\documentclass[fleqn,usenatbib]{mnras}

\usepackage{newtxtext,newtxmath}

\usepackage{graphicx}	
\usepackage[svgnames]{xcolor}
\usepackage{booktabs,caption}
\usepackage{threeparttable}
\usepackage{microtype}
\usepackage{gensymb}
\usepackage{pdflscape}
\usepackage{subcaption}
\usepackage{amsfonts}
\usepackage[normalem]{ulem}

\DeclareUnicodeCharacter{2212}{-}

\title[WISDOM XXIX.\ CO Tully-Fisher]{WISDOM Project -- XXIX.\ Catalogue of CO observations of nearby galaxies and the CO Tully-Fisher relation}

\author[P.\ Dominiak et al.]{Pandora Dominiak,$^{1}$\thanks{E-mail: pandora.dominiak@physics.ox.ac.uk}
Martin Bureau,$^{1}$\thanks{E-mail: martin.bureau@physics.ox.ac.uk}, Sel\c{c}uk Topal$^{2}$, Michele Cappellari$^{1}$, Woorak Choi$^{3}$,
\newauthor{Timothy A.\ Davis$^{4}$, Jindra Gensior$^{5}$, Marie Korsaga$^{6,7}$, Fu-Heng Liang$^{8}$, Ilaria Ruffa$^{9,4}$, Jacob Elford$^{10}$,}
\newauthor{Thomas G.\ Williams$^{11}$, Hengyue Zhang$^{1}$}\\
$^{1}$Sub-department of Astrophysics, Department of Physics, University of Oxford, Denys Wilkinson Building, Keble Road, Oxford, OX1~3RH, UK\\
$^{2}$Department of Physics, Van Yuzuncu Yil University, Van 65080, Turkey\\
$^{3}$Department of Physics and Astronomy, McMaster University, Hamilton, ON L8S~4M1, Canada\\
$^{4}$Cardiff Hub for Astrophysics Research \& Technology, Department of Physics \& Astronomy, Cardiff University, Queens Buildings, The Parade, Cardiff, CF243AA, UK\\
$^{5}$Institute for Astronomy, University of Edinburgh, Royal Observatory, Blackford Hill, Edinburgh EH9~3HJ, UK\\
$^{6}$Instituto de Astrof\'isica de Andaluc\'ia (CSIC), PO Box 3004, 18080 Granada, Spain\\ 
$^{7}$Laboratoire de Physique et de Chimie de l’Environnement, Université Joseph Ki-Zerbo, 03 BP 7021, Ouaga 03, Burkina Faso\\
$^{8}$Astronomisches Rechen-Institut, Zentrum f\"{u}r Astronomie der Universit\"{a}t Heidelberg, M\"{o}nchhofstra\ss e 12-14, D-69120 Heidelberg, Germany\\
$^{9}$INAF, Arcetri Astrophysical Observatory, Largo Enrico Fermi 5, I-50125 Florence, Italy\\
$^{10}$Instituto de Estudios Astrofísicos, Facultad de Ingeniería y Ciencias, Universidad Diego Portales, Av.\ Ejército Libertador 441, Santiago, Chile\\
$^{11}$UK ALMA Regional Centre Node, Jodrell Bank Centre for Astrophysics, Department of Physics and Astronomy, The University of Manchester,\\
Oxford Road, Manchester M13~9PL, UK\\
}

\date{Accepted XXX. Received YYY; in original form ZZZ}

\pubyear{2026}

\begin{document}
\label{firstpage}
\pagerange{\pageref{firstpage}--\pageref{lastpage}}
\maketitle

\begin{abstract}
We present Tully-Fisher relations (TFRs) that use carbon monoxide (CO) as the kinematic tracer and \textit{Wide-Field Infrared Survey Explorer} (WISE) $W1$-band and Two Micron All Sky Survey (2MASS) $K_\text{s}$-band magnitudes as tracers of stellar luminosity. CO data were obtained using the Atacama Large Millimeter/submillimeter Array for the mm-Wave Interferometric Survey of Dark Object Masses (WISDOM) project for a sample of $102$ galaxies located at redshifts $z \lesssim 0.03$, with $81$ detections. The best-fitting reverse $W1$-band and $K_\text{s}$-band CO TFRs are $M_{W1}=(-7.18\pm0.93)\left[\log\left(\frac{W_{50}/\sin i}{\text{km~s}^{-1}}\right)-2.6\right]+(-24.04\pm0.13)$ and $M_{K_\text{s}}=(-5.95\pm0.62)\left[\log\left(\frac{W_{50}/\sin i}{\text{km~s}^{-1}}\right)-2.6\right]+(-24.01\pm0.10)$, respectively, where $W_{50}$ is the velocity width of each integrated CO spectrum at $50\%$ of its maximum, $i$ is the galaxy inclination and $M$ is the galaxy absolute magnitude. Our results are in excellent agreement with those of Tiley et al. (2016b), and together they constitute a robust local benchmark for the CO TFR. We however measure a shallower slope and a smaller zero-point offset compared to the relation of Tully \& Pierce (2000). Due to the morphological diversity of our sample, we construct separate late-type and early-type galaxy (ETG) CO TFRs, finding that ETGs are $1.40\pm0.39$~mag fainter at $W1$ band and $0.89\pm0.28$~mag fainter at $K_\text{s}$ band at a given velocity width. We also generate and include a comprehensive catalogue of WISDOM CO data products for use in future molecular gas studies.
\end{abstract}



\section{Introduction}
\label{section:Introduction}

Scaling relations between the luminosity (or mass) and kinematics of galaxies are central to our understanding of galaxy formation and evolution. Among these, the Tully-Fisher relation (TFR; \citealt{Tully_Fisher_1977}) remains one of the most fundamental, linking a galaxy's luminosity and rotational velocity. As the former traces the stellar mass of a galaxy while the latter traces its enclosed dynamical mass, the relation provides insights into the distribution of both baryons and dark matter within galaxies \citep[e.g.][]{Zwaan_1995, McGaugh_1998, McGaugh_2000}, it constrains stellar mass-to-light ratios ($M/L$; e.g.\ \citealt{Bell_deJong_2001}) and it serves as a test of galaxy formation and cosmological models \citep[e.g.][]{vdBosch_2000, Yang_2003, Trachternach_2009, McGaugh_2012, Ponomareva_2017}. In addition, the TFR can be used as a distance indicator, extending the cosmic distance ladder beyond the local Universe \citep[e.g.][]{Pierce_1988, Sakai_2000, Springob_2007, Lelli_2016_ApJL}. Indeed, by comparing the absolute magnitude of a galaxy expected from the TFR with its observed apparent magnitude, the distance to the galaxy can be estimated independently of its redshift. 

Traditionally, the TFR has been derived for late-type, gas-rich galaxies using emission from the hyperfine transition of atomic hydrogen (H\,\textsc{i}) at $21$~cm  \citep[e.g.][]{Tully_2000, Verheijen_2001, Meyer_2008, Ponomareva_2021}. However, direct H\,\textsc{i} detections of individual galaxies within large-area untargeted surveys are rare beyond the nearby Universe \citep[e.g.][]{Giovanelli_2005, Serra_2015, Xi_2021, de_Blok_2024}. Recently, targeted deep observations have yielded eleven detections at redshifts $0.25<z<0.384$ \citep{Jarvis_2025}, a single detection at $z=0.376$ (\citealt{Fernandez_2016}; although this result has recently been called into question by \citealt{Heywood_2024}) and six detections at $0.38<z<0.5$ \citep{Xi_2024}. In exceptional cases H\,\textsc{i} can be detected at $z\approx0.7$, but only through the use of strong gravitational lensing \citep{Lawrie_2025}. A previously claimed strongly lensed H\,\textsc{i} detection at $z\approx1.3$ \citep{Chakraborty_2023} was subsequently challenged by \citet{Deane_2024}, highlighting the difficulty of H\,\textsc{i} detections at such redshifts. H\,\textsc{i} is also problematic when attempting to probe the TFR of early-type galaxies (ETGs). Indeed, extended atomic gas is less frequent in ETGs \citep[e.g.][]{Young_2014, Young_2018}, and when present it is often kinematically disturbed \citep[e.g.][]{Williams_2010, Serra_2012}, preventing reliable determinations of rotational velocities and therefore undermining the construction of reliable TFRs. When extended regularly rotating H\,\textsc{i} discs are present, ETG TFRs can be defined, but samples remain small \citep[e.g.][]{denHeijer_2015}.

When comparing TFR differences across galaxy morphological types, it is crucial that the same velocity tracer is used for all sample galaxies, as different velocity tracers will likely introduce different systematic biases \citep[e.g.][]{Williams_2010}. Thus, the use of carbon monoxide (CO) as an alternative kinematic tracer in TFR studies has grown considerably in recent years \citep[e.g.][]{Davis_2011, MASSIVE_III, Tiley_2016, Topal_2018}. CO is observed across the full Hubble sequence, with large surveys mapping CO in a substantial fraction of nearby spirals [e.g.\ Berkeley-Illinois-Maryland Association Survey of Nearby Galaxies (BIMA SONG), \citealt{Helfer_2003}; Extragalactic Database for Galaxy Evolution and Calar Alto Legacy Integral Field Area Survey (EDGE-CALIFA), \citealt{Utomo_2017}; Extended CO Legacy Database for GASS (xCOLD GASS), \citealt{Saintonge_2017}; CO Multi-line Imaging of Nearby Galaxies (COMING), \citealt{Sorai_2019}] and targeted studies demonstrating routine CO detections in $\approx20\%$ of ETGs (e.g.\ ATLAS$^{\rm 3D}$, \citealt{Young_2011}; MASSIVE, \citealt{MASSIVE_I} and \citealt{MASSIVE_XI}; KILOGAS, Davis et al.\ in prep.). Moreover, contrary to H\,\textsc{i}, CO is observable across a wide range of galaxy environments, from brightest cluster galaxies (where H\,\textsc{i} is often depleted due to environmental processes such as ram-pressure stripping and tidal interactions; e.g.\ \citealt{Chung_2009b,Rose_2019,Ruffa_2019a,Brown_2021,Dunne_2021}) to isolated field galaxies \citep[e.g.][]{Young_2011, Atlas3d_X}. In addition, CO remains detectable up to $z\geq6$ \citep[e.g.][]{Carilli_2013, Novak_2019, Decarli_2022}, $z=6.9$ being the highest redshift for which CO is detected \citep{Venemans_2017,Kaasinen_2024}.

Early studies \citep[e.g.][]{Dickey_1992, Schoniger_1994}
showed that CO line widths correlate well with rotation velocities in spiral galaxies, suggesting that CO could yield TFRs with a comparable slope to those derived using H\,\textsc{i}. The use of CO as a kinematic tracer to derive a TFR for nearby ETGs was first demonstrated by \citet{Davis_2011}. Subsequent studies extended this approach to larger and more diverse galaxy samples in the local Universe (e.g.\ \citealt{Tiley_2016}; Topal et al.\ in prep.), to massive ETGs \citep{MASSIVE_III} and to galaxies at intermediate redshifts \citep{Topal_2018}. These studies collectively demonstrate that CO TFRs yield slopes, zero-points and scatters consistent with those of the classical relations, thus providing a robust alternative to H\,\textsc{i} for rotation-based scaling relations.

In this work, we use the Atacama Large Millimeter/submillimeter Array (ALMA) observations from the mm-Wave Interferometric Survey of Dark Object Masses (WISDOM) project \citep[e.g.][]{WISDOM_I, Zhang_2024, Dominiak_2025} to construct a new CO TFR calibration for the local Universe ($z\leq0.032$). The WISDOM sample comprises {$102$ galaxies ($81$ of which are detected in CO), comparable in size to previous CO TFR samples (e.g.\ \citealt{Tiley_2016} had $84$ galaxies in their final sub-sample). These galaxies were observed predominantly in the $^{12}\text{CO(2–1)}$ line with high spatial and spectral resolution, providing one of the largest interferometric datasets of molecular gas in nearby galaxies to date. ALMA's exquisite spatial resolution allows us to measure the emission's spatial extent and to identify and exclude targets with non-circular and/or disturbed motions, whereas its spectral resolution ensures precise velocity widths with unambiguous profile shapes.

The WISDOM targets are morphologically diverse, selected as those most likely to host regular dynamically cold molecular gas discs. The presence of these discs was established from prior single-dish or interferometric CO observations, or from regular dust (a reliable proxy for molecular gas) structures in \textit{Hubble Space Telescope} (\textit{HST}) images. As a result, the WISDOM survey provides a large and homogeneous set of observations suitable for constructing a CO TFR. In addition, the widespread presence of dust discs in the sample enables robust inclination measurements, improving the accuracy of the velocity deprojections, a long-standing issue in TFR studies. Using methods analogous to those of  \citet{Tiley_2016} and \citet{Topal_2018}, we measure the galaxy rotation velocities from fits to the integrated CO profiles and derive the luminosities in the near-infrared $W1$- ($3.4$~$\mu$m) and $K_{\rm s}$- ($2.15$~$\mu$m) bands from \textit{Wide-Field Infrared Survey Explorer} (\textit{WISE}) and Two Micron All Sky Survey (2MASS; \citealt{2MASS}) images, respectively, as they best trace the old stellar populations dominating stellar mass. Alongside the TFRs, we also compile a comprehensive catalogue of CO data products (moment maps, position-velocity diagrams and integrated spectra) for all the WISDOM sample galaxies, to serve as a lasting resource for molecular gas and dynamical studies. The CO data cubes and aforementioned data products will be made available on the WISDOM project website\footnote{\url{https://wisdom-project.org}}.

The structure of this paper is as follows. In Section~\ref{section:Data} we discuss the creation of CO data products from the interferometric data and list the properties of our CO data cubes, serving as the catalogue portion of this paper. In Section~\ref{section:TFR data} we discuss the data necessary for the TFR portion of this study. This includes the near-infrared photometric data, the measurements associated with the fits to our CO data, and the deprojection of our CO velocity widths. In Section~\ref{section:results} we present the TFRs of our parent sample of galaxies and various sub-samples. In particular, we explore the impacts of imposing CO data quality and inclination criteria to create more homogenous sub-samples, and look at TFR differences as a function of galaxy morphology. In Section~\ref{section:Discussion} we discuss our results and compare them against literature results. We conclude briefly in Section~\ref{section:conclusion}. All WISDOM CO data products are shown in Appendix~\ref{section:Appendix A}.

\section{ALMA CO Observations}
\label{section:Data}

\subsection{CO data}
\label{subsection:CO data}

The parent sample of the WISDOM project catalogue presented in this paper consists of $102$ galaxies for which ALMA CO data were obtained.

The $^{12}\text{CO(2–1)}$ emission line of $60$ galaxies was observed using Band 6 of the $7$-m Atacama Compact Array (ACA) as a part of projects 2016.2.00046.S (PI: Bureau), 2019.1.00363.S (PI: Davis), 2022.1.01173.S (PI: Liang) and 2023.1.01084.S (PI: Liang). Those data were calibrated using the \textsc{Common Astronomy Software Applications} (\textsc{casa}) package (using the version appropriate for the data depending on the ALMA cycle; \citealt{McMullin}) and its ALMA pipeline. The imaging steps used \textsc{casa} version 6.7.2. 

For these galaxies, data cubes were created adopting the same method as previous works in the WISDOM series \citep[e.g.][]{WISDOM_X,Ruffa_2023,Dominiak_2026}. Specifically, for each galaxy we first measured the continuum emission in the continuum spectral windows and the line-free channels of the line spectral window and removed it from the line emission using the \textsc{casa} task \texttt{uvcontsub}. We then cleaned the continuum-subtracted data using the \texttt{tclean} task to a threshold appropriate for the data set, typically $\approx3$ times the root-mean-square (RMS) noise determined from the dirty continuum-subtracted data cube. The data were then imaged using the \texttt{tclean} task, adopting the kinematic Local Standard of Rest (LSRK) as the spectral reference frame, and natural weighting (Briggs weighting with a robust parameter of $2$), to achieve the best possible emission signal-to-noise ratio. Each data cube was initially imaged using $10$~km~s$^{-1}$ channels, however in cases where signal-to-noise ratio was low the channel width was increased in increments of $10$~km~s$^{-1}$ until a satisfactory signal-to-noise ratio was achieved.

One third ($21/60$) of these ACA sample galaxies were not detected in $^{12}\text{CO(2–1)}$, and the basic properties of these data cubes are listed in Table~\ref{table:non-detected cubes}. The synthesised beam sizes listed are the full-widths at half-maximum (FWHM) along the major and minor axes of the beam. Non-detection is determined through a visual inspection, by a lack of emission in the cube after binning to a channel width of $50$~km~s$^{-1}$. In three galaxies (NGC0193, NGC3078 and UGC7115) there appears to be some faint emission in the centre, however this emission does not exhibit the typical velocity profile expected from a galaxy and each has a very high velocity dispersion ($>200$~km~s$^{-1}$). We therefore classify these galaxies as non-detections, as the emission does not permit meaningful scientific analyses. In these cases the RMS noise was measured away from the galaxy centre.

\begin{table}
\begin{threeparttable}
    \centering
    \caption{Properties of the data cubes of galaxies with no CO detection.}
    \label{table:non-detected cubes}
    \begin{tabular}{lcc}
        \hline
         Galaxy & Beam & RMS noise \\
        & (arcsec$^{2}$) & (mJy~beam$^{-1}$~channel$^{-1}$) \\
         \hline
         3C296 & $8.5\times6.0$ & $4.9$ \\
         MCG\_-02-36-002 & $8.8\times4.4$ & $2.8$ \\
         NGC0193 & $7.7\times4.9$ & $3.4$ \\
         NGC1052 & $7.4\times4.6$ & $4.3$\\ 
         NGC1461 & $7.4\times4.5$ & $4.4$\\
         NGC1700 & $8.2\times5.5$ & $3.6$\\
         NGC3078 & $6.5\times4.8$ & $3.5$ \\
         NGC3379 & $9.0\times4.9$ & $5.7$ \\
         NGC3414 & $8.7\times4.6$ & $4.5$ \\
         NGC4233 & $7.7\times4.9$ & $4.0$ \\
         NGC4342 & $8.0\times4.6$ & $2.9$ \\
         NGC4350 & $9.0\times4.8$ & $5.0$ \\
         NGC4866 & $8.1\times5.5$ & $4.2$ \\
         NGC5077 & $11.0\times4.4\phantom{0}$ & $3.7$ \\
         NGC5127 & $7.6\times6.5$ & $4.2$ \\
         NGC5845 & $7.7\times5.3$ & $2.8$\\
         NGC7457 & $7.8\times6.6$ & $3.7$ \\
         NGC7768 & $7.3\times6.5$ & $3.7$\\
         PGC046123 & $8.2\times5.5$ & $5.3$ \\
         PGC214240 & $7.9\times5.0$ & $3.4$ \\
         UGC7115 & $6.9\times6.7$ & $3.6$ \\
         \hline
    \end{tabular}
    \begin{tablenotes}[flushleft]
    \item \textit{Note:} All cubes listed have a channel width of $50$~km~s$^{-1}$.
    \end{tablenotes}
\end{threeparttable}
\end{table}

The remaining $39/60$ ACA sample galaxies yielded detections in $^{12}\text{CO(2–1)}$. These detections resulted in $28$ data cubes with $10$~km~s$^{-1}$ channels, $8$ cubes with $20$~km~s$^{-1}$ channels and $3$ cubes with $30$~km~s$^{-1}$ channels. These cubes have synthesised beam sizes (full widths at half maximum measured along the kinematic major-axis of the CO emission) ranging from $4.0$ to $8.2$~arcsec. For all cubes we chose a pixel size ensuring a sampling of $3$ -- $5$ spaxels across the synthesised beam and a total spatial extent equal to the $20$~per~cent primary beam response. The properties of these $39$ cubes are summarised in Table~\ref{tab:observational CO data}. 

\begin{table*}
\begin{threeparttable}
    \centering
    \caption{Observational parameters of the CO data of all CO-detected galaxies.}
    \label{tab:observational CO data}
	\begin{tabular}{l@{\hskip12pt}lrcccccccc}
    \hline
    Galaxy & Type & Array & Channels & Ref. & Beam & RMS noise & $W_{50,\text{obs}}$ & $v_{0,\text{obs}}$ & $\langle v\rangle_\text{obs}$ & $(S_\text{CO}\Delta v)_\text{obs}$ \\
    & & & (km~s$^{-1}$) & & (arcsec$^{2}$) & (mJy) & (km~s$^{-1}$) & (km~s$^{-1}$) & (km~s$^{-1}$) & (Jy~km~s$^{-1}$) \\
    (1) & (2) & (3) & (4) & (5) & (6) & (7) & (8) & (9) & (10) & (11) \\
    \hline
    ESO383-G035 & Sab & $7$-m & $10$ & TW & $6.4\phantom{0} \times 5.5\phantom{0}$ & $8.0$ & $220\pm\phantom{0}7$ & $2335\pm\phantom{0}4$ & $2342.8\pm\phantom{0}1.8$ & $\phantom{00}32.4\pm\phantom{0}0.8$ \\ 
    FRL49 & E-S0 & $12$-m & $15$ & L22 & $0.22 \times 0.16$ & $0.8$ & $375\pm 11$ & $5890\pm \phantom{0}5$ & $5900.7\pm \phantom{0}1.0$ & $\phantom{0}479.7 \pm \phantom{0}3.4$ \\
    FRL1146 & Sc & $12$-m & $20$ & TW & $0.39\times0.34$ & $0.8$ & $420\pm 14$ & $9180\pm\phantom{0}7$ & $9192.3 \pm 16.7$ & $\phantom{0}110.6 \pm 15.9$ \\
    IC0843 & S0 & $7$-m & $20$ & TW & $7.7\phantom{0} \times 6.1\phantom{0}$ & $3.8$ & $760\pm14$ & $7210\pm\phantom{0}7$ & $7229.5\pm\phantom{0}8.6$ & $\phantom{00}22.9\pm\phantom{0}0.8$ \\
    IC1481 & Sd & $7$-m & $10$ & TW & $7.6\phantom{0} \times 4.7\phantom{0}$ & $9.2$ & $170\pm\phantom{0}7$ & $6010\pm\phantom{0}4$ & $6007.2\pm\phantom{0}1.3$ & $\phantom{00}72.0\pm\phantom{0}1.0$ \\
    IC4329A & S0-a & $7$-m & $30$ & TW & $8.3\phantom{0} \times 4.5\phantom{0}$ & $2.3$ & $\phantom{0}90 \pm 21$ & $4630 \pm 11$ & $4693.2 \pm \phantom{0}8.2$ & $\phantom{00}11.0 \pm \phantom{0}0.5$ \\
    MRK0567 & Sc & $12$-m & $15$ & D22 & $0.16 \times 0.12$ & $1.1$ & $195\pm 11$ & $9415\pm\phantom{0}5$ & $9413.1\pm\phantom{0} 0.4$ & $\phantom{0}167.1 \pm \phantom{0}1.0$ \\
    NGC0017 & S0-a & $7$-m & $10$ & TW & $7.1\phantom{0} \times 5.5\phantom{0}$ & $4.0$ & $320\pm\phantom{0}7$ & $5695 \pm \phantom{0}4$ & $5701.9 \pm \phantom{0}0.2$ & $\phantom{0}454.1 \pm\phantom{0} 0.8$ \\
    NGC0289$^\dagger$& SBbc & $7$-m & $10$ & TW & $6.6\phantom{0} \times 4.9\phantom{0}$ & $3.1$ & $160\pm \phantom{0}7$ & $1615 \pm \phantom{0}4$ & $1626.6 \pm\phantom{0} 0.1$ & $\phantom{0}880.8 \pm\phantom{0} 0.6$ \\
    NGC0383 & E-S0 & $12$-m & $10$ & Z25 & $0.18\times0.10$ & $0.4$ & $510\pm\phantom{0}7$ & $4931\pm \phantom{0}4$ & $4938.3\pm \phantom{0}0.7$ & $\phantom{0}109.5\pm\phantom{0} 0.4$ \\
    NGC0404 & E-S0 & $12$-m & $\phantom{1}2$ & D20 & $0.05 \times 0.03$ & $0.8$ & $\phantom{0}24\pm \phantom{0}1$ & $\ -57 \pm \phantom{0}1$ & $\ -54.4 \pm\phantom{0} 0.1$ & $\phantom{0}156.3 \pm \phantom{0}0.7$ \\
    NGC0449 & SBa & $12$-m & $10$ & D22 & $0.85\times0.51$ & $3.9$ & $220\pm \phantom{0}7$ & $4686\pm \phantom{0}4$ & $4693.0 \pm\phantom{0} 1.0$ & $\phantom{00}67.5\pm\phantom{0} 1.0$ \\
    NGC0524 & S0-a & $12$-m & $10$ & S19 & $0.38 \times 0.28$ & $0.5$ & $300\pm \phantom{0}7$ & $2381\pm\phantom{0}4$ & $2388.1\pm\phantom{0} 0.4$ & $\phantom{00}37.8\pm \phantom{0}0.1$ \\
    NGC0541 & E & $7$-m & $10$ & TW & $7.0\phantom{0} \times 4.7\phantom{0}$ & $4.6$ & $220\pm\phantom{0}7$ & $5315 \pm \phantom{0}4$ & $ 5334.2 \pm\phantom{0} 4.4$ & $\phantom{000}4.1 \pm\phantom{0} 0.2$ \\
    NGC0612 & S0-a & $12$-m & $10$ & R23 & $0.21 \times 0.18$ & $1.0$ & $770\pm \phantom{0}7$ & $8629\pm\phantom{0}4$ & $8644.6\pm\phantom{0} 1.8$ & $\phantom{0}289.2\pm \phantom{0}1.9$\\
    NGC0708 & E & $12$-m & $10$ & N21 & $0.09 \times 0.08$ & $0.4$ & $310\pm \phantom{0}7$ & $4760\pm \phantom{0}4$ & $4756.5\pm\phantom{0}1.0$ & $\phantom{00}42.3\pm\phantom{0} 0.3$ \\
    NGC0863$^\dagger$ & Sa & $7$-m & $10$ & TW & $7.7\phantom{0} \times 5.4\phantom{0}$ & $9.0$ & $220 \pm \phantom{0}7$ & $7715 \pm \phantom{0}4$ & $7722.2 \pm \phantom{0}0.6$ & $\phantom{00}67.7 \pm\phantom{0} 0.7$ \\
    NGC1084$^\dagger$ & Sc & $7$-m & $30$ & TW & $7.7\phantom{0} \times 5.0\phantom{0}$ & $3.5$ & $150\pm 21$ & $1350 \pm 11$ & $1368.0 \pm \phantom{0}0.1$ & $\phantom{0}505.7 \pm \phantom{0}0.5$ \\
    NGC1194 & S0-a & $12$-m & $20$ & E23 & $0.30 \times 0.23$ & $0.3$ & $380\pm 14$ & $4008\pm\phantom{0}7$ & $4046.1\pm \phantom{0}3.5$ & $\phantom{000}8.6\pm\phantom{0} 0.2$ \\
    NGC1300$^\dagger$ & Sbc & $12$-m & $10$ & TW & $0.82 \times 0.48$ & $0.8$ & $190\pm\phantom{0}7$ & $1540\pm\phantom{0}4$ & $1543.4\pm\phantom{0} 0.2$ & $\phantom{0}468.5\pm \phantom{0}1.1$ \\
    NGC1358 & S0-a & $7$-m & $20$ & TW & $7.6\phantom{0} \times 4.1\phantom{0}$ & $5.3$ & $160 \pm 14$ & $3950 \pm \phantom{0}7$ & $3962.9 \pm \phantom{0}4.2$ & $\phantom{000}6.4 \pm \phantom{0}0.4$ \\
    NGC1387 & E-S0 & $12$-m & $10$ & D26 & $0.42 \times 0.35$ & $0.8$ & $140\pm\phantom{0}7$ & $1265\pm\phantom{0}4$ & $1268.5\pm \phantom{0}0.2$ & $\phantom{0}387.5\pm\phantom{0} 1.3$ \\
    NGC1400 & E & $7$-m & $10$ & TW & $7.2\phantom{0} \times 4.2\phantom{0}$ & $9.9$ & $410 \pm \phantom{0}7$ & $\phantom{0}550 \pm \phantom{0}4$ & $\phantom{0}553.0 \pm \phantom{0}1.0$ & $\phantom{0}115.5 \pm \phantom{0}0.9$ \\
    NGC1448$^\dagger$ & Sc & $7$-m & $20$ & TW & $7.5\phantom{0} \times 5.6\phantom{0}$ & $3.2$ & $220 \pm 14$ & $1100 \pm \phantom{0}7$ & $1142.1 \pm \phantom{0}0.2$ & $1024.9 \pm \phantom{0}1.6$ \\
    NGC1493 & SBc & $7$-m & $10$ & TW & $9.2\phantom{0} \times 3.9\phantom{0}$ & $7.2$ & $\phantom{0}10 \pm \phantom{0}7$ & $1030 \pm \phantom{0}4$ & $1033.9 \pm\phantom{0} 0.1$ & $\phantom{00}58.8 \pm\phantom{0} 0.5$ \\
    NGC1497 & S0 & $7$-m & $10$ & TW & $7.0\phantom{0} \times 6.2\phantom{0}$ & $5.0$ & $570 \pm \phantom{0}7$ & $6100 \pm \phantom{0}4$ & $6109.0 \pm \phantom{0}2.1$ & $\phantom{00}41.5 \pm\phantom{0} 0.5$ \\
    NGC1574 & E-S0 & $12$-m & $10$ & R23 & $0.20 \times 0.14$ & $0.7$ & $220\pm\phantom{0}7$ & $1018\pm\phantom{0}4$ & $1020.8\pm \phantom{0}1.5$ & $\phantom{00}30.0\pm \phantom{0}0.6$ \\
    NGC2110 & E-S0 & $12$-m & $20$ & E23 & $0.74 \times 0.43$ & $1.2$ & $406\pm14$ & $2373\pm\phantom{0}7$ & $2380.0\pm \phantom{0}1.4$ & $\phantom{00}56.3\pm\phantom{0} 0.6$ \\
    NGC2179 & S0-a & $7$-m & $10$ & TW & $7.5\phantom{0} \times 4.2\phantom{0}$ & $3.1$ & $390 \pm \phantom{0}7$ & $2830 \pm \phantom{0}4$ & $2830.0 \pm \phantom{0}2.1$ & $\phantom{00}26.1 \pm \phantom{0}0.4$ \\
    NGC2187A$^\dagger$ & S0-a & $7$-m & $20$ & TW & $7.1\phantom{0} \times 6.4\phantom{0}$ & $3.5$ & $400 \pm 14$ & $3910 \pm \phantom{0}7$ & $3899.3 \pm \phantom{0}2.2$ & $\phantom{0}362.9 \pm \phantom{0}5.5$ \\
    NGC2903 & Sbc & $12$-m & $10$ & TW & $0.12\times0.10$ & $0.3$ & $150\pm\phantom{0}7$ & $\phantom{0}530\pm\phantom{0}4$ & $\phantom{0}548.8\pm \phantom{0}0.1$ & $1242.1\pm\phantom{0}0.9$ \\
    NGC2911 & S0 & $7$-m & $10$ & TW & $6.9\phantom{0} \times 5.5\phantom{0}$ & $6.8$ & $500\pm\phantom{0} 7$ & $3195 \pm\phantom{0} 4$ & $3190.0 \pm\phantom{0} 1.9$ & $\phantom{00}41.3 \pm \phantom{0}0.5$ \\
    NGC2964 & Sbc & $7$-m & $10$ & TW & $7.5\phantom{0} \times 6.5\phantom{0}$ & $22.6\phantom{0}$ & $170\pm \phantom{0}7$ & $1320 \pm \phantom{0}4$ & $1323.9 \pm \phantom{0}0.6$ & $\phantom{0}454.5 \pm \phantom{0}4.4$ \\
    NGC2974 & E & $7$-m & $10$ & TW & $6.6\phantom{0} \times 4.6\phantom{0}$ & $9.5$ & $520\pm \phantom{0}7$ & $1805\pm\phantom{0}4$ & $1817.8 \pm \phantom{0}7.5$ & $\phantom{00}26.7 \pm \phantom{0}1.2$ \\
    NGC3021$^\dagger$ & Sbc & $7$-m & $10$ & TW & $10.5\phantom{0} \times 5.7\phantom{0}\phantom{0}$ & $10.7\phantom{0}$ & $260\pm\phantom{0}7$ & $1515\pm\phantom{0}4$ & $1523.1\pm \phantom{0}0.3$ & $\phantom{0}501.3\pm \phantom{0}1.4$ \\
    NGC3156 & S0 & $7$-m & $20$ & TW & $6.3\phantom{0} \times 5.5\phantom{0}$ & $5.3$ & $120\pm14$ & $1310\pm\phantom{0} 7$ & $1325.8\pm \phantom{0}1.7$ & $\phantom{00}10.8 \pm \phantom{0}0.4$ \\
    NGC3169 & Sa & $12$-m & $10$ & D22 & $0.70\times0.51$ & $1.1$ & $ 467\pm\phantom{0}7$ & $1231\pm\phantom{0}4$ & $1236.7\pm \phantom{0}0.5$ & $\phantom{0}846.2\pm \phantom{0}2.4$ \\
    NGC3277 & Sab & $7$-m & $10$ & TW & $7.6\phantom{0} \times 6.3\phantom{0}$ & $6.4$ & $230\pm\phantom{0}7$ & $1400\pm \phantom{0}4$ & $1406.0 \pm\phantom{0} 0.4$ & $\phantom{0}135.9 \pm\phantom{0} 0.6$ \\
    NGC3351$^\dagger$ & Sb & $12$-m & $10$ & E23 & $0.29 \times 0.14$ & $3.2$ & $190\pm\phantom{0}6$ & $\phantom{0}787\pm\phantom{0}3$ & $\phantom{0}796.9\pm \phantom{0}0.4$ & $\phantom{0}285.3\pm \phantom{0}1.6$ \\
    NGC3368 & Sab & $12$-m & $\phantom{1}2$ & D22 & $0.21 \times 0.20$ & $0.9$ & $180\pm\phantom{0}1$ & $\phantom{0}884\pm\phantom{0}1$ & $ \phantom{0}909.3\pm \phantom{0}0.1$ & $\phantom{0}878.3\pm \phantom{0}0.5$ \\
    NGC3393$^\dagger$ & SBa & $12$-m & $10$ & L23 & $0.56 \times 0.45$ & $ 0.4$ & $190\pm\phantom{0}7$ & $3681\pm\phantom{0}4$ & $3686.4\pm\phantom{0}0.4$ & $\phantom{0}103.0\pm\phantom{0} 0.6$ \\
    NGC3607 & E-S0 & $12$-m & $10$ & D22 & $0.60\times0.50$ & $1.2$ & $520 \pm\phantom{0}7$ & $1184\pm\phantom{0}4$ & $1194.2\pm \phantom{0}0.7$ & $\phantom{0}164.6\pm \phantom{0}0.7$ \\
    NGC3810$^\dagger$ & Sc & $7$-m & $10$ & TW & $7.3\phantom{0} \times 5.0\phantom{0}$ & $6.3$ & $160\pm\phantom{0}7$ & $1005\pm \phantom{0}4$ & $1005.0 \pm \phantom{0}0.1$ & $\phantom{0}638.2 \pm \phantom{0}0.7$ \\
    NGC3862 & E & $12$-m & $10$ & E23 & $0.69 \times 0.53$ & $1.9$ & $199\pm\phantom{0}7$ & $6366\pm\phantom{0}4\,$ & $6367.9\pm \phantom{0}1.1$ & $\phantom{00}15.3\pm\phantom{0} 0.3$ \\
    NGC4030$^\dagger$ & Sbc & $7$-m & $10$ & TW & $7.9\phantom{0} \times 4.5\phantom{0}$ & $5.8$ & $330\pm \phantom{0}7$ & $1450\pm \phantom{0}4$ & $1452.8 \pm \phantom{0}0.1$ & $1415.7 \pm \phantom{0}1.0$ \\
    NGC4061 & E & $12$-m & $10$ & D22 & $0.15\times0.12$ & $0.3$ & $690\pm\phantom{0}7$ & $7187\pm\phantom{0}4$ & $ 7197.6\pm \phantom{0}1.4$ & $\phantom{00}23.6\pm \phantom{0}0.2$ \\
    NGC4203 & E-S0 & $7$-m & $20$ & TW & $7.4\phantom{0} \times 6.8\phantom{0}$ & $6.9$ & $100\pm 14$ & $\phantom{0}960\pm \phantom{0}7$ & $\phantom{0}997.4 \pm\phantom{0} 2.4$ & $\phantom{00}33.5 \pm \phantom{0}0.9$ \\
    NGC4212$^\dagger$ & Sc & $7$-m & $10$ & TW & $7.2\phantom{0} \times 5.7\phantom{0}$ & $8.8$ & $160\pm \phantom{0}7$ & $\ -65\pm \phantom{0}4$ & $\ -69.1 \pm\phantom{0} 0.2$ & $\phantom{0}563.9 \pm \phantom{0}1.9$ \\ 
    NGC4235 & Sa & $7$-m & $10$ & TW & $8.1\phantom{0} \times 5.1\phantom{0}$ & $10.3\phantom{0}$ & $240\pm \phantom{0}7$ & $2235\pm\phantom{0}4$ & $2238.8\pm\phantom{0} 2.5$ & $\phantom{00}34.9\pm \phantom{0}1.1$ \\
    NGC4245 & S0-a & $7$-m & $10$ & TW & $8.7\phantom{0} \times 5.4\phantom{0}$ & $7.9$ & $190\pm\phantom{0} 7$ & $\phantom{0}880\pm\phantom{0}4$ & $\phantom{0}887.3 \pm\phantom{0} 0.7$ & $\phantom{00}65.4 \pm \phantom{0}0.6$ \\
    NGC4261 & E & $12$-m & $15$ & R23 & $0.20 \times 0.15$ & $0.7$ & $462\pm11$ & $2237\pm\phantom{0}5$ & $2250.2\pm\phantom{0}9.7$ & $\phantom{000}4.3\pm \phantom{0}0.2$ \\
    \hline
    \end{tabular}
    \begin{tablenotes}[flushleft] 
    \item \textit{Notes.} The data in this table are available in a machine-readable format online. (1) Galaxy name. $^\dagger$ indicates a galaxy for which the CO emission extends close to or beyond the primary beam FWHM, and for which some of the measured quantities, particularly the velocity width and integrated flux, may be underestimated. (2) Galaxy morphological type from HyperLEDA. (3) More extended ALMA array used for the CO data cube. ``12-m" indicates 12-m data combined with ACA data when available, while ``7-m" indicates ACA data only. (4) Cube channel width. (5) Source of CO data. TW: this work; D17: \cite{Davis2017}; D18: \cite{WISDOM_III}; D20: \cite{Davis_2020}; D22: \cite{WISDOM_X}; D26: \citet{Dominiak_2026}; E23: \cite{Elford_2023}; L22: \cite{Lelli_2022}; N21: \cite{North_2021}; R23: \cite{Ruffa_2023}; S19: \cite{WISDOM_IV}; S21: \cite{Smith_2021}; and Z25: \cite{Zhang_2025}. (6) Cube synthesised beam size (full widths at half maximum). (7) RMS noise of the data cube, measured per beam per channel in the line-free channels. (8) Integrated spectrum velocity width at $50\%$ of the peak. (9) Central velocity, defined as the midpoint between the (edge) velocities at which the profile reaches $50\%$ of the peak, reported in the LSRK reference frame. (10) Flux-weighted mean velocity, also in the LSRK reference frame. (11) Integrated CO flux.
    \end{tablenotes}
\end{threeparttable}
\end{table*}

\begin{table*}
\begin{threeparttable}
    \centering
	 \ContinuedFloat
	\caption[]{Continued.}
	\begin{tabular}{l@{\hskip24pt}lrcccccccc}
    \hline
    Galaxy & Type & Array & Channels & Ref. & Beam & RMS noise & $W_{50,\text{obs}}$ & $v_{0,\text{obs}}$ & $\langle v\rangle_\text{obs}$ & $(S_\text{CO}\Delta v)_\text{obs}$ \\
    & & & (km~s$^{-1}$) & & (arcsec$^{2}$) & (mJy) & (km~s$^{-1}$) & (km~s$^{-1}$) & (km~s$^{-1}$) & (Jy~km~s$^{-1}$) \\
    (1) & (2) & (3) & (4) & (5) & (6) & (7) & (8) & (9) & (10) & (11) \\
    \hline
    NGC4274$^\dagger$ & SBab & $12$-m & $10$ & TW & $0.30\times0.21$ & $0.3$ & $340\pm\phantom{0}7$ & $\phantom{0}915\pm\phantom{0}4$ & $\phantom{0}912.5\pm\phantom{0}0.4$ & $\phantom{0}695.6\pm \phantom{0}2.2$ \\
    NGC4278 & E & $7$-m & $20$ & TW & $7.1\phantom{0} \times 6.3\phantom{0}$ & $4.4$ & $\phantom{0}80 \pm 14$ & $\phantom{0}790 \pm \phantom{0}7$ & $\phantom{0}741.0 \pm \phantom{0}4.0 $ & $\phantom{00}18.7 \pm\phantom{0} 0.5$ \\
    NGC4429 & S0-a & $12$-m & $10$ & D18 & $0.23\times0.16$ & $0.6$ & $450\pm\phantom{0}7$ & $1094\pm\phantom{0}4$ & $1110.0\pm \phantom{0}0.5$ & $\phantom{00}92.0\pm \phantom{0}0.3$ \\
    NGC4435 & S0 & $12$-m & $10$ & D22 & $0.25\times0.23$ & $0.9$ & $410\pm\phantom{0}7$ & $\phantom{0}788\pm\phantom{0}4$ & $\phantom{0}789.5\pm \phantom{0}0.4$ & $\phantom{00}88.2\pm \phantom{0}0.3$ \\
    NGC4438$^\dagger$ & Sa & $12$-m & $10$ & D22 & $0.60\times0.53$ & $1.4$ & $418\pm\phantom{0}7$ & $\ -33 \pm\phantom{0}4$ & $\ -19.9\pm \phantom{0}0.3$ & $\phantom{0}906.3 \pm \phantom{0}1.8$ \\
    NGC4501$^\dagger$ & Sb & $12$-m & $15$ & D22 & $0.69\times0.59$ & $1.2$ & $222\pm 11$ & $2272\pm\phantom{0}5$ & $2268.6\pm \phantom{0}0.3$ & $1166.3\pm\phantom{0} 4.1$ \\
    NGC4569$^\dagger$ & Sab & $12$-m & $10$ & TW & $0.21\times0.15$ & $0.6$ & $220\pm\phantom{0}7$ & $-225\pm\phantom{0}4\,$ & $-226.9\pm \phantom{0}0.2\,$ & $2877.4\pm\phantom{0} 6.3$ \\
    NGC4697 & E & $12$-m & $10$ & D17 & $0.56\times0.53$ & $1.1$ & $420\pm\phantom{0}7$ & $1241\pm\phantom{0}4$ & $1248.4\pm \phantom{0}1.6$ & $\phantom{00}13.2\pm \phantom{0}0.2$ \\
    NGC4826 & SABa & $12$-m & $\phantom{1}2$ & D22 & $0.18 \times 0.14$ & $3.4$ & $220\pm\phantom{0}1$ & $\phantom{0}409\pm\phantom{0}1$ & $\phantom{0}405.7\pm \phantom{0}0.1$ & $\phantom{0}700.1\pm \phantom{0}0.7$ \\
    NGC4845 & SABa & $7$-m & $10$ & TW & $7.5\phantom{0} \times 4.5\phantom{0}$ & $7.3$ & $350\pm \phantom{0}7$ & $1090 \pm \phantom{0}4$ & $1094.0\pm\phantom{0} 0.1$ & $\phantom{0}963.6\pm\phantom{0} 0.7$ \\
    NGC5064 & Sb & $12$-m & $20$ & D22 & $0.08\times0.04$ & $0.7$ & $520\pm 14$ & $2966\pm\phantom{0}7$ & $2977.0\pm \phantom{0}1.0$ & $1302.4\pm\phantom{0} 8.5$ \\
    NGC5084 & S0 & $12$-m & $\phantom{1}2$ & TW & $0.59 \times 0.53$ & $2.4$ & $496\pm\phantom{0}1$ & $1715 \pm \phantom{0}1$ & $1702.9\pm \phantom{0}5.0$ & $\phantom{000}3.3\pm \phantom{0}0.1$ \\
    NGC5248$^\dagger$ & SABb & $12$-m & $10$ & TW & $0.17\times0.17$ & $0.4$ & $190\pm\phantom{0}7$ & $1160\pm\phantom{0}4$ & $1162.2\pm \phantom{0}0.1$ & $1061.8\pm \phantom{0}1.1$ \\ 
    NGC5273 & S0 & $7$-m & $10$ & TW & $8.4\phantom{0} \times 6.4\phantom{0}$ & $9.6$ & $120 \pm \phantom{0}7$ & $1085\pm\phantom{0}4$ & $1092.4\pm \phantom{0}0.9$ & $\phantom{00}36.6 \pm\phantom{0} 0.8$ \\
    NGC5427$^\dagger$ & SABc & $7$-m & $10$ & TW & $7.3\phantom{0} \times 4.5\phantom{0}$ & $6.9$ & $120\pm\phantom{0}7$ & $2715\pm\phantom{0}4$ & $2716.4\pm \phantom{0}0.1$ & $\phantom{0}280.6\pm \phantom{0}0.5$ \\
    NGC5746$^\dagger$ & Sb & $7$-m & $10$ & TW & $8.6\phantom{0} \times 4.6\phantom{0}$ & $1.6$ & $150\pm\phantom{0}7$ & $1720\pm\phantom{0}4$ & $1727.2\pm\phantom{0} 0.6$ & $\phantom{0}104.1\pm \phantom{0}0.8$ \\
    NGC5765b & SABb & $12$-m & $10$ & D22 & $0.39\times0.27$ & $0.8$ & $264 \pm\phantom{0}7$ & $8078 \pm \phantom{0}4$ & $8079.9 \pm\phantom{0} 0.2$ & $\phantom{0}133.4 \pm\phantom{0} 0.2$ \\
    NGC5806 & Sb & $12$-m & $10$ & D22 & $0.33\times0.27$ & $1.0$ & $320\pm\phantom{0}7$ & $1367 \pm\phantom{0}4$ & $1373.1\pm \phantom{0}0.3$ & $\phantom{0}171.4\pm \phantom{0}0.7$ \\
    NGC5995 & SABa & $12$-m & $10$ & TW & $0.31\times0.23$ & $0.7$ & $440\pm \phantom{0}7$ & $7415\pm\phantom{0}4$ & $7418.9\pm\phantom{0} 0.6$ & $\phantom{0}391.2 \pm \phantom{0}1.6$ \\
    NGC6500$^\dagger$ & Sab & $7$-m & $10$ & TW & $8.0\phantom{0} \times 6.2\phantom{0}$ & $9.2$ & $440\pm\phantom{0}7$ & $3005\pm\phantom{0}4$ & $3016.1\pm\phantom{0}1.7$ & $\phantom{00}96.0\pm \phantom{0}1.1$ \\
    NGC6753 & Sb & $12$-m & $10$ & D22 & $0.21\times0.10$ & $0.9$ & $390\pm\phantom{0}7$ & $3151 \pm\phantom{0}4$ & $3157.7 \pm\phantom{0} 0.3$ & $2780.3\pm \phantom{0}6.2$  \\
    NGC6958 & E & $12$-m & $10$ & TW & $0.18\times0.09$ & $1.1$ & $440\pm\phantom{0}7$ & $2689\pm\phantom{0}4$ & $2700.8\pm\phantom{0}3.1$ & $\phantom{00}54.7\pm \phantom{0}1.1$ \\
    NGC7049$^\dagger$ & S0 & $7$-m & $20$ & TW & $7.4\phantom{0} \times 5.0\phantom{0}$ & $2.9$ & $620\pm14$ & $2220\pm \phantom{0}7$ & $2243.5\pm\phantom{0} 4.7$ & $\phantom{00}79.0 \pm \phantom{0}1.9$ \\
    NGC7052 & E & $12$-m & $15$ & S21 & $0.13\times0.10$ & $0.4$ & $690\pm 11$ & $4628\pm\phantom{0}5$ & $4631.2\pm \phantom{0}1.1$ & $\phantom{00}39.4\pm \phantom{0}0.2$ \\
    NGC7172 & Sa & $12$-m & $10$ & D22 & $0.15 \times 0.12$ & $0.2$ & $620\pm\phantom{0}7$ & $2509 \pm\phantom{0}4$ & $2514.1\pm\phantom{0} 0.3$ & $\phantom{0}660.2\pm \phantom{0}0.8$ \\
    NGC7424$^\dagger$ & Sc & $7$-m & $10$ & TW & $7.1\phantom{0} \times 4.9\phantom{0}$ & $7.5$ & $\phantom{0}10\pm\phantom{0}7$ & $\phantom{0}930\pm\phantom{0}4$ & $\phantom{0}937.1 \pm \phantom{0}0.5$ & $\phantom{00}35.6\pm \phantom{0}1.0$ \\
    PGC006240 & E-S0 & $7$-m & $10$ & TW & $7.9\phantom{0} \times 5.9\phantom{0}$ & $7.5$ & $350\pm\phantom{0}7$ & $7890\pm\phantom{0}4$ & $7876.1\pm \phantom{0}1.1$ & $\phantom{0}364.9\pm \phantom{0}3.3$ \\
    PGC043387 & E & $12$-m & $13$ & D22 & $0.76 \times 0.52$ & $1.1$ & $483\pm\phantom{0}9$ & $6705\pm\phantom{0}5$ & $6711.7\pm \phantom{0}1.8$ & $\phantom{00}24.3\pm\phantom{0} 0.3$ \\
    UGC1214 & S0-a & $7$-m & $30$ & TW & $8.0\phantom{0} \times 4.7\phantom{0}$ & $2.5$ & $180\pm 21$ & $5085\pm 11$ & $5077.8\pm \phantom{0}2.1$ & $\phantom{00}27.1\pm\phantom{0} 0.6$ \\
    UGC1395$^\dagger$ & Sb & $7$-m & $10$ & TW & $7.2\phantom{0} \times 4.7\phantom{0}$ & $6.2$ & $190\pm\phantom{0}7$ & $5060\pm \phantom{0}4$ & $5069.9\pm \phantom{0}1.0$ & $\phantom{00}46.3\pm \phantom{0}0.8$ \\
    \hline
    \end{tabular}
\end{threeparttable}
\end{table*}

In addition to the $39$ galaxies with ACA low angular resolution detections, we include another $42$ galaxies with 12-m array higher angular resolution observations, all part of previous WISDOM studies (and thus all detected in CO; e.g.\ \citealt{WISDOM_X, Elford_2023}). The associated data cubes use data obtained as part of the aforementioned projects as well as projects 2013.1.00493.S (PI: Bureau), 2015.1.00419.S (PI: Davis), 2015.1.00466.S (PI: Onishi), 2015.1.00598.S (PI: Bureau), 2016.1.00437.S (PI: Davis), 2016.1.00839.S (PI: Onishi), 2016.1.01553.S (PI: Bureau), 2016.2.00053.S (PI: Liu), 2017.1.00277.S (PI: Onishi), 2017.1.00391.S (PI: North), 2017.1.00904.S (PI: Smith) and 2018.1.00397.S (PI: Smith). In most cases ($39/42$) the observations targeted the $^{12}\text{CO(2–1)}$ line using ALMA Band 6, but in a few cases ($3/42$; NGC3351, NGC4429 and NGC4826) the $^{12}\text{CO(3–2)}$ line was targeted using ALMA Band 7. The majority of these cubes were used in previous WISDOM studies (see column~5 in Table~\ref{tab:observational CO data}), so the robust weightings, cleaning thresholds and channel widths vary. The weighting and cleaning threshold choices primarily affect the angular resolutions and signal-to-noise ratios of the resulting cubes, but they are unlikely to significantly affect the total recovered flux densities and line widths, as they do not alter the calibrated visibilities \citep{Briggs_1995}. Consequently, for reasonable choices, the integrated fluxes are expected to be robust. Having said that, interferometric observations may miss extended emission, particularly when the CO distribution exceeds the primary beam or the maximum recoverable scale of the array (all the WISDOM galaxies were observed with a single central pointing in each array). In such cases some of our quantities, such as integrated flux, may be underestimated. We have identified these galaxies using a dagger alongside their names in column~1 of Tables~\ref{tab:observational CO data} and \ref{tab:CO data}. Four of the resulting cubes have $2$~km~s$^{-1}$ channels, $23$ cubes $10$~km~s$^{-1}$ channels and $9$ cubes $10<{\rm channels}\leq20$~km~s$^{-1}$. The cubes have synthesised beam sizes ranging from $0.03$ to $0.84$~arcsec. A subset of the high-resolution WISDOM cubes were originally provided in the barycentric spectral reference frame. To ensure a homogeneous catalogue, these were converted to LSRK using the \textsc{casa} task \texttt{imreframe}. Consequently, the systemic velocities reported in this work may differ by up to $20$~km~s$^{-1}$ from those previously published for these galaxies, due to the differing spectral reference frames. The properties of these $42$ high-resolution cubes are also summarised in Table~\ref{tab:observational CO data}. 

Integrated-intensity, luminosity-weighted mean line-of-sight velocity and luminosity-weighted line-of-sight velocity dispersion maps were created for all CO-detected sample galaxies using the \textsc{pymakeplots} package\footnote{\url{https://github.com/TimothyADavis/pymakeplots}}, which implements the standard masked-moment technique of \citet{Dame_2011}. For the ACA data, we created data cubes with and without primary-beam correction. The cubes without primary-beam correction were first spatially and spectrally smoothed, using typically a boxcar spatial kernel $1.5$ times the width of the synthesised beam and a boxcar spectral kernel $4$ times the channel width. The smoothed data cubes were then clipped to a specific threshold (typically $3$ -- $8$ times the RMS noise of the line-free channels) to create a binary mask. These masks were then applied to the unsmoothed primary-beam corrected cubes to create moment maps.
Major-axis position-velocity diagrams (PVDs) were created by taking cuts through the masked primary-beam corrected cubes at kinematic position angles determined by eye, as the position angles of the molecular gas distributions do not always align with the optical position angles. The resulting moment maps and PVDs of the ACA sample galaxies are shown in Fig.~\ref{fig:CO ACA moment maps}. 

For the $12$-m data, we primarily had access to data cubes without primary-beam correction (27/42). For these, we followed the same procedure as above, but the binary masks were created from and applied to the same cubes. In addition, the position angle from previous kinematic modelling work was adopted when available. The resulting moment maps and PVDs of the 12-m sample galaxies are shown in Fig.~\ref{fig:CO high res moment maps}.

\subsection{Catalogue measurements}
\label{subsection:Catalogue measurements}

For each galaxy, in addition to the moment maps and PVD, we report a set of standard catalogue parameters derived from the CO data in Table~\ref{tab:observational CO data}. These parameters are derived from the integrated spectrum, constructed by summing, for each velocity channel, the flux density over all spatial pixels within the region of detected emission, as defined by the mask used in moment map creation. The noise in each velocity bin was calculated using the same method as \citet{Liang_2023}, whereby for a given velocity bin the \textsc{pymakeplots} mask is applied to each of the line-free channel of the data cube, the flux within each masked region is summed, and the RMS of these sums is adopted as the uncertainty on the flux in that velocity bin. The integrated spectrum is shown alongside the moment maps and PVD in Figs.~\ref{fig:CO ACA moment maps} and \ref{fig:CO high res moment maps}.

Throughout this paper, the subscript ``obs'' denotes quantities measured directly from the observed integrated spectrum, distinguishing them from quantities derived from the line profile fit used in the TFR analysis (Section~\ref{subsection:velocity widths}). In column~11 of Table~\ref{tab:observational CO data} we list the integrated galaxy flux ($(S_\text{CO}\Delta v)_\text{obs}$), obtained by summing the integrated spectrum over the channels containing line emission. The uncertainty on the integrated flux was calculated using standard error propagation (i.e.\ summing the squared noise of each channel, taking the square root, and multiplying by the channel width). We assume that the noise between velocity channels is not correlated.

In column~8 we list the width of the integrated CO spectrum at $50\%$ of its maximum ($W_{50,\text{obs}}$).  We traverse the spectrum from the low- and high-velocity ends, respectively, and select the first channel on each side whose flux exceeds $50\%$ of the peak value. The width is then taken as the difference between the central velocities of those two edge channels. Since the edges of $W_{50,\text{obs}}$ are identified at the centres of the edge channels, each edge has an uncertainty of half a channel width. As such, the uncertainty on $W_{50,\text{obs}}$ is calculated as $\sigma_{W_{50,\text{obs}}}=\Delta v /\sqrt{2}$, where $\Delta v$ is the channel width. This represents the minimum uncertainty imposed by the spectral resolution alone, and it does not account for possible additional uncertainty arising from the finite signal-to-noise ratio of the observations.

In column~9 we list the mean velocity ($v_{0,\text{obs}}$), defined as the midpoint between the two velocities defining $W_{50,\text{obs}}$. Using standard error propagation, the uncertainty on $v_{0,\text{obs}}$ is calculated as $\sigma_{v_{0,\text{obs}}}=\Delta v /(2\sqrt{2})$. This should likewise be regarded as a lower limit.

Lastly, in column~10 we list the flux-weighted mean velocity, defined as 
\begin{equation}
    \langle v \rangle_\text{obs}=\frac{\sum_i S_i v_i}{\sum_iS_i}\quad,
\end{equation}
where $S_i$ and $v_i$ are the flux density and velocity in the $i^\text{th}$ channel, respectively. This provides an estimate of the centroid of the line profile that accounts for the full flux distribution across the spectrum. The uncertainty is obtained by propagating the uncertainty in each velocity channel, assuming that the noise between channels is uncorrelated.


\subsection{Optical data}
\label{subsection:optical data}

To compare the CO and optical extents of the galaxies, in Figs.~\ref{fig:CO ACA moment maps} and \ref{fig:CO high res moment maps} we also overlay the CO contours of each galaxy over a red-band optical image from the Digitized Sky Survey-2 (DSS), obtained from the European Southern Observatory archive\footnote{\url{https://archive.eso.org/dss/dss}}.

\begin{table*}
\begin{threeparttable}
    \centering
	\caption{Tully-Fisher relation data of all CO-detected galaxies.}
	\label{tab:CO data}
	\begin{tabular}{l@{\hskip12pt}cccccccl}
    \hline
    Galaxy & $v_0$ & $W_{50}$ & $S_\text{CO}\Delta v$ & $i_{b/a}$ & $i_{\text{dust}}$ & $i_{\text{CO}}$ & $W_{50}/\sin i$ & Sample \\
    & (km~s$^{-1}$) & (km~s$^{-1}$) & (Jy~km~s$^{-1}$) & (deg) & (deg) & (deg) & (km~s$^{-1}$) & \\
    (1) & (2) & (3) & (4) & (5) & (6) & (7) & (8) & (9) \\
    \hline
    ESO383-G035 & $2332\phantom{.0}\pm1\phantom{.0}$ & $242.5\pm\phantom{0}2.5$ & $\phantom{00}31.9\pm\phantom{0}0.9$ & $57\pm\phantom{0}3$ & $-$ & $-$ & $289\phantom{.0}\pm\phantom{0}10\phantom{.0}$ & I A B \\ 
    FRL49 & $5886.6\pm0.5$ & $391.6\pm\phantom{0}1.1$ & $\phantom{0}461.3 \pm\phantom{0}3.6$ & $40^\dagger\phantom{\pm00}$ & $57\pm12$ & $64\pm1$ & $435.7\pm\phantom{00}3.9$ & I A B \\
    FRL1146 & $9170\phantom{.0}\pm4\phantom{.0}$ & $435\phantom{.0}\pm23\phantom{.0}$ & $\phantom{00}90\phantom{.0} \pm 11\phantom{.0}$ & $60\pm\phantom{0}3$ & $72\pm\phantom{0}4$ & $-$ & $458\phantom{.0}\pm\phantom{0}26\phantom{.0}$ & I A B \\
    IC0843 & $7198\phantom{.0}\pm6\phantom{.0}$ & $783\phantom{.0}\pm12\phantom{.0}$ & $\phantom{00}22.7\pm\phantom{0}1.0$ & $68\pm\phantom{0}4$ & $79\pm\phantom{0}2$ & $-$ & $797\phantom{.0}\pm\phantom{0}14\phantom{.0}$ & I A B \\
    IC1481 & $5994\phantom{.0}\pm2\phantom{.0}$ & $219.3\pm\phantom{0}4.0$ & $\phantom{00}76.4\pm\phantom{0}1.2$ & $28\pm20$ & $-$ & $-$ & $467\phantom{.0}\pm310\phantom{.0}$ & I \\
    IC4329A & $4773\phantom{.4}\pm4\phantom{.1}$ & $498\phantom{.4}\pm18\phantom{.0}$ & $\phantom{00}12.5\pm\phantom{0}0.6$ & $58\pm\phantom{0}3$ & $84\pm\phantom{0}2$ & $-$ & $501\phantom{.1}\pm\phantom{0}18\phantom{.0}$ & I A B \\
    MRK0567 & $9422.0\pm0.3$ & $222.2\pm\phantom{0}0.6$ & $\phantom{0}158.0\pm\phantom{0}1.0$ & $26^\dagger\phantom{\pm00}$ & $19\pm\phantom{0}6$ & $-$ & $683\phantom{.0}\pm210\phantom{.0}$ & I A \\
    NGC0017 & $5699.0\pm0.2$ & $335.5\pm\phantom{0}0.8$ & $\phantom{0}444.2\pm\phantom{0}1.0$ & $74\pm\phantom{0}2$ & $20\pm\phantom{0}9$ & $-$ & $981\phantom{.0}\pm420\phantom{.0}$ & I A \\
    NGC0289$^\dagger$ & $1625.0\pm0.1$ & $144.0\pm\phantom{0}0.1$ & $\phantom{0}759.2\pm\phantom{0}0.5$ & $43\pm\phantom{0}3$ & $-$ & $-$ & $211\phantom{.0}\pm\phantom{0}12\phantom{.0}$ & I \\
    NGC0383 & $4930.6\pm0.2$ & $523.2\pm\phantom{0}0.5$ & $\phantom{0}108.2\pm \phantom{0}0.5$ & $36\pm\phantom{0}7$ & $37\pm\phantom{0}6$ & $38\pm3$ & $850\phantom{.0}\pm\phantom{0}57\phantom{.0}$ & I A B \\
    NGC0404 & $\ -55.7\pm0.1$ & $\phantom{0}26.9\pm\phantom{0}0.1$ & $\phantom{0}153.4 \pm\phantom{0}0.8$ & $15\pm15$ & $16\pm\phantom{0}6$ & $37\pm2$ & $\phantom{0}44.7\pm\phantom{00}2.1$ & $\times$ \\
    NGC0449 & $4684.7\pm0.6$ & $233.2\pm\phantom{0}1.6$ & $\phantom{00}65.6\pm \phantom{0}1.0$ & $58\pm\phantom{0}3$ & $70\pm\phantom{0}4$ & $-$ & $248.2\pm\phantom{00}6.5$ & I A B \\
    NGC0524 & $2382.1\pm0.2$ & $306.0\pm\phantom{0}0.4$ & $\phantom{00}37.6 \pm \phantom{0}0.2$ & $13\pm13$ & $13\pm\phantom{0}8$ & $20\pm7$ & $895\phantom{.0}\pm300\phantom{.0}$ & I A \\
    NGC0541 & $5319\phantom{.0}\pm2\phantom{.0}$ & $239.8\pm\phantom{0}3.5$ & $\phantom{000}3.1\pm\phantom{0}0.2$ & $50\pm\phantom{0}3$ & $29\pm10$ & $-$ & $495\phantom{.0}\pm160\phantom{.0}$ & I A \\
    NGC0612 & $8625.8\pm0.3$ & $790.0\pm\phantom{0}0.8$ & $\phantom{0}269.9\pm\phantom{0} 2.3$ & $29\pm\phantom{0}9$ & $79\pm\phantom{0}2$ & $81\pm1$ & $799.8\pm\phantom{00}2.4$ & I A B \\
    NGC0708 & $4762.8\pm0.7$ & $354.5\pm\phantom{0}1.9$ & $\phantom{00}41.7 \pm\phantom{0}0.4$ & $69\pm\phantom{0}3$ & $73\pm\phantom{0}3$ & $77\pm3$ & $363.8\pm\phantom{00}4.8$ & I A B \\
    NGC0863$^\dagger$ & $7715.3\pm0.3$ & $226.4\pm\phantom{0}0.7$ & $\phantom{00}65.4 \pm\phantom{0}0.9$ & $25\pm10$ & $19\pm14$ & $-$ & $695\phantom{.0}\pm490\phantom{.0}$ & I A \\
    NGC1084$^\dagger$ & $1356.0\pm0.1$ & $206.5\pm\phantom{0}0.1$ & $\phantom{0}458.7\pm\phantom{0}0.1$ & $50\pm\phantom{0}2$ & $67\pm\phantom{0}2$ & $-$ & $224.3\pm\phantom{00}3.3$ & I A B \\
    NGC1194 & $4006\phantom{.0}\pm1\phantom{.0}$ & $408.9\pm\phantom{0}3.1$ & $\phantom{000}7.5\pm\phantom{0}0.2$ & $67\pm\phantom{0}3$ & $81\pm\phantom{0}3$ & $-$ & $414.0\pm\phantom{00}4.7$ & I A B \\
    NGC1300$^\dagger$ & $1538.0\pm0.2$ & $211.8\pm\phantom{0}0.5$ & $\phantom{0}459.6\pm\phantom{0}1.2$  & $61\pm\phantom{0}3$ & $55\pm\phantom{0}4$ & $-$ & $259\phantom{.0}\pm\phantom{0}13\phantom{.0}$ & I A B \\
    NGC1358 & $3954\phantom{.0}\pm2\phantom{.0}$ & $189.7\pm\phantom{0}6.4$ & $\phantom{000}6.2\pm\phantom{0}0.3$ & $59\pm\phantom{0}3$ & $46\pm\phantom{0}5$ & $-$ & $264\phantom{.0}\pm\phantom{0}24\phantom{.0}$ & I A B \\
    NGC1387 & $1263.6\pm0.1$ & $143.3\pm\phantom{0}0.1$ & $\phantom{0}387.8\pm\phantom{0}1.3$ & $11\pm11$ & $16\pm\phantom{0}6$ & $14\pm1$ & $592\phantom{.0}\pm\phantom{0}42\phantom{.0}$ & I A \\
    NGC1400 & $\phantom{1}547.9\pm0.1$ & $426.5\pm\phantom{0}0.2$ & $\phantom{0}117.3\pm\phantom{0}0.6$ & $29\pm\phantom{0}5$ & $27\pm\phantom{0}8$ & $-$ & $939\phantom{.0}\pm260\phantom{.0}$ & I A \\
    NGC1448$^\dagger$ & $1122.3\pm0.1$ & $221.2\pm\phantom{0}0.4$ & $1185.7\pm\phantom{0}1.4$ & $89\pm\phantom{0}1$ & $82\pm\phantom{0}2$ & $-$ & $223.4\pm\phantom{00}1.2$ & I \\
    NGC1493 & $1028.5\pm0.1$ & $\phantom{0}23.5\pm\phantom{0}0.2$ & $\phantom{00}61.7\pm\phantom{0}0.5$ & $22\pm\phantom{0}7$ & $25\pm\phantom{0}7$ & $-$ & $\phantom{0}56\phantom{.0}\pm\phantom{0}15\phantom{.0}$ & $\times$ \\
    NGC1497 & $6096.7\pm0.5$ & $588.8\pm\phantom{0}0.9$ & $\phantom{00}43.1 \pm\phantom{0}0.6$ & $65\pm\phantom{0}3$ & $70\pm\phantom{0}4$ & $-$ & $627\phantom{.0}\pm\phantom{0}16\phantom{.0}$ & I A B \\
    NGC1574 & $1016.6\pm0.5$ & $226.4\pm\phantom{0}0.9$ & $\phantom{00}30.0 \pm\phantom{0}0.6$ & $25\pm\phantom{0}6$ & $28\pm\phantom{0}4$ & $27\pm4$ & $499\phantom{.0}\pm\phantom{0}68\phantom{.0}$ & I A \\
    NGC2110 & $2376\phantom{.0}\pm1\phantom{.0}$ & $403.5\pm\phantom{0}3.8$ & $\phantom{00}55.1\pm\phantom{0}0.7$ & $25\pm\phantom{0}6$ & $45\pm\phantom{0}3$ & $-$ & $571\phantom{.0}\pm\phantom{0}30\phantom{.0}$ & I A B \\
    NGC2179 & $2829.8\pm0.8$ & $413.6\pm\phantom{0}1.4$ & $\phantom{00}25.5 \pm\phantom{0}0.5$ & $45\pm\phantom{0}3$ & $51\pm\phantom{0}3$ & $-$ & $532\phantom{.0}\pm\phantom{0}23\phantom{.0}$ & I A B \\
    NGC2187A$^\dagger$ & $3919.8\pm0.4$ & $449.7\pm\phantom{0}0.8$ & $\phantom{0}232.3\pm\phantom{0}3.1$ & $64\pm\phantom{0}4$ & $63\pm\phantom{0}2$ & $-$ & $504.7\pm\phantom{00}9.0$ & I A B \\
    NGC2903 & $\phantom{0}565.9\pm0.1$ & $237.8\pm\phantom{0}0.1$ & $1186.9\pm\phantom{0}1.1$ & $24\pm\phantom{0}5$ & $60\pm\phantom{0}8$ & $-$ & $275\phantom{.0}\pm\phantom{0}22\phantom{.0}$ & I A B \\
    NGC2911 & $3188\phantom{.0}\pm1\phantom{.0}$ & $521.7\pm\phantom{0}2.7$ & $\phantom{00}42.1\pm\phantom{0}0.6$ & $57\pm\phantom{0}3$ & $-$ & $-$ & $622\phantom{.0}\pm\phantom{0}21\phantom{.0}$ & I A B \\
    NGC2964 & $1316.1\pm0.5$ & $173.4\pm\phantom{0}1.3$ &  $\phantom{0}473.4\pm\phantom{0}4.7$ & $43\pm\phantom{0}5$ & $63\pm\phantom{0}2$ & $-$ & $194.6\pm\phantom{00}3.8$ & I A B \\
    NGC2974 & $1807\phantom{.0}\pm6\phantom{.0}$ & $519\phantom{.0}\pm15\phantom{.0}$ & $\phantom{00}25.9\pm\phantom{0}1.2$ & $57\pm\phantom{0}4$ & $60\pm\phantom{0}2$ & $-$ & $600\phantom{.0}\pm\phantom{0}21\phantom{.0}$ & I A B \\
    NGC3021$^\dagger$ & $1515.2\pm0.3$ & $273.2\pm\phantom{0}0.6$ & $\phantom{0}513.9\pm\phantom{0}1.7$ & $56\pm\phantom{0}5$ & $63\pm\phantom{0}3$ & $-$ & $306.6\pm\phantom{00}8.2$ & I A B \\
    NGC3156 & $1310\phantom{.0}\pm1\phantom{.0}$ & $140.9\pm\phantom{0}3.2$ & $\phantom{00}11.3\pm\phantom{0}0.4$ & $64\pm\phantom{0}5$ & $-$ & $-$ & $156.8\pm\phantom{00}7.6$ & I A B \\
    NGC3169 & $1218.3\pm0.2$ & $466.0\pm\phantom{0}0.5$ & $\phantom{0}756.7 \pm\phantom{0}2.5$ & $53\pm\phantom{0}6$ & $67\pm\phantom{0}3$ & $-$ & $506\phantom{.0}\pm\phantom{0}11\phantom{.0}$ & I A B \\
    NGC3277 & $1404.5\pm0.2$ & $235.4\pm\phantom{0}0.4$ & $\phantom{0}134.3\pm\phantom{0}0.7$ & $42\pm\phantom{0}5$ & $32\pm\phantom{0}6$ & $-$ & $444\phantom{.0}\pm\phantom{0}74\phantom{.0}$ & I A B \\
    NGC3351$^\dagger$ & $\phantom{1}786.9\pm0.2$ & $203.5\pm\phantom{0}0.4$ & $\phantom{0}259.6\pm\phantom{0}1.6$ & $53\pm\phantom{0}4$ & $40\pm\phantom{0}6$ & $-$ & $317\phantom{.0}\pm\phantom{0}40\phantom{.0}$ & I A B \\
    NGC3368 & $\phantom{1}907.0\pm0.1$ & $222.6\pm\phantom{0}0.1$ & $\phantom{0}816.4\pm\phantom{0}0.6$ & $50\pm\phantom{0}5$ & $50\pm\phantom{0}4$ & $-$ & $291\phantom{.0}\pm\phantom{0}17\phantom{.0}$ & I \\
    NGC3393$^\dagger$ & $3688.3\pm0.3$ & $231.6\pm\phantom{0}1.5$ & $\phantom{00}99.5\pm\phantom{0}0.6$ & $30\pm\phantom{0}7$ & $40\pm\phantom{0}5$ & $-$ & $353\phantom{.0}\pm\phantom{0}37\phantom{.0}$ & I A B \\
    NGC3607 & $1181.5\pm0.3$ & $537.5\pm\phantom{0}0.5$ & $\phantom{0}166.8 \pm\phantom{0}0.8$ & $32\pm13$ & $51\pm\phantom{0}2$ & $-$ & $692\phantom{.0}\pm\phantom{0}20\phantom{.0}$ & I A B \\
    NGC3810$^\dagger$ & $1002.0\pm0.1$ & $199.9\pm\phantom{0}0.1$ & $\phantom{0}614.4\pm\phantom{0}0.8$ & $49\pm\phantom{0}3$ & $48\pm\phantom{0}3$ & $-$ & $269\phantom{.0}\pm\phantom{0}13\phantom{.0}$ & I A B \\
    NGC3862 & $6363.0\pm0.6$ & $216.6\pm\phantom{0}1.3$ & $\phantom{00}15.2 \pm\phantom{0}0.3$ & $\phantom{0}5\pm\phantom{0}5$ & $16\pm\phantom{0}9$ & $-$ & $786\phantom{.0}\pm431\phantom{.0}$ & I A \\
    NGC4030$^\dagger$ & $1452.4\pm0.1$ & $329.4\pm\phantom{0}0.1$ & $1375.8\pm\phantom{0}1.0$ & $46\pm\phantom{0}3$ & $40\pm\phantom{0}4$ & $-$ & $513\phantom{.0}\pm\phantom{0}43\phantom{.0}$ & I A B \\
    NGC4061 & $7189.8\pm0.6$ & $694.2\pm\phantom{0}1.2$ & $\phantom{00}23.6\pm\phantom{0}0.2$ & $34\pm\phantom{0}7$ & $60\pm\phantom{0}2$ & $-$ & $802\phantom{.0}\pm\phantom{0}16\phantom{.0}$ & I A B \\
    NGC4203 & $1076\phantom{.0}\pm2\phantom{.0}$ & $351.1\pm\phantom{0}1.3$ & $\phantom{00}21.0\pm\phantom{0}0.6$ & $25\pm10$ & $30\pm\phantom{0}6$ & $-$ & $702\phantom{.0}\pm127\phantom{.0}$ & I A B \\
    NGC4212$^\dagger$ & $\ -74.4\pm0.3$ & $179.1\pm\phantom{0}0.8$ & $\phantom{0}546.1\pm\phantom{0}2.1$ & $55\pm\phantom{0}3$ & $57\pm\phantom{0}3$ & $-$ & $213.6\pm\phantom{00}7.3$ & I A B \\
    NGC4235 & $2234.2\pm2.0$ & $246.3\pm\phantom{0}4.5$ & $\phantom{00}35.3\pm\phantom{0}1.3$ & $87\pm\phantom{0}3$ & $82\pm\phantom{0}2$ & $-$ & $248.7\pm\phantom{00}4.7$ & I A B \\
    \hline
    \end{tabular}
    \begin{tablenotes}[flushleft] 
    \item \textit{Notes.} The data in this table are available in a  machine-readable format online. (1) Galaxy name. $^\dagger$ indicates a galaxy for which the CO emission extends close to or beyond the primary beam FWHM, and for which some of the derived quantities, particularly the velocity width and integrated flux, may be underestimated. (2) Central velocity, derived from the best fit to the integrated spectrum and reported in the LSRK reference frame. (3) Velocity width at $50\%$ of the peak, derived from the best fit to the integrated spectrum. (4) Integrated CO flux, derived from the best fit to the integrated spectrum. (5) Inclination derived from the galaxy HyperLEDA optical axis ratio. Uncertainties are derived using the optical axis ratio uncertainties listed in HyperLEDA. $^\dagger$ indicates galaxies for which $K$-band axis ratios from 2MASS were used instead (for which there is no listed uncertainty). (6) Inclination derived from ellipse fitting to unsharp-masked optical \textit{HST} images. The ellipse parameters were adjusted by eye and the uncertainties listed bracket the range of inclinations for which the ellipse remains an acceptable visual match. (7) Inclination derived from kinematic modelling of the CO data cube in the literature. The uncertainties are the average of the listed upper and lower $3\sigma$ uncertainties. (8) Inclination-corrected velocity width at $50\%$ of peak, derived using the velocity width in column~3 and the `best' inclination, selected based on availability in the following order of preference: $i_{\text{CO}}$, $i_{\text{dust}}$ and $i_{b/a}$. Uncertainties are propagated in the standard manner. (9) Sample membership of each galaxy. I: initial sample; A: sub-sample A; B: sub-sample B; $\times$: excluded.
    \end{tablenotes}
\end{threeparttable}
\end{table*}

\begin{table*}
\begin{threeparttable}
    \centering
    \ContinuedFloat
	\caption[]{Continued.}
	\begin{tabular}{l@{\hskip22pt}cccccccl}
    \hline
    Galaxy & $v_0$ & $W_{50}$ & $S_\text{CO}\Delta v$ & $i_{b/a}$ & $i_{\text{dust}}$ & $i_{\text{CO}}$ & $W_{50}/\sin i$ & Sample \\
    & (km~s$^{-1}$) & (km~s$^{-1}$) & (Jy~km~s$^{-1}$) & (deg) & (deg) & (deg) & (km~s$^{-1}$) & \\
    (1) & (2) & (3) & (4) & (5) & (6) & (7) & (8) & (9) \\
    \hline
    NGC4245 & $\phantom{1}880.4\pm0.3$ & $204.6\pm\phantom{0}0.4$ & $\phantom{00}66.4\pm\phantom{0}0.6$ & $52\pm\phantom{0}3$ & $40\pm\phantom{0}6$ & $-$ & $318\phantom{.0}\pm\phantom{0}40\phantom{.0}$ & I A B \\
    NGC4261 & $\ 2157\phantom{.0}\pm8\phantom{.0}$ & $864\phantom{.0}\pm38\phantom{.0}$ & $\phantom{000}3.7 \pm\phantom{0}0.2$ & $40\pm\phantom{0}4$ & $61\pm\phantom{0}3$ & $61\pm3$ & $988\phantom{.0}\pm\phantom{0}52\phantom{.0}$ & I A B \\
    NGC4274$^\dagger$ & $\phantom{0}913.4\pm0.1$ & $357.5\pm\phantom{0}0.4$ & $\phantom{0}693.8\pm\phantom{0}2.1$ & $64\pm\phantom{0}2$ & $68\pm\phantom{0}5$ &  $-$ & $386\phantom{.0}\pm\phantom{0}14\phantom{.0}$ & I A B \\
    NGC4278 & $\phantom{1}629\phantom{.0}\pm2\phantom{.0}$ & $411.9\pm\phantom{3}6.1$ & $\phantom{00}12.9\pm\phantom{0}0.4$ & $13\pm11$ & $57\pm\phantom{0}3$ & $-$ & $491\phantom{.0}\pm\phantom{0}18\phantom{.0}$ & I A B \\
    NGC4429 & $1094.3\pm0.3$ & $455.9\pm\phantom{0}0.5$ & $\phantom{00}90.3\pm\phantom{0}0.3$ & $71\pm\phantom{0}4$ & $65\pm\phantom{0}1$ & $67\pm1$ & $495.3\pm\phantom{00}3.7$ & I A B \\
    NGC4435 & $\phantom{0}790.5\pm0.1$ & $415.7\pm\phantom{0}0.3$ & $\phantom{00}89.6\pm\phantom{0}0.3$ & $49\pm\phantom{0}5$ & $71\pm\phantom{0}2$ & $-$ & $439.7\pm\phantom{00}5.3$ & I A B \\
    NGC4438$^\dagger$ & $\ -38.6\pm0.1$ & $441.6\pm\phantom{0}0.3$ & $\phantom{0}899.2\pm\phantom{0}2.1$ & $67\pm\phantom{0}4$ & $65\pm\phantom{0}4$ & $-$ & $487\phantom{.0}\pm\phantom{0}16\phantom{.0}$ & I A B \\
    NGC4501$^\dagger$ & $2265.4\pm0.3$ & $205.1\pm\phantom{0}0.5$ & $ 1257.6\pm\phantom{0}4.5$ & $62\pm\phantom{0}3$ & $63\pm\phantom{0}3$ & $-$ & $230.2\pm\phantom{00}6.2$ & I \\
    NGC4569$^\dagger$ & $-254.7\pm0.2\,$ & $175.1\pm\phantom{0}0.2$ & $2383.2\pm\phantom{0}3.9$ & $68\pm\phantom{0}3$ & $69\pm\phantom{0}3$ & $-$ & $187.6\pm\phantom{00}3.8$ & I \\
    NGC4697 & $1243.5\pm0.4$ & $432.9\pm\phantom{0}0.7$ & $\phantom{00}13.7\pm\phantom{0} 0.2$ & $59\pm\phantom{0}4$ & $76\pm\phantom{0}1$ & $76\pm1$ & $446.2\pm\phantom{00}2.1$ & I A B \\
    NGC4826 & $\phantom{0}413.5\pm0.1$ & $235.1\pm\phantom{0}0.1$ & $\phantom{0}717.2\pm\phantom{0}0.9$ & $61\pm\phantom{0}3$ & $64\pm\phantom{0}3$ & $-$ & $261.6\pm\phantom{00}6.7$ & I A B \\
    NGC4845 & $1095.0\pm0.1$ & $362.7\pm\phantom{0}0.1$ & $\phantom{0}956.1 \pm\phantom{0}0.8$ & $85\pm\phantom{0}2$ & $80\pm\phantom{0}1$ & $-$ & $368.3\pm\phantom{00}1.1$ & I A B \\
    NGC5064 & $2960.9\pm0.2$ & $543.3\pm\phantom{0}0.5$ & $1273.3\pm\phantom{0}7.4$ & $60\pm\phantom{0}2$ & $69\pm\phantom{0}2$ & $71\pm2$ & $574.6\pm\phantom{00}6.9$ & I A B \\
    NGC5084 & $1728\phantom{.0}\pm3\phantom{.0}$ & $525\phantom{.0}\pm\phantom{0}8\phantom{.0}$ & $\phantom{000}4.3 \pm\phantom{0}0.2$ & $90\phantom{\pm000}$ & $-$ & $-$ & $563\phantom{.0}\pm\phantom{0}14\phantom{.0}$ & I A B \\
    NGC5248$^\dagger$ & $1160.7\pm0.1$ & $207.5\pm\phantom{0}0.1$ & $1032.1\pm\phantom{0}1.1$ & $56\pm\phantom{0}4$ & $50\pm\phantom{0}6$ & $-$ & $271\phantom{.0}\pm\phantom{0}24\phantom{.0}$ & I A B \\
    NGC5273 & $1086.3\pm0.8$ & $123.8\pm\phantom{0}2.3$ & $\phantom{0}36.2 \pm\phantom{0}0.8$ & $53\pm\phantom{0}4$ & $26\pm10$ & $-$ & $282\phantom{.0}\pm100\phantom{.0}$ & I A \\
    NGC5427$^\dagger$ & $2711.8\pm0.1$ & $123.0\pm\phantom{0}0.2$ & $\phantom{0}282.5\pm\phantom{0}0.5$ & $25\pm15$ & $26\pm\phantom{0}8$ & $-$ & $281\phantom{.0}\pm\phantom{0}80\phantom{.0}$ & I A \\
    NGC5746$^\dagger$ & $1724.6\pm0.7$ & $183.8\pm\phantom{0}2.8$ & $\phantom{0}101.0\pm\phantom{0}0.9$ & $90\phantom{\pm000} $ & $85\pm\phantom{0}2$ & $-$ & $184.5\pm\phantom{00}2.9$ & $\times$ \\
    NGC5765b & $8081.2\pm0.2$ & $282.2\pm\phantom{0}0.7$ & $\phantom{0}133.8 \pm \phantom{0}0.3$ & $49\pm\phantom{0}8$ & $43\pm\phantom{0}5$ & $-$ & $414\phantom{.0}\pm\phantom{0}39\phantom{.0}$ & I A B \\
    NGC5806 & $1366.4\pm0.2$ & $345.6\pm\phantom{0}0.6$ & $\phantom{0}163.2 \pm\phantom{0}0.7$ & $59\pm\phantom{0}2$ & $59\pm\phantom{0}5$ & $-$ & $403\phantom{.0}\pm\phantom{0}21\phantom{.0}$ & I A B \\
    NGC5995 & $7413.5\pm0.4$ & $463.8\pm1.0$ & $\phantom{0}396.1\pm\phantom{0}1.8$ & $40\pm\phantom{0}3$ & $59\pm\phantom{0}6$ & $-$ & $541\phantom{.0}\pm\phantom{0}34\phantom{.0}$ & I A B \\
    NGC6500$^\dagger$ & $3001.8\pm0.7$ & $447.5\pm\phantom{0}1.3$ & $\phantom{00}86.6 \pm\phantom{0}1.1$ & $38\pm\phantom{0}9$ & $53\pm\phantom{0}3$ & $-$ & $560\phantom{.0}\pm\phantom{0}22\phantom{.0}$ & I A B \\
    NGC6753 & $3148.9\pm0.1$ & $410.3\pm\phantom{0}0.3$ &  $2780.1\pm\phantom{0}7.4$ & $30\pm\phantom{0}7$ & $34\pm10$ & $-$ & $734\phantom{.0}\pm190\phantom{.0}$ & I A B \\
    NGC6958 & $2690.6\pm0.8$ & $450.3\pm\phantom{0}2.1$ &  $\phantom{00}55.0\pm\phantom{0}1.3$ & $42\pm\phantom{0}4$ & $45\pm\phantom{0}6$ & $-$ & $637\phantom{.0}\pm\phantom{0}67\phantom{.0}$ & I A B \\
    NGC7049$^\dagger$ & $2217.7\pm0.8$ & $644.4\pm\phantom{0}1.6$ & $\phantom{00}79.6\pm\phantom{0}1.6$ & $52\pm\phantom{0}4$ & $55\pm\phantom{0}3$ & $-$ & $787\phantom{.0}\pm\phantom{0}29\phantom{.0}$ & I A B \\
    NGC7052 & $4627.1\pm0.9$ & $701.4\pm\phantom{0}0.5$ & $\phantom{00}39.4\pm\phantom{0}0.2$ & $68\pm\phantom{0}3$ & $73\pm\phantom{0}2$ & $75\pm1$ & $726.1\pm\phantom{00}3.4$ & I A B \\
    NGC7172 & $2509.1\pm0.1$ & $637.6\pm\phantom{0}0.4$ &  $\phantom{0}651.5\pm\phantom{0}1.0$ & $50\pm\phantom{0}4$ & $83\pm\phantom{0}1$ & $-$ & $642.4\pm\phantom{00}1.4$ & I A B \\
    NGC7424$^\dagger$ & $\phantom{0}931.5\pm0.5$ & $\phantom{0}33.1\pm\phantom{0}1.1$ & $\phantom{00}34.6 \pm\phantom{0}1.0$ & $33\pm\phantom{0}4$ & $-$ & $-$ & $\phantom{0}60.8\pm\phantom{00}6.8$ & $\times$ \\
    PGC006240 & $7877\phantom{.0}\pm1\phantom{.0}$ & $345.8\pm\phantom{0}4.0$ & $\phantom{0}367.1\pm\phantom{0}3.9$ & $63\pm\phantom{0}4$ & $28\pm\phantom{0}6$ & $-$ & $737\phantom{.0}\pm150\phantom{.0}$ & I A \\
    PGC043387 & $6704.0\pm0.3$ & $500.5\pm\phantom{0}0.7$ & $\phantom{00}24.3\pm\phantom{0}0.3$ & $47\pm\phantom{0}5$ & $47\pm\phantom{0}5$ & $-$ & $684\phantom{.0}\pm\phantom{0}56\phantom{.0}$ & I A B \\
    UGC1214 & $5066\phantom{.0}\pm3\phantom{.0}$ & $225.2\pm\phantom{0}8.3$ & $\phantom{00}29.3\pm\phantom{0}0.3$ & $25\pm10$ & $18\pm\phantom{0}3$ & $-$ & $729\phantom{.0}\pm120\phantom{.0}$ & I \\
    UGC1395$^\dagger$ & $5062.2\pm0.7$ & $200.0\pm\phantom{0}1.7$ & $\phantom{00}46.5 \pm\phantom{0}0.8$ & $55\pm\phantom{0}3$ & $51\pm\phantom{0}5$ & $-$ & $257\phantom{.0}\pm\phantom{0}18\phantom{.0}$ & I A B \\
    \hline
    \end{tabular}
\end{threeparttable}
\end{table*}


\section{Tully-Fisher Relation Data}
\label{section:TFR data}


\subsection{Velocity widths}
\label{subsection:velocity widths}

To characterise the rotational velocity of each CO-detected sample galaxy, we adopt the width of the integrated CO spectrum at $50\%$ of its maximum ($W_{50}$). As in previous studies, we fit the integrated spectrum and derive $W_{50}$ from the fit rather than measure it directly from the data, as done in the previous Section \citep[e.g.][]{Tiley_2016, Topal_2018}. We use one of two functions to fit each spectrum. When the integrated spectrum has the typical double-horned shape expected of a rotating disc, we use a Gaussian Double Peak function (i.e.\ a central parabola flanked by two symmetric half-Gaussians equidistant from a central velocity; Eq.~A2 of \citealt{Tiley_2016}) 
\begin{equation}
    f(v) = 
    \begin{cases}
        A_\text{G}\times \exp\Big(\frac{-[v-(v_0-w)]^2}{2\sigma^2}\Big)\quad\quad v<v_0-w \\
        A_\text{C} + a(v-v_0)^2\quad\quad\quad\quad v_0-w\leq v\leq v_0+w, \\
        A_\text{G}\times \exp\Big(\frac{-[v-(v_0+w)]^2}{2\sigma^2}\Big)\quad\quad v>v_0+w \\
    \end{cases}
\end{equation}
where $v_0$ is the central velocity, $w$ is the half velocity width of the central parabola, $\sigma$ is the velocity width of the Gaussian edges, $A_\text{G}$ is the amplitude of the two half-Gaussians, $A_\text{C}$ is the amplitude of the central parabola, and $a=(A_\text{G}-A_\text{C})/w^2$.
This has been shown to be the most accurate fitting function, minimising potential biases as a function of the amplitude-to-noise ratio of the spectrum, the galaxy inclination and the galaxy rotation velocity \citep{Tiley_2016}. In cases where the spectrum is not double-horned, we use a single Gaussian function. Spectra of that shape are typically observed when a galaxy has an incomplete or spatially limited CO distribution, the inclination of the galaxy is low and/or the galaxy is of a particularly late morphological type (with a solid-body rather than flat rotation curve). 

To fit the functions to the integrated spectra, we adopt the Levenberg-Marquardt least-squares minimisation algorithm, using the \textsc{mpfit} code \citep{Markwardt_2009} implemented as a Python routine in the \texttt{astrolibpy}\footnote{\url{https://github.com/segasai/astrolibpy/tree/master}} library. The best-fitting parameters are defined as those that minimise the reduced $\chi^2$ 
\begin{equation}
    \chi^2_\text{red}\equiv\frac{1}{\text{DOF}}\sum_i\left(\frac{[f(v_i)-F(v_i)]^2}{\sigma^2_{i,\text{RMS}}}\right)\quad,
\end{equation}
where $f(v_i)$ and $F(v_i)$ are the modelled and the observed flux density in the $i^\text{th}$ velocity bin $v_i$, respectively, DOF the number of degrees of freedom of the fit and $\sigma_{i,\text{RMS}}$ the RMS noise of the spectrum in the $i^\text{th}$ velocity bin. In previous studies which primarily used single-dish data, $\sigma_\text{RMS}$ was taken as a constant, defined as the RMS noise in the line-free channels of the integrated spectrum. However, for interferometric data using masked data cubes, it is more appropriate to consider the (varying) RMS of each velocity bin, which we calculated in Section~\ref{subsection:CO data}. For each sample galaxy, the best fit overlaid on the integrated spectrum is shown in Appendix~\ref{section:Appendix A}, alongside the moment maps and PVD of the galaxy.  

To avoid converging to local minima, we run the \textsc{mpfit} code several times with a variety of initial guesses. For certain integrated spectra, the central parabola of the best-fitting Gaussian Double Peak function has a larger amplitude than that of the half-Gaussians. We allow these profiles as long as $2A_\text{C}/3\leq A_\text{G}$. If this condition is not met we choose to fit the spectrum with a single Gaussian function irrespective of $\chi_\text{red}^2$ \citep[see][]{Tiley_2016, Topal_2018}.

The velocity width at $50\%$ of the peak of the Gaussian Double Peak function ($W_{50}$) can be calculated in an analytical manner as
\begin{equation}    W_{50}=2\,(w+\sqrt{2\ln2}\,\sigma)\quad.
\end{equation}
For a single Gaussian function (i.e.\ when $w=0$), this reduces to
\begin{equation}
    W_{50}=2\sqrt{2\ln2}\,\sigma\quad.
\end{equation} 
The best-fitting $W_{50}$ and $v_0$, and their associated uncertainties, are listed in columns~3 and 2 of Table~\ref{tab:CO data}, respectively.

The associated integrated fluxes are obtained by integrating the best-fitting model profiles over velocity and are listed in column~4 of Table~\ref{tab:CO data}, along with their uncertainties. For the Gaussian Double Peak function the integrated flux has the analytical form
\begin{equation}
    S_\text{CO}\Delta v=2w\bigg(\frac{2}{3}A_\text{C}+\frac{1}{3}A_\text{G}\bigg)+\sqrt{2\pi}\,A_\text{G}\sigma\quad,
\end{equation}
whereas for the single Gaussian profile this reduces to the usual analytical form 
\begin{equation}
    S_\text{CO}\Delta v=\sqrt{2\pi}\,A_\text{G}\sigma\quad. 
\end{equation}

The uncertainties of all fitted quantities, and thus $v_0$, $W_{50}$ and $S_\text{CO}\Delta v$, were estimated using a Monte Carlo method and $250$ realisations of each spectrum, in which the best-fitting model was perturbed by adding Gaussian noise with a standard deviation equal to the per-channel uncertainties. Each perturbed model was then refit and the standard deviation of each fitted quantity distribution taken as its uncertainty. These Monte Carlo uncertainties quantify the effect of statistical noise on the fitted parameters, but they do not capture possible systematic effects associated with the specific model adopted. For example, asymmetric profiles or residual baseline artefacts could bias $W_{50}$ and $v_0$ in manners that are not reflected in the quoted statistical uncertainties. We note, however, that only a small fraction of the galaxies in our sample exhibit strongly asymmetric profiles. For this small number of galaxies, we performed exploratory fits in which the two Gaussian peaks were permitted to have different amplitudes, but the derived $W_{50}$ were broadly consistent with those obtained from our adopted symmetric model. Furthermore, \citet{Tiley_2016} compared several alternative profile parameterisations, and found that the Gaussian Double Peak profile generally yields the most reliable line widths.


\subsection{Photometry}
\label{subsection:photometry}

In this paper we use two photometric measures as proxies of the luminous masses of the galaxies. First, as done in many previous TFR studies, we use the total absolute $K$-band ($2.15$~$\mu$m) magnitude as a proxy of the bulk of the stellar mass. In particular, we use the \texttt{k\_m\_ext} quantity from the 2MASS Extended Source Catalog (XSC), which is the integrated apparent $K_\text{s}$-band magnitude measured within a Kron aperture with an extrapolated surface brightness profile, ensuring the recovery of flux lost as a result of background noise. Despite the near-infrared wavelength, these measurements are nevertheless subject to dust extinction. As such, we calculate the extinction-corrected total absolute $K_\text{s}$-band magnitude as
\begin{equation}
    M_{K_\text{s}}=m_{K_\text{s}}-A_{K_\text{s}}-\mu\quad, 
\label{eq1}
\end{equation}
where $m_{K_\text{s}}$ is the aforementioned integrated apparent $K_\text{s}$-band magnitude, $A_{K_\text{s}}$ the extinction in the 2MASS $K_\text{s}$-band calculated from the dust maps of \citet{Schlafly_Finkbeiner_2011} taken from the NASA/IPAC Infrared Science Archive\footnote{\url{https://irsa.ipac.caltech.edu/applications/DUST/}} and $\mu$ is the distance modulus taken as the \texttt{modbest} parameter from HyperLEDA\footnote{\url{http://atlas.obs-hp.fr/hyperleda/}}.

Second, we use the \textit{WISE} Band~1 ($3.4$~$\mu$m) magnitudes ($W1$) as another probe of stellar mass. We consider both the $K_\text{s}$ and the $W1$ bands, as they are both commonly adopted in TFR studies. Using both bands allows direct comparisons with the literature, where results are based on either band. Specifically, we adopt the \texttt{w1gmag} quantity from the AllWISE Source Catalog\footnote{\url{http://wise2.ipac.caltech.edu/docs/release/allwise/}}, which is the $W1$ apparent magnitude $m_{W1}$ measured within an elliptical aperture with a size, shape and orientation based on that reported in the 2MASS XSC. We calculate the extinction-corrected total absolute $W1$ magnitudes as in Eq.~\ref{eq1}, with the extinctions $A_{W1}$ also taken from the NASA/IPAC Infrared Science Archive and the same distance moduli. 

The $K_\text{s}$- and $W1$-band apparent magnitudes, as well as the extinctions and the distance moduli, were available for $80$ of the $81$ CO-detected galaxies in our sample, with the exception of NGC0404 which is a nearby low-luminosity dwarf galaxy, and are listed in Table~\ref{tab:photometry}. As such, we remove NGC0404 from subsequent analyses. The $K_\text{s}$ and $W1$ bands are known to track similar stellar populations and are strongly correlated with one another \citep[e.g.][]{Jarrett_2013, Cluver_2014}. We show this correlation in Fig.~\ref{fig:band correlation}, demonstrating that there is little systematic difference between the two measures. The Spearman correlation coefficient is $r_\text{s}=0.97$, indicating a highly significant correlation ($p\ll0.001$, where $p$ is the probability of obtaining such a correlation by chance under the null hypothesis of no correlation), in line with Topal et al.\ (in prep.). The average $K_\text{s}-W1$ colour of our CO-detected sample galaxies (i.e.\ the average difference between the integrated $K_\text{s}$- and $W1$-band apparent magnitudes) is $0.07\pm0.03$ mag (where the stated uncertainty represents the standard error on the mean), with a scatter of $\approx0.3$~mag. While a colour of zero is consistent with this result within $3\sigma$, the measured mean colour suggests the possibility of a small systematic difference between the two bands, further motivating the use of both for direct comparisons with the literature.

\begin{figure}
    \centering
    \includegraphics[width=\linewidth]{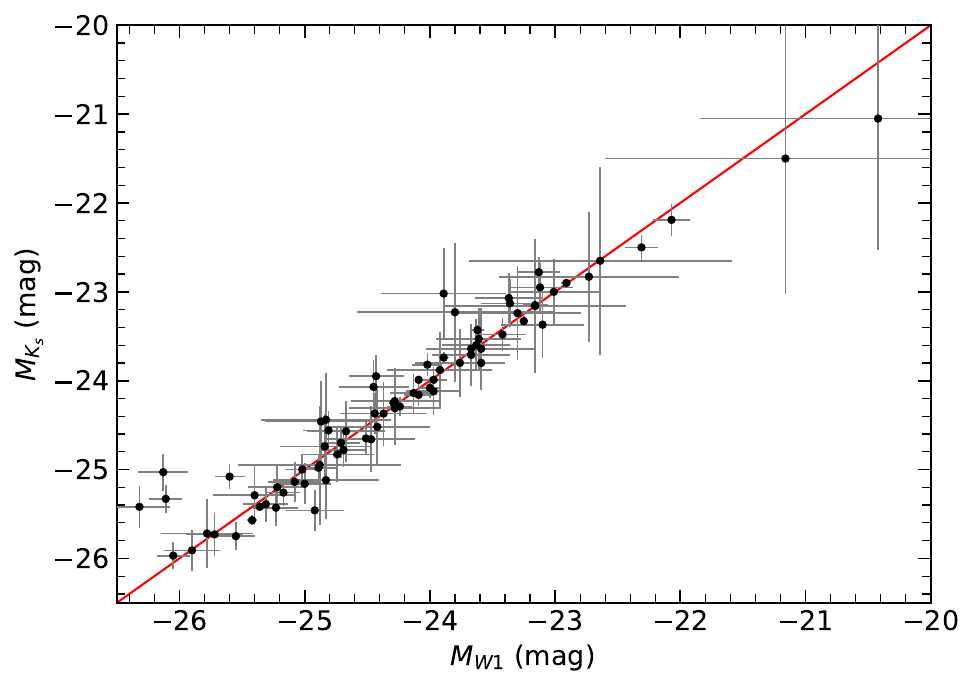}
    \caption{Correlation between the total absolute $W1$- and $K_\text{s}$-band magnitudes of our sample of galaxies. The solid red line shows the $1:1$ relation.}
    \label{fig:band correlation}
\end{figure}

\begin{table*}
\begin{threeparttable}
    \centering
	\caption{Photometric data of all CO-detected galaxies.}
	\label{tab:photometry}
	\begin{tabular}{l@{\hskip4pt}c@{\hskip8pt}c@{\hskip8pt}r@{\hskip8pt}c@{\hskip8pt}c@{\hskip6pt}c@{\hskip8pt}c@{\hskip8pt}c@{\hskip8pt}c@{\hskip8pt}c}
    \hline
    Galaxy & $D_{25}$ & $b/a$ & PA & $m_{W1}$ & $A_{W1}$ & $m_{K_\text{s}}$ &  $A_{K_\text{s}}$ & $\mu$ & $M_{W1}$ & $M_{K_\text{s}}$ \\
    & (arcsec) &  & (deg) & (mag) & (mag) & (mag) & (mag) & (mag) & (mag) & (mag) \\
    (1) & (2) & (3) & (4) & (5) & (6) & (7) & (8) & (9) & (10) & (11) \\
    \hline
    ESO383-G035 & $\phantom{0}61\pm\phantom{0}7$ & $0.58\pm0.04$ & $116$ & $8.702\pm0.005$ & $0.011$ & $9.582\pm0.017$ & $0.019$ & $32.58\pm0.49$ & $-23.89\pm0.50$ & $-23.02\pm0.51$ \\ 
    FRL49 & $\phantom{0}29^\dagger\phantom{\pm00}$ & $0.80^\dagger \phantom{\pm0.00}$ & $160$ & $8.534\pm0.005$ & $0.012$ & $9.644\pm0.015$ & $0.020$ & $34.65\pm0.19$ & $-26.13\pm0.20$ & $-25.03\pm0.21$ \\
    FRL1146 & $\phantom{0}39\pm\phantom{0}5$ & $0.52\pm0.04$ & $17$ & $9.619\pm0.006$ & $0.076$ & $10.449\pm0.039\phantom{0}$ & $0.125$ & $35.65\pm0.12$ & $-26.11\pm0.13$ & $-25.33\pm0.16$ \\
    IC0843 & $\phantom{0}64\pm15$ & $0.49\pm0.06$ & $135$ & $10.312\pm0.006\phantom{0}$ & $0.003$ & $10.228\pm0.024\phantom{0}$ & $0.005$ & $35.20\pm0.14$ & $-24.89\pm0.15$ & $-24.98\pm0.16$ \\
    IC1481 & $\phantom{0}43\pm12$ & $0.89\pm0.12$ & $45$ & $10.627\pm0.006\phantom{0}$ & $0.015$ & $10.623\pm0.037\phantom{0}$ & $0.024$ & $34.74\pm0.18$ & $-24.13\pm0.19$ & $-24.14\pm0.22$ \\
    IC4329A & $\phantom{0}95\pm\phantom{0}7$ & $0.56\pm0.04$ & $46$ & $7.898\pm0.005$ & $0.011$ & $8.805\pm0.013$ & $0.018$ & $34.21\pm0.23$ & $-26.32\pm0.24$ & $-25.42\pm0.24$ \\
    MRK0567 & $\phantom{0}25^\dagger\phantom{\pm00}$ & $0.90^\dagger\phantom{\pm0.00}$ & $70$ & $10.148\pm0.006\phantom{0}$ & $0.005$ & $10.664\pm0.029\phantom{0}$ & $0.008$ & $35.74\pm0.11$ & $-25.60\pm0.12$ & $-25.08\pm0.14$\\
    NGC0017 & $\phantom{0}69\pm10$ & $0.34\pm0.02$ & $32$ & $9.825\pm0.006$ & $0.005$ & $10.074\pm0.030\phantom{0}$ & $0.008$ & $34.63\pm0.19$ & $-24.81\pm0.20$ & $-24.56\pm0.22$ \\
    NGC0289 & $203\pm\phantom{0}9$ & $0.74\pm0.03$ & $132$ & $7.960\pm0.006$ & $0.004$ & $7.997\pm0.027$ & $0.006$ & $31.59\pm0.26$ & $-23.63\pm0.27$ & $-23.60\pm0.29$  \\
    NGC0383 & $144\pm10$ & $0.83\pm0.06$ & $8$ & $8.486\pm0.007$ & $0.014$ & $8.481\pm0.023$ & $0.022$ & $34.37\pm0.21$ & $-25.90\pm0.22$ & $-25.91\pm0.23$ \\
    NGC0404 & $203\pm14$ & $1.00_{-0.12}^{+0.00}\phantom{00}$ & $-$ & $-$ & $-$ & $-$ & $-$ & $-$ & $-$ & $-$ \\
    NGC0449 & $\phantom{0}41\pm\phantom{0}5$ & $0.56\pm0.04$ & $78$ & $10.929\pm0.006\phantom{0}$ & $0.006$ & $11.173\pm0.052\phantom{0}$ & $0.018$ & $34.28\pm0.22$ & $-23.36\pm0.23$ & $-23.13\pm0.27$ \\
    NGC0524 & $203\pm19$ & $1.00_{-0.09}^{+0.00}\phantom{00}$ & $-$ & $7.215\pm0.006$ & $0.016$ & $7.163\pm0.012$ & $0.026$ & $32.56\pm0.03$ & $-25.36\pm0.04$ & $-25.42\pm0.04$ \\
    NGC0541 & $131\pm\phantom{0}9$ & $0.69\pm0.03$ & $69$ & $9.453\pm0.006$ & $0.009$ & $9.295\pm0.031$ & $0.014$ & $34.44\pm0.20$ & $-25.00\pm0.21$ & $-25.16\pm0.23$ \\
    NGC0612 & $\phantom{0}89\pm\phantom{0}6$ & $0.89\pm0.06$ & $160$ & $9.492\pm0.006$ & $0.004$ & $9.575\pm0.028$ & $0.006$ & $35.54\pm0.12$ & $-26.05\pm0.13$ & $-25.97\pm0.15$ \\
    NGC0708 & $154\pm14$ & $0.48\pm0.03$ & $38$ & $8.859\pm0.008$ & $0.017$ & $8.574\pm0.026$ & $0.028$ & $33.67\pm0.41$ & $-24.83\pm0.42$ & $-25.12\pm0.44$ \\
    NGC0863 & $\phantom{0}76\pm\phantom{0}7$ & $0.91\pm0.06$ & $176$ & $9.738\pm0.006$ & $0.007$ & $9.540\pm0.024$ & $0.011$ & $35.28\pm0.14$ & $-25.55\pm0.15$ & $-25.75\pm0.16$ \\
    NGC1084 & $203\pm14$ & $0.66\pm0.03$ & $40$ & $7.628\pm0.006$ & $0.005$ & $7.930\pm0.023$ & $0.008$ & $30.99\pm0.26$ & $-23.37\pm0.27$ & $-23.07\pm0.28$ \\
    NGC1194 & $\phantom{0}93\pm\phantom{0}9$ & $0.50\pm0.03$ & $139$ & $9.361\pm0.005$ & $0.014$ & $9.758\pm0.029$ & $0.024$ & $33.80\pm0.27$ & $-24.45\pm0.28$ & $-24.07\pm0.30$ \\
    NGC1300 & $353\pm24$ & $0.51\pm0.04$ & $105$ & $7.831\pm0.008$ & $0.006$ & $7.564\pm0.047$ & $0.009$ & $30.92\pm0.32$ & $-23.10\pm0.33$ & $-23.37\pm0.37$ \\
    NGC1358 & $128\pm\phantom{0}9$ & $0.55\pm0.04$ & $15$ & $9.033\pm0.006$ & $0.011$ & $8.948\pm0.032$ & $0.018$ & $33.76\pm0.28$ & $-24.74\pm0.29$ & $-24.83\pm0.31$\\
    NGC1387 & $169\pm13$ & $1.00_{-0.07}^{+0.00}\phantom{00}$ & $-$ & $7.446\pm0.006$ & $0.002$ & $7.425\pm0.018$ & $0.004$ & $31.41\pm0.10$ & $-23.97\pm0.11$ & $-23.99\pm0.12$ \\
    NGC1400 & $169\pm\phantom{0}8$ & $0.89\pm0.04$ & $45$ & $7.855\pm0.006$ & $0.012$ & $7.811\pm0.018$ & $0.020$ & $32.08\pm0.09$ & $-24.24\pm0.10$ & $-24.29\pm0.11$ \\ 
    NGC1448 & $477\pm22$ & $0.19\pm0.01$ & $42$ & $7.517\pm0.005$ & $0.003$ & $7.660\pm0.026$ & $0.004$ & $31.40\pm0.03$ & $-23.89\pm0.04$ & $-23.74\pm0.06$\\ 
    NGC1493 & $208\pm10$ & $0.93\pm0.04$ & $93$ & $9.142\pm0.008$ & $0.002$ & $8.806\pm0.090$ & $0.003$ & $30.30\pm1.43$ & $-21.16\pm1.44$ & $-21.50\pm1.52$ \\
    NGC1497 & $\phantom{0}74\pm\phantom{0}7$ & $0.52\pm0.04$ & $63$ & $9.535\pm0.006$ & $0.041$ & $9.475\pm0.028$ & $0.068$ & $34.80\pm0.17$ & $-25.31\pm0.18$ & $-25.39\pm0.20$ \\ 
    NGC1574 & $237\pm11$ & $0.92\pm0.04$ & $14$ & $7.033\pm0.006$ & $0.003$ & $7.109\pm0.023$ & $0.005$ & $31.47\pm0.21$ & $-24.44\pm0.22$ & $-24.37\pm0.23$ \\
    NGC2110 & $109\pm10$ & $0.92\pm0.04$ & $175$ & $7.711\pm0.006$ & $0.071$ & $8.144\pm0.019$ & $0.116$ & $32.47\pm0.51$ & $-24.83\pm0.52$ & $-24.44\pm0.53$ \\
    NGC2179 & $120\pm\phantom{0}8$ & $0.72\pm0.03$ & $165$ & $9.267\pm0.006$ & $0.012$ & $9.224\pm0.027$ & $0.019$ & $32.84\pm0.43$ & $-23.59\pm0.44$ & $-23.64\pm0.46$\\
    NGC2187A & $213\pm20$ & $0.48\pm0.06$ & $75$ & $9.984\pm0.006$ & $0.018$ & $9.961\pm0.051$ & $0.029$ & $33.64\pm0.30$ & $-23.67\pm0.31$ & $-23.71\pm0.35$\\ 
    NGC2903 & $721\pm33$ & $0.92\pm0.03$ & $22$ & $5.835\pm0.006$ & $0.006$ & $6.036\pm0.018$ & $0.010$ & $29.85\pm0.11$ & $-24.02\pm0.12$ & $-23.82\pm0.13$ \\
    NGC2911 & $208\pm14$ & $0.62\pm0.04$ & $135$ & $8.897\pm0.009$ & $0.006$ & $8.711\pm0.033$ & $0.010$ & $33.36\pm0.34$ & $-24.47\pm0.35$ & $-24.66\pm0.37$ \\
    NGC2964 & $177\pm12$ & $0.74\pm0.05$ & $97$ & $7.781\pm0.005$ & $0.004$ & $8.355\pm0.014$ & $0.006$ & $31.58\pm0.77$ & $-23.80\pm0.78$ & $-23.23\pm0.78$ \\
    NGC2974 & $213\pm10$ & $0.62\pm0.06$ & $44$ & $6.788\pm0.005$ & $0.010$ & $6.255\pm0.005$ & $0.017$ & $31.70\pm0.22$ & $-24.92\pm0.23$ & $-25.46\pm0.23$ \\
    NGC3021 & $\phantom{0}81\pm13$ & $0.59\pm0.07$ & $110$ & $8.987\pm0.006$ & $0.003$ & $9.178\pm0.015$ & $0.004$ & $32.60\pm0.04$ & $-23.62\pm0.05$ & $-23.43\pm0.06$ \\
    NGC3156 & $114\pm11$ & $0.54\pm0.06$ & $50$ & $9.663\pm0.006$ & $0.007$ & $9.548\pm0.037$ & $0.011$ & $31.73\pm0.14$ & $-22.07\pm0.15$ & $-22.19\pm0.18$ \\ 
    NGC3169 & $262\pm24$ & $0.62\pm0.07$ & $49$ & $7.235\pm0.006$ & $0.006$ & $7.283\pm0.021$ & $0.010$ & $31.52\pm0.03$ & $-24.29\pm0.04$ & $-24.25\pm0.05$ \\
    NGC3277 & $128\pm\phantom{0}9$ & $0.76\pm0.05$ & $29$ & $9.023\pm0.006$ & $0.005$ & $8.926\pm0.017$ & $0.008$ & $31.75\pm0.71$ & $-22.73\pm0.72$ & $-22.83\pm0.73$ \\
    NGC3351 & $435\pm20$ & $0.62\pm0.06$ & $11$ & $6.742\pm0.008$ & $0.005$ & $6.665\pm0.036$ & $0.009$ & $29.99\pm0.02$ & $-23.25\pm0.03$ & $-23.33\pm0.06$ \\
    NGC3368 & $499\pm35$ & $0.66\pm0.06$ & $5$ & $6.286\pm0.008$ & $0.005$ & $6.320\pm0.023$ & $0.008$ & $29.95\pm0.02$ & $-23.67\pm0.03$ & $-23.64\pm0.04$ \\
    NGC3393 & $117\pm\phantom{0}8$ & $0.87\pm0.06$ & $13$ & $8.955\pm0.006$ & $0.014$ & $9.059\pm0.039$ & $0.023$ & $33.61\pm0.30$ & $-24.67\pm0.31$ & $-24.57\pm0.34$ \\
    NGC3607 & $274\pm25$ & $0.87\pm0.10$ & $120$ & $7.082\pm0.006$ & $0.004$ & $6.994\pm0.023$ & $0.006$ & $31.77\pm0.17$ & $-24.69\pm0.18$ & $-24.78\pm0.19$ \\
    NGC3810 & $199\pm14$ & $0.68\pm0.03$ & $29$ & $7.777\pm0.010$ & $0.008$ & $7.952\pm0.028$ & $0.014$ & $30.89\pm0.25$ & $-23.12\pm0.26$ & $-22.95\pm0.28$ \\
    NGC3862 & $\phantom{0}66\pm17$ & $1.00_{-0.01}^{+0.00}\phantom{00}$ & $-$ & $9.680\pm0.007$ & $0.004$ & $9.489\pm0.035$ & $0.007$ & $34.91\pm0.17$ & $-25.23\pm0.18$ & $-25.43\pm0.21$ \\
    NGC4030 & $228\pm11$ & $0.71\pm0.03$ & $21$ & $7.221\pm0.007$ & $0.005$ & $7.330\pm0.015$ & $0.008$ & $32.06\pm0.35$ & $-24.84\pm0.36$ & $-24.74\pm0.37$ \\
    NGC4061 & $\phantom{0}55\pm\phantom{0}8$ & $0.85\pm0.06$ & $1$ & $9.907\pm0.007$ & $0.007$ & $9.763\pm0.027$ & $0.011$ & $35.32\pm0.03$ & $-25.42\pm0.04$ & $-25.57\pm0.06$ \\
    NGC4203 & $203\pm\phantom{0}9$ & $0.92\pm0.06$ & $10$ & $7.463\pm0.008$ & $0.002$ & $7.406\pm0.014$ & $0.004$ & $30.88\pm0.18$ & $-23.42\pm0.19$ & $-23.48\pm0.19$ \\
    NGC4212 & $169\pm12$ & $0.60\pm0.04$ & $74$ & $8.021\pm0.006$ & $0.006$ & $8.375\pm0.010$ & $0.010$ & $31.14\pm0.16$ & $-23.13\pm0.17$ & $-22.78\pm0.17$ \\
    NGC4235 & $262\pm18$ & $0.20\pm0.02$ & $49$ & $8.397\pm0.005$ & $0.004$ & $8.396\pm0.028$ & $0.006$ & $31.55\pm0.72$ & $-23.16\pm0.73$ & $-23.16\pm0.75$ \\ 
    NGC4245 & $154\pm11$ & $0.63\pm0.04$ & $146$ & $8.312\pm0.006$ & $0.004$ & $8.311\pm0.019$ & $0.006$ & $30.95\pm1.04$ & $-22.64\pm1.05$ & $-22.65\pm1.06$ \\
    NGC4261 & $256\pm12$ & $0.79\pm0.04$ & $160$ & $7.350\pm0.007$ & $0.003$ & $7.263\pm0.028$ & $0.005$ & $32.52\pm0.12$ & $-25.17\pm0.13$ & $-25.26\pm0.15$ \\
    NGC4274 & $218\pm15$ & $0.47\pm0.03$ & $102$ & $7.021\pm0.006$ & $0.004$ & $7.032\pm0.020$ & $0.007$ & $31.39\pm0.34$ & $-24.37\pm0.35$ & $-24.37\pm0.36$ \\
    NGC4278 & $173\pm16$ & $0.98_{-0.05}^{+0.02}\phantom{00}$ & $23$ & $7.245\pm0.006$ & $0.006$ & $7.184\pm0.011$ & $0.009$ & $31.33\pm0.11$ & $-24.09\pm0.12$ & $-24.16\pm0.12$ \\
    \hline
    \end{tabular}
    \begin{tablenotes}[flushleft]
          \item \textit{Notes.} The data in this table are available in a  machine-readable format online. (1) Galaxy name. (2) Galaxy diameter at the $25$~mag~arcsec$^{-2}$ isophote in the $B$-band from HyperLeda. $\dagger$ indicates galaxies with no measurement, in which case the diameter at the $20$~mag~arcsec$^{-2}$ isophote in the $K_\text{s}$-band from 2MASS is listed instead (quantity \texttt{r\_k20fe} of the 2MASS XSC). (3) Galaxy axis ratio from HyperLEDA. $^\dagger$ indicates galaxies for which the axis ratio of the $3\sigma$ $K_\text{s}$-band isophote is listed instead (quantity \texttt{k\_ba} of the 2MASS XSC). (4) Position angle of the galaxy major axis at the $25$~mag~arcsec$^{-2}$ isophote in the $B$-band from HyperLEDA (no uncertainty available). (5) Galaxy total apparent $W1$-band Vega magnitude from the AllWISE Source Catalog. (6) $W1$-band extinction from the NASA/IPAC Infrared Science Archive. (7) Galaxy total apparent $K_\text{s}$-band Vega magnitude from the 2MASS XSC. (8) $K_\text{s}$-band extinction from the NASA/IPAC Infrared Science Archive. (9) Galaxy distance modulus from HyperLEDA (\texttt{modbest} parameter). (10) Galaxy extinction-corrected total absolute $W1$-band Vega magnitude. (11) Galaxy extinction-corrected total absolute $K_\text{s}$-band Vega magnitude.
    \end{tablenotes}
\end{threeparttable}
\end{table*}

\begin{table*}
\begin{threeparttable}
    \centering
	\ContinuedFloat
	\caption[]{Continued.}
	\begin{tabular}{l@{\hskip12pt}c@{\hskip8pt}c@{\hskip8pt}r@{\hskip8pt}c@{\hskip8pt}c@{\hskip8pt}c@{\hskip8pt}c@{\hskip8pt}c@{\hskip8pt}c@{\hskip8pt}c}
    \hline
    Galaxy & $D_{25}$ & $b/a$ & PA & $m_{W1}$ & $A_{W1}$ & $m_{K_\text{s}}$ &  $A_{K_\text{s}}$ & $\mu$ & $M_{W1}$ & $M_{K_\text{s}}$ \\
    &  &  & (deg) & (mag) & (mag) & (mag) & (mag) & (mag) & (mag) & (mag) \\
    (1) & (2) & (3) & (4) & (5) & (6) & (7) & (8) & (9) & (10) & (11) \\
    \hline
    NGC4429 & $315\pm15$ & $0.46\pm0.04$ & $98$ & $6.738\pm0.006$ & $0.006$ & $6.783\pm0.023$ & $0.010$ & $30.65\pm0.41$ & $-23.92\pm0.42$ & $-23.88\pm0.43$ \\
    NGC4435 & $181\pm\phantom{0}8$ & $0.71\pm0.05$ & $12$ & $7.506\pm0.006$ & $0.006$ & $7.297\pm0.016$ & $0.009$ & $31.09\pm0.04$ & $-23.59\pm0.05$ & $-23.80\pm0.06$ \\
    NGC4438 & $547\pm25$ & $0.44\pm0.05$ & $27$ & $7.258\pm0.006$ & $0.005$ & $7.265\pm0.023$ & $0.009$ & $30.26\pm0.35$ & $-23.01\pm0.36$ & $-23.00\pm0.37$ \\
    NGC4501 & $523\pm36$ & $0.50\pm0.05$ & $138$ & $6.160\pm0.008$ & $0.007$ & $6.267\pm0.017$ & $0.012$ & $30.24\pm0.02$ & $-24.09\pm0.03$ & $-23.99\pm0.04$ \\
    NGC4569 & $547\pm25$ & $0.42\pm0.05$ & $22$ & $6.615\pm0.008$ & $0.009$ & $6.581\pm0.027$ & $0.015$ & $30.37\pm0.35$ & $-23.76\pm0.36$ & $-23.80\pm0.38$ \\
    NGC4697 & $425\pm29$ & $0.59\pm0.04$ & $79$ & $6.442\pm0.008$ & $0.006$ & $6.367\pm0.027$ & $0.009$ & $30.44\pm0.08$ & $-24.00\pm0.09$ & $-24.08\pm0.11$ \\
    NGC4826 & $628\pm29$ & $0.51\pm0.04$ & $114$ & $5.317\pm0.017$ & $0.008$ & $5.330\pm0.017$ & $0.013$ & $28.22\pm0.02$ & $-22.91\pm0.04$ & $-22.90\pm0.04$ \\
    NGC4845 & $330\pm15$ & $0.22\pm0.01$ & $82$ & $7.717\pm0.005$ & $0.004$ & $7.794\pm0.023$ & $0.006$ & $31.32\pm0.33$ & $-23.61\pm0.34$ & $-23.53\pm0.35$ \\
    NGC5064 & $161\pm\phantom{0}7$ & $0.52\pm0.02$ & $37$ & $7.967\pm0.006$ & $0.030$ & $8.014\pm0.018$ & $0.049$ & $33.16\pm0.22$ & $-25.22\pm0.23$ & $-25.20\pm0.24$ \\
    NGC5084 & $600\pm28$ & $0.25\pm0.02$ & $80$ & $7.110\pm0.006$ & $0.022$ & $7.058\pm0.027$ & $0.036$ & $31.97\pm0.64$ & $-24.88\pm0.65$ & $-24.95\pm0.67$ \\
    NGC5248 & $244\pm17$ & $0.59\pm0.05$ & $120$ & $7.190\pm0.006$ & $0.005$ & $7.253\pm0.033$ & $0.008$ & $30.48\pm0.50$ & $-23.30\pm0.51$ & $-23.24\pm0.53$ \\
    NGC5273 & $137\pm10$ & $0.66\pm0.05$ & $6$ & $8.850\pm0.007$ & $0.002$ & $8.665\pm0.024$ & $0.003$ & $31.16\pm0.12$ & $-22.31\pm0.13$ & $-22.50\pm0.14$ \\ 
    NGC5427 & $218\pm15$ & $0.91\pm0.08$ & $11$ & $8.614\pm0.008$ & $0.005$ & $8.590\pm0.046$ & $0.009$ & $32.89\pm0.36$ & $-24.28\pm0.37$ & $-24.31\pm0.41$ \\
    NGC5746 & $435\pm20$ & $0.15\pm0.01$ & $172$ & $6.773\pm0.005$ & $0.008$ & $6.878\pm0.017$ & $0.012$ & $32.16\pm0.32$ & $-25.40\pm0.33$ & $-25.29\pm0.34$ \\
    NGC5765b & $\phantom{0}49\pm14$ & $0.68\pm0.09$ & $144$ & $10.429\pm0.006\phantom{0}$ & $0.007$ & $10.455\pm0.039\phantom{0}$ & $0.012$ & $35.44\pm0.13$ & $-25.02\pm0.14$ & $-25.00\pm0.17$ \\
    NGC5806 & $181\pm\phantom{0}8$ & $0.54\pm0.02$ & $172$ & $8.429\pm0.006$ & $0.010$ & $8.449\pm0.022$ & $0.016$ & $31.58\pm0.09$ & $-23.16\pm0.10$ & $-23.15\pm0.11$ \\
    NGC5995 & $\phantom{0}61\pm\phantom{0}7$ & $0.78\pm0.04$ & $147$ & $8.495\pm0.005$ & $0.030$ & $9.393\pm0.019$ & $0.049$ & $35.24\pm0.14$ & $-26.78\pm0.15$ & $-25.90\pm0.16$ \\
    NGC6500 & $125\pm17$ & $0.79\pm0.09$ & $45$ & $9.102\pm0.007$ & $0.017$ & $9.154\pm0.029$ & $0.028$ & $33.36\pm0.34$ & $-24.28\pm0.35$ & $-24.23\pm0.37$ \\ 
    NGC6753 & $177\pm12$ & $0.87\pm0.06$ & $27$ & $7.450\pm0.007$ & $0.013$ & $7.523\pm0.025$ & $0.021$ & $33.22\pm0.36$ & $-25.78\pm0.37$ & $-25.72\pm0.39$ \\
    NGC6958 & $147\pm10$ & $0.78\pm0.04$ & $93$ & $8.499\pm0.006$ & $0.009$ & $8.400\pm0.018$ & $0.014$ & $32.91\pm0.41$ & $-24.42\pm0.42$ & $-24.52\pm0.43$ \\
    NGC7049 & $239\pm17$ & $0.68\pm0.05$ & $64$ & $7.299\pm0.007$ & $0.011$ & $7.246\pm0.022$ & $0.017$ & $32.37\pm0.20$ & $-25.08\pm0.21$ & $-25.14\pm0.22$ \\
    NGC7052 & $109\pm10$ & $0.49\pm0.03$ & $66$ & $8.560\pm0.006$ & $0.023$ & $8.574\pm0.020$ & $0.039$ & $34.26\pm0.22$ & $-25.72\pm0.23$ & $-25.73\pm0.24$ \\
    NGC7172 & $151\pm10$ & $0.66\pm0.05$ & $104$ & $7.905\pm0.006$ & $0.005$ & $8.317\pm0.023$ & $0.008$ & $32.77\pm0.44$ & $-24.87\pm0.45$ & $-24.46\pm0.46$ \\
    NGC7424 & $455\pm31$ & $0.85\pm0.04$ & $8$ & $9.885\pm0.009$ & $0.002$ & $9.254\pm0.066$ & $0.003$ & $30.30\pm1.41$ & $-20.42\pm1.42$ & $-21.05\pm1.48$ \\
    PGC006240 & $\phantom{0}66\pm11$ & $0.55\pm0.05$ & $109$ & $10.606\pm0.006\phantom{0}$ & $0.004$ & $10.617\pm0.034\phantom{0}$ & $0.006$ & $35.31\pm0.14$ & $-24.71\pm0.15$ & $-24.70\pm0.17$ \\
    PGC043387 & $\phantom{0}50\pm\phantom{0}6$ & $0.72\pm0.05$ & $150$ & $10.549\pm0.006\phantom{0}$ & $0.002$ & $10.419\pm0.031\phantom{0}$ & $0.004$ & $35.06\pm0.15$ & $-24.51\pm0.16$ & $-24.65\pm0.18$ \\
    UGC1214 & $\phantom{0}63\pm\phantom{0}7$ & $0.91\pm0.06$ & $57$ & $9.900\pm0.006$ & $0.004$ & $10.385\pm0.030\phantom{0}$ & $0.007$ & $34.33\pm0.21$ & $-24.43\pm0.22$ & $-23.95\pm0.24$ \\
    UGC1395 & $\phantom{0}74\pm\phantom{0}7$ & $0.60\pm0.04$ & $159$ & $10.396\pm0.007\phantom{0}$ & $0.014$ & $10.254\pm0.064\phantom{0}$ & $0.022$ & $34.35\pm0.21$ & $-23.97\pm0.22$ & $-24.12\pm0.27$
    \\
    \hline
    \end{tabular}
\end{threeparttable}
\end{table*}


\subsection{Inclination}
\label{subsection:inclination}

For use in the TFR, the velocity widths obtained in Section~\ref{subsection:velocity widths} must be deprojected (i.e.\ corrected for the galaxy inclinations). It is important to use reliable inclination measures, as inclination uncertainties contribute greatly to the deprojected velocity width uncertainties, and reliable inclinations lead to reduced total scatters around the best-fitting TFRs \citep{Davis_2011}. Ideally, we should be using a methodology that directly measures the inclination of the molecular gas. When available, we thus use inclinations obtained through kinematic modelling of the CO data ($i_{\text{CO}}$; column~7 of Table~\ref{tab:CO data}). However, the number of galaxies with existing kinematic modelling is rather small.

When no such modelling is available, we attempt to obtain inclination estimates from ellipse fitting to the dust lanes visible in \textit{HST} images ($i_\text{dust}$; column~6 of Table~\ref{tab:CO data}). This method has been shown to yield reliable inclination measures (e.g.\ \citealt{Davis_2011}, who used the method for ETGs at distances $<40$~Mpc, similar to our WISDOM sample), as the dust is typically spatially coincident with the molecular gas \citep[e.g.][]{Young_2008, Crocker_2008, Bolatto_2013}. Unsharp-masked optical images, preferably in a blue band for which the dust is more prominent, were thus created and ellipses fit by eye. As each dust disc is assumed to be infinitely thin, the inclination is obtained from the usual expression
\begin{equation}
    i_\text{dust}=\cos^{-1}(q)
\end{equation}
\citep{Holmberg_1958}, where $q$ is the minor-to-major axis ratio of the best-fitting ellipse. We were able to derive inclinations for the vast majority of our sample galaxies using this method.

When no dust lane is visible in \textit{HST} images, we estimate the galaxy inclination using the axis ratio of the $25$~mag~arcsec$^{-2}$ $B$-band isophote taken from HyperLEDA. These axis ratios can be converted into inclinations  ($i_{a/b}$) using the standard expression
\begin{equation}
    i_{b/a}=\cos^{-1}\left(\sqrt{\frac{q^2-q_0^2}{1-q_0^2}}\right)
\end{equation}
\citep{Holmberg_1958}, where $q$ is taken from HyperLEDA and $q_0$ is the intrinsic axis ratio (when the galaxy is seen edge-on). As $q_0$ is morphology dependent, we choose $q_0=0.34$ for ETGs and $q_0=0.20$ for late-type galaxies (LTGs), following the methodologies of \citet{Davis_2011} and \cite{Tiley_2016}. The galaxy morphological types were also obtained from HyperLEDA and are listed in Table~\ref{tab:observational CO data}. Inclinations obtained using this method are listed in column~5 of Table~\ref{tab:CO data}. 

We note that in some galaxies the molecular gas distribution appears significantly misaligned with respect to the stellar disc, as indicated by different projected position angles (see e.g.\ NGC1497 and NGC5084). In these cases, the dust lanes appear aligned with the stellar distribution rather than the CO emission, suggesting that even a dust-derived inclination may not accurately represent the inclination of the molecular gas. Such misalignments could therefore introduce systematic uncertainties into the inclination-corrected velocity widths, potentially contributing additional scatter to the TFRs. Quantifying any inclination mismatch and its impact on the TFRs would require detailed kinematic modelling (for the galaxies without existing kinematically derived inclinations), which is beyond the scope of this work, but fortunately differences of projected position angles do not imply equally large differences of inclinations and misaligned objects are rare.


\section{Results}
\label{section:results}

\subsection{Data sub-samples}

Of the $81$ galaxies with detected CO emission, one does not have the necessary photometric data available (NGC0404) and three are clear outliers in the TFRs. All four are excluded from the TFR fits. Two of these outliers (NGC1493 and NGC7424) have unusually narrow Gaussian-shaped integrated profiles, with $\log(W_{50}/\sin i)<1.9$. They lie in a region of parameter space that is poorly sampled by the main dataset and therefore exert disproportionate leverage on the fitted relations without being supported by a statistically representative population.
The third of these outliers (NGC5746) has an unusually small velocity width for its (very bright) absolute magnitude, likely because the ACA primary beam is much smaller than the galaxy and therefore does not probe the flat part of the rotation curve. This source was also identified as highly influential in the regressions, with disproportionate impacts on the inferred scatters. 
We refer to the remaining sample of $77$ CO-detected galaxies as the \textit{initial} sample, as it represents the most complete sample of galaxies in this study. However, we can and do impose additional criteria to create more homogenous samples. We explore these below.

\subsubsection{Integrated spectrum shapes}

When studying the CO TFR, we must ensure that the CO emission of each galaxy reaches a sufficiently large radius to trace the asymptotically flat region of the galaxy's rotation curve. As the rotational velocity serves as a proxy of the total dynamical mass of the galaxy, any underestimation of this rotational velocity will result in systematic biases (and increased scatters) of the TFRs. As can be seen from our moment maps in Appendix~\ref{section:Appendix A}, the CO extents of the galaxies vary greatly in proportion to their optical (i.e.\ stellar) extents. However, the integrated spectra of kinematic tracers exhibiting boxy or double-horned shapes likely extend beyond the turnover radii of the rotation curves, and thus reliably probe the outer parts of the galaxies \citep{Davis_2011}. As such, in addition to analysing the entire initial sample of galaxies, we will also analyse galaxy subsets which omit all the integrated spectra that do not meet this criterion, i.e.\ single Gaussian profiles. This eliminates an additional $4$ galaxies from our ACA sample (IC1481, NGC0289, NGC1448 and UGC1214) and $3$ galaxies from our $12$-m sample (NGC3368, NGC4501 and NGC4569), bringing the total number of CO-detected sample galaxies to $70$. We will refer to this galaxy sub-sample as \textit{sub-sample A}. 

It is worth noting that while such a profile-shape selection reduces the risk of measuring velocities that do not trace the flat parts of the rotation curves, it cannot eliminate the effects of non-circular motions entirely. In some galaxies the CO emission appears to be confined to the stellar bar, where non-circular streaming motions may contribute to the observed velocity widths. In such cases, the measured $W_{50}$ may not provide an accurate estimate of the circular rotation velocities, potentially introducing additional scatter into the TFRs. Furthermore, some galaxies, such as NGC1497, exhibit apparent misalignments between the molecular gas distribution and the stellar disc, suggesting that the gas does not trace the same plane as the stars.


\subsubsection{Inclinations}

Given that the velocities need to be deprojected (i.e.\ corrected for inclination) for TFR work, the uncertainties of the inclinations translate to (additional) uncertainties in the deprojected velocities. In addition to the restriction based on the shape of the velocity profile discussed above, we therefore create another subset by eliminating galaxies whose inclination uncertainties are large, i.e.\ (face-on) galaxies with inclinations $i<30\degree$ \citep[e.g.][]{Pierce_1988,Tiley_2016}. This eliminates $7$ more targets from our ACA sample (NGC0017, NGC0541, NGC0863, NGC1400, NGC5273, NGC5427 and PGC006240) and $5$ more from our $12$-m data sample (MRK0567, NGC0524, NGC1387, NGC1574 and NGC3607), bringing the total number of CO-detected sample galaxies to $58$. We will refer to this galaxy sub-sample, which imposes both the velocity profile shape restriction and the inclination cut-off, as \textit{sub-sample B}.


\subsection{Tully-Fisher relation}
\label{sec:TFR}

The general form of the Tully-Fisher relation we adopt is
\begin{equation}
    M_{W1}=a\left[\log\left(\frac{W_{50}/\sin i}{\text{km~s}^{-1}}\right)-2.6\right]+b\quad, \label{eq:8; main TFR}
\end{equation}
where $a$ is the slope of the best-fitting relation (formally dimensionless but expressed in units of mag~dex$^{-1}$), $b$ the best-fitting zero-point (formally dimensionless but expressed in units of mag) and $2.6$ the pivot point (based on the mean of $\log(W_{50}/\sin i)\approx2.65$; formally dimensionless but expressed in units of dex), chosen to reduce the covariance between the slope and the zero-point. We use an analogous form of the relation when considering the $K_\text{s}$-band data. 

The slope and zero-point of the TFR together define both the scaling and normalisation of the coupling between stellar mass and dynamical mass. The slope sets how strongly luminosity, and hence stellar mass, varies with rotational velocity. This effectively encodes how the stellar component traces the underlying gravitational potential across the galaxy population. The zero-point fixes the normalisation of this relation, corresponding to the luminosity at a given velocity, and therefore reflects the typical stellar mass associated with a given dynamical mass. Consistency of both the slope and zero-point across different samples indicates that they are tracing a common underlying relation between stellar mass and rotational velocity, reflecting a shared scaling between the stellar component and the total gravitational mass probed by the galaxy kinematics.

We fit the TFR with a straight line using the Levenberg-Marquardt least-squares algorithm \textsc{mpfit} \citep{Markwardt_2009}. We first perform a forward fit, which in our case implies that the deprojected velocity is the independent variable and the absolute magnitude is the dependent variable. This fit formally minimises
\begin{equation}
    \chi^2_\text{for}\equiv\sum_j\frac{1}{\sigma_j^2}\left\{M_{W1}-\left(a\left[\log\left(\frac{W_{50}/\sin i}{\text{km~s}^{-1}}\right)-2.6\right]+b\right)\right\}^2\quad,
\end{equation}
where the sum is over all sample galaxies and $\sigma_j$ is the effective uncertainty, defined as
\begin{equation}
    \sigma_j^2\equiv\sigma_{M_{W1}, j}^2+a^2\sigma_{\log(W_{50}/\sin i),j}^2+\sigma_\text{int}^2\quad,
\end{equation}
where $\sigma_{M_{W1},j}$ and $\sigma_{\log(W_{50}/\sin i),j}$ are the uncertainties of the absolute magnitude and the logarithm of the deprojected velocity of a given galaxy, respectively, and $\sigma_\text{int}$ quantifies the intrinsic scatter of the relation.

The absolute magnitude uncertainties are obtained by propagating the uncertainties of the apparent magnitudes and the adopted distance moduli, assuming negligible uncertainty on the Galactic extinction corrections. The uncertainties of $\log(W_{50}/\sin i)$ are obtained by first propagating the uncertainties of the fitted velocity widths and adopted inclinations to the inclination-corrected velocity widths, and then propagating these into logarithmic space using standard first-order error propagation. This yields symmetric $1\sigma$ uncertainties that are used in the TFR fits. It is however worth noting that the uncertainties of $\log(W_{50}/\sin i)$ are not constant across the sample, with galaxies that have larger inclination-corrected velocity widths generally having smaller uncertainties in logarithmic space. In practice, the observational uncertainties do not fully account for the scatter about the TFR, so $\sigma_\text{int}$ is introduced to account for the variance of the relation not captured by measurement uncertainties. The value of $\sigma_\text{int}$ is determined by adjusting it until the reduced $\chi_\text{red,for}^2\equiv\chi^2_\text{for}/\text{DOF}$ equals $1$. By doing so, the effective uncertainties used in the fit are rescaled to match the observed scatter of the data, ensuring that the formal parameter uncertainties returned by the fit are realistic. Without accounting for this intrinsic scatter, the uncertainties on both would be underestimated. The quoted uncertainties on the slope and zero-point are the formal $1\sigma$ uncertainties returned by the \textsc{mpfit} algorithm after the intrinsic scatter has been adjusted such that $\chi^2_\text{red,for}=1$.

Given a fit, we can also calculate the total scatter of the relation as
\begin{equation}
    \sigma_\text{tot}^2=\frac{\chi_\text{for}^2}{\sum_j(1/\sigma_j^2)}\quad.
\end{equation}
This measures the dispersion of the data around the relation and thus quantifies the relations' quality. Unlike the intrinsic scatter, which attempts to quantify the underlying astrophysical scatter after accounting for measurement uncertainties, the total scatter includes both intrinsic variations and observational uncertainties.

As the velocity uncertainties are typically (much) larger than the absolute magnitude uncertainties, slopes determined from forward fits are often biased (shallower than the true slopes) and less accurate \citep[e.g.][]{Willick_1994, Willick_1995}. In such cases, reverse fits are preferable, where the absolute magnitude is the independent variable and the deprojected velocity is the dependent variable. Such fits minimise 
\begin{equation}
    \chi^2_\text{rev}\equiv\sum_j\frac{1}{\zeta^2_j}\left[\log\left(\frac{W_{50}/\sin i}{\text{km~s}^{-1}}\right)-2.6-(A\,M_{W1}+B)\right]^2\quad,
\end{equation}
where $A$ is the slope of the best-fitting relation (formally dimensionless but expressed in units of dex~mag$^{-1}$), $B$ the best-fitting zero-point (formally dimensionless but expressed in units of dex) and $\zeta_j$ the effective uncertainty, defined as
\begin{equation}
    \zeta_j^2\equiv\sigma^2_{\log(W_{50}/\sin i),j}+A^2\sigma^2_{M_{W1},j}+\zeta_\text{int}^2\quad,
\end{equation}
where $\zeta_\text{int}$ is the intrinsic scatter adjusted such that $\chi_\text{red,rev}^2\equiv\chi_\text{rev}^2/\text{DOF}=1$. The total scatter of the relation is then
\begin{equation}
    \zeta_\text{tot}^2=\frac{\chi_\text{rev}^2}{\sum_j(1/\zeta_j^2)}\quad.
\end{equation}
The reverse-fit parameters can be redefined to enable a direct comparison to the forward-fit parameters\footnote{\citet{Williams_2010} carry out an equivalent derivation, but they list the redefined intrinsic scatter and total scatter as $\sigma_\text{int}'\equiv A\,\zeta_\text{int}$ and $\sigma_\text{tot}'\equiv A\,\zeta_\text{tot}$. This appears to be a typographical error (see Eq.~\ref{eq 15} for the correct expressions). Nonetheless, their reported best-fitting scatters appear sensible.}:
\begin{equation}
    a'\equiv\frac{1}{A},\quad b'\equiv-\frac{B}{A},\quad\sigma'_\text{int}\equiv\frac{\zeta_\text{int}}{A}\quad\text{and}\quad\sigma'_\text{tot}\equiv\frac{\zeta_\text{tot}}{A}\quad.
    \label{eq 15}
\end{equation}

The $W1$- and $K_\text{s}$-band TFRs are shown in Fig.~\ref{fig:CO TFR} for the initial sample, sub-sample A and sub-sample B. Both forward and reverse fits are shown, the latter redefined as above to allow direct comparisons. The benchmark relation of \citet{Tully_2000} in the local Universe is also shown, with a zero-point calibrated using known Cepheid calibrator galaxies. It is worth noting that \citet{Tully_2000} use H\,\textsc{i} as their velocity tracer, although it has been shown that slopes calculated using CO are generally consistent within the uncertainties with (albeit sometimes slightly shallower than) slopes calculated using H\,\textsc{i} \citep{Tiley_2016}. The relation of \citet{Tully_2000} was determined in the $K'$-band but is shown on both the $W1$- and $K_\text{s}$-band TFR plots, assuming that the $K'$ and the $W1$ bands, and the $K'$ and the $K_\text{s}$ bands, are equivalent to one another, respectively (i.e.\ assuming a $K'-W1$ and a $K'-K_\text{s}$ colour of $0$, respectively). In addition, the relation of \citet{Tiley_2016} is shown on the $W1$-band plots, whereas the relation of \citet{Topal_2018} is shown on the $K_\text{s}$-band plots, both in their respective band. Comparisons of our fits with literature results are discussed at more length in Section~\ref{section:Discussion}. The best-fitting parameters of all the fits are listed in Table~\ref{tab:results}. When reporting reverse fits, we list the redefined parameters $a'$, $b'$, $\sigma'_\text{int}$ and $\sigma'_\text{tot}$ (rather than the parameters $A$, $B$, $\zeta_\text{int}$ and $\zeta_\text{tot}$).

\begin{table*}
    \begin{threeparttable}
        \centering
        \caption{Best-fitting parameters of the $W1$- and $K_\text{s}$-band CO TFRs.}
        \label{tab:results}
        \begin{tabular}{lllllccc}
             \hline
             Infrared band & Sample & Fit & Slope & Zero-point & $\sigma_\text{int}$ & $\sigma_\text{tot}$ & Offset \\
             & & & (mag~dex$^{-1}$) & (mag) & (mag) & (mag) & (mag) \\
             \hline
             WISE $W1$ & Initial & Forward & $-3.35\pm0.44$ & $-24.19\pm 0.09$ & $0.69$ & $0.76$ & $-$ \\
             & & Reverse & $-7.18\pm0.93$ & $-24.04\pm0.13$ & $1.01$ & $1.10$ & $-$ \\
             & & Fixed (TP00) & $-8.78$ & $-23.97\pm 0.16$ & $1.25$ & $1.35$ & $0.08$ \\
             & & Fixed (Ti16) & $-7.10$ & $-24.04\pm 0.13$ & $0.99$ & $1.09$ & $-0.07\phantom{-}$ \\
             & Sub-sample A & Forward & $-3.73\pm 0.50$ & $-24.14\pm 0.10$ & $0.71$ & $0.78$ & $-$\\
             & & Reverse & $-7.73\pm 1.02$ & $-23.90\pm 0.15$ & $1.02$ & $1.11$ & $-$ \\ 
             & & Fixed (TP00) & $-8.78$ & $-23.84 \pm 0.15$ & $1.15$ & $1.26$ & $0.21$ \\
             & & Fixed (Ti16) & $-7.10$ & $-23.94\pm 0.12$ & $0.93$ & $1.03$ & $0.03$ \\
             & Sub-sample B & Forward & $-3.81\pm 0.57$ & $-24.18\pm 0.11$ & $0.74$ & $0.79$ & $-$ \\
             & & Reverse & $-7.85\pm 1.11$ & $-23.99\pm 0.16$ & $1.05$ & $1.10$ & $-$ \\ 
             & & Fixed (TP00) & $-8.78$ & $-23.94\pm 0.16$ & $1.17$ & $1.23$ & $0.11$ \\
             & & Fixed (Ti16) & $-7.10$ & $-24.02\pm 0.13$ & $0.95$ & $1.01$ & $-0.05\phantom{-}$ \\
             & LTGs & Forward & $-3.97\pm 0.75$ & $-24.34\pm 0.13$ & $0.66$ & $0.74$ & $-$ \\
             & & Reverse & $-8.20\pm 1.50$ & $-24.65\pm 0.21$ & $0.95$ & $1.04$ & $-$ \\
             & & Common slope & $-8.76\pm1.25$ & $-24.69\pm0.21$ & $1.05$ & $1.18$ & $-$ \\
             & & Fixed (Ti16) & $-7.10$ & $-24.57\pm0.15$ & $0.82$ & $0.92$ & $-$ \\
             & ETGs & Forward & $-3.55\pm0.76$ & $-24.05\pm0.17$ & $0.72$ & $0.77$ & $-$ \\
             & & Reverse & $-9.43\pm2.12$ & $-23.19\pm0.38$ & $1.18$ & $1.28$ & $-$ \\
             & & Common slope & $-8.76\pm1.25$ & $-23.29\pm0.27$ & $1.05$ & $1.18$ & $-$ \\
              \hline
             2MASS $K_\text{s}$ & Initial & Forward & $-3.54 \pm 0.38$ & $-24.11 \pm 0.08$ & $0.57$ & $0.66$ & $-$ \\
             & & Reverse & $-5.95\pm 0.62$ & $-24.01\pm 0.10$ & $0.74$ & $0.84$ & $-$ \\ 
             & & Fixed (TP00) & $-8.78$ & $-23.89\pm 0.15$ & $1.16$ & $1.27$ & $0.15$ \\
             & & Fixed (To18) & $-8.43$ & $ -23.91\pm 0.14$ & $1.10$ & $1.21$ & $0.40$ \\
             & Sub-sample A & Forward & $-4.04\pm 0.43$ & $-24.05\pm 0.08$ & $0.56$ & $0.65$ & $-$ \\
             & & Reverse & $-6.30\pm 0.64$ & $-23.91\pm 0.11$ & $0.70$ & $0.80$ & $-$ \\
             & & Fixed (TP00) & $-8.78$ & $-23.77\pm 0.14$ & $1.03$ & $1.14$ & $0.29$ \\
             & & Fixed (To18) & $-8.43$ & $-23.79\pm 0.13$ & $0.97$ & $1.09$ & $0.52$ \\
             & Sub-sample B & Forward & $-4.14 \pm 0.47$ & $-24.08\pm 0.09$ & $0.57$ & $0.65$ & $-$ \\
             & & Reverse & $-6.34\pm 0.68$ & $-23.97\pm 0.11$ & $0.71$ & $0.78$ & $-$ \\
             & & Fixed (TP00) & $-8.78$ & $-23.85\pm 0.15$ & $1.04$ & $1.10$ & $0.20$ \\
             & & Fixed (To18) & $-8.43$ & $-23.87\pm 0.14$ & $0.98$ & $1.05$ & $0.44$ \\
             & LTGs & Forward & $-3.75\pm 0.64$ & $-24.18\pm 0.11$ & $0.54$ & $0.63$ & $-$ \\
             & & Reverse & $-6.80\pm 1.11$ & $ -24.40 \pm 0.16$ & $0.73$ & $0.83$ & $-$ \\
             & & Common slope & $-7.11\pm0.87$ & $-24.42\pm0.16$ & $0.78$ & $0.90$ & $-$\\
             & ETGs & Forward & $-3.74\pm0.67$ & $-24.02\pm0.15$ & $0.60$ & $0.68$ & $-$ \\
             & & Reverse & $-7.52\pm1.40$ & $-23.46\pm0.26$ & $0.86$ & $0.96$ & $-$ \\
             & & Common slope & $-7.11\pm0.87$ & $-23.53\pm0.20$ & $0.78$ & $0.90$ & $-$\\
             \hline
        \end{tabular}
        \begin{tablenotes}[flushleft]
        \item \textit{Notes.} All results are derived using a pivot point of $2.6$. The zero-point offsets are measured as ours$-$theirs. As discussed in Section~\ref{sec:TFR}, the reverse-fit parameters listed are the redefined quantities $a'$, $b'$, $\sigma'_\text{int}$ and $\sigma'_\text{tot}$.
    \end{tablenotes}
    \end{threeparttable}
\end{table*}

\begin{figure*}
    \centering
    \includegraphics[width=0.49\linewidth]{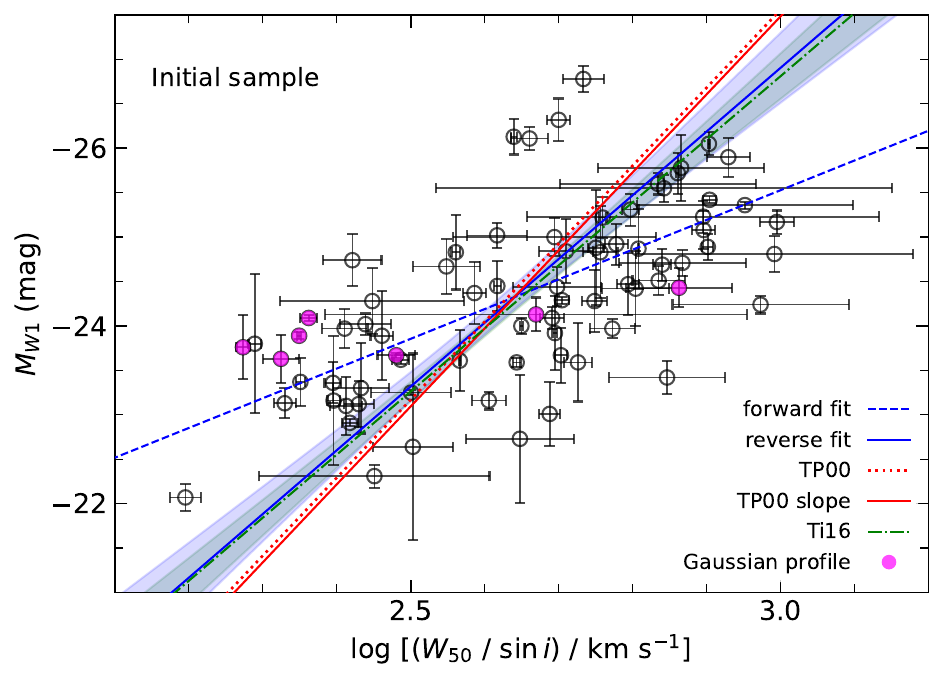}
    \includegraphics[width=0.49\linewidth]{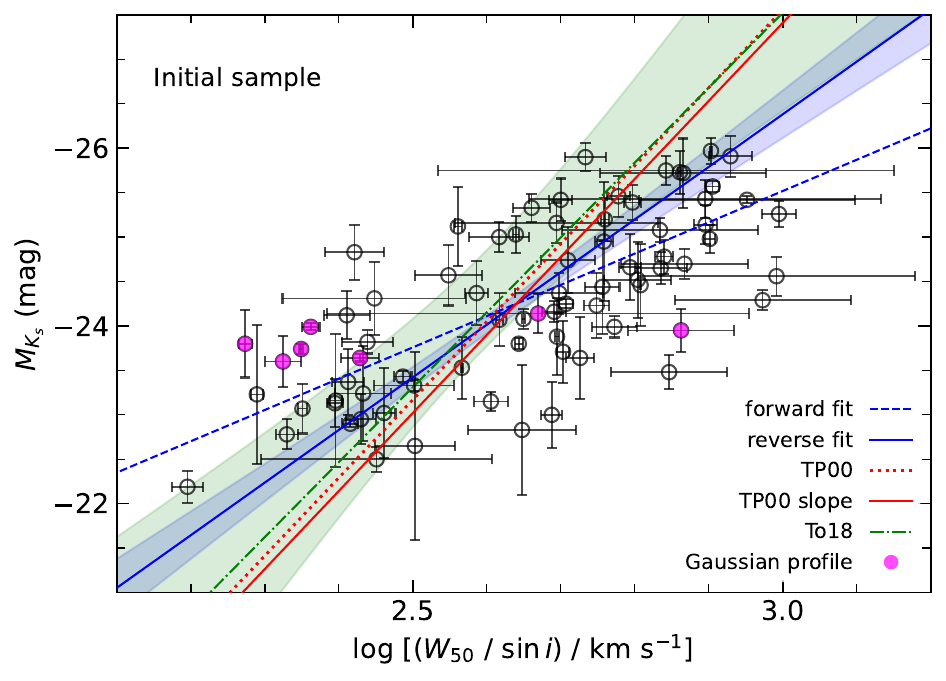}\\
    \includegraphics[width=0.49\linewidth]{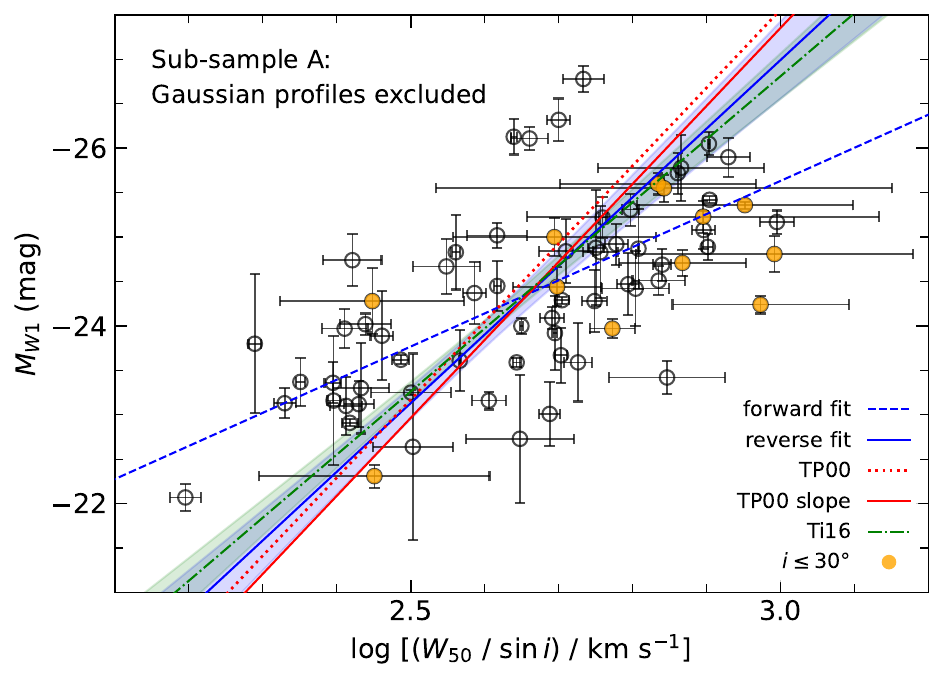} 
    \includegraphics[width=0.49\linewidth]{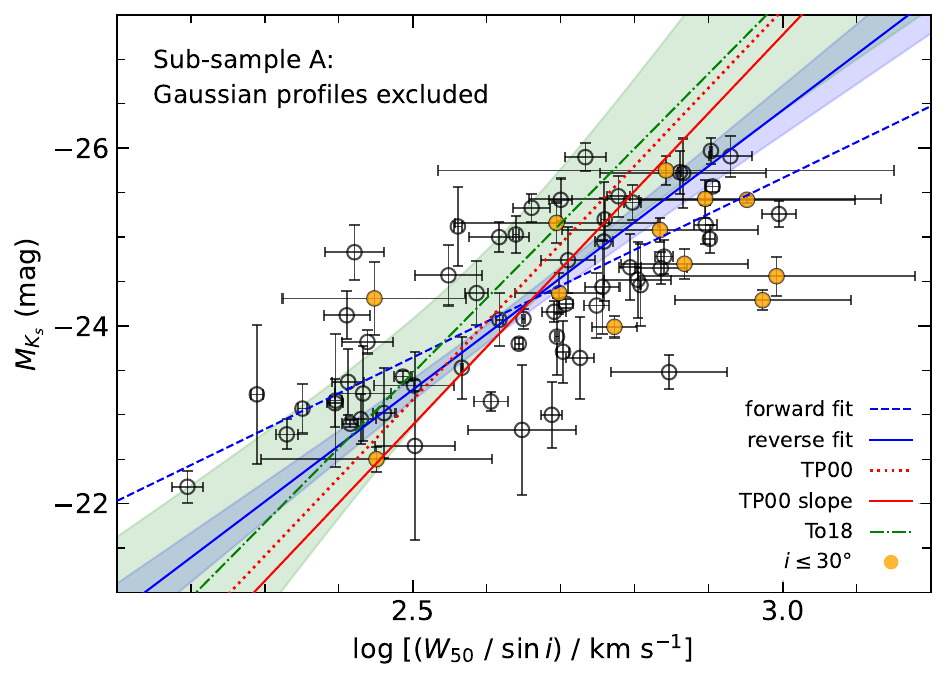}\\
    \includegraphics[width=0.49\linewidth]{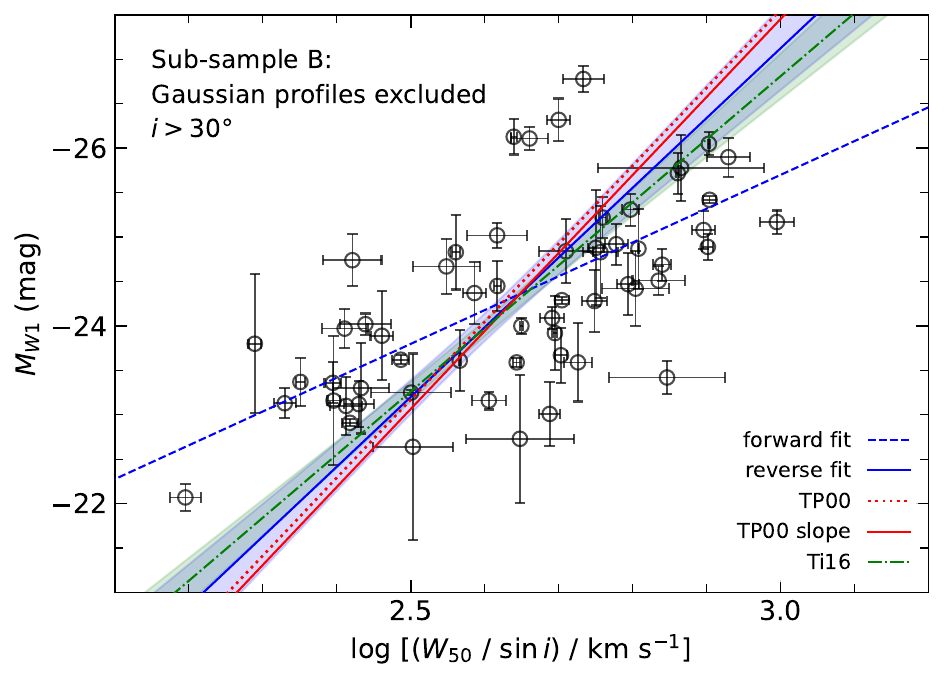}
    \includegraphics[width=0.49\linewidth]{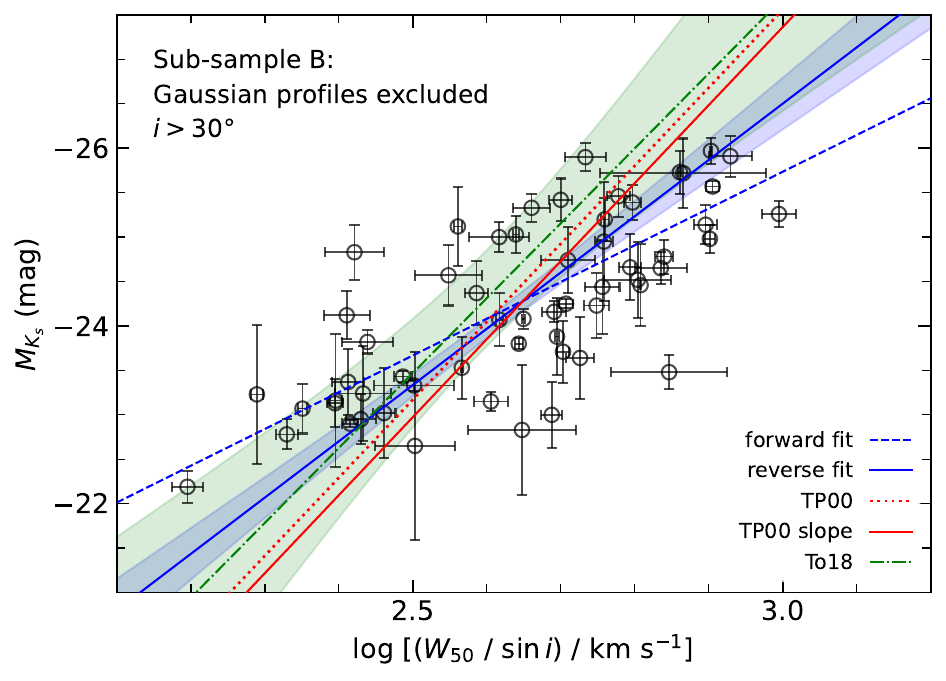}
    \caption{CO Tully–Fisher relations. \textbf{Top-left:} $W1$-band CO TFR of the initial sample galaxies. Filled magenta data points indicate galaxies with a single Gaussian profile. \textbf{Top-right:} As before, but for the $K_\text{s}$-band CO TFR. \textbf{Middle-left:} $W1$-band CO TFR of the sub-sample A galaxies, with single Gaussian profile galaxies removed. Filled orange data points indicate galaxies with inclinations less than $30\degree$. \textbf{Middle-right:} As before, but for the $K_\text{s}$-band CO TFR. \textbf{Bottom-left:} $W1$-band CO TFR of the sub-sample B galaxies, with single Gaussian profile and face-on ($i\leq30\degree$) galaxies removed. \textbf{Bottom-right:} As before, but for the $K_\text{s}$-band CO TFR. The blue dashed and solid lines show the best forward fit and reverse fit, respectively. The blue shaded band indicates the $1\sigma$ uncertainty around the best reverse fit. The dotted red line is the $K'$-band TFR of \citet{Tully_2000}, plotted for each band assuming a $K'-W1$ and a $K'-K_\text{s}$ colour of $0$, respectively. The solid red line shows the reverse fit to our data when the slope is fixed to that of \citet{Tully_2000}. The green dot-dashed line on the $W1$-band CO TFR plots is the $W1$-band TFR of \citet{Tiley_2016}, whereas the green dot-dashed line on the $K_\text{s}$-band CO TFR plots is the $K_\text{s}$-band TFR of \citet{Topal_2018}; the green shaded band indicates the $1\sigma$ uncertainty around both of their best fit, respectively.}
    \label{fig:CO TFR}
\end{figure*}


\subsection{Galaxy morphology}
\label{subsection:morphology results}

Separate TFRs were fit for different morphological types, to examine whether galaxy morphology introduces systematic variations in the relation. Previous studies \citep[e.g.][]{Masters_2008, Williams_2010, Davis_2011, denHeijer_2015} have shown that the TFR can vary systematically with morphology, with ETG TFRs typically offset to fainter magnitudes at a given rotation velocity (see detailed discussion in Section~\ref{subsection:morphology discussion}). By constructing morphology-dependent fits, we test whether similar trends are present in our samples and quantify the extent of any systematic difference. 

We performed the fits on our initial sample of $77$ galaxies. Of these, $20$ are classified as ellipticals, $19$ as lenticulars and $38$ as spirals. We group the lenticulars and ellipticals together as ETGs, as all ellipticals in our sample have high stellar masses (and thus higher luminosities and velocity widths) and are thus at larger deprojected velocities than our adopted pivot point. The addition of lenticulars to the ellipticals extends the data over the entire velocity range of the sample, thus offering more reliable determinations of the slopes and zero-points of the relations. 

The $W1$- and $K_\text{s}$-band ETG and LTG TFRs are shown in Fig.~\ref{fig:morphology TFR}. In the same figure we show the best-fitting relations of other studies which are most similar to ours methodologically, yielding the most appropriate benchmarks against which to compare our results. These are \citet{Tully_2000} and \citet{Tiley_2016} for our LTGs and \citet{Williams_2010} and \citet{Davis_2011} for our ETGs. All best-fitting TFRs are listed in Table~\ref{tab:results}.

It is worth highlighting that different studies choose different pivot points for their TFRs, depending on their data (see Eq.~\ref{eq:8; main TFR}). To ensure a fair comparison, we thus recalculate the zero-points of other studies to our pivot point in the following manner:
\begin{equation}
\begin{split}
    & b_\text{pivot}=b+a\,(2.6-x_0)\quad\text{and}\\
    & \sigma^2_{b_\text{pivot}}=\sigma_b^2+\sigma_a^2\,(2.6-x_0)^2 \quad,
\end{split}
\end{equation}
where $a$, $b$, $\sigma_a$ and $\sigma_b$ are the original (literature) slope, zero-point and their respective uncertainty, respectively, $x_0$ is the original pivot point, $b_\text{pivot}$ is the new zero-point and $\sigma_{b_\text{pivot}}$ is its uncertainty adjusted to our pivot point of $2.6$. The adjusted zero-points of the aforementioned studies are listed in Table~\ref{tab:literature}, along with other key information.

\begin{table*}
    \centering
    \begin{threeparttable}
    \caption{Best-fitting TFR parameters of literature studies.}
    \label{tab:literature}
    \begin{tabular}{lllllcccl}
        \hline
         Work & Sample & Slope $a$ & Zero-point $b$ & Pivot & Tracer & Velocity measure & Band & Adjusted zero-point $b_\text{pivot}$ \\
         & & (mag~dex$^{-1}$) & (mag) & (dex) & & & & (mag)\\
         \hline
         \citet{Tully_2000} & Spirals & $-8.78$ & $-23.17$ & $2.5$ & H\,\textsc{i} & $W_{20}$ & $K'$ & $−24.05$ \\
         \citet{Williams_2010} & S0s & $-9.13\pm0.87$ & $-23.98\pm0.20$ & $2.4$ & Stars & $v_\text{model}$ & $K_\text{s}$ & $-23.06\pm0.22$ \\
         \citet{Davis_2011} & ETGs & $-8.38\pm0.70$ & $-23.12\pm0.09$ & $2.6$ & CO & $W_{20}$ & $K_\text{s}$ & $-23.12\pm0.09$ \\
         \citet{Tiley_2016} & Spirals & $-7.1\phantom{0}\pm0.6$ & $-23.83\pm0.09$ & $2.58$ & CO & $W_{50}$ & $W1$ & $-23.97\pm0.09$ \\
         \citet{Topal_2018} & $z\leq0.3$ & $-8.4\phantom{3}\pm2.9$ & $-23.47\pm0.54$ & $2.5$ & CO & $W_{50}$ & $K_\text{s}$ & $-24.31\pm0.61$ \\
         \hline
    \end{tabular}
    \end{threeparttable}
\end{table*}

\begin{figure}
    \centering
    \includegraphics[width=\linewidth]{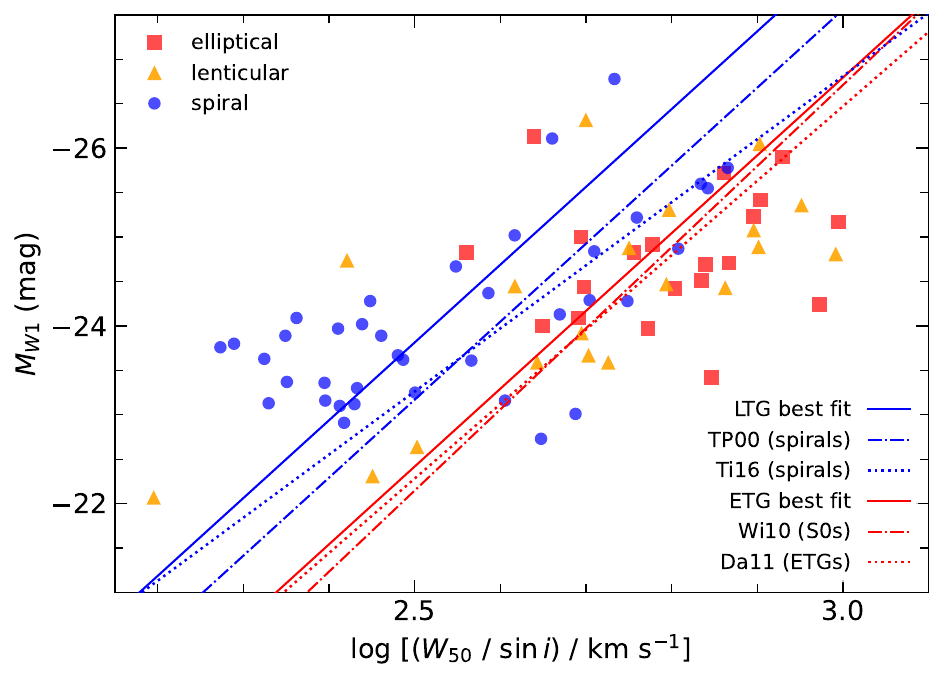} \\
    \includegraphics[width=\linewidth]{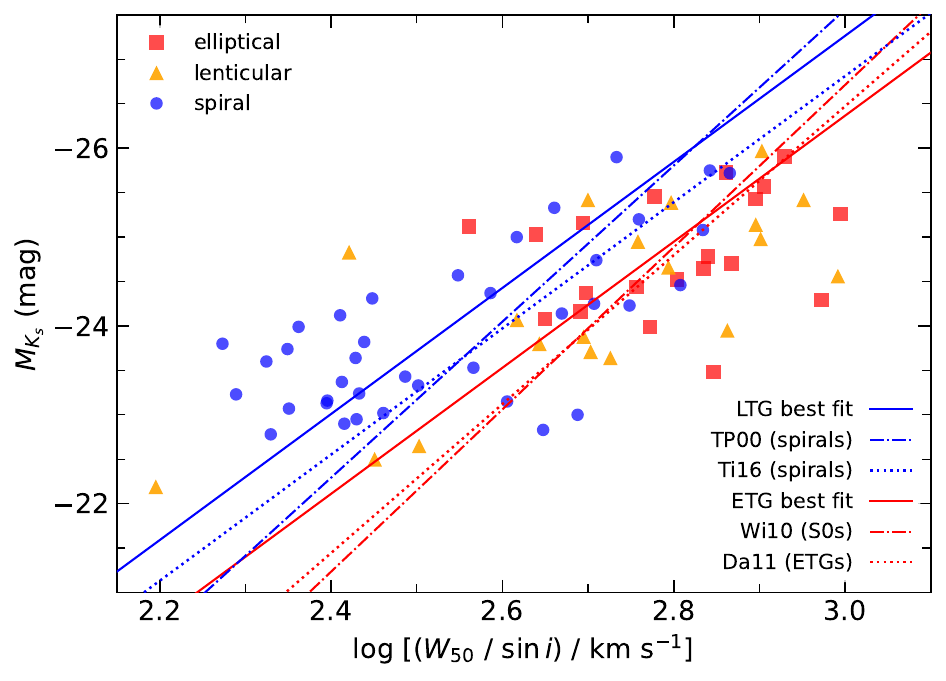}
    \caption{As Fig.~\ref{fig:CO TFR}, but for galaxies in the initial sample categorised by morphological type. \textbf{Top:} $W1$-band CO TFR. \textbf{Bottom:} $K_\text{s}$-band CO TFR. Blue circles indicate spiral galaxies (LTGs), orange triangles lenticular galaxies and red squares elliptical galaxies. Lenticulars and ellipticals together comprise ETGs. The best-fitting TFRs are coloured as the data points, blue lines showing fits for LTGs, red lines for ETGs. Our reverse fits are shown as solid lines. Morphology-dependent reverse TFR fits from the literature are also shown. For LTGs these are \citeauthor{Tully_2000} (\citeyear{Tully_2000}; blue dot-dashed line) and \citeauthor{Tiley_2016} (\citeyear{Tiley_2016}; blue dotted line). For ETGs these are \citeauthor{Williams_2010} (\citeyear{Williams_2010}; red dot-dashed line) and \citeauthor{Davis_2011} (\citeyear{Davis_2011}; red dotted line). All are plotted assuming $K'-W1$ and $K'-K_\text{s}$ (and thus $K_\text{s}-W1$) colours of $0$. Error bars have been omitted for clarity.}
    \label{fig:morphology TFR}
\end{figure}


\section{Discussion}
\label{section:Discussion}

Given their common use in previous TFR studies, and their reduced susceptibility to selection biases, to facilitate direct comparison with the literature we will focus our discussion on the reverse fits.

\subsection{Literature Comparisons}

\subsubsection{Tiley et al.\ (2016b)}
\label{subsubsec: Tiley}

\citet{Tiley_2016} established the exact method used in this work. Use of the same kinematic tracer and the same velocity width measure ($W_{50}$ of fits to integrated spectra rather than $W_{50}$ of integrated spectra) as us makes this work ideal to compare to our results.

\citet{Tiley_2016} studied the $W1$-band TFR of $207$ CO Legacy Database for the GALEX Arecibo SDSS Survey (COLD GASS; \citealt{Saintonge_2011}) galaxies at redshifts $0.025<z<0.05$. Using a reverse fit, they found a $W1$-band slope of $−7.1\pm0.6$, entirely consistent (within the uncertainties) with the slopes derived for all three of our samples ($-7.18\pm0.93$, $-7.73\pm1.02$ and $-7.85\pm1.11$). To fairly compare zero-points, we carried out reverse fits with the slope set to that of \citet{Tiley_2016}, yielding $W1$-band zero-points of $-24.04\pm0.13$, $-23.94\pm0.12$ and $-24.02\pm0.13$, in full agreement with those of \citeauthor{Tiley_2016} (\citeyear{Tiley_2016}; $-23.97\pm0.09$ when converted to our pivot point of $2.6$). 

It is worth noting that $93\%$ of the galaxies in our sample are closer than those of \citet{Tiley_2016}, i.e.\ closer than $z=0.025$. Taken together, the two studies thus cover the entire extent of the local Universe, and their excellent agreement with one another demonstrates that we have successfully established a robust local benchmark for $W1$-band CO TFRs.

\subsubsection{Topal et al. (2018)}
\label{subsubsec: Topal}

\citet{Topal_2018} studied the $K_\text{s}$-band TFR of $25$ mostly late-type galaxies at redshifts $0.05<z<0.3$. All galaxies in our sample, as well as those of \citet{Tiley_2016}, are closer than those of \citet{Topal_2018}. They found a reverse-fit slope of $-8.43\pm2.86$ which, partly owing to the large uncertainty, is consistent with the slopes derived for all three of our samples ($-5.95\pm0.62$, $-6.30\pm0.64$ and $-6.34\pm0.68$). To investigate potential zero-point differences, we again carry out reverse fits with the slope fixed, now to that of \citet{Topal_2018}. Again, our zero-points ($-23.91\pm0.14$, $-23.79\pm0.13$ and $-23.87\pm0.14$) are in broad agreement with that of \citeauthor{Topal_2018} (\citeyear{Topal_2018}; $-24.31\pm0.61$ when converted to our pivot point of $2.6$). These results bolster the conclusions drawn by \citet{Topal_2018}, who compared their results to those of \citet{Tiley_2016}, namely that there does not appear to be any evolution of the $K_\text{s}$-band CO TFR since $z\approx0.3$.

Given that $z\approx0.3$ corresponds to a lookback time of $\approx3.5$~Gyr, this result is consistent with expectations. Indeed, over this period, massive galaxies are expected to have undergone relatively little structural evolution, while their near-infrared luminosities evolve only slowly. The absence of measurable evolution therefore supports the view that the relation has remained largely unchanged over the past $\approx3.5$~Gyr, and is in line with a growing body of literature results showing relatively stable slope and zero-point of the $K$-band TFR up to such intermediate redshifts \citep[e.g.][]{Conselice_2005, Flores_2006, Tiley_2016b, Tiley_2019}. 

Notwithstanding this formal consistency, it is worth noting that our measurements favour slopes different from those of \citet{Topal_2018}. Relative to our uncertainties, a slope as steep as $-8.43$ is disfavoured at the $\approx4\sigma$ level, indicating that our best-fitting relation has a shallower slope, although it remains statistically consistent with that of \citet{Topal_2018} once their uncertainties are also taken into account. Similarly, our data suggest a best-fitting zero-point $\approx0.45$~mag fainter than that of \citet{Topal_2018} for the same slope, also an $\approx3\sigma$ offset when judged against our uncertainties. Given that we use the same tracer, velocity width definition, photometric band, pivot point and fitting function as \citet{Topal_2018}, these differences are not attributable to our choice of methodology. Rather, they likely represent systematic differences emerging from sample differences. This tension, while not significant once \citeauthor{Topal_2018}'s (\citeyear{Topal_2018}) larger uncertainties are taken into account, nevertheless motivates further large-scale studies of the $K_\text{s}$-band CO TFR, particularly to probe whether there could be a potential evolution since intermediate redshifts.

\subsubsection{Tully \& Pierce (2000)}

Given that the work of \citet{Tully_2000} is commonly treated as a standard reference within TFR work, we will compare it to our relations in each photometric band assuming a $K'-W1$ and a $K'-K_\text{s}$ colour of $0$, respectively (and thus $K_\text{s}-W1=0$), i.e.\ that results in all three bands are directly comparable. It is also plotted as such in Fig.~\ref{fig:CO TFR}.

The $W1$-band reverse fits of all three of our samples have slopes shallower than the reverse $K'$-band fit of \citeauthor{Tully_2000} (\citeyear{Tully_2000}; $-8.78$). However, as the samples become more homogenous (from initial sample, to sub-sample A to sub-sample B), their slopes become steeper and more closely approximate that of \citet{Tully_2000}. In fact, the \citet{Tully_2000} slope is statistically consistent with the slope of our sub-sample B ($−7.85\pm1.11$). The slopes of our $K_\text{s}$-band reverse fits are consistently shallower than those of our $W1$-band reverse fits (for all three of our samples), so our $K_\text{s}$-band slopes depart even more from the fit of \citet{Tully_2000}. It is however worth noting that \citet{Tully_2000} do not list the uncertainty of their slope, making a direct comparison difficult. Using CO velocity widths and near-infrared photometry, \citet{Tiley_2016} also reported a slope shallower than that of \citet{Tully_2000}.

To compare the zero-points of our TFRs to that of \citet{Tully_2000}, we again carried out reverse fits with the slope fixed, now to that of \citet{Tully_2000}. Of the three $W1$-band samples, both the initial sample and sub-sample B have zero-points ($-23.97\pm0.16$ and $-23.94\pm0.16$) which are in agreement with that of \citeauthor{Tully_2000} (\citeyear{Tully_2000}; $-24.05$ when converted to our pivot point of $2.6$), whereas the zero-point for sub-sample A ($-23.84\pm0.15$) is marginally inconsistent.
All three of our $K_\text{s}$-band samples have zero-points ($-23.89\pm0.15$, $-23.77\pm0.14$ and $-23.85\pm0.15$) which are not in agreement with that of \citet{Tully_2000}. However, \citet{Tully_2000} again do not list the uncertainty of their zero-point, so it is hard to gauge how significant those disagreements are. Given the modest offsets and unspecified uncertainties, it is possible that the initial sample and sub-sample B are in fact in agreement with \citet{Tully_2000} in the $K_\text{s}$-band.

One reason why we might in general expect to find an offset between our work and that of \citeauthor{Tully_2000} relates to the choice of kinematic tracer. The \citet{Tully_2000} TFR is built from H\,\textsc{i} integrated spectra of late-type spirals. H\,\textsc{i} almost always reaches well into the flat parts of the rotation curves, whereas CO can be centrally concentrated (corroborated by our data; see Appendix~\ref{section:Appendix A}). If the CO emission does not extend beyond the turnover radius of a rotation curve, the resulting velocity width will underestimate the true asymptotic rotation velocity, leading to a systematic offset of the inferred zero-point. This effect arises from the more compact radial extent of CO relative to H\,\textsc{i}, rather than any intrinsic limitation of CO as a kinematic tracer. In practice, however, this bias is mitigated by our selection of galaxies with double-horned or boxy CO profiles.

\subsection{Galaxy morphology}
\label{subsection:morphology discussion}

As mentioned in Section~\ref{subsection:morphology results}, it is essential to look at the morphological breakdown of our galaxy samples. Indeed, a comparison of TFRs between LTGs and ETGs can provide valuable insight into the connection between stellar populations, $M/L$ and the dynamical evolution of galactic discs \citep[e.g.][]{Masters_2008, Williams_2010}. In particular, a difference between the TFR zero-points of spiral and S0 galaxies could indicate a passive fading of spirals into lenticulars following a cessation of star formation \citep[e.g.][]{Bedregal_2006}. Alternatively, a spiral -- S0 offset could be the result of mass distribution differences, driven by the more concentrated mass distribution of S0 galaxies \citep[e.g.][]{Williams_2010}. Thus, the degree to which the TFR changes (or not) across Hubble types constrains the mechanisms by which galaxies quench star formation and transform morphologically. Additionally, systematic variations between morphological types within the LTG population itself are well-established. Earlier-type spirals (Sa -- Sb) typically possess larger bulge fractions, redder colours and higher $M/L$ than late-type (Sc -- Sd) systems, producing measurable differences of both TFR slope and zero-point \citep[e.g.][]{Pierce_1988, Masters_2008}. These variations indicate that the TFR can also be sensitive to the detailed star-formation history and structural composition of galactic discs.


Previous observational studies seem to collectively suggest that ETG TFRs are truly offset from spiral TFRs in the $K$-band, but the derived magnitude of the offset varies. \citet{Bedregal_2006} analysed S0 galaxies using stellar kinematics and found that they are $0.8$ -- $1.2$~mag fainter than spirals at a given velocity. \citet{DeRijcke_2007} reported a comparable offset of $\approx1.2$~mag, whilst \citet{Davis_2011} reported an offset of $0.98\pm0.22$~mag between their ETG CO TFR and a spiral CO TFR established using the line widths of \citet{Chung_2009}. In contrast, \citet{denHeijer_2015} reported a smaller offset of $0.5$ -− $0.7$~mag when looking at ETGs with extended, regular H\,\textsc{i} discs measured at large radii. However, these studies employ different kinematic tracers and velocity definitions, and are based on samples with varying selection criteria, meaning that the reported offsets are not strictly directly comparable. Whether this offset persists when the relation is expressed in terms of stellar or baryonic mass rather than luminosity is less clear. \citet{Williams_2010} reported a $\approx0.2$~dex offset for both their stellar-mass and their dynamical-mass TFRs, and concluded that the offset of the baryonic relation is likely even larger. This suggests that S0s are systematically smaller and/or more concentrated than spirals and therefore spirals may undergo a systematic contraction as they transform into lenticulars. On the other hand, \citet{denHeijer_2015} reported no significant offset between their baryonic-mass ETG TFR and the baryonic-mass spiral TFR of \citet{Noordermeer_2007}. These works suggest that whilst some of the apparent offset may arise from differences of the stellar populations and/or photometric bandpasses, there are also likely fundamental structural and/or dynamical differences between morphological types.

Nevertheless, when comparing the TFRs of different morphological samples, it is essential to use consistent kinematic tracers. The use of different velocity measures, e.g.\ H\,\textsc{i} integrated spectrum line widths for spirals and stellar or molecular gas rotation curves for ETGs, can introduce systematic biases that mimic intrinsic luminosity offsets. \citet{Williams_2010} showed that when circular velocities are derived homogeneously for both ETGs and LTGs, the two TFRs are offset by a more modest $0.53\pm0.15$~mag, nevertheless significantly different from $0$ given the stated uncertainties.

Most past works that probed the morphological dependence of the TFR used the $K$-band, so we will begin our discussion by considering our $K_\text{s}$-band results. Our LTG slope of $-6.80\pm1.11$ is consistent with our ETG slope of $-7.52\pm1.40$. Our LTG zero-point of $-24.40\pm0.16$ is $0.93\pm0.31$~mag brighter than our ETG zero-point of $−23.46\pm0.26$, which is statistically significant. 

As the TFR slopes are consistent within the uncertainties, we can obtain a more robust zero-point offset by constraining both slopes to be equal. We do so by following the methodology of \citet{Williams_2010} and simultaneously reverse fitting two straight lines, of the usual form for the LTGs and the following form for ETGs:
\begin{equation}
    \log\left(\frac{W_{50}/\sin i}{\text{km~s}^{-1}}\right)=A\,M_{W1}+B+2.6+\Delta\quad,
    \label{eq:17; S0 TFR}
\end{equation}
where $\Delta$ represents the reverse-fit zero-point offset between the two relations (formally dimensionless but expressed in units of dex). This offset can then be converted back to a forward-fit equivalent offset $\delta\equiv-\Delta/A$. Using this method, we derive a common slope of $-7.11\pm0.87$, a LTG zero-point of $-24.42\pm0.16$ and an ETG zero-point offset of $0.89\pm0.28$~mag toward fainter magnitudes, in agreement with our previous estimate. 

Considering the $W1$ band, we derive a LTG slope of $-8.20\pm1.50$ and an ETG slope of $-9.43\pm2.12$. Whilst these slopes differ from each other somewhat more than in the $K_\text{s}$ band, they are still just statistically consistent. Our $W1$-band LTG zero-point of $-24.65\pm0.21$ is $1.46\pm0.43$~mag brighter than the ETG zero-point of $-23.19\pm0.38$, which is statistically significant. As the slopes of the two relations are in rough agreement, we again perform a common-slope fit to our $W1$-band data. Using this method, we derive a common slope of 
$-8.76\pm1.25$, a LTG zero-point of $-24.69\pm0.21$ and an ETG zero-point offset of $1.40\pm0.39$~mag toward fainter magnitudes. 

Our $K_\text{s}$-band offset is consistent within the uncertainties with the offsets derived by \citet{Bedregal_2006} and \citet{Davis_2011}. Our $W1$-band offset is comparably larger, even larger than that of \citet{DeRijcke_2007}.
This result is somewhat surprising, as \citet{Williams_2010} suggested that a significant driver behind these observed offsets could be systematic effects due to different tracers and/or methodologies. While our use of a homogeneous approach suggests that such systematics are unlikely to fully account for the observed offsets, residual systematics related to sample selection may still be present. In particular, when comparing the zero-point of \citeauthor{Tiley_2016} (\citeyear{Tiley_2016}; $-23.97\pm0.09$), whose sample is dominated by spiral galaxies, with that of our LTG sub-sample, we find a mild tension between the two results. This can be further examined by refitting our $W1$-band LTG sample with the slope fixed to that of \citet{Tiley_2016}, which yields a zero-point of $-24.58\pm 0.15$ still significantly offset from that of \citet{Tiley_2016}.

This suggests that the close agreement between the \citet{Tiley_2016} relation and the results presented in Section~\ref{subsubsec: Tiley}, derived from a morphologically mixed sample, is not driven by the LTG population alone, but is instead mediated by the inclusion of ETGs which act to lower the overall zero-point. As such, our LTG sample may be systematically brighter at fixed velocity relative to that of \citet{Tiley_2016}, leading to the larger-than-expected offset. Alternatively, this discrepancy may reflect differences in how well the kinematic tracer probes the outer rotation curves. Spiral galaxies are typically more extended than ETGs, and if the CO emission does not reach the flat parts of the rotation curves, the resulting line widths will underestimate the true asymptotic rotation velocities. This would shift galaxies towards lower velocities at a fixed luminosity, producing an apparent increase of the zero-point offset. Another possibility is that there are systematic differences between the inclinations adopted for ETGs and LTGs. As the velocity widths are deprojected using these inclinations, even modest systematic biases would directly translate into systematic TFR zero-point offsets. In addition, ETGs typically host older stellar populations, with higher stellar $M/L$ than spirals. Although the use of the $K_\text{s}$ and $W1$ bands minimises stellar population effects, residual stellar $M/L$ differences between ETGs and LTGs may contribute to the observed offset. Distinguishing between these possibilities will require larger homogeneous samples together with detailed dynamical modelling and stellar population analyses, and is beyond the scope of the present work.

\subsection{TFR scatter}

As shown in Table~\ref{tab:results}, the reverse $W1$-band CO TFR of our initial sample has an intrinsic scatter of $\sigma_\text{int}=1.01$, higher than expected when compared to \citeauthor{Davis_2011} (\citeyear{Davis_2011}; $\sigma_\text{int}=0.36$) and \citeauthor{Tiley_2016} (\citeyear{Tiley_2016}; $\sigma_\text{int}=0.73$). The reverse $K_\text{s}$-band initial sample CO TFR has an intrinsic scatter of $\sigma_\text{int}=0.74$, smaller than our $W1$-band scatter but still rather large. 

One explanation for this large scatter may be our choice of inclination measure. Very few of our sample galaxies have inclinations derived from kinematic modelling, which is the gold standard for inclinations. As shown by \citet{Davis_2011}, the intrinsic scatter of the TFR derived using dust inclinations only is $\sigma_\text{int}=0.66$, almost double that of the TFR derived using inclination measures dominated by interferometric measurements ($\sigma_\text{int}=0.36$). 

It also appears that the scatter is at least somewhat morphology-dependent. Previous studies have typically focused on a particular morphology, such as ETGs \citep[e.g.][]{Davis_2011} or spirals \citep[e.g.][]{Tiley_2016}. When dividing our sample galaxies by morphological types, the LTGs have a notably smaller scatter in both bands ($\sigma_\text{int}=0.95$ in $W1$ and $0.73$ in $K_\text{s}$) compared to ETGs ($\sigma_\text{int}=1.18$ in $W1$ and $0.86$ in $K_\text{s}$). A potential reason for the larger scatter of ETGs is that they tend to have more concentrated CO distributions, leading to potentially biased velocity widths and thus higher scatter. Therefore, when all galaxies are combined together into a single sample, ETGs drive the resulting scatter. Furthermore, by construction we expect our ETG sample to have a larger scatter, as it consists of both elliptical and lenticular galaxies which will span a broader range of structural properties than typical spiral galaxy samples, including variations of bulge-to-disc ratio and stellar $M/L$. These differences affect the underlying gravitational potentials and in turn the shapes of the rotation curves that the line widths trace.

It is also worth noting that the intrinsic scatter does not significantly decrease as the galaxy sample is made more homogeneous (i.e.\ from the initial sample to sub-sample A to sub-sample B), in either band ($\sigma_\text{int}=1.01$, $1.02$ and $1.05$, respectively, in the $W1$ band and $\sigma_\text{int}=0.74$, $0.69$ and $0.71$, respectively, in the $K_\text{s}$ band). While sub-sample A, which excludes galaxies with single Gaussian CO integrated spectra, yields a modest reduction of the scatter in the $K_\text{s}$ band, the effect is small and likely reflects the limited number of such systems in the sample. This suggests that while profile shape is a useful selection criterion, it is not the dominant driver of the observed scatter. Indeed, even a galaxy with a double-horned profile may have a molecular gas distribution that does not extend beyond the turnover radius of its rotation curve, for example if the gas is confined to a ring. Sub-sample B additionally excludes galaxies with inclinations $i<30\degree$, but this does not result in any further reduction in scatter. This is perhaps unsurprising, as the large uncertainties associated with these nearly face-on galaxies reduce their weights in the $\chi^2$ minimisation, and hence their influence on the derived TFRs. Their removal therefore has relatively little effect on the fitted slopes, zero-points and scatters. This indicates that excluding low-inclination galaxies alone is insufficient to reduce the overall scatter of the relation. In other words, while the formal precision of the dust-based inclinations may be adequate, their accuracy in tracing the true disc orientation is not universal across a galaxy sample. This suggests that the reliability of inclination measurements is more important than the application of simple inclination cuts, and that such cuts should be applied with caution when robust inclination estimates are available.


\section{Conclusions}
\label{section:conclusion}

In this paper we have presented a new CO TFR calibration for the local Universe ($z\lesssim0.03$), based on a sample of galaxies observed by the WISDOM Project. Using this sample we probed the $W1$- and $K_\text{s}$-band CO TFRs of an initial sample of $77$ galaxies, a sub-sample of $70$ galaxies with double-horned integrated spectra and a sub-sample of $58$ relatively edge-on galaxies. The main results are as follows.
\begin{enumerate}
    \item The best-fitting reverse $W1$- and $K_\text{s}$-band CO TFRs of our initial sample are\\
    $M_{W1}=(−7.18\pm0.93)\left[\log\left(\frac{W_{50}/\sin i}{\text{km~s}^{-1}}\right)-2.6\right]+(−24.04\pm0.13)$\\
    and\\
    $M_{K_\text{s}}=(−5.95\pm0.62)\left[\log\left(\frac{W_{50}/\sin i}{\text{km~s}^{-1}}\right)-2.6\right]+(−24.01\pm0.10)$. 
    \item The slopes and zero-points of the best-fitting CO TFRs of our sub-samples that exclude galaxies with single Gaussian profiles and low inclinations are statistically consistent with those of the initial sample. Excluding these galaxies does not substantively decrease the intrinsic scatters of the relations.
    \item Our results are in excellent agreement with the work of \citet{Tiley_2016}, who used the same methodology and kinematic tracer. Taken together, these two studies cover the entire extent of the local Universe and establish a robust local benchmark for the $W1$-band CO TFR.
    \item Our results are in broad agreement with those of \citet{Topal_2018} and indicate no significant evolution of either the zero-point or slope of the CO TFR since $z=0.3$.
    \item Our best-fitting CO TFRs are in moderate tension with other studies such as that of \citet{Tully_2000}, suggesting a shallower TFR slope; there is marginal agreement between our $W1$-band results and the relation of \citet{Tully_2000}. However, due to the use of different kinematic tracers, analysis methods and sample galaxy morphologies, we cannot assign these differences to specific astrophysical phenomena rather than methodological differences. 
    \item Due to the morphological variety of our sample galaxies, we constructed a LTG CO TFR for $38$ spiral galaxies and an ETG CO TFR for $20$ elliptical and $19$ lenticular galaxies. By simultaneously fitting the two galaxy samples with a common slope, we infer that ETGs are $0.89\pm0.28$~mag fainter in the $K_\text{s}$ band and $1.40\pm0.39$~mag fainter in the $W1$ band than LTGs of the same rotation velocity. The offset in the $K_\text{s}$ band is broadly consistent with other literature results, whereas the offset in the $W1$ band is larger than expected. 
    \item We have generated and included a comprehensive catalogue of WISDOM CO data products for use in future molecular gas studies. 
\end{enumerate}

Looking ahead, several avenues remain for improving CO TFR studies. A key priority is the use of spatially resolved kinematic modelling to derive more accurate molecular gas disc inclinations, thereby reducing systematic uncertainties and the scatter of the relationship. Additionally, increasing the sample size, especially at higher redshifts and across a wider range of galaxy morphologies, will allow for tighter constraints on potential evolutionary trends and the role of morphological diversity in shaping the TFR. Finally, combining CO data with complementary tracers such as H\,\textsc{i} will allow to extend observations to larger radii and quantify the true extent to which CO robustly traces the flat parts of rotation curves. Together, these steps will further refine the CO TFR as a precise and physically informative probe of galaxy dynamics and evolution.


\section*{Acknowledgements}

MB was supported by STFC consolidated grant ‘Astrophysics at Oxford’ ST/K00106X/1 and ST/W000903/1 and STFC Small Award grant UKRI1181. FHL acknowledges funding from the European Research Council’s starting grant ERC StG-101077573 (`ISM-METALS'). ST acknowledges the support of the Scientific and Technological Research Council of Türkiye (TUBITAK) 2219 International Postdoctoral Research Fellowship Program for Turkish Citizens. JG gratefully acknowledges funding via STFC grant ST/Y001133/1. MK acknowlegdes financial support from the grants PID2021-123930OB-C21 and PID2024-155817OB-I00 funded by MICIU AEI/10.13039/501100011033 and by ERDF/EU, and from the SAFE project funded by the European Union under Grant Agreement No. 101148426.


\section*{Data Availability}

All CO data products will are available at \url{https://wisdom-project.org}. The observations underlying this article are available in the ALMA archive, at \url{https://almascience.eso.org/asax/}, and in the ESO Archive, at \url{https://archive.eso.org/dss/dss}. 


\bibliographystyle{mnras}
\bibliography{BIB} 


\appendix
\section{Moment maps, position-velocity diagrams, integrated spectra and optical images}
\label{section:Appendix A}

\begin{figure*}
    \centering
    \begin{subfigure}{.34\textwidth}
    \centering
    \includegraphics[height=5.9cm]{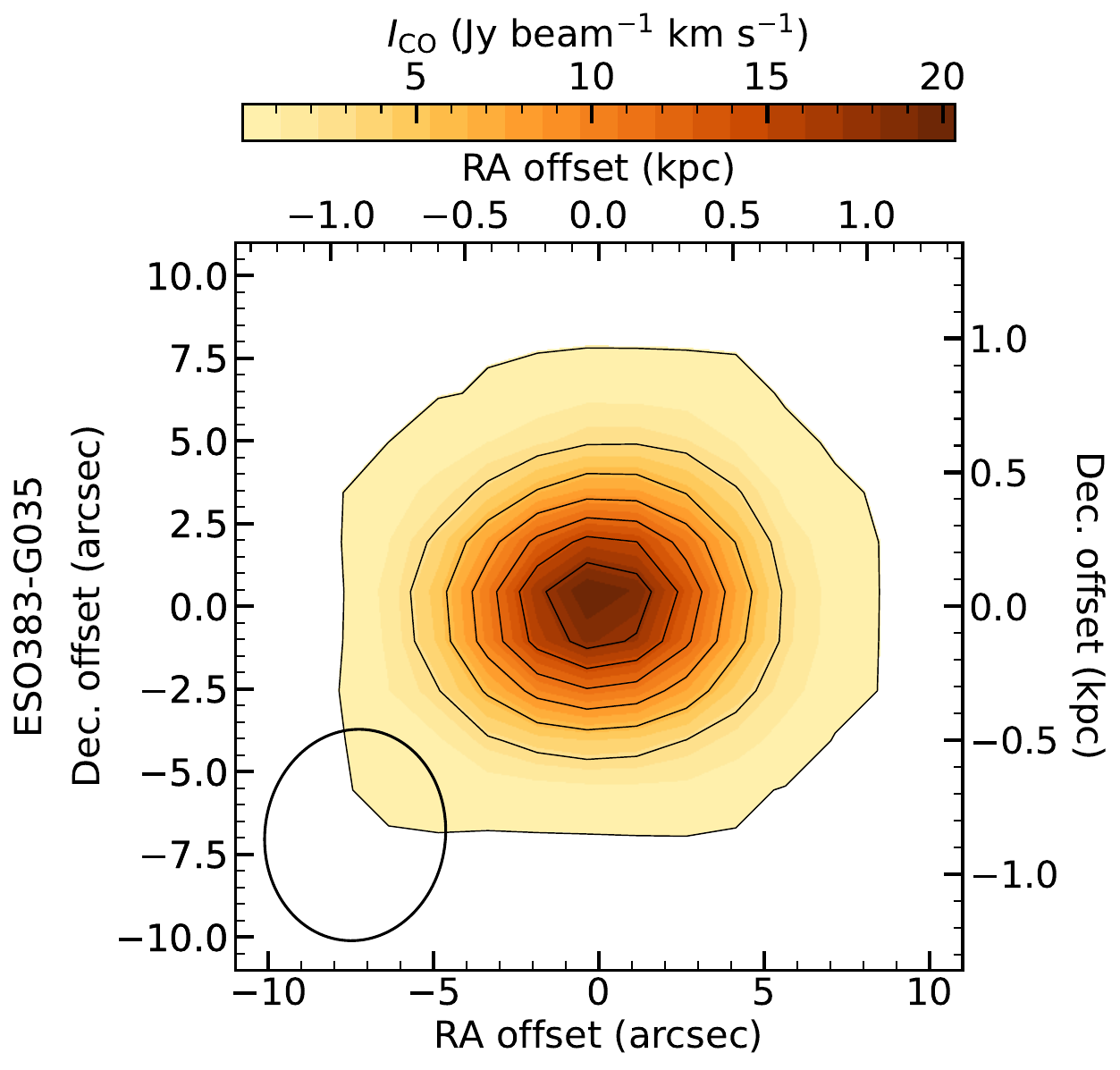}
    \end{subfigure}
    \begin{subfigure}{.32\linewidth}
    \centering
    \includegraphics[height=5.9cm]{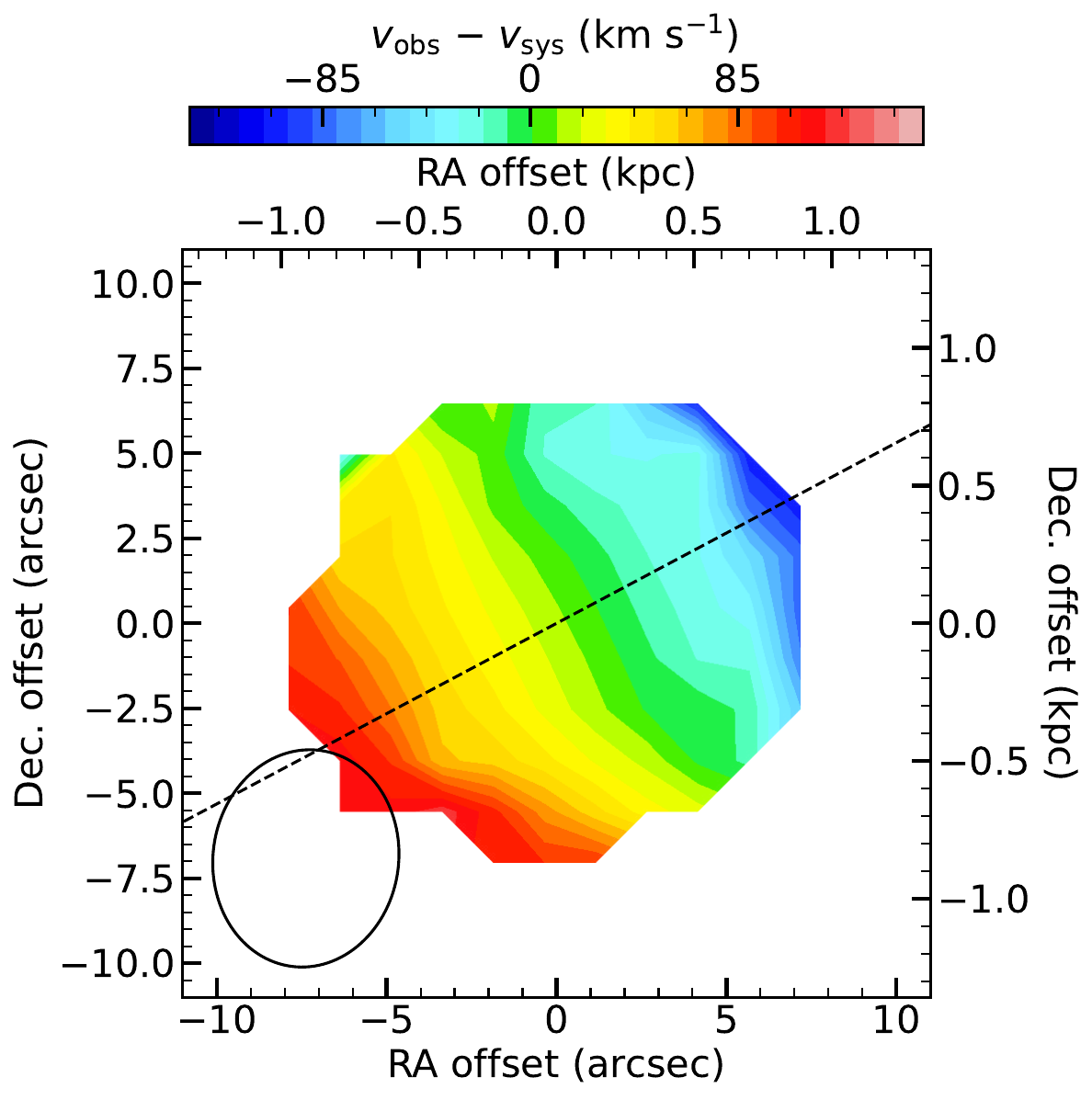}
    \end{subfigure}
    \begin{subfigure}{.32\linewidth}
    \centering
    \includegraphics[height=5.9cm]{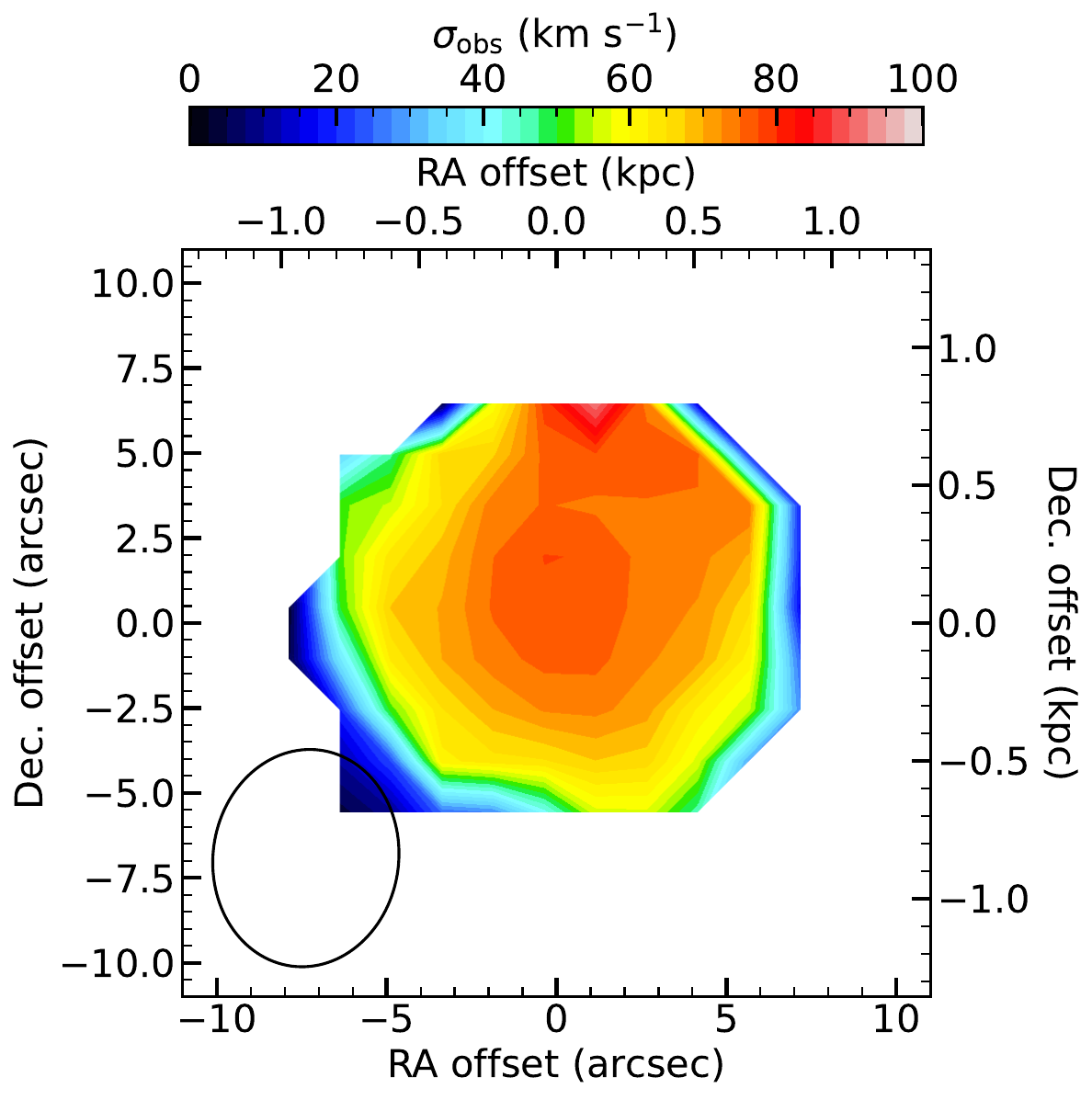}
    \end{subfigure}
    \\
    \begin{subfigure}{.34\linewidth}
    \centering
    \includegraphics[height=5.2cm]{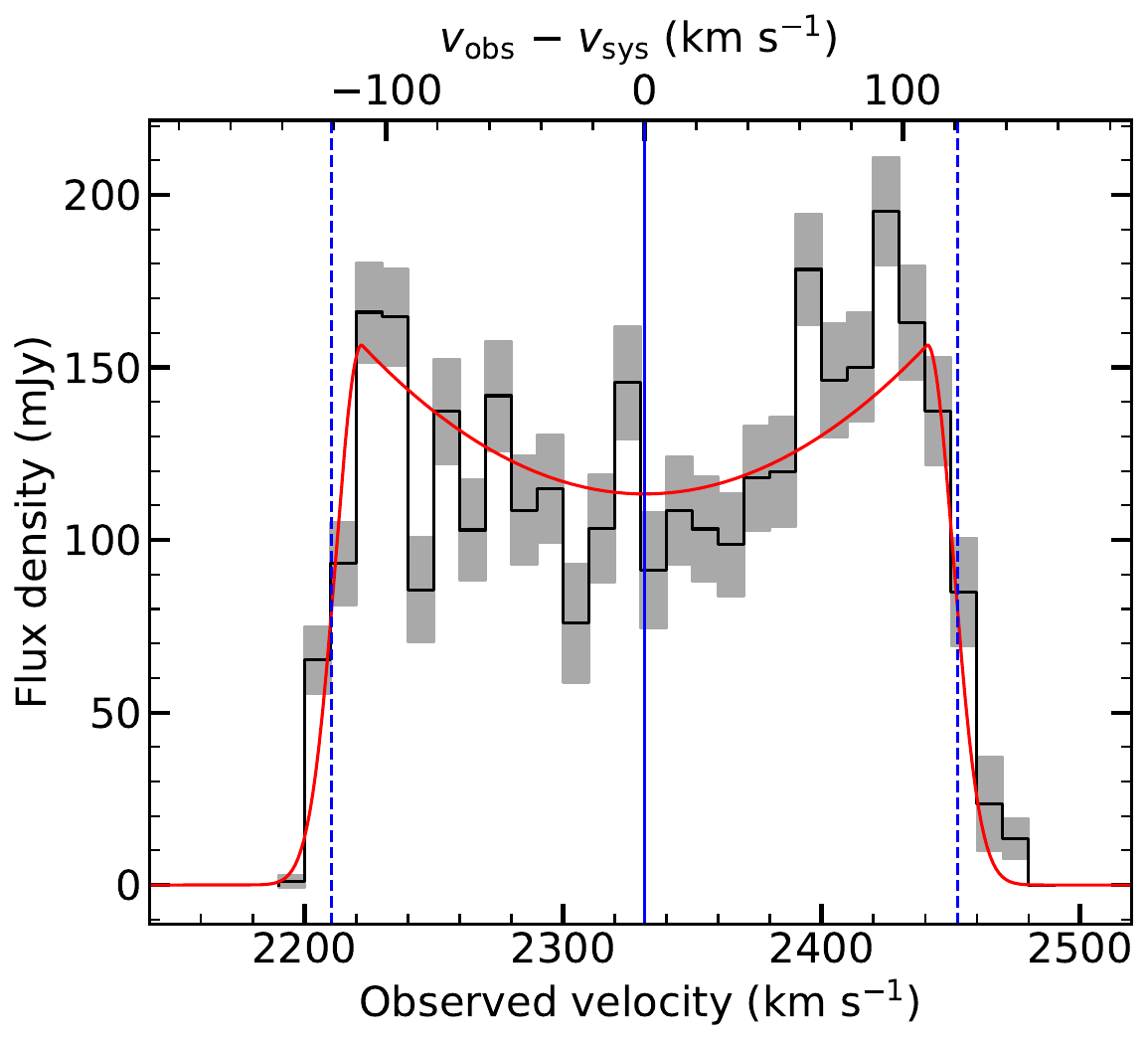}
    \end{subfigure}
    \begin{subfigure}{.32\linewidth}
    \centering
    \includegraphics[height=4.8cm]{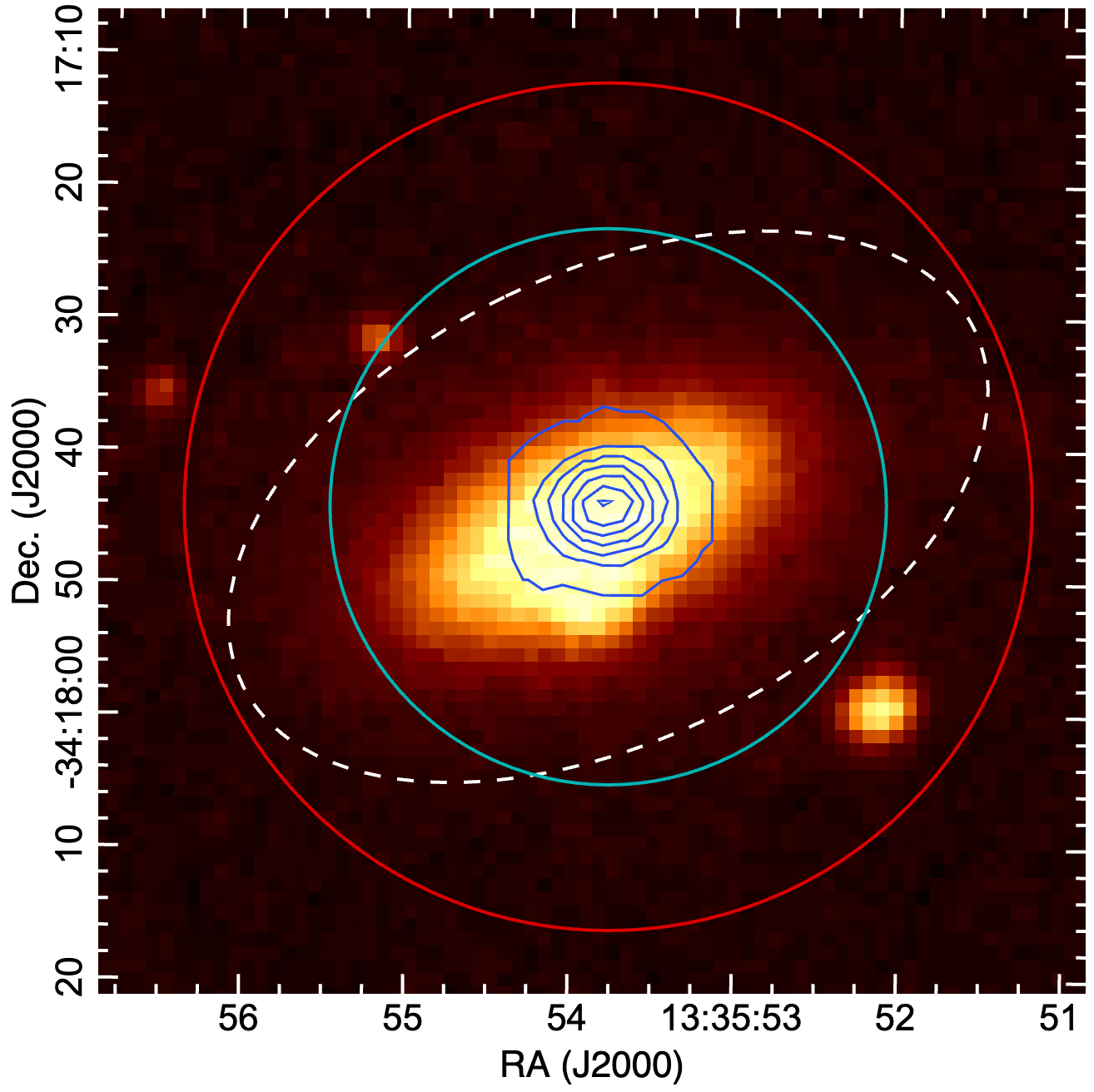}
    \end{subfigure}
    \begin{subfigure}{.33\linewidth}
    \centering
    \includegraphics[height=5.2cm]{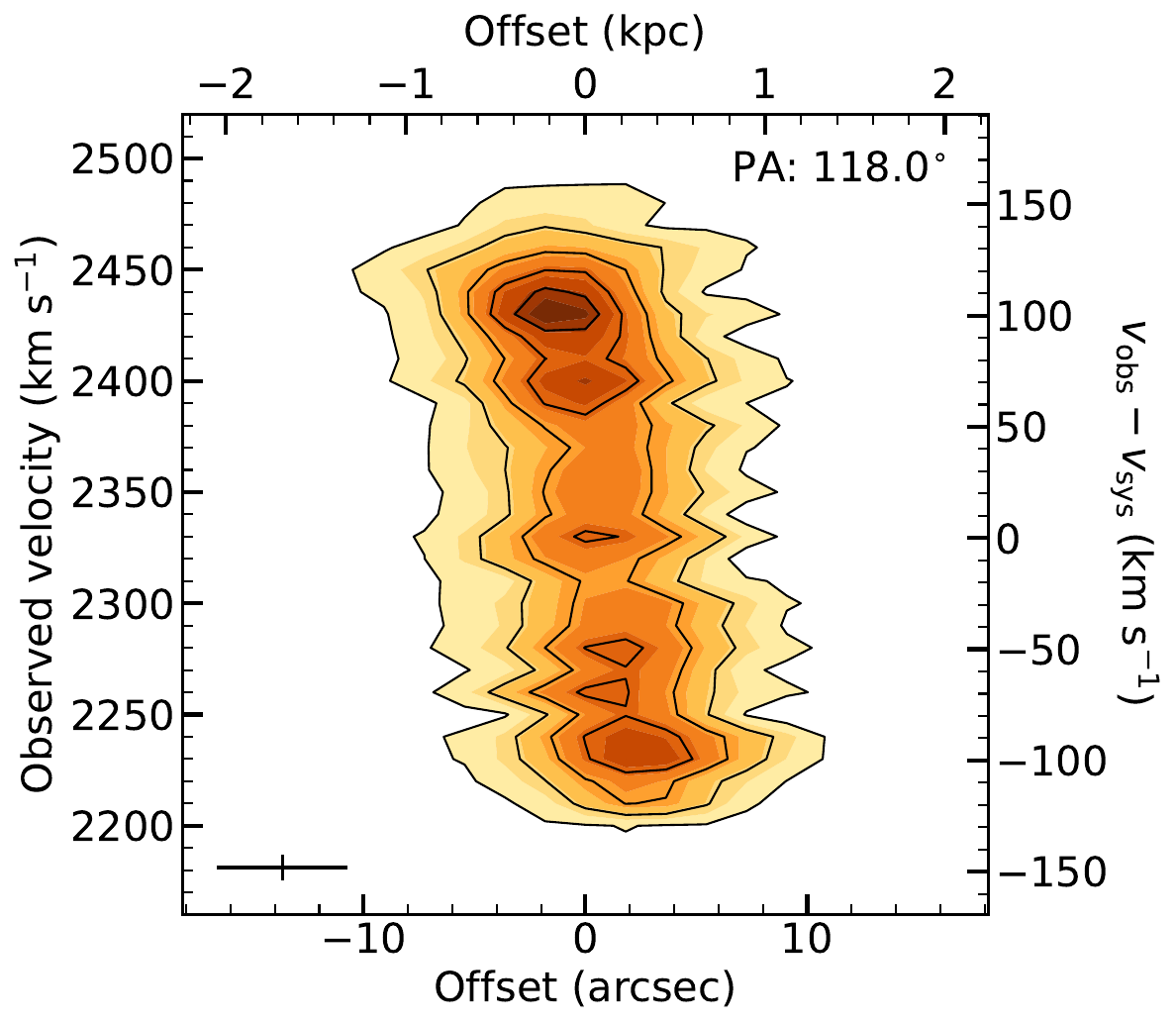}
    \end{subfigure}
     \\

    \caption{$^{12}$CO(2–1) data products derived from our $7$-m ACA ALMA data. \textbf{Top-left:} zeroth-moment (integrated-intensity) map. The solid blue circle indicates the FWHM of the primary beam, i.e.\ the field of view. The solid red circle indicates the extent where the sensitivity is below $20$~per~cent of the primary beam gain, i.e.\ the maximal extent of the data cube. We note that these circles are not always visible, as their extents may lie beyond the displayed field of view, which has been chosen to best highlight the CO emission. The galaxy name is displayed to the left of the map. \textbf{Top-middle:} first-moment (intensity-weighted mean line-of-sight velocity) map. The dashed black line overlaid indicates the assumed kinematic major-axis position angle. \textbf{Top-right:} second-moment (intensity-weighted line-of-sight velocity dispersion) map. The solid black open ellipse in the bottom-left corner of each moment map shows the synthesised beam. \textbf{Bottom-left:} integrated spectrum. The grey shading shows the statistical uncertainties. The solid red line shows the best-fitting function (see Section~\ref{subsection:velocity widths}). The vertical solid blue line shows the best-fitting central velocity $v_0$, whilst the vertical dashed blue lines show the width at $50\%$ of the peak $W_{50}$. \textbf{Bottom-middle:} Digitized Sky Survey-2 red-band optical image of the whole galaxy. The solid blue contours overlaid show the $^{12}$CO(2–1) ALMA emission. The solid cyan circle indicates the primary beam and the solid red circle indicates the cube extent. The dashed white ellipse indicates the optical $D_{25}$ diameter of the galaxy, representing the diameter of the isophote where the surface brightness is $25$~mag~arcsec$^{-2}$ in the $B$-band. \textbf{Bottom-right:} kinematic major-axis position-velocity diagram. The black cross in the bottom-left corner shows the channel width of the cube vertically and the size of the synthesised beam projected along the kinematic major-axis horizontally. The position angle along which the PVD is taken is listed in the upper-right corner.}
    \label{fig:CO ACA moment maps}
\end{figure*}

\begin{figure*}
    \ContinuedFloat
    \centering
    \begin{subfigure}{.34\linewidth}
    \centering
    \includegraphics[height=6.0cm]{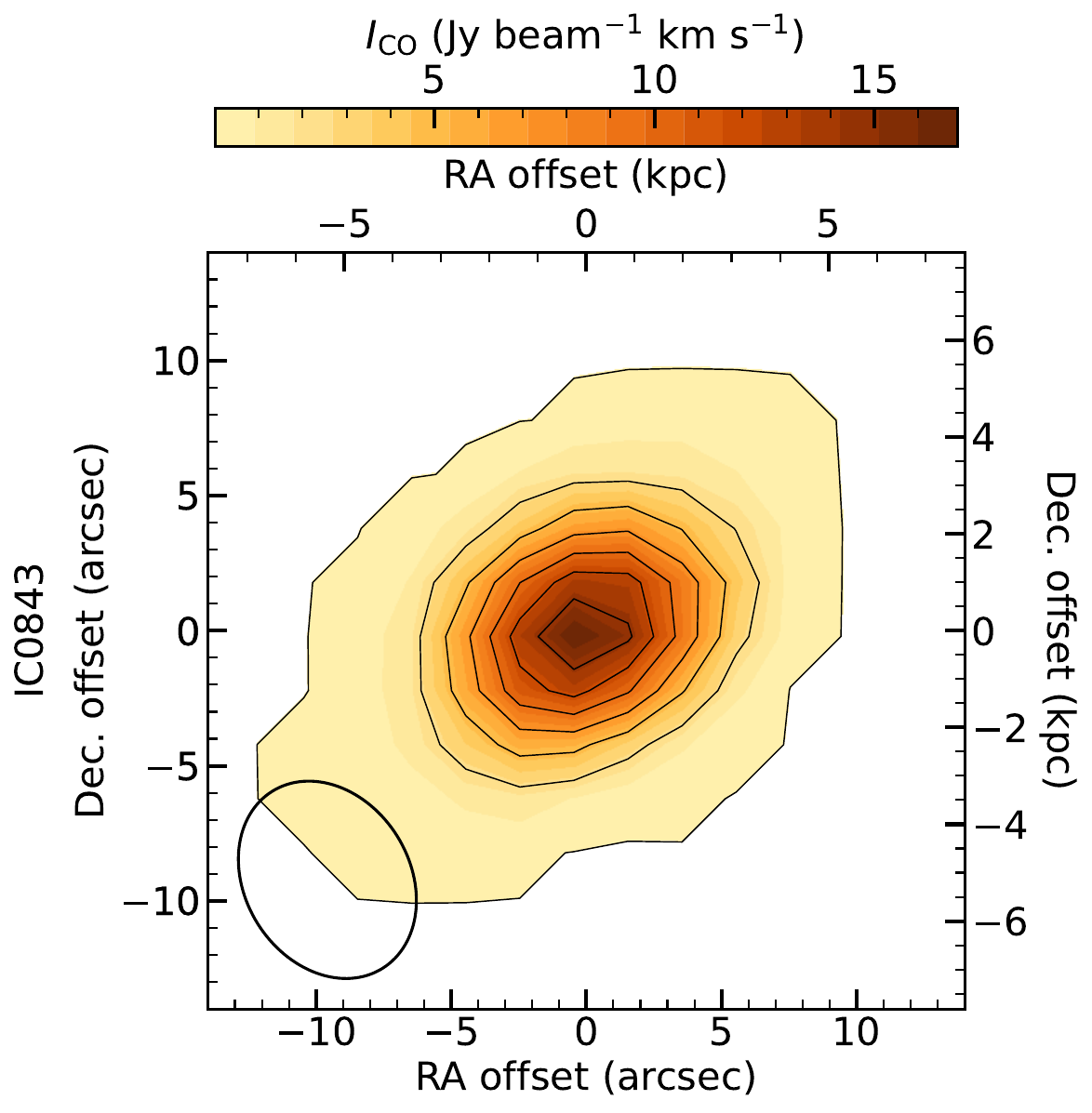}
    \end{subfigure}
    \begin{subfigure}{.32\linewidth}
    \centering
    \includegraphics[height=6.0cm]{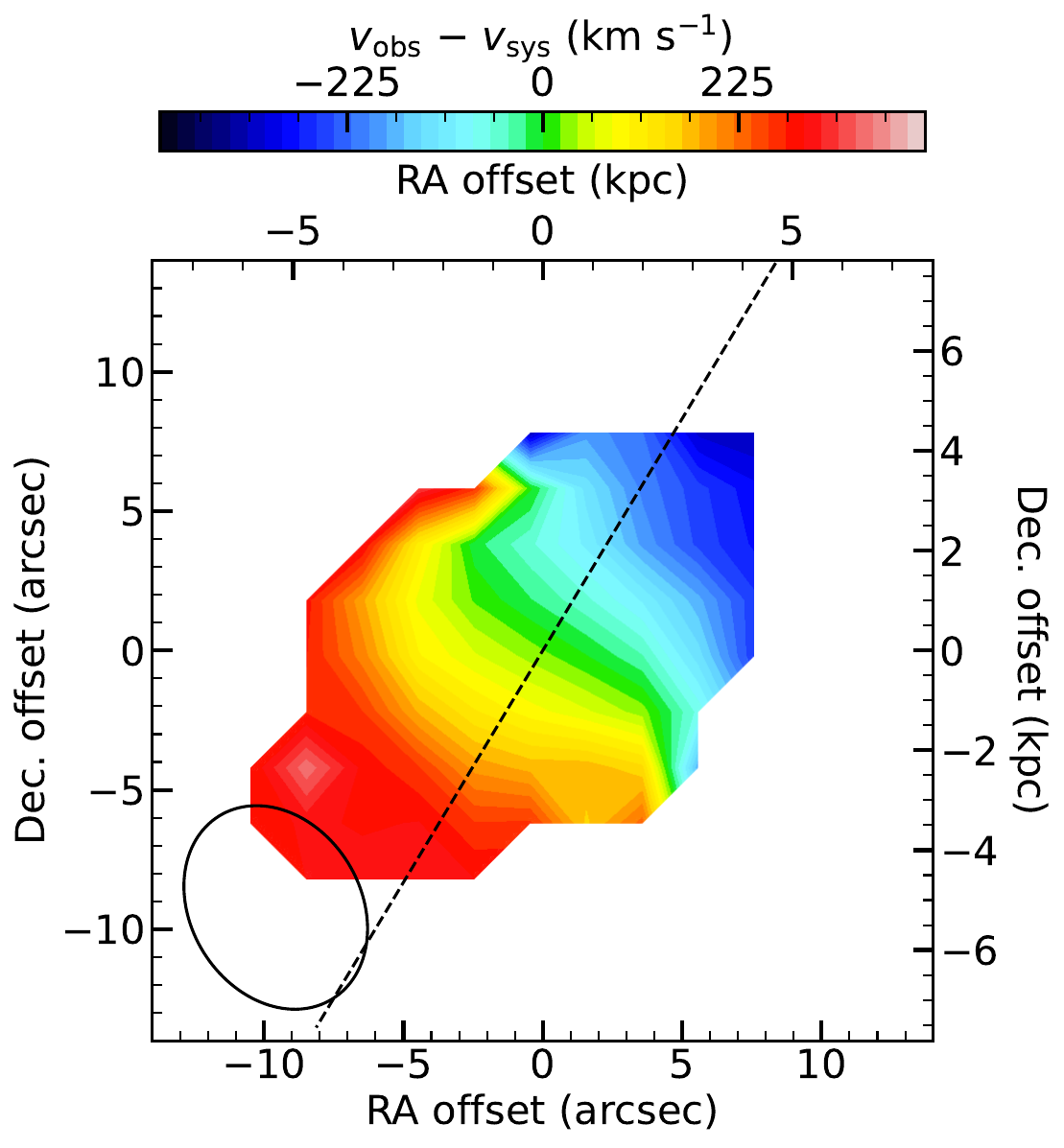}
    \end{subfigure}
    \begin{subfigure}{.32\linewidth}
    \centering
    \includegraphics[height=6.0cm]{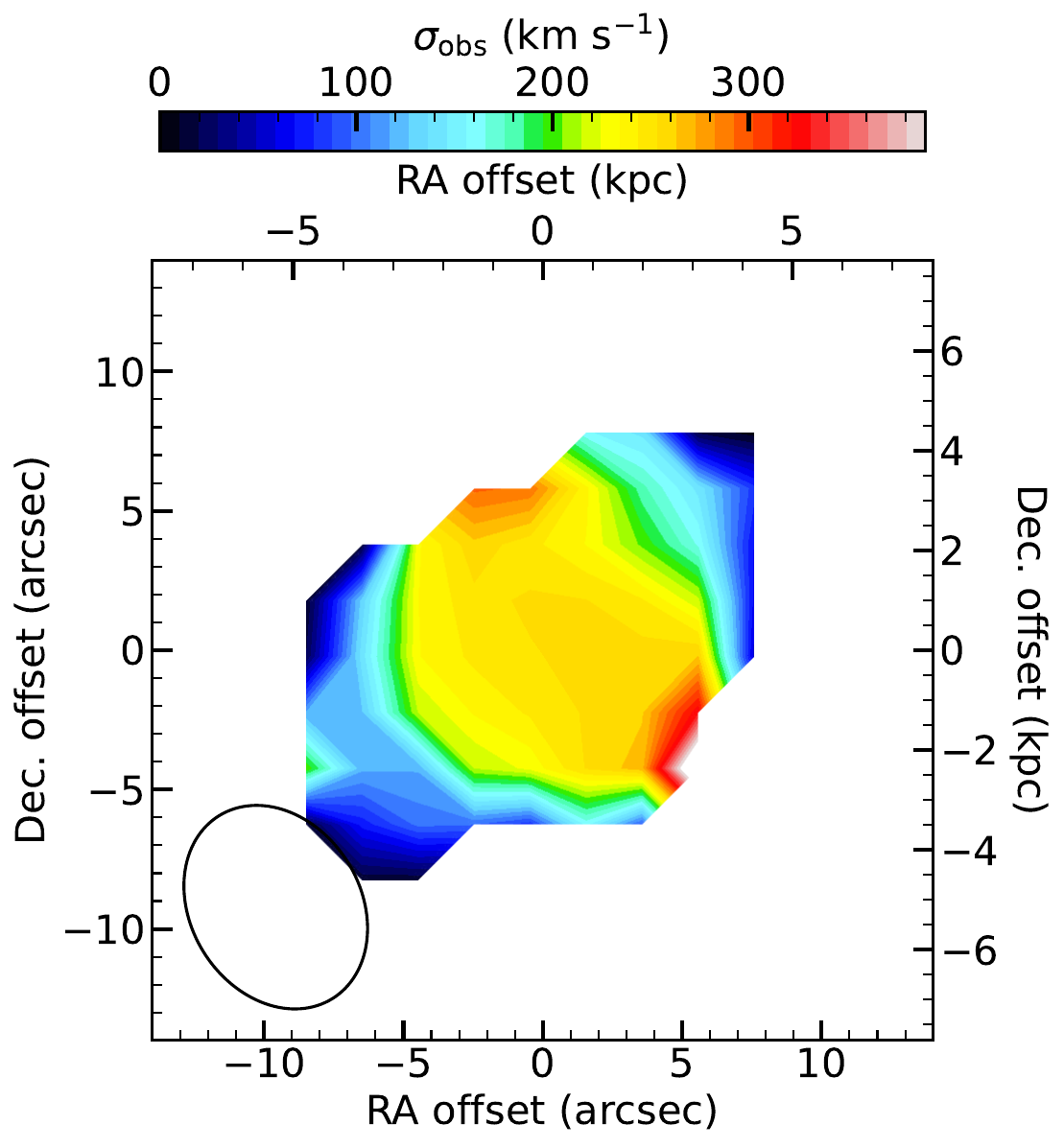}
    \end{subfigure}
    \\
    \begin{subfigure}{.34\linewidth}
    \centering
    \includegraphics[height=5.2cm]{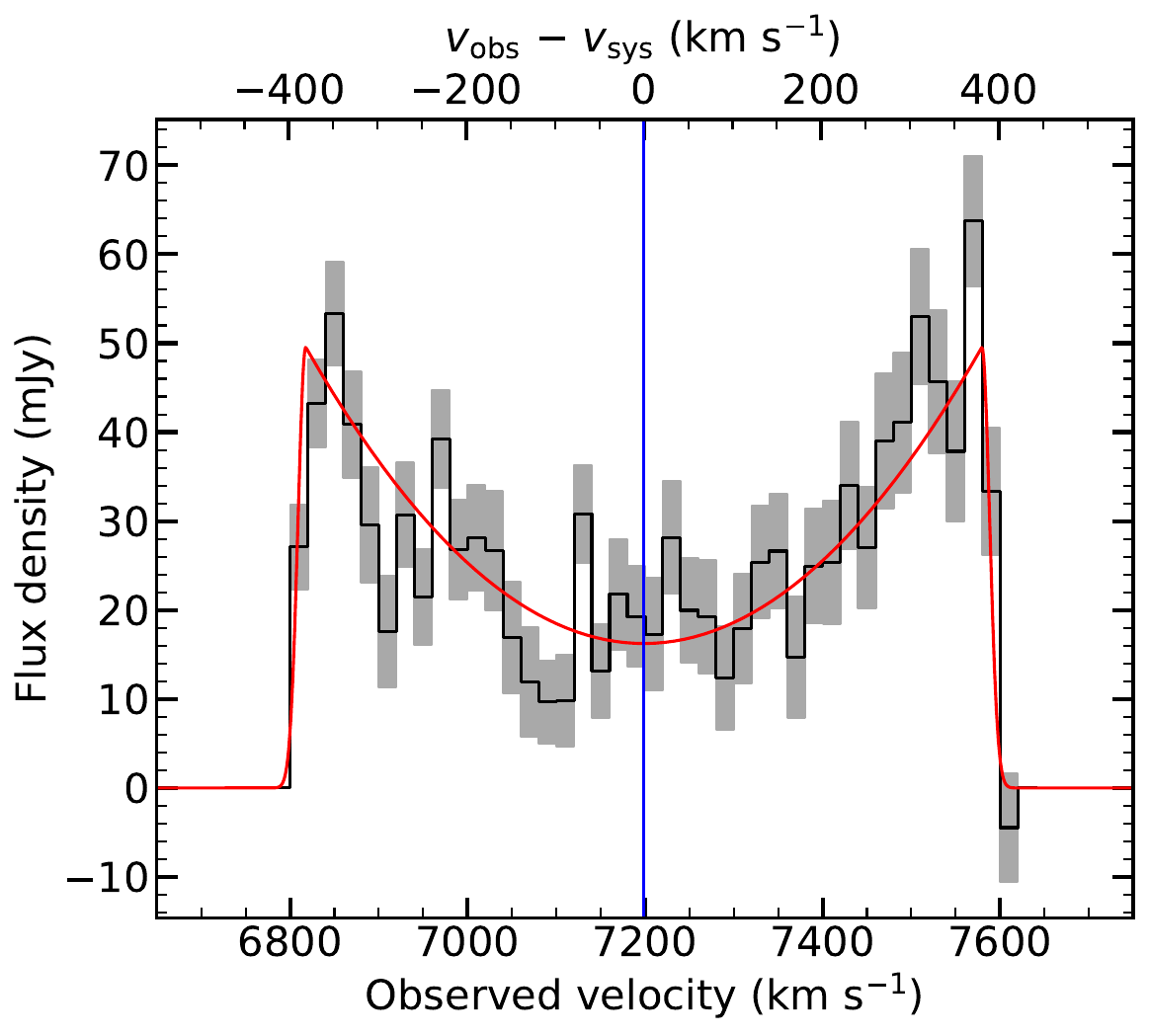}
    \end{subfigure}
    \begin{subfigure}{.32\linewidth}
    \centering
    \includegraphics[height=4.8cm]{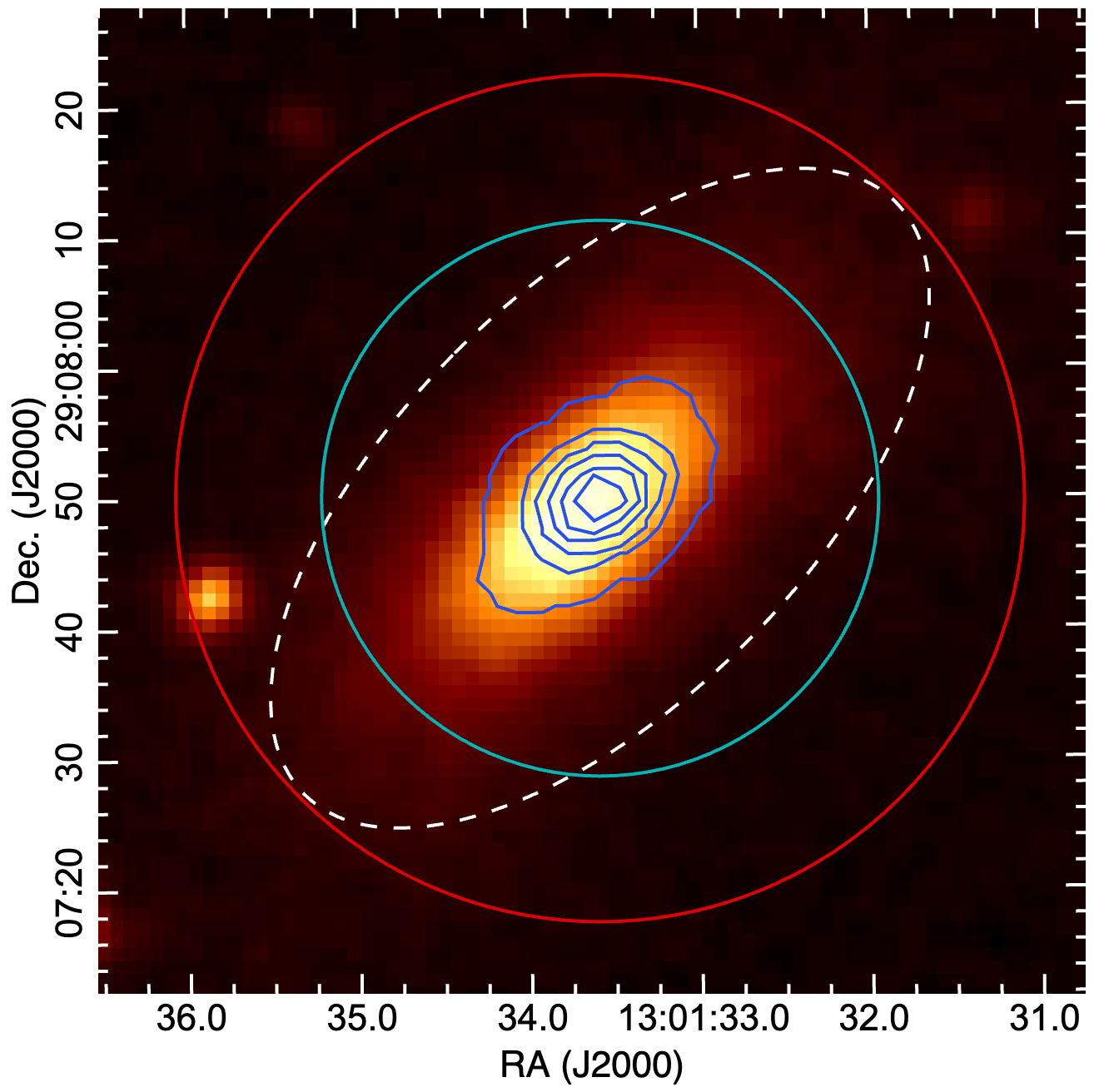}
    \end{subfigure}
    \begin{subfigure}{.33\linewidth}
    \centering
    \includegraphics[height=5.2cm]{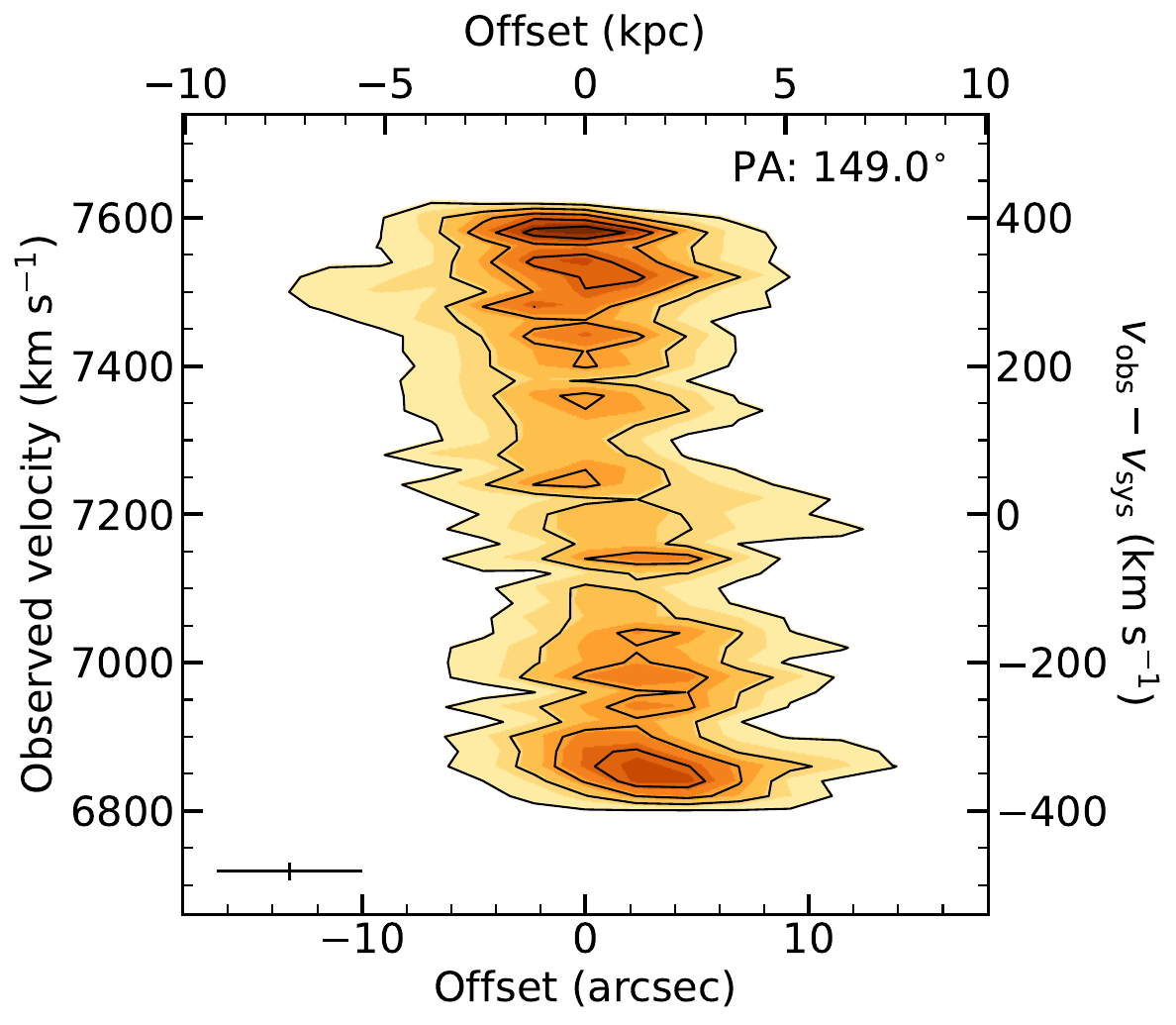}
    \end{subfigure}
    \\
    \begin{subfigure}{.34\linewidth}
    \centering
    \includegraphics[height=6.0cm]{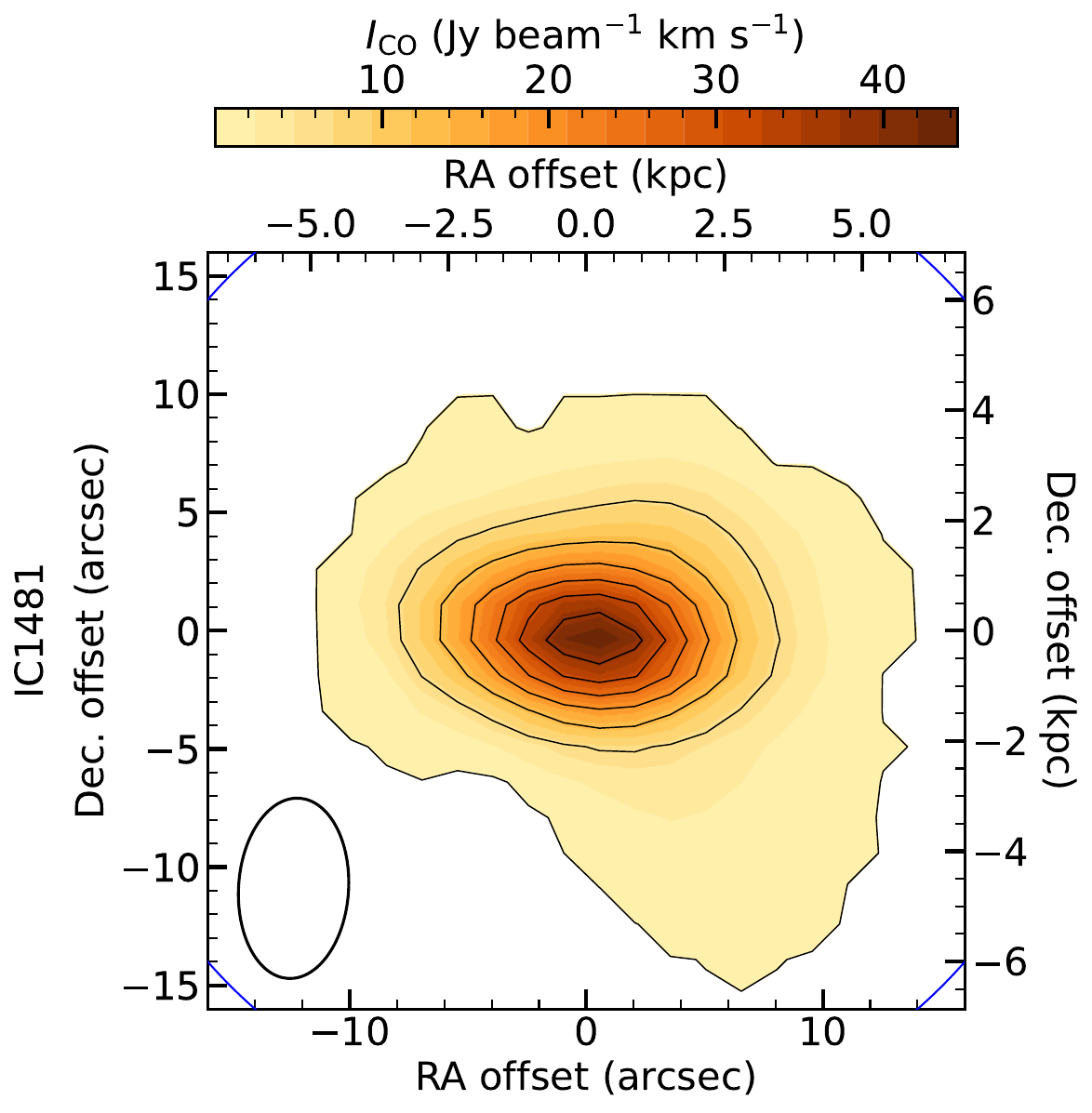}
    \end{subfigure}
    \begin{subfigure}{.32\linewidth}
    \centering
    \includegraphics[height=6.0cm]{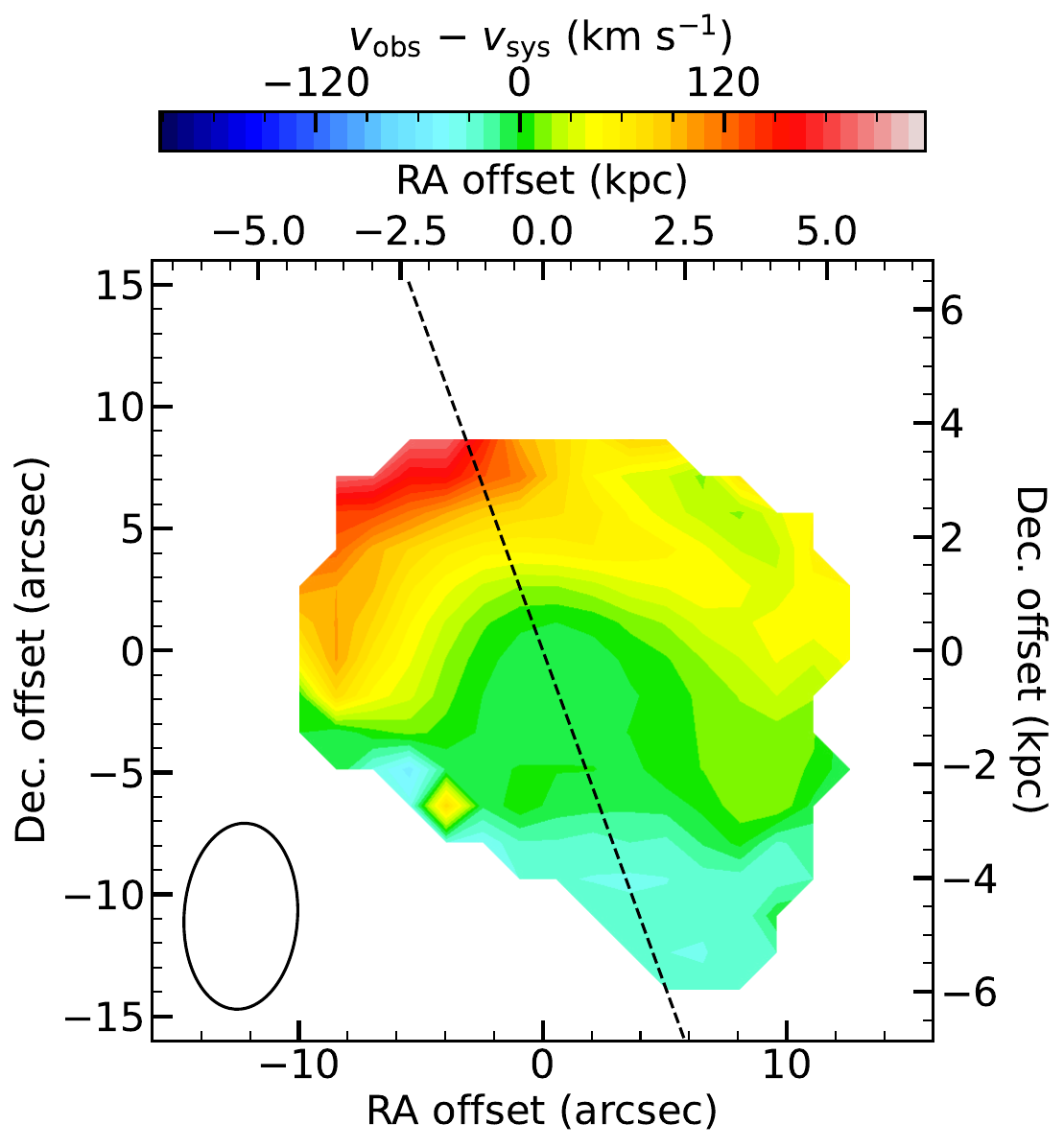}
    \end{subfigure}
    \begin{subfigure}{.32\linewidth}
    \centering
    \includegraphics[height=6.0cm]{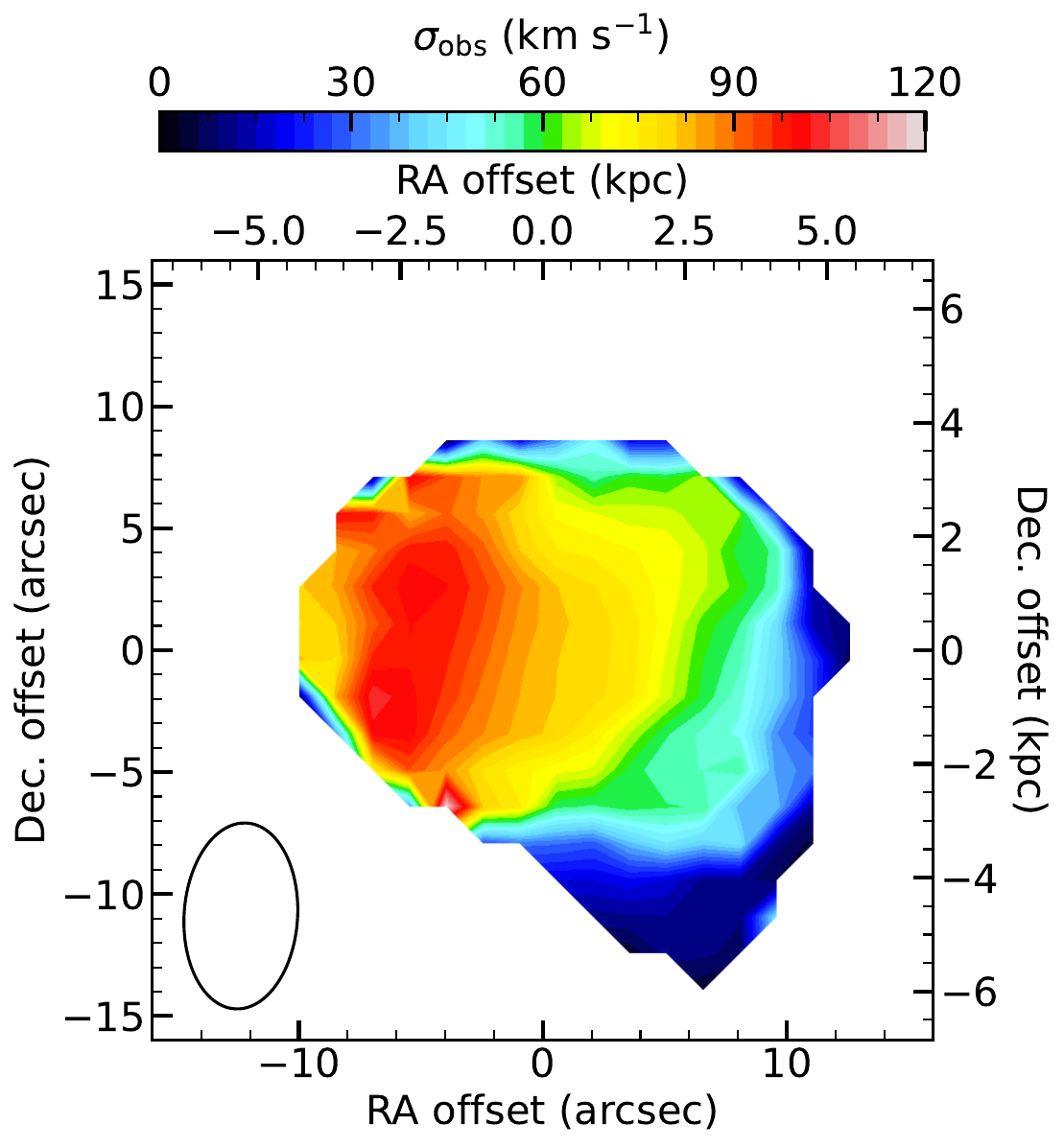}
    \end{subfigure}
    \\
    \begin{subfigure}{.34\linewidth}
    \centering
    \includegraphics[height=5.3cm]{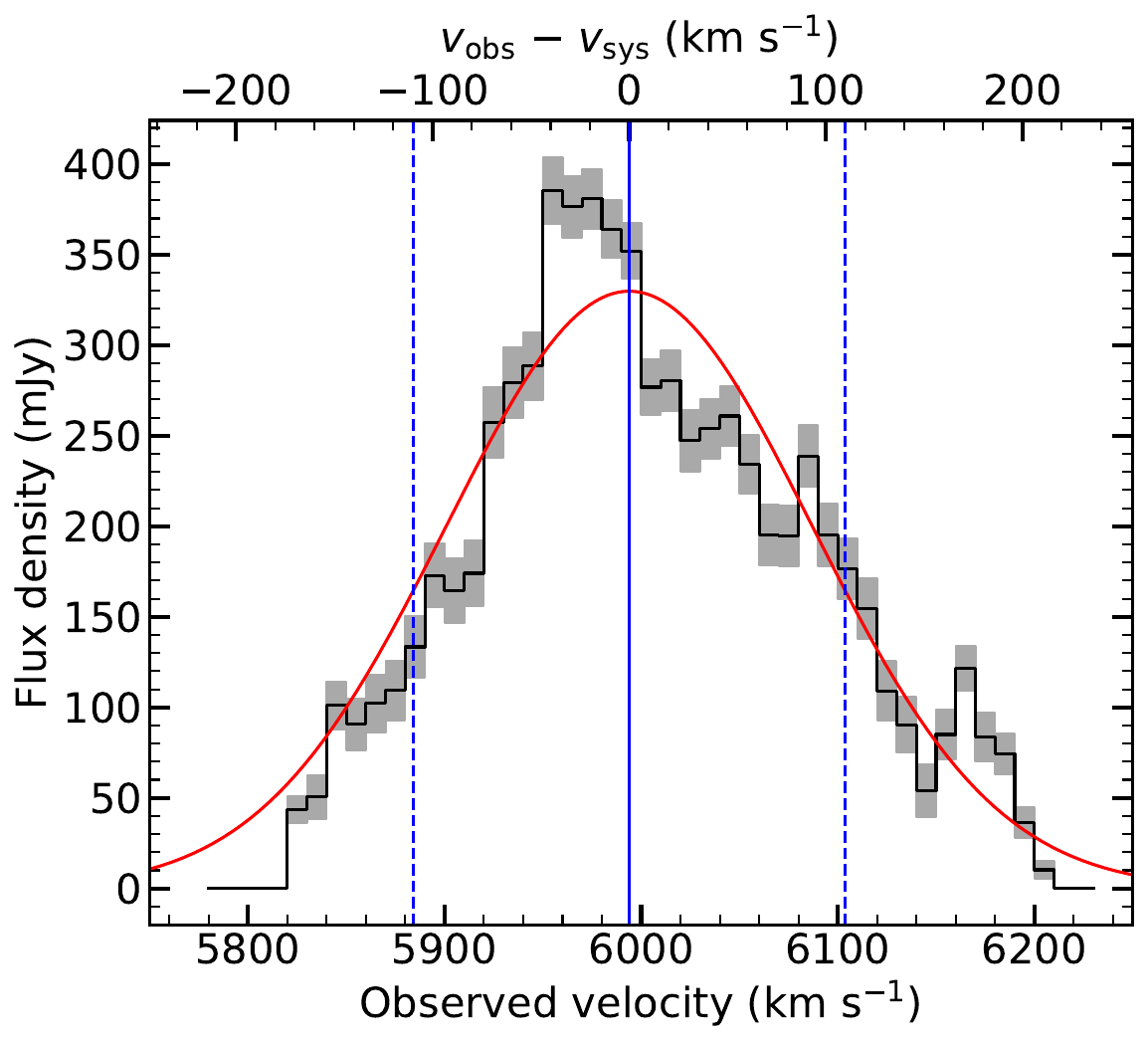}
    \end{subfigure}
    \begin{subfigure}{.32\linewidth}
    \centering
    \includegraphics[height=4.8cm]{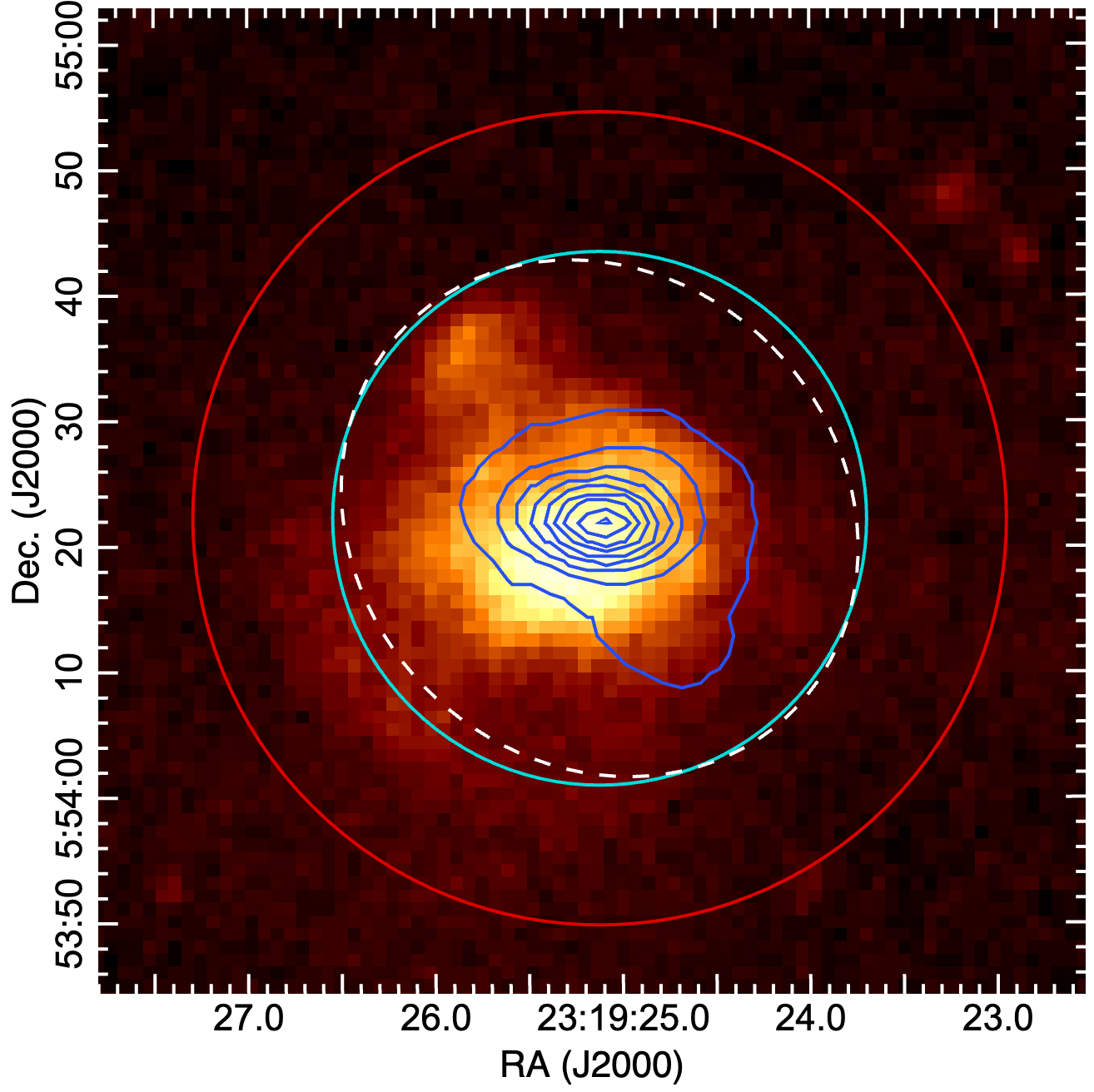}
    \end{subfigure}
    \begin{subfigure}{.33\linewidth}
    \centering
    \includegraphics[height=5.3cm]{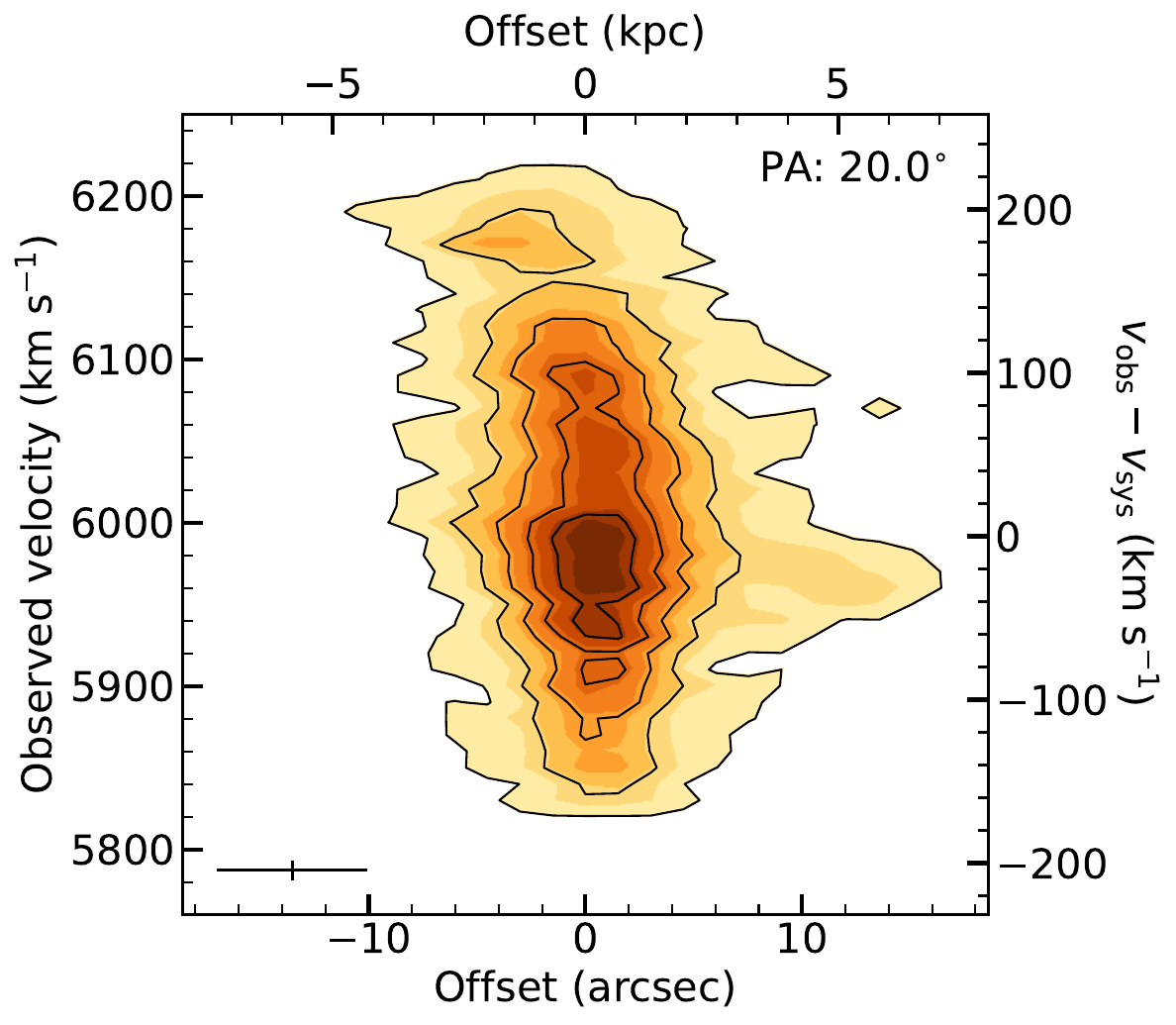}
    \end{subfigure}
    \caption[]{continued.}
\end{figure*}

\begin{figure*}
    \ContinuedFloat
    \centering
    \begin{subfigure}{.34\linewidth}
    \centering
    \includegraphics[height=6.0cm]{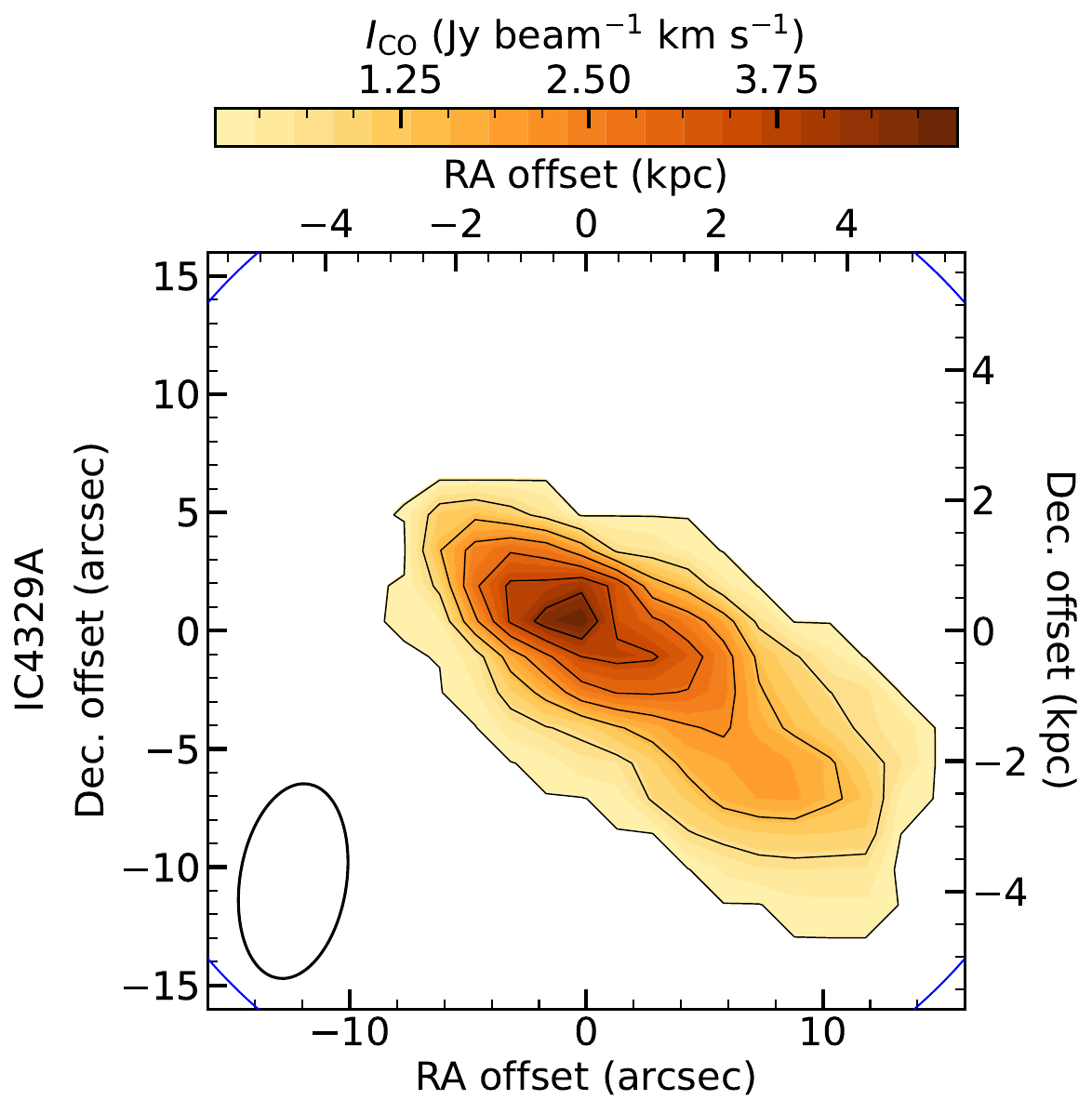}
    \end{subfigure}
    \begin{subfigure}{.32\linewidth}
    \centering
    \includegraphics[height=6.0cm]{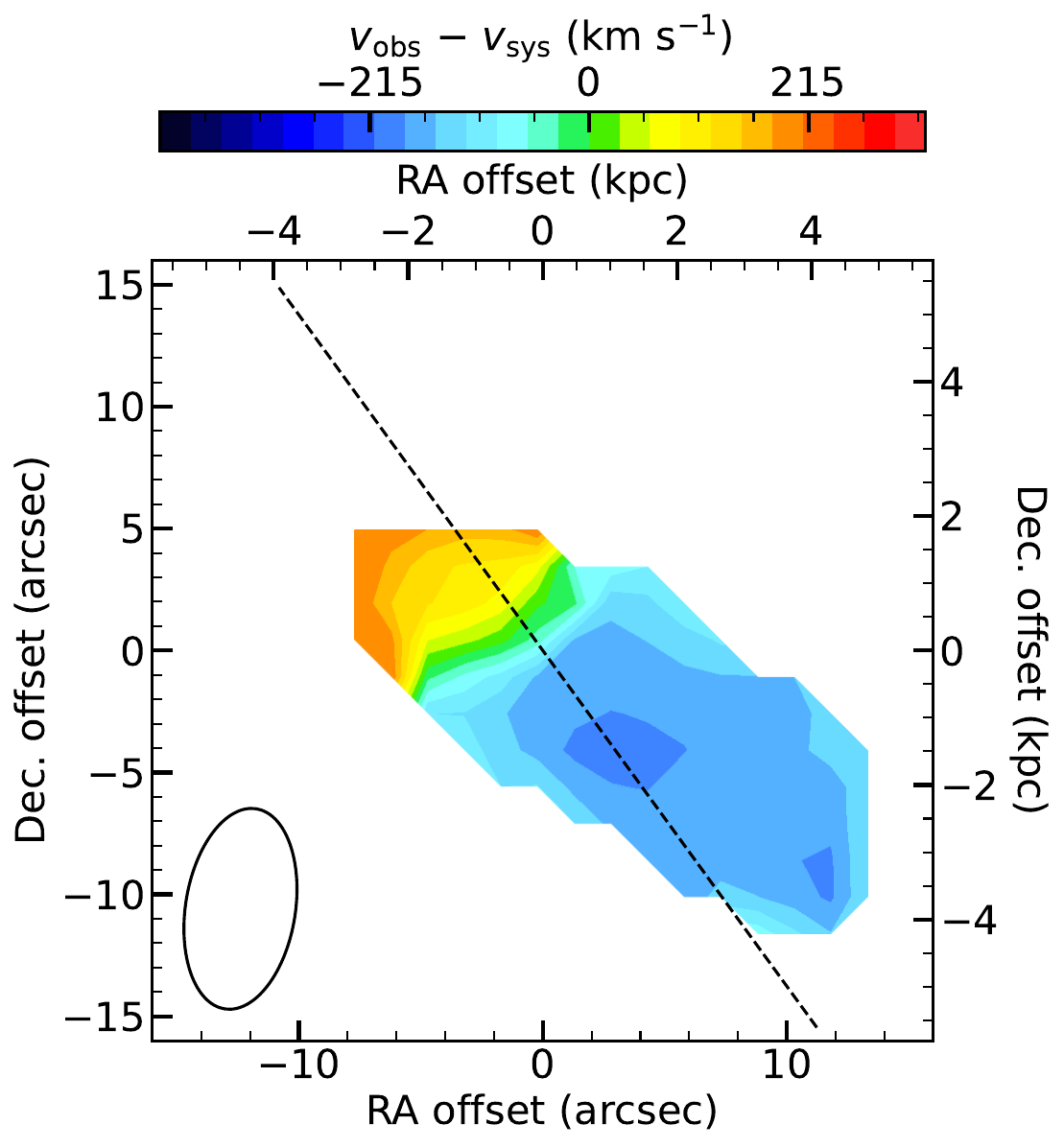}
    \end{subfigure}
    \begin{subfigure}{.32\linewidth}
    \centering
    \includegraphics[height=6.0cm]{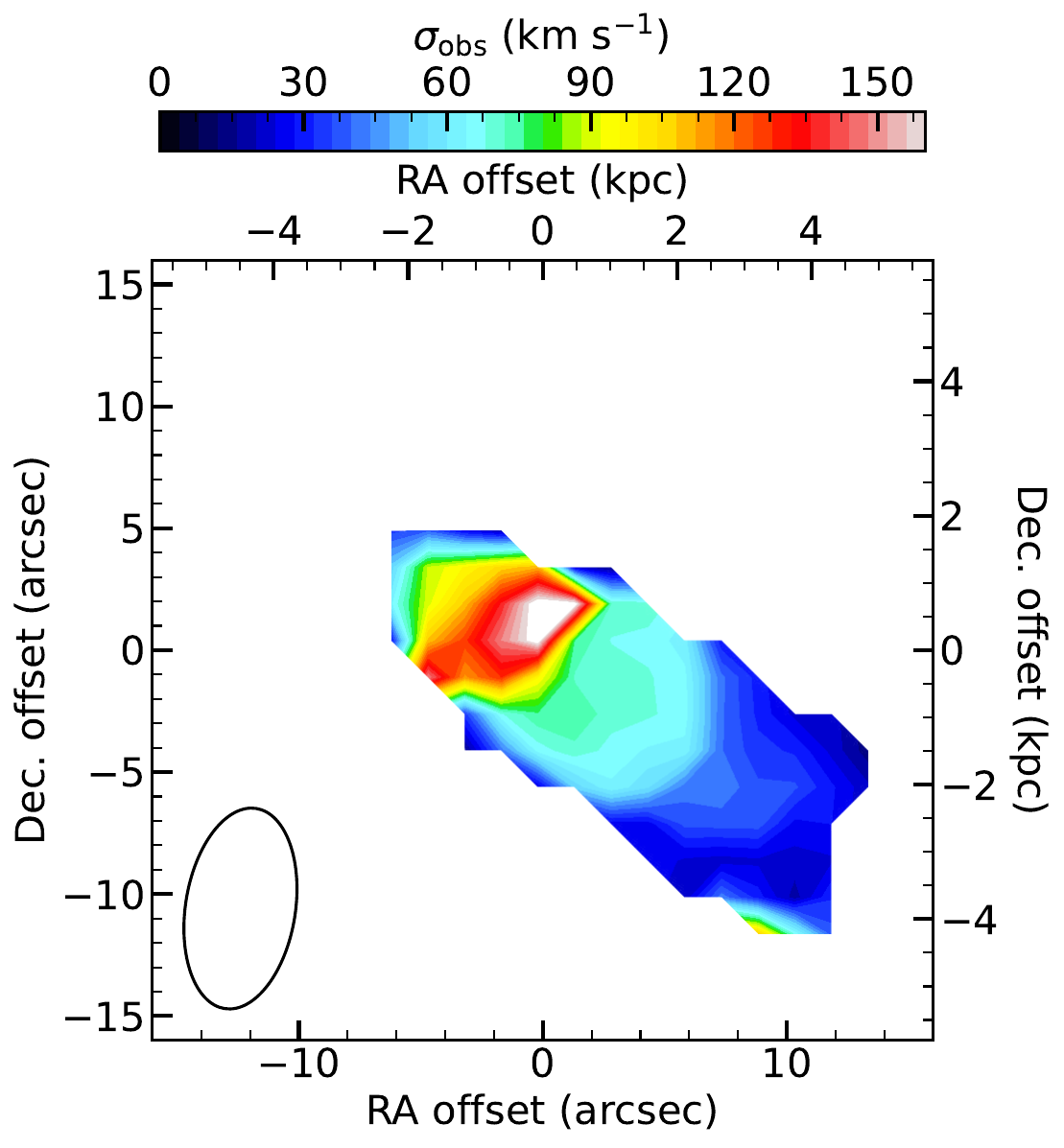}
    \end{subfigure}
    \\
    \begin{subfigure}{.34\linewidth}
    \centering
    \includegraphics[height=5.2cm]{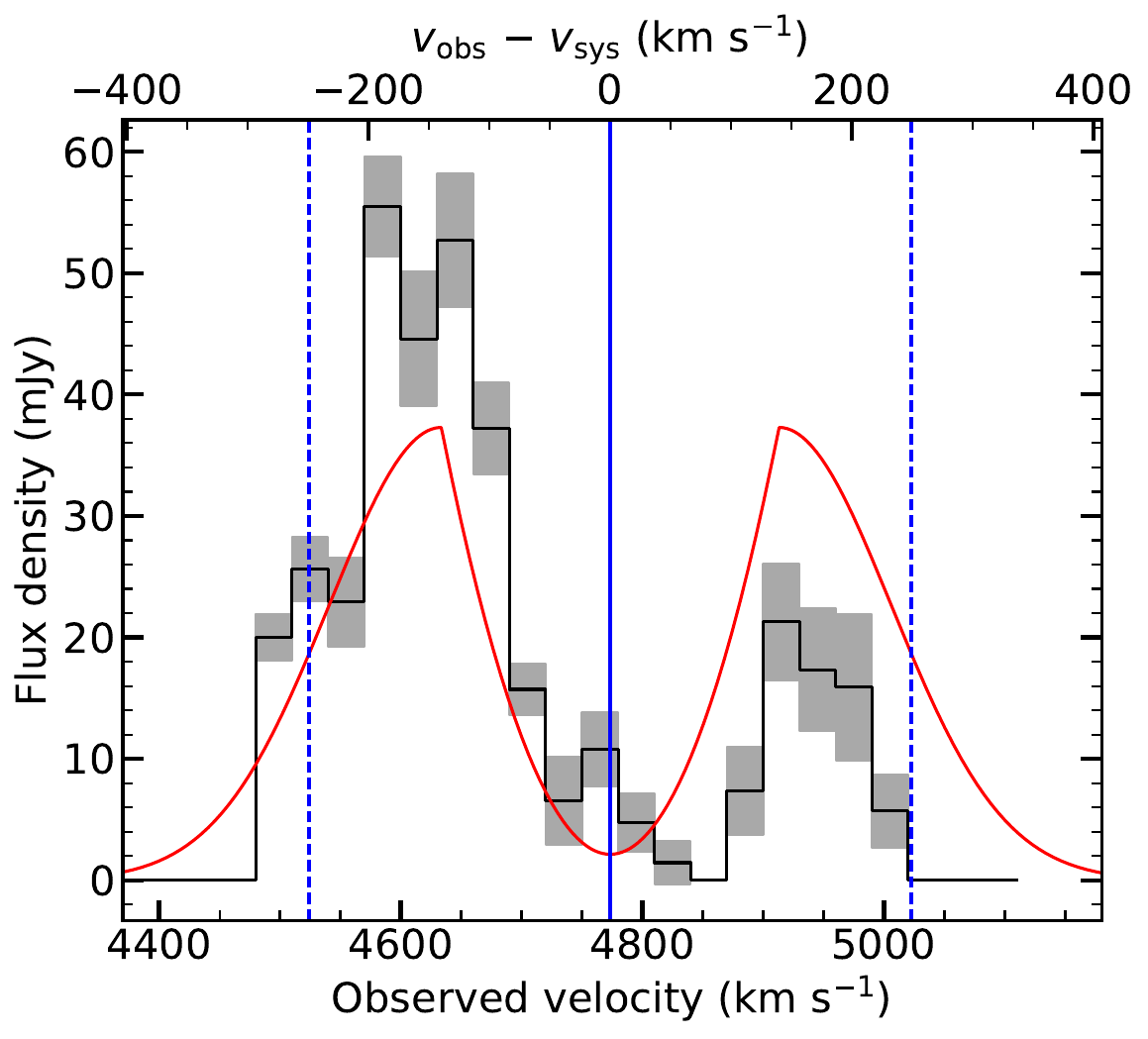}
    \end{subfigure}
    \begin{subfigure}{.32\linewidth}
    \centering
    \includegraphics[height=4.8cm]{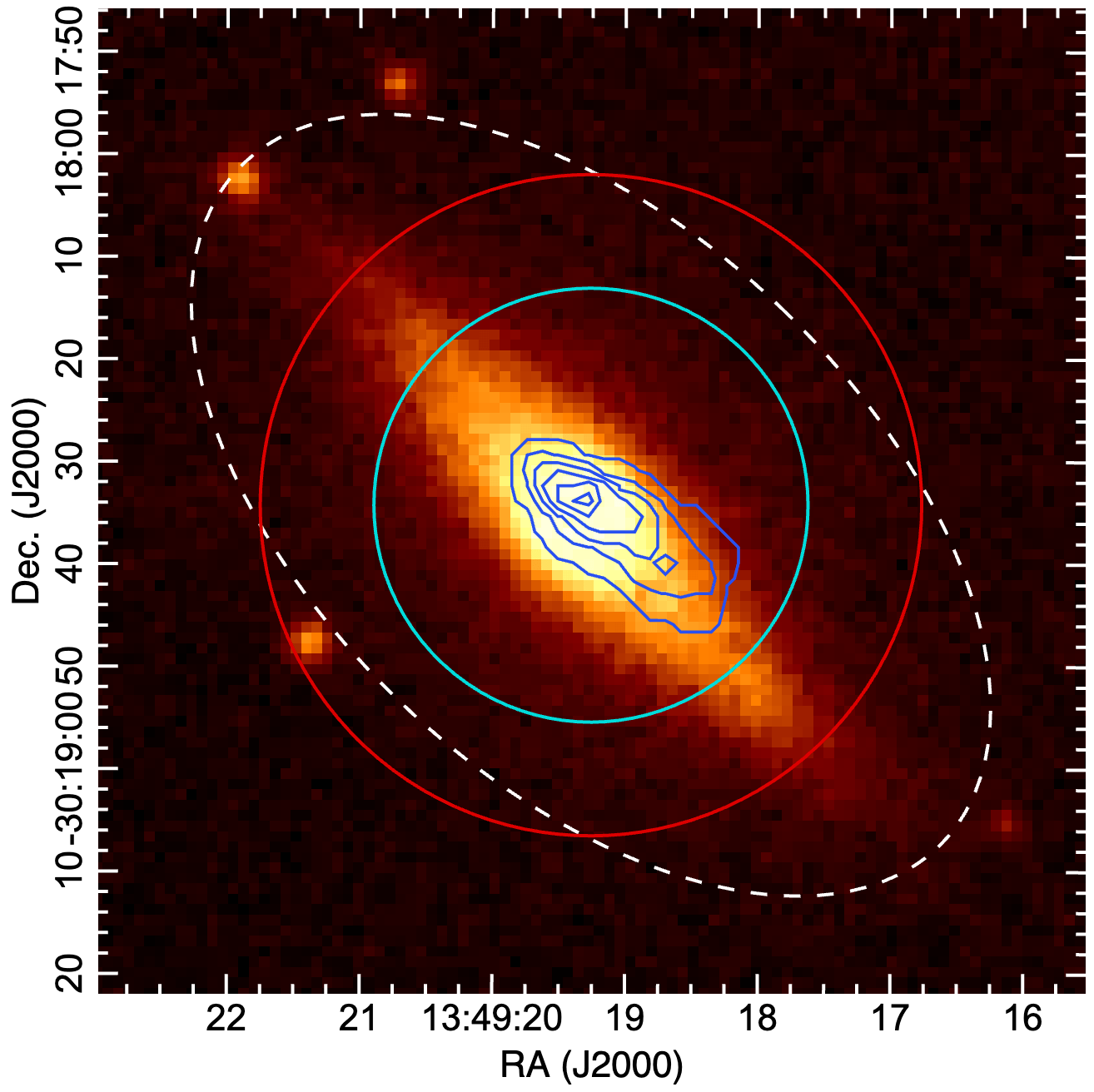}
    \end{subfigure}
    \begin{subfigure}{.33\linewidth}
    \centering
    \includegraphics[height=5.2cm]{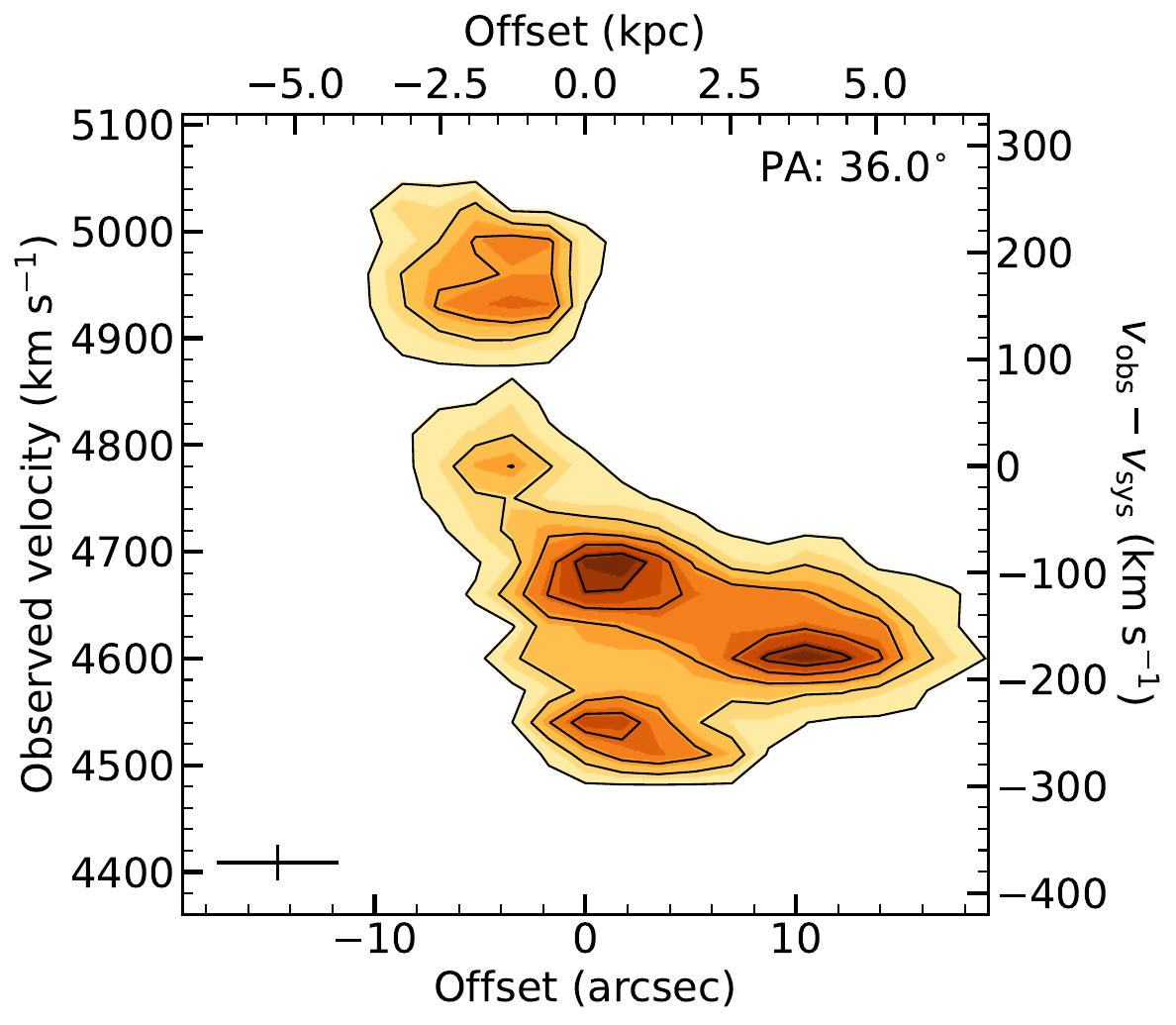}
    \end{subfigure}
    \\
    \begin{subfigure}{.34\linewidth}
    \centering
    \includegraphics[height=6.0cm]{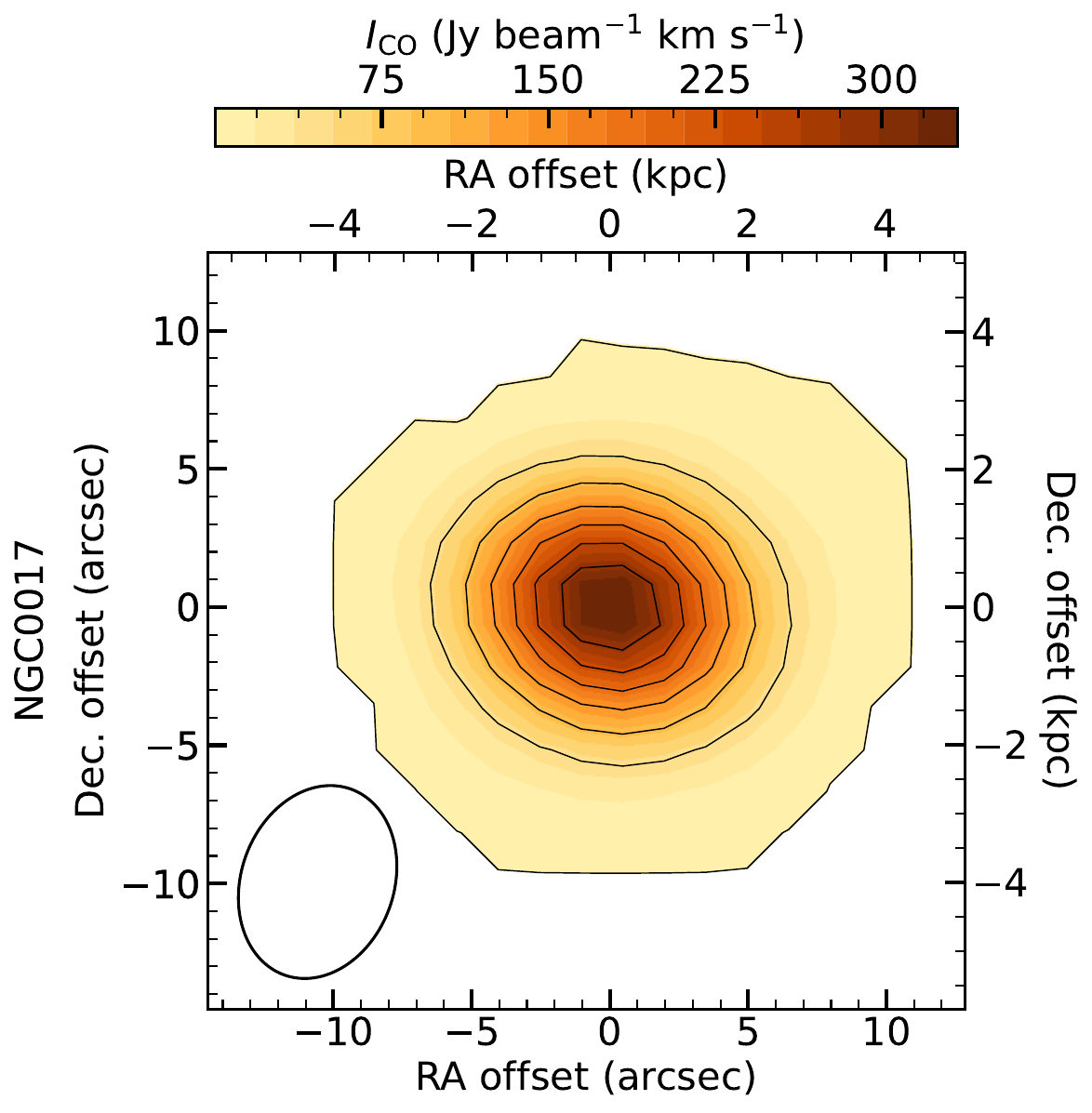}
    \end{subfigure}
    \begin{subfigure}{.32\linewidth}
    \centering
    \includegraphics[height=6.0cm]{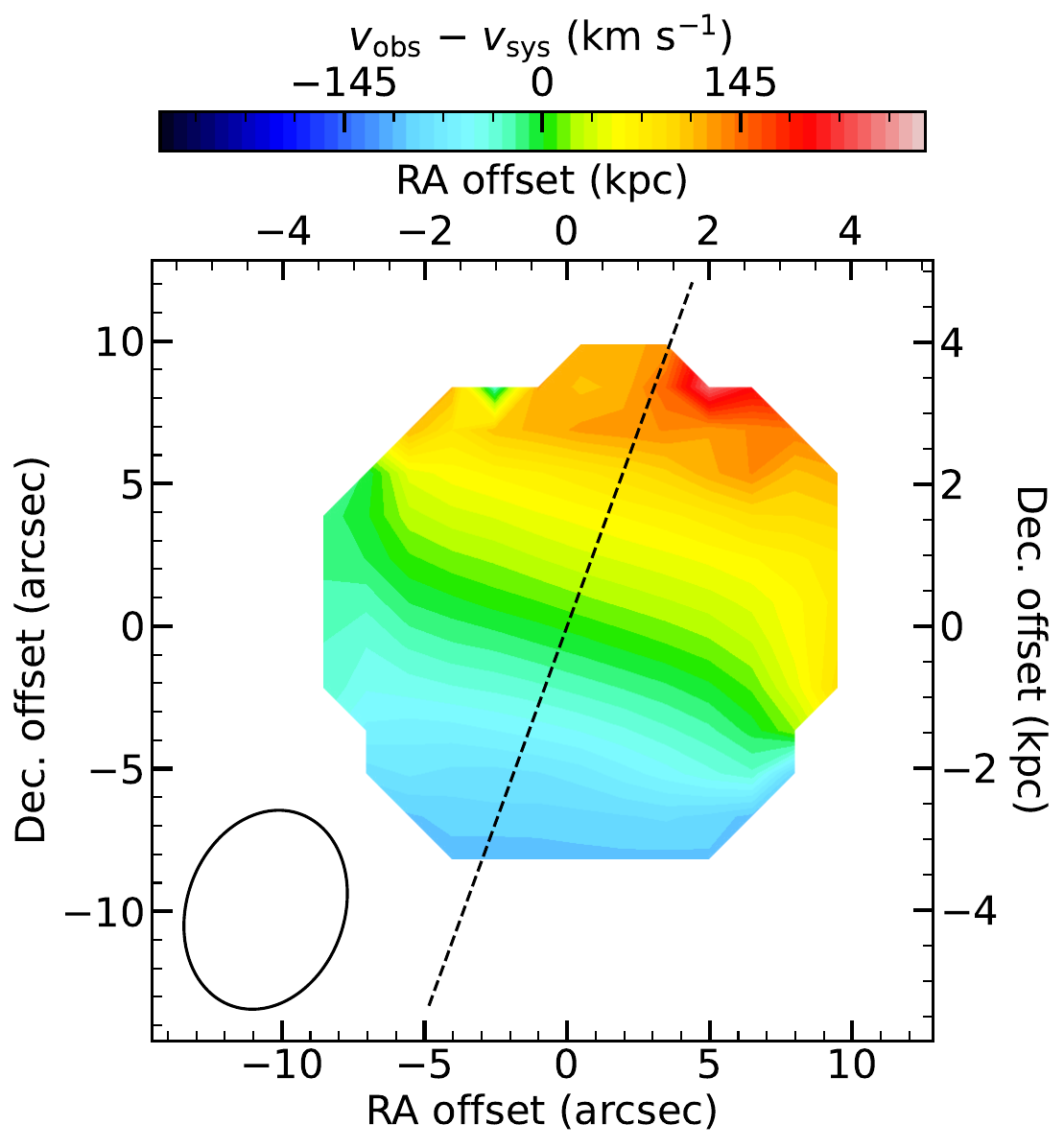}
    \end{subfigure}
    \begin{subfigure}{.32\linewidth}
    \centering
    \includegraphics[height=6.0cm]{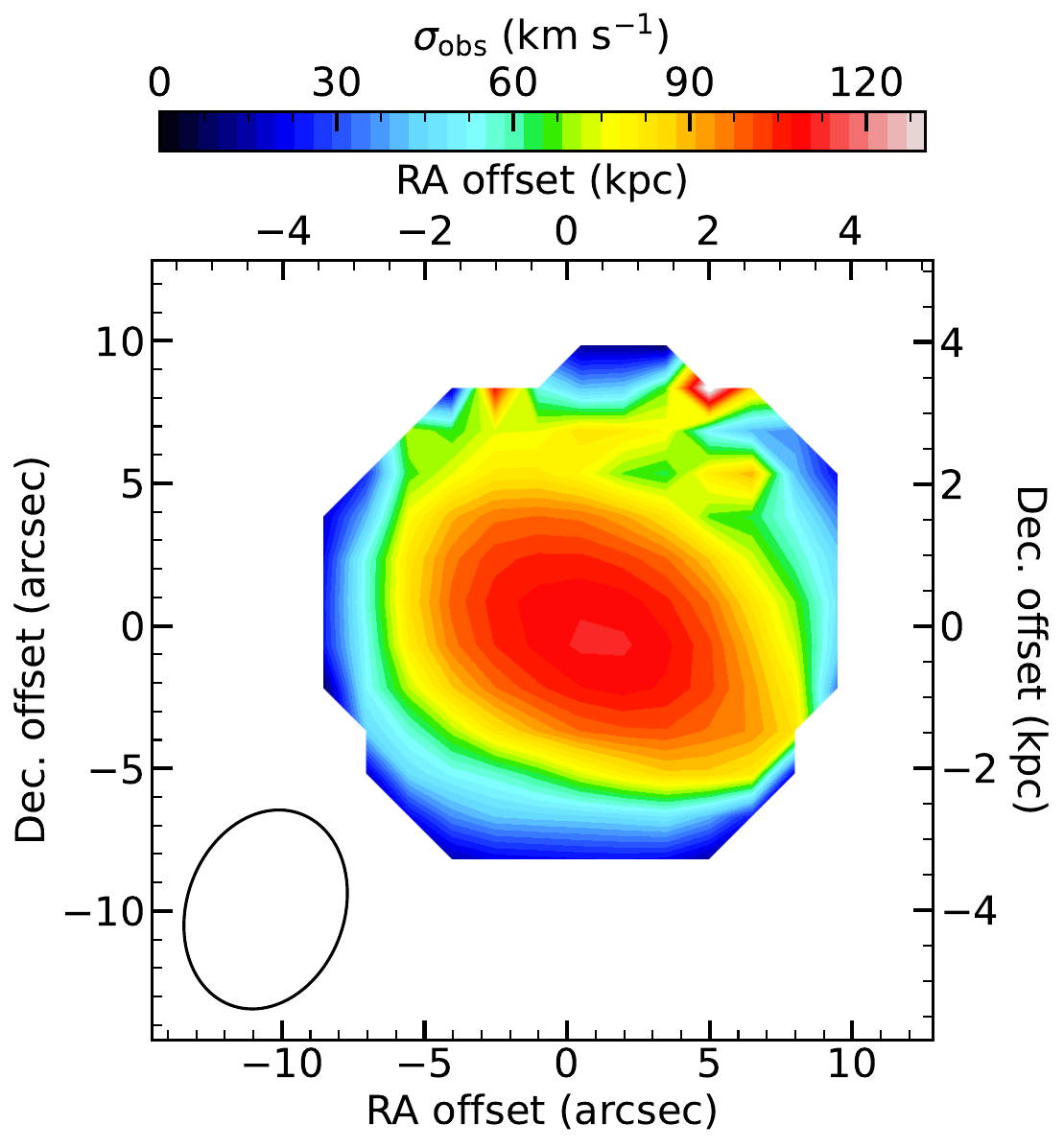}
    \end{subfigure}
    \\
    \begin{subfigure}{.34\linewidth}
    \centering
    \includegraphics[height=5.2cm]{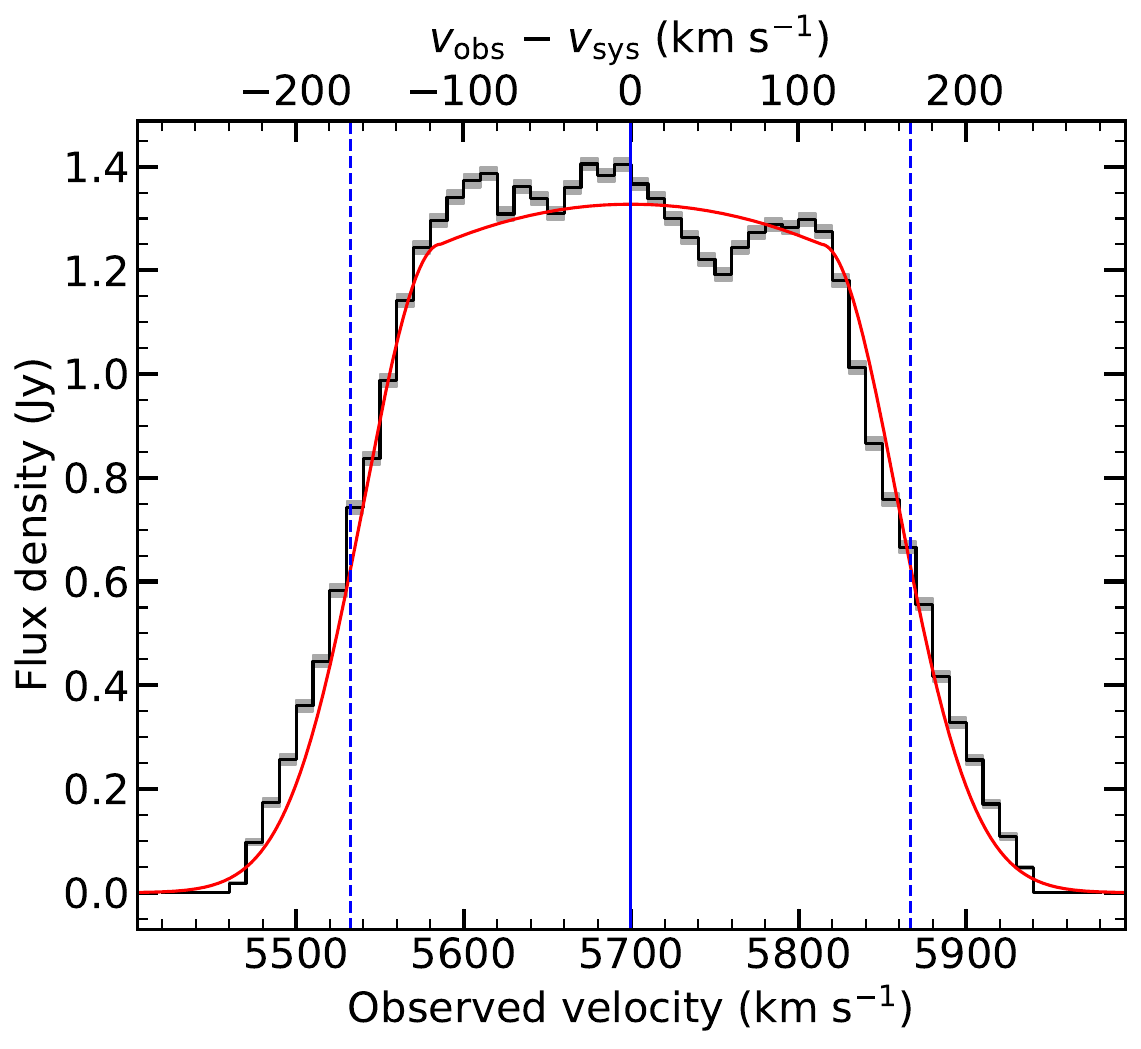}
    \end{subfigure}
    \begin{subfigure}{.32\linewidth}
    \centering
    \includegraphics[height=4.8cm]{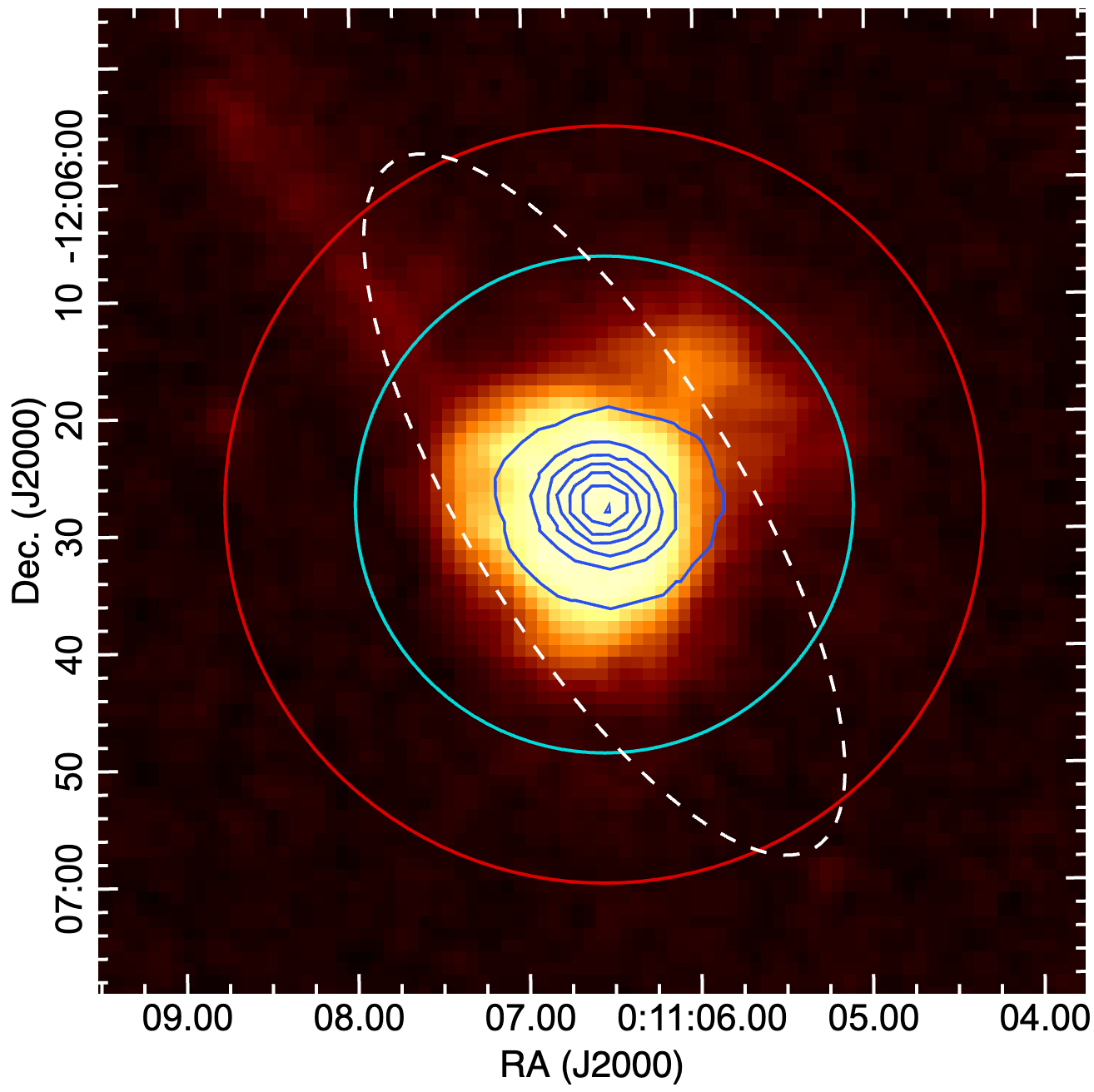}
    \end{subfigure}
    \begin{subfigure}{.33\linewidth}
    \centering
    \includegraphics[height=5.2cm]{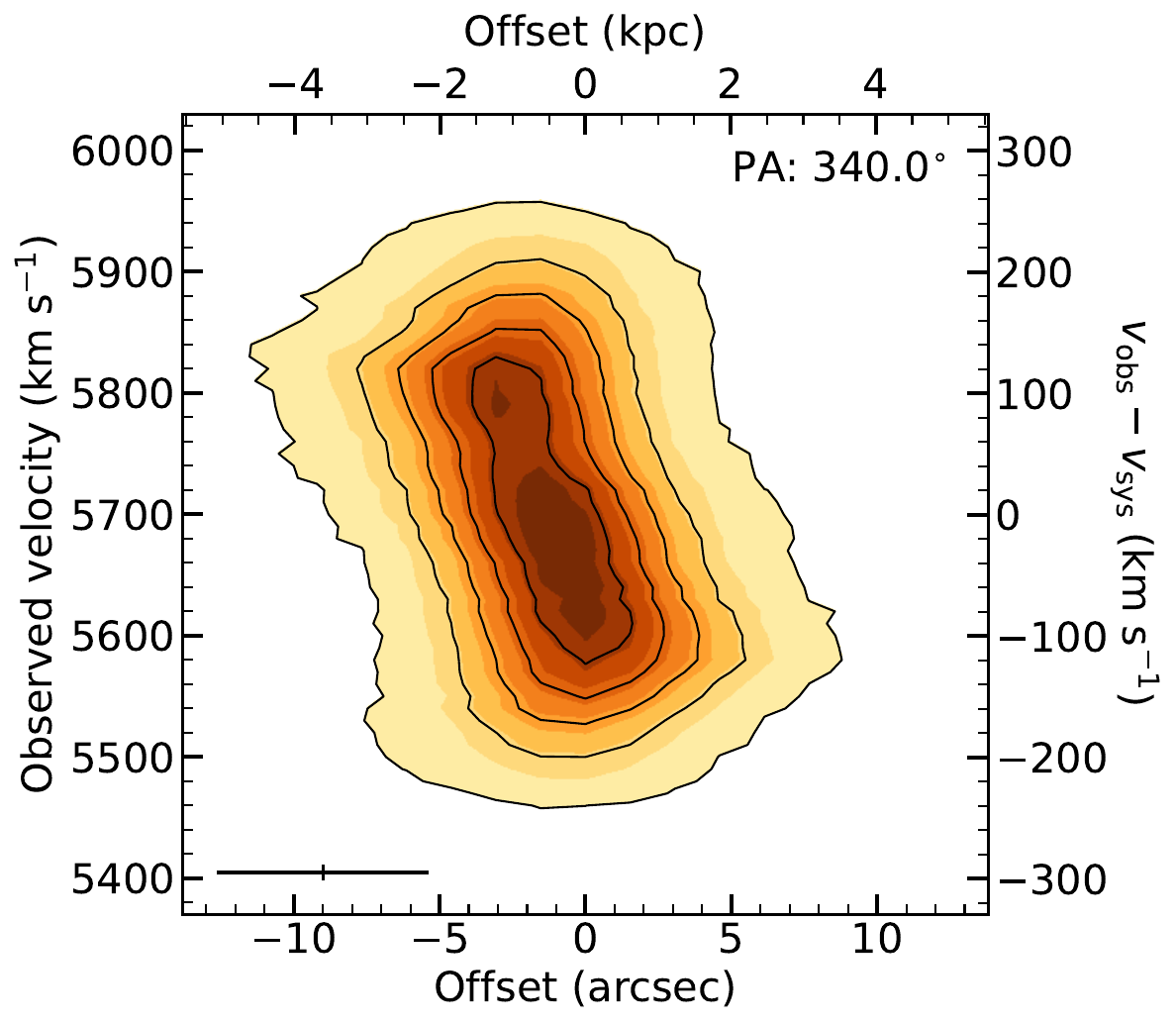}
    \end{subfigure}
    \caption[]{continued.}
\end{figure*}

\begin{figure*}
    \ContinuedFloat
    \begin{subfigure}{.34\linewidth}
    \centering
    \includegraphics[height=6.0cm]{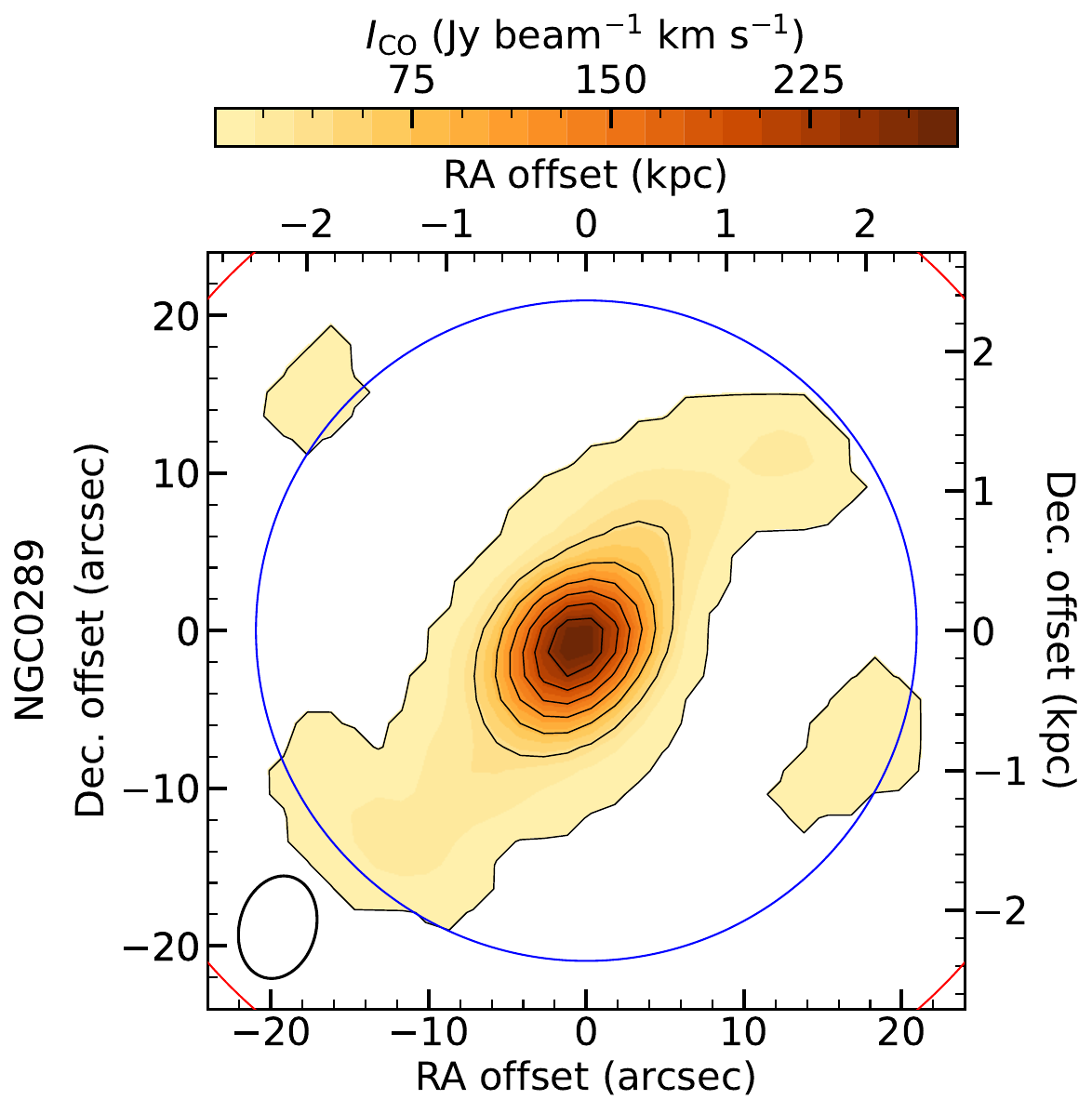}
    \end{subfigure}
    \begin{subfigure}{.32\linewidth}
    \centering
    \includegraphics[height=6.0cm]{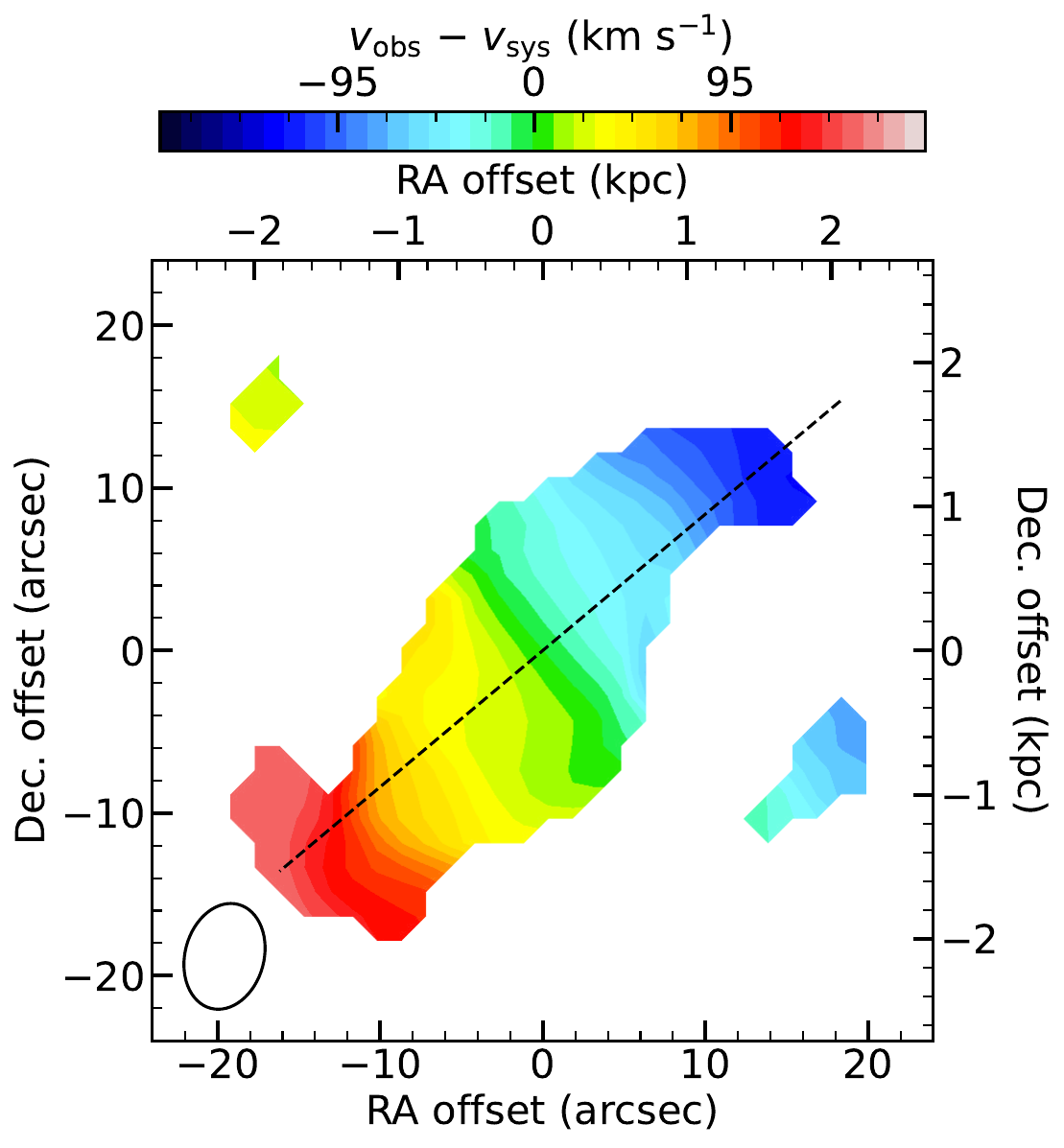}
    \end{subfigure}
    \begin{subfigure}{.32\linewidth}
    \centering
    \includegraphics[height=6.0cm]{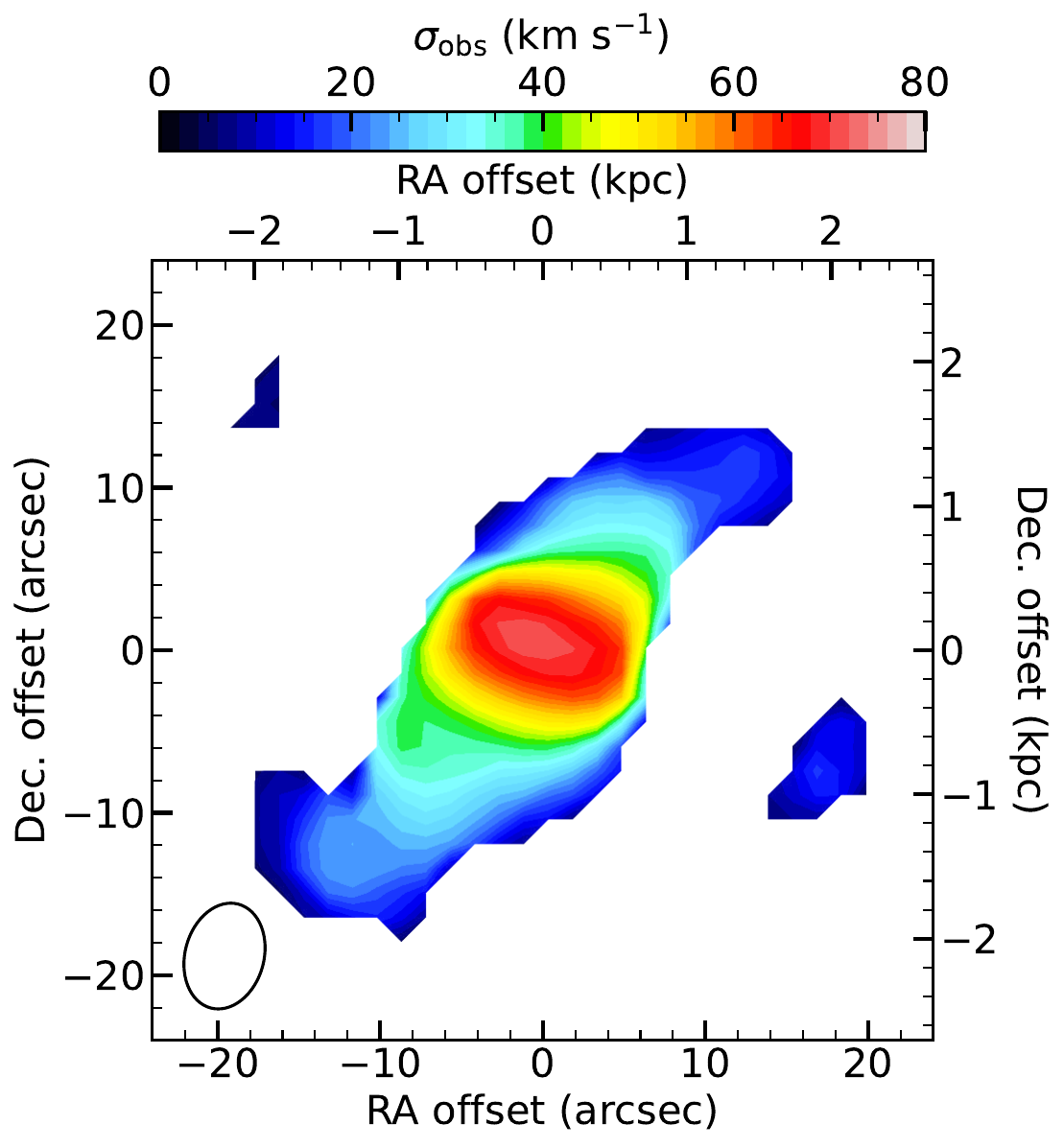}
    \end{subfigure}
    \\
    \begin{subfigure}{.34\linewidth}
    \centering
    \includegraphics[height=5.2cm]{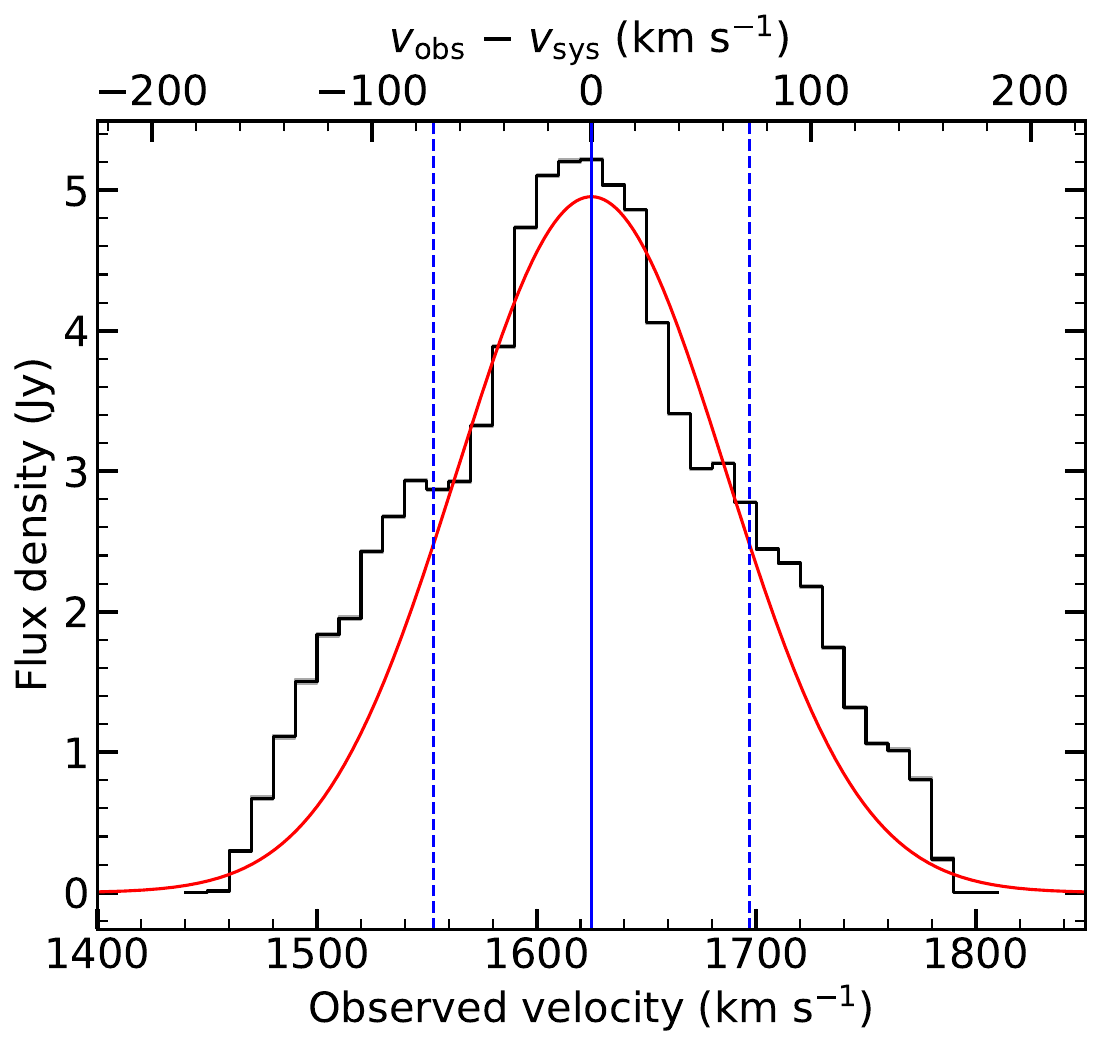}
    \end{subfigure}
    \begin{subfigure}{.32\linewidth}
    \centering
    \includegraphics[height=4.8cm]{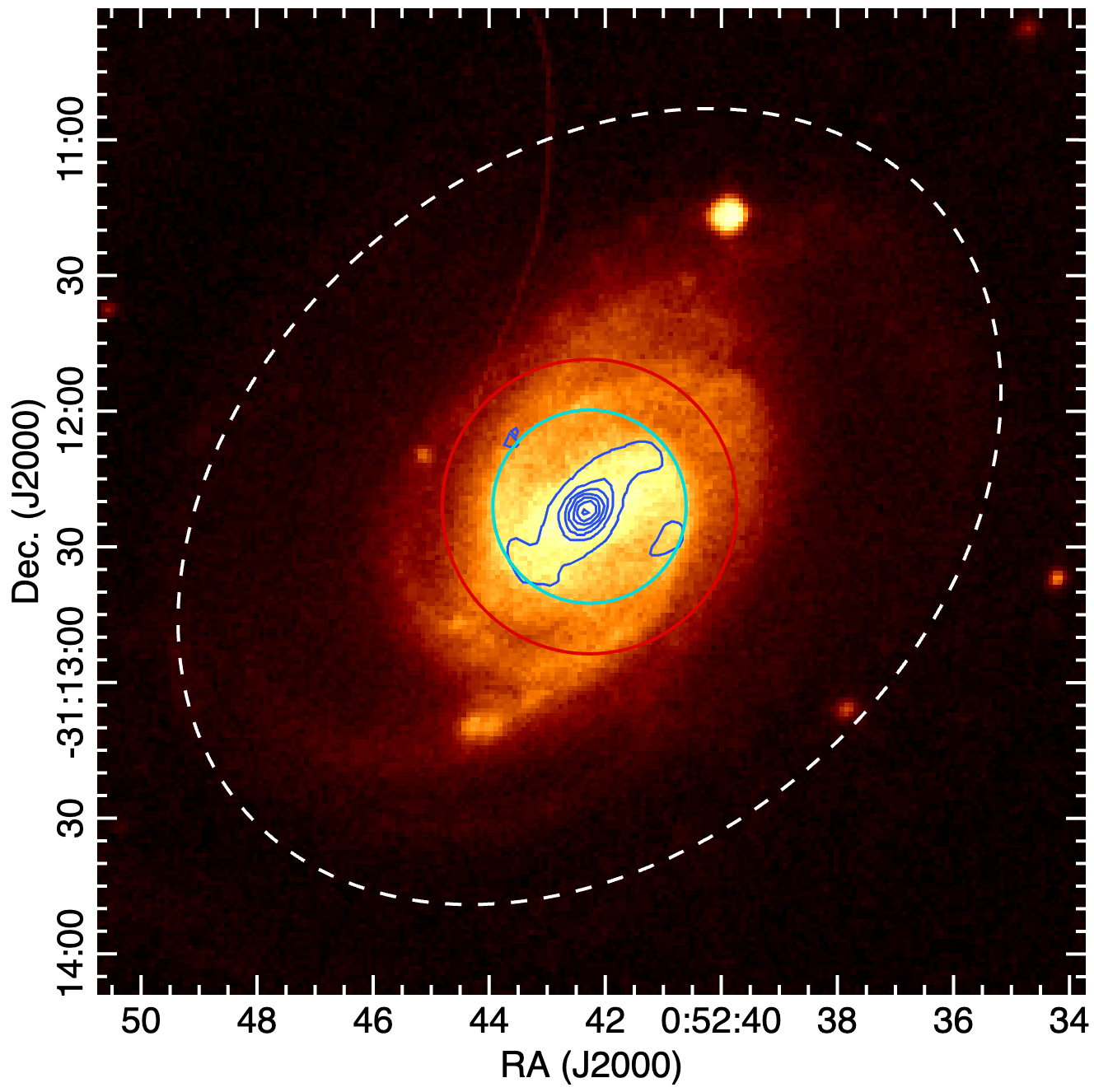}
    \end{subfigure}
    \begin{subfigure}{.33\linewidth}
    \centering
    \includegraphics[height=5.2cm]{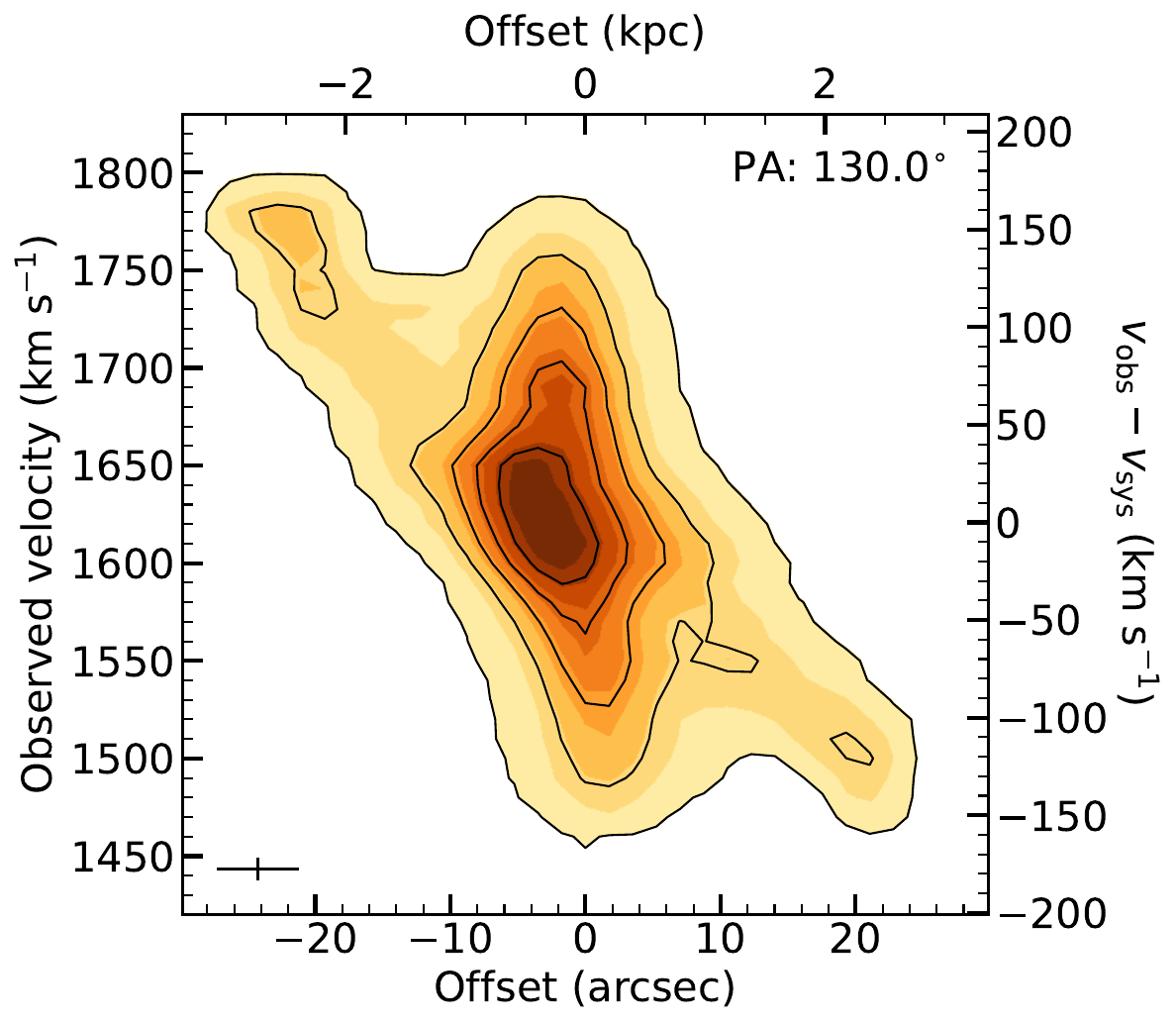}
    \end{subfigure}
    \\
    \begin{subfigure}{.34\linewidth}
    \centering
    \includegraphics[height=5.9cm]{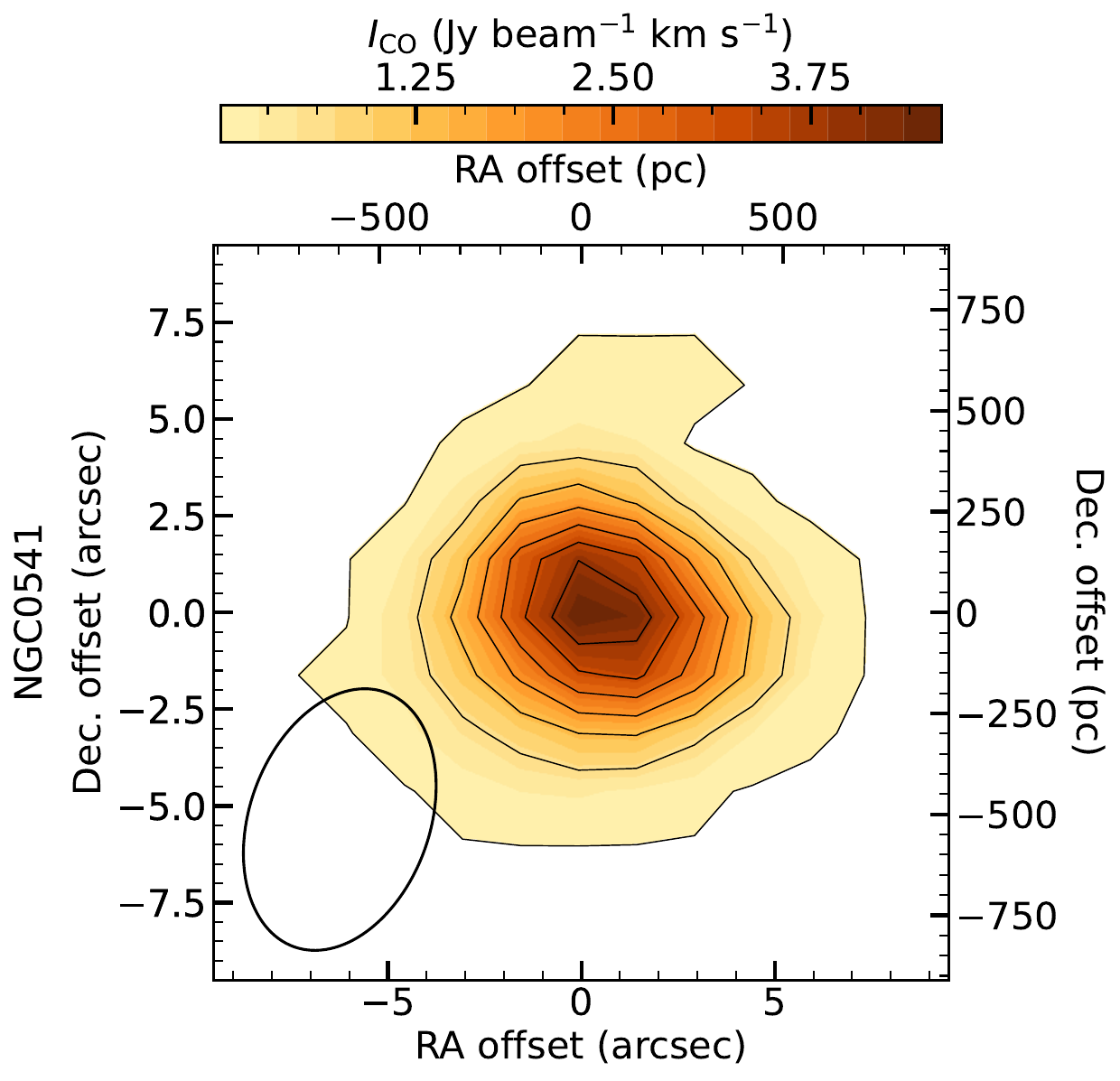}
    \end{subfigure}
    \begin{subfigure}{.32\linewidth}
    \centering
    \includegraphics[height=5.9cm]{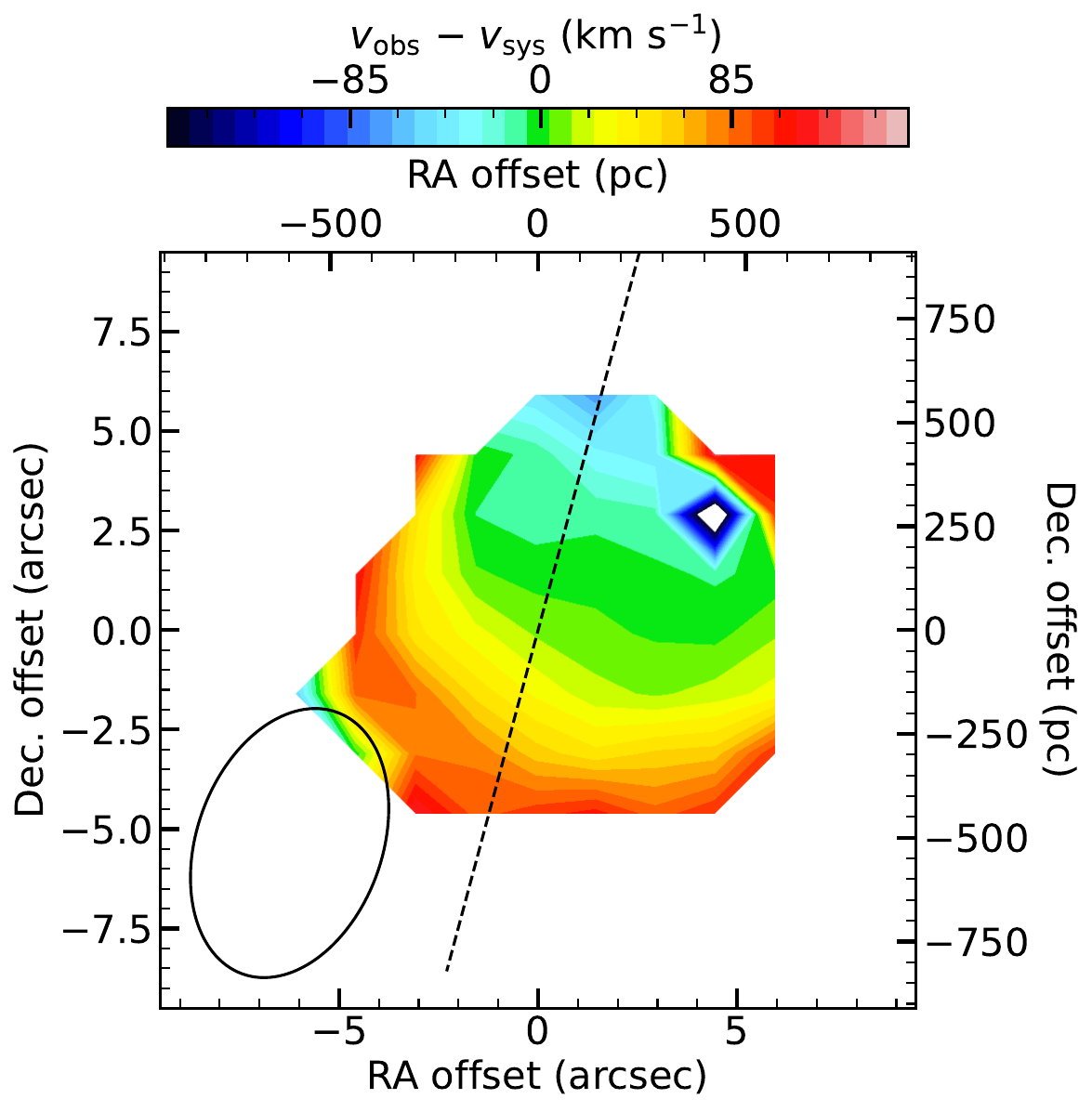}
    \end{subfigure}
    \begin{subfigure}{.32\linewidth}
    \centering
    \includegraphics[height=5.9cm]{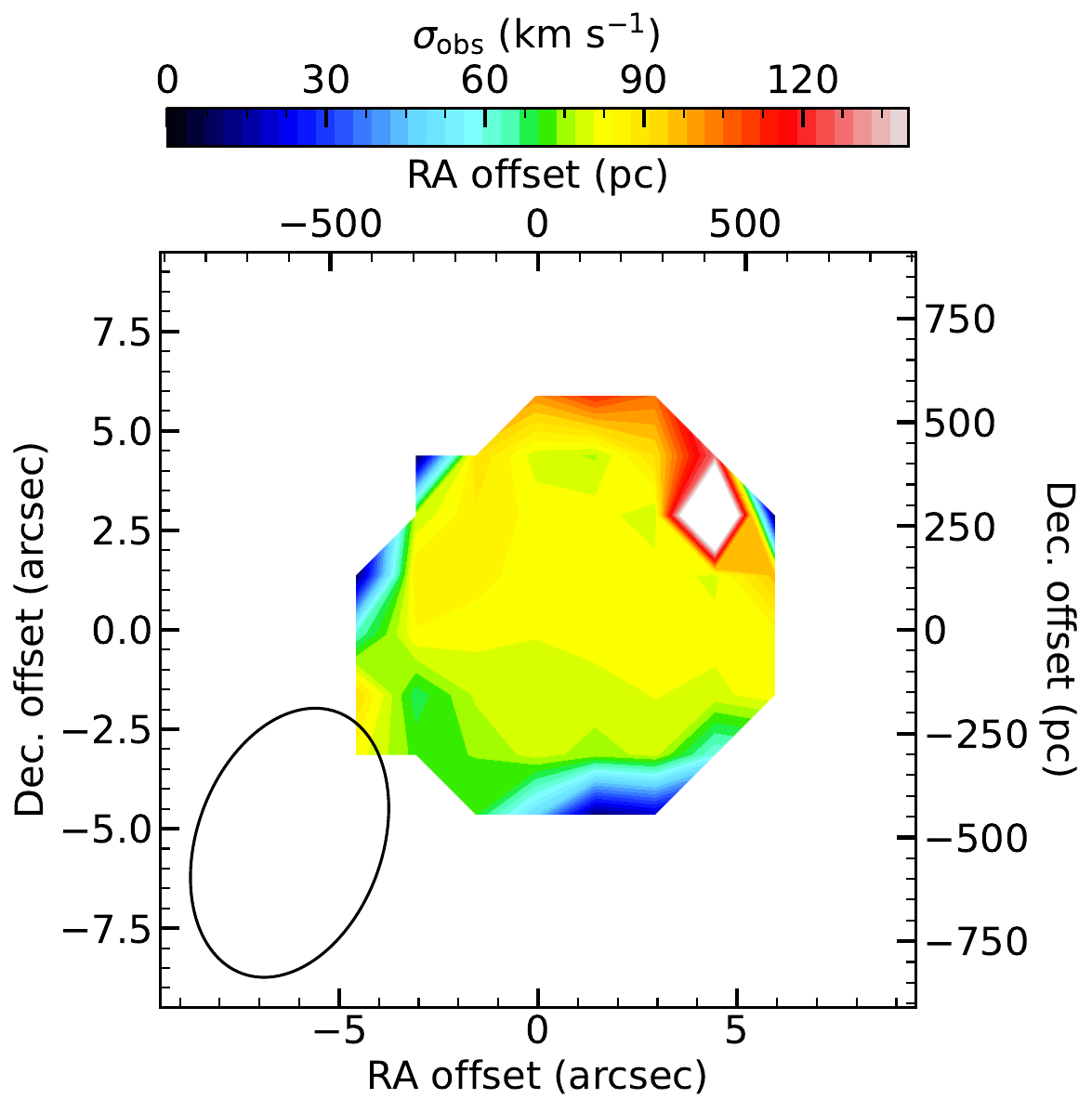}
    \end{subfigure}
    \\
    \begin{subfigure}{.34\linewidth}
    \centering
    \includegraphics[height=5.2cm]{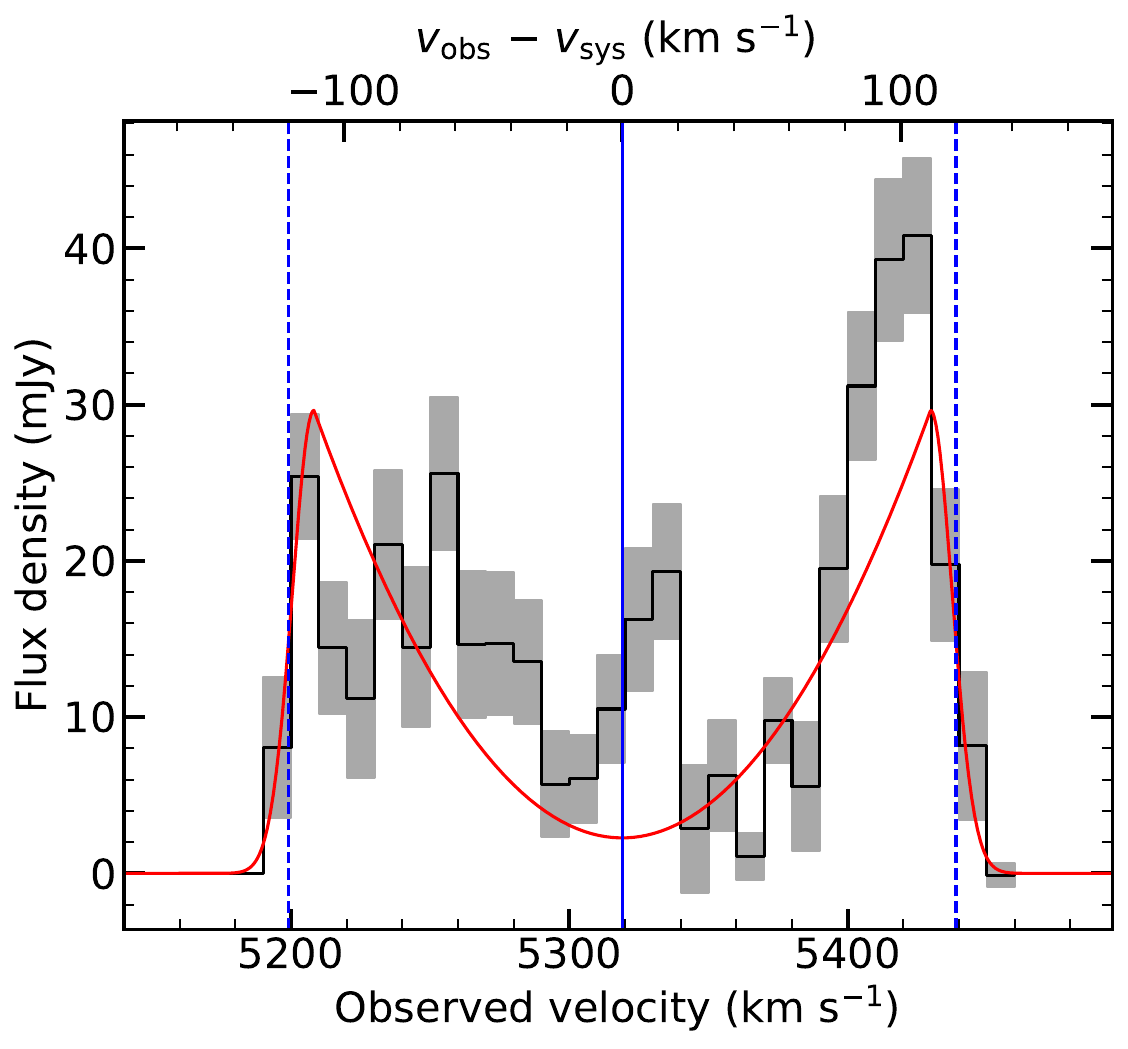}
    \end{subfigure}
    \begin{subfigure}{.32\linewidth}
    \centering
    \includegraphics[height=4.8cm]{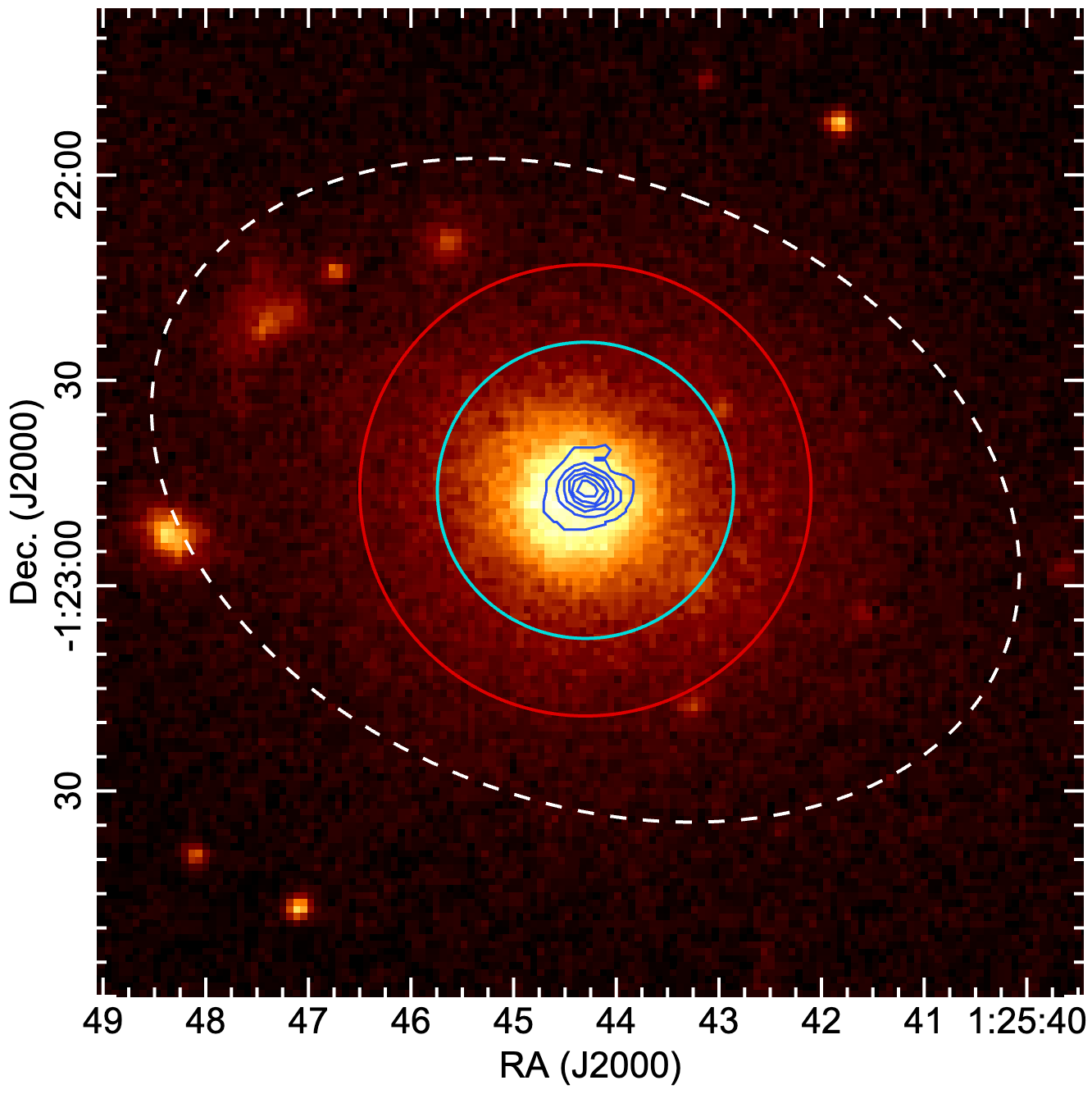}
    \end{subfigure}
    \begin{subfigure}{.33\linewidth}
    \centering
    \includegraphics[height=5.2cm]{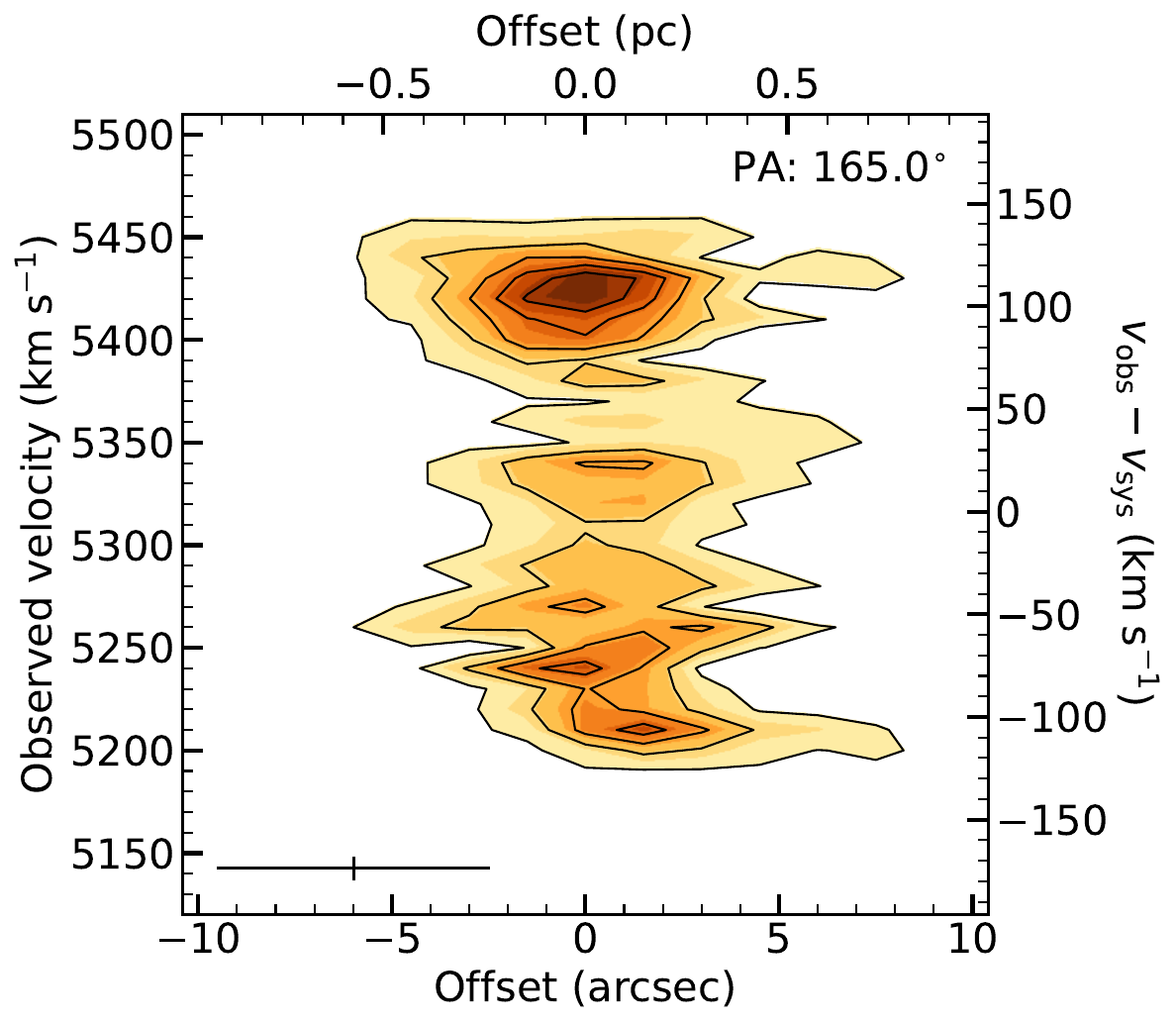}
    \end{subfigure}
    \caption[]{continued.}
\end{figure*}

\begin{figure*}
    \ContinuedFloat
    \centering
    \begin{subfigure}{.34\linewidth}
    \centering
    \includegraphics[height=6.0cm]{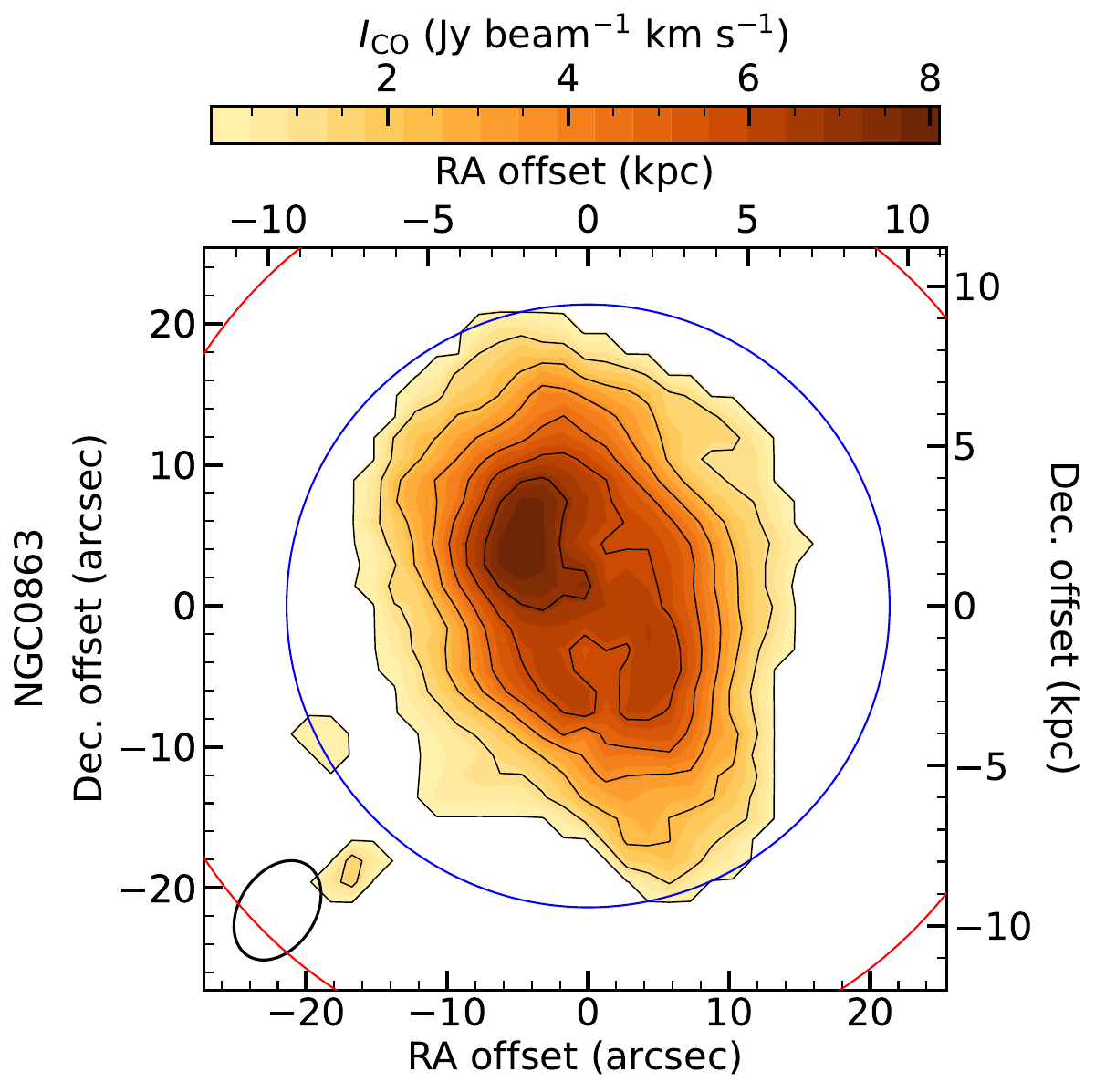}
    \end{subfigure}
    \begin{subfigure}{.32\linewidth}
    \centering
    \includegraphics[height=6.0cm]{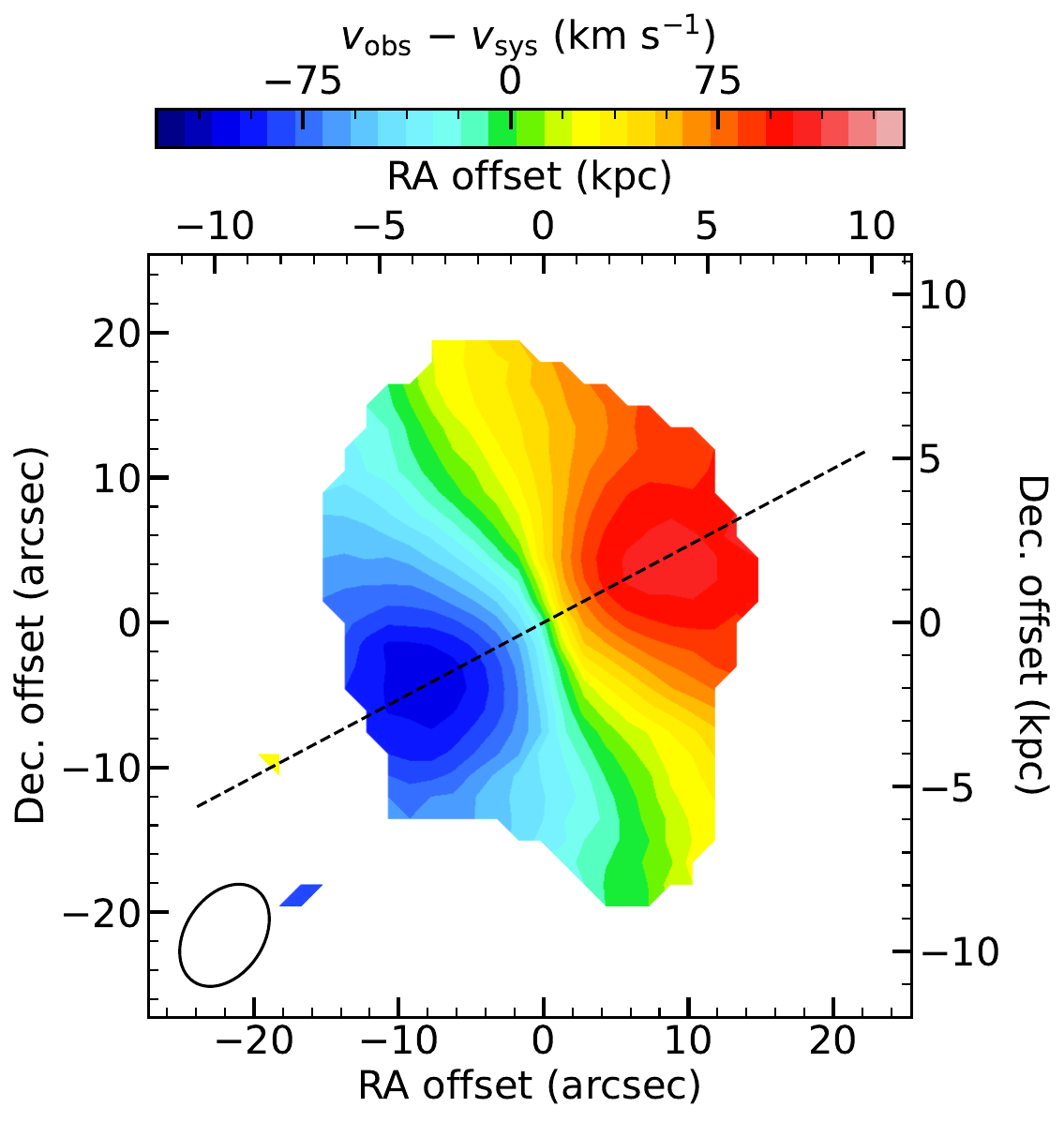}
    \end{subfigure}
    \begin{subfigure}{.32\linewidth}
    \centering
    \includegraphics[height=6.0cm]{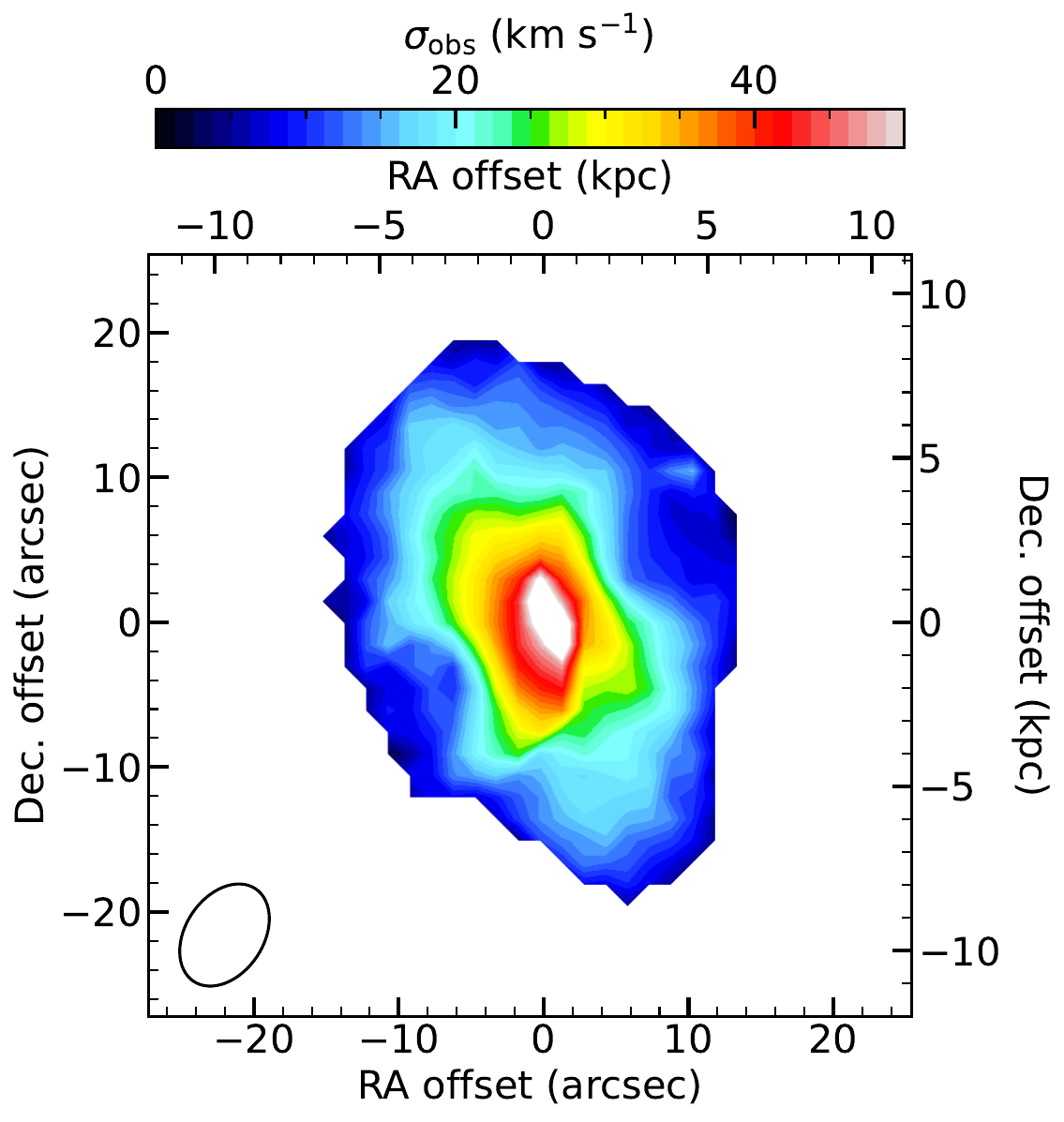}
    \end{subfigure}
    \\
    \begin{subfigure}{.34\linewidth}
    \centering
    \includegraphics[height=5.2cm]{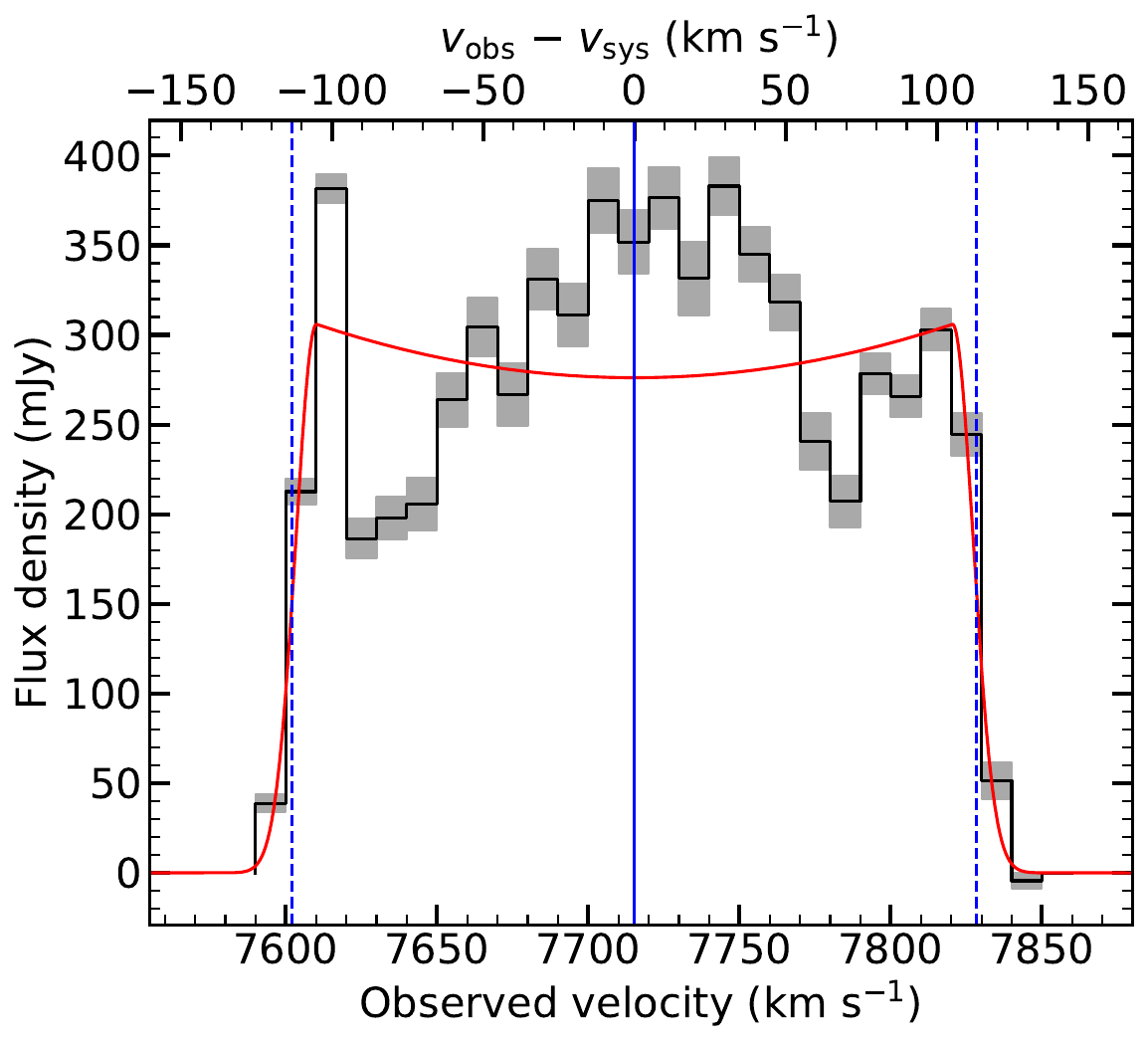}
    \end{subfigure}
    \begin{subfigure}{.32\linewidth}
    \centering
    \includegraphics[height=4.8cm]{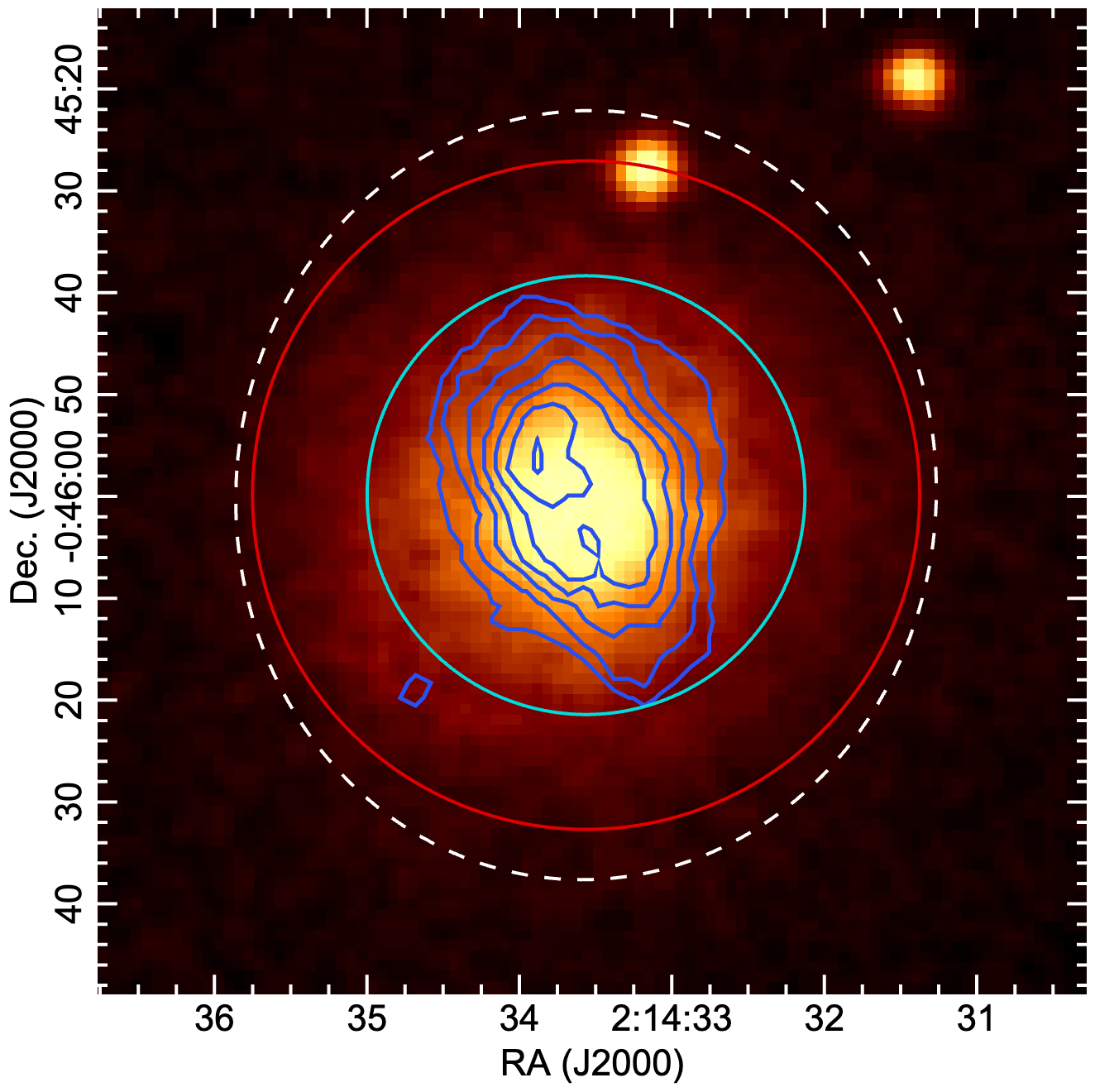}
    \end{subfigure}
    \begin{subfigure}{.33\linewidth}
    \centering
    \includegraphics[height=5.2cm]{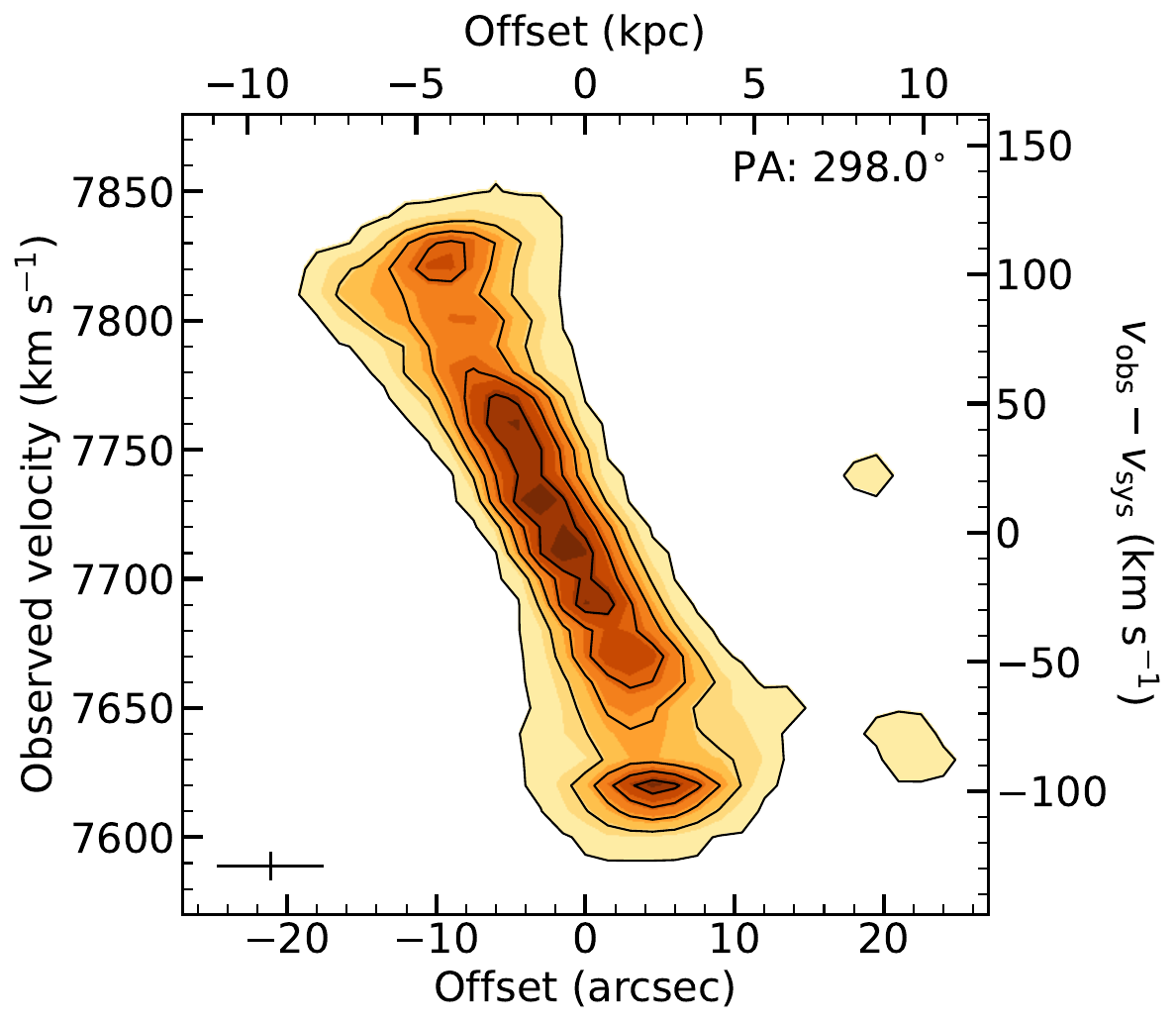}
    \end{subfigure}
    \\
    \begin{subfigure}{.34\linewidth}
    \centering
    \includegraphics[height=6.0cm]{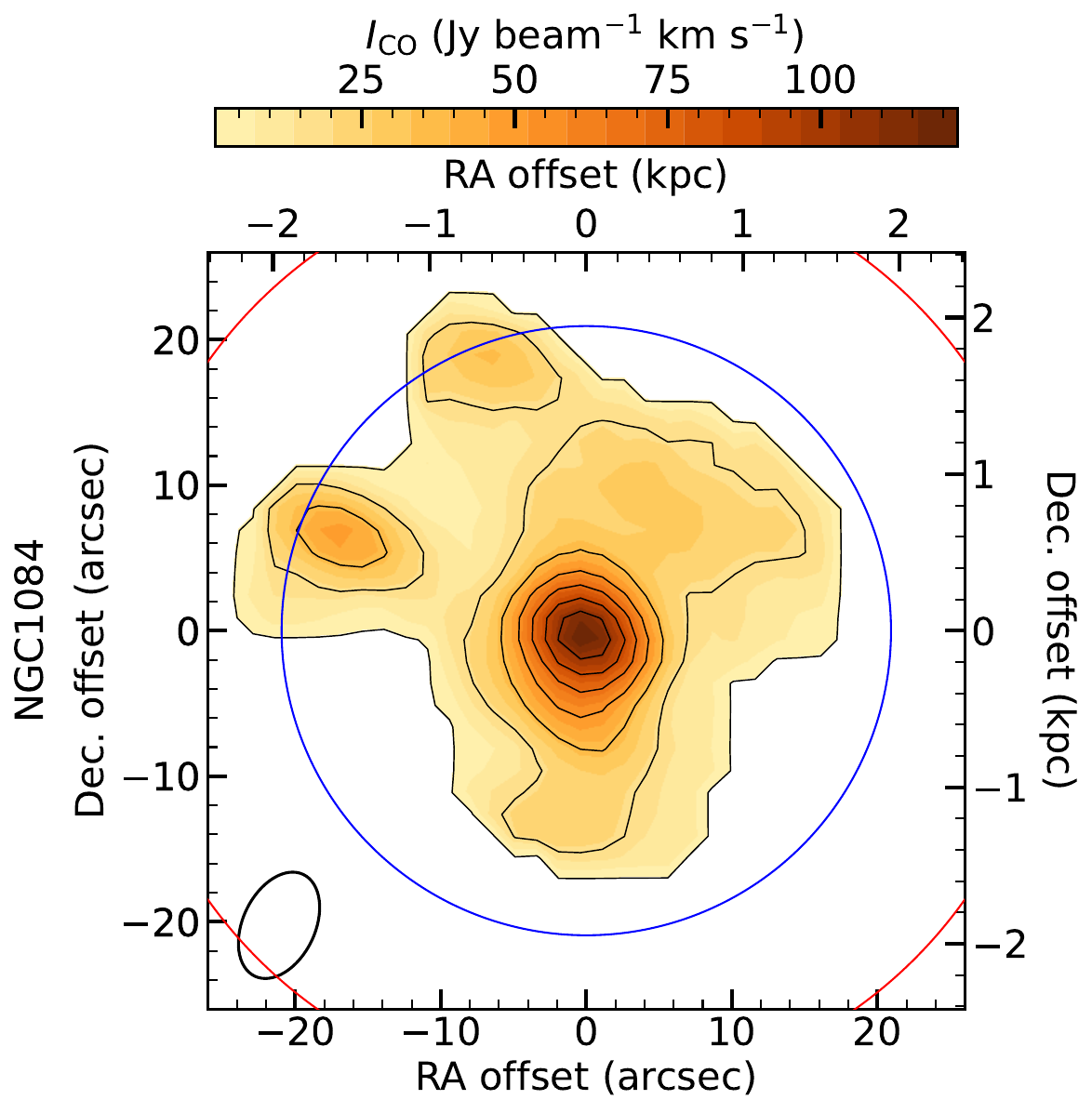}
    \end{subfigure}
    \begin{subfigure}{.32\linewidth}
    \centering
    \includegraphics[height=6.0cm]{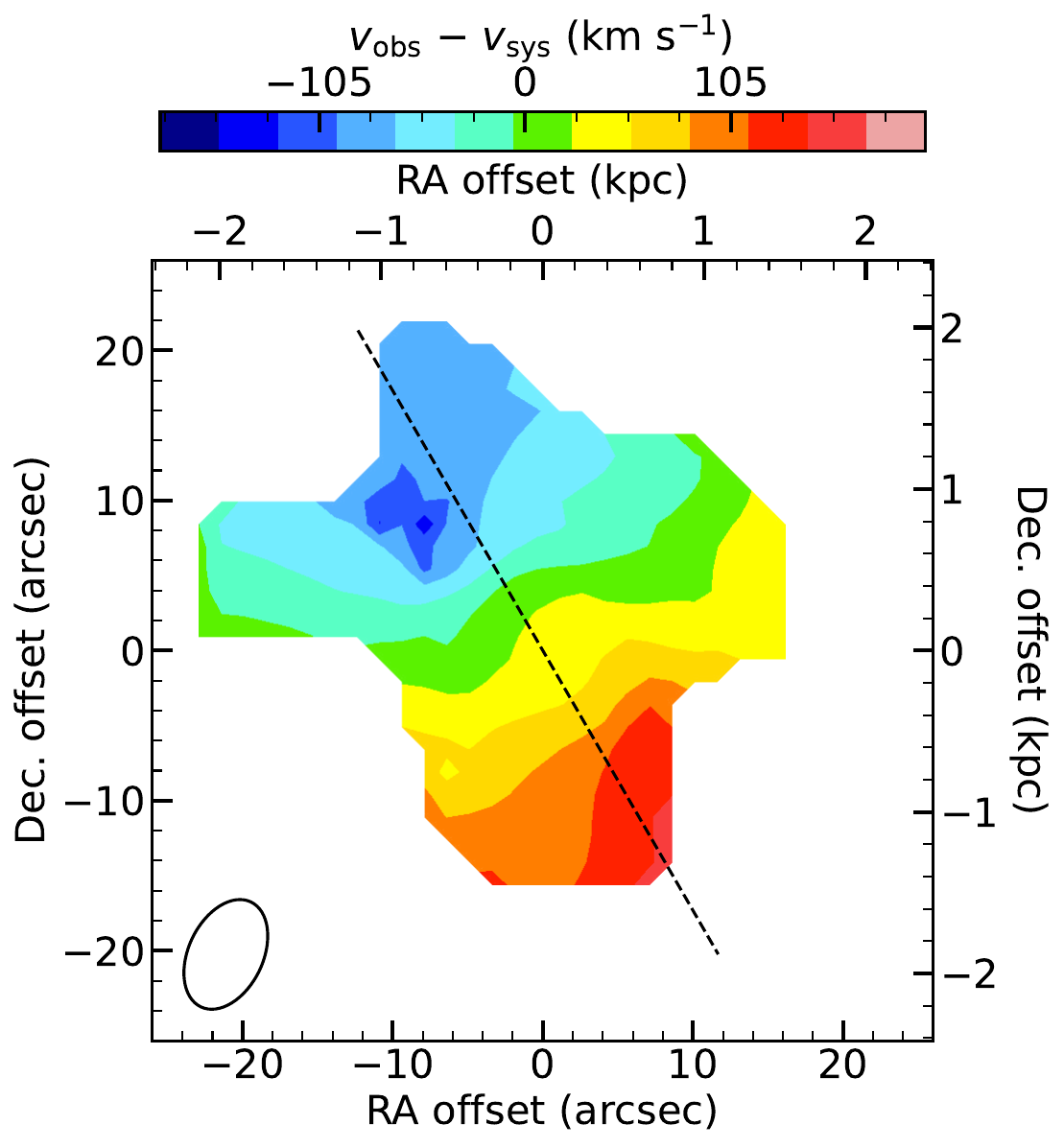}
    \end{subfigure}
    \begin{subfigure}{.32\linewidth}
    \centering
    \includegraphics[height=6.0cm]{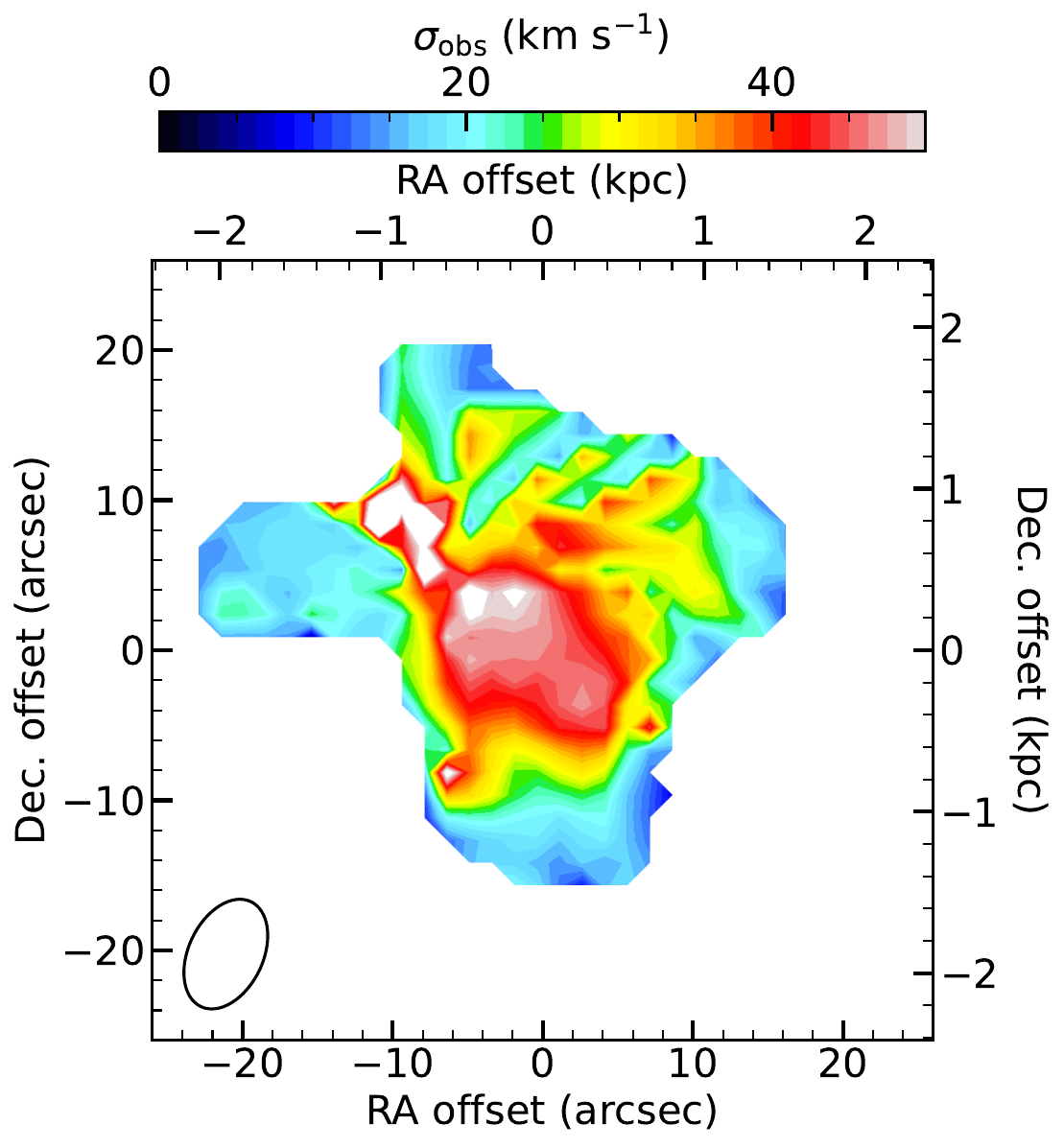}
    \end{subfigure}
    \\
    \begin{subfigure}{.34\linewidth}
    \centering
    \includegraphics[height=5.2cm]{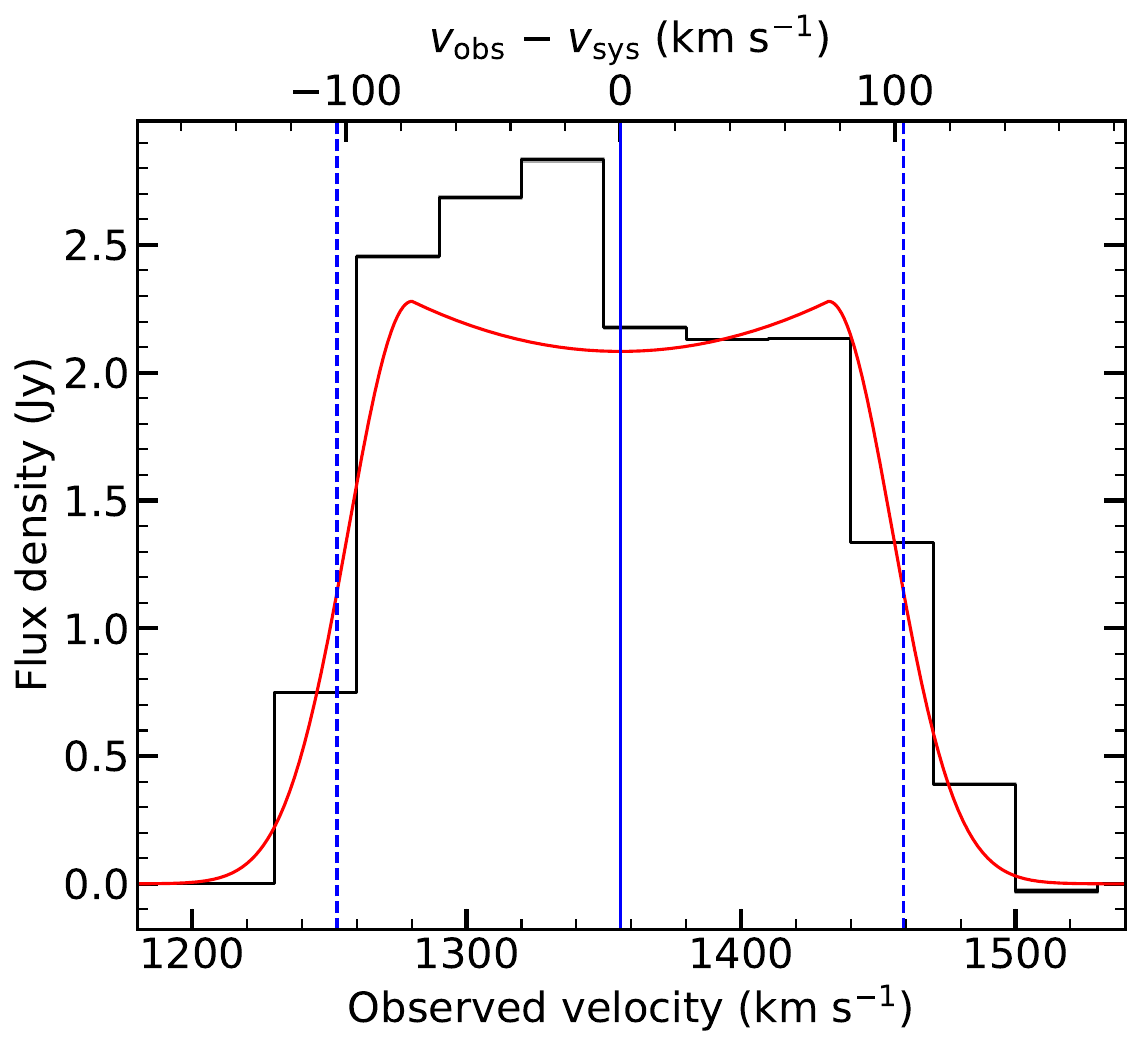}
    \end{subfigure}
    \begin{subfigure}{.32\linewidth}
    \centering
    \includegraphics[height=4.8cm]{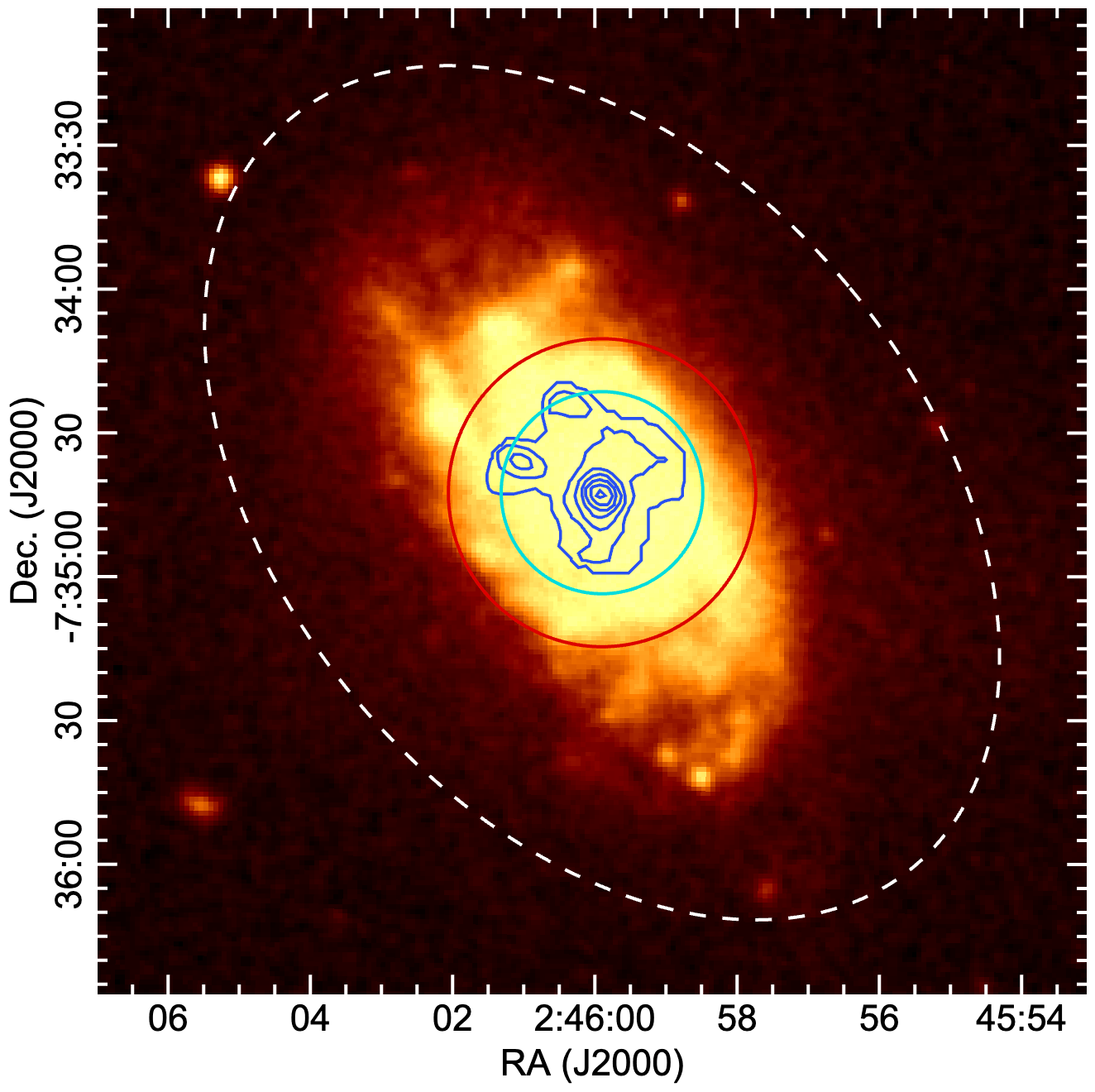}
    \end{subfigure}
    \begin{subfigure}{.33\linewidth}
    \centering
    \includegraphics[height=5.2cm]{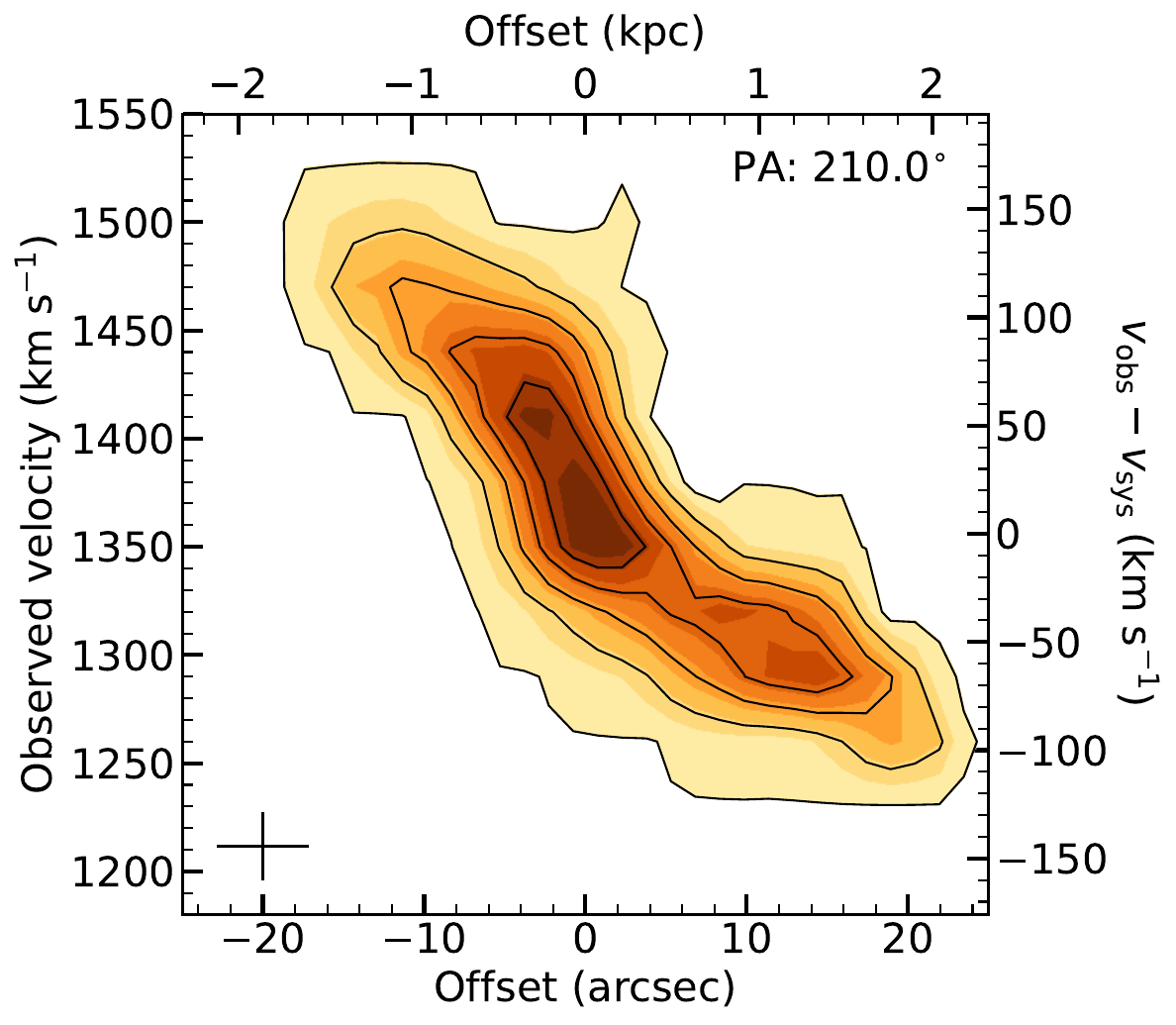}
    \end{subfigure}
    \caption[]{continued.}
\end{figure*}

\begin{figure*}
    \ContinuedFloat
    \centering
    \begin{subfigure}{.34\linewidth}
    \centering
    \includegraphics[height=6.0cm]{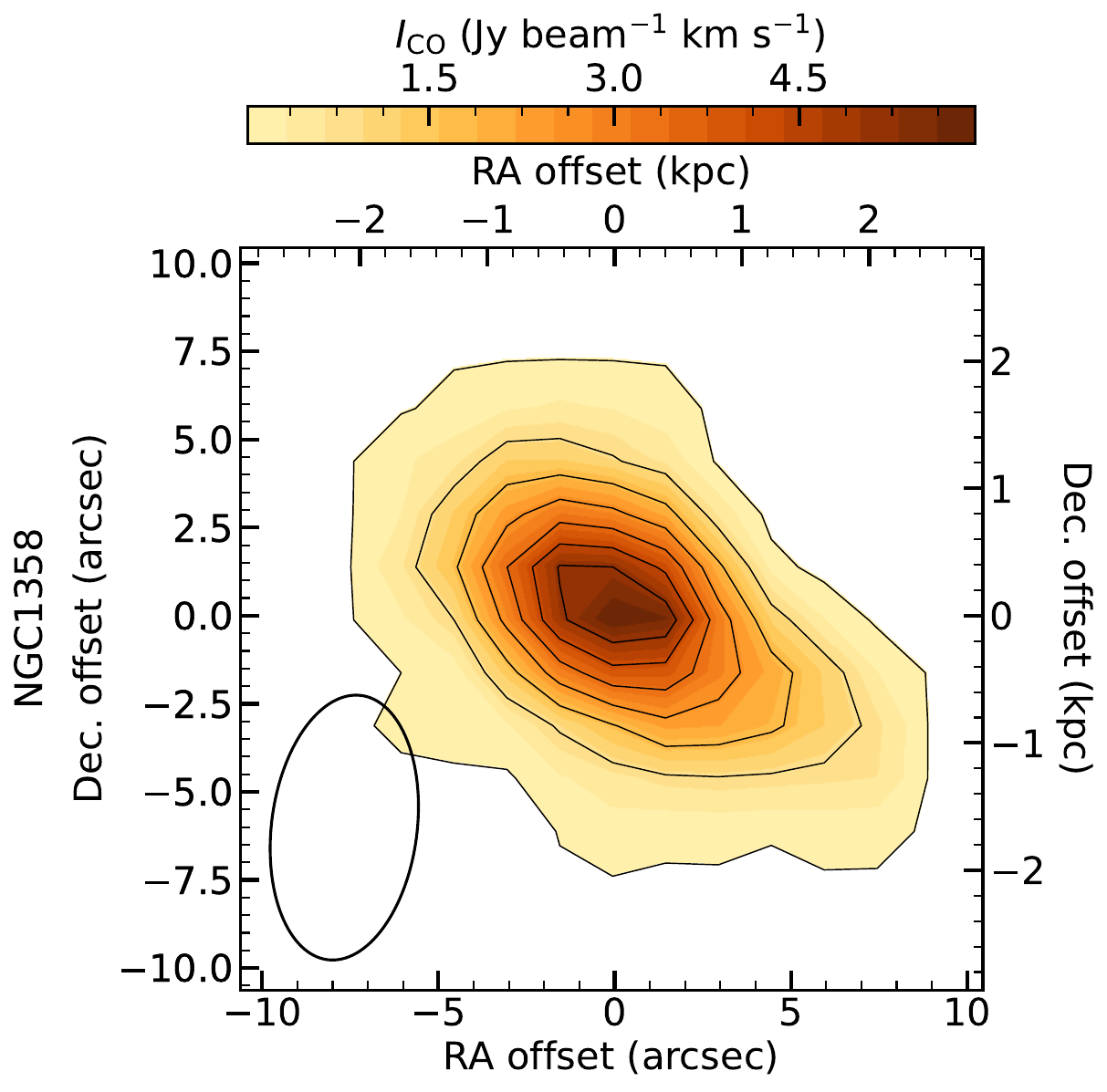}
    \end{subfigure}
    \begin{subfigure}{.32\linewidth}
    \centering
    \includegraphics[height=6.0cm]{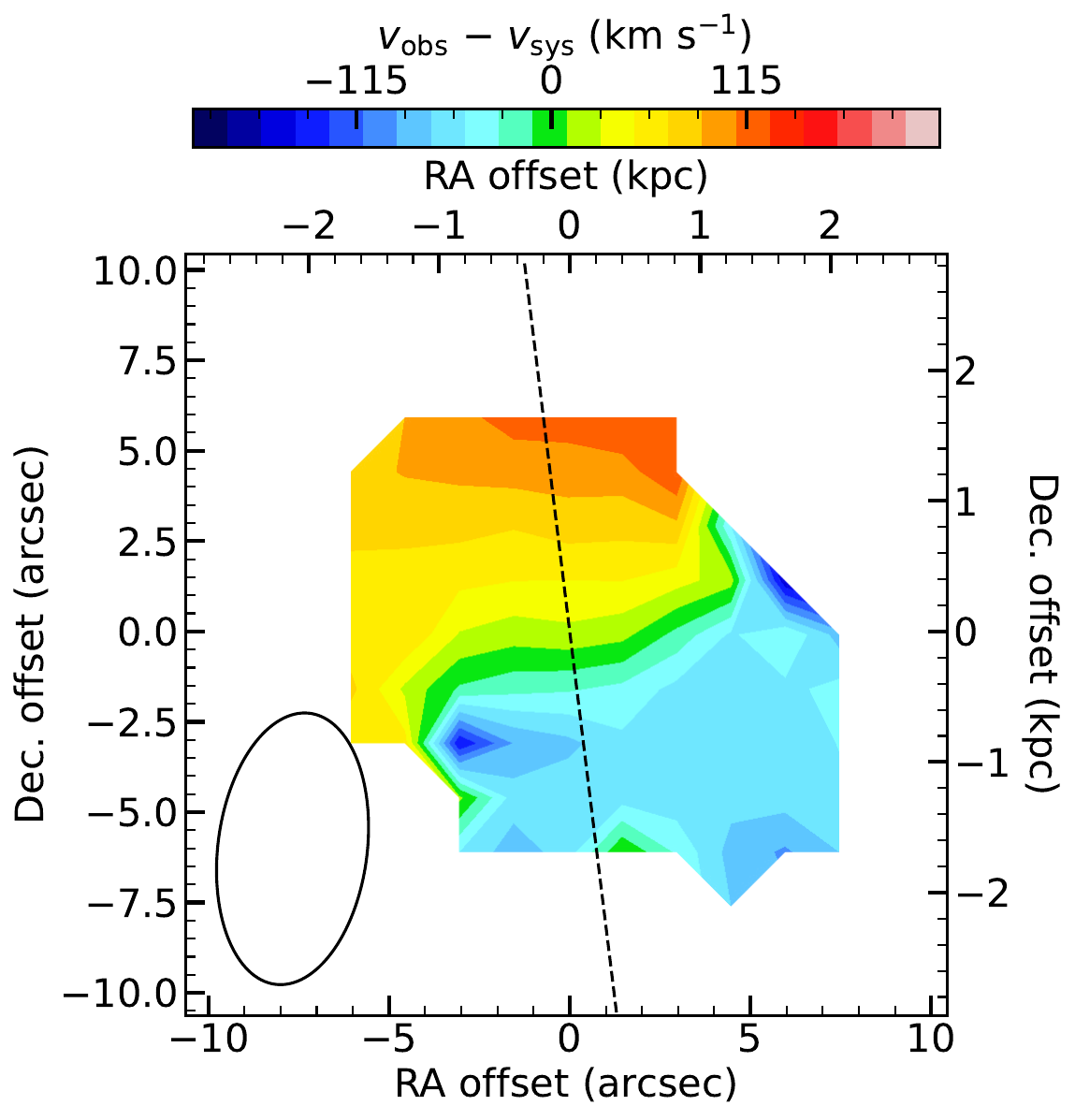}
    \end{subfigure}
    \begin{subfigure}{.32\linewidth}
    \centering
    \includegraphics[height=6.0cm]{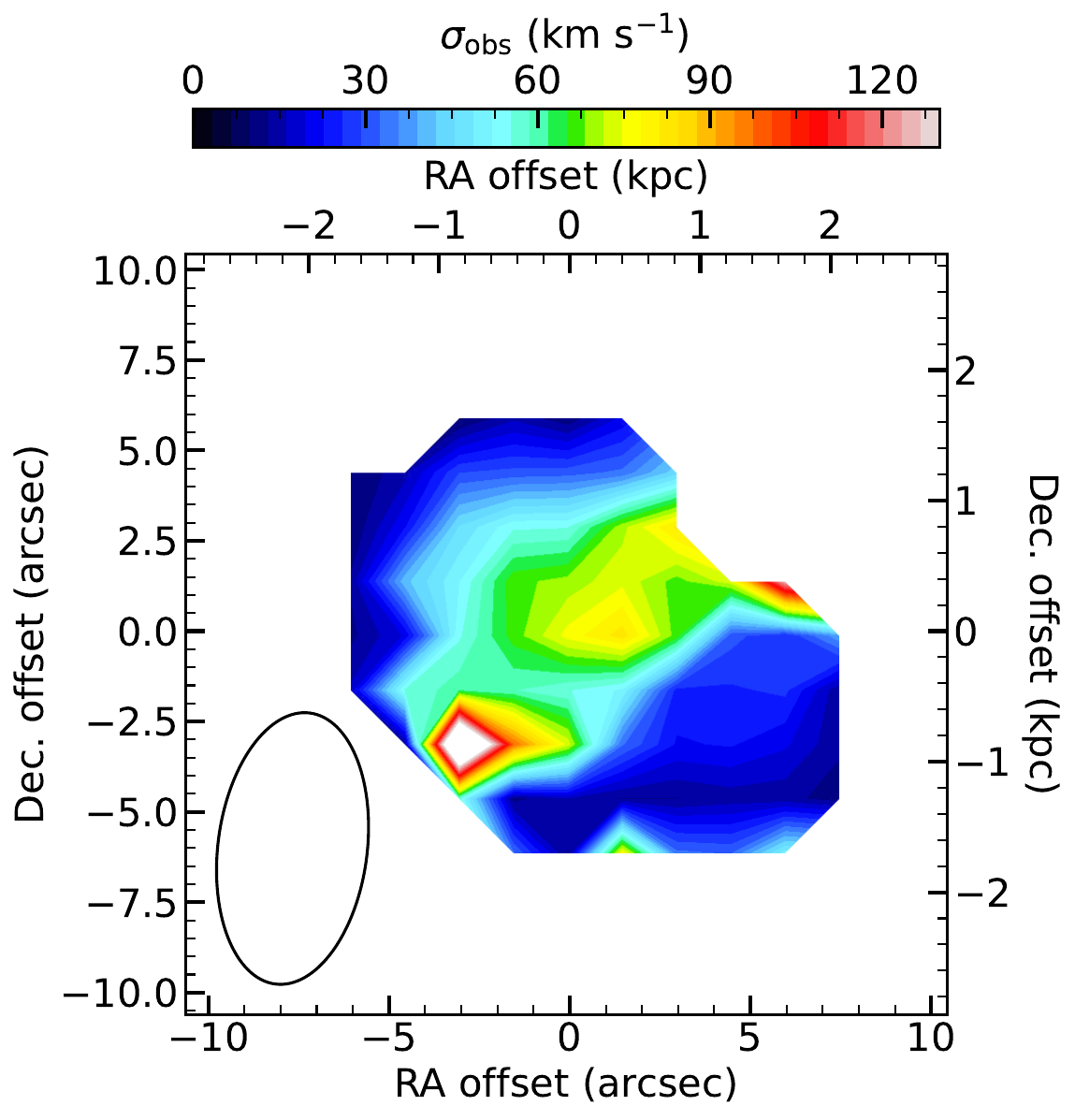}
    \end{subfigure}
    \\
    \begin{subfigure}{.34\linewidth}
    \centering
    \includegraphics[height=5.2cm]{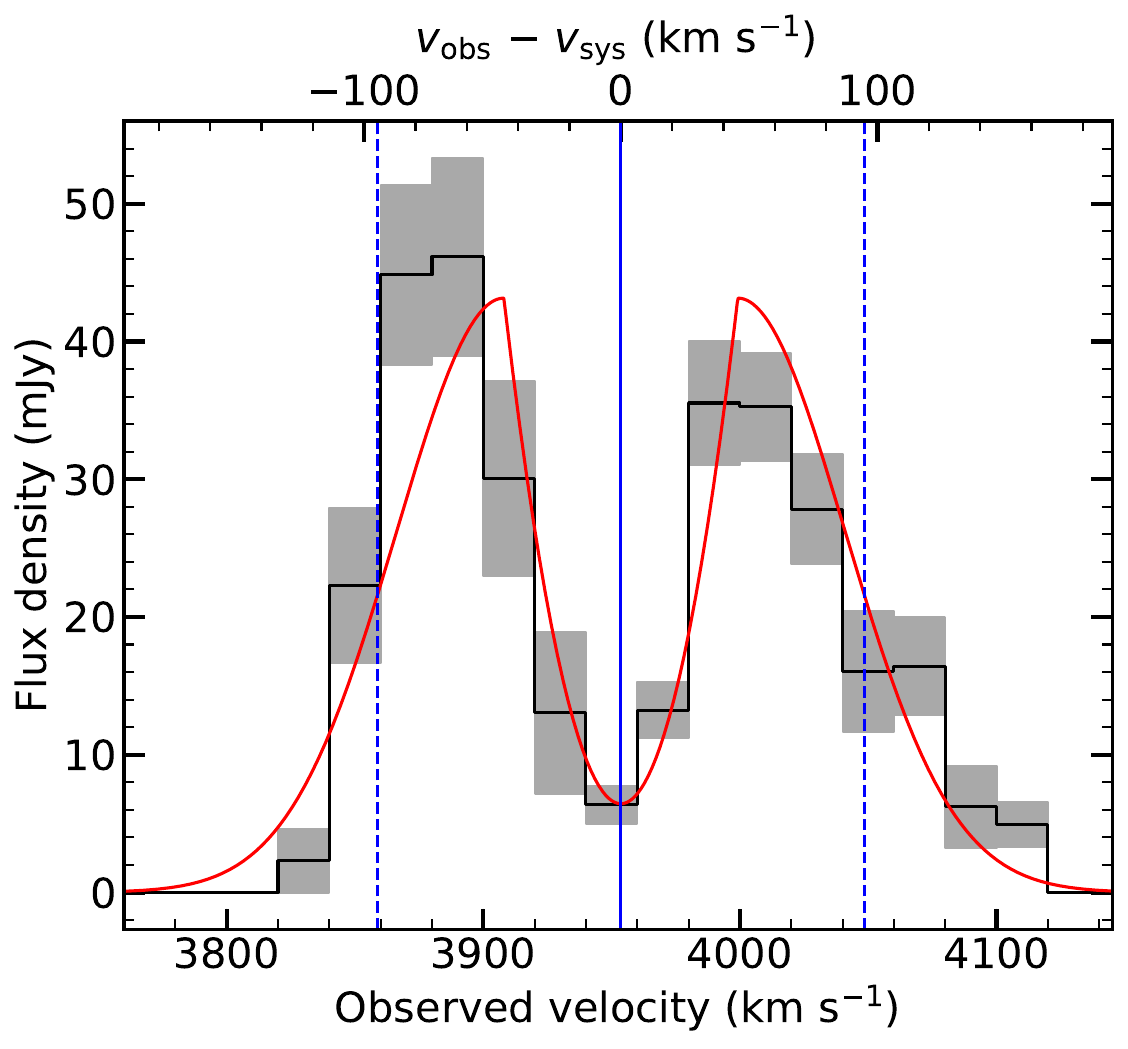}
    \end{subfigure}
    \begin{subfigure}{.32\linewidth}
    \centering
    \includegraphics[height=4.8cm]{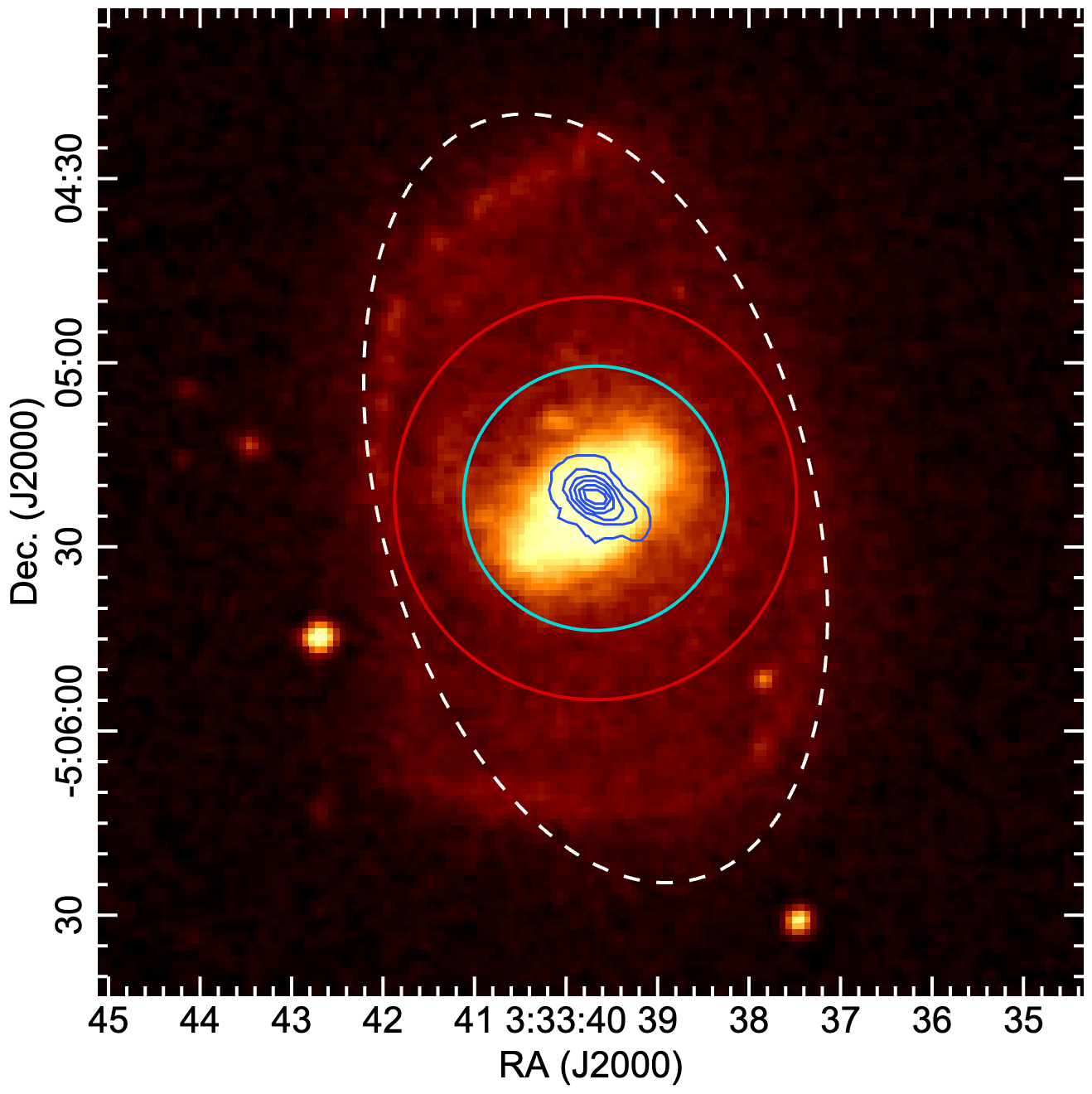}
    \end{subfigure}
    \begin{subfigure}{.33\linewidth}
    \centering
    \includegraphics[height=5.2cm]{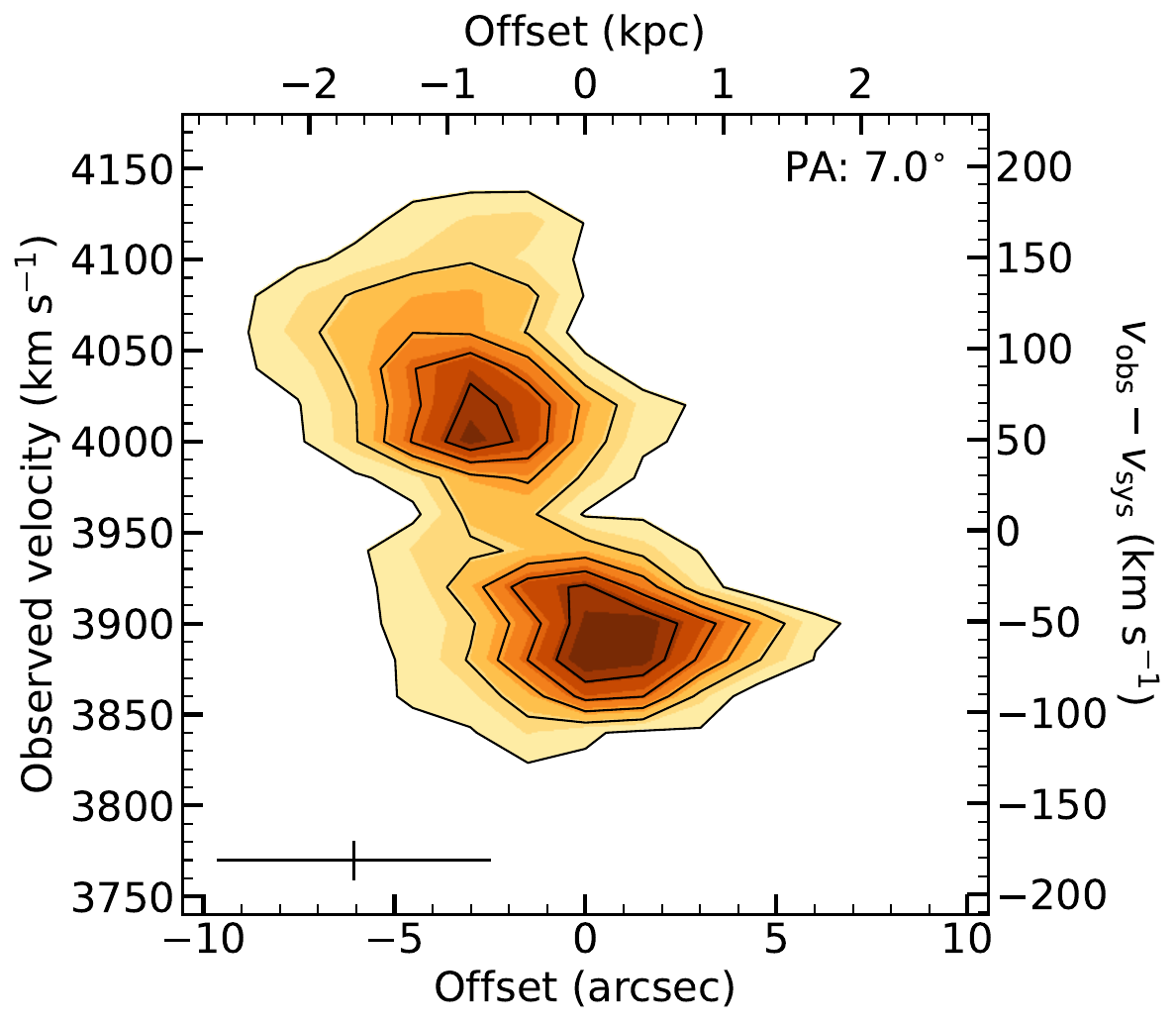}
    \end{subfigure}
    \\
    \begin{subfigure}{.34\linewidth}
    \centering
    \includegraphics[height=6.0cm]{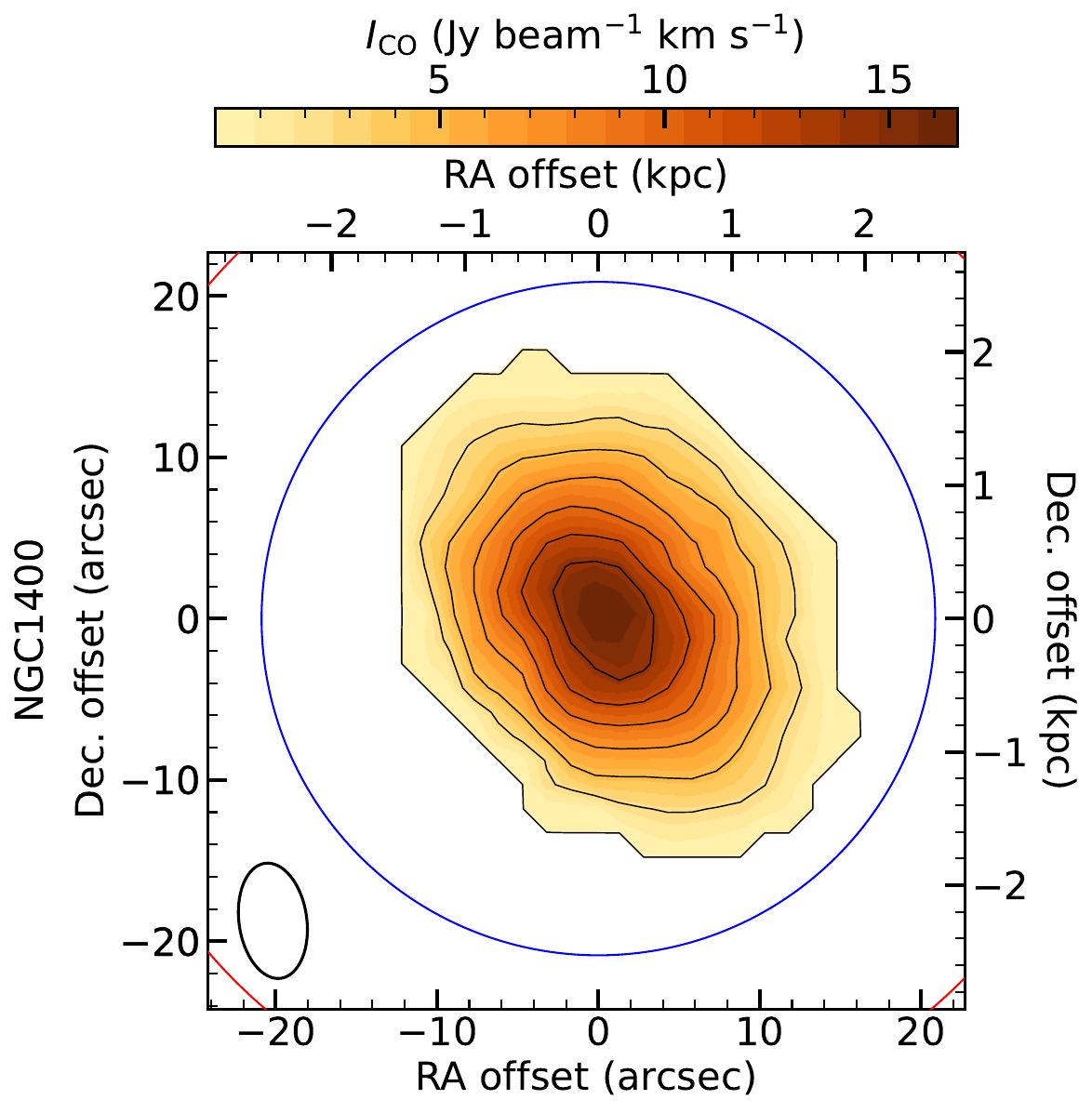}
    \end{subfigure}
    \begin{subfigure}{.32\linewidth}
    \centering
    \includegraphics[height=6.0cm]{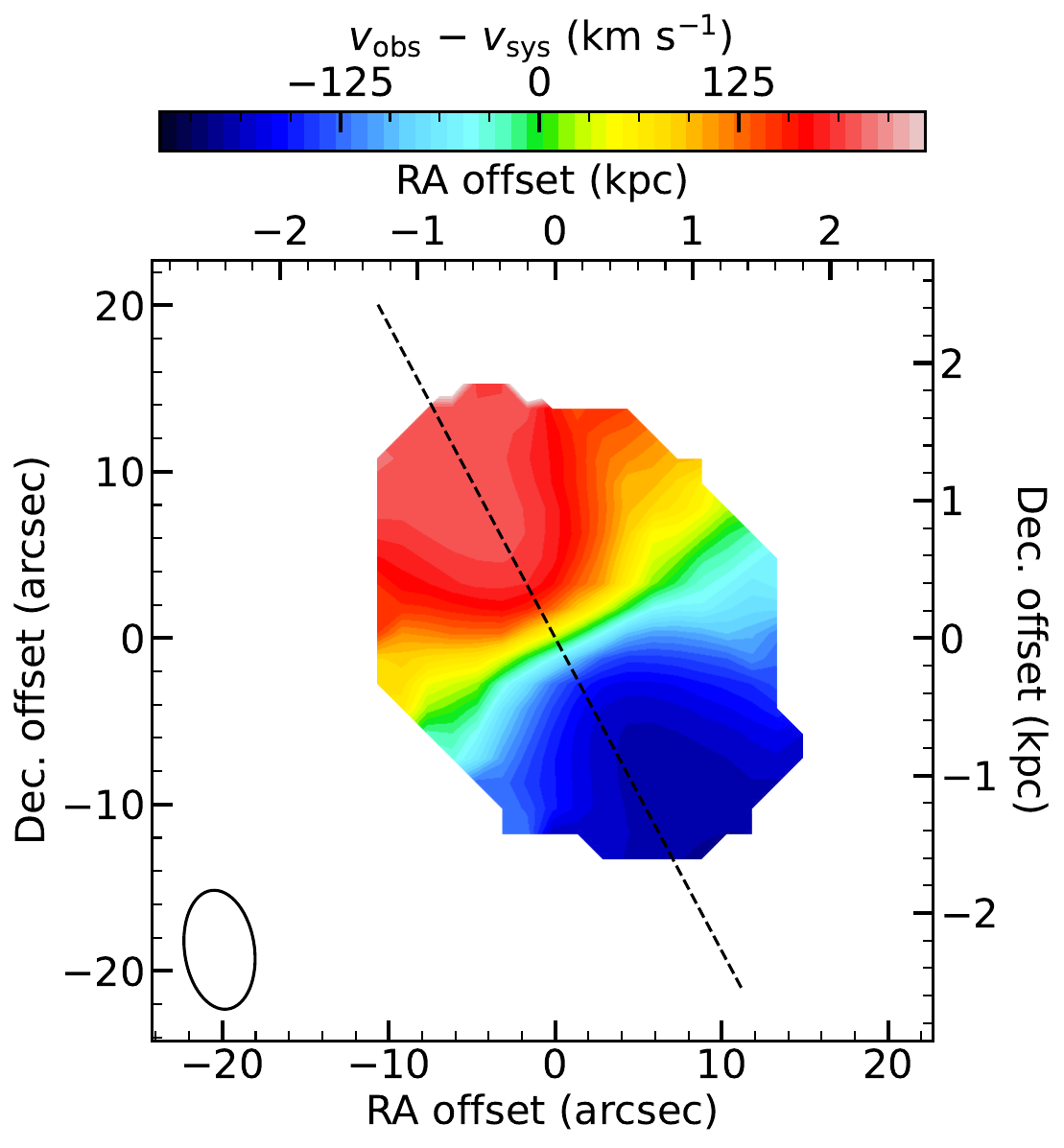}
    \end{subfigure}
    \begin{subfigure}{.32\linewidth}
    \centering
    \includegraphics[height=6.0cm]{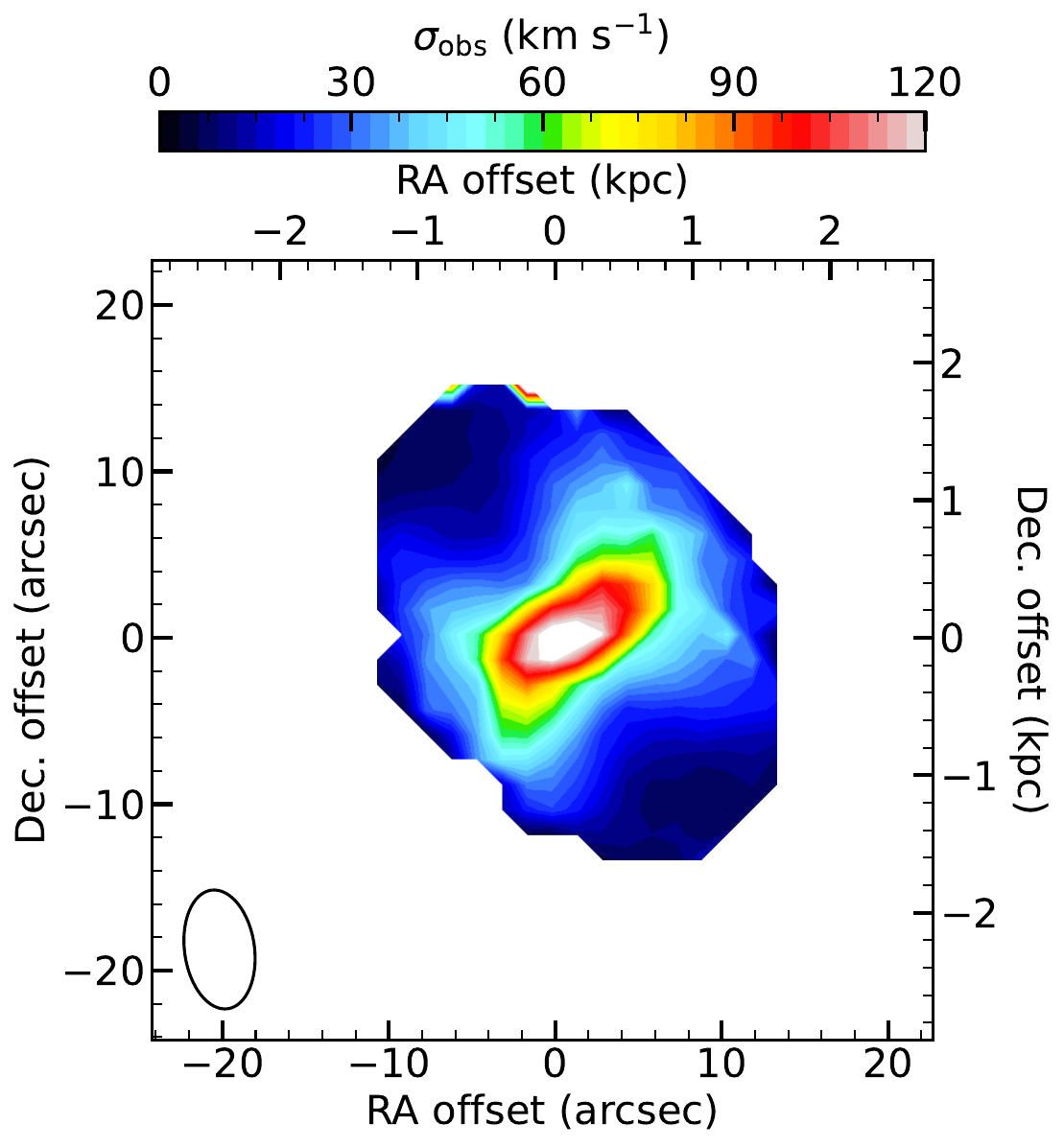}
    \end{subfigure}
    \\
    \begin{subfigure}{.34\linewidth}
    \centering
    \includegraphics[height=5.2cm]{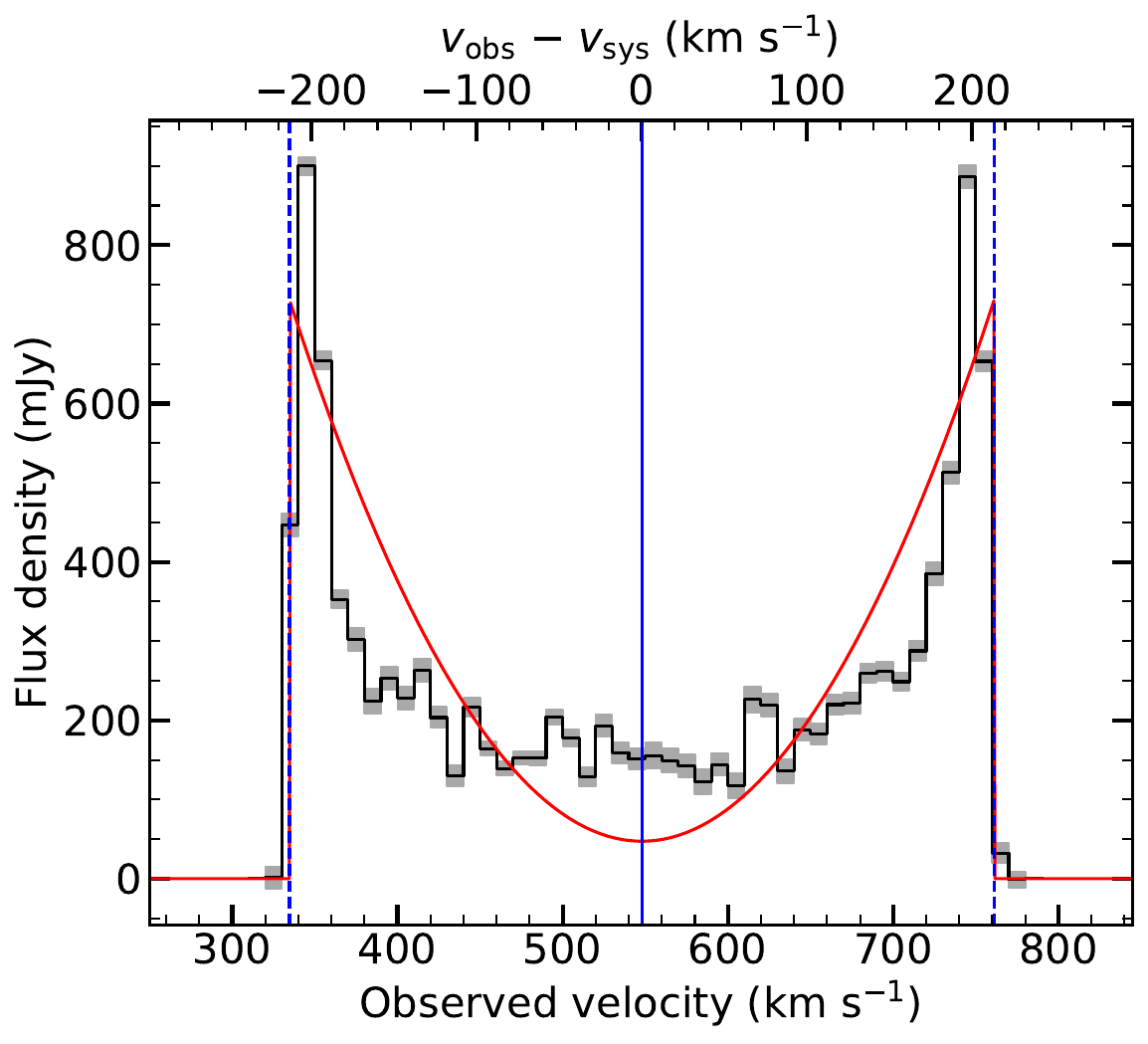}
    \end{subfigure}
    \begin{subfigure}{.32\linewidth}
    \centering
    \includegraphics[height=4.8cm]{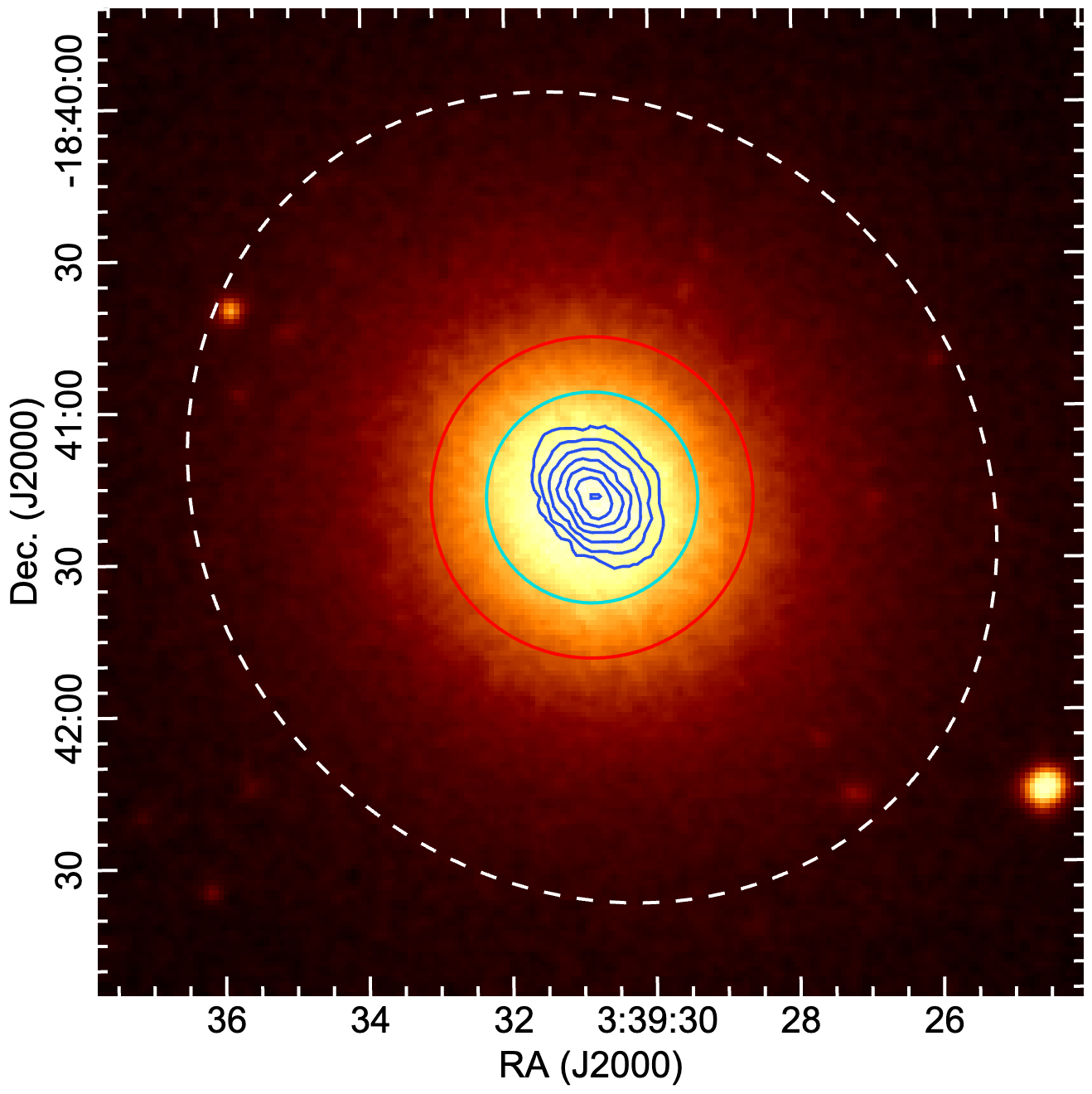}
    \end{subfigure}
    \begin{subfigure}{.33\linewidth}
    \centering
    \includegraphics[height=5.2cm]{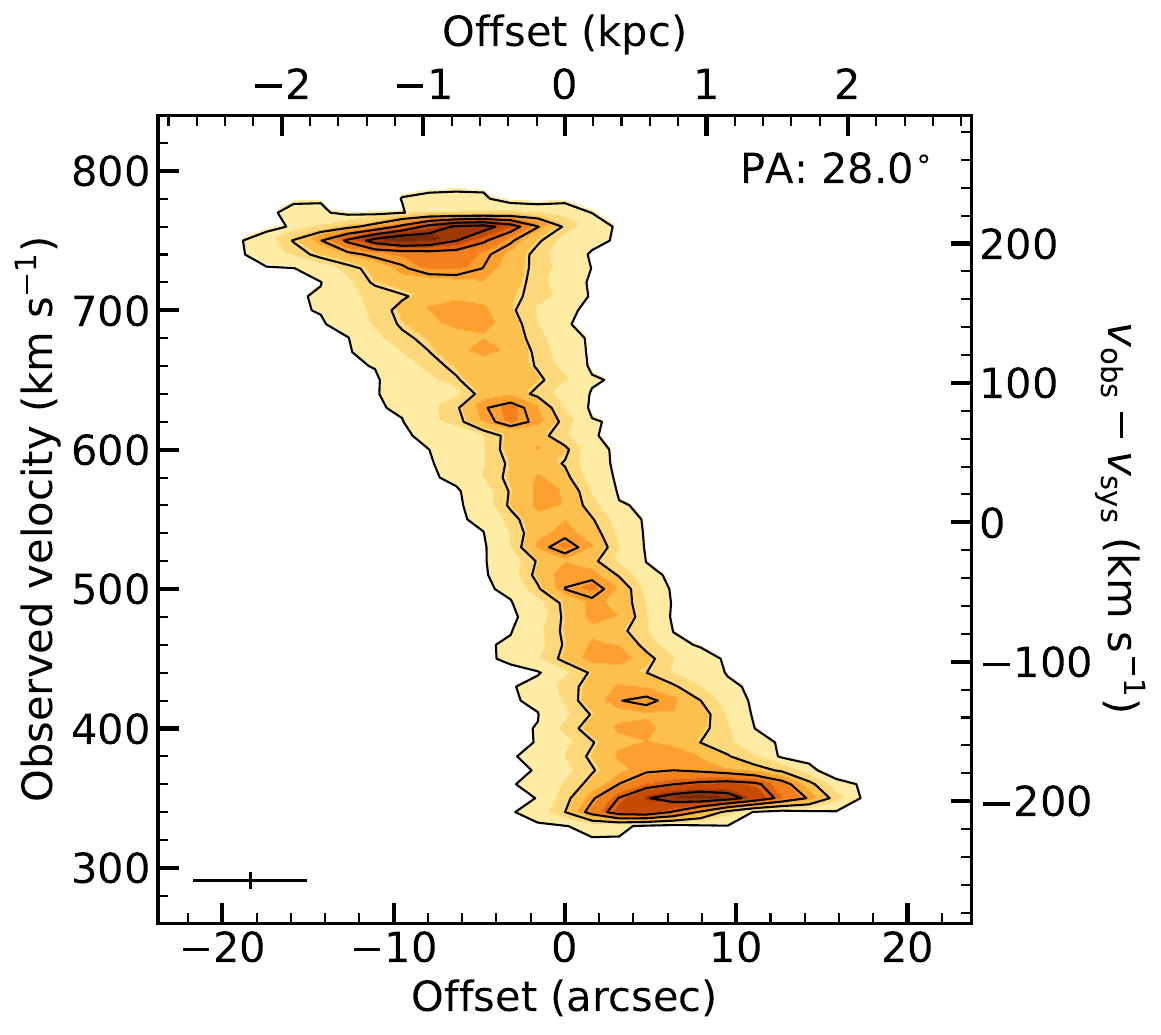}
    \end{subfigure}
    \caption[]{continued.}
\end{figure*}

\begin{figure*}
    \ContinuedFloat
    \centering
    \begin{subfigure}{.34\linewidth}
    \centering
    \includegraphics[height=6.0cm]{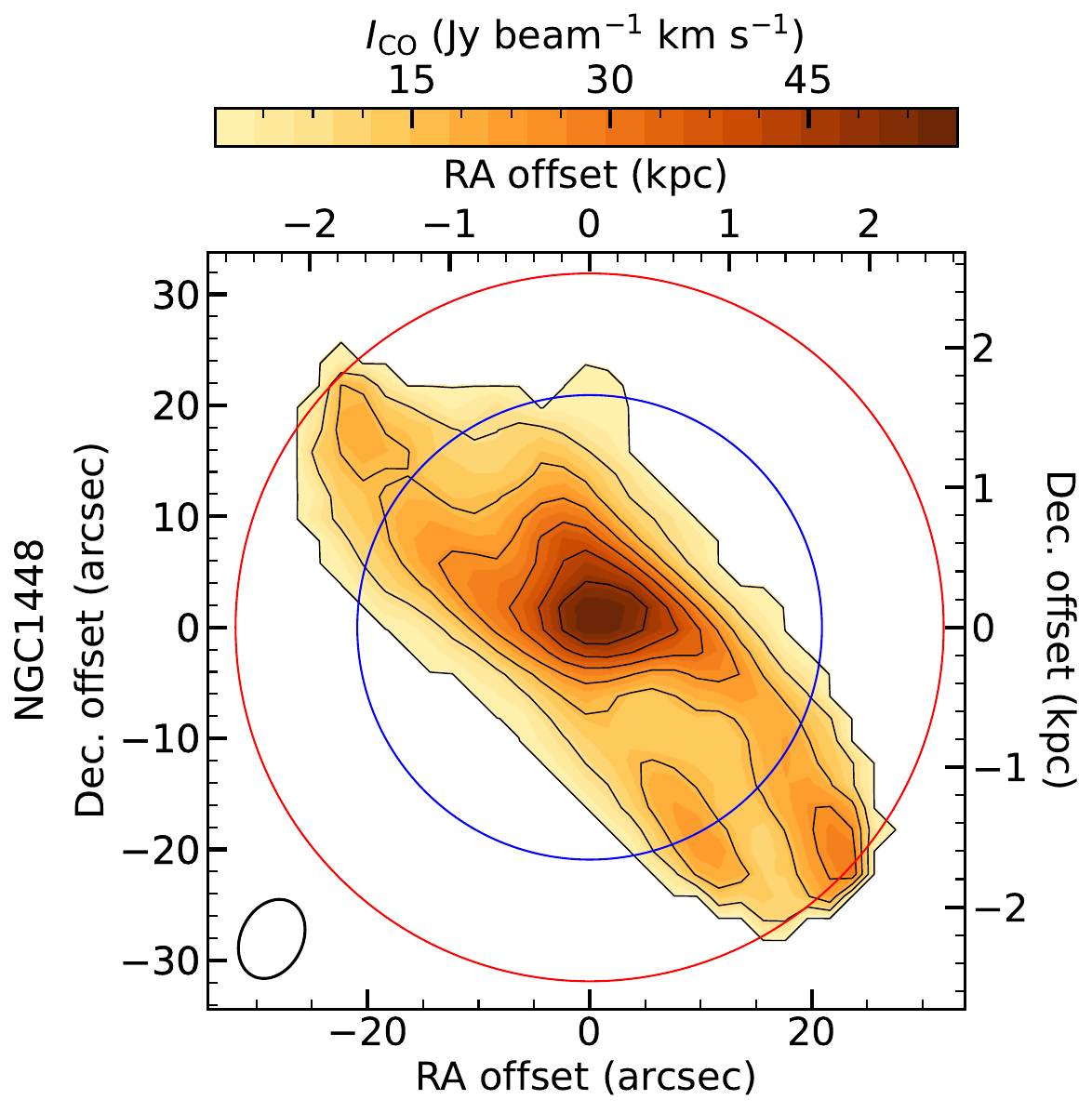}
    \end{subfigure}
    \begin{subfigure}{.32\linewidth}
    \centering
    \includegraphics[height=6.0cm]{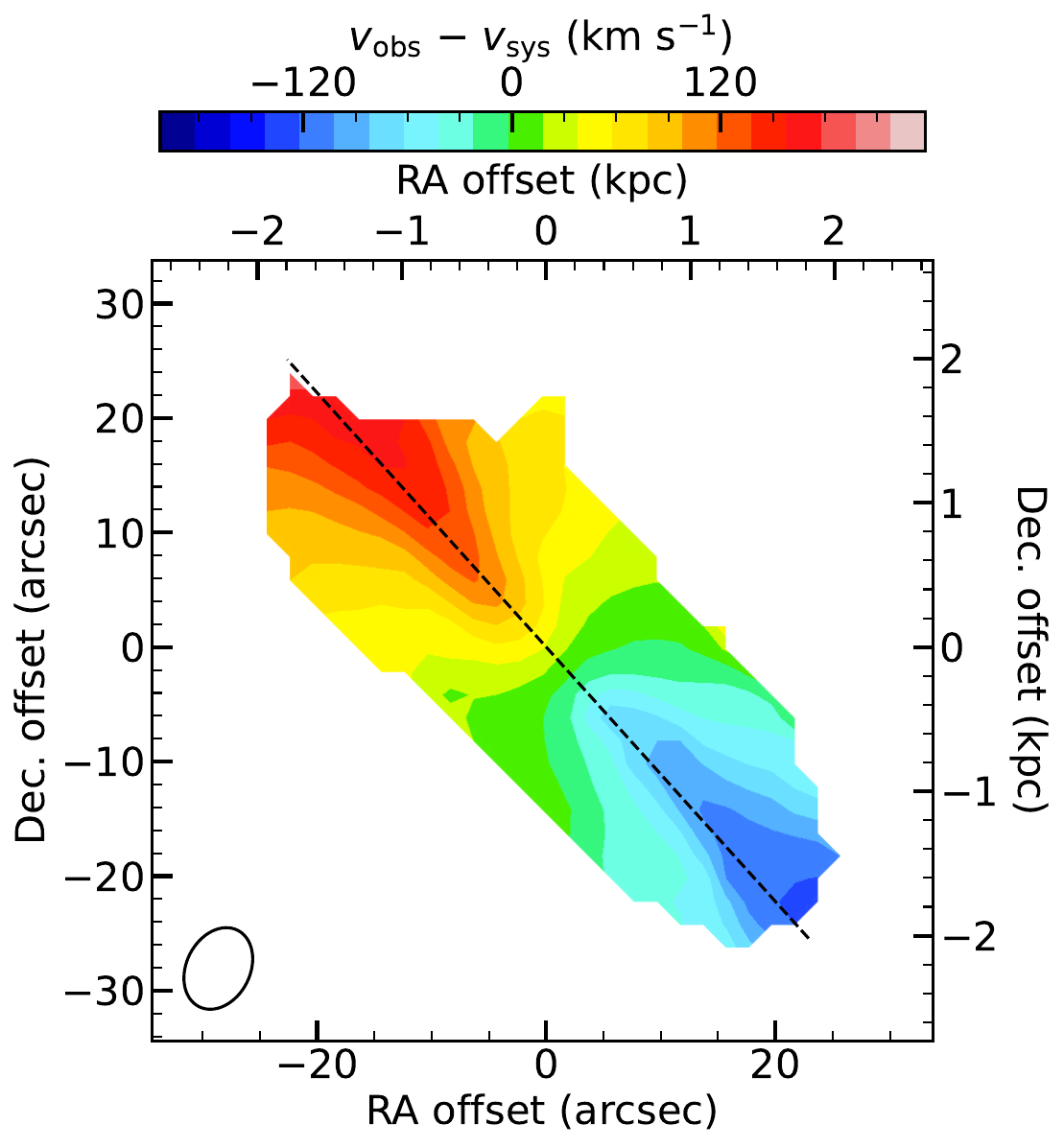}
    \end{subfigure}
    \begin{subfigure}{.32\linewidth}
    \centering
    \includegraphics[height=6.0cm]{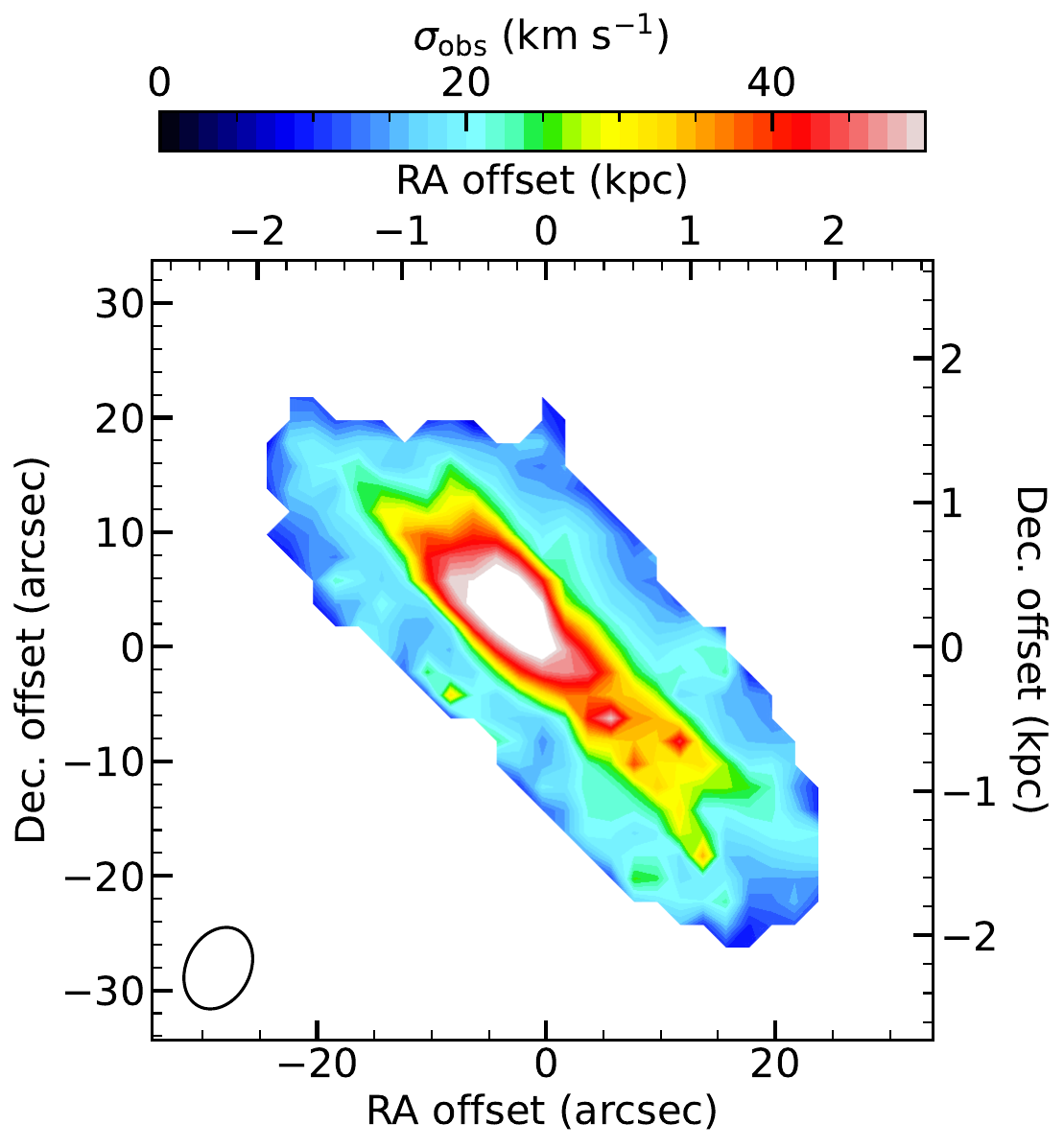}
    \end{subfigure}
    \\
    \begin{subfigure}{.34\linewidth}
    \centering
    \includegraphics[height=5.2cm]{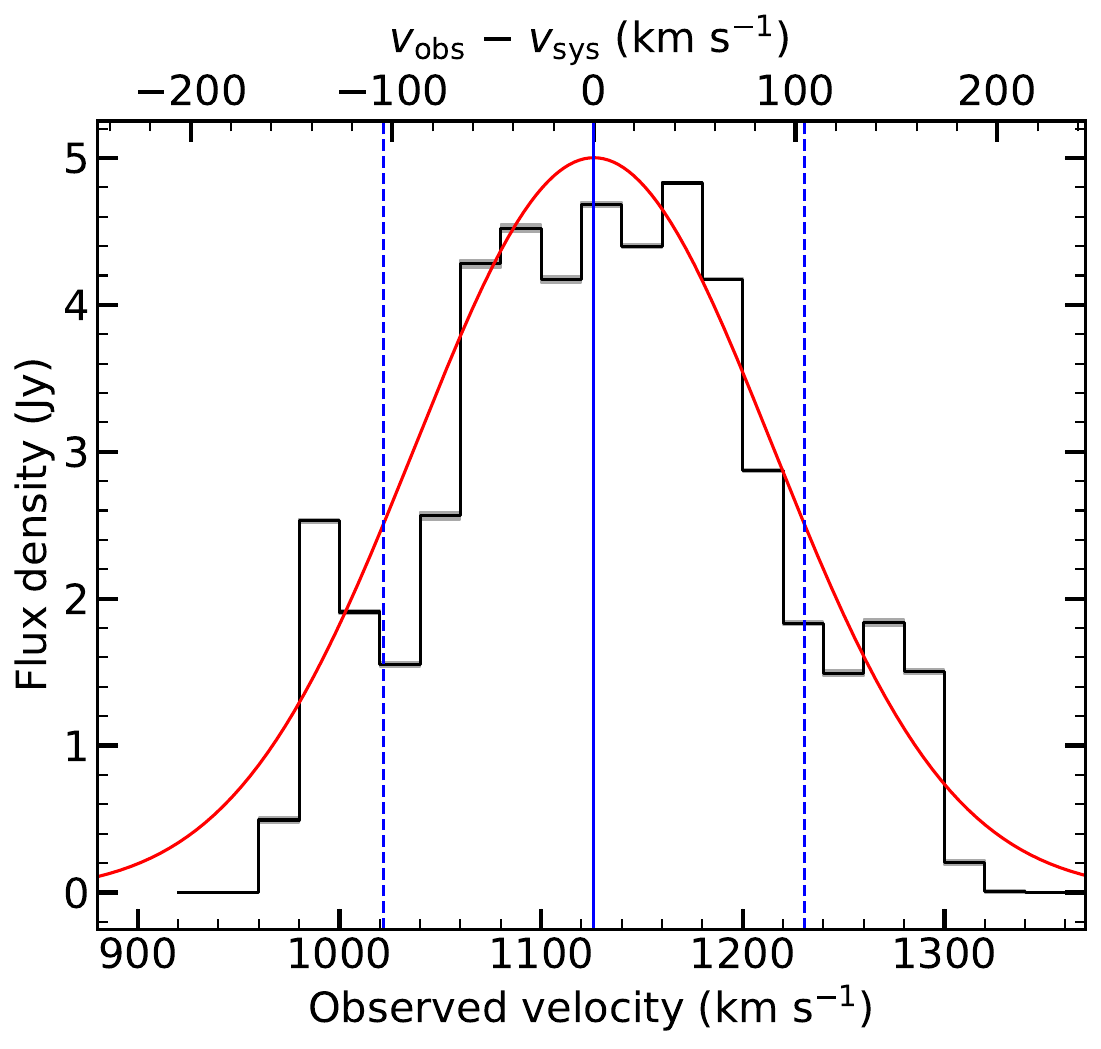}
    \end{subfigure}
    \begin{subfigure}{.32\linewidth}
    \centering
    \includegraphics[height=4.8cm]{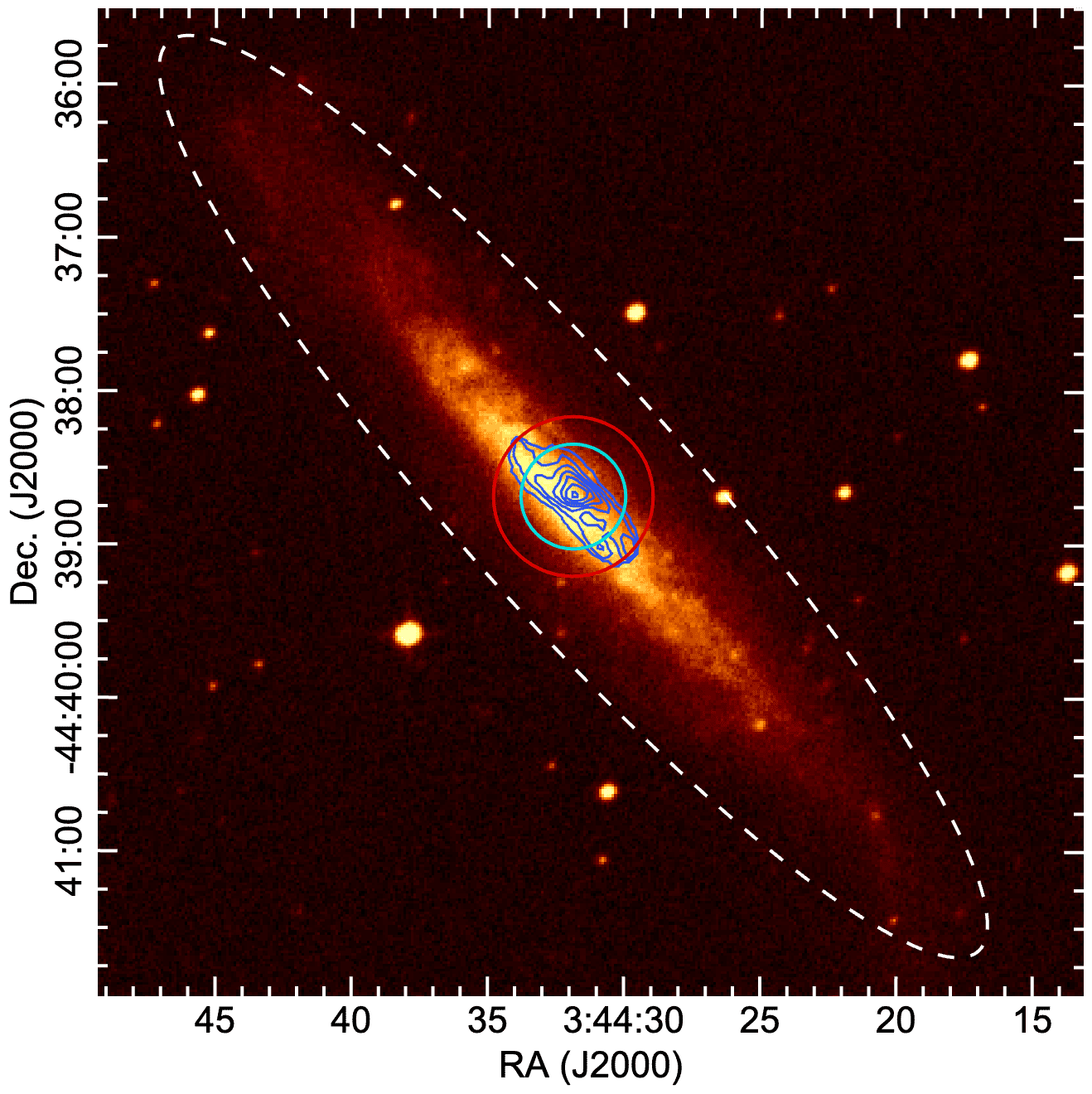}
    \end{subfigure}
    \begin{subfigure}{.33\linewidth}
    \centering
    \includegraphics[height=5.2cm]{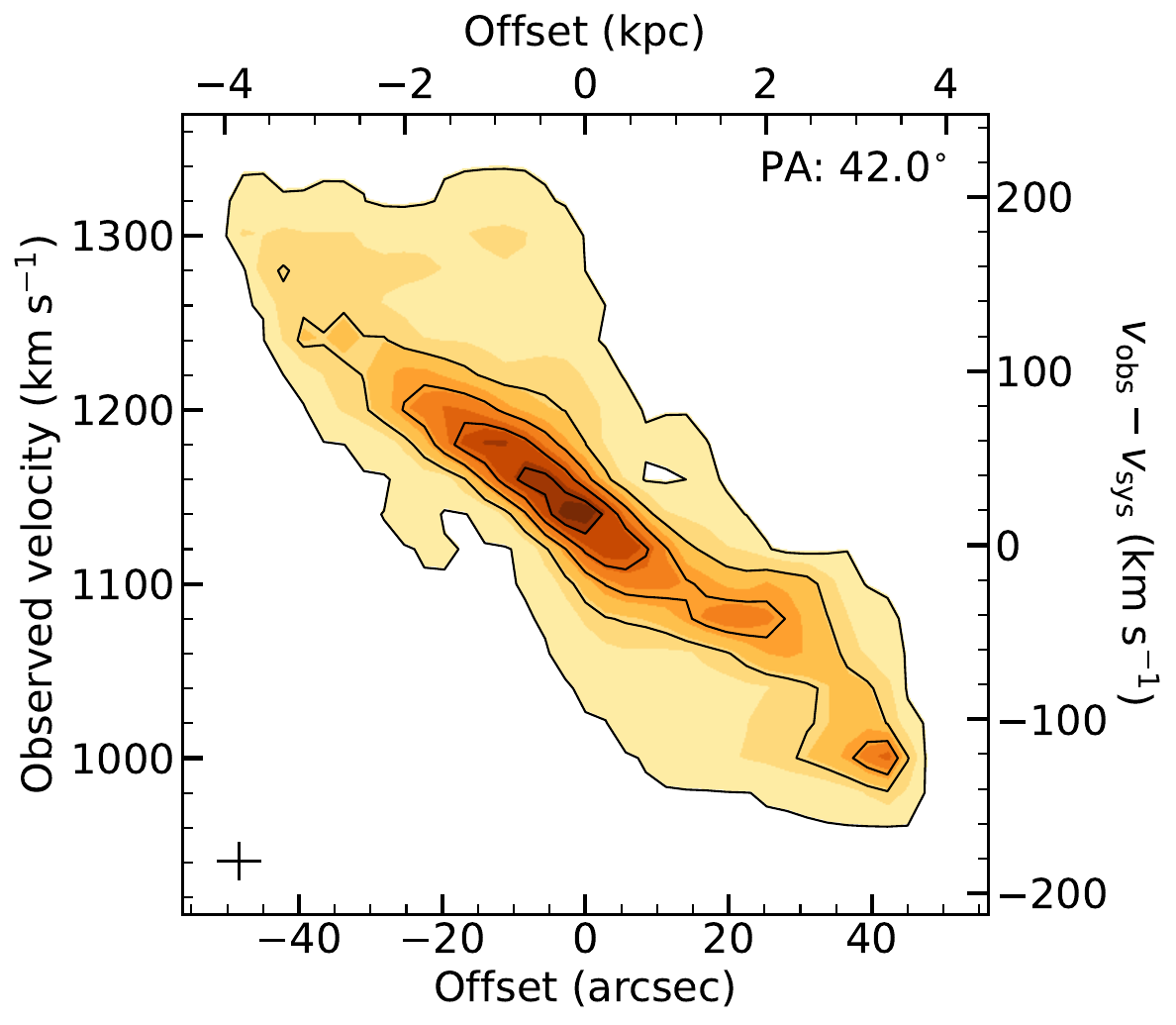}
    \end{subfigure}
    \\
    \begin{subfigure}{.34\linewidth}
    \centering
    \includegraphics[height=6.0cm]{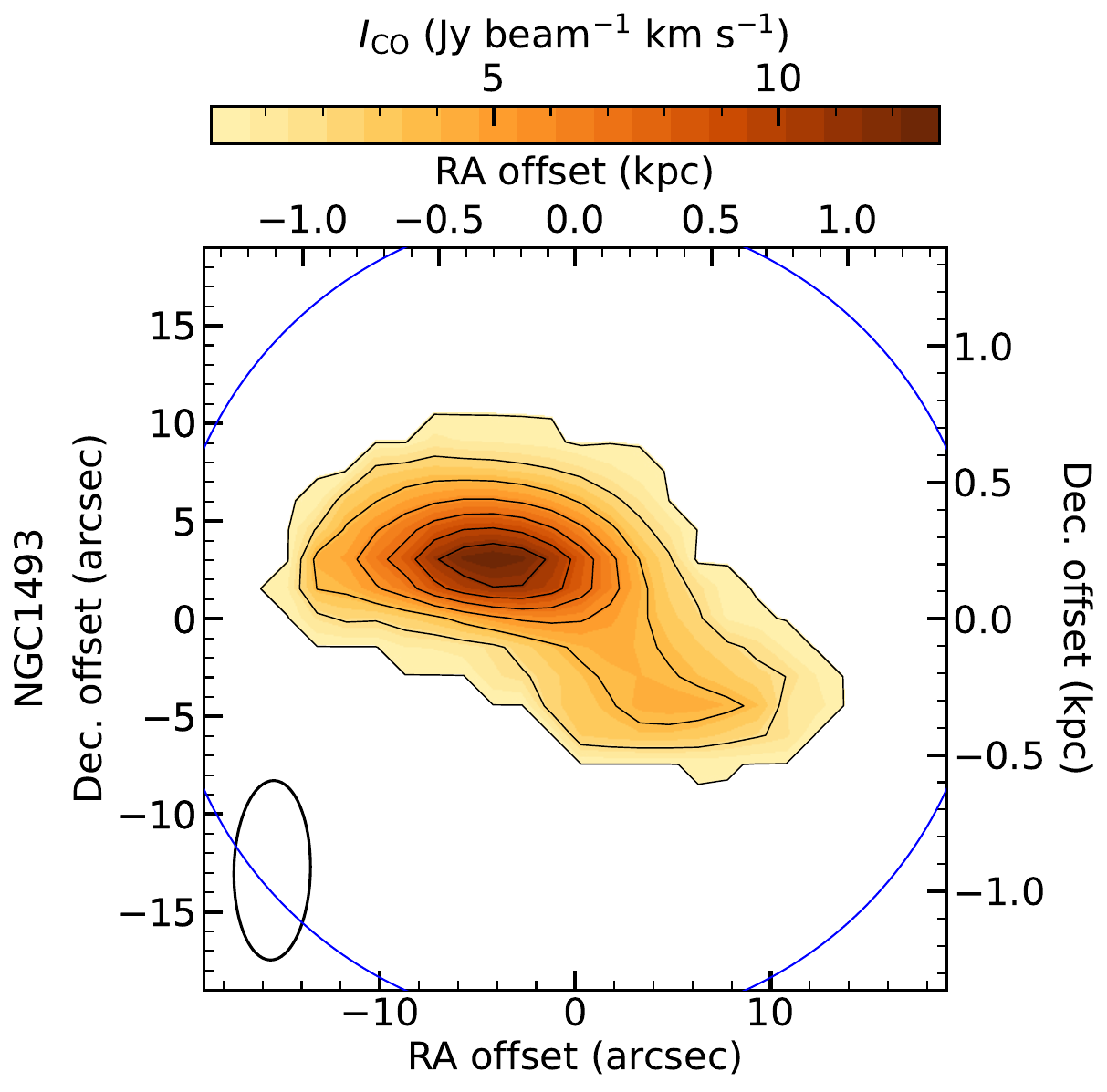}
    \end{subfigure}
    \begin{subfigure}{.32\linewidth}
    \centering
    \includegraphics[height=6.0cm]{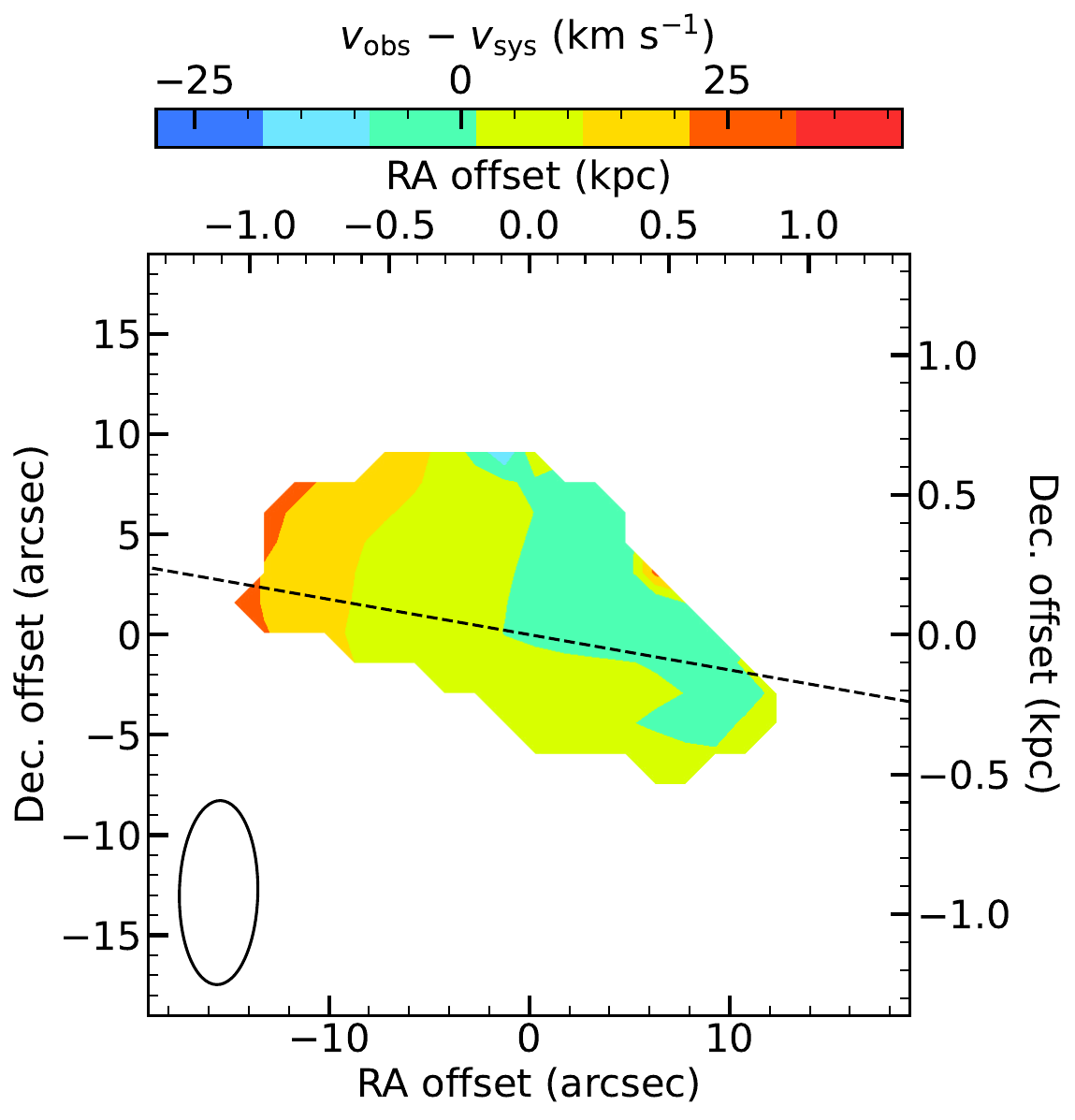}
    \end{subfigure}
    \begin{subfigure}{.32\linewidth}
    \centering
    \includegraphics[height=6.0cm]{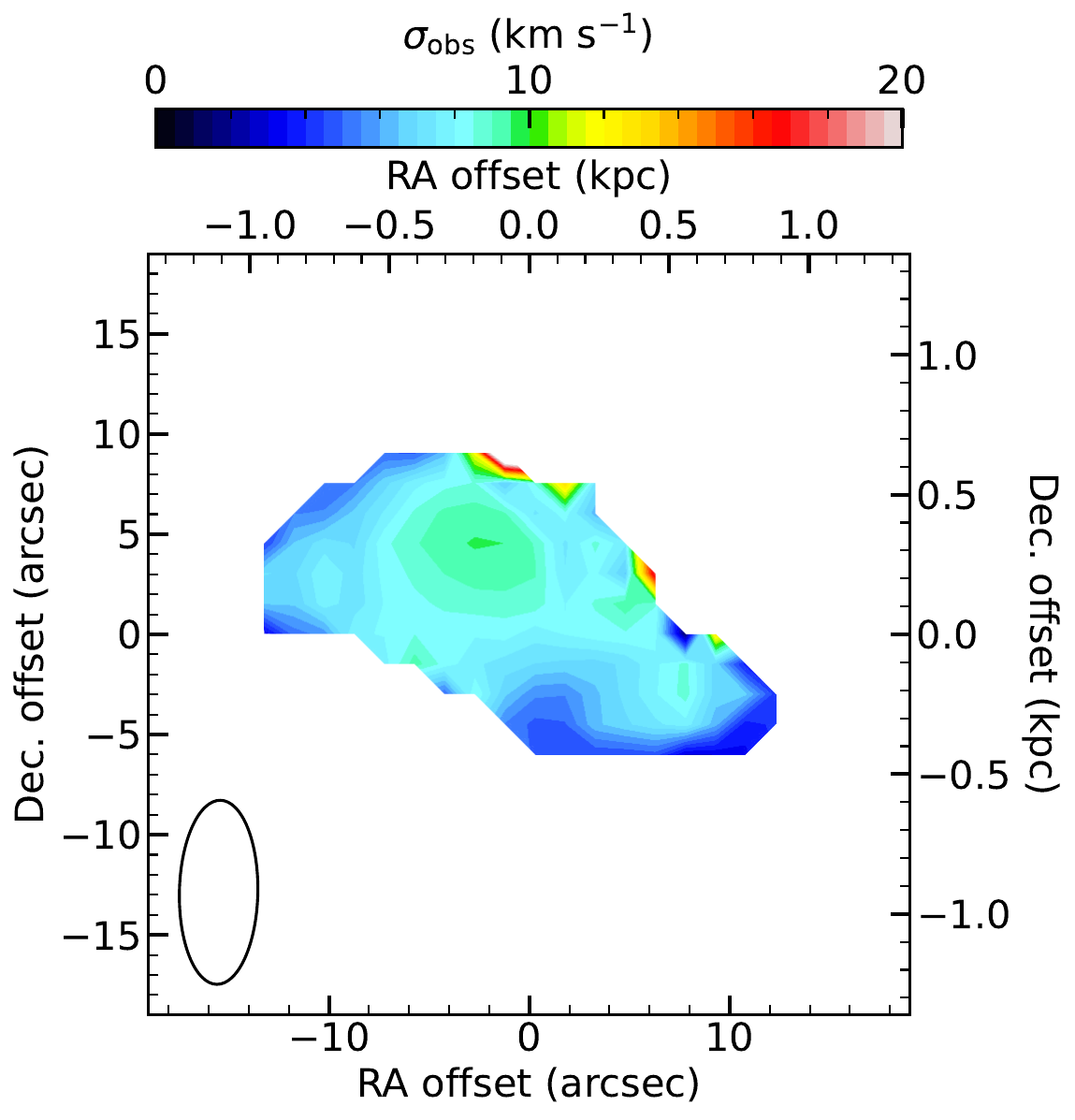}
    \end{subfigure}
    \\
    \begin{subfigure}{.34\linewidth}
    \centering
    \includegraphics[height=5.2cm]{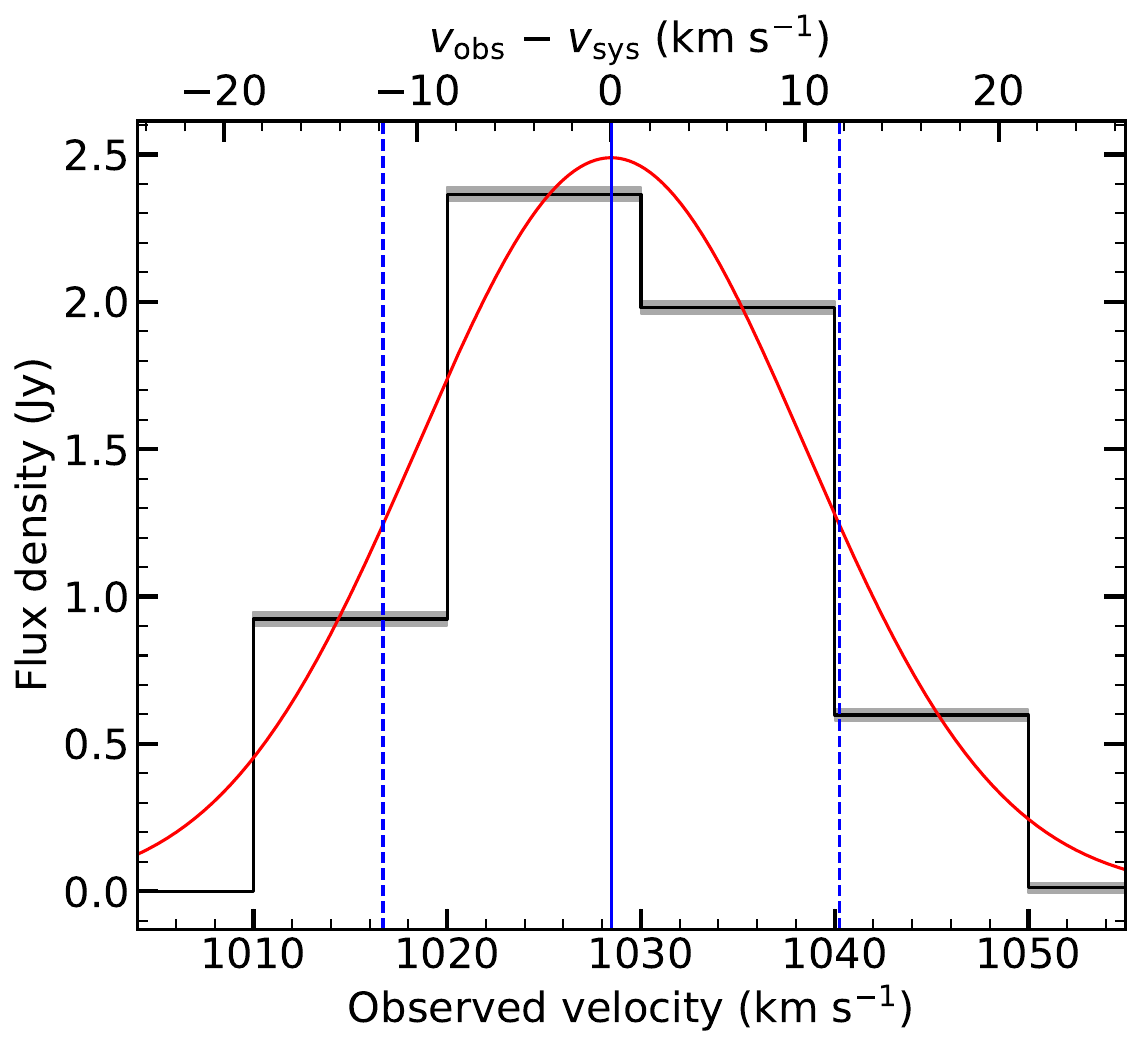}
    \end{subfigure}
    \begin{subfigure}{.32\linewidth}
    \centering
    \includegraphics[height=4.8cm]{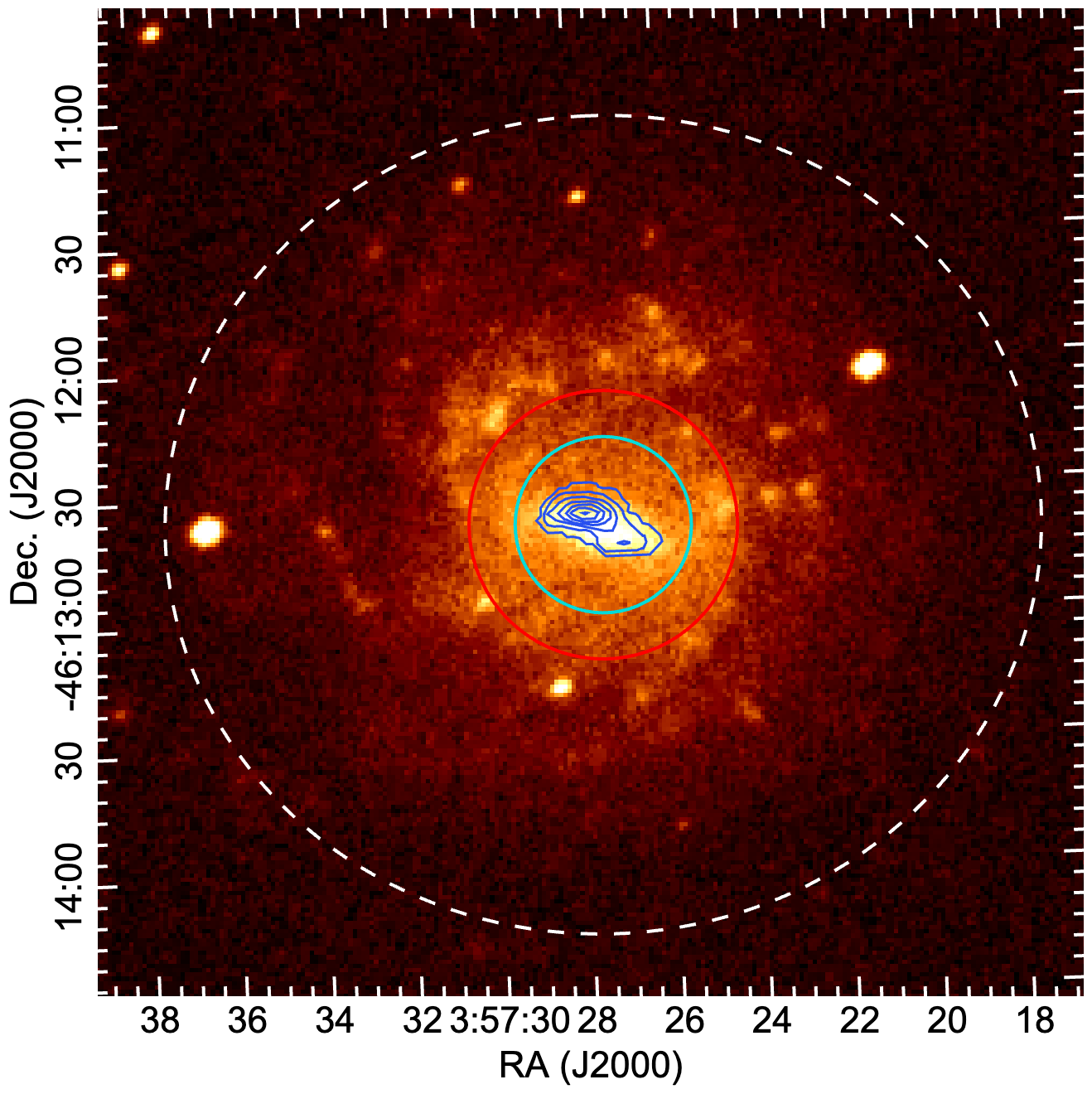}
    \end{subfigure}
    \begin{subfigure}{.33\linewidth}
    \centering
    \includegraphics[height=5.2cm]{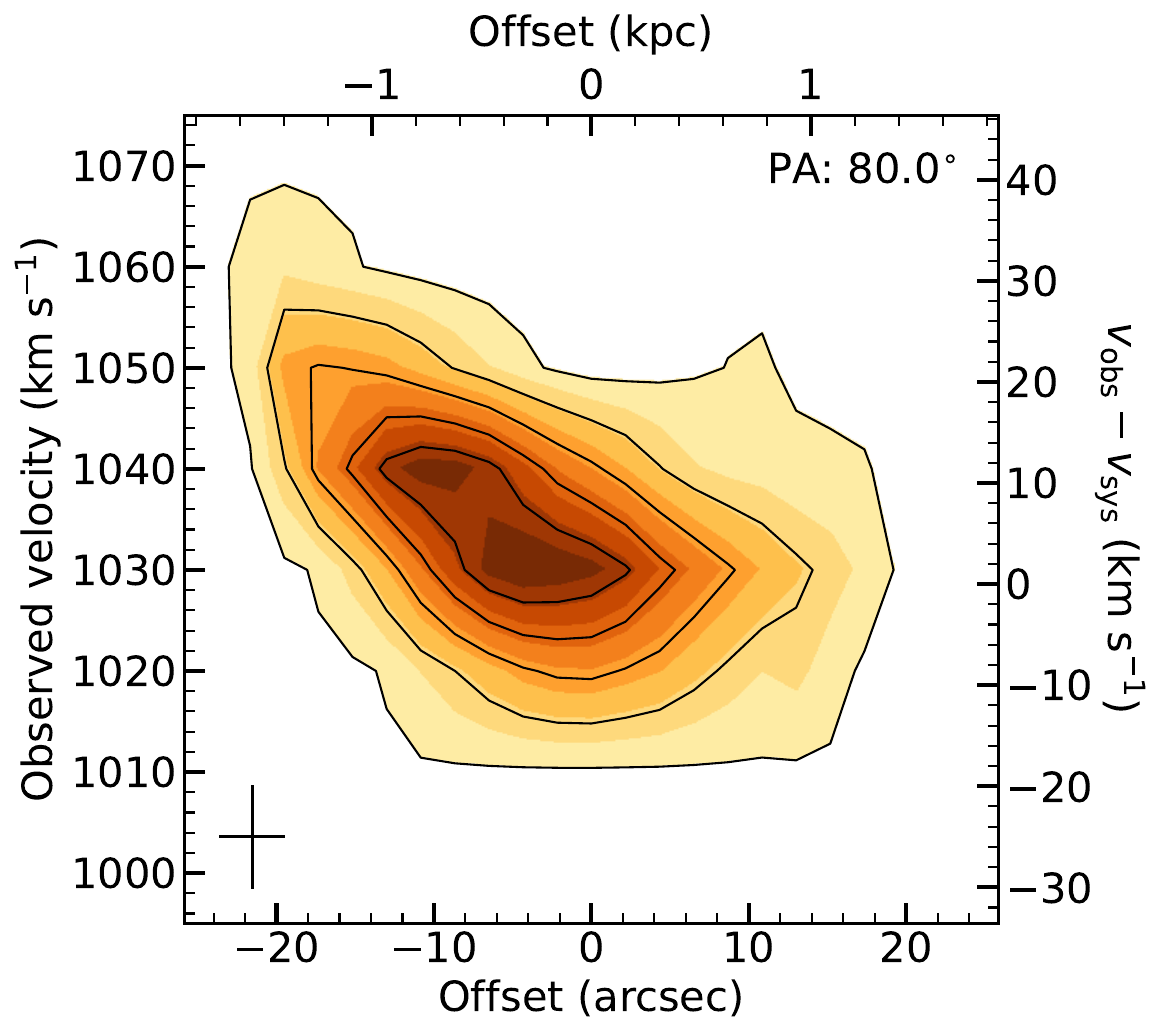}
    \end{subfigure}
    \caption[]{continued.}
\end{figure*}

\begin{figure*}
    \ContinuedFloat
    \centering
    \begin{subfigure}{.34\linewidth}
    \centering
    \includegraphics[height=6.0cm]{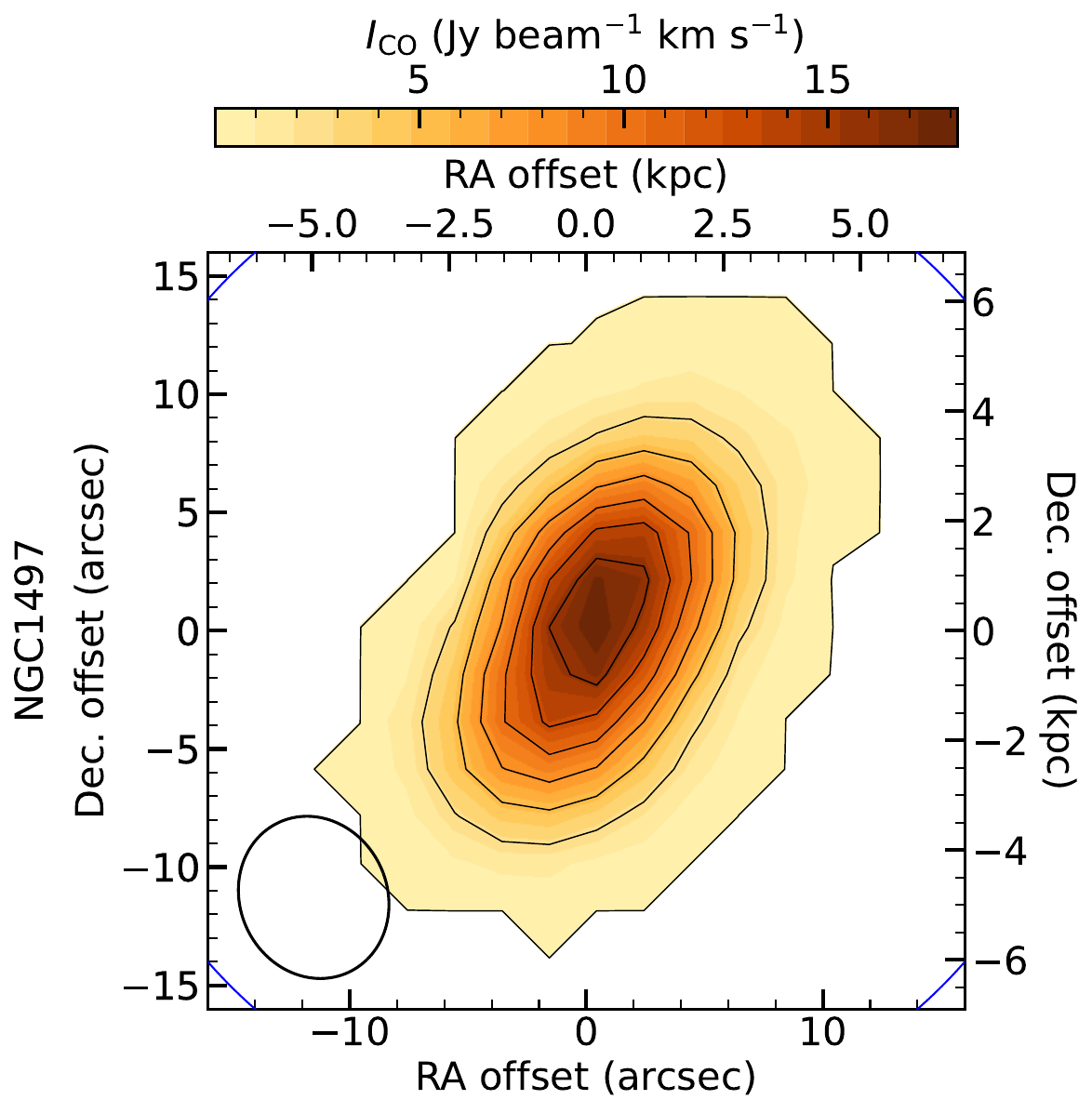}
    \end{subfigure}
    \begin{subfigure}{.32\linewidth}
    \centering
    \includegraphics[height=6.0cm]{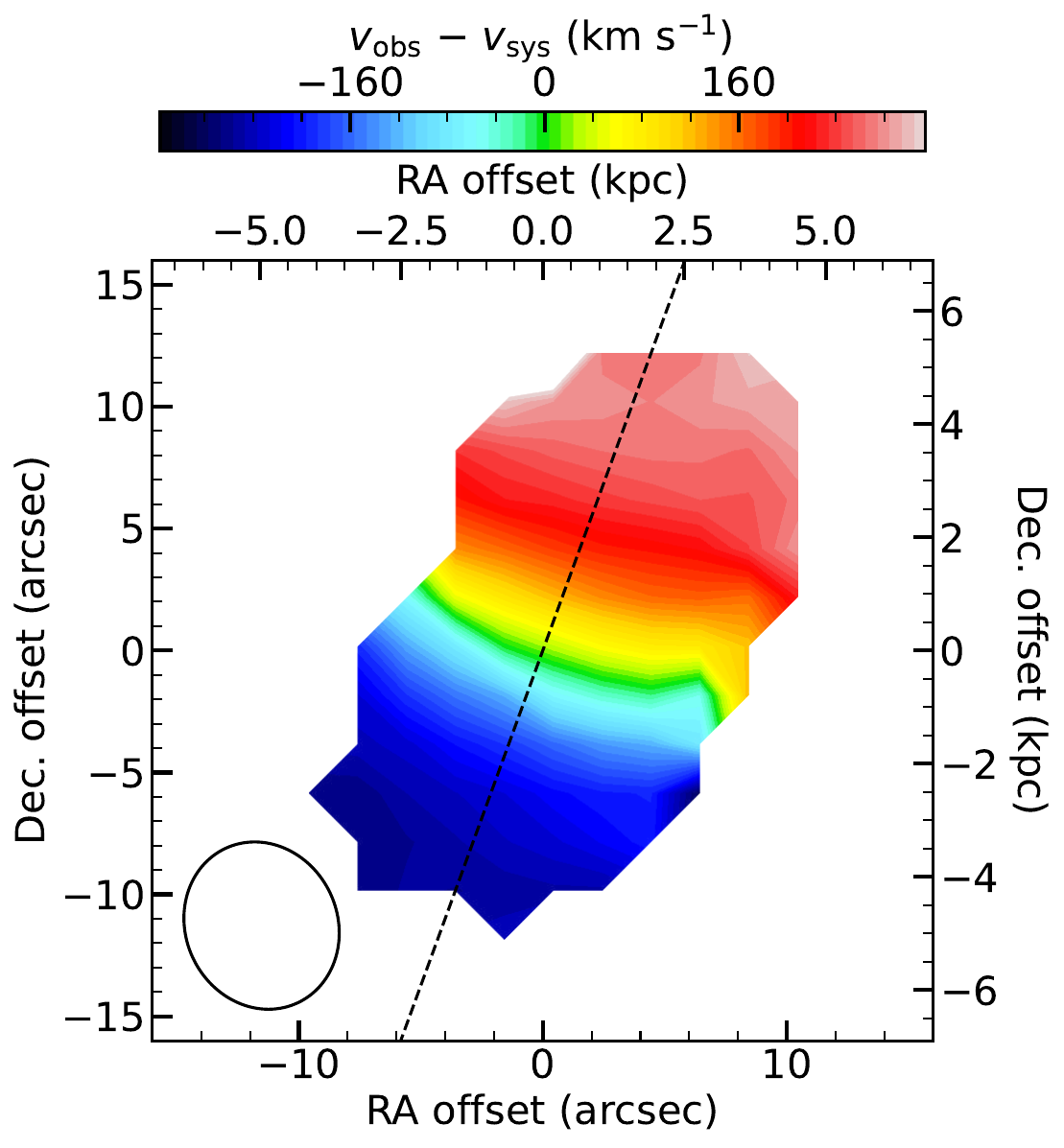}
    \end{subfigure}
    \begin{subfigure}{.32\linewidth}
    \centering
    \includegraphics[height=6.0cm]{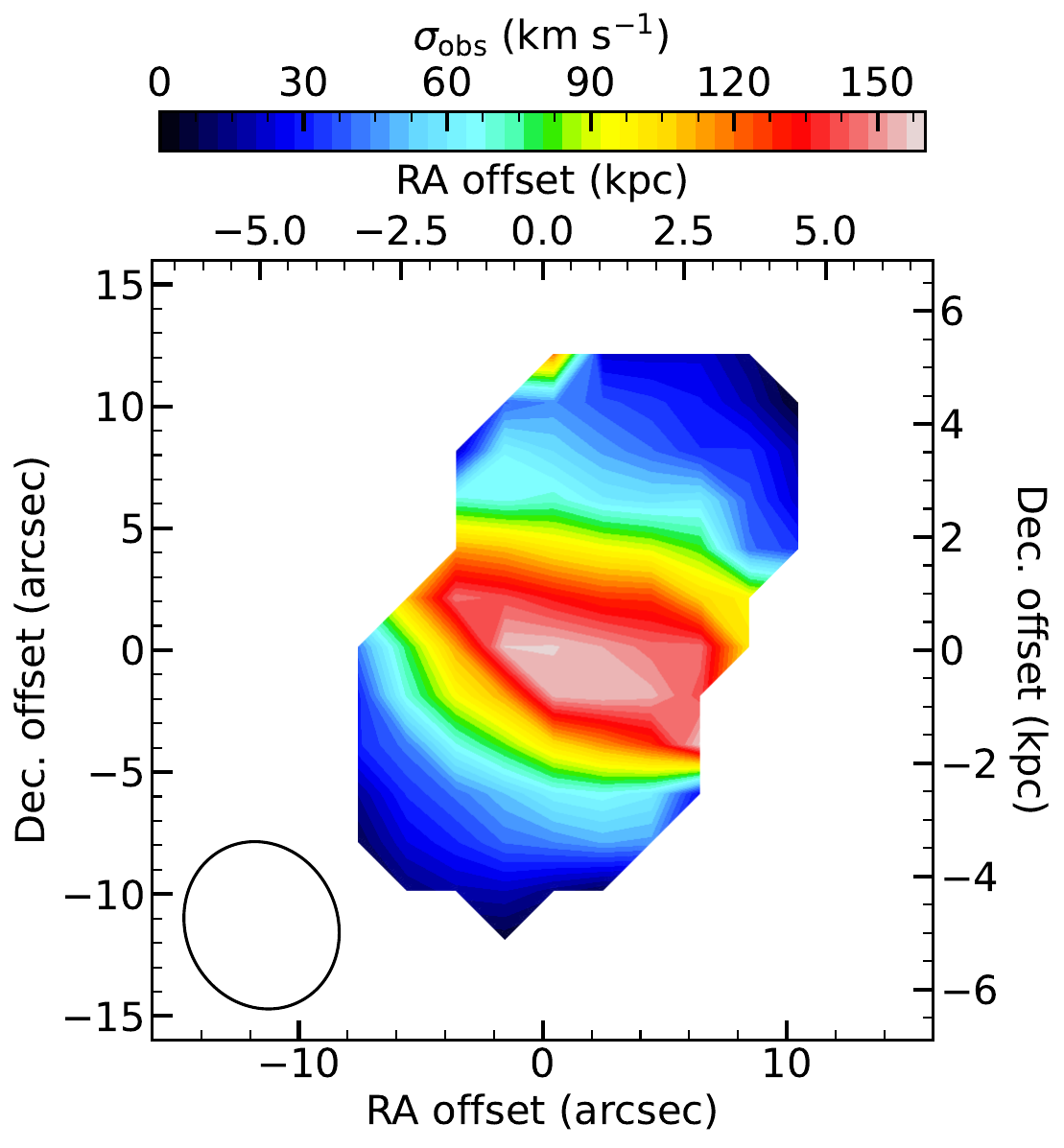}
    \end{subfigure}
    \\
    \begin{subfigure}{.34\linewidth}
    \centering
    \includegraphics[height=5.2cm]{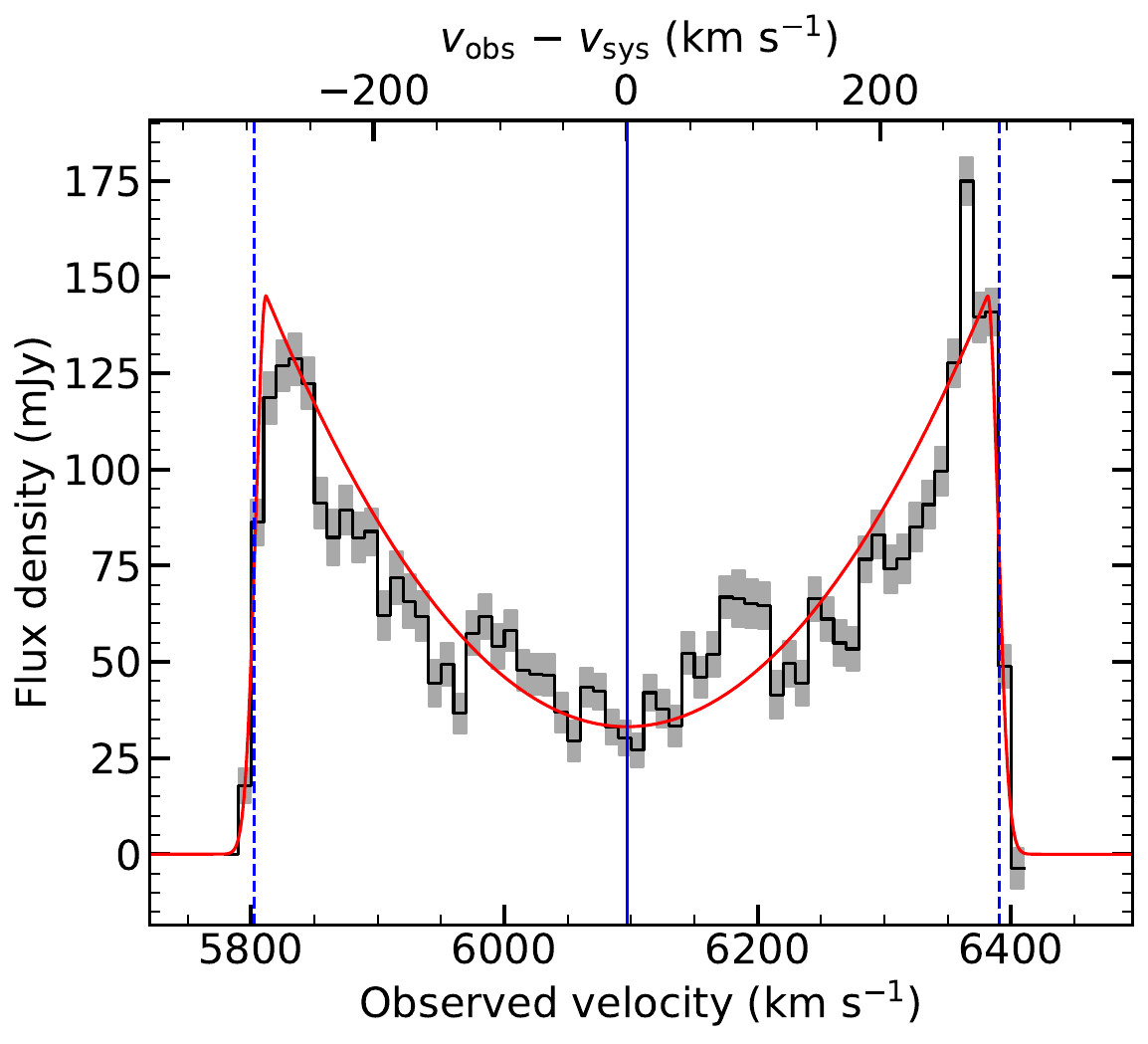}
    \end{subfigure}
    \begin{subfigure}{.32\linewidth}
    \centering
    \includegraphics[height=4.8cm]{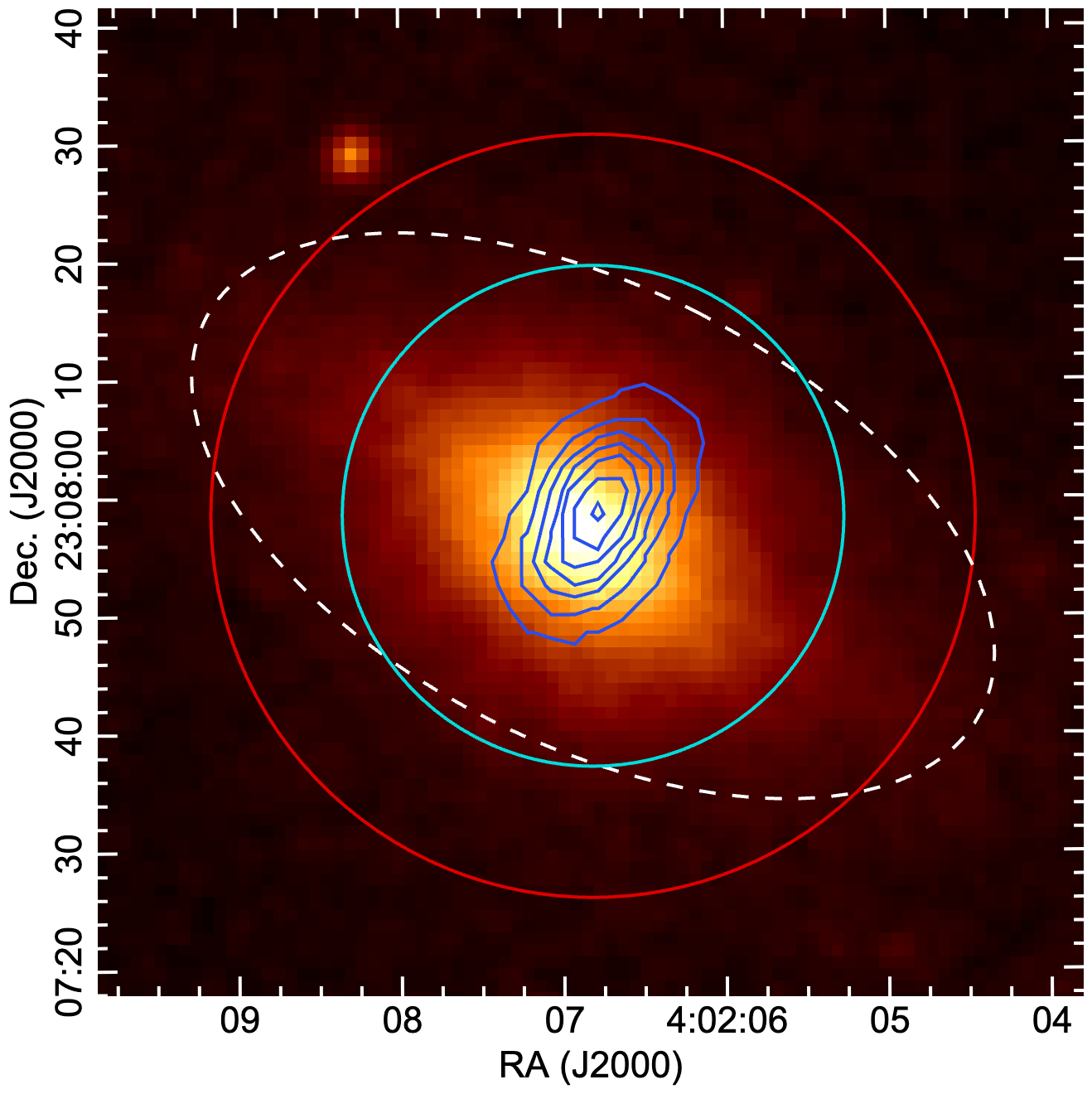}
    \end{subfigure}
    \begin{subfigure}{.33\linewidth}
    \centering
    \includegraphics[height=5.2cm]{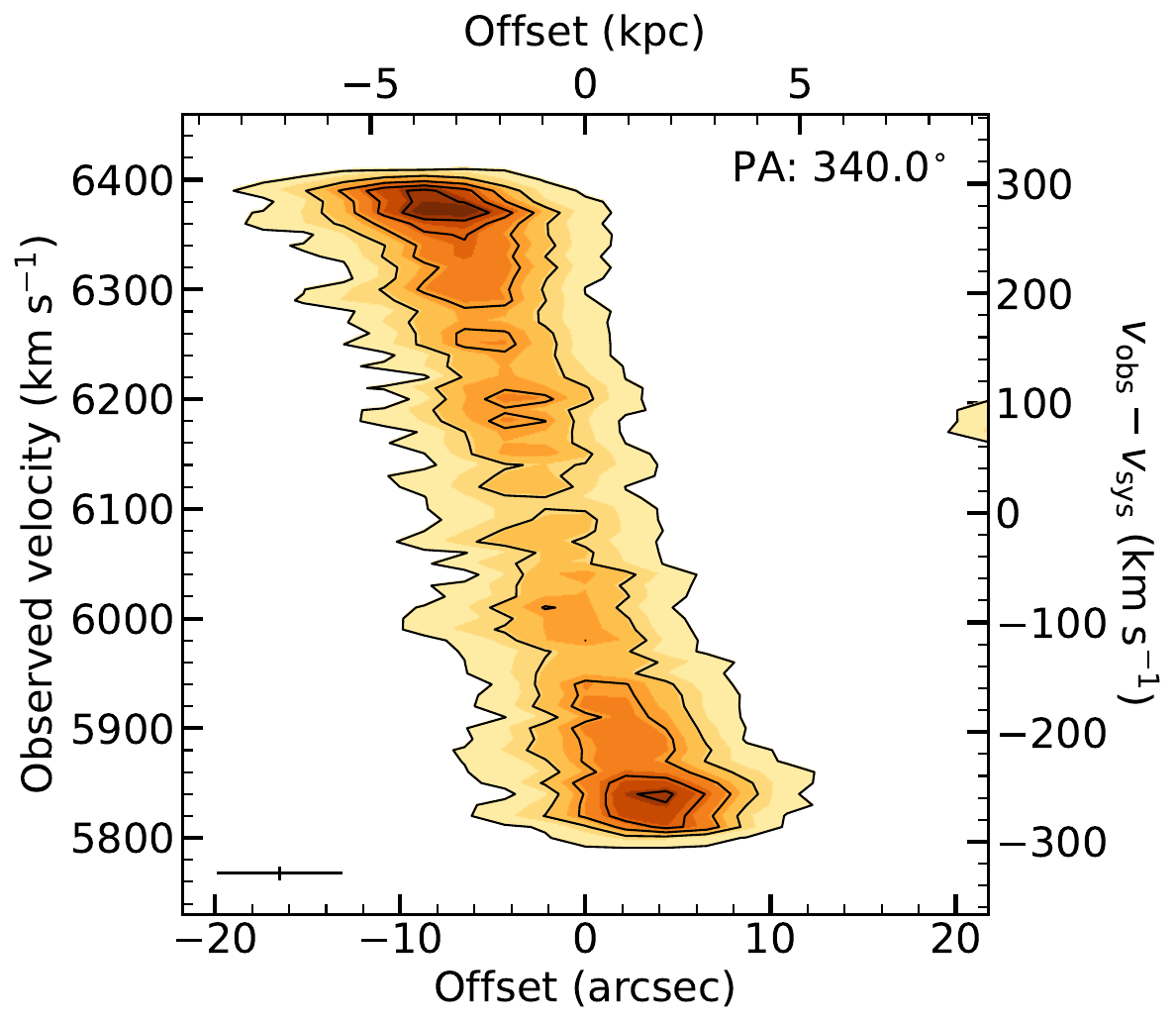}
    \end{subfigure}
    \\
    \begin{subfigure}{.34\linewidth}
    \centering
    \includegraphics[height=6.0cm]{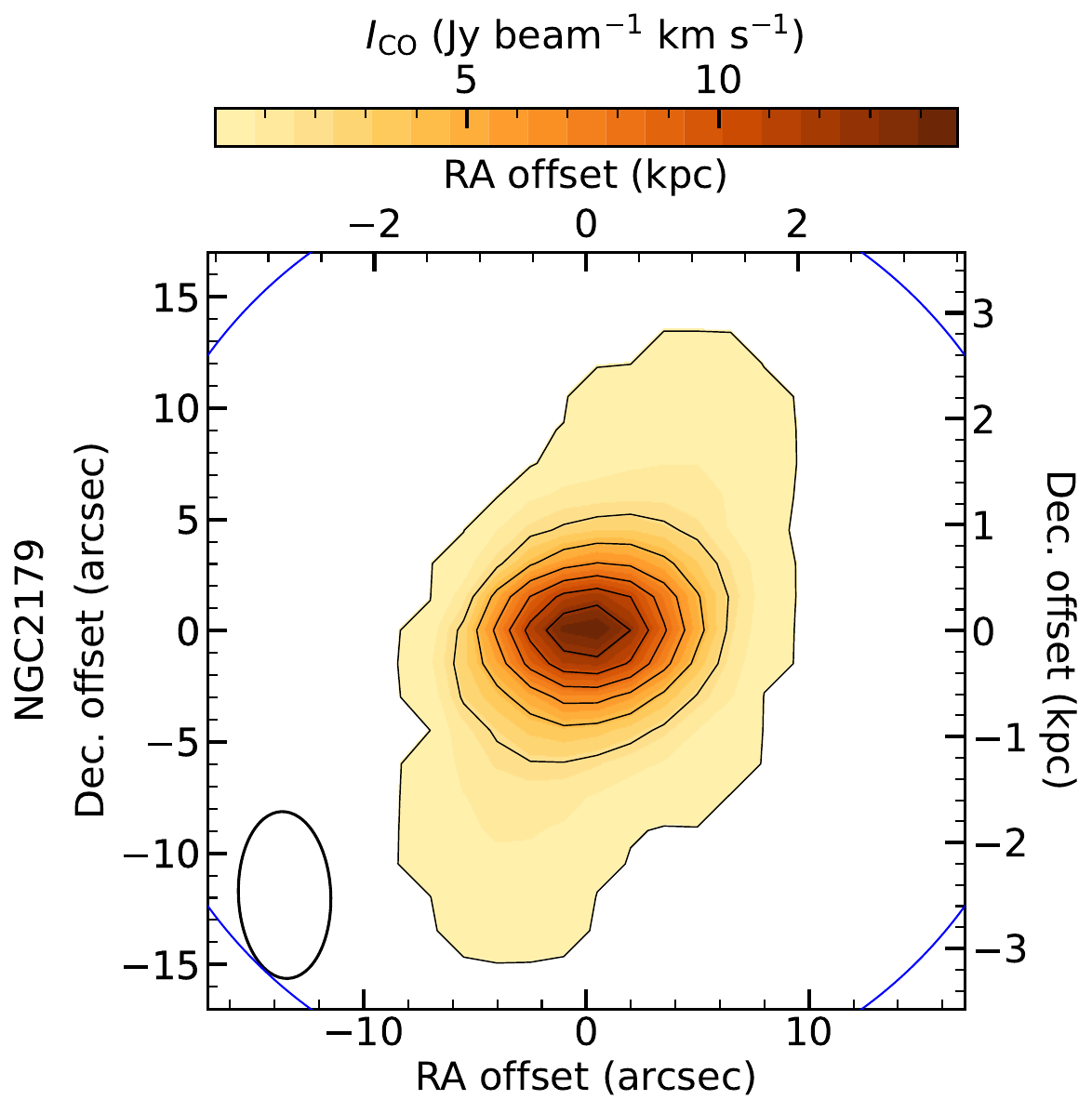}
    \end{subfigure}
    \begin{subfigure}{.32\linewidth}
    \centering
    \includegraphics[height=6.0cm]{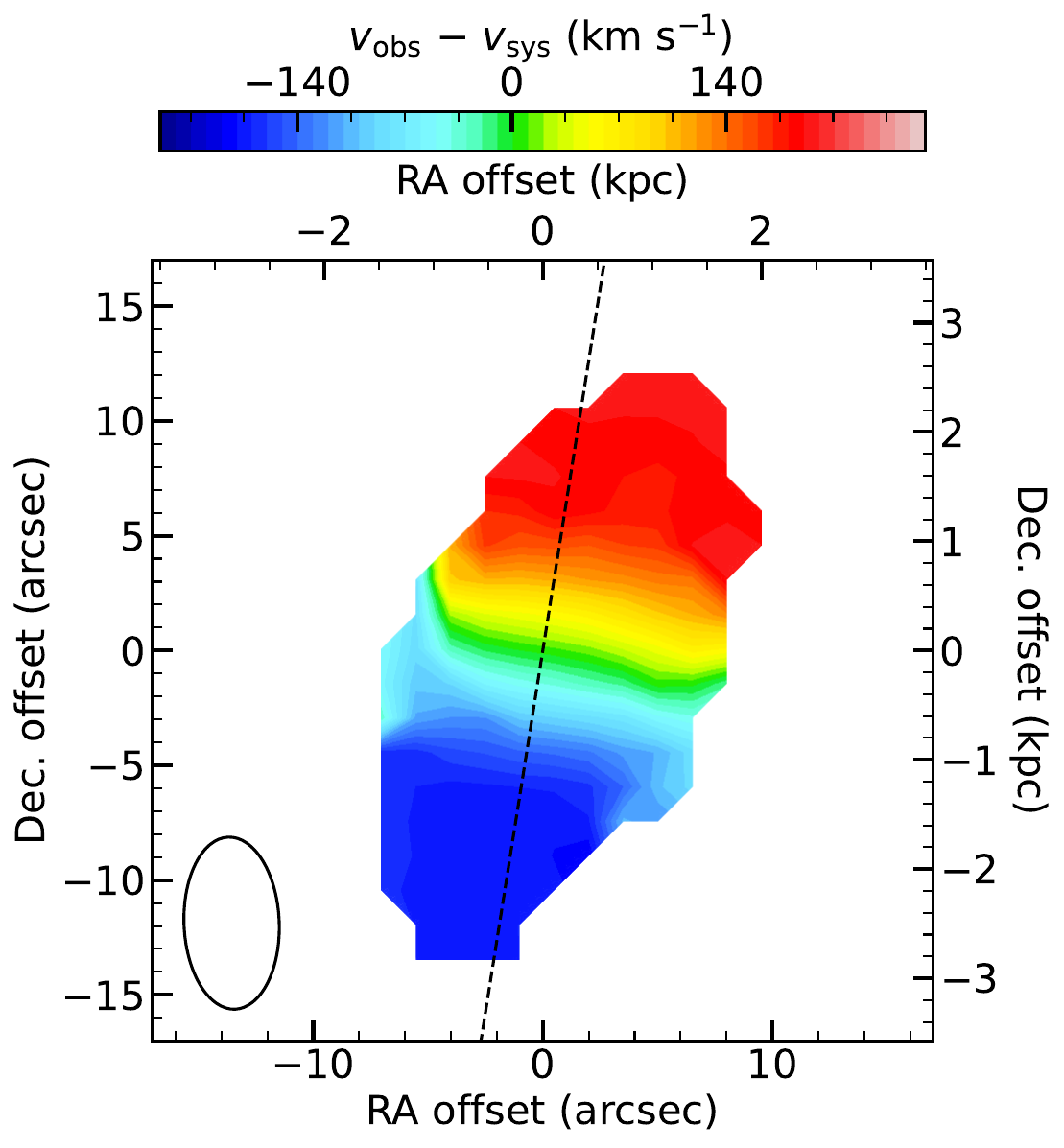}
    \end{subfigure}
    \begin{subfigure}{.32\linewidth}
    \centering
    \includegraphics[height=6.0cm]{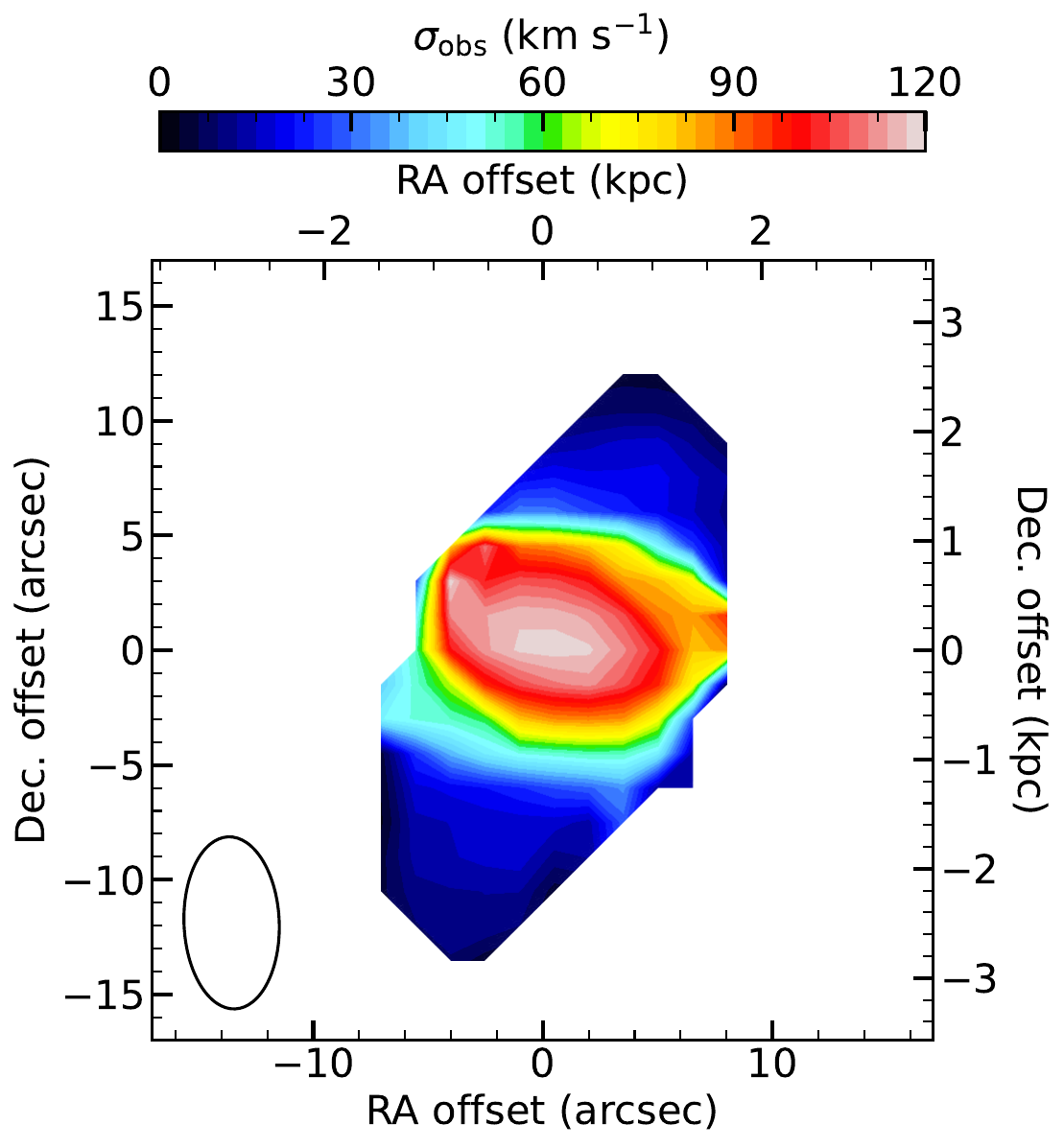}
    \end{subfigure}
    \\
    \begin{subfigure}{.34\linewidth}
    \centering
    \includegraphics[height=5.2cm]{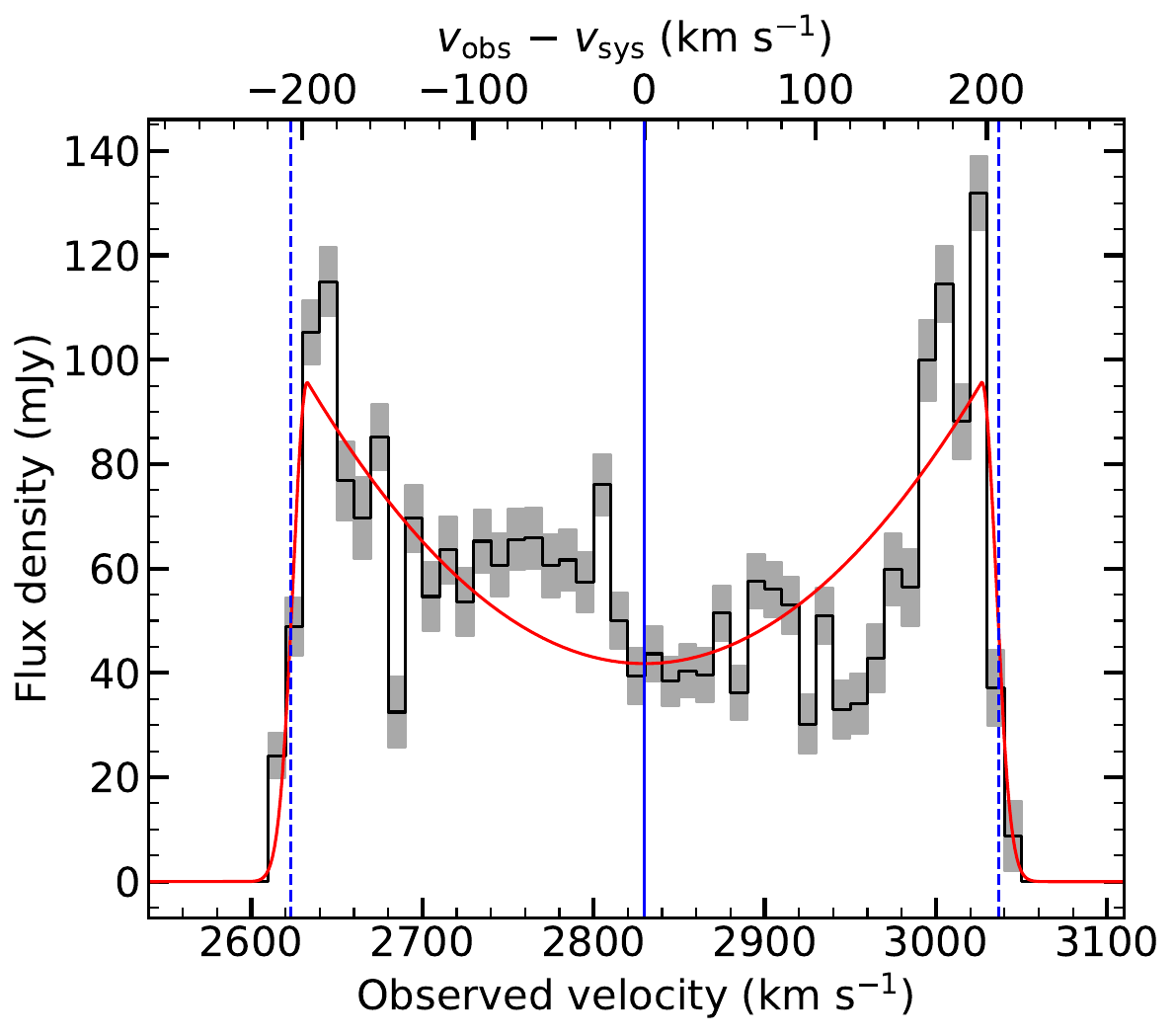}
    \end{subfigure}
    \begin{subfigure}{.32\linewidth}
    \centering
    \includegraphics[height=4.8cm]{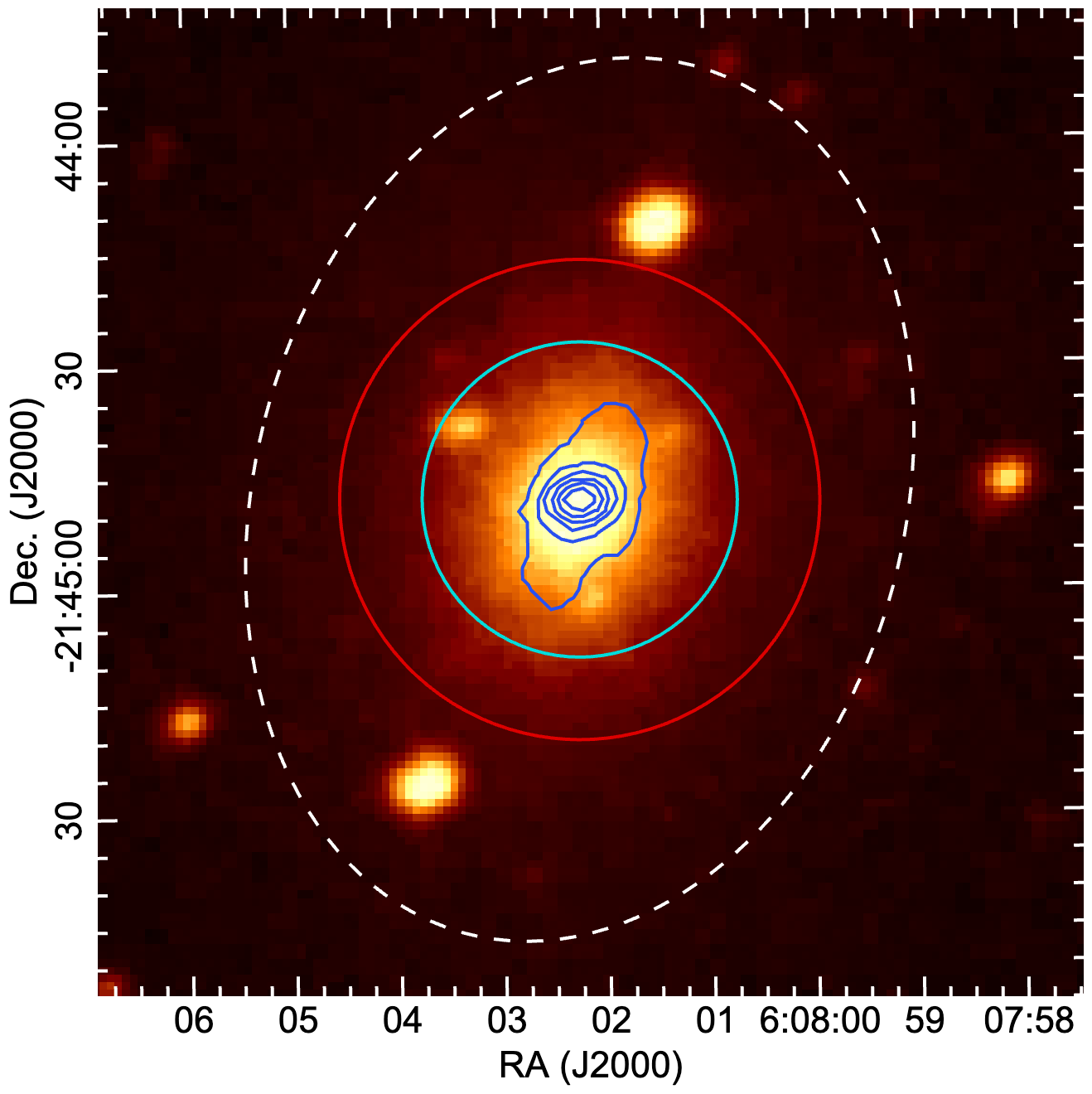}
    \end{subfigure}
    \begin{subfigure}{.33\linewidth}
    \centering
    \includegraphics[height=5.2cm]{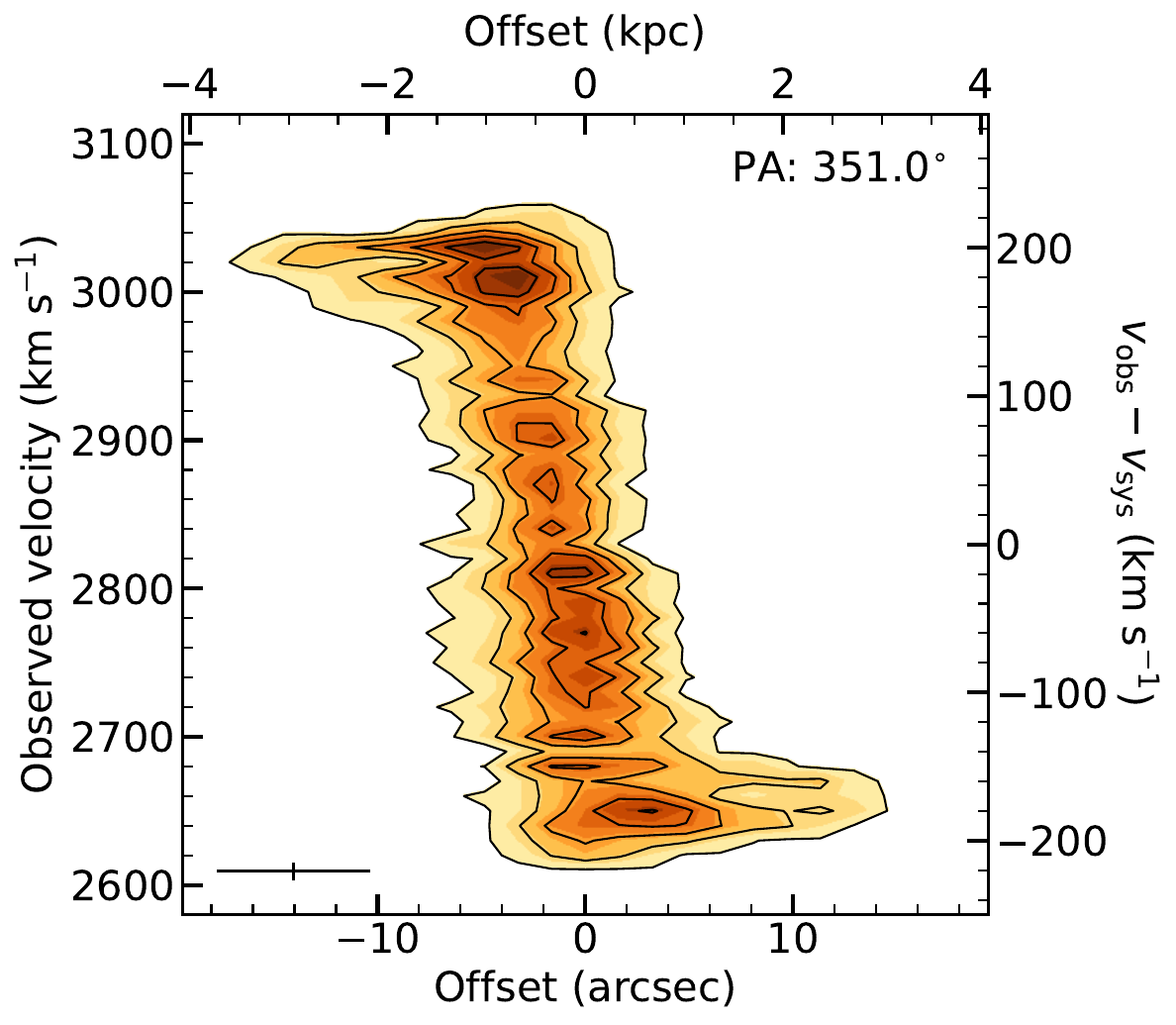}
    \end{subfigure}
    \caption[]{continued.}
\end{figure*}

\begin{figure*}
    \ContinuedFloat
    \centering
    \begin{subfigure}{.34\linewidth}
    \centering
    \includegraphics[height=6.0cm]{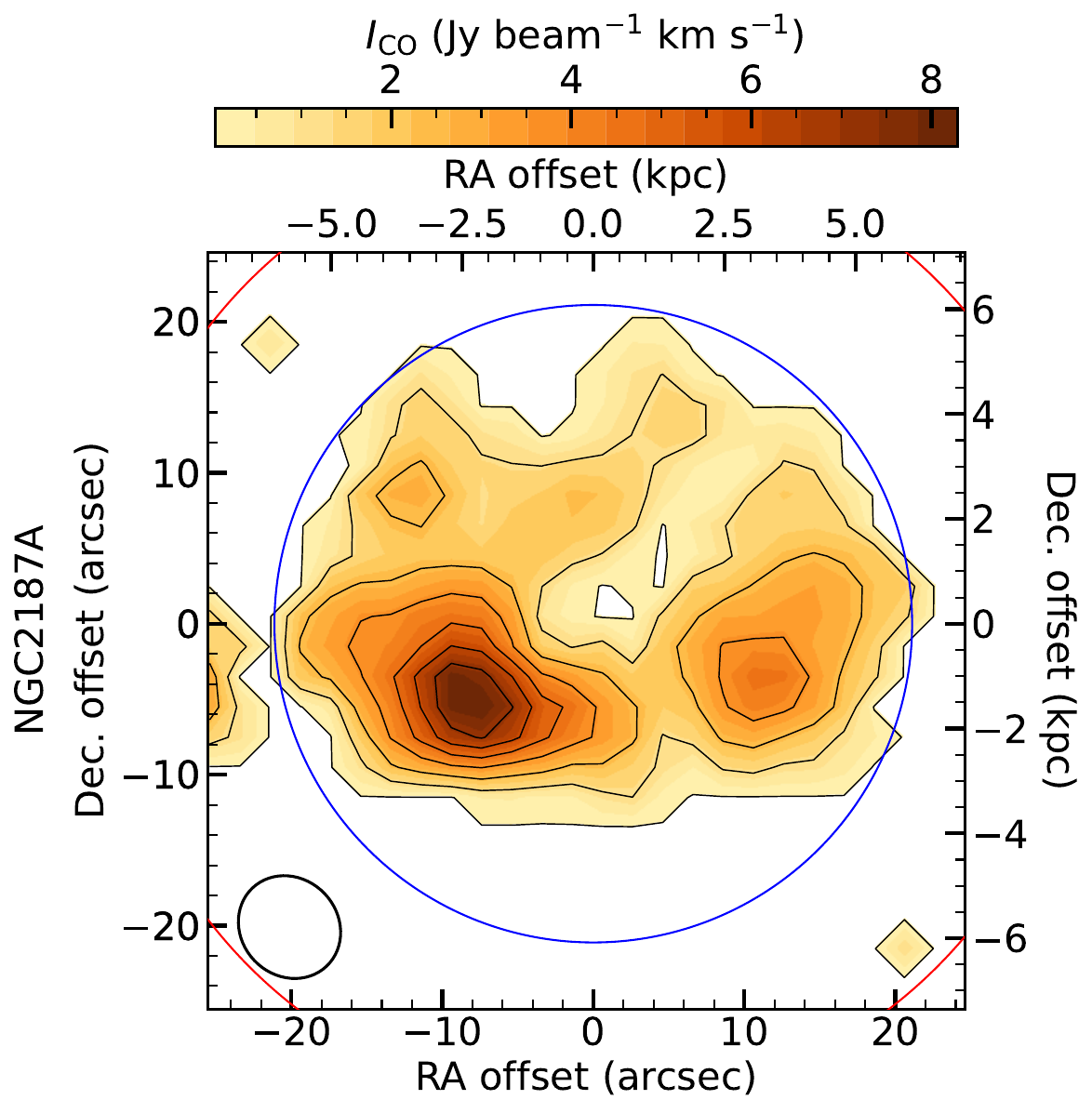}
    \end{subfigure}
    \begin{subfigure}{.32\linewidth}
    \centering
    \includegraphics[height=6.0cm]{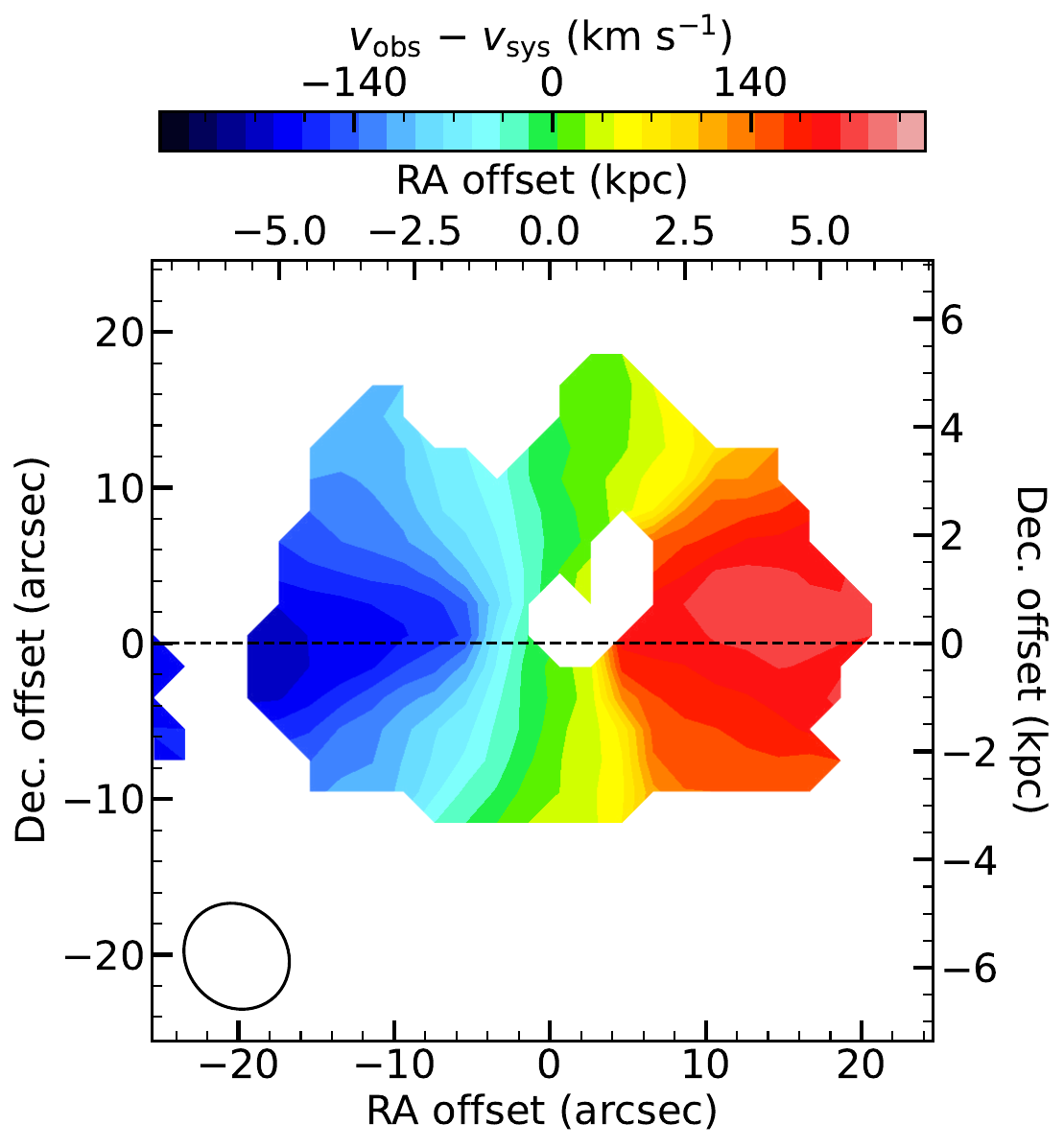}
    \end{subfigure}
    \begin{subfigure}{.32\linewidth}
    \centering
    \includegraphics[height=6.0cm]{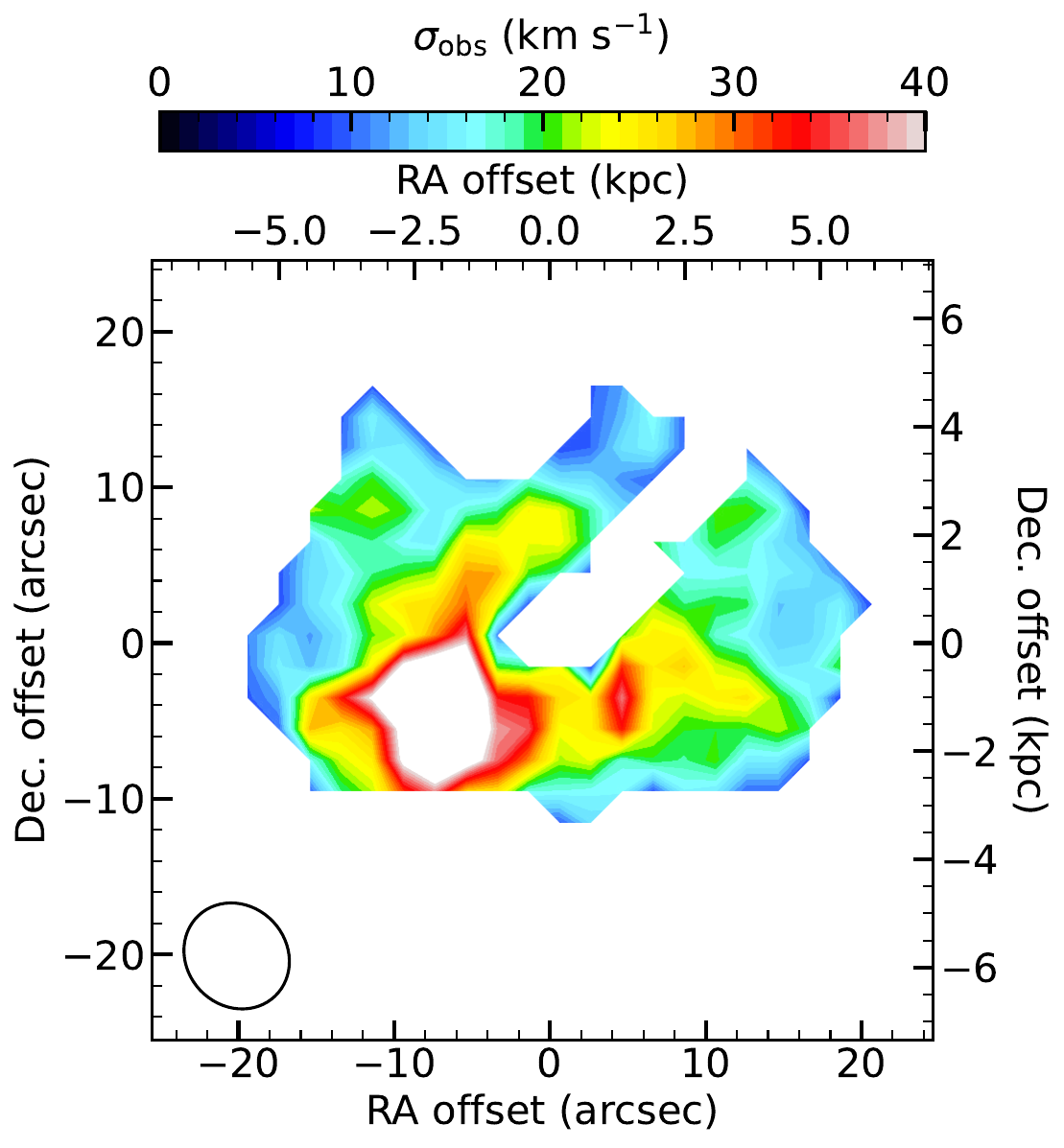}
    \end{subfigure}
    \\
    \begin{subfigure}{.34\linewidth}
    \centering
    \includegraphics[height=5.2cm]{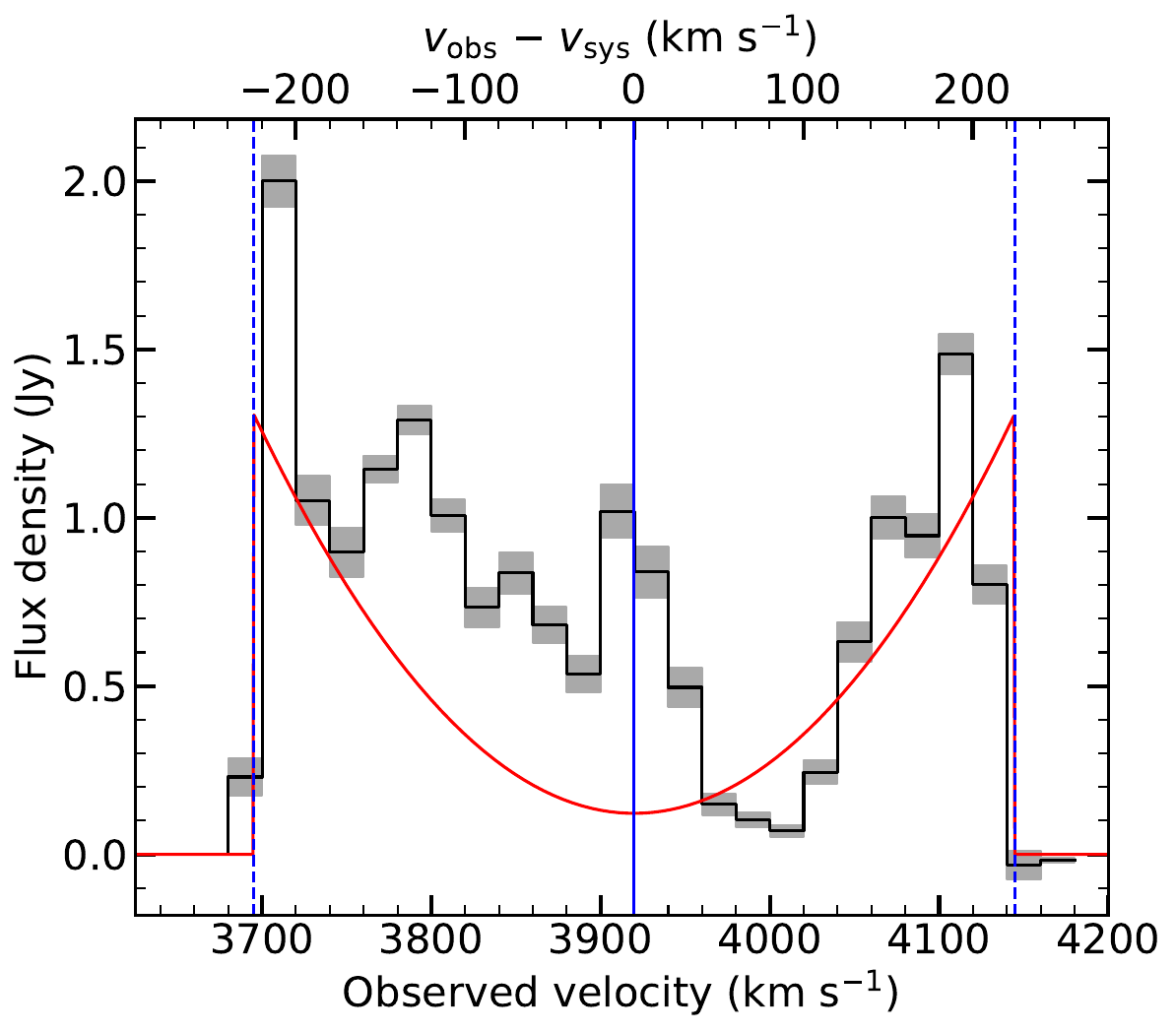}
    \end{subfigure}
    \begin{subfigure}{.32\linewidth}
    \centering
    \includegraphics[height=4.8cm]{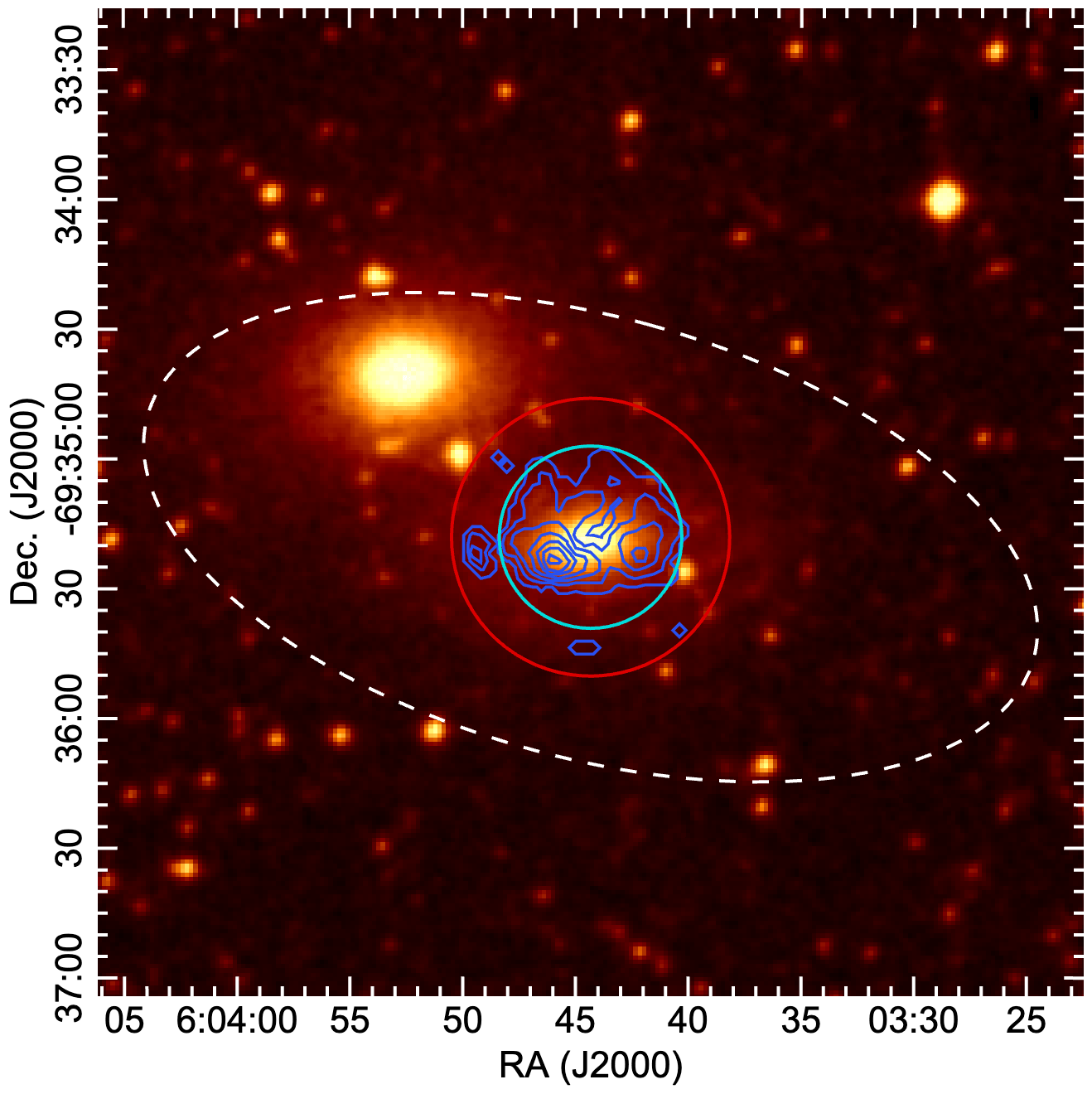}
    \end{subfigure}
    \begin{subfigure}{.33\linewidth}
    \centering
    \includegraphics[height=5.2cm]{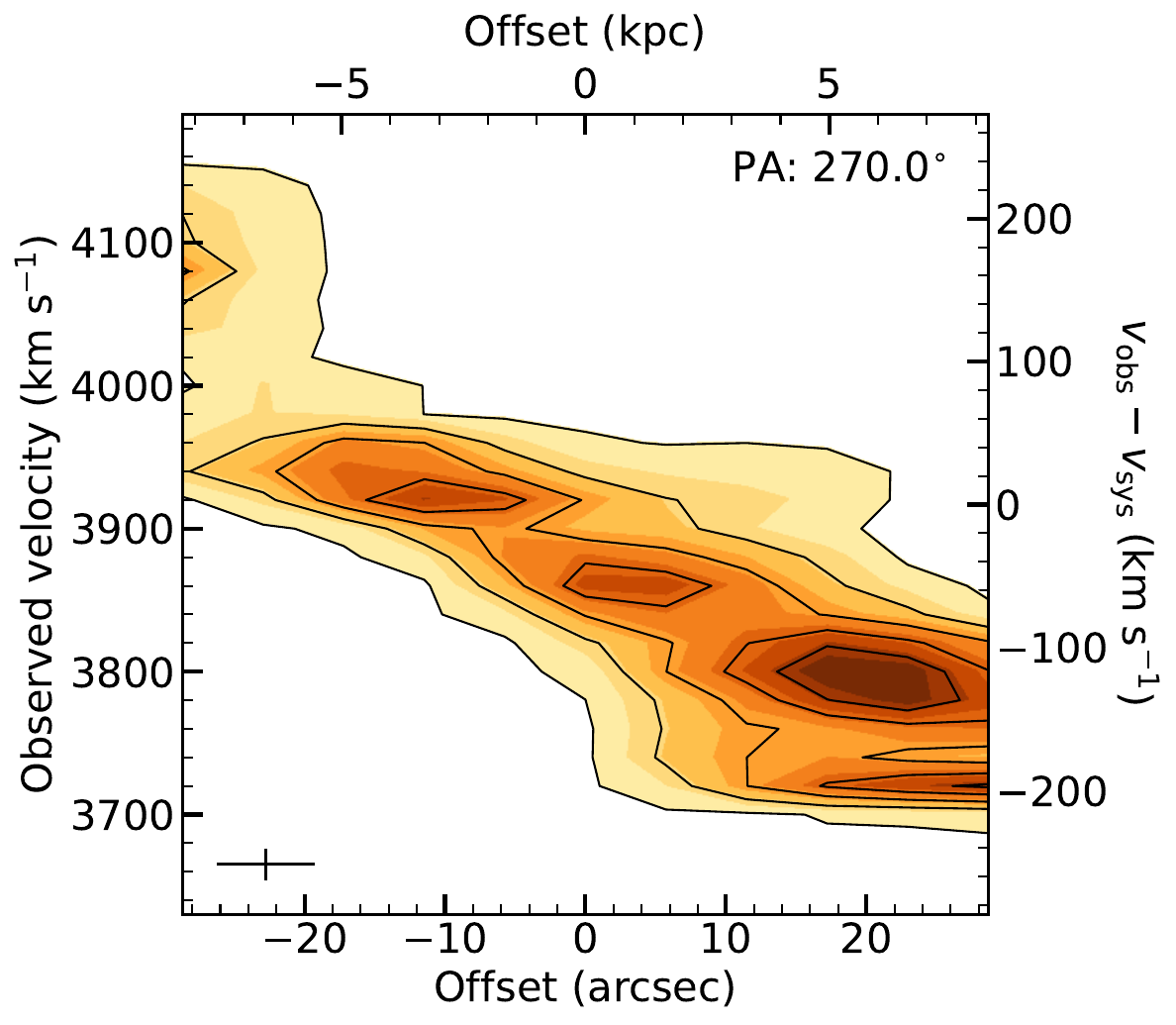}
    \end{subfigure}
    \\
    \begin{subfigure}{.34\linewidth}
    \centering
    \includegraphics[height=6.0cm]{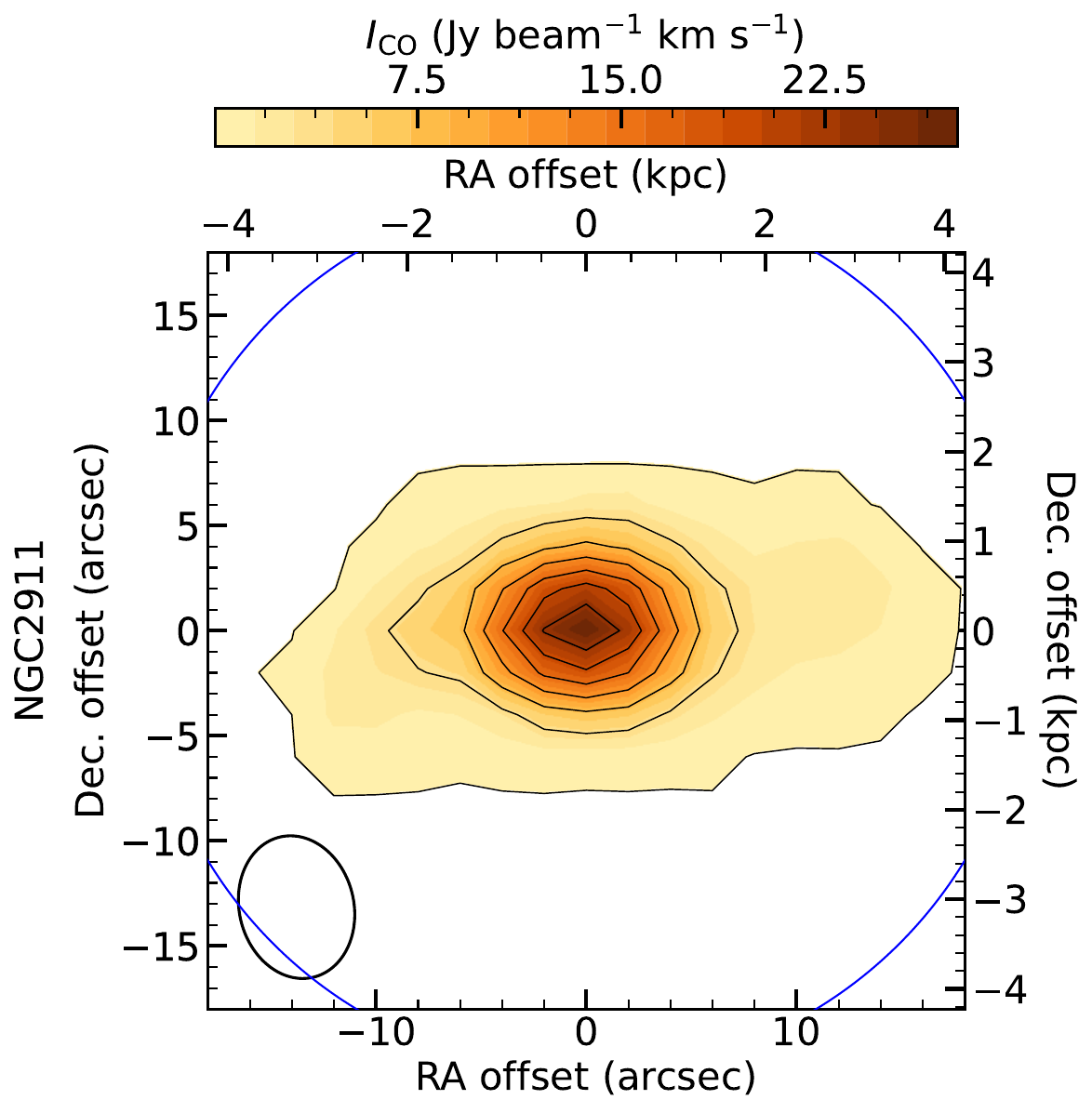}
    \end{subfigure}
    \begin{subfigure}{.32\linewidth}
    \centering
    \includegraphics[height=6.0cm]{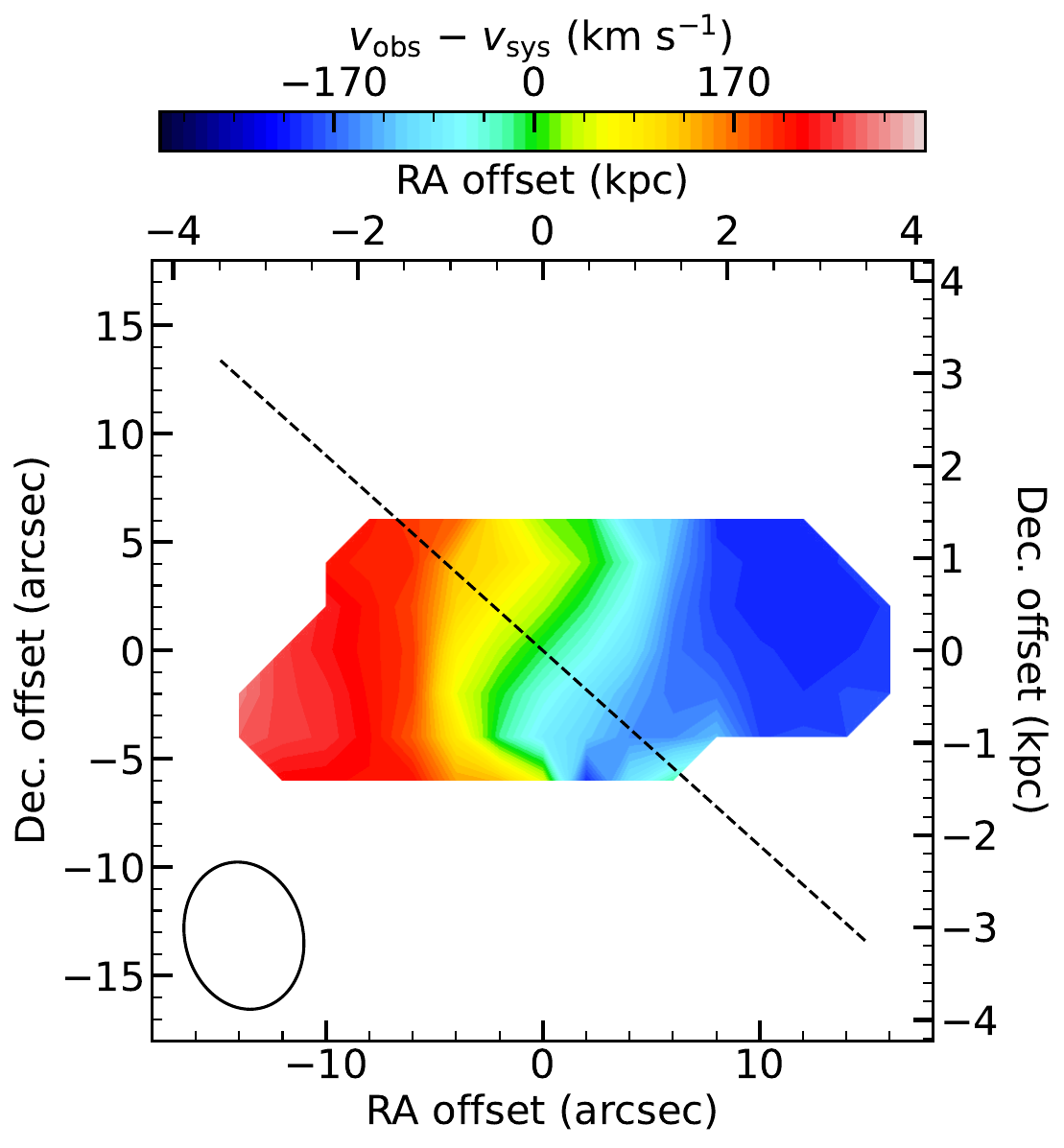}
    \end{subfigure}
    \begin{subfigure}{.32\linewidth}
    \centering
    \includegraphics[height=6.0cm]{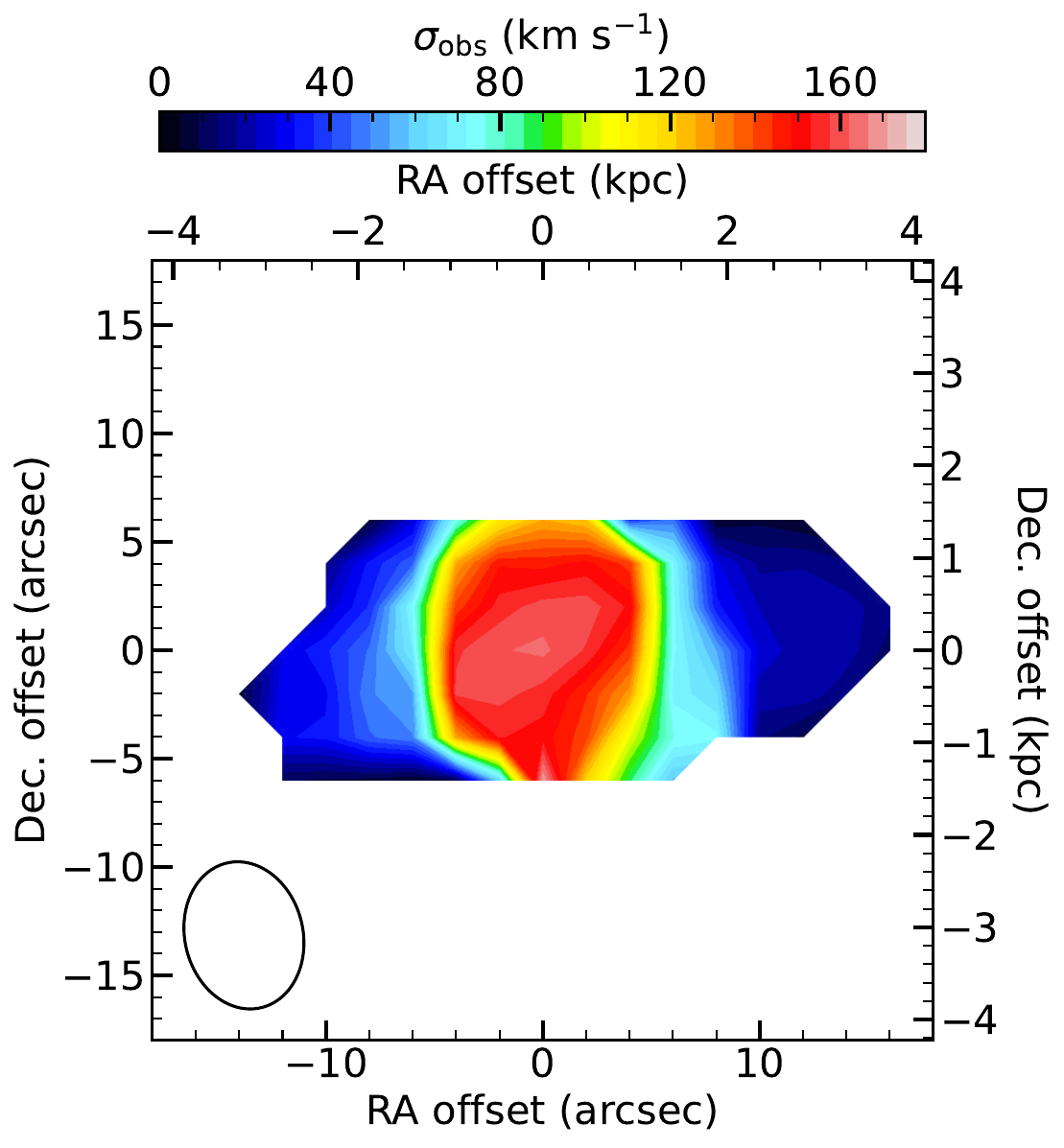}
    \end{subfigure}
    \\
    \begin{subfigure}{.34\linewidth}
    \centering
    \includegraphics[height=5.2cm]{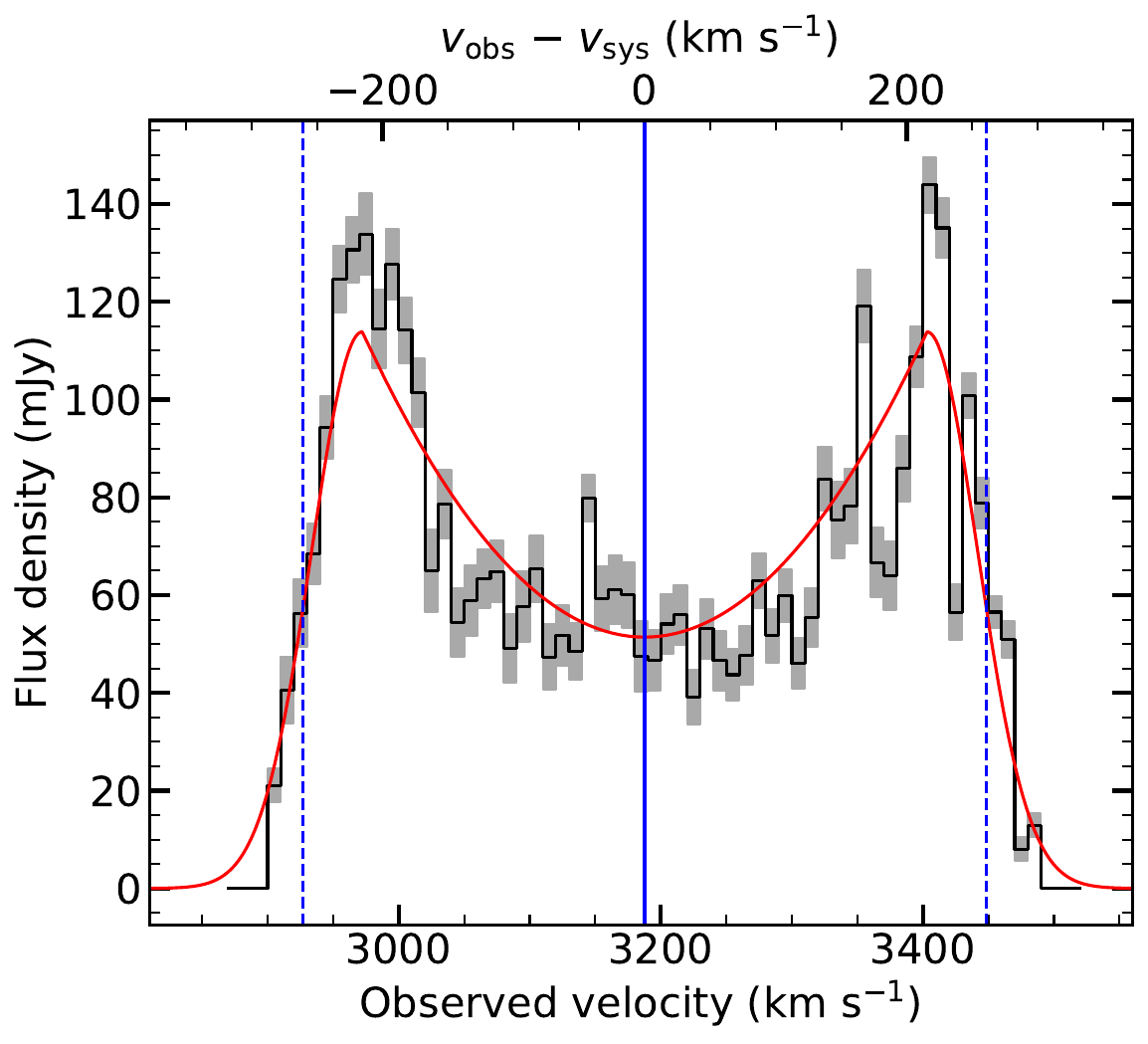}
    \end{subfigure}
    \begin{subfigure}{.32\linewidth}
    \centering
    \includegraphics[height=4.8cm]{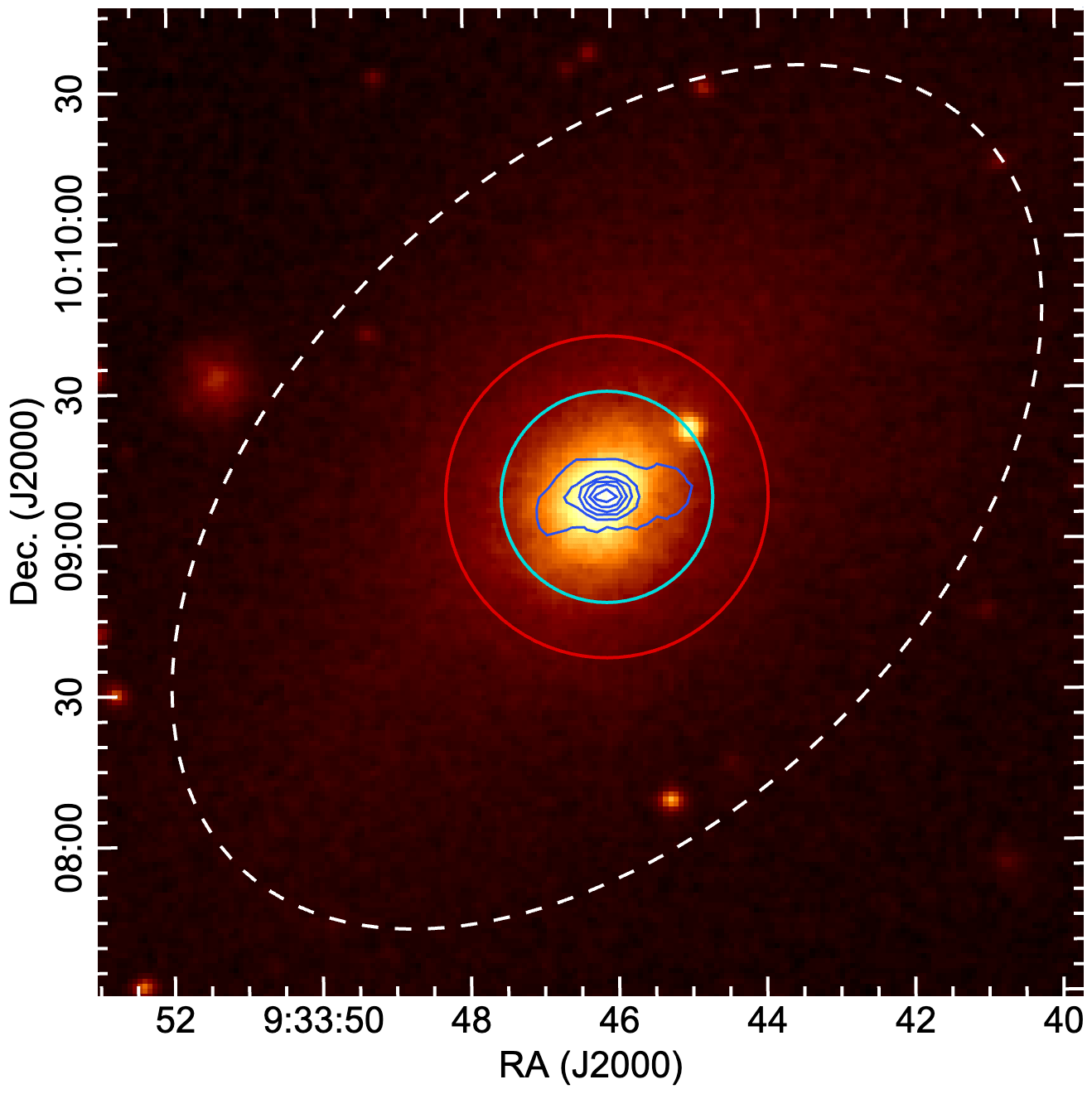}
    \end{subfigure}
    \begin{subfigure}{.33\linewidth}
    \centering
    \includegraphics[height=5.2cm]{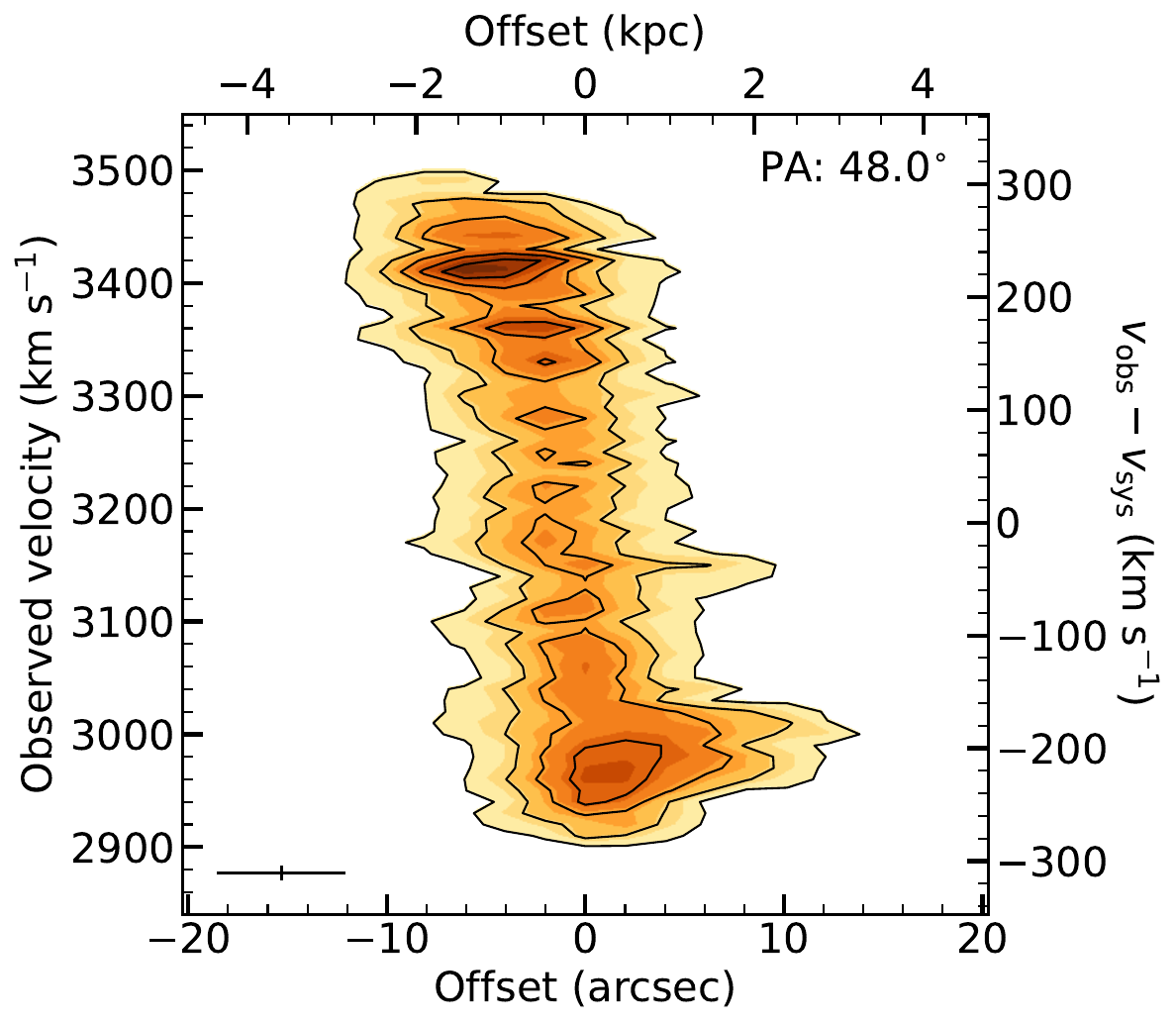}
    \end{subfigure}
    \caption[]{continued.}
\end{figure*}

\begin{figure*}
    \ContinuedFloat
    \centering
    \begin{subfigure}{.34\linewidth}
    \centering
    \includegraphics[height=5.9cm]{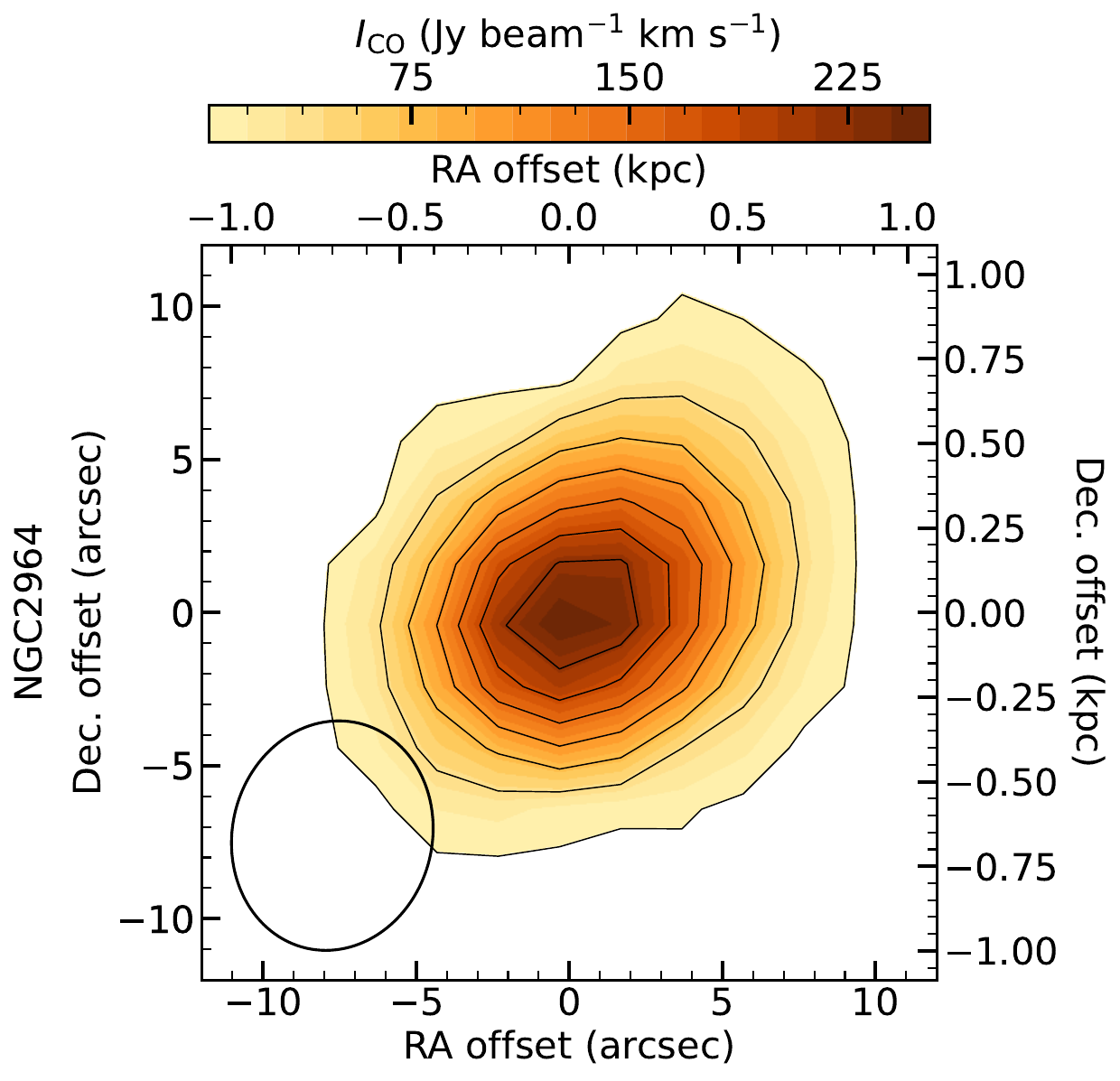}
    \end{subfigure}
    \begin{subfigure}{.32\linewidth}
    \centering
    \includegraphics[height=5.9cm]{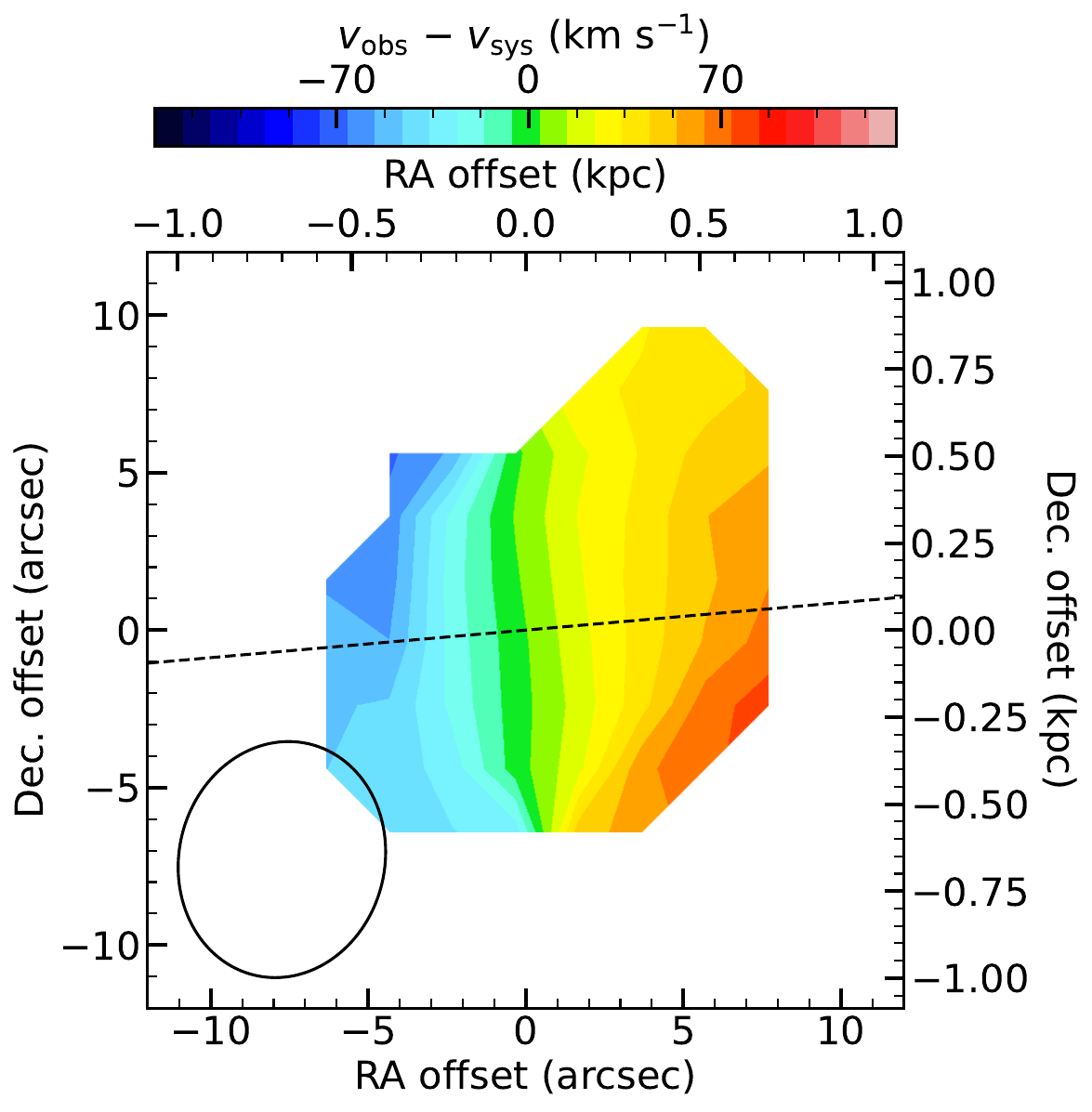}
    \end{subfigure}
    \begin{subfigure}{.32\linewidth}
    \centering
    \includegraphics[height=5.9cm]{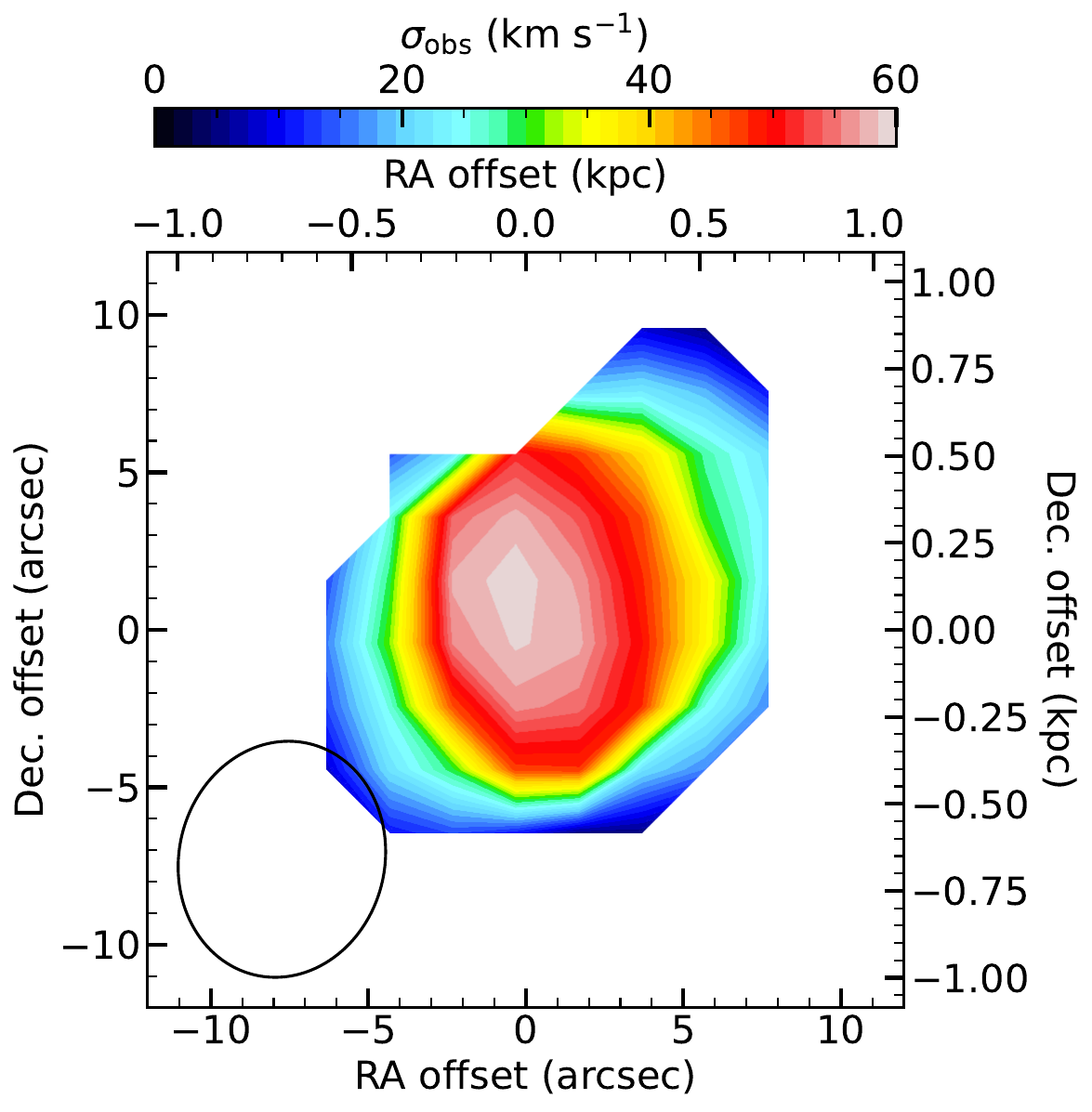}
    \end{subfigure}
    \\
    \begin{subfigure}{.34\linewidth}
    \centering
    \includegraphics[height=5.2cm]{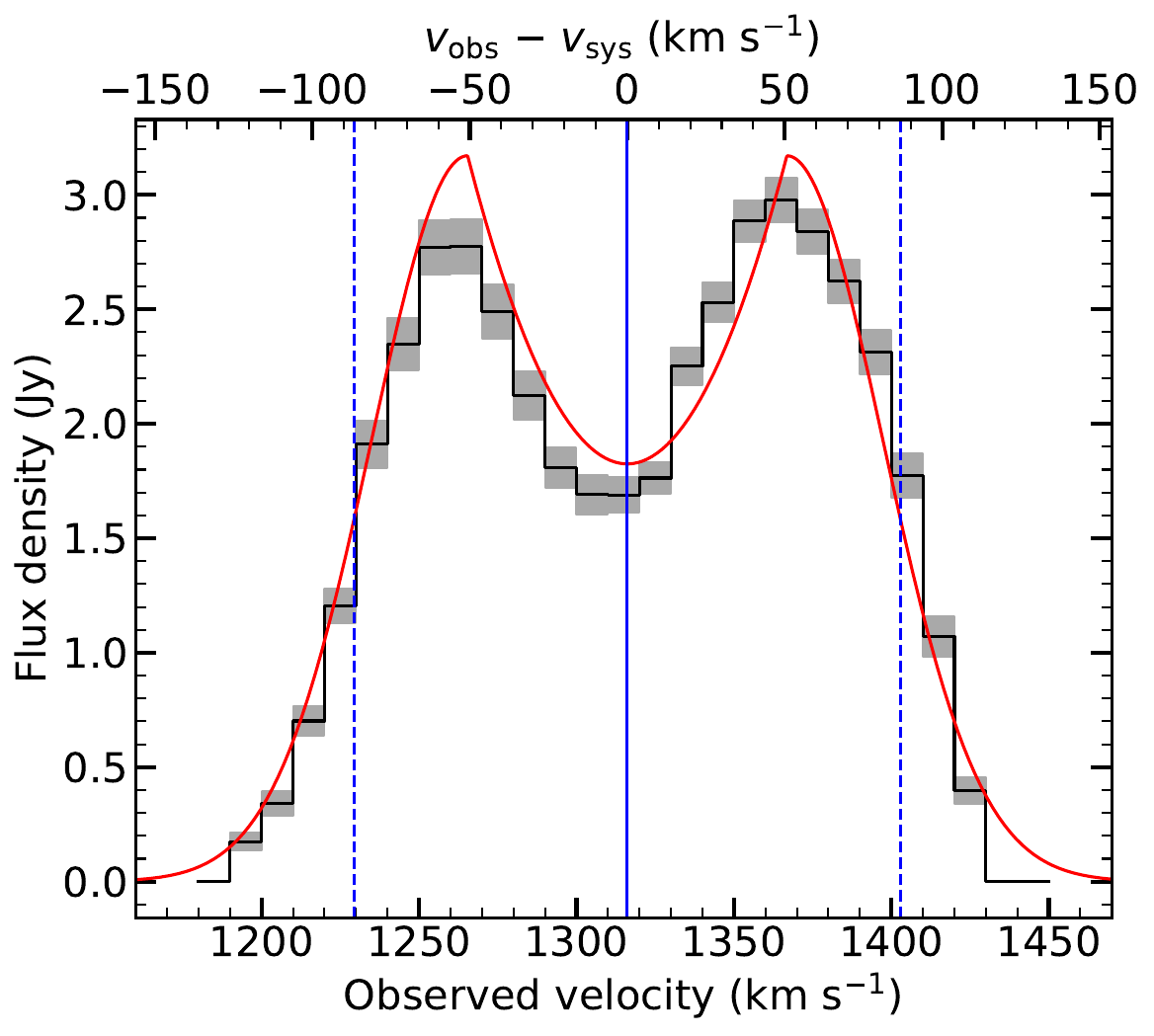}
    \end{subfigure}
    \begin{subfigure}{.32\linewidth}
    \centering
    \includegraphics[height=4.8cm]{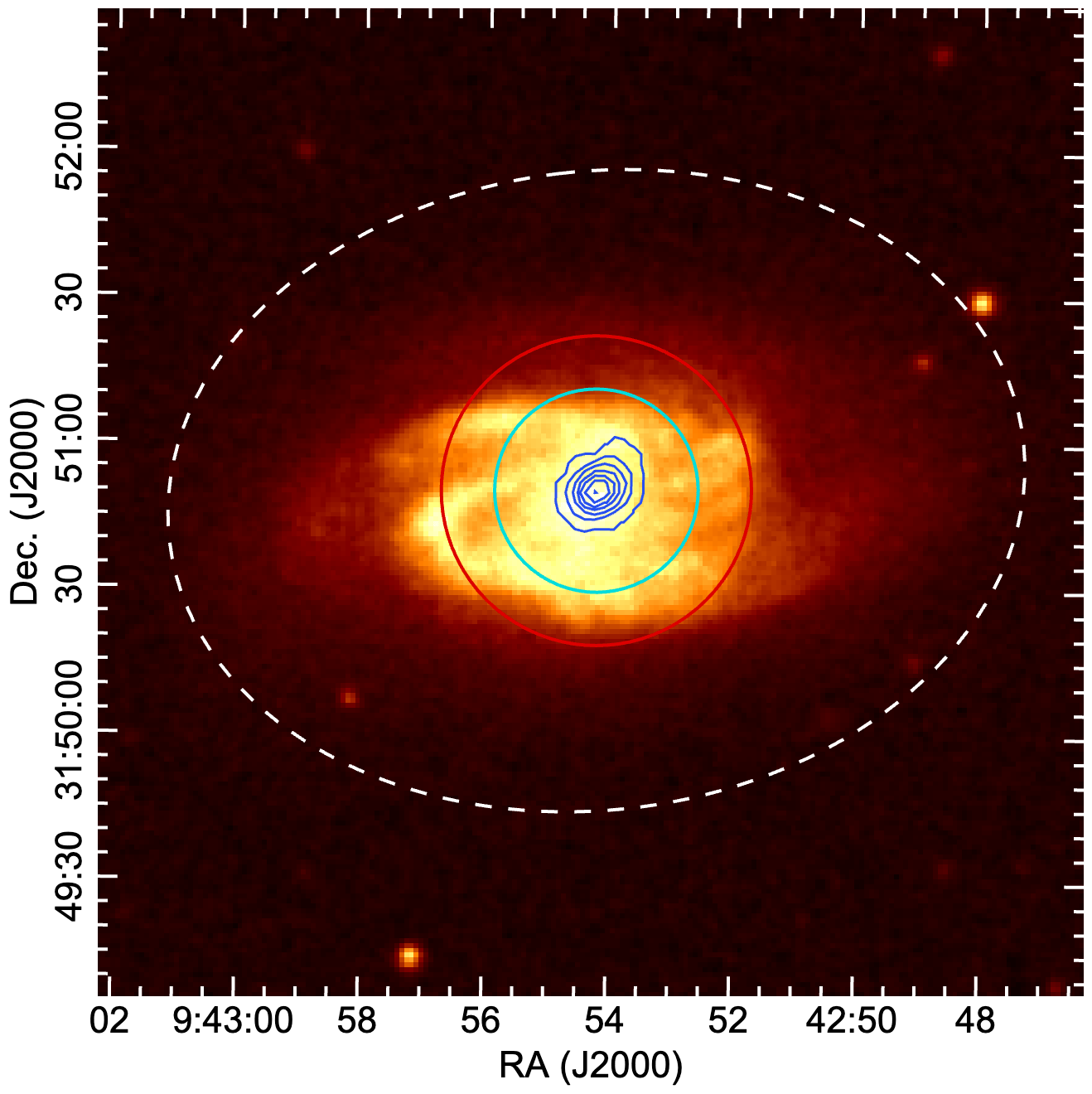}
    \end{subfigure}
    \begin{subfigure}{.33\linewidth}
    \centering
    \includegraphics[height=5.2cm]{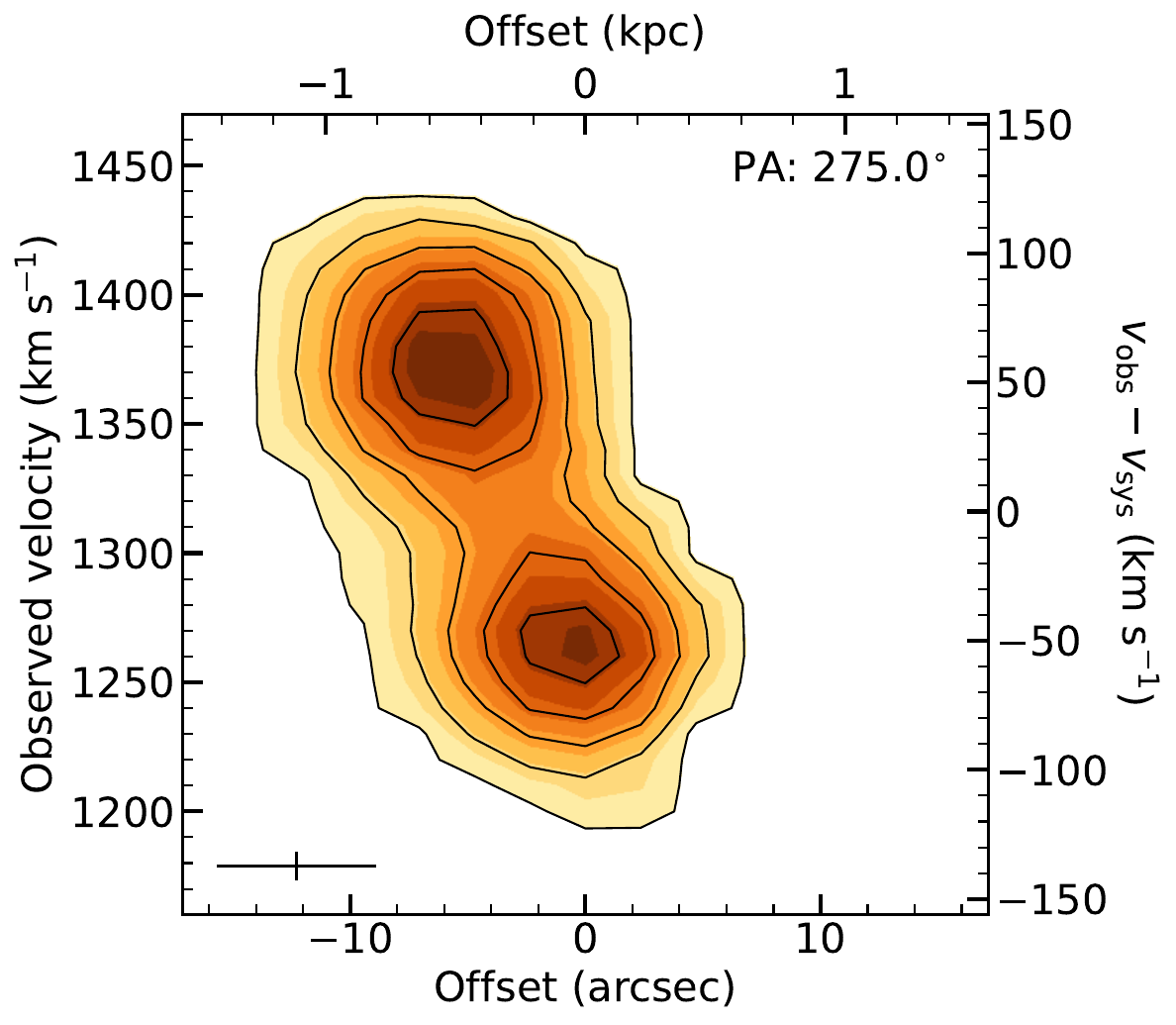}
    \end{subfigure}
    \\
    \begin{subfigure}{.34\linewidth}
    \centering
    \includegraphics[height=6.0cm]{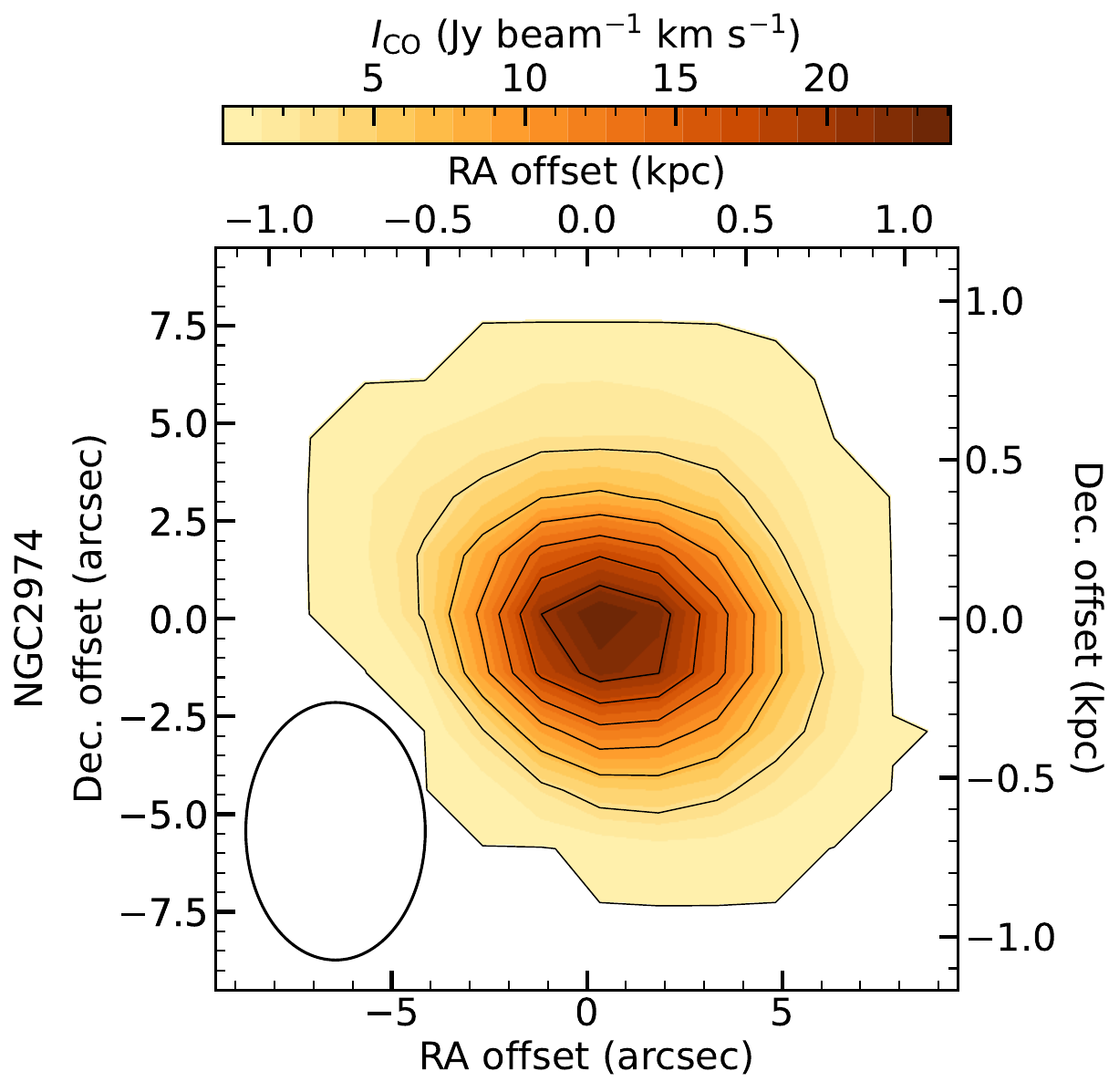}
    \end{subfigure}
    \begin{subfigure}{.32\linewidth}
    \centering
    \includegraphics[height=6.0cm]{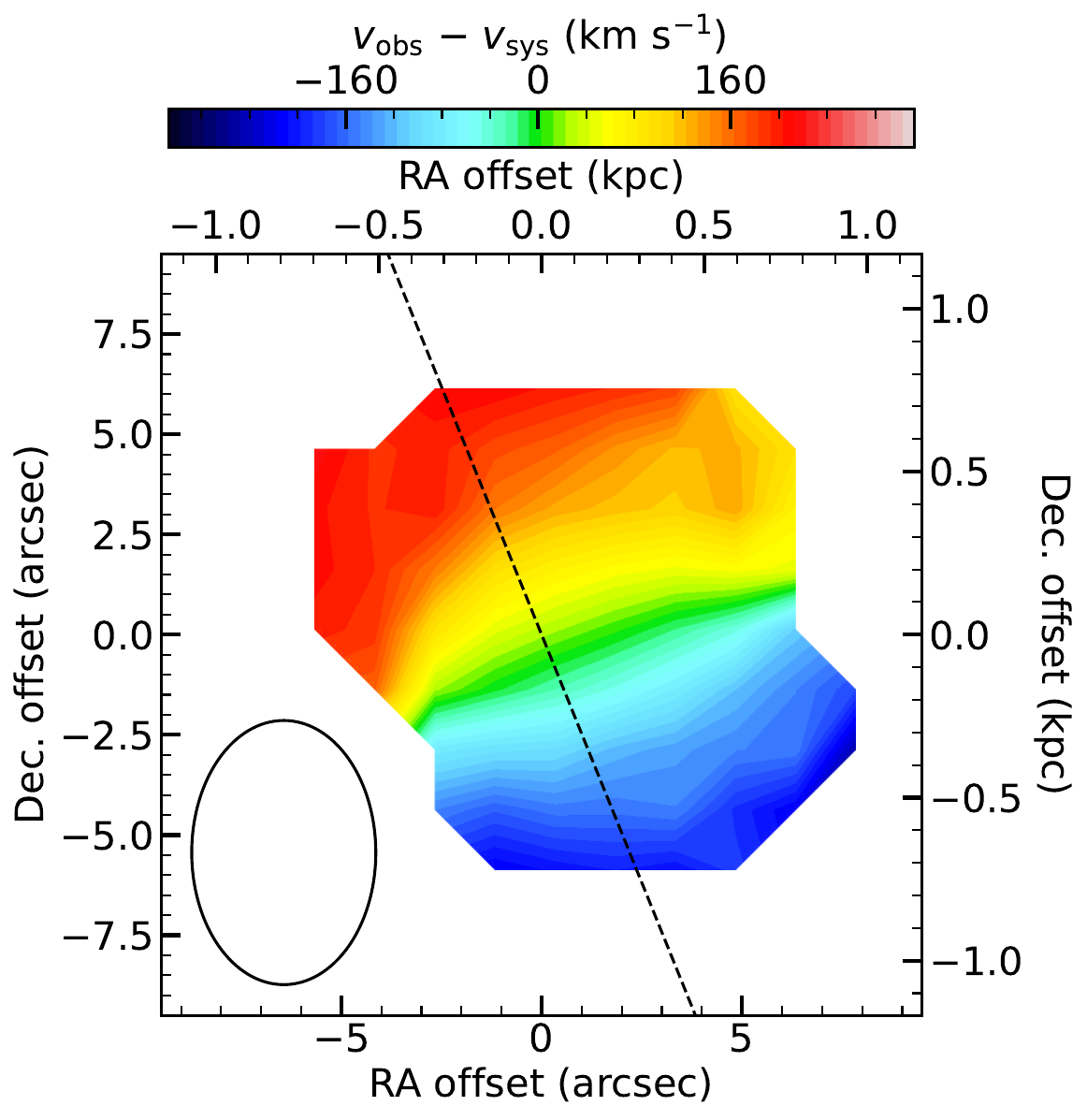}
    \end{subfigure}
    \begin{subfigure}{.32\linewidth}
    \centering
    \includegraphics[height=6.0cm]{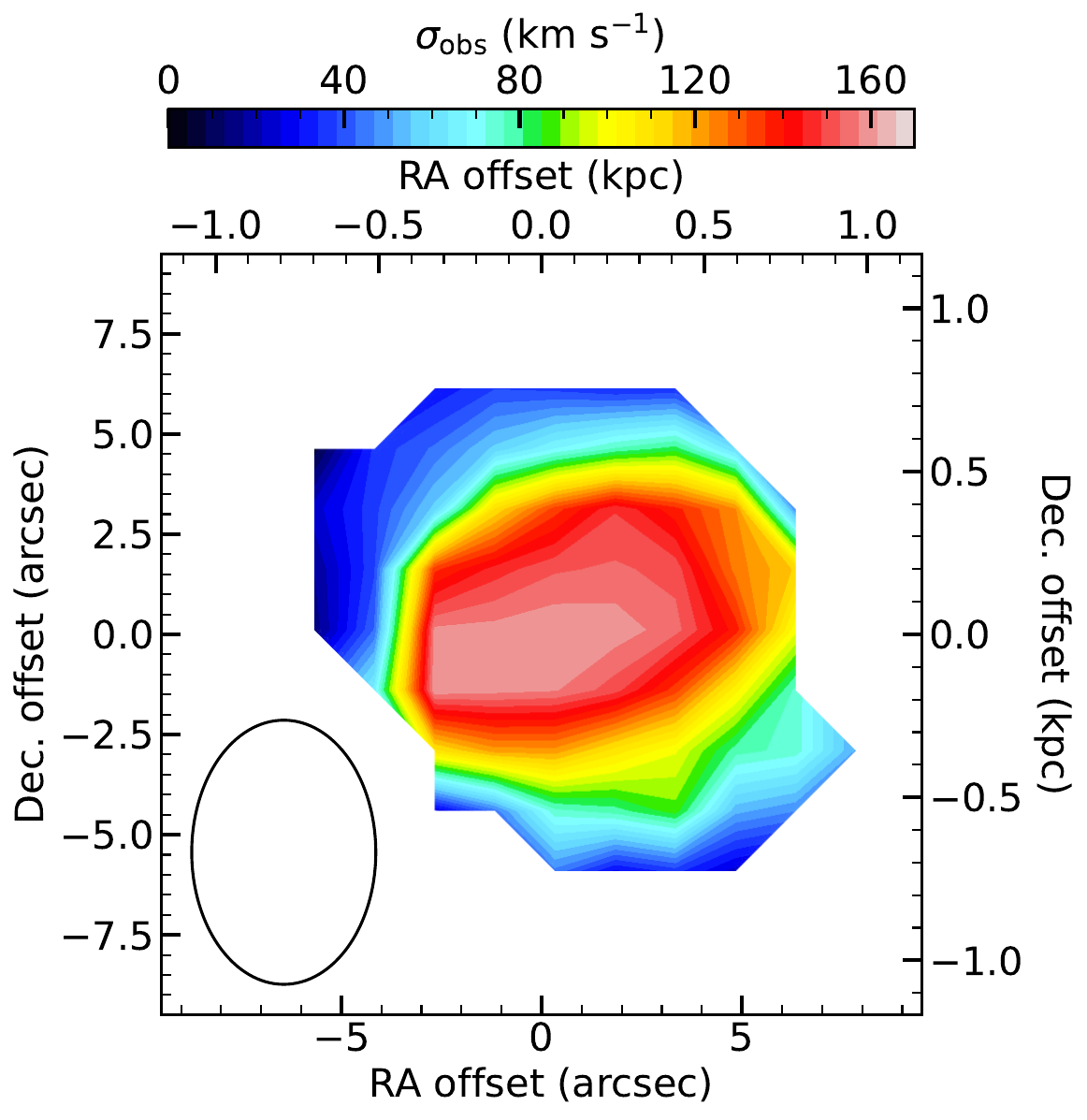}
    \end{subfigure}
    \\
    \begin{subfigure}{.34\linewidth}
    \centering
    \includegraphics[height=5.2cm]{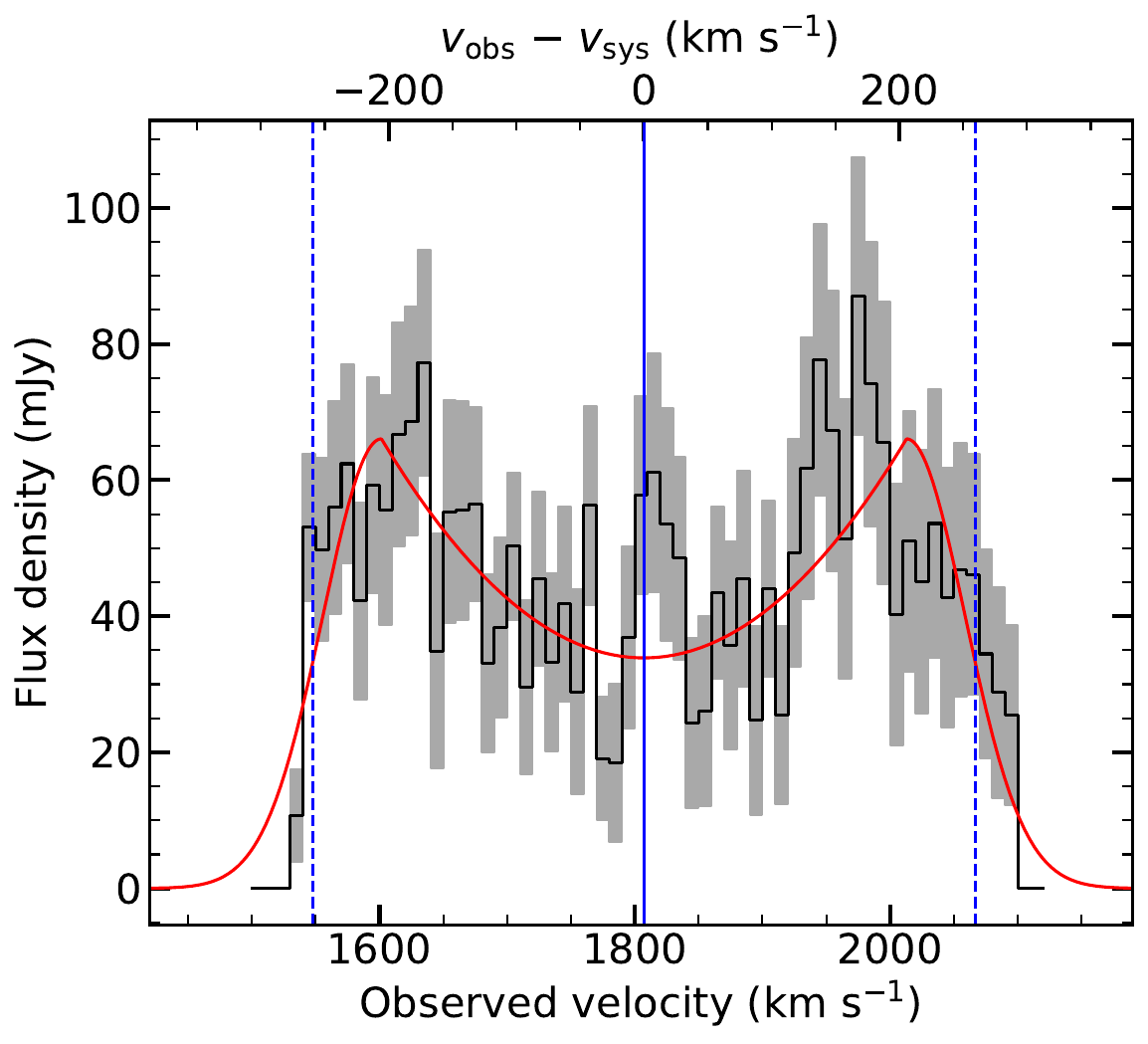}
    \end{subfigure}
    \begin{subfigure}{.32\linewidth}
    \centering
    \includegraphics[height=4.8cm]{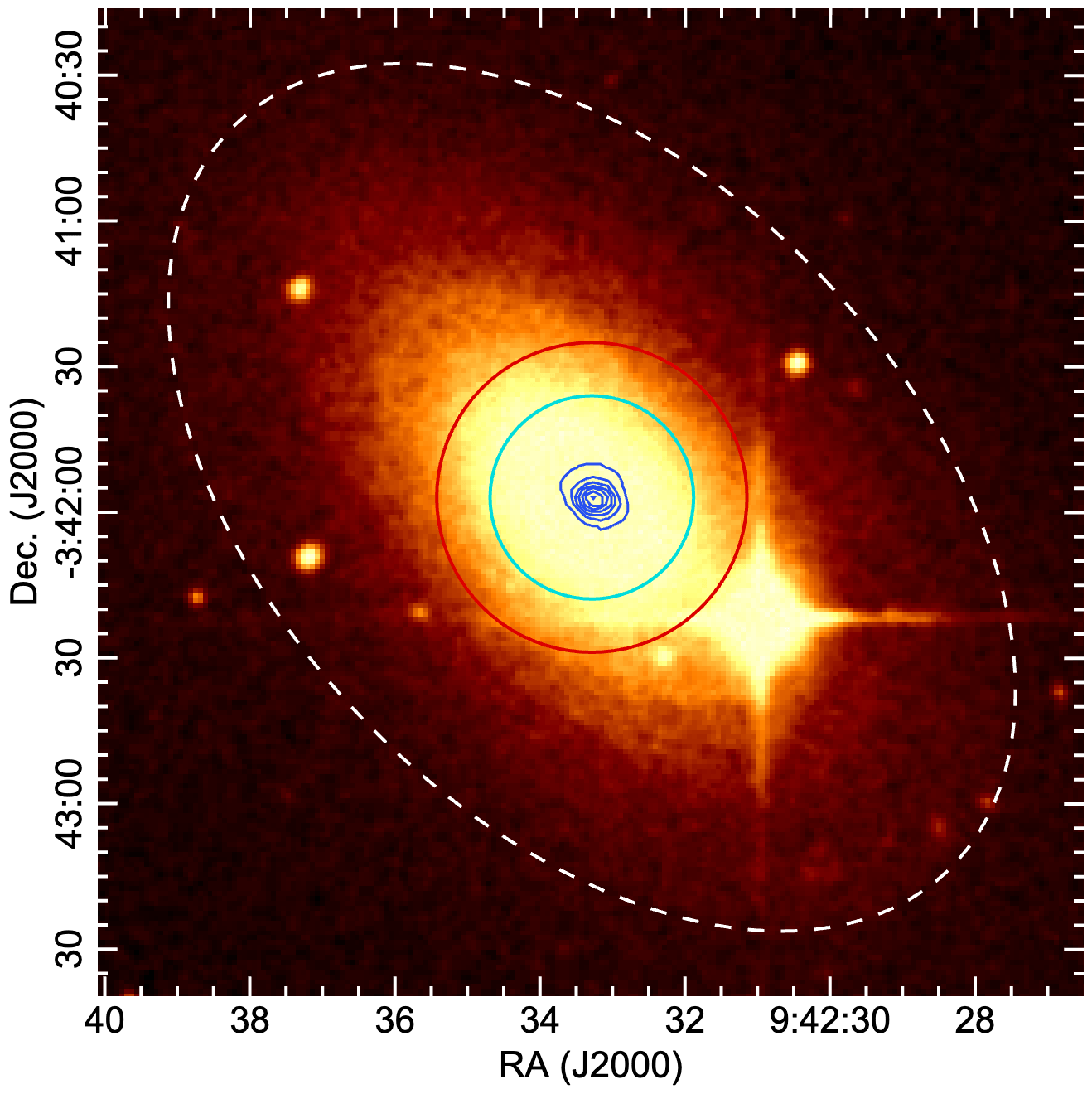}
    \end{subfigure}
    \begin{subfigure}{.33\linewidth}
    \centering
    \includegraphics[height=5.2cm]{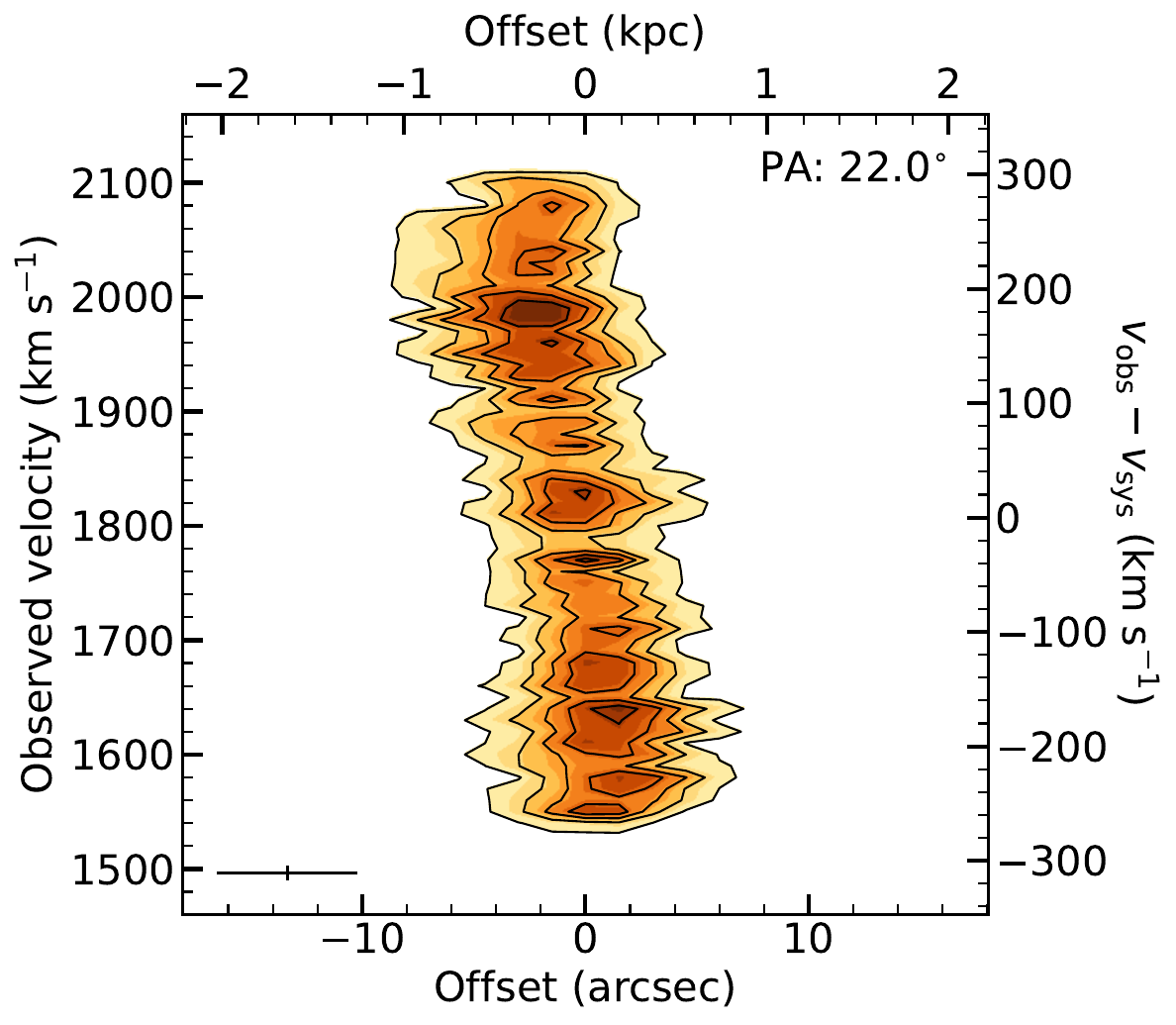}
    \end{subfigure}
    \caption[]{continued.}
\end{figure*}

\begin{figure*}
    \ContinuedFloat
    \centering
    \begin{subfigure}{.34\linewidth}
    \centering
    \includegraphics[height=6.0cm]{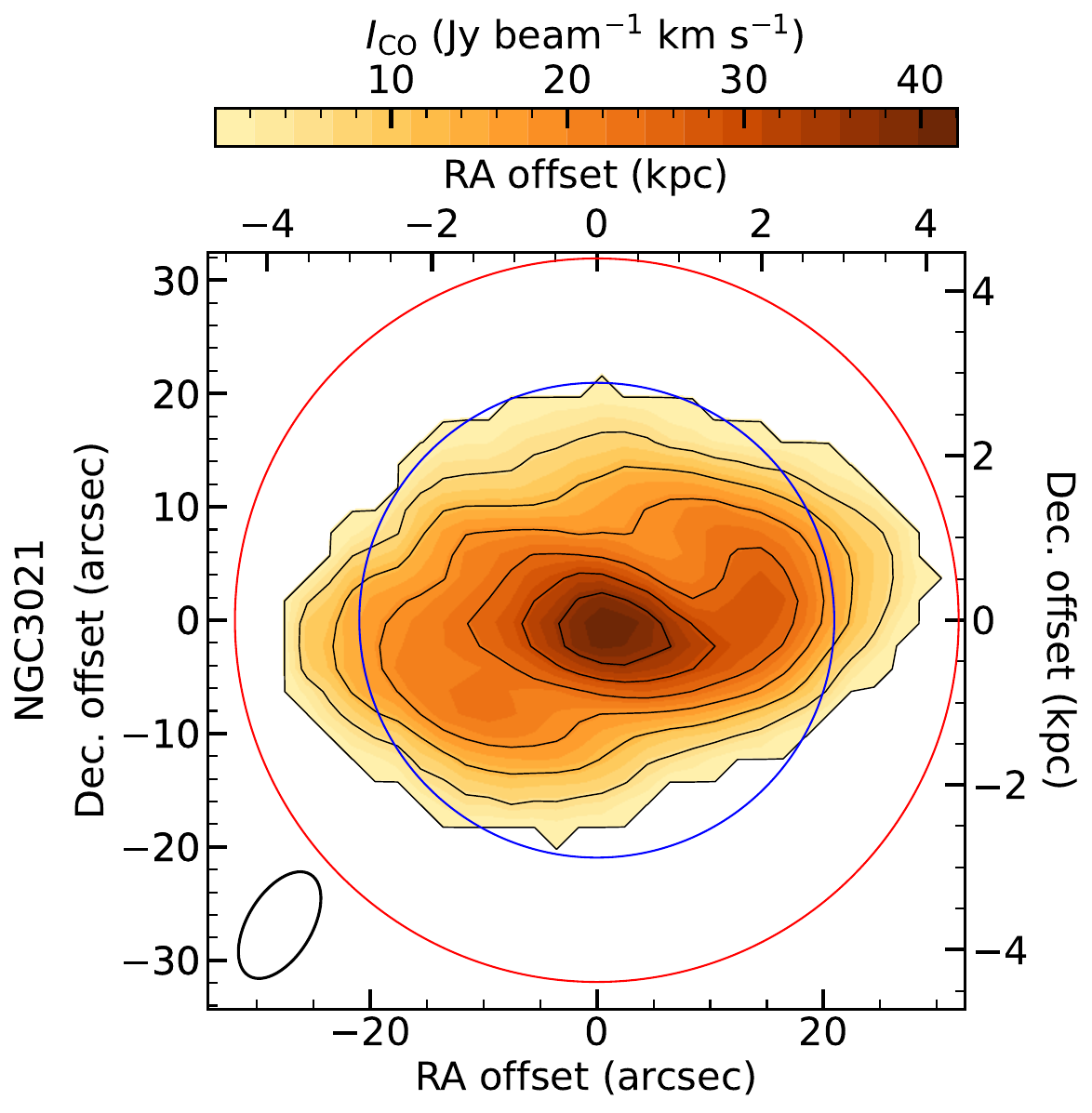}
    \end{subfigure}
    \begin{subfigure}{.32\linewidth}
    \centering
    \includegraphics[height=6.0cm]{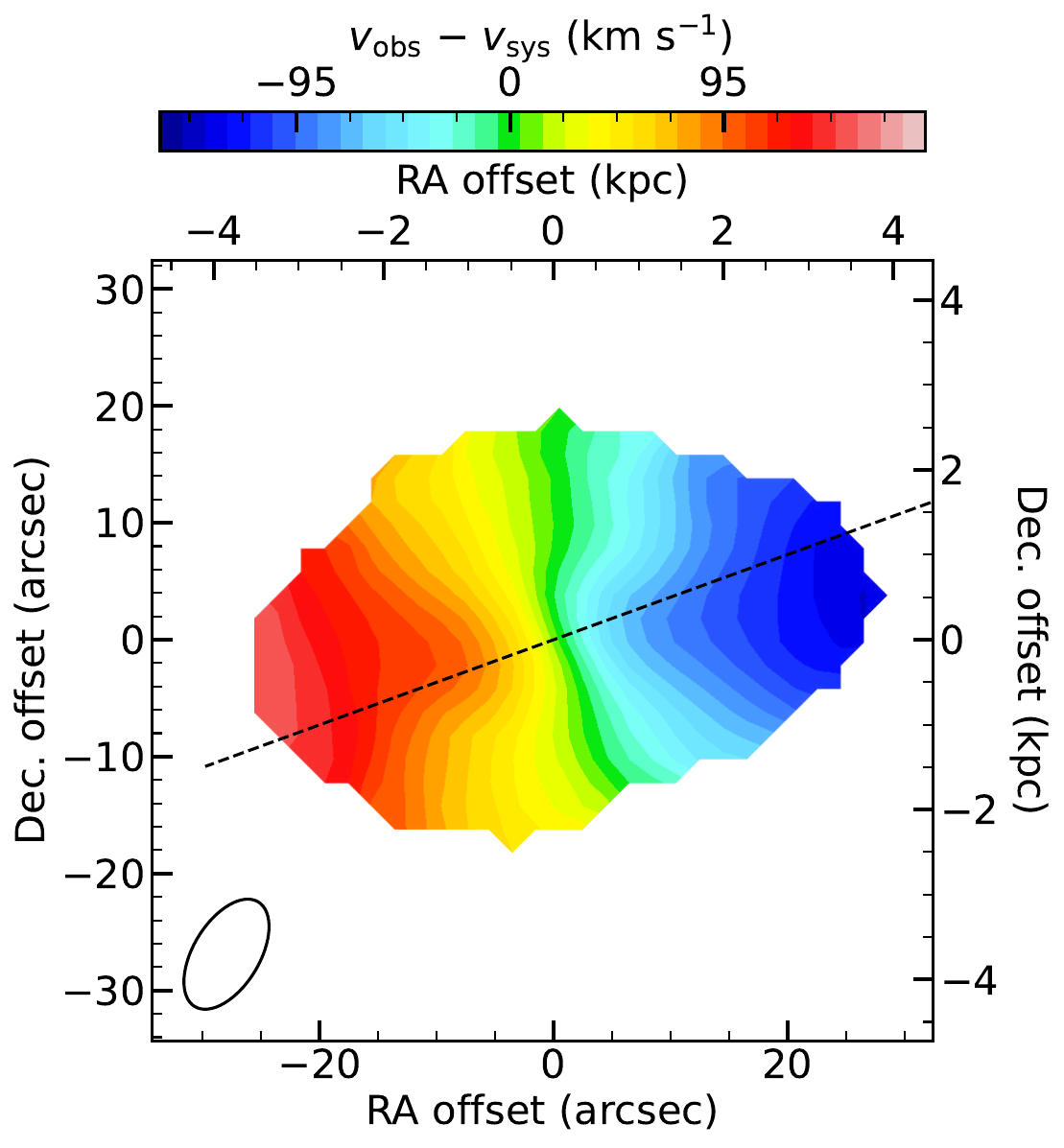}
    \end{subfigure}
    \begin{subfigure}{.32\linewidth}
    \centering
    \includegraphics[height=6.0cm]{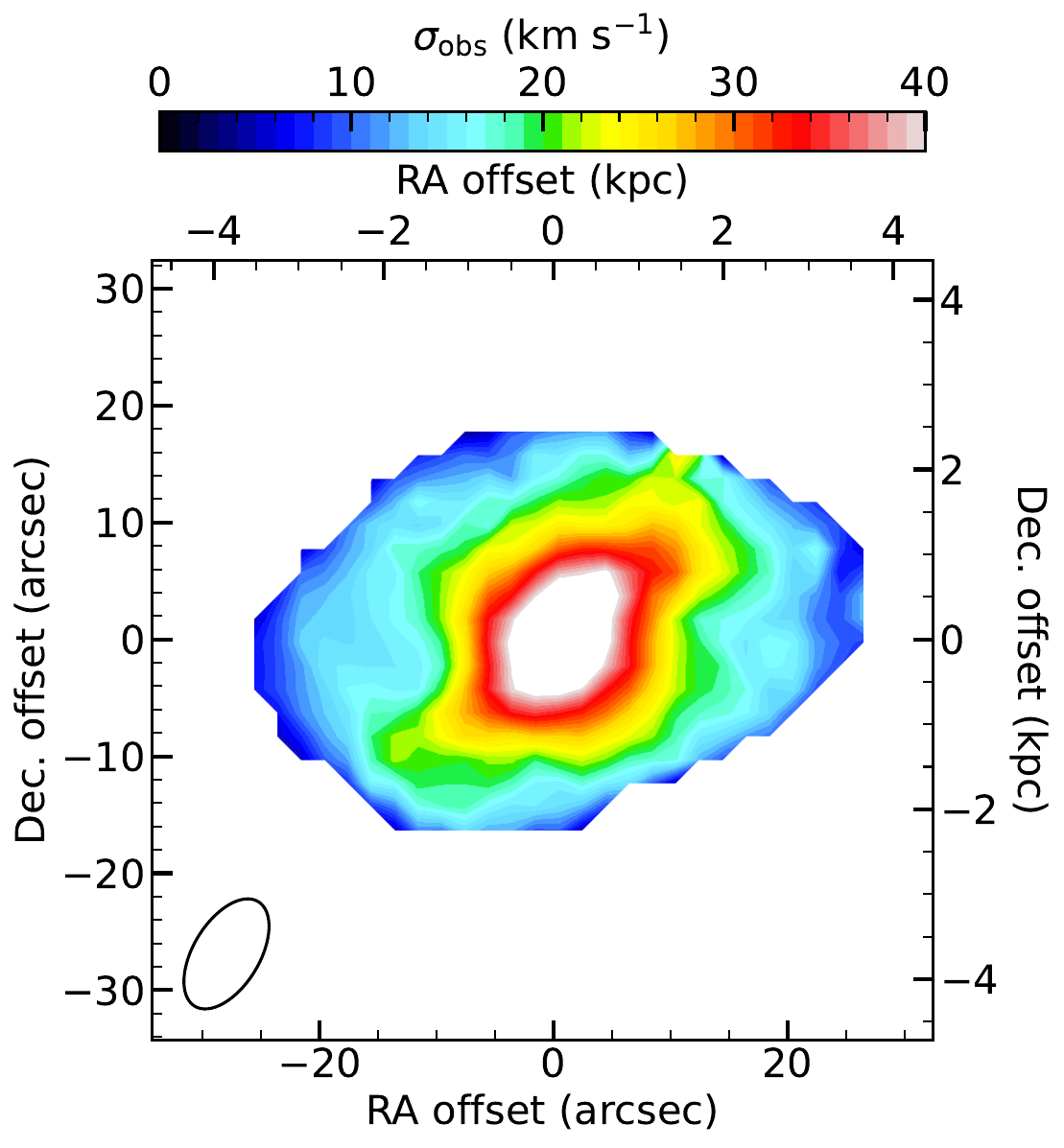}
    \end{subfigure}
    \\
    \begin{subfigure}{.34\linewidth}
    \centering
    \includegraphics[height=5.2cm]{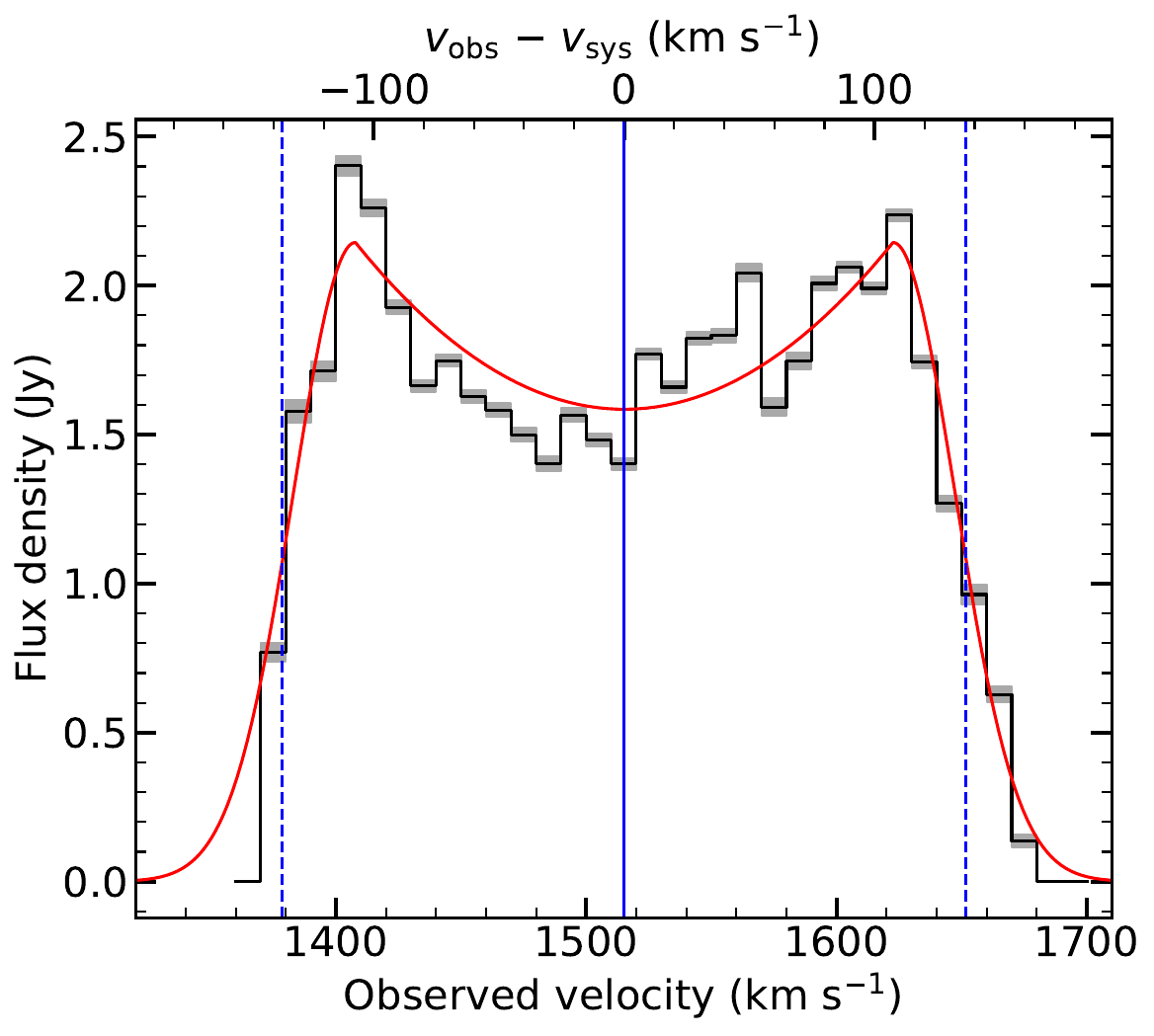}
    \end{subfigure}
    \begin{subfigure}{.32\linewidth}
    \centering
    \includegraphics[height=4.8cm]{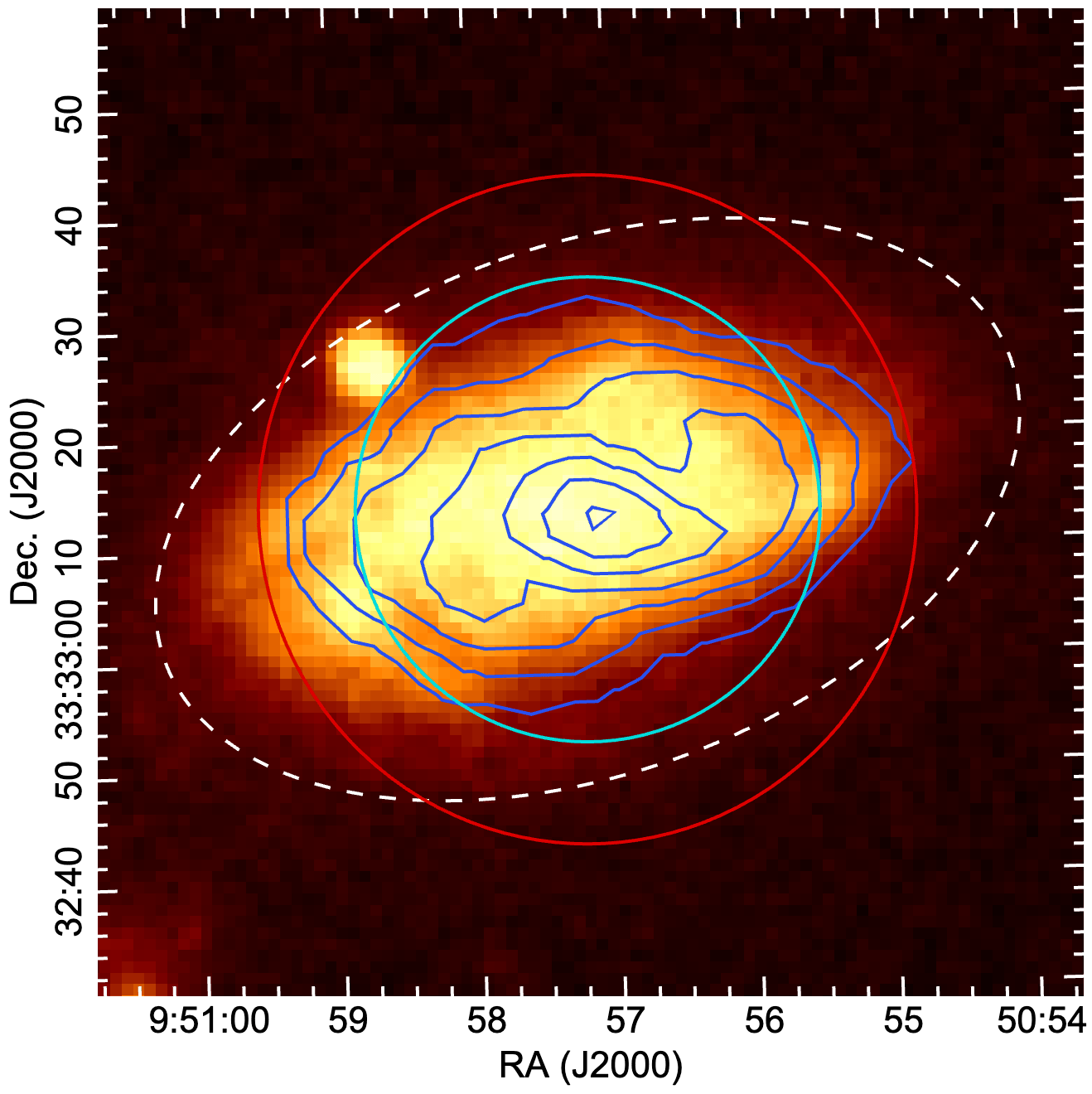}
    \end{subfigure}
    \begin{subfigure}{.33\linewidth}
    \centering
    \includegraphics[height=5.2cm]{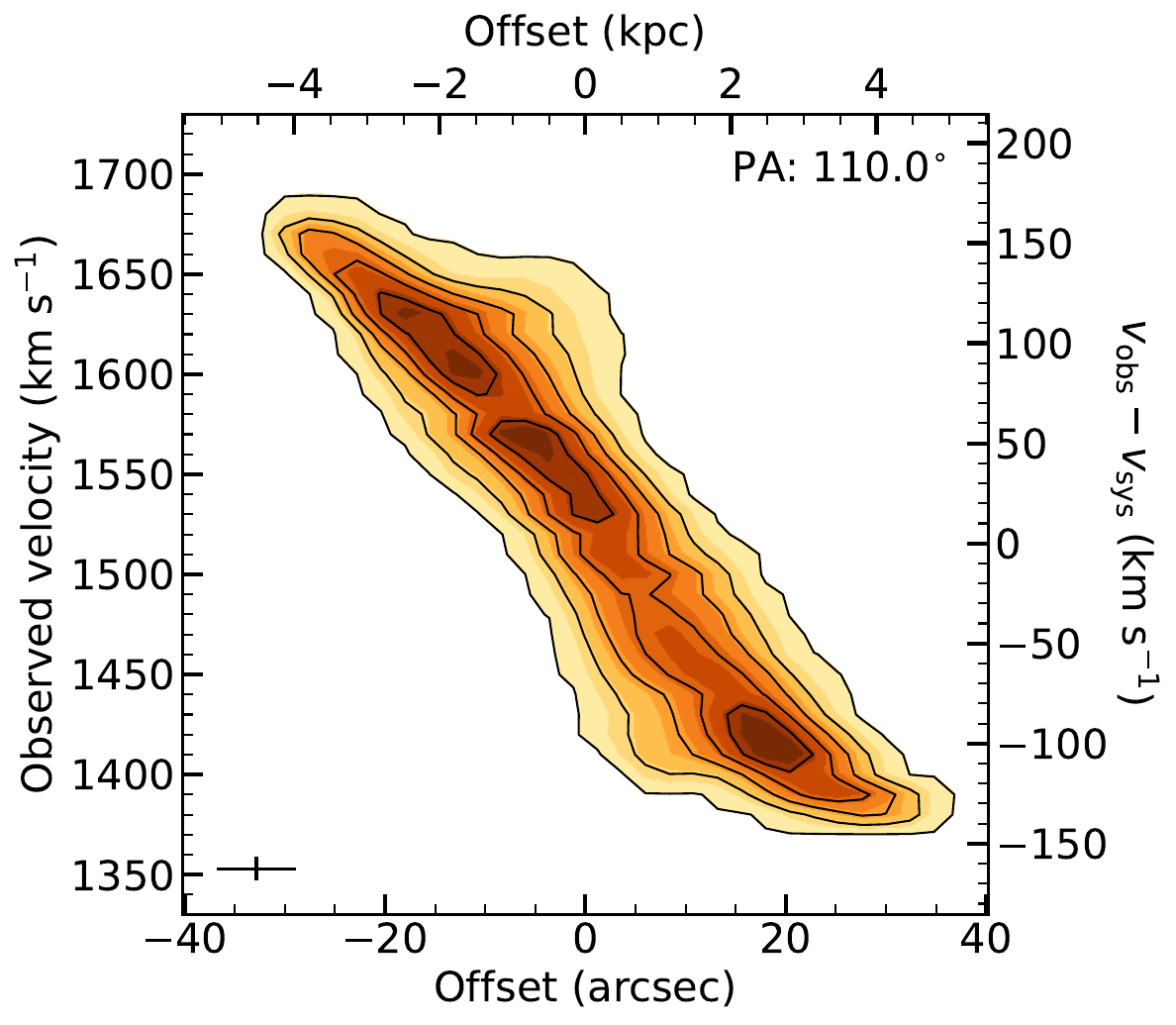}
    \end{subfigure}
    \\
    \begin{subfigure}{.34\linewidth}
    \centering
    \includegraphics[height=5.85cm]{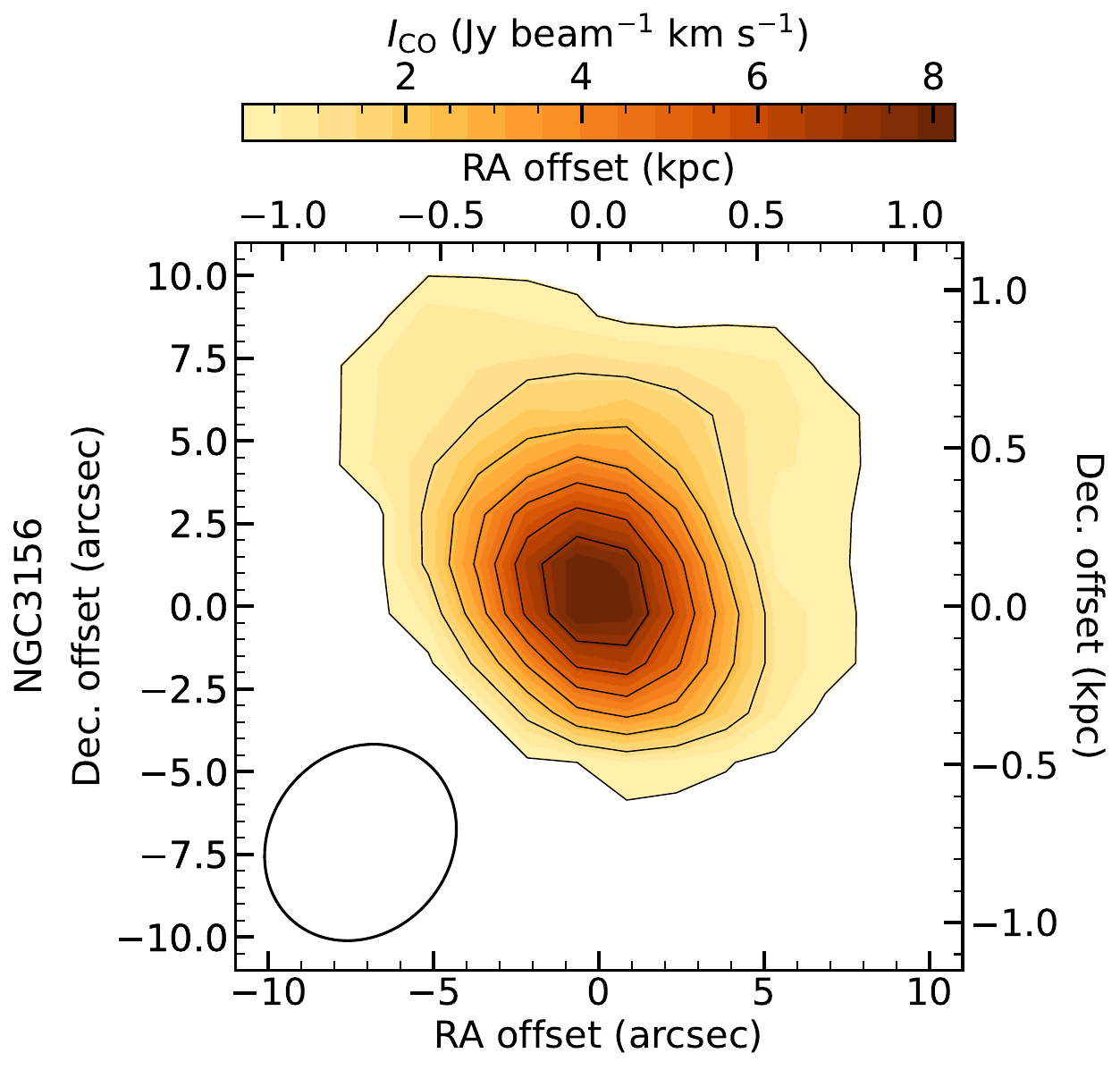}
    \end{subfigure}
    \begin{subfigure}{.32\linewidth}
    \centering
    \includegraphics[height=5.85cm]{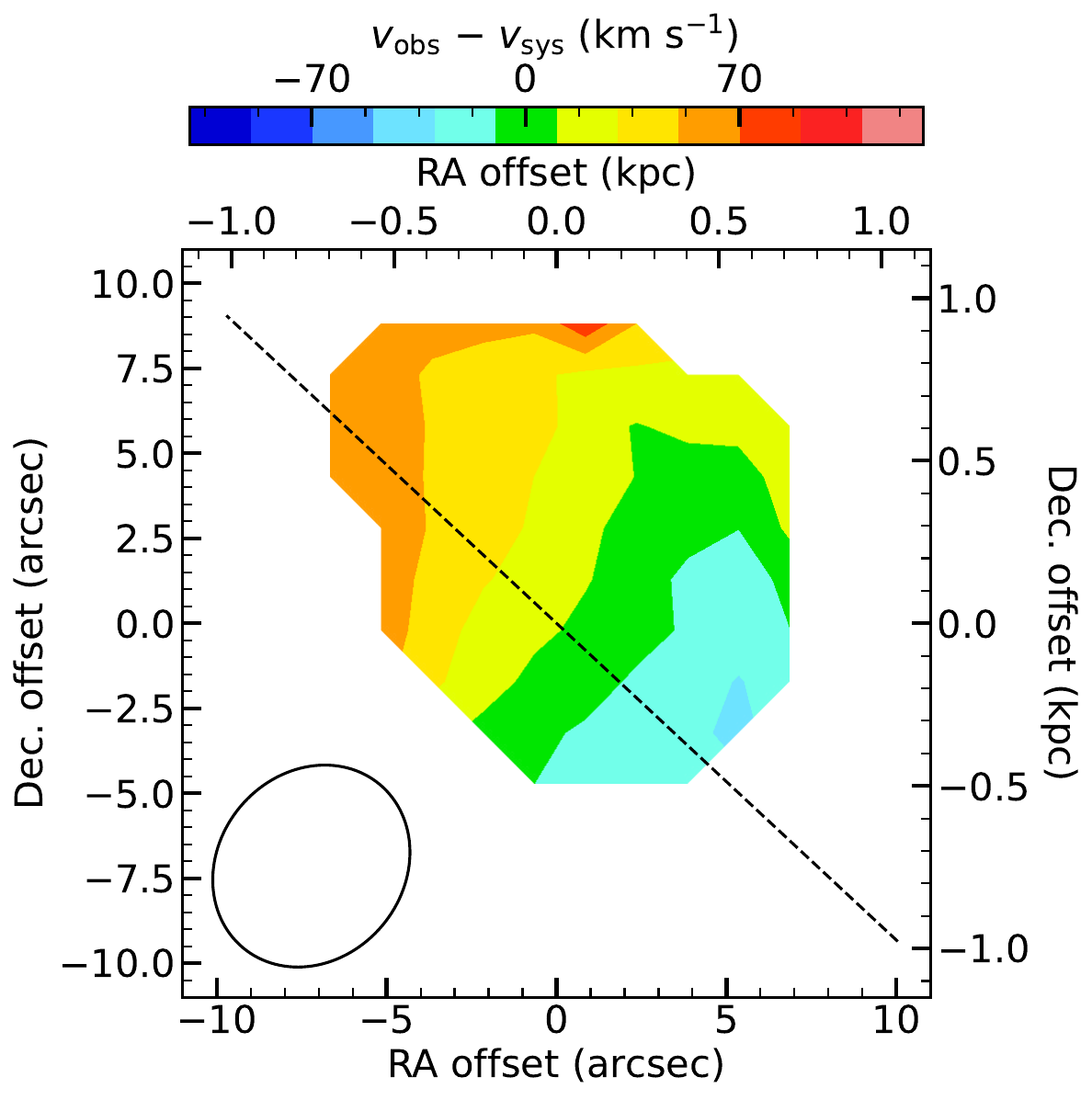}
    \end{subfigure}
    \begin{subfigure}{.32\linewidth}
    \centering
    \includegraphics[height=5.85cm]{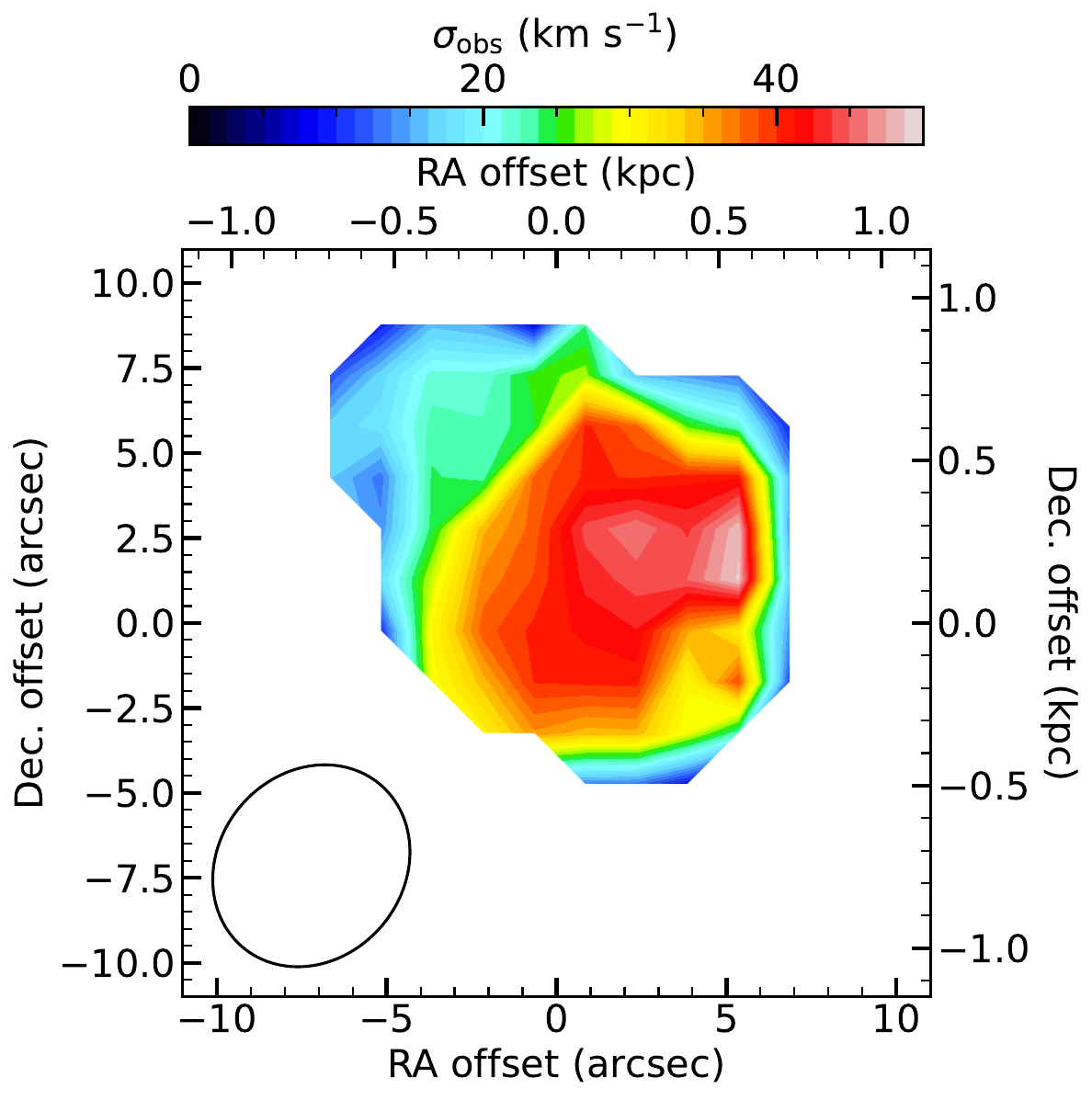}
    \end{subfigure}
    \\
    \begin{subfigure}{.34\linewidth}
    \centering
    \includegraphics[height=5.2cm]{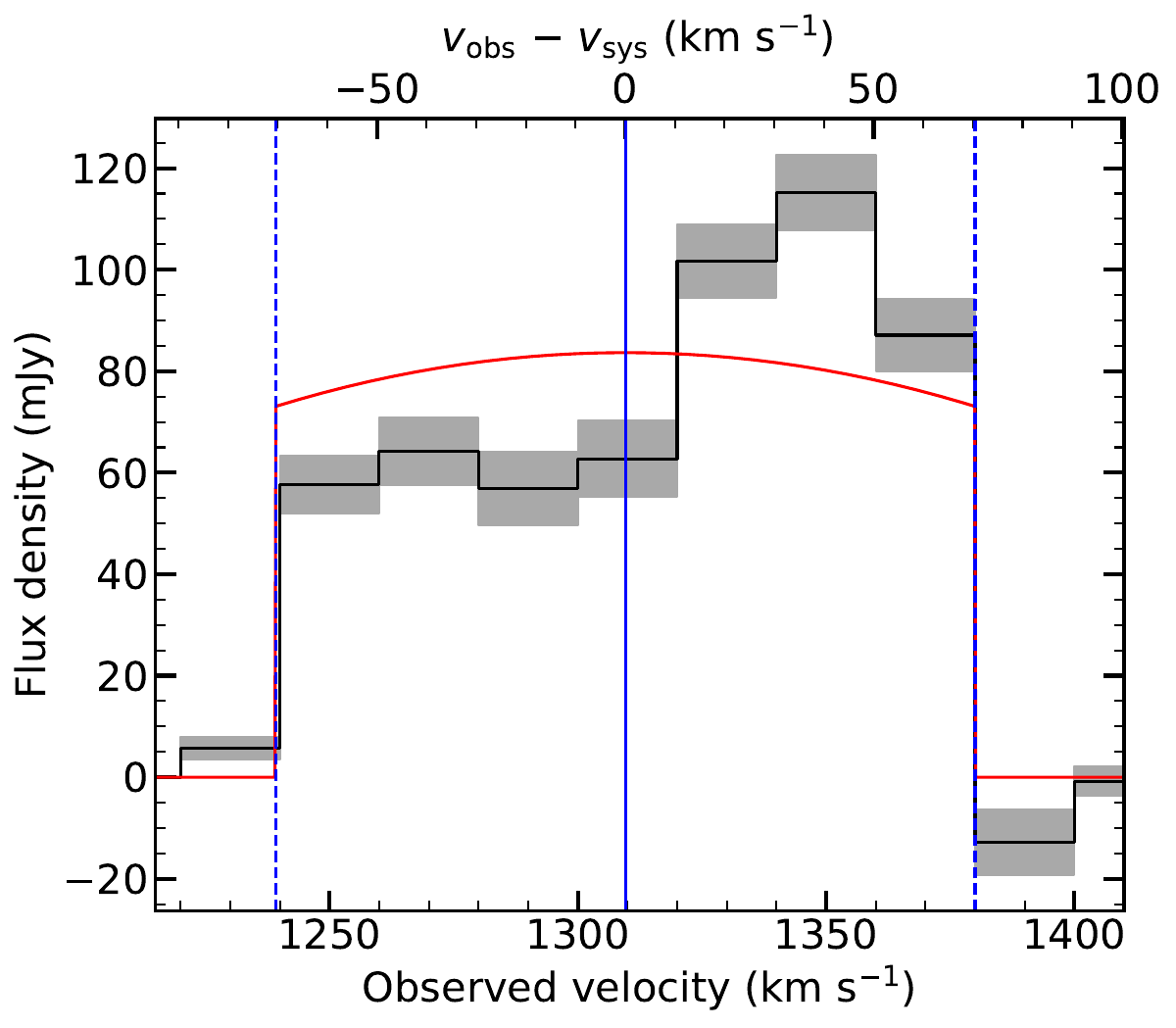}
    \end{subfigure}
    \begin{subfigure}{.32\linewidth}
    \centering
    \includegraphics[height=4.8cm]{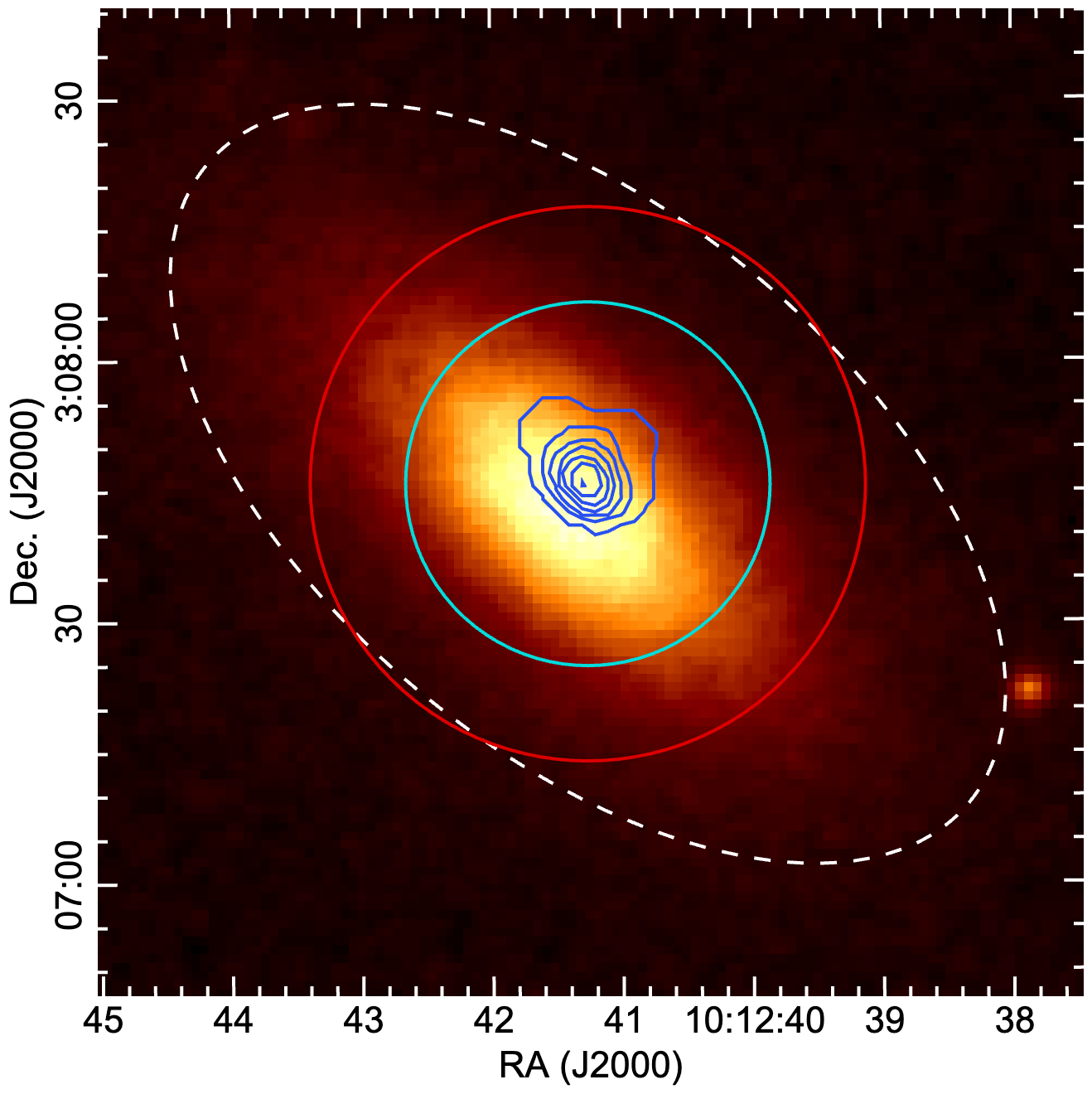}
    \end{subfigure}
    \begin{subfigure}{.33\linewidth}
    \centering
    \includegraphics[height=5.2cm]{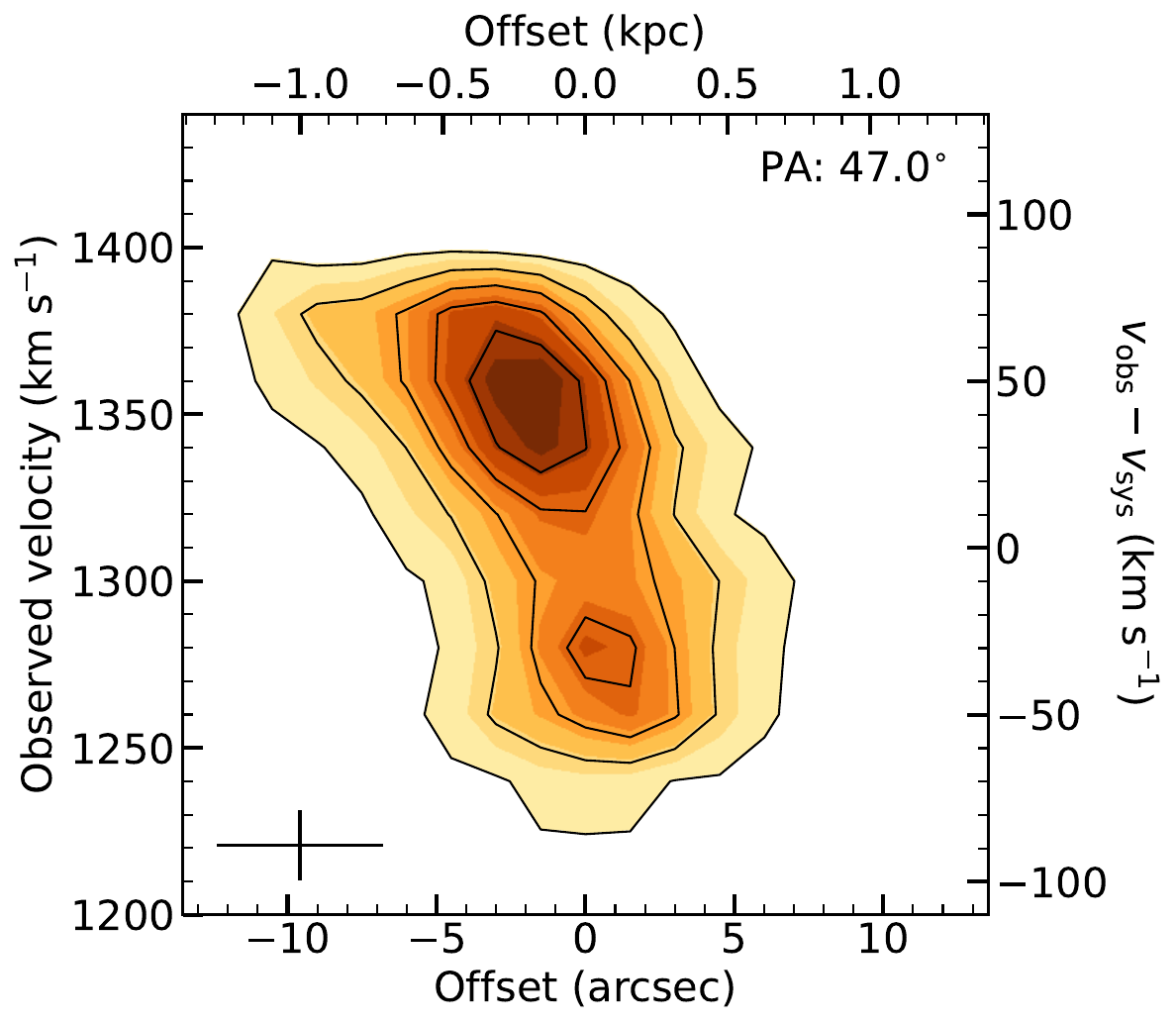}
    \end{subfigure}
    \caption[]{continued.}
\end{figure*}

\begin{figure*}
    \ContinuedFloat
    \centering
    \begin{subfigure}{.34\linewidth}
    \centering
    \includegraphics[height=6.0cm]{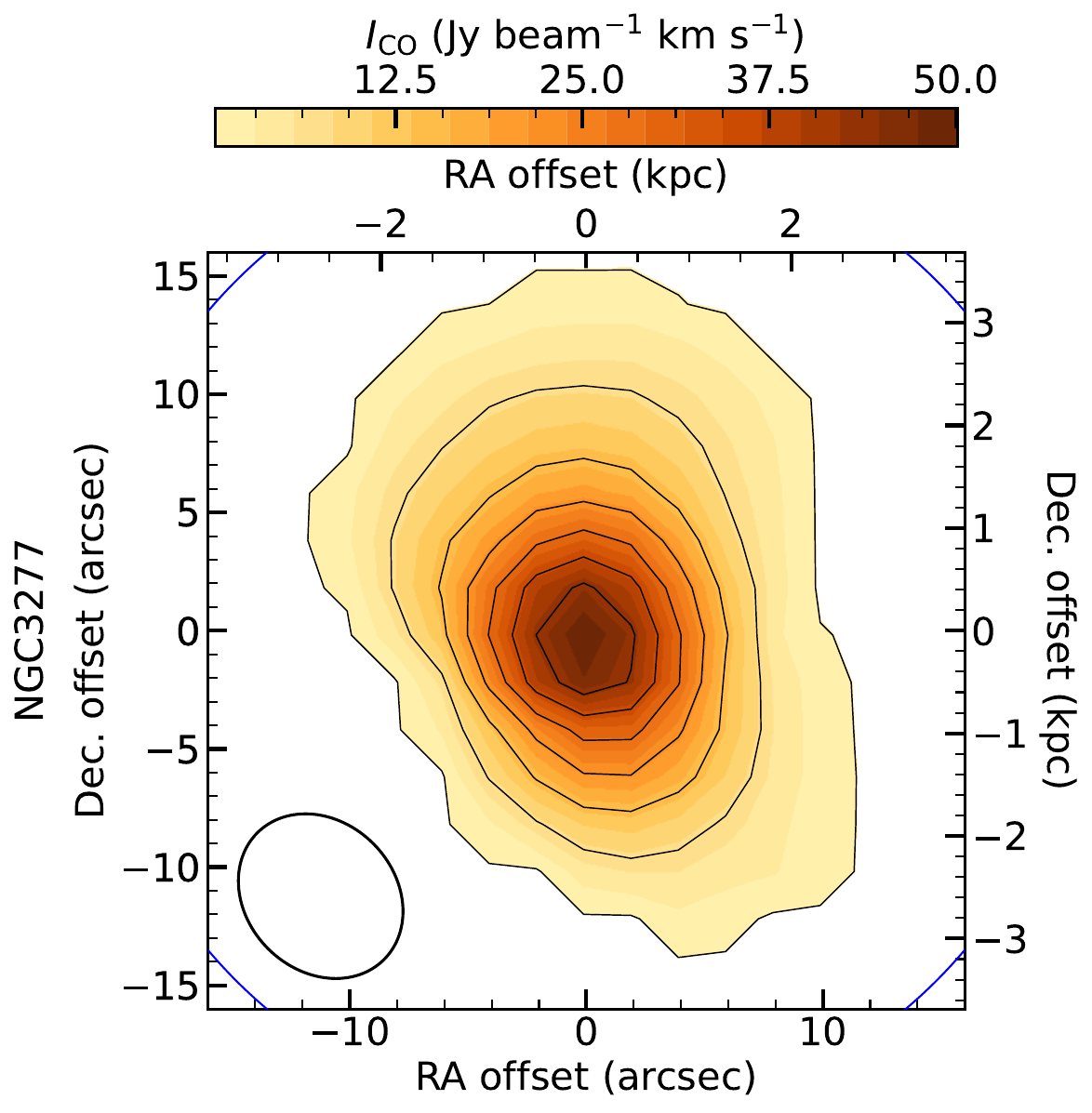}
    \end{subfigure}
    \begin{subfigure}{.32\linewidth}
    \centering
    \includegraphics[height=6.0cm]{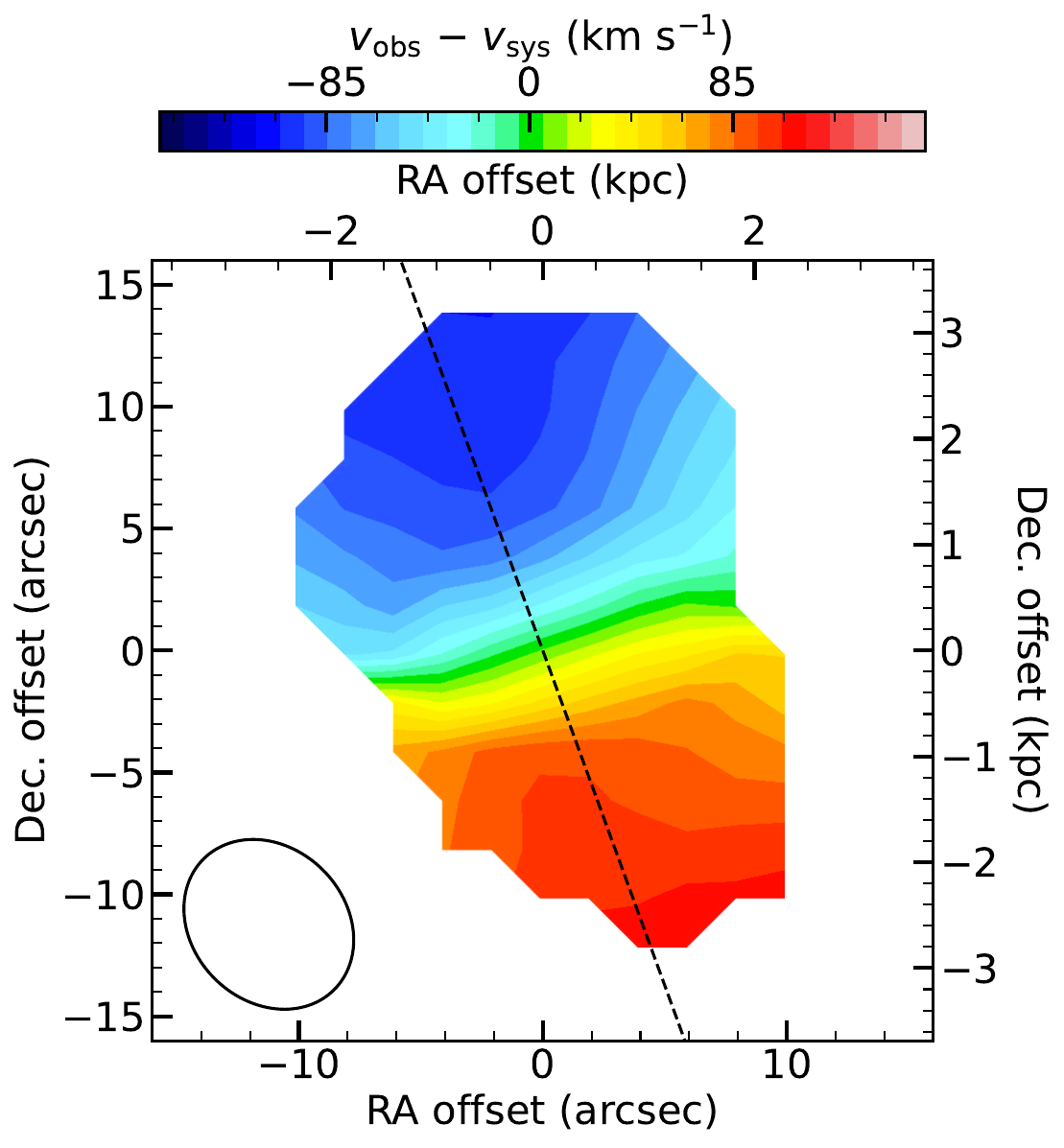}
    \end{subfigure}
    \begin{subfigure}{.32\linewidth}
    \centering
    \includegraphics[height=6.0cm]{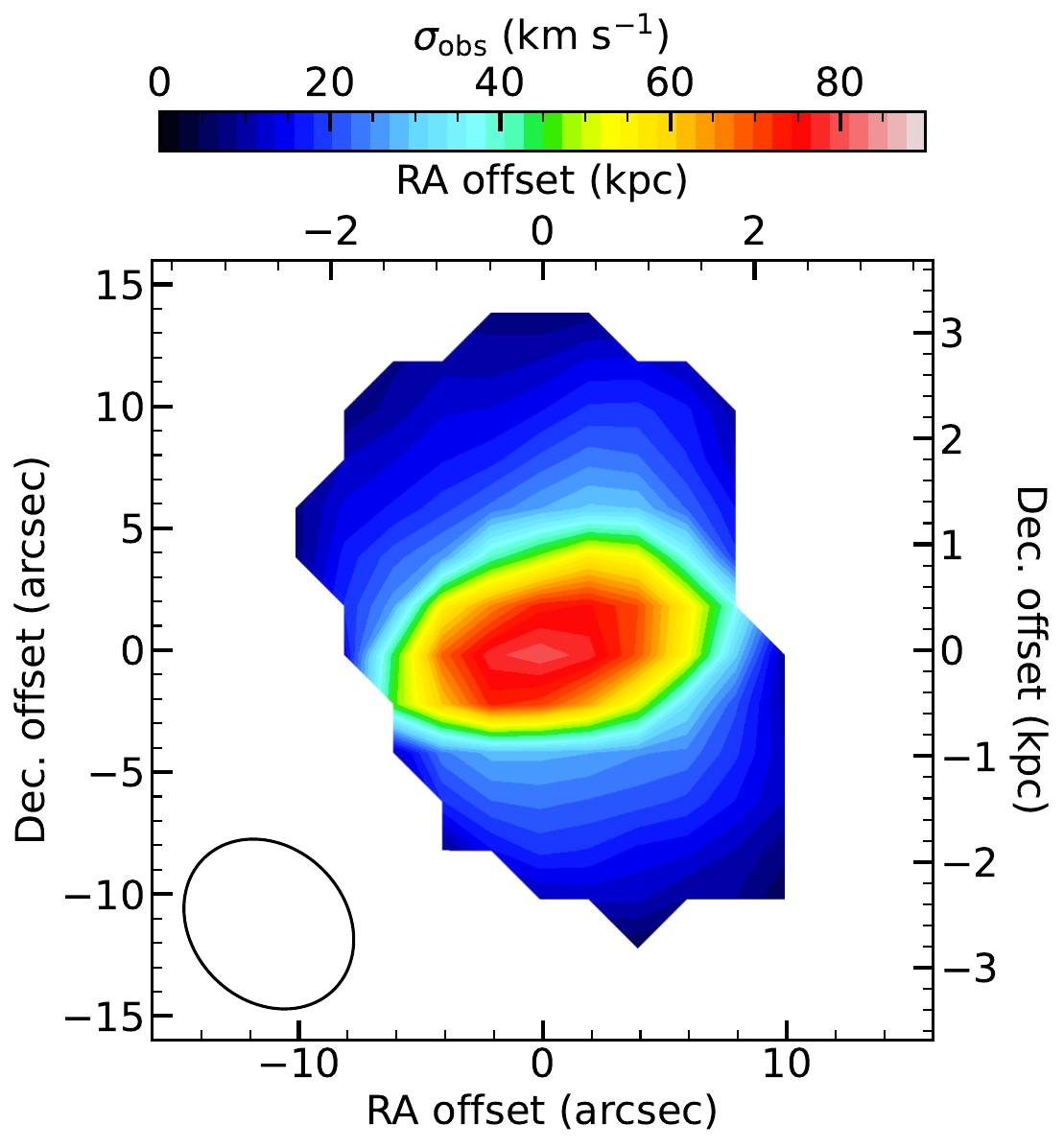}
    \end{subfigure}
    \\
    \begin{subfigure}{.34\linewidth}
    \centering
    \includegraphics[height=5.2cm]{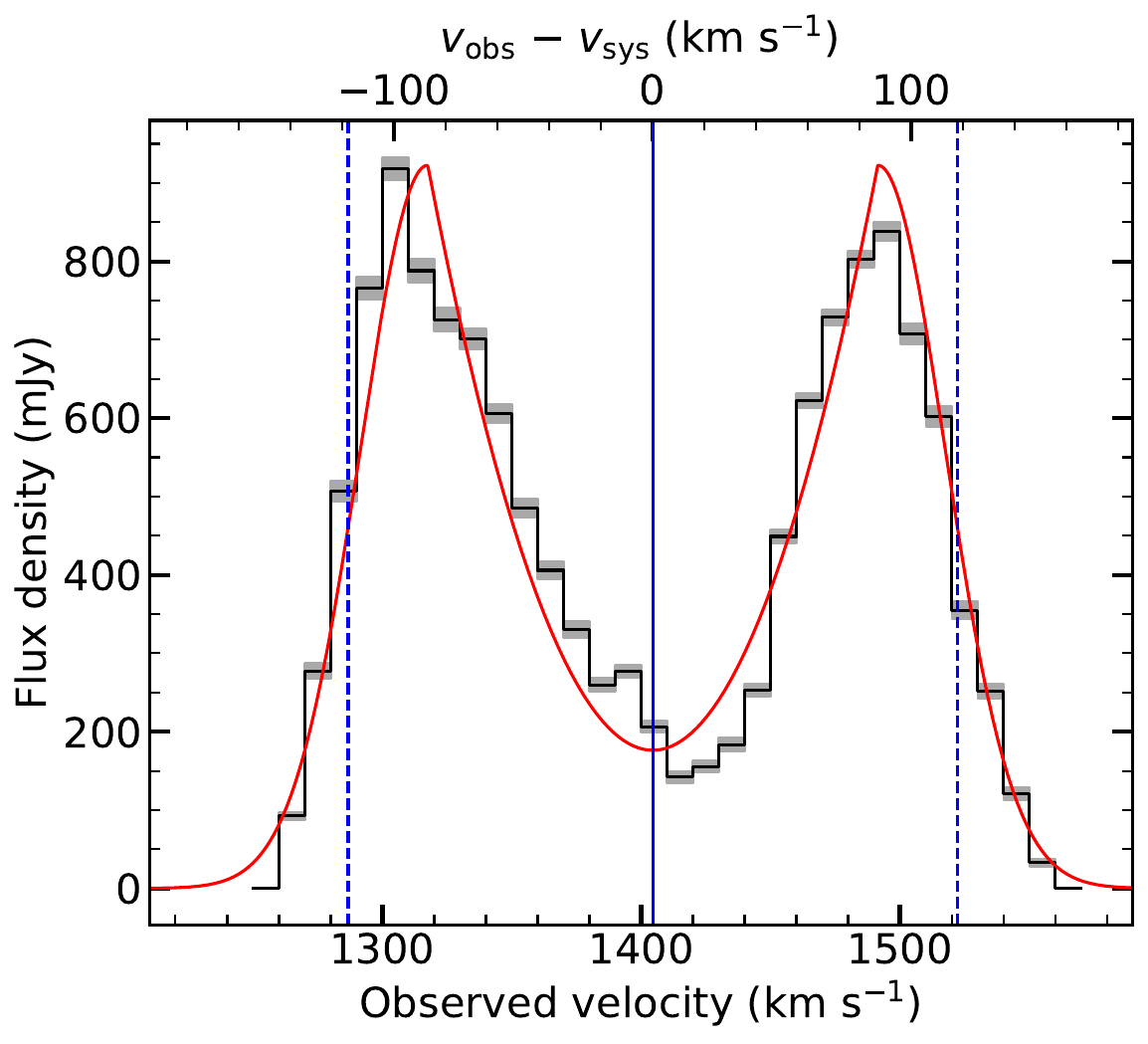}
    \end{subfigure}
    \begin{subfigure}{.32\linewidth}
    \centering
    \includegraphics[height=4.8cm]{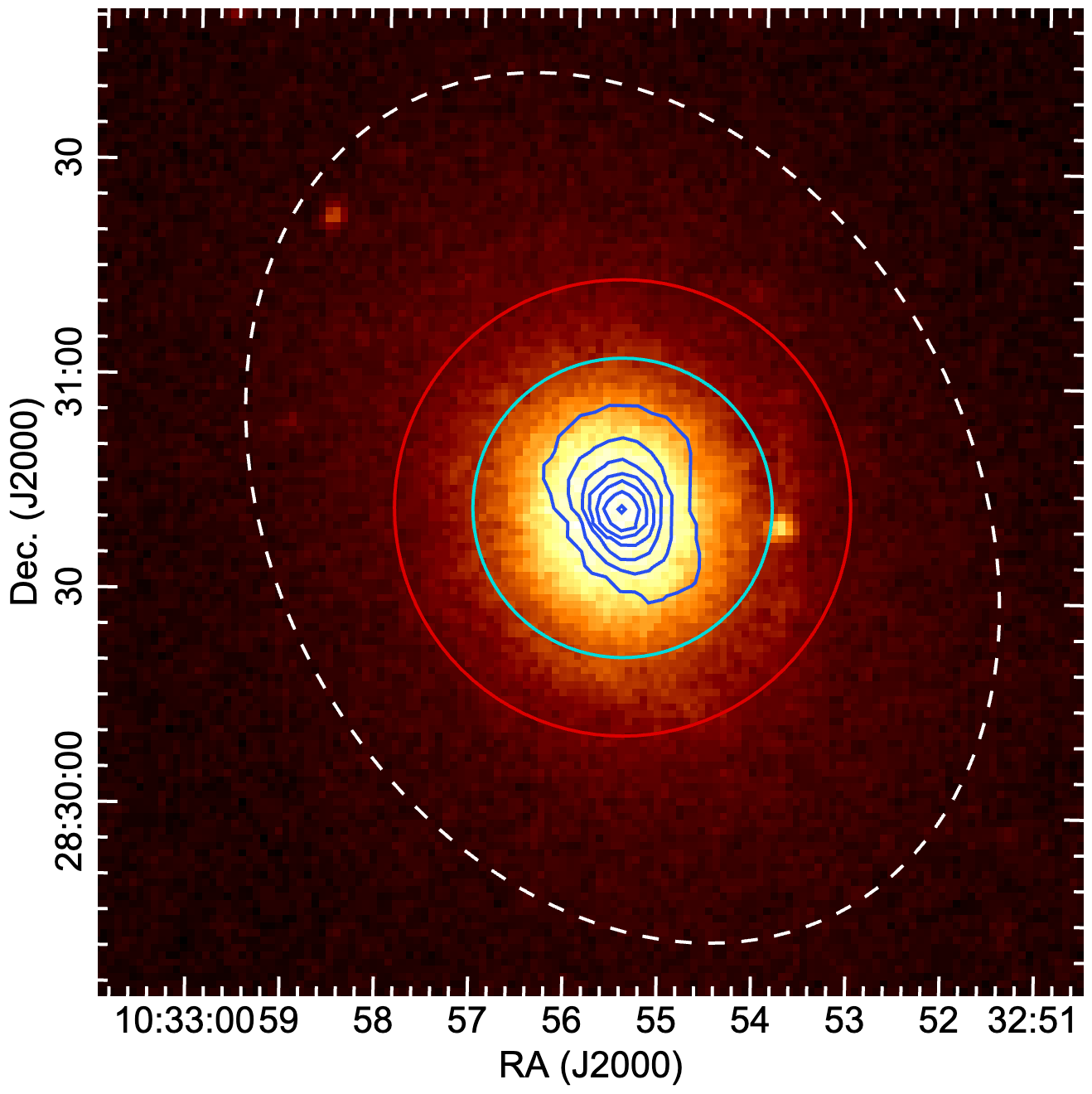}
    \end{subfigure}
    \begin{subfigure}{.33\linewidth}
    \centering
    \includegraphics[height=5.2cm]{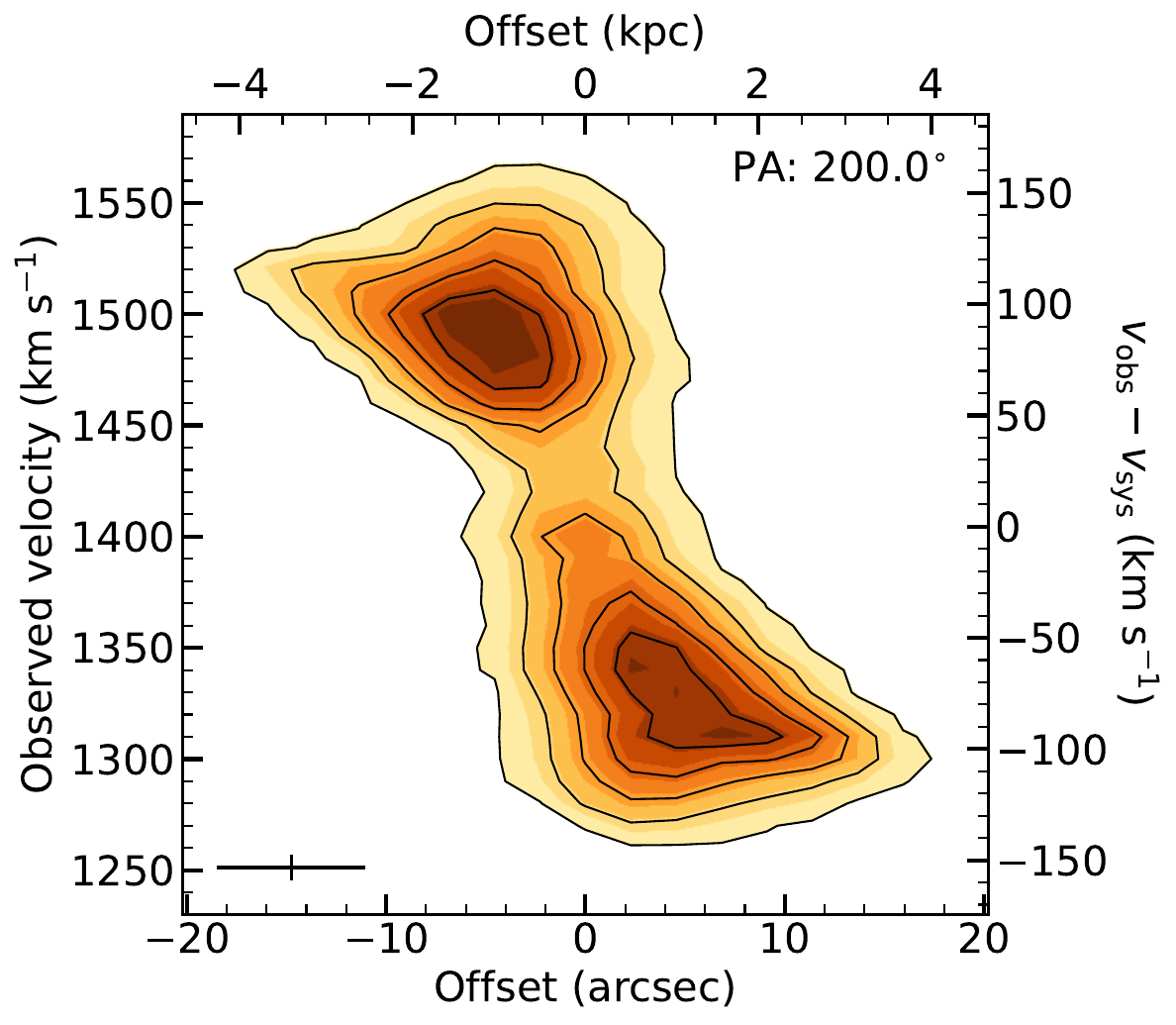}
    \end{subfigure}
    \\
    \begin{subfigure}{.34\linewidth}
    \centering
    \includegraphics[height=6.0cm]{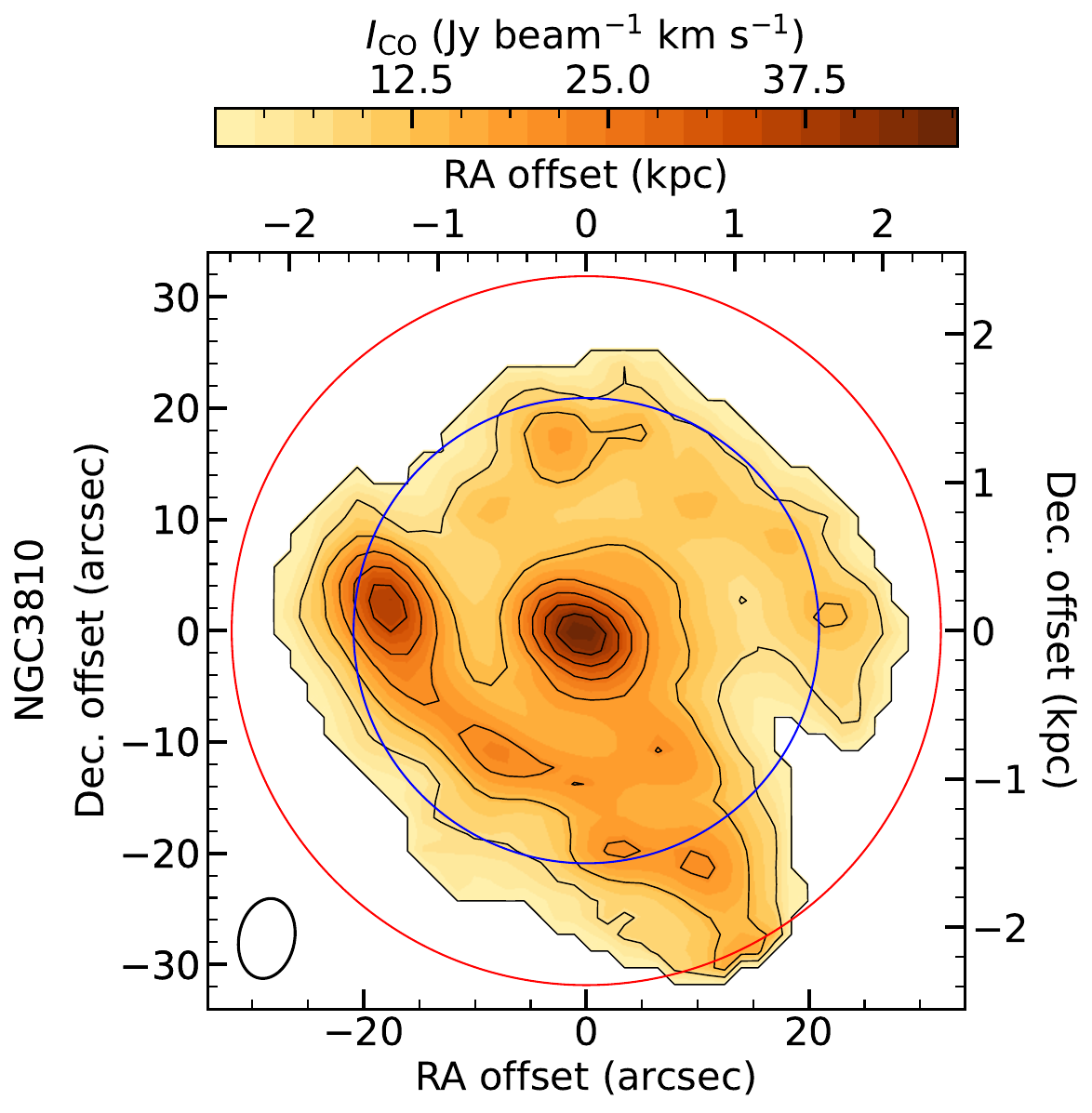}
    \end{subfigure}
    \begin{subfigure}{.32\linewidth}
    \centering
    \includegraphics[height=6.0cm]{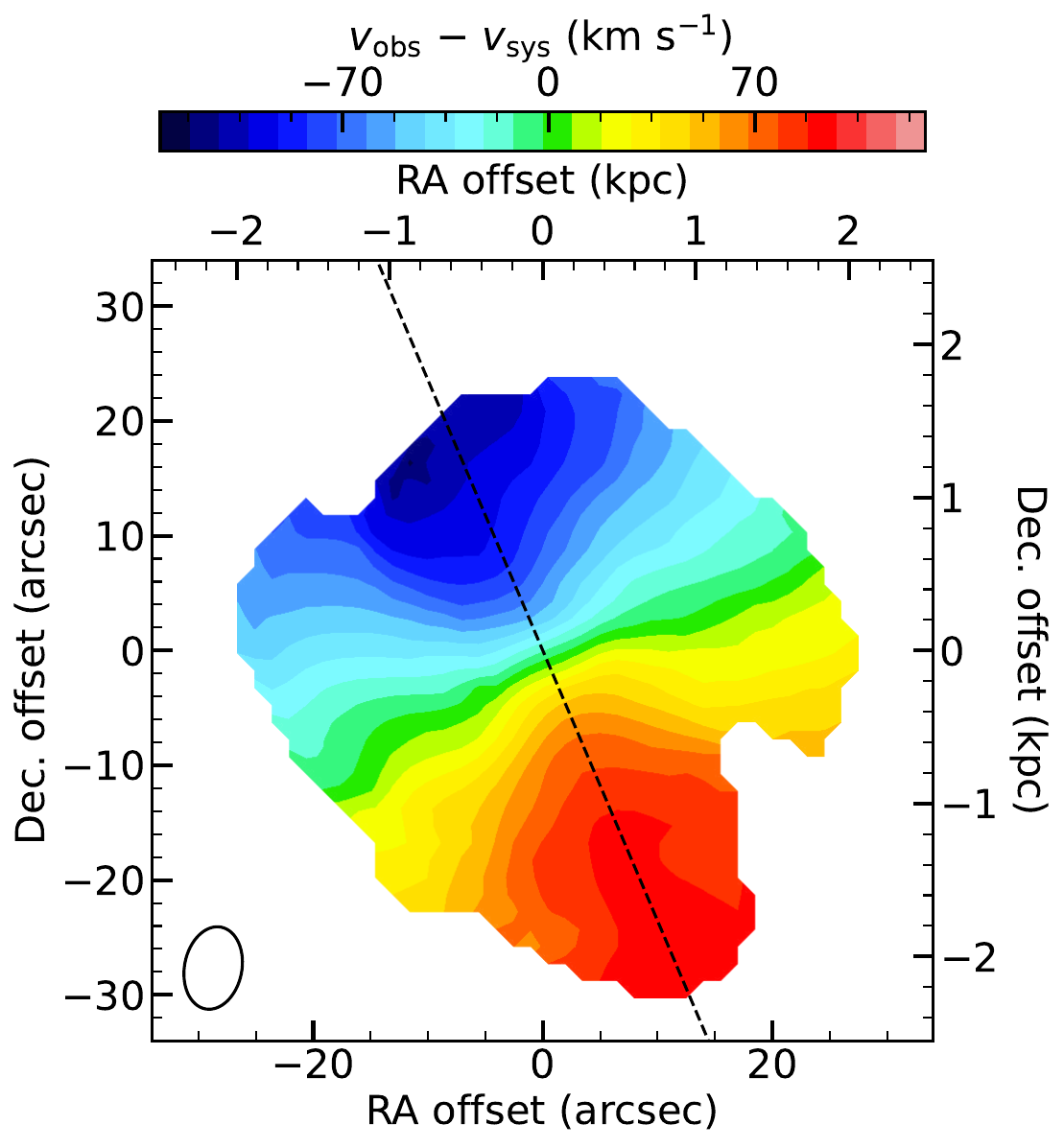}
    \end{subfigure}
    \begin{subfigure}{.32\linewidth}
    \centering
    \includegraphics[height=6.0cm]{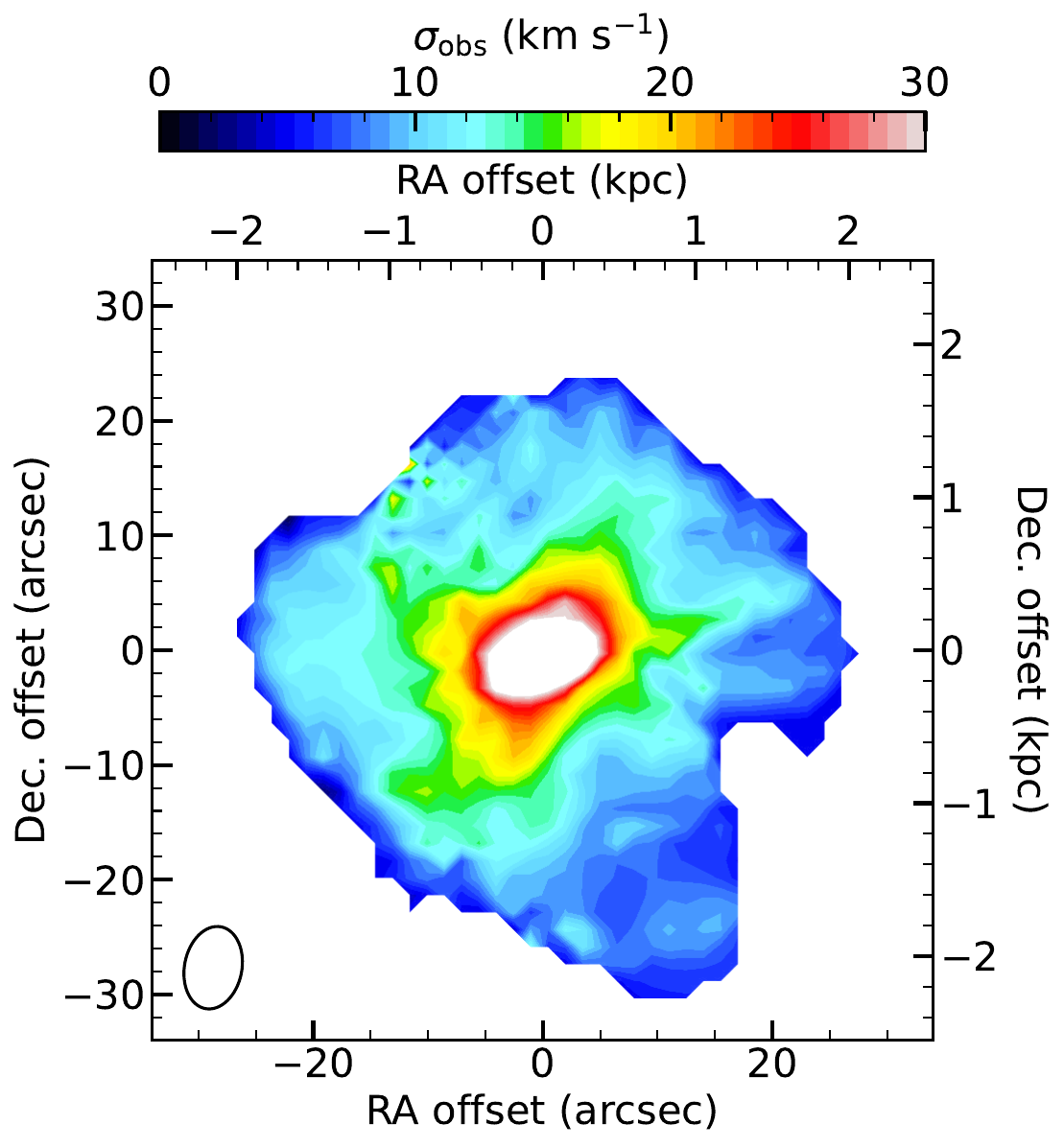}
    \end{subfigure}
    \\
    \begin{subfigure}{.34\linewidth}
    \centering
    \includegraphics[height=5.2cm]{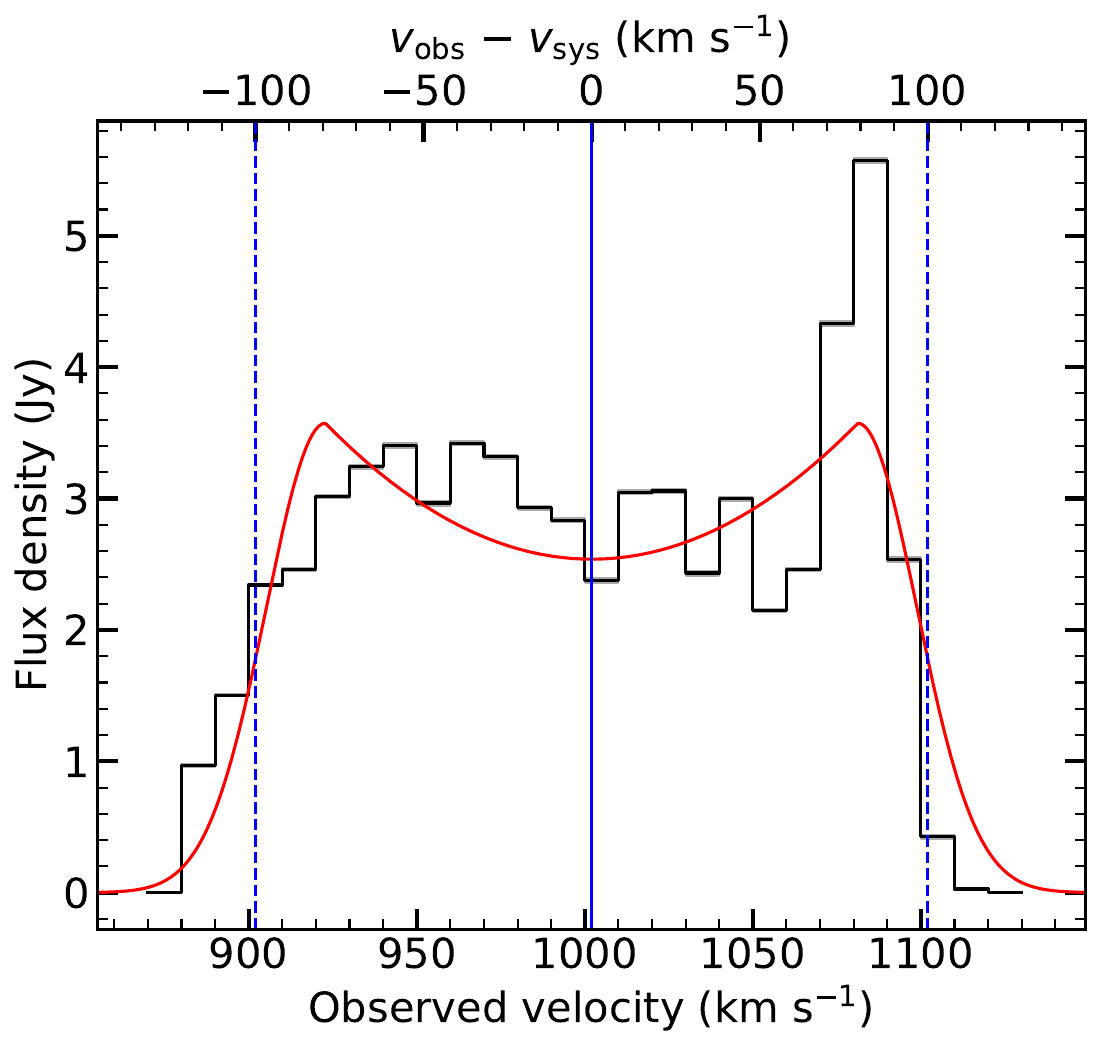}
    \end{subfigure}
    \begin{subfigure}{.32\linewidth}
    \centering
    \includegraphics[height=4.8cm]{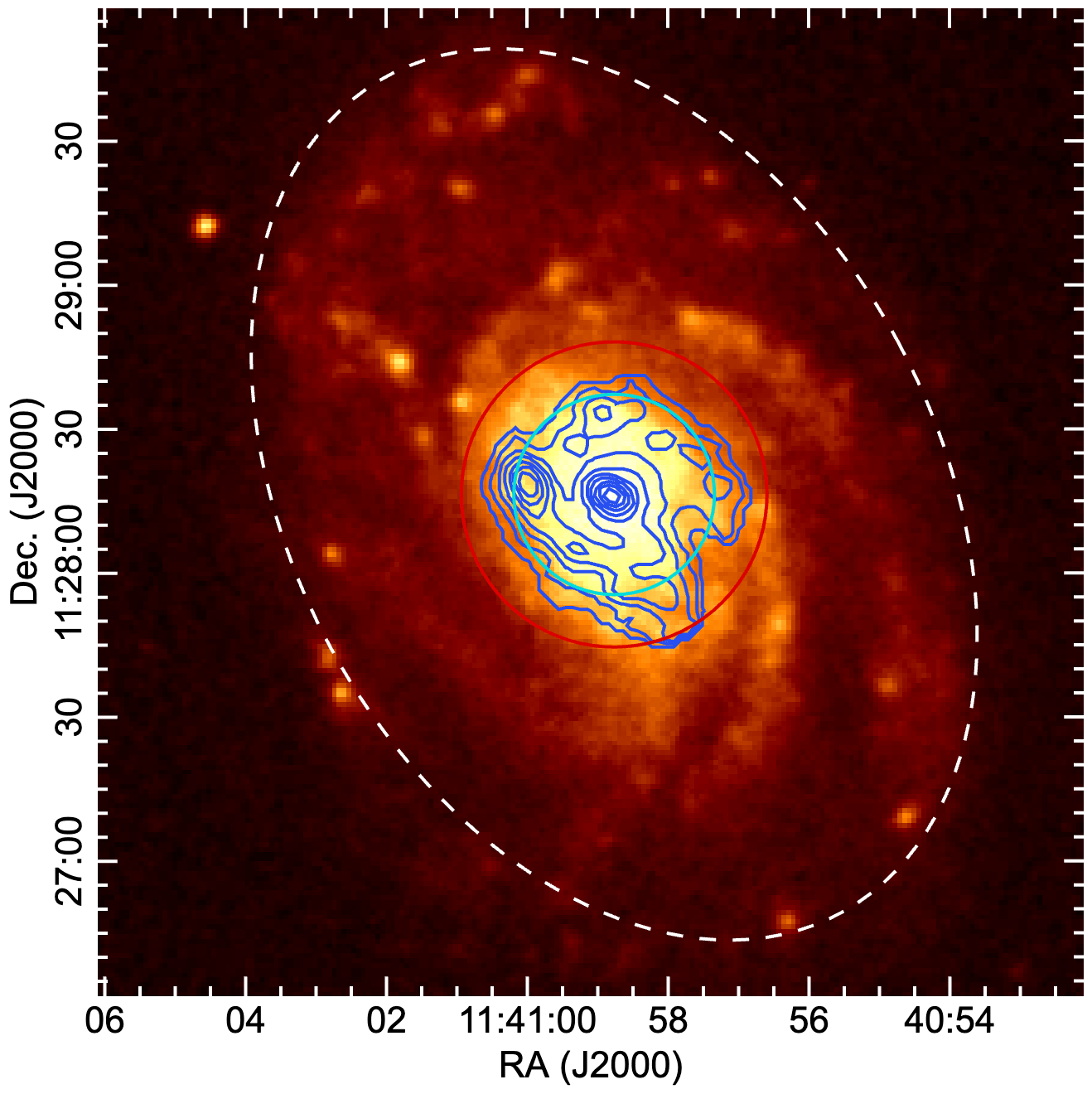}
    \end{subfigure}
    \begin{subfigure}{.33\linewidth}
    \centering
    \includegraphics[height=5.2cm]{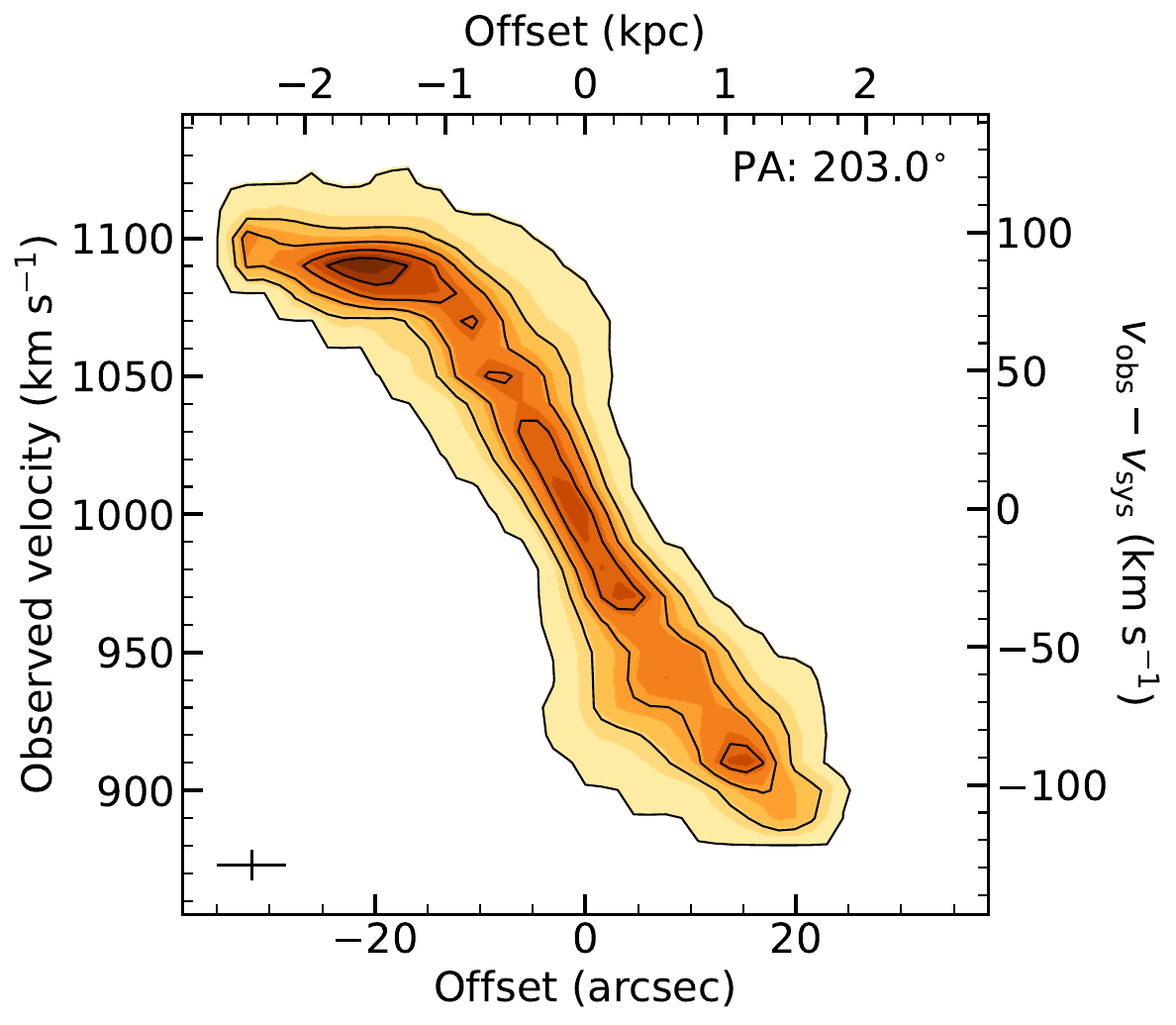}
    \end{subfigure}
    \caption[]{continued.}
\end{figure*}

\begin{figure*}
    \ContinuedFloat
    \centering
    \begin{subfigure}{.34\linewidth}
    \centering
    \includegraphics[height=6.0cm]{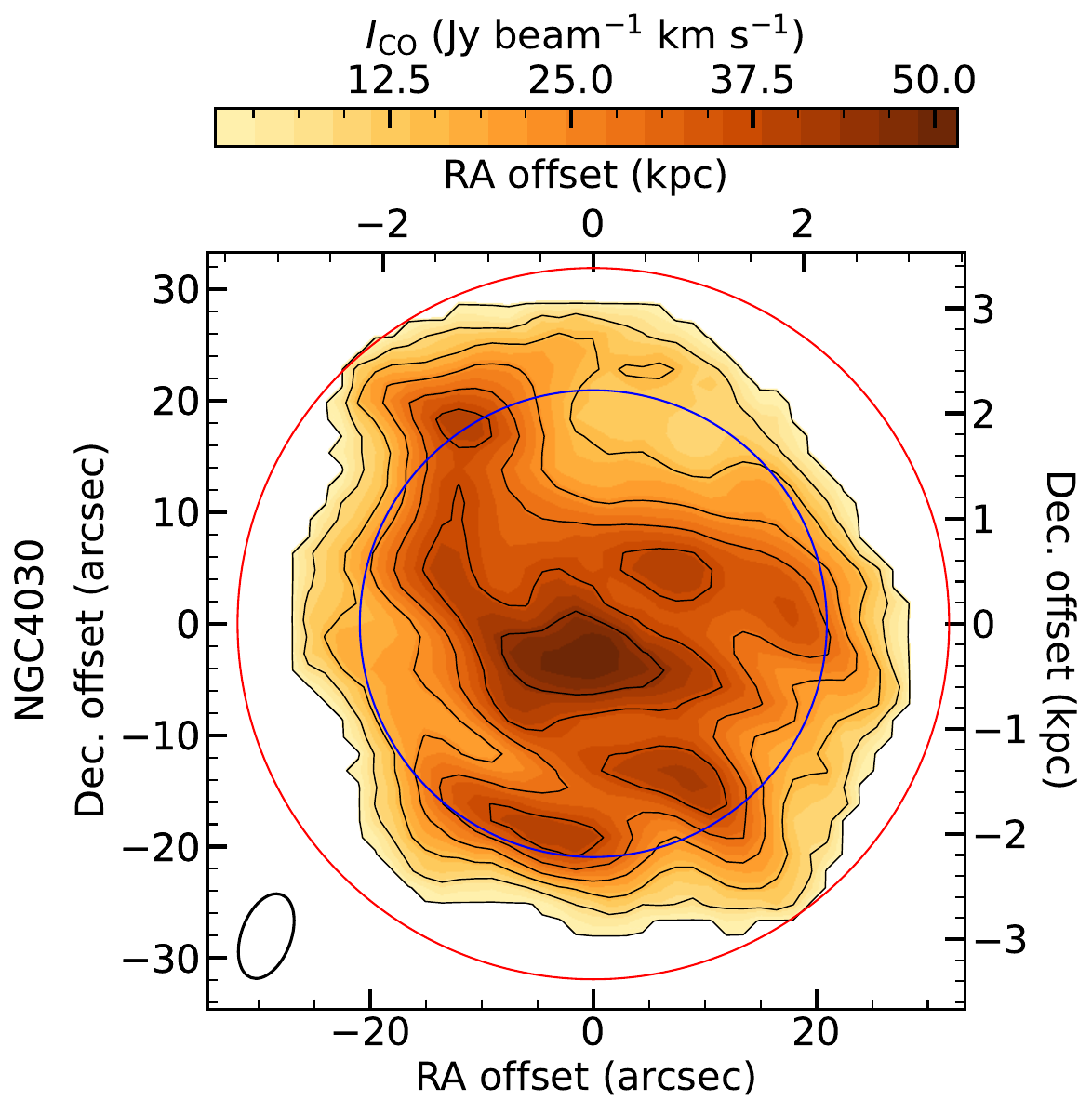}
    \end{subfigure}
    \begin{subfigure}{.32\linewidth}
    \centering
    \includegraphics[height=6.0cm]{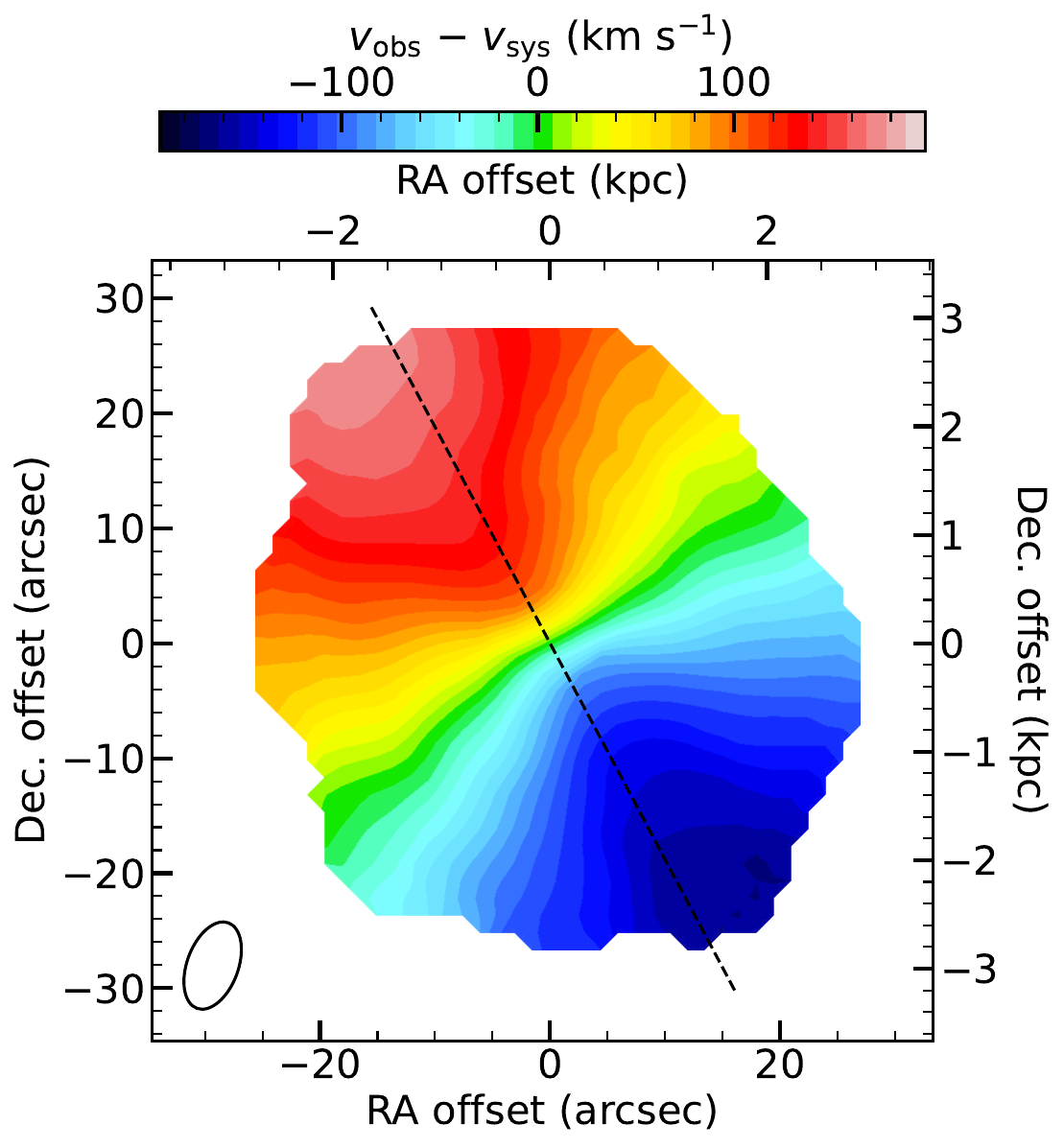}
    \end{subfigure}
    \begin{subfigure}{.32\linewidth}
    \centering
    \includegraphics[height=6.0cm]{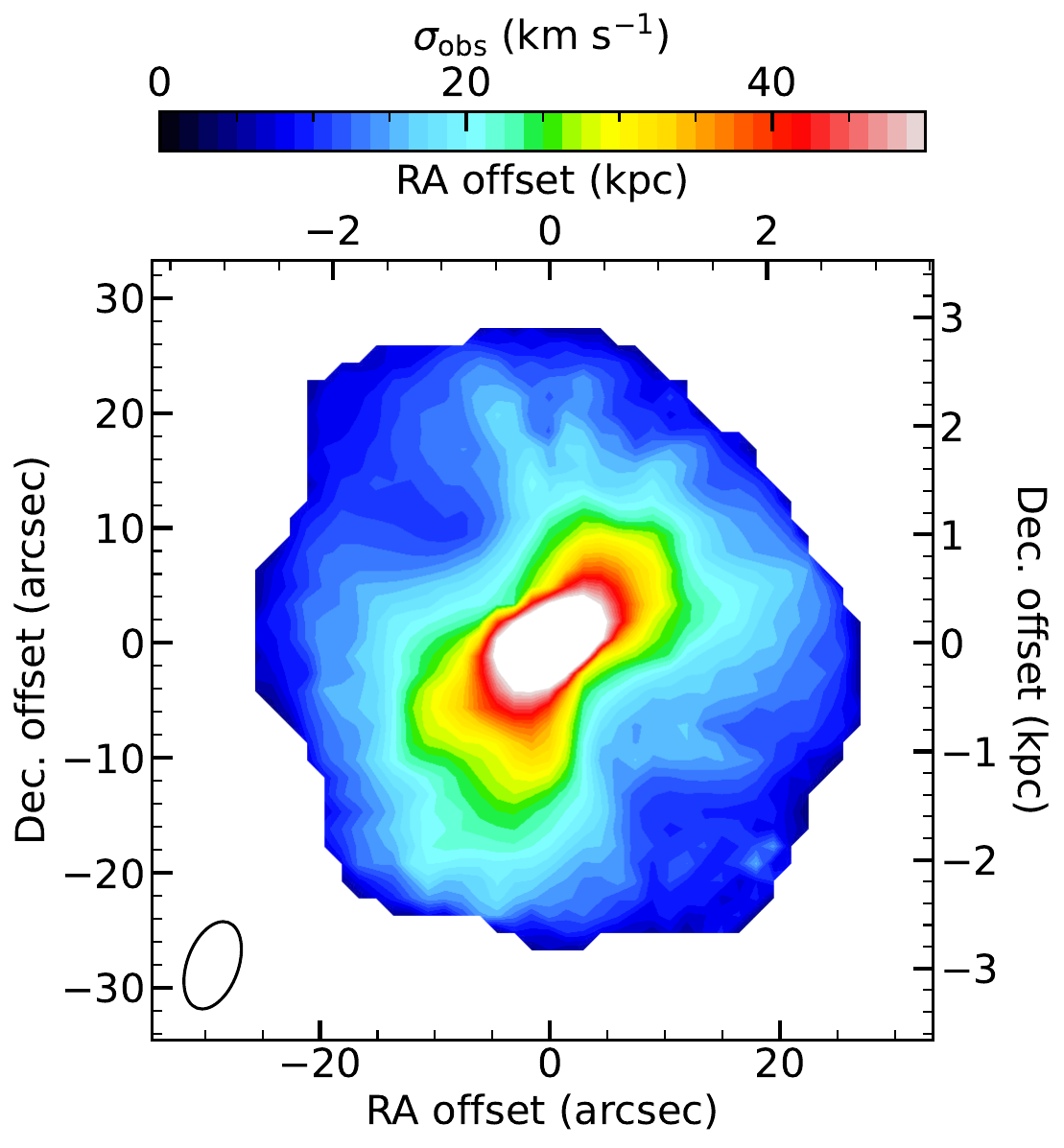}
    \end{subfigure}
    \\
    \begin{subfigure}{.34\linewidth}
    \centering
    \includegraphics[height=5.2cm]{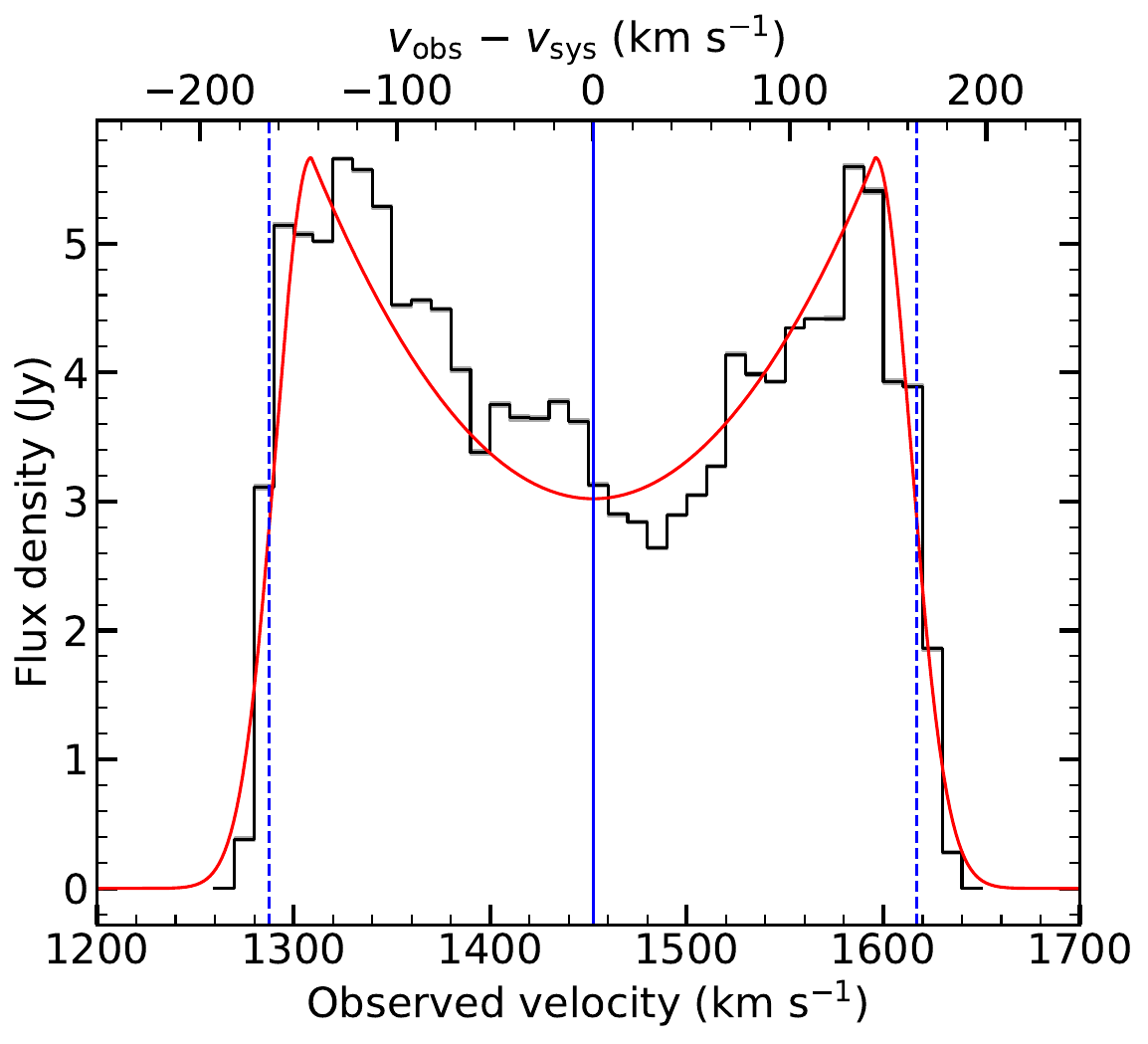}
    \end{subfigure}
    \begin{subfigure}{.32\linewidth}
    \centering
    \includegraphics[height=4.8cm]{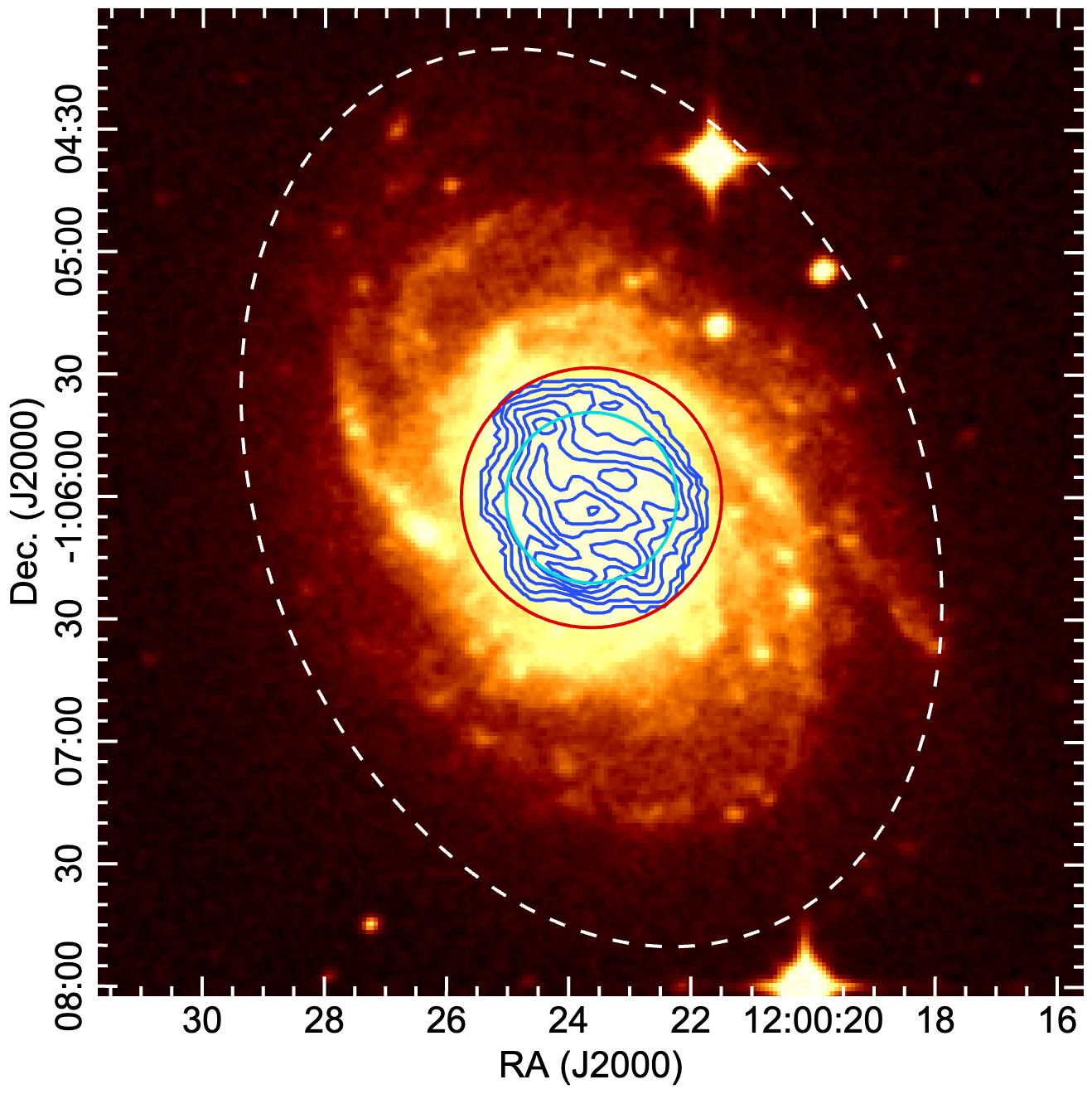}
    \end{subfigure}
    \begin{subfigure}{.33\linewidth}
    \centering
    \includegraphics[height=5.2cm]{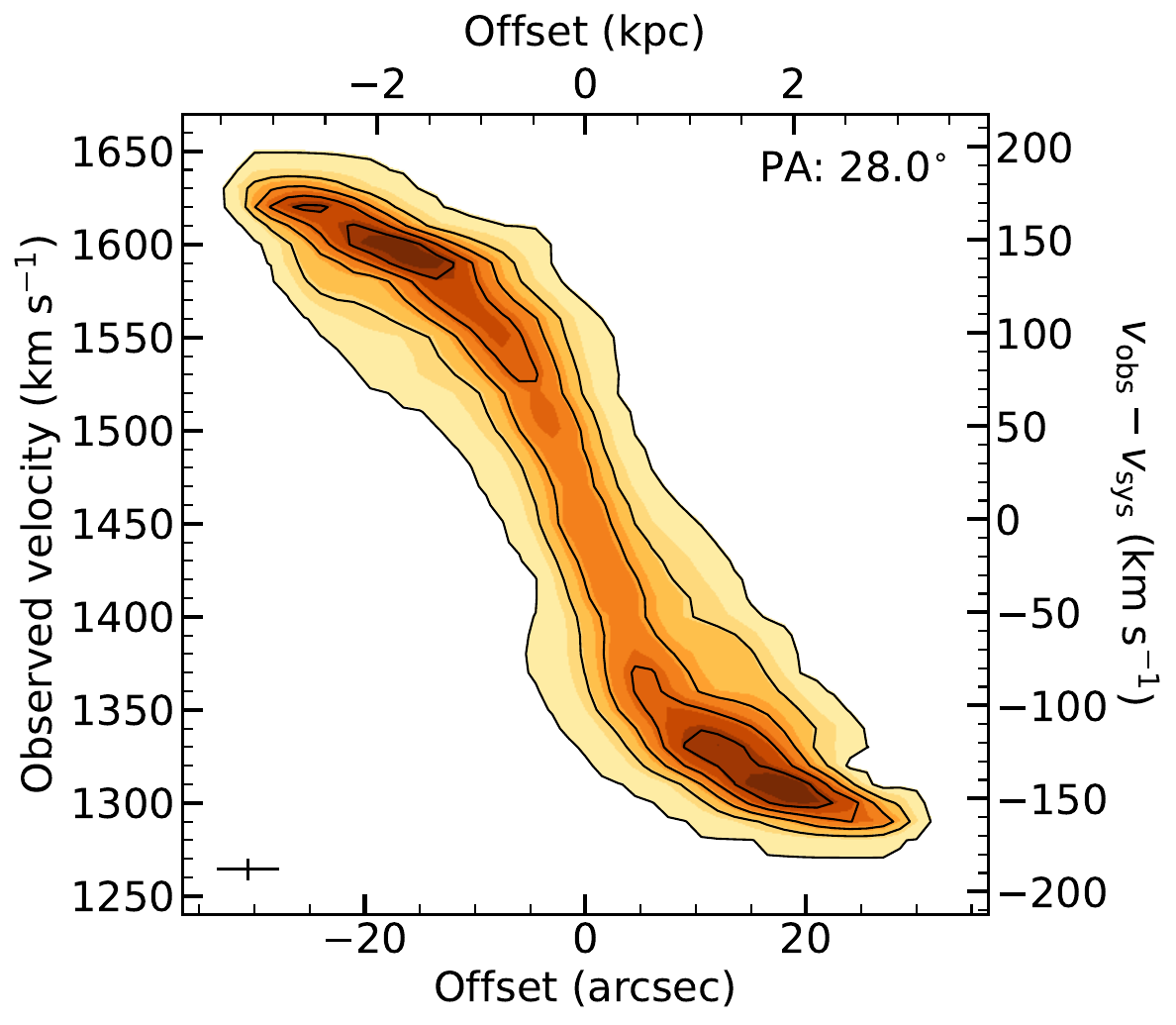}
    \end{subfigure}
    \\
    \begin{subfigure}{.34\linewidth}
    \centering
    \includegraphics[height=6.0cm]{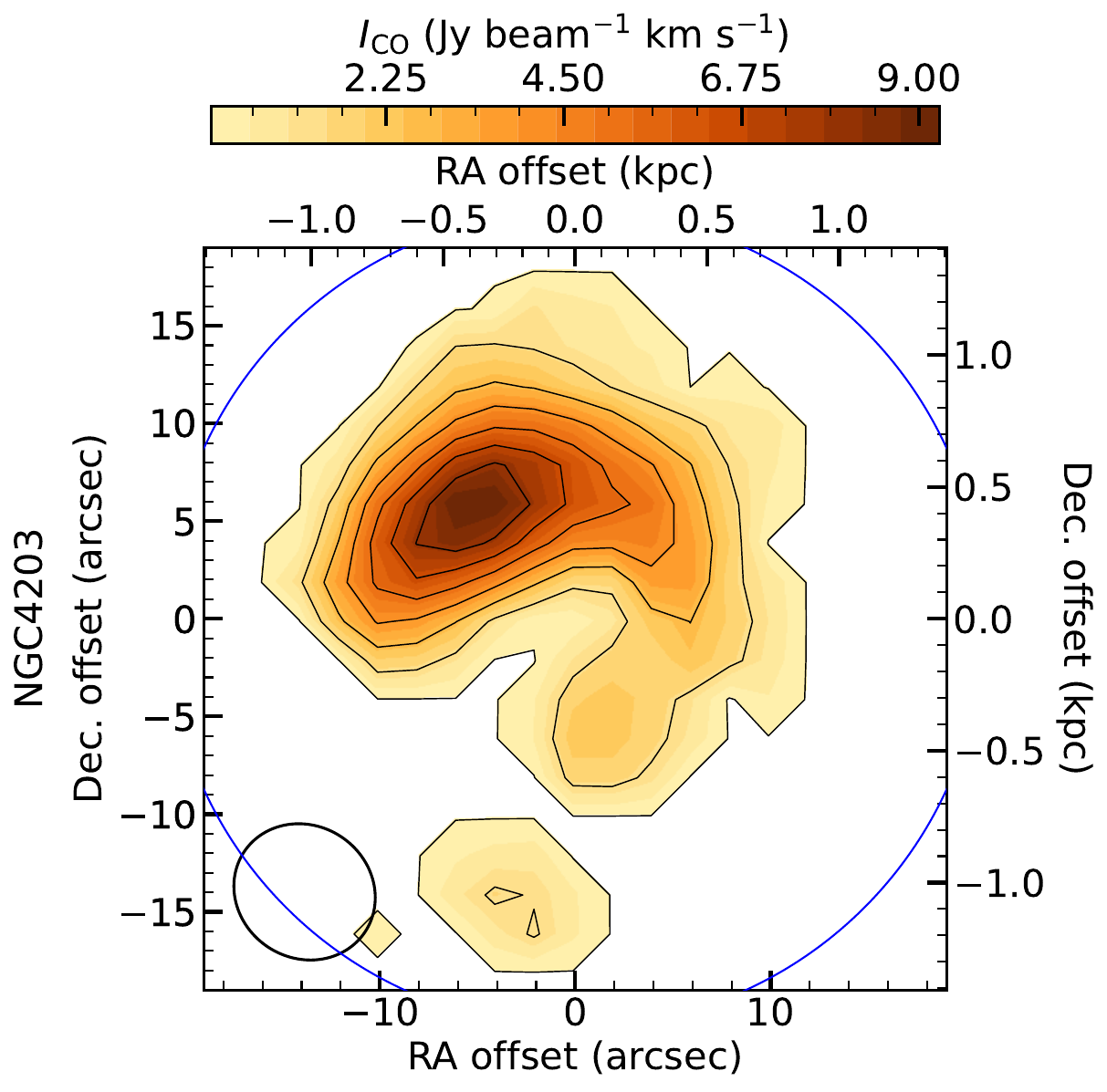}
    \end{subfigure}
    \begin{subfigure}{.32\linewidth}
    \centering
    \includegraphics[height=6.0cm]{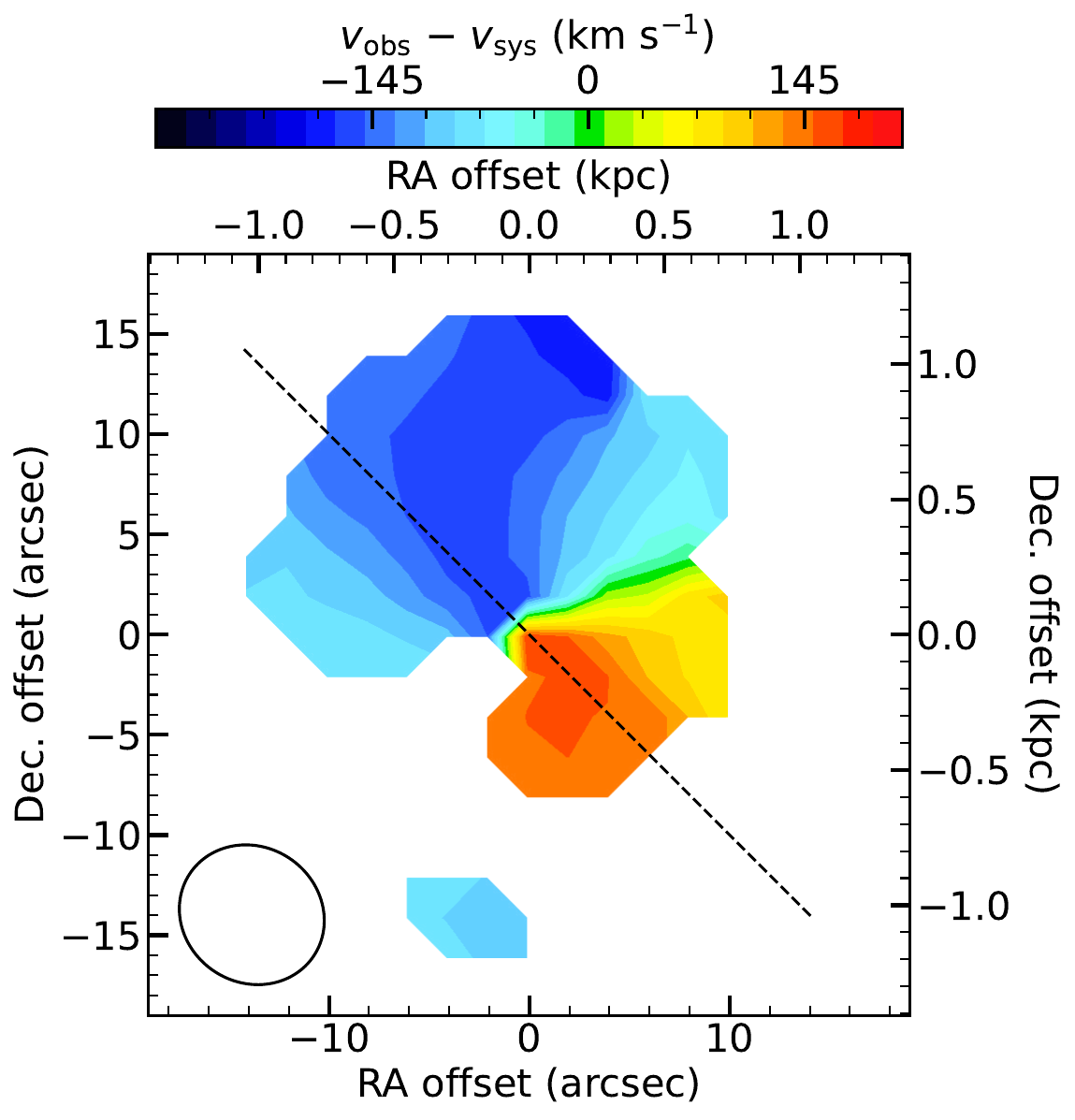}
    \end{subfigure}
    \begin{subfigure}{.32\linewidth}
    \centering
    \includegraphics[height=6.0cm]{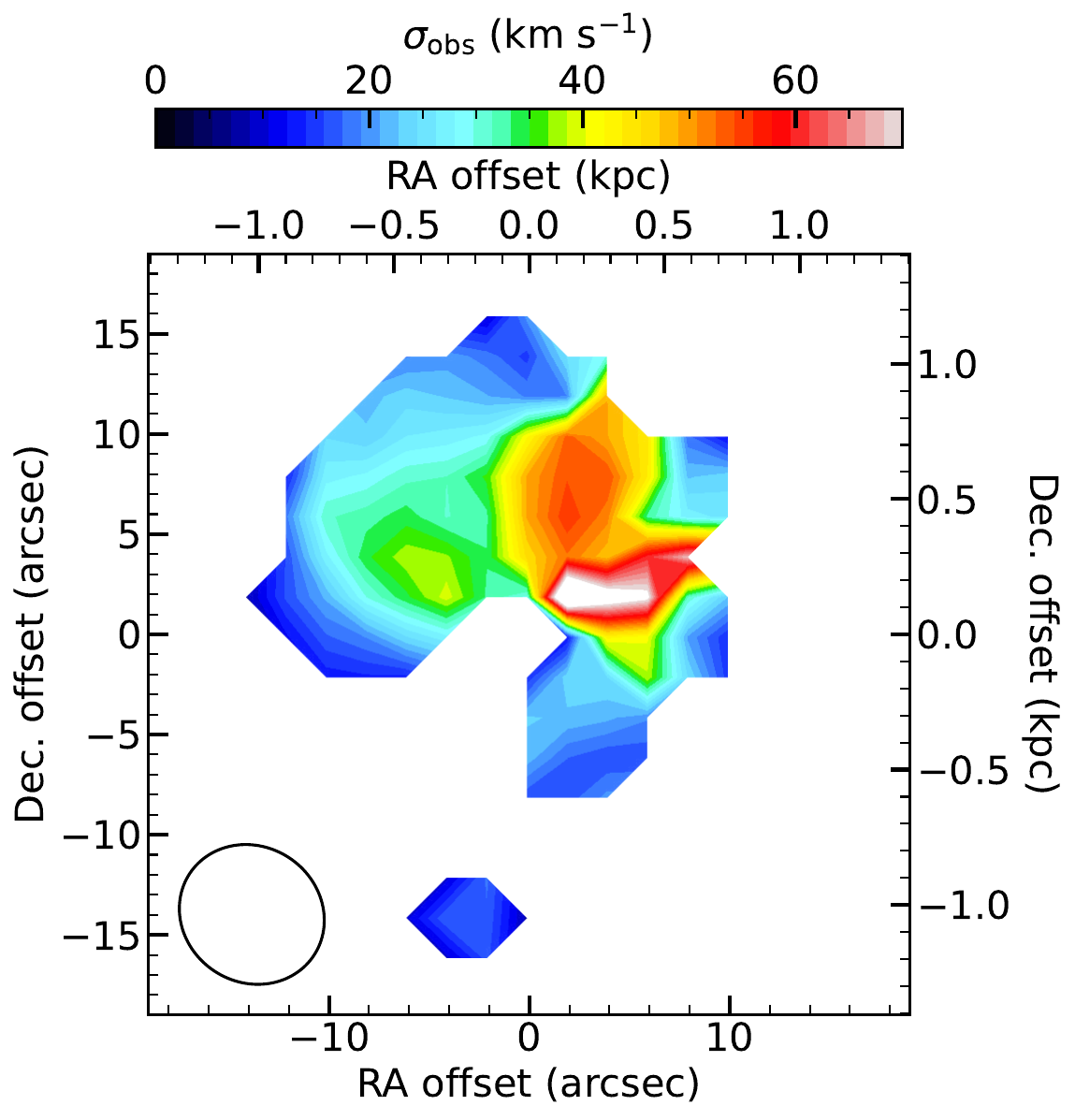}
    \end{subfigure}
    \\
    \begin{subfigure}{.34\linewidth}
    \centering
    \includegraphics[height=5.2cm]{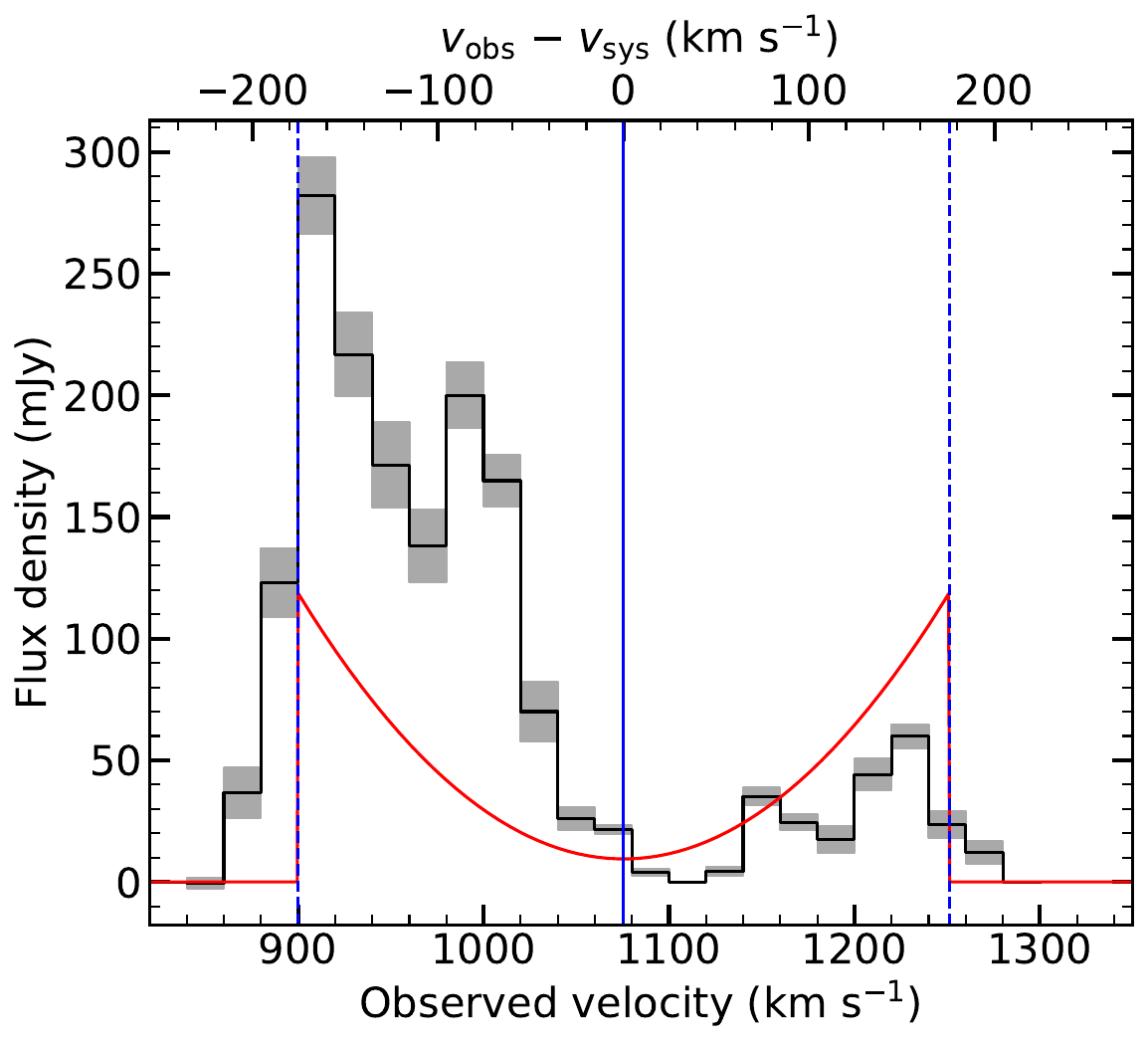}
    \end{subfigure}
    \begin{subfigure}{.32\linewidth}
    \centering
    \includegraphics[height=4.8cm]{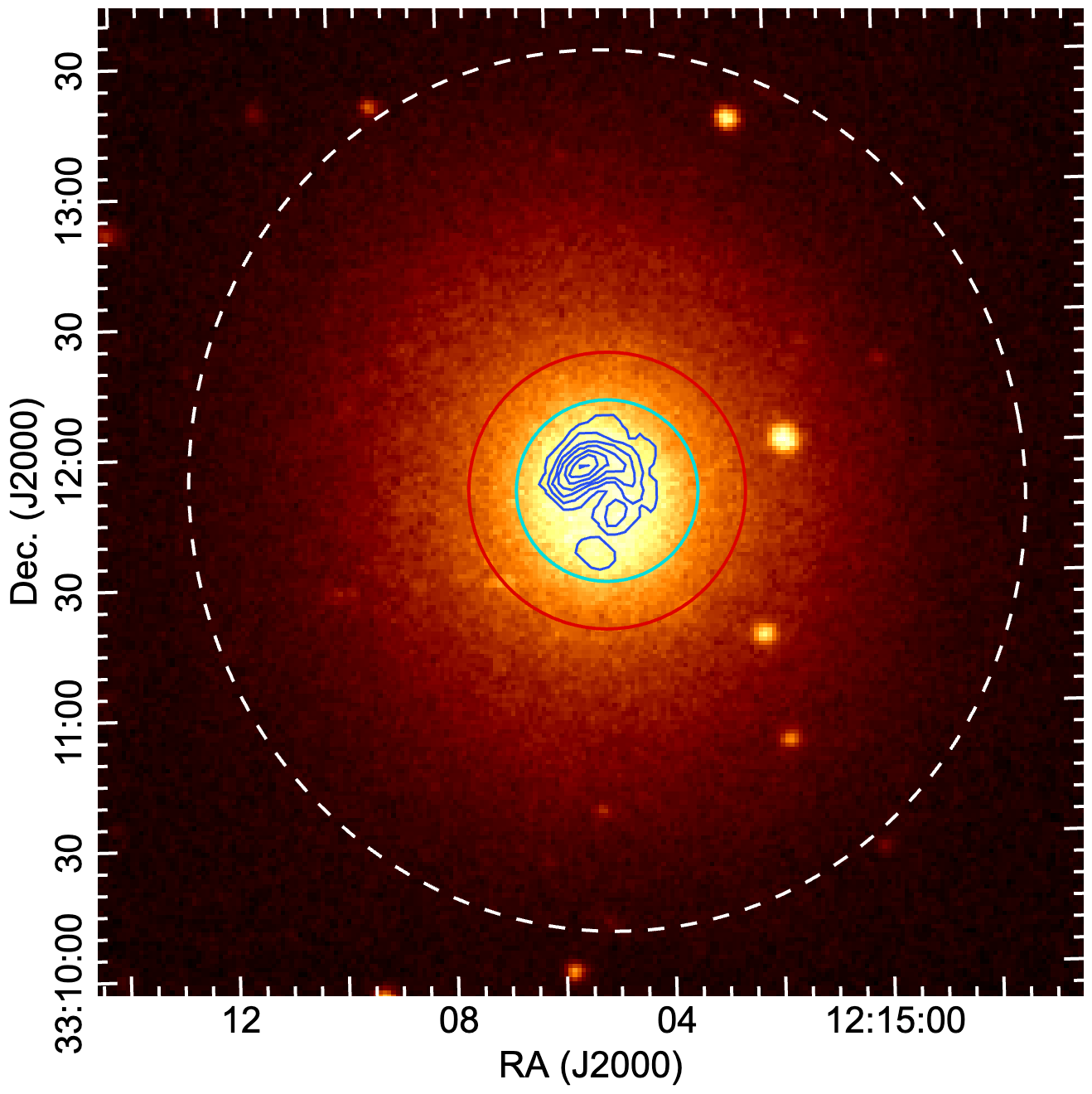}
    \end{subfigure}
    \begin{subfigure}{.33\linewidth}
    \centering
    \includegraphics[height=5.2cm]{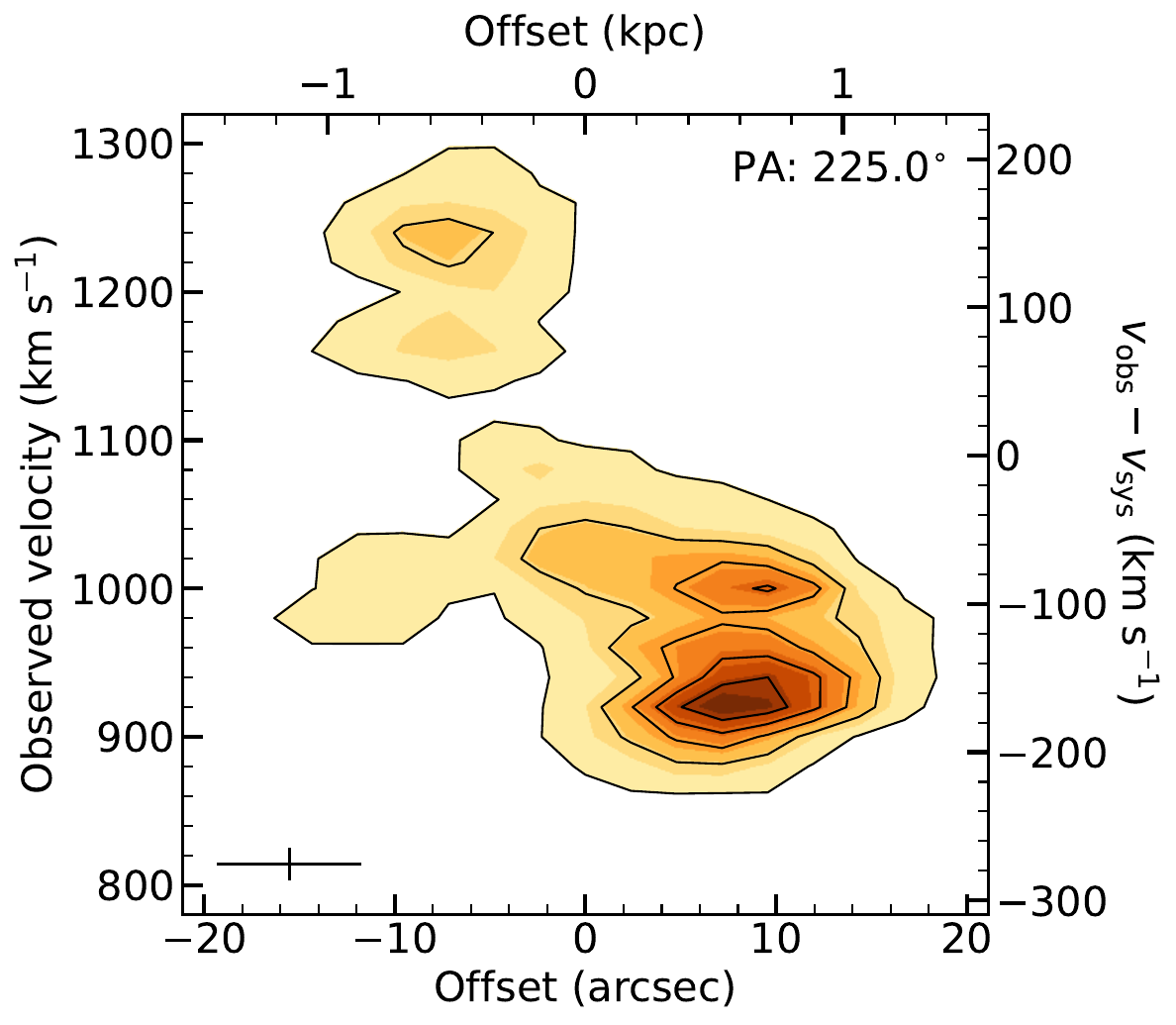}
    \end{subfigure}
    \caption[]{continued.}
\end{figure*}

\begin{figure*}
    \ContinuedFloat
    \centering
    \begin{subfigure}{.34\linewidth}
    \centering
    \includegraphics[height=6.0cm]{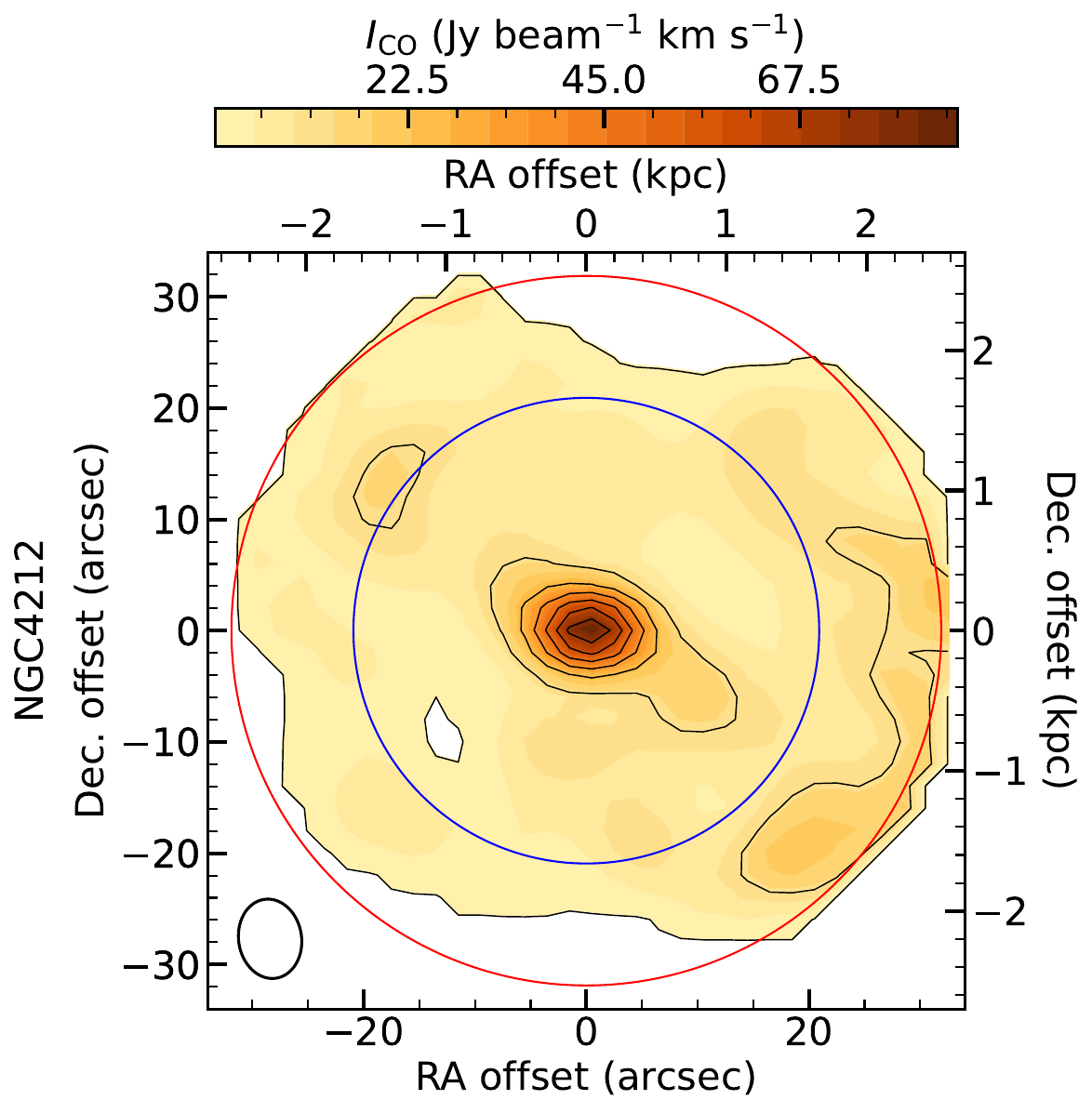}
    \end{subfigure}
    \begin{subfigure}{.32\linewidth}
    \centering
    \includegraphics[height=6.0cm]{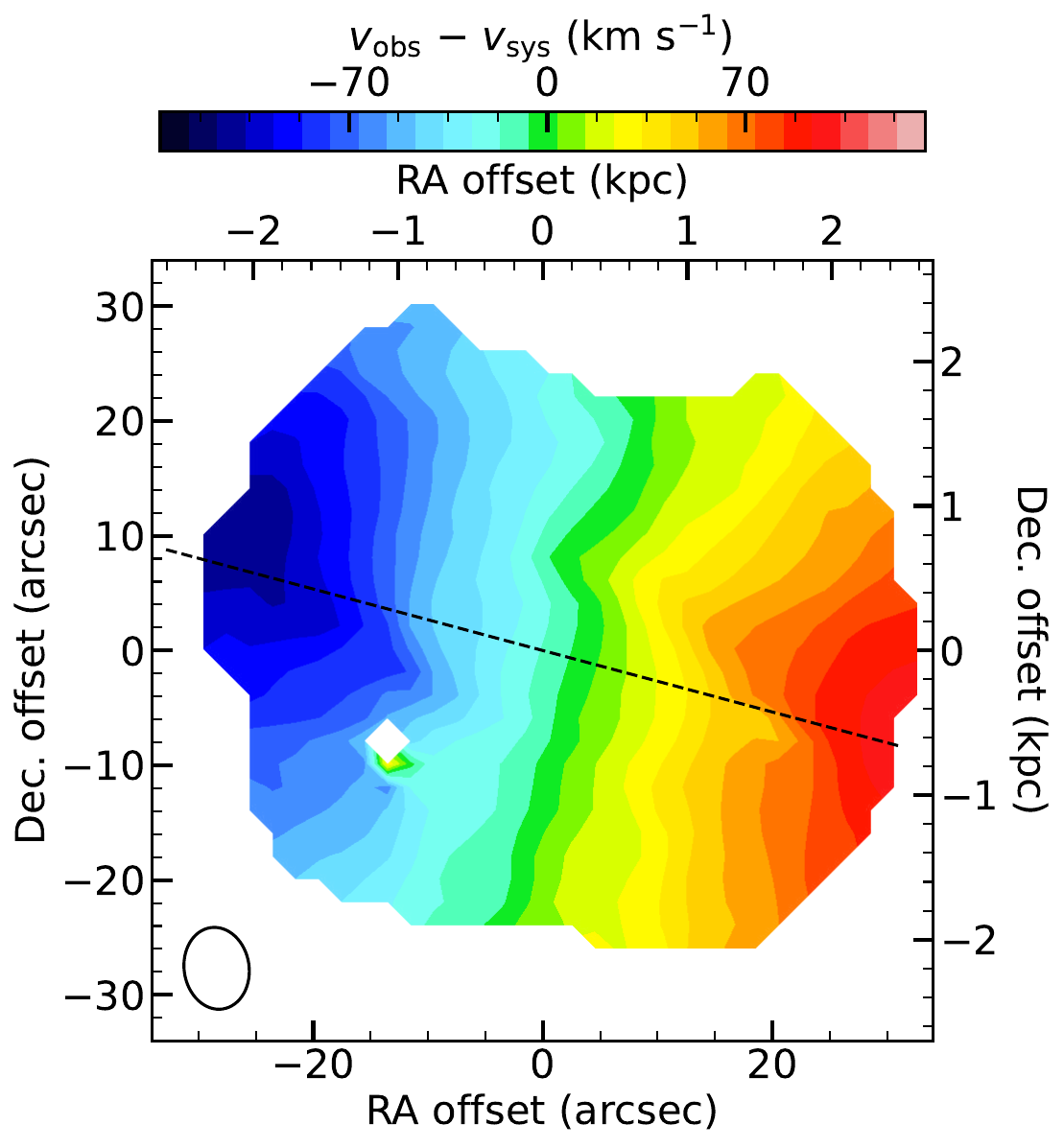}
    \end{subfigure}
    \begin{subfigure}{.32\linewidth}
    \centering
    \includegraphics[height=6.0cm]{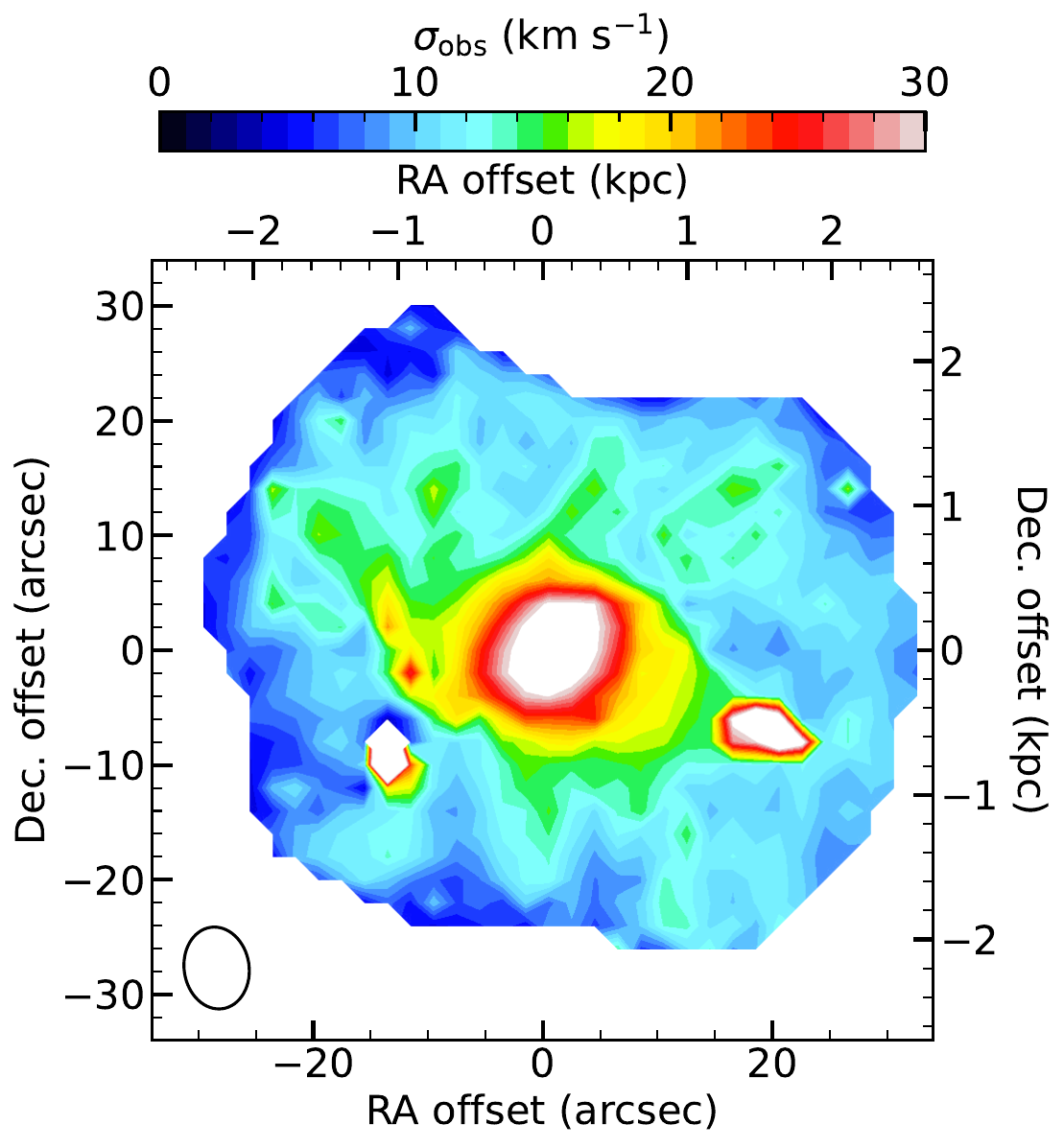}
    \end{subfigure}
    \\
    \begin{subfigure}{.34\linewidth}
    \centering
    \includegraphics[height=5.2cm]{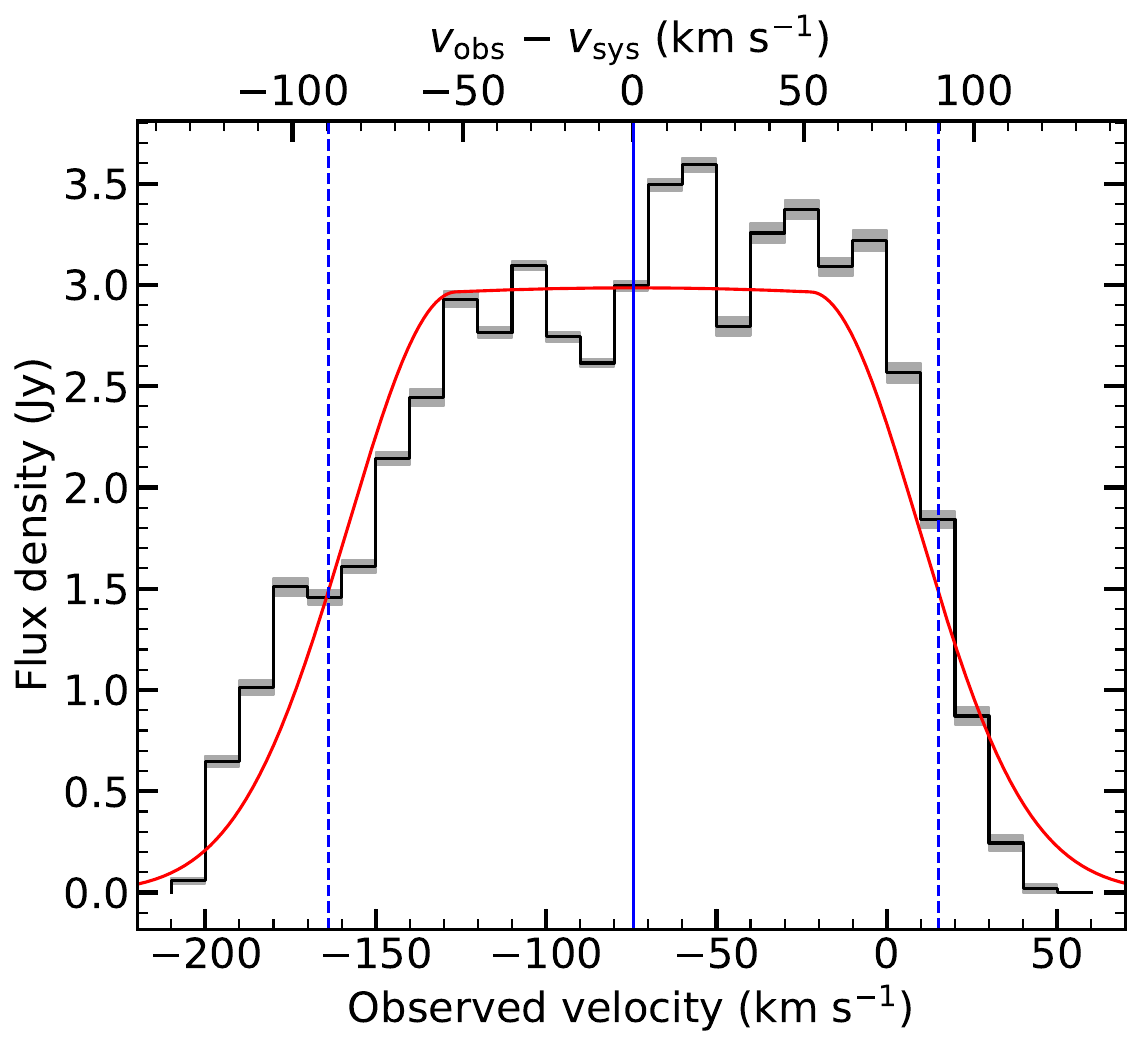}
    \end{subfigure}
    \begin{subfigure}{.32\linewidth}
    \centering
    \includegraphics[height=4.8cm]{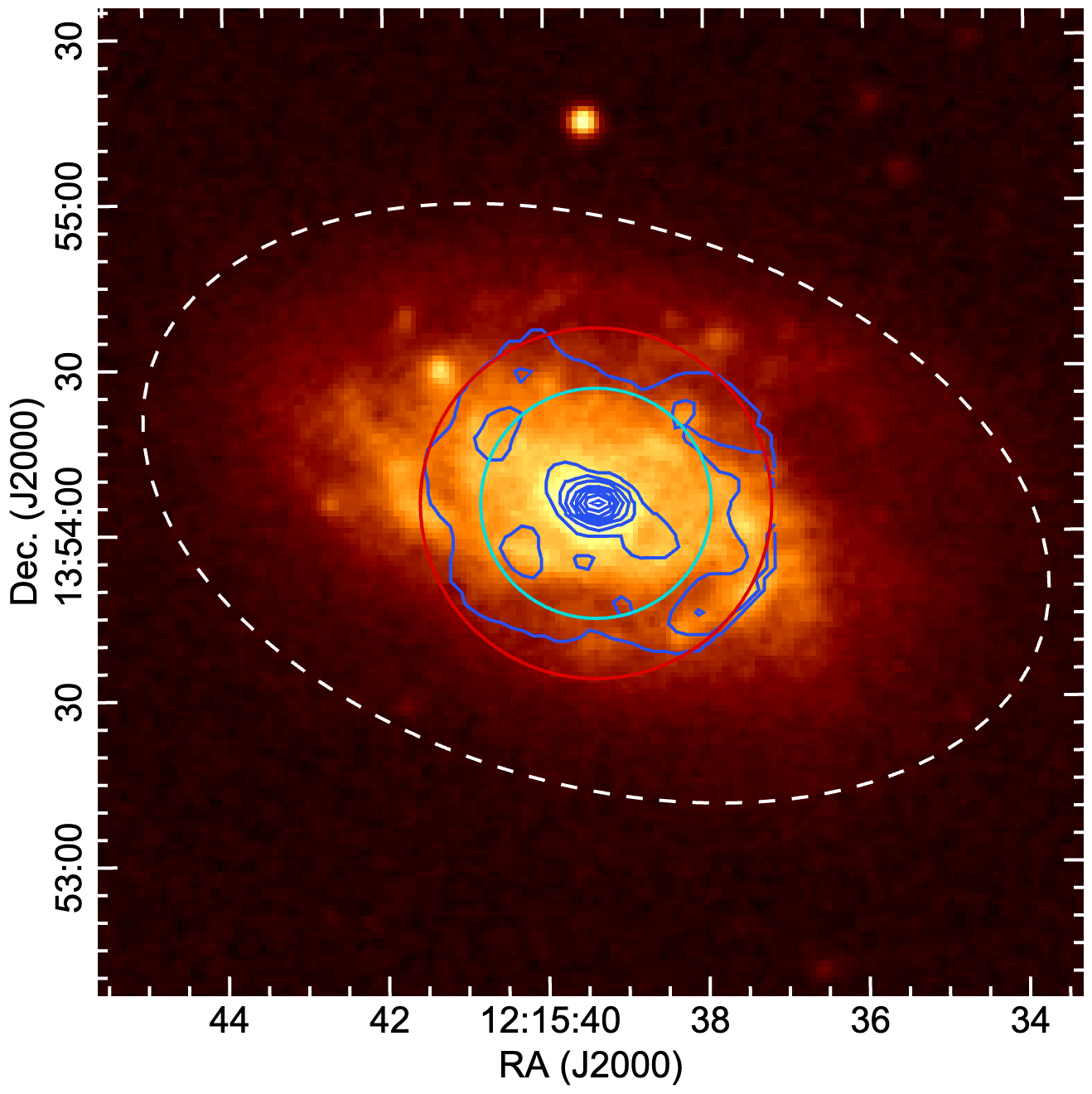}
    \end{subfigure}
    \begin{subfigure}{.33\linewidth}
    \centering
    \includegraphics[height=5.2cm]{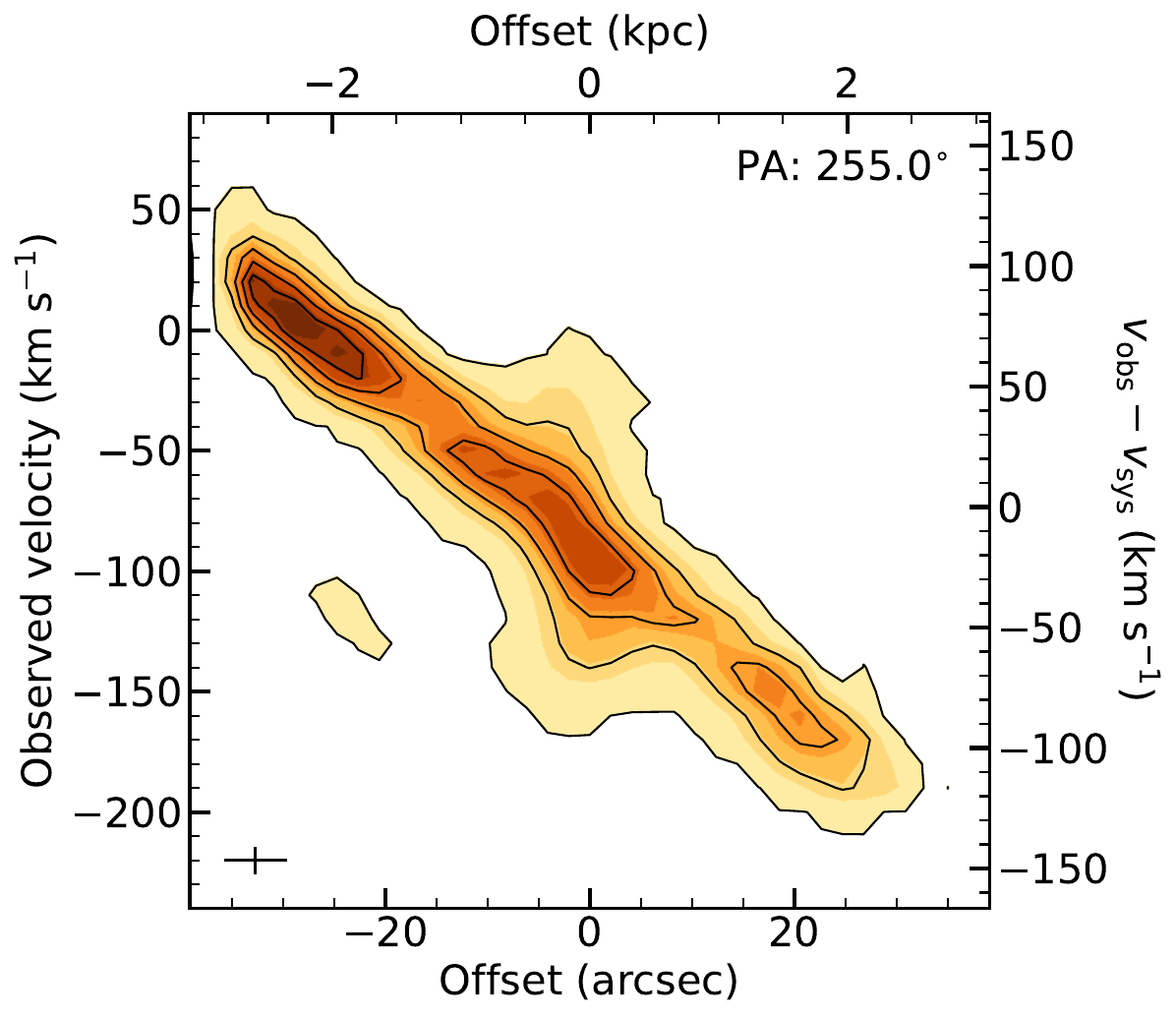}
    \end{subfigure}
    \\
    \begin{subfigure}{.34\linewidth}
    \centering
    \includegraphics[height=5.85cm]{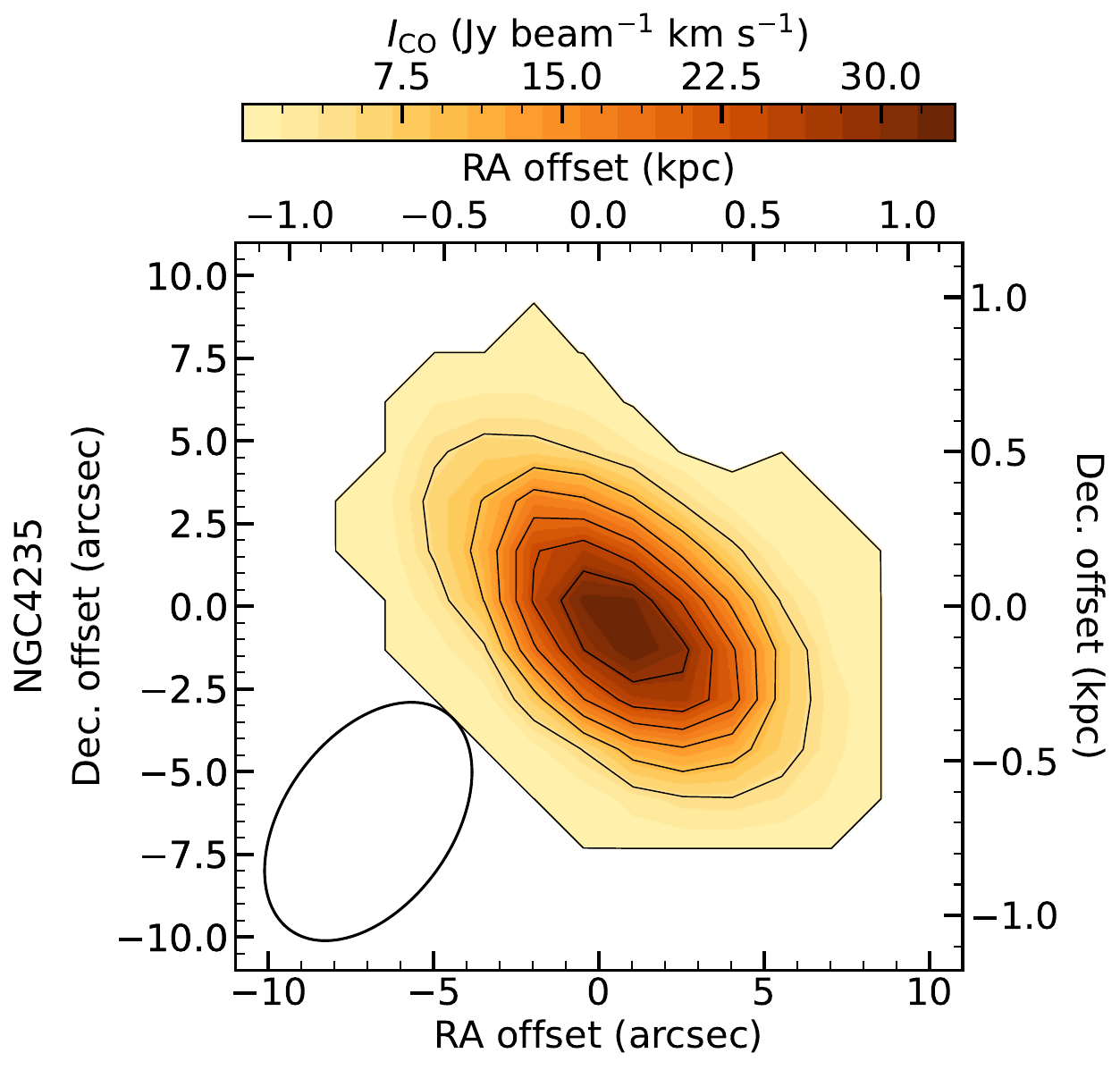}
    \end{subfigure}
    \begin{subfigure}{.32\linewidth}
    \centering
    \includegraphics[height=5.85cm]{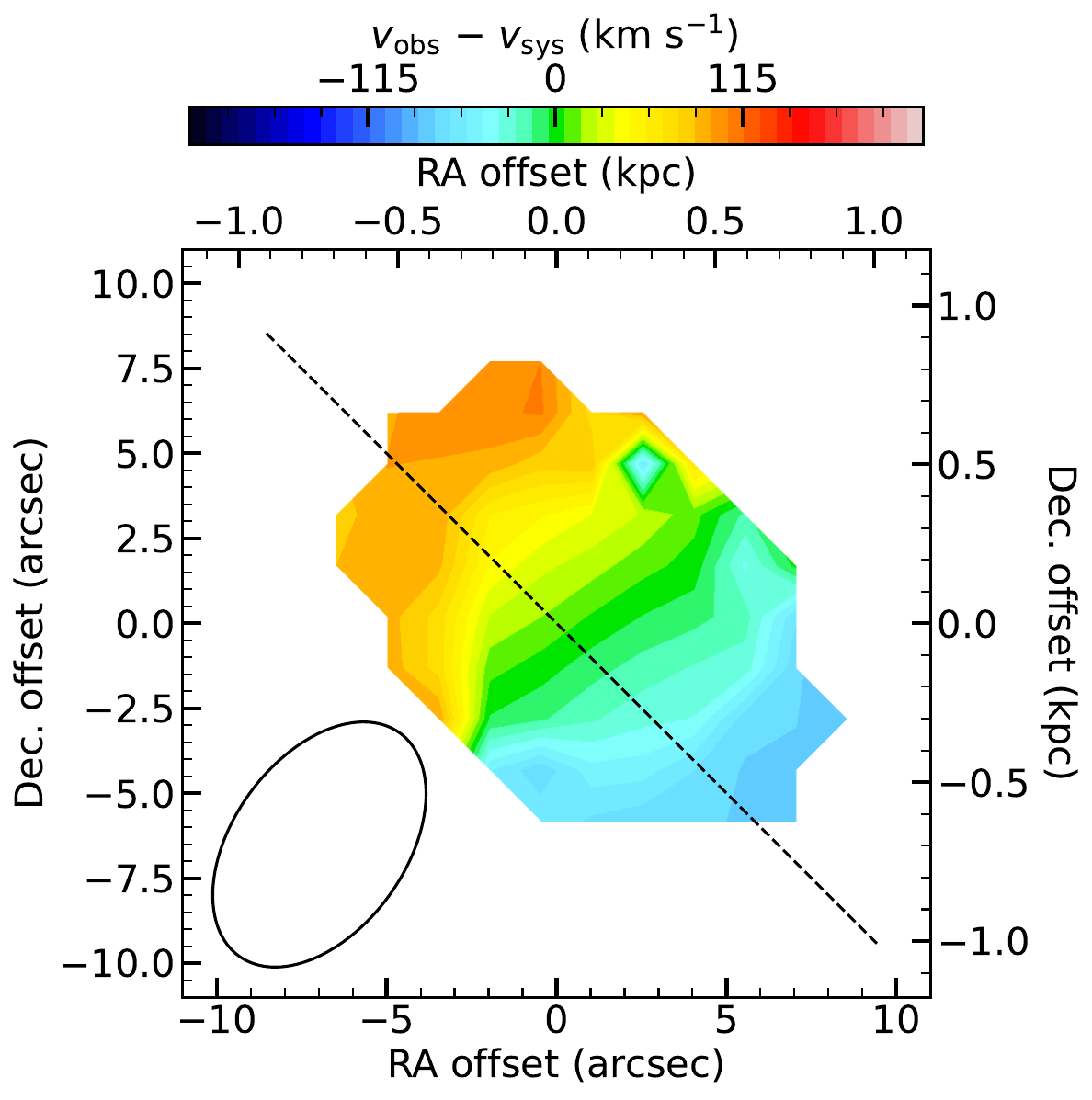}
    \end{subfigure}
    \begin{subfigure}{.32\linewidth}
    \centering
    \includegraphics[height=5.85cm]{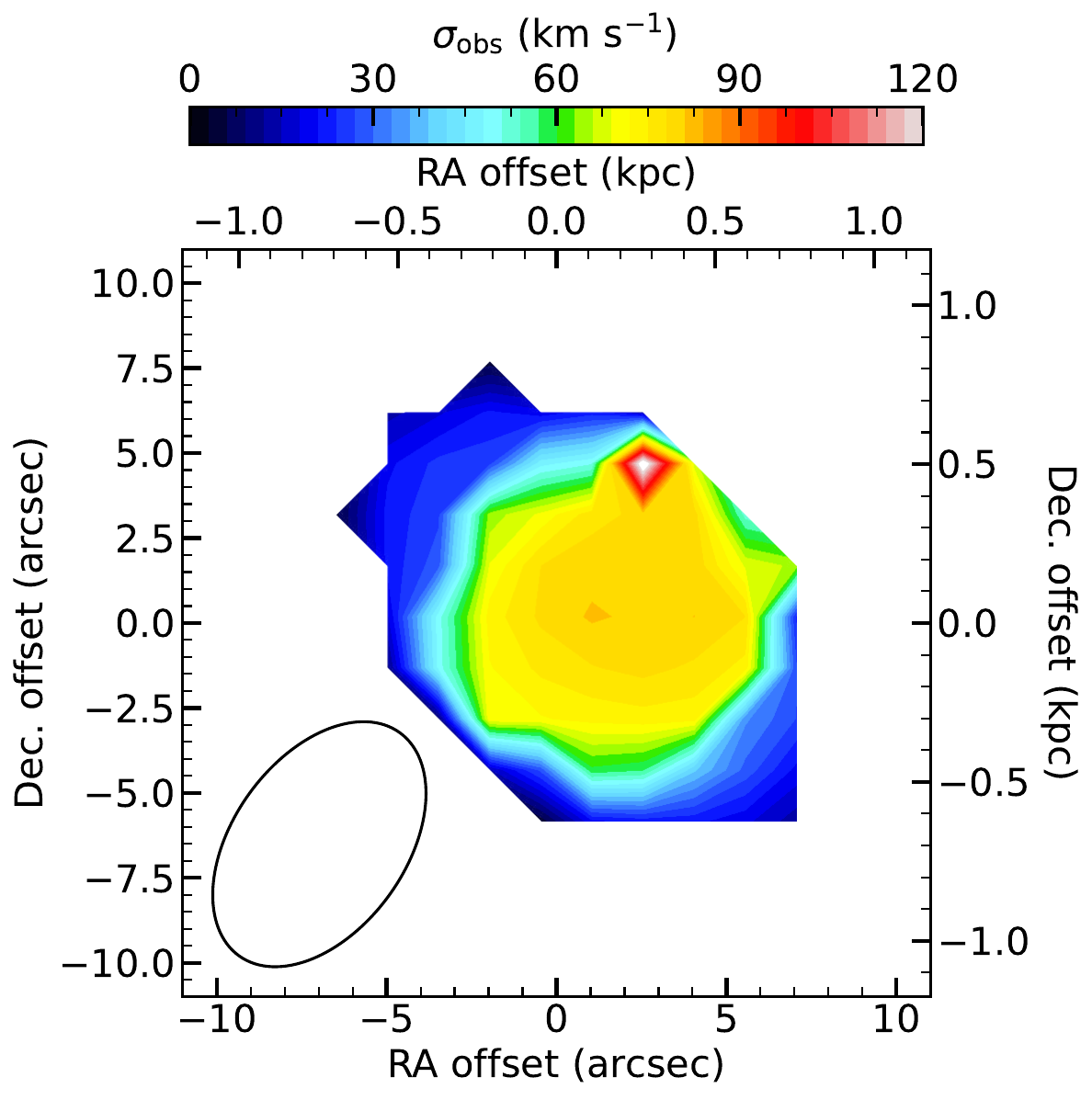}
    \end{subfigure}
    \\
    \begin{subfigure}{.34\linewidth}
    \centering
    \includegraphics[height=5.2cm]{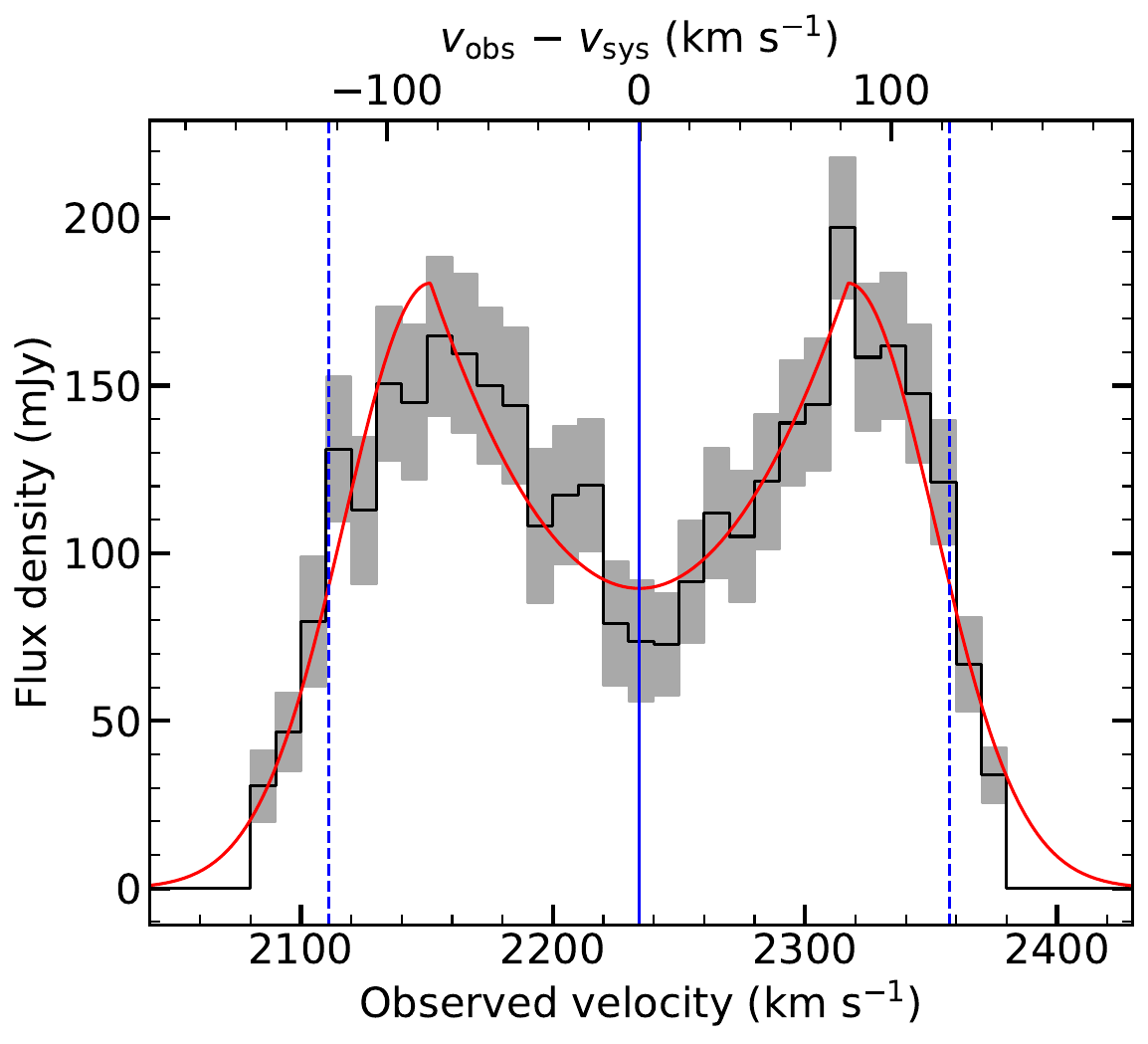}
    \end{subfigure}
    \begin{subfigure}{.32\linewidth}
    \centering
    \includegraphics[height=4.8cm]{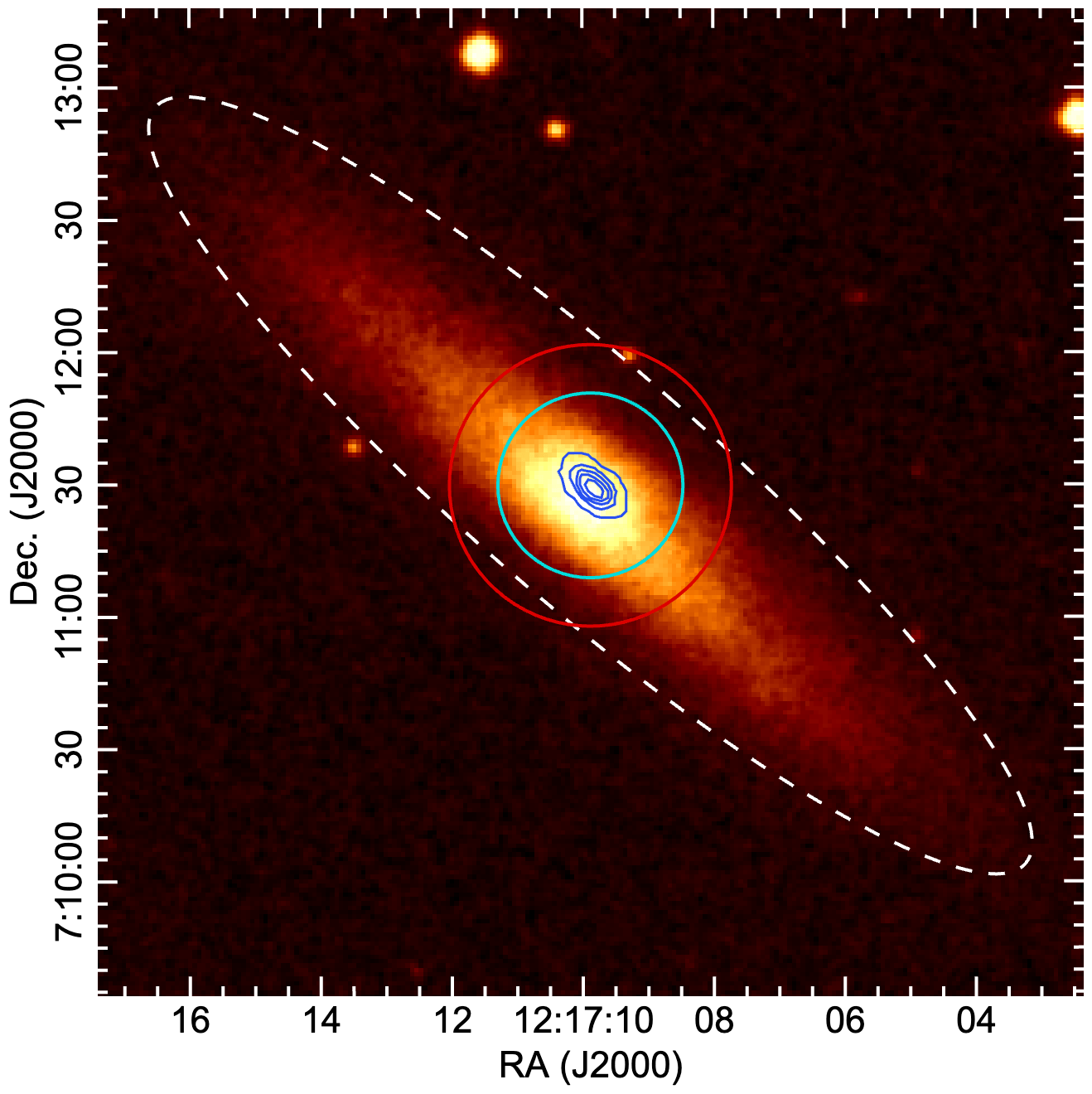}
    \end{subfigure}
    \begin{subfigure}{.33\linewidth}
    \centering
    \includegraphics[height=5.2cm]{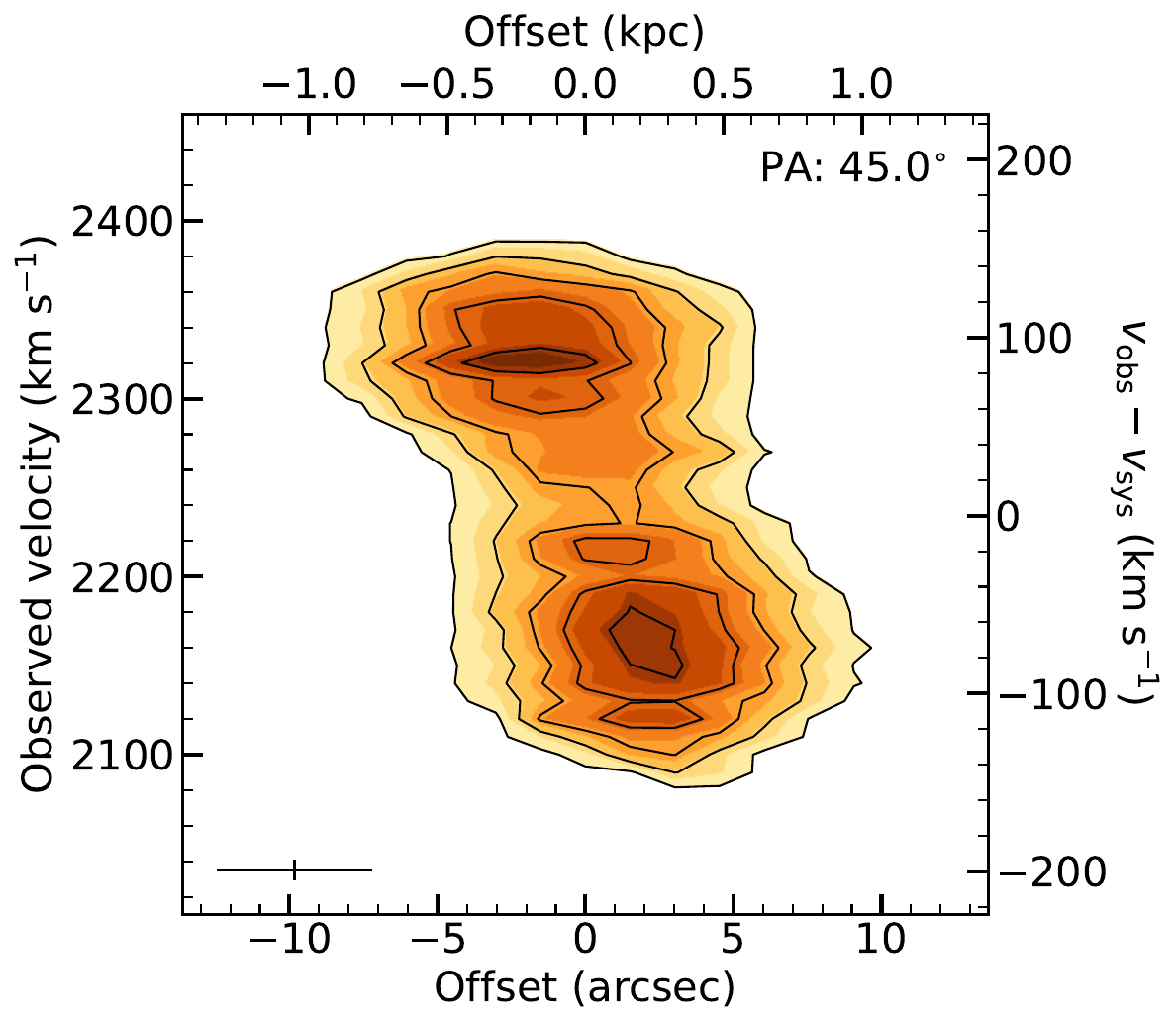}
    \end{subfigure}
    \caption[]{continued.}
\end{figure*}

\begin{figure*}
    \ContinuedFloat
    \centering
    \begin{subfigure}{.34\linewidth}
    \centering
    \includegraphics[height=6.0cm]{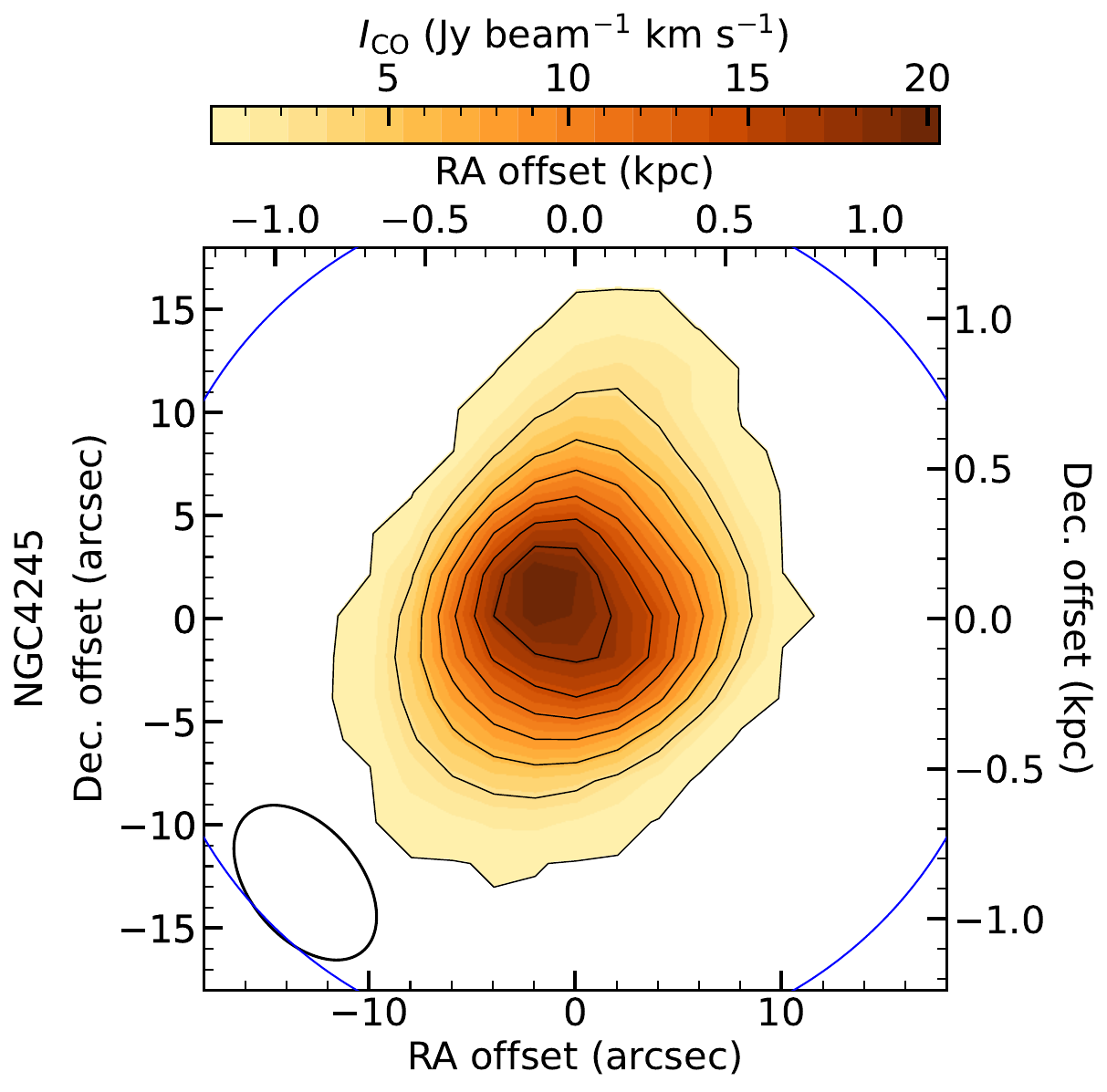}
    \end{subfigure}
    \begin{subfigure}{.32\linewidth}
    \centering
    \includegraphics[height=6.0cm]{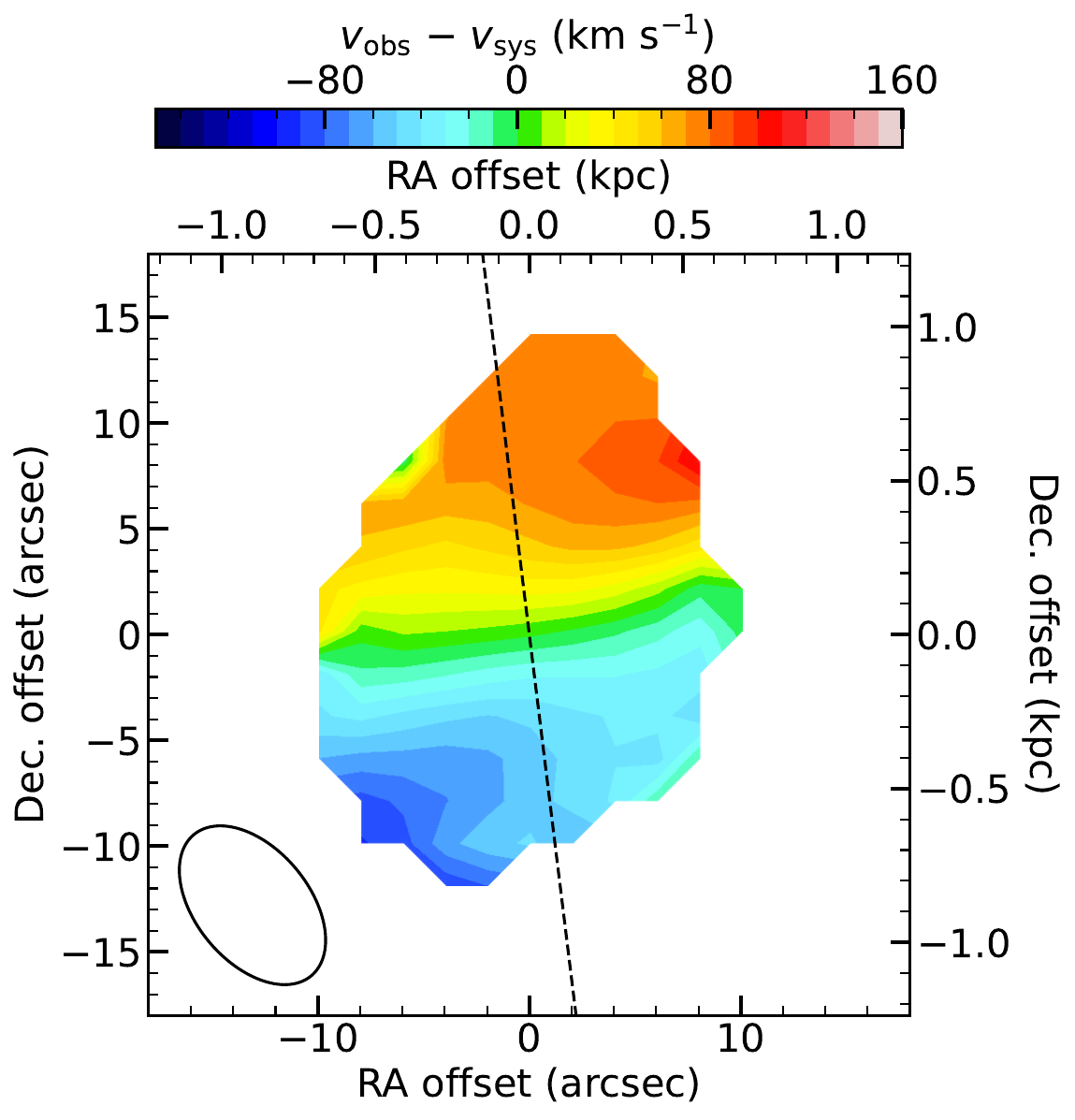}
    \end{subfigure}
    \begin{subfigure}{.32\linewidth}
    \centering
    \includegraphics[height=6.0cm]{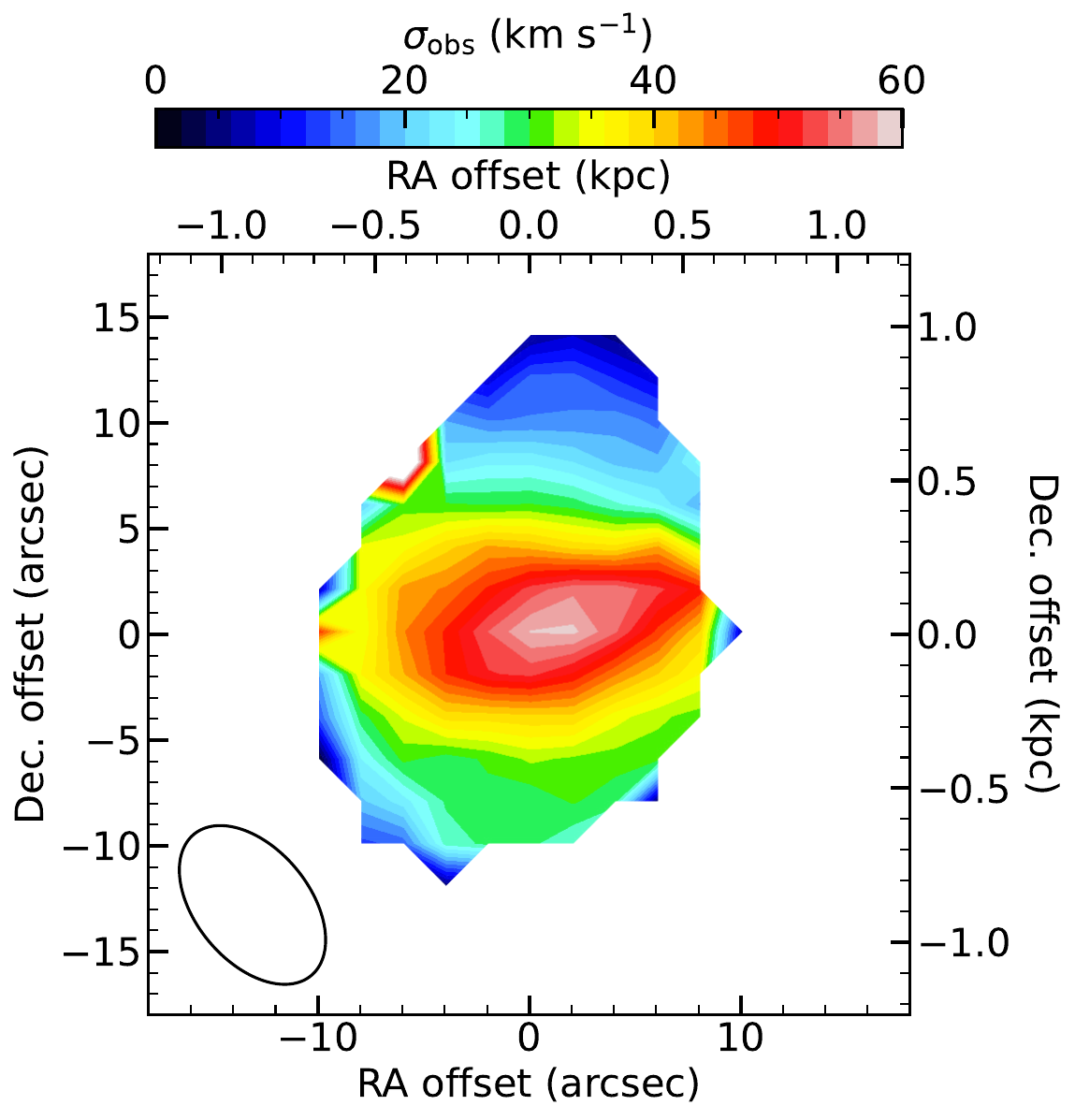}
    \end{subfigure}
    \\
    \begin{subfigure}{.34\linewidth}
    \centering
    \includegraphics[height=5.2cm]{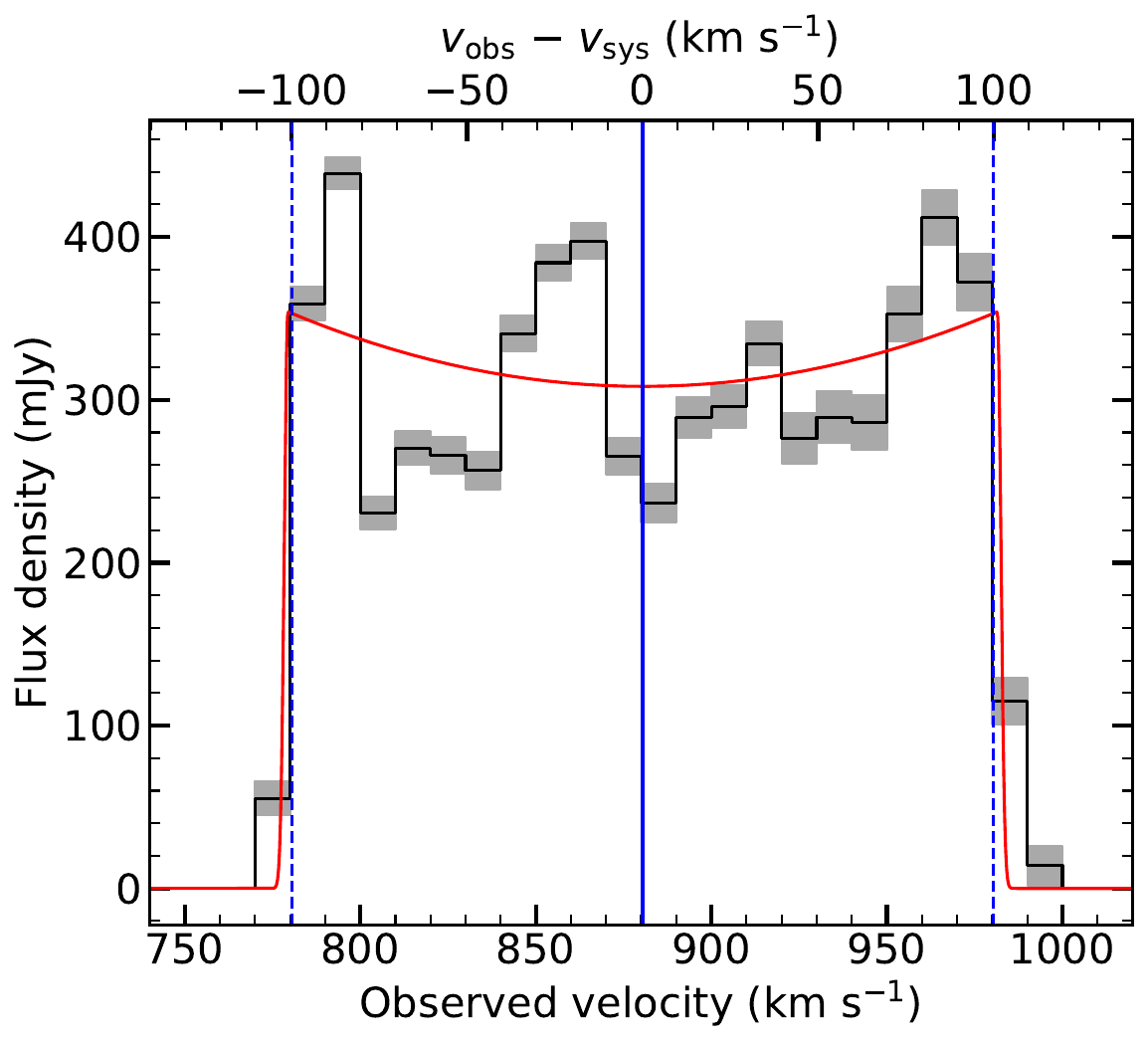}
    \end{subfigure}
    \begin{subfigure}{.32\linewidth}
    \centering
    \includegraphics[height=4.8cm]{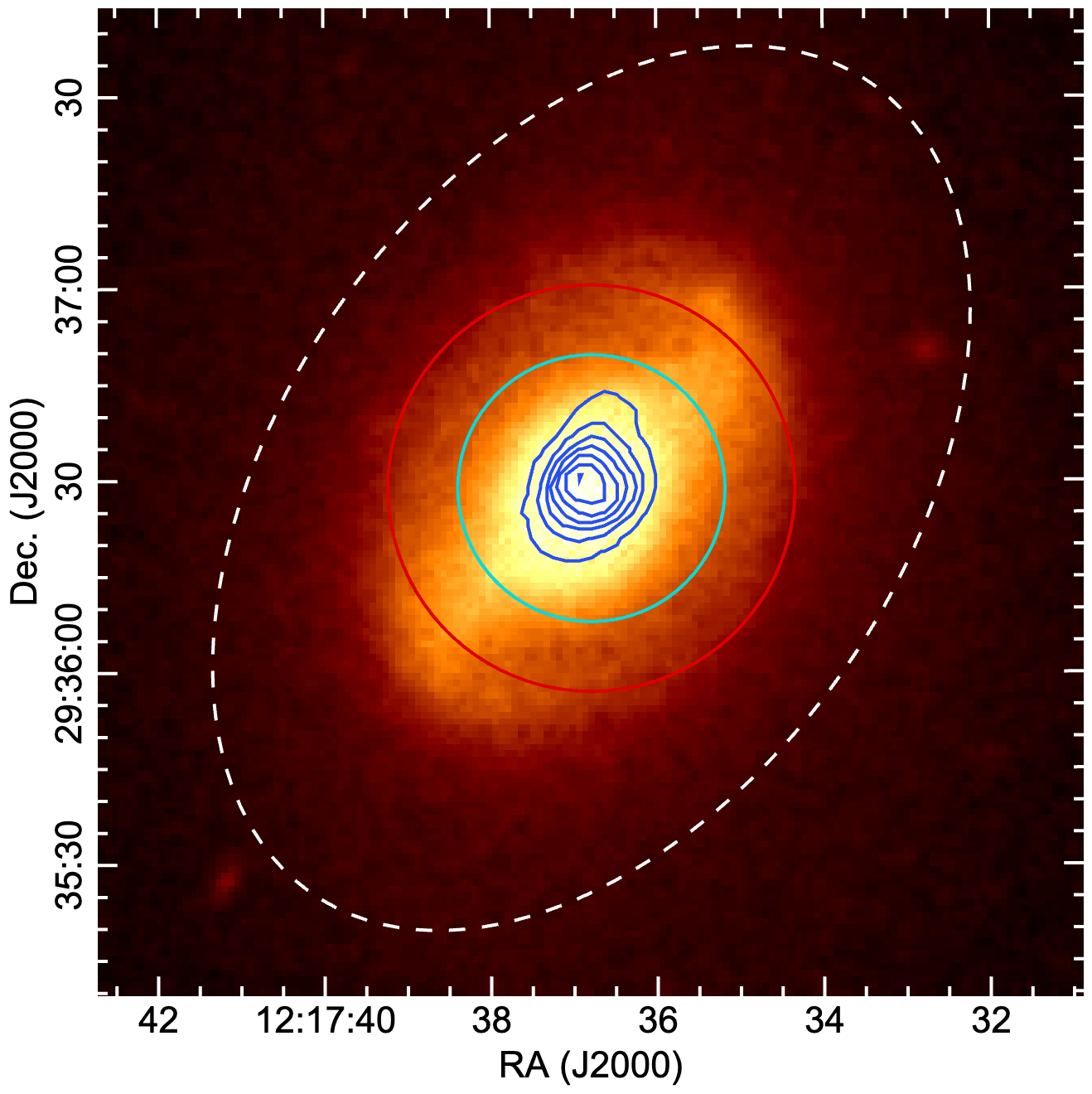}
    \end{subfigure}
    \begin{subfigure}{.33\linewidth}
    \centering
    \includegraphics[height=5.2cm]{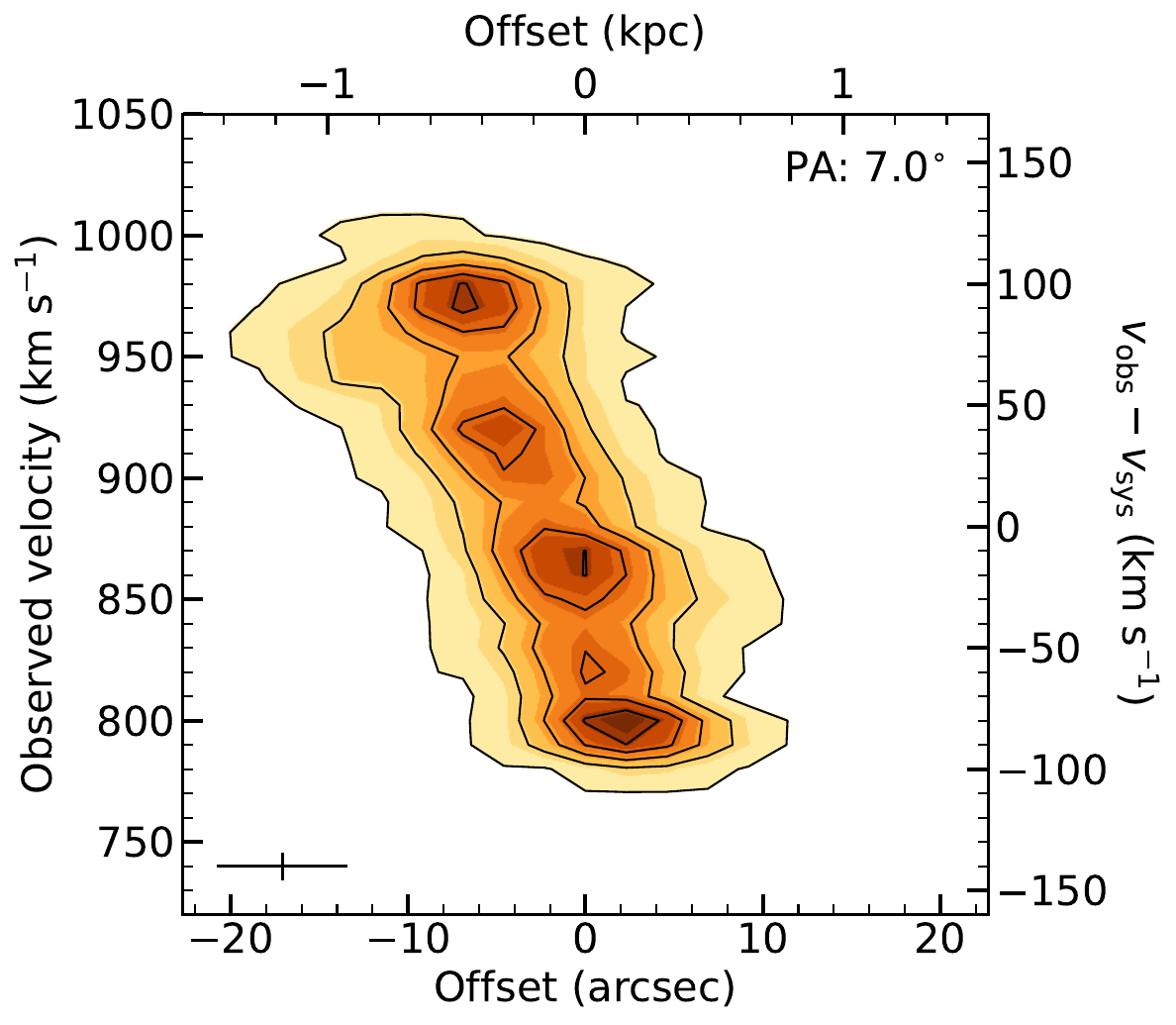}
    \end{subfigure}
    \\
    \begin{subfigure}{.34\linewidth}
    \centering
    \includegraphics[height=6.0cm]{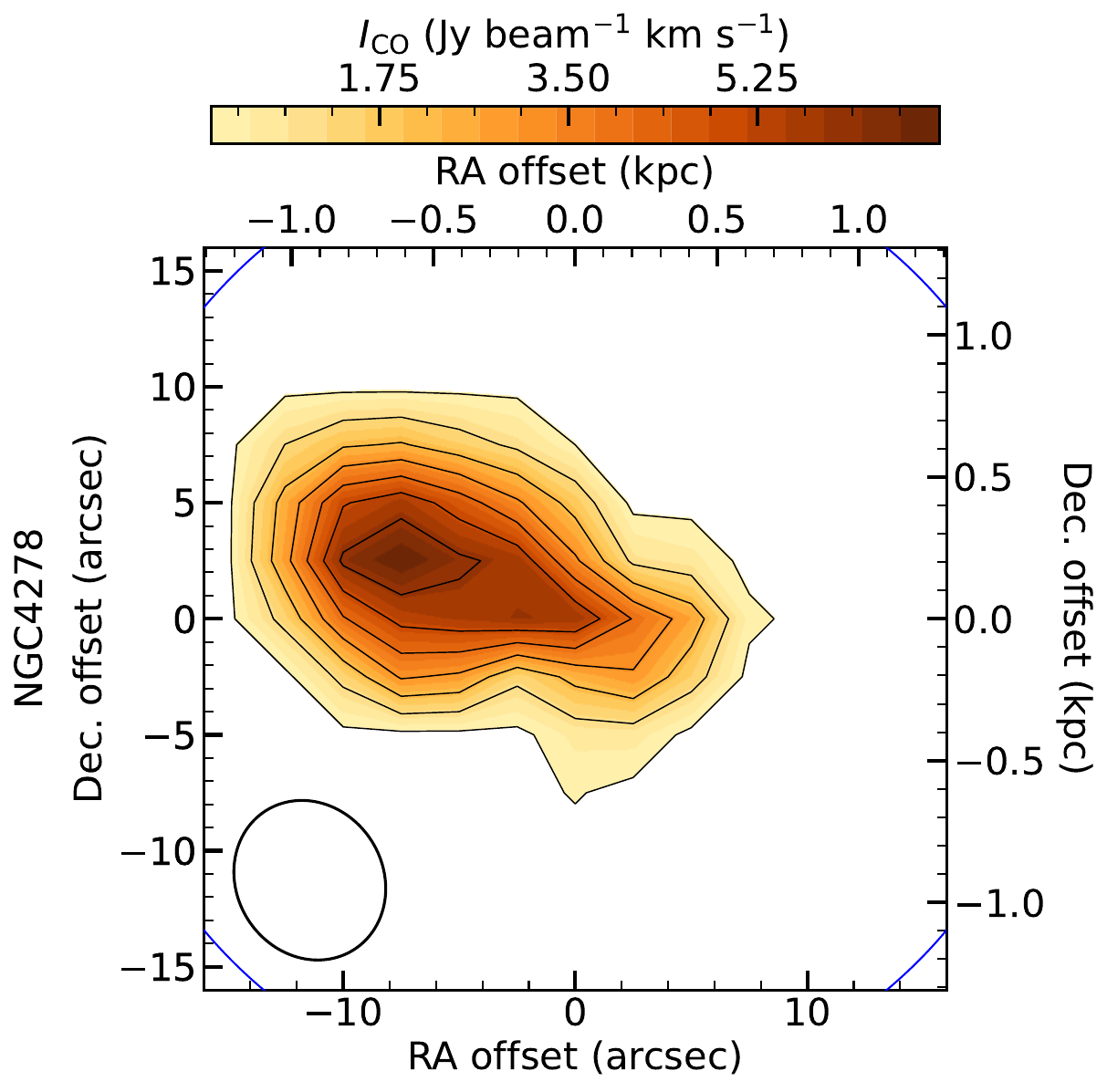}
    \end{subfigure}
    \begin{subfigure}{.32\linewidth}
    \centering
    \includegraphics[height=6.0cm]{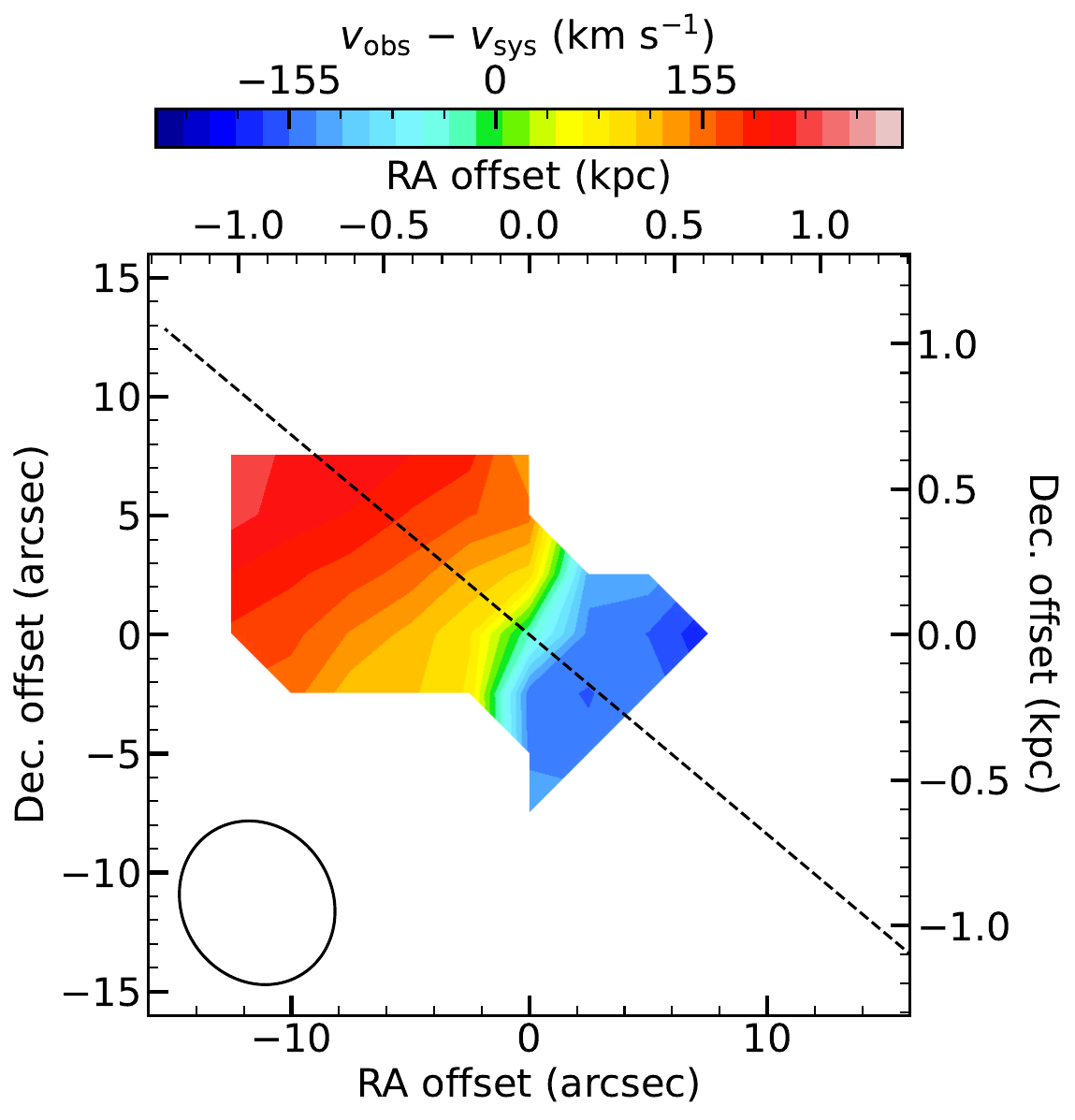}
    \end{subfigure}
    \begin{subfigure}{.32\linewidth}
    \centering
    \includegraphics[height=6.0cm]{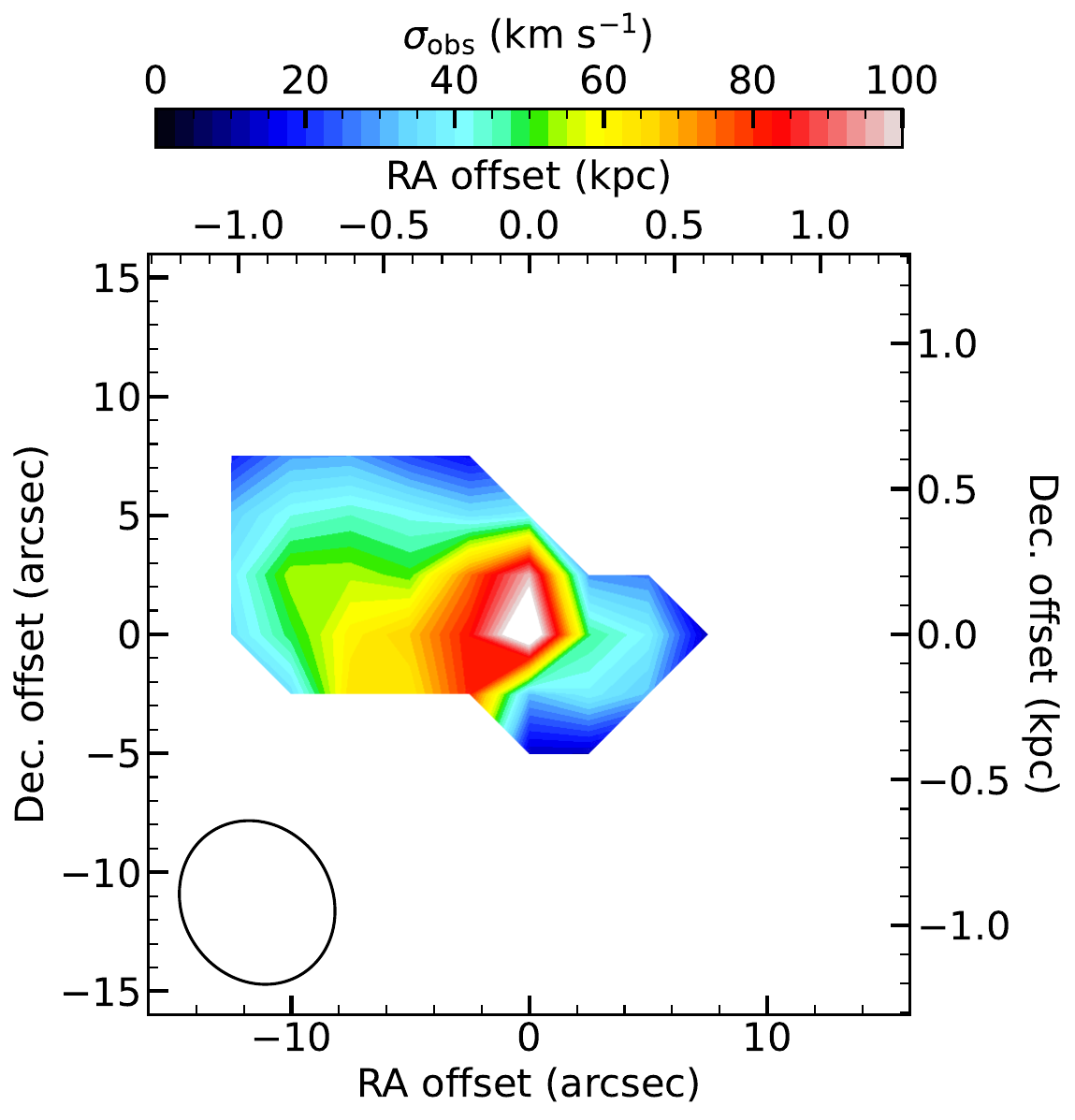}
    \end{subfigure}
    \\
    \begin{subfigure}{.34\linewidth}
    \centering
    \includegraphics[height=5.2cm]{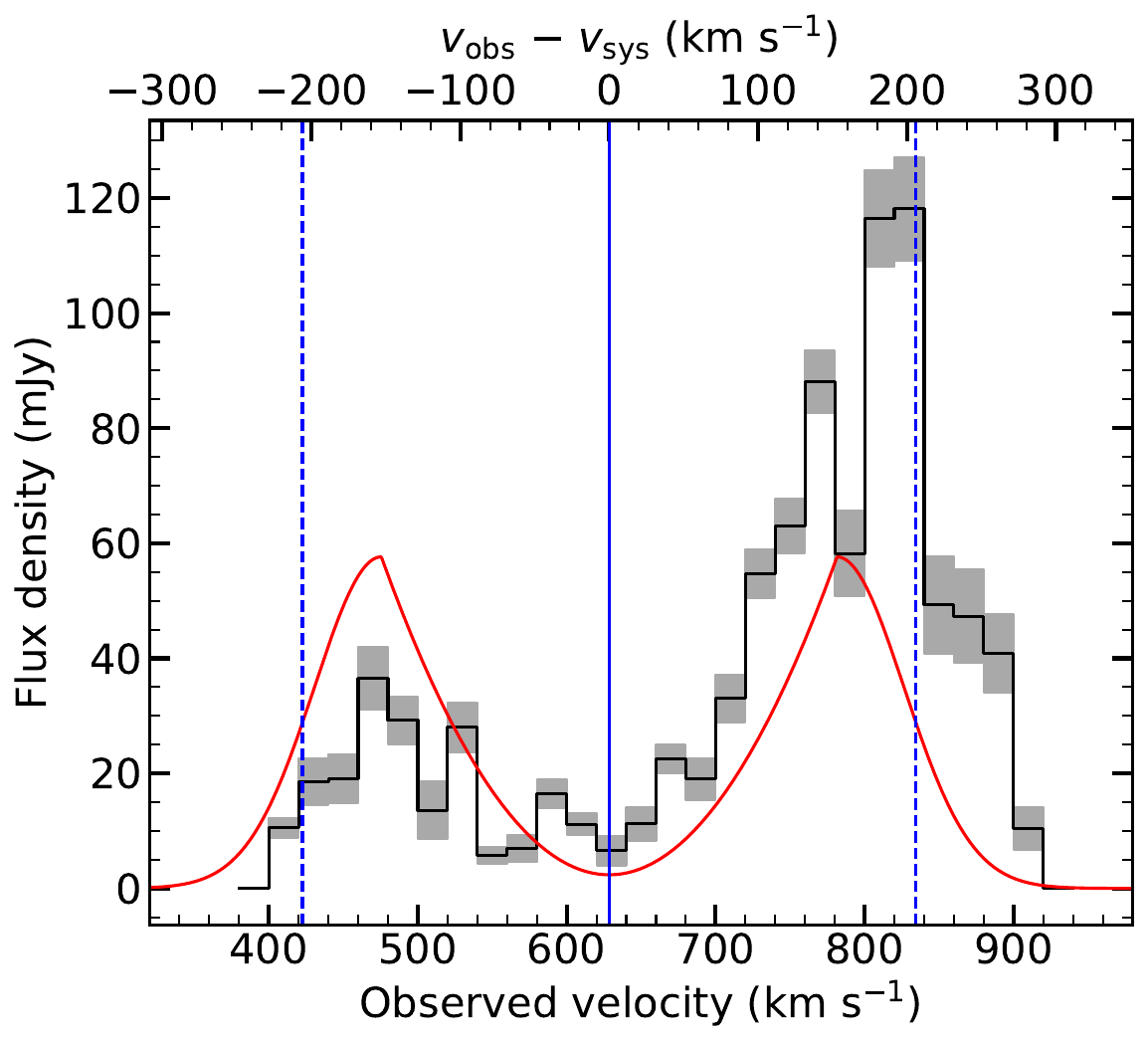}
    \end{subfigure}
    \begin{subfigure}{.32\linewidth}
    \centering
    \includegraphics[height=4.8cm]{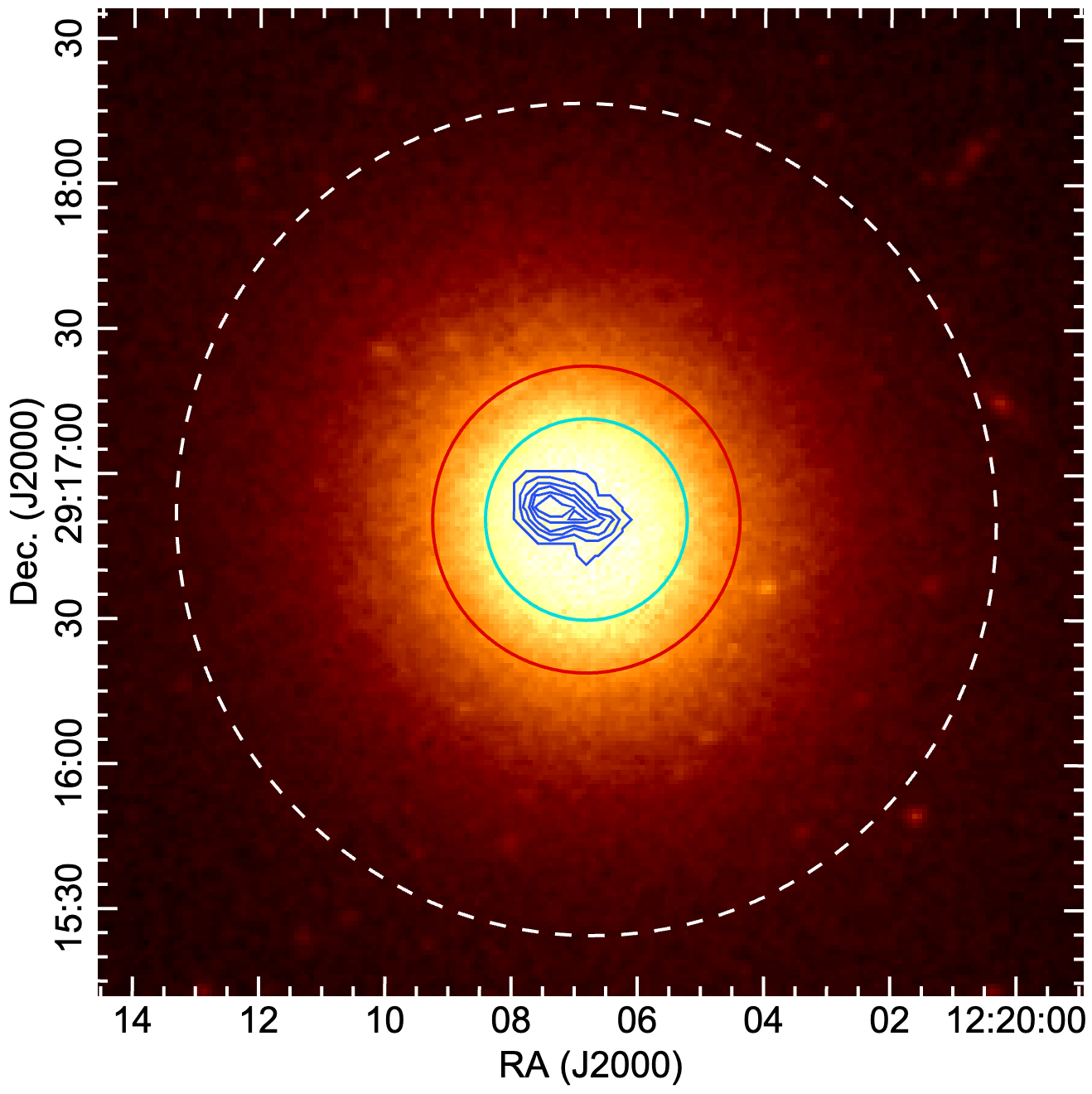}
    \end{subfigure}
    \begin{subfigure}{.33\linewidth}
    \centering
    \includegraphics[height=5.2cm]{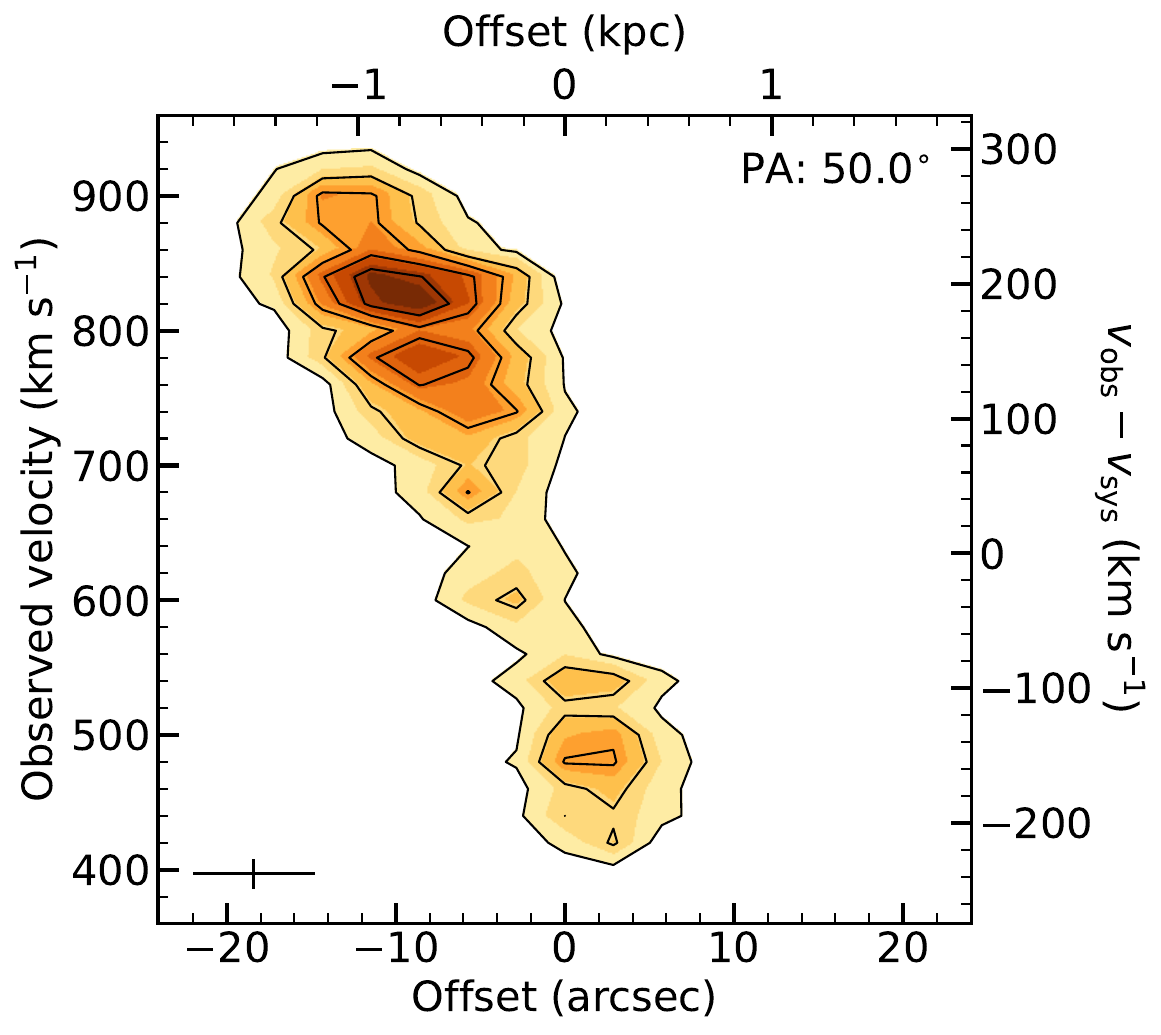}
    \end{subfigure}
    \caption[]{continued.}
\end{figure*}

\begin{figure*}
    \ContinuedFloat
    \centering
    \begin{subfigure}{.34\linewidth}
    \centering
    \includegraphics[height=6.0cm]{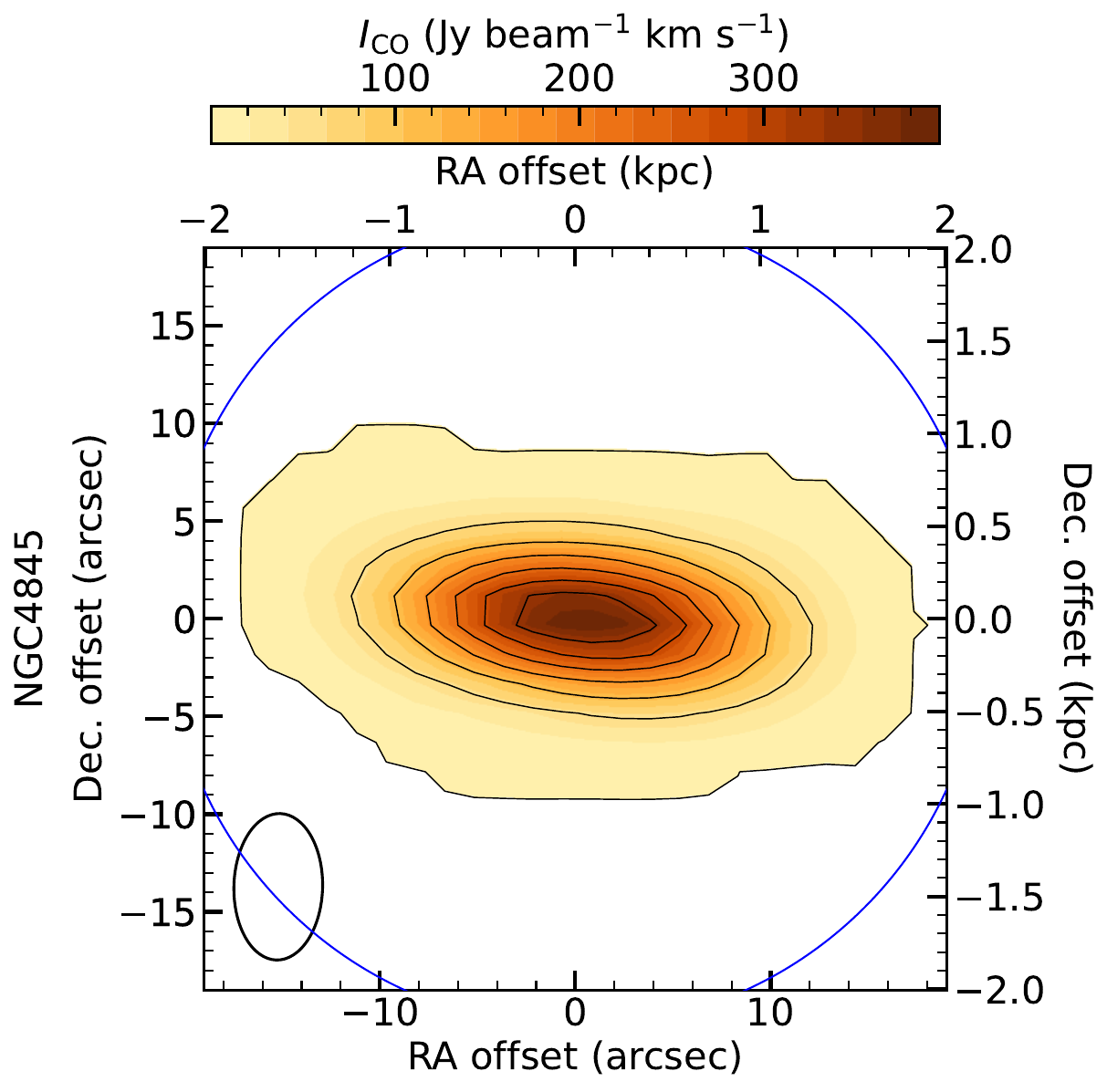}
    \end{subfigure}
    \begin{subfigure}{.32\linewidth}
    \centering
    \includegraphics[height=6.0cm]{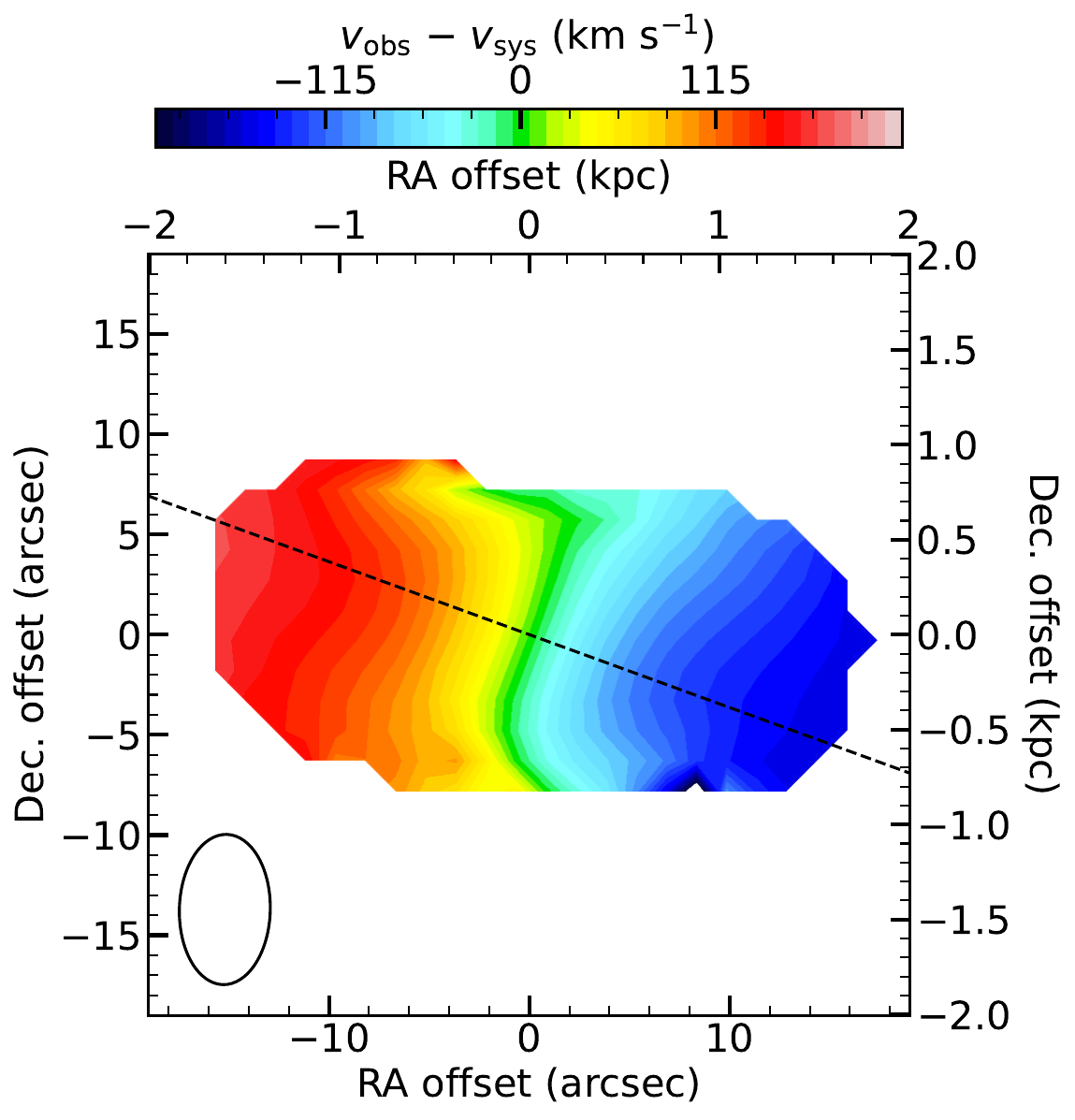}
    \end{subfigure}
    \begin{subfigure}{.32\linewidth}
    \centering
    \includegraphics[height=6.0cm]{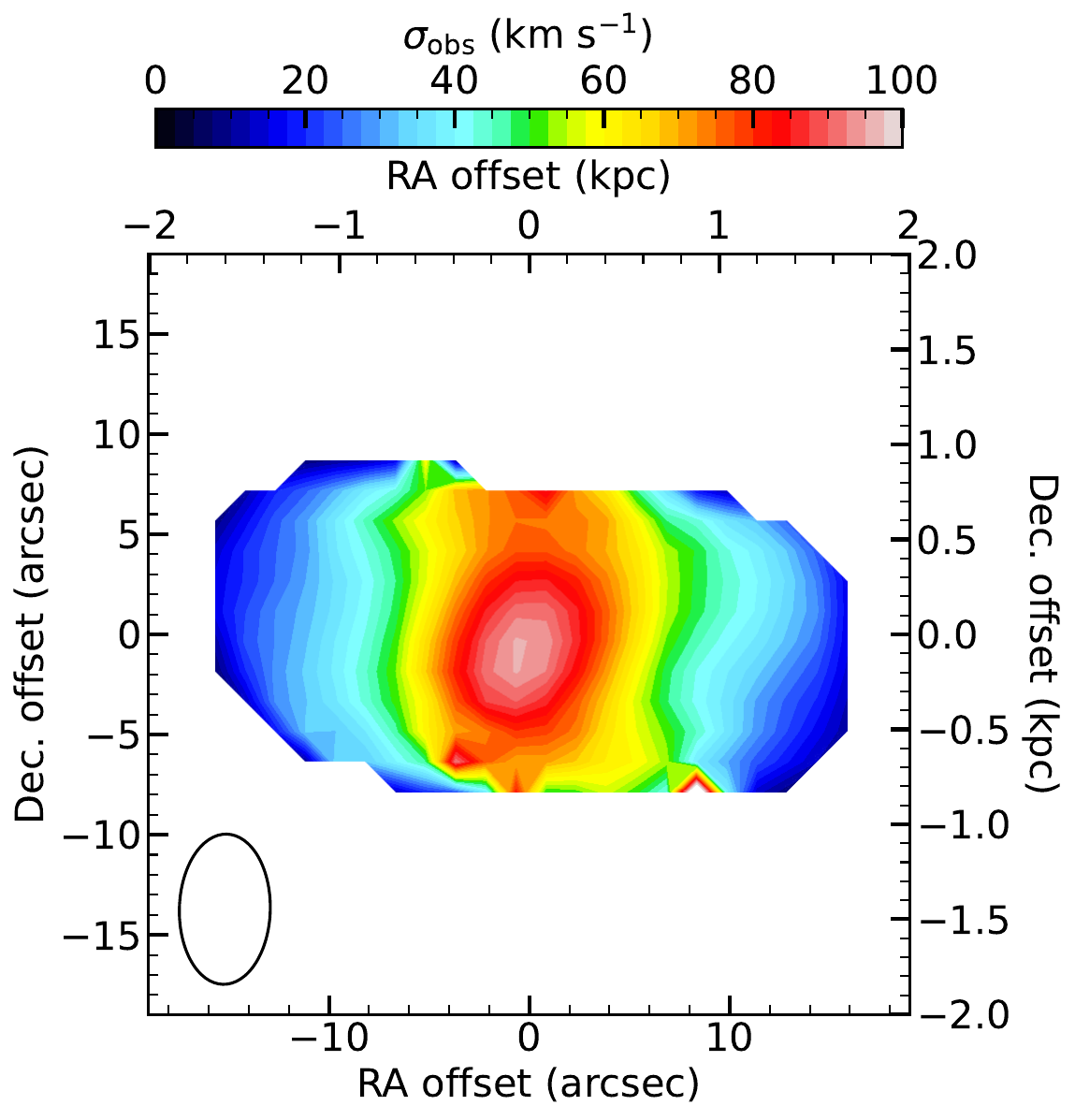}
    \end{subfigure}
    \\
    \begin{subfigure}{.34\linewidth}
    \centering
    \includegraphics[height=5.2cm]{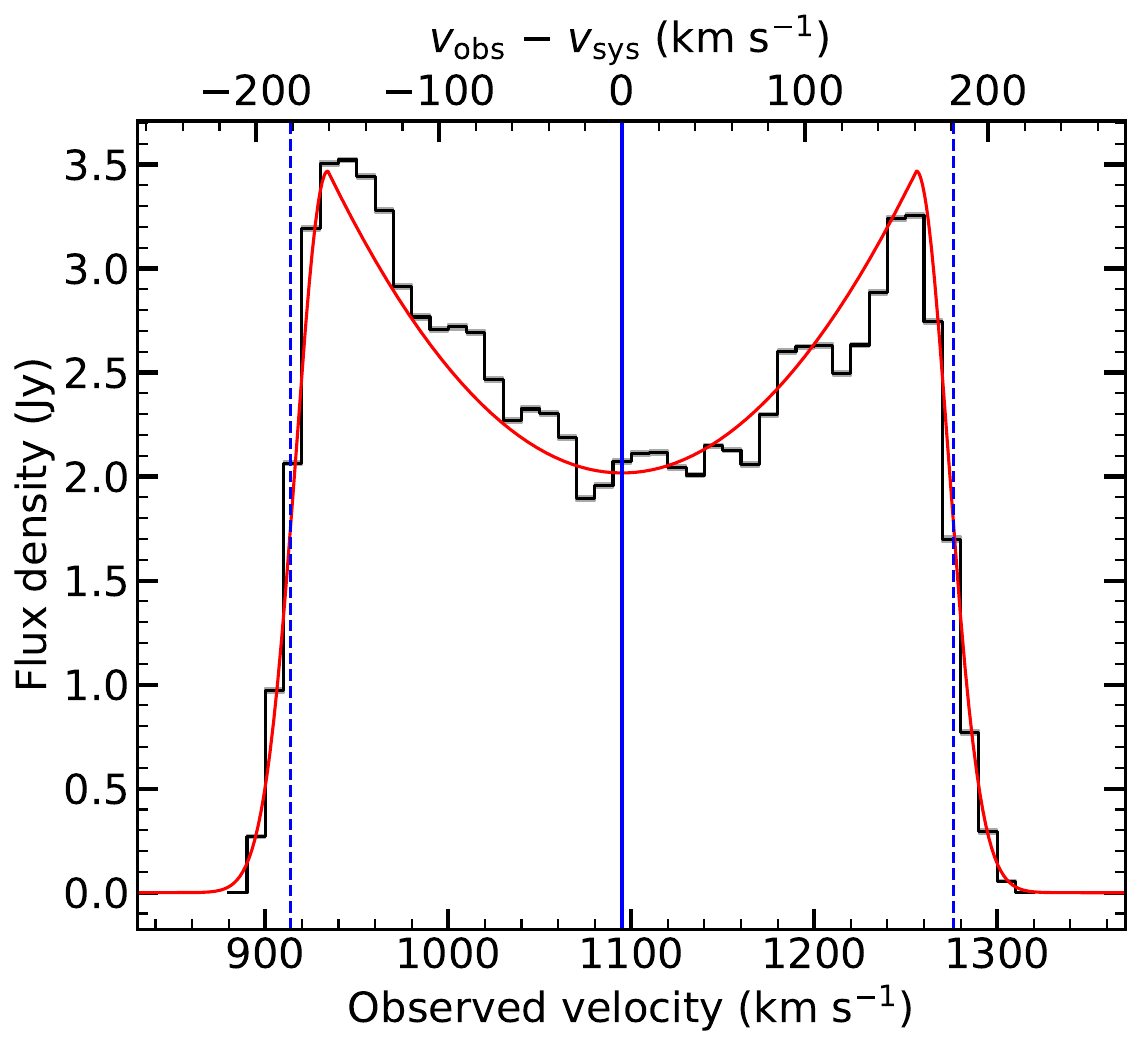}
    \end{subfigure}
    \begin{subfigure}{.32\linewidth}
    \centering
    \includegraphics[height=4.8cm]{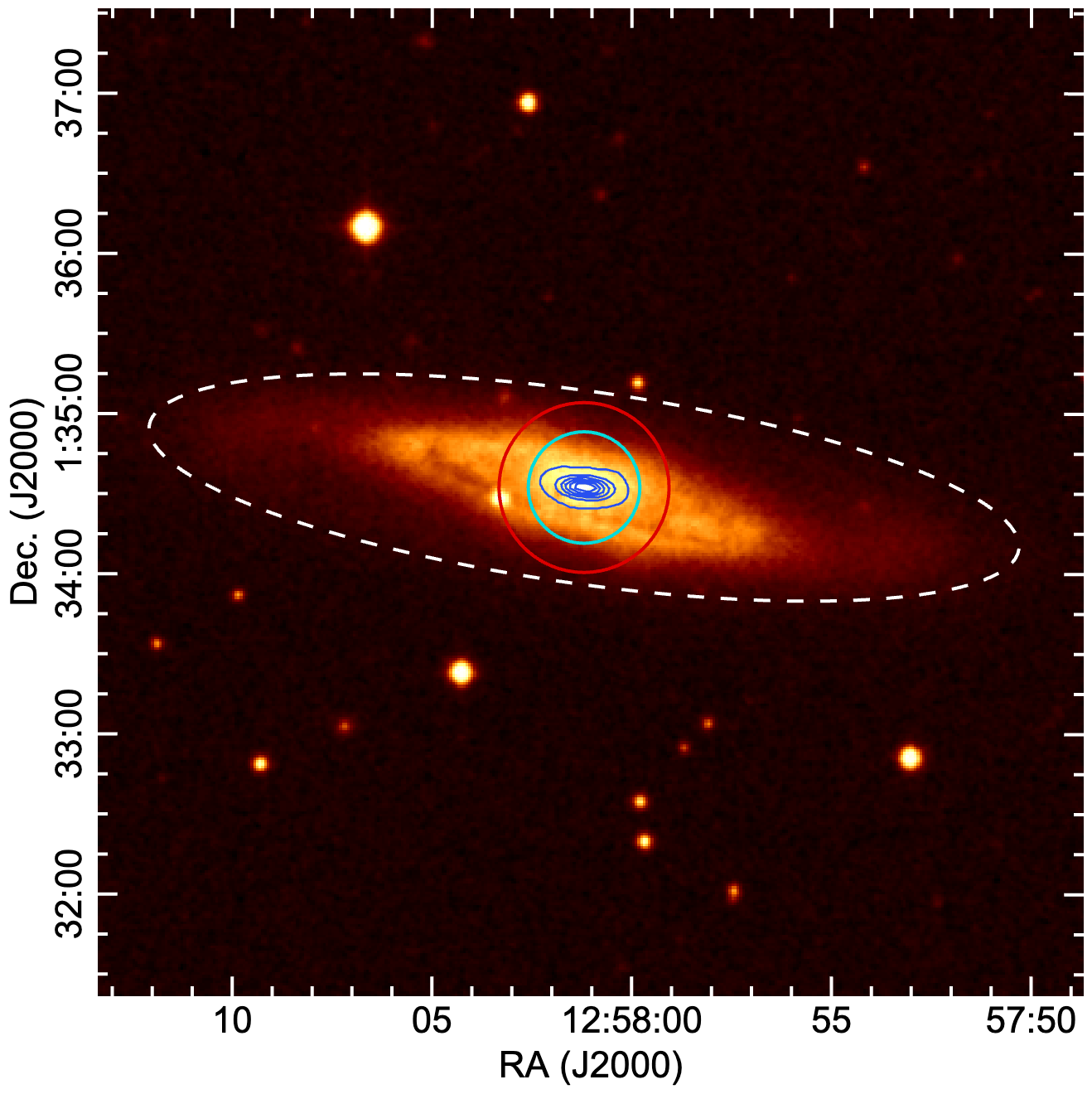}
    \end{subfigure}
    \begin{subfigure}{.33\linewidth}
    \centering
    \includegraphics[height=5.2cm]{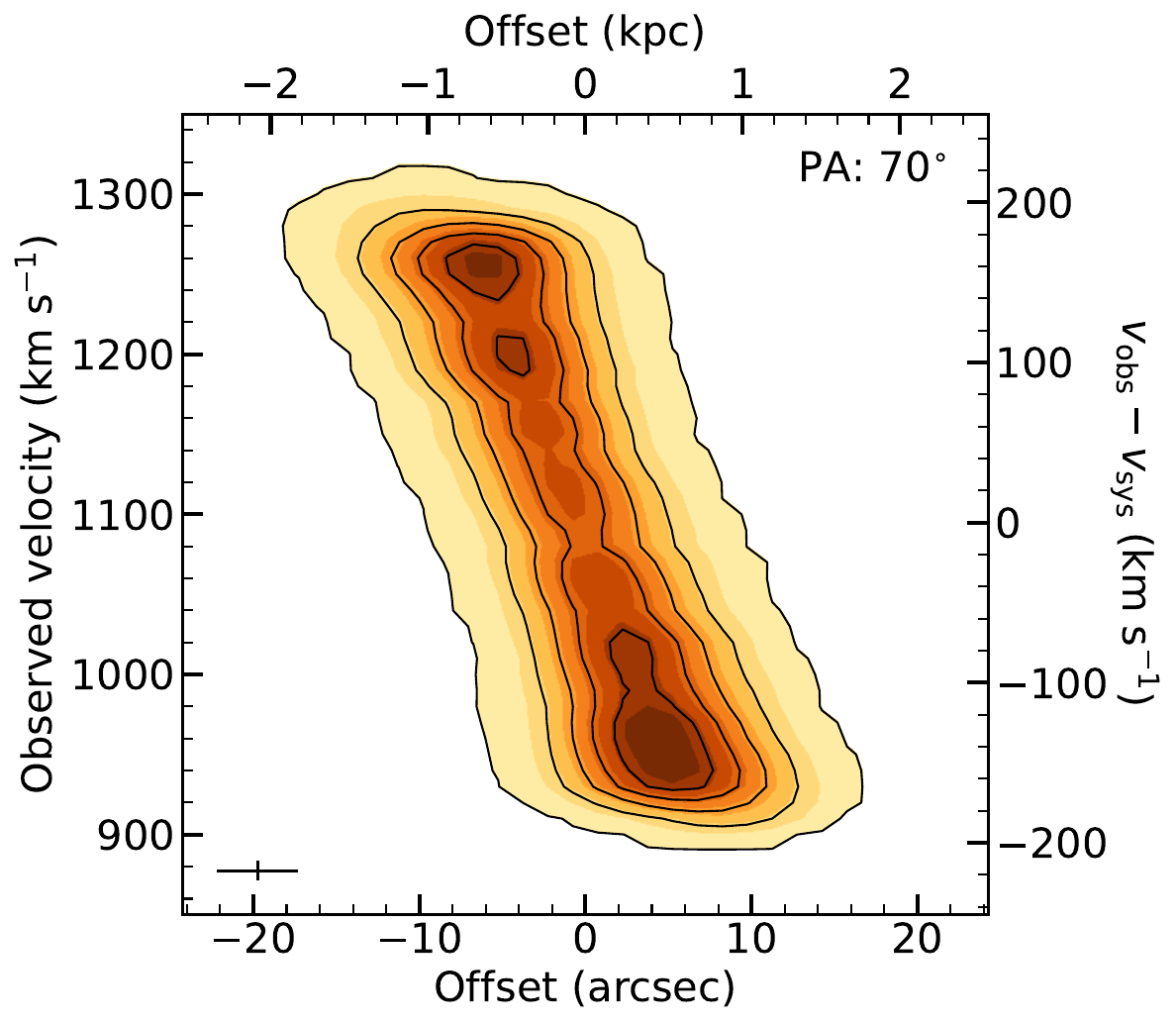}
    \end{subfigure}
    \\
    \begin{subfigure}{.34\linewidth}
    \centering
    \includegraphics[height=5.9cm]{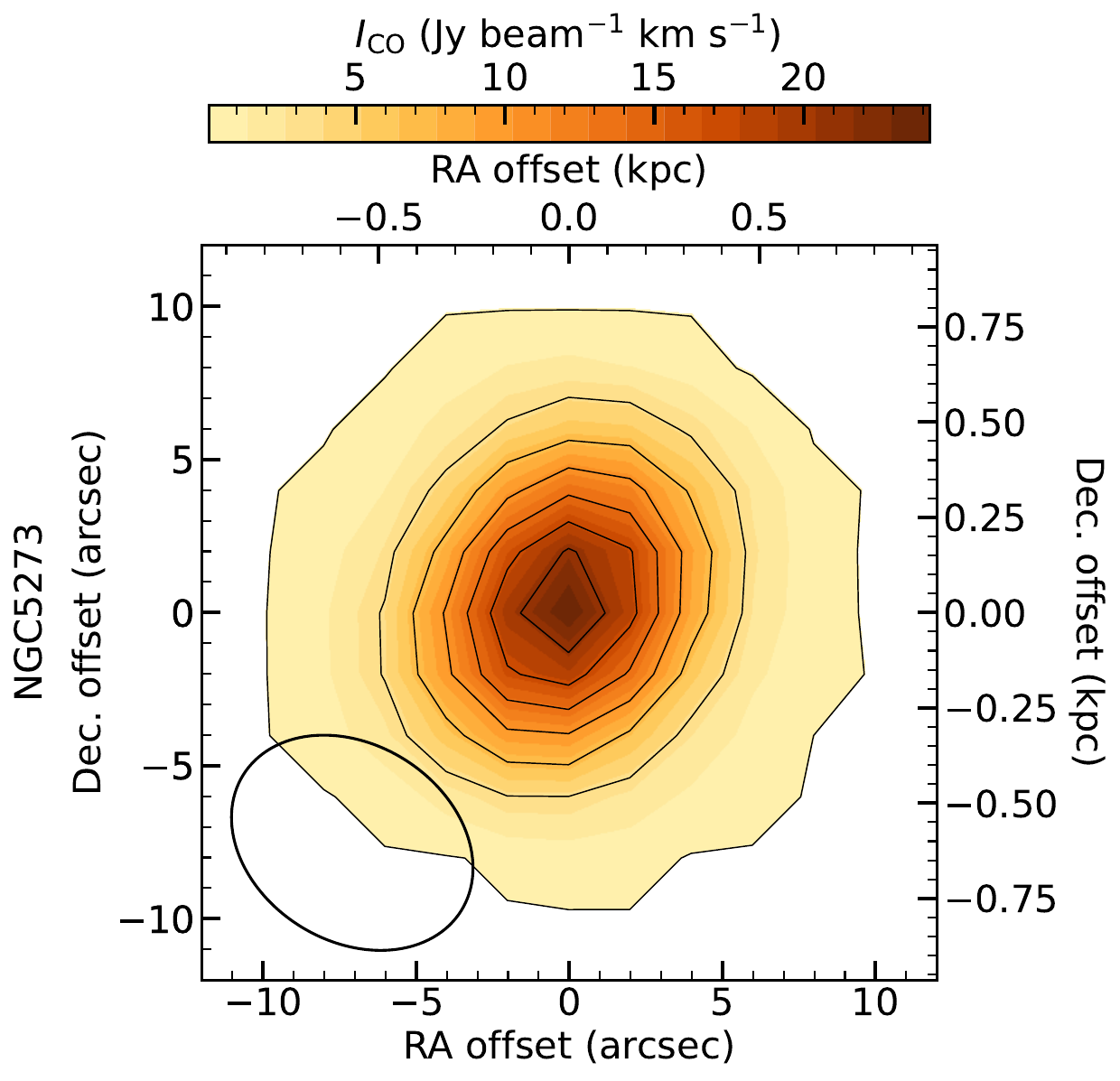}
    \end{subfigure}
    \begin{subfigure}{.32\linewidth}
    \centering
    \includegraphics[height=5.9cm]{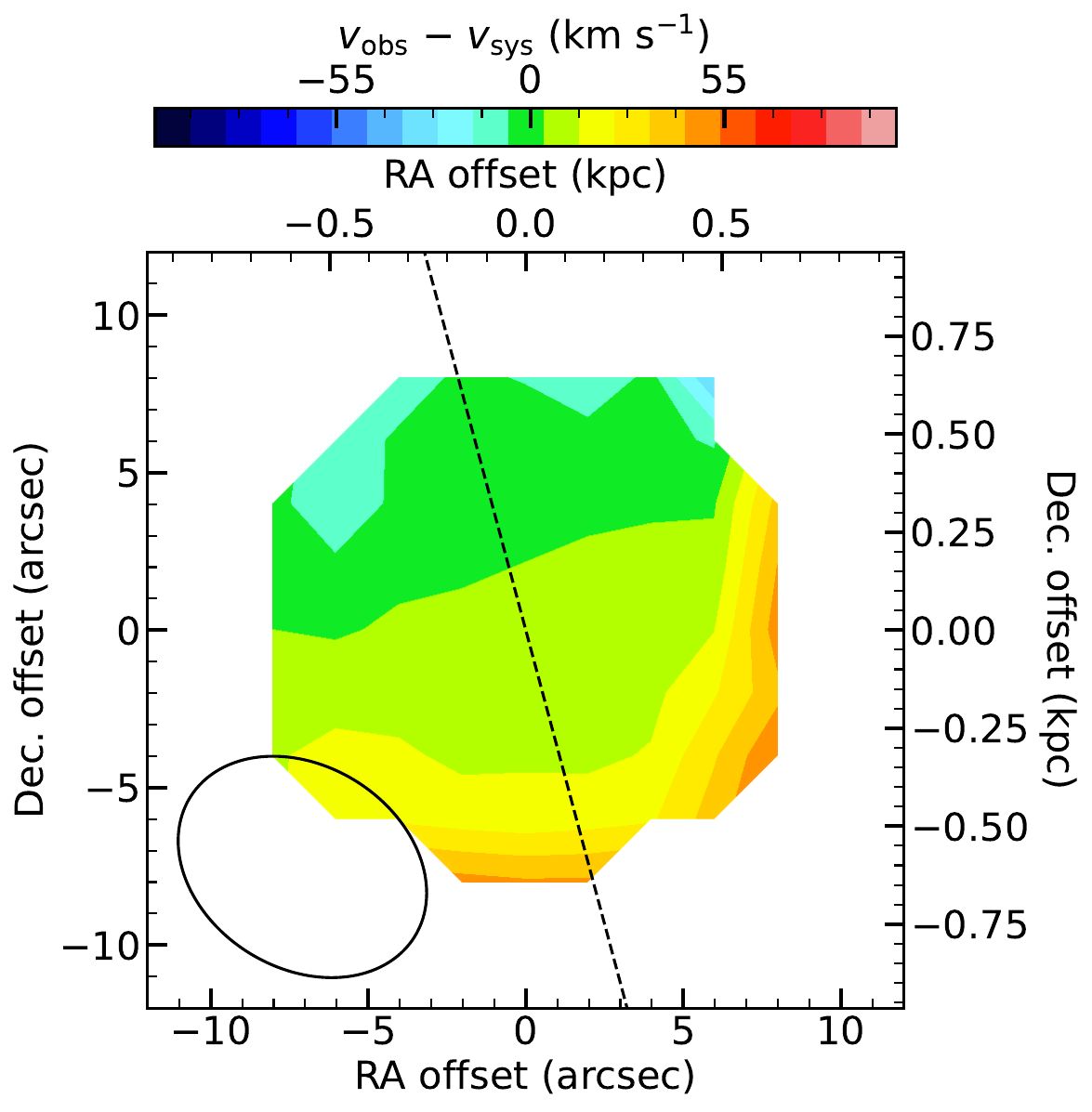}
    \end{subfigure}
    \begin{subfigure}{.32\linewidth}
    \centering
    \includegraphics[height=5.9cm]{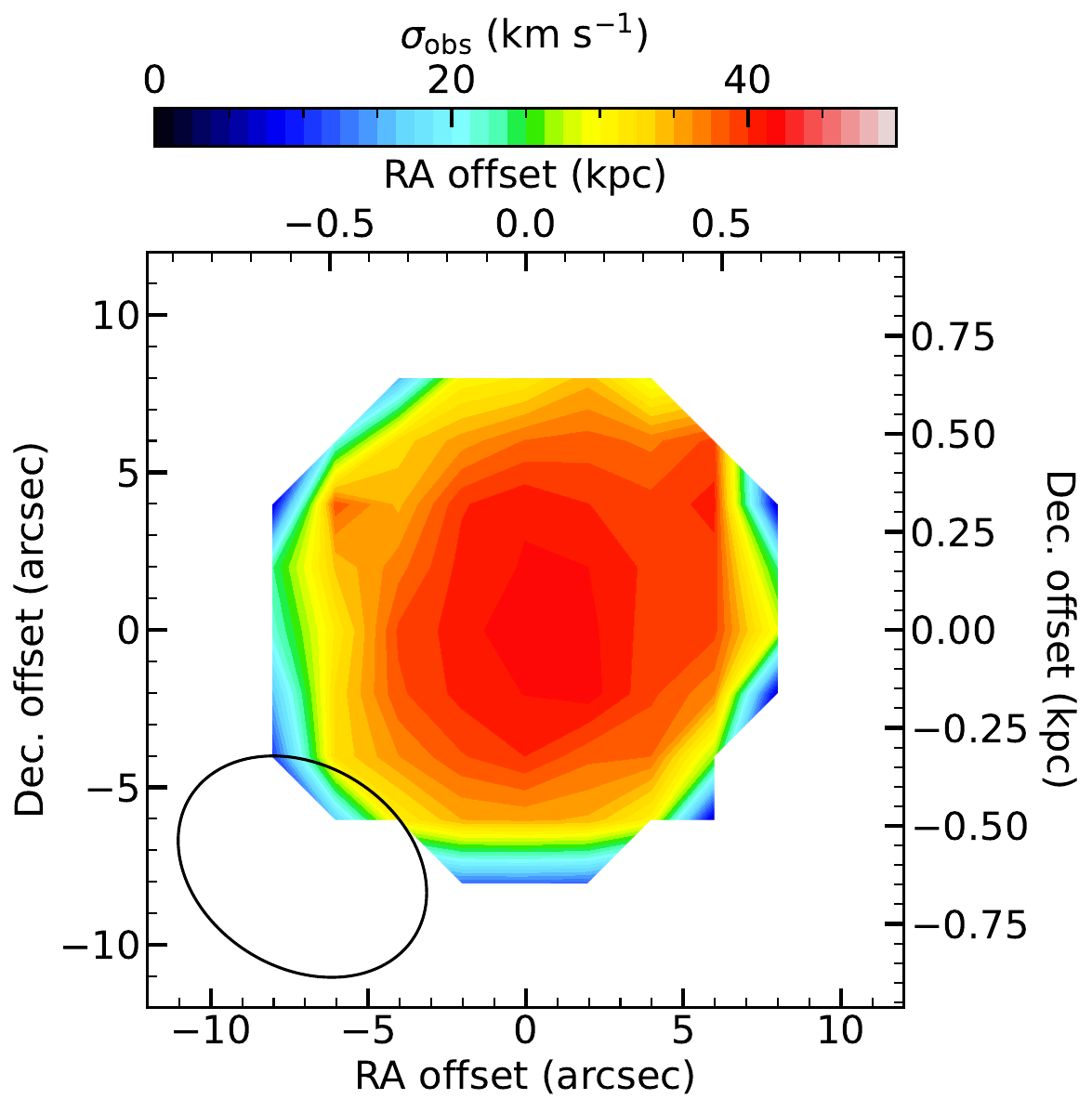}
    \end{subfigure}
    \\
    \begin{subfigure}{.34\linewidth}
    \centering
    \includegraphics[height=5.2cm]{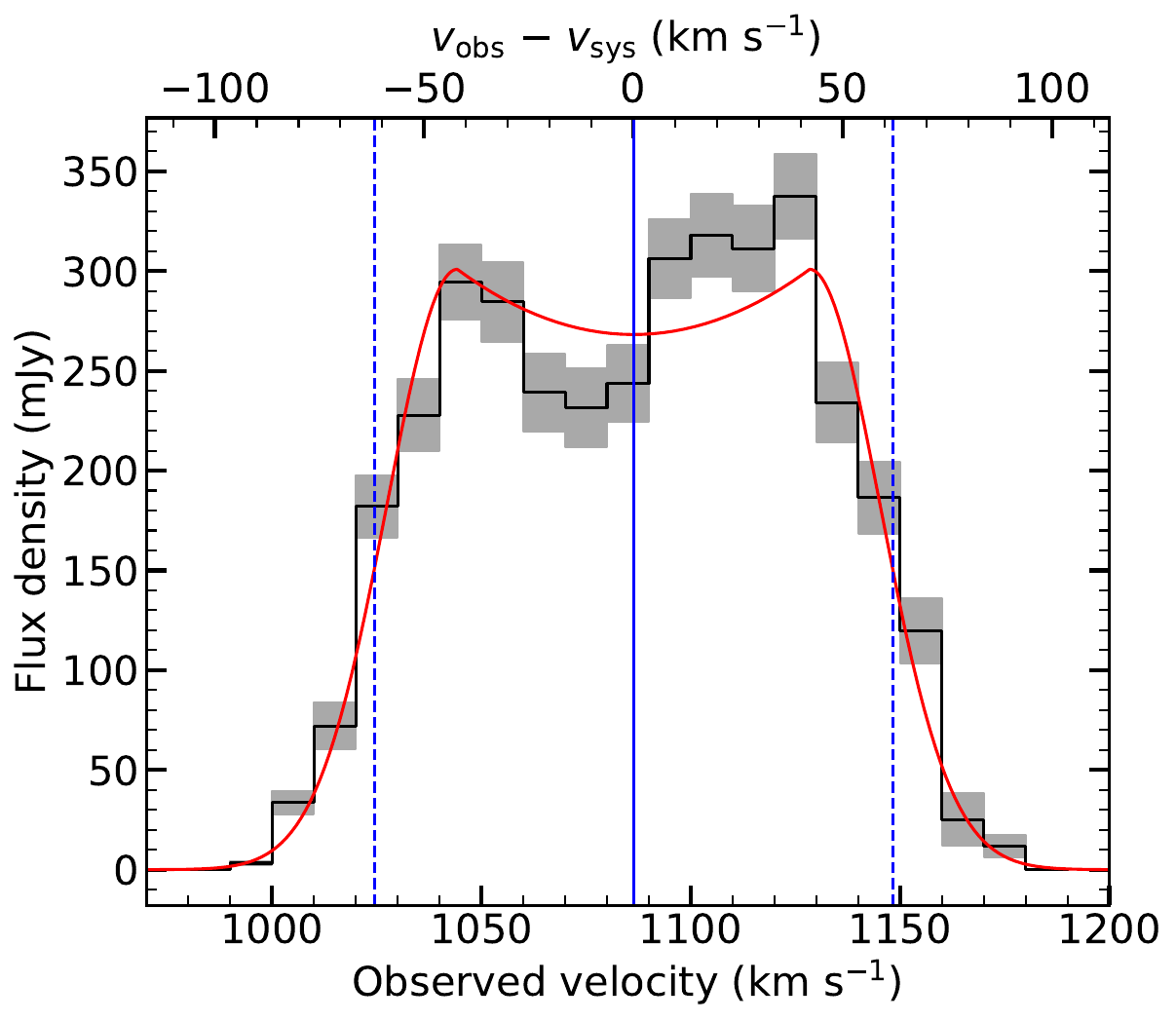}
    \end{subfigure}
    \begin{subfigure}{.32\linewidth}
    \centering
    \includegraphics[height=4.8cm]{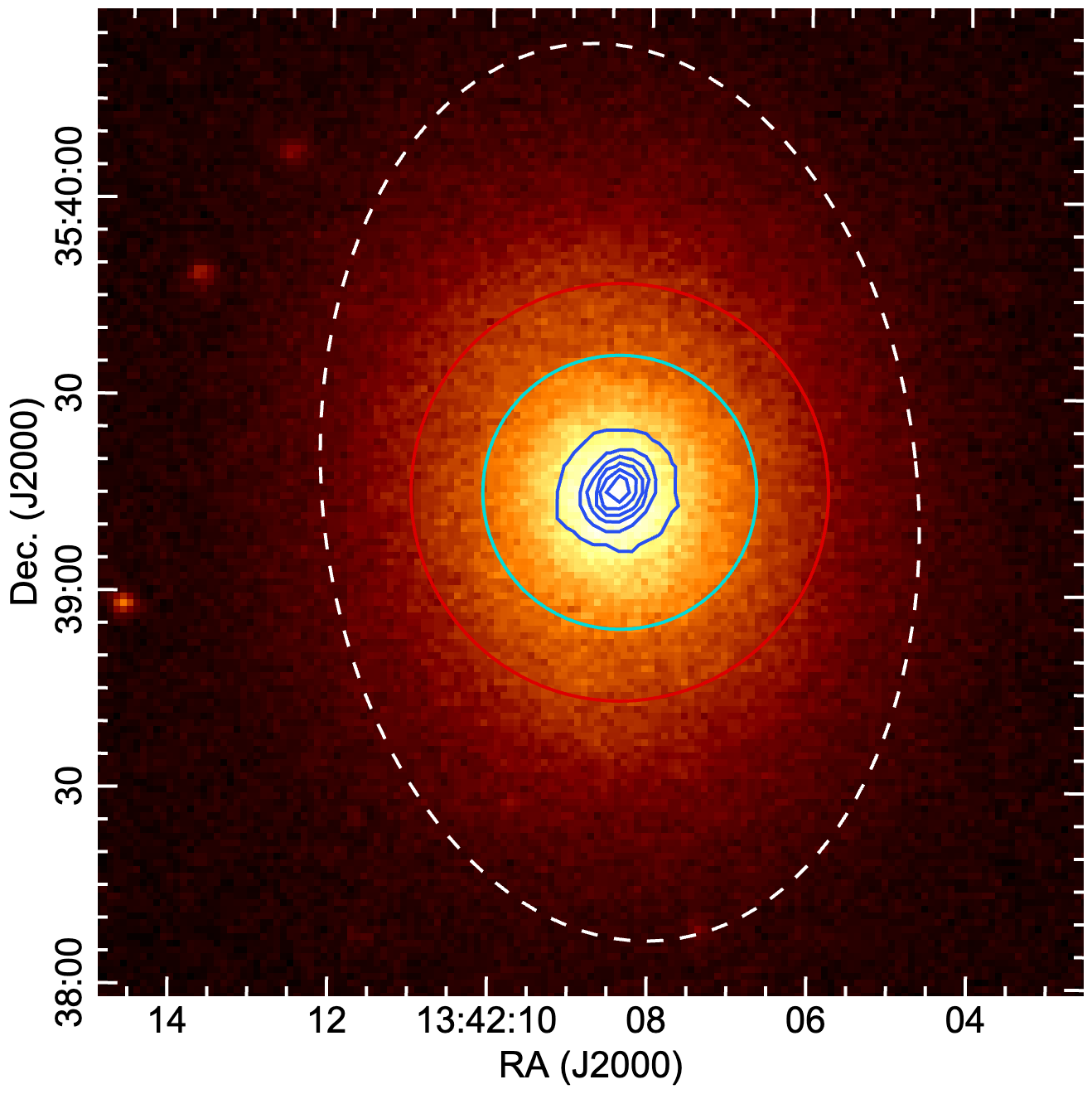}
    \end{subfigure}
    \begin{subfigure}{.33\linewidth}
    \centering
    \includegraphics[height=5.2cm]{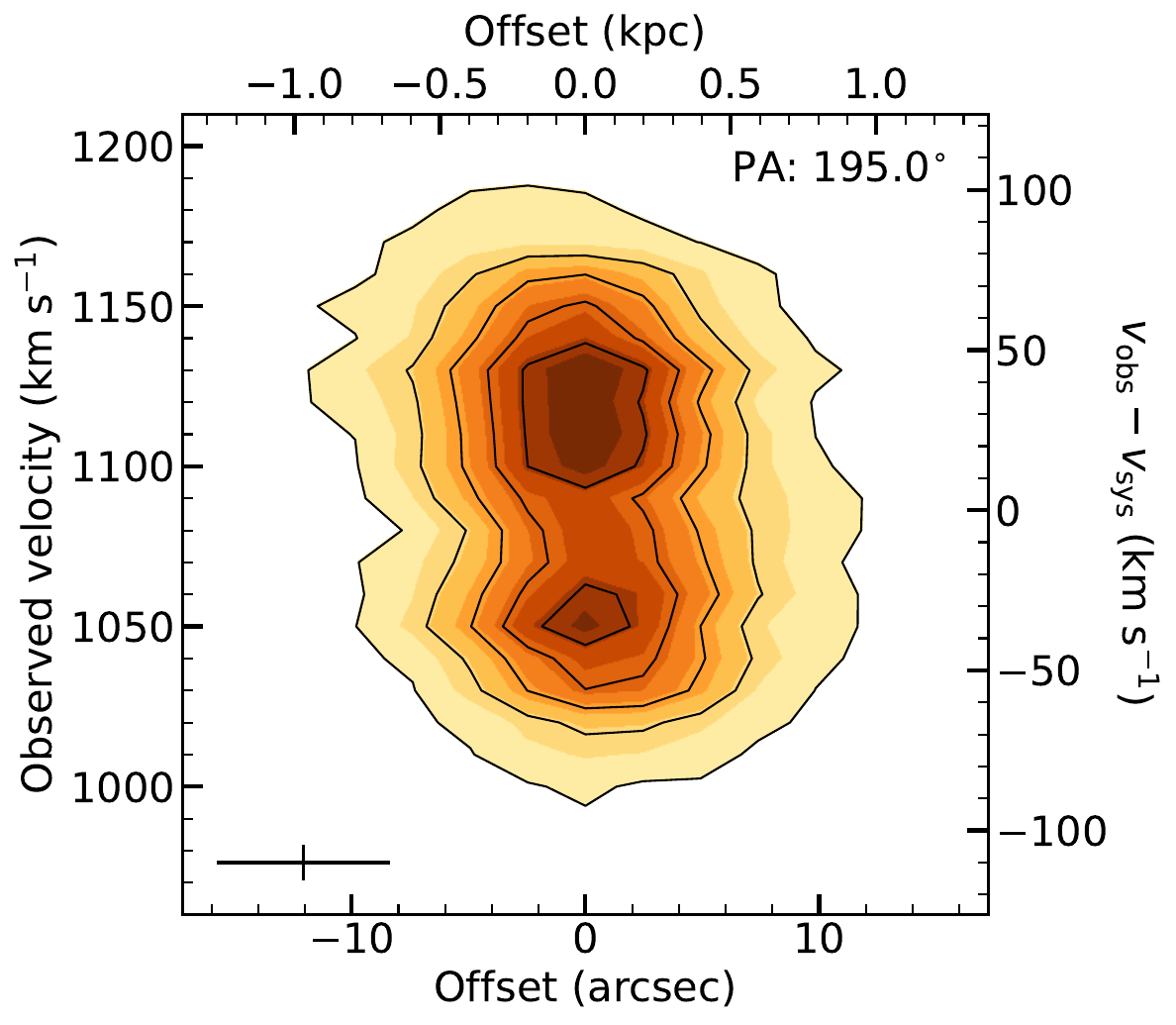}
    \end{subfigure}
    \caption[]{continued.}
\end{figure*}

\begin{figure*}
    \ContinuedFloat
    \centering
    \begin{subfigure}{.34\linewidth}
    \centering
    \includegraphics[height=6.0cm]{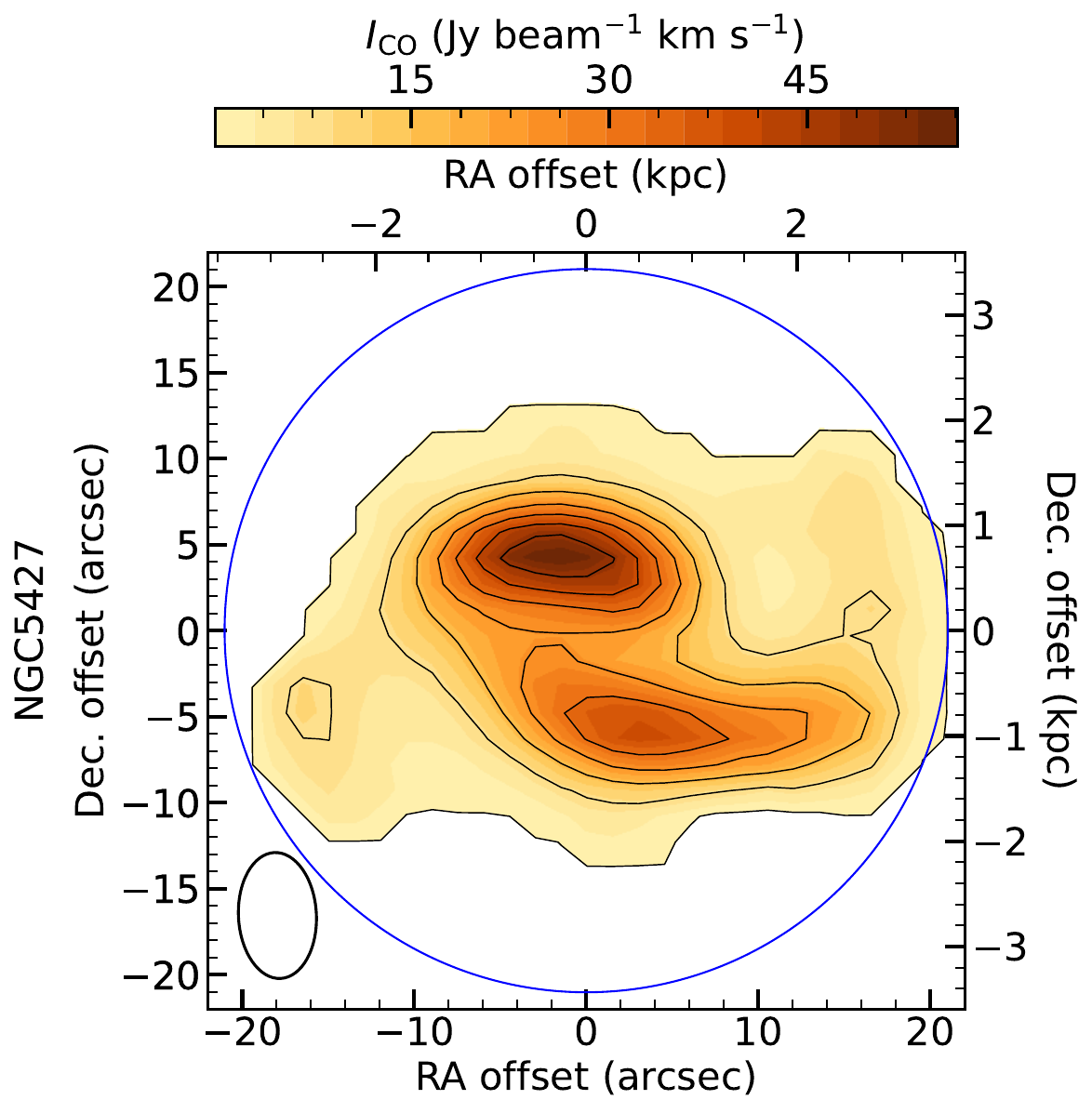}
    \end{subfigure}
    \begin{subfigure}{.32\linewidth}
    \centering
    \includegraphics[height=6.0cm]{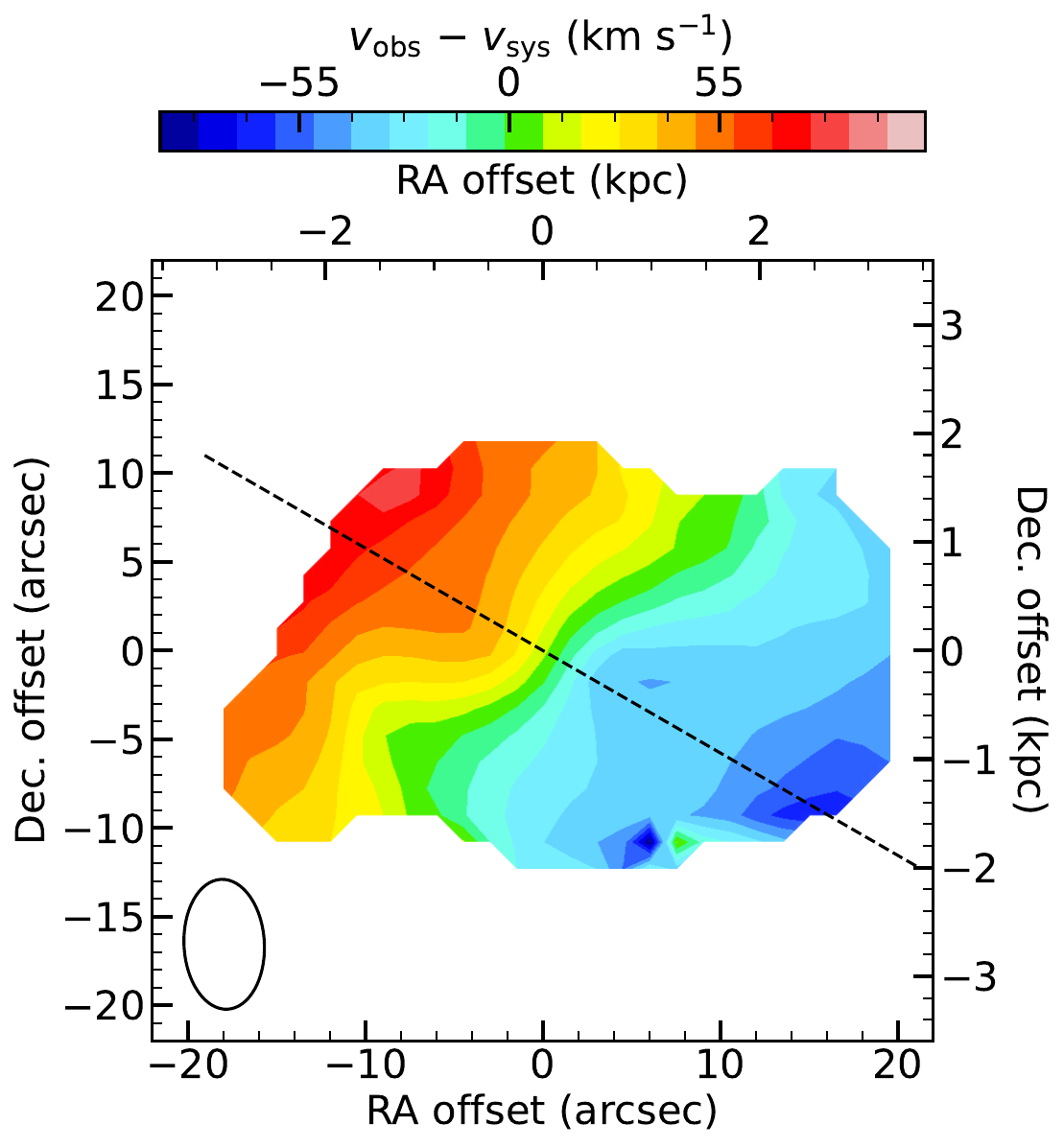}
    \end{subfigure}
    \begin{subfigure}{.32\linewidth}
    \centering
    \includegraphics[height=6.0cm]{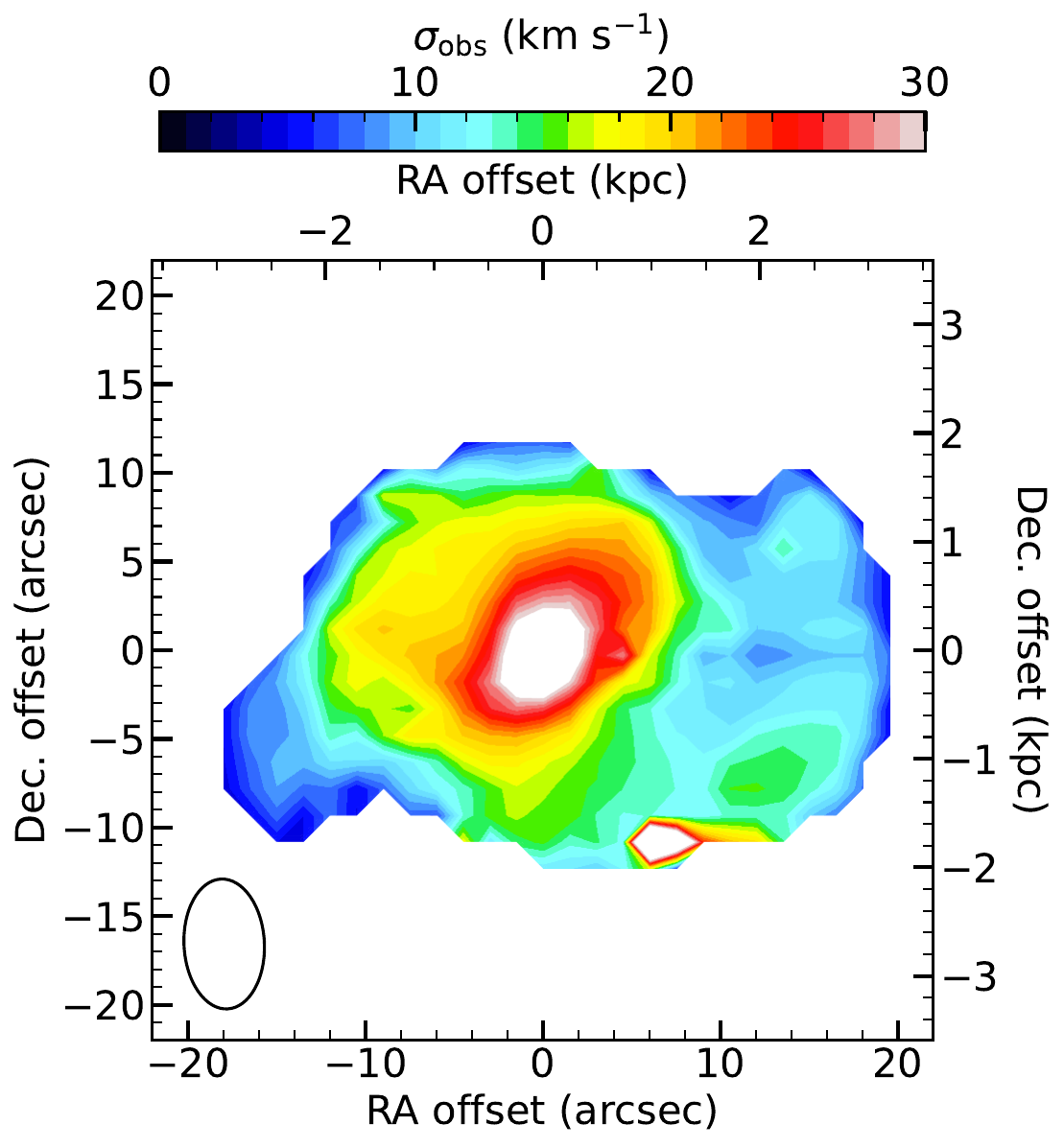}
    \end{subfigure}
    \\
    \begin{subfigure}{.34\linewidth}
    \centering
    \includegraphics[height=5.2cm]{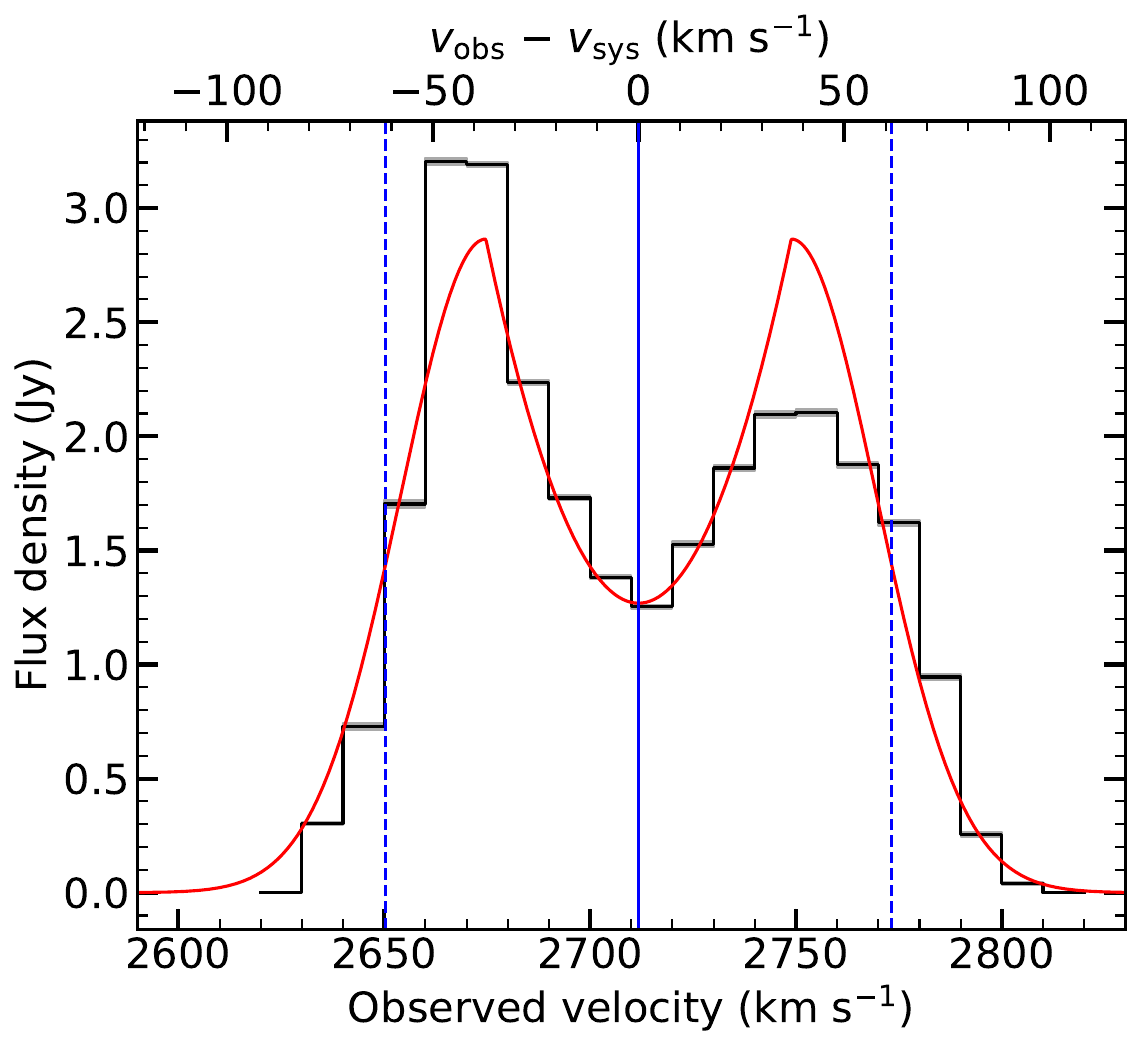}
    \end{subfigure}
    \begin{subfigure}{.32\linewidth}
    \centering
    \includegraphics[height=4.8cm]{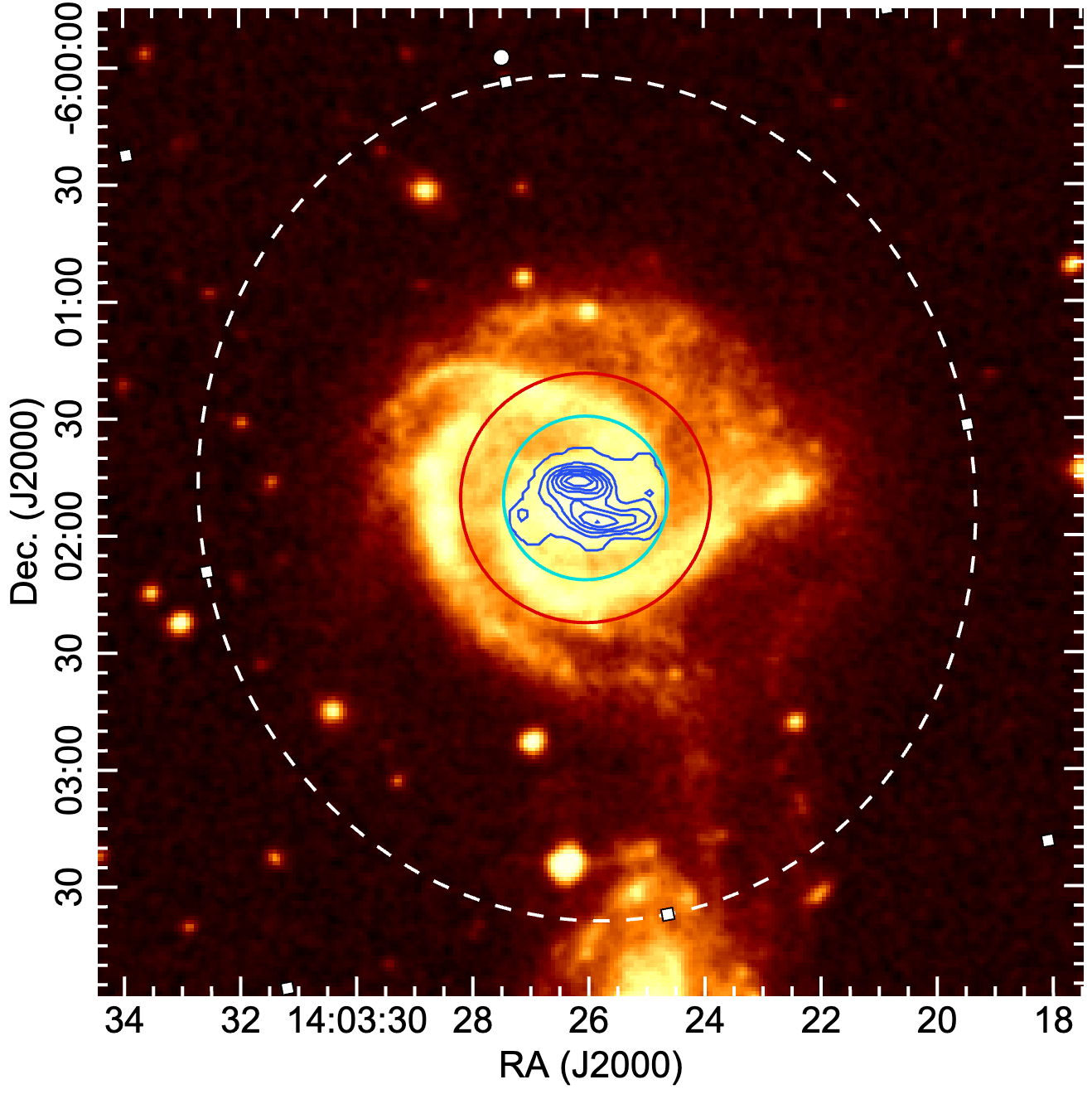}
    \end{subfigure}
    \begin{subfigure}{.33\linewidth}
    \centering
    \includegraphics[height=5.2cm]{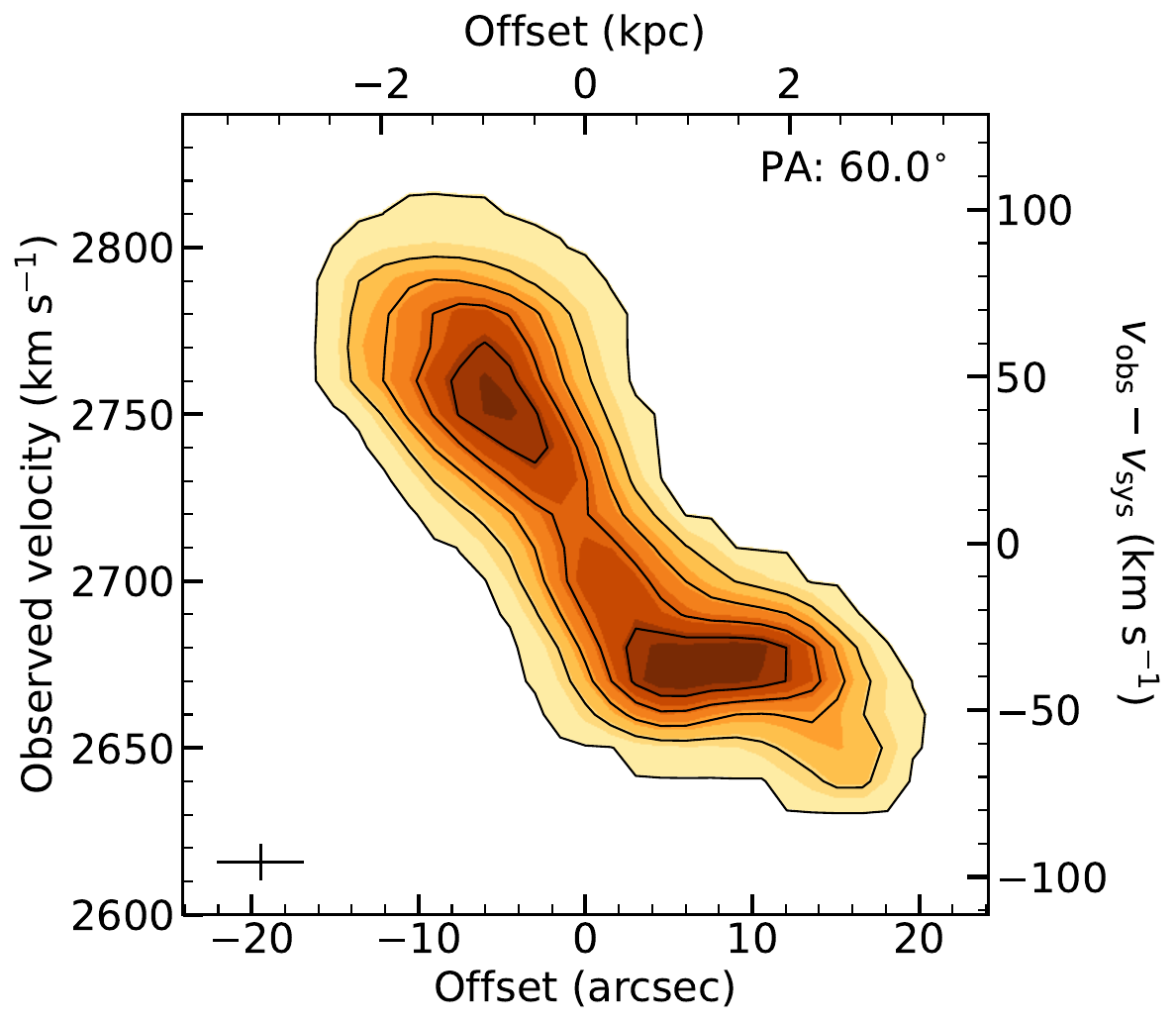}
    \end{subfigure}
    \\
    \begin{subfigure}{.34\linewidth}
    \centering
    \includegraphics[height=6.0cm]{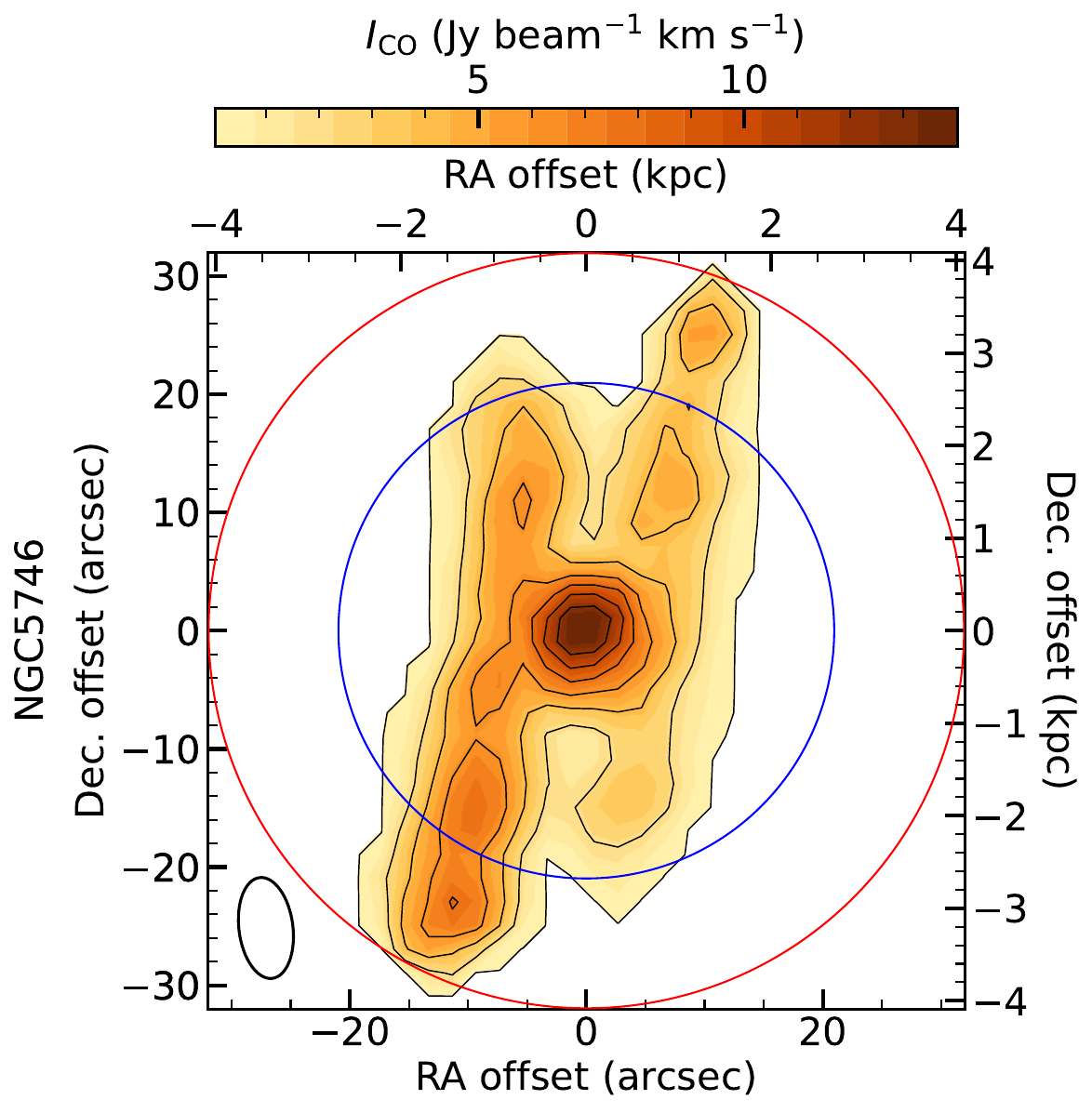}
    \end{subfigure}
    \begin{subfigure}{.32\linewidth}
    \centering
    \includegraphics[height=6.0cm]{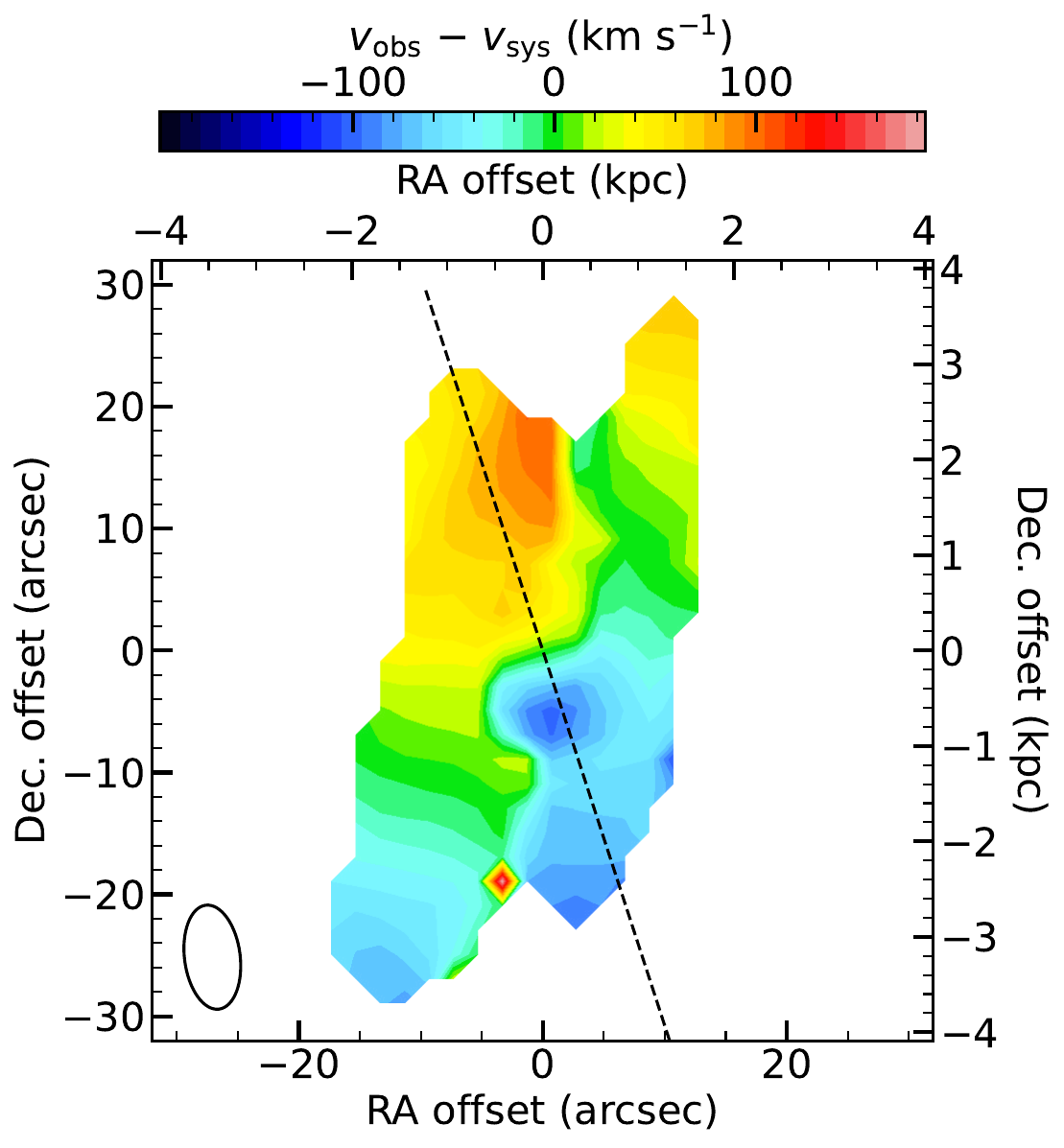}
    \end{subfigure}
    \begin{subfigure}{.32\linewidth}
    \centering
    \includegraphics[height=6.0cm]{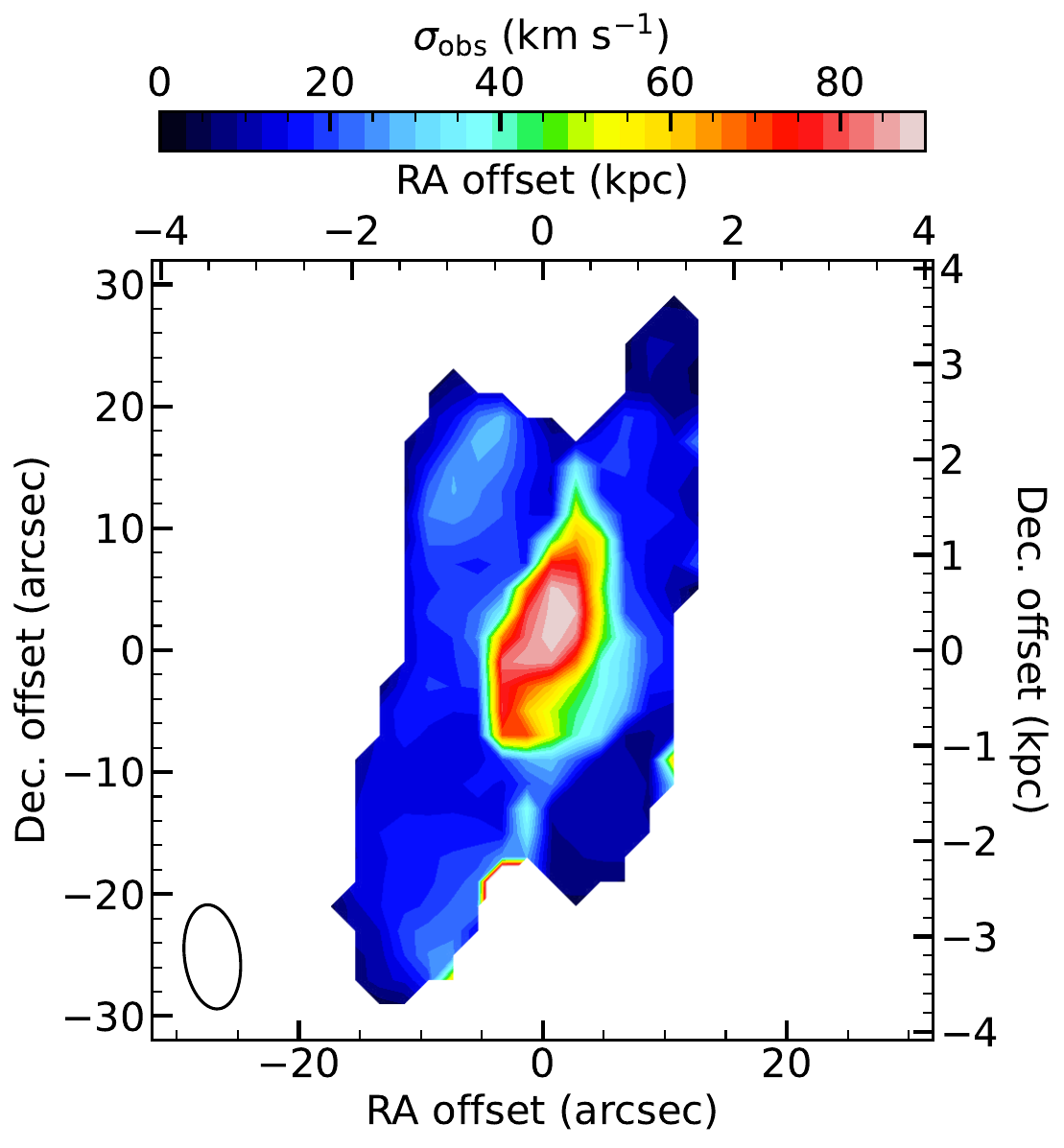}
    \end{subfigure}
    \\
    \begin{subfigure}{.34\linewidth}
    \centering
    \includegraphics[height=5.2cm]{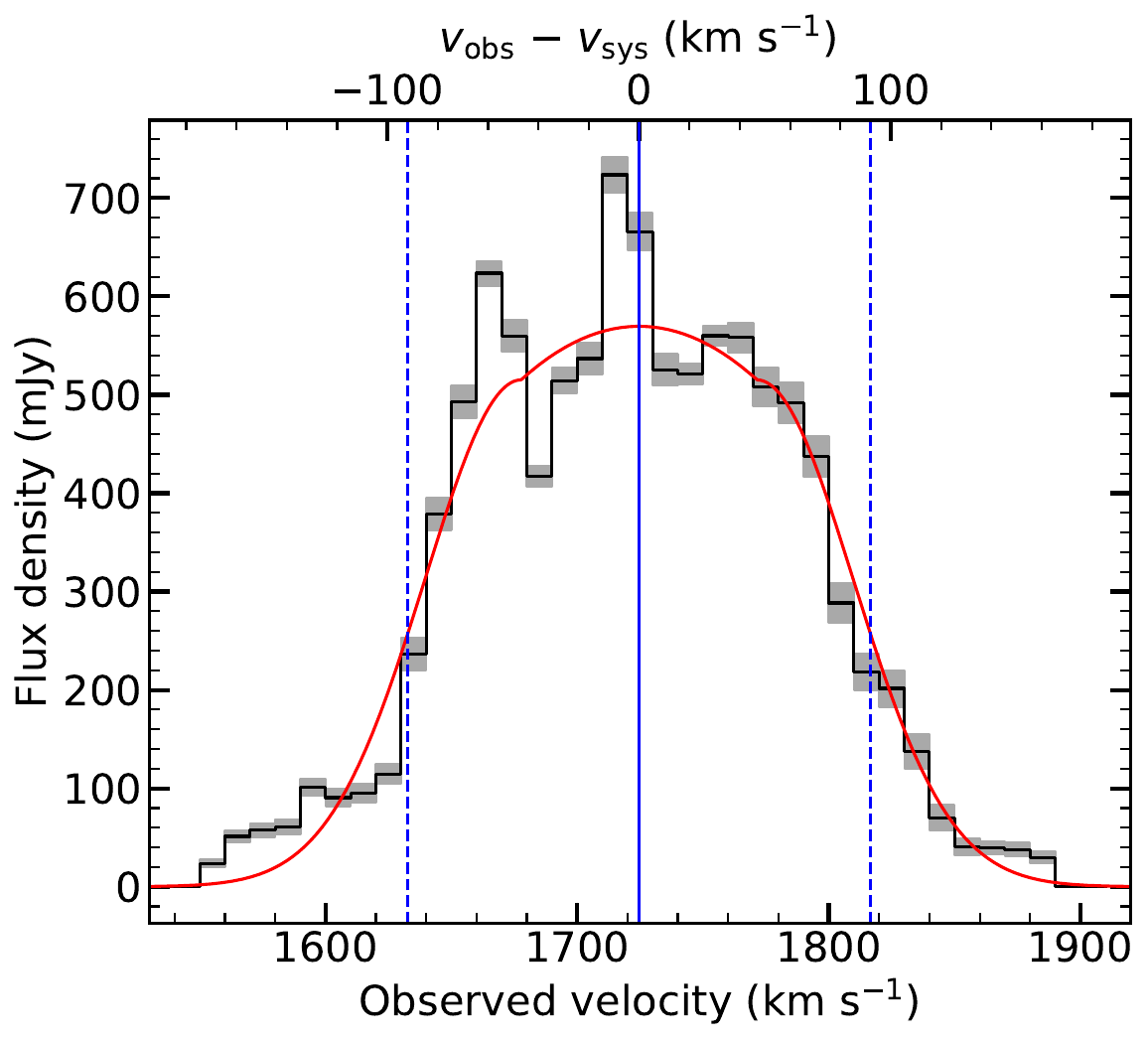}
    \end{subfigure}
    \begin{subfigure}{.32\linewidth}
    \centering
    \includegraphics[height=4.8cm]{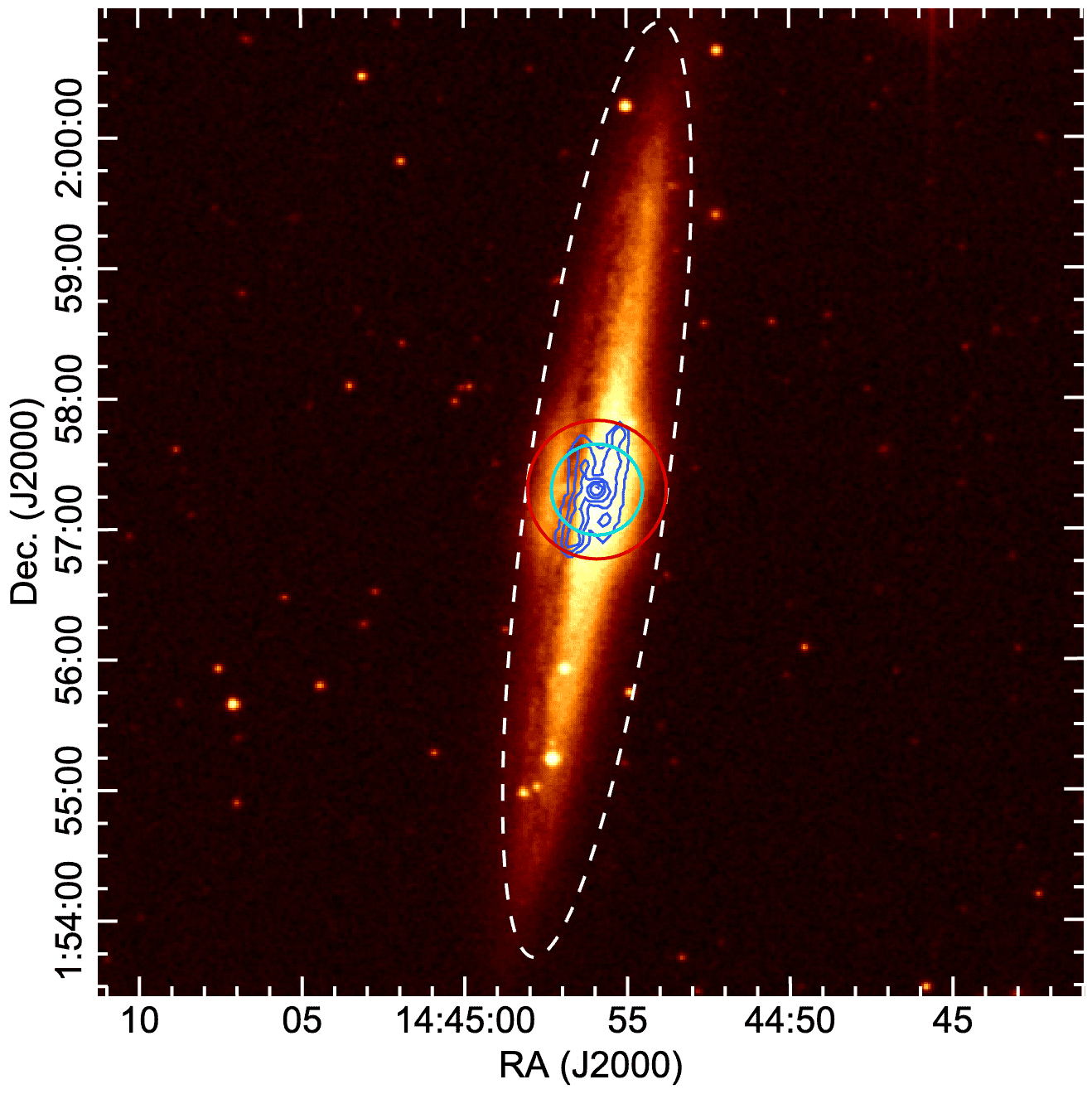}
    \end{subfigure}
    \begin{subfigure}{.33\linewidth}
    \centering
    \includegraphics[height=5.2cm]{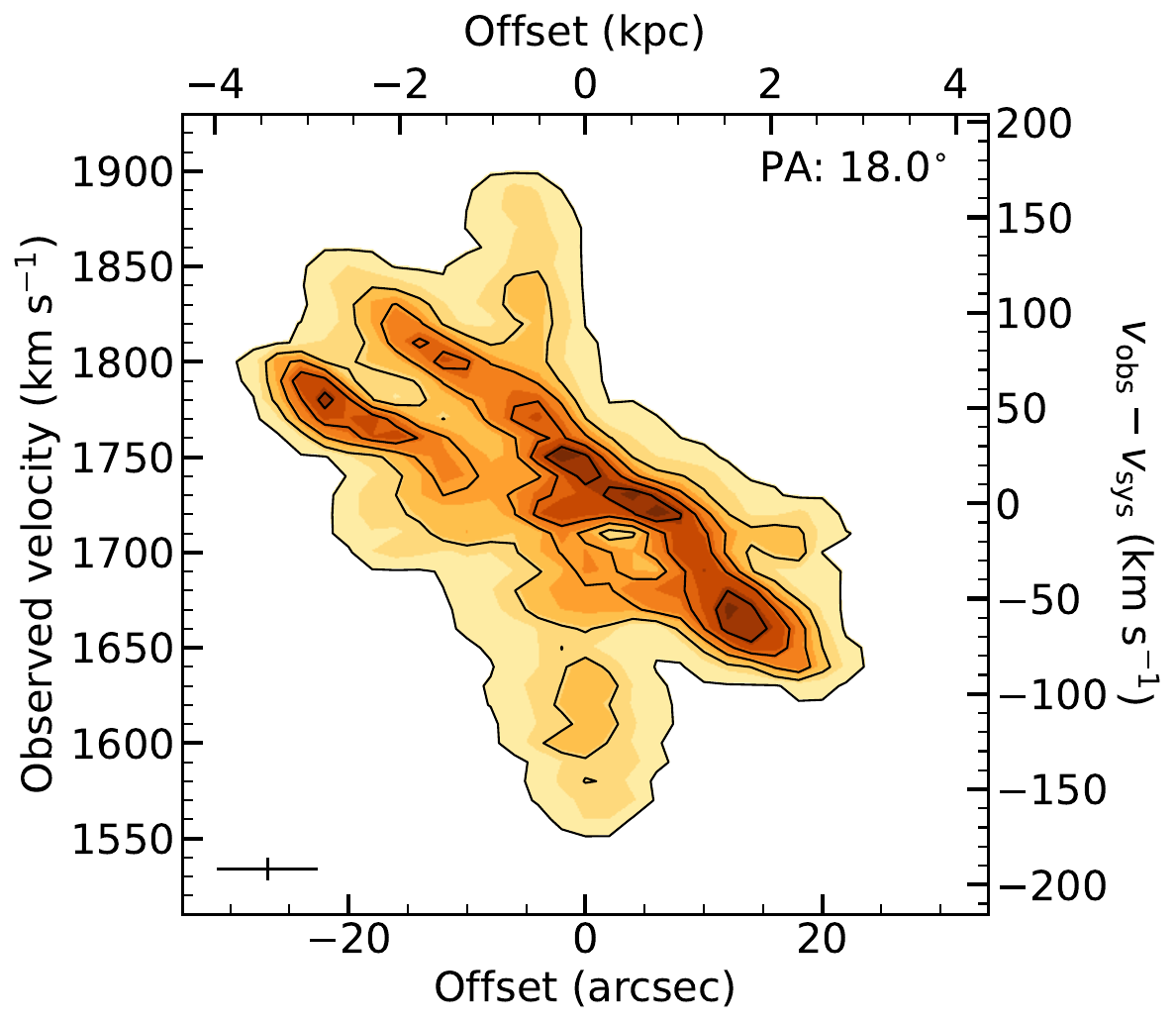}
    \end{subfigure}
    \caption[]{continued.}
\end{figure*}

\begin{figure*}
    \ContinuedFloat
    \centering
    \begin{subfigure}{.34\linewidth}
    \centering
    \includegraphics[height=6.0cm]{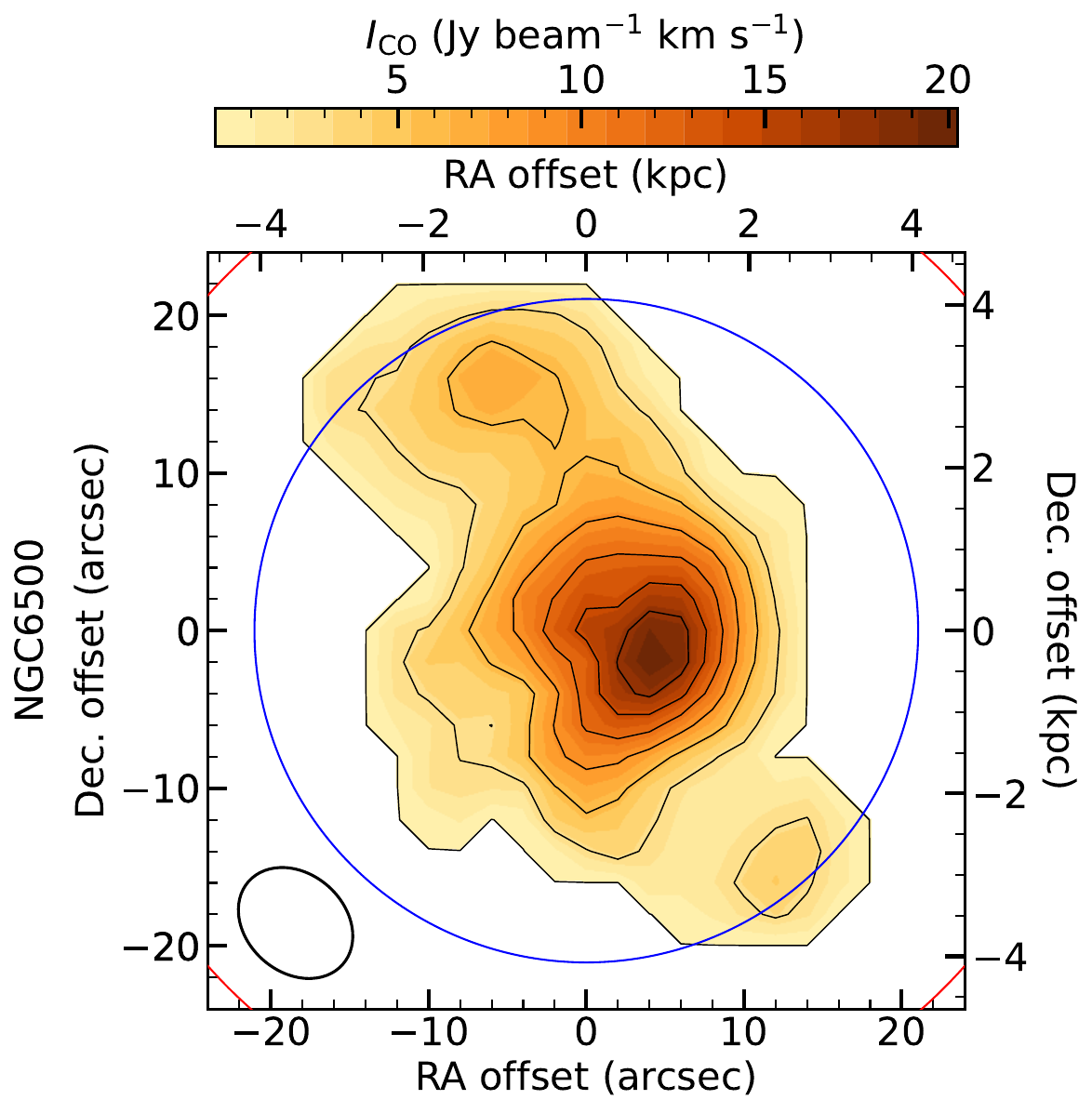}
    \end{subfigure}
    \begin{subfigure}{.32\linewidth}
    \centering
    \includegraphics[height=6.0cm]{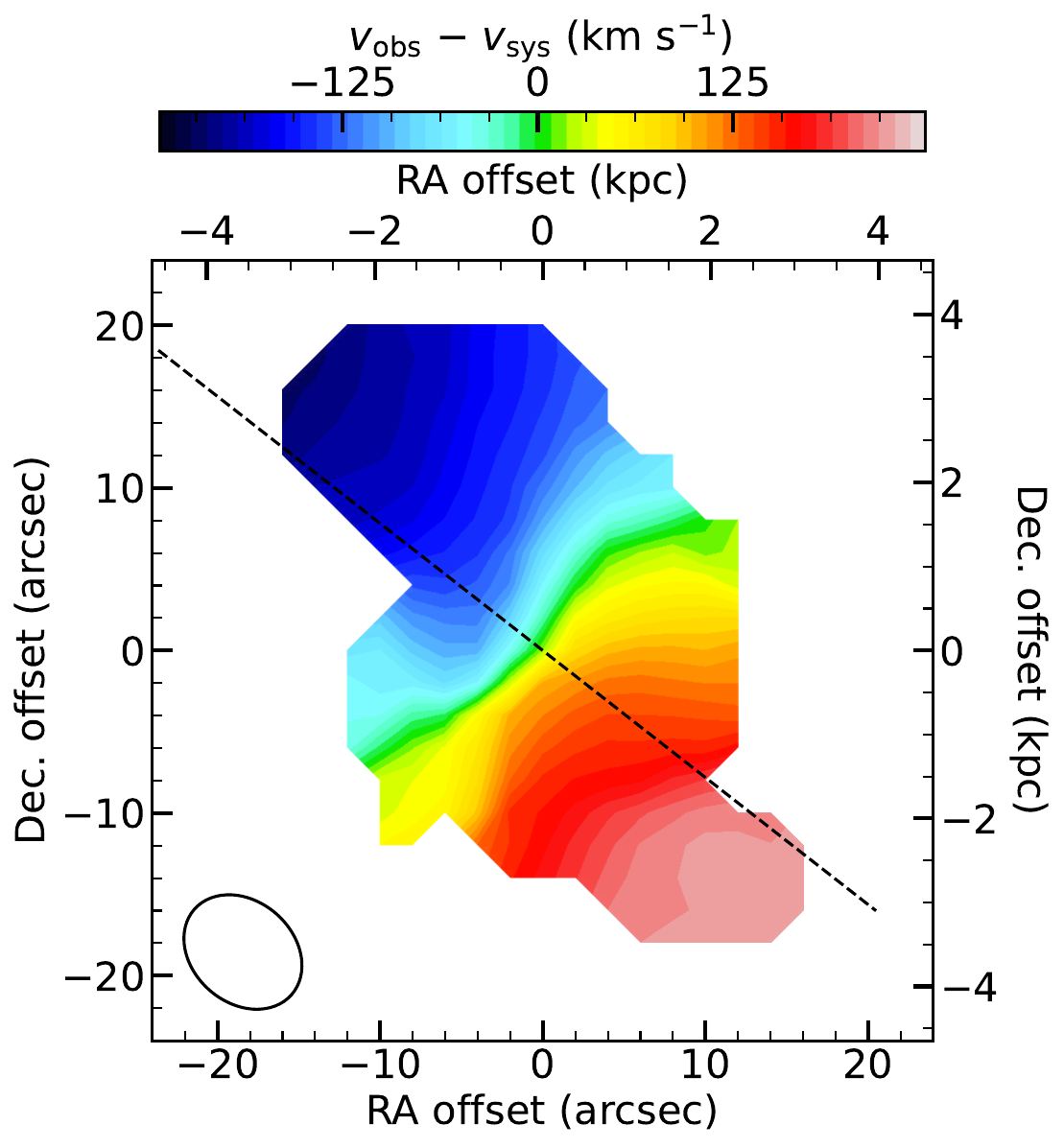}
    \end{subfigure}
    \begin{subfigure}{.32\linewidth}
    \centering
    \includegraphics[height=6.0cm]{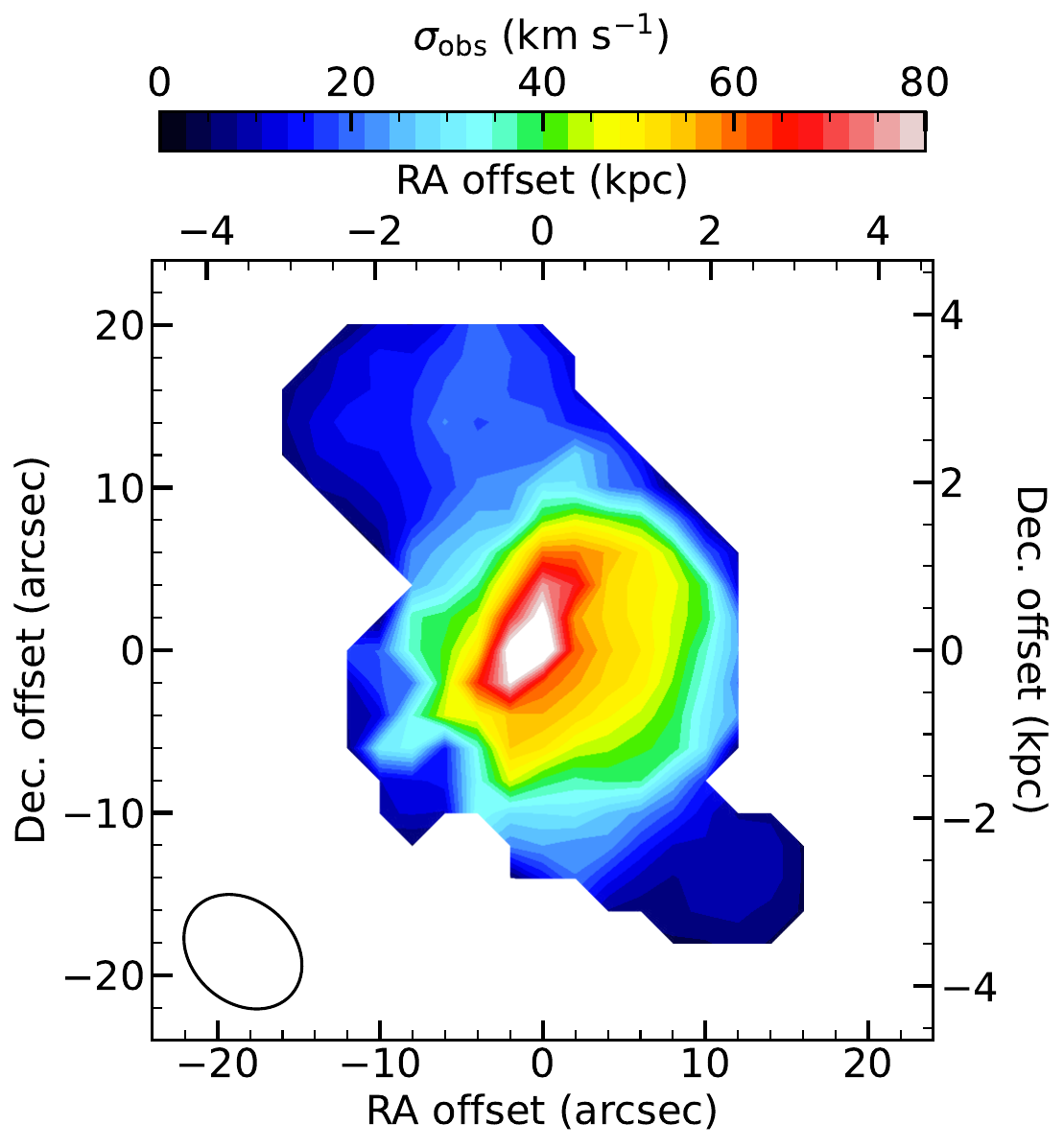}
    \end{subfigure}
    \\
    \begin{subfigure}{.34\linewidth}
    \centering
    \includegraphics[height=5.2cm]{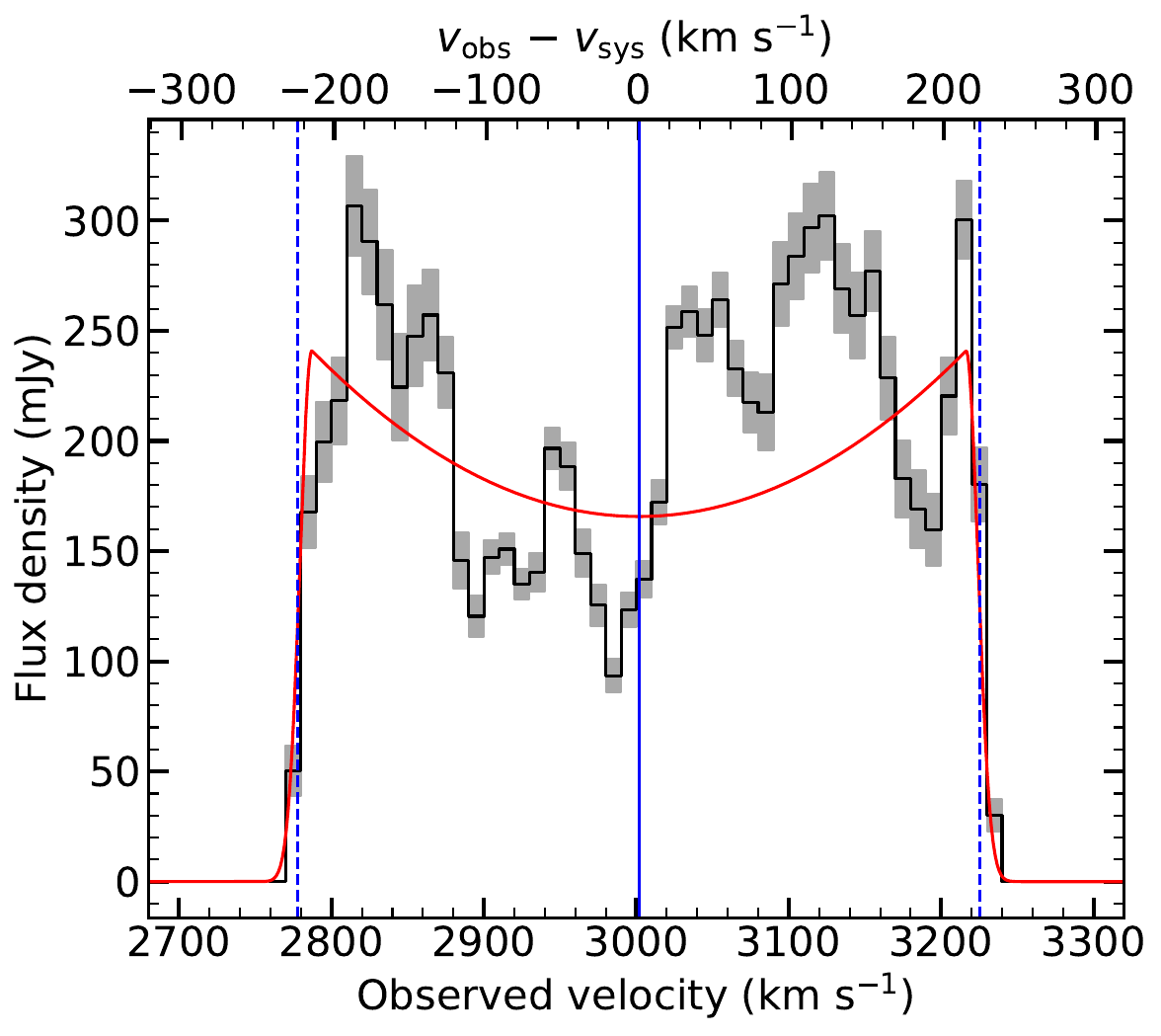}
    \end{subfigure}
    \begin{subfigure}{.32\linewidth}
    \centering
    \includegraphics[height=4.8cm]{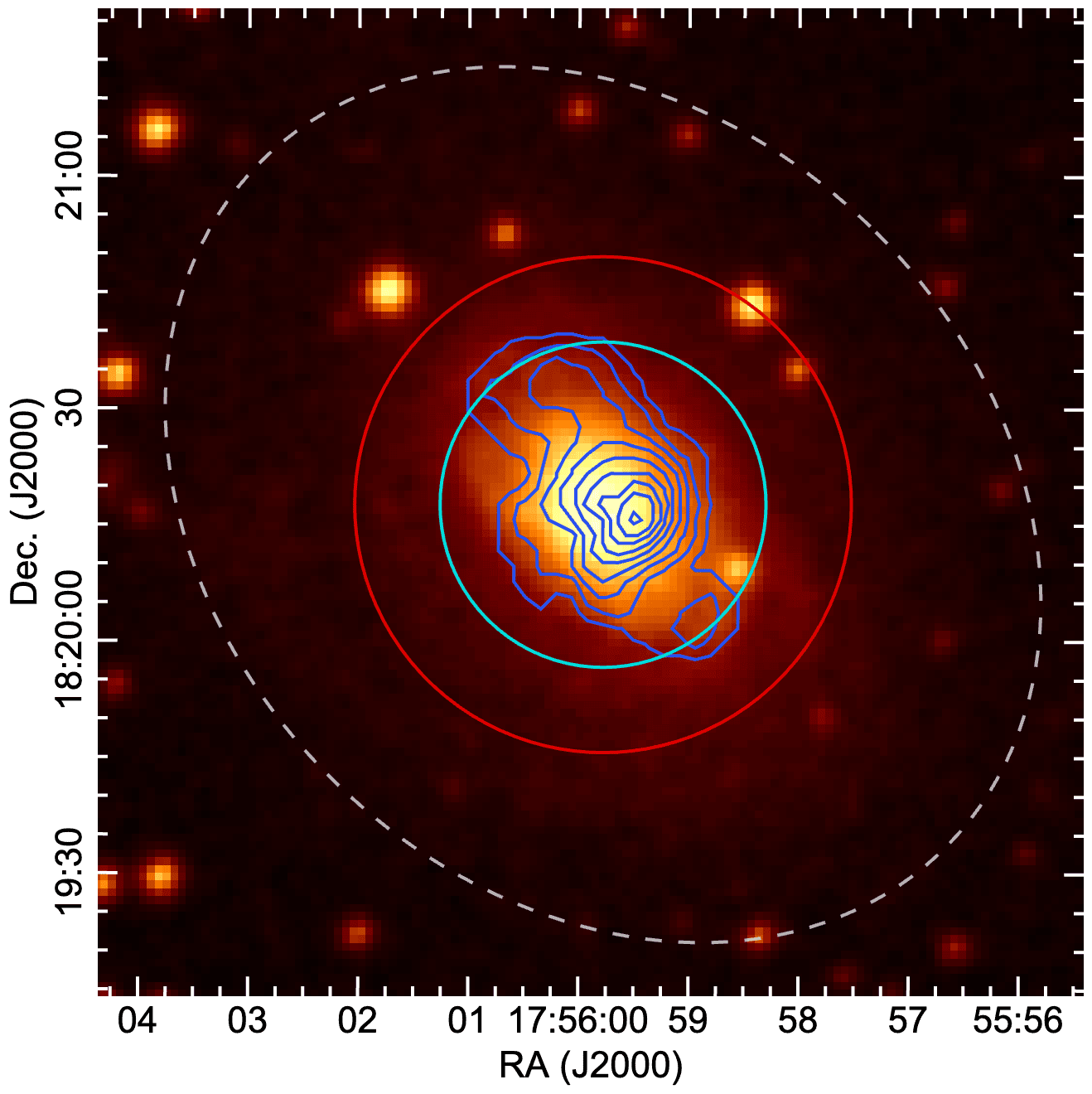}
    \end{subfigure}
    \begin{subfigure}{.33\linewidth}
    \centering
    \includegraphics[height=5.2cm]{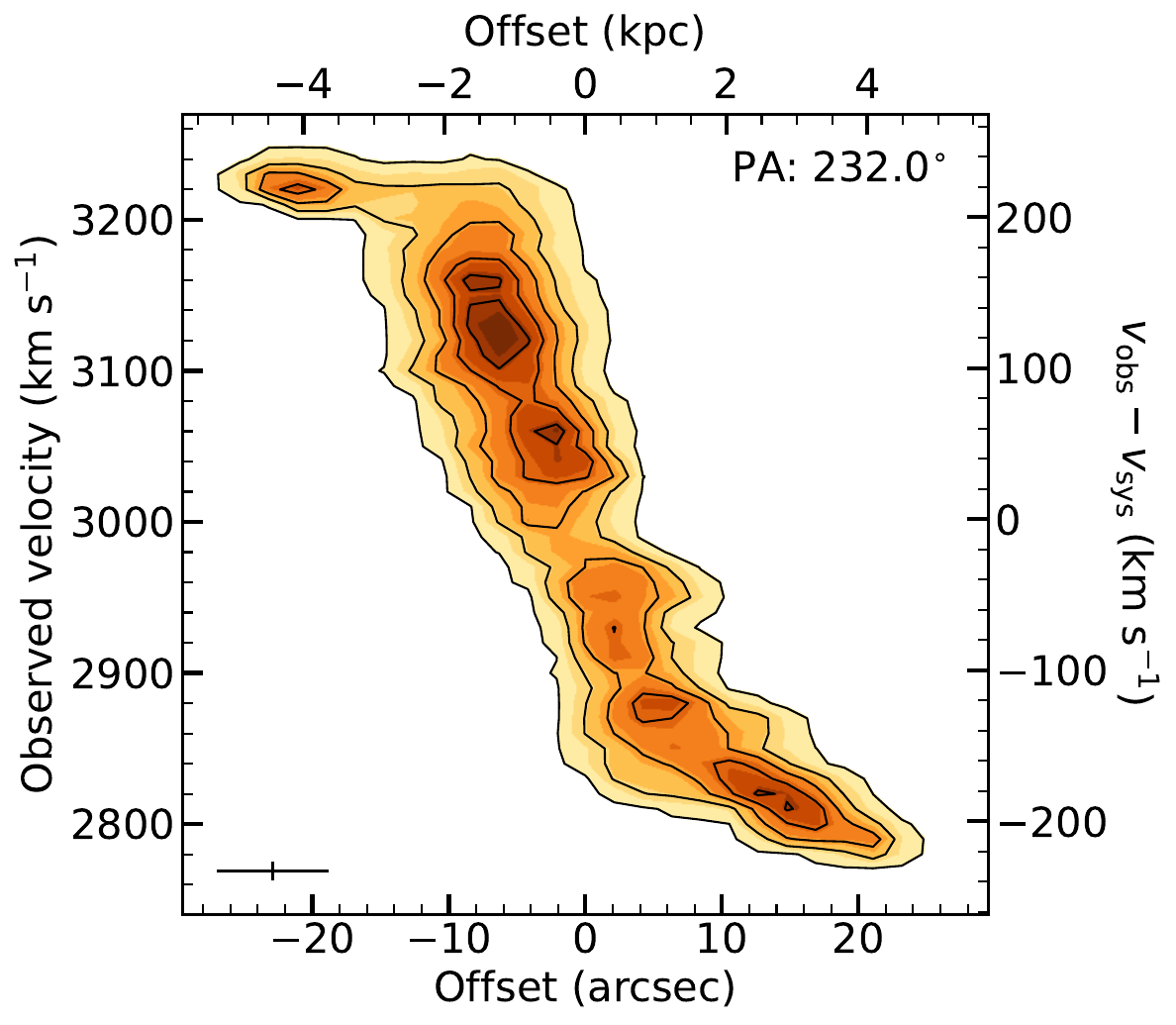}
    \end{subfigure}
    \\
    \begin{subfigure}{.34\linewidth}
    \centering
    \includegraphics[height=6.0cm]{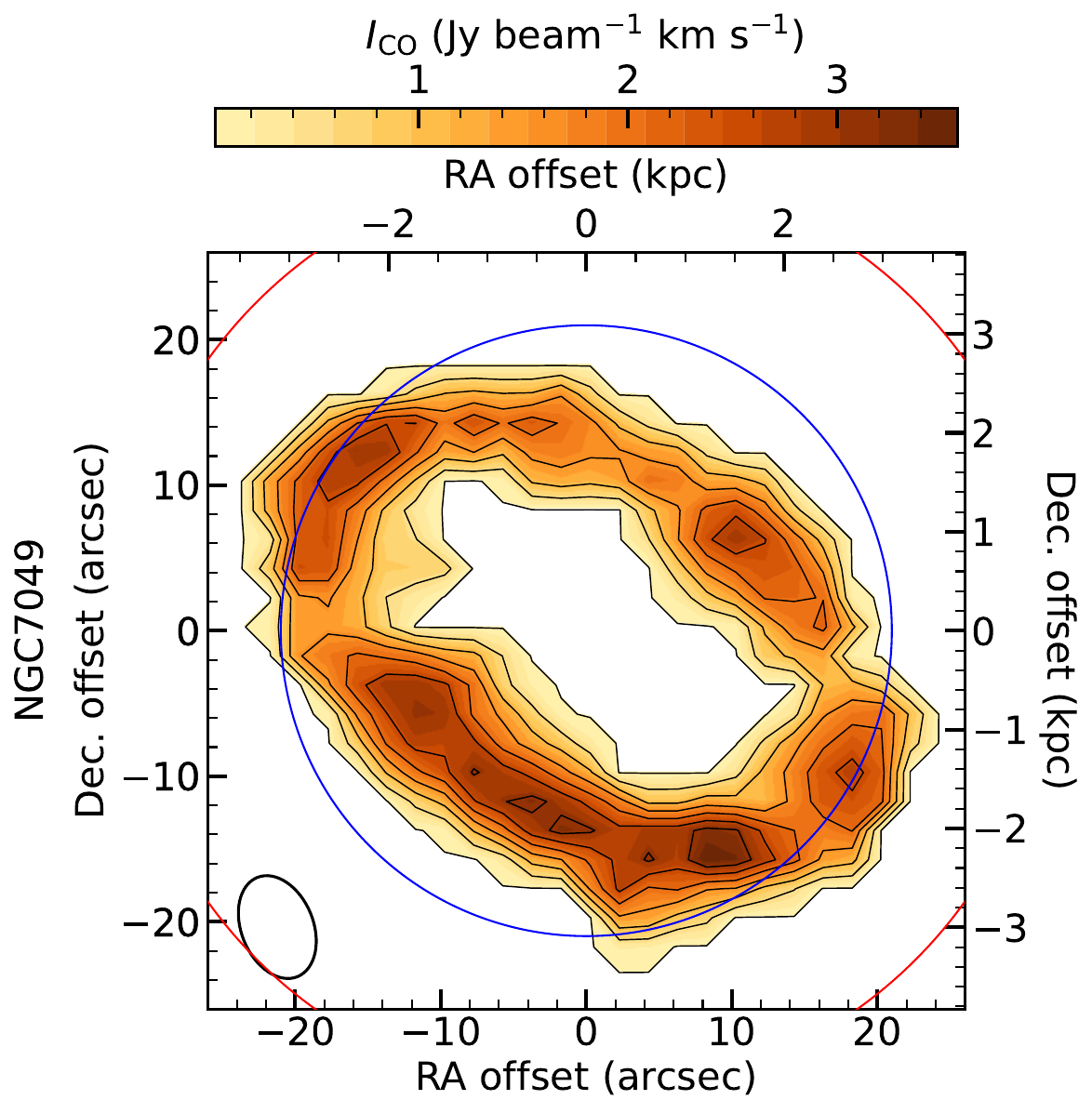}
    \end{subfigure}
    \begin{subfigure}{.32\linewidth}
    \centering
    \includegraphics[height=6.0cm]{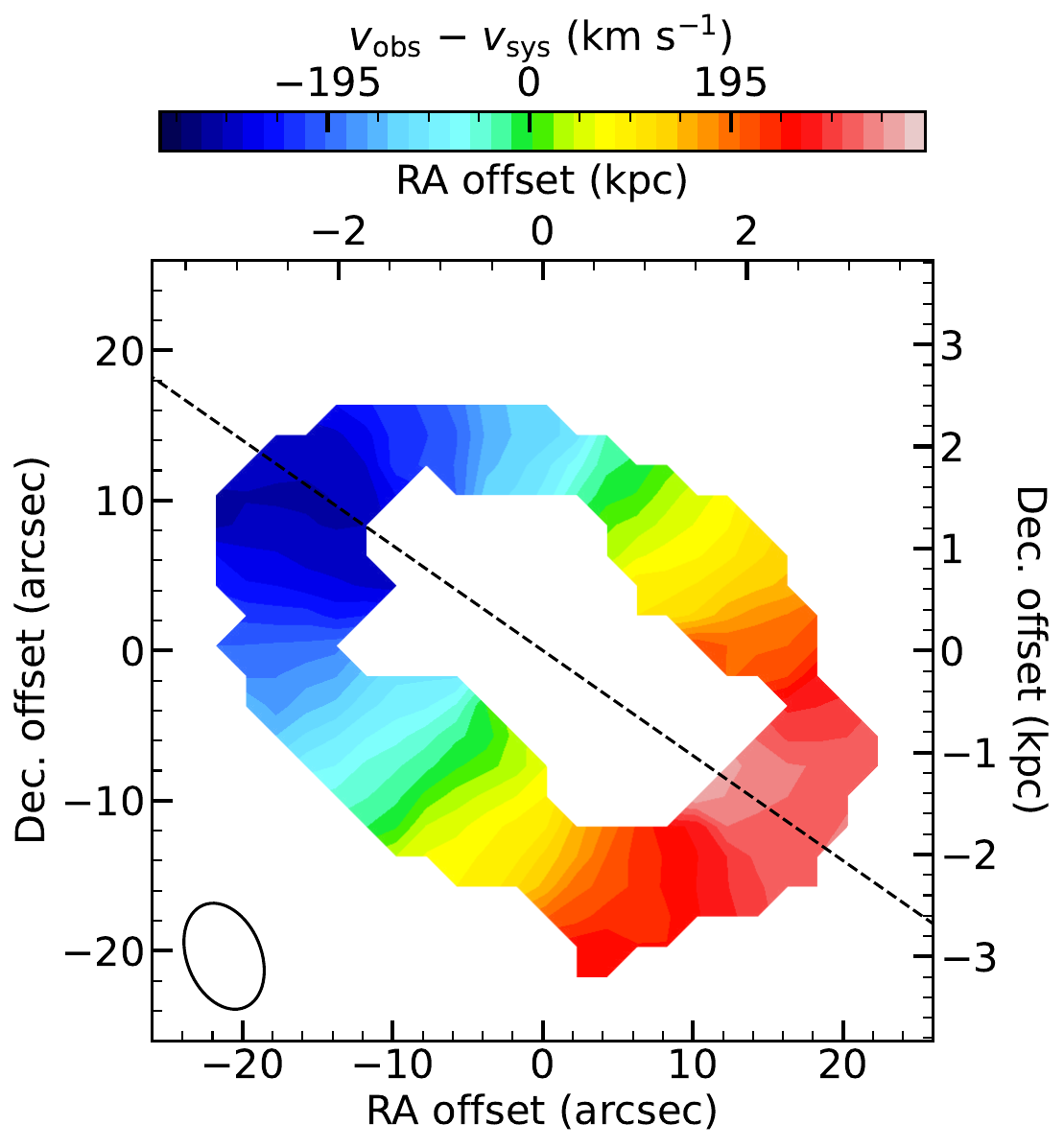}
    \end{subfigure}
    \begin{subfigure}{.32\linewidth}
    \centering
    \includegraphics[height=6.0cm]{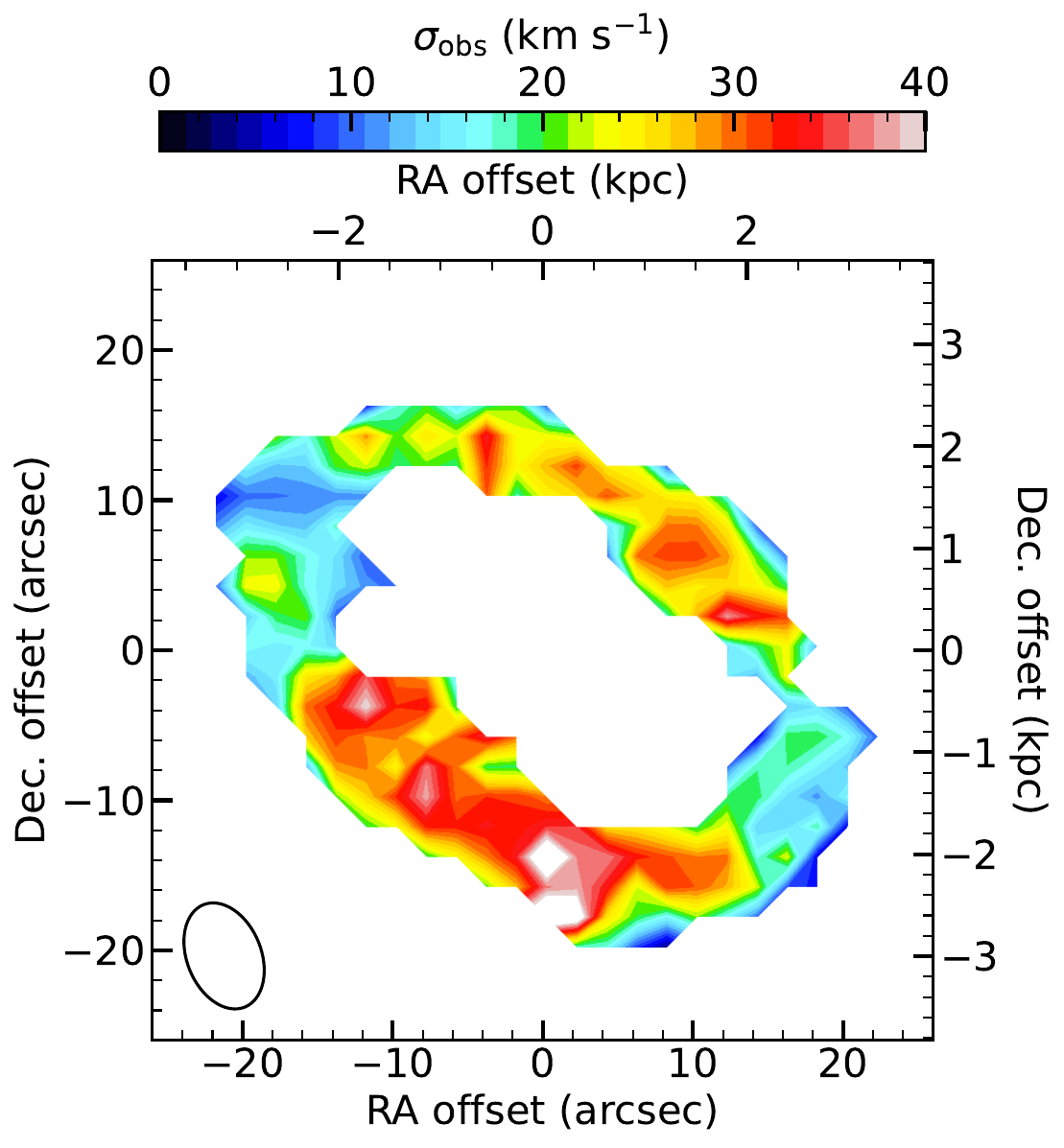}
    \end{subfigure}
    \\
    \begin{subfigure}{.34\linewidth}
    \centering
    \includegraphics[height=5.2cm]{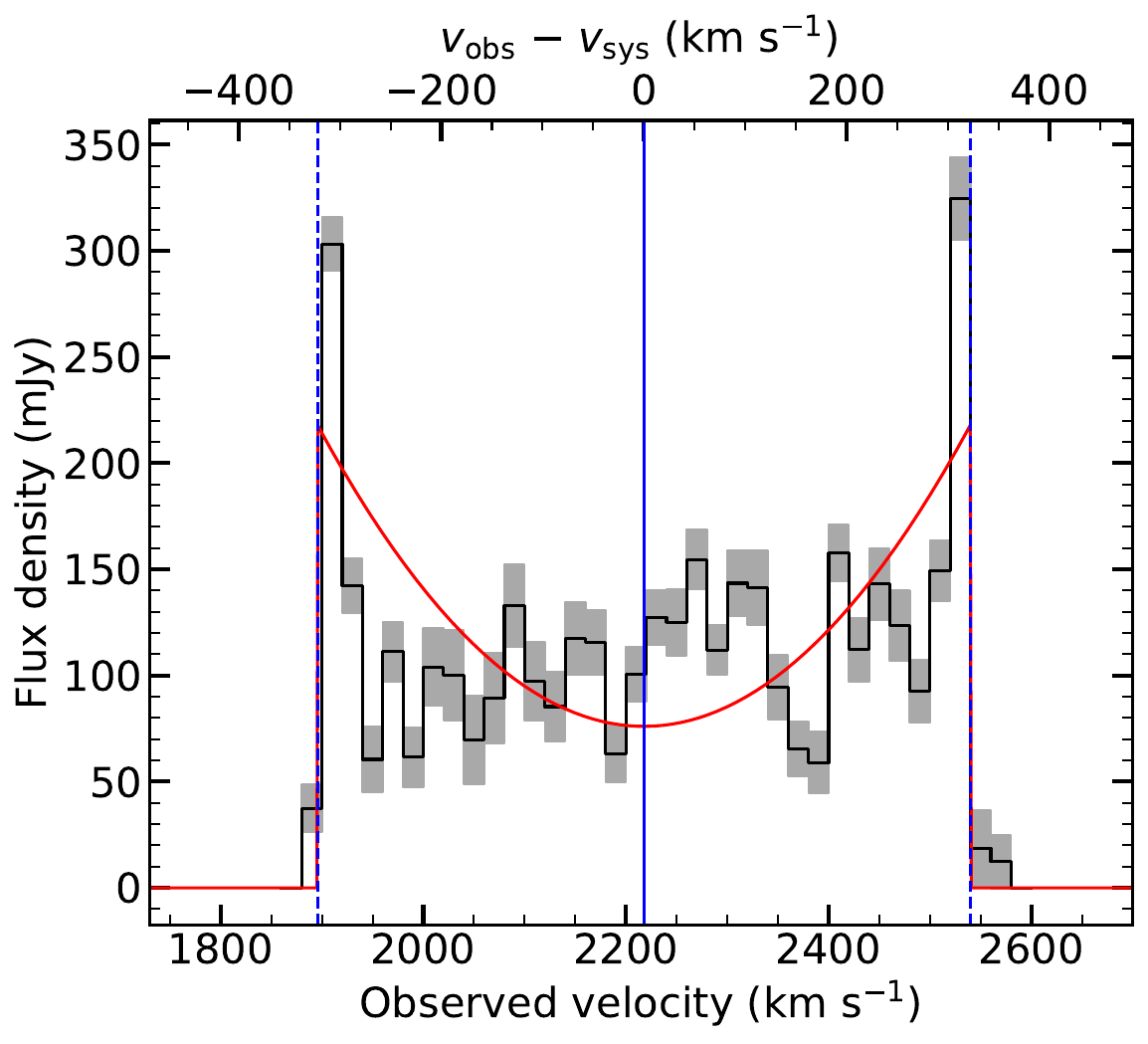}
    \end{subfigure}
    \begin{subfigure}{.32\linewidth}
    \centering
    \includegraphics[height=4.8cm]{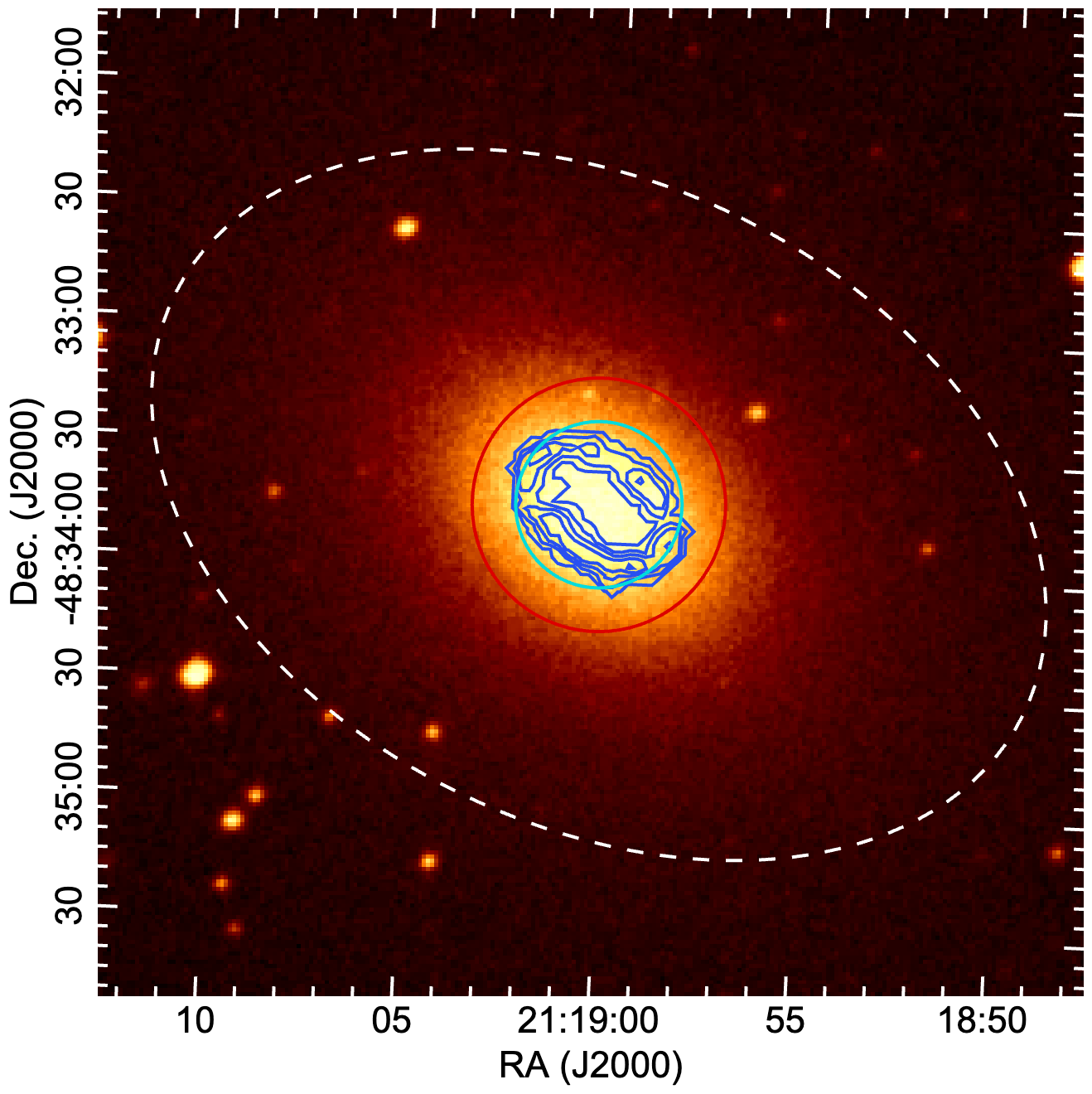}
    \end{subfigure}
    \begin{subfigure}{.33\linewidth}
    \centering
    \includegraphics[height=5.2cm]{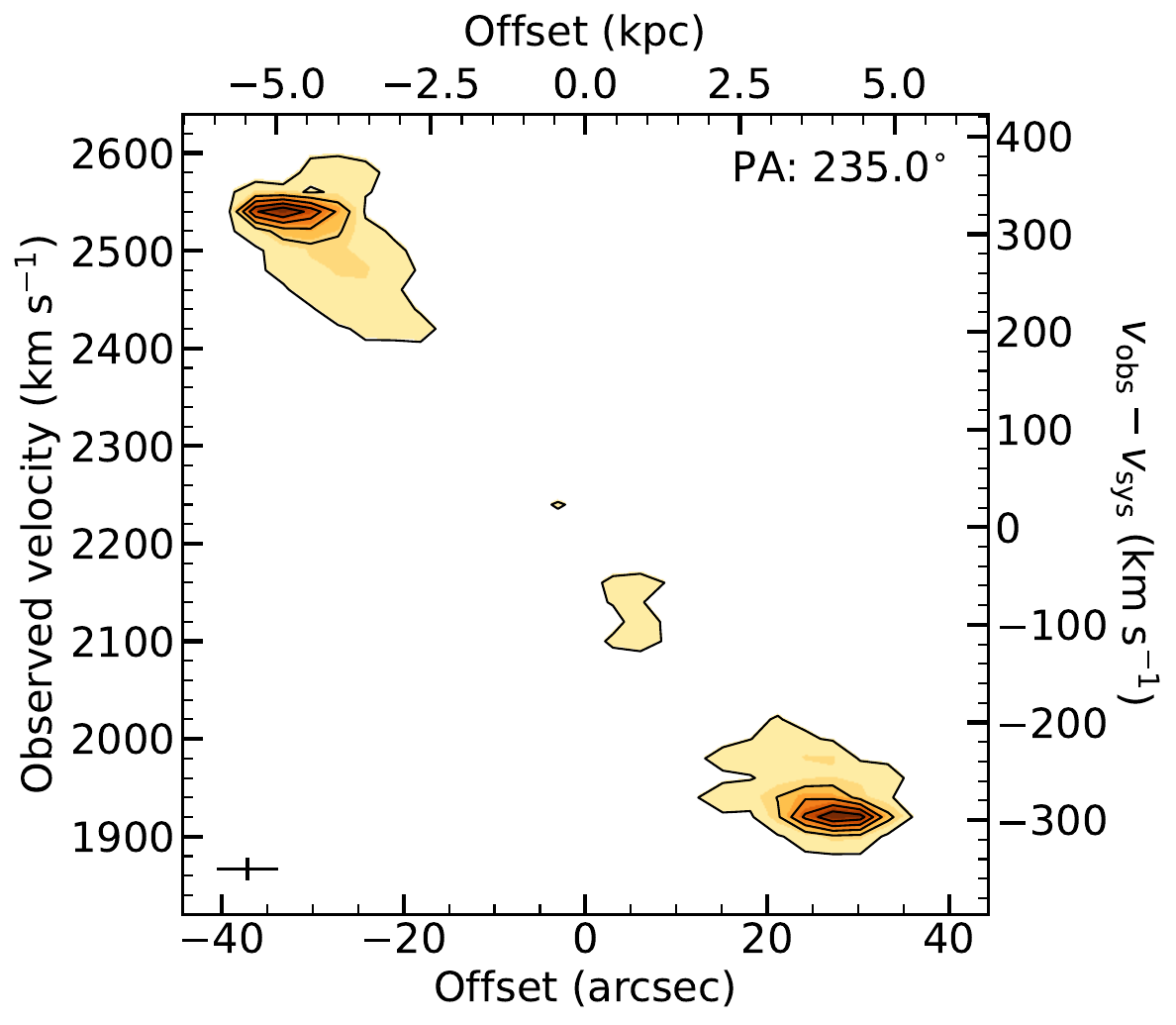}
    \end{subfigure}
    \caption[]{continued.}
\end{figure*}

\begin{figure*}
    \ContinuedFloat
    \centering
    \begin{subfigure}{.34\linewidth}
    \centering
    \includegraphics[height=6.0cm]{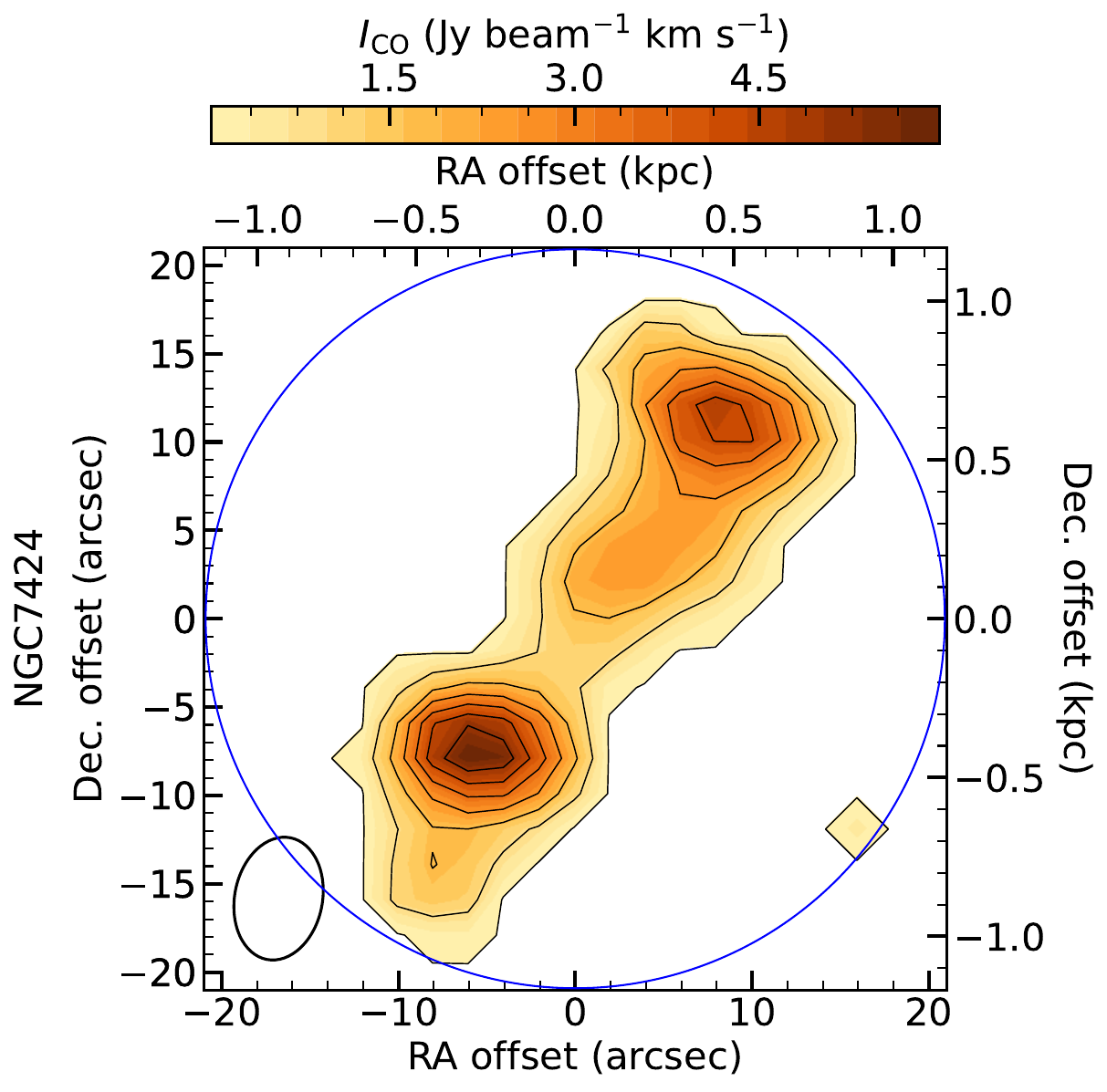}
    \end{subfigure}
    \begin{subfigure}{.32\linewidth}
    \centering
    \includegraphics[height=6.0cm]{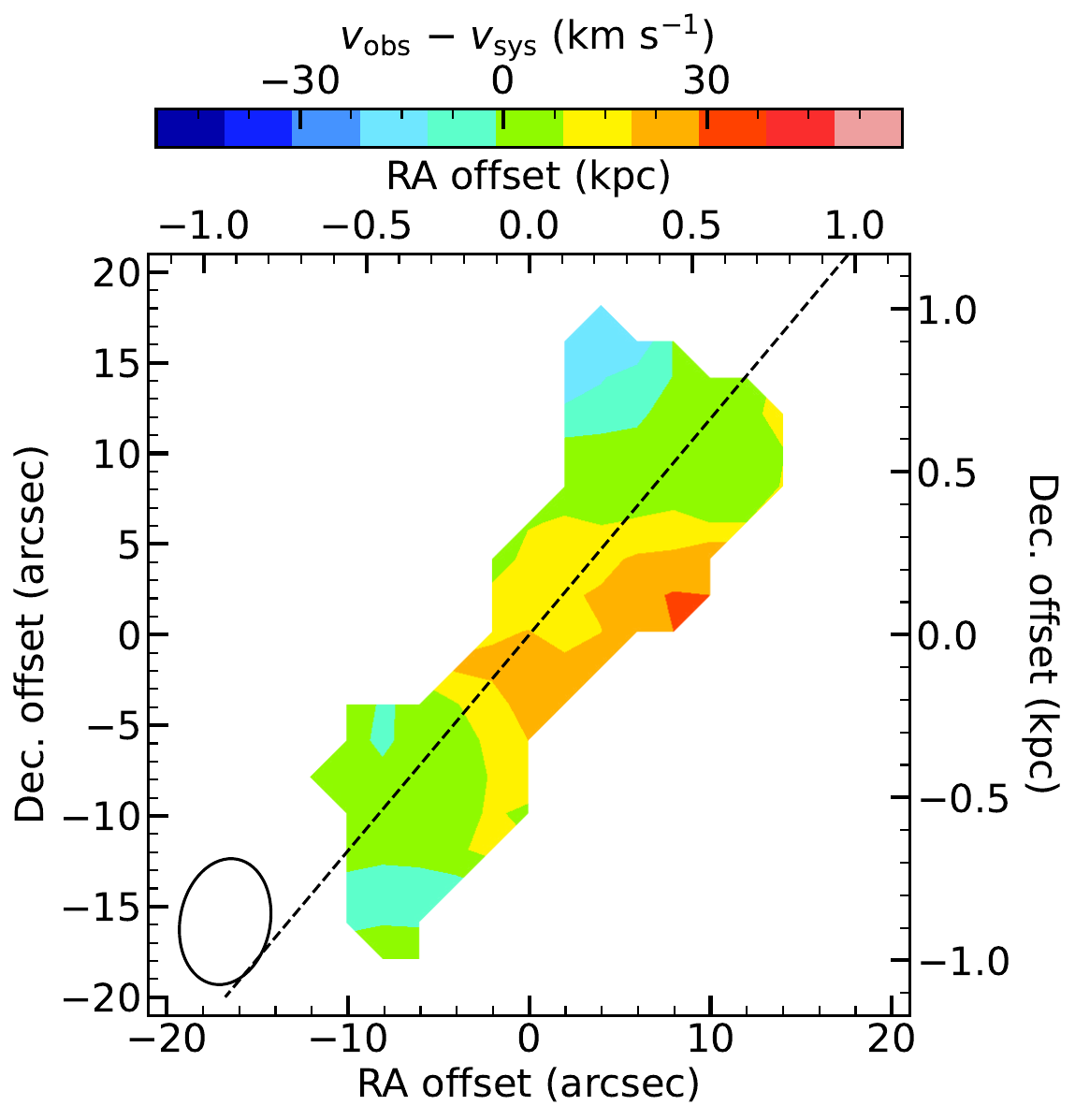}
    \end{subfigure}
    \begin{subfigure}{.32\linewidth}
    \centering
    \includegraphics[height=6.0cm]{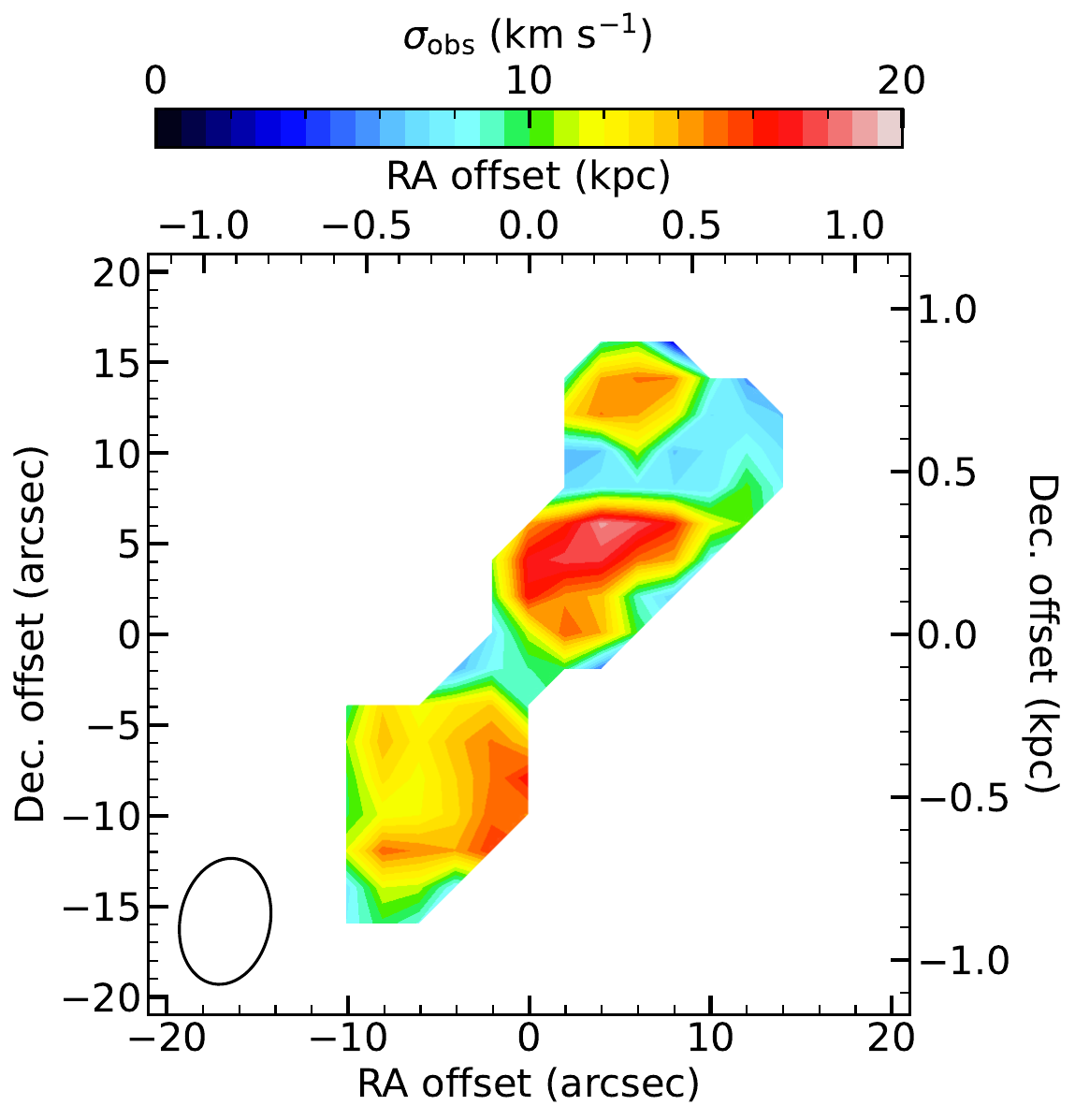}
    \end{subfigure}
    \\
    \begin{subfigure}{.34\linewidth}
    \centering
    \includegraphics[height=5.2cm]{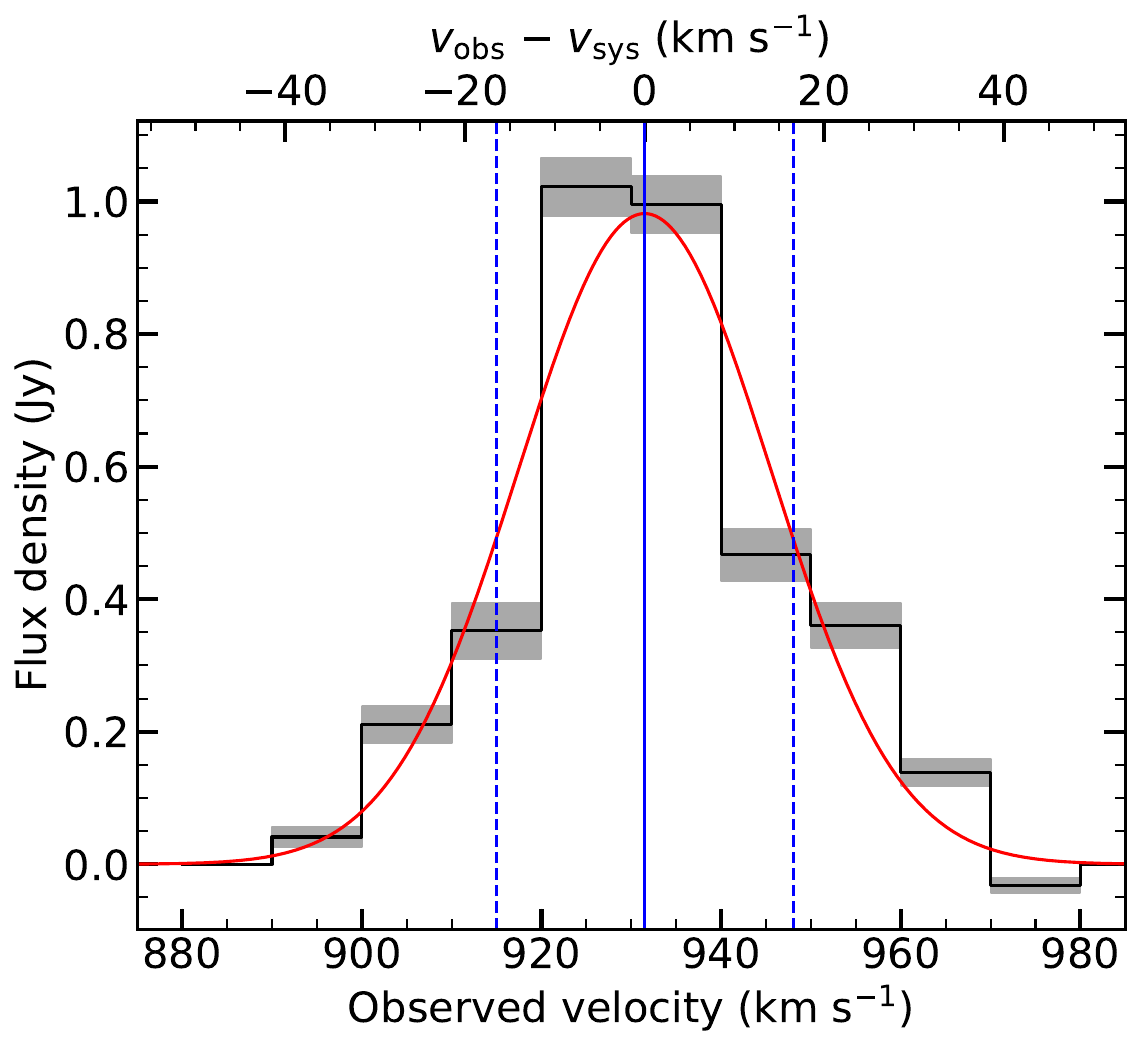}
    \end{subfigure}
    \begin{subfigure}{.32\linewidth}
    \centering
    \includegraphics[height=4.8cm]{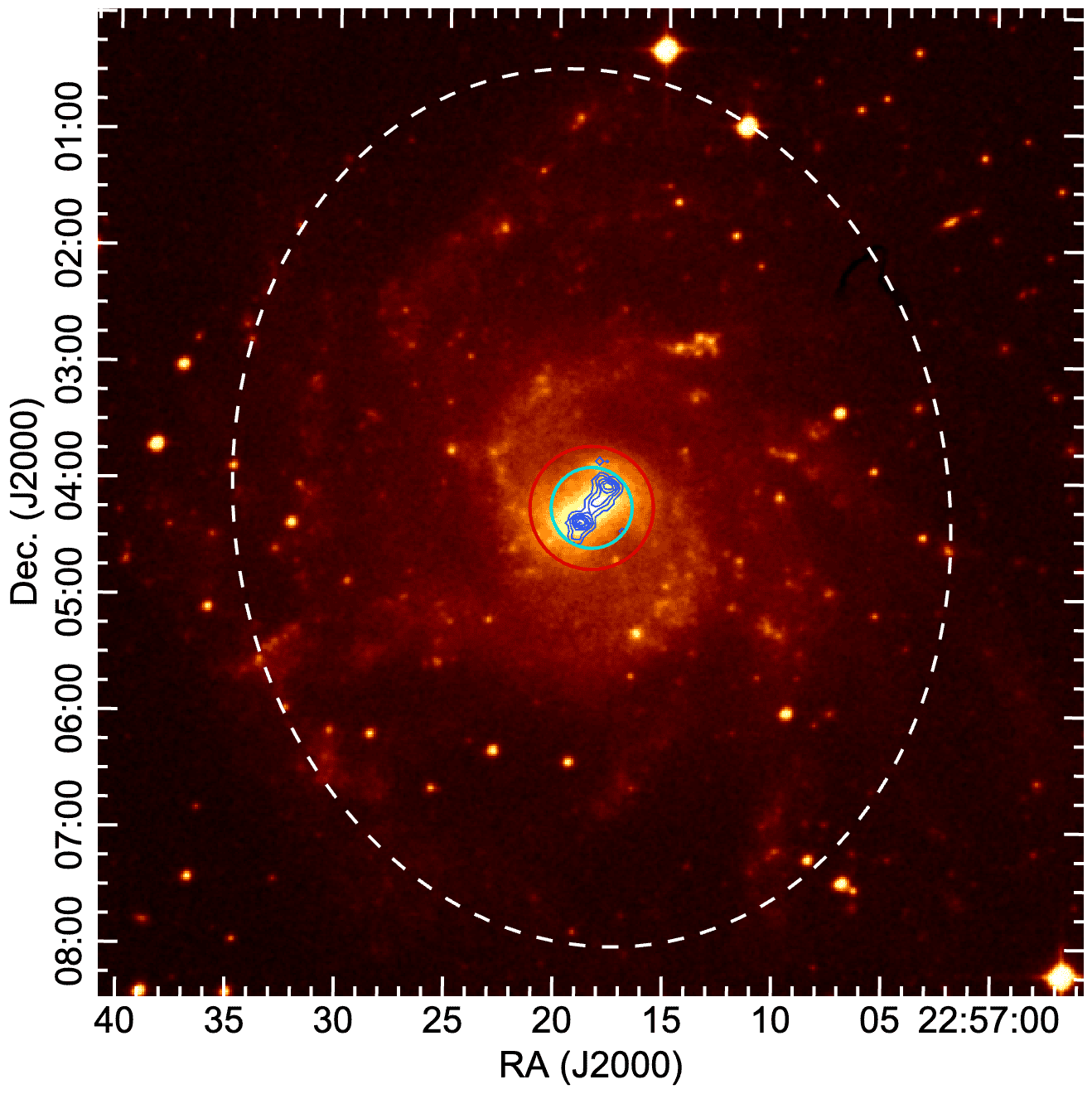}
    \end{subfigure}
    \begin{subfigure}{.33\linewidth}
    \centering
    \includegraphics[height=5.2cm]{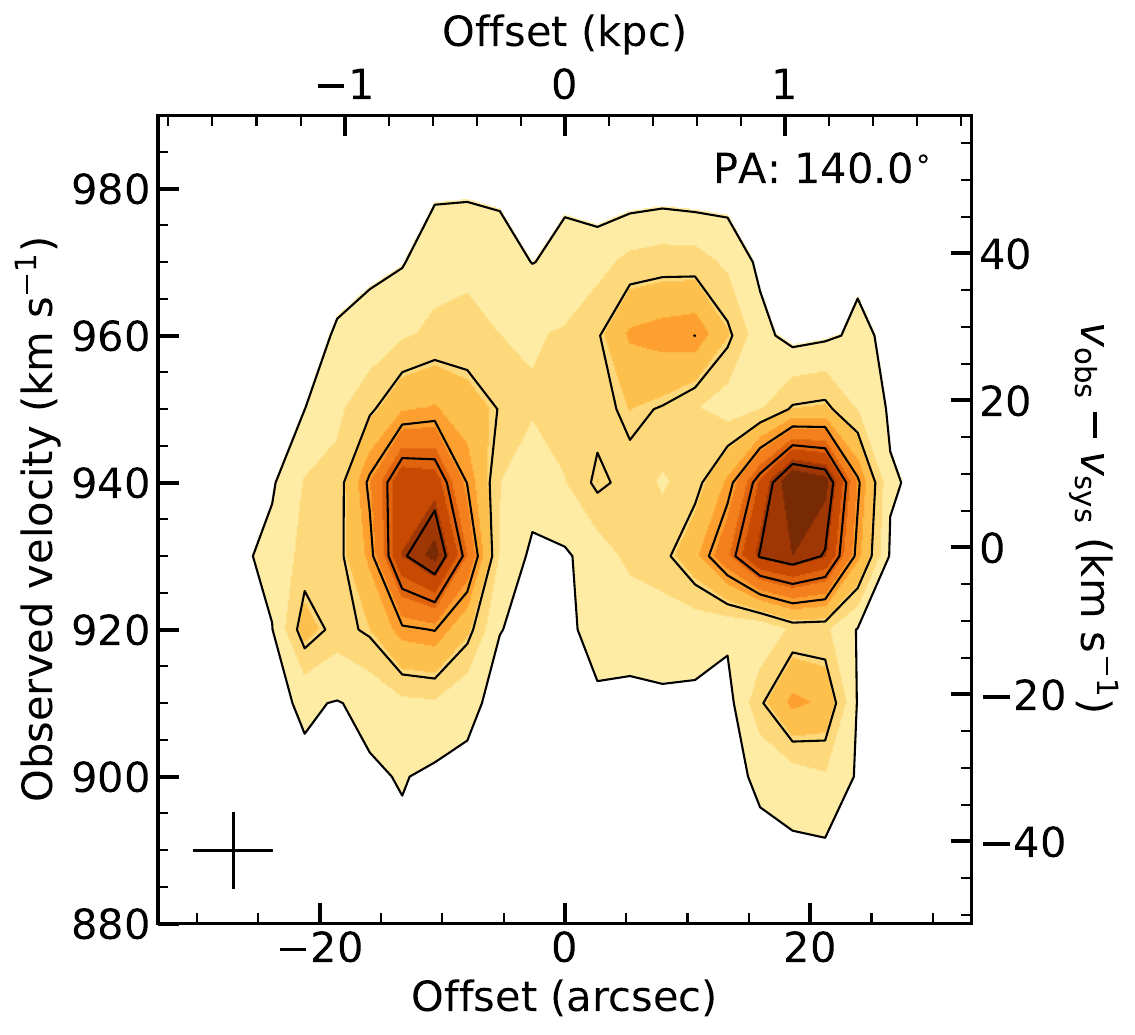}
    \end{subfigure}
    \\
    \begin{subfigure}{.34\linewidth}
    \centering
    \includegraphics[height=6.0cm]{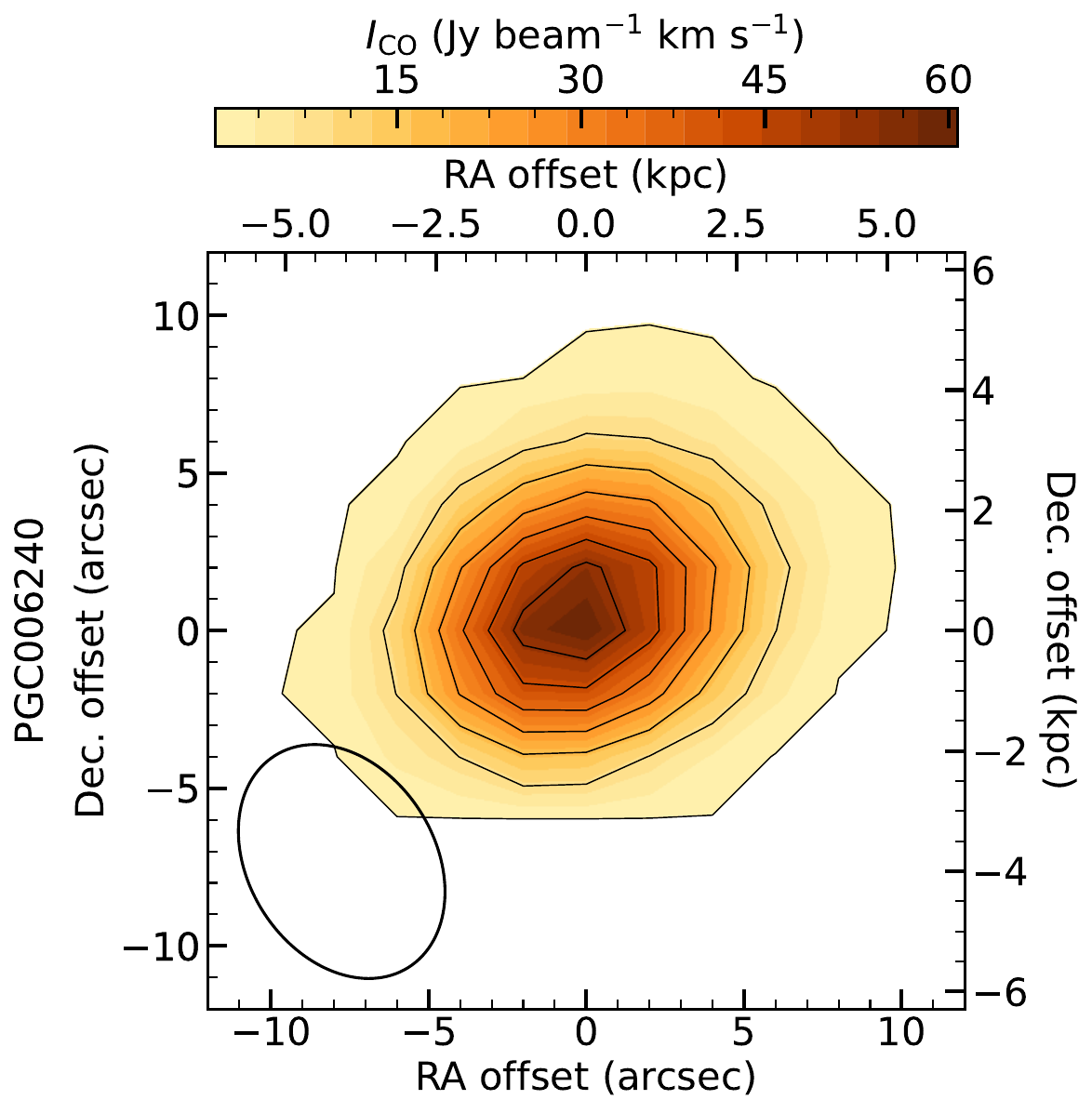}
    \end{subfigure}
    \begin{subfigure}{.32\linewidth}
    \centering
    \includegraphics[height=6.0cm]{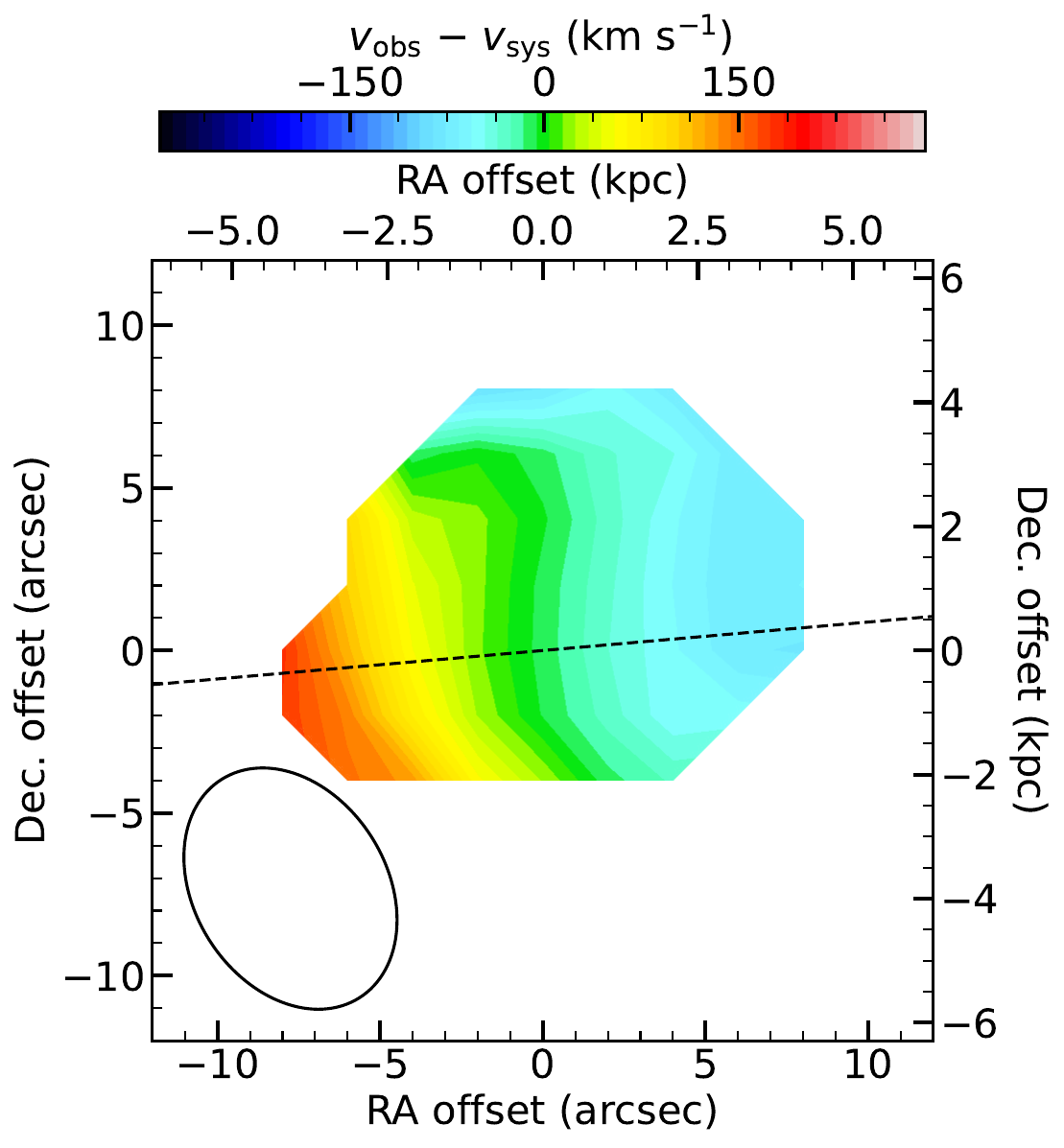}
    \end{subfigure}
    \begin{subfigure}{.32\linewidth}
    \centering
    \includegraphics[height=6.0cm]{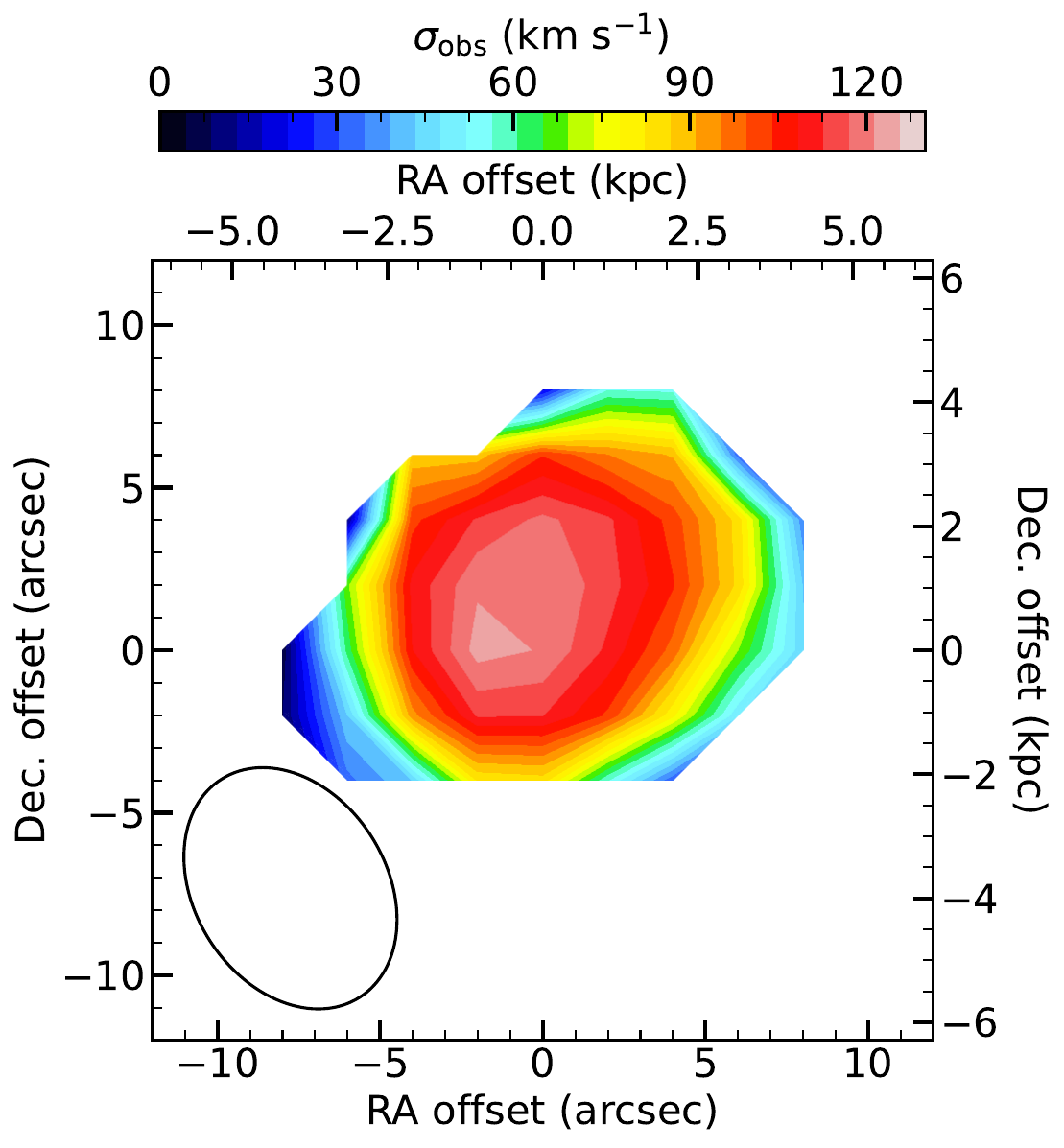}
    \end{subfigure}
    \\
    \begin{subfigure}{.34\linewidth}
    \centering
    \includegraphics[height=5.2cm]{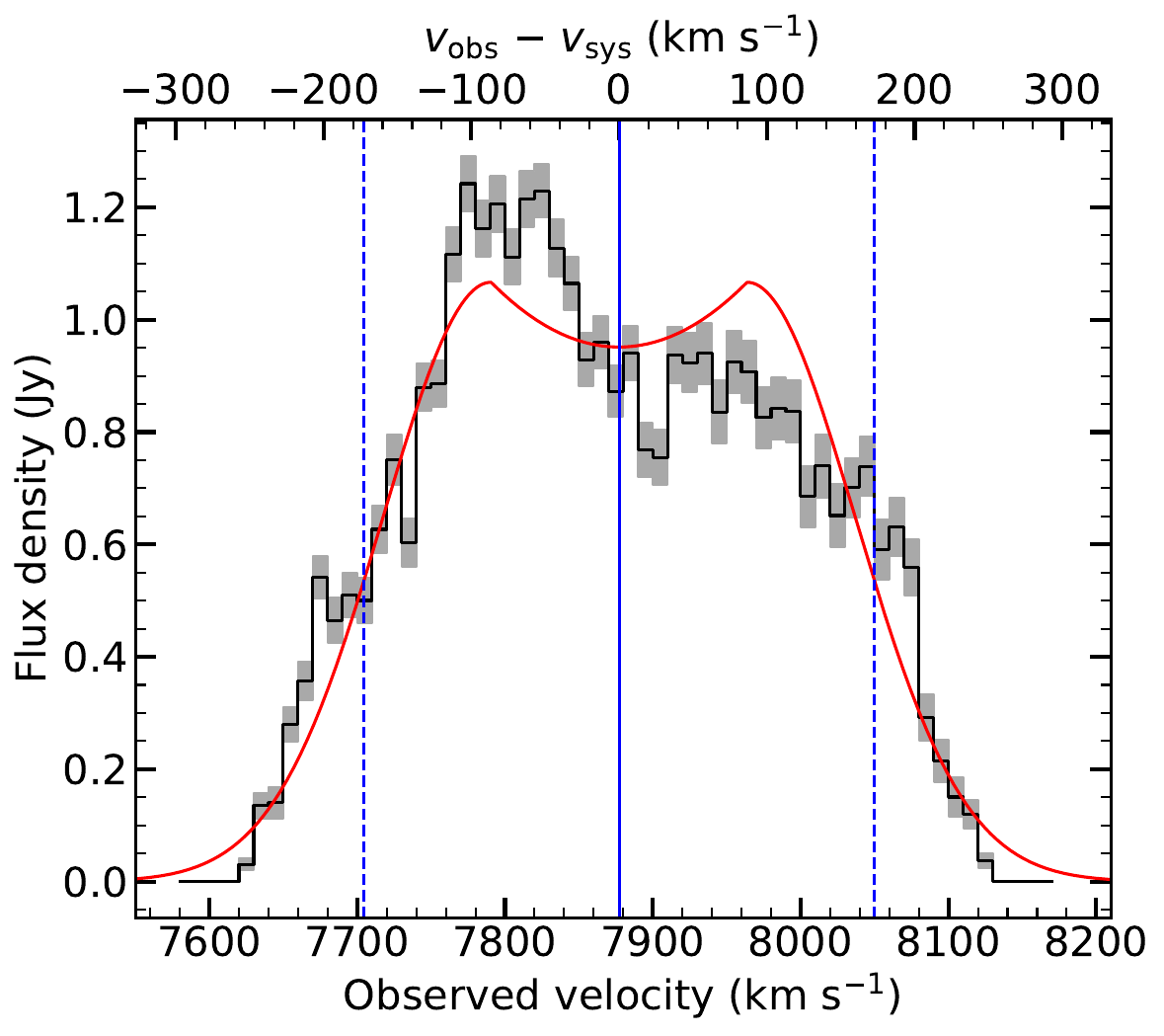}
    \end{subfigure}
    \begin{subfigure}{.32\linewidth}
    \centering
    \includegraphics[height=4.8cm]{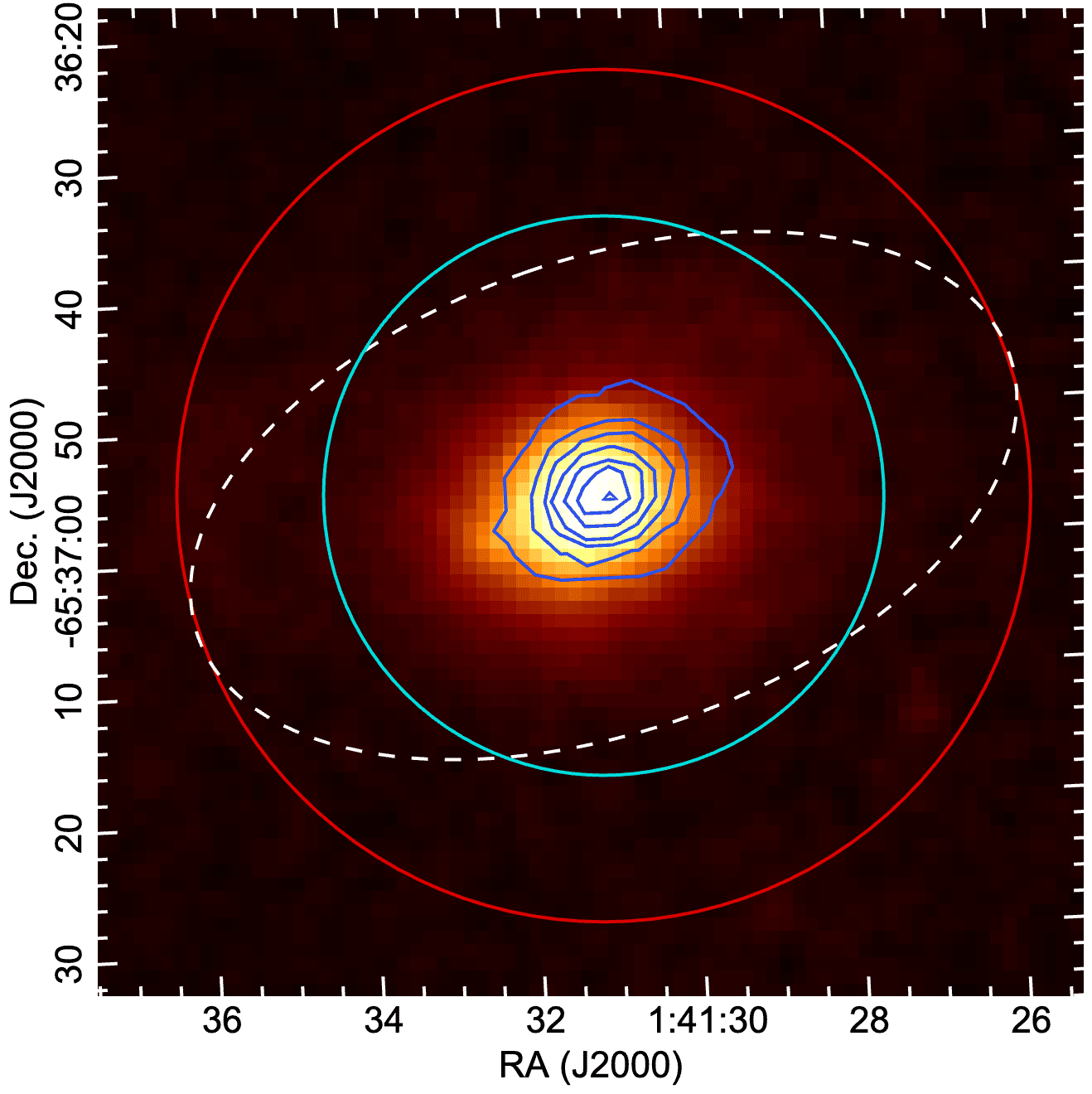}
    \end{subfigure}
    \begin{subfigure}{.33\linewidth}
    \centering
    \includegraphics[height=5.2cm]{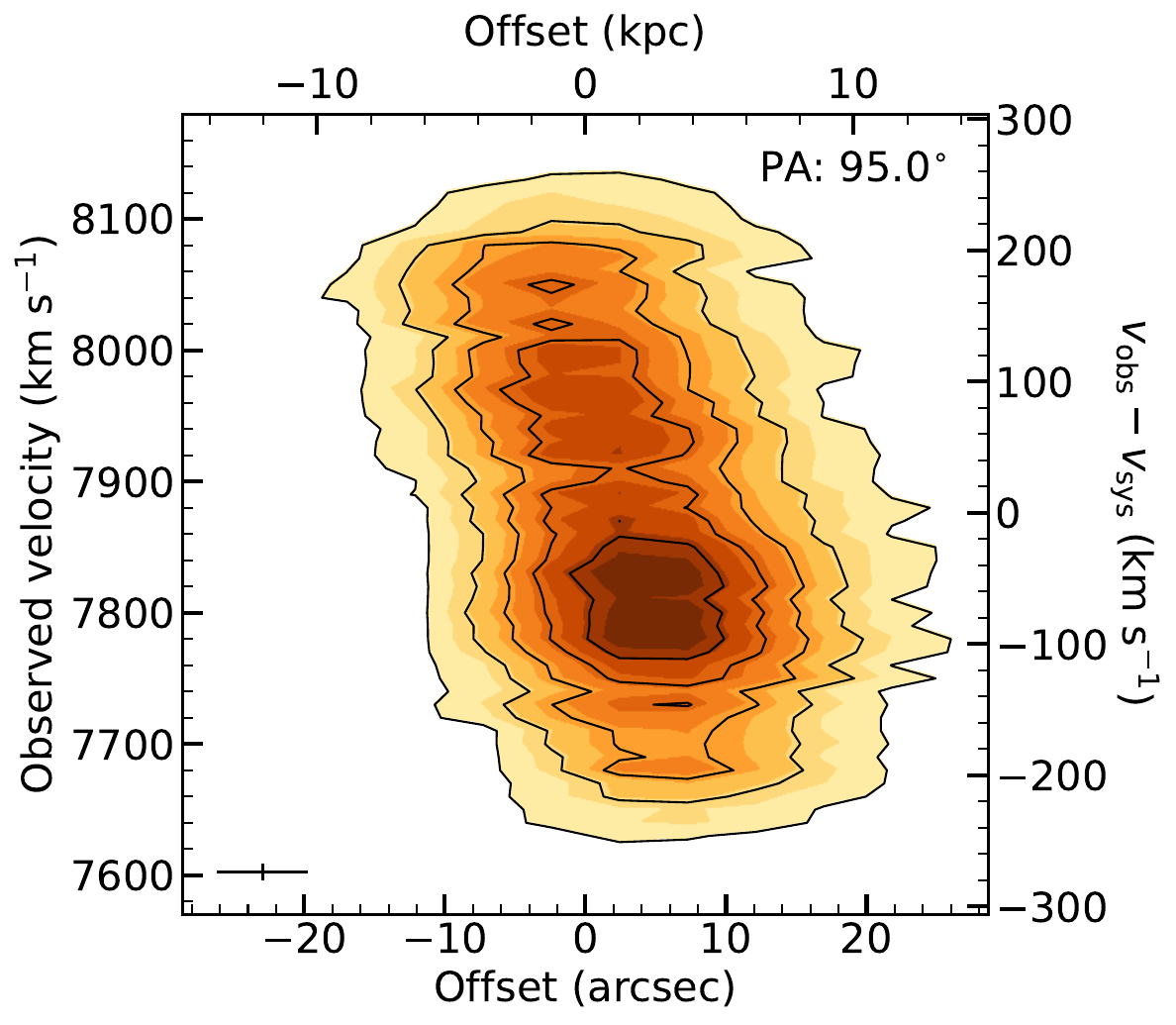}
    \end{subfigure}
    \caption[]{continued.}
\end{figure*}

\begin{figure*}
    \ContinuedFloat
    \centering
    \begin{subfigure}{.34\linewidth}
    \centering
    \includegraphics[height=6.0cm]{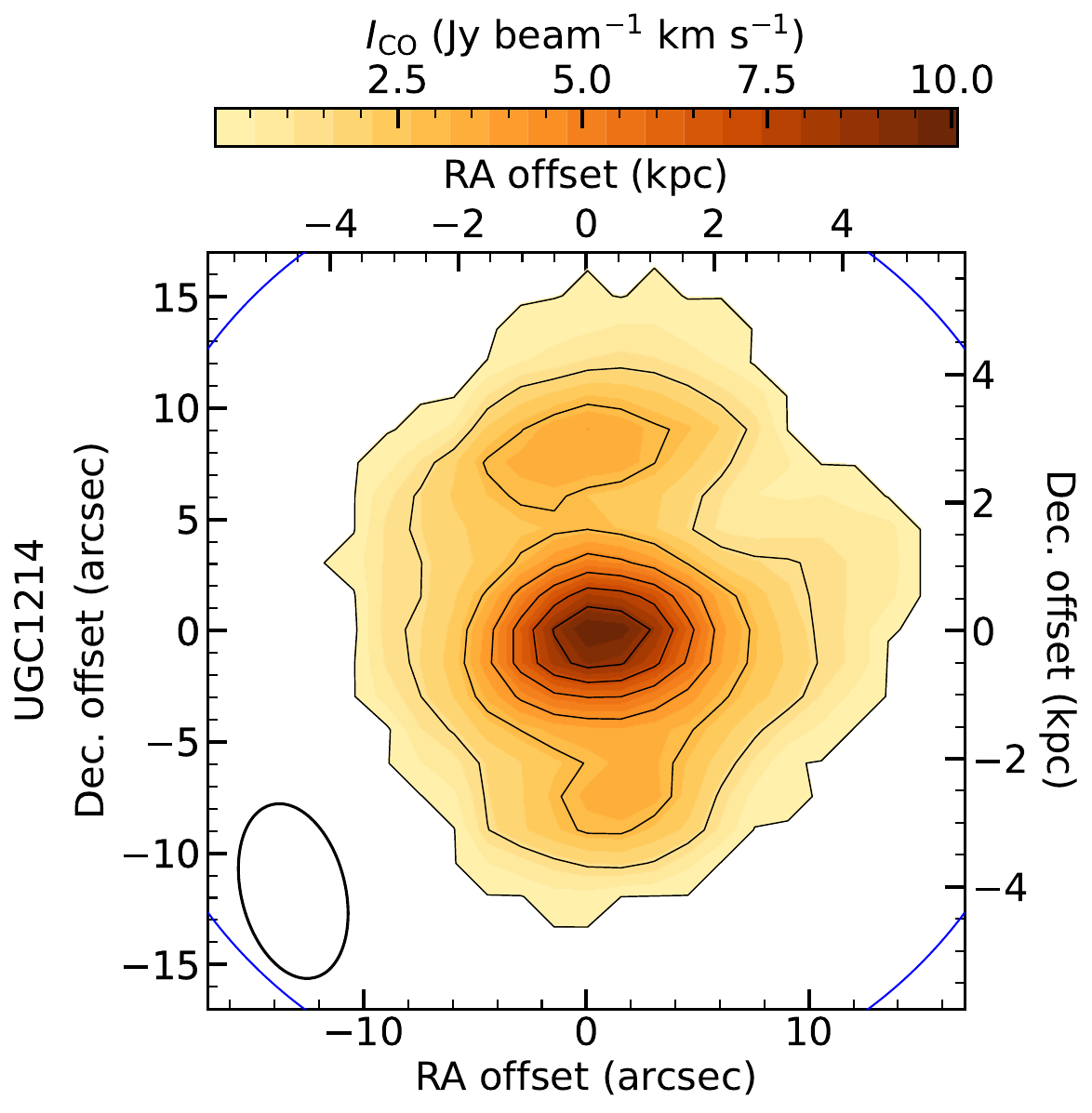}
    \end{subfigure}
    \begin{subfigure}{.32\linewidth}
    \centering
    \includegraphics[height=6.0cm]{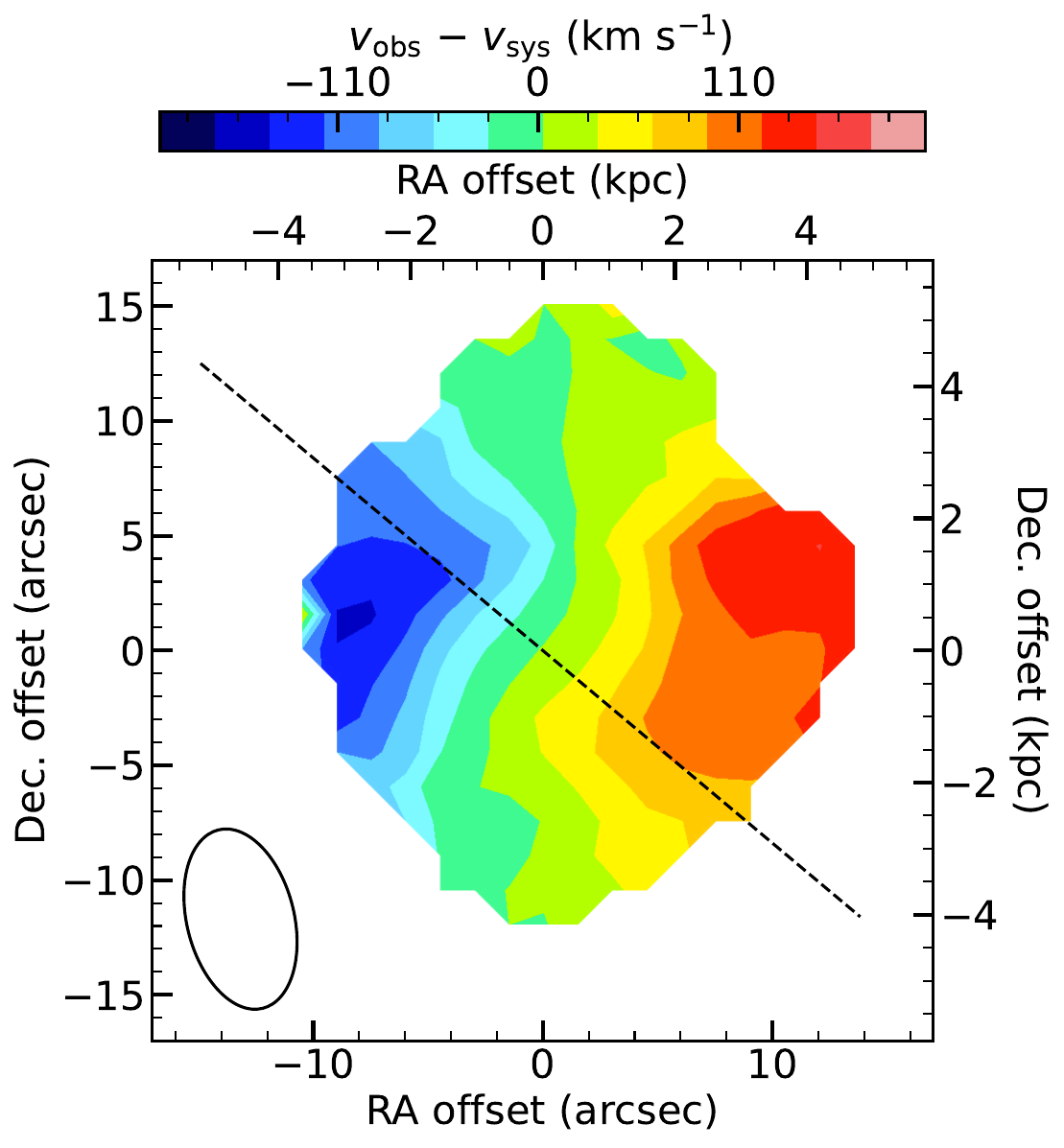}
    \end{subfigure}
    \begin{subfigure}{.32\linewidth}
    \centering
    \includegraphics[height=6.0cm]{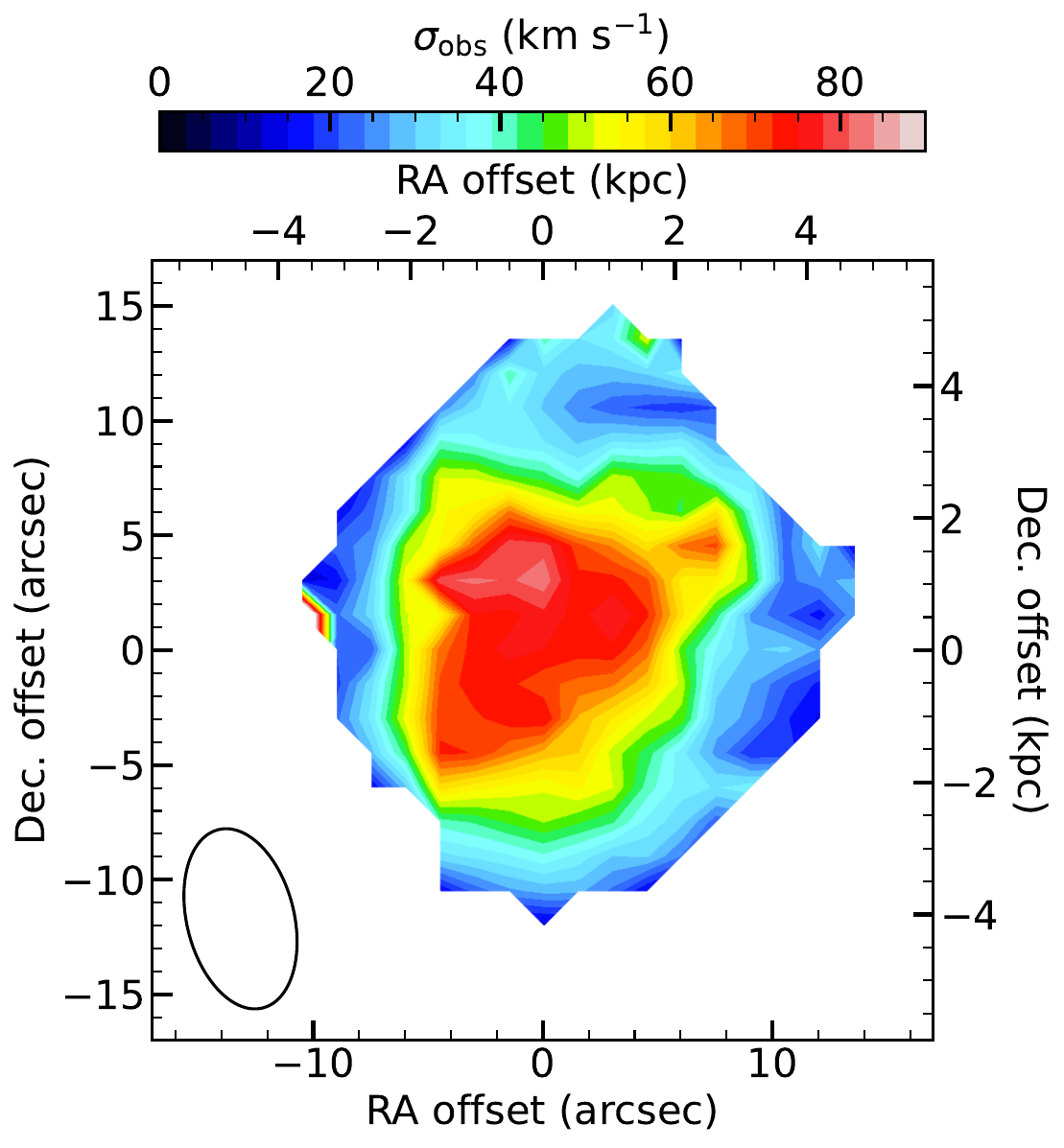}
    \end{subfigure}
    \\  
    \begin{subfigure}{.34\linewidth}
    \centering
    \includegraphics[height=5.2cm]{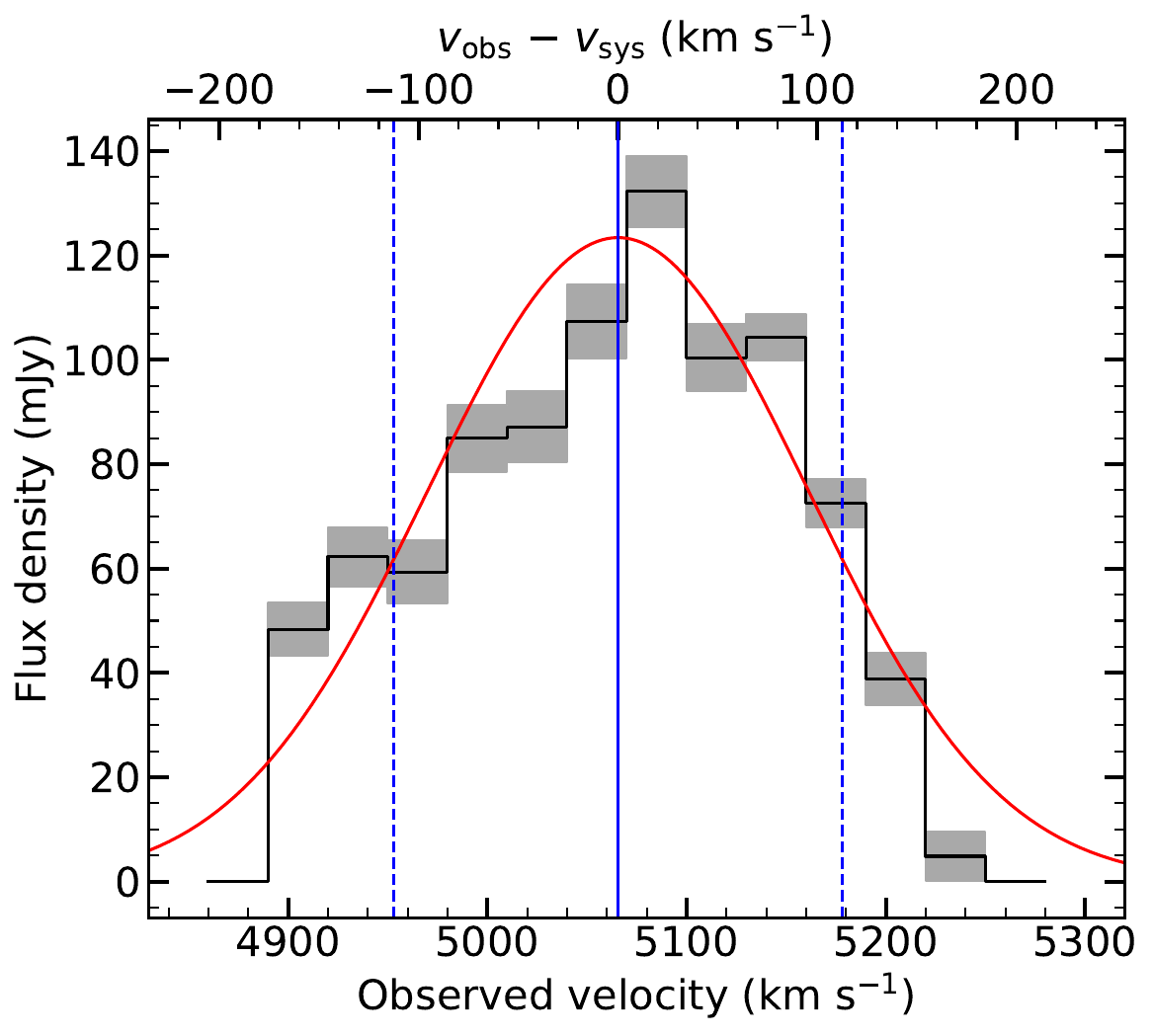}
    \end{subfigure}
    \begin{subfigure}{.32\linewidth}
    \centering
    \includegraphics[height=4.8cm]{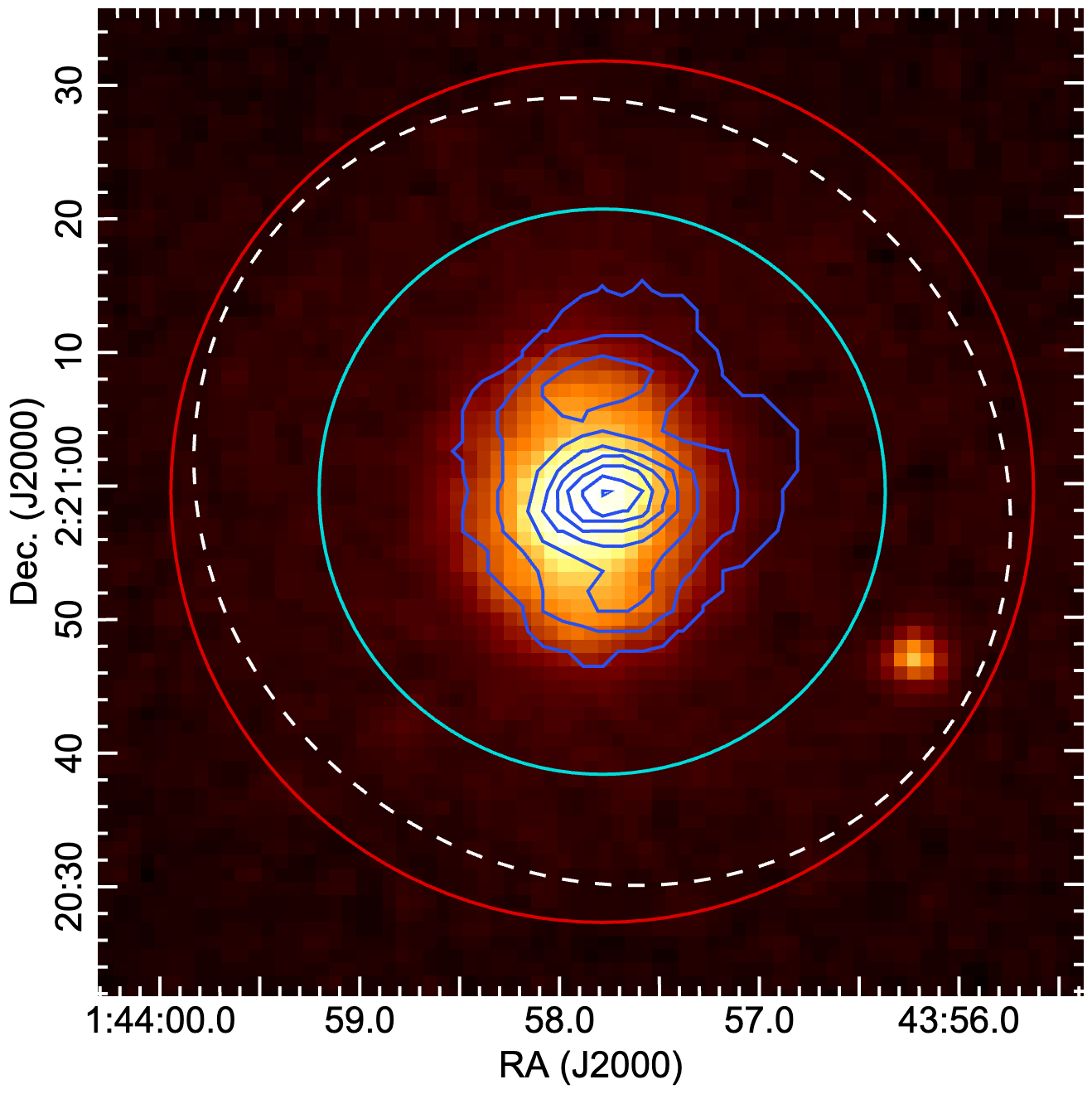}
    \end{subfigure}
    \begin{subfigure}{.33\linewidth}
    \centering
    \includegraphics[height=5.2cm]{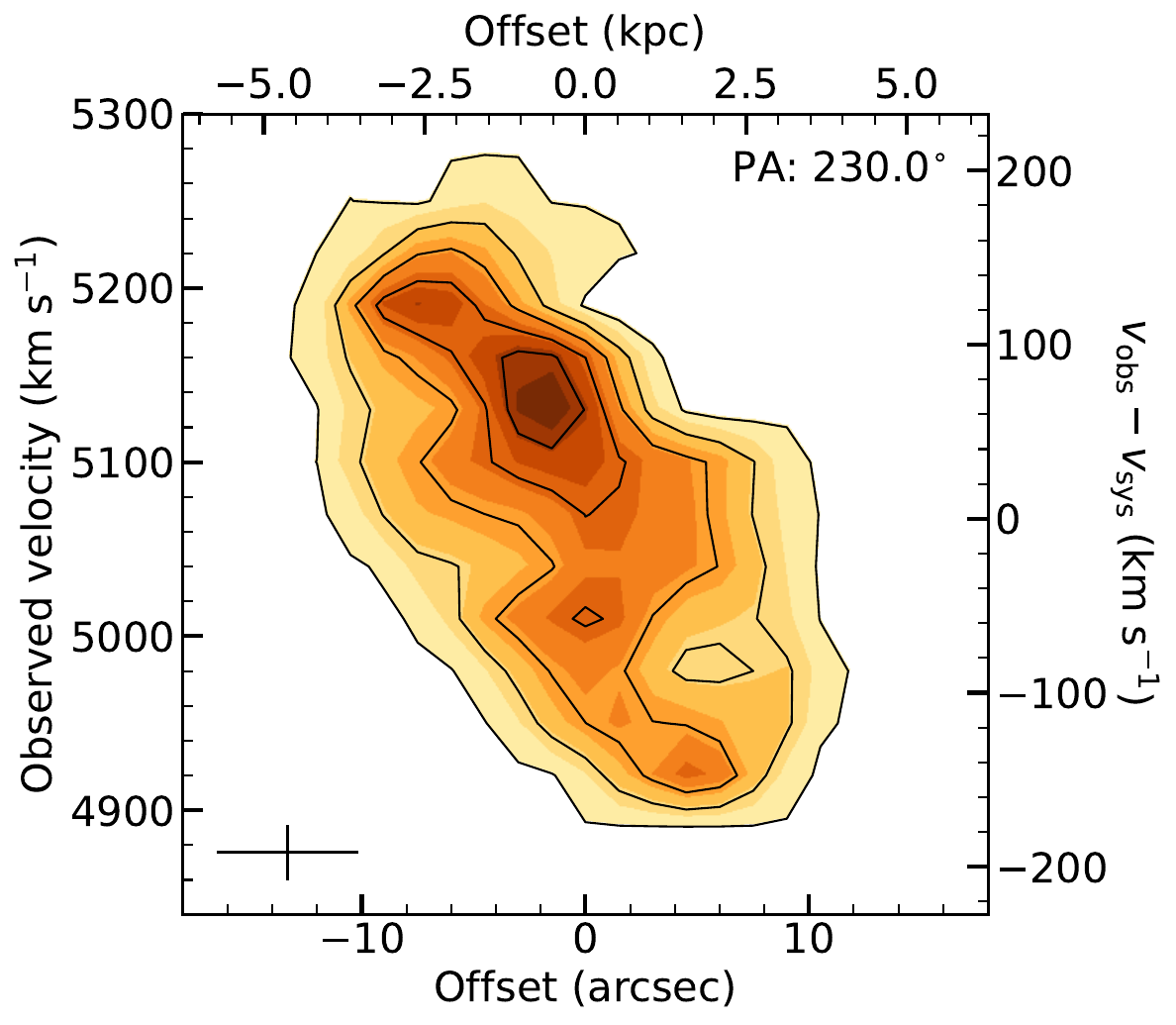}
    \end{subfigure}\\
    \begin{subfigure}{.34\linewidth}
    \centering
    \includegraphics[height=6.0cm]{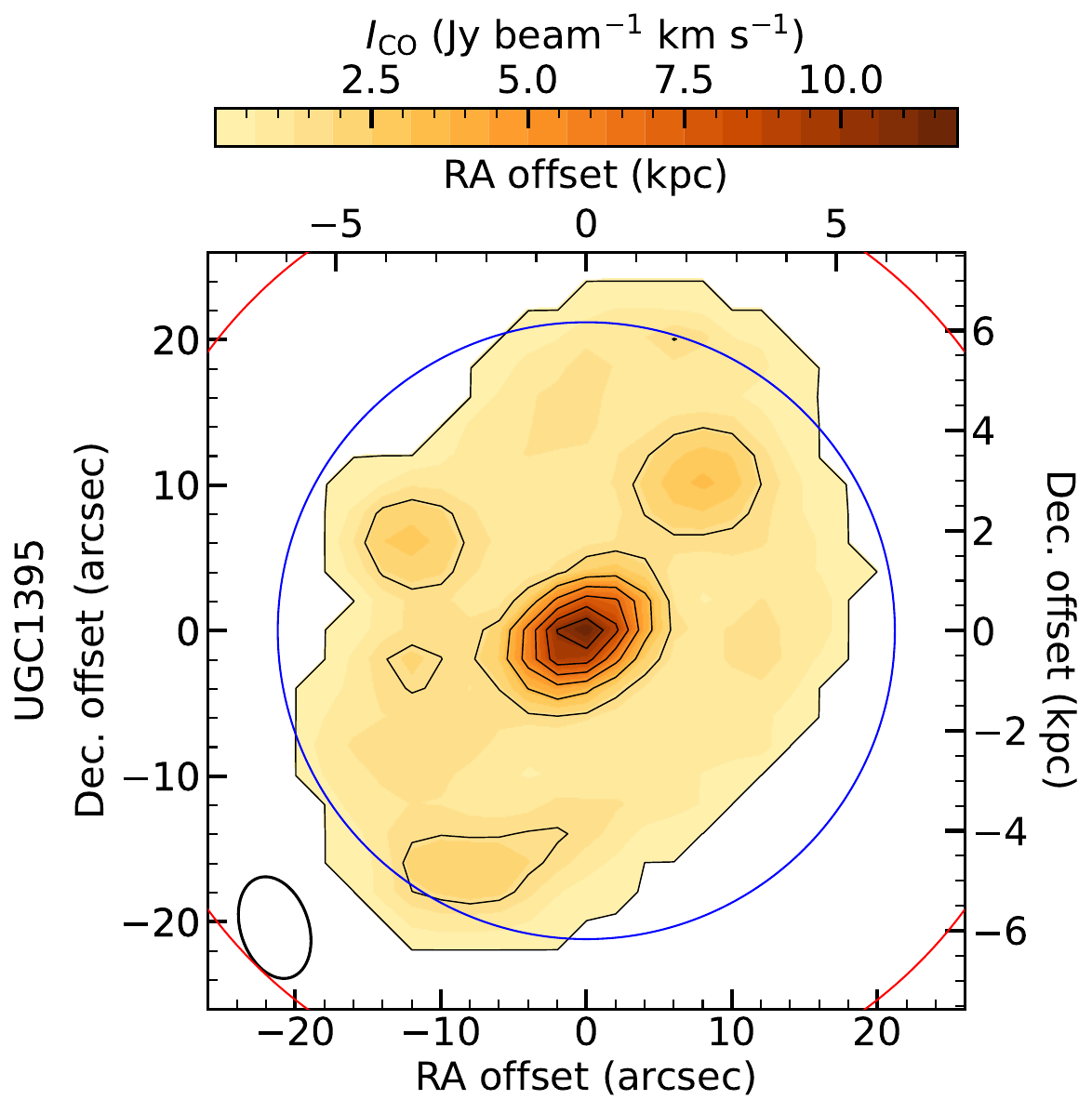}
    \end{subfigure}
    \begin{subfigure}{.32\linewidth}
    \centering
    \includegraphics[height=6.0cm]{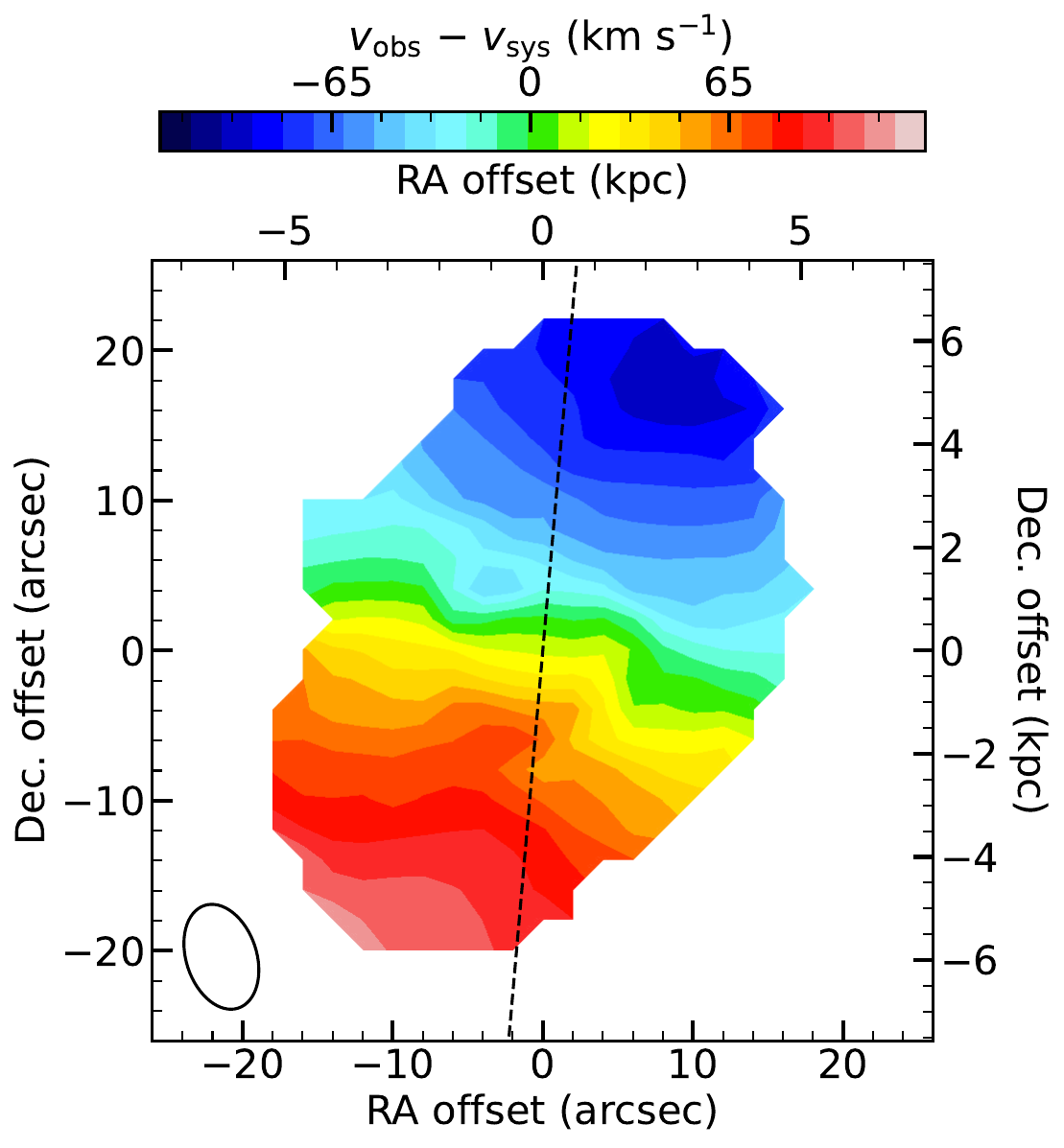}
    \end{subfigure}
    \begin{subfigure}{.32\linewidth}
    \centering
    \includegraphics[height=6.0cm]{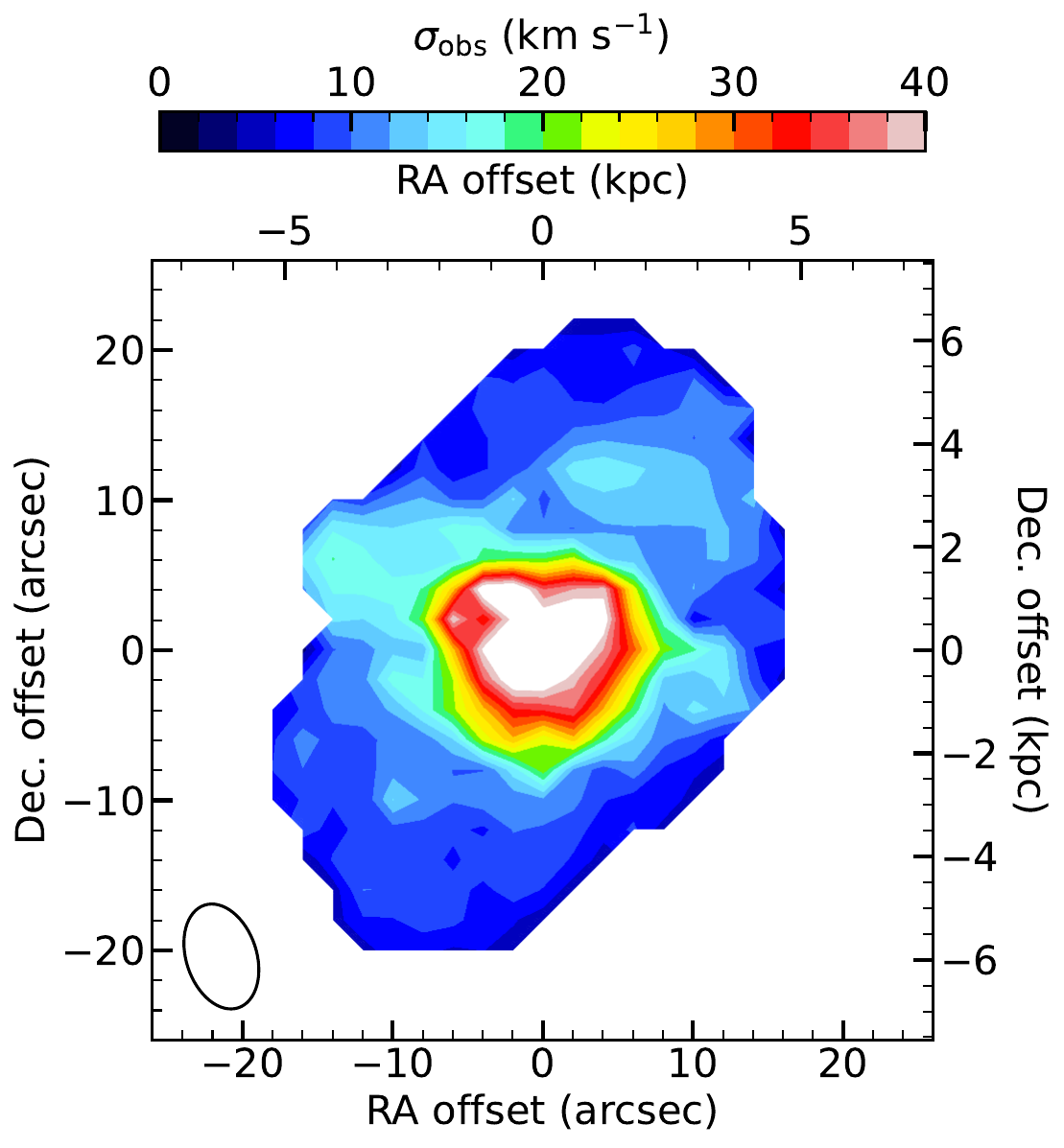}
    \end{subfigure}
    \\
    \begin{subfigure}{.34\linewidth}
    \centering
    \includegraphics[height=5.2cm]{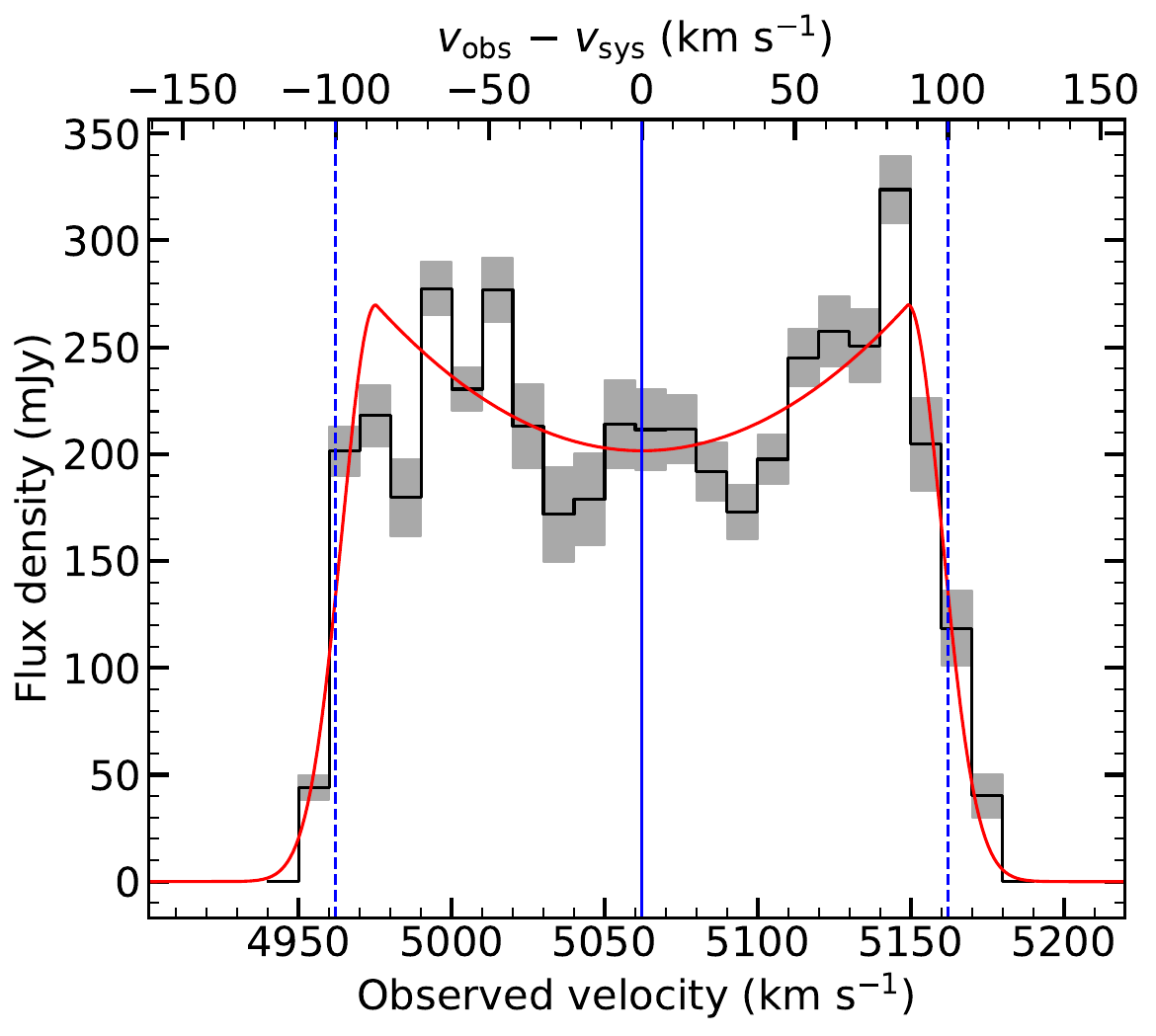}
    \end{subfigure}
    \begin{subfigure}{.32\linewidth}
    \centering
    \includegraphics[height=4.8cm]{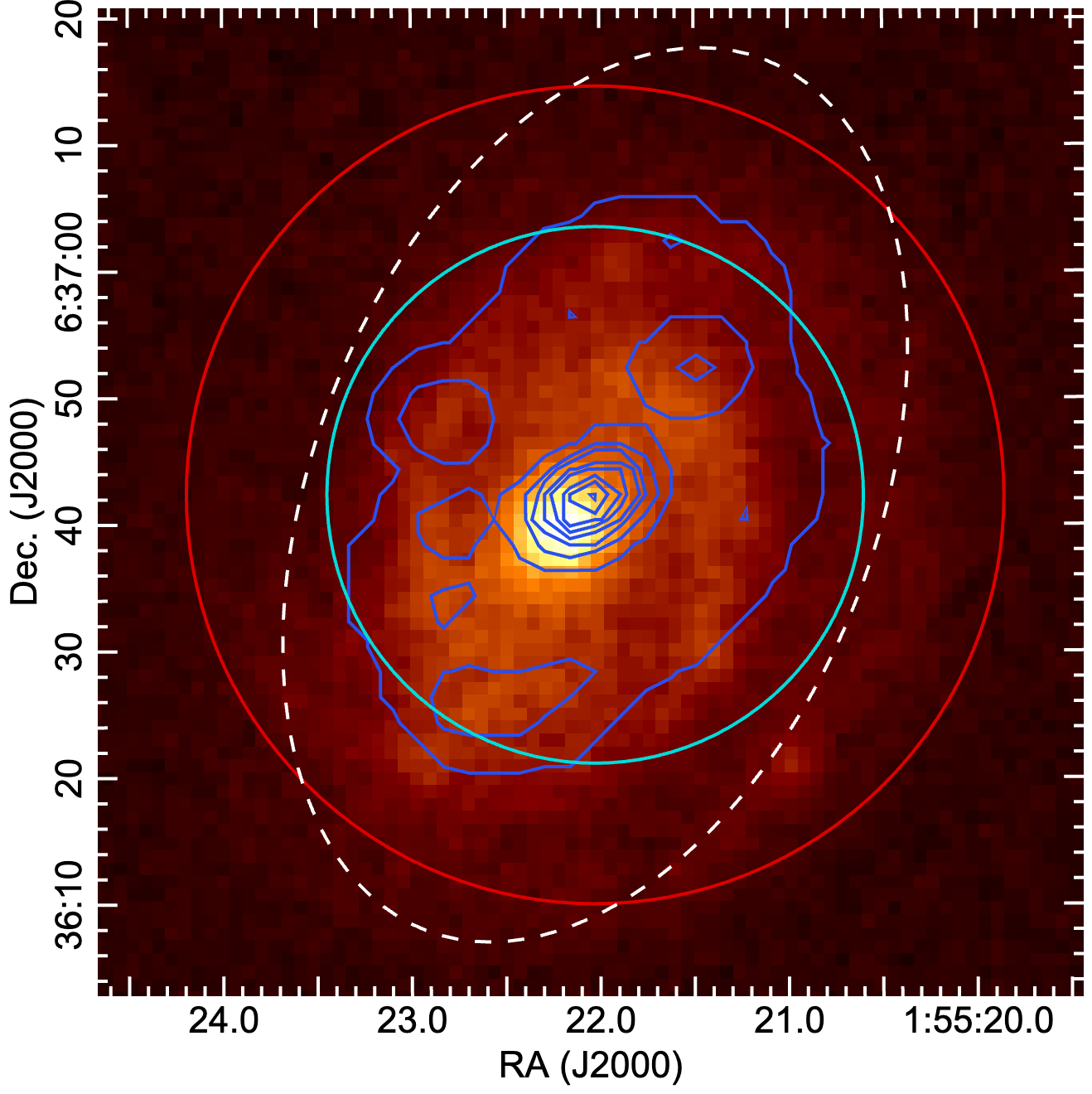}
    \end{subfigure}
    \begin{subfigure}{.33\linewidth}
    \centering
    \includegraphics[height=5.2cm]{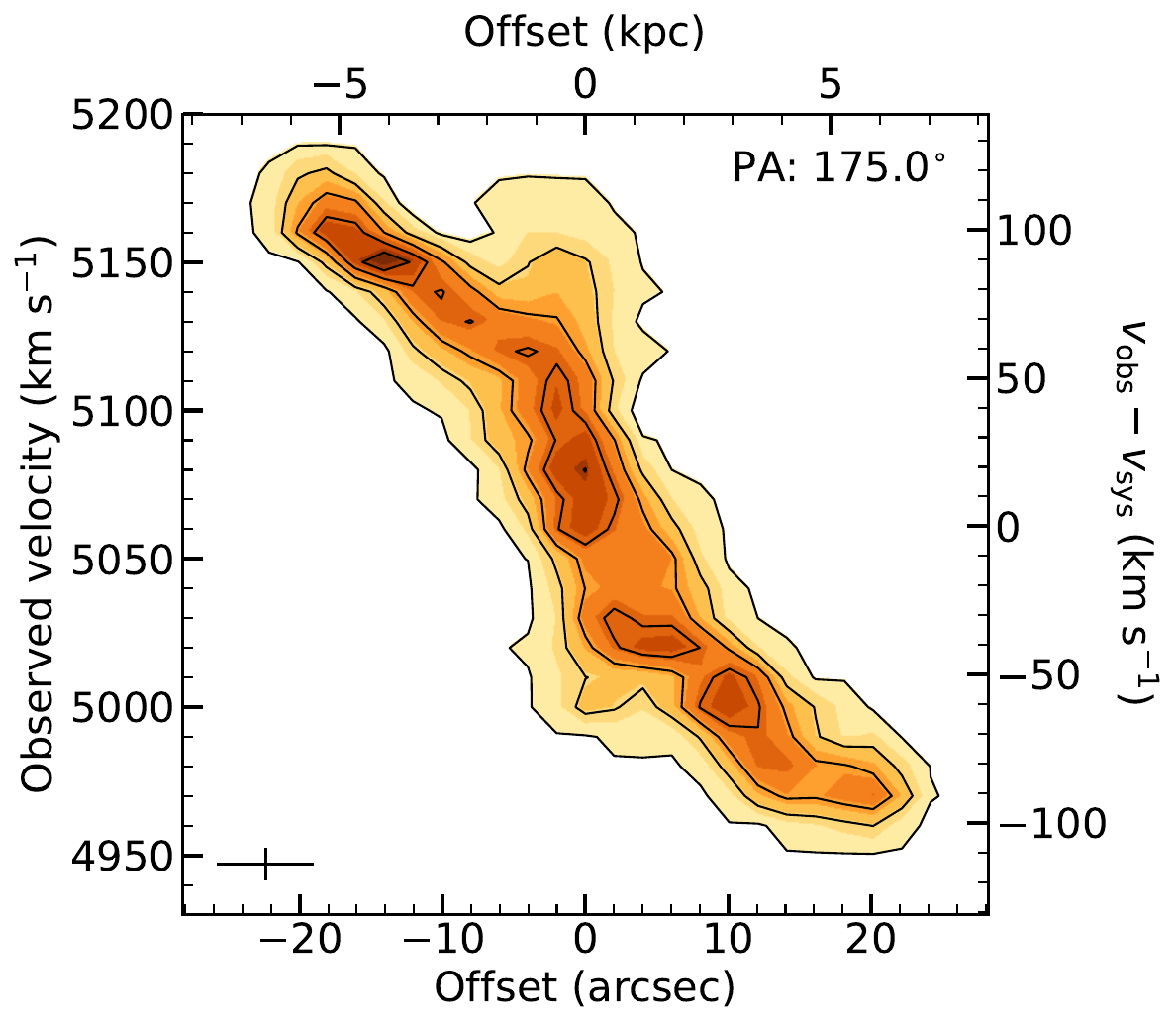}
    \end{subfigure}
    \caption[]{continued.}
\end{figure*}


\begin{figure*}
    \begin{subfigure}{.34\linewidth}
    \includegraphics[height=6.0cm]{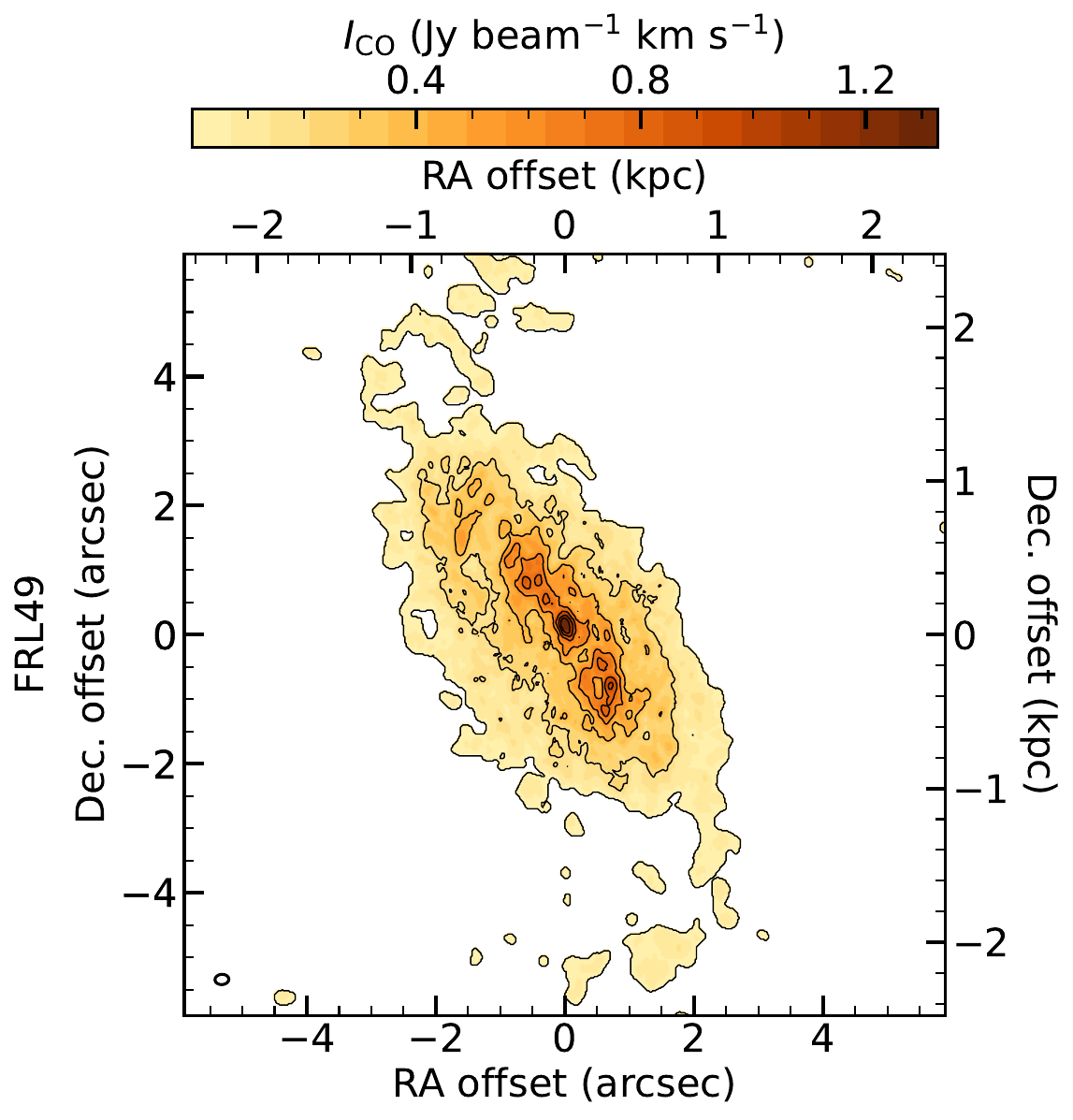}
    \end{subfigure}
    \begin{subfigure}{.32\linewidth}
    \centering
    \includegraphics[height=6.0cm]{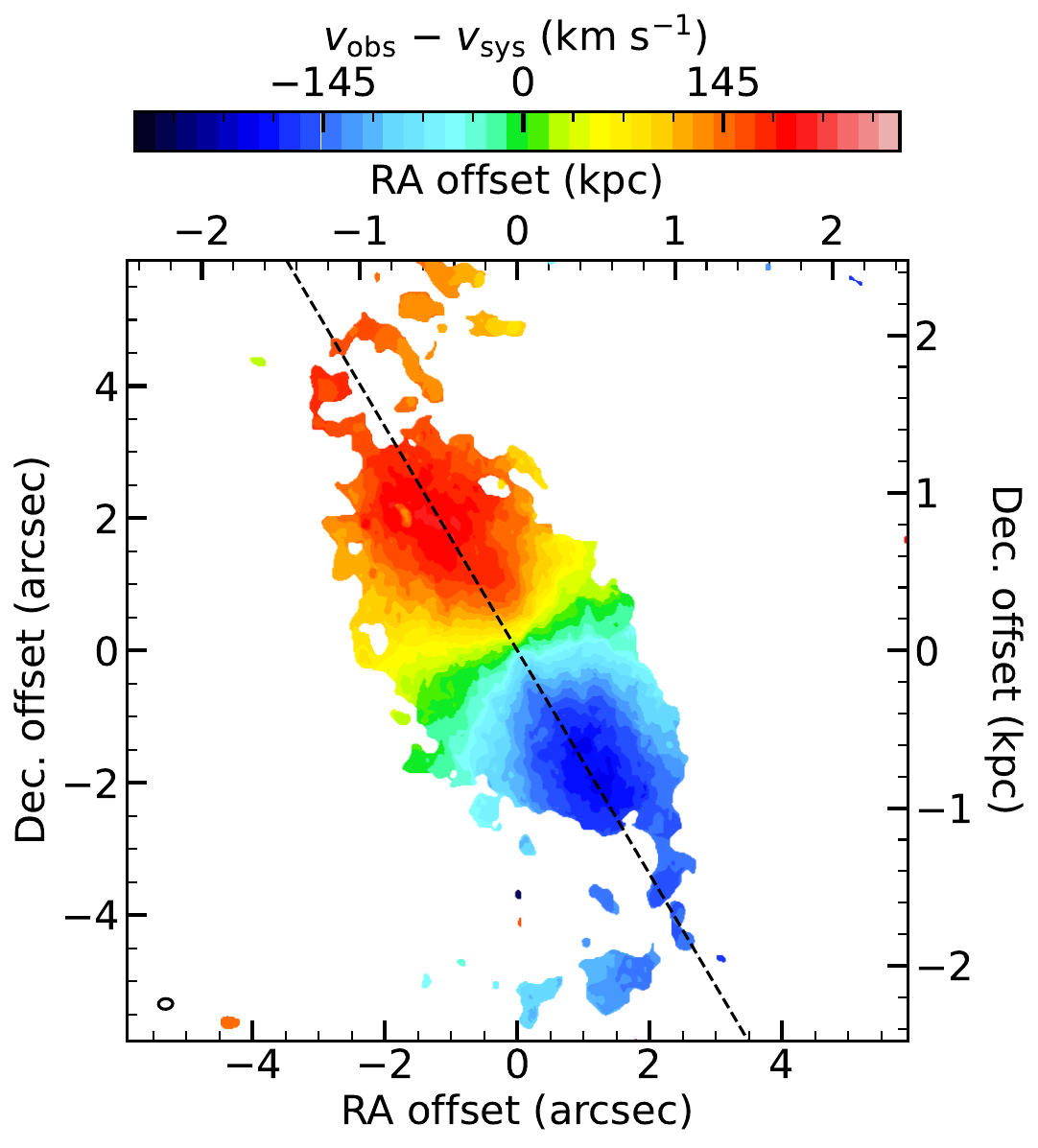}
    \end{subfigure}
    \begin{subfigure}{.32\linewidth}
    \centering
    \includegraphics[height=6.0cm]{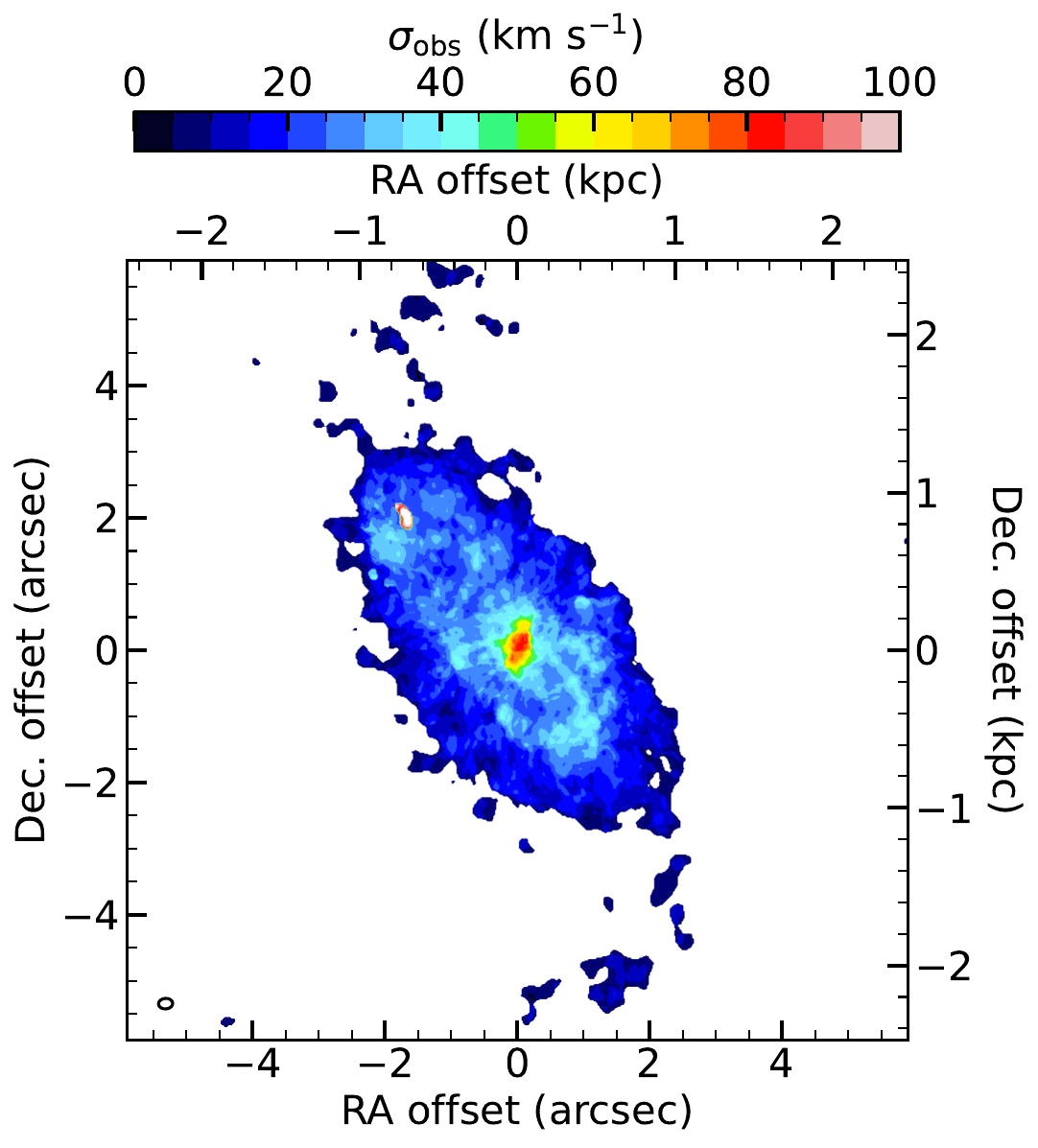}
    \end{subfigure}
    \\
    \begin{subfigure}{.33\linewidth}
    \centering
    \includegraphics[height=5.2cm]{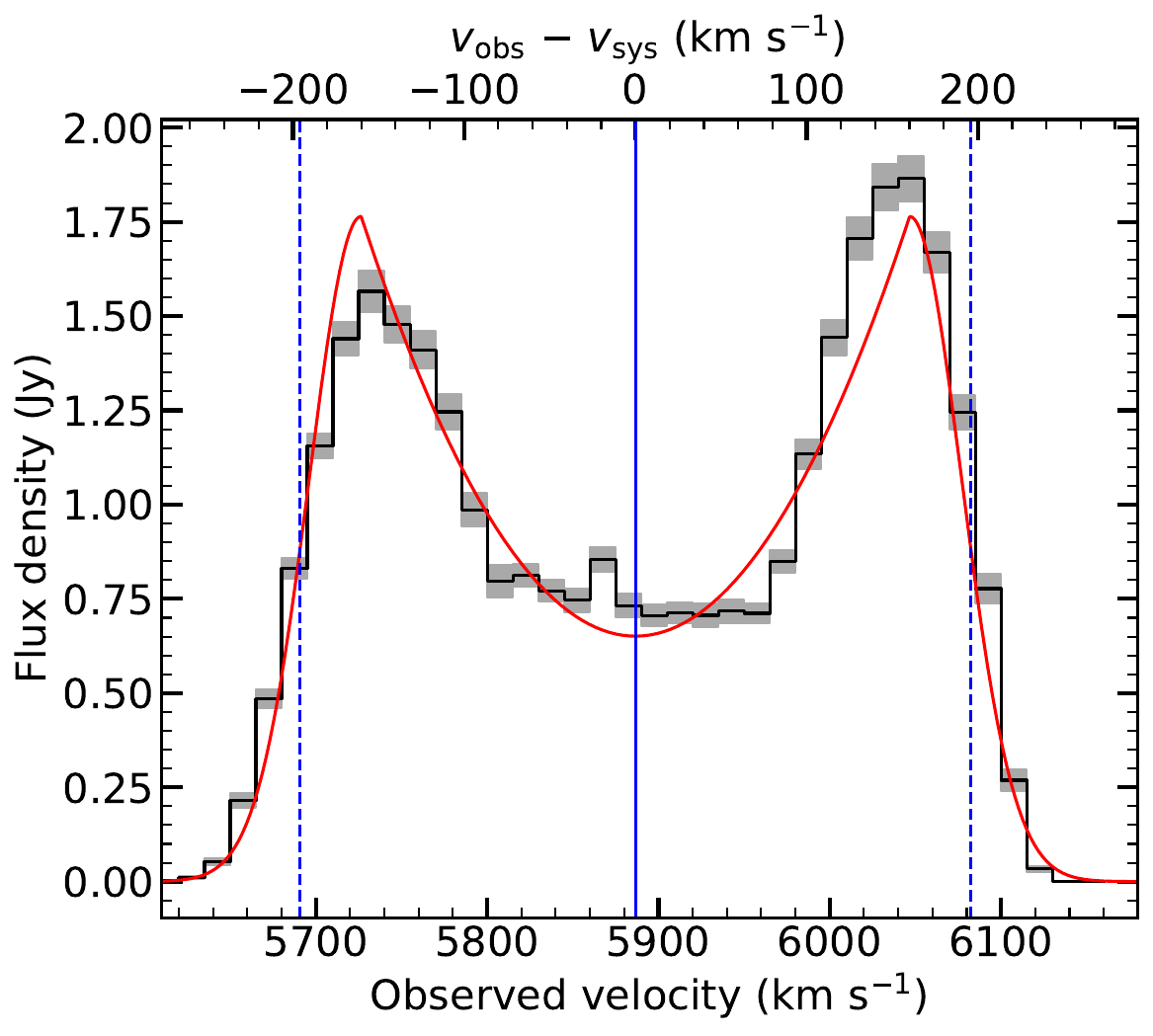}
    \end{subfigure}
    \begin{subfigure}{.32\linewidth}
    \centering
    \includegraphics[height=4.8cm]{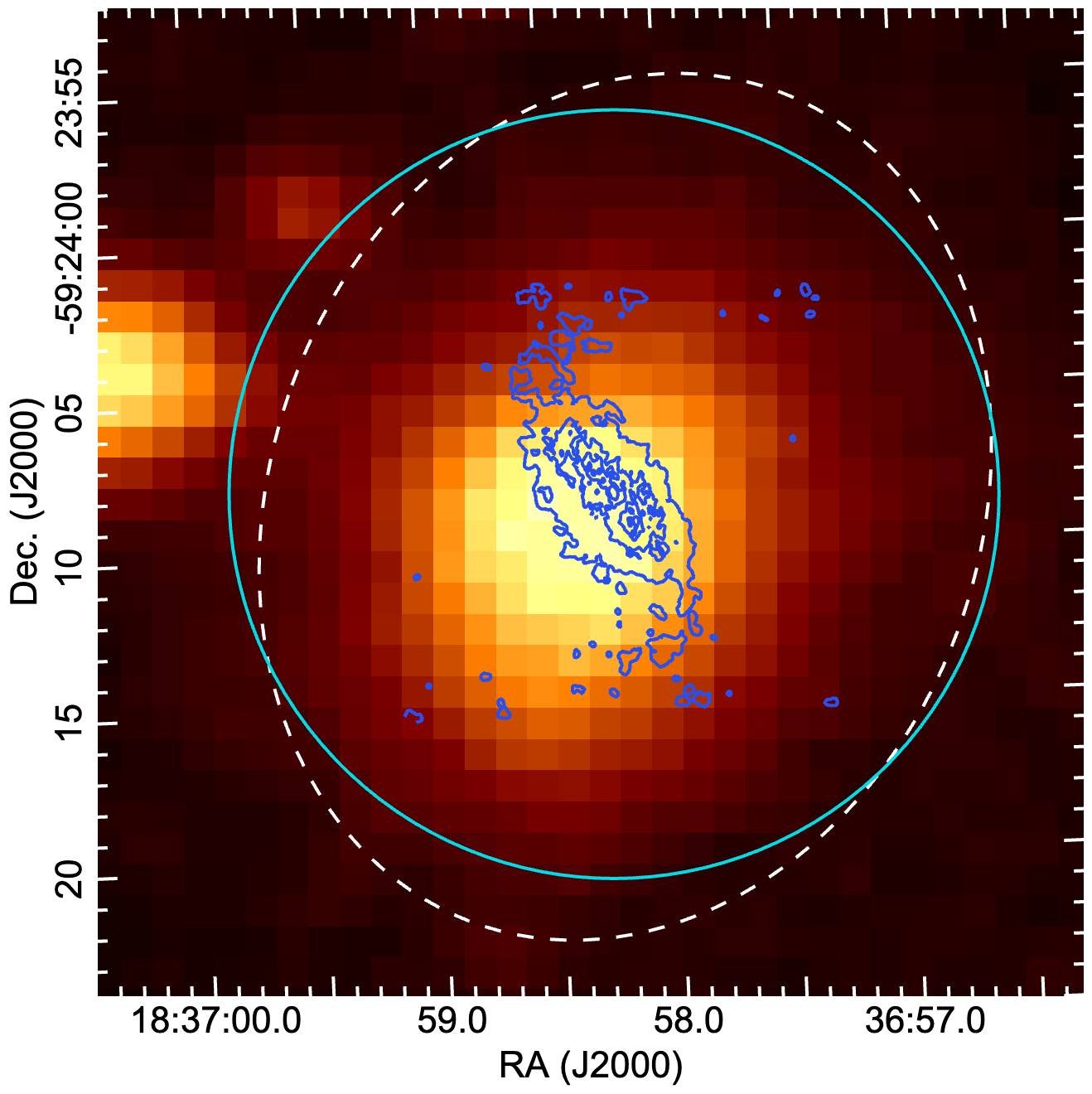}
    \end{subfigure}
    \begin{subfigure}{.33\linewidth}
    \centering
    \includegraphics[height=5.2cm]{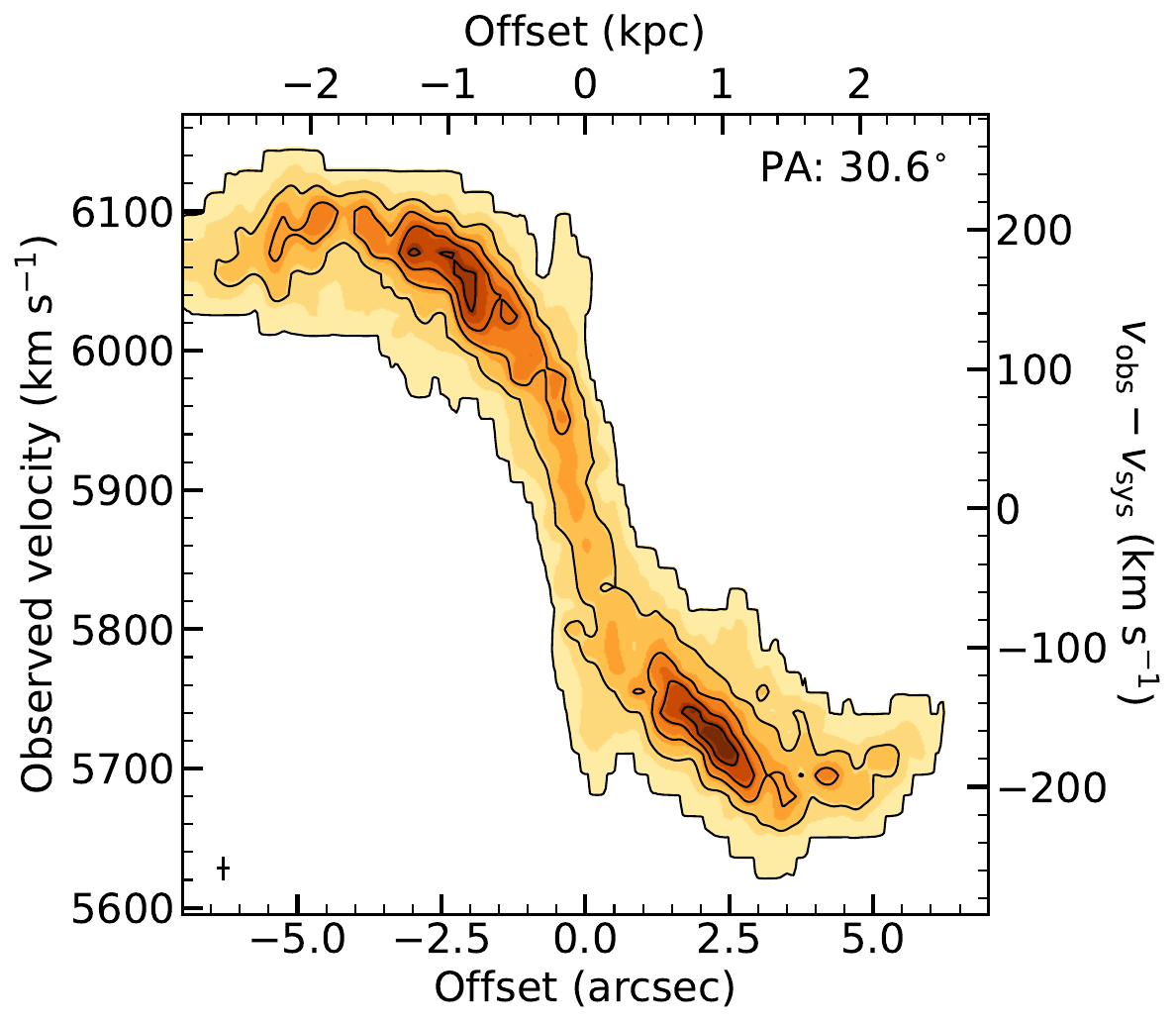}
    \end{subfigure}
    \begin{subfigure}{.34\linewidth}
    \includegraphics[height=6.0cm]{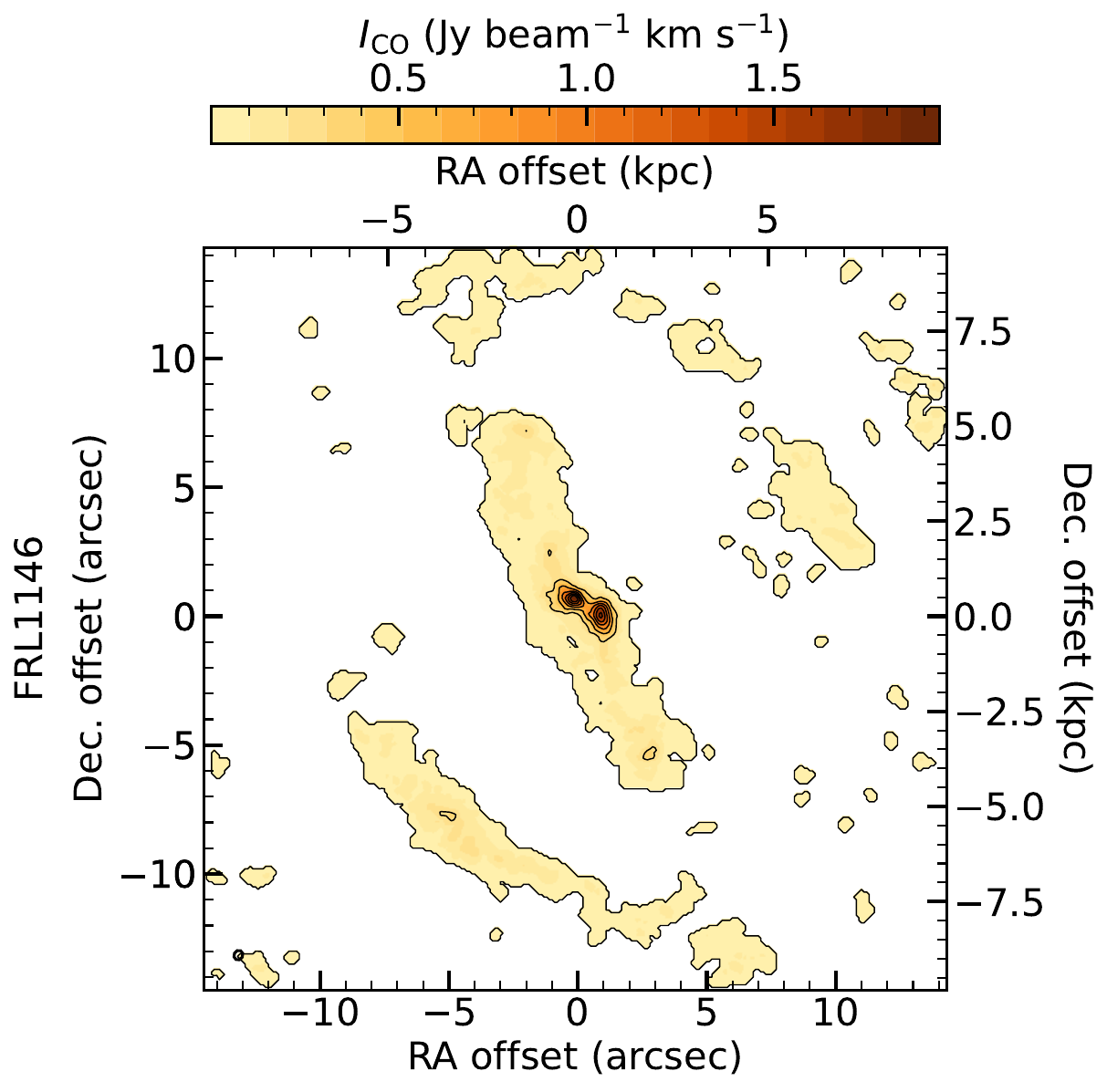}
    \end{subfigure}
    \begin{subfigure}{.32\linewidth}
    \centering
    \includegraphics[height=6.0cm]{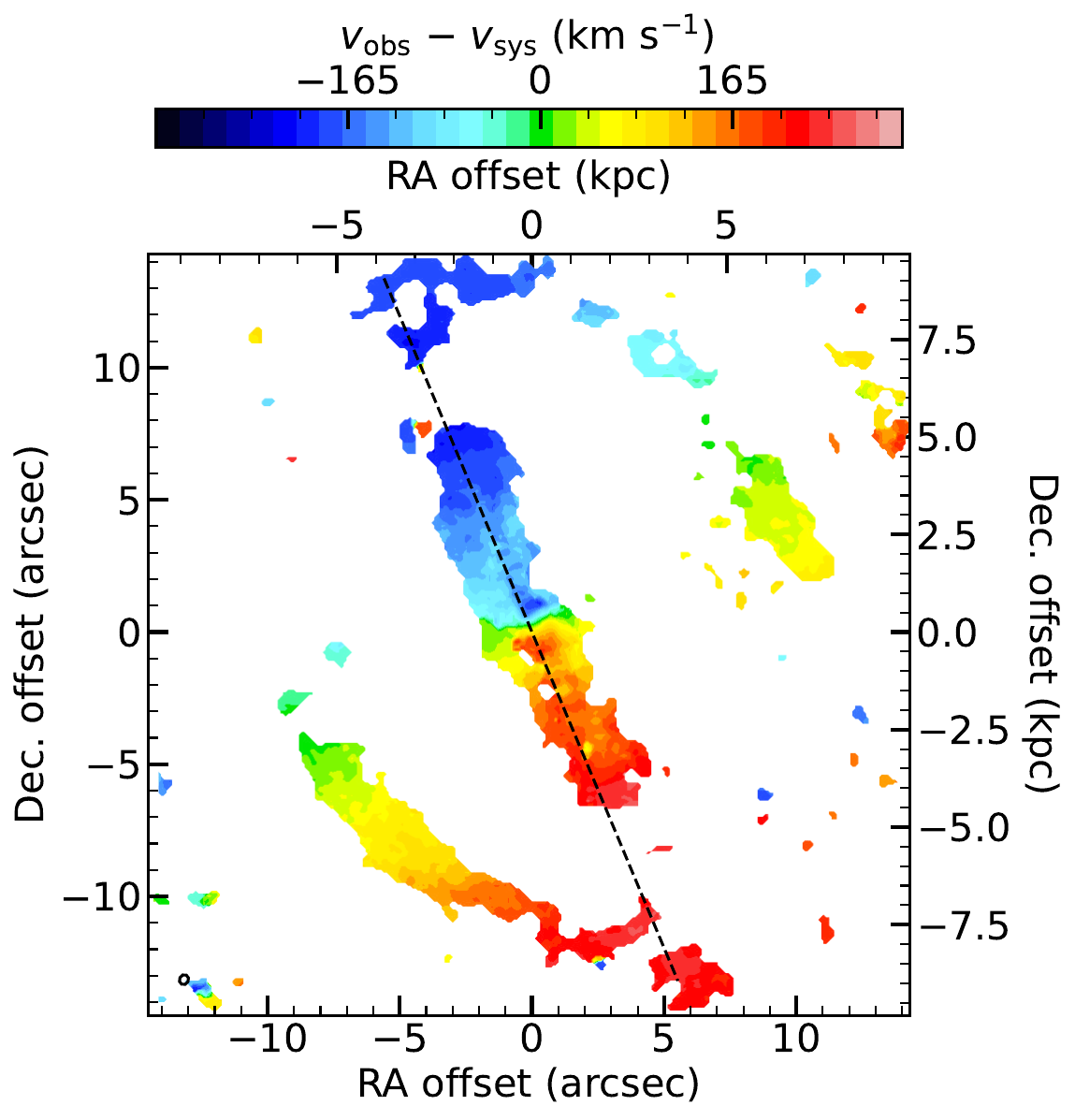}
    \end{subfigure}
    \begin{subfigure}{.32\linewidth}
    \centering
    \includegraphics[height=6.0cm]{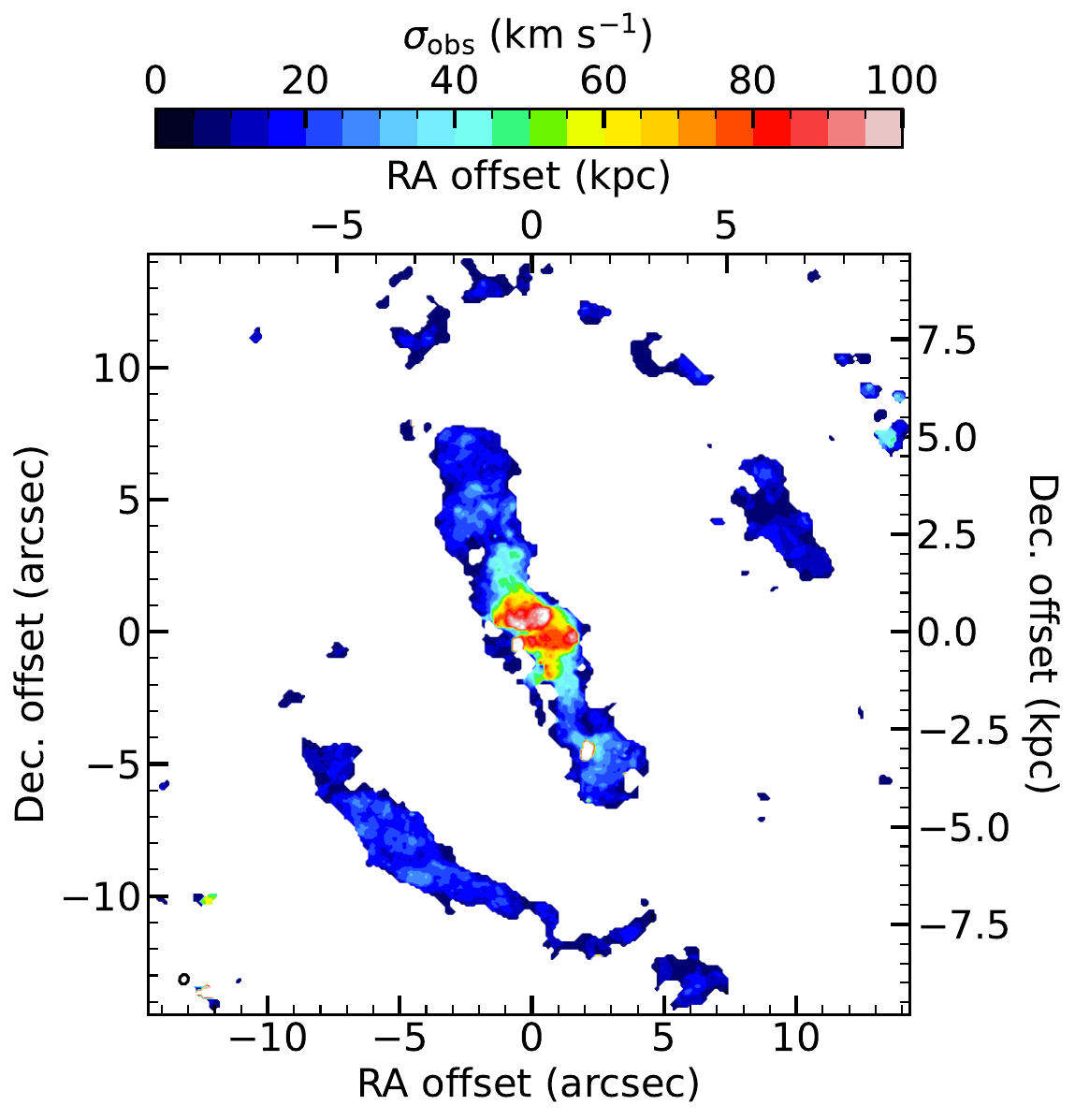}
    \end{subfigure}
    \\
    \begin{subfigure}{.34\linewidth}
    \centering
    \includegraphics[height=5.2cm]{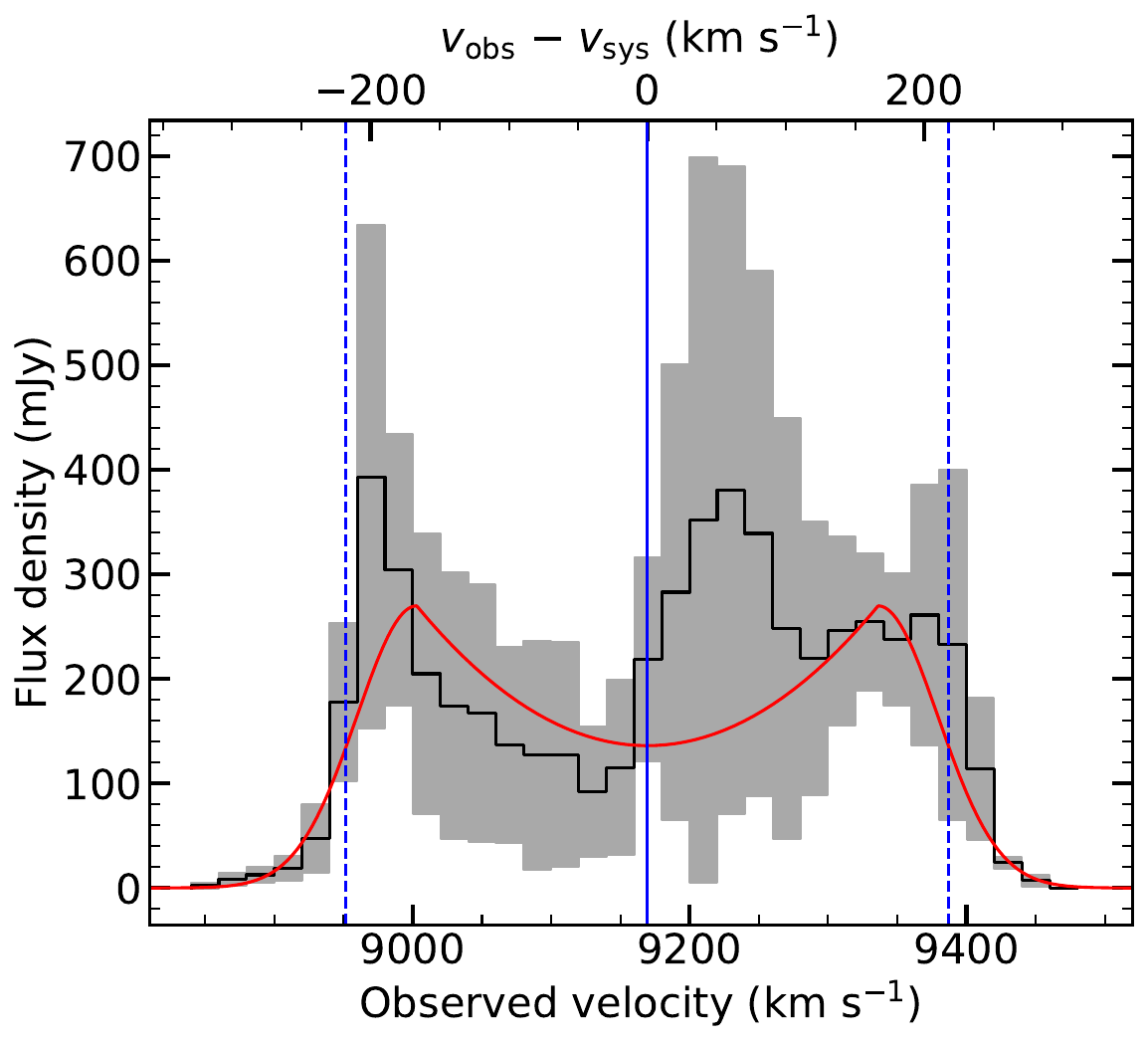}
    \end{subfigure}
    \begin{subfigure}{.32\linewidth}
    \centering
    \includegraphics[height=4.8cm]{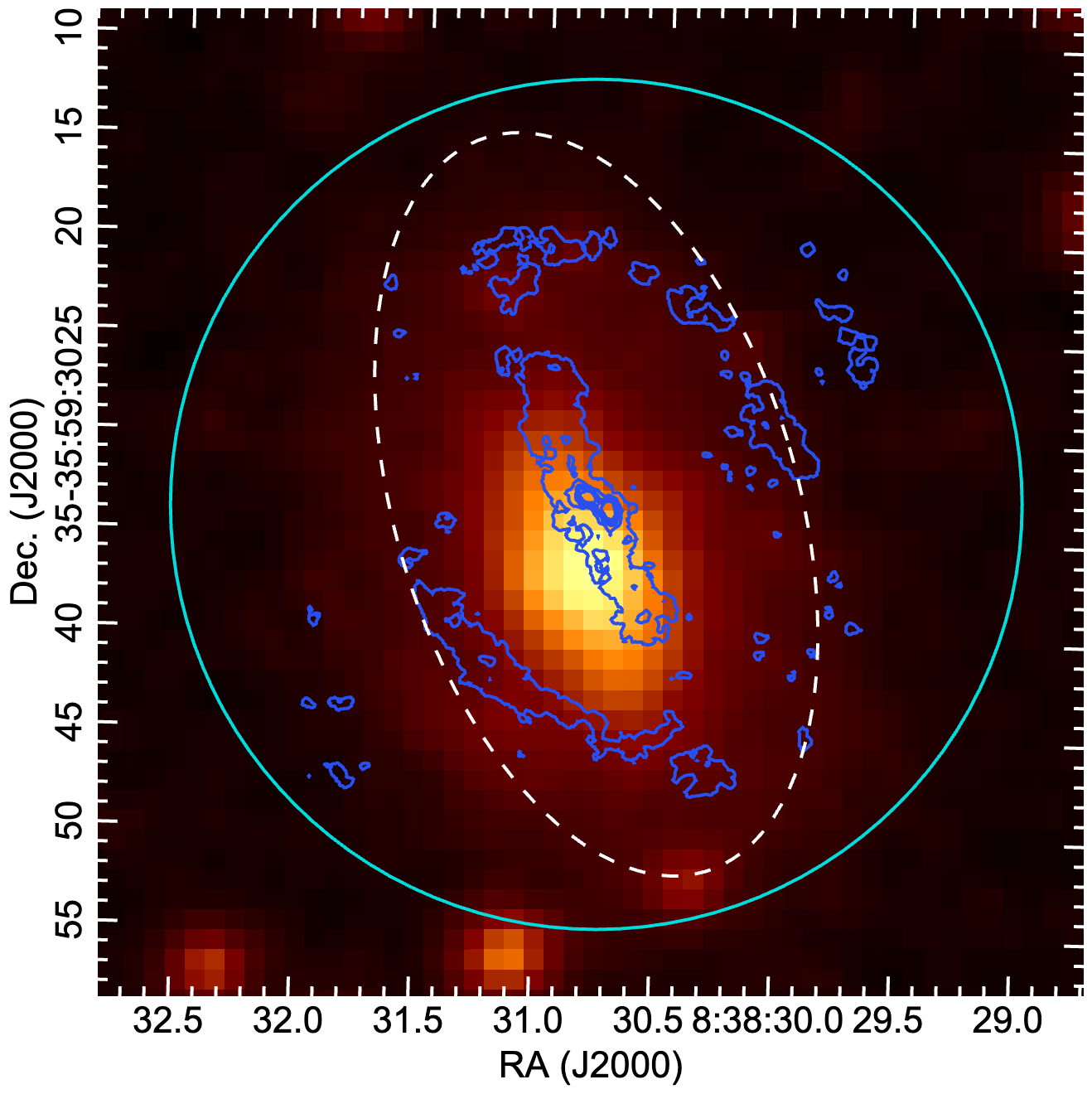}
    \end{subfigure}
    \begin{subfigure}{.33\linewidth}
    \centering
    \includegraphics[height=5.2cm]{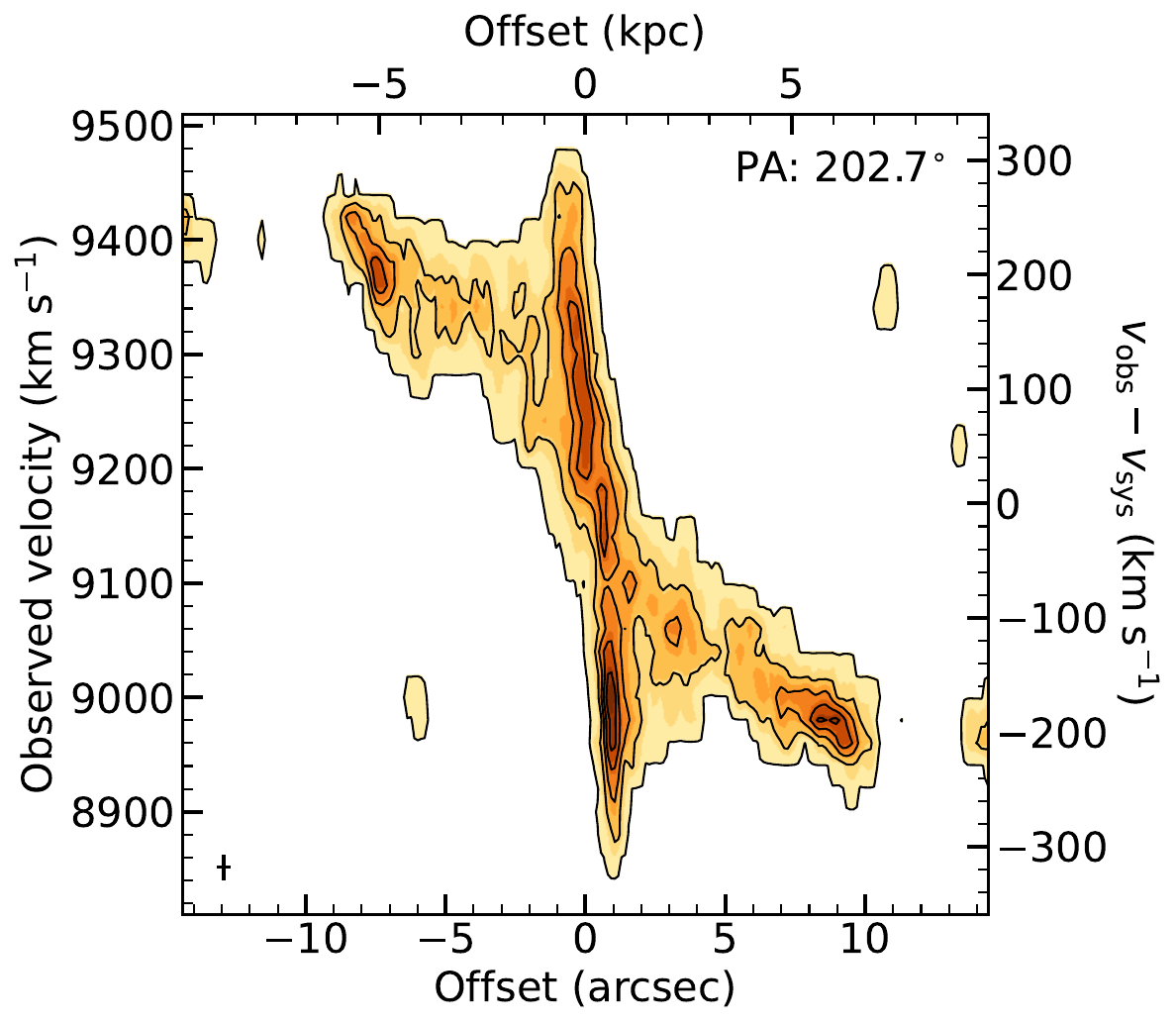}
    \end{subfigure}
    \caption{As Fig.~\ref{fig:CO ACA moment maps}, but for our high-resolution data. }
    \label{fig:CO high res moment maps}
\end{figure*}

\begin{figure*}
    \ContinuedFloat
    \begin{subfigure}{.34\linewidth}
    \includegraphics[height=6.0cm]{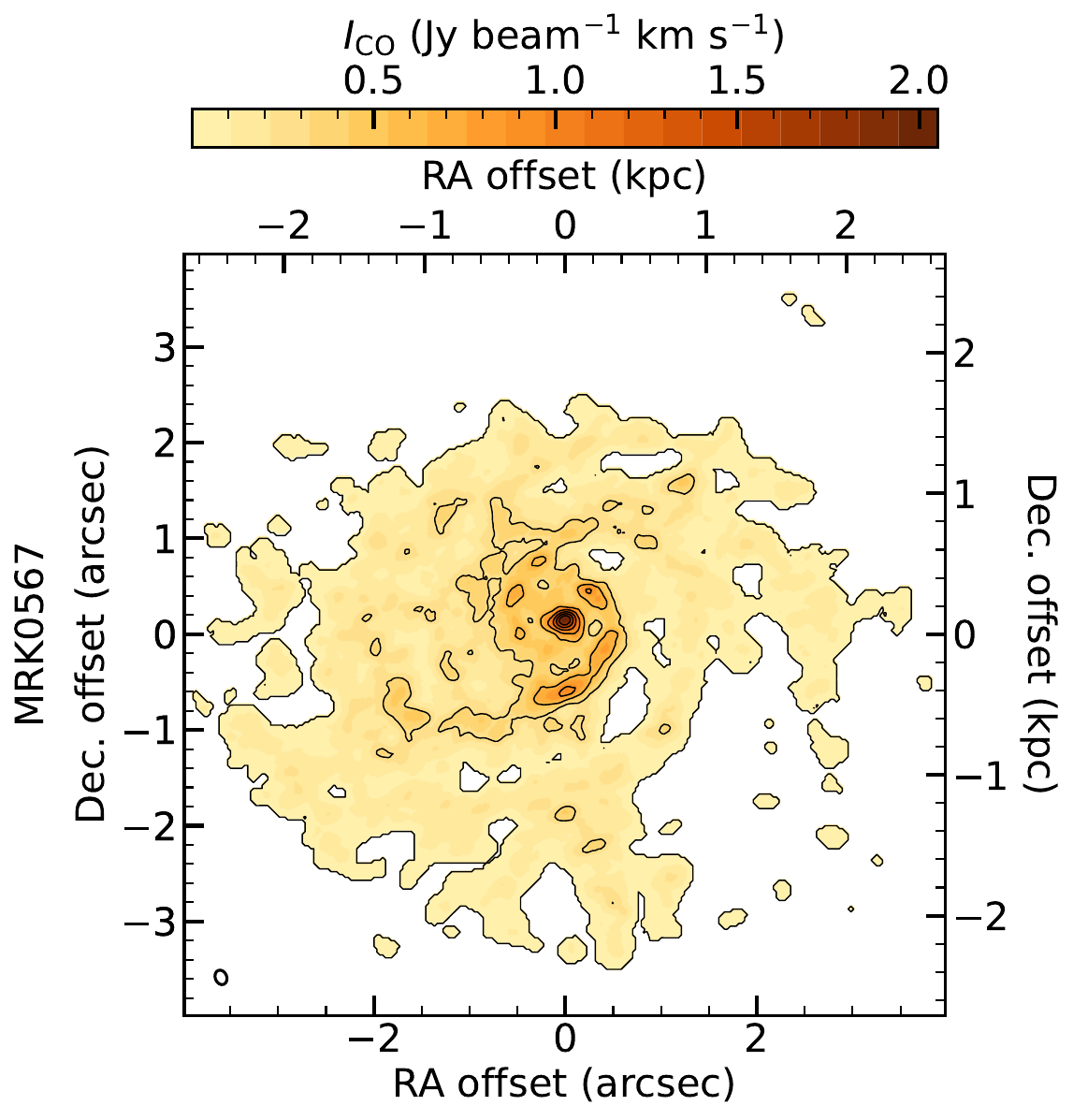}
    \end{subfigure}
    \begin{subfigure}{.32\linewidth}
    \centering
    \includegraphics[height=6.0cm]{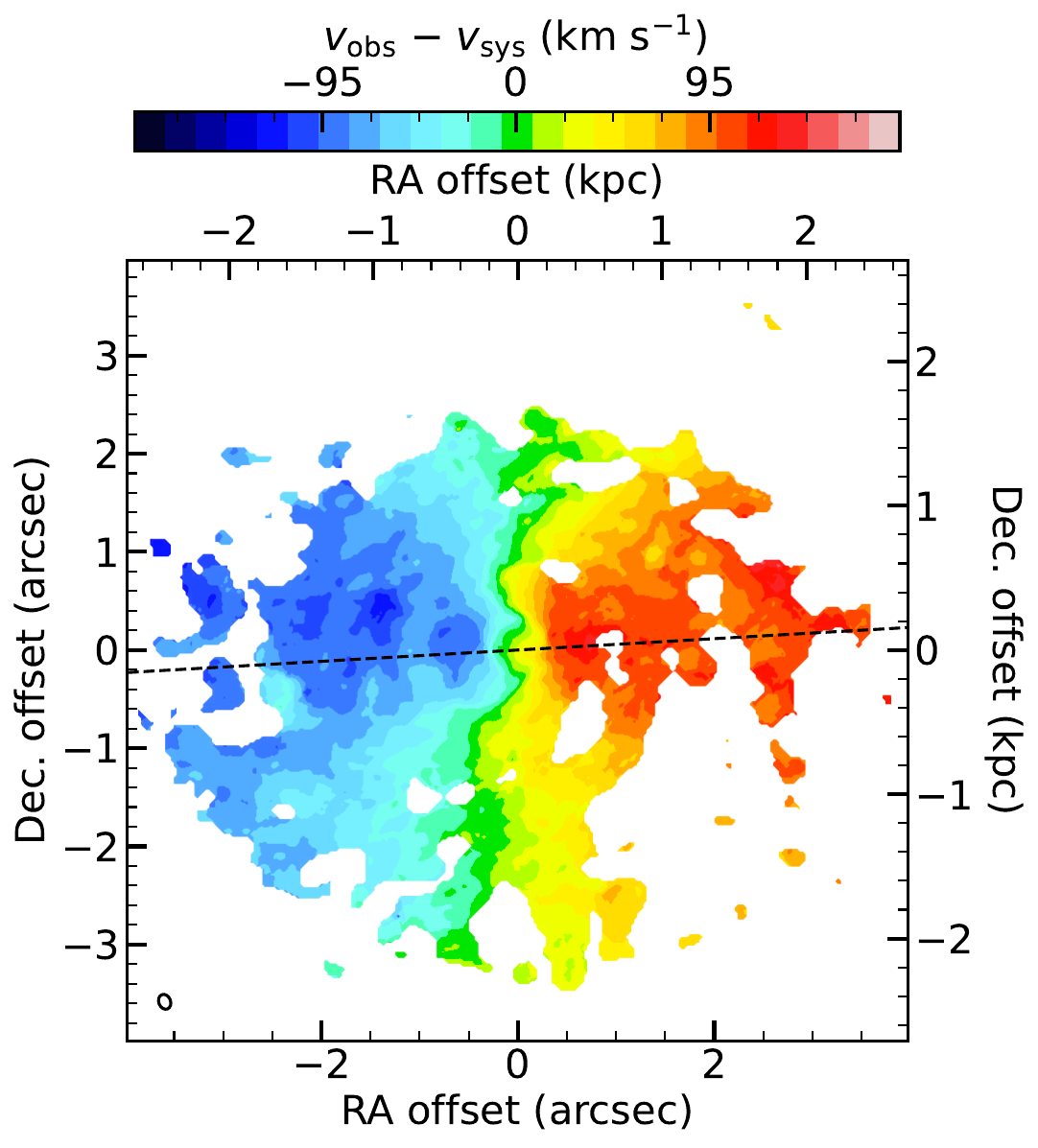}
    \end{subfigure}
    \begin{subfigure}{.32\linewidth}
    \centering
    \includegraphics[height=6.0cm]{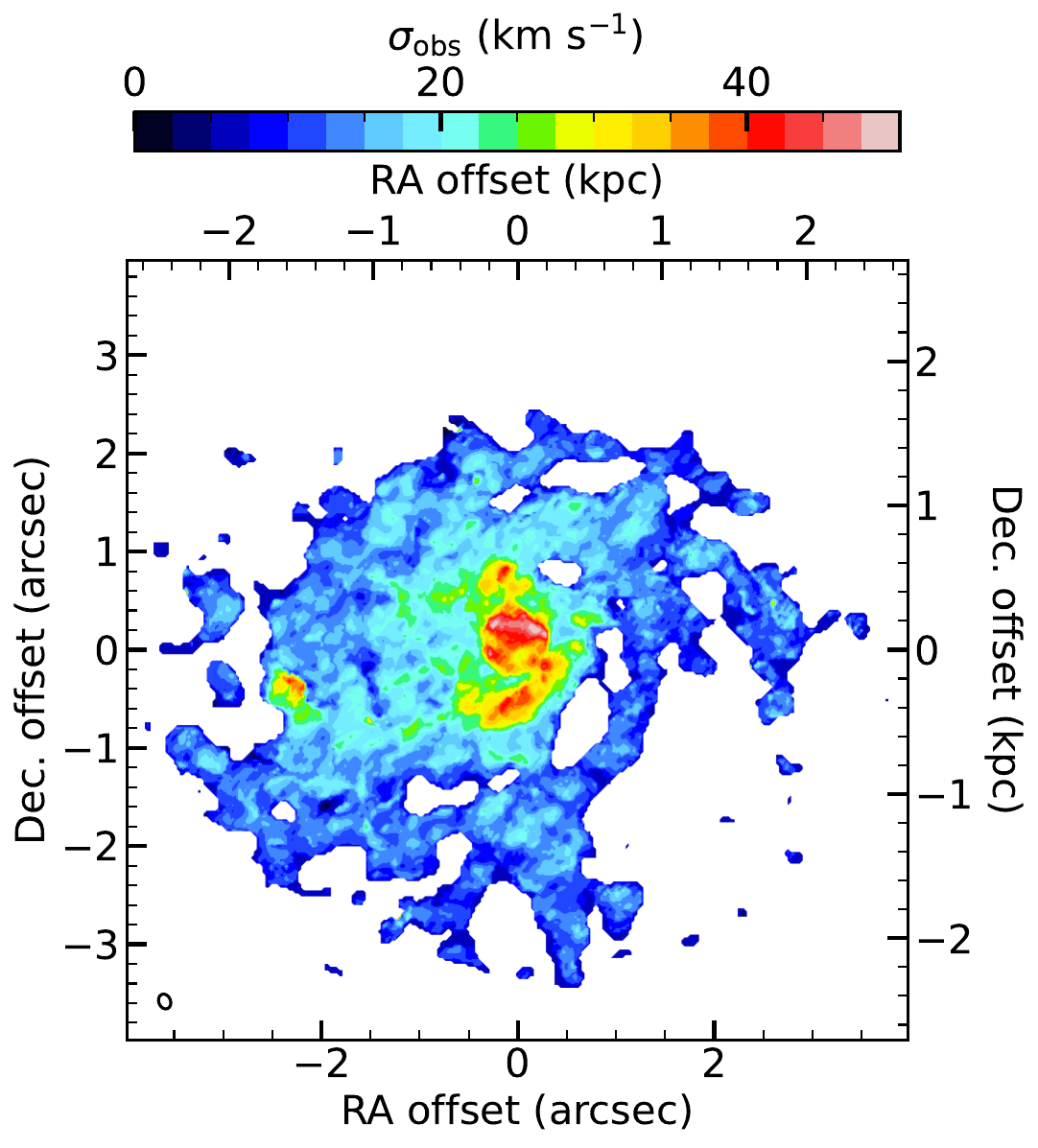}
    \end{subfigure}
    \\
    \begin{subfigure}{.34\linewidth}
    \centering
    \includegraphics[height=5.2cm]{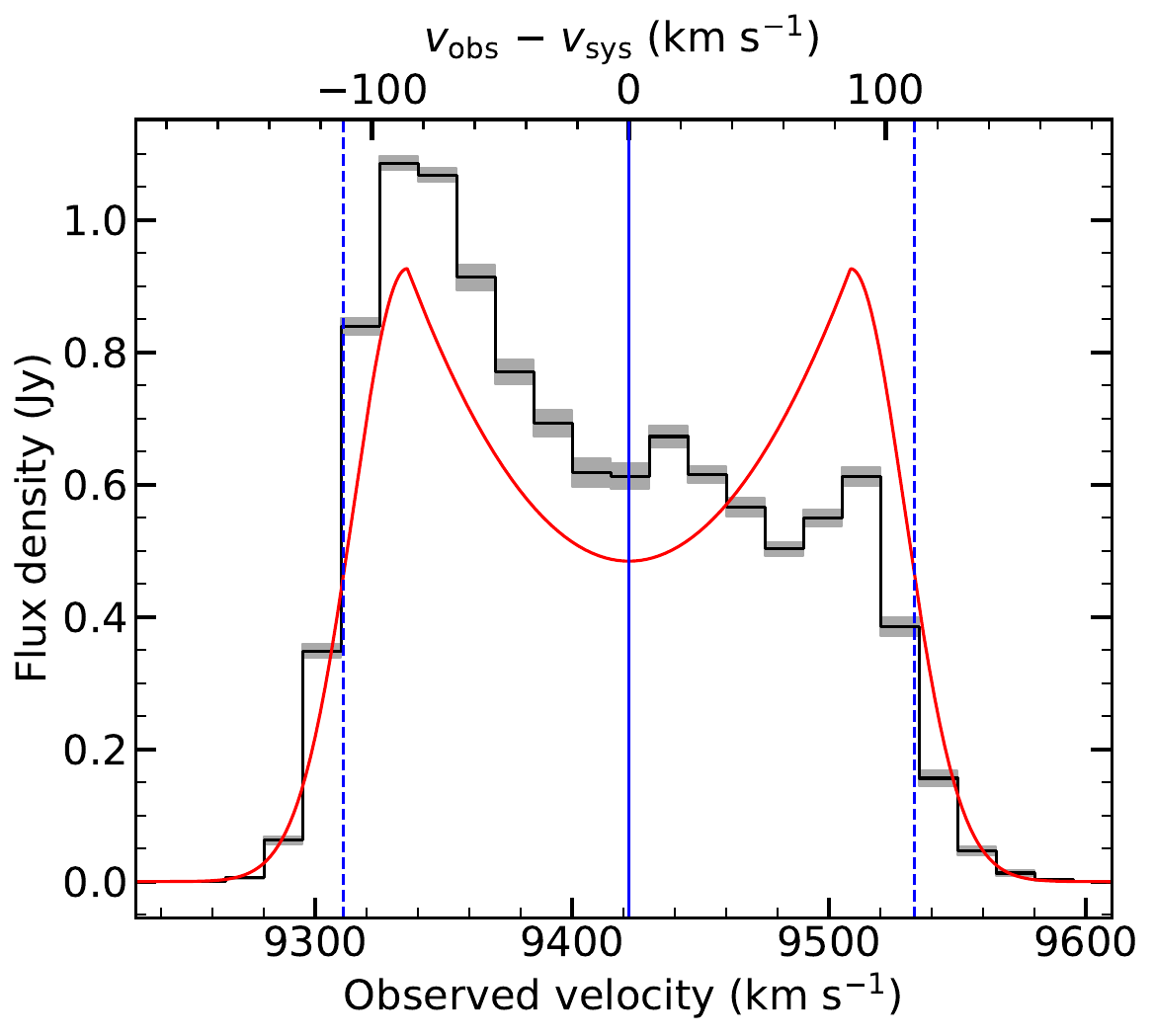}
    \end{subfigure}
    \begin{subfigure}{.32\linewidth}
    \centering
    \includegraphics[height=4.8cm]{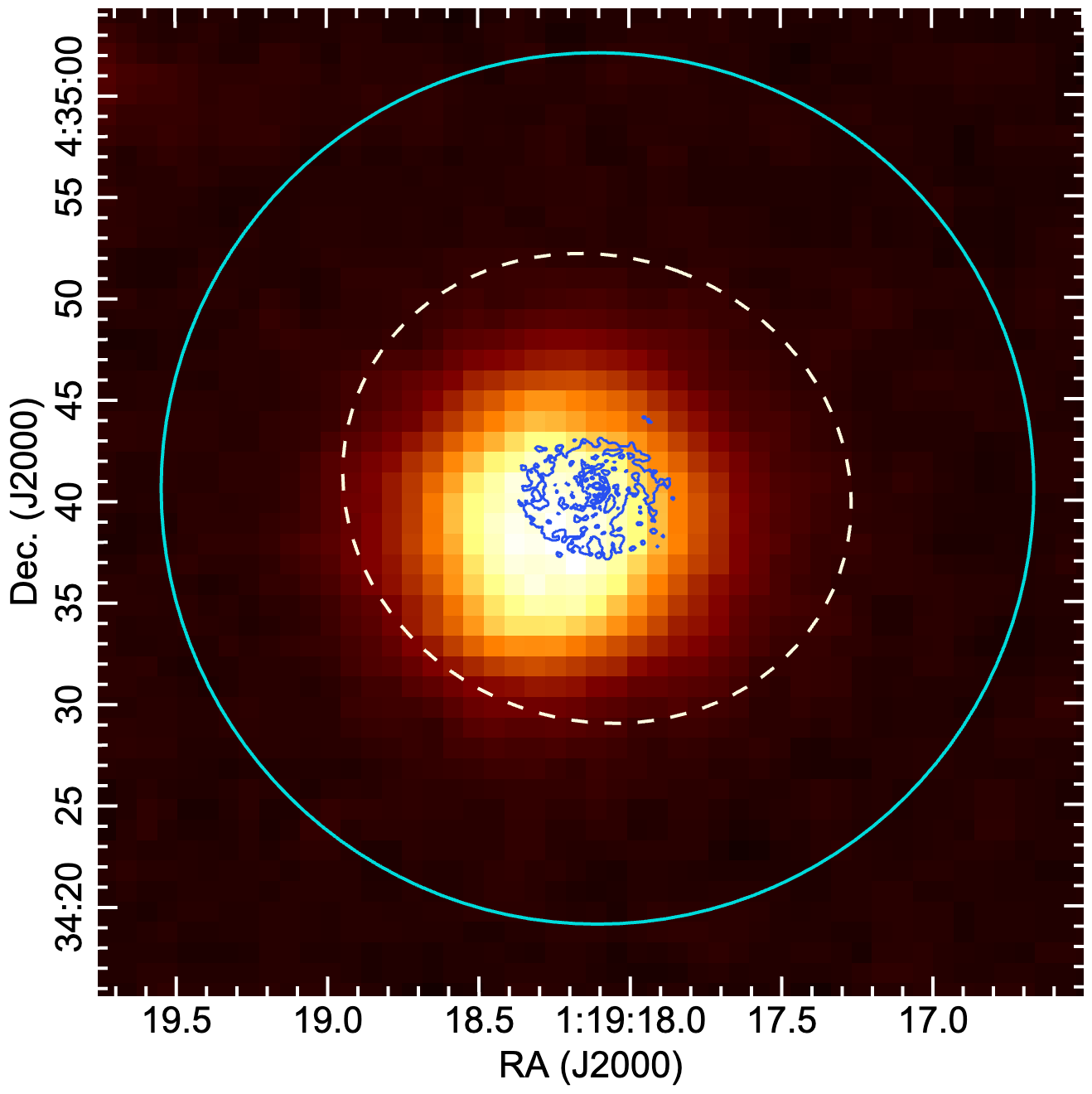}
    \end{subfigure}
    \begin{subfigure}{.33\linewidth}
    \centering
    \includegraphics[height=5.2cm]{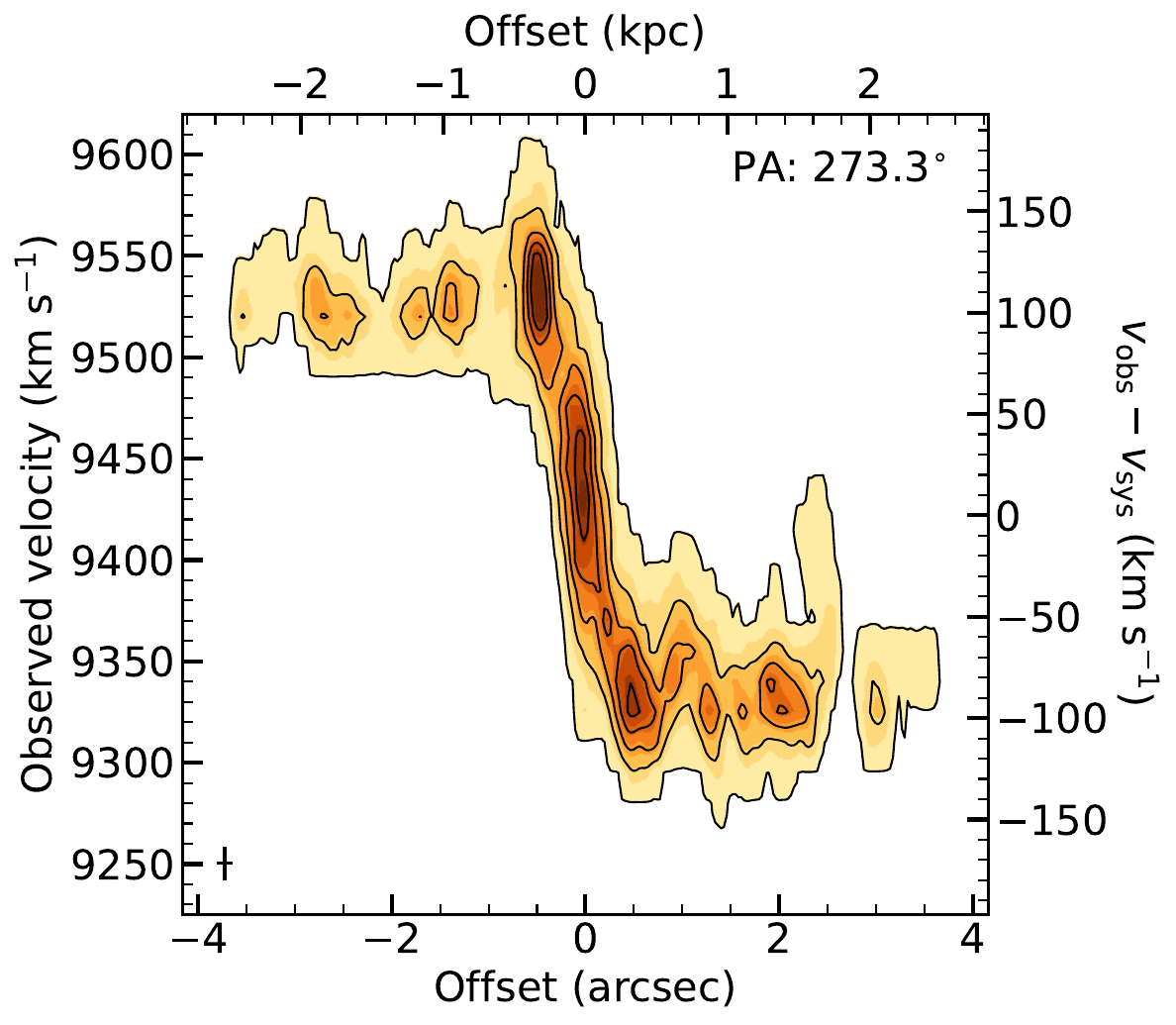}
    \end{subfigure}
    \\
    \begin{subfigure}{.34\linewidth}
    \centering
    \includegraphics[height=6.0cm]{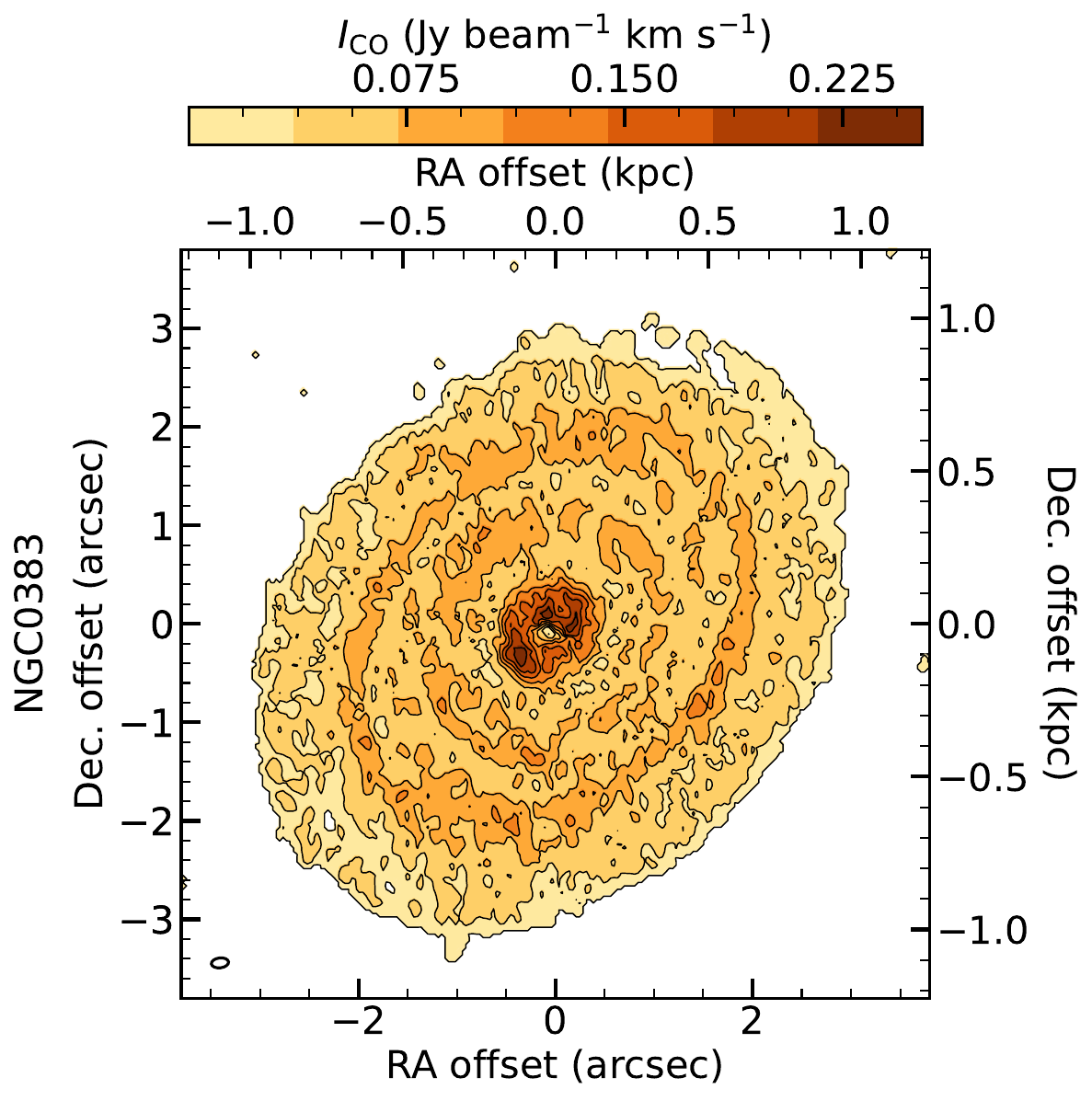}
    \end{subfigure}
    \begin{subfigure}{.32\linewidth}
    \centering
    \includegraphics[height=6.0cm]{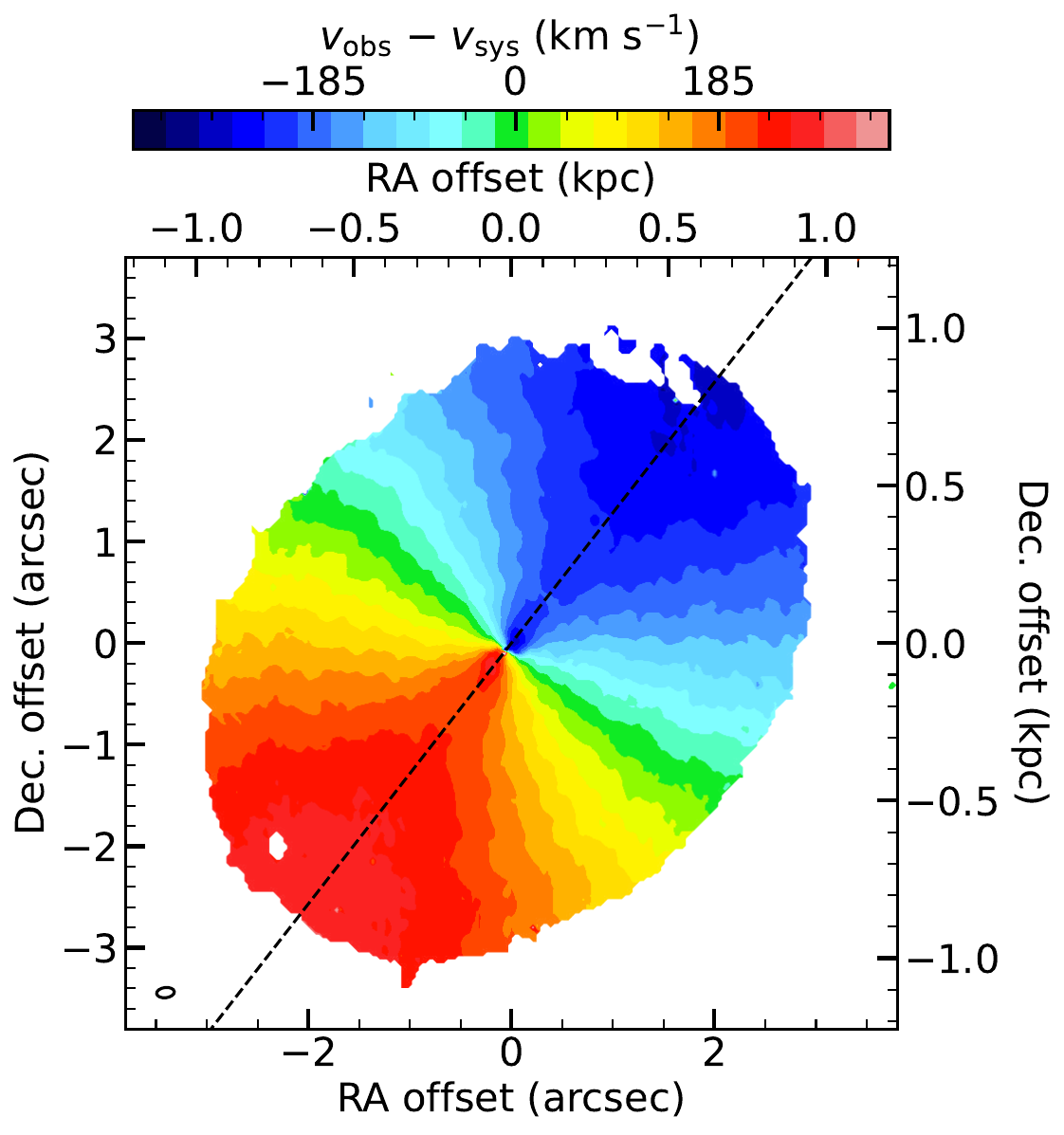}
    \end{subfigure}
    \begin{subfigure}{.32\linewidth}
    \centering
    \includegraphics[height=6.0cm]{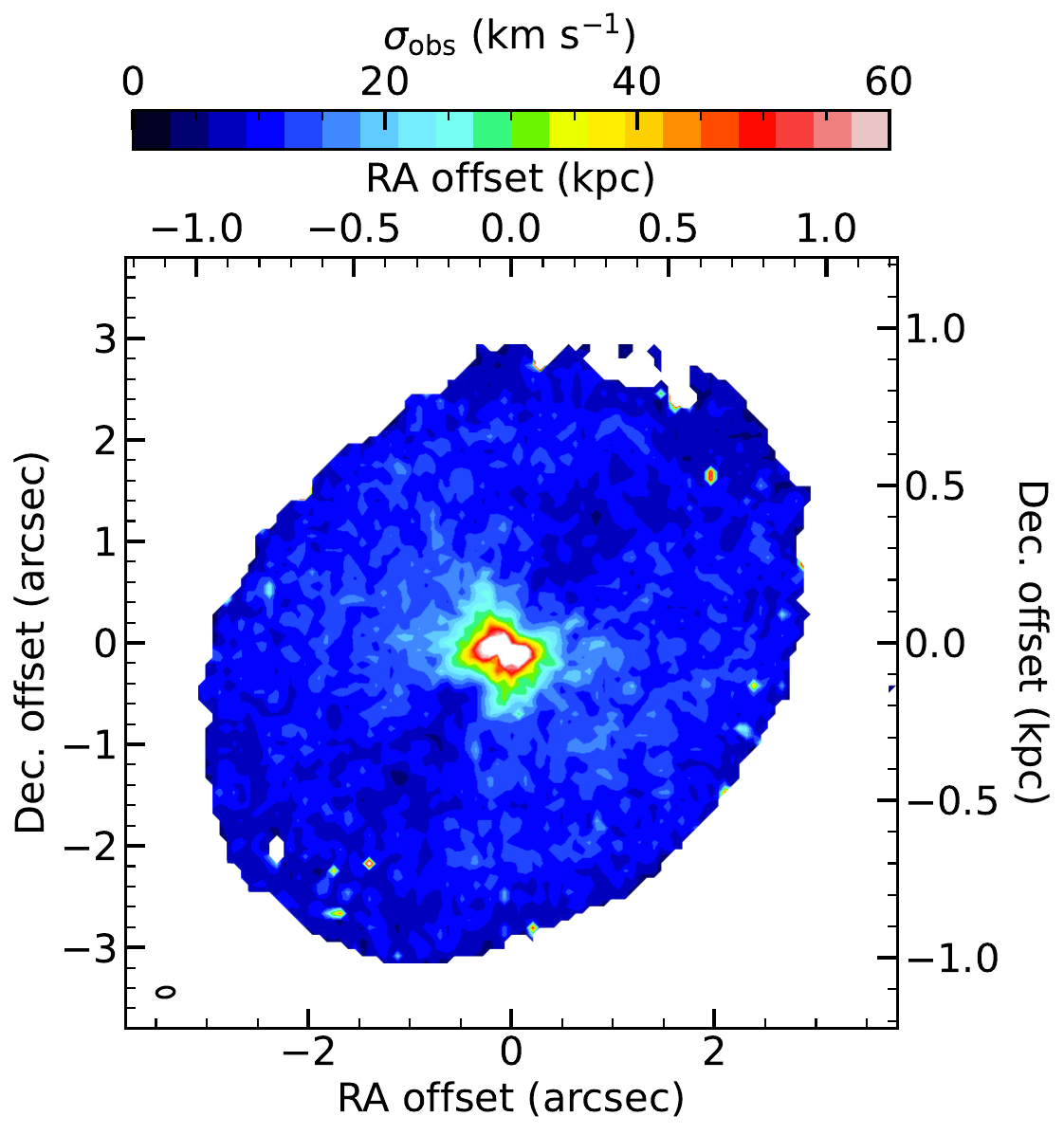}
    \end{subfigure}
    \\
    \begin{subfigure}{.34\linewidth}
    \centering
    \includegraphics[height=5.2cm]{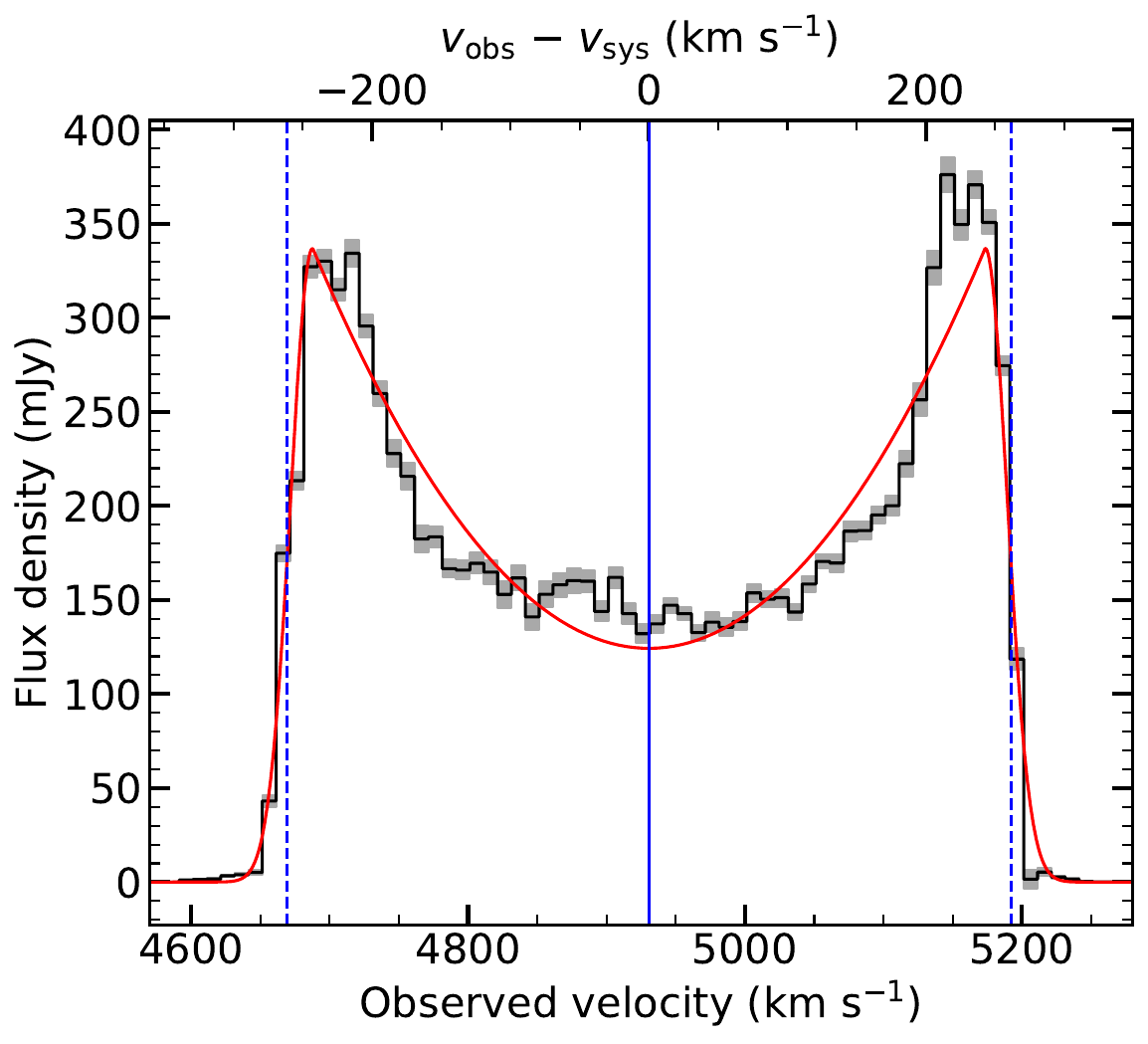}
    \end{subfigure}
    \begin{subfigure}{.32\linewidth}
    \centering
    \includegraphics[height=4.8cm]{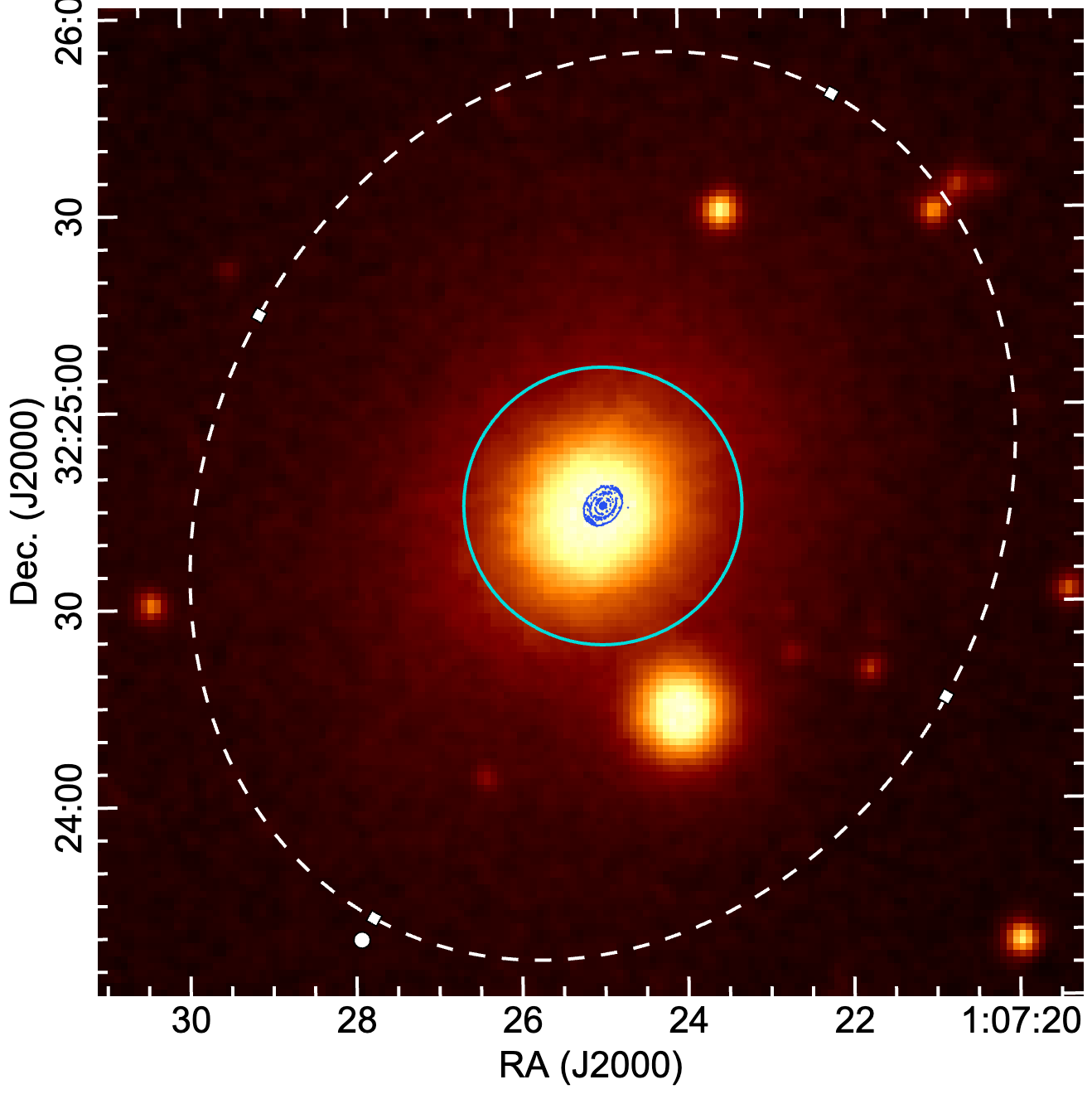}
    \end{subfigure}
    \begin{subfigure}{.33\linewidth}
    \centering
    \includegraphics[height=5.2cm]{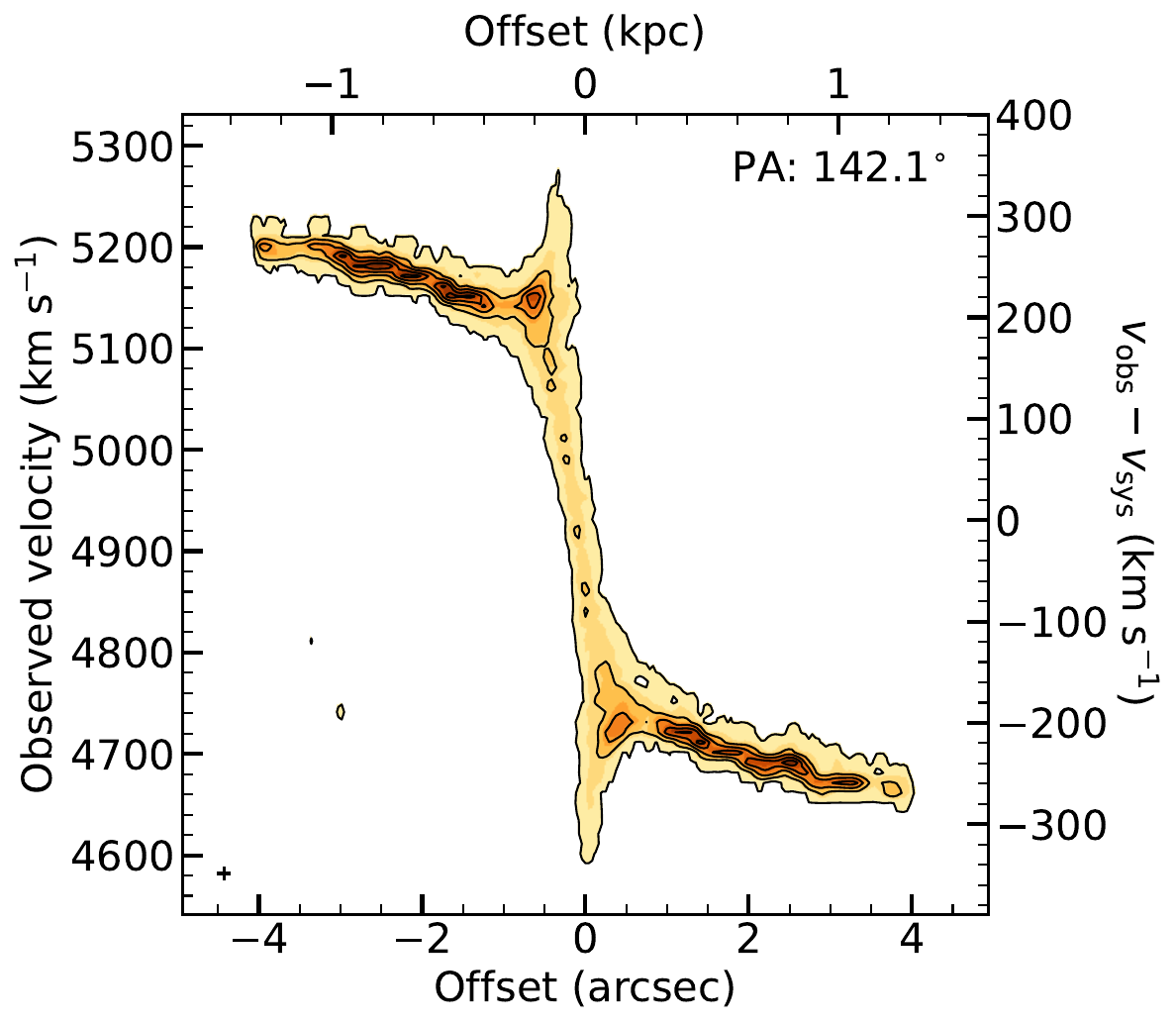}
    \end{subfigure}
    \\
    \caption[]{continued.}
\end{figure*}

\begin{figure*}
    \ContinuedFloat
    \begin{subfigure}{.34\linewidth}
    \includegraphics[height=6.0cm]{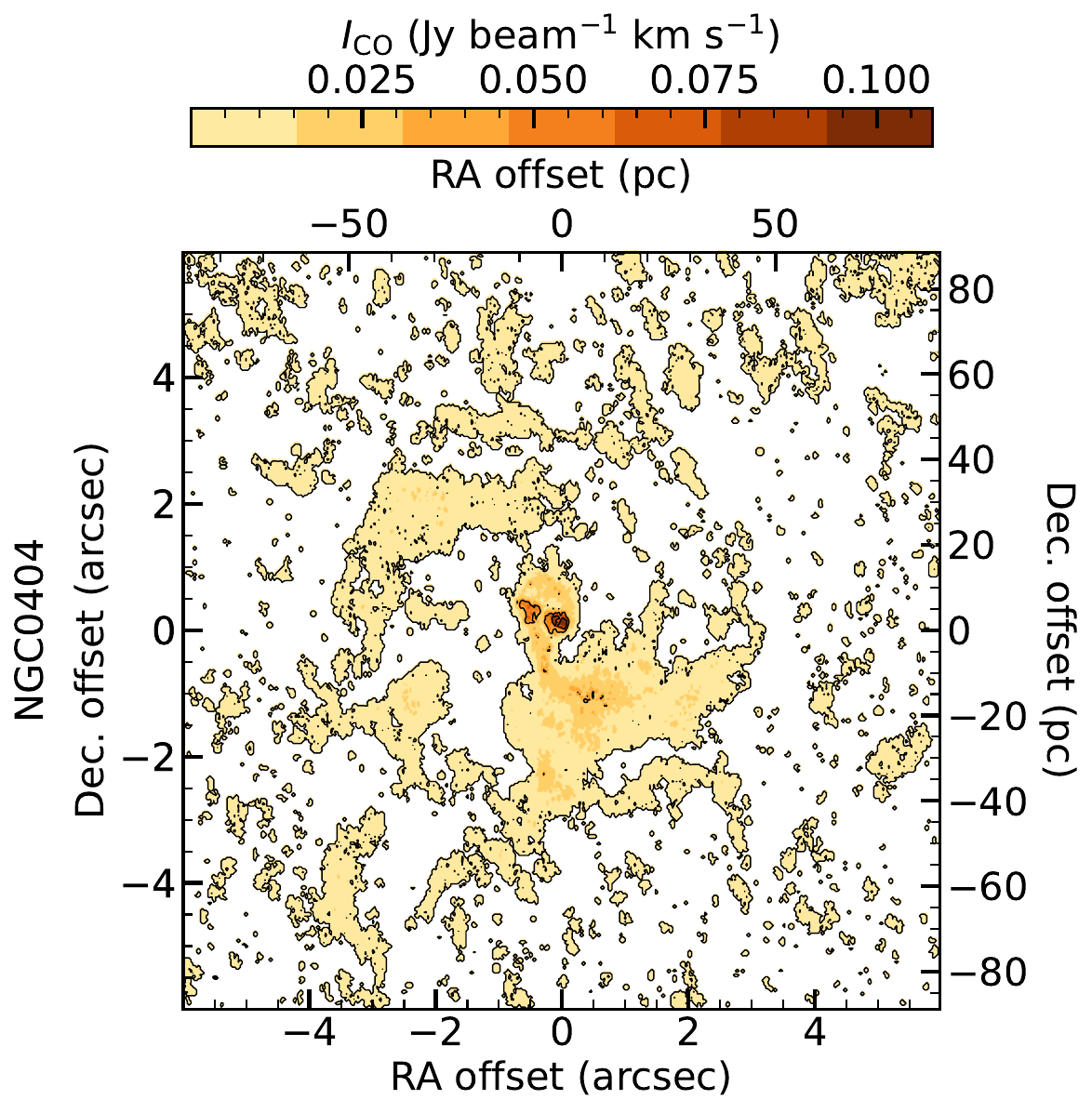}
    \end{subfigure}
    \begin{subfigure}{.32\linewidth}
    \centering
    \includegraphics[height=6.0cm]{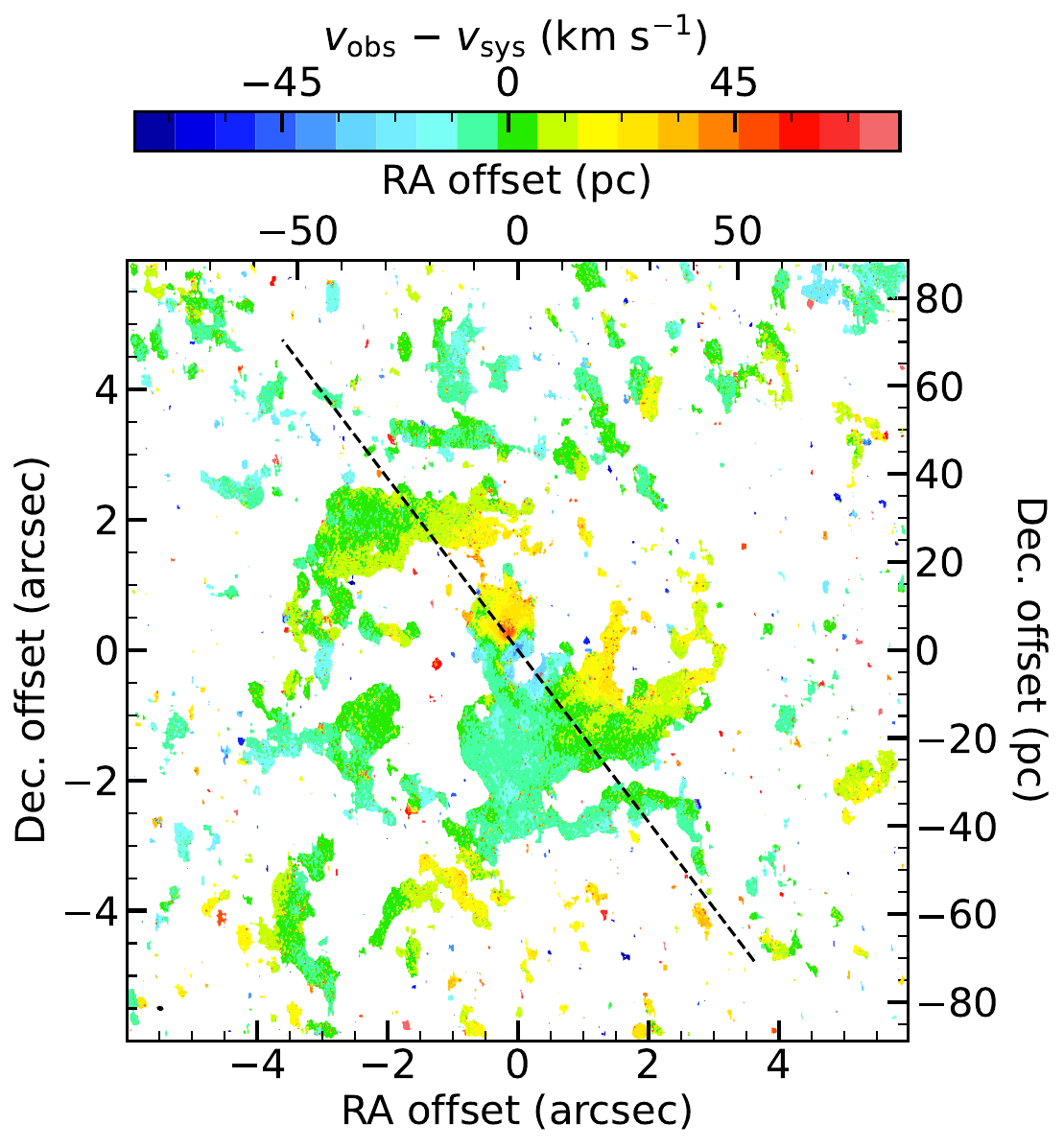}
    \end{subfigure}
    \begin{subfigure}{.32\linewidth}
    \centering
    \includegraphics[height=6.0cm]{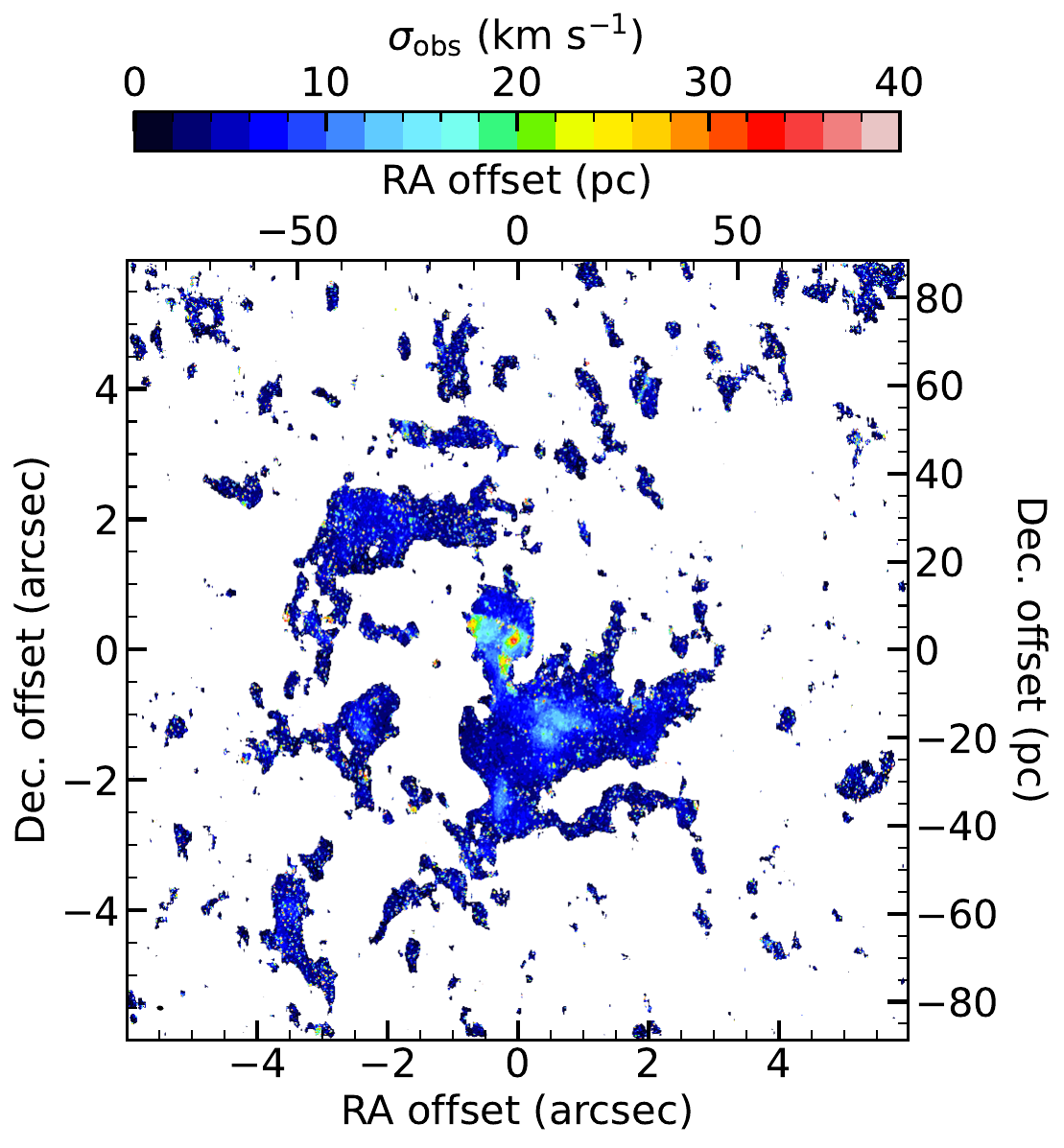}
    \end{subfigure}
    \\
    \begin{subfigure}{.34\linewidth}
    \centering
    \includegraphics[height=5.2cm]{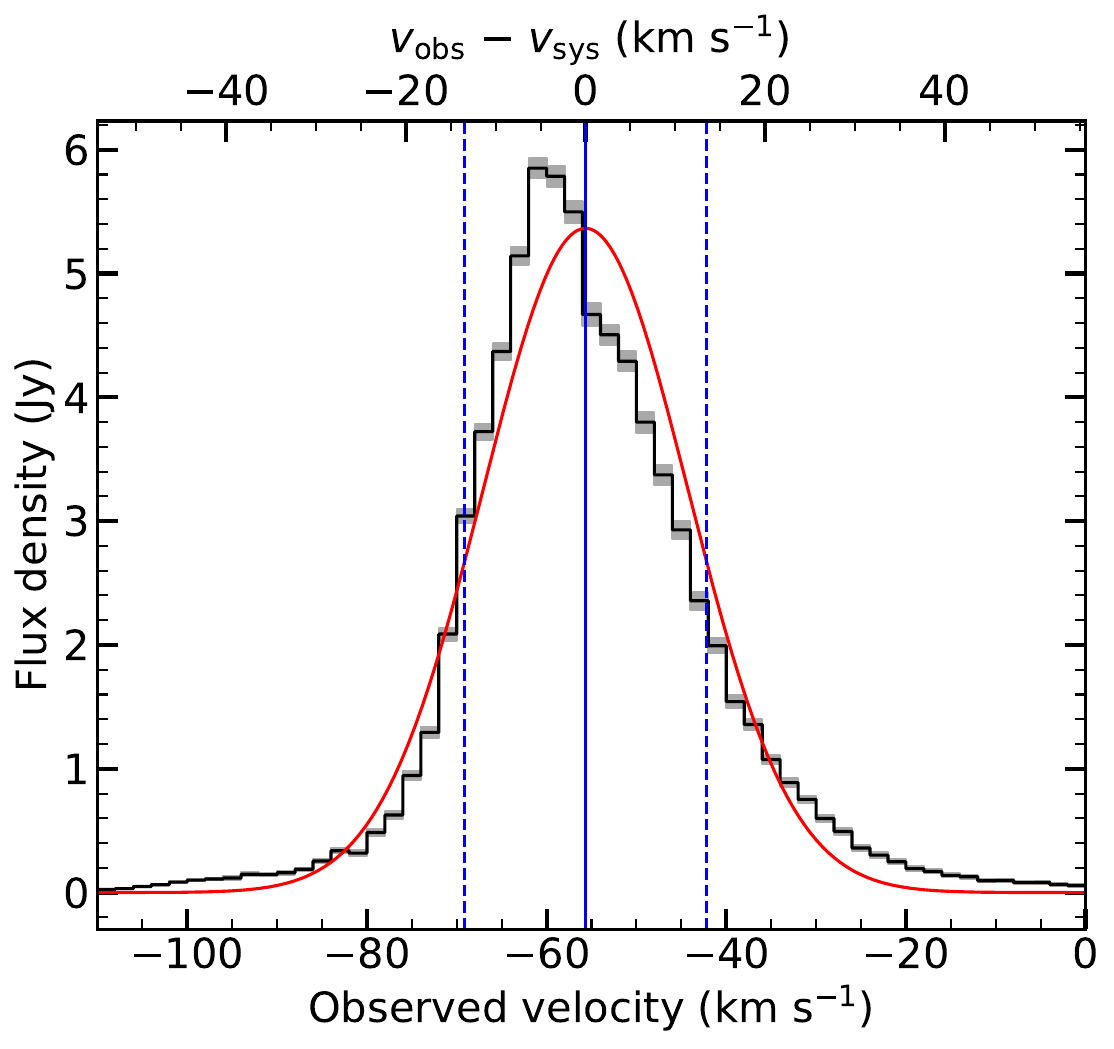}
    \end{subfigure}
    \begin{subfigure}{.32\linewidth}
    \centering
    \includegraphics[height=4.8cm]{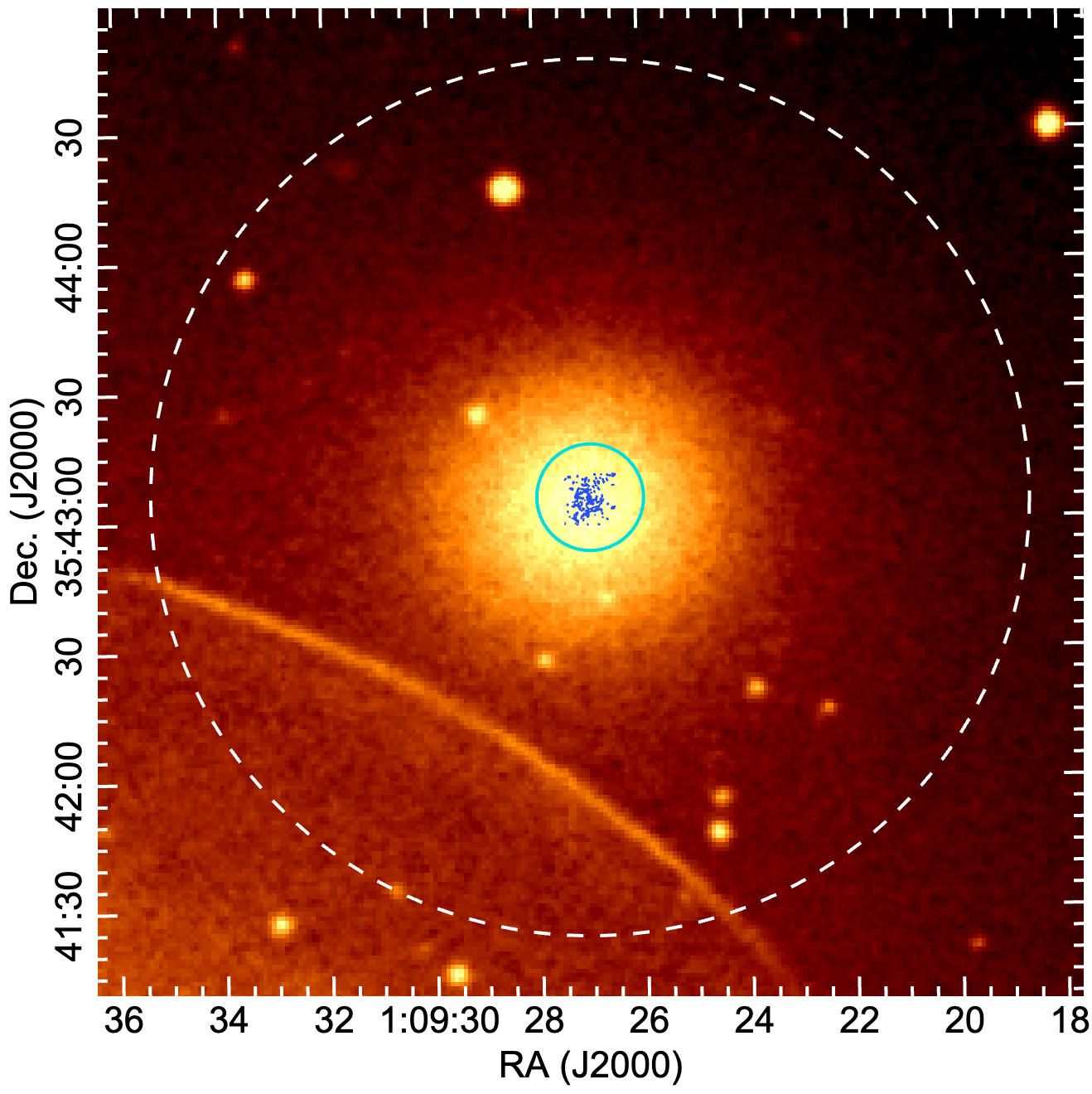}
    \end{subfigure}
    \begin{subfigure}{.33\linewidth}
    \centering
    \includegraphics[height=5.2cm]{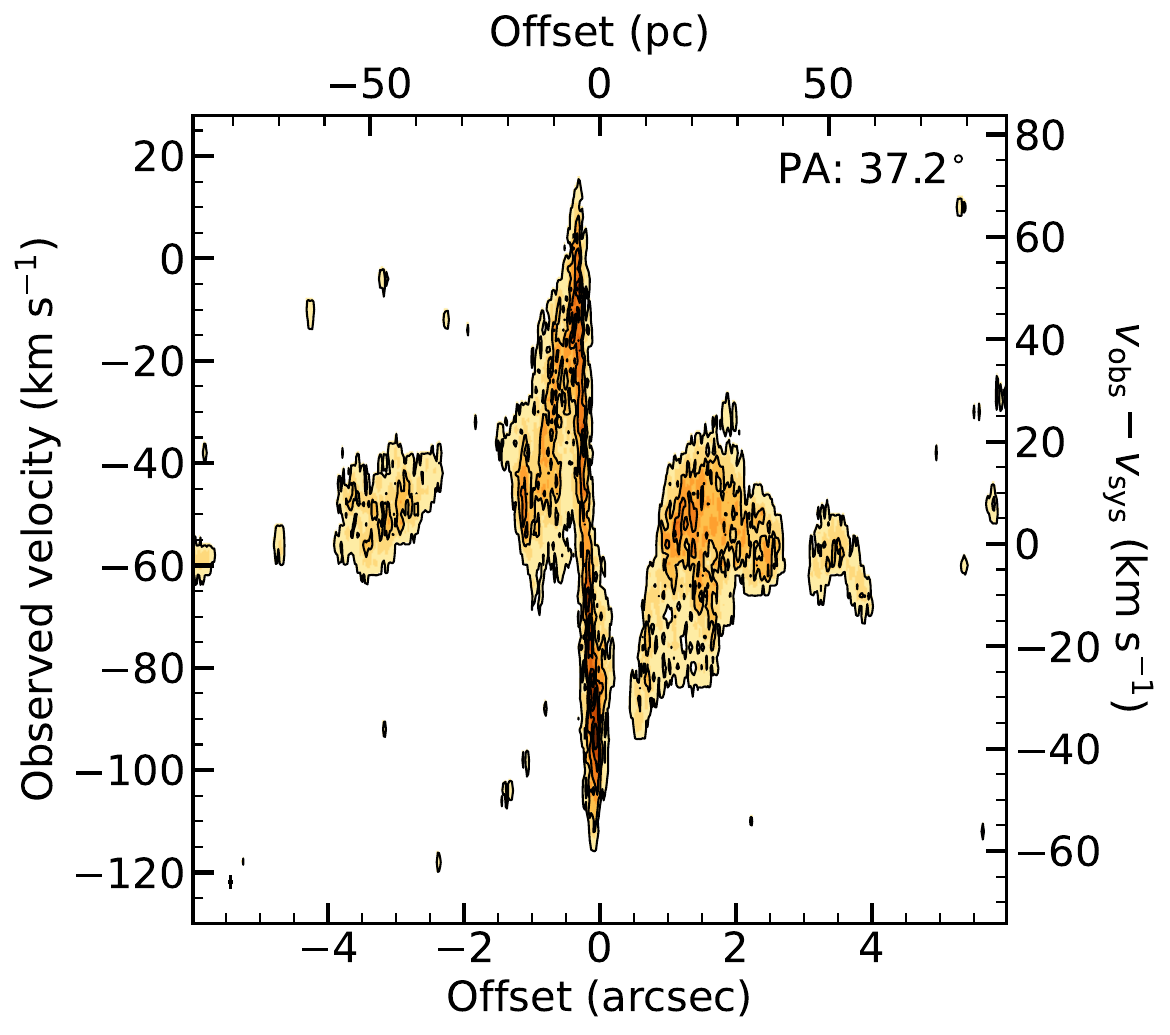}
    \end{subfigure}
    \\
    \begin{subfigure}{.34\linewidth}
    \includegraphics[height=6.0cm]{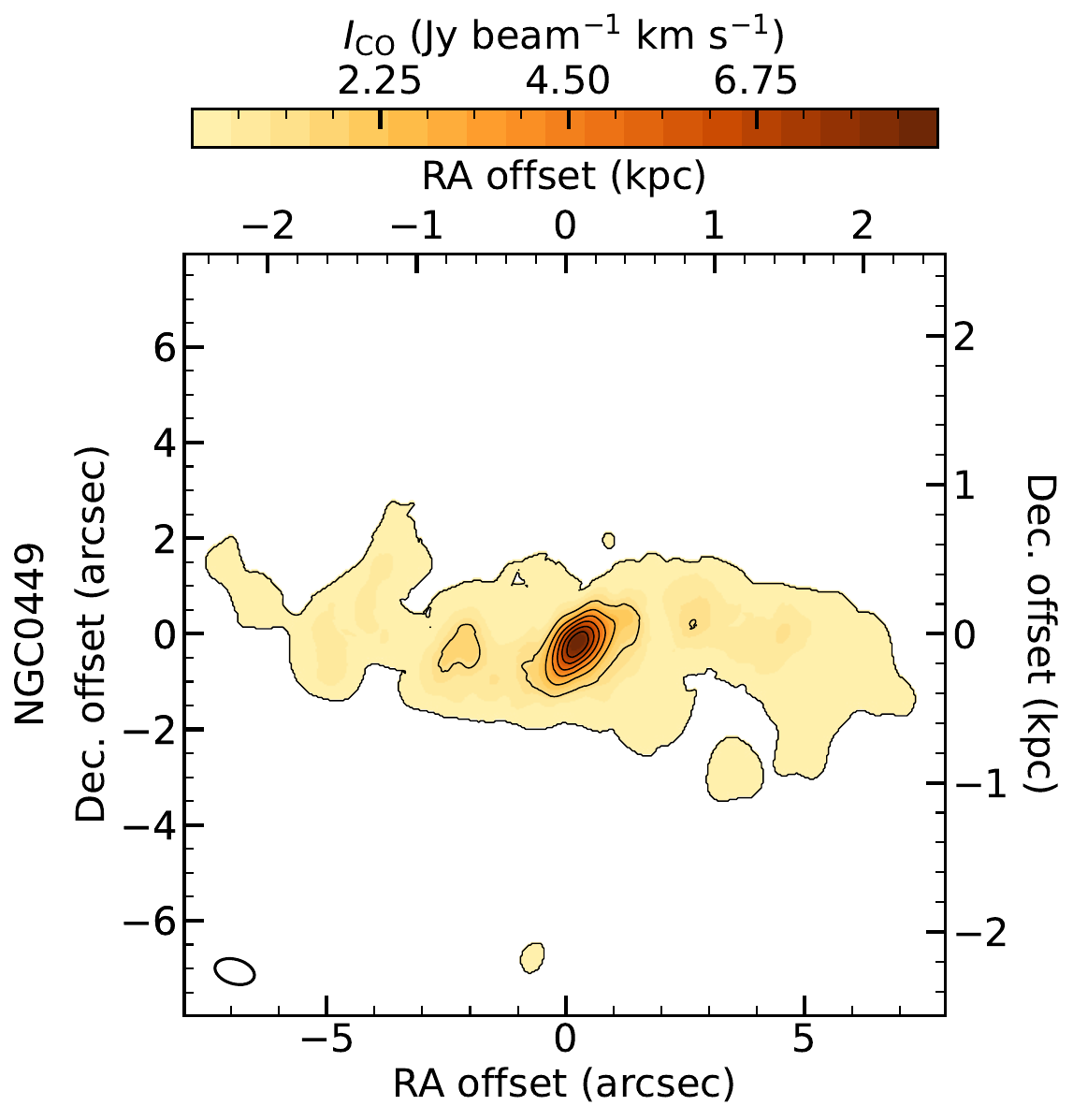}
    \end{subfigure}
    \begin{subfigure}{.32\linewidth}
    \centering
    \includegraphics[height=6.0cm]{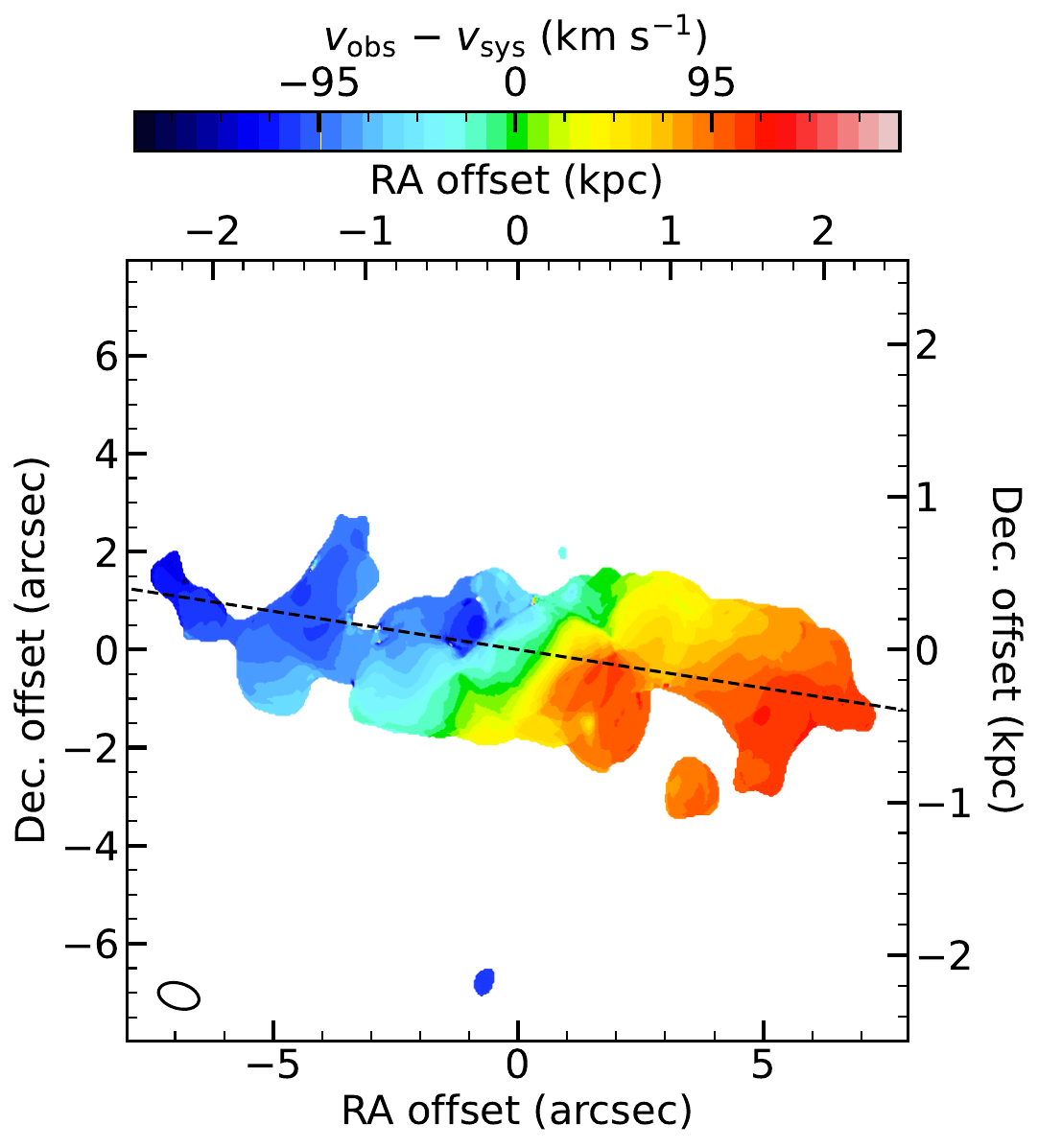}
    \end{subfigure}
    \begin{subfigure}{.32\linewidth}
    \centering
    \includegraphics[height=6.0cm]{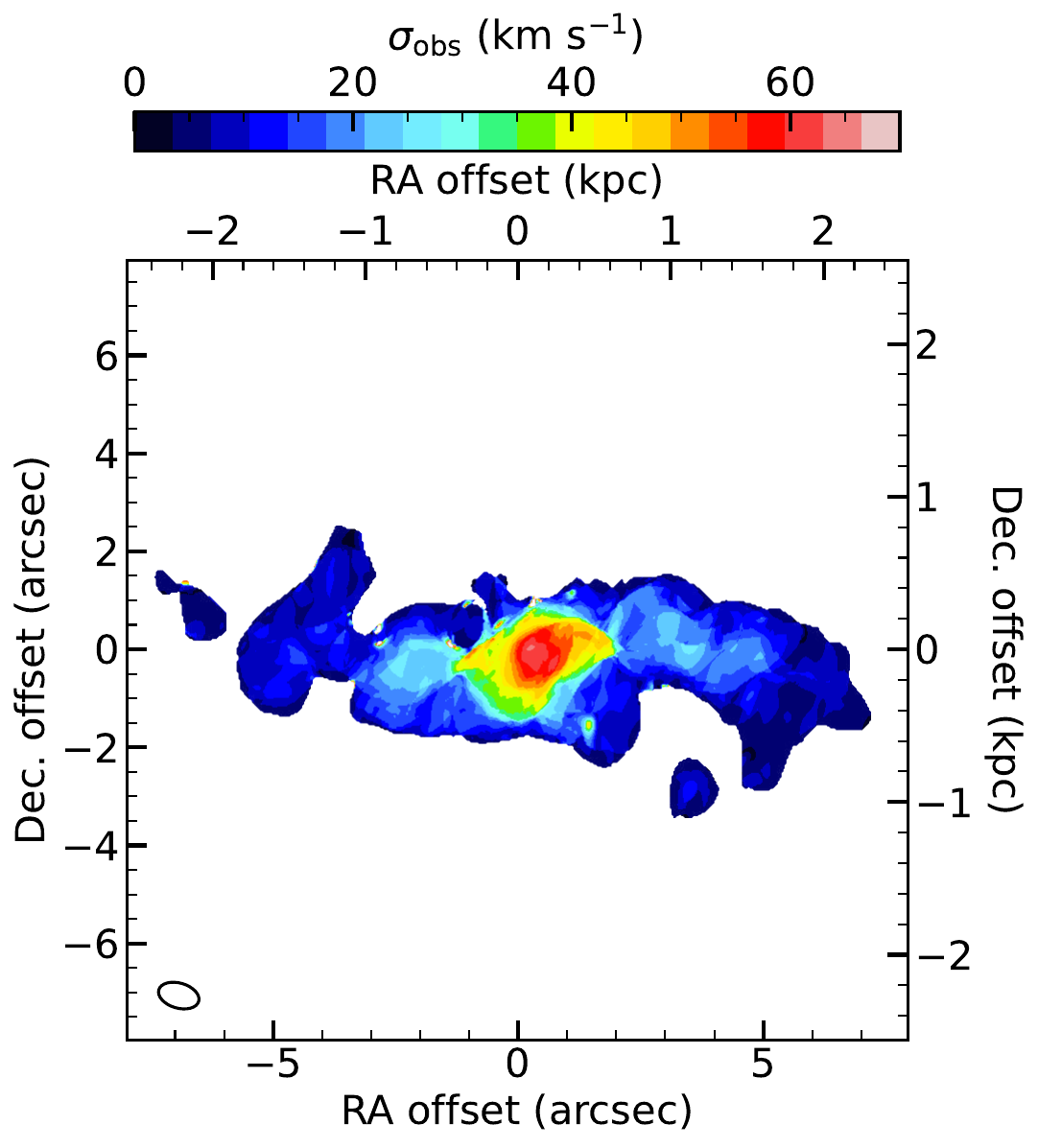}
    \end{subfigure}
    \\
    \begin{subfigure}{.34\linewidth}
    \centering
    \includegraphics[height=5.2cm]{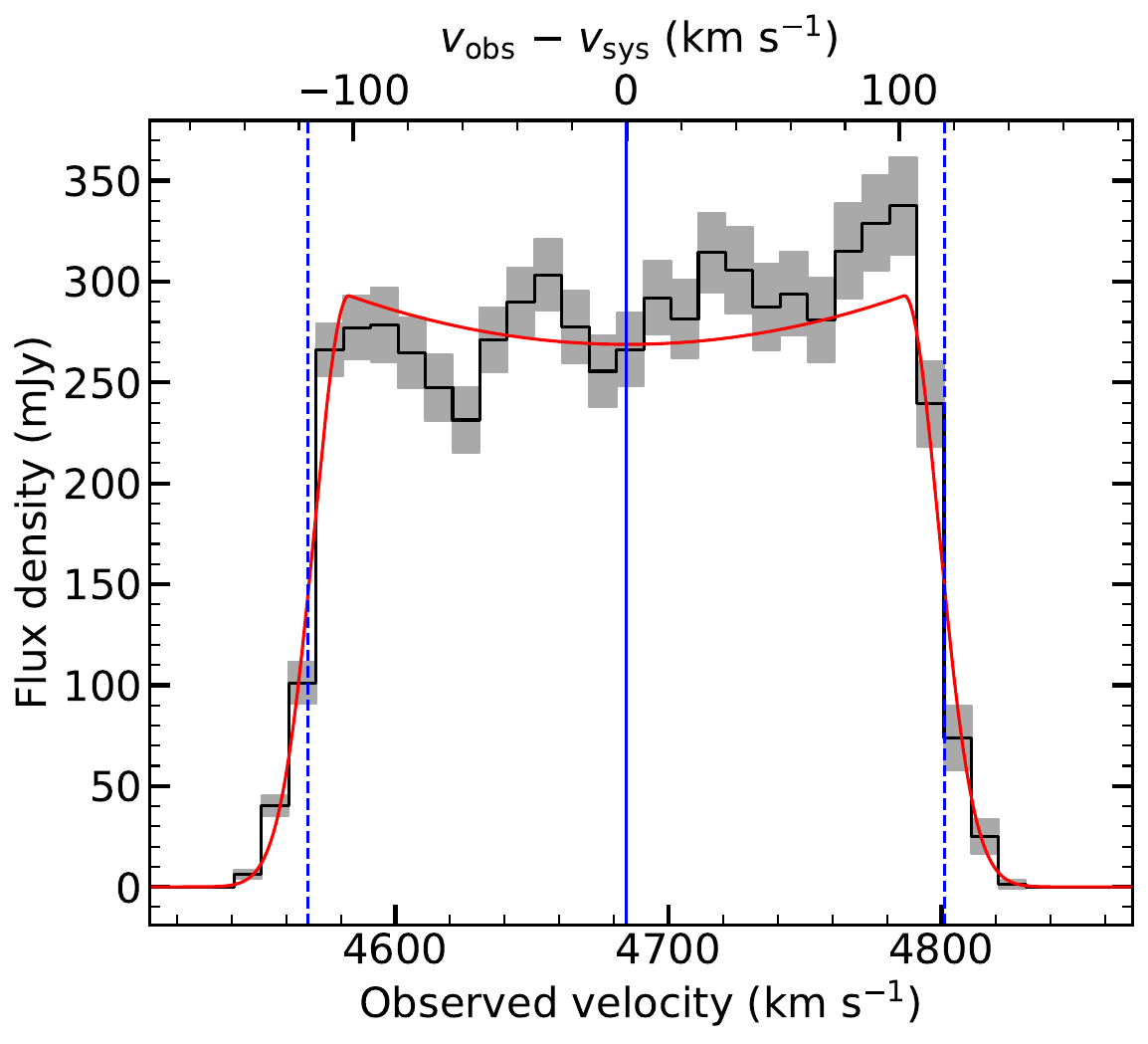}
    \end{subfigure}
    \begin{subfigure}{.32\linewidth}
    \centering
    \includegraphics[height=4.8cm]{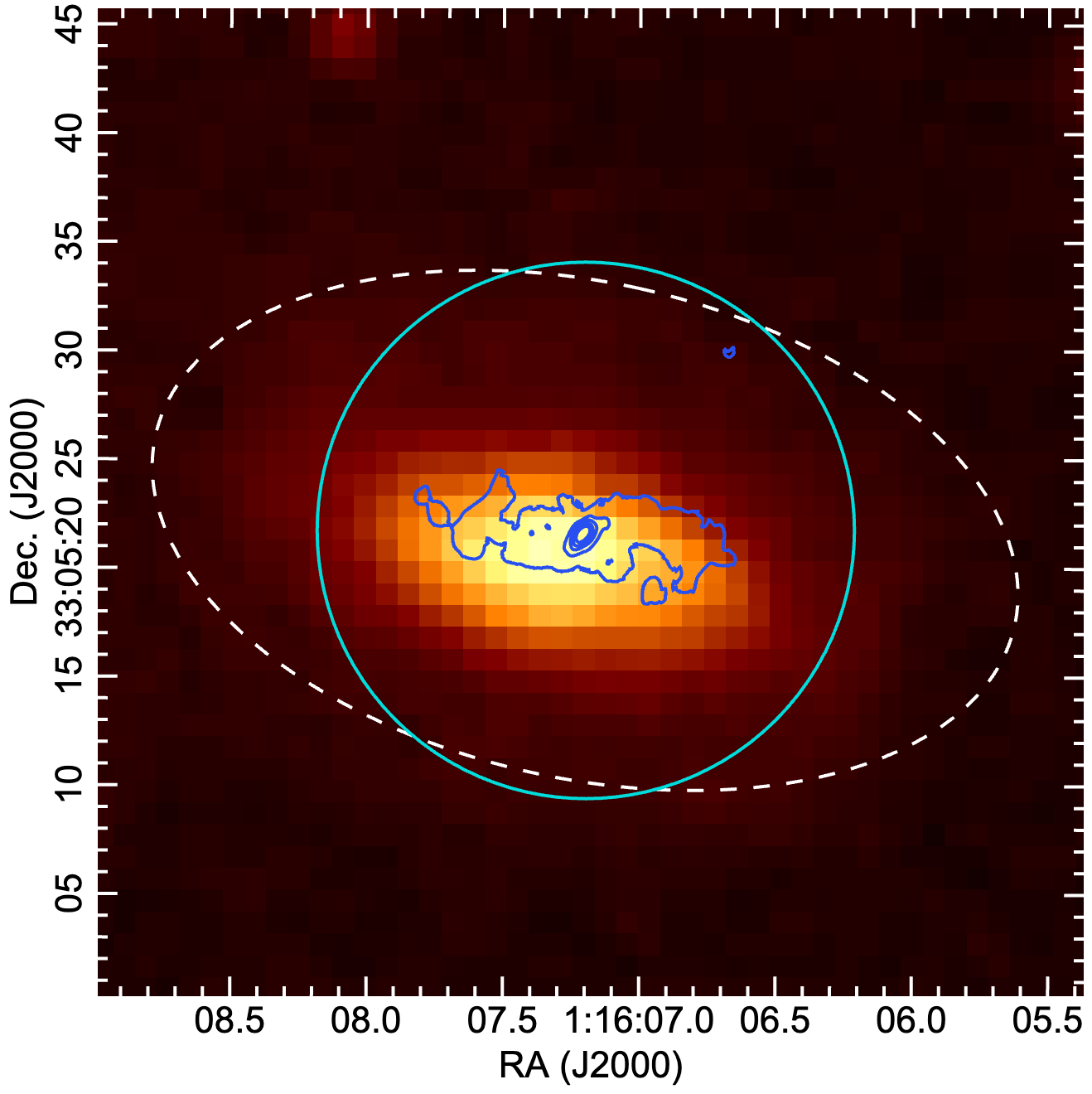}
    \end{subfigure}
    \begin{subfigure}{.33\linewidth}
    \centering
    \includegraphics[height=5.2cm]{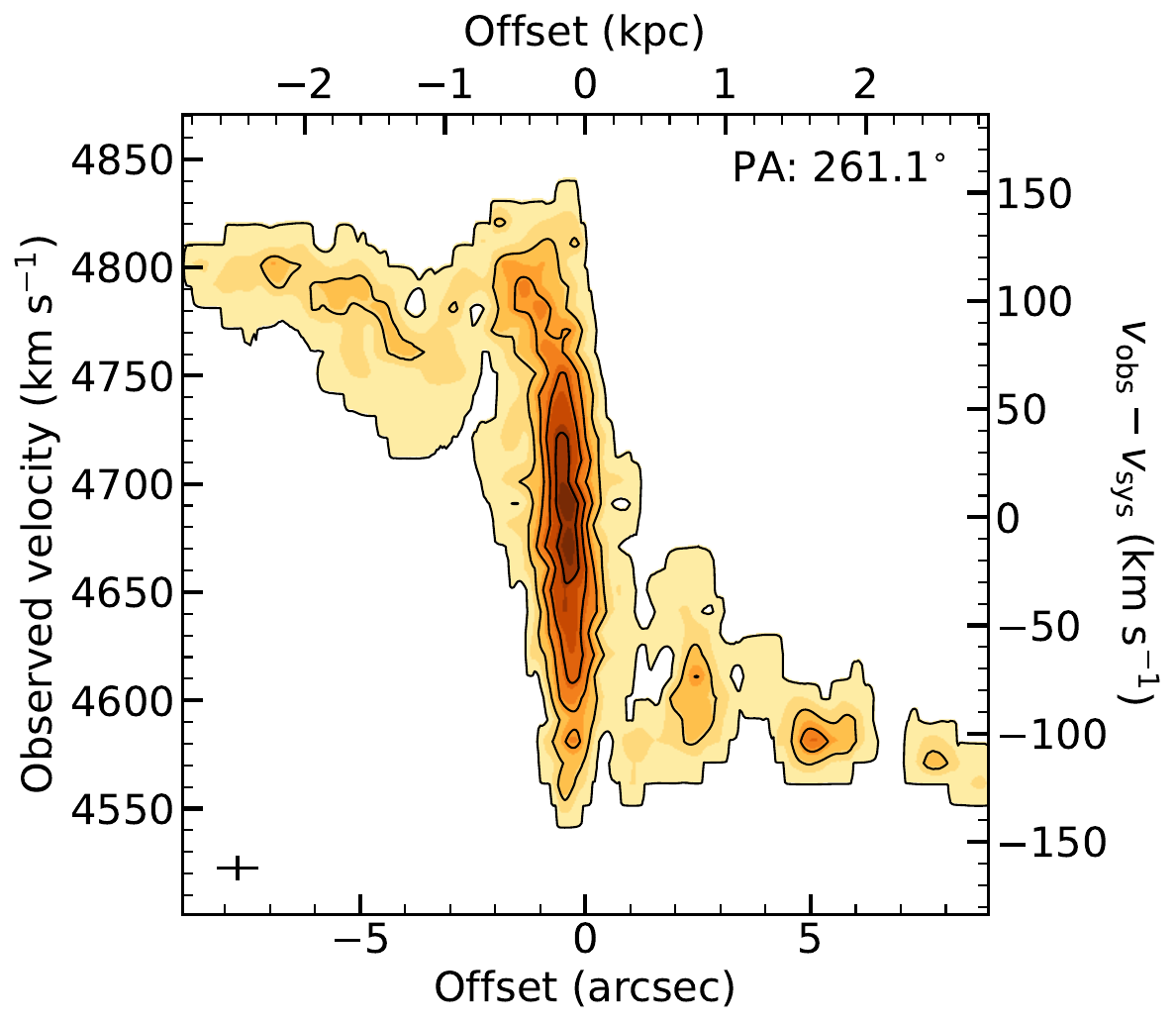}
    \end{subfigure}
    \caption[]{continued.}
\end{figure*}

\begin{figure*}
    \ContinuedFloat
    \begin{subfigure}{.34\linewidth}
    \includegraphics[height=6.0cm]{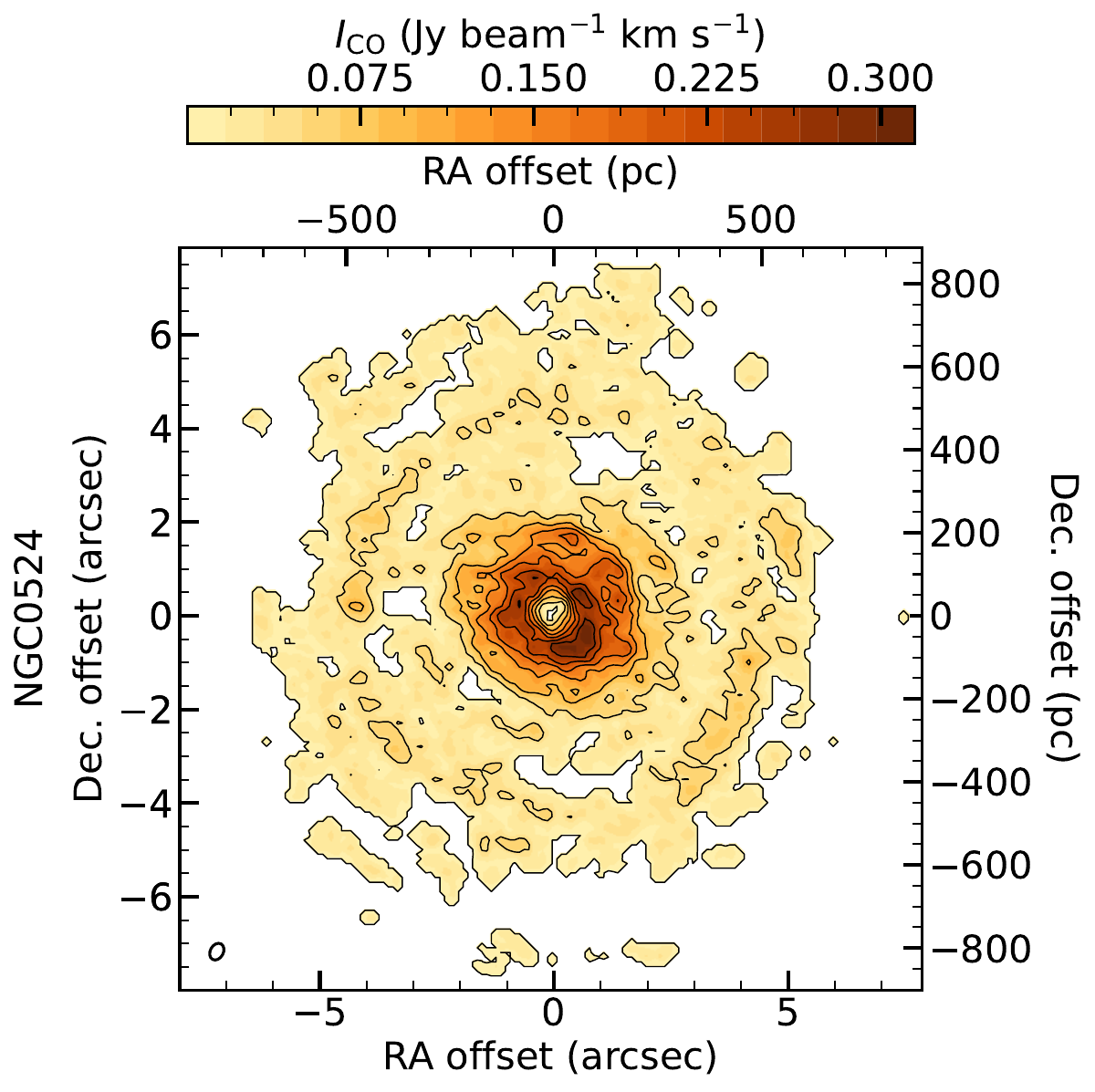}
    \end{subfigure}
    \begin{subfigure}{.32\linewidth}
    \centering
    \includegraphics[height=6.0cm]{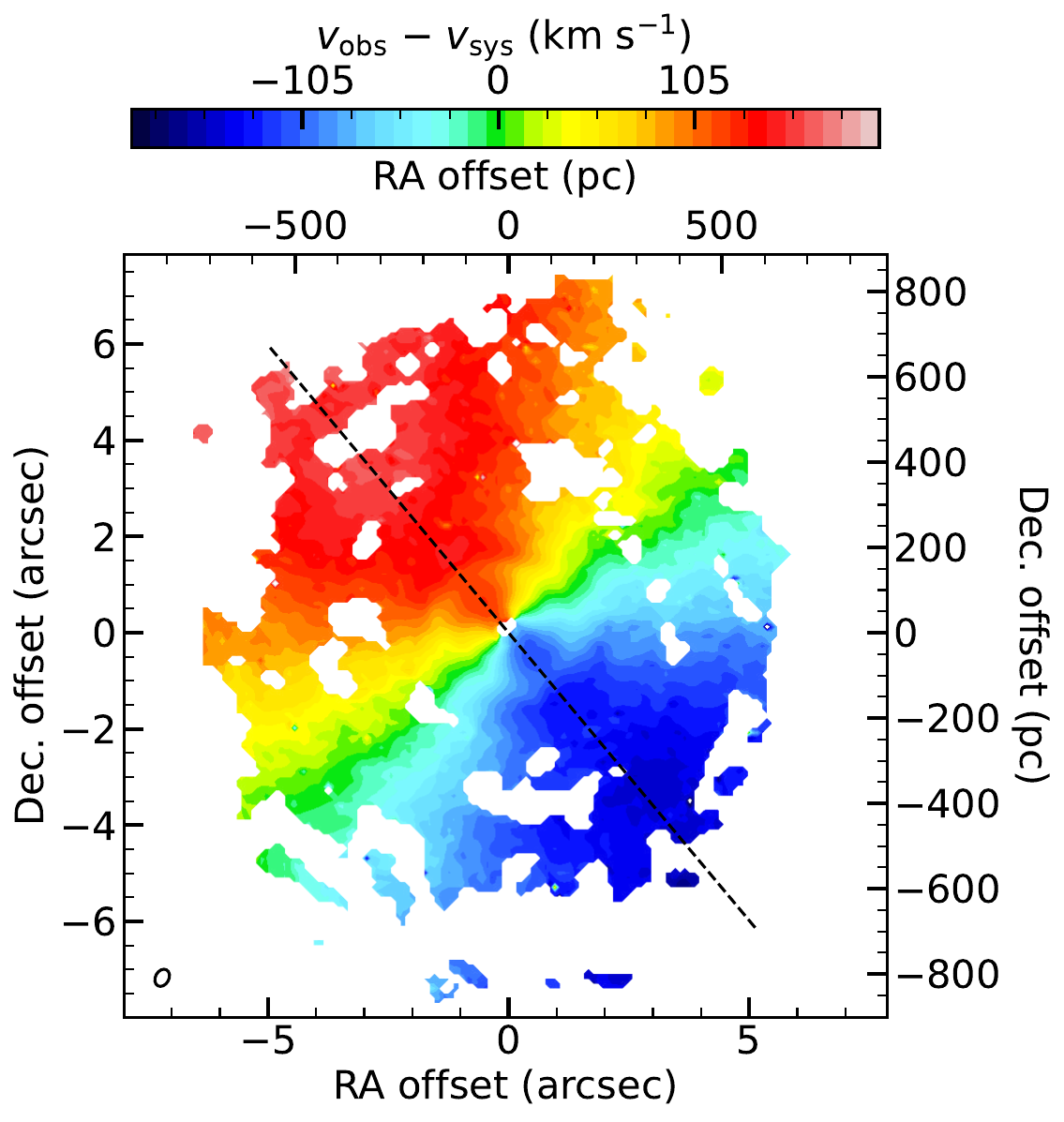}
    \end{subfigure}
    \begin{subfigure}{.32\linewidth}
    \centering
    \includegraphics[height=6.0cm]{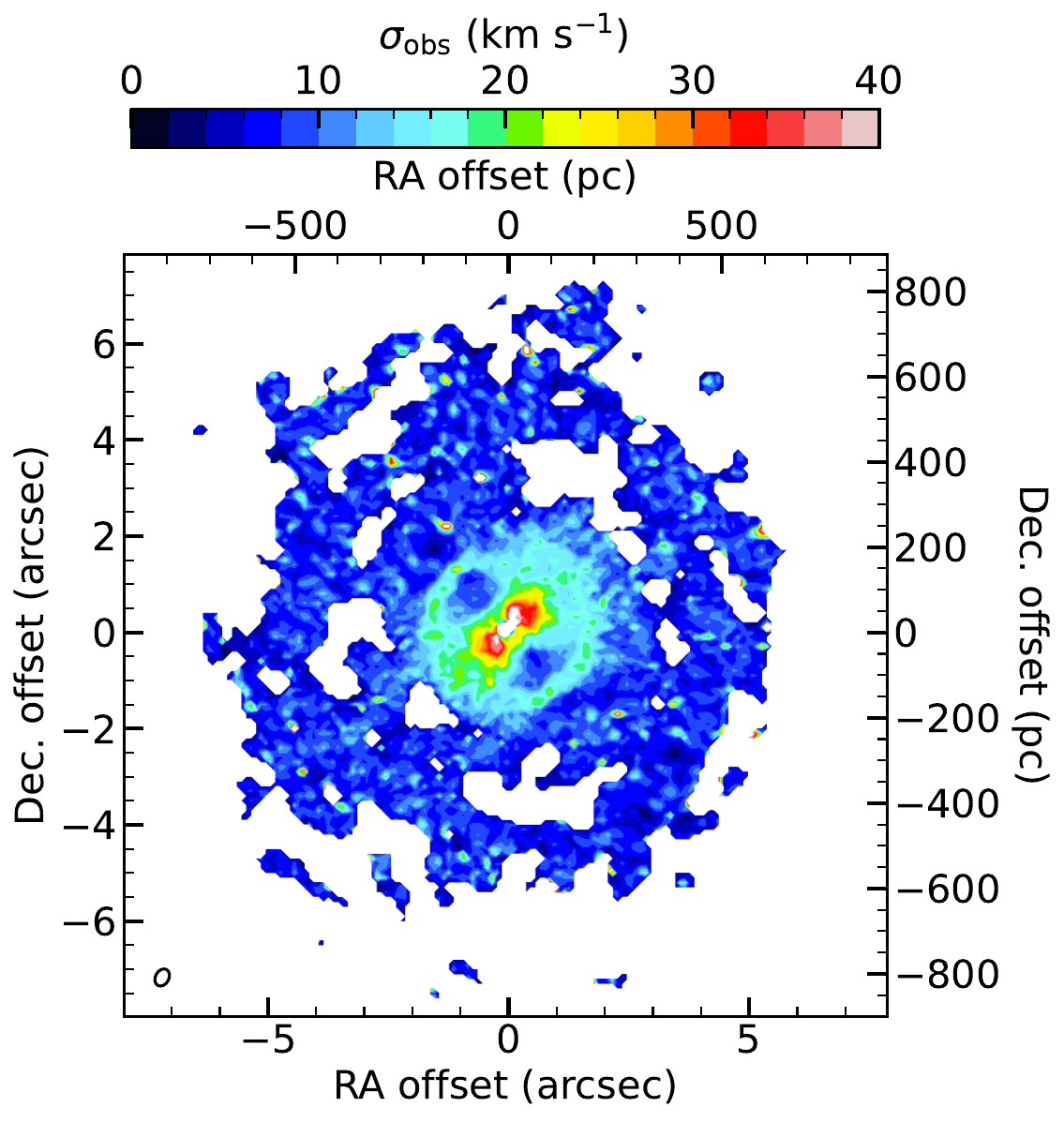}
    \end{subfigure}
    \\
    \begin{subfigure}{.34\linewidth}
    \centering
    \includegraphics[height=5.2cm]{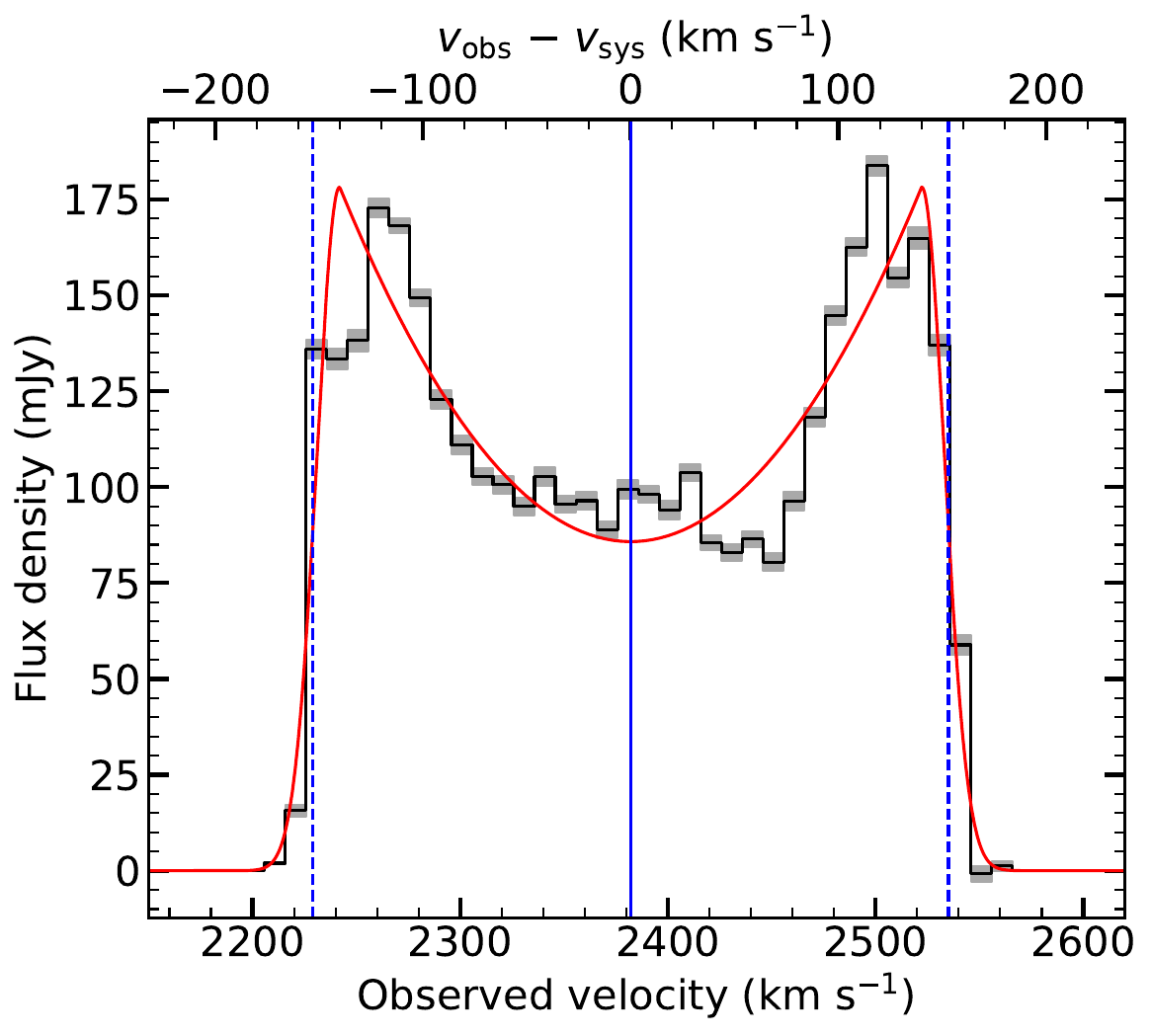}
    \end{subfigure}
    \begin{subfigure}{.32\linewidth}
    \centering
    \includegraphics[height=4.8cm]{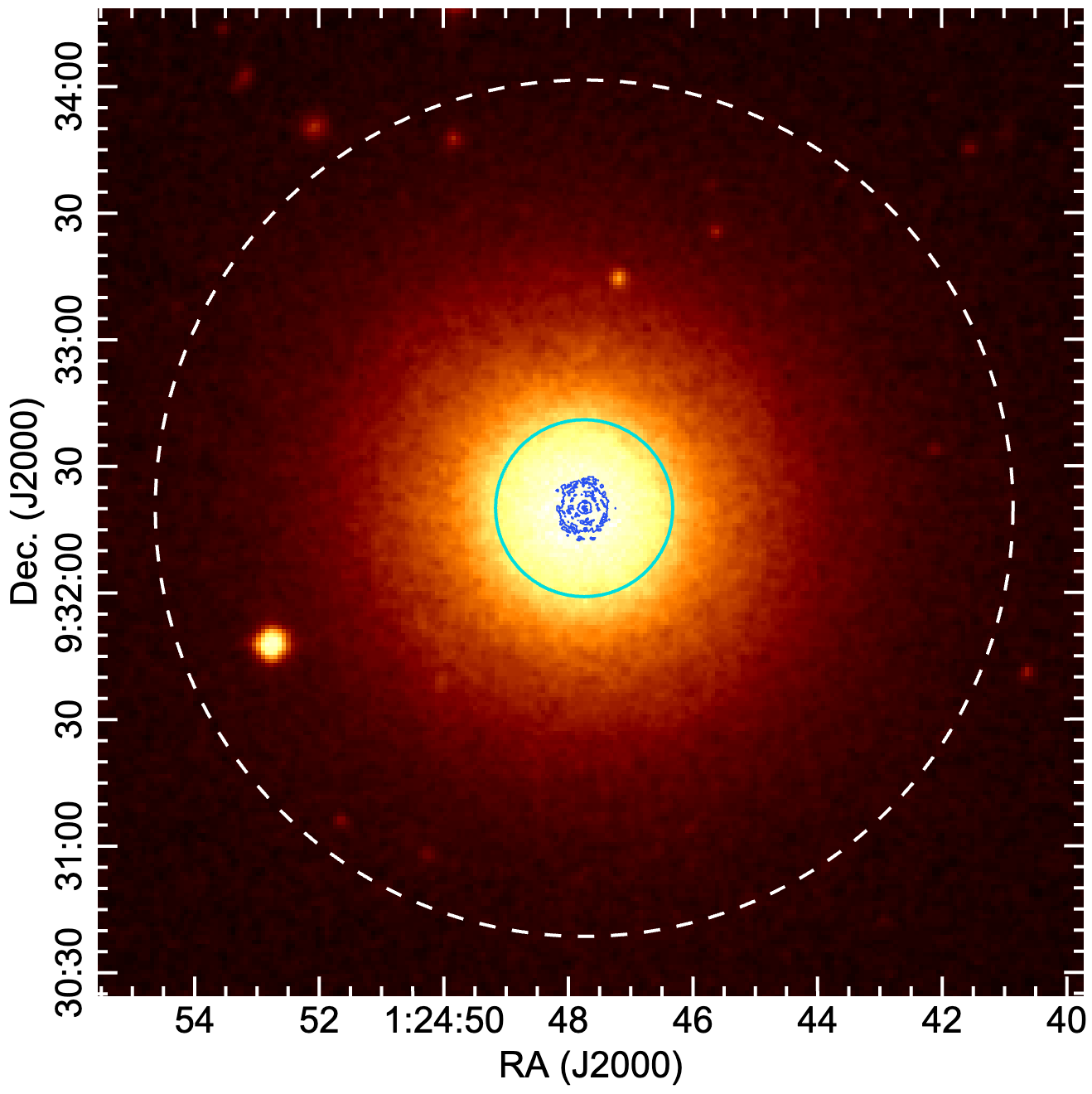}
    \end{subfigure}
    \begin{subfigure}{.33\linewidth}
    \centering
    \includegraphics[height=5.2cm]{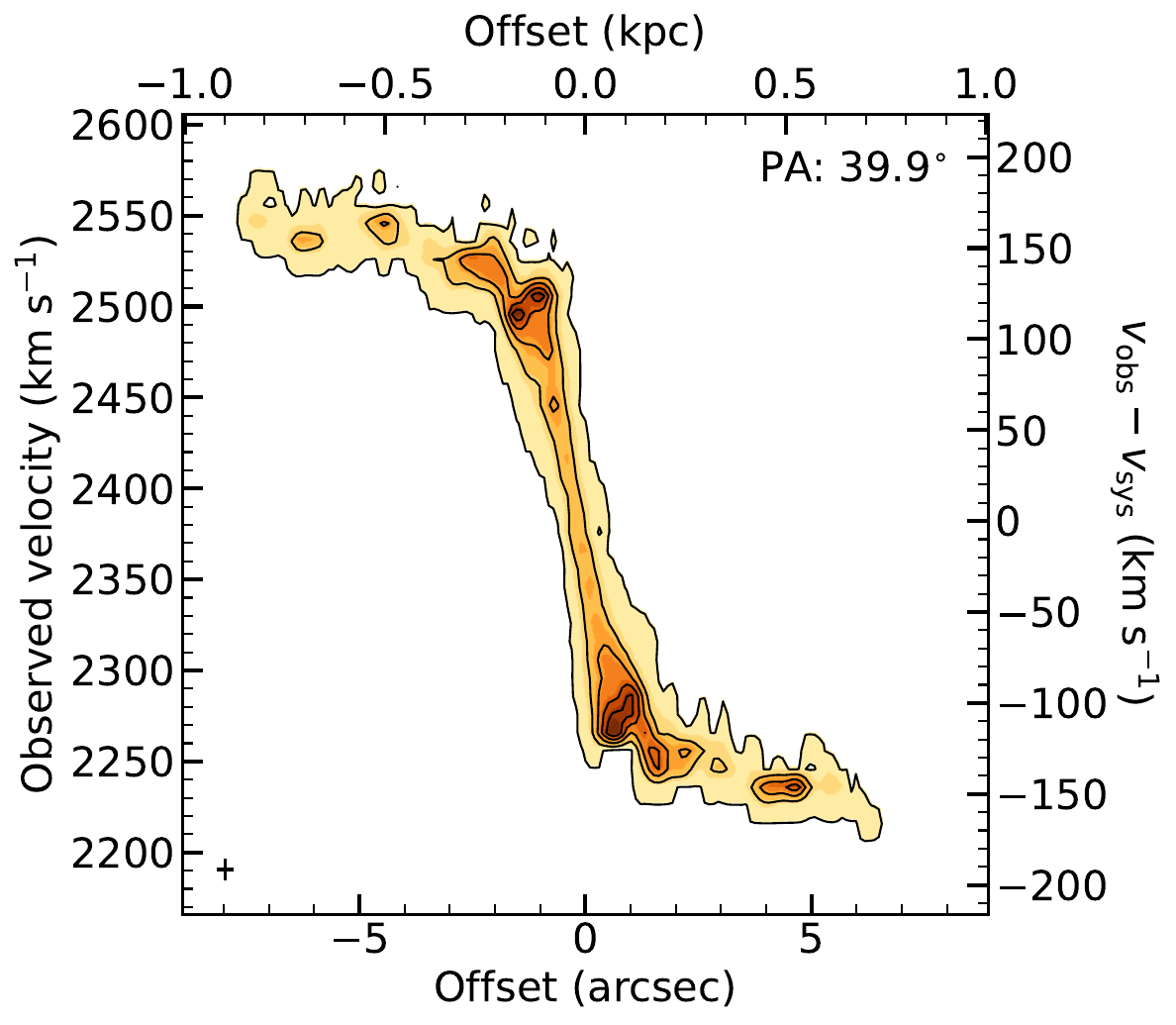}
    \end{subfigure}
    \\
    \begin{subfigure}{.34\linewidth}
    \includegraphics[height=6.0cm]{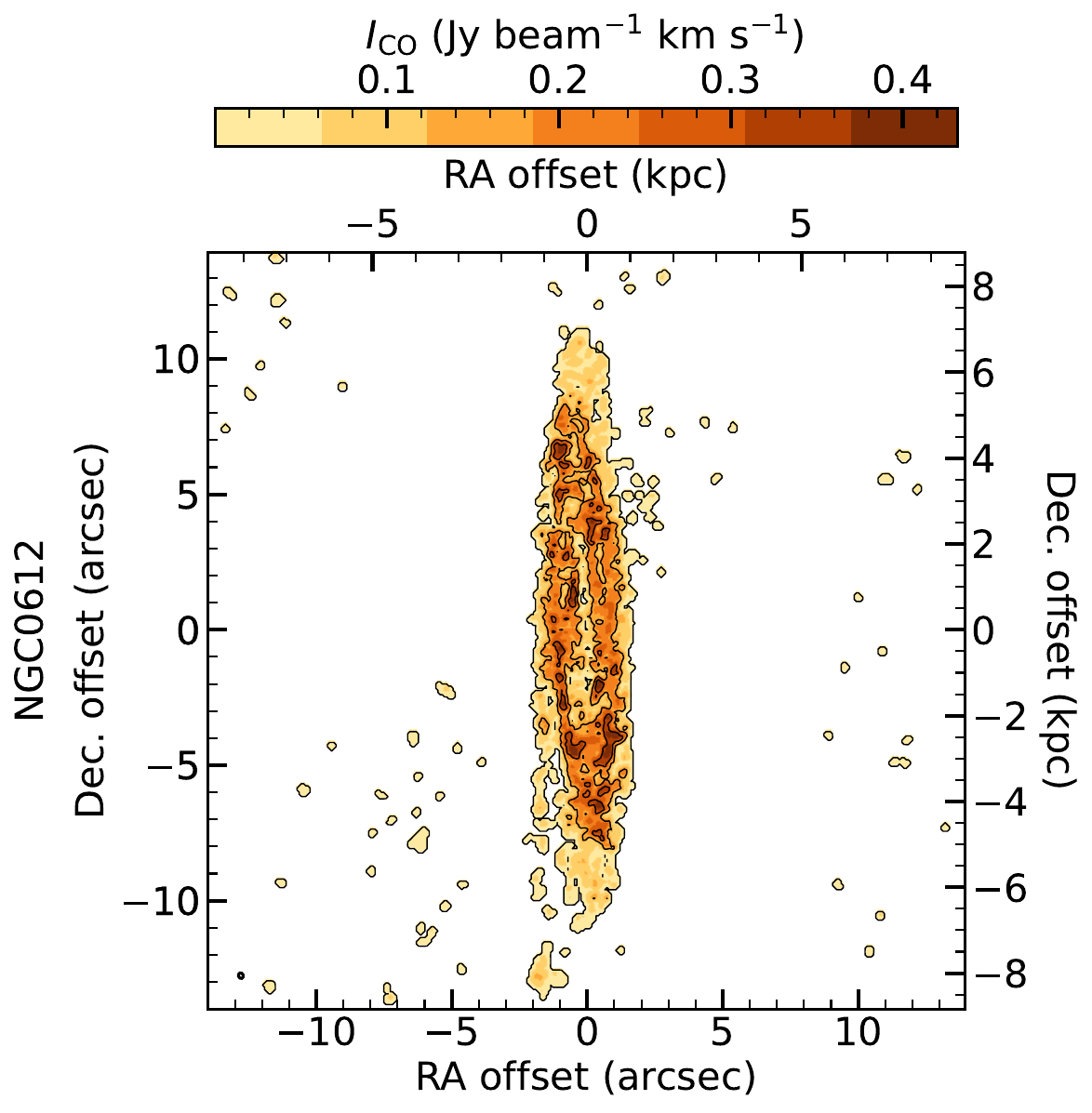}
    \end{subfigure}
    \begin{subfigure}{.32\linewidth}
    \centering
    \includegraphics[height=6.0cm]{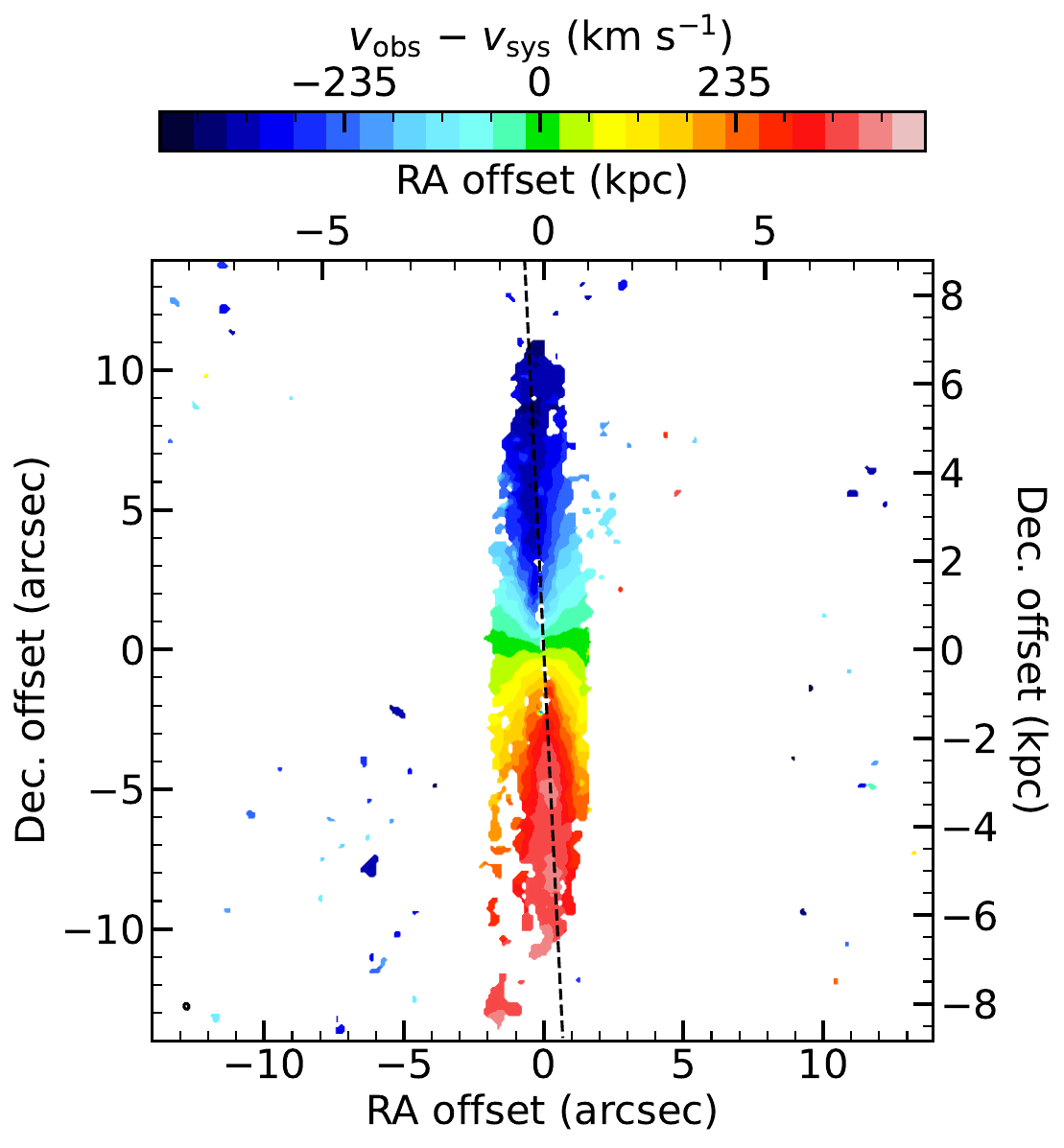}
    \end{subfigure}
    \begin{subfigure}{.32\linewidth}
    \centering
    \includegraphics[height=6.0cm]{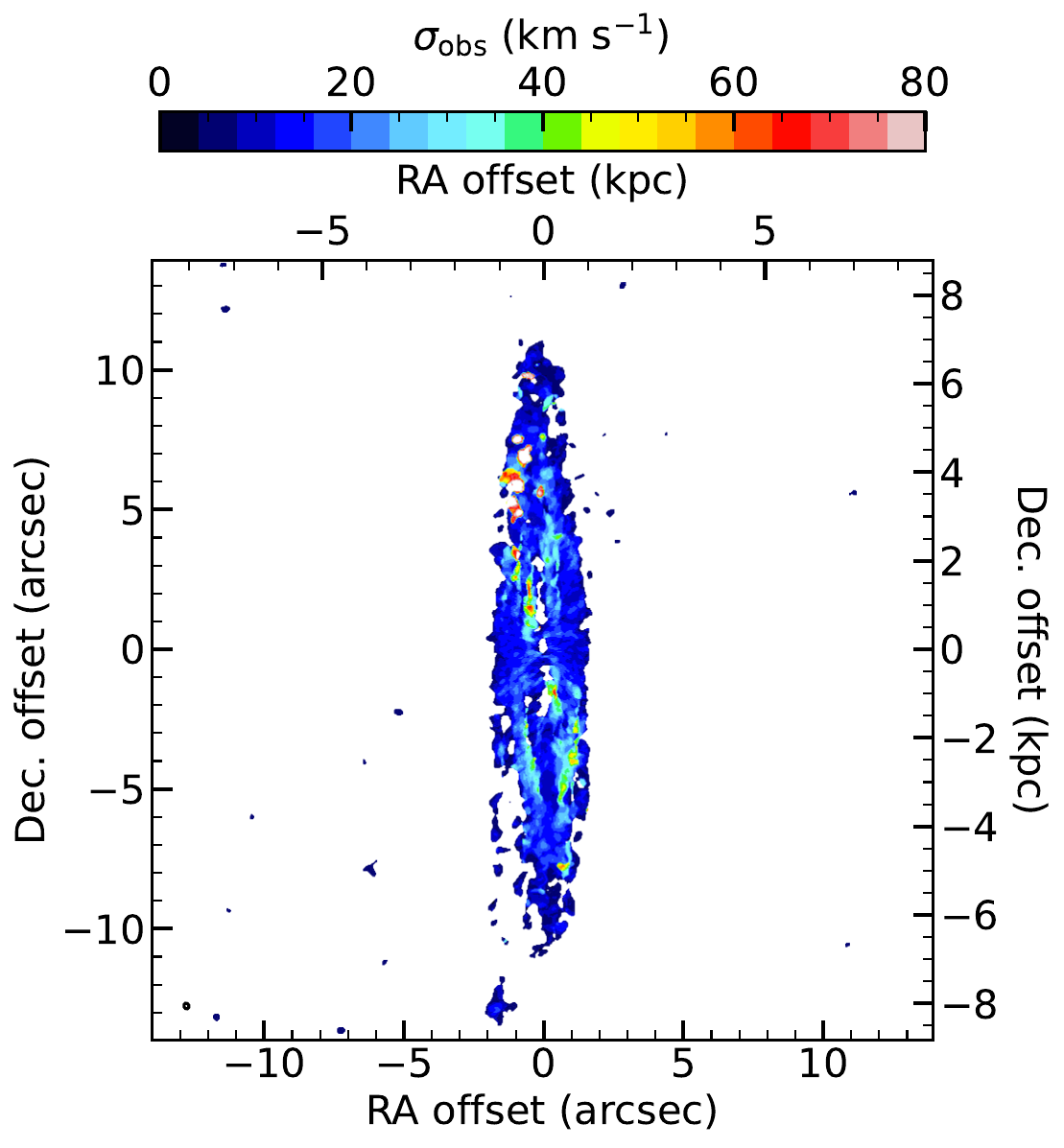}
    \end{subfigure}
    \\
    \begin{subfigure}{.34\linewidth}
    \centering
    \includegraphics[height=5.2cm]{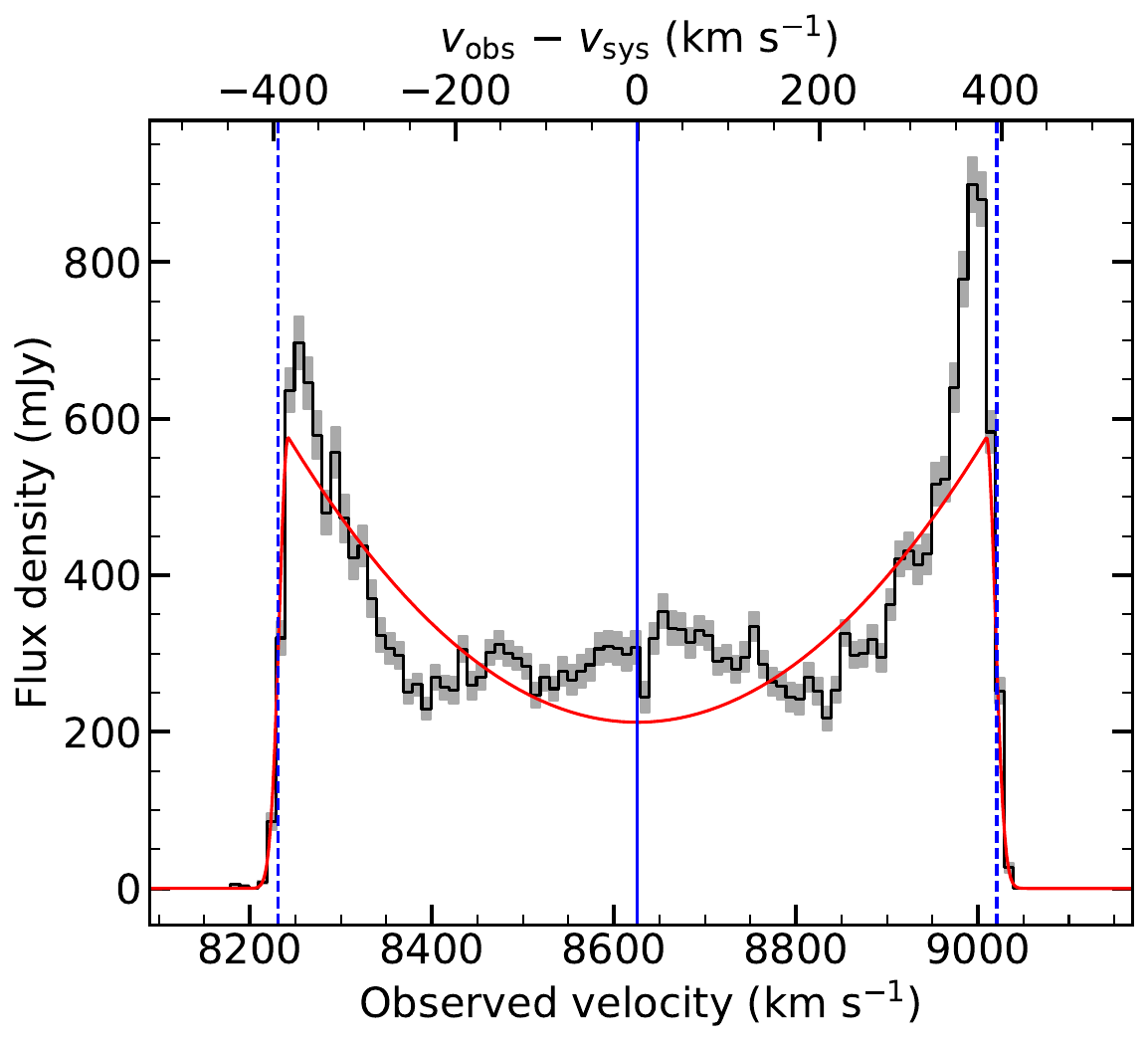}
    \end{subfigure}
    \begin{subfigure}{.32\linewidth}
    \centering
    \includegraphics[height=4.8cm]{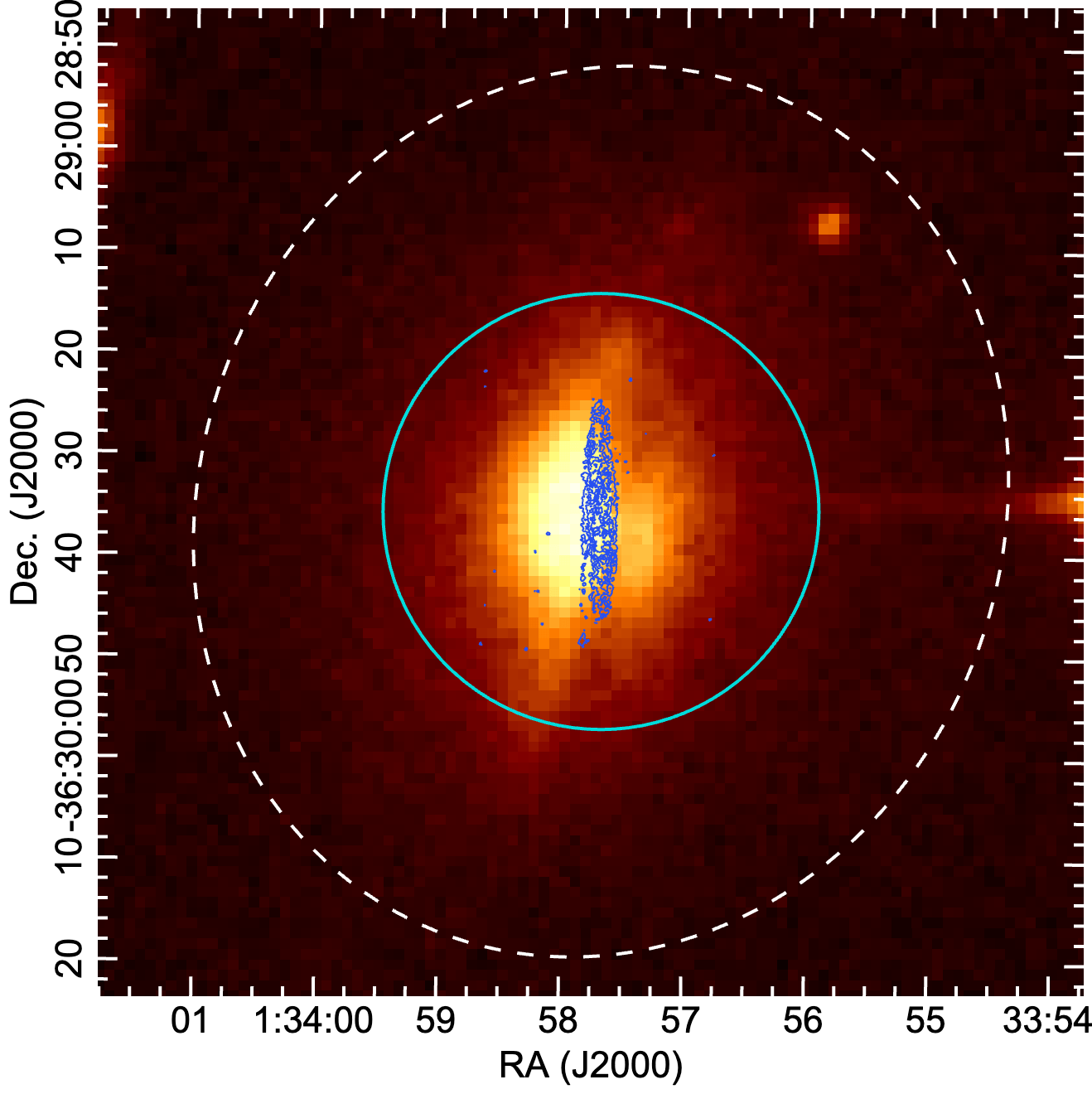}
    \end{subfigure}
    \begin{subfigure}{.33\linewidth}
    \centering
    \includegraphics[height=5.2cm]{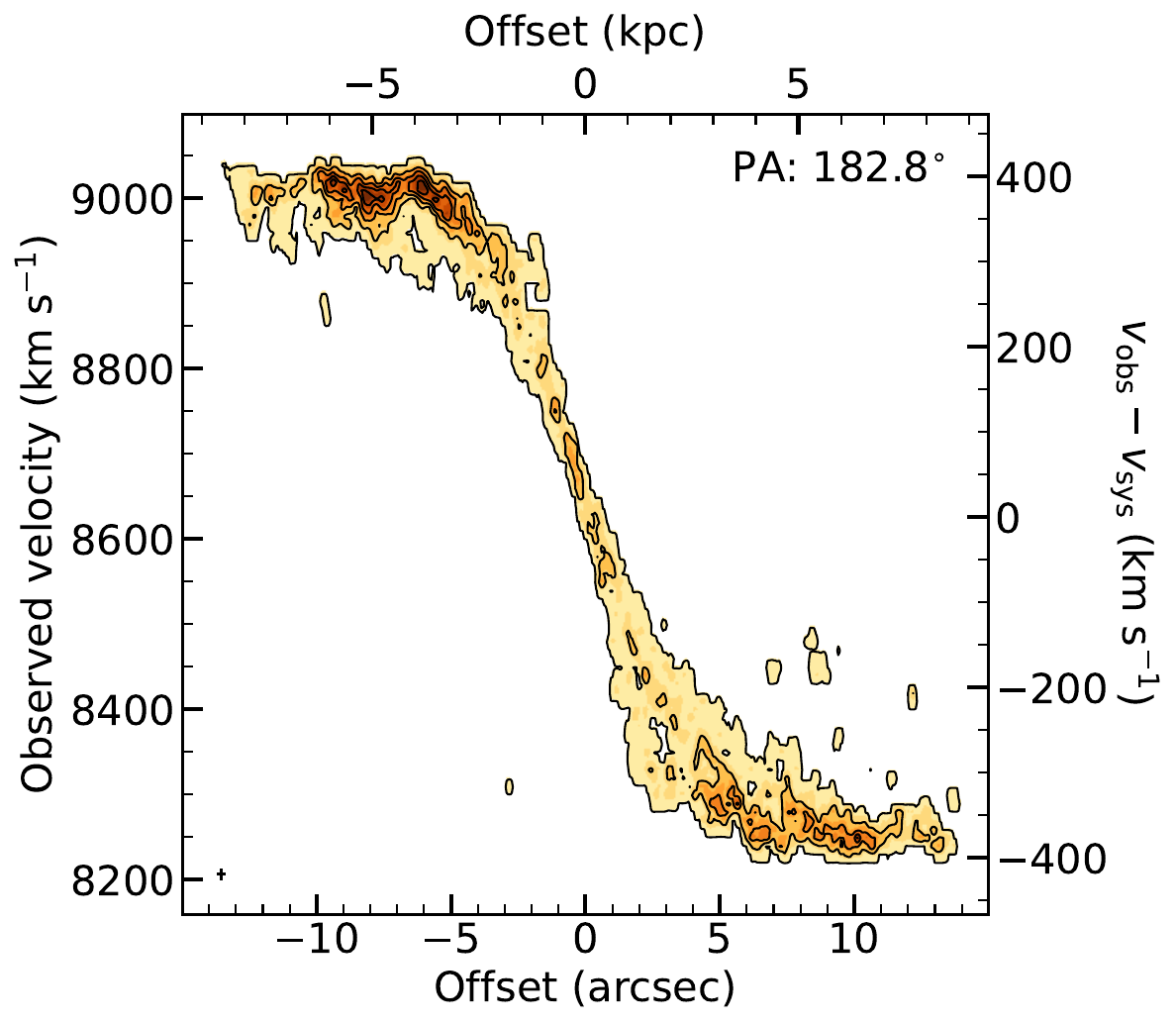}
    \end{subfigure}
    \caption[]{continued.}
\end{figure*}

\begin{figure*}
    \ContinuedFloat
    \begin{subfigure}{.34\linewidth}
    \includegraphics[height=6.0cm]{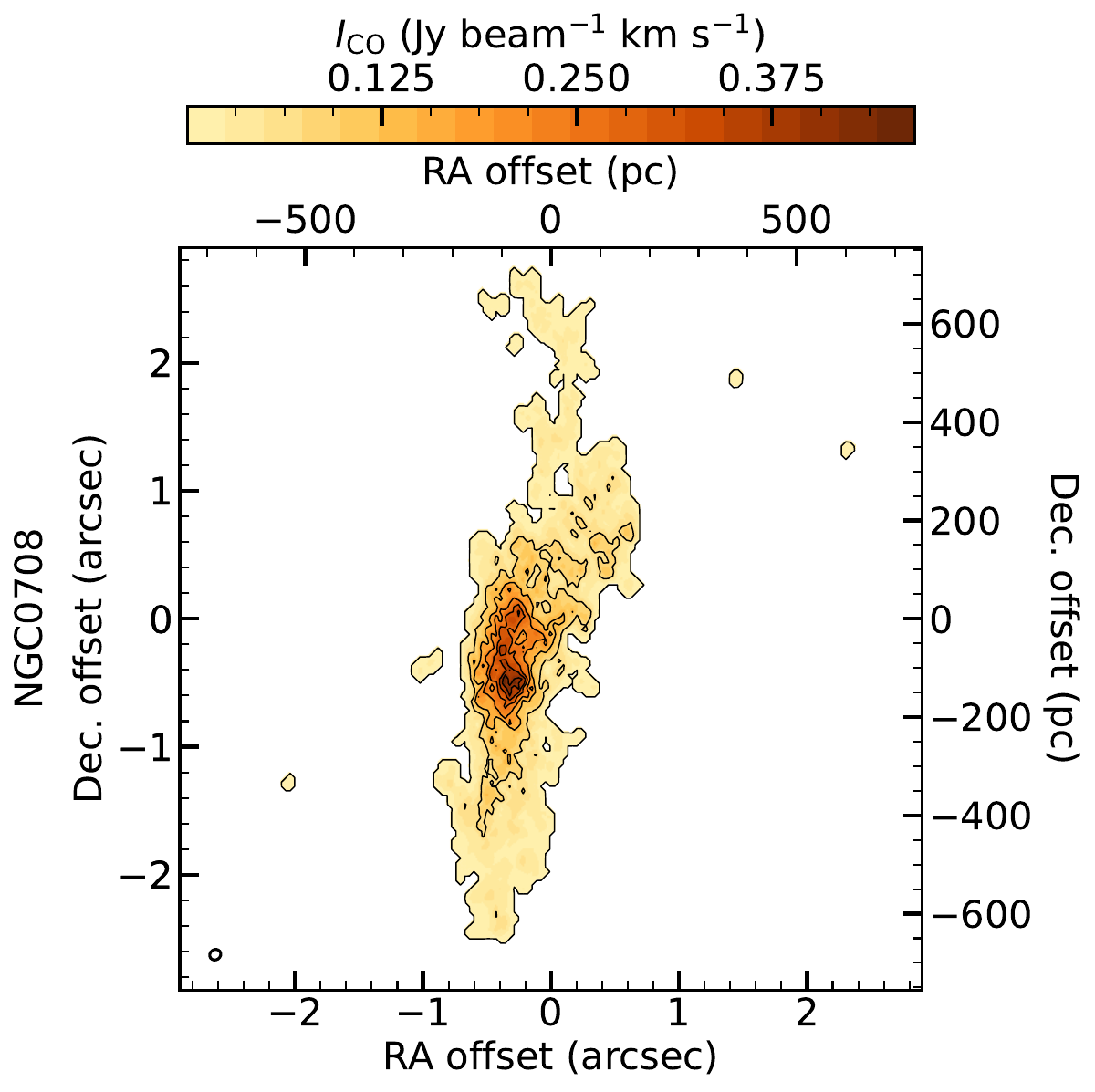}
    \end{subfigure}
    \begin{subfigure}{.32\linewidth}
    \centering
    \includegraphics[height=6.0cm]{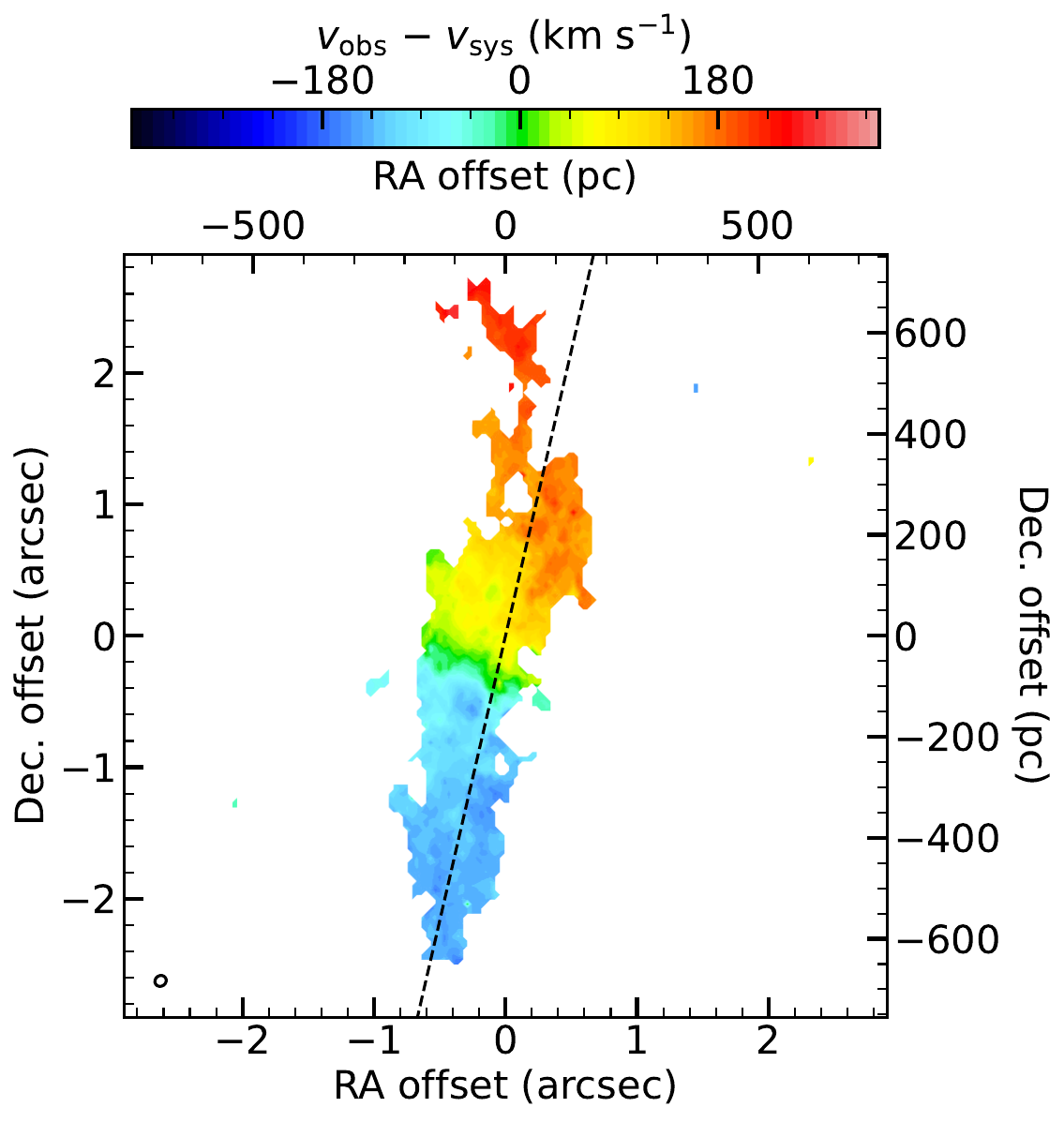}
    \end{subfigure}
    \begin{subfigure}{.32\linewidth}
    \centering
    \includegraphics[height=6.0cm]{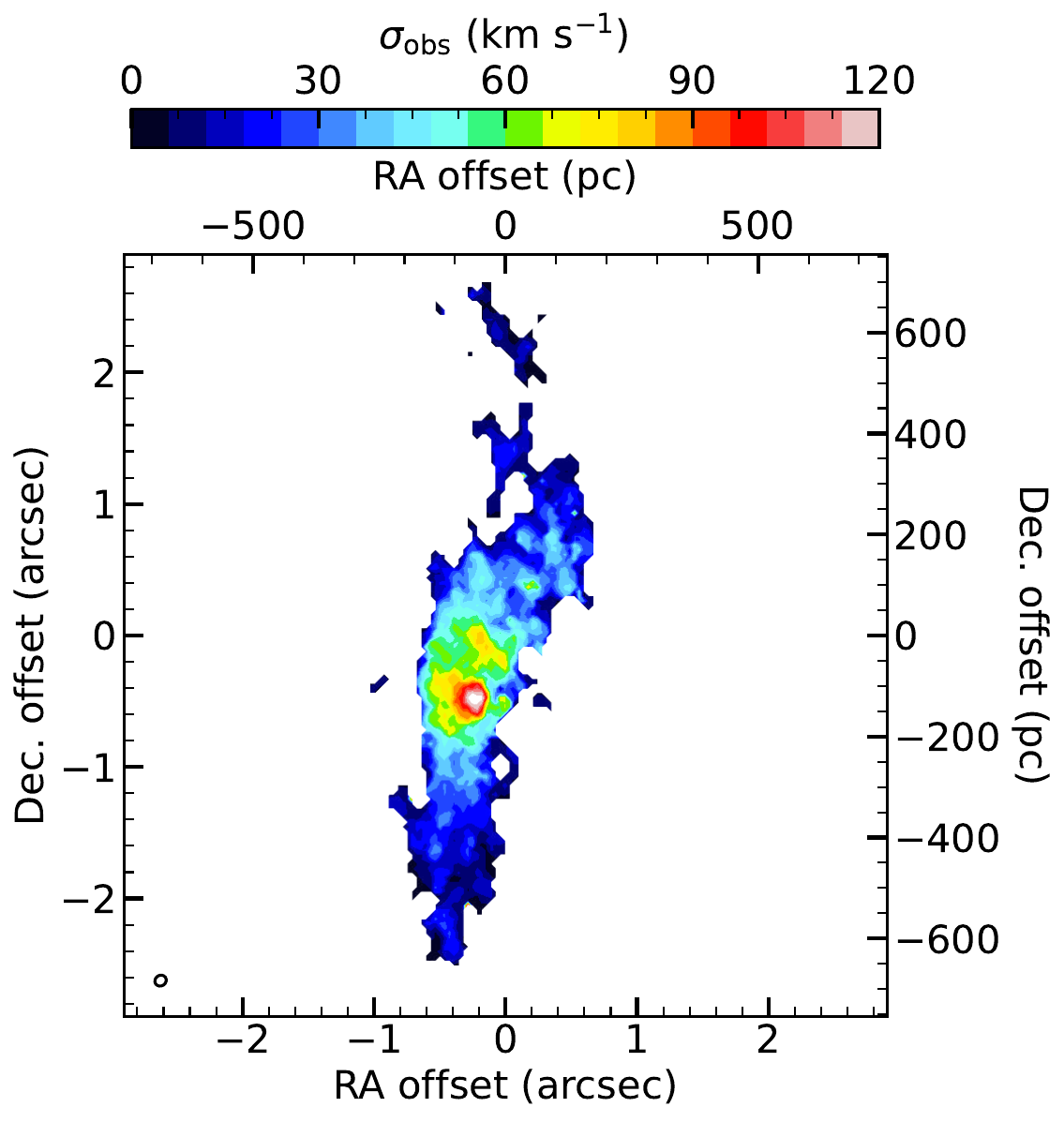}
    \end{subfigure}
    \\
    \begin{subfigure}{.34\linewidth}
    \centering
    \includegraphics[height=5.2cm]{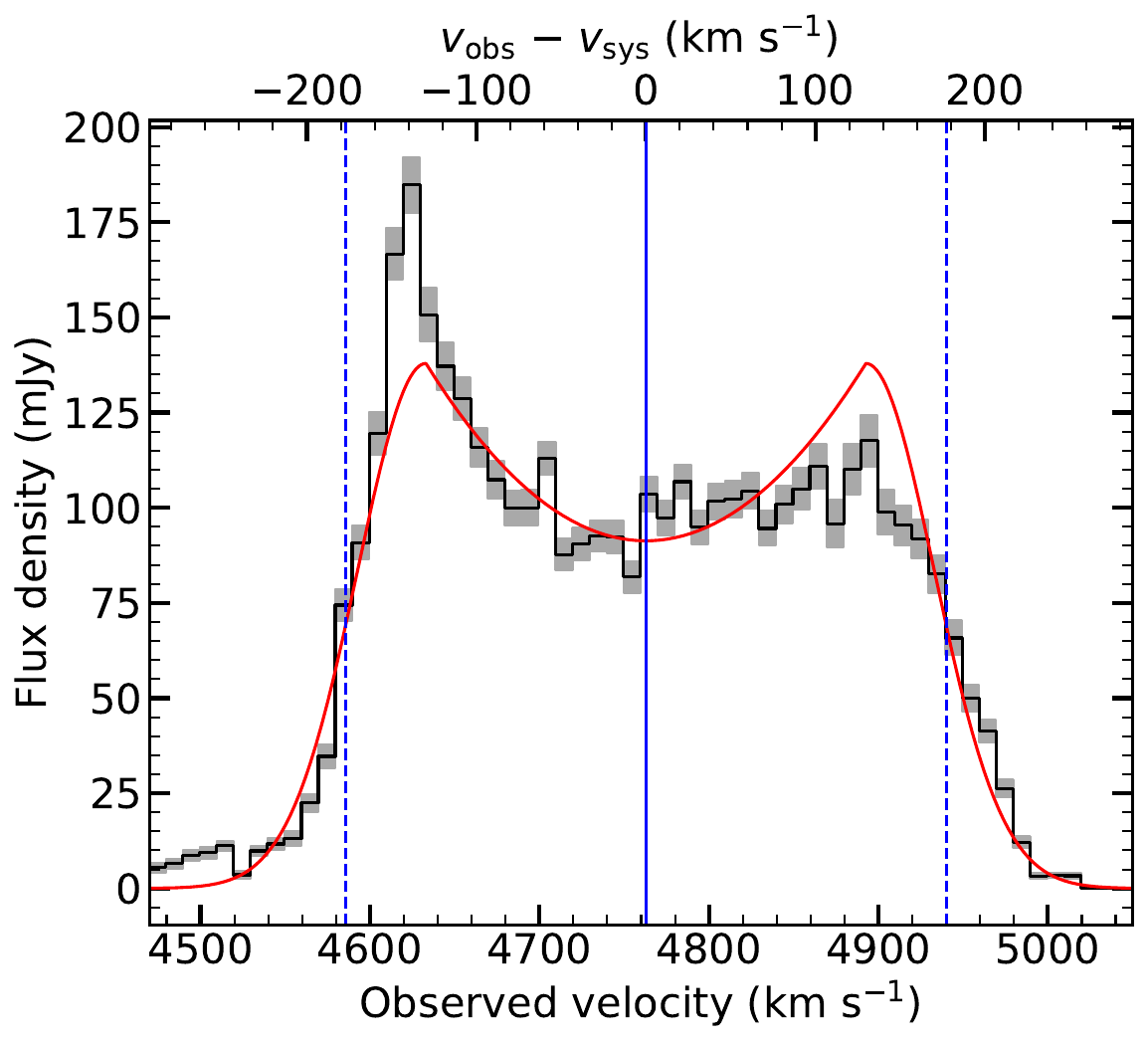}
    \end{subfigure}
    \begin{subfigure}{.32\linewidth}
    \centering
    \includegraphics[height=4.8cm]{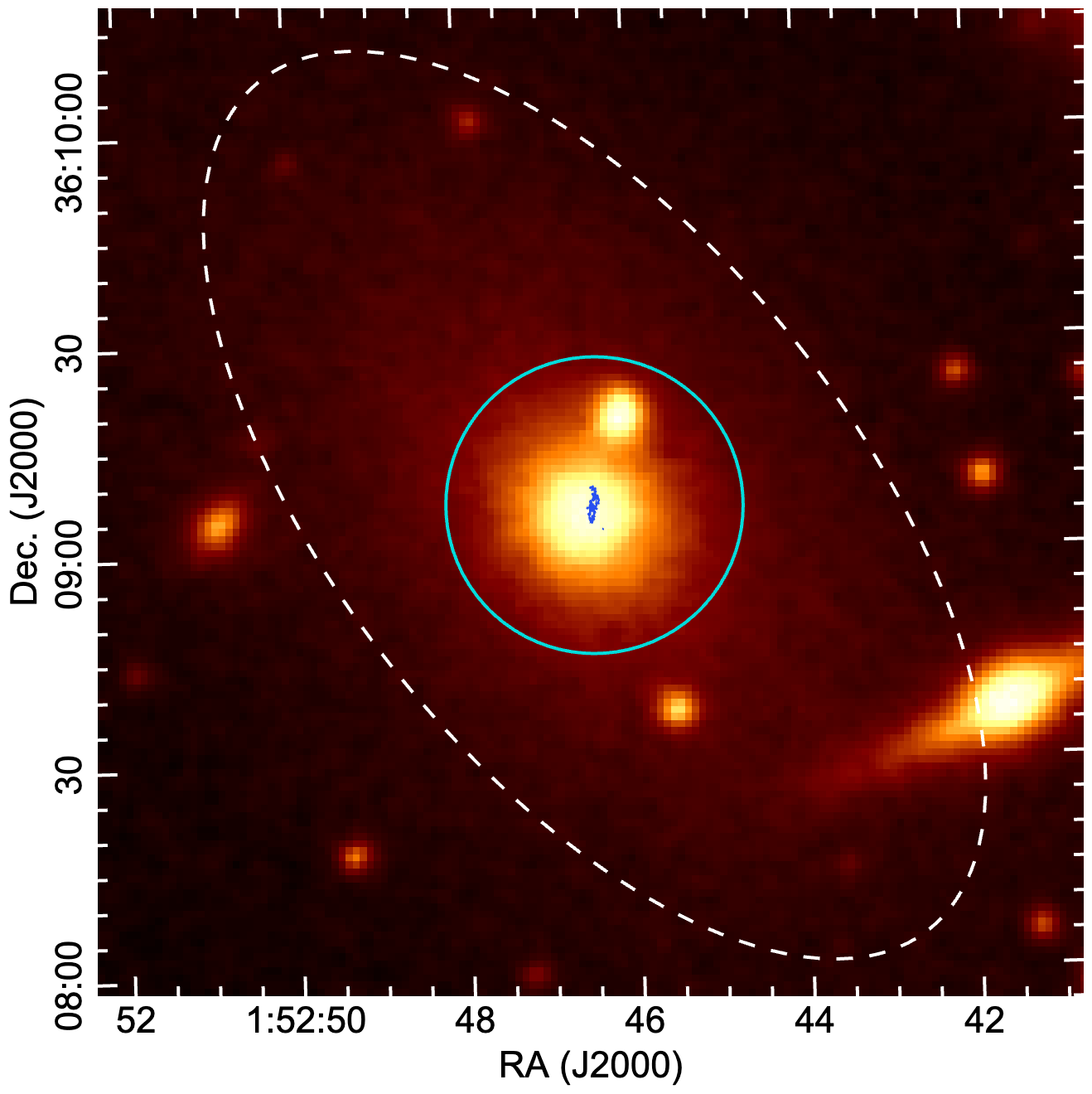}
    \end{subfigure}
    \begin{subfigure}{.33\linewidth}
    \centering
    \includegraphics[height=5.2cm]{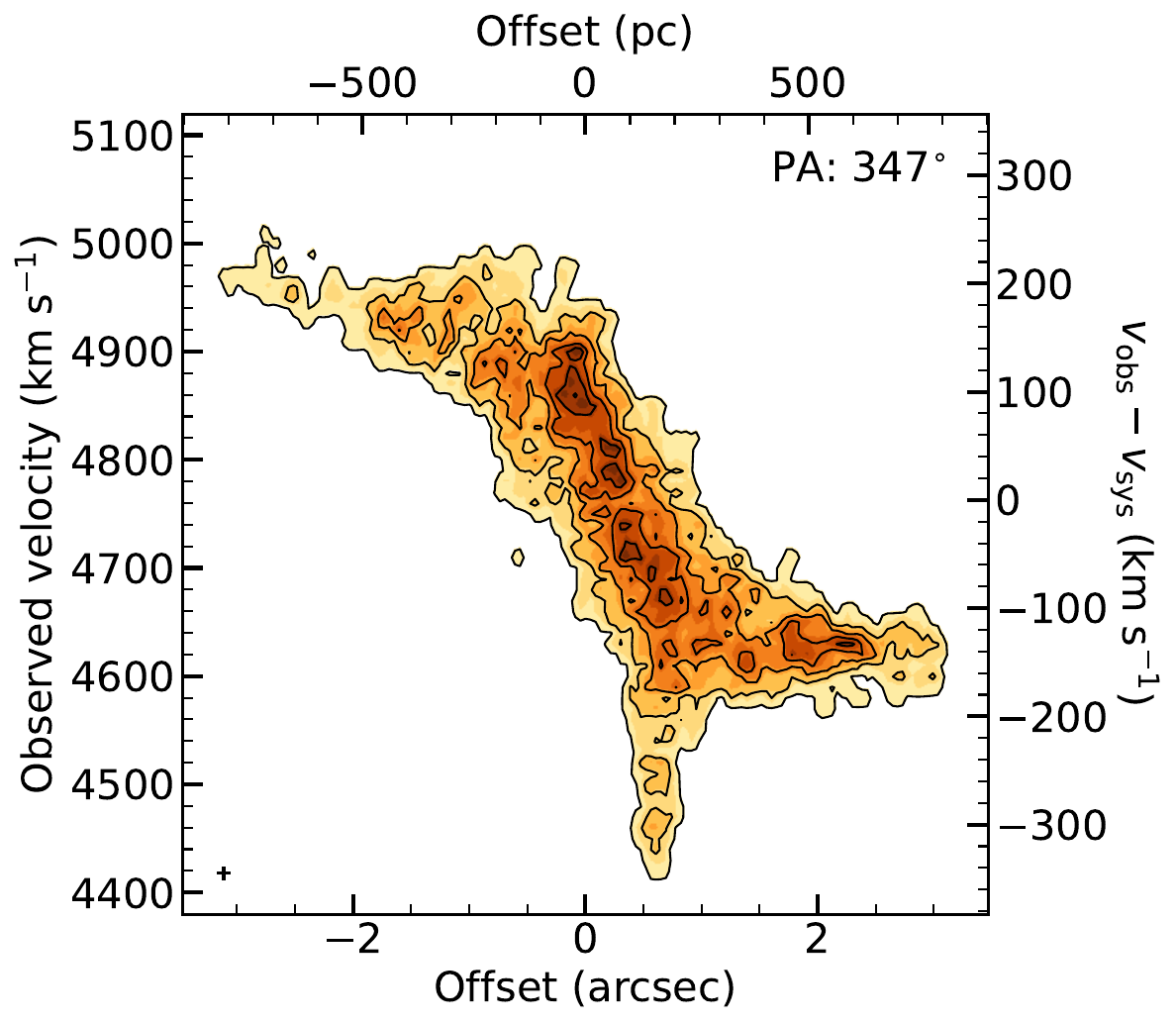}
    \end{subfigure}
    \\
    \begin{subfigure}{.34\linewidth}
    \includegraphics[height=6.0cm]{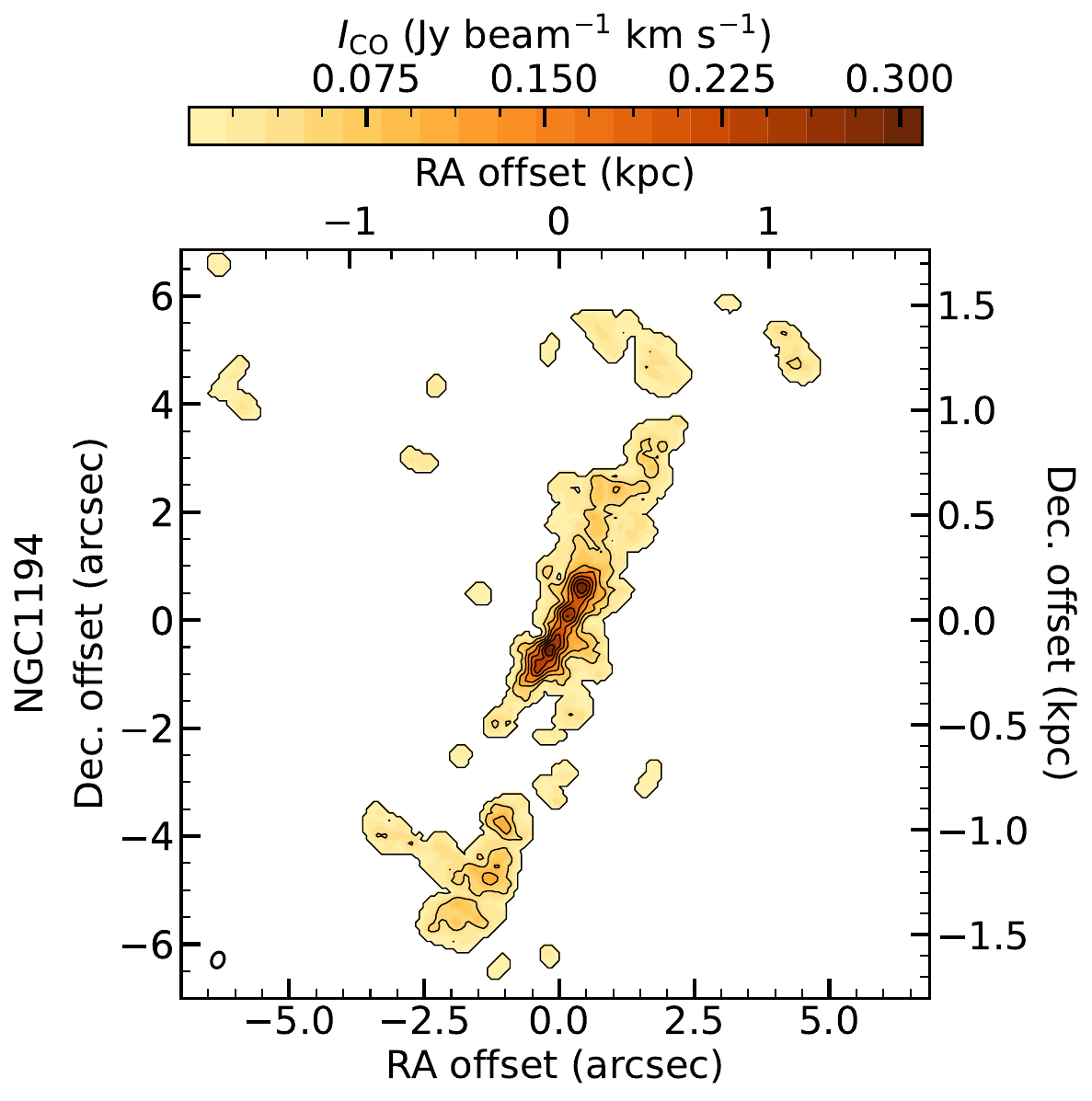}
    \end{subfigure}
    \begin{subfigure}{.32\linewidth}
    \centering
    \includegraphics[height=6.0cm]{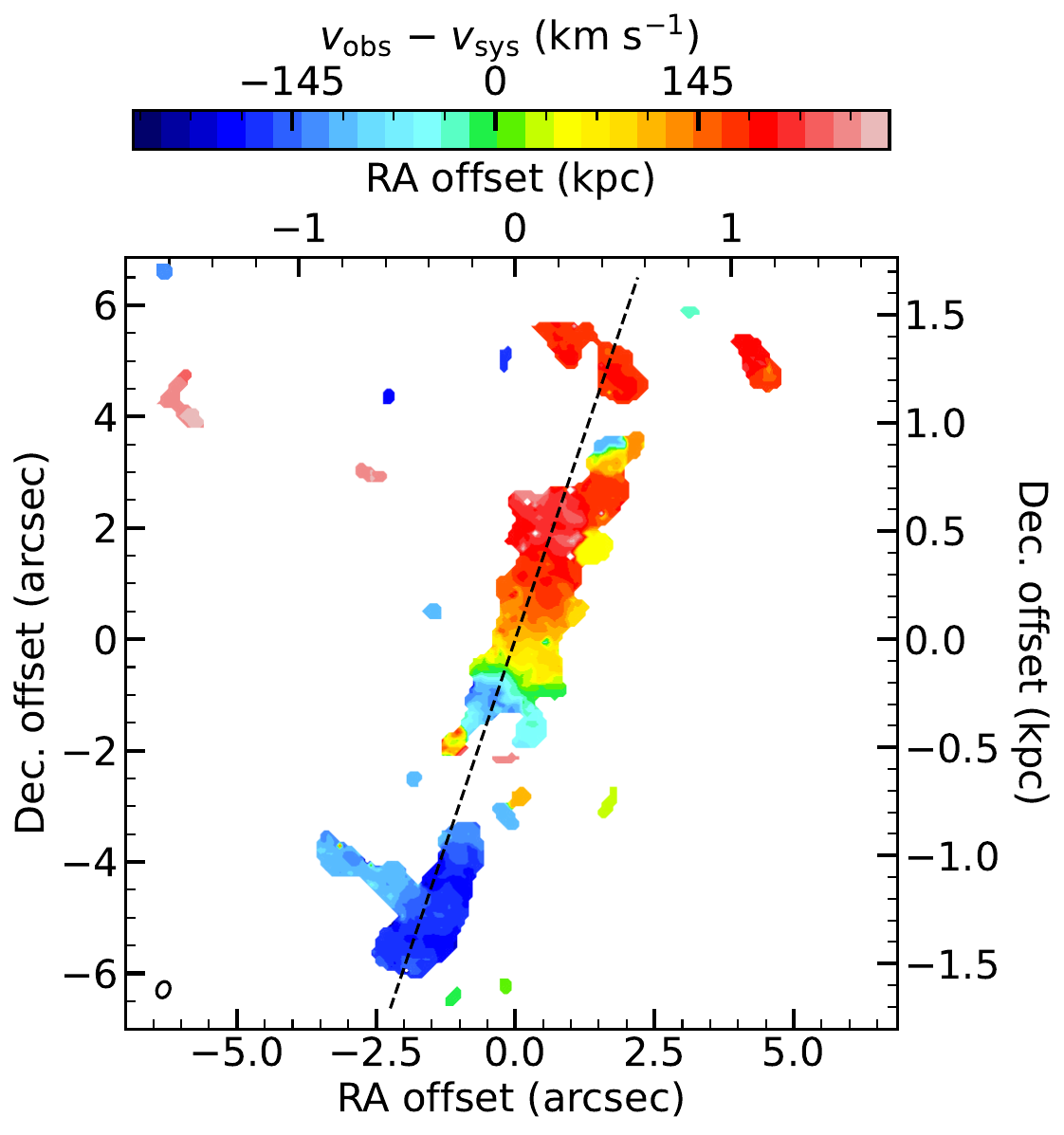}
    \end{subfigure}
    \begin{subfigure}{.32\linewidth}
    \centering
    \includegraphics[height=6.0cm]{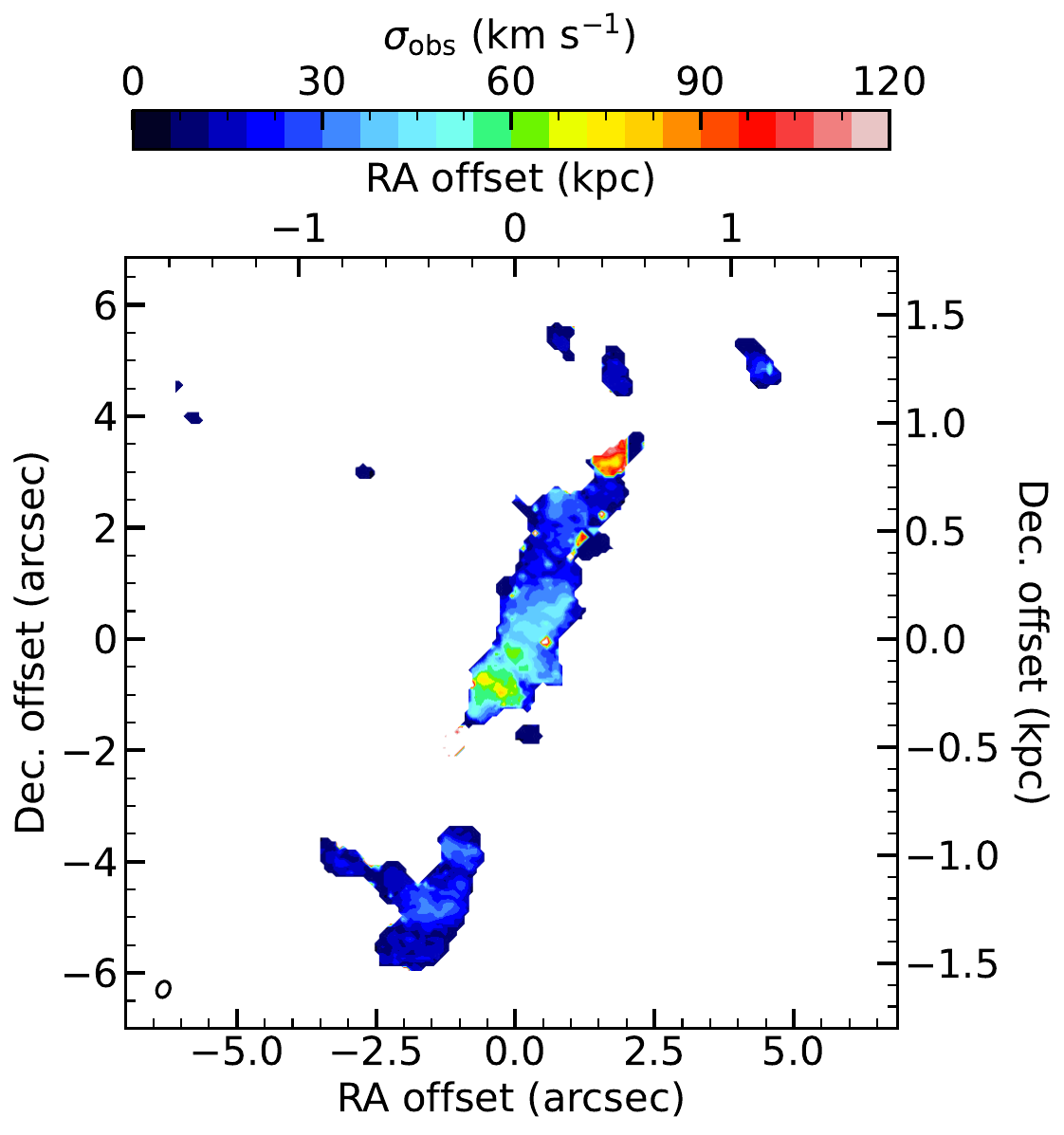}
    \end{subfigure}
    \\
    \begin{subfigure}{.34\linewidth}
    \centering
    \includegraphics[height=5.2cm]{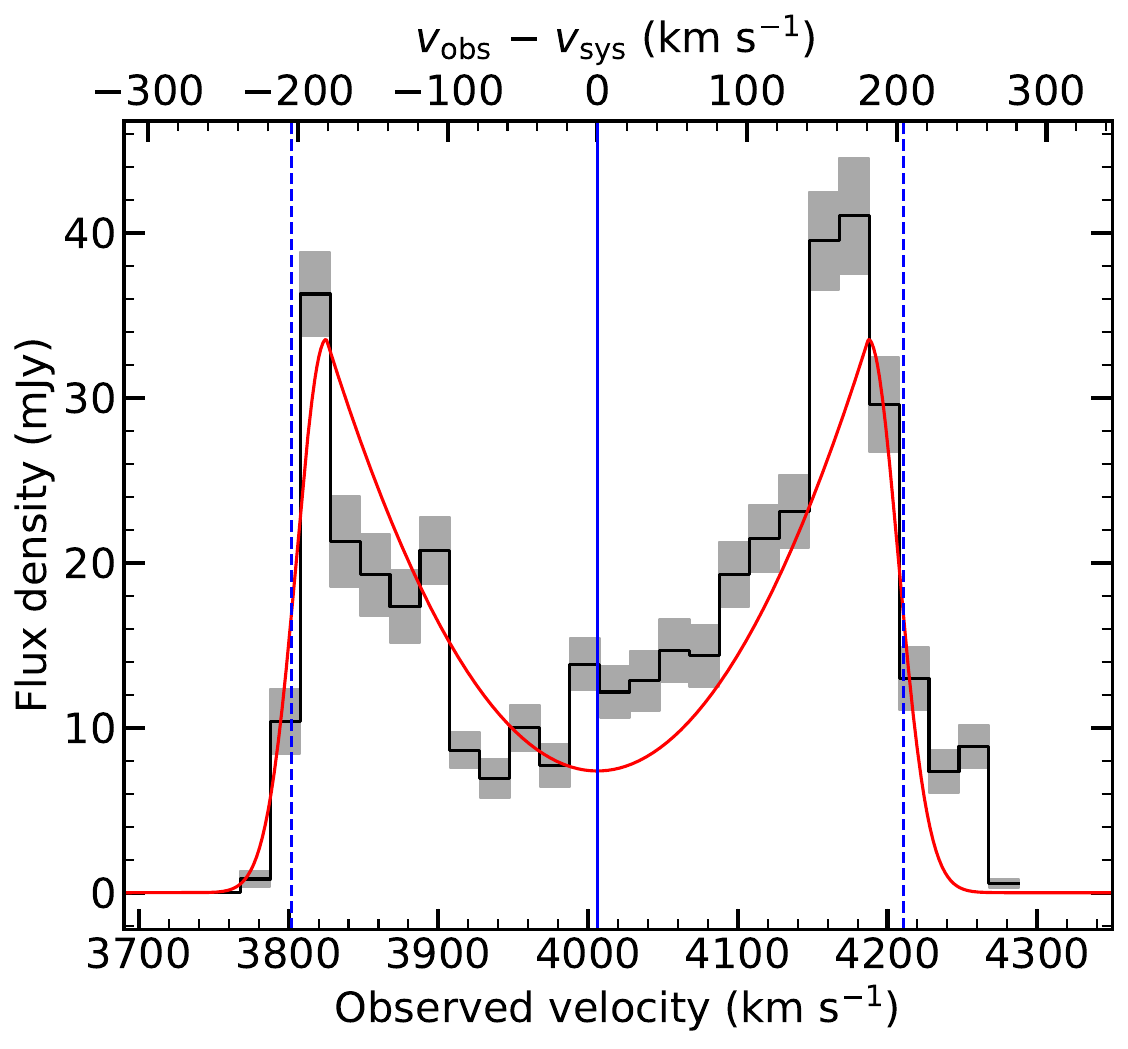}
    \end{subfigure}
    \begin{subfigure}{.32\linewidth}
    \centering
    \includegraphics[height=4.8cm]{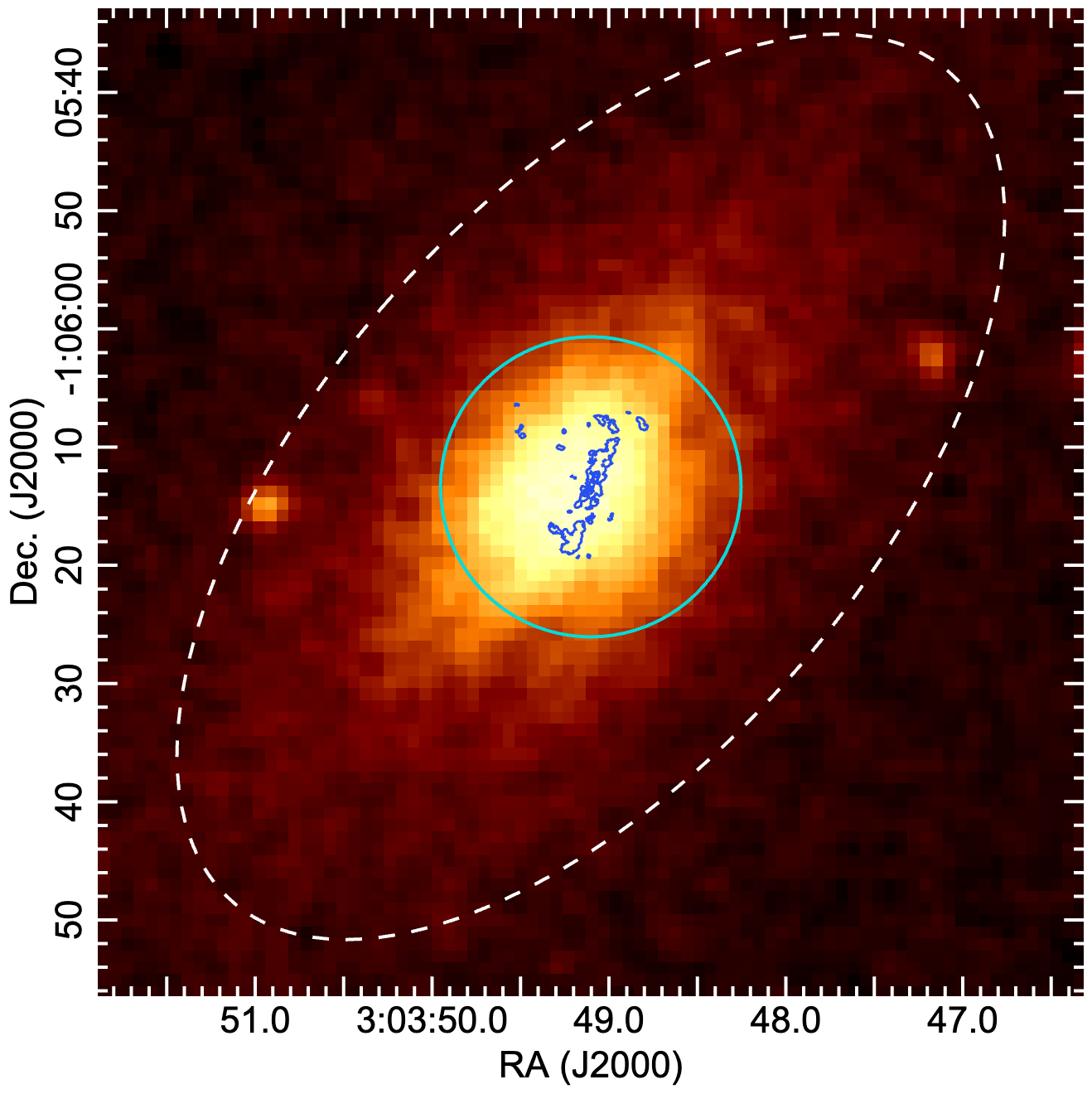}
    \end{subfigure}
    \begin{subfigure}{.33\linewidth}
    \centering
    \includegraphics[height=5.2cm]{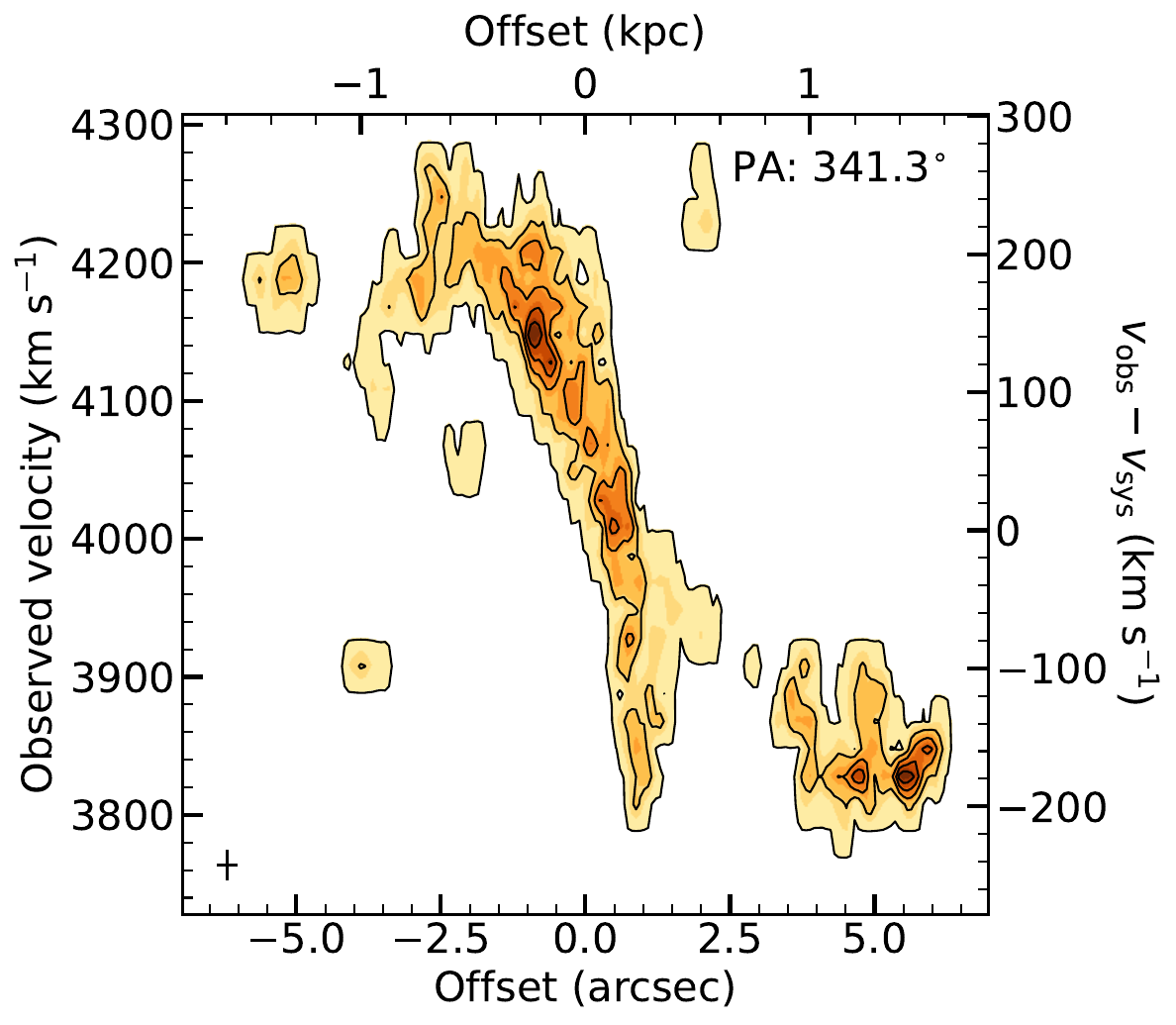}
    \end{subfigure}
    \caption[]{continued.}
\end{figure*}

\begin{figure*}
    \ContinuedFloat
    \centering
    \begin{subfigure}{.34\linewidth}
    \includegraphics[height=5.9cm]{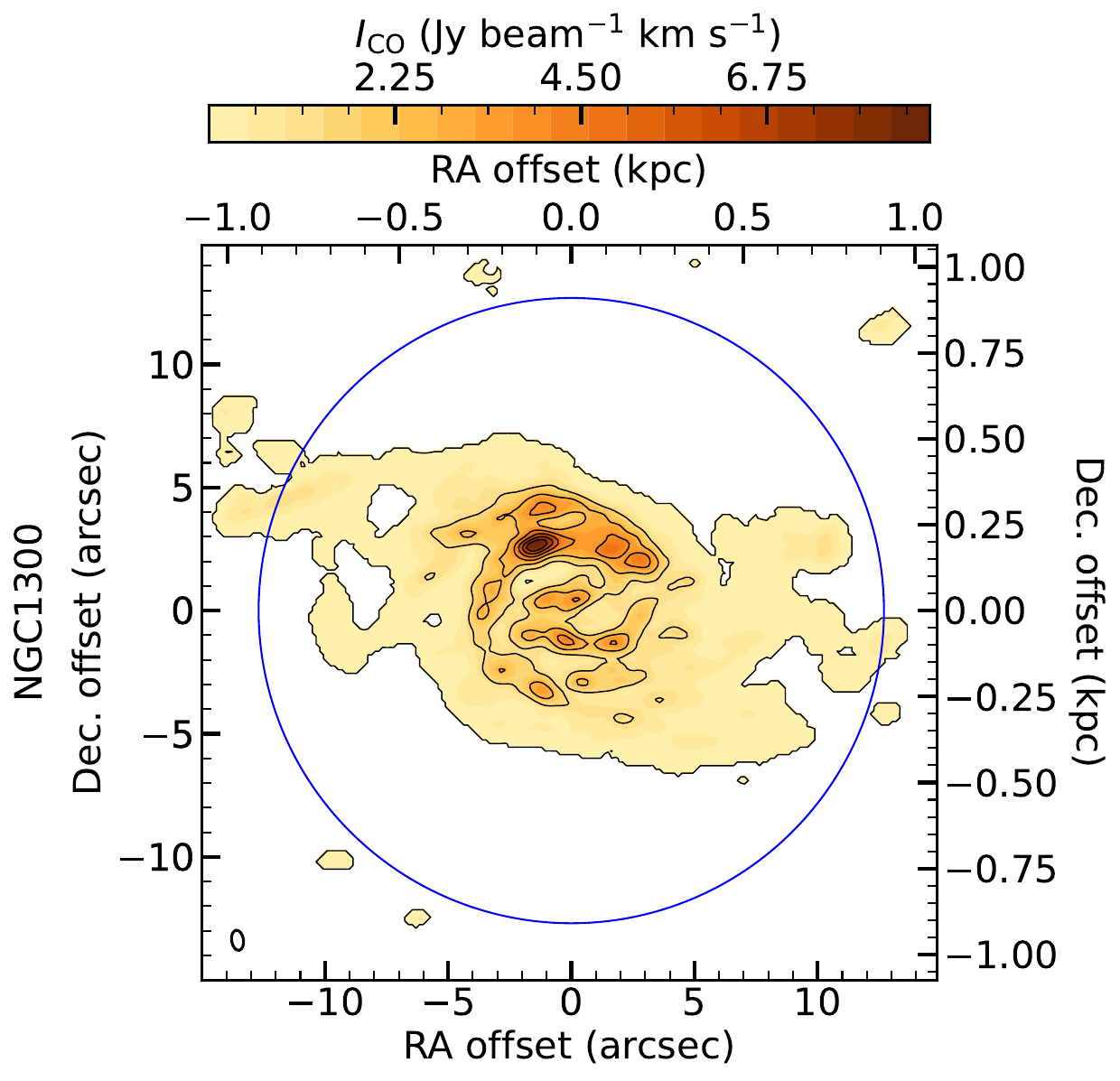}
    \end{subfigure}
    \begin{subfigure}{.32\linewidth}
    \centering
    \includegraphics[height=5.9cm]{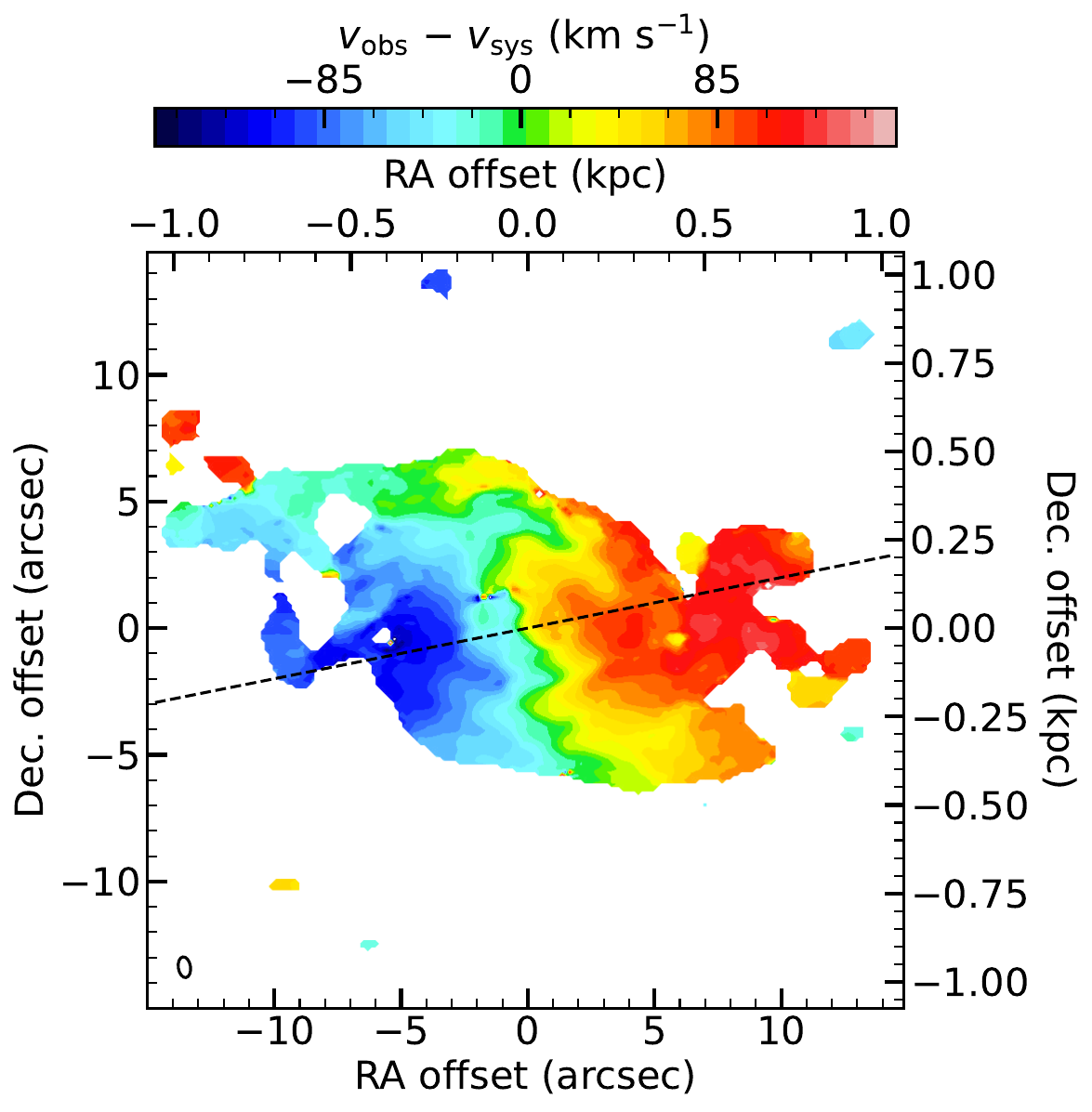}
    \end{subfigure}
    \begin{subfigure}{.32\linewidth}
    \centering
    \includegraphics[height=5.9cm]{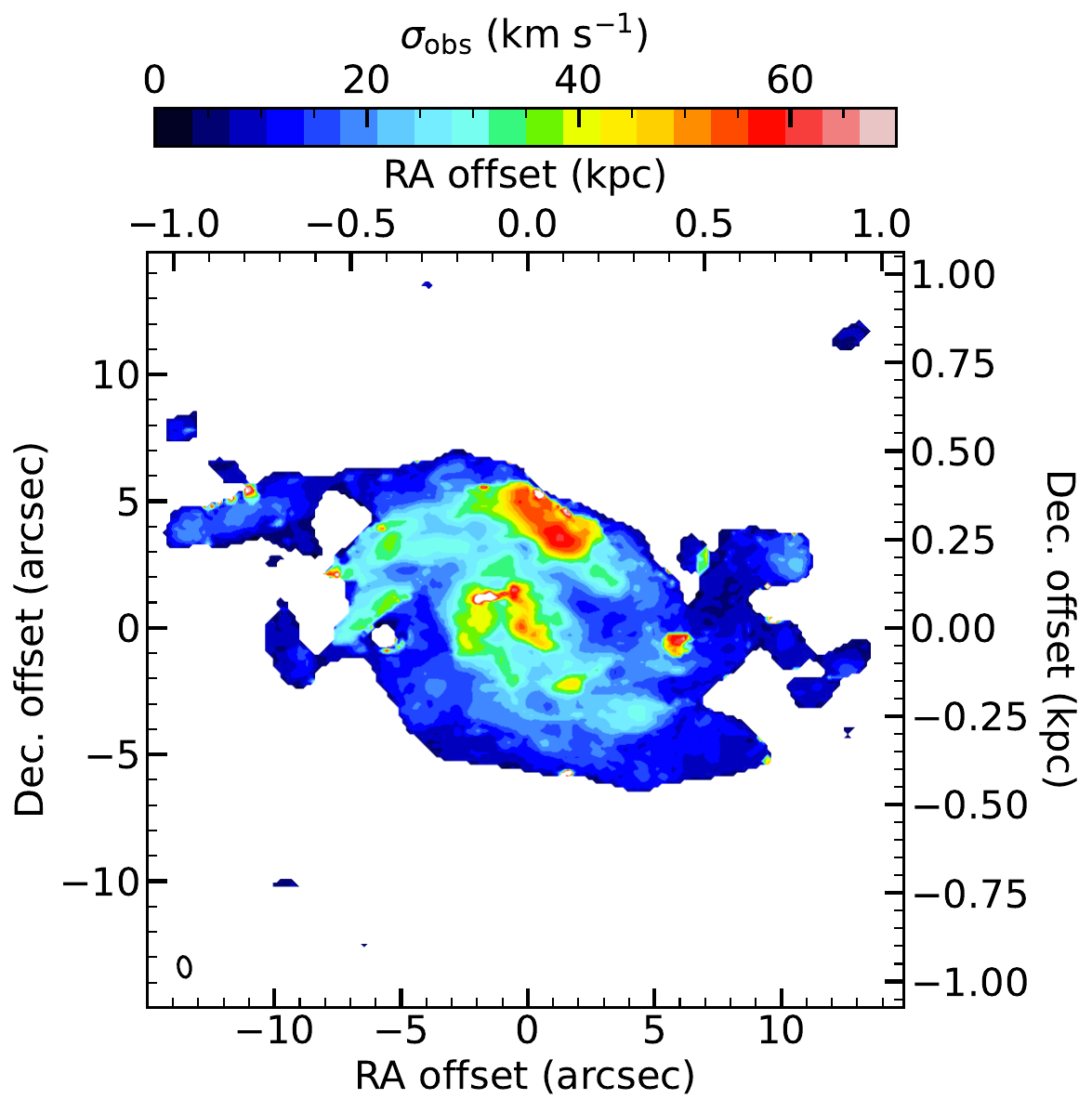}
    \end{subfigure}
    \\
    \begin{subfigure}{.34\linewidth}
    \centering
    \includegraphics[height=5.2cm]{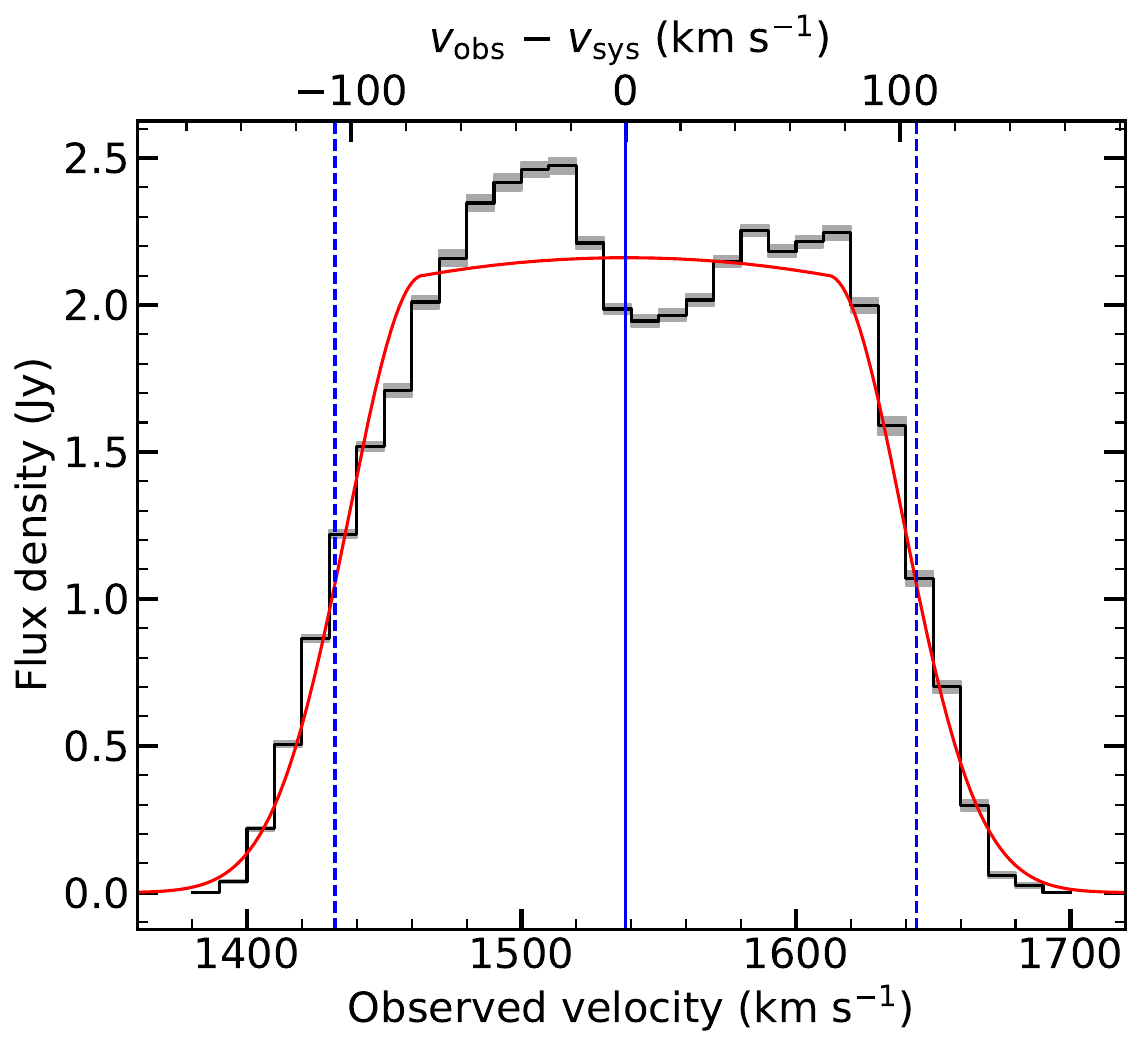}
    \end{subfigure}
    \begin{subfigure}{.32\linewidth}
    \centering
    \includegraphics[height=4.8cm]{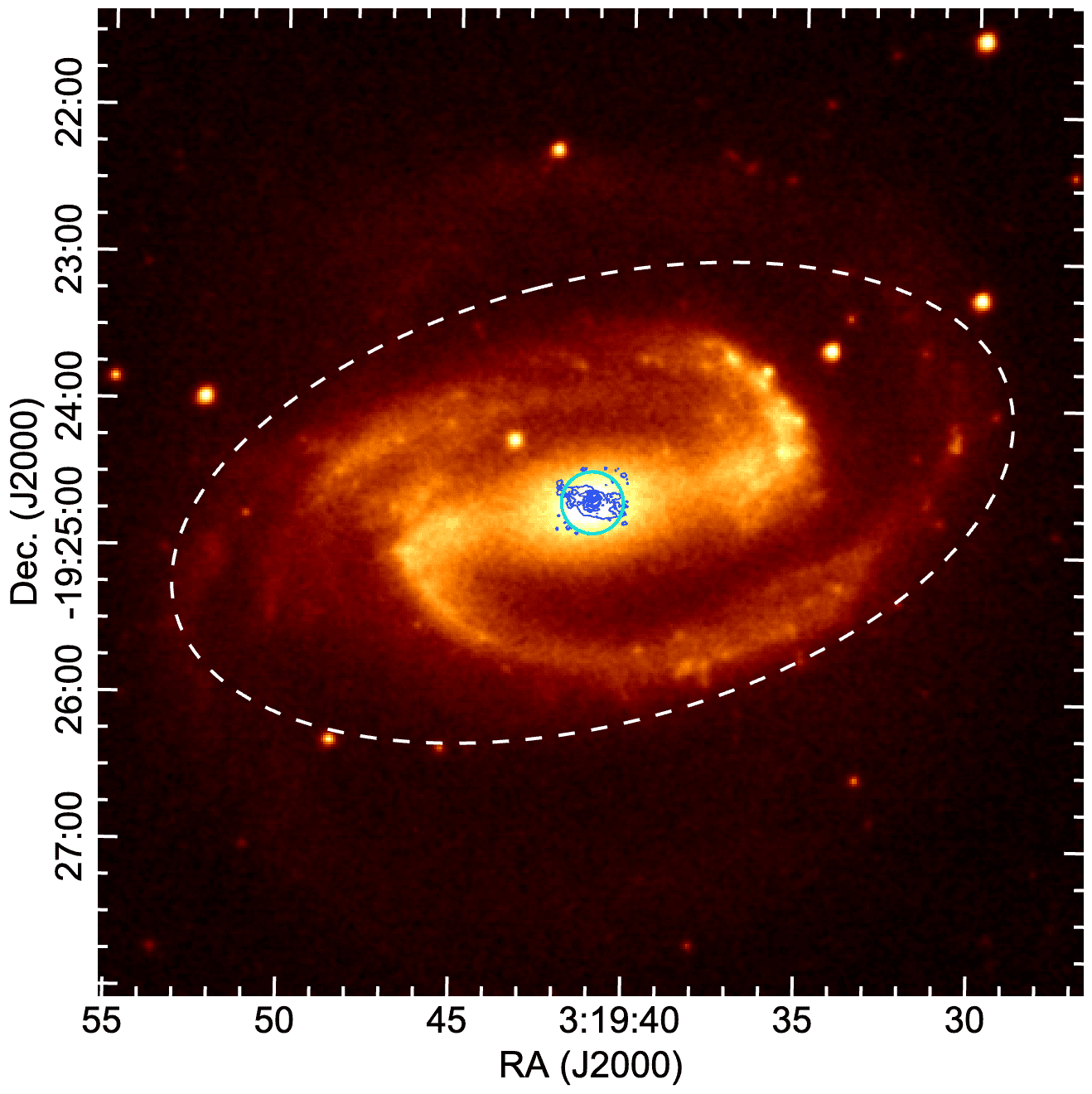}
    \end{subfigure}
    \begin{subfigure}{.33\linewidth}
    \centering
    \includegraphics[height=5.2cm]{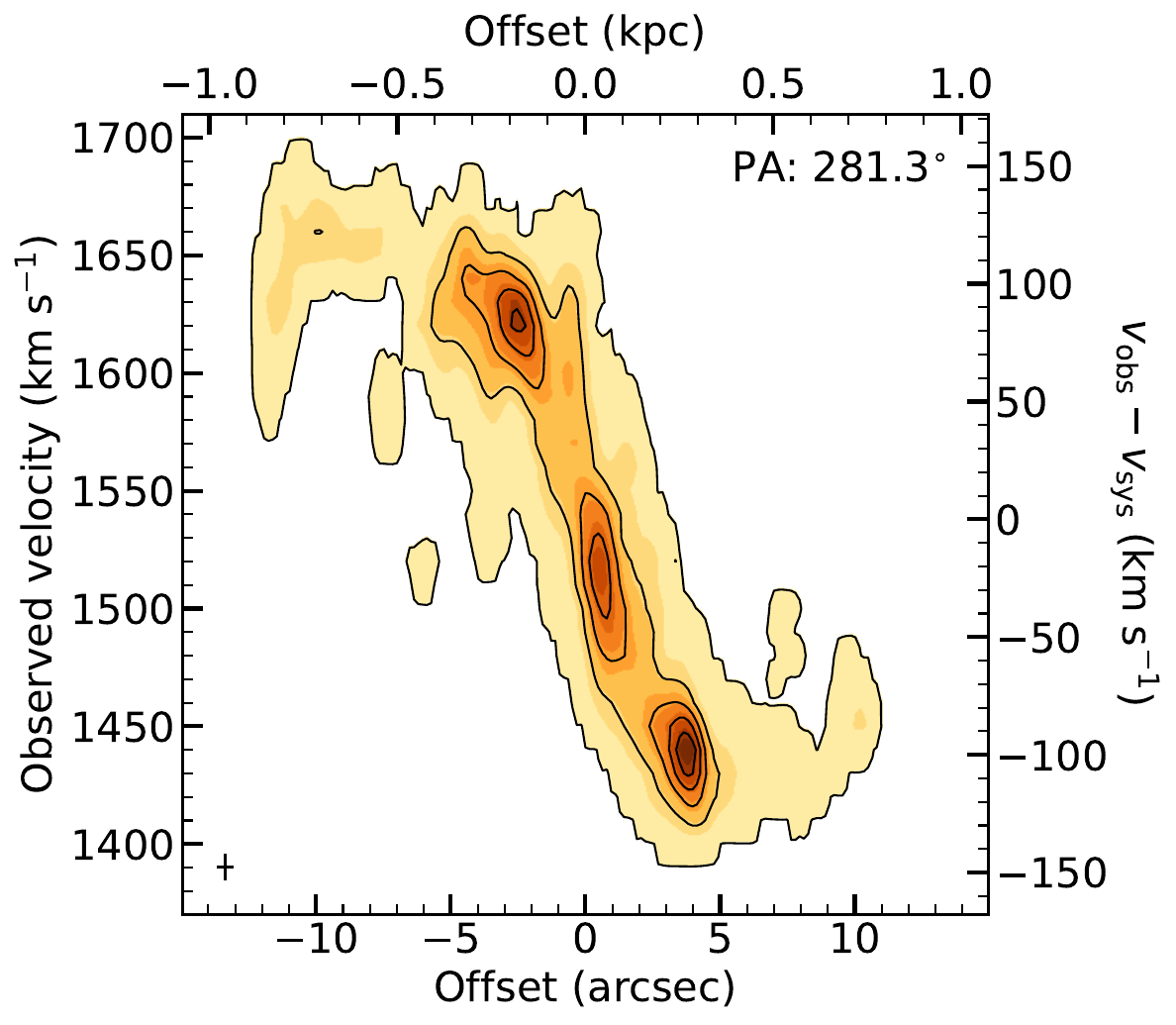}
    \end{subfigure}
    \begin{subfigure}{.34\linewidth}
    \includegraphics[height=6.0cm]{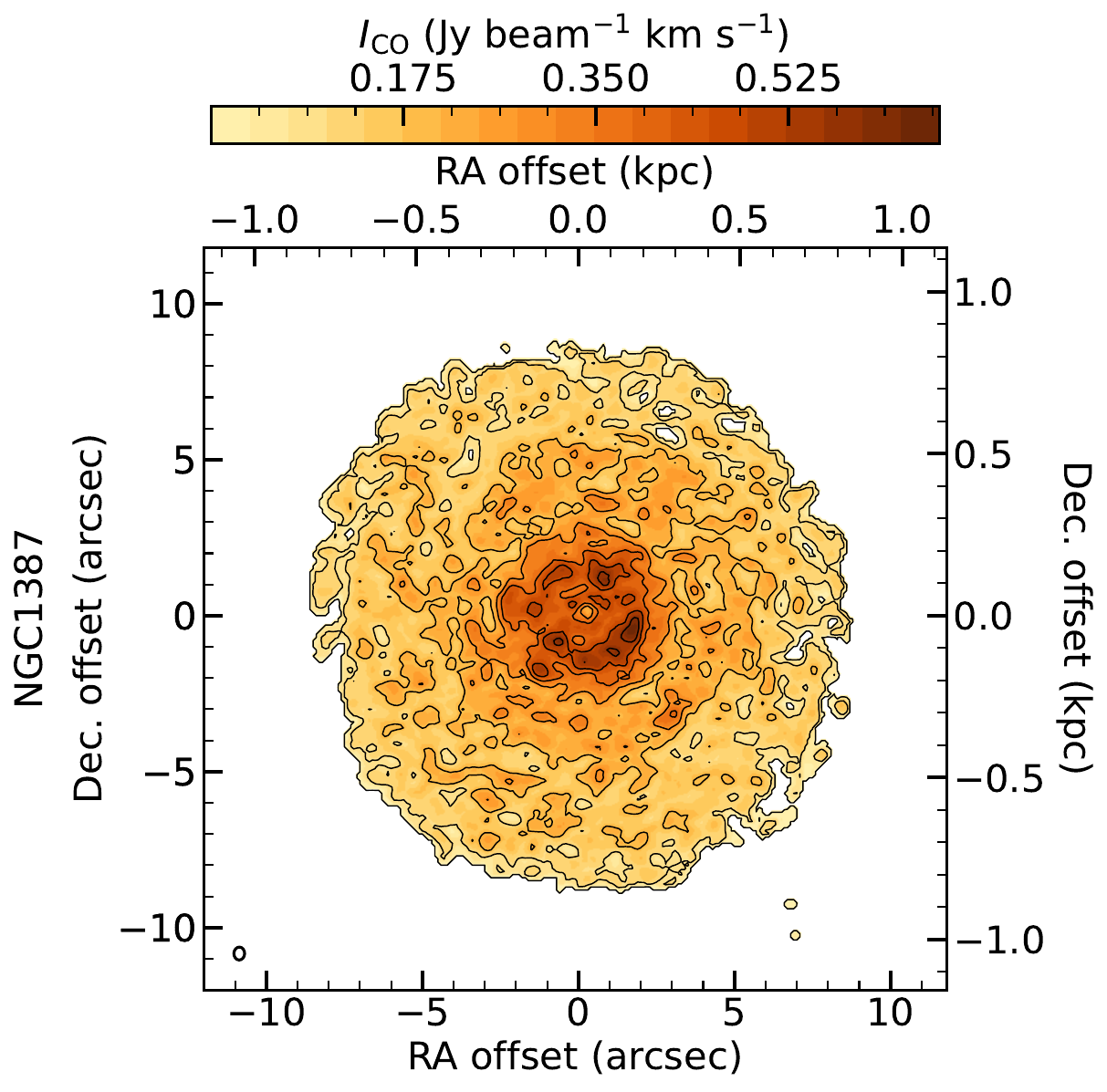}
    \end{subfigure}
    \begin{subfigure}{.32\linewidth}
    \centering
    \includegraphics[height=6.0cm]{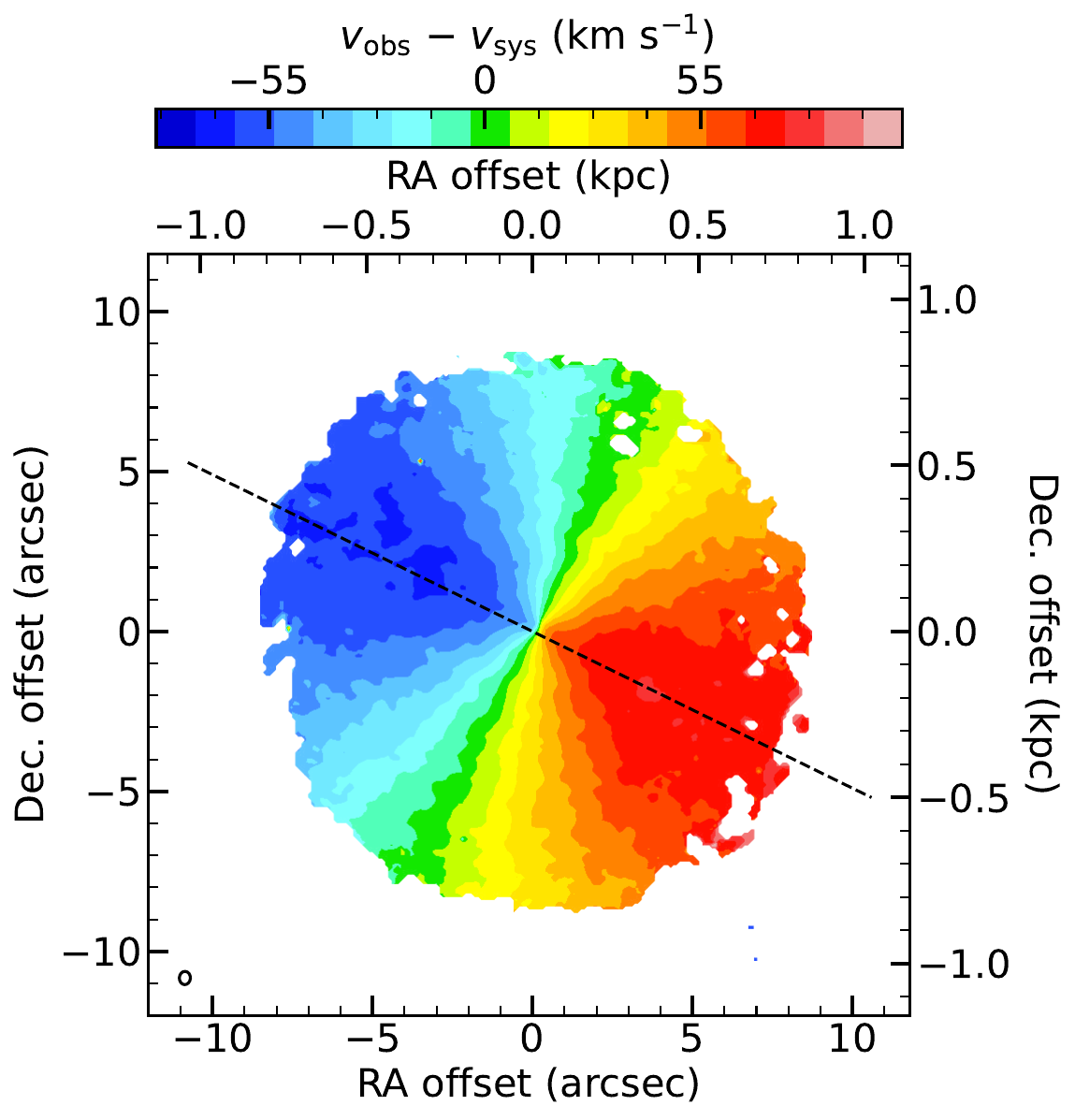}
    \end{subfigure}
    \begin{subfigure}{.32\linewidth}
    \centering
    \includegraphics[height=6.0cm]{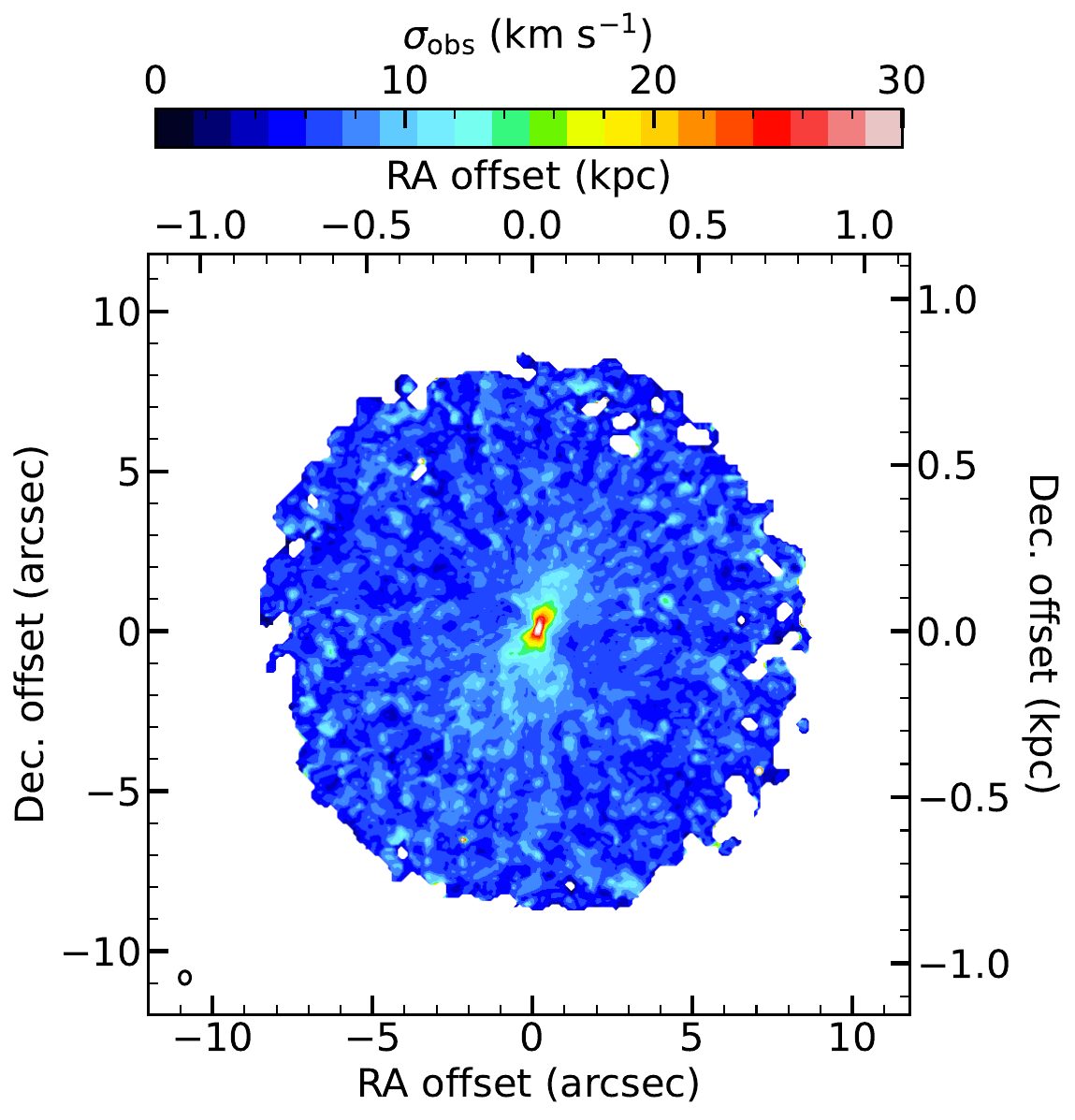}
    \end{subfigure}
    \\
    \begin{subfigure}{.34\linewidth}
    \centering
    \includegraphics[height=5.2cm]{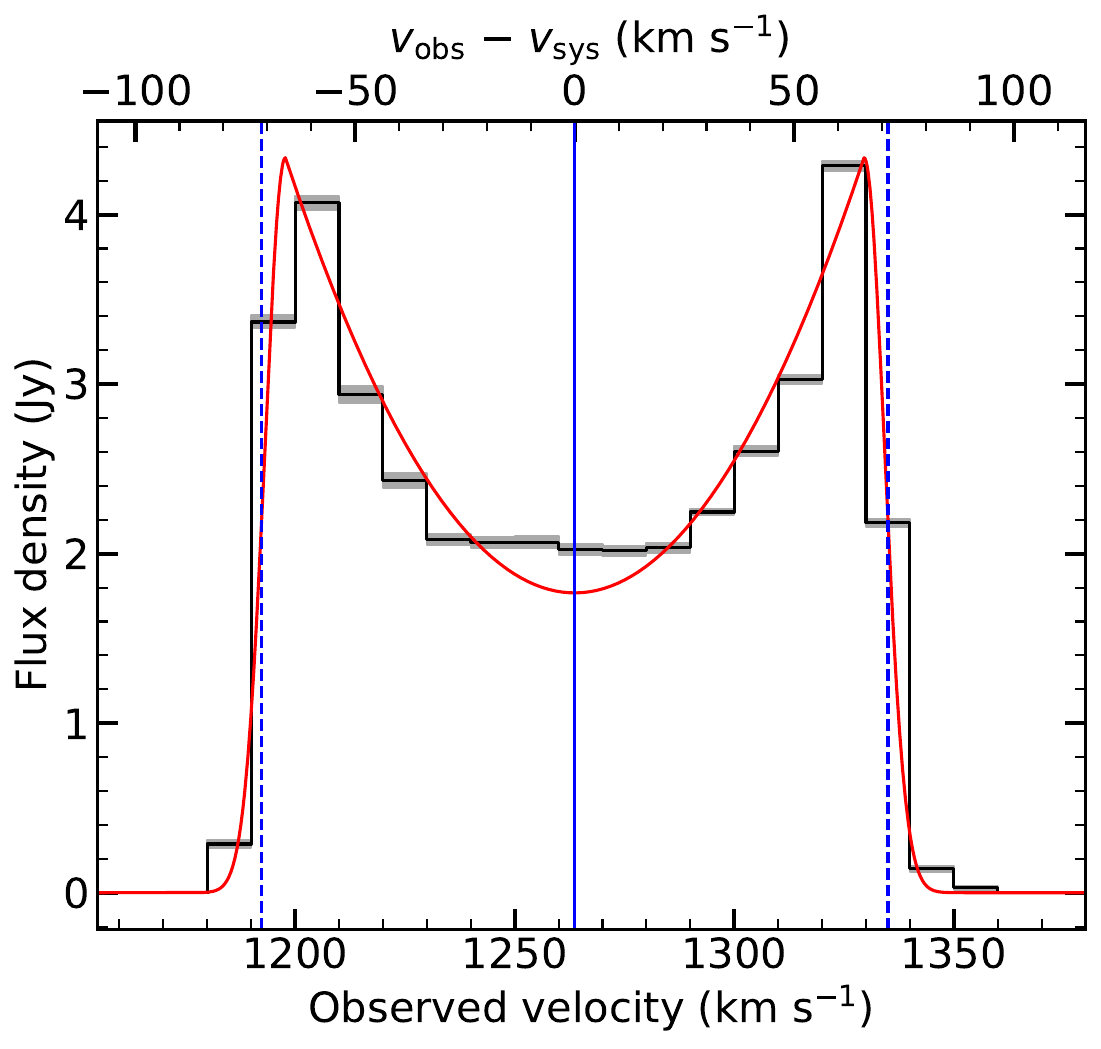}
    \end{subfigure}
    \begin{subfigure}{.32\linewidth}
    \centering
    \includegraphics[height=4.8cm]{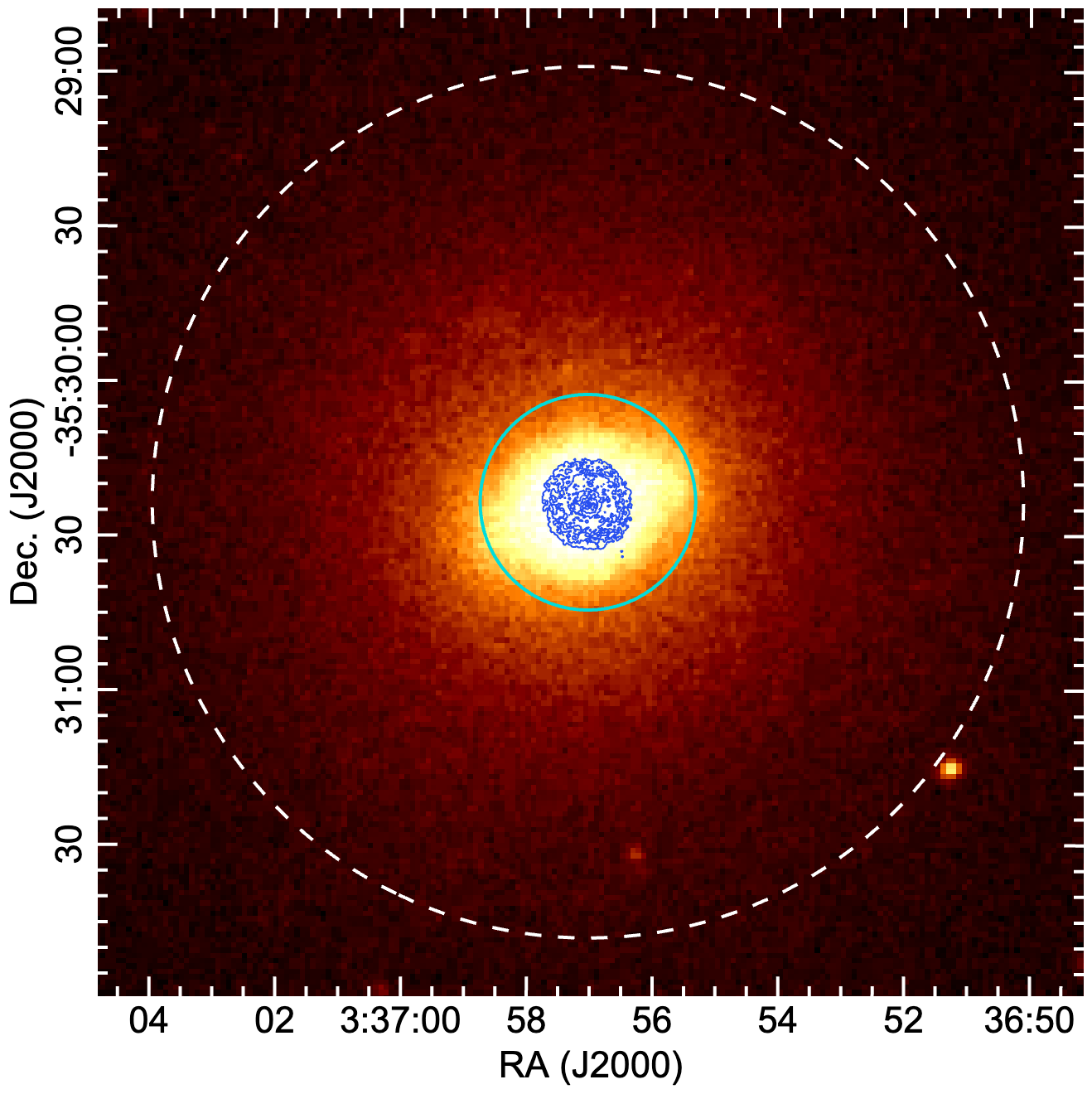}
    \end{subfigure}
    \begin{subfigure}{.33\linewidth}
    \centering
    \includegraphics[height=5.2cm]{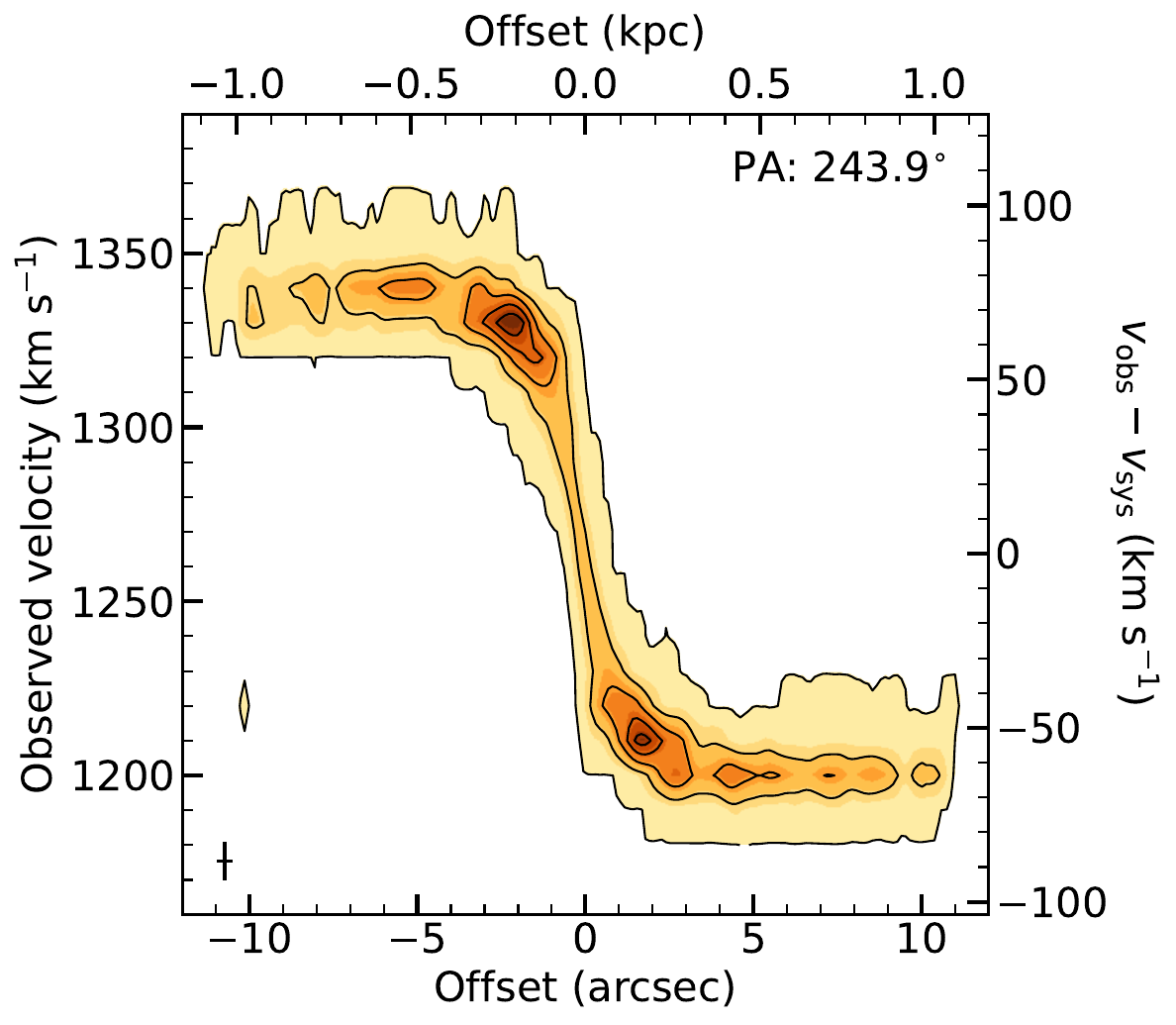}
    \end{subfigure}
    \caption[]{continued.}
\end{figure*}

\begin{figure*}
    \ContinuedFloat
    \centering
    \begin{subfigure}{.34\linewidth}
    \includegraphics[height=5.9cm]{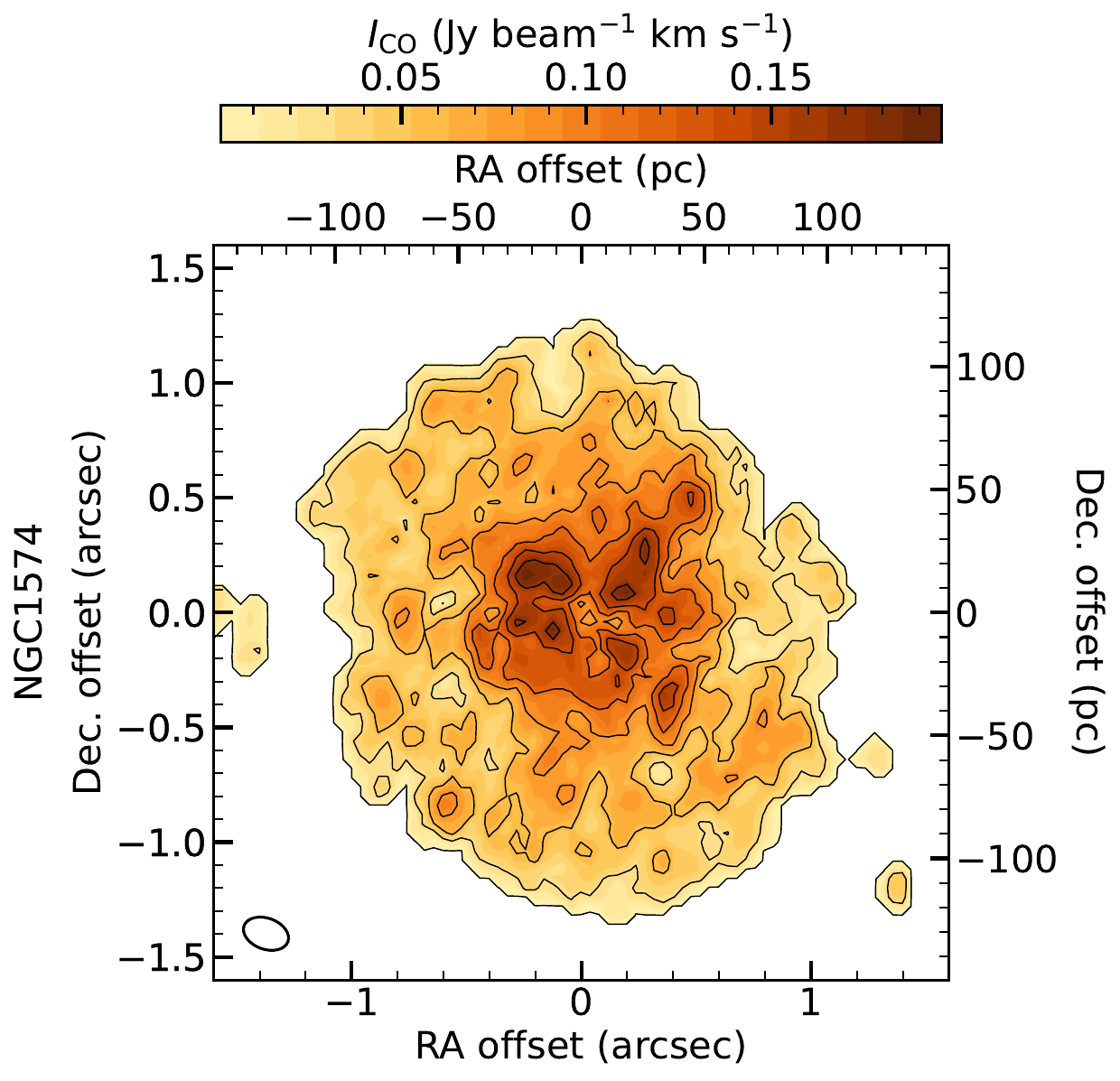}
    \end{subfigure}
    \begin{subfigure}{.32\linewidth}
    \centering
    \includegraphics[height=5.9cm]{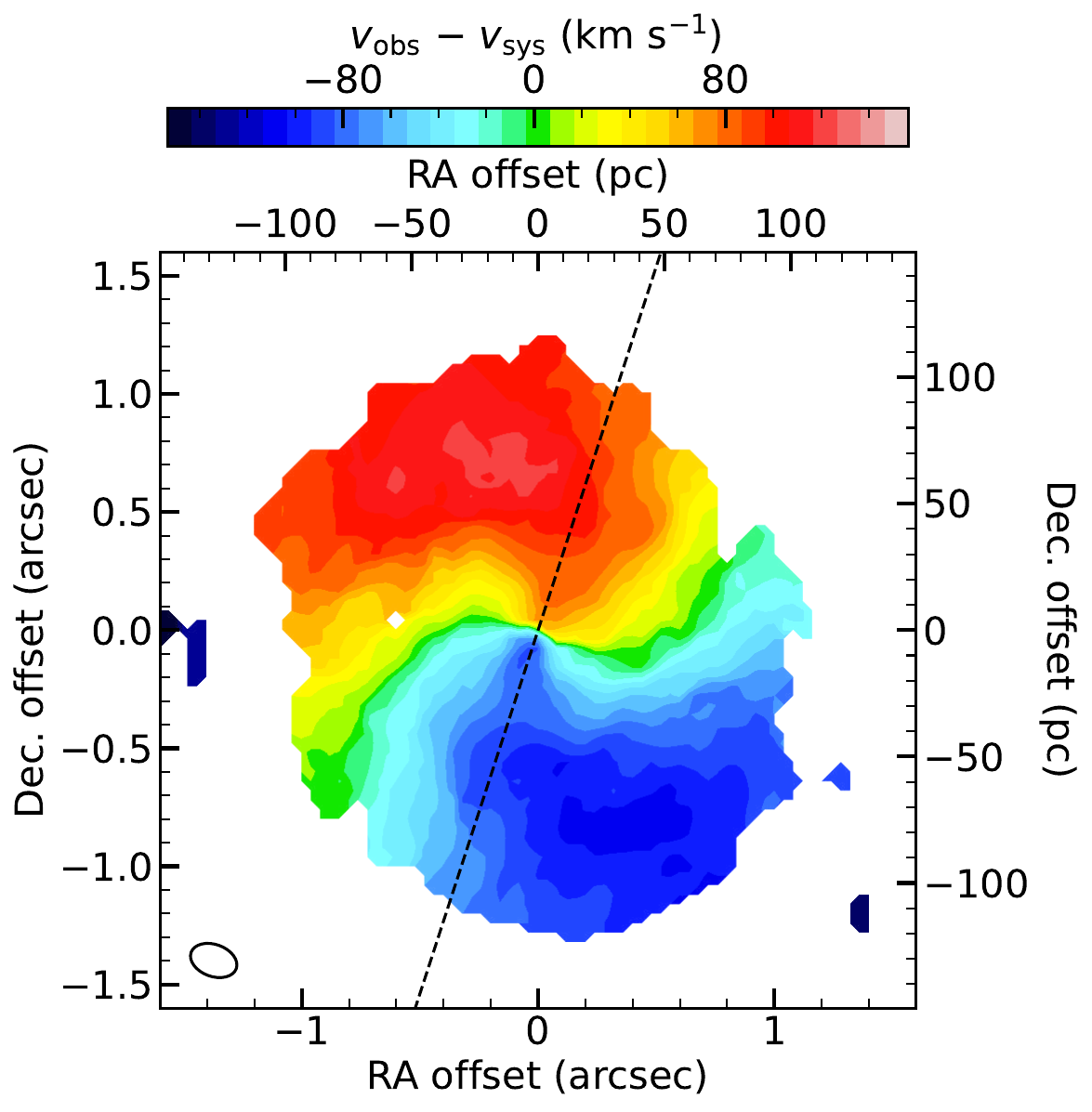}
    \end{subfigure}
    \begin{subfigure}{.32\linewidth}
    \centering
    \includegraphics[height=5.9cm]{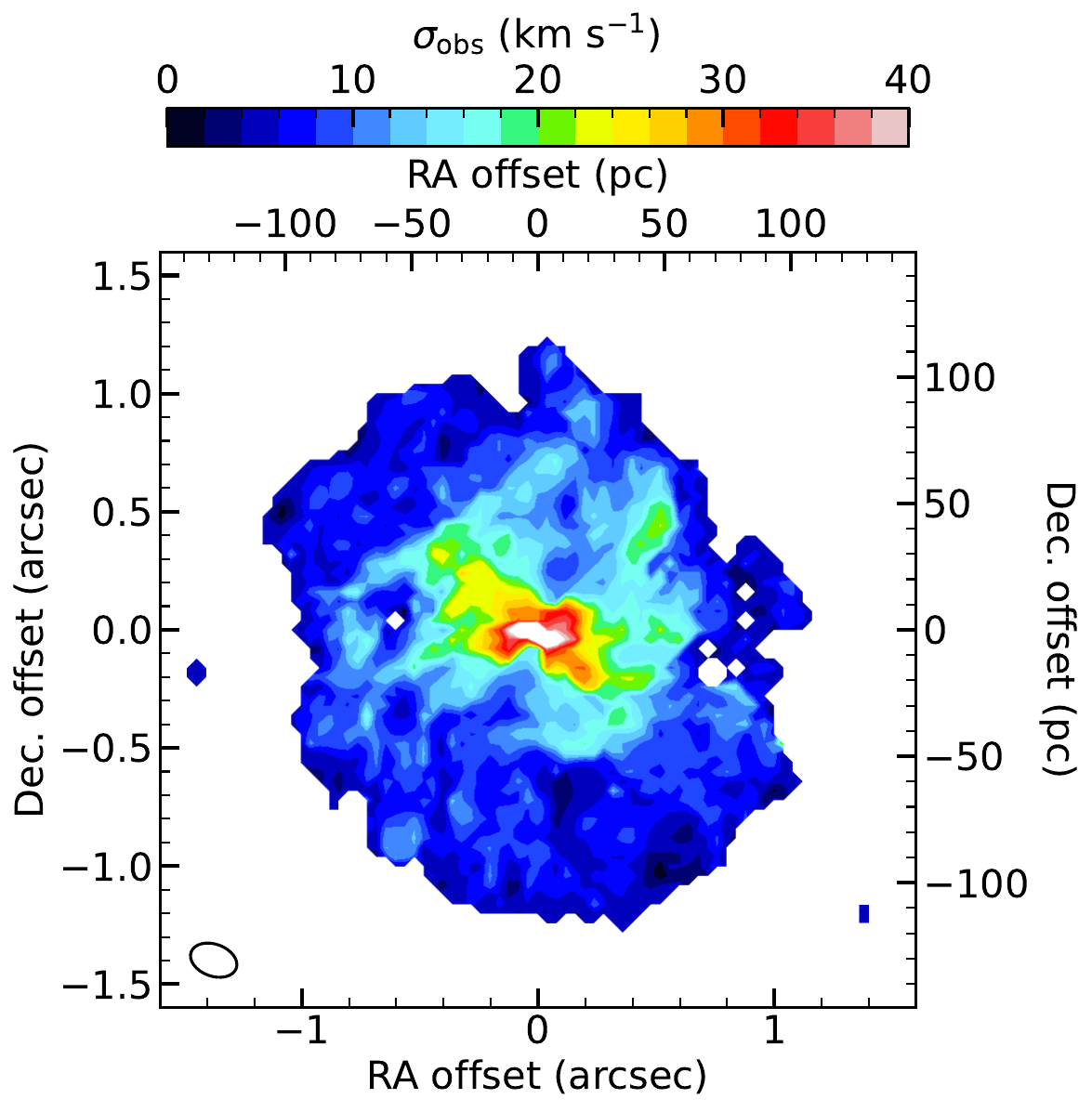}
    \end{subfigure}
    \\
    \begin{subfigure}{.34\linewidth}
    \centering
    \includegraphics[height=5.2cm]{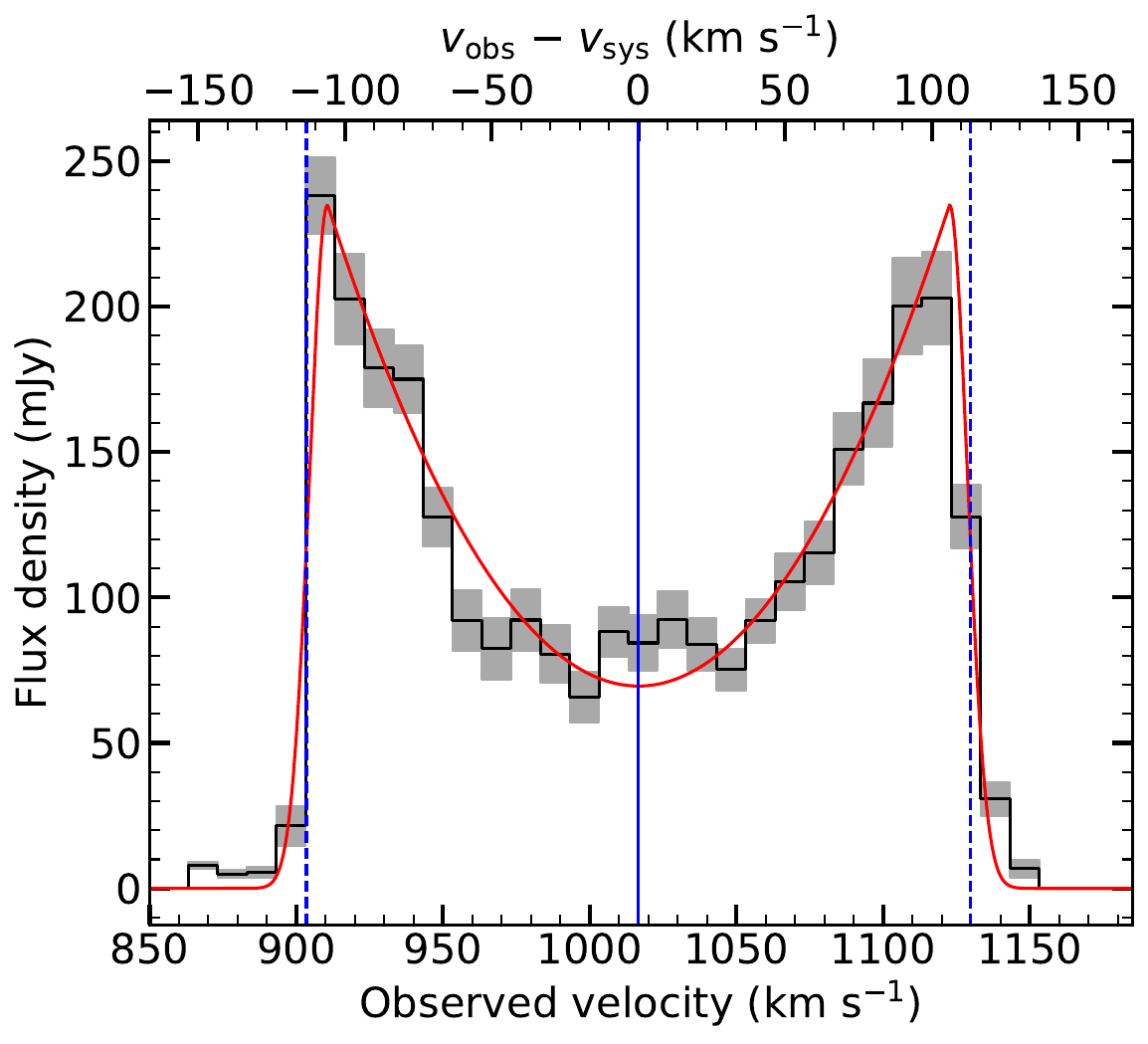}
    \end{subfigure}  
    \begin{subfigure}{.32\linewidth}
    \centering
    \includegraphics[height=4.8cm]{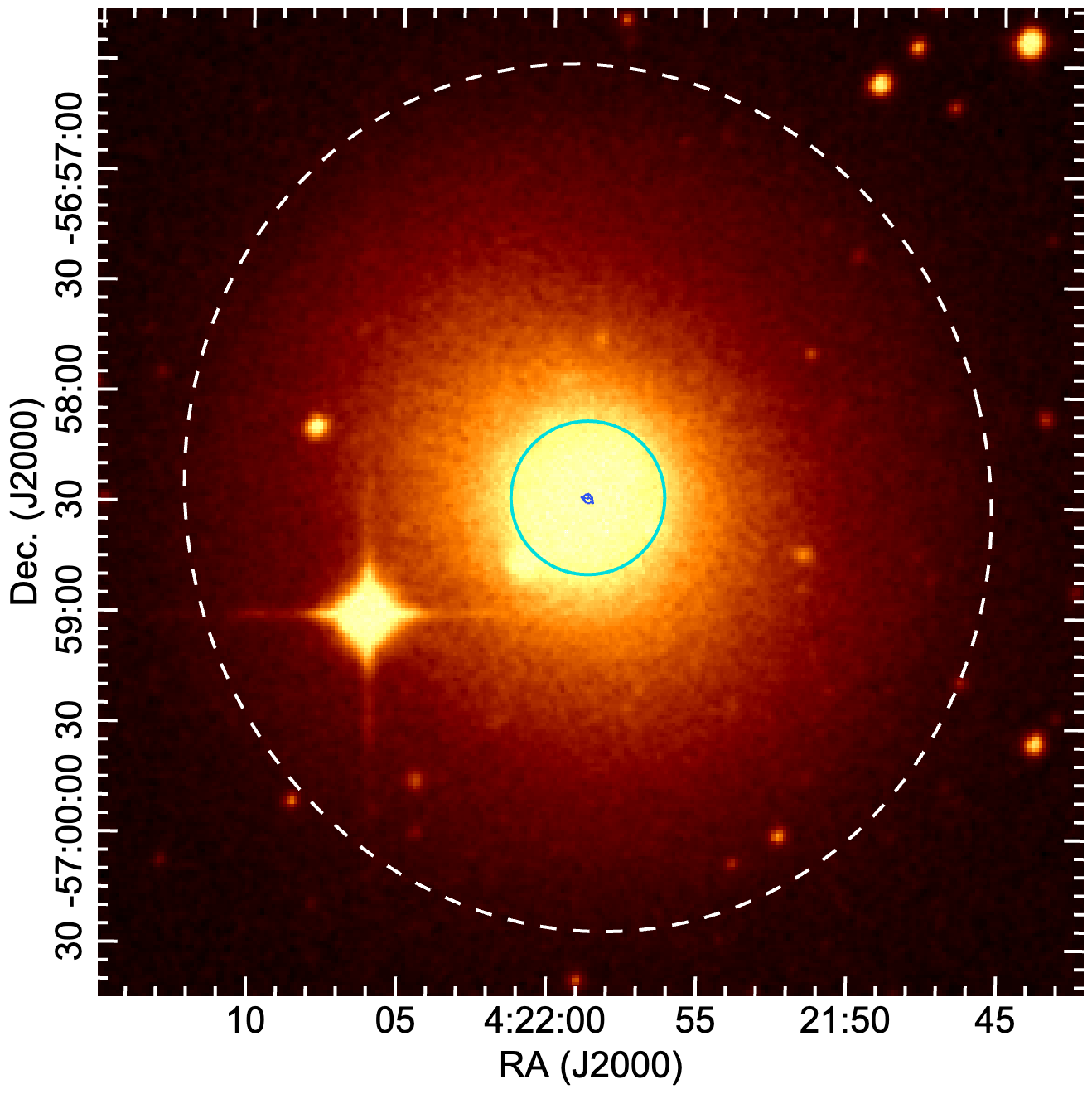}
    \end{subfigure}
    \begin{subfigure}{.33\linewidth}
    \centering
    \includegraphics[height=5.2cm]{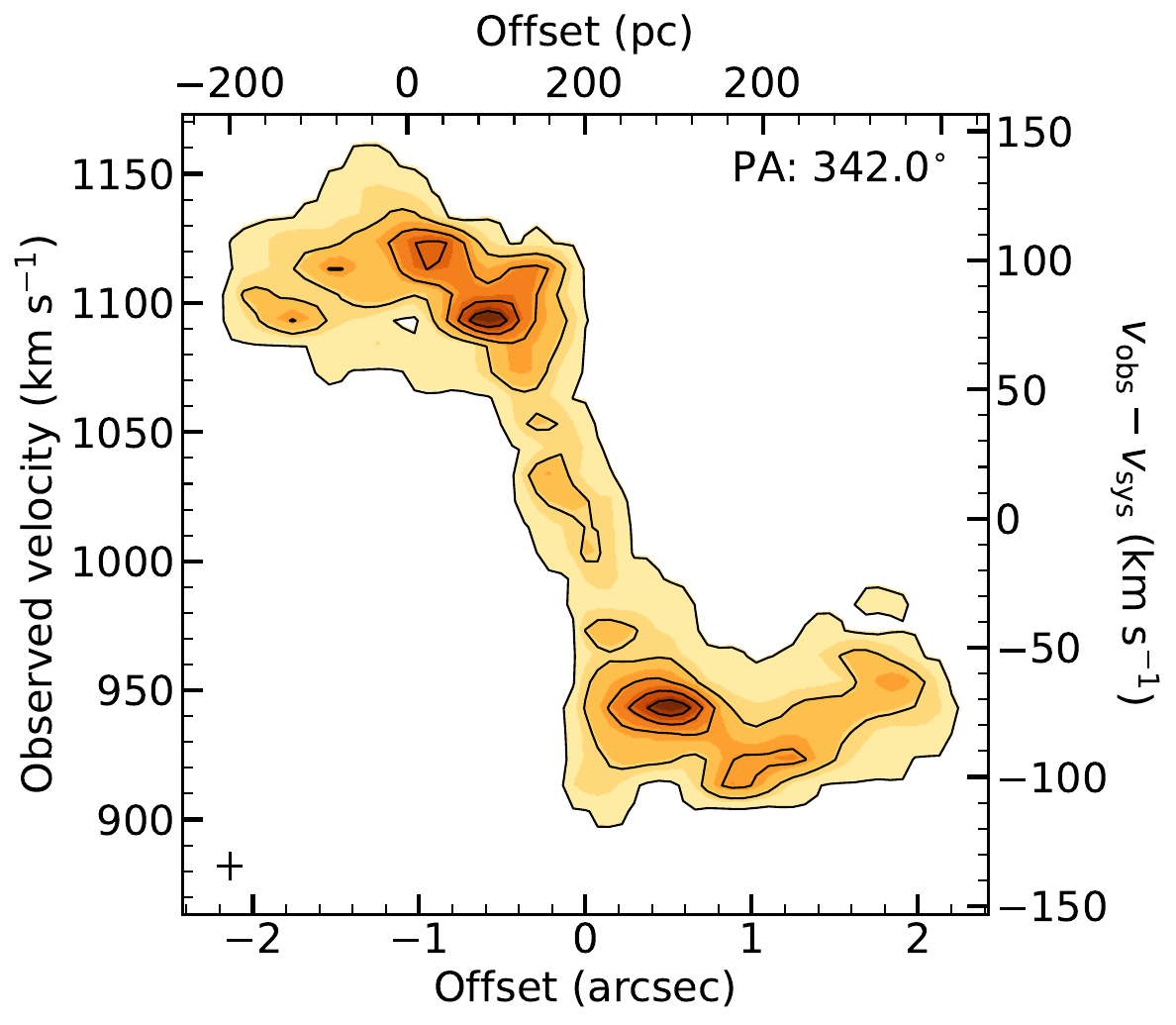}
    \end{subfigure}
    \begin{subfigure}{.34\linewidth}
    \centering
    \includegraphics[height=6.0cm]{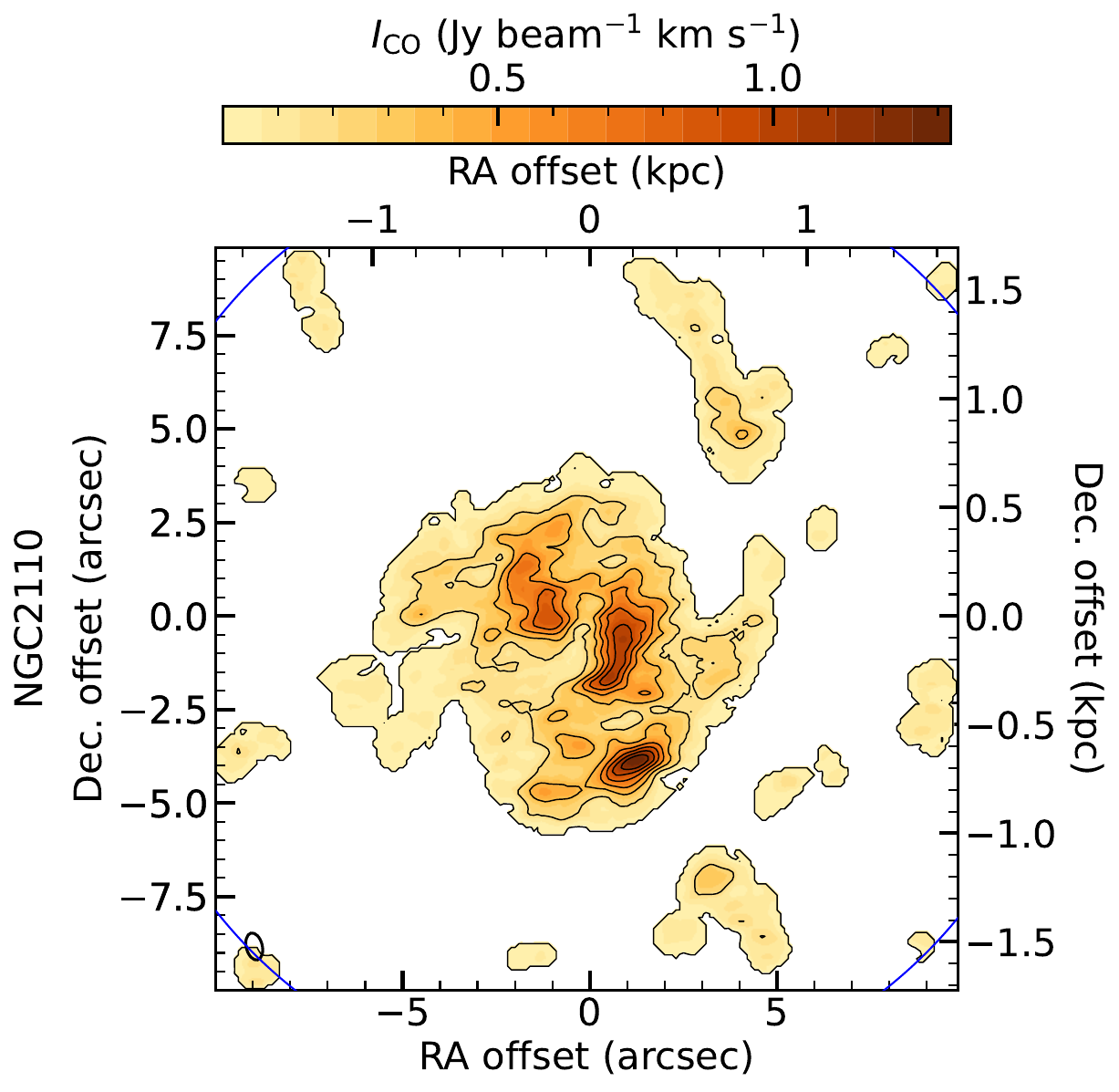}
    \end{subfigure}
    \begin{subfigure}{.32\linewidth}
    \centering
    \includegraphics[height=6.0cm]{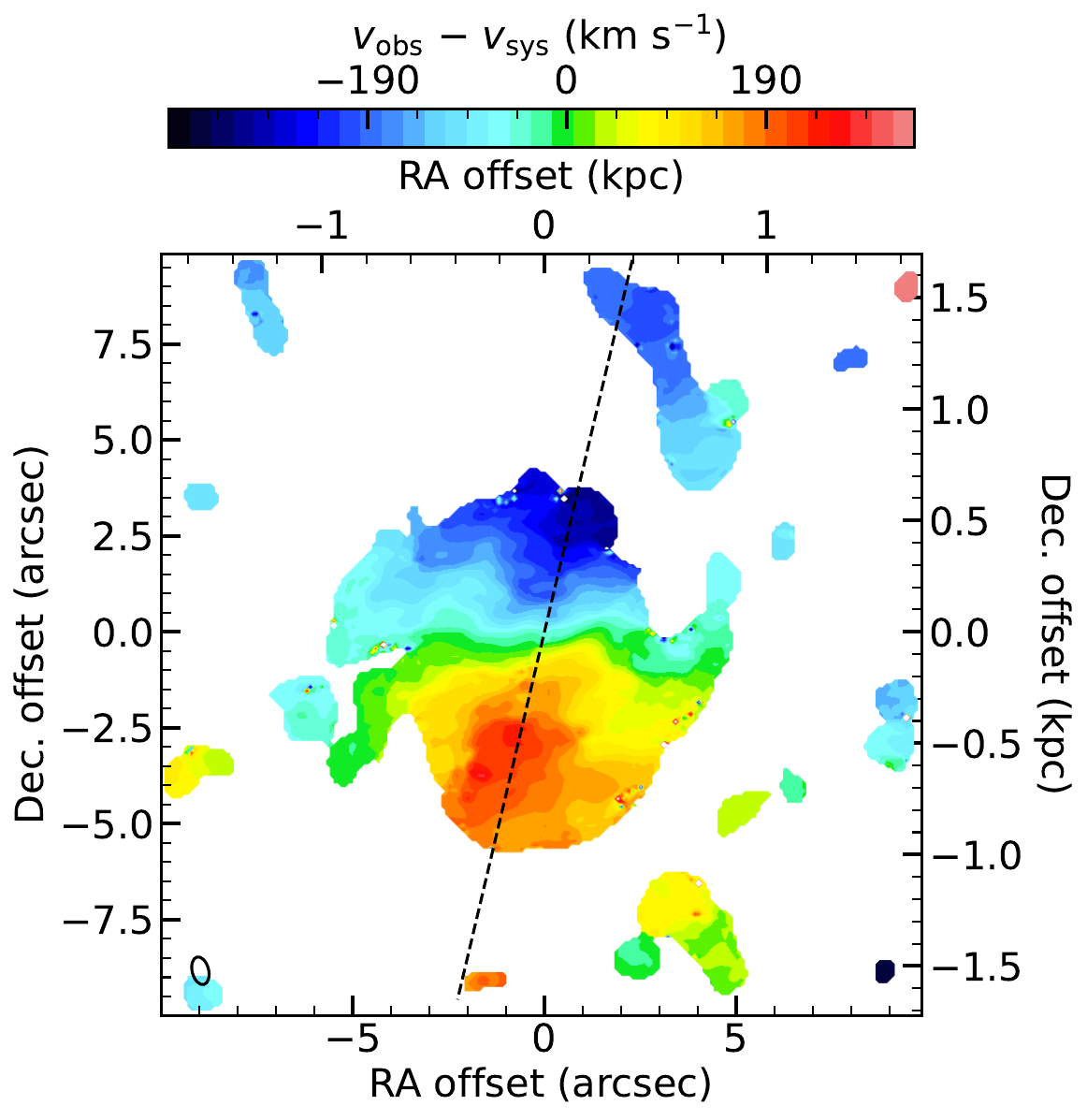}
    \end{subfigure}
    \begin{subfigure}{.32\linewidth}
    \centering
    \includegraphics[height=6.0cm]{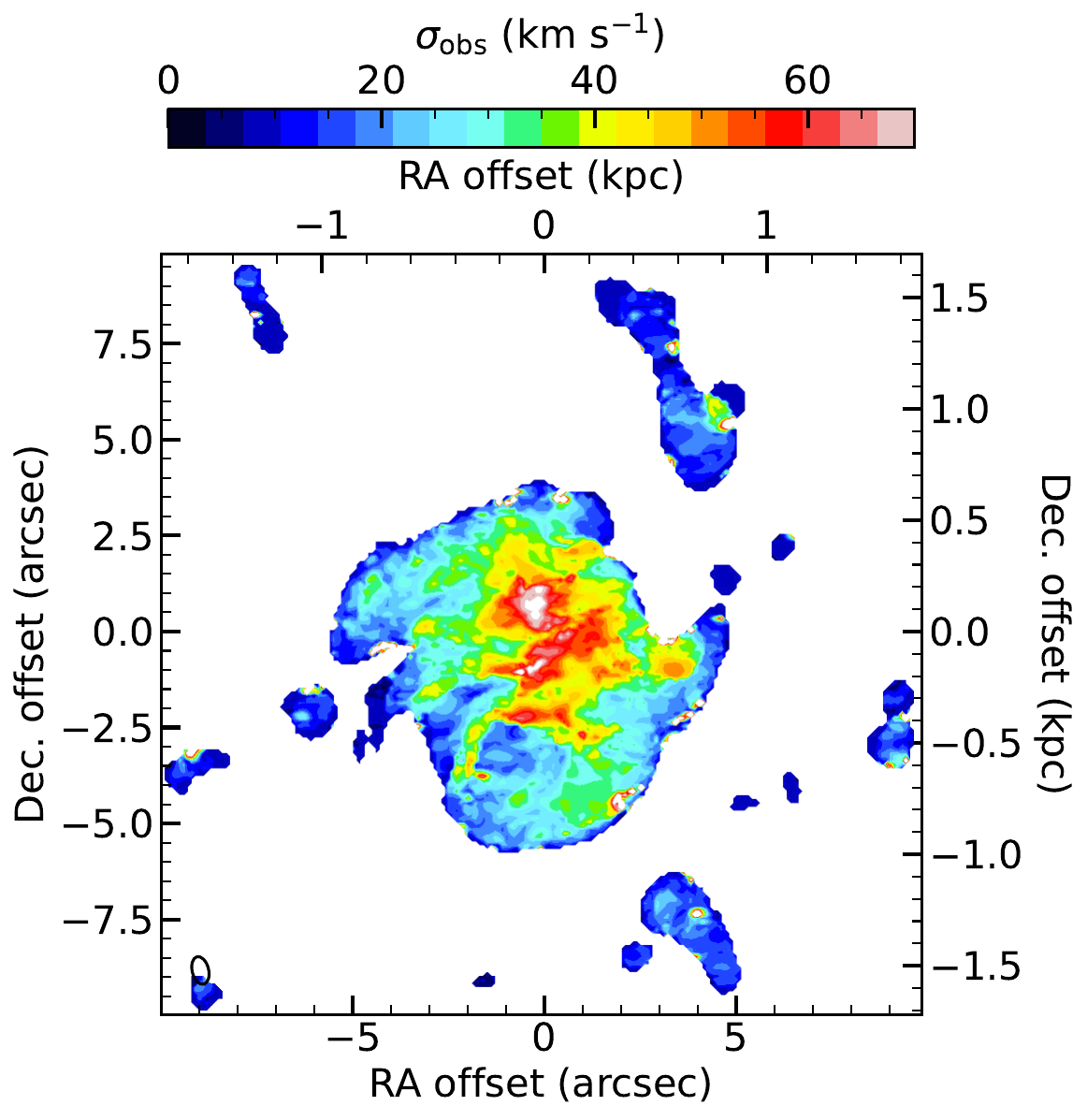}
    \end{subfigure}
    \\
    \begin{subfigure}{.34\linewidth}
    \centering
    \includegraphics[height=5.2cm]{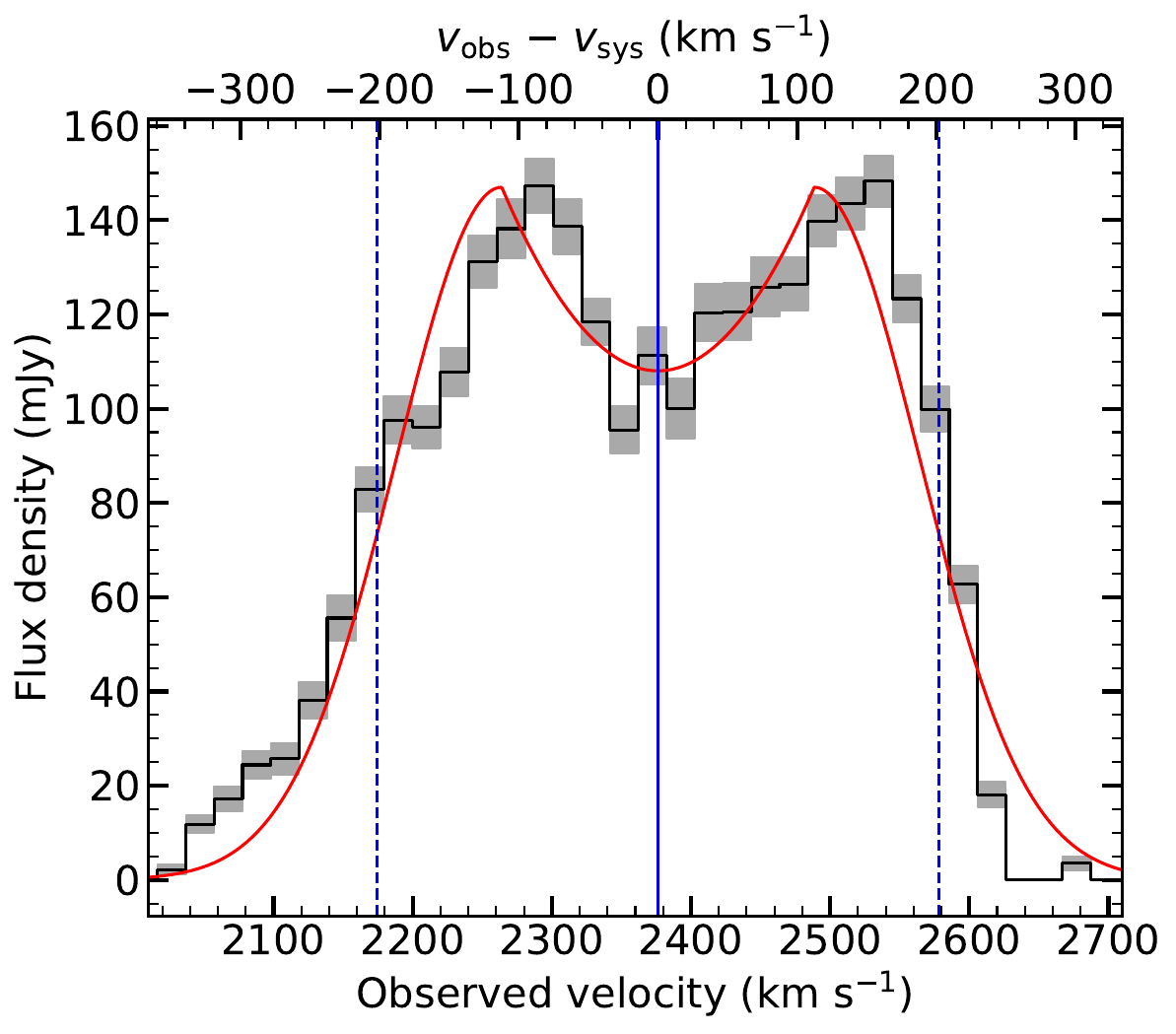}
    \end{subfigure} 
    \begin{subfigure}{.32\linewidth}
    \centering
    \includegraphics[height=4.8cm]{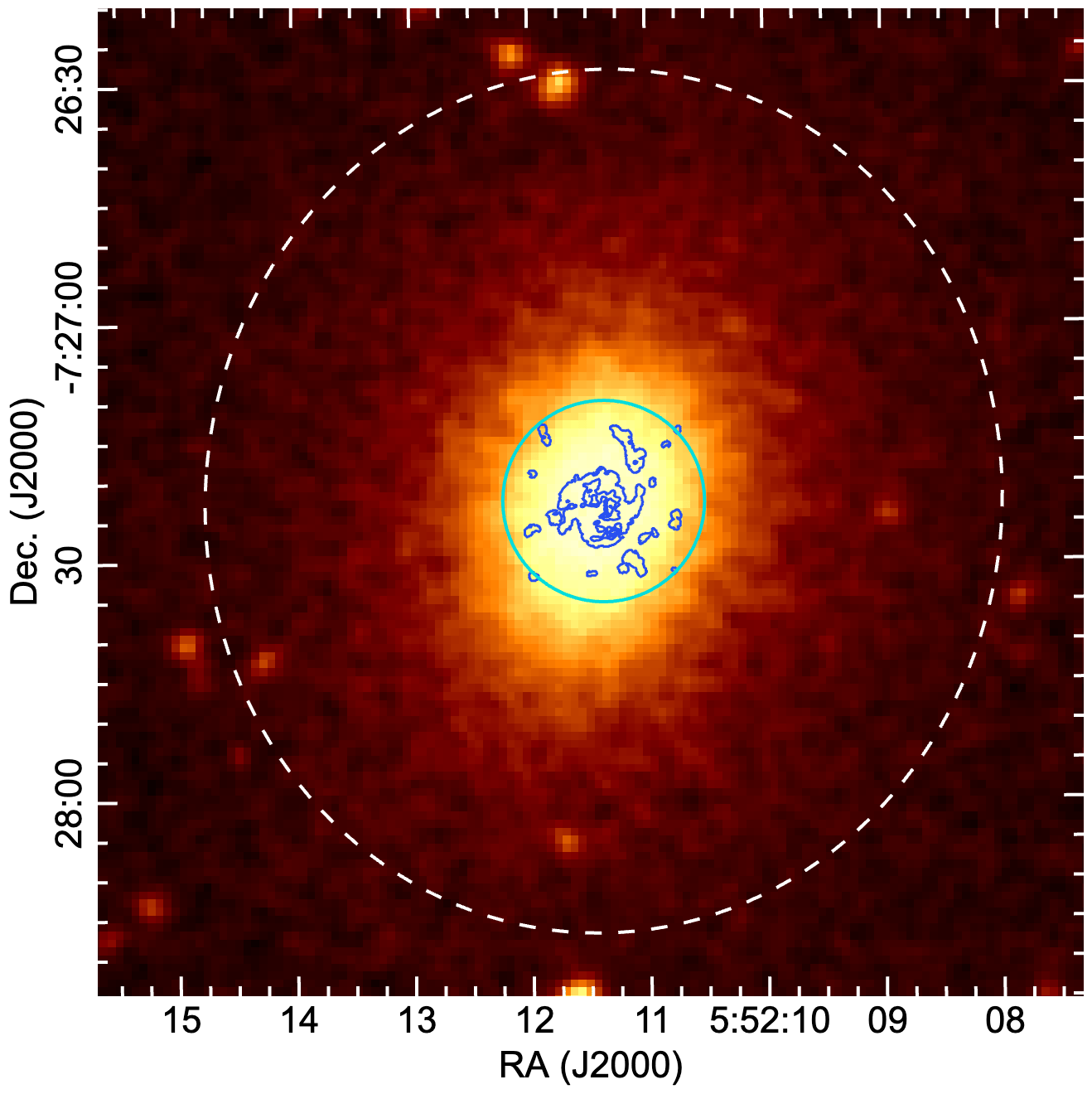}
    \end{subfigure}
    \begin{subfigure}{.33\linewidth}
    \centering
    \includegraphics[height=5.2cm]{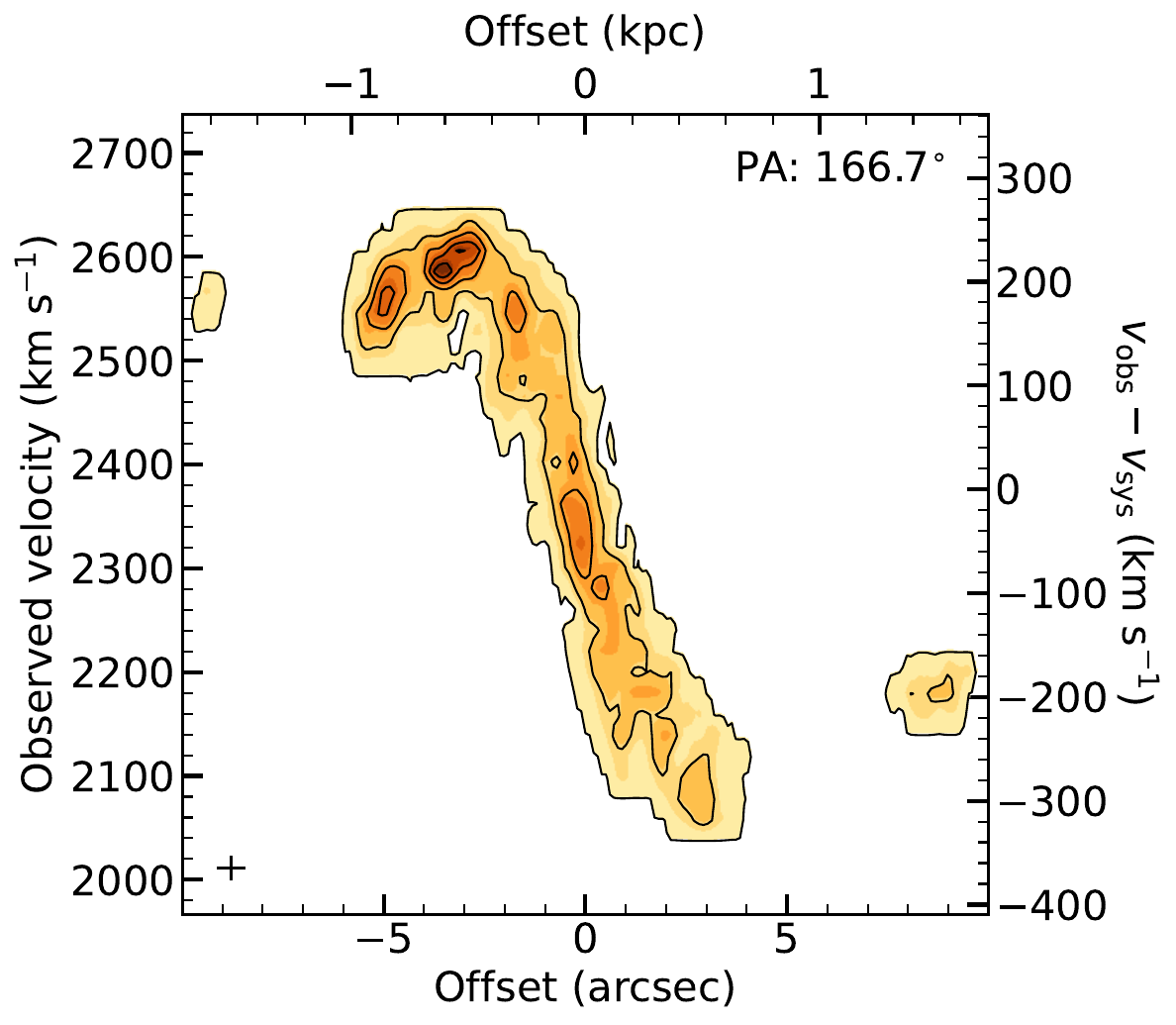}
    \end{subfigure} 
    \caption[]{continued.}
\end{figure*}

\begin{figure*}
    \ContinuedFloat
    \centering
    \begin{subfigure}{.34\linewidth}
    \includegraphics[height=6.0cm]{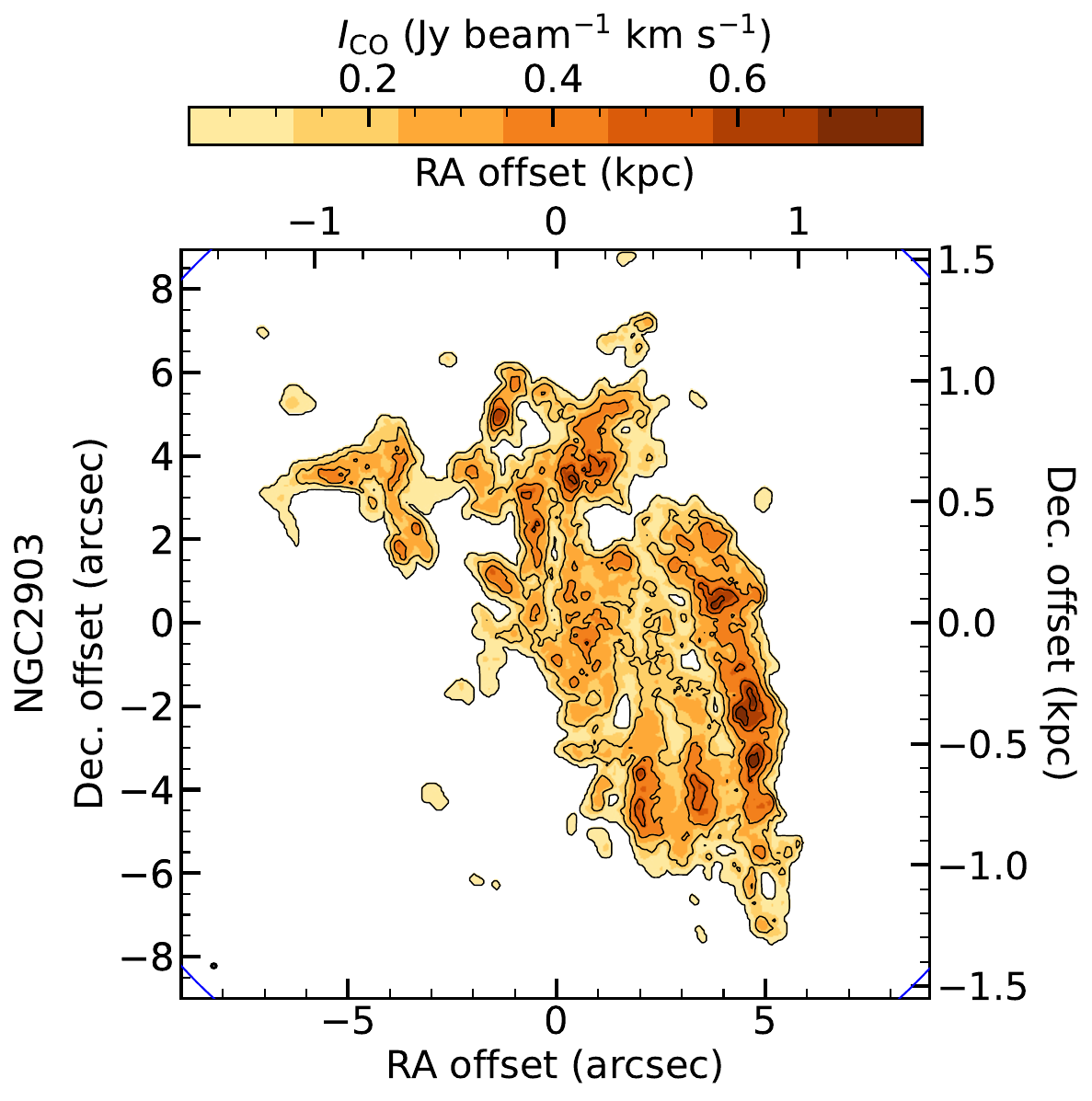}
    \end{subfigure}
    \begin{subfigure}{.32\linewidth}
    \centering
    \includegraphics[height=6.0cm]{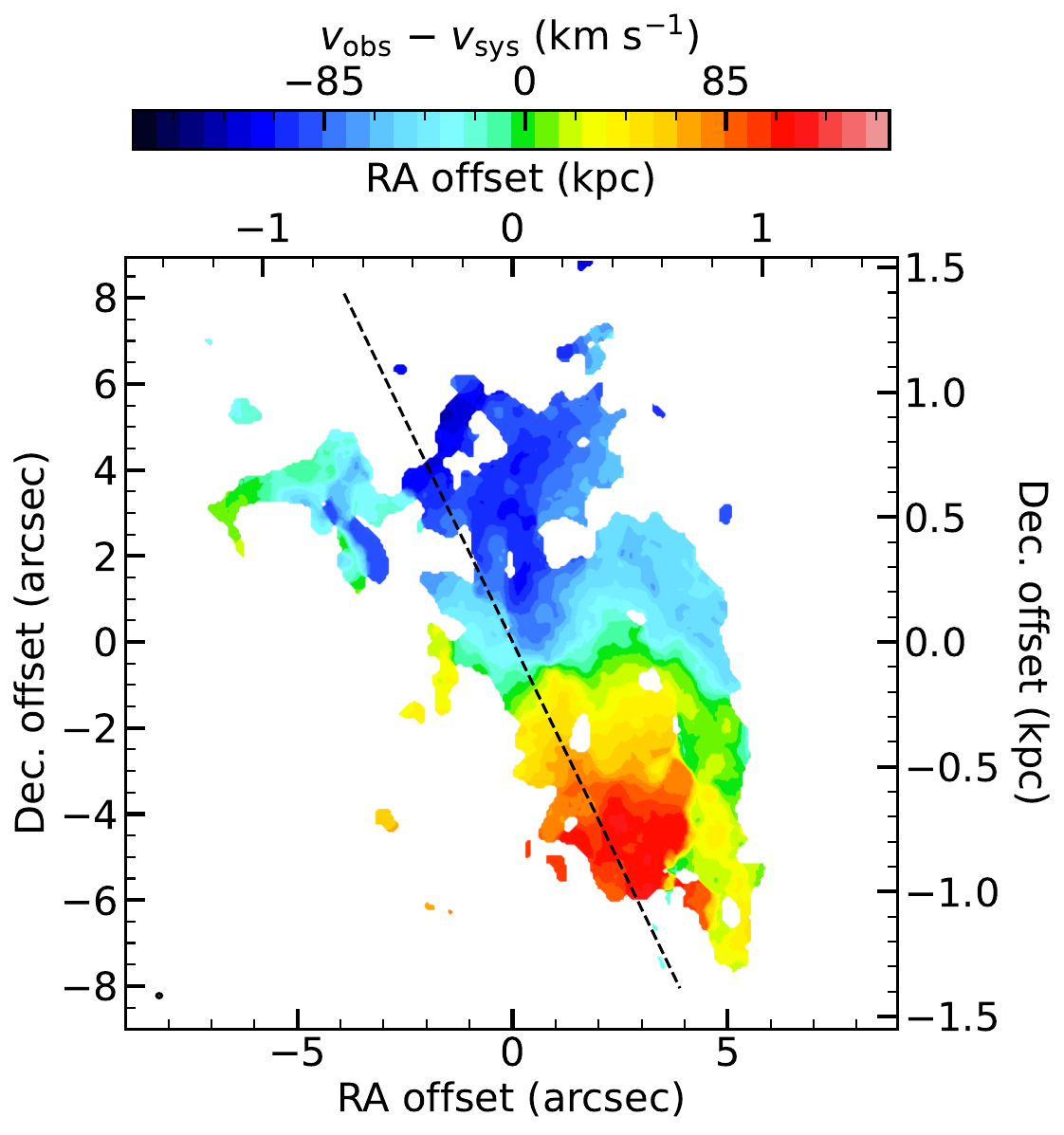}
    \end{subfigure}
    \begin{subfigure}{.32\linewidth}
    \centering
    \includegraphics[height=6.0cm]{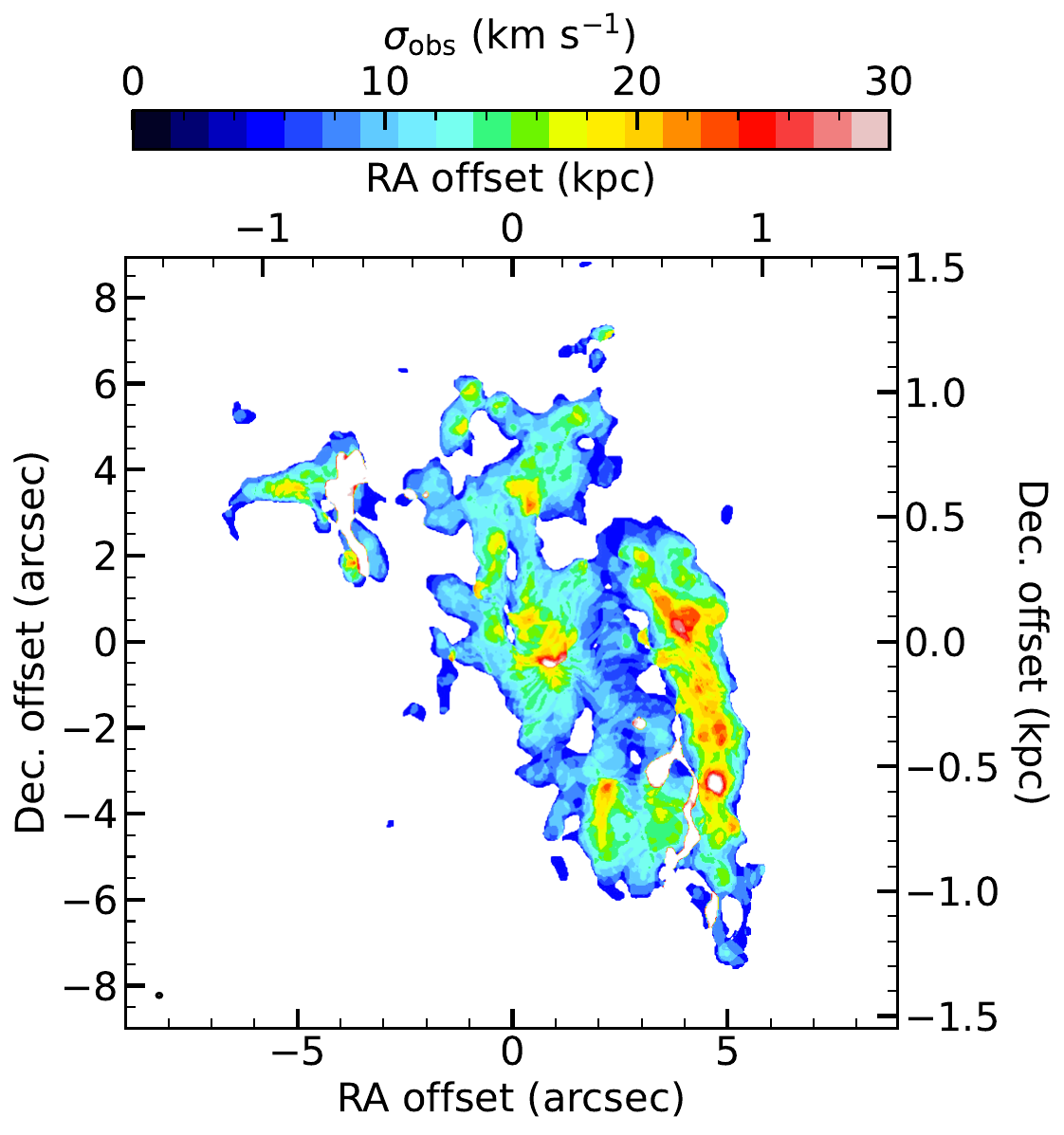}
    \end{subfigure}
    \\
    \begin{subfigure}{.34\linewidth}
    \centering
    \includegraphics[height=5.2cm]{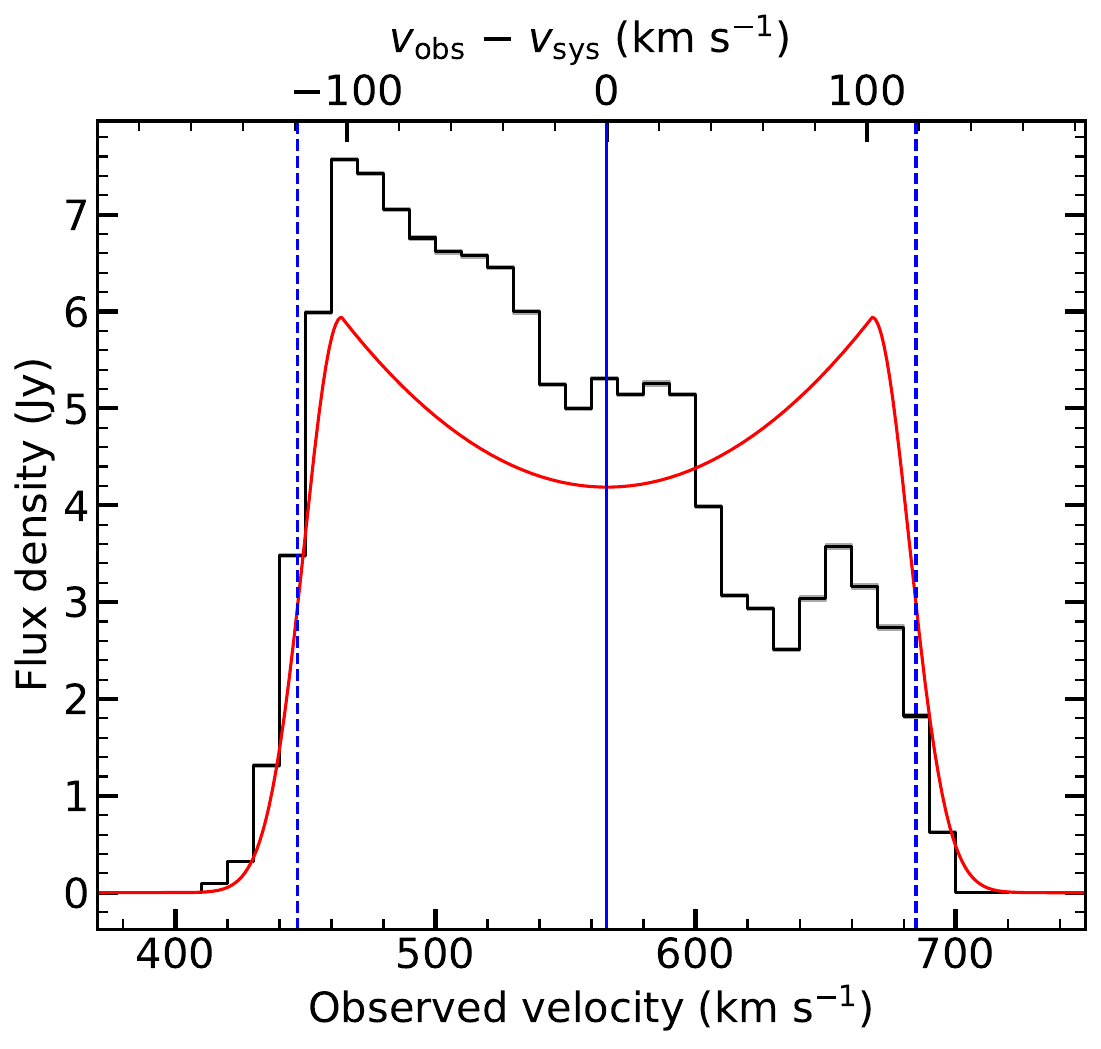}
    \end{subfigure}  
    \begin{subfigure}{.32\linewidth}
    \centering
    \includegraphics[height=4.8cm]{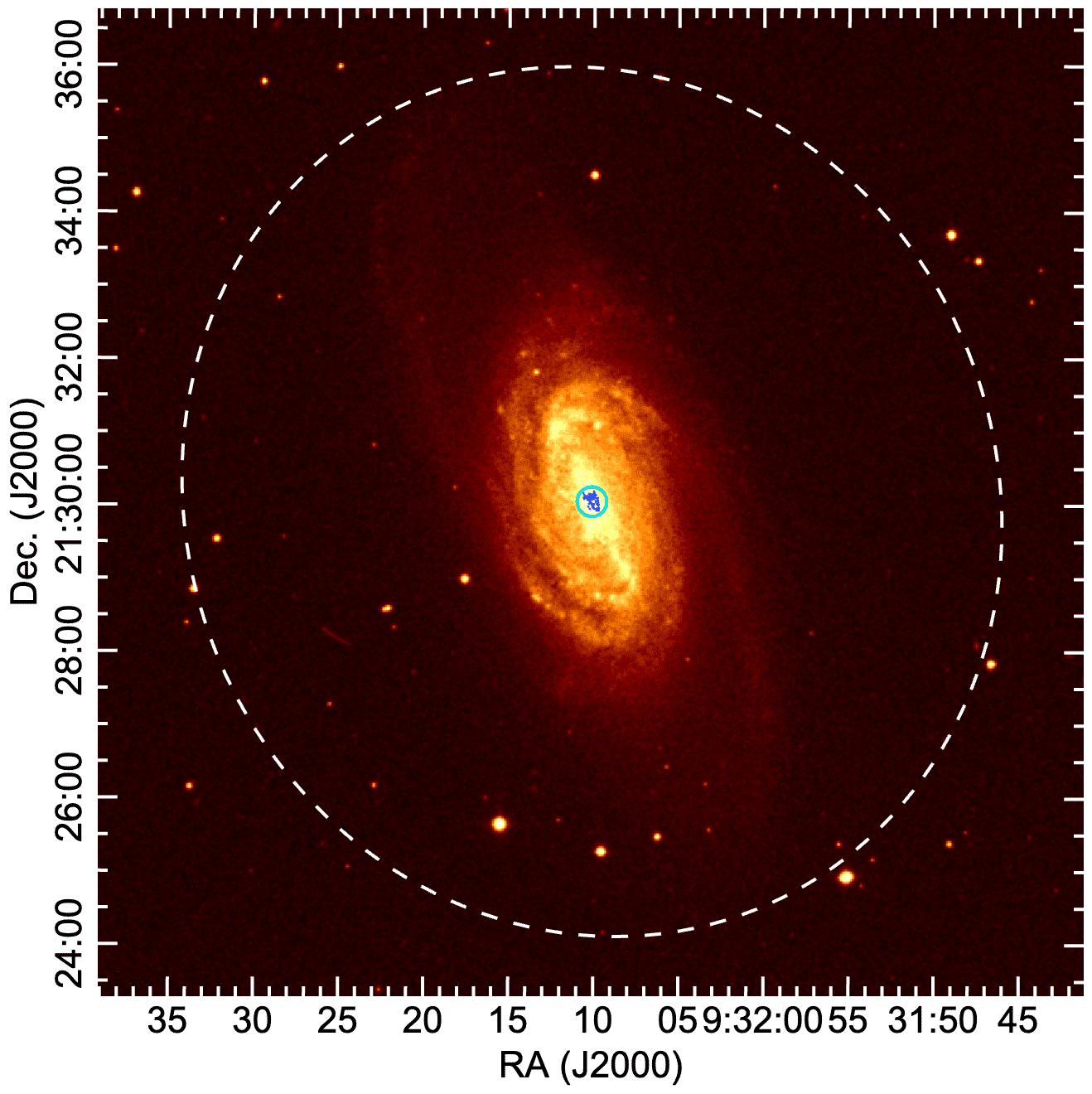}
    \end{subfigure}
    \begin{subfigure}{.33\linewidth}
    \centering
    \includegraphics[height=5.2cm]{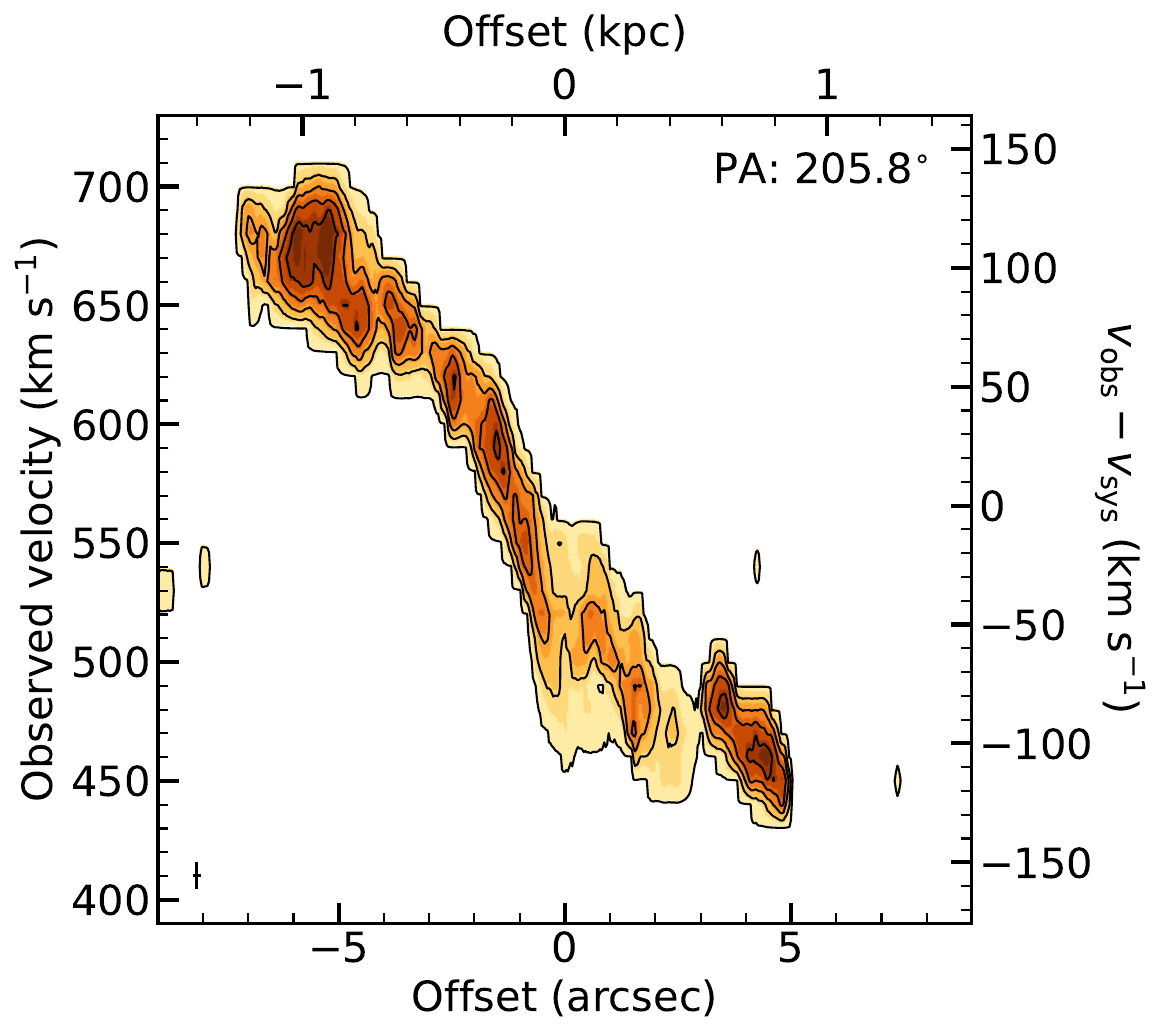}
    \end{subfigure}
    \begin{subfigure}{.34\linewidth}
    \includegraphics[height=6.0cm]{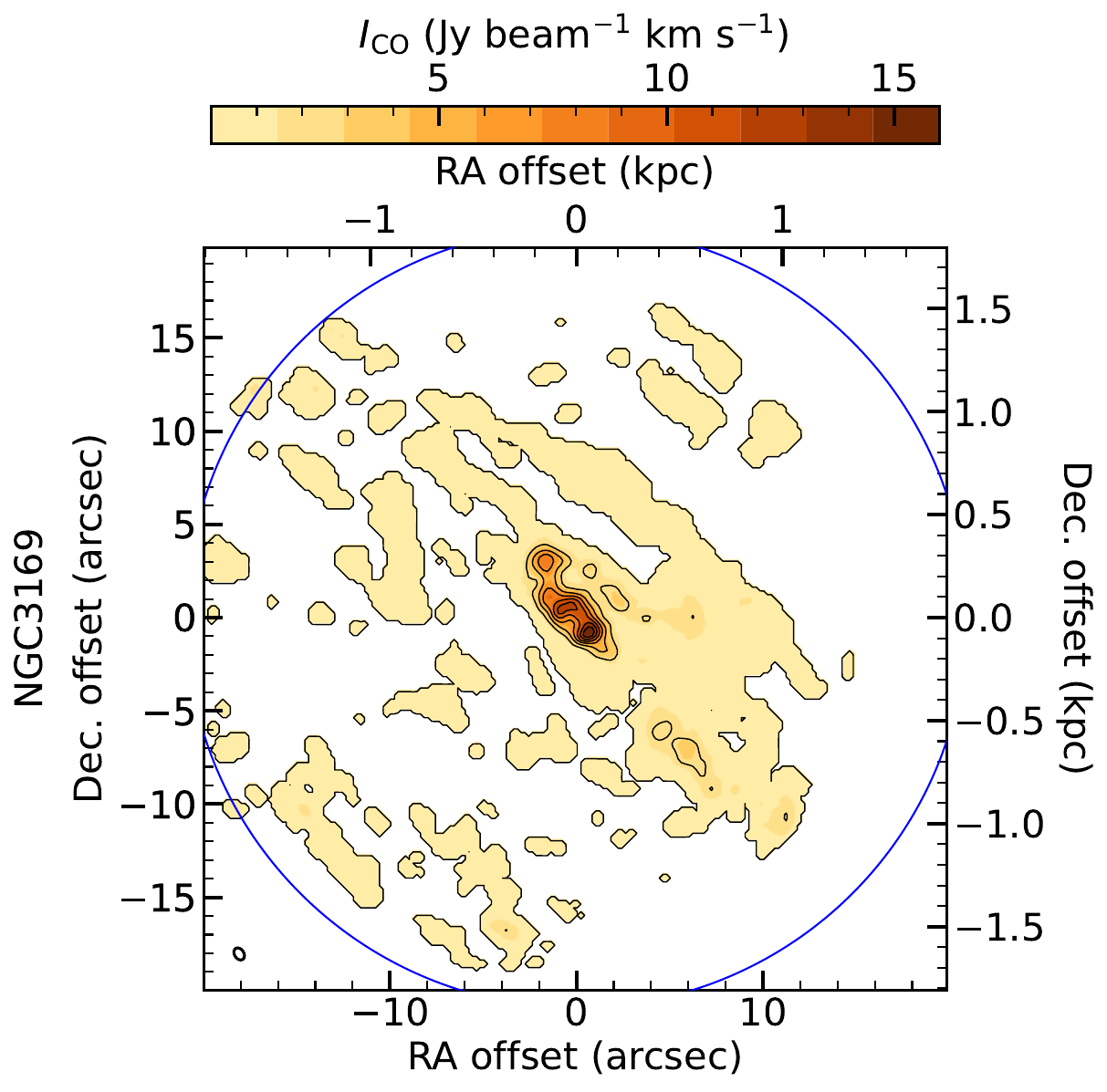}
    \end{subfigure}
    \begin{subfigure}{.32\linewidth}
    \centering
    \includegraphics[height=6.0cm]{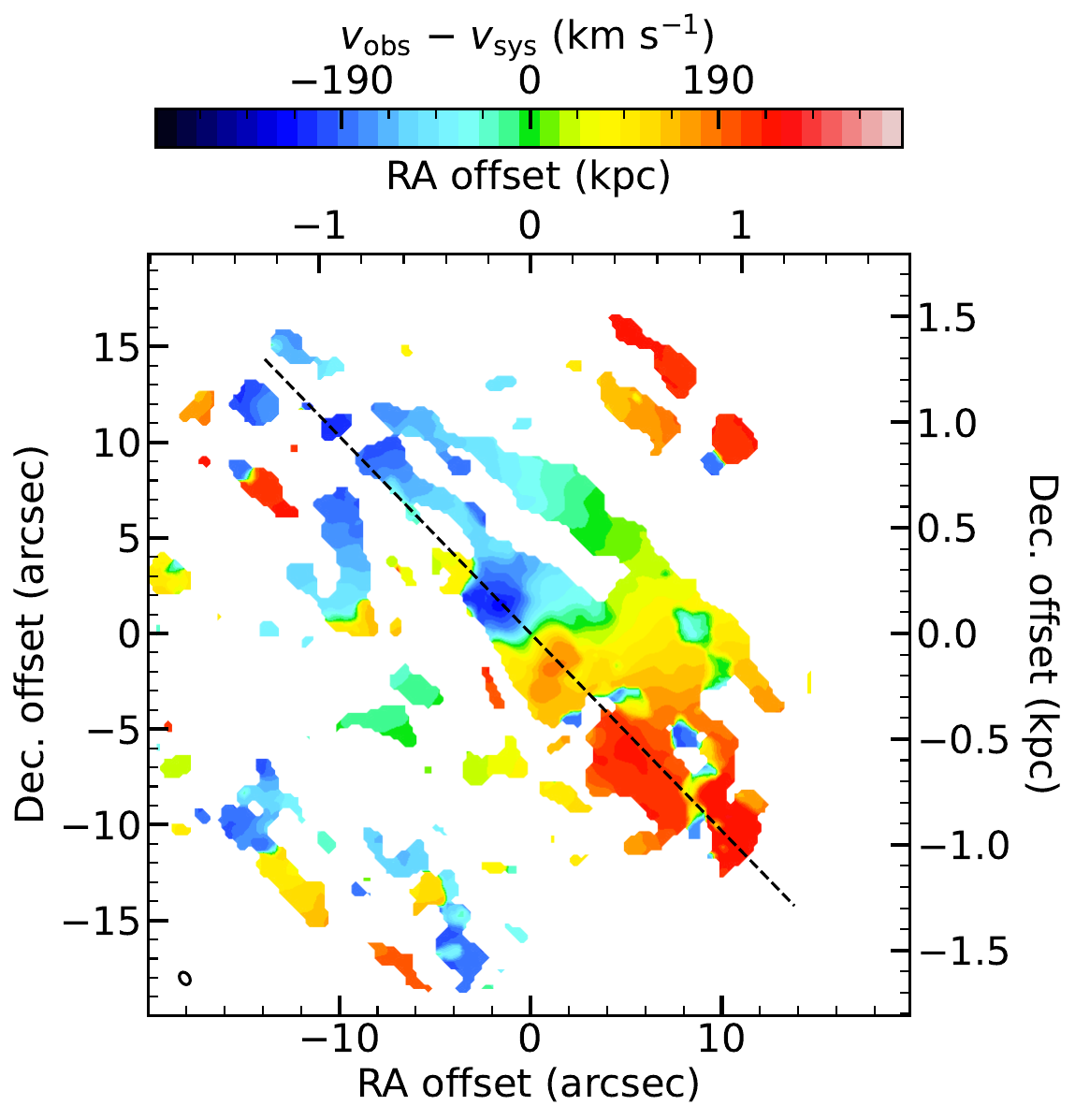}
    \end{subfigure}
    \begin{subfigure}{.32\linewidth}
    \centering
    \includegraphics[height=6.0cm]{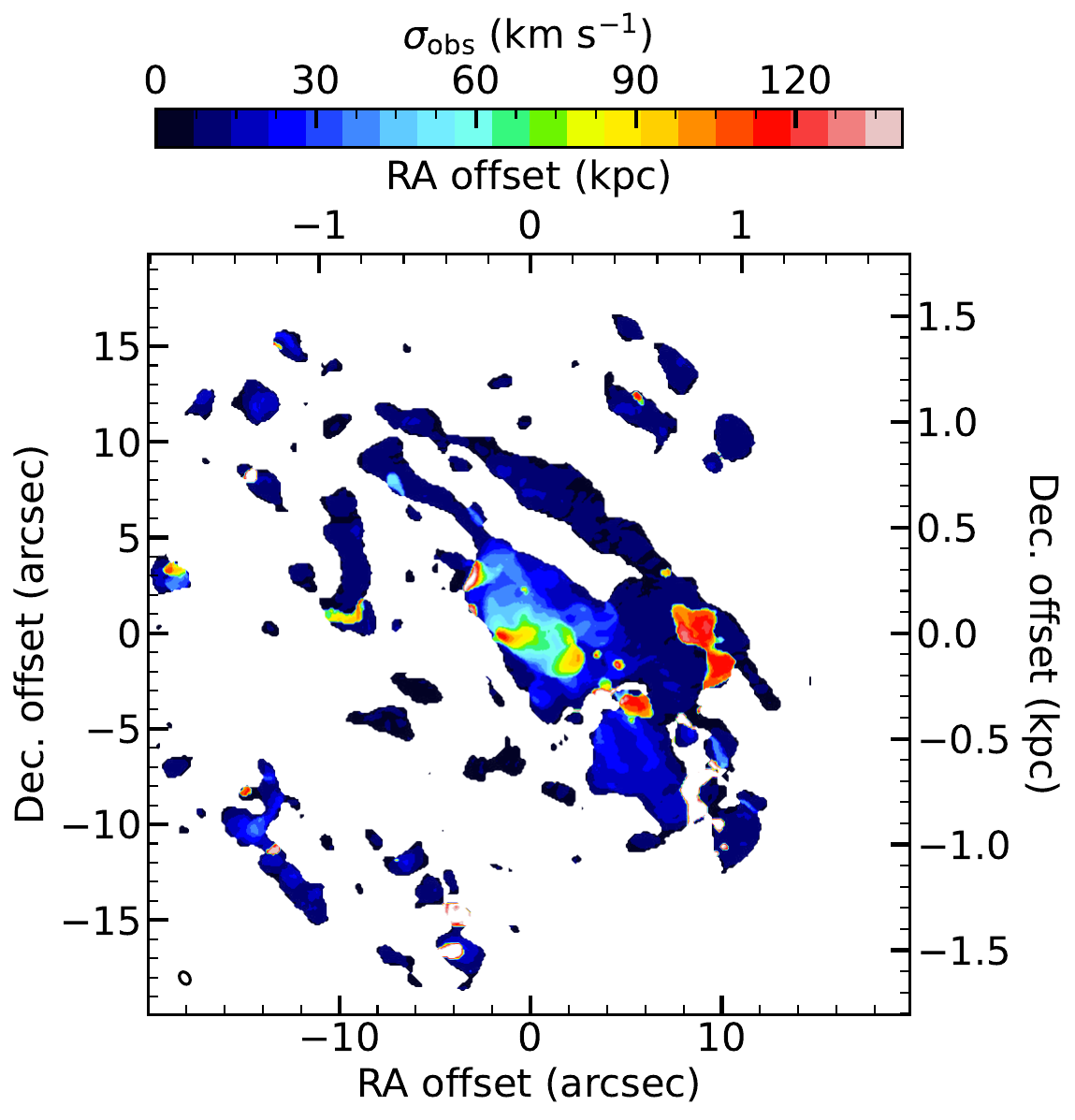}
    \end{subfigure}
    \\
    \begin{subfigure}{.34\linewidth}
    \centering
    \includegraphics[height=5.2cm]{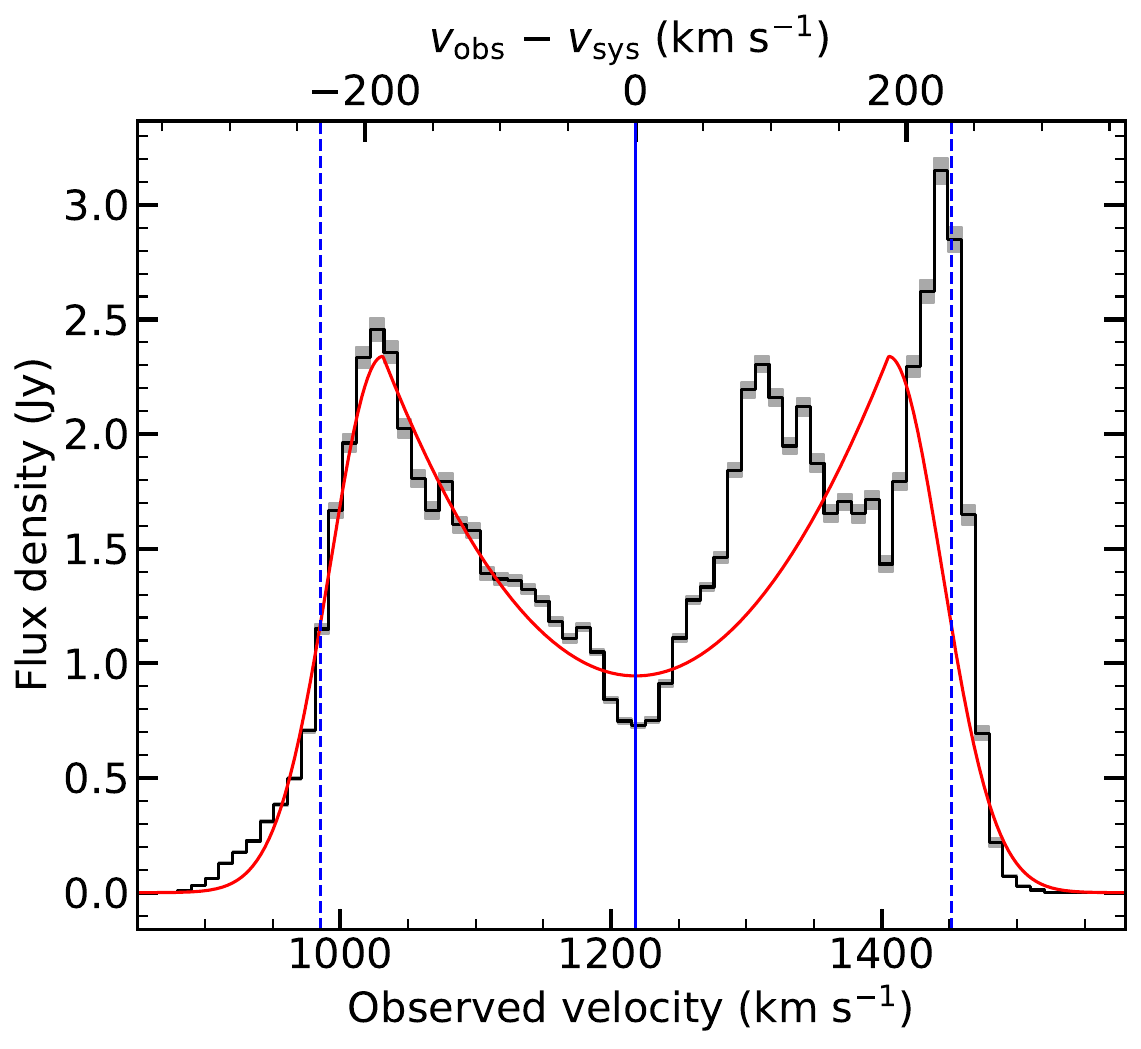}
    \end{subfigure}
    \begin{subfigure}{.32\linewidth}
    \centering
    \includegraphics[height=4.8cm]{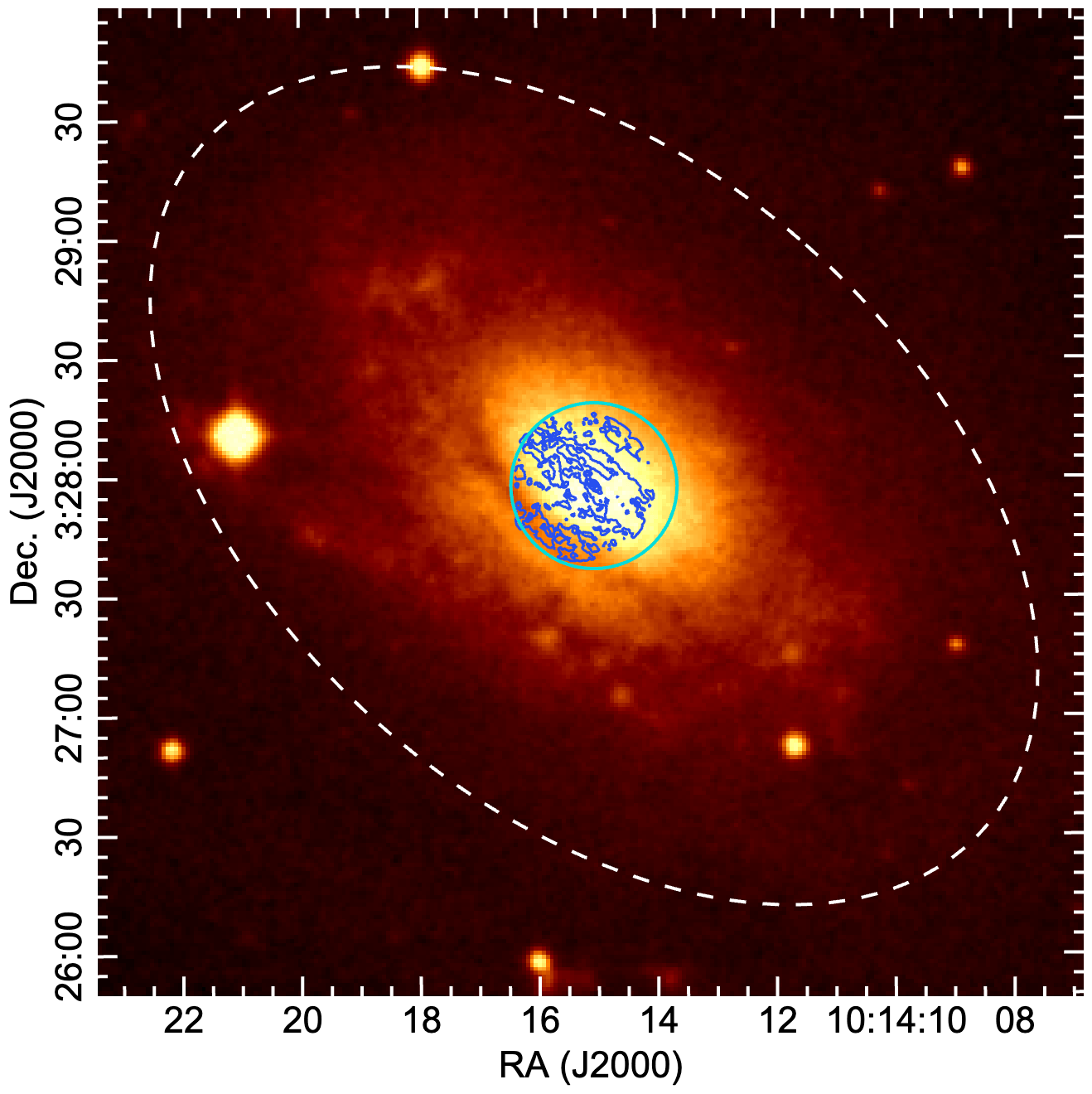}
    \end{subfigure}
    \begin{subfigure}{.33\linewidth}
    \centering
    \includegraphics[height=5.2cm]{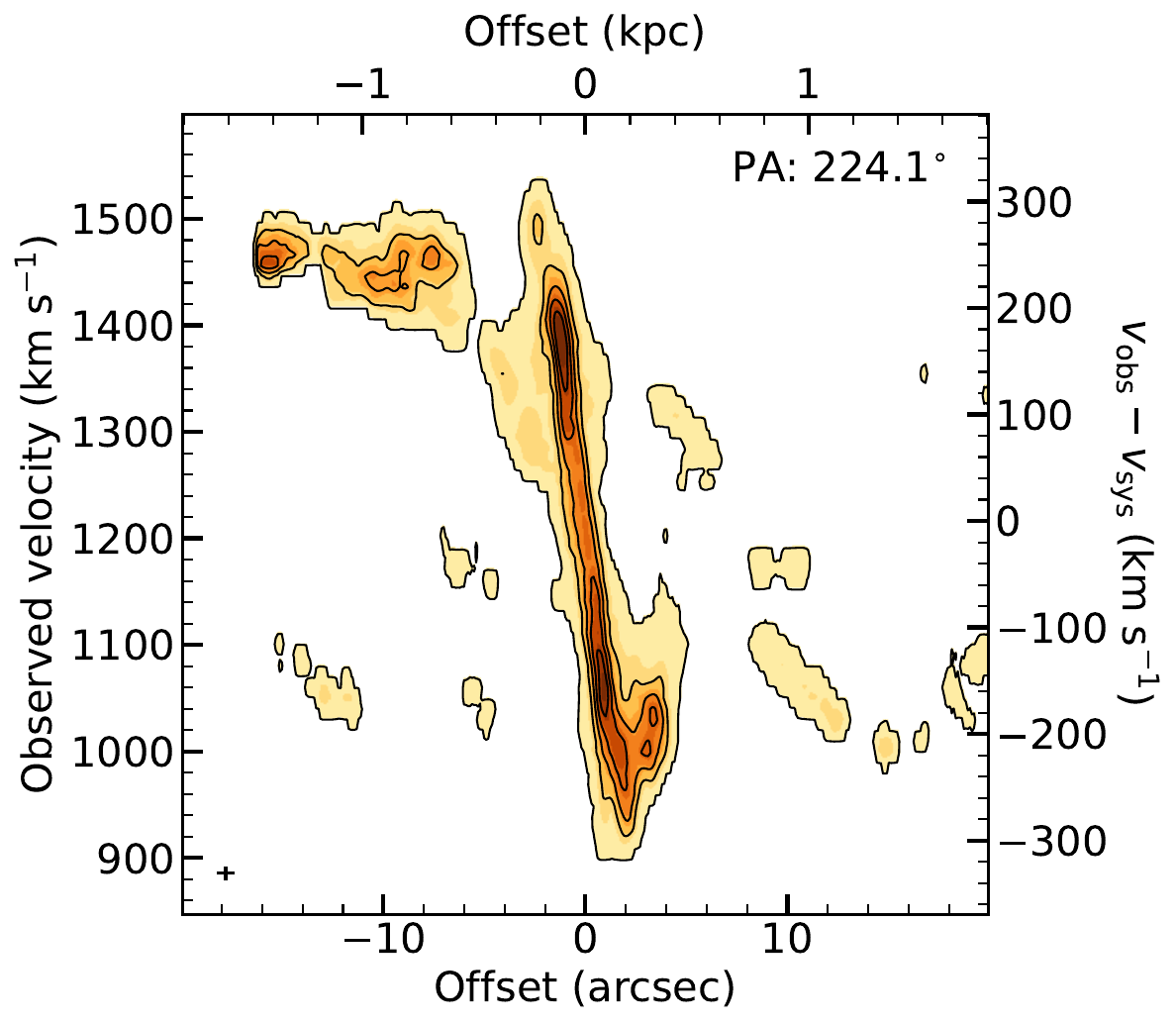}
    \end{subfigure}
    \caption[]{continued.}
\end{figure*}

\begin{figure*}
    \ContinuedFloat
    \begin{subfigure}{.34\linewidth}
    \includegraphics[height=5.9cm]{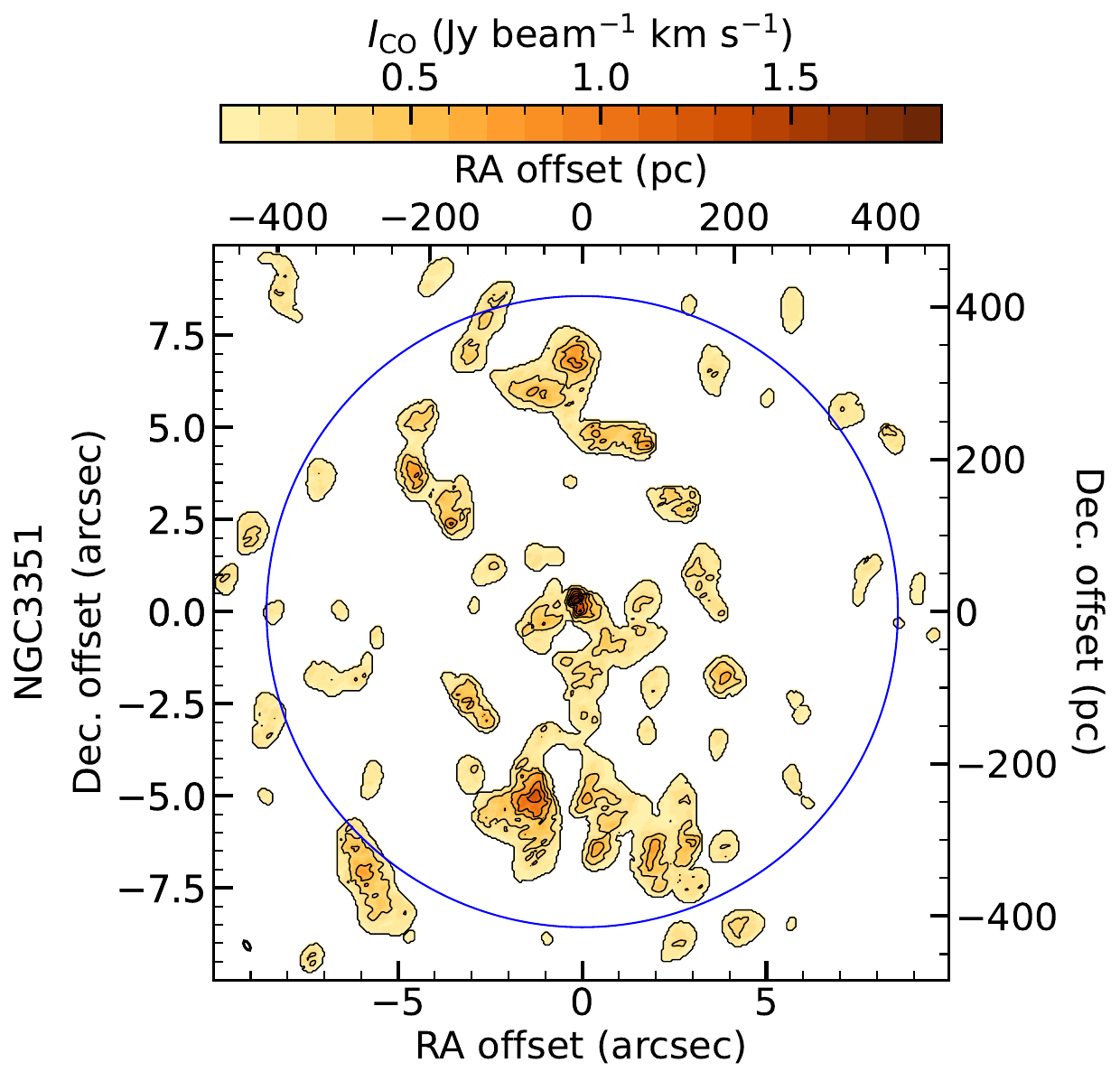}
    \end{subfigure}
    \begin{subfigure}{.32\linewidth}
    \centering
    \includegraphics[height=5.9cm]{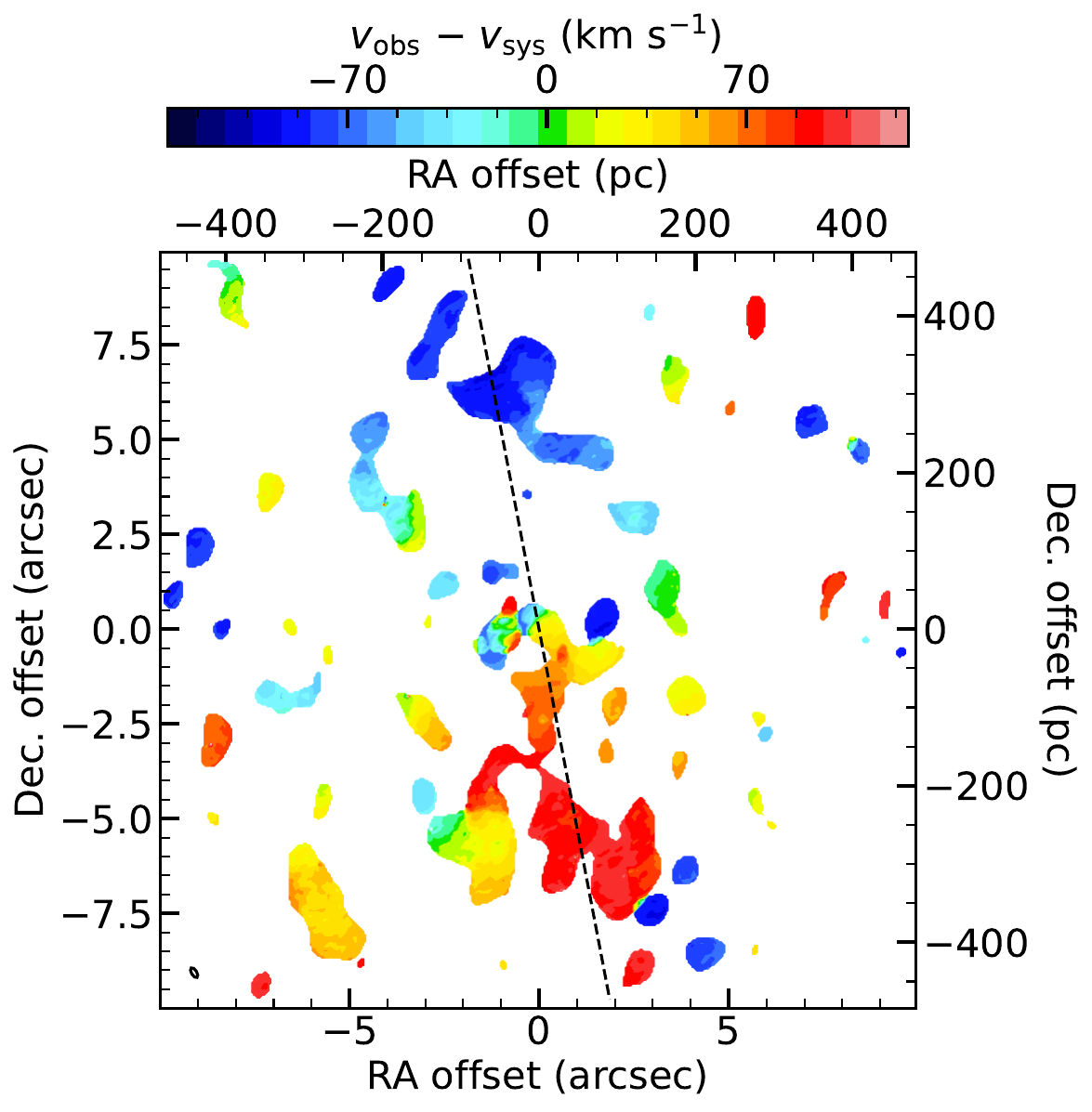}
    \end{subfigure}
    \begin{subfigure}{.32\linewidth}
    \centering
    \includegraphics[height=5.9cm]{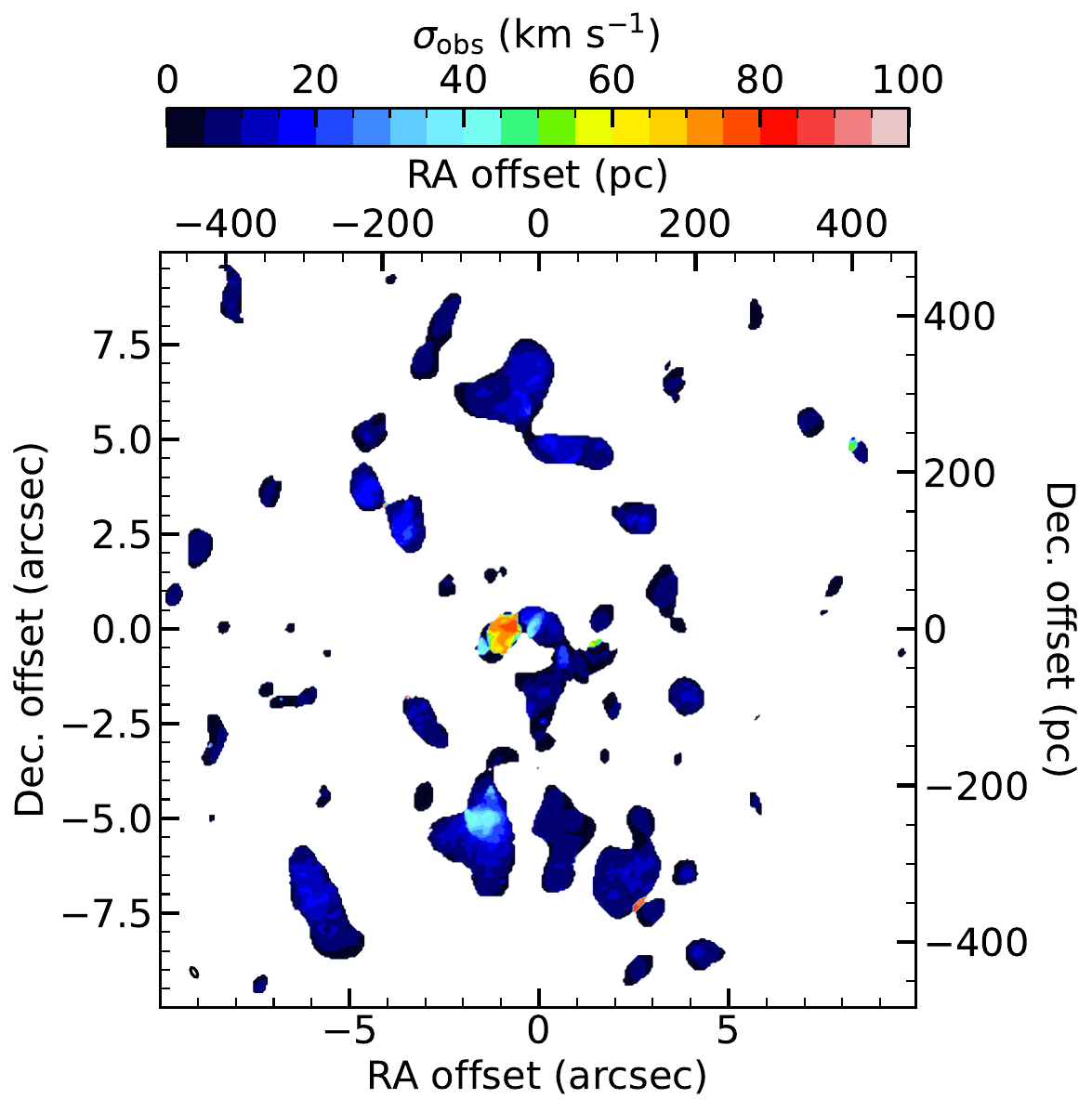}
    \end{subfigure}
    \\
    \begin{subfigure}{.34\linewidth}
    \centering
    \includegraphics[height=5.2cm]{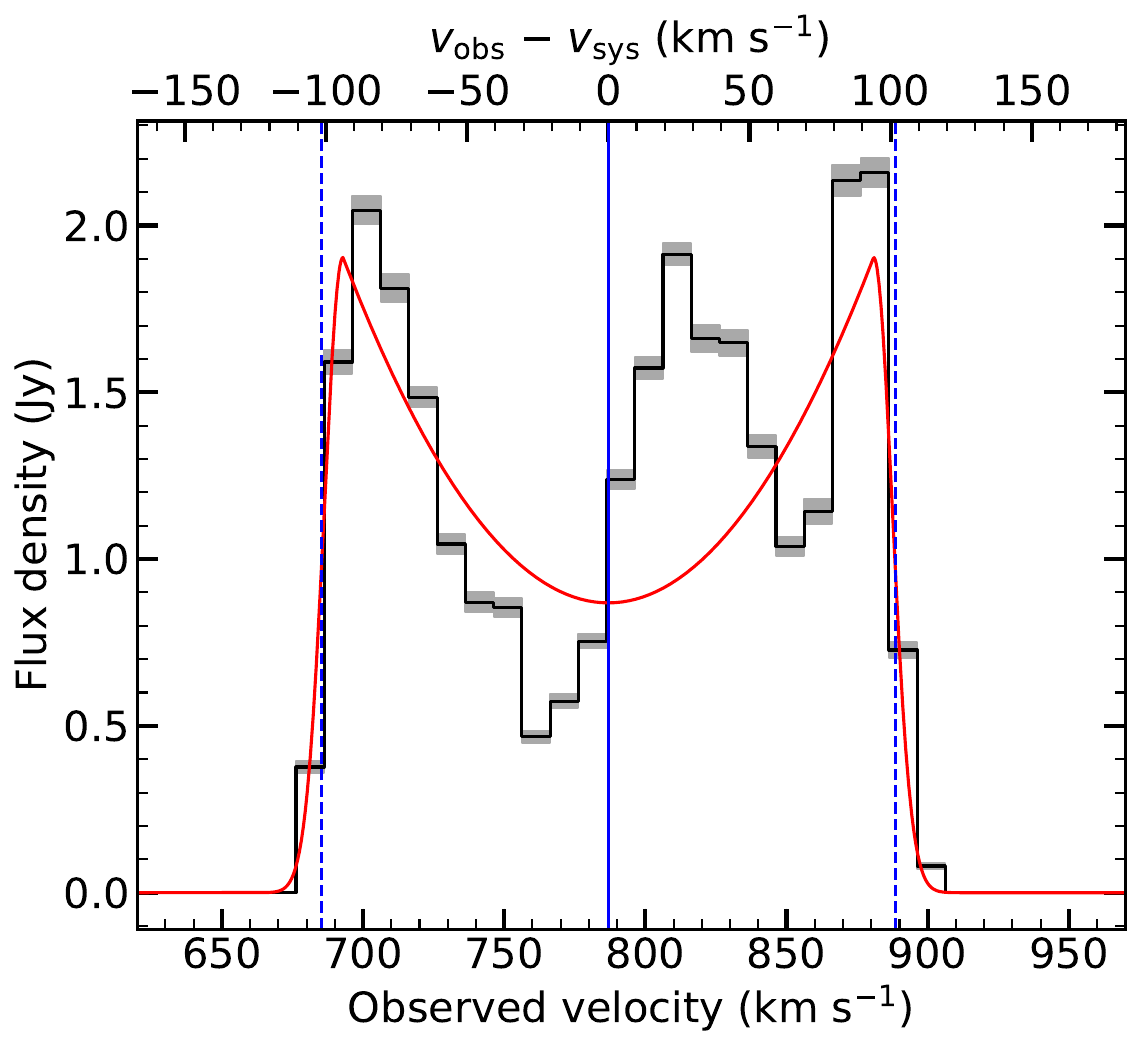}
    \end{subfigure}
    \begin{subfigure}{.32\linewidth}
    \centering
    \includegraphics[height=4.8cm]{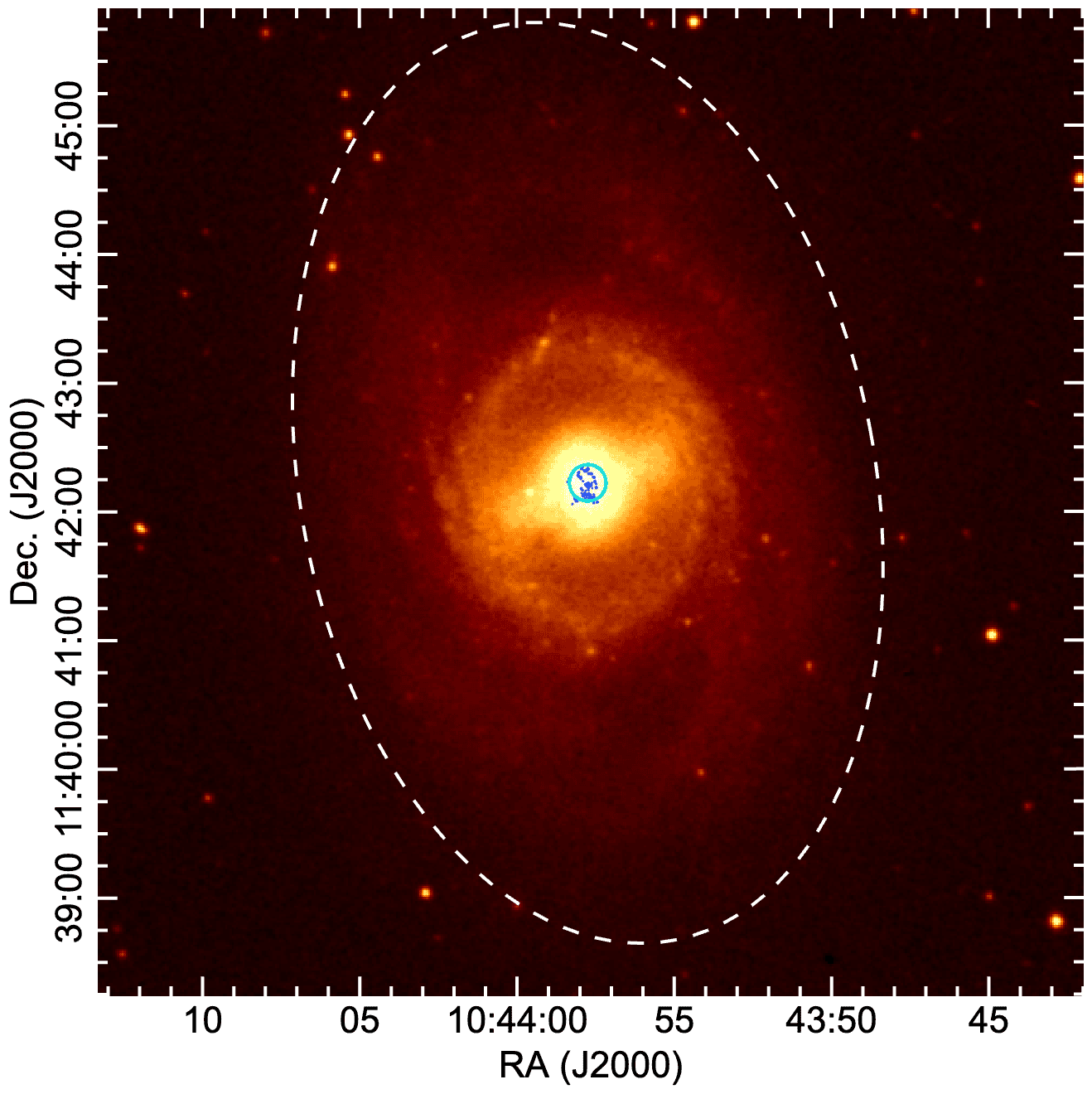}
    \end{subfigure}
    \begin{subfigure}{.33\linewidth}
    \centering
    \includegraphics[height=5.2cm]{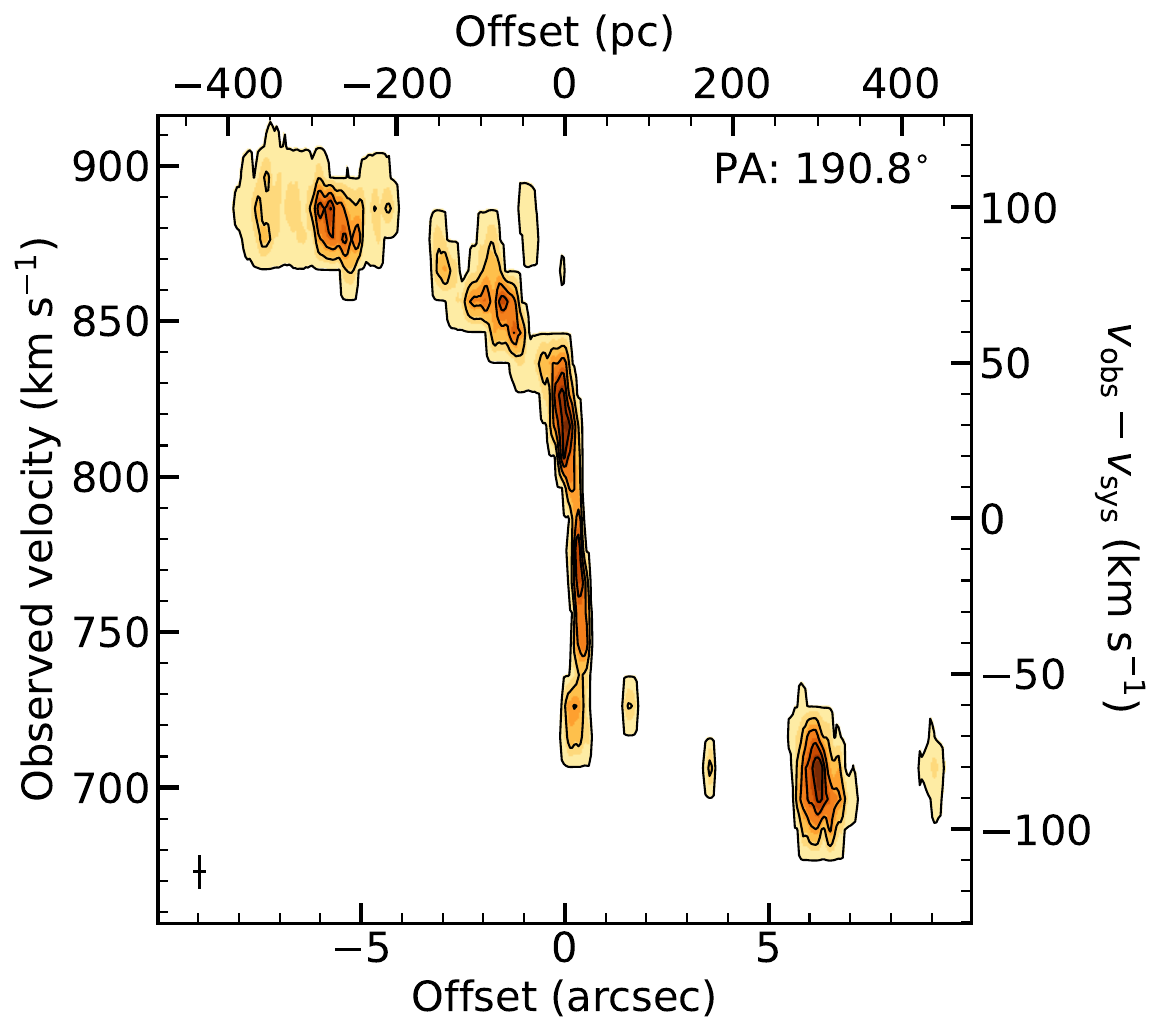}
    \end{subfigure}
    \begin{subfigure}{.34\linewidth}
    \centering
    \includegraphics[height=5.9cm]{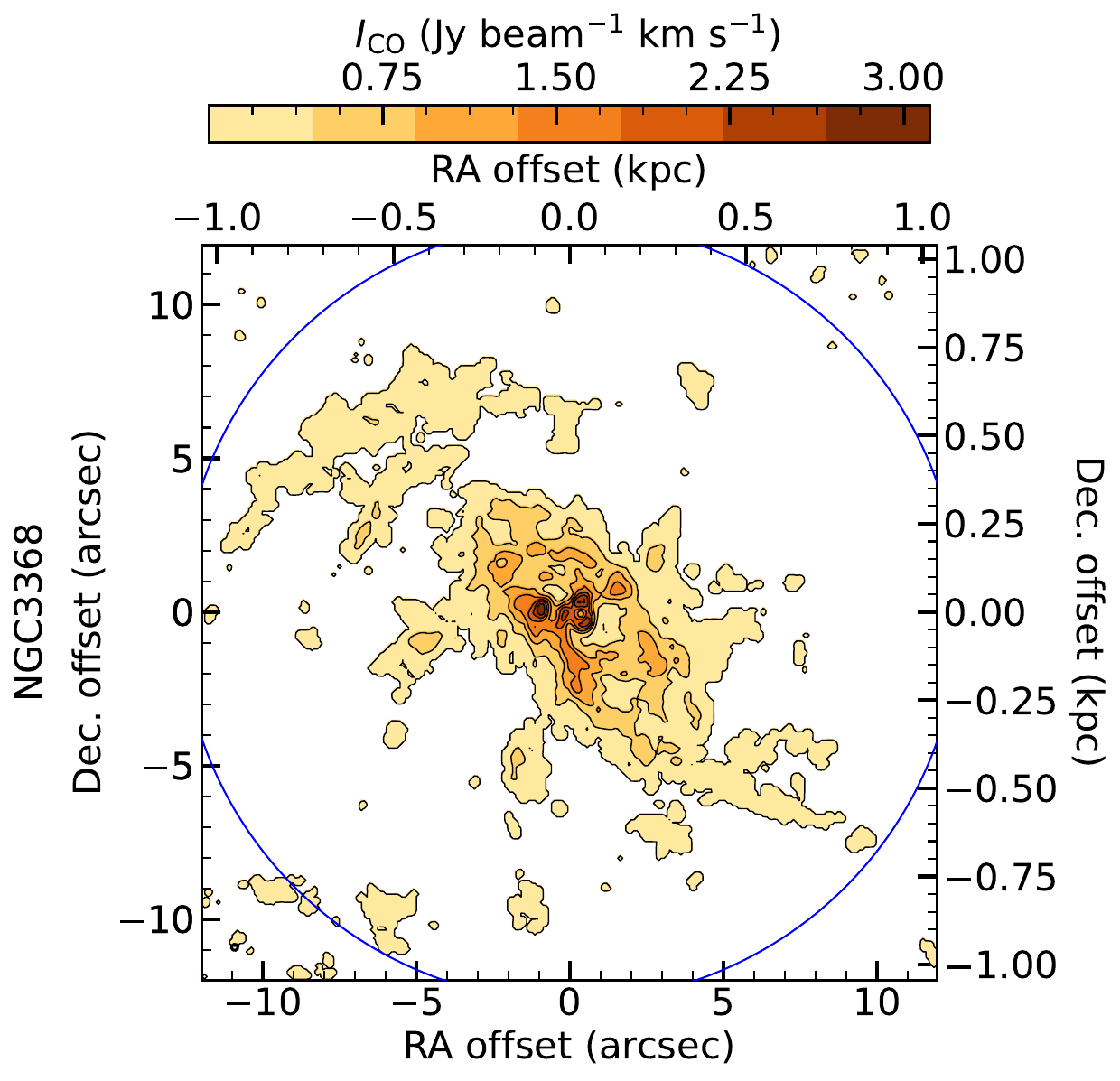}
    \end{subfigure}
    \begin{subfigure}{.32\linewidth}
    \centering
    \includegraphics[height=5.9cm]{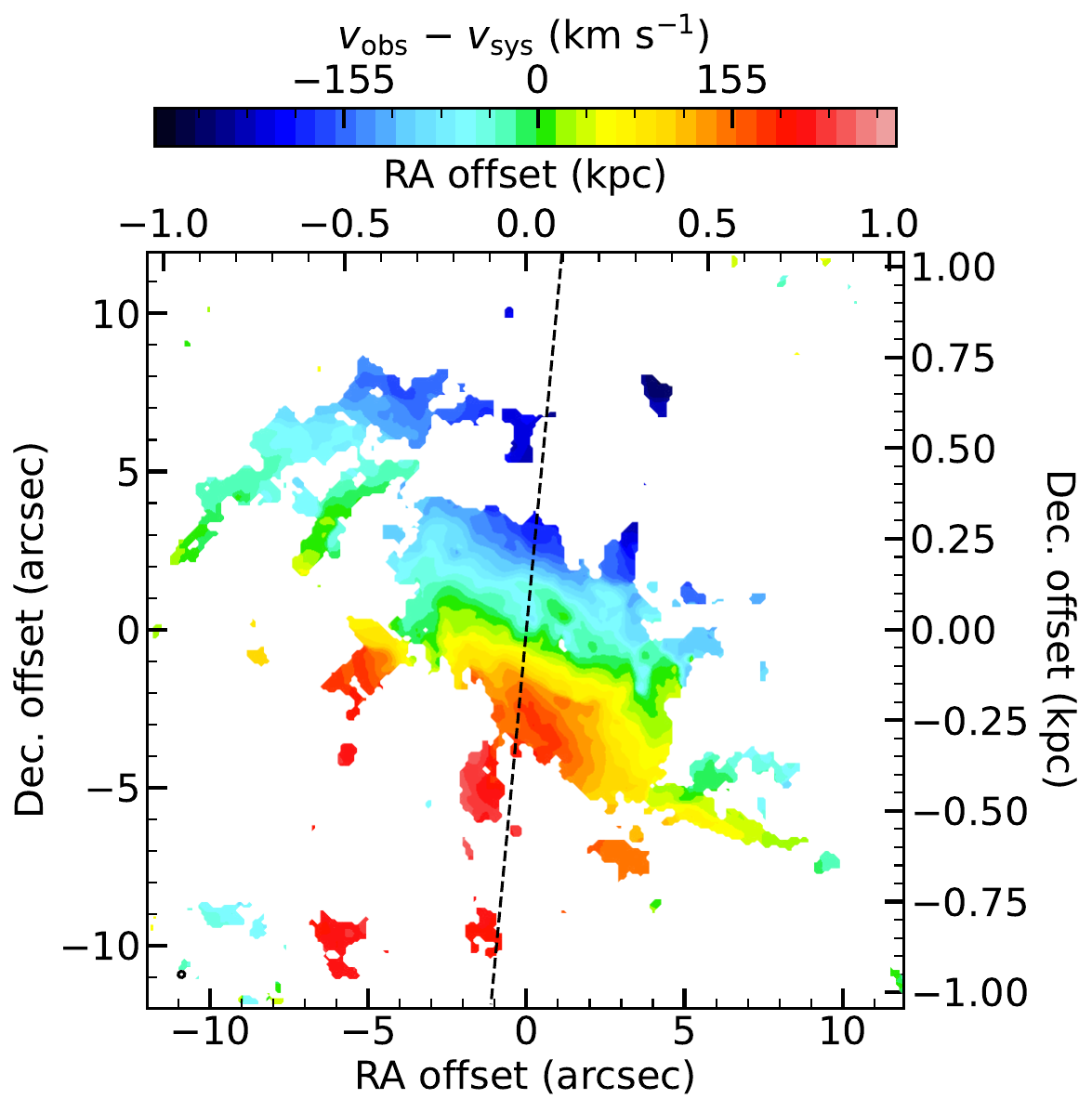}
    \end{subfigure}
    \begin{subfigure}{.32\linewidth}
    \centering
    \includegraphics[height=5.9cm]{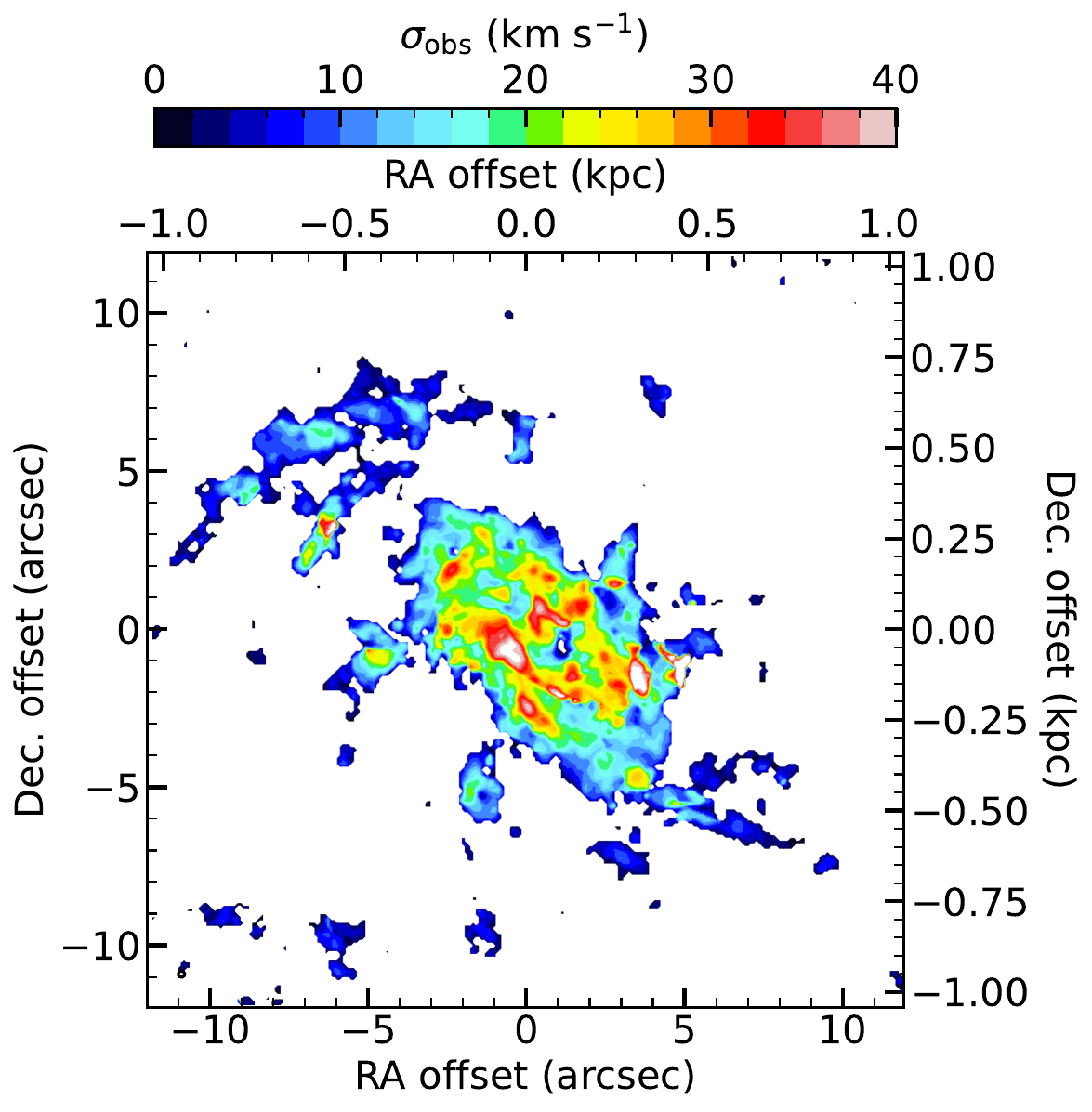}
    \end{subfigure}
    \\
    \begin{subfigure}{.34\linewidth}
    \centering
    \includegraphics[height=5.2cm]{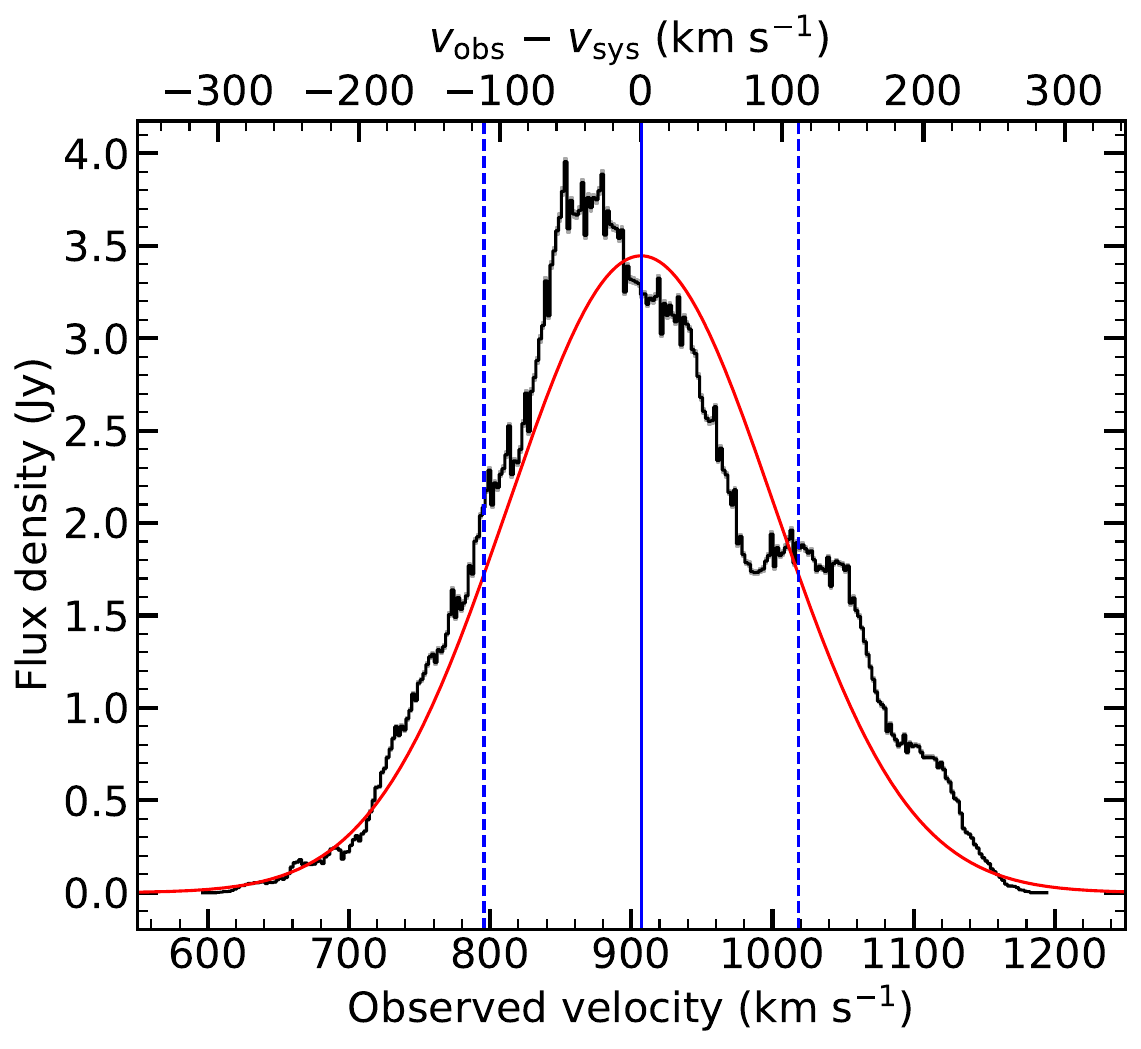}
    \end{subfigure}
    \begin{subfigure}{.32\linewidth}
    \centering
    \includegraphics[height=4.8cm]{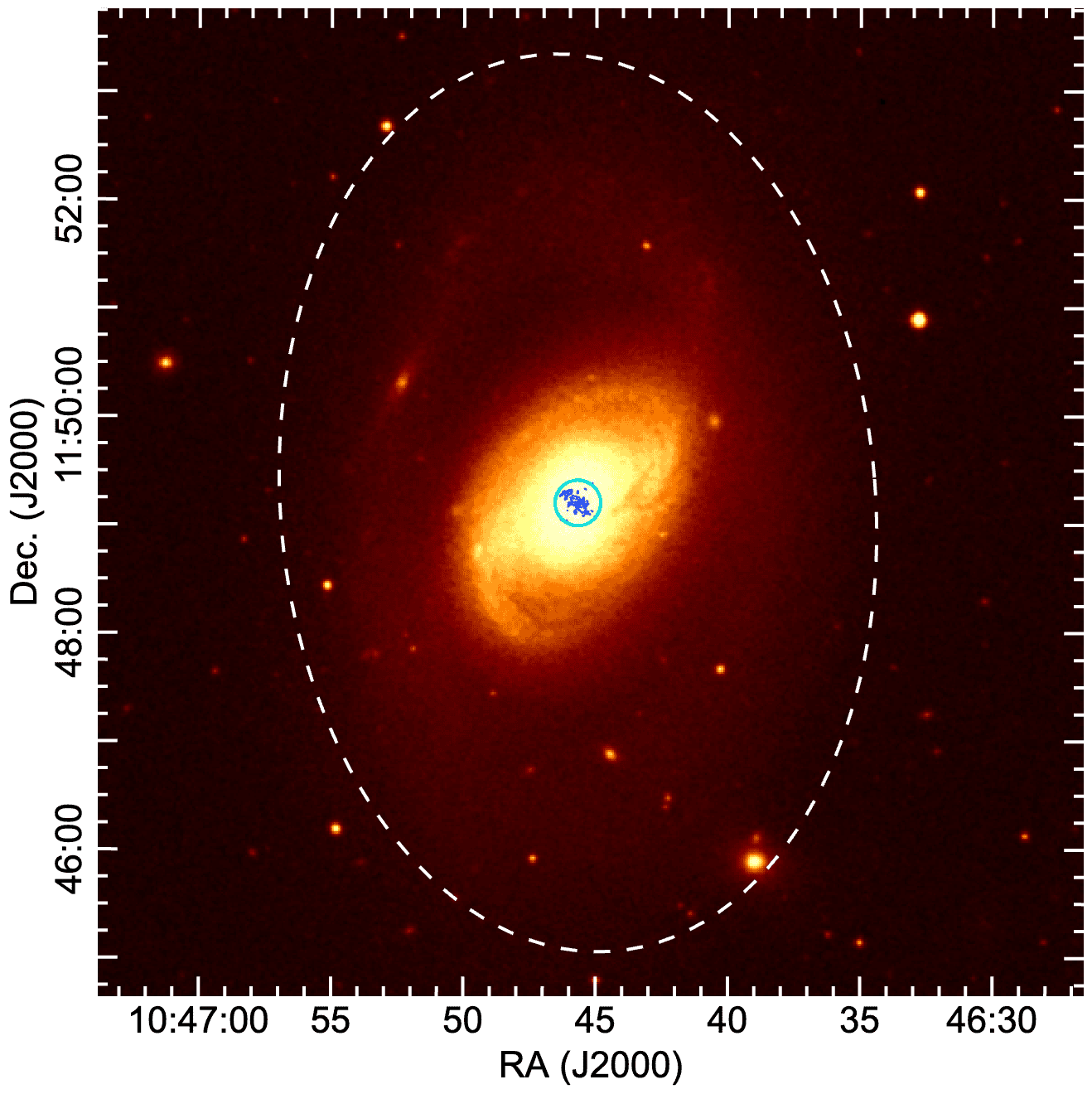}
    \end{subfigure}
    \begin{subfigure}{.33\linewidth}
    \centering
    \includegraphics[height=5.2cm]{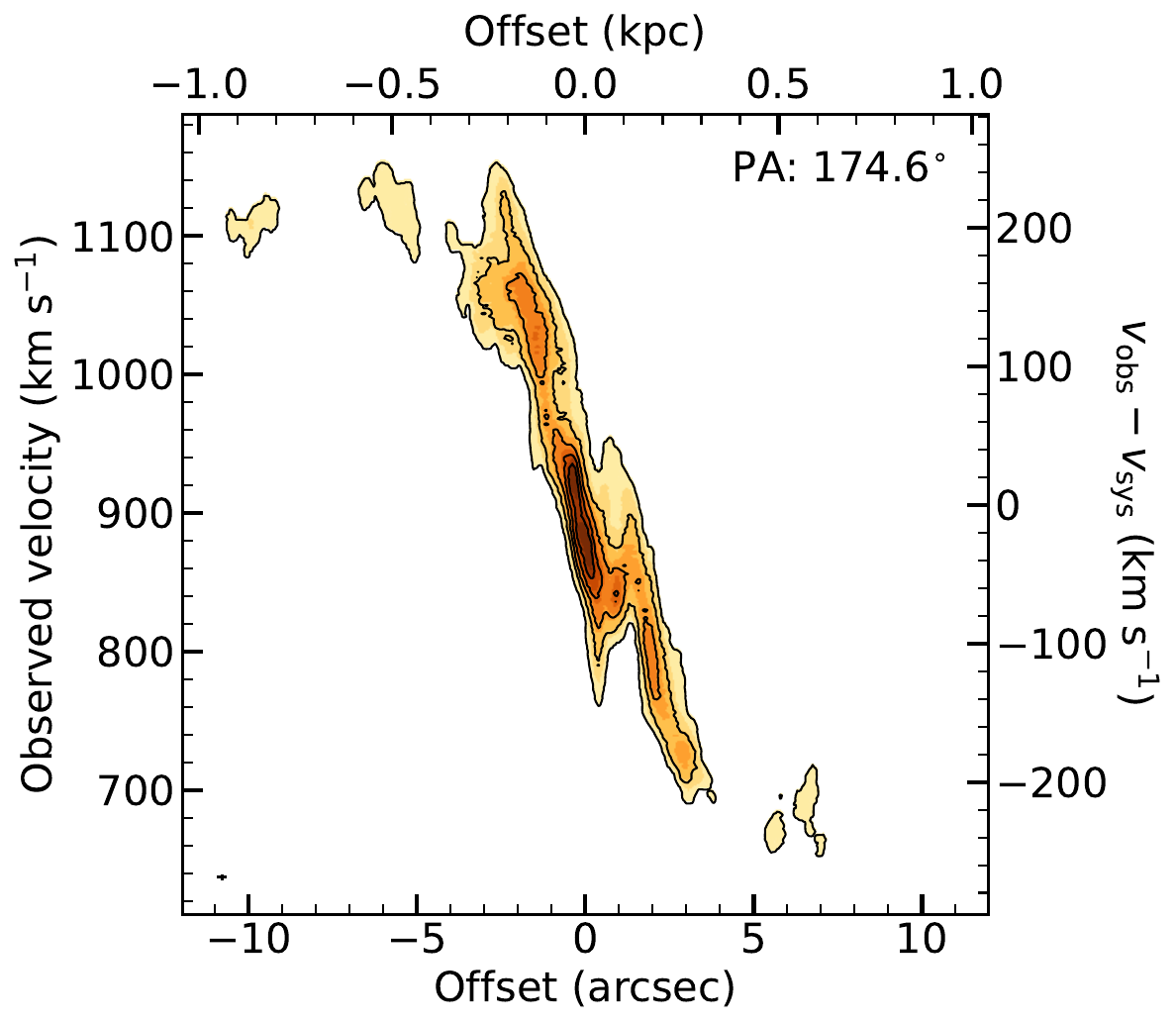}
    \end{subfigure}
    \caption[]{continued.}
\end{figure*}

\begin{figure*}
    \ContinuedFloat
    \begin{subfigure}{.34\linewidth}
    \includegraphics[height=6.0cm]{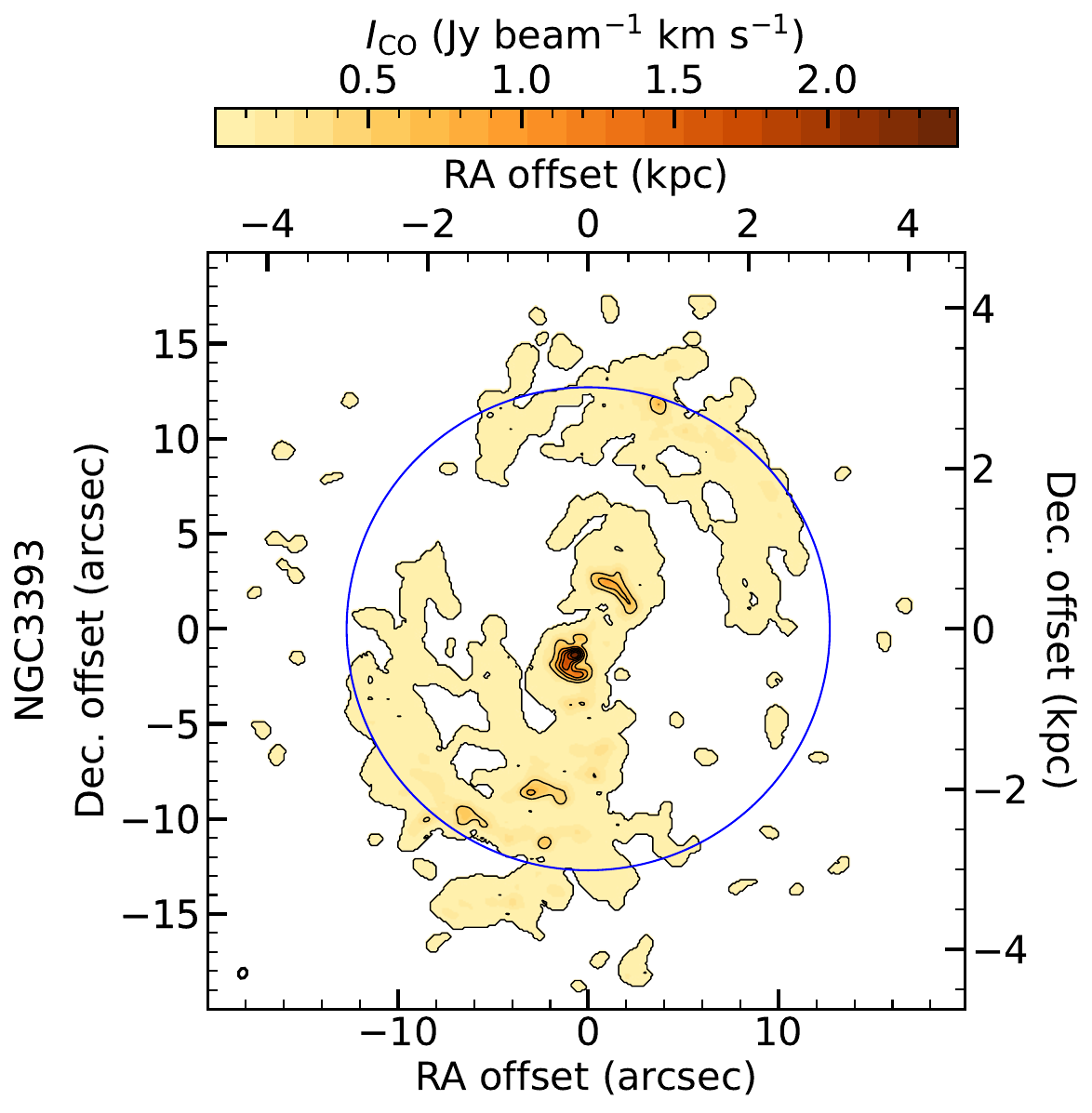}
    \end{subfigure}
    \begin{subfigure}{.32\linewidth}
    \centering
    \includegraphics[height=6.0cm]{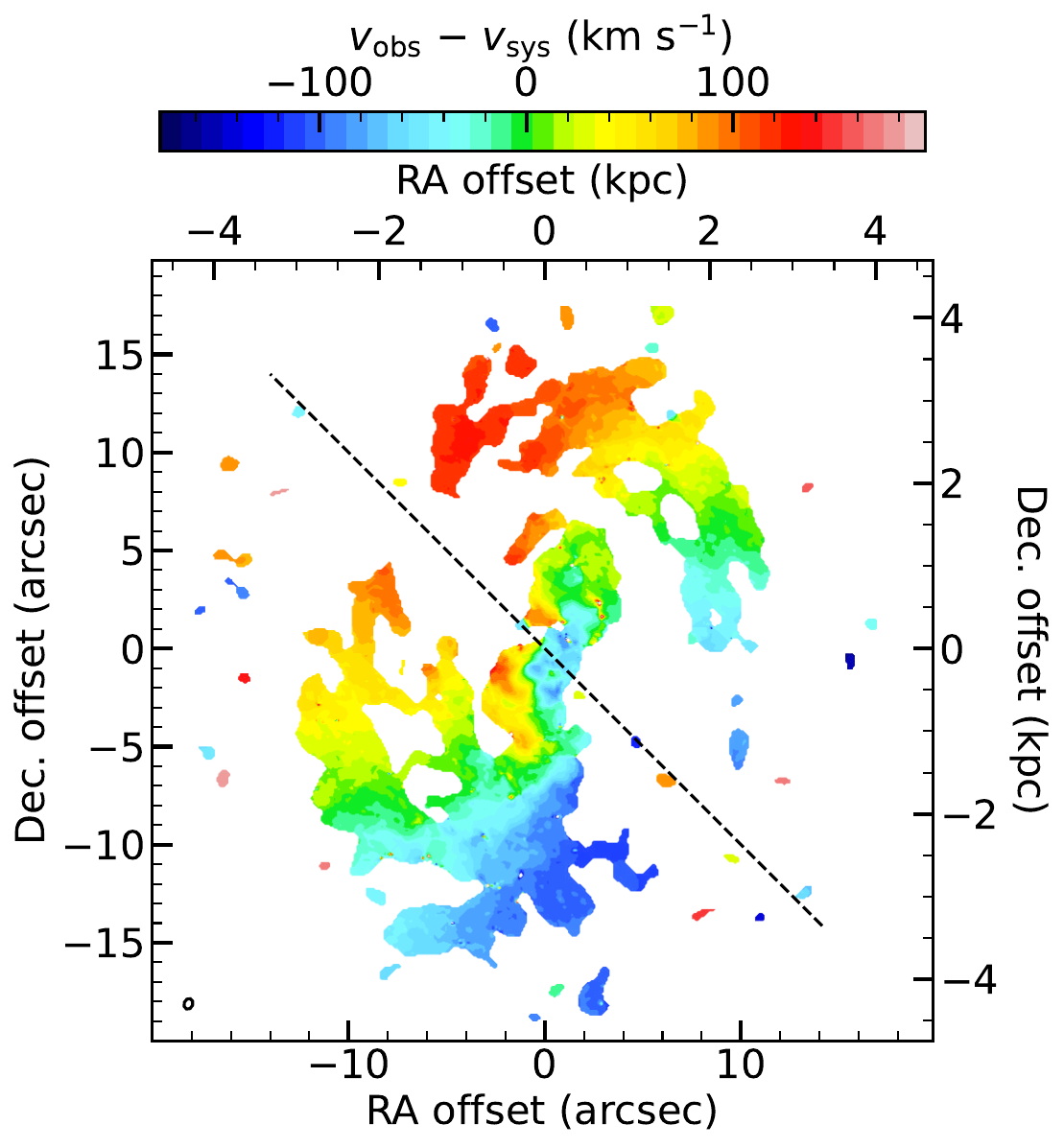}
    \end{subfigure}
    \begin{subfigure}{.32\linewidth}
    \centering
    \includegraphics[height=6.0cm]{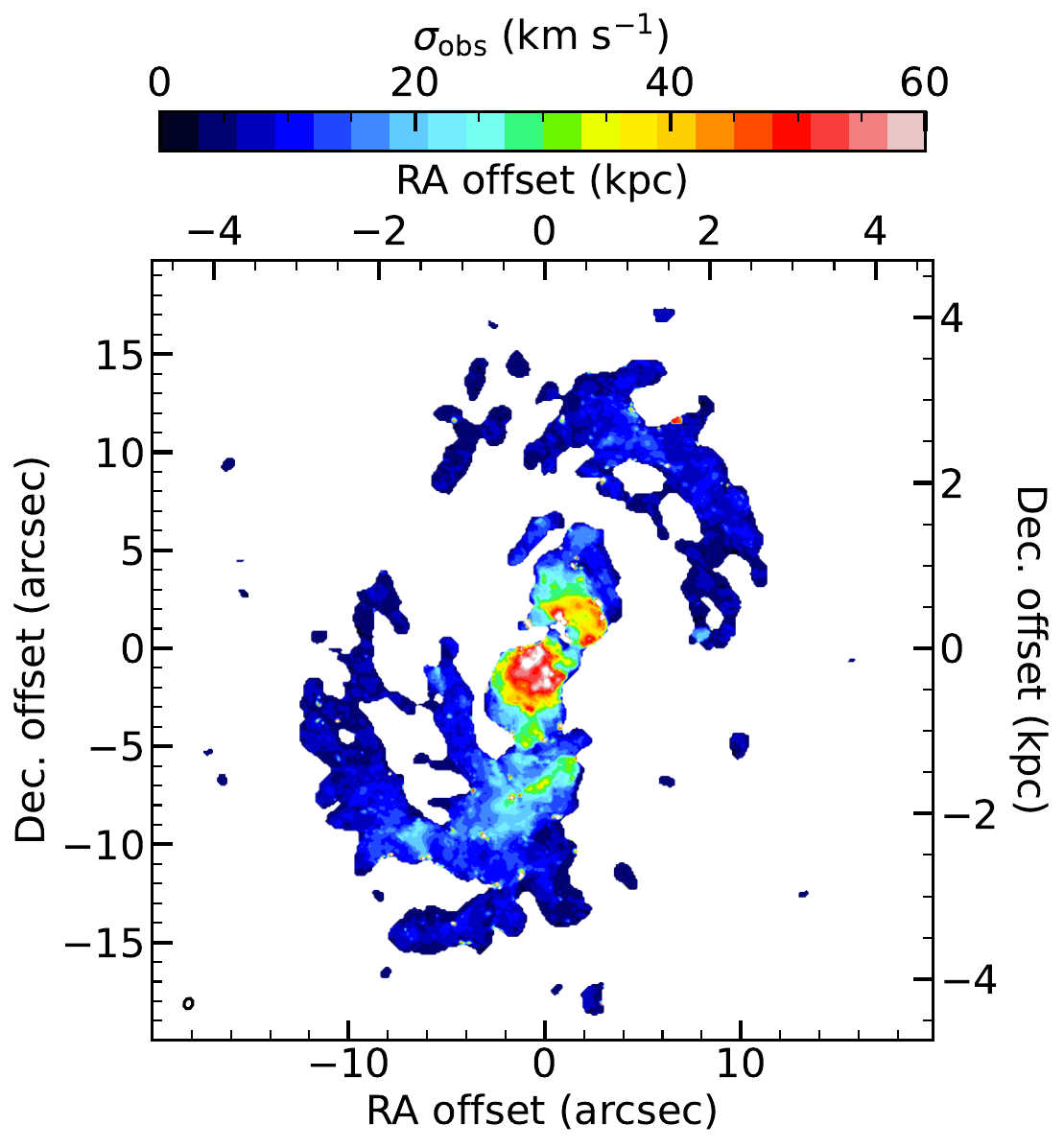}
    \end{subfigure}
    \\
    \begin{subfigure}{.34\linewidth}
    \centering
    \includegraphics[height=5.2cm]{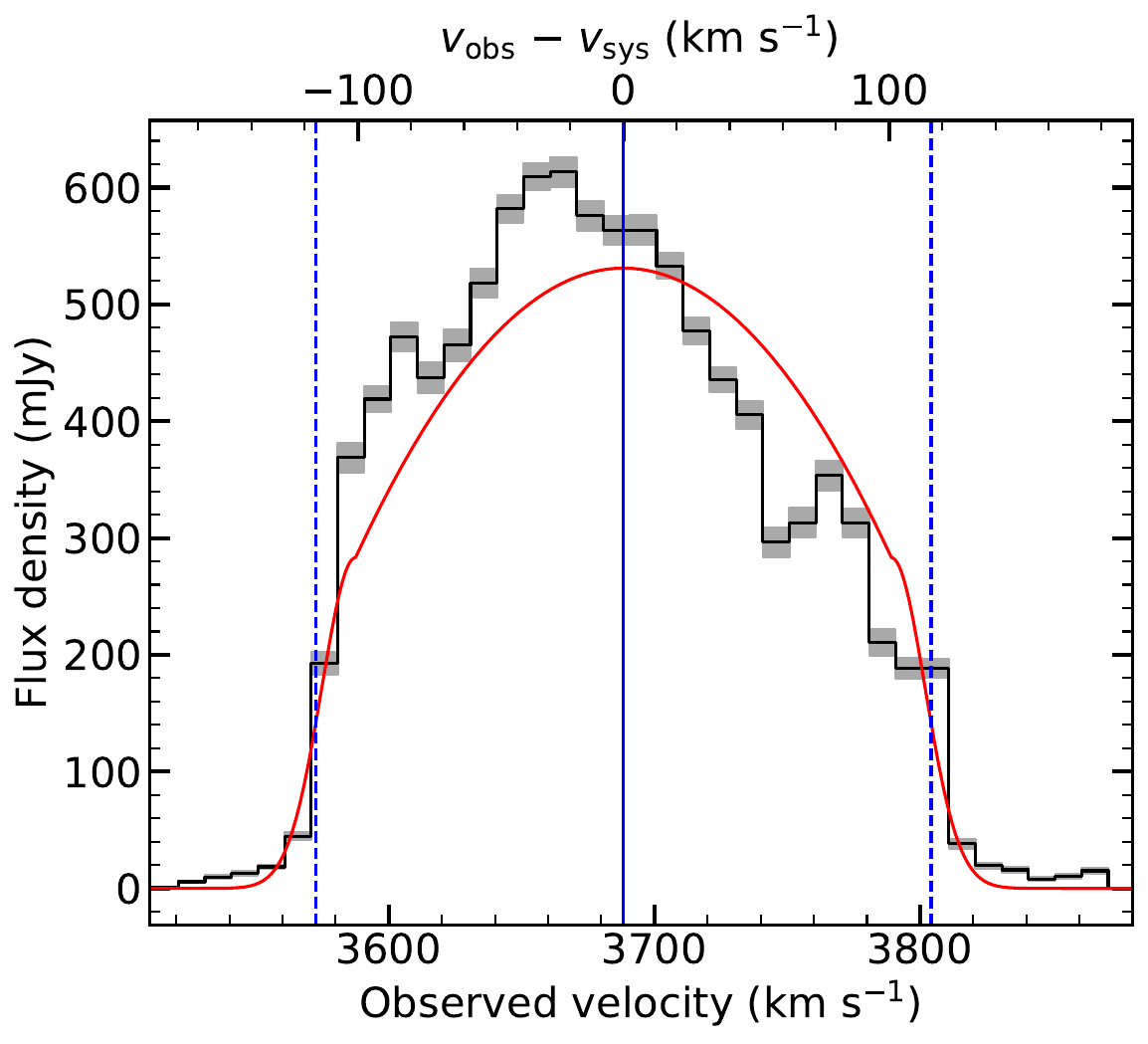}
    \end{subfigure}
    \begin{subfigure}{.32\linewidth}
    \centering
    \includegraphics[height=4.8cm]{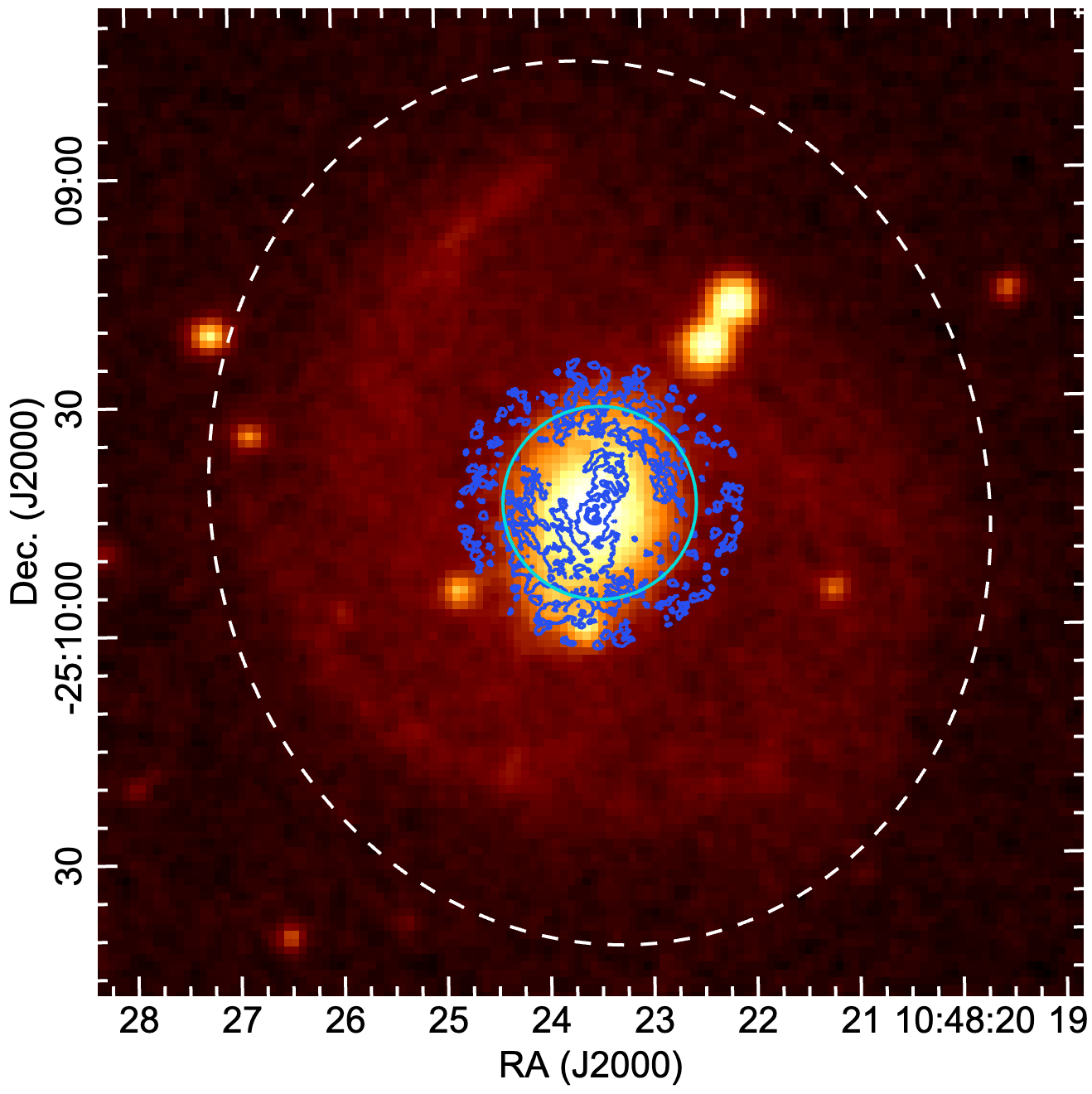}
    \end{subfigure}
    \begin{subfigure}{.33\linewidth}
    \centering
    \includegraphics[height=5.2cm]{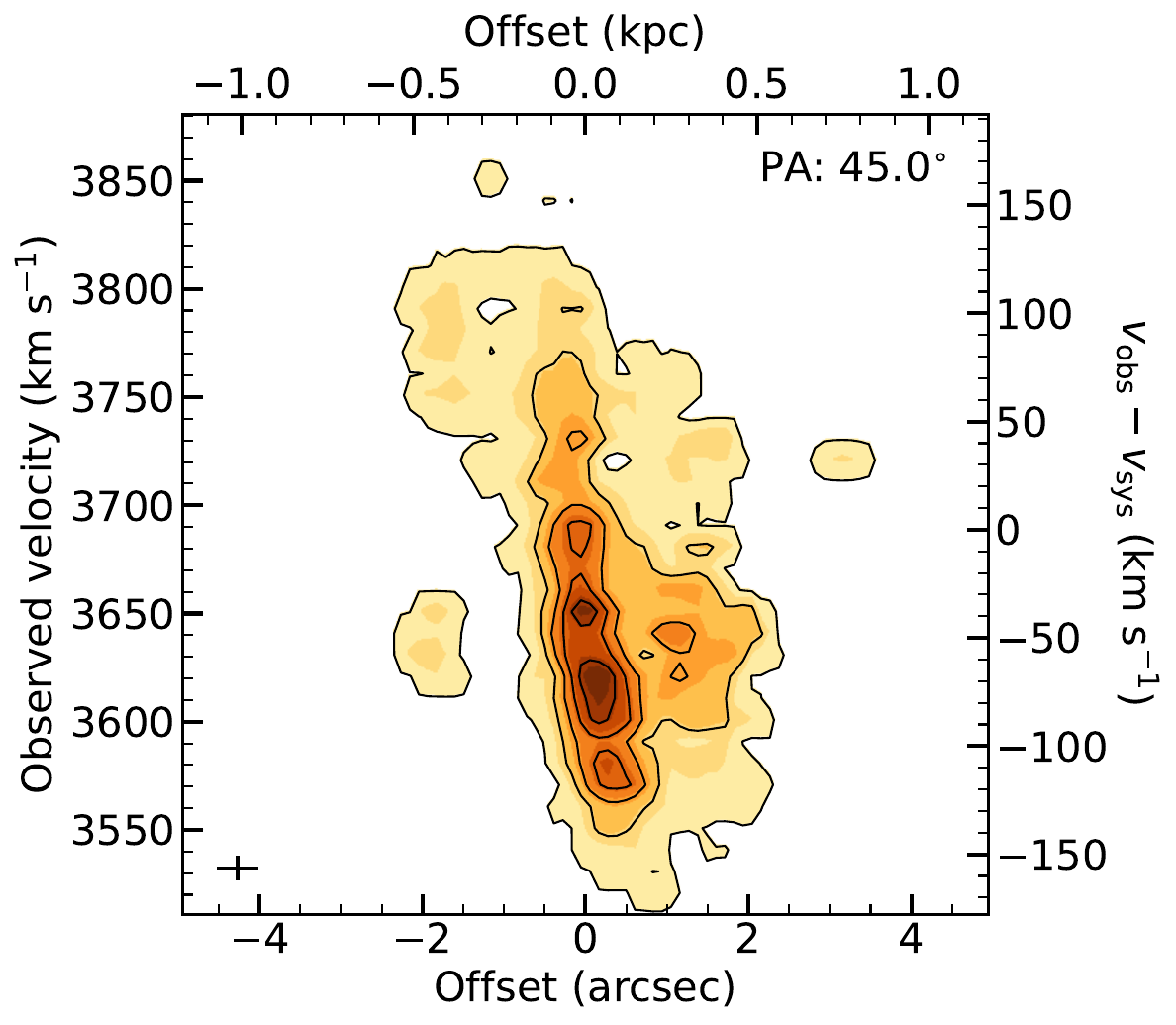}
    \end{subfigure}
    \begin{subfigure}{.34\linewidth}
    \includegraphics[height=6.0cm]{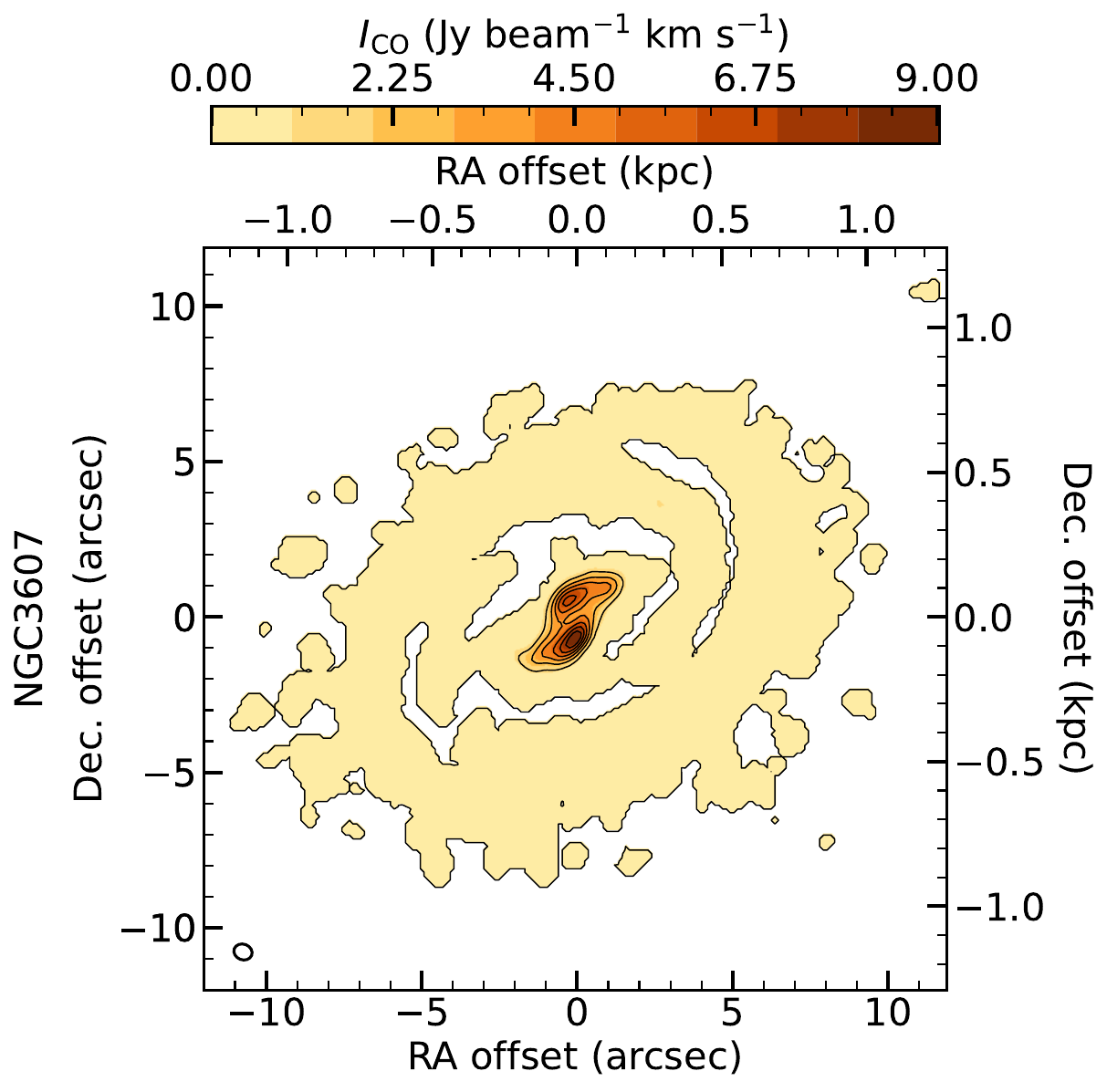}
    \end{subfigure}
    \begin{subfigure}{.32\linewidth}
    \centering
    \includegraphics[height=6.0cm]{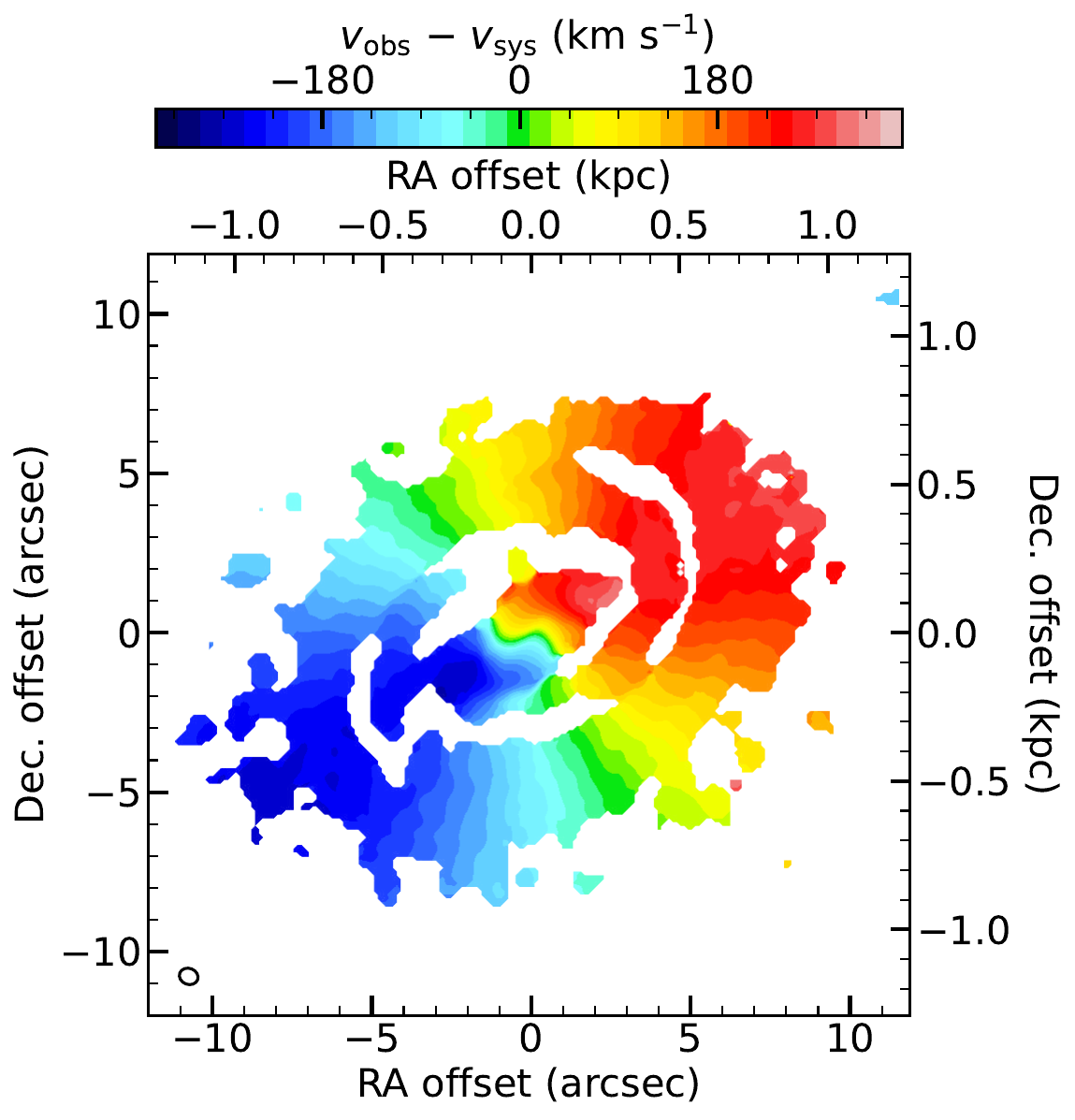}
    \end{subfigure}
    \begin{subfigure}{.32\linewidth}
    \centering
    \includegraphics[height=6.0cm]{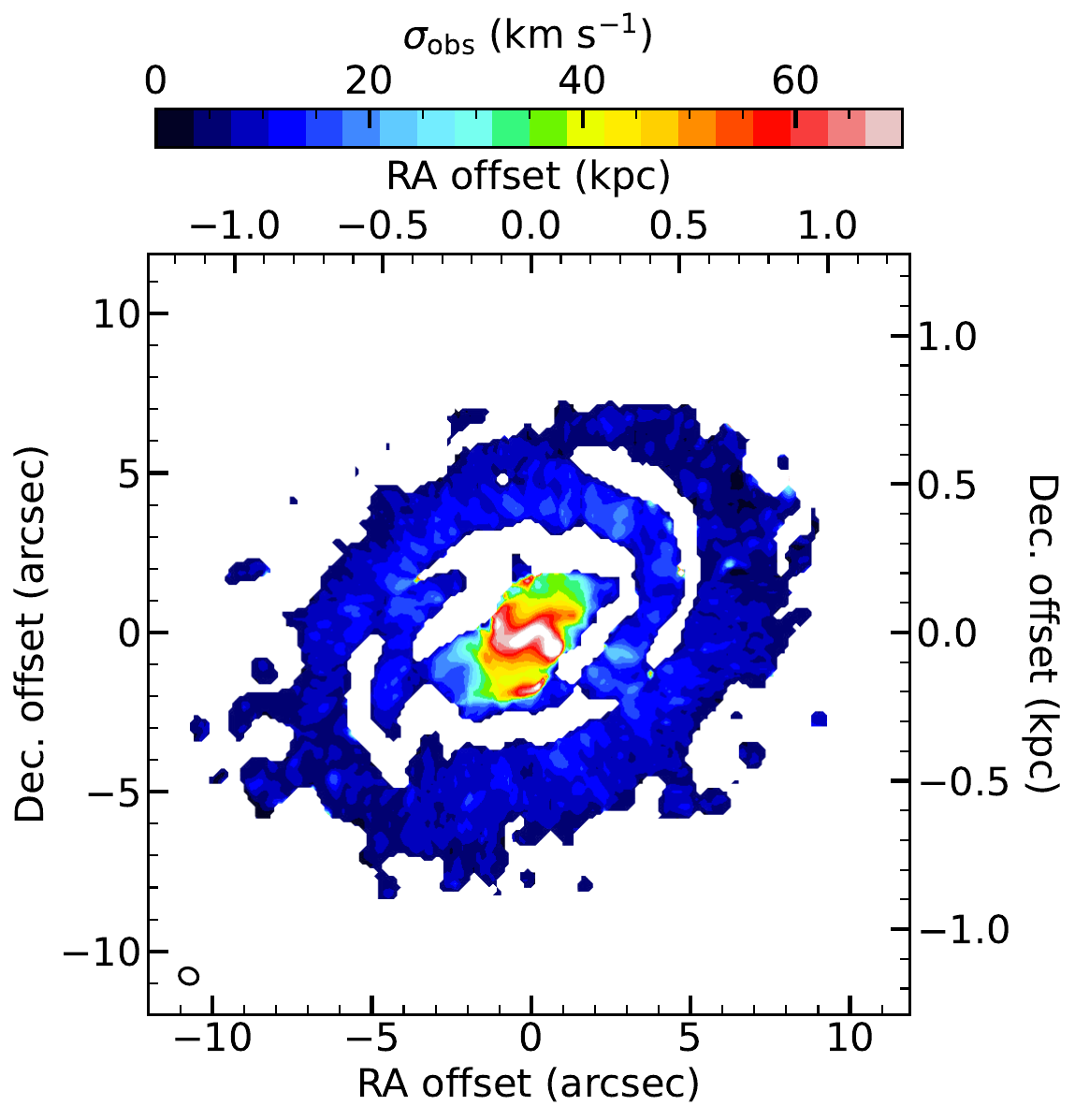}
    \end{subfigure}
    \\
    \begin{subfigure}{.34\linewidth}
    \centering
    \includegraphics[height=5.2cm]{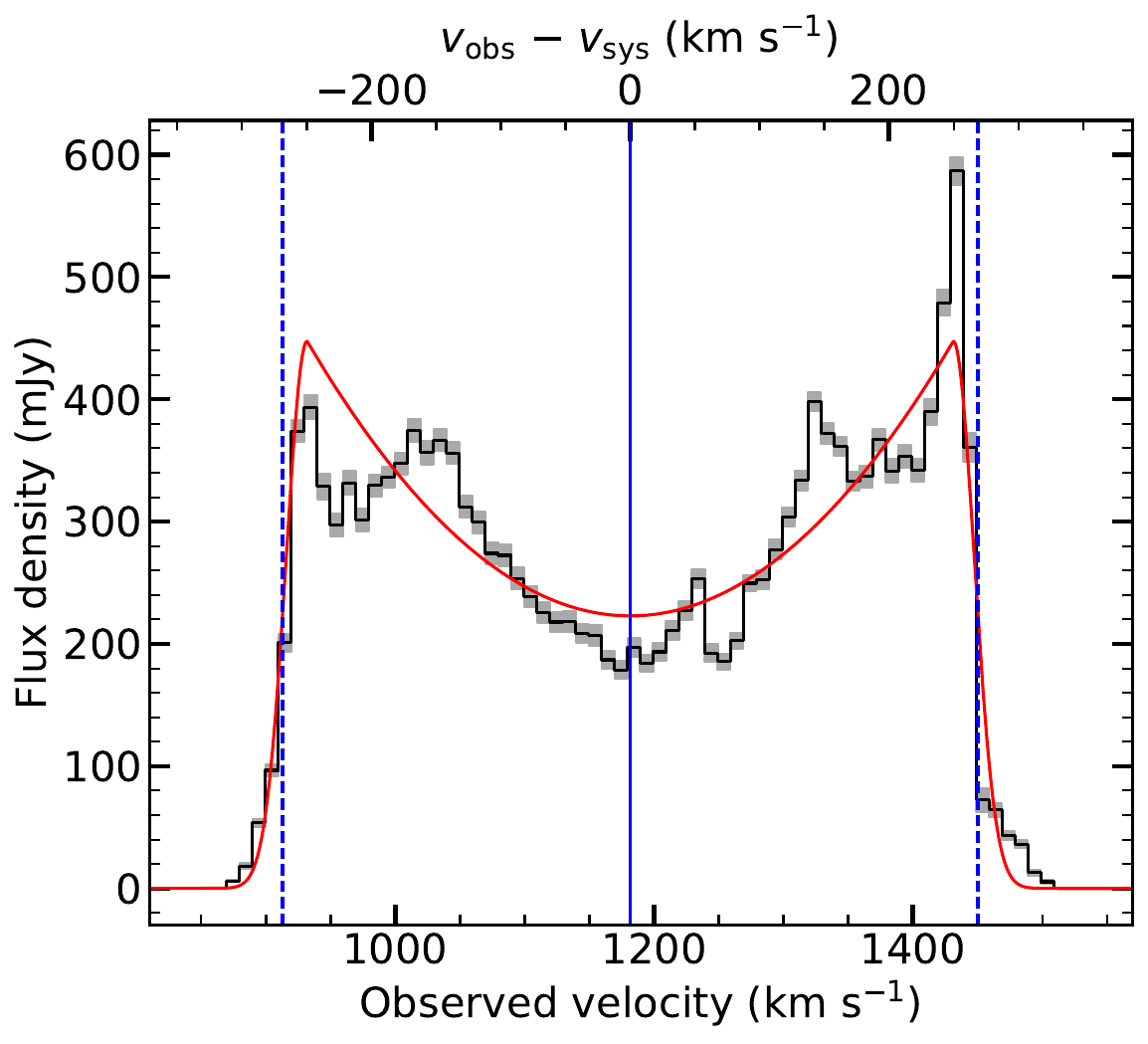}
    \end{subfigure}
    \begin{subfigure}{.32\linewidth}
    \centering
    \includegraphics[height=4.8cm]{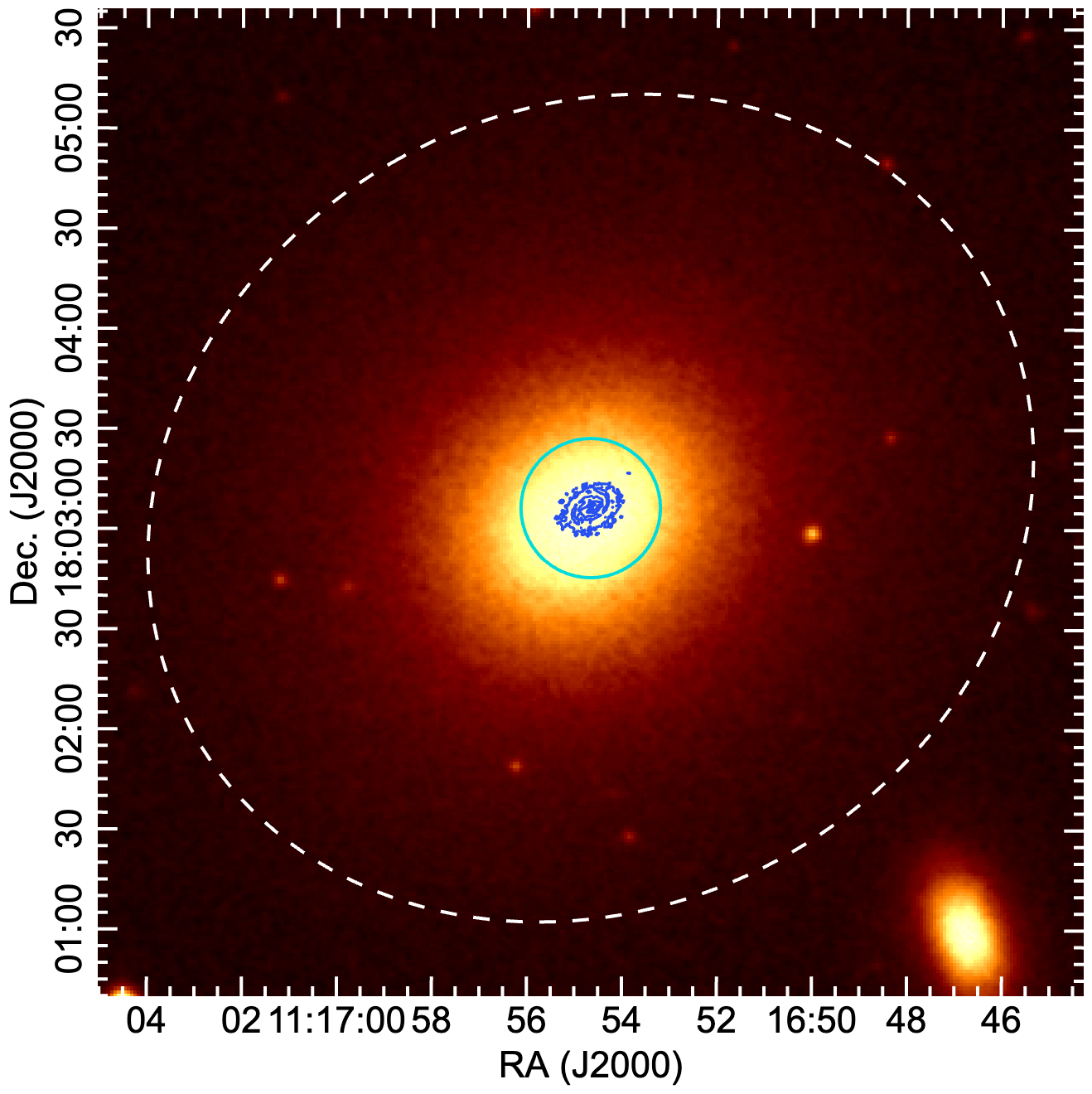}
    \end{subfigure}
    \begin{subfigure}{.33\linewidth}
    \centering
    \includegraphics[height=5.2cm]{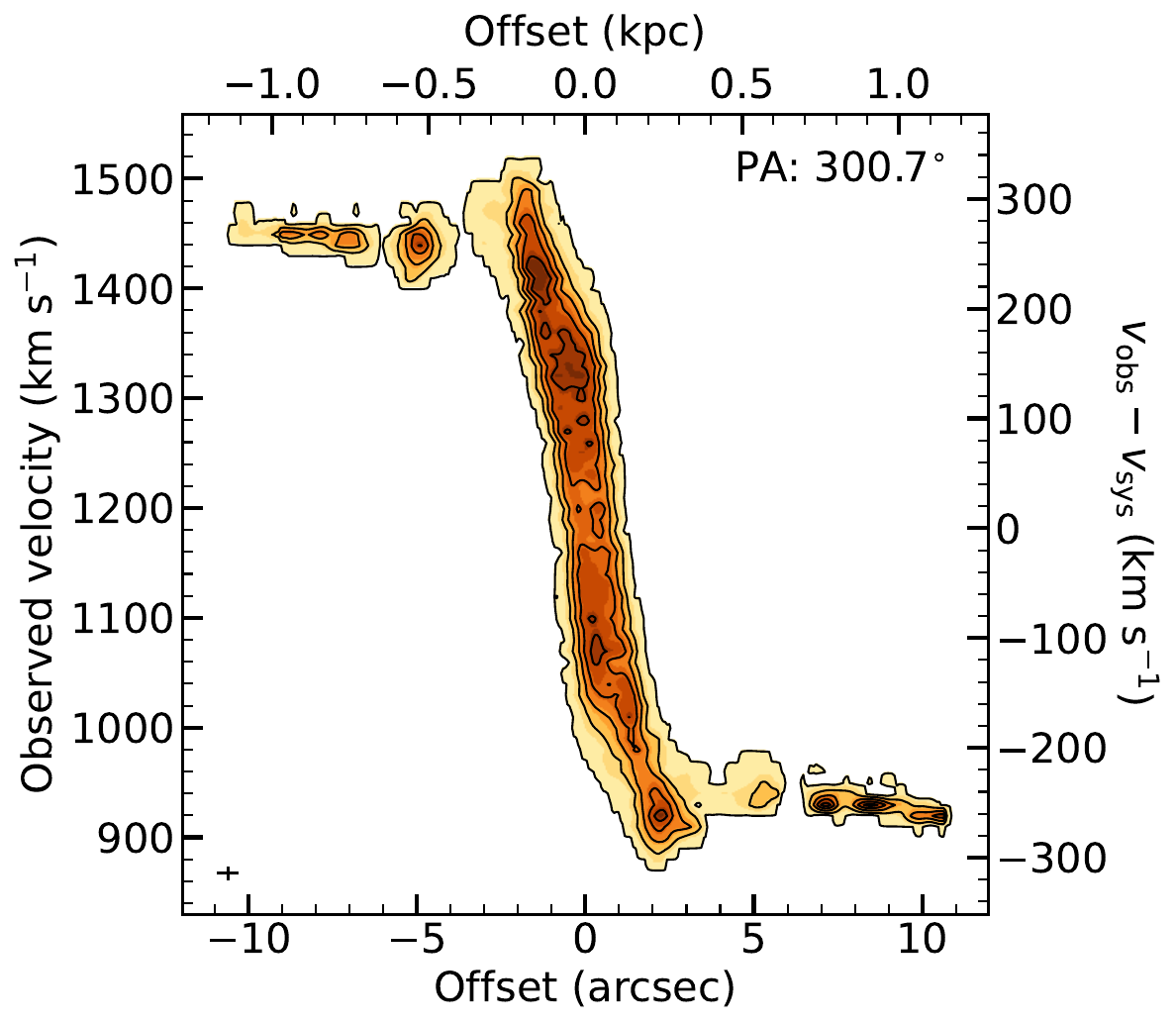}
    \end{subfigure}
    \caption[]{continued.}
\end{figure*}

\begin{figure*}
    \ContinuedFloat
    \begin{subfigure}{.34\linewidth}
    \centering
    \includegraphics[height=5.9cm]{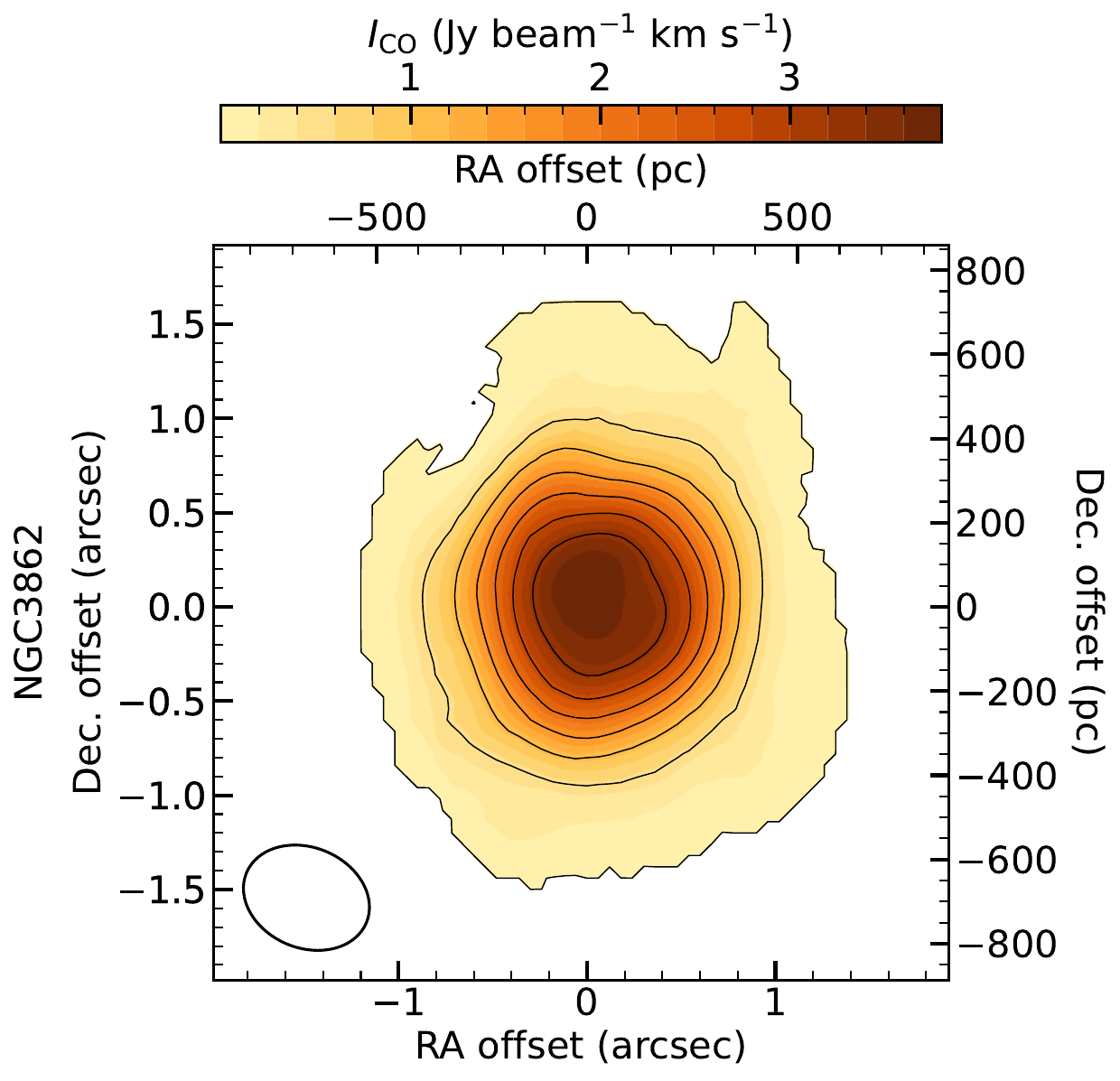}
    \end{subfigure}
    \begin{subfigure}{.32\linewidth}
    \centering
    \includegraphics[height=5.9cm]{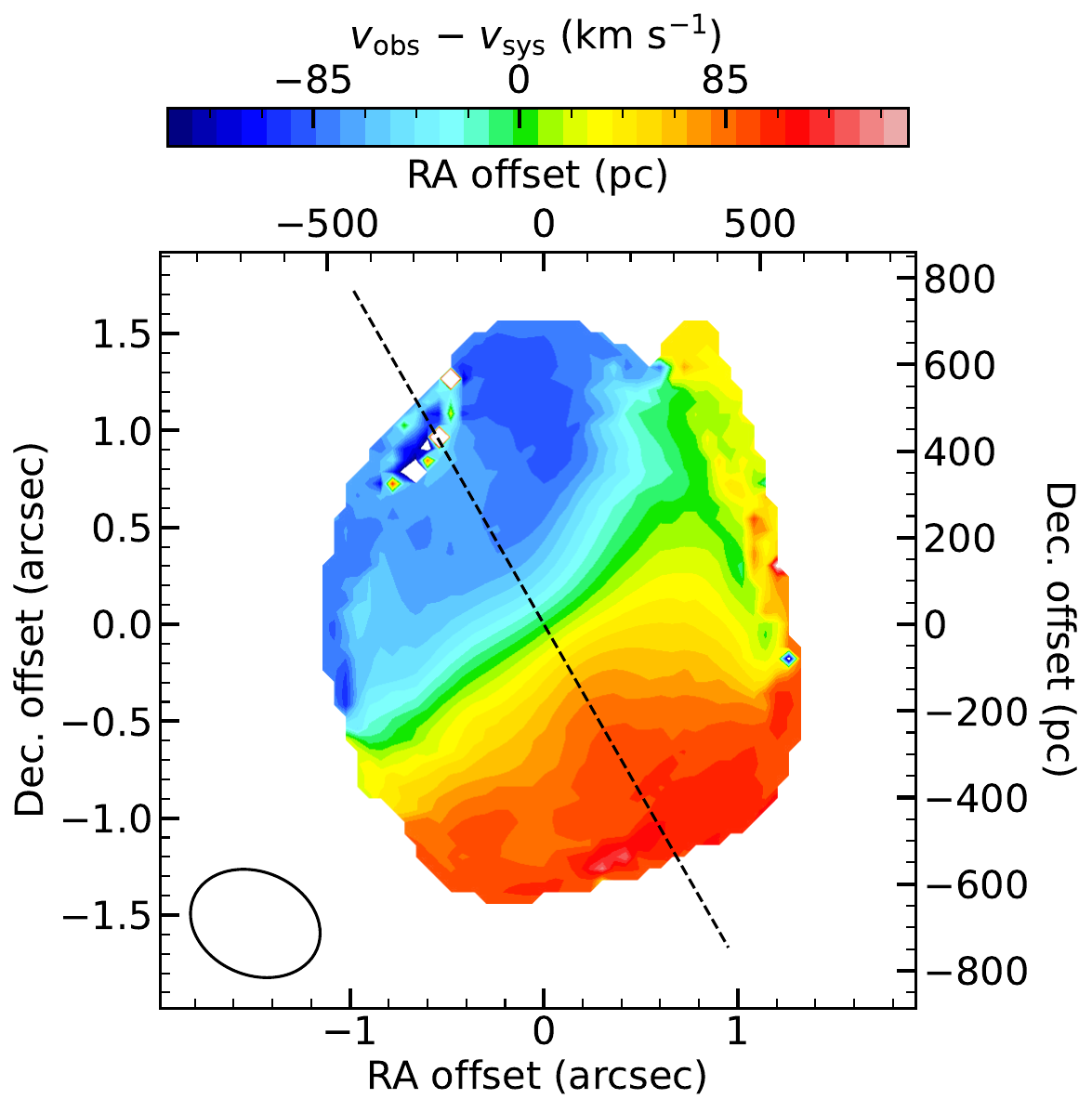}
    \end{subfigure}
    \begin{subfigure}{.32\linewidth}
    \centering
    \includegraphics[height=5.9cm]{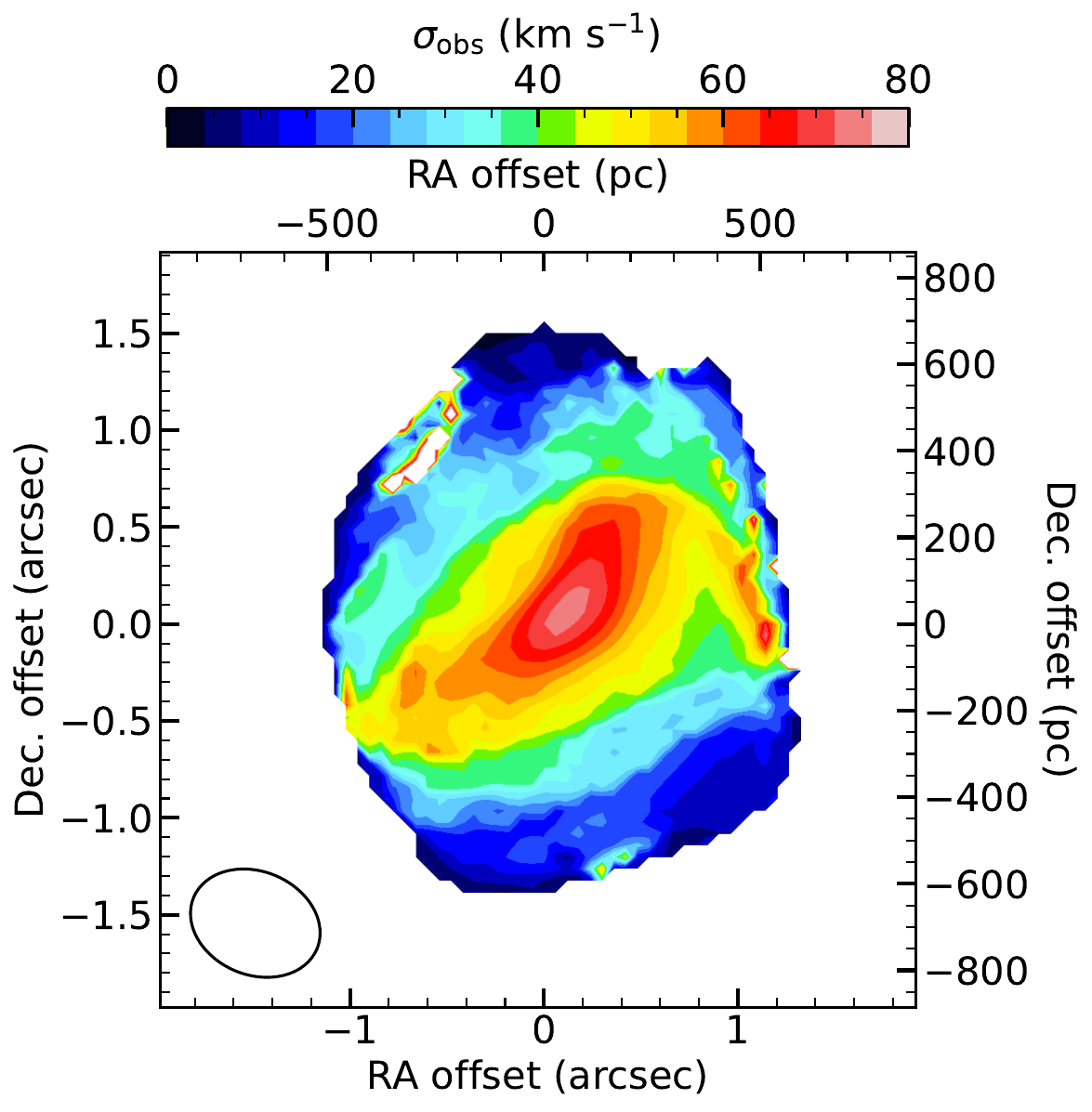}
    \end{subfigure}
    \\
    \begin{subfigure}{.34\linewidth}
    \centering
    \includegraphics[height=5.2cm]{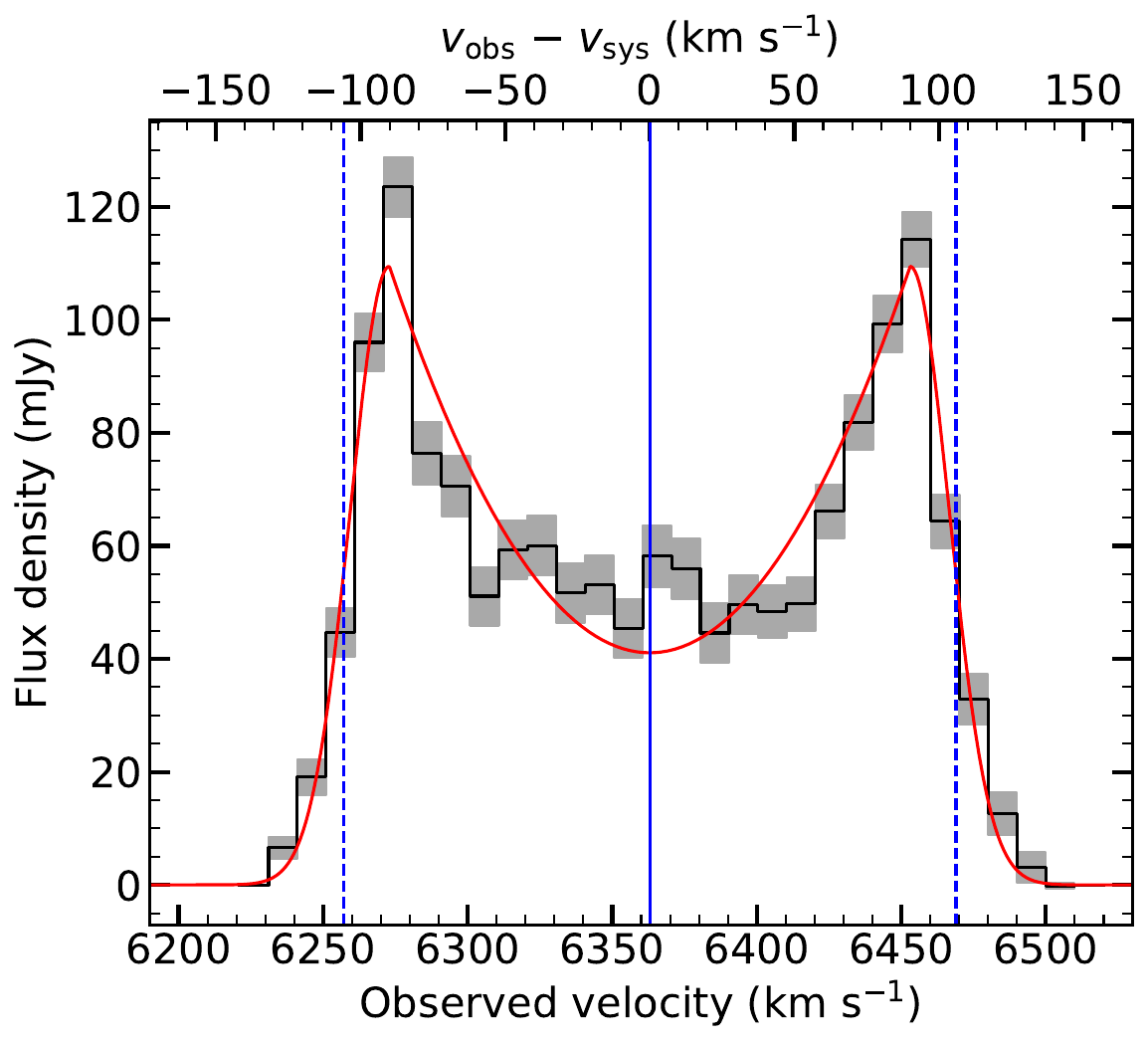}
    \end{subfigure}
    \begin{subfigure}{.32\linewidth}
    \centering
    \includegraphics[height=4.8cm]{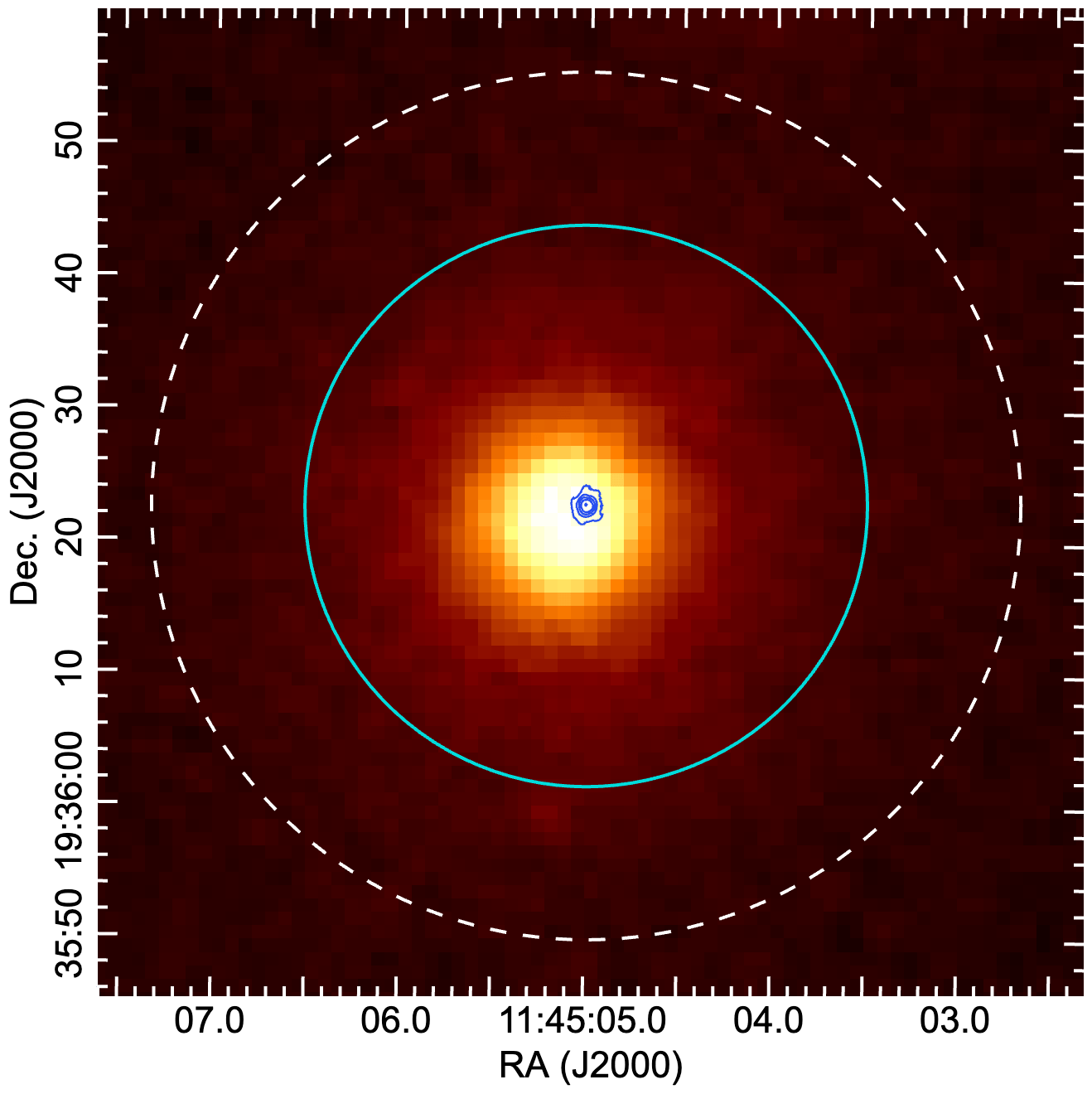}
    \end{subfigure}
    \begin{subfigure}{.33\linewidth}
    \centering
    \includegraphics[height=5.2cm]{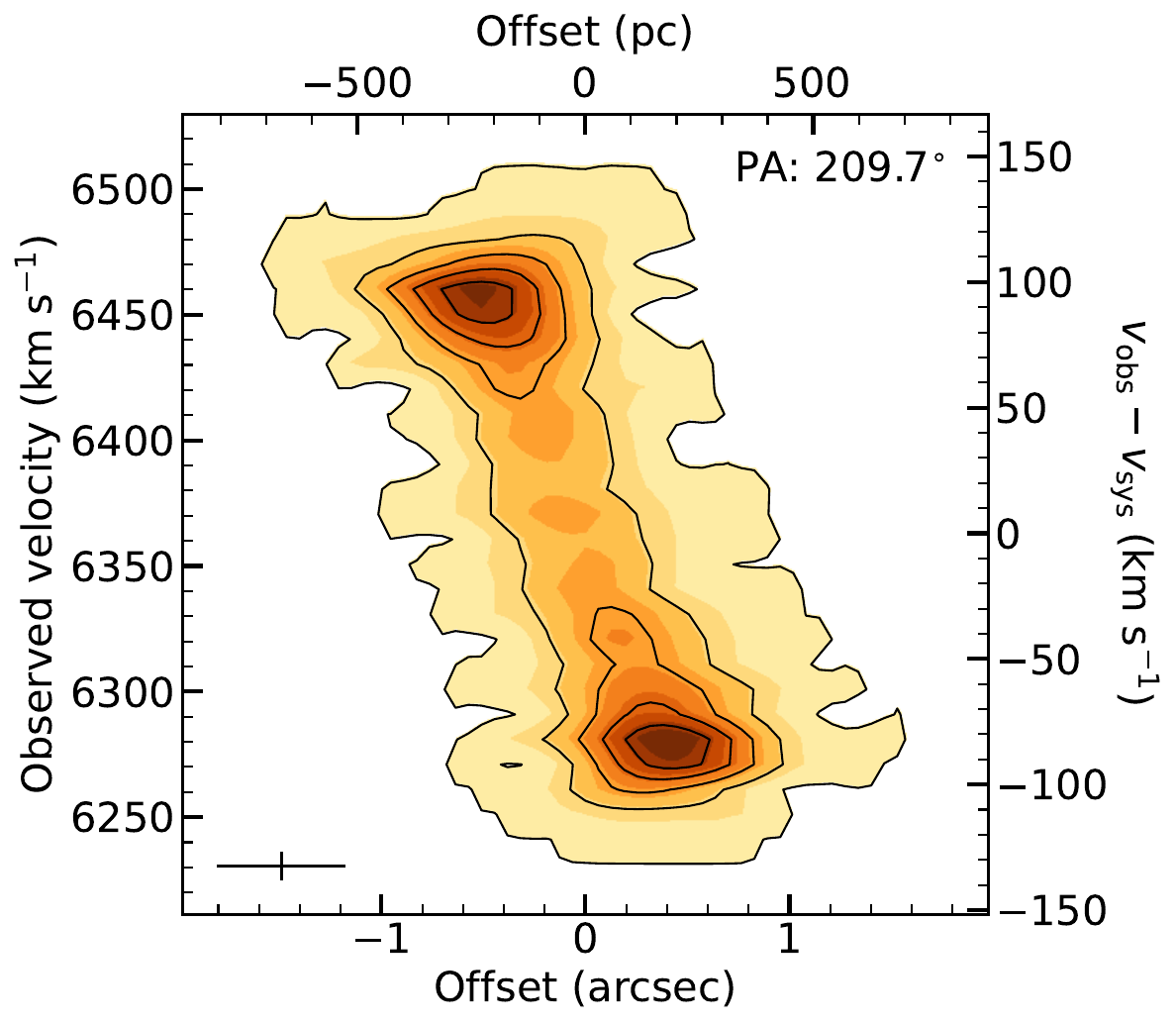}
    \end{subfigure}
    \begin{subfigure}{.34\linewidth}
    \centering
    \includegraphics[height=6.0cm]{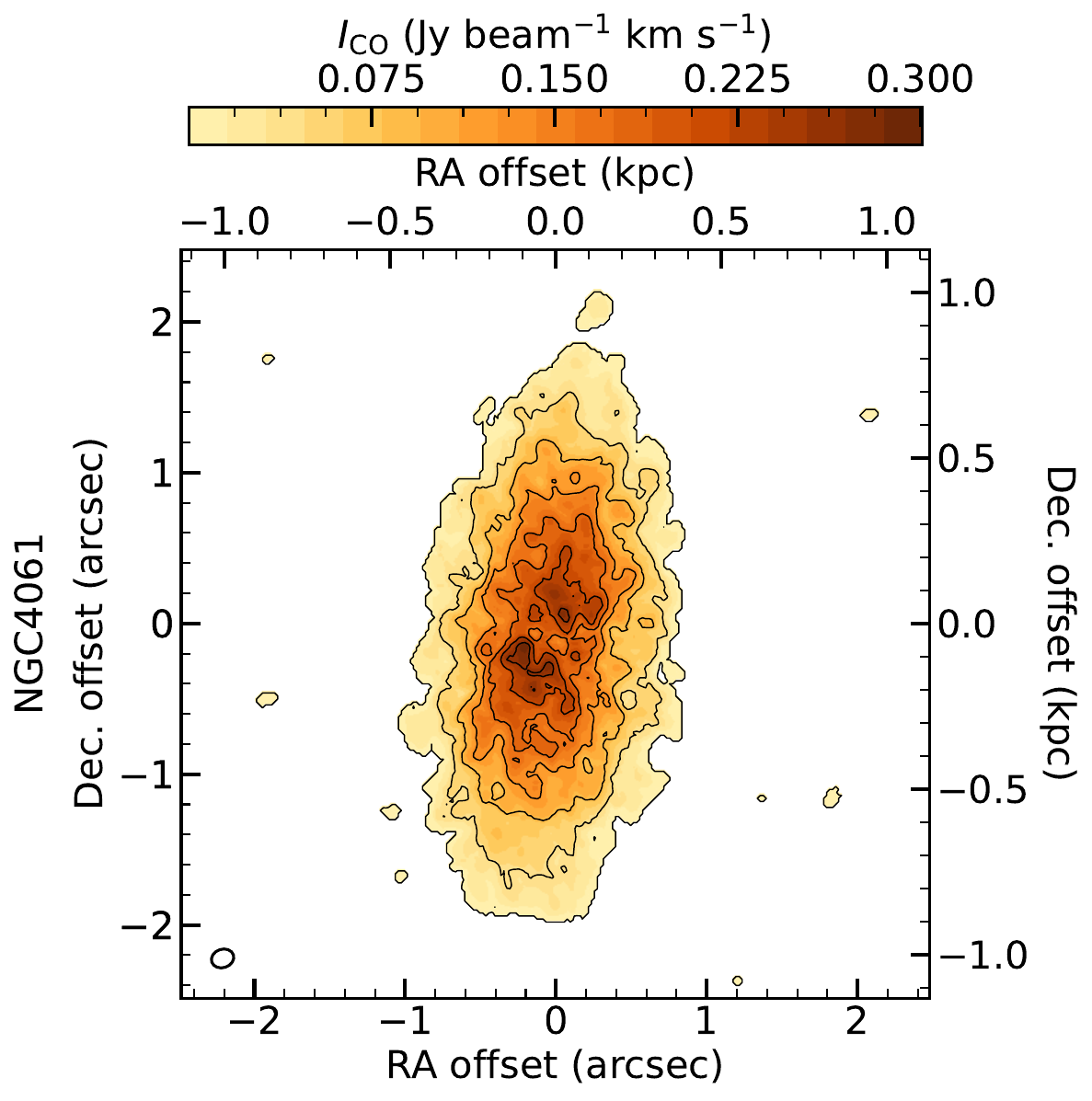}
    \end{subfigure}
    \begin{subfigure}{.32\linewidth}
    \centering
    \includegraphics[height=6.0cm]{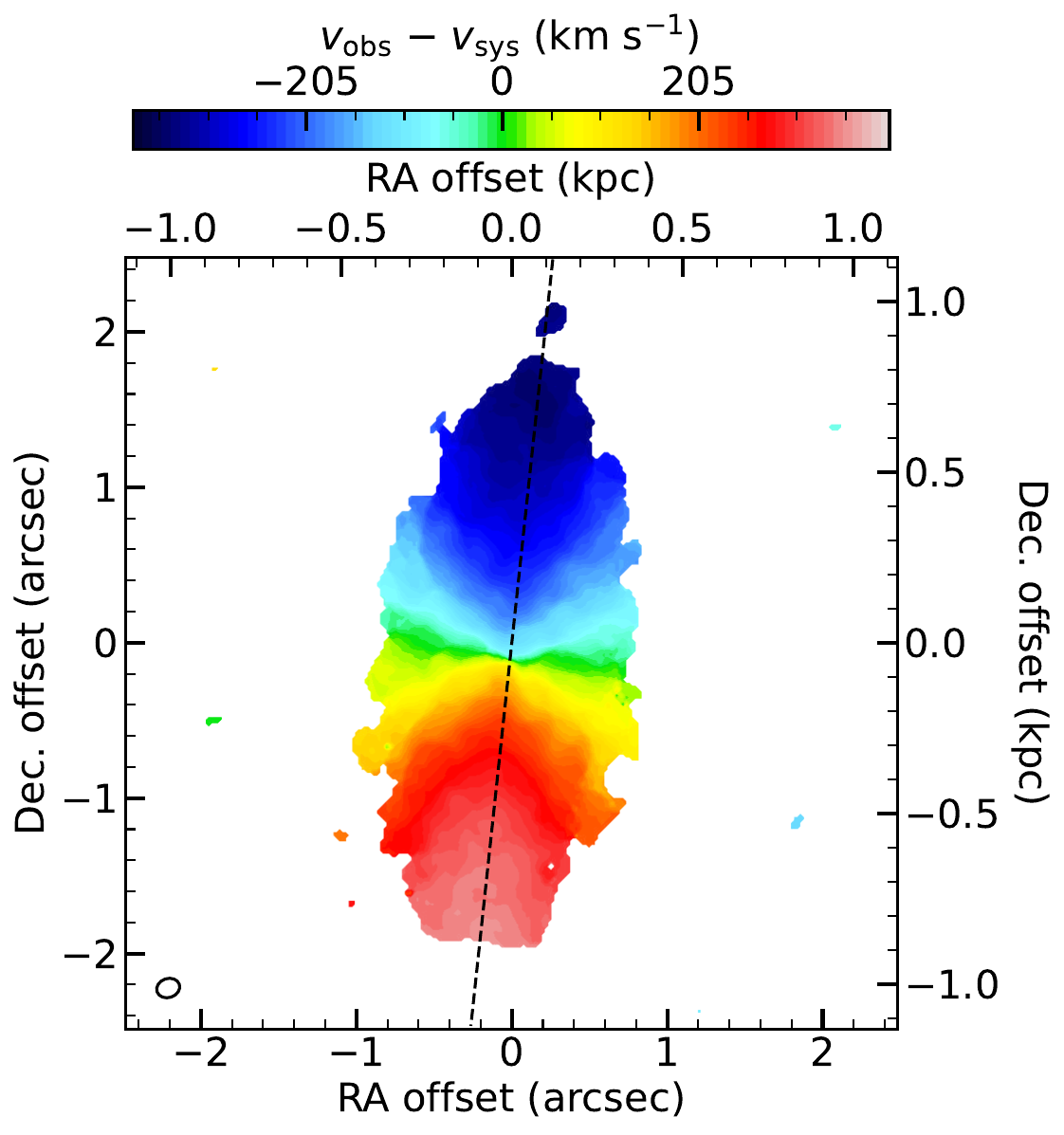}
    \end{subfigure}
    \begin{subfigure}{.32\linewidth}
    \centering
    \includegraphics[height=6.0cm]{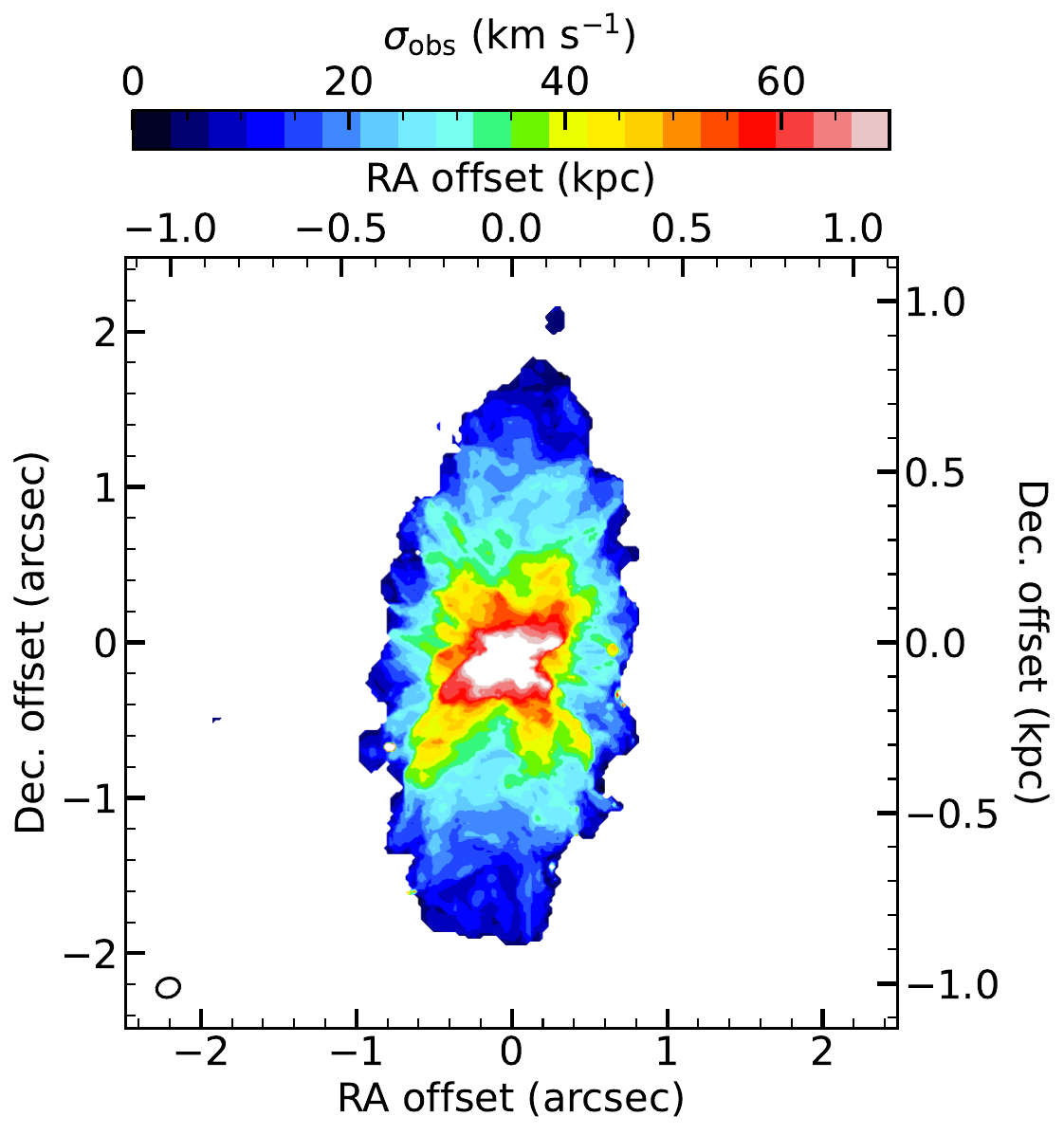}
    \end{subfigure}
    \\
    \begin{subfigure}{.34\linewidth}
    \centering
    \includegraphics[height=5.2cm]{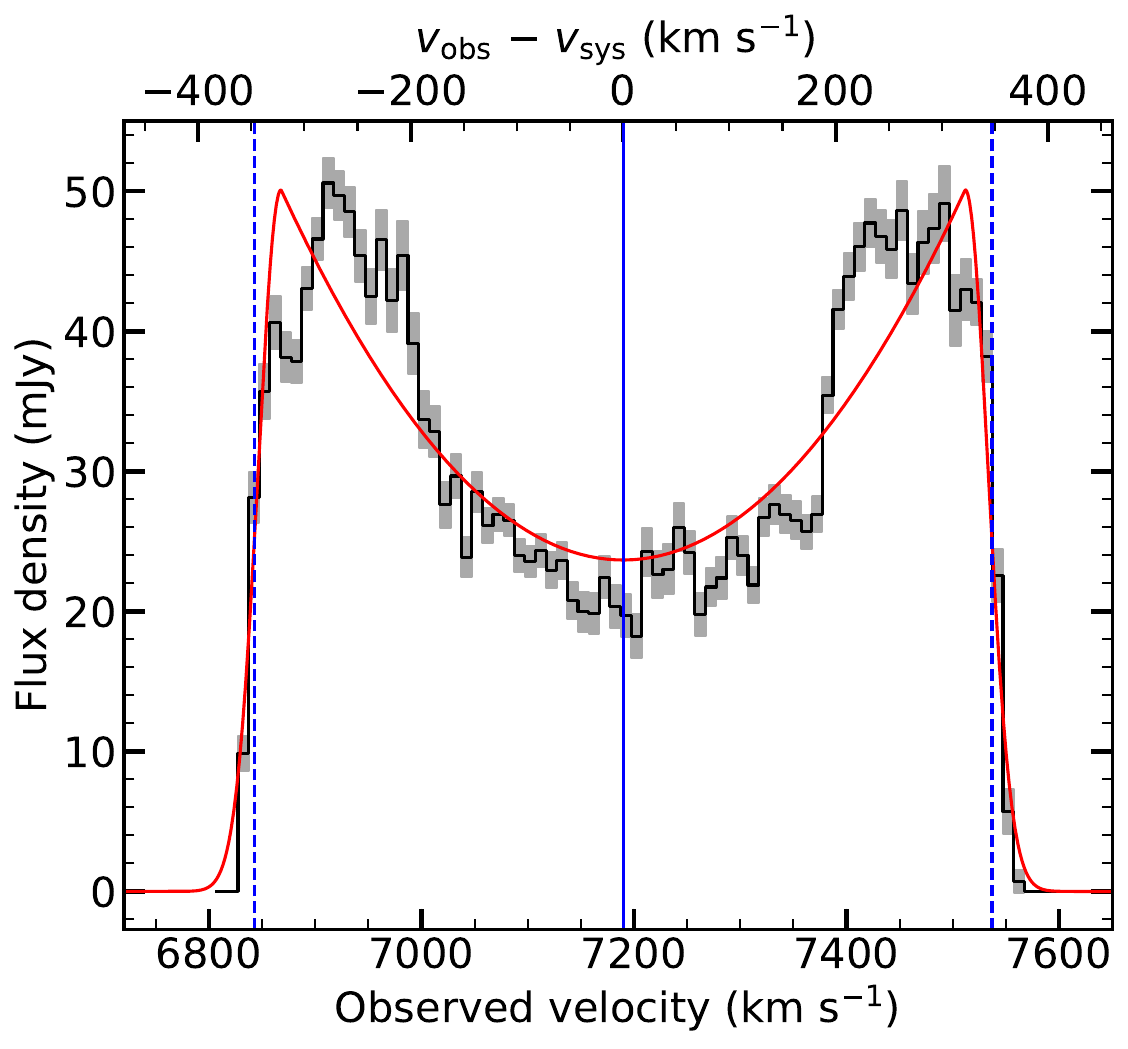}
    \end{subfigure}
    \begin{subfigure}{.32\linewidth}
    \centering
    \includegraphics[height=4.8cm]{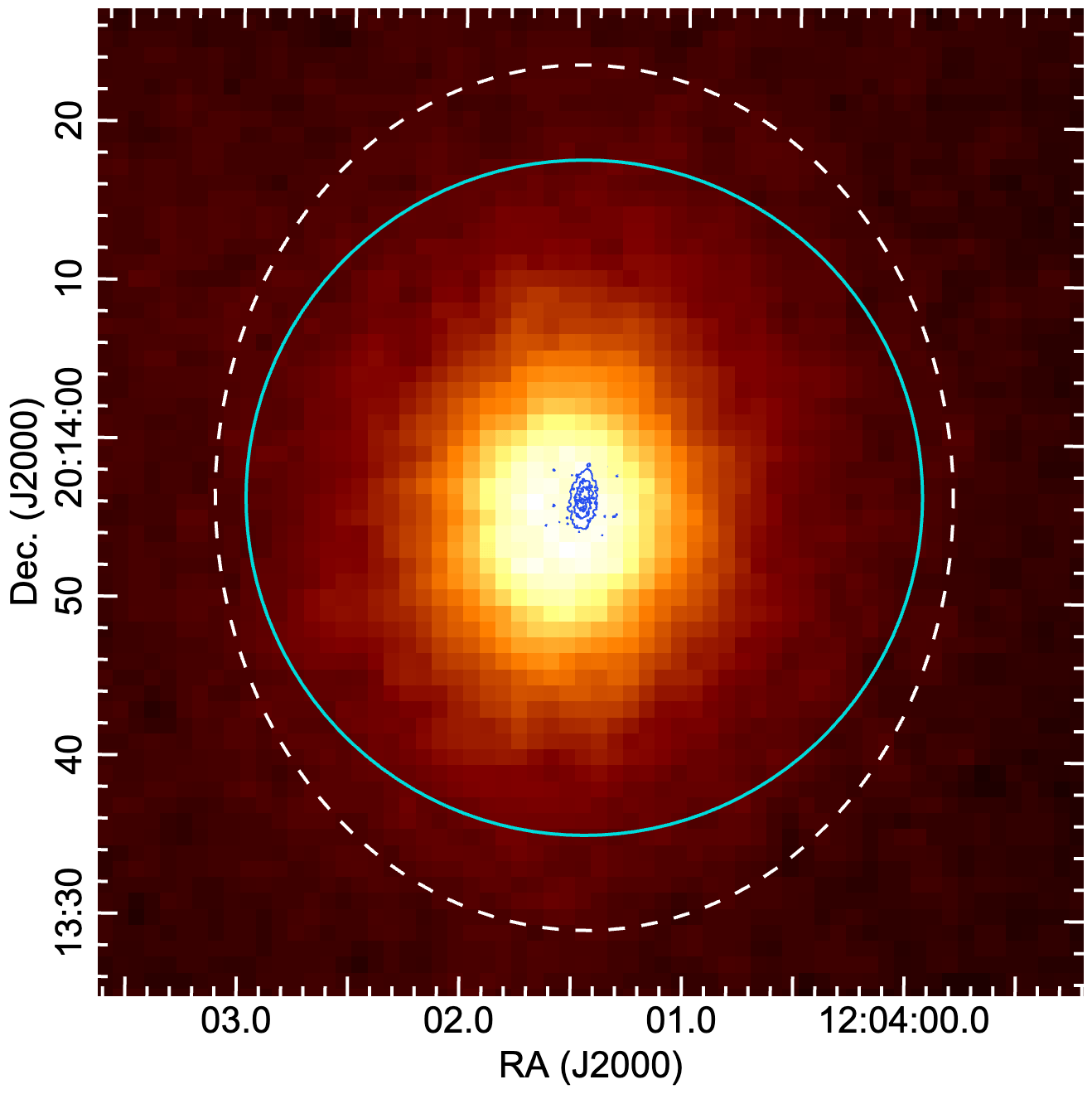}
    \end{subfigure}
    \begin{subfigure}{.33\linewidth}
    \centering
    \includegraphics[height=5.2cm]{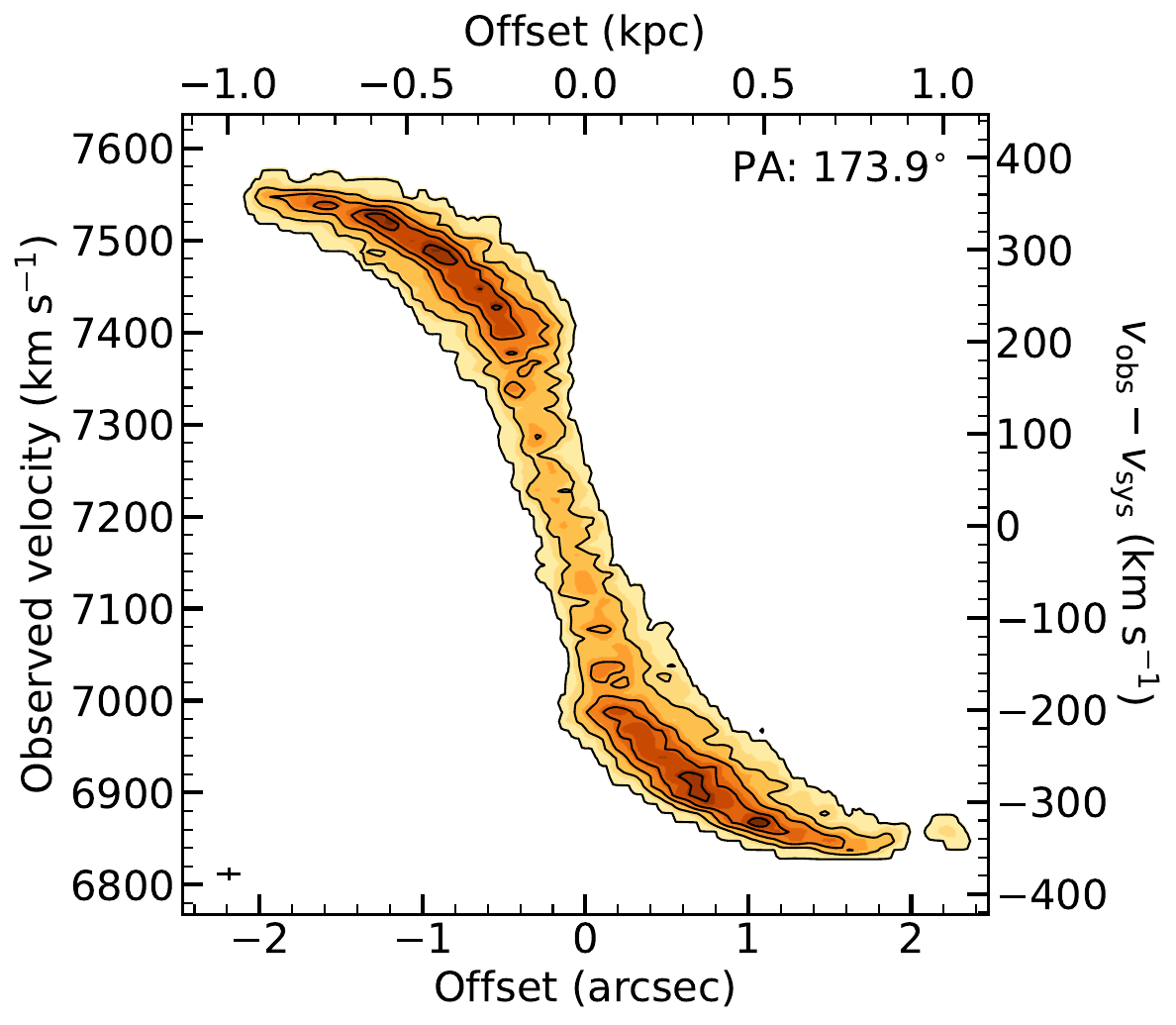}
    \end{subfigure}
    \caption[]{continued.}
\end{figure*}

\begin{figure*}
    \ContinuedFloat
    \begin{subfigure}{.34\linewidth}
    \includegraphics[height=5.85cm]{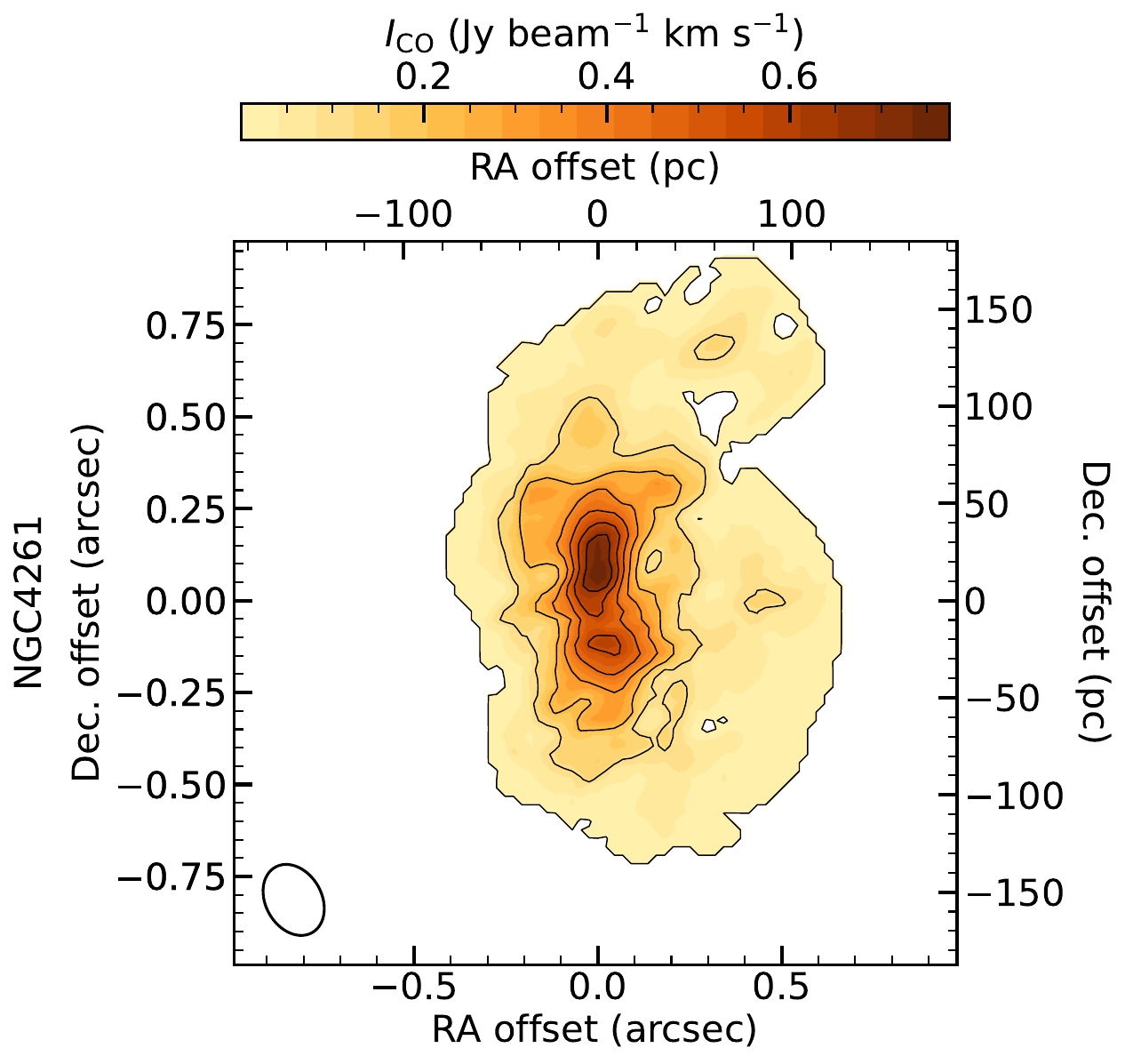}
    \end{subfigure}
    \begin{subfigure}{.32\linewidth}
    \centering
    \includegraphics[height=5.85cm]{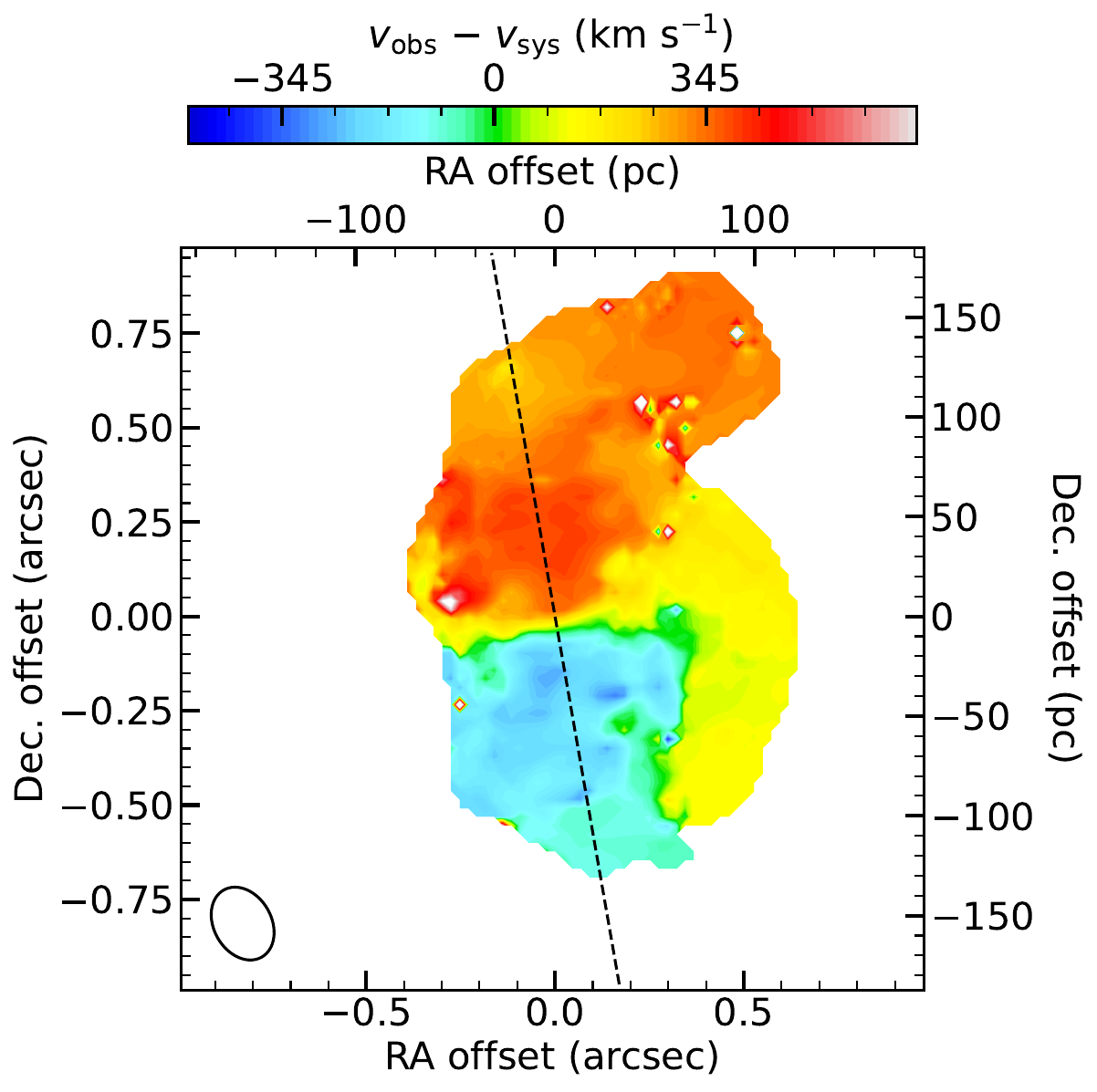}
    \end{subfigure}
    \begin{subfigure}{.32\linewidth}
    \centering
    \includegraphics[height=5.85cm]{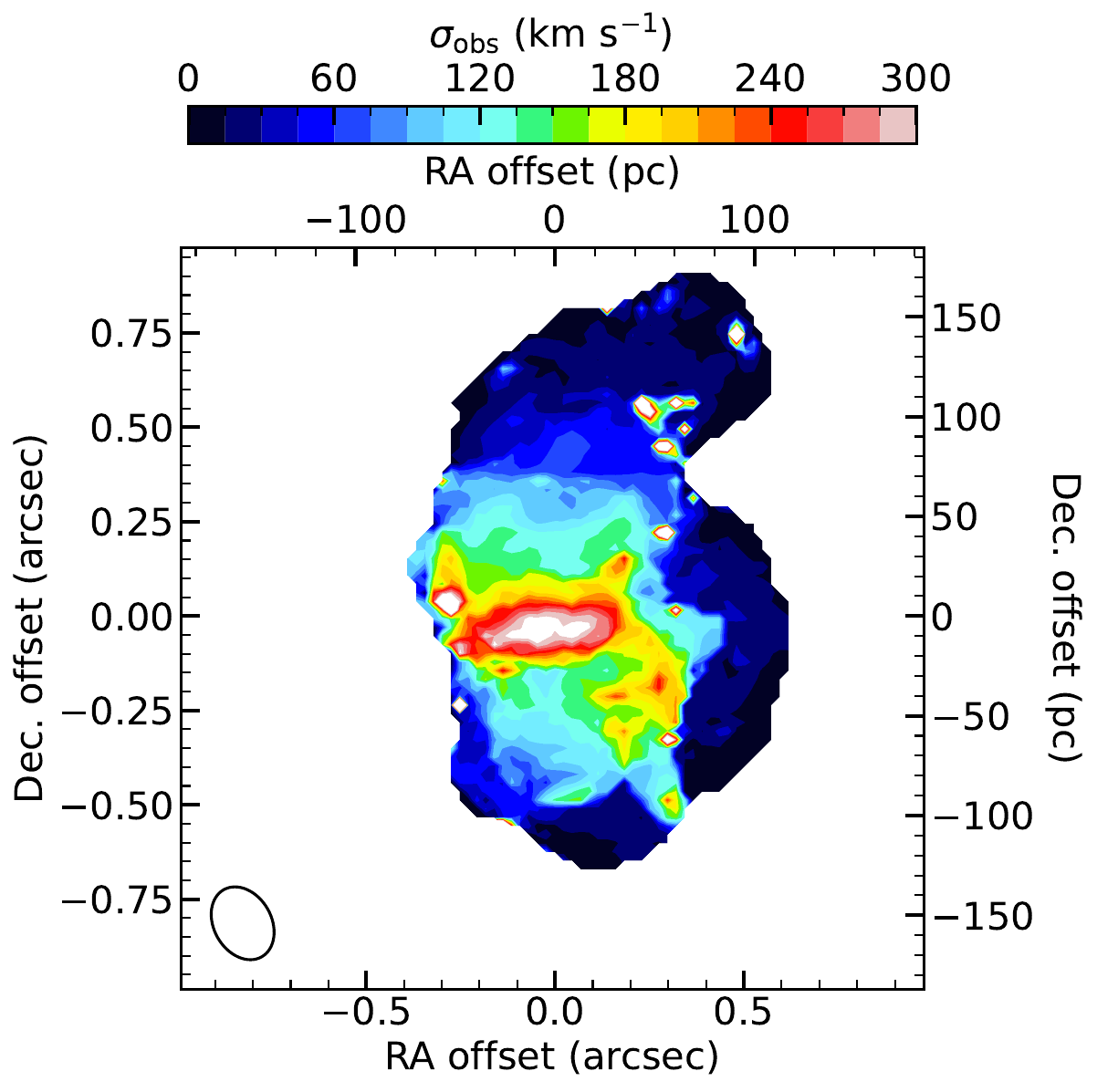}
    \end{subfigure}
    \\
    \begin{subfigure}{.34\linewidth}
    \centering
    \includegraphics[height=5.2cm]{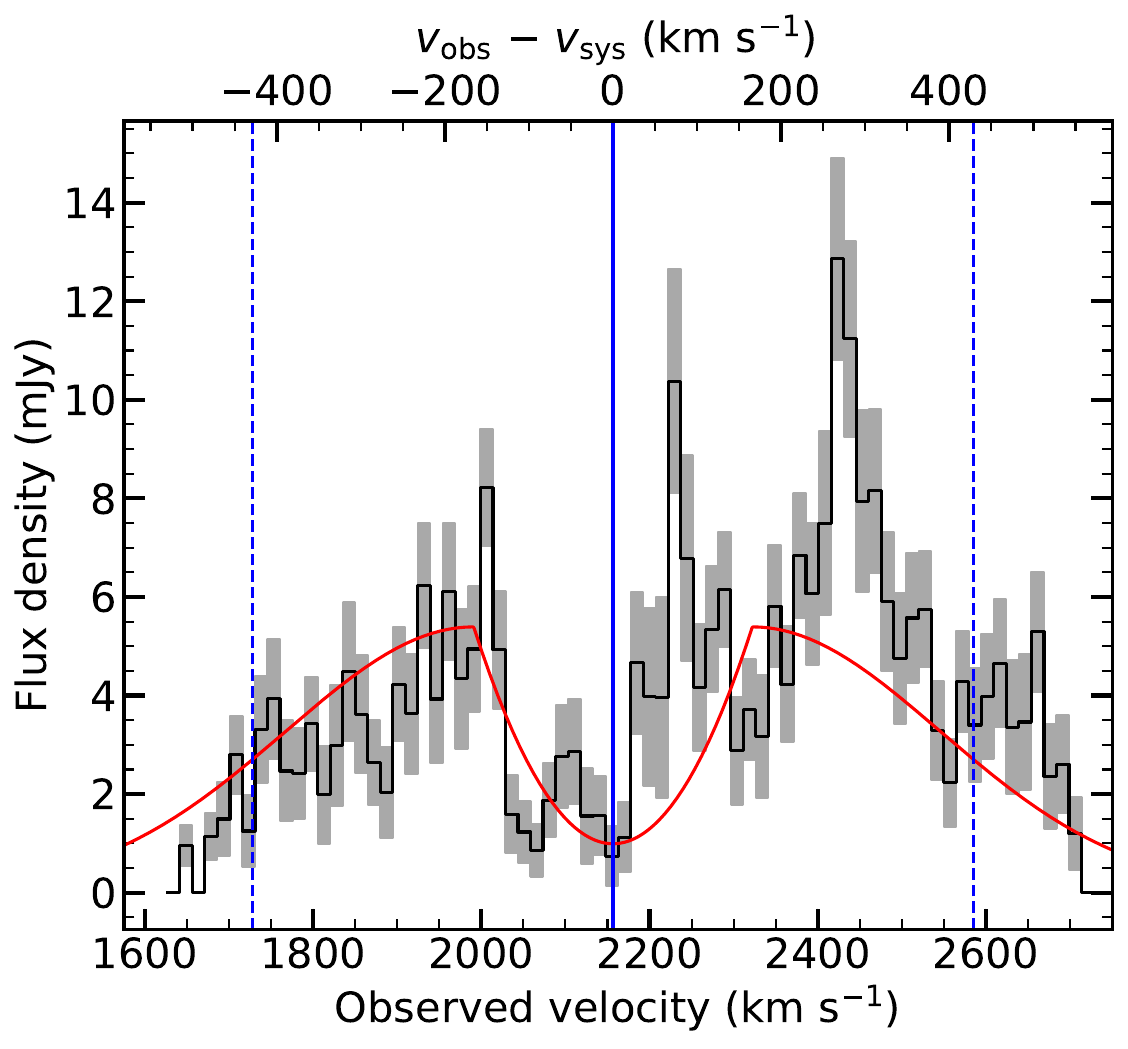}
    \end{subfigure}
    \begin{subfigure}{.32\linewidth}
    \centering
    \includegraphics[height=4.8cm]{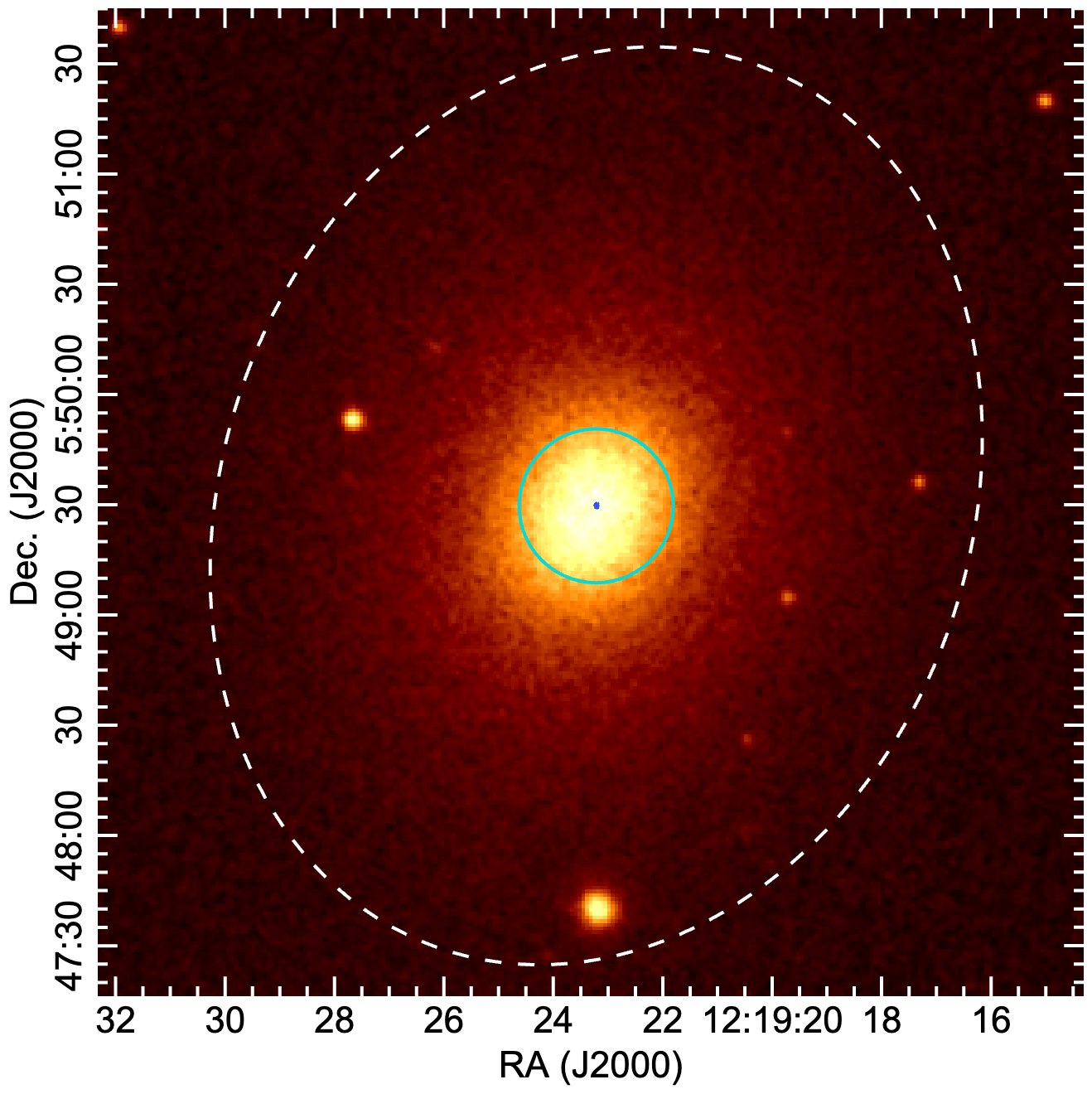}
    \end{subfigure}
    \begin{subfigure}{.33\linewidth}
    \centering
    \includegraphics[height=5.2cm]{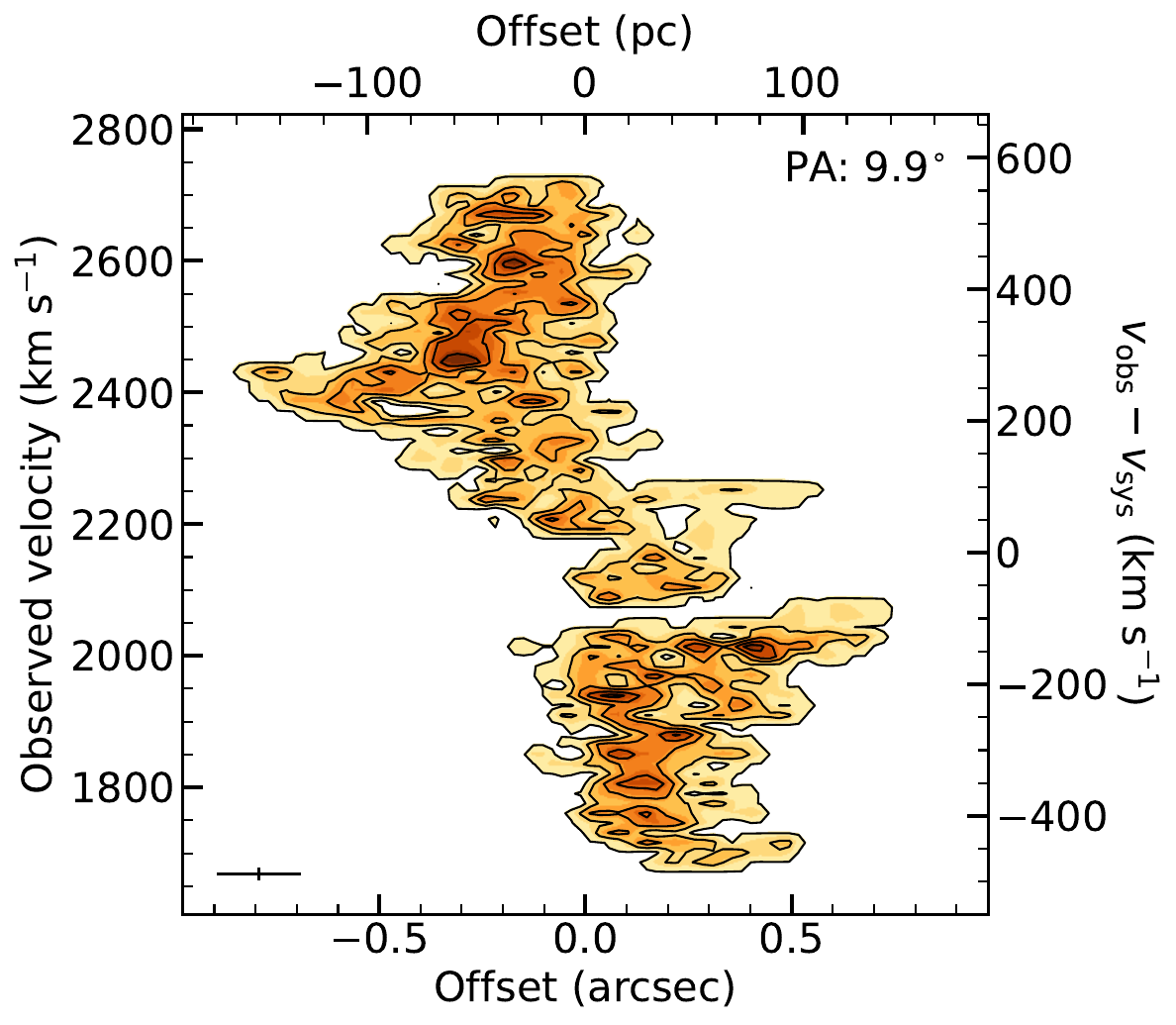}
    \end{subfigure}
    \\
    \begin{subfigure}{.34\linewidth}
    \includegraphics[height=6.0cm]{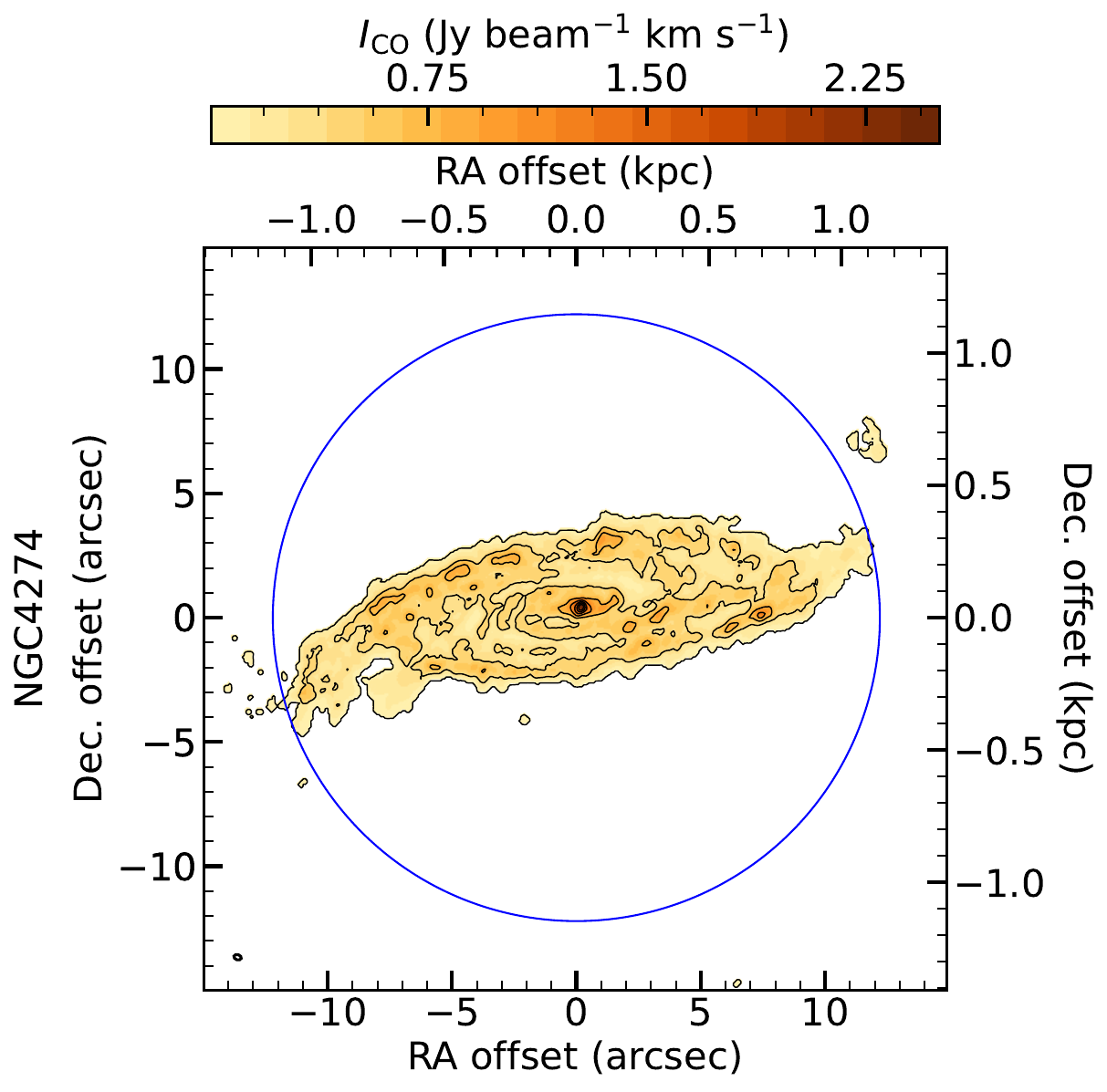}
    \end{subfigure}
    \begin{subfigure}{.32\linewidth}
    \centering
    \includegraphics[height=6.0cm]{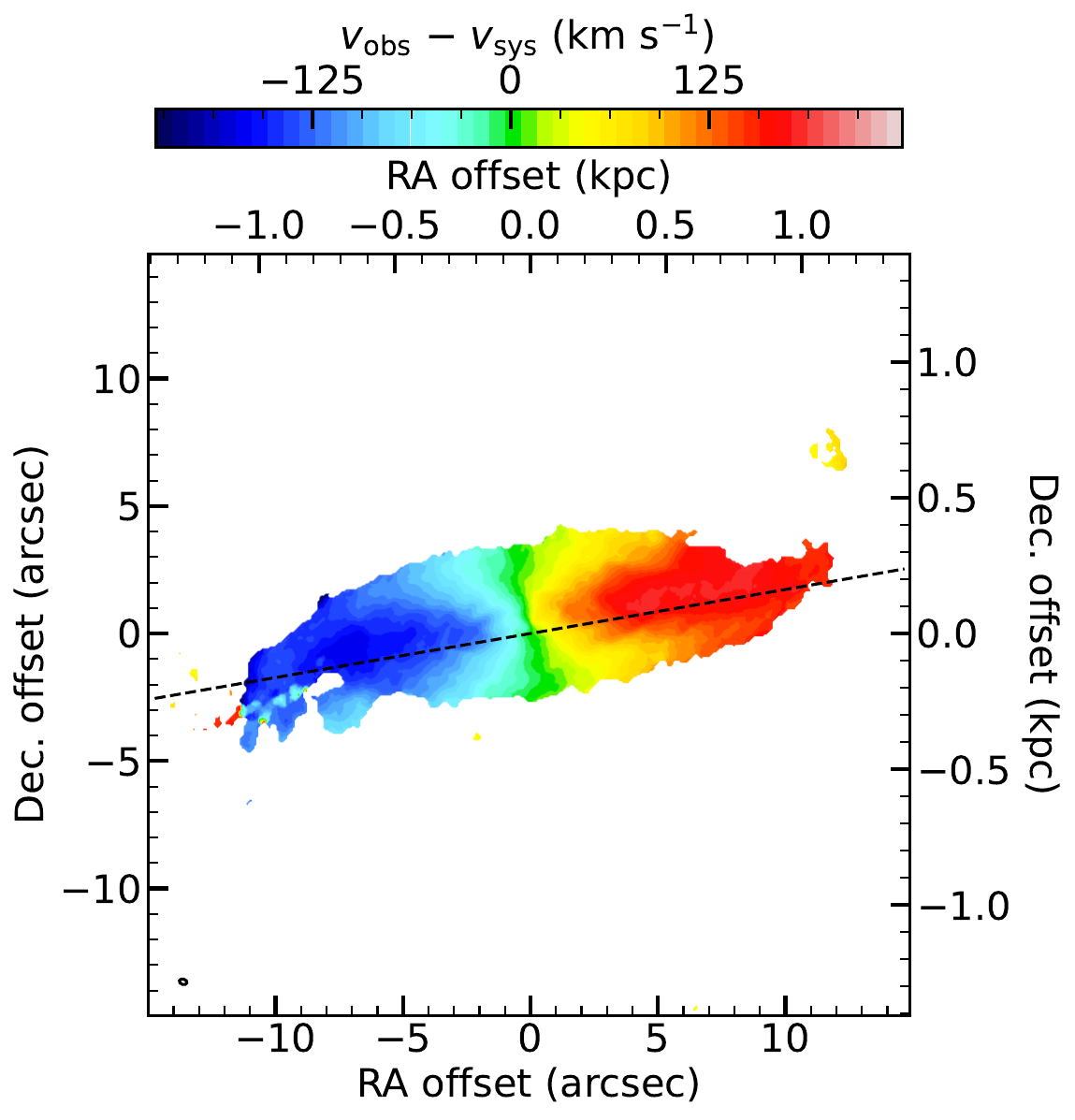}
    \end{subfigure}
    \begin{subfigure}{.32\linewidth}
    \centering
    \includegraphics[height=6.0cm]{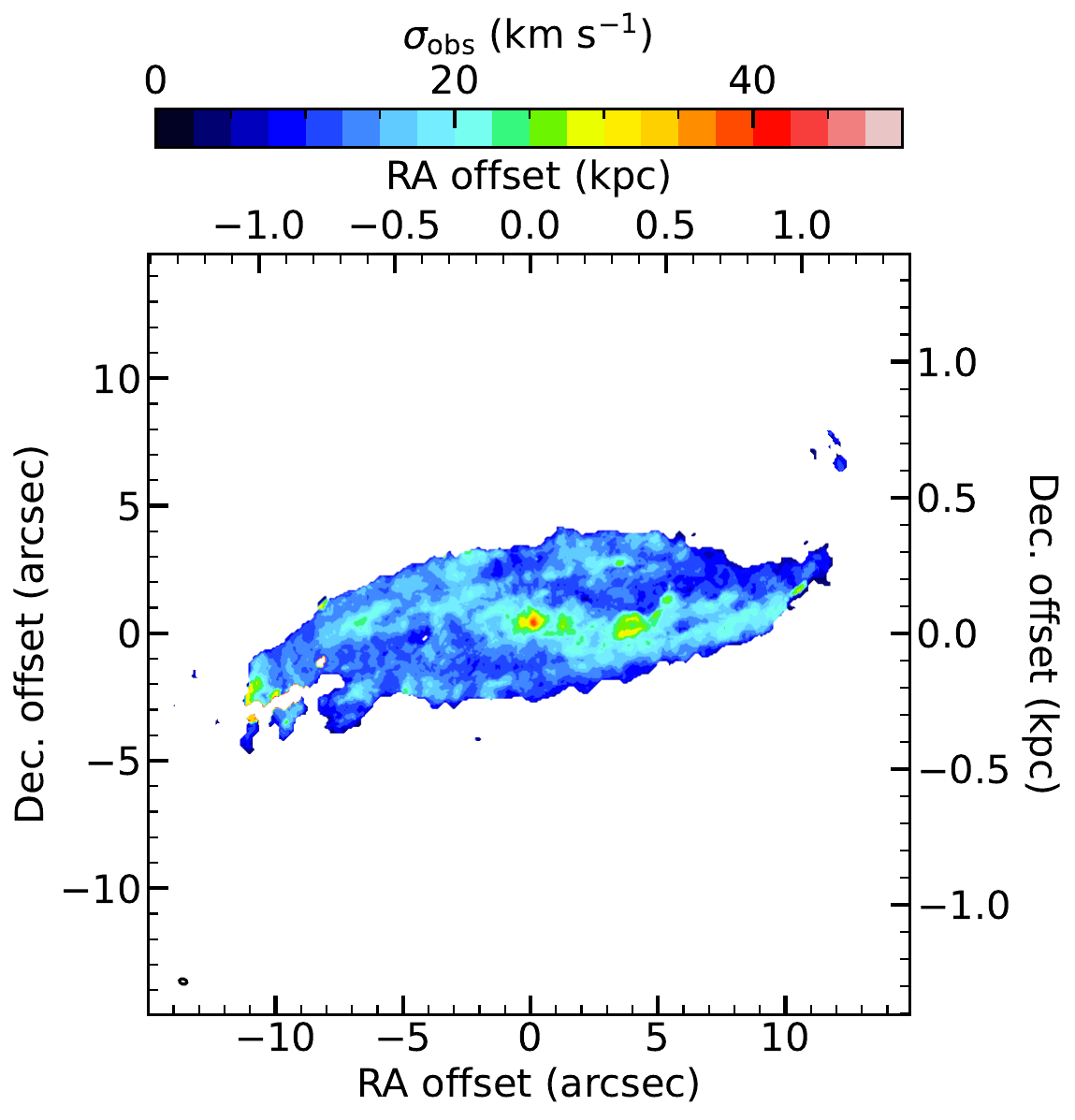}
    \end{subfigure}
    \\
    \begin{subfigure}{.34\linewidth}
    \centering
    \includegraphics[height=5.2cm]{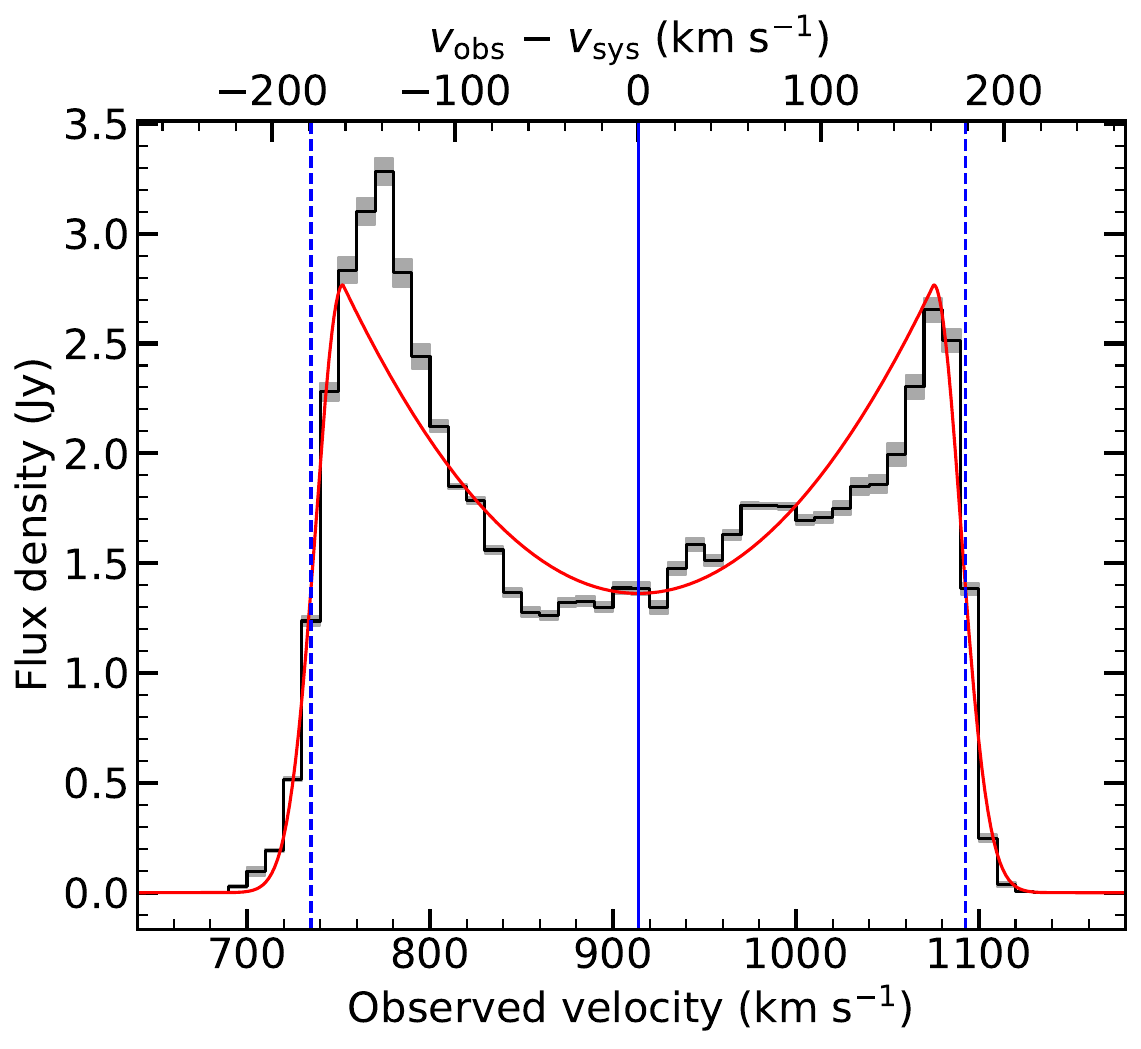}
    \end{subfigure}
    \begin{subfigure}{.32\linewidth}
    \centering
    \includegraphics[height=4.8cm]{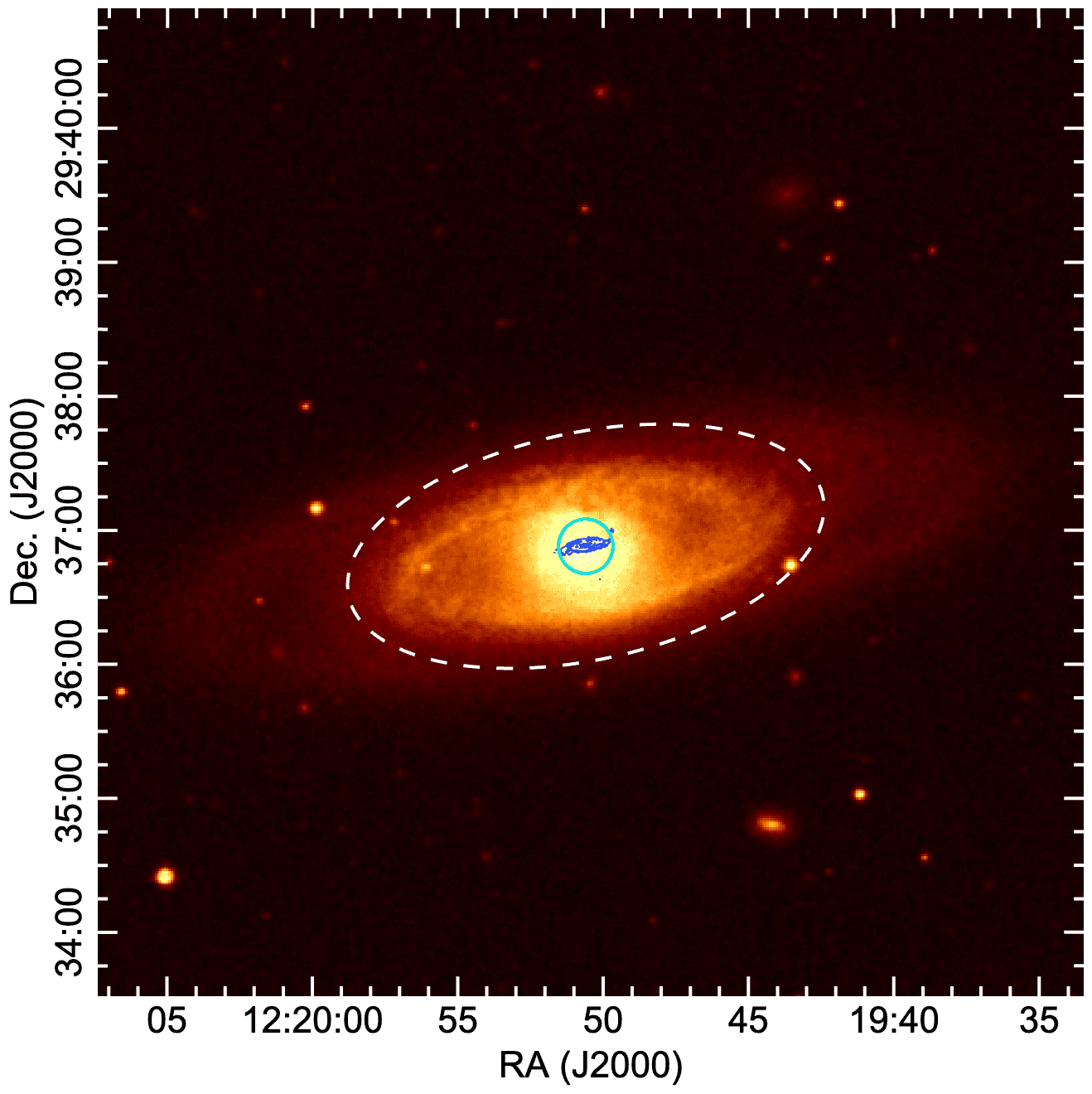}
    \end{subfigure}
    \begin{subfigure}{.33\linewidth}
    \centering
    \includegraphics[height=5.2cm]{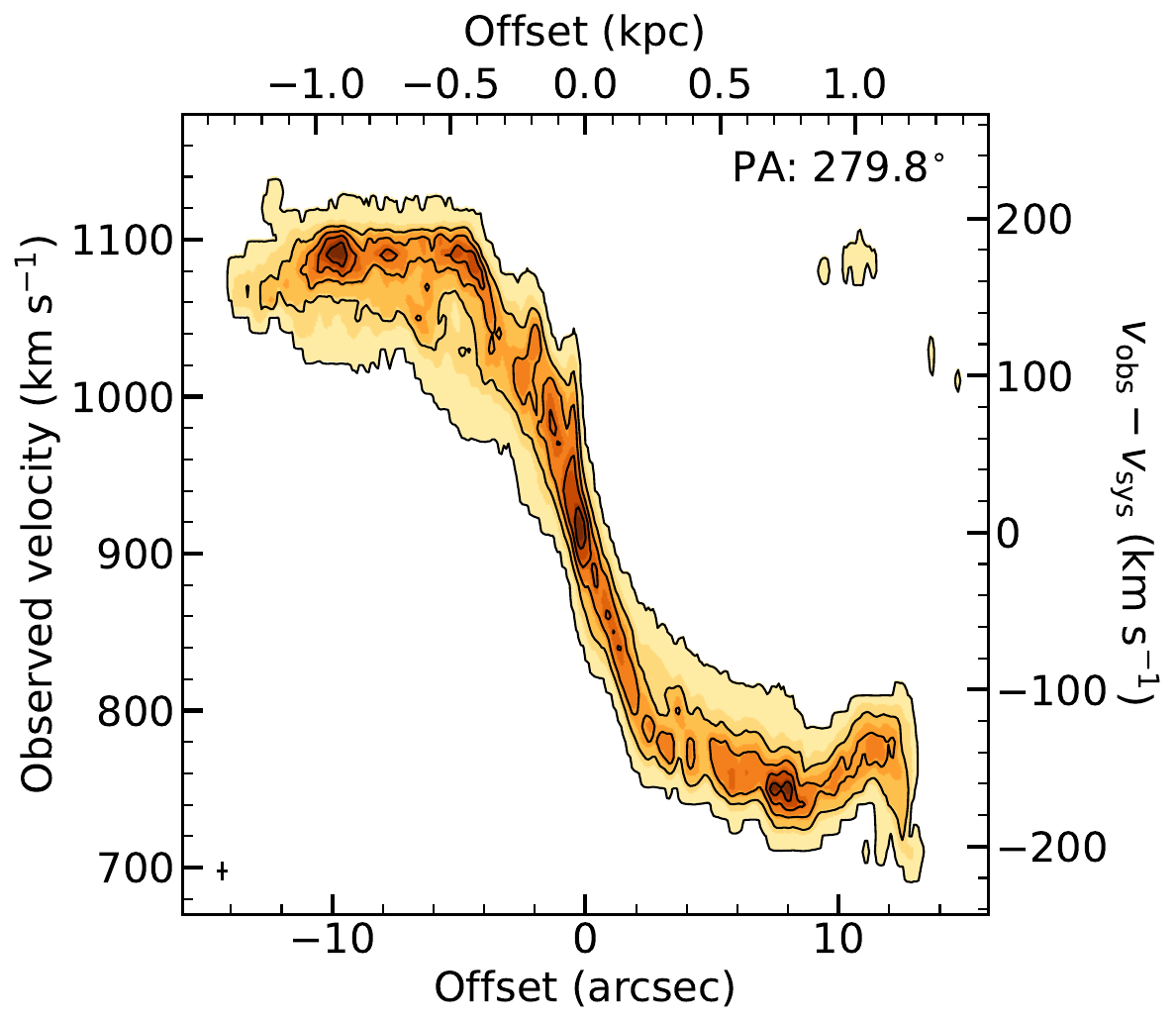}
    \end{subfigure}
    \caption[]{continued.}
\end{figure*}

\begin{figure*}
    \ContinuedFloat
    \begin{subfigure}{.34\linewidth}
    \includegraphics[height=6.0cm]{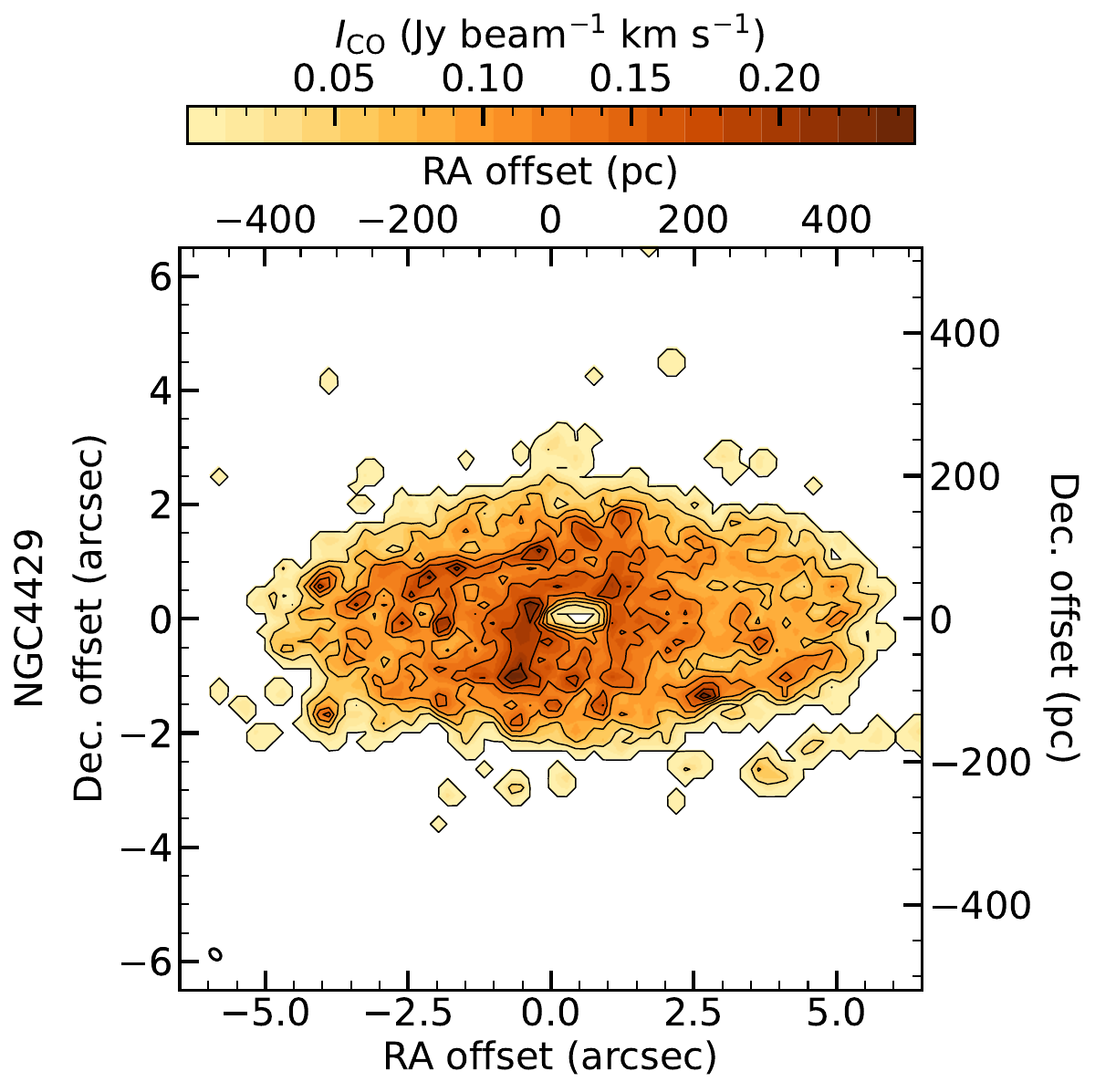}
    \end{subfigure}
    \begin{subfigure}{.32\linewidth}
    \centering
    \includegraphics[height=6.0cm]{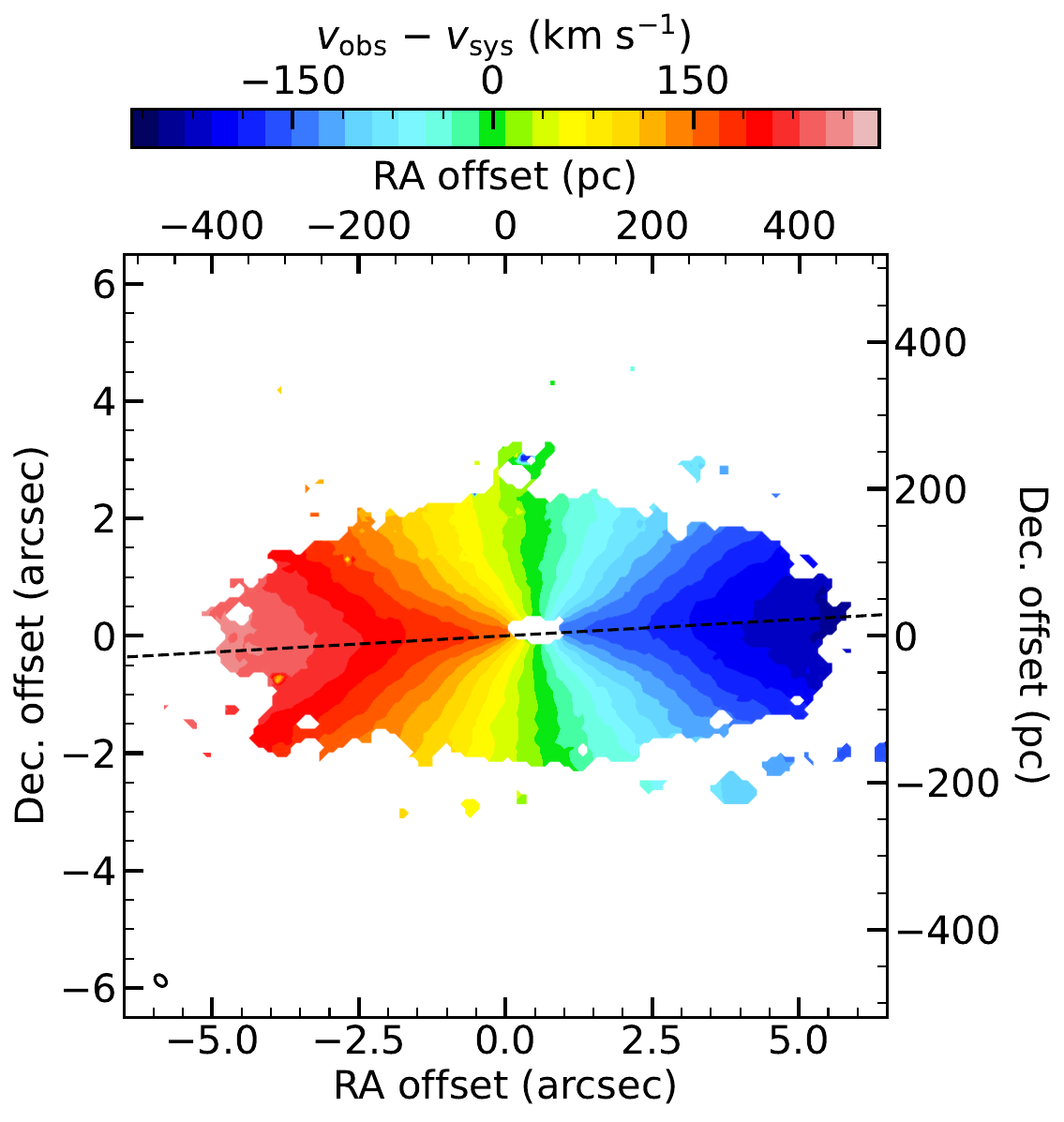}
    \end{subfigure}
    \begin{subfigure}{.32\linewidth}
    \centering
    \includegraphics[height=6.0cm]{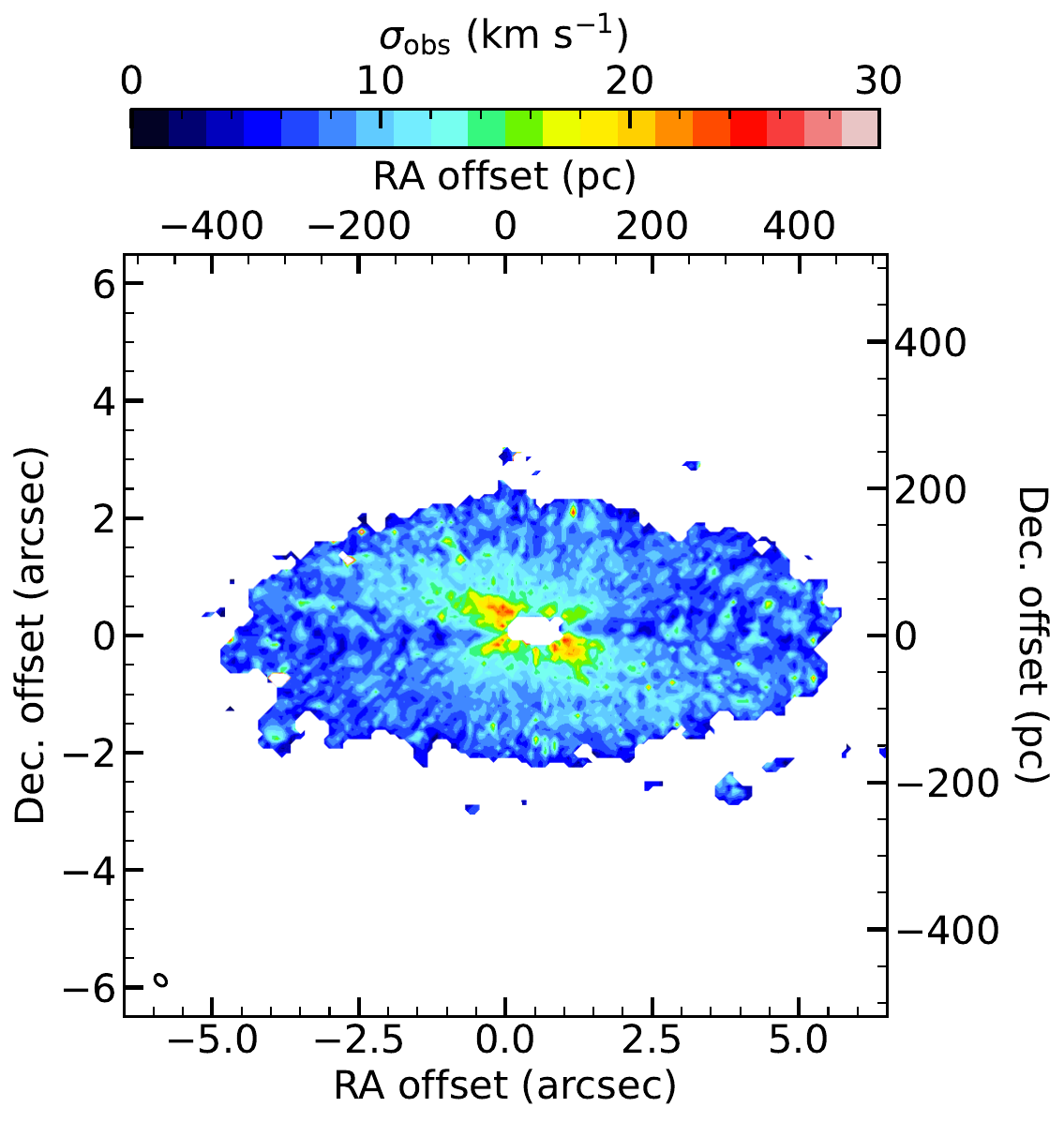}
    \end{subfigure}
    \\
    \begin{subfigure}{.34\linewidth}
    \centering
    \includegraphics[height=5.2cm]{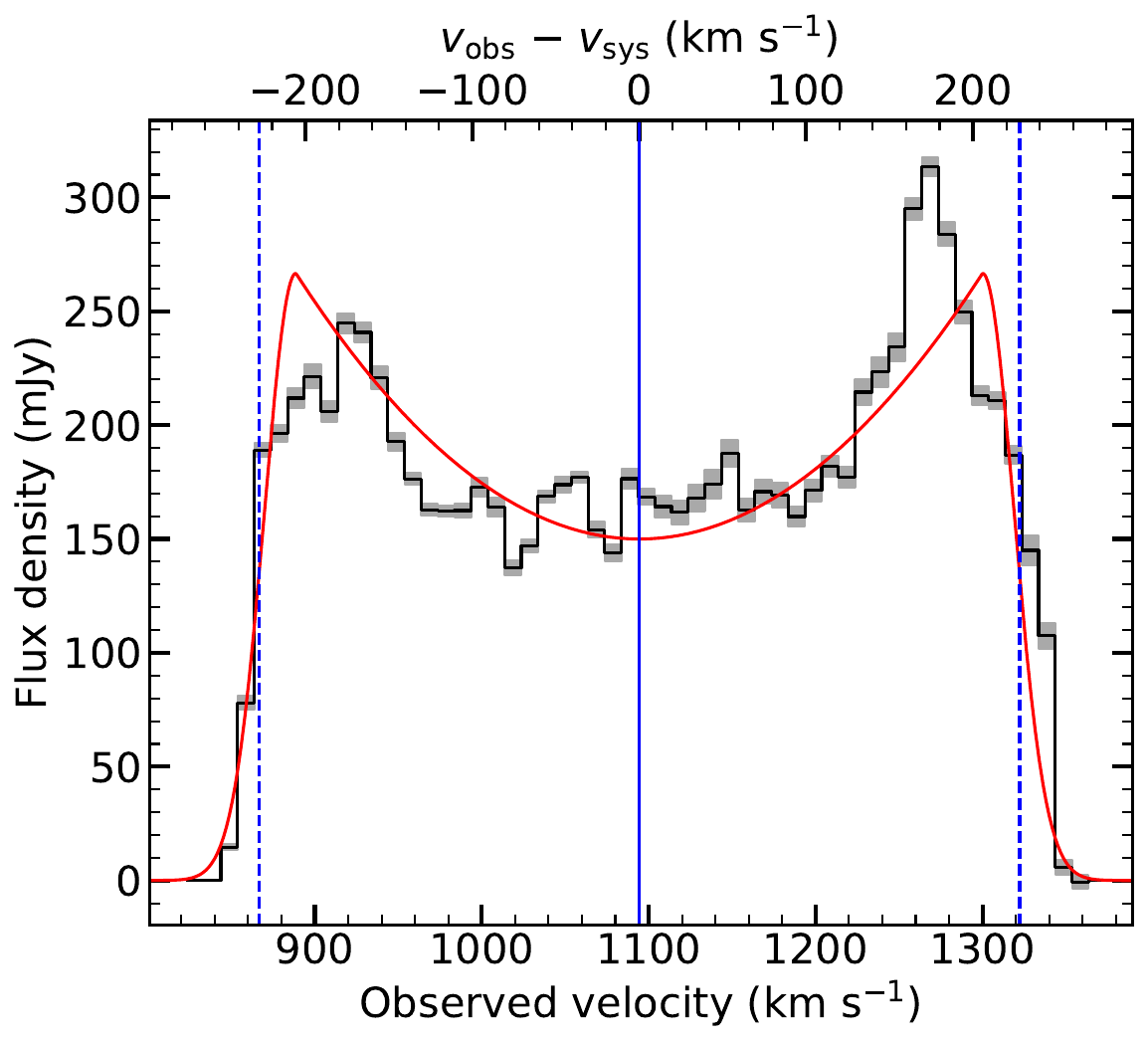}
    \end{subfigure}
    \begin{subfigure}{.32\linewidth}
    \centering
    \includegraphics[height=4.8cm]{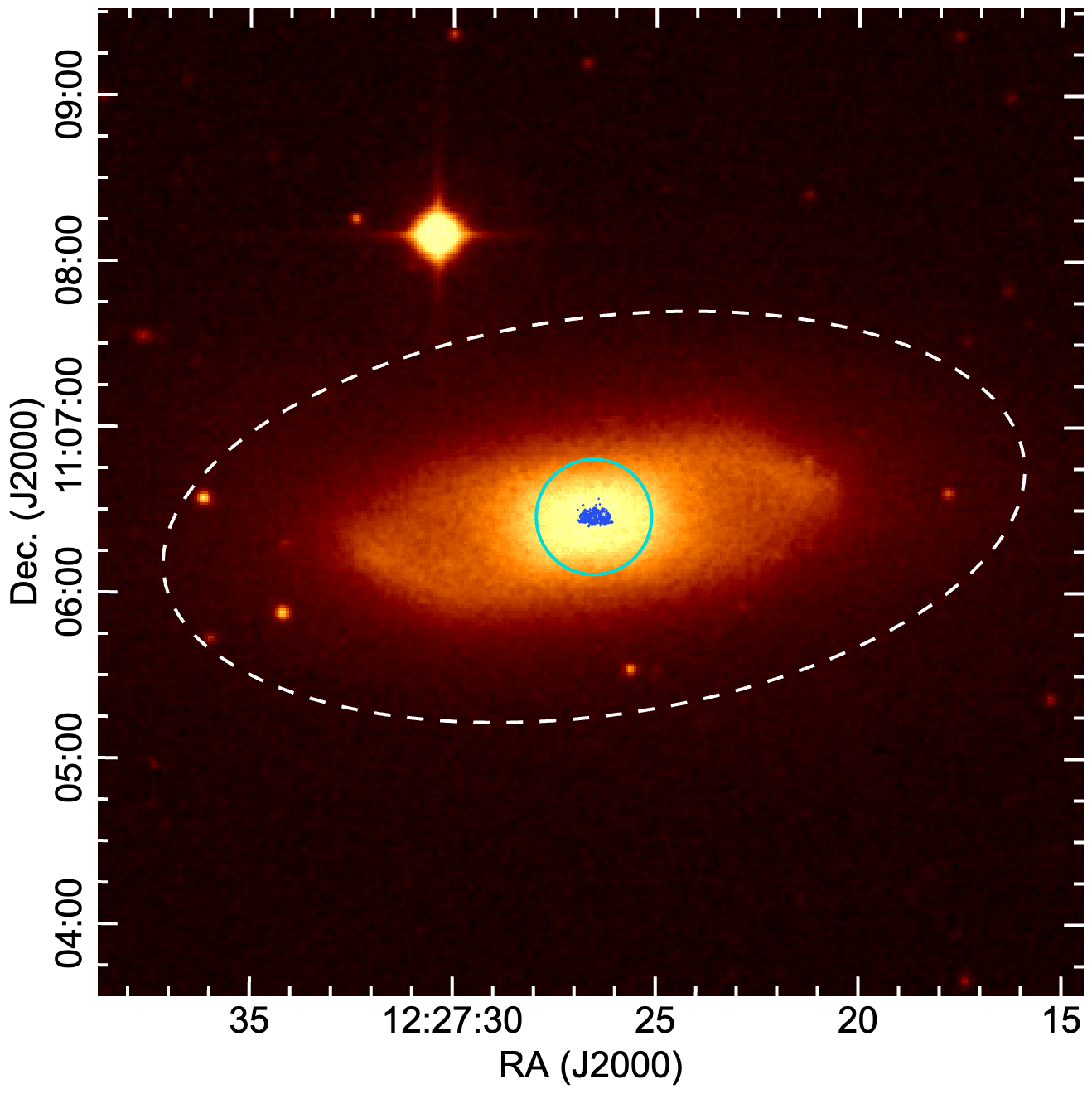}
    \end{subfigure}
    \begin{subfigure}{.33\linewidth}
    \centering
    \includegraphics[height=5.2cm]{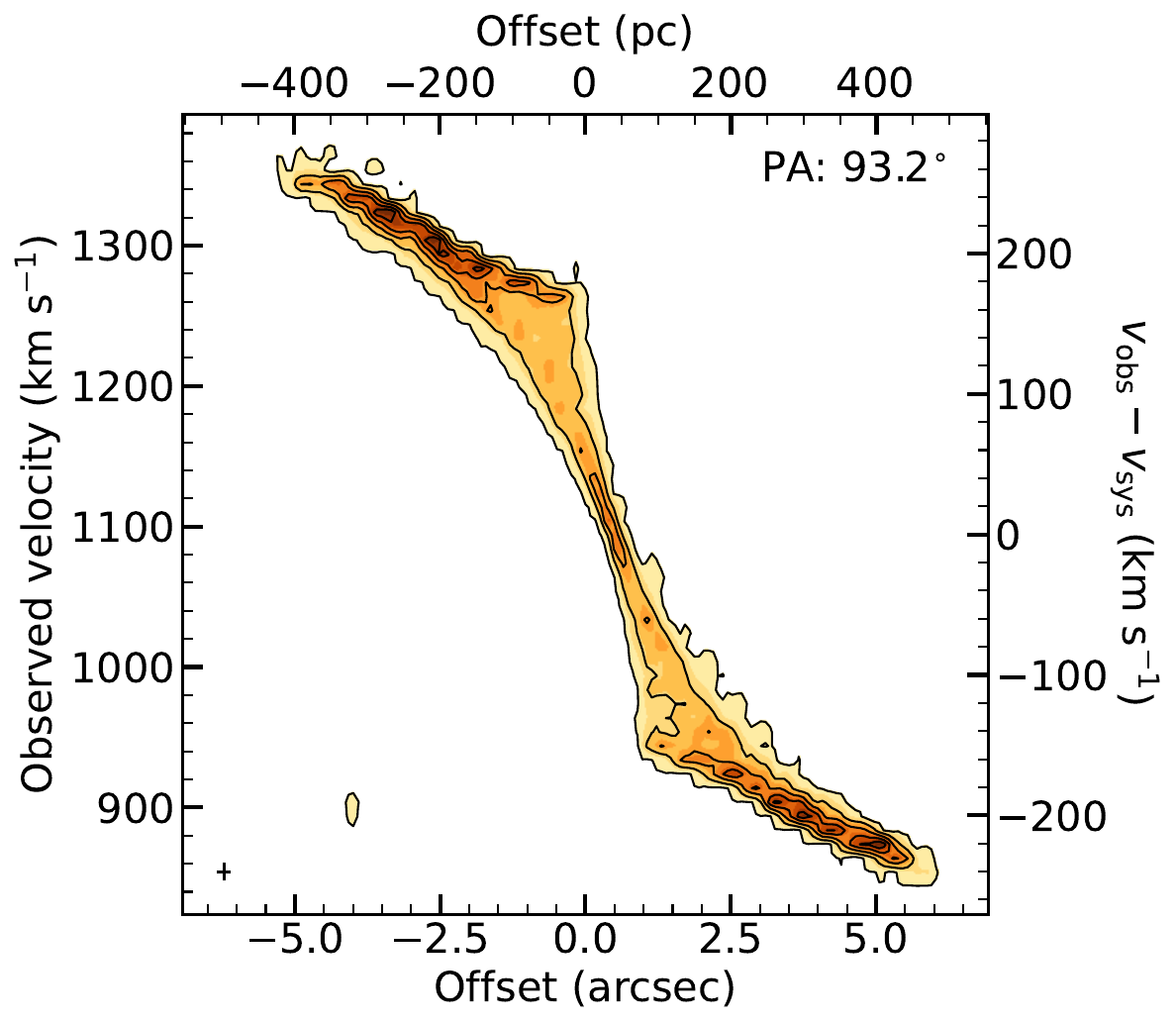}
    \end{subfigure}
    \begin{subfigure}{.34\linewidth}
    \centering
    \includegraphics[height=6.0cm]{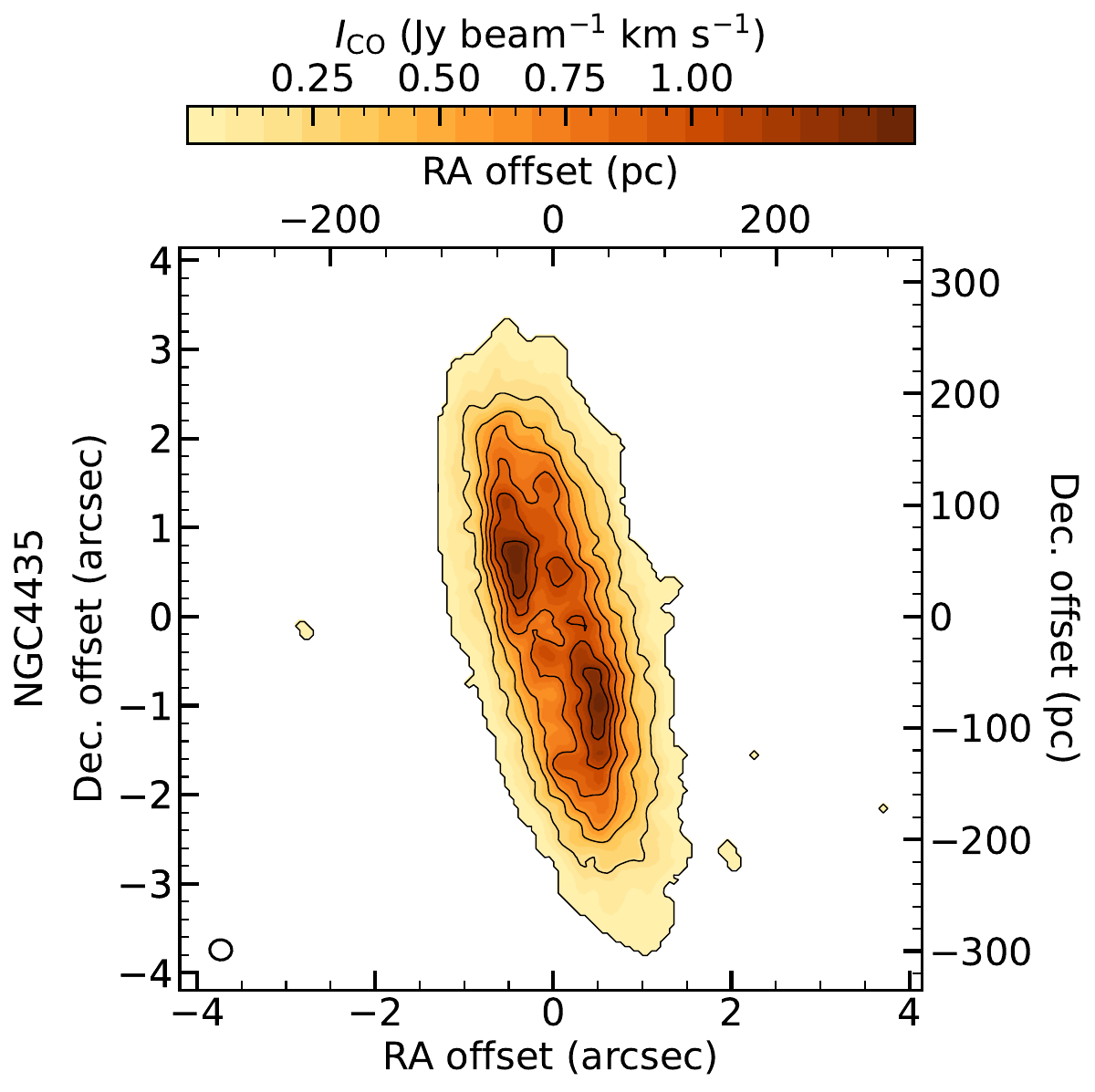}
    \end{subfigure}
    \begin{subfigure}{.32\linewidth}
    \centering
    \includegraphics[height=6.0cm]{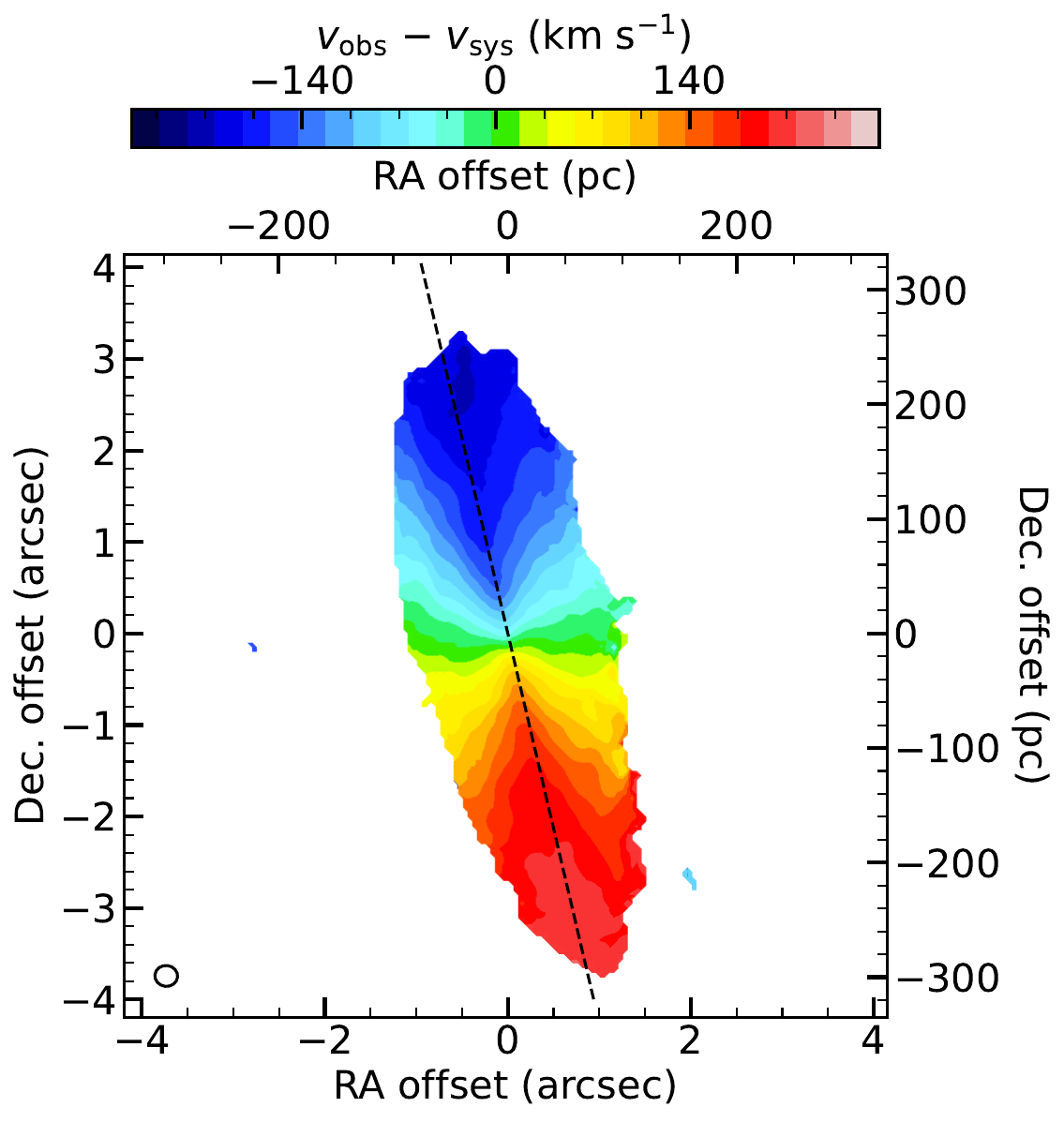}
    \end{subfigure}
    \begin{subfigure}{.32\linewidth}
    \centering
    \includegraphics[height=6.0cm]{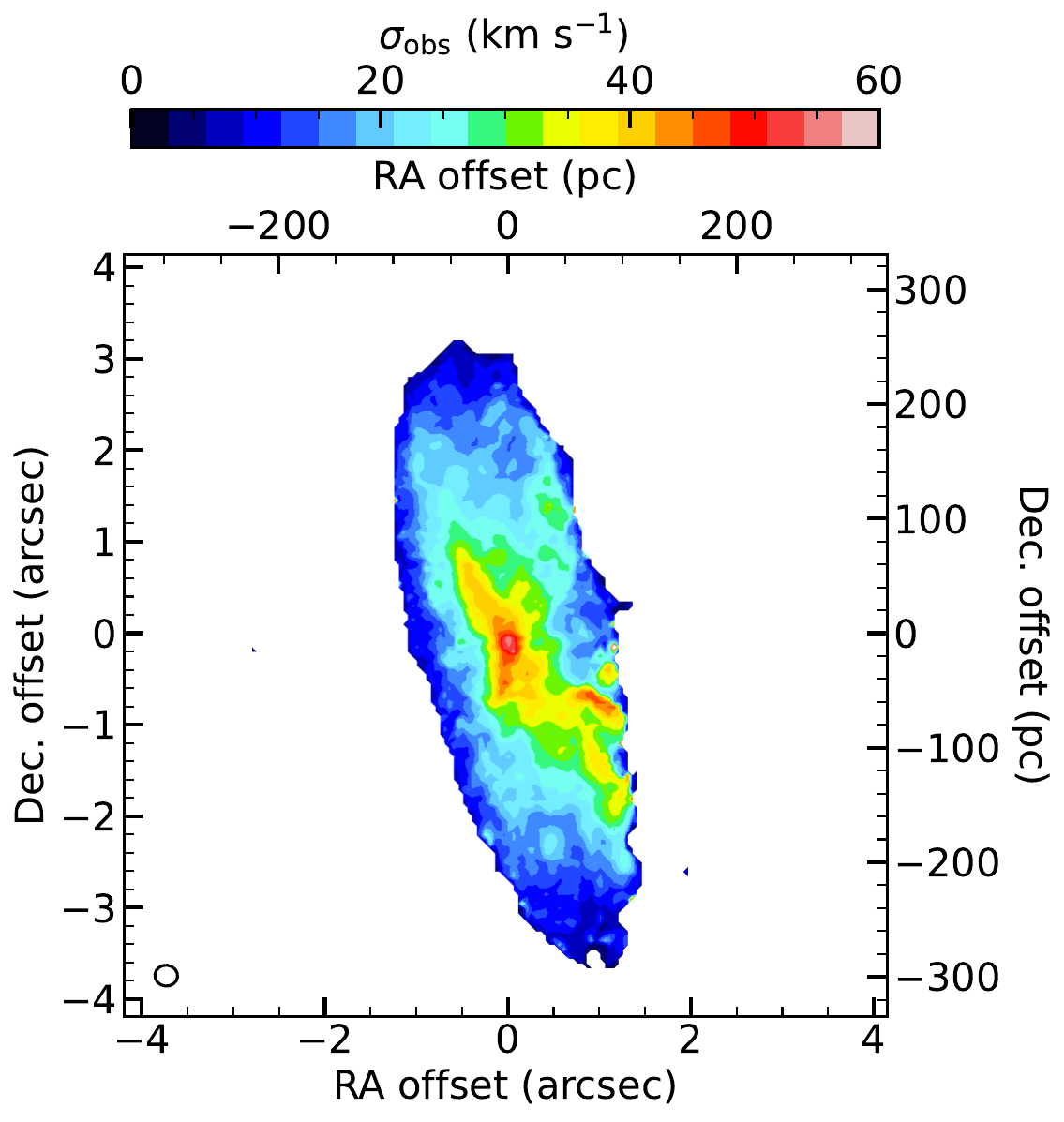}
    \end{subfigure}
    \\
    \begin{subfigure}{.34\linewidth}
    \centering
    \includegraphics[height=5.2cm]{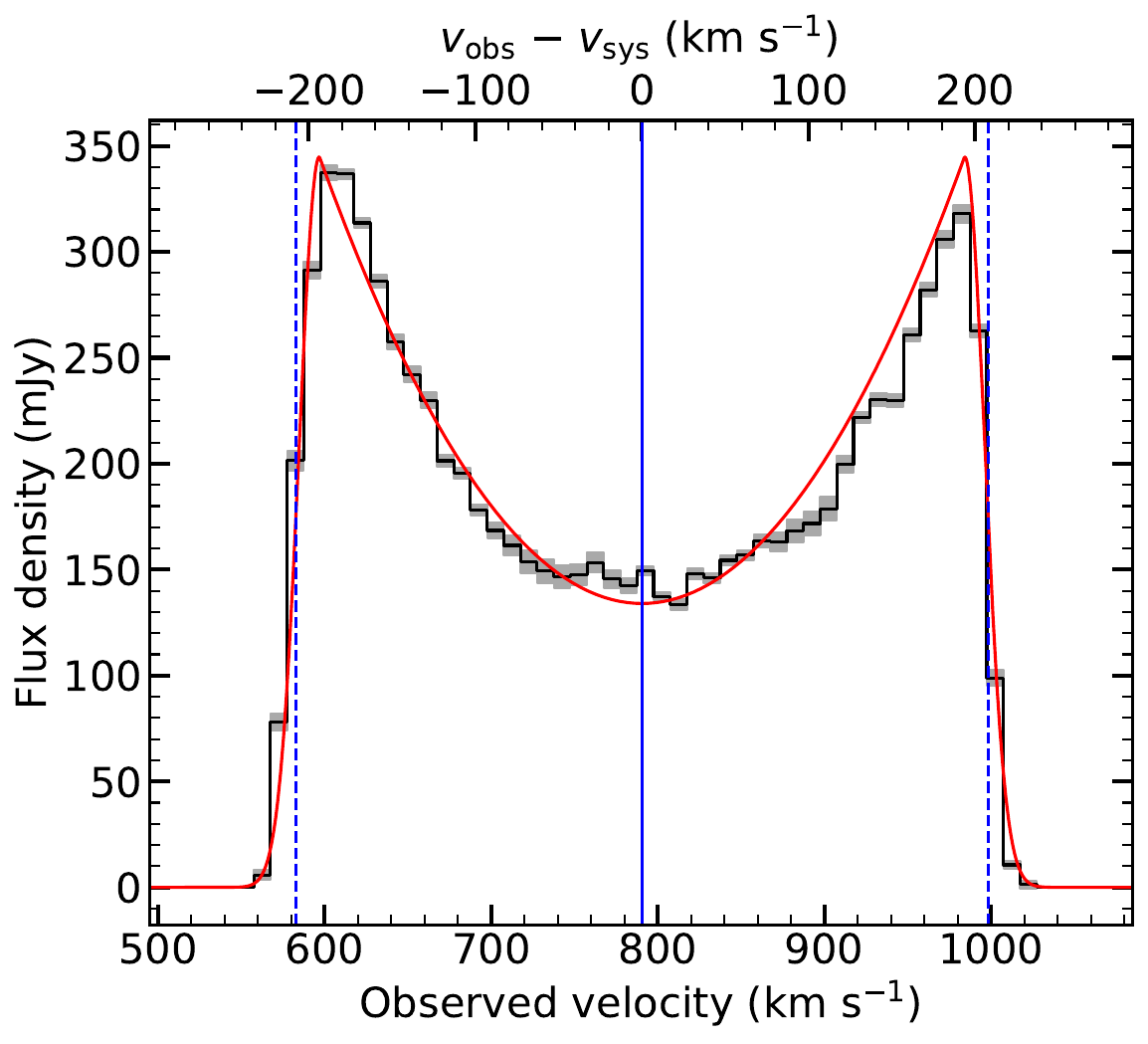}
    \end{subfigure}
    \begin{subfigure}{.32\linewidth}
    \centering
    \includegraphics[height=4.8cm]{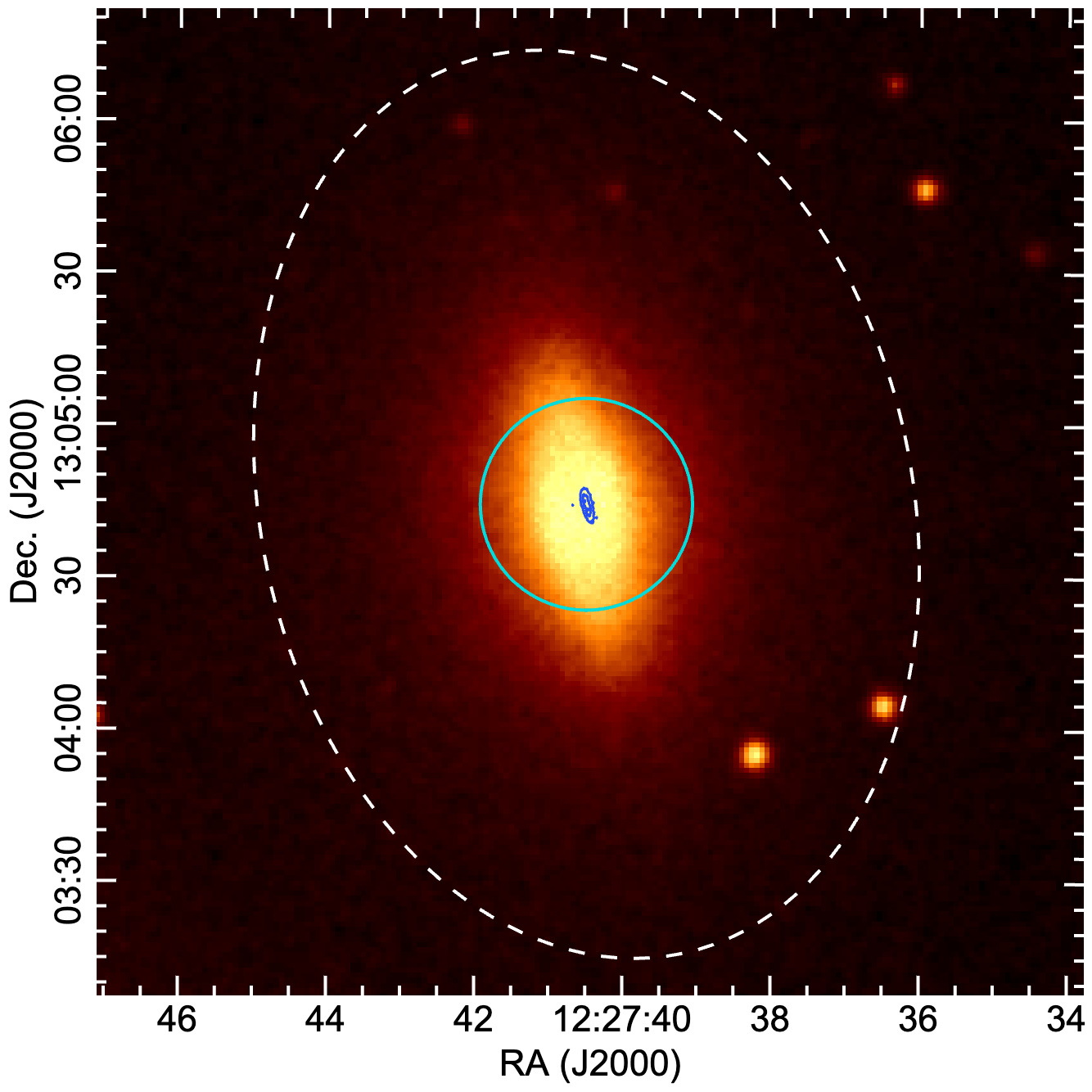}
    \end{subfigure}
    \begin{subfigure}{.33\linewidth}
    \centering
    \includegraphics[height=5.2cm]{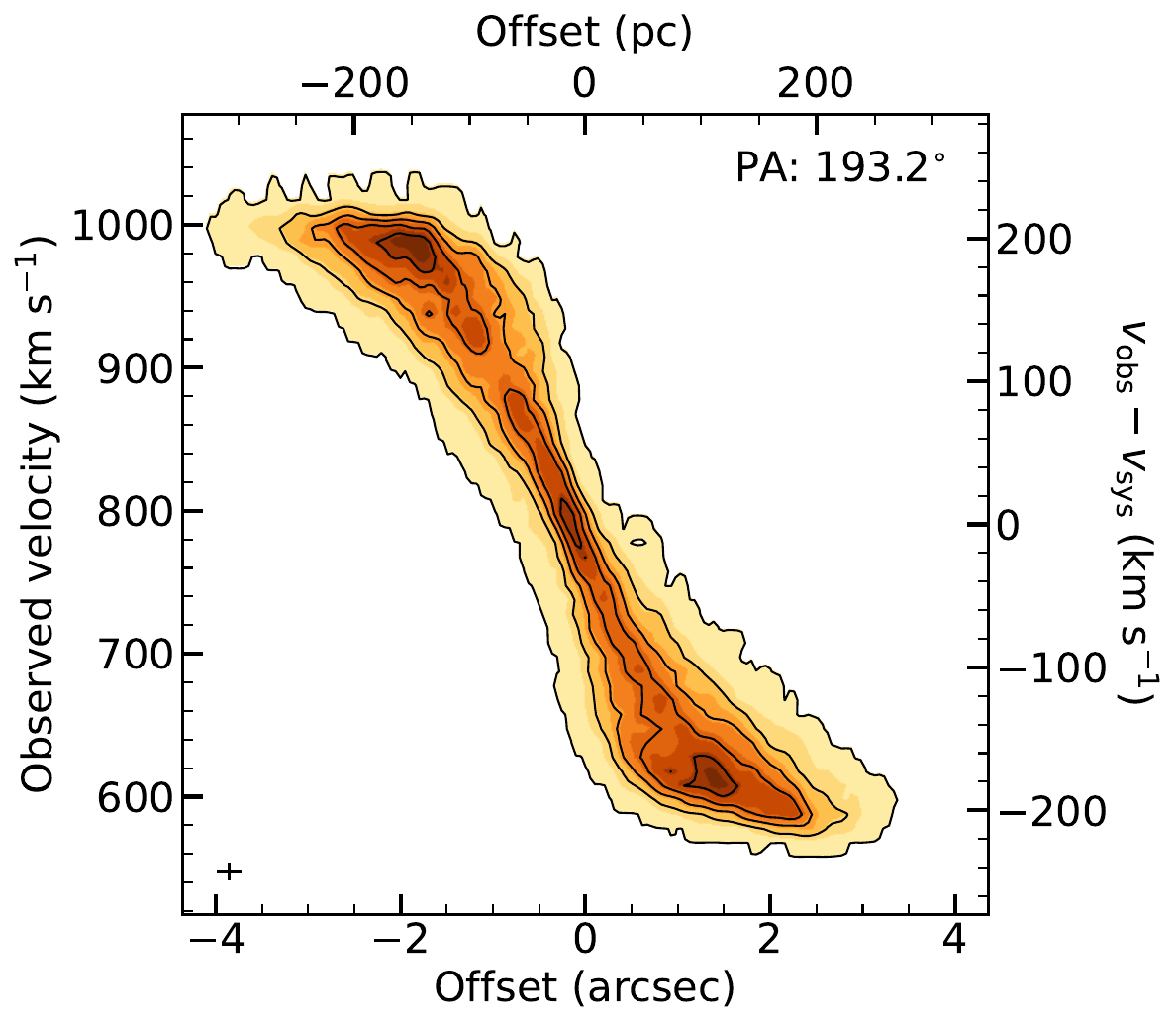}
    \end{subfigure}
    \caption[]{continued.}
\end{figure*}

\begin{figure*}
    \ContinuedFloat
    \begin{subfigure}{.34\linewidth}
    \centering
    \includegraphics[height=6.0cm]{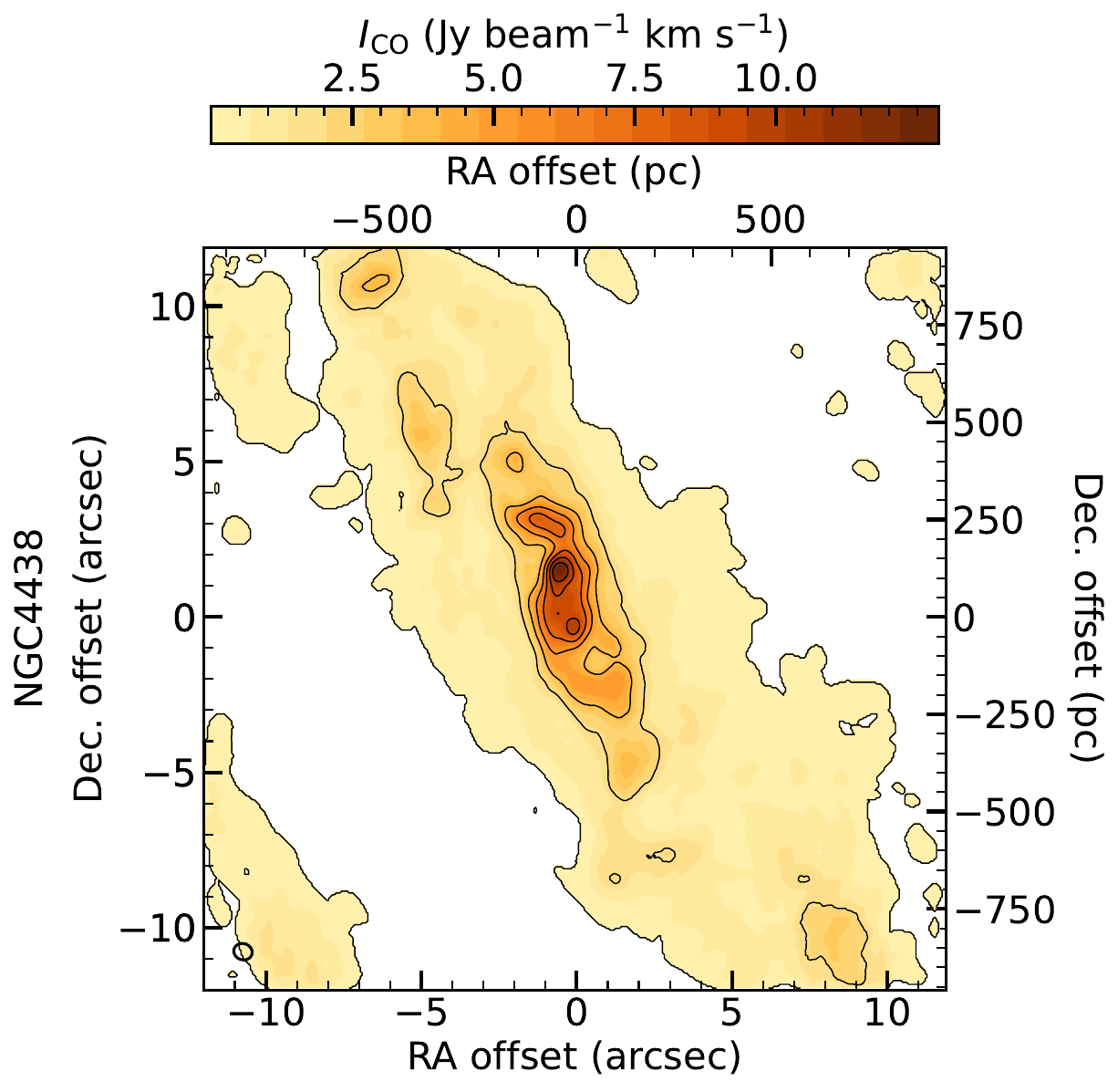}
    \end{subfigure}
    \begin{subfigure}{.32\linewidth}
    \centering
    \includegraphics[height=6.0cm]{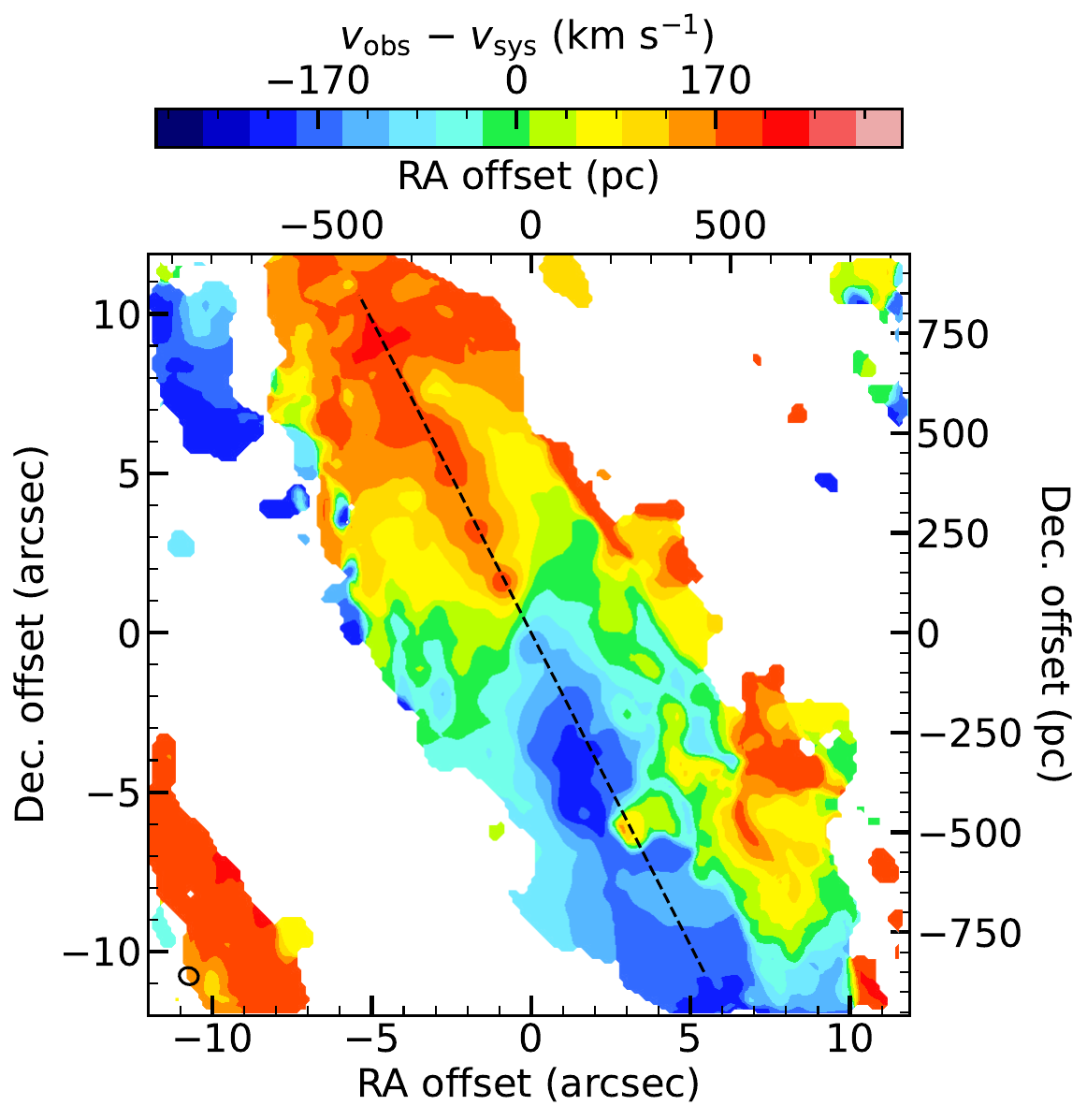}
    \end{subfigure}
    \begin{subfigure}{.32\linewidth}
    \centering
    \includegraphics[height=6.0cm]{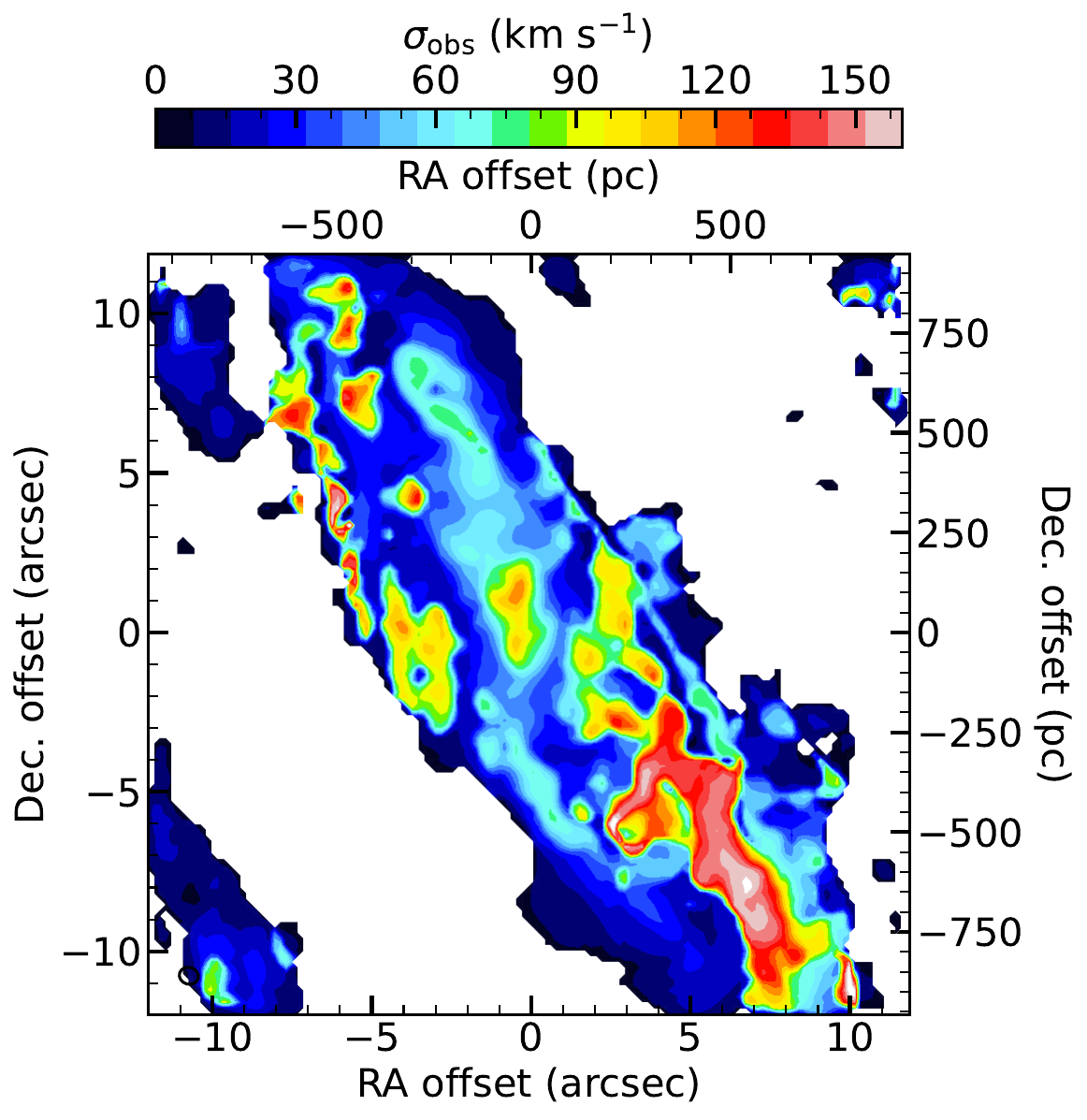}
    \end{subfigure}
    \\
    \begin{subfigure}{.34\linewidth}
    \centering
    \includegraphics[height=5.2cm]{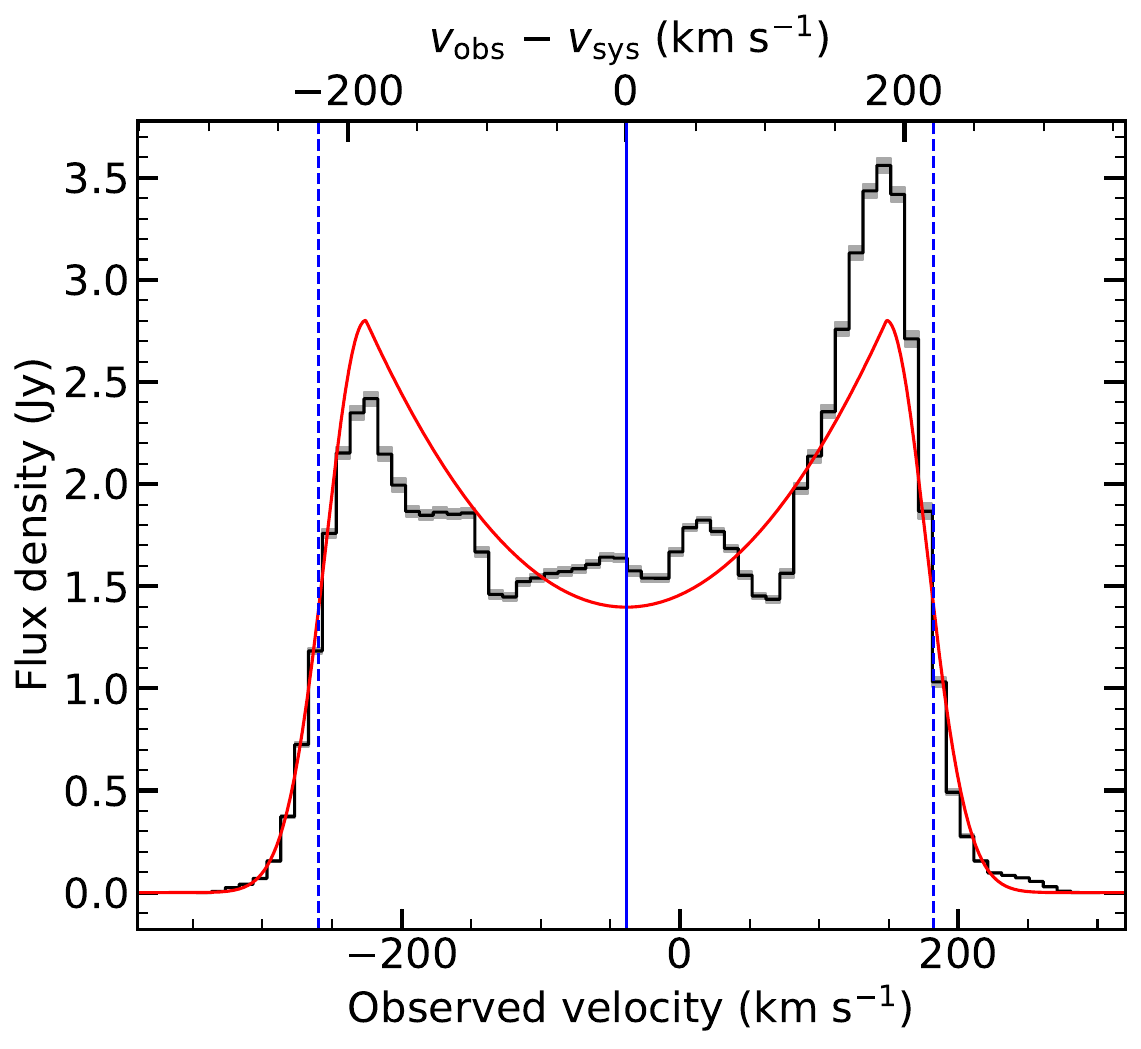}
    \end{subfigure}
    \begin{subfigure}{.32\linewidth}
    \centering
    \includegraphics[height=4.8cm]{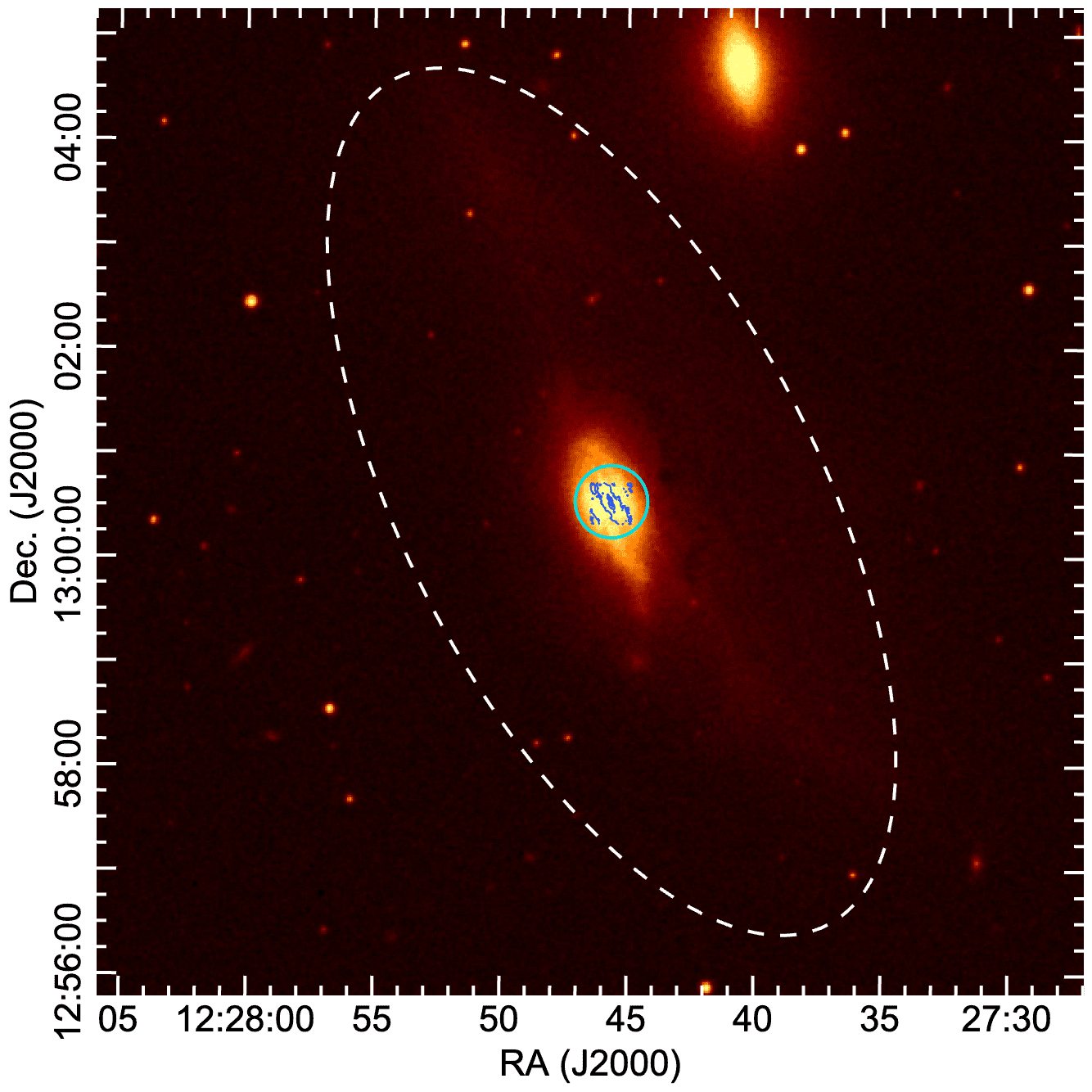}
    \end{subfigure}
    \begin{subfigure}{.33\linewidth}
    \centering
    \includegraphics[height=5.2cm]{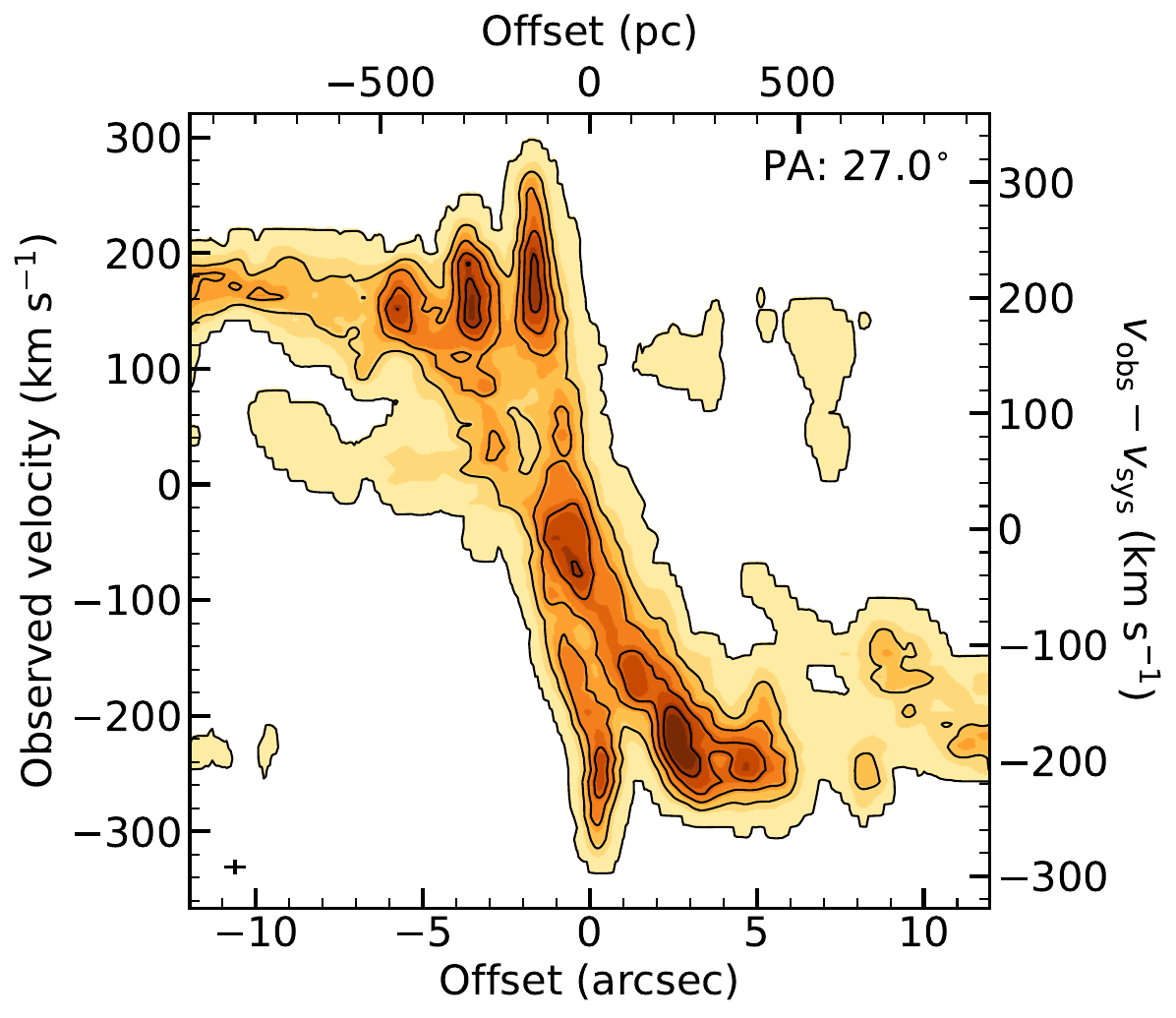}
    \end{subfigure}
    \begin{subfigure}{.34\linewidth}
    \centering
    \includegraphics[height=6.0cm]{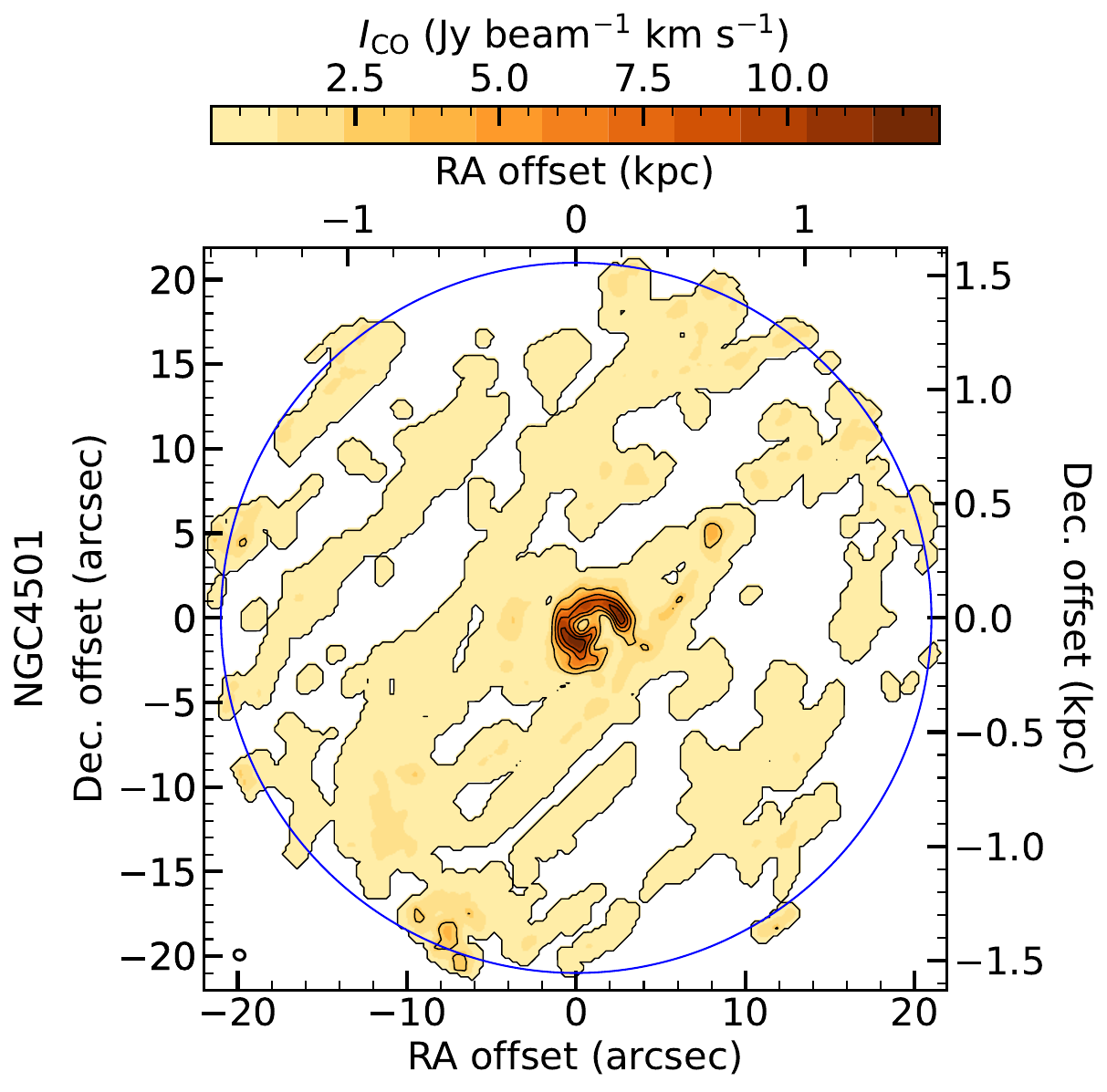}
    \end{subfigure}
    \begin{subfigure}{.32\linewidth}
    \centering
    \includegraphics[height=6.0cm]{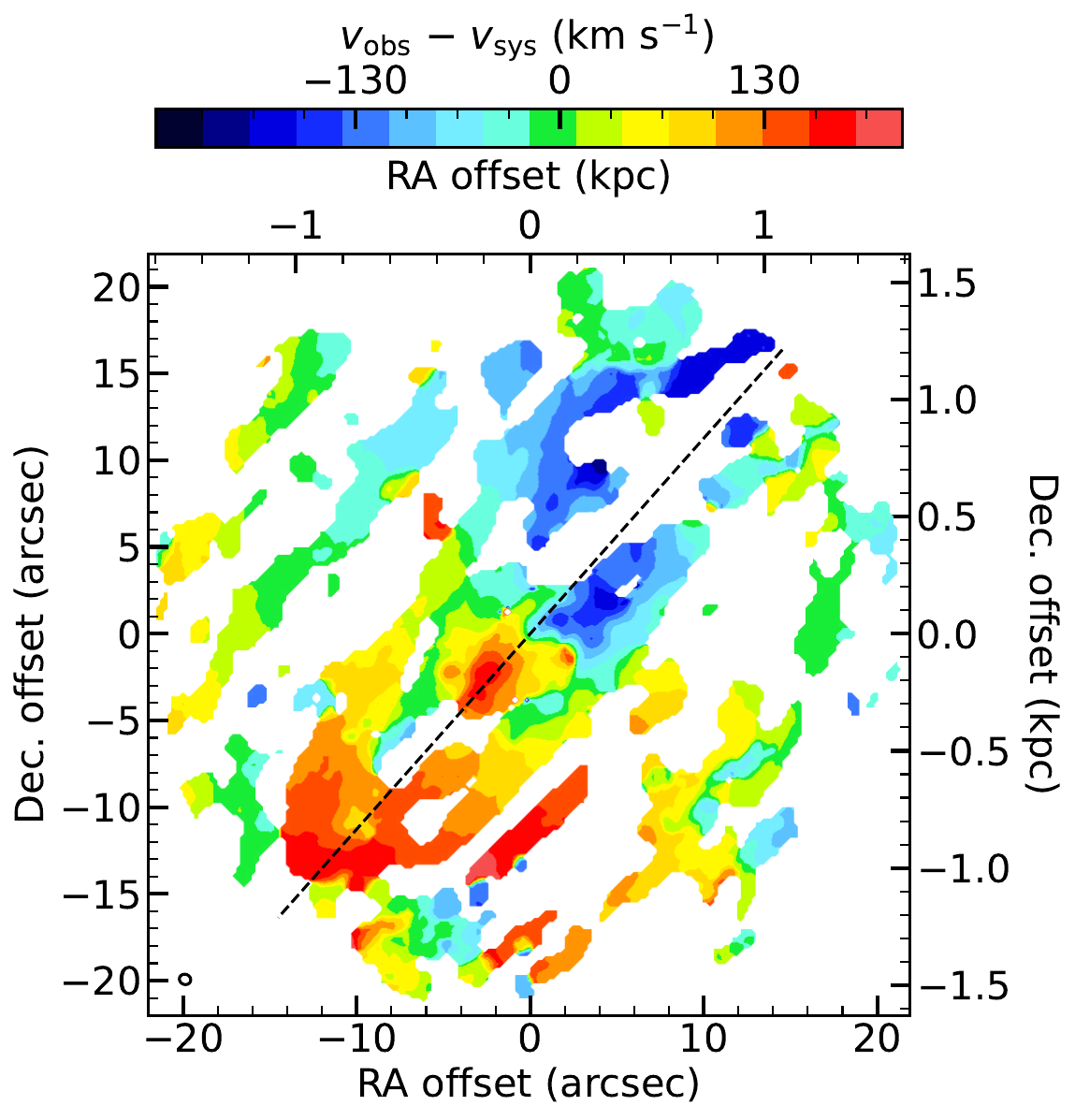}
    \end{subfigure}
    \begin{subfigure}{.32\linewidth}
    \centering
    \includegraphics[height=6.0cm]{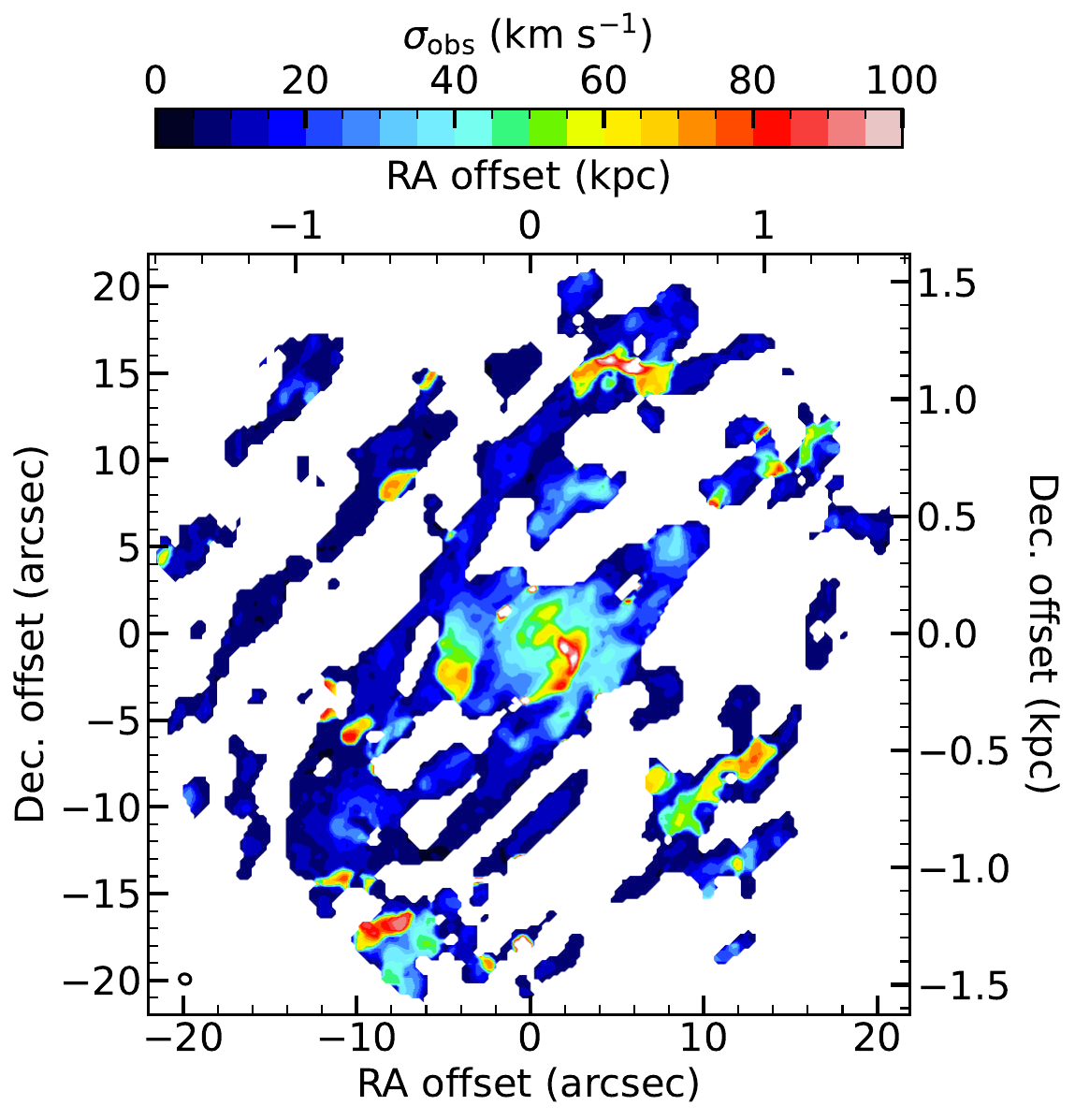}
    \end{subfigure}
    \\
    \begin{subfigure}{.34\linewidth}
    \centering
    \includegraphics[height=5.2cm]{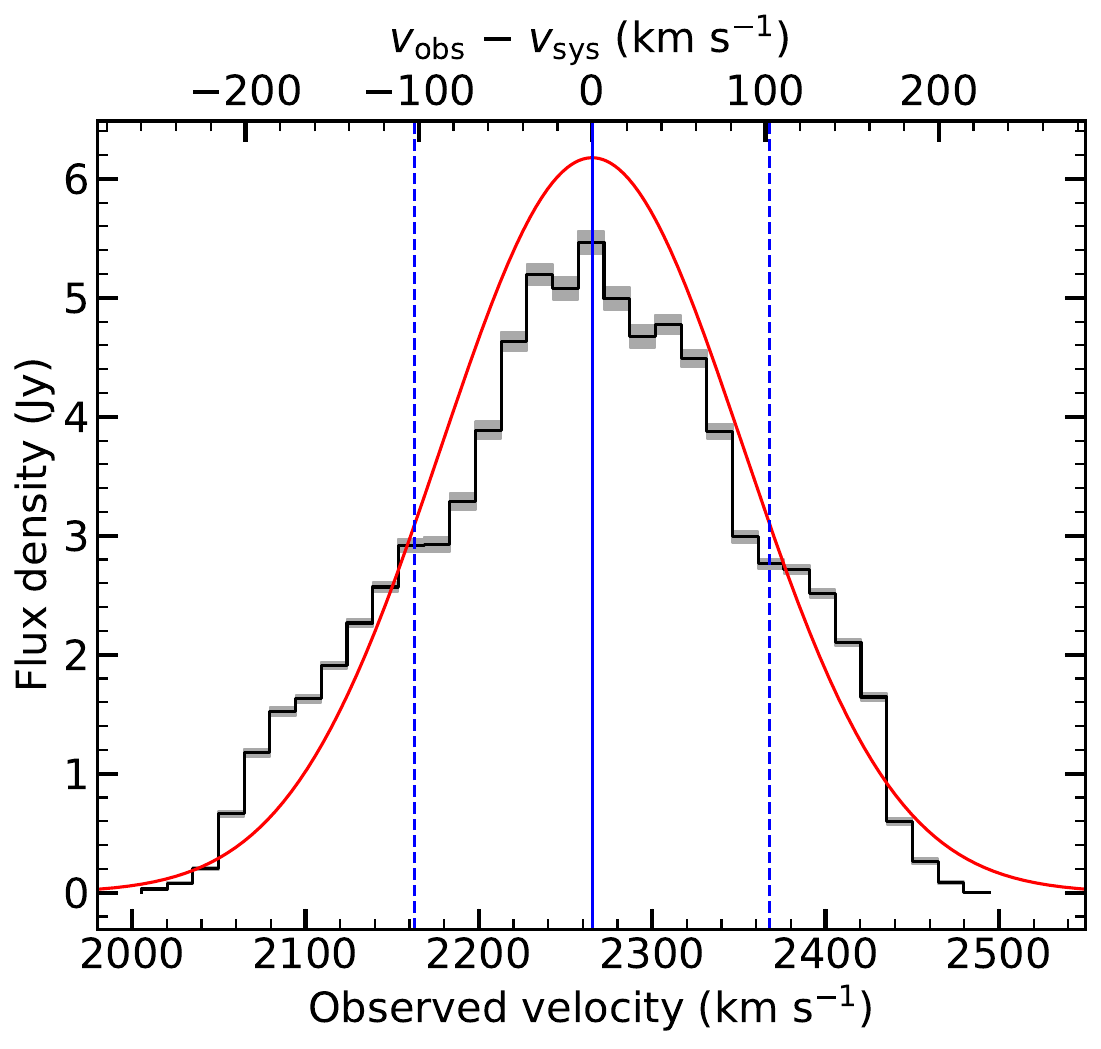}
    \end{subfigure}
    \begin{subfigure}{.32\linewidth}
    \centering
    \includegraphics[height=4.8cm]{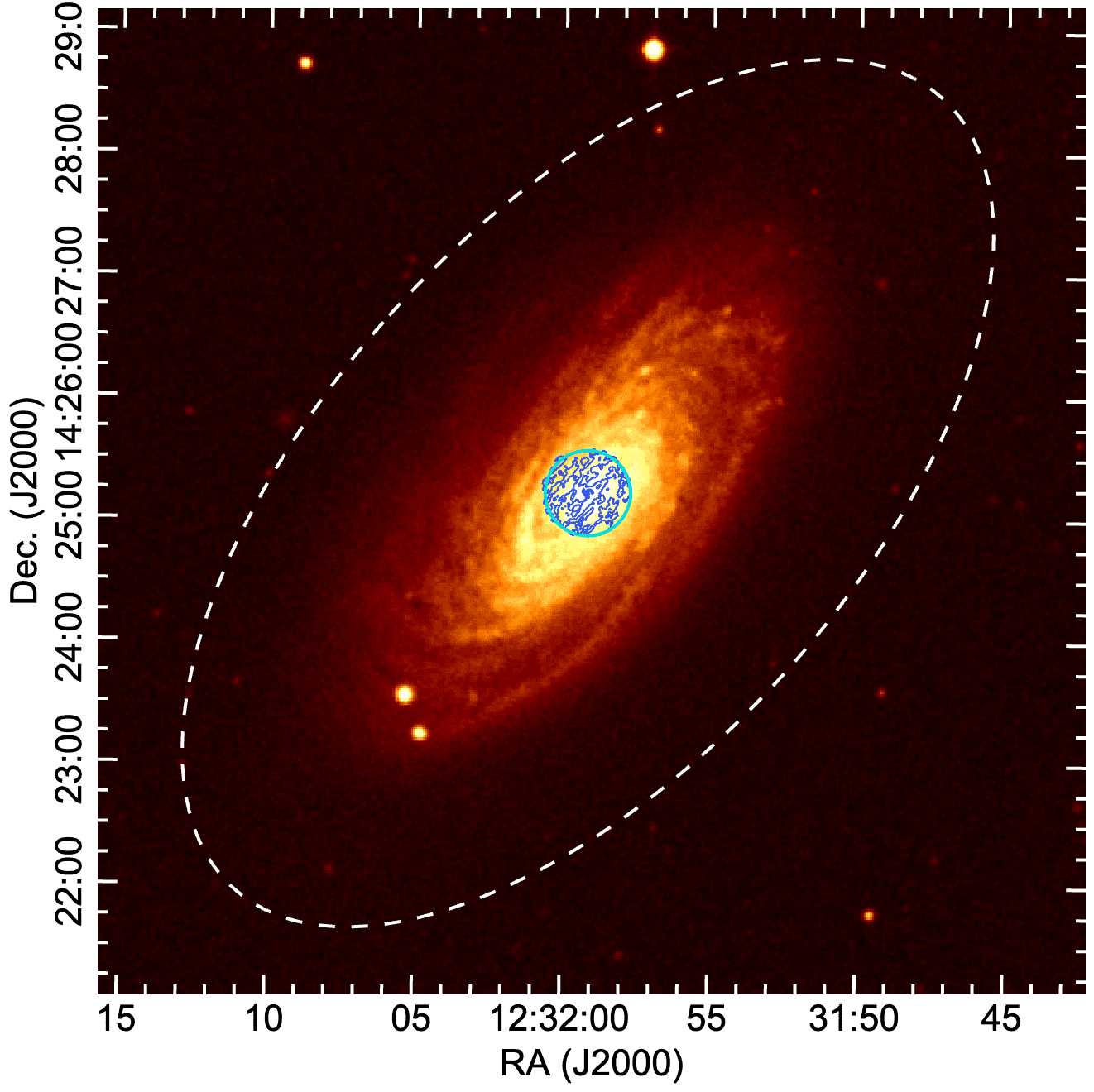}
    \end{subfigure}
    \begin{subfigure}{.33\linewidth}
    \centering
    \includegraphics[height=5.2cm]{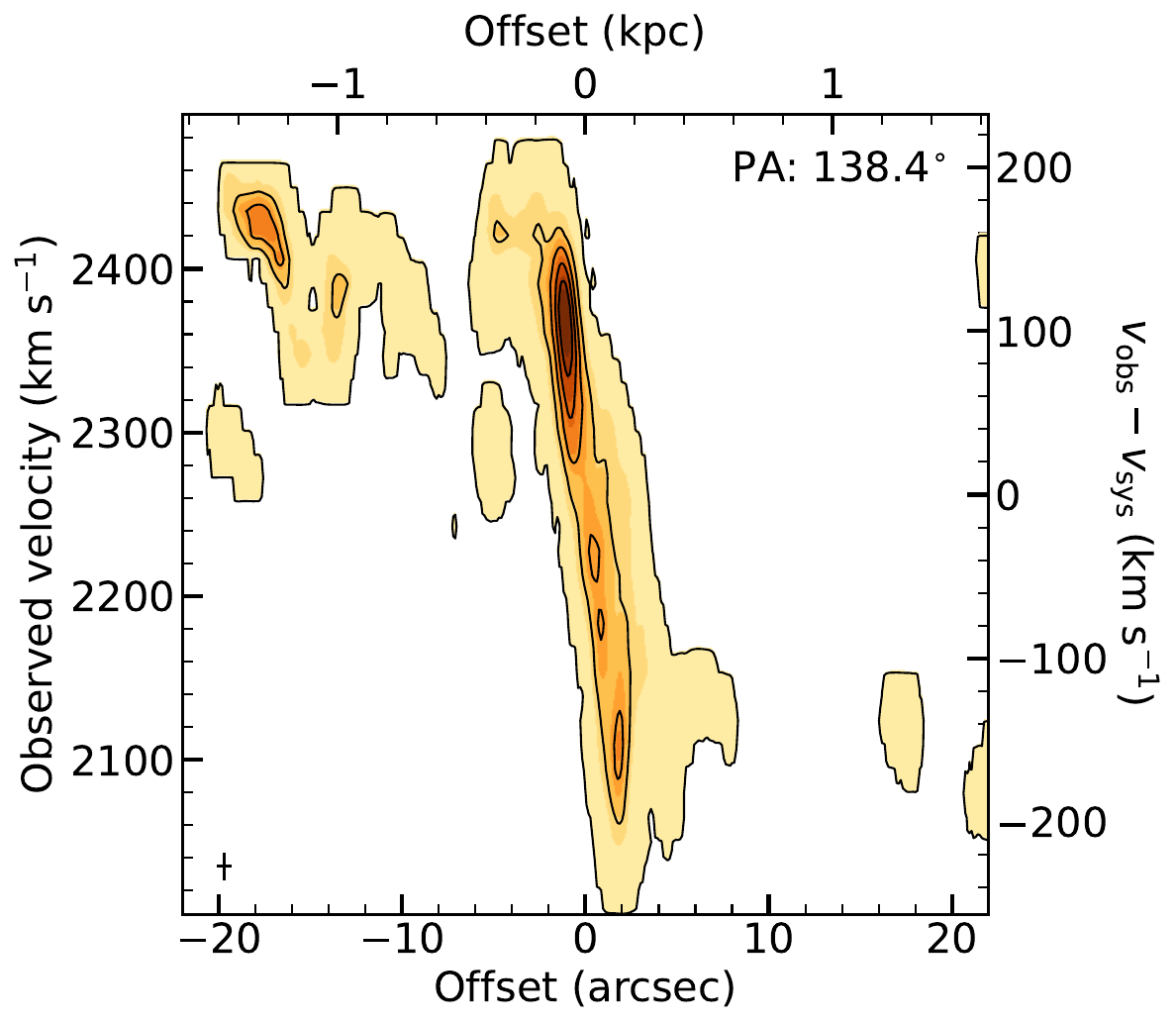}
    \end{subfigure}
    \caption[]{continued.}
\end{figure*}

\begin{figure*}
    \ContinuedFloat
    \begin{subfigure}{.34\linewidth}
    \centering
    \includegraphics[height=6.0cm]{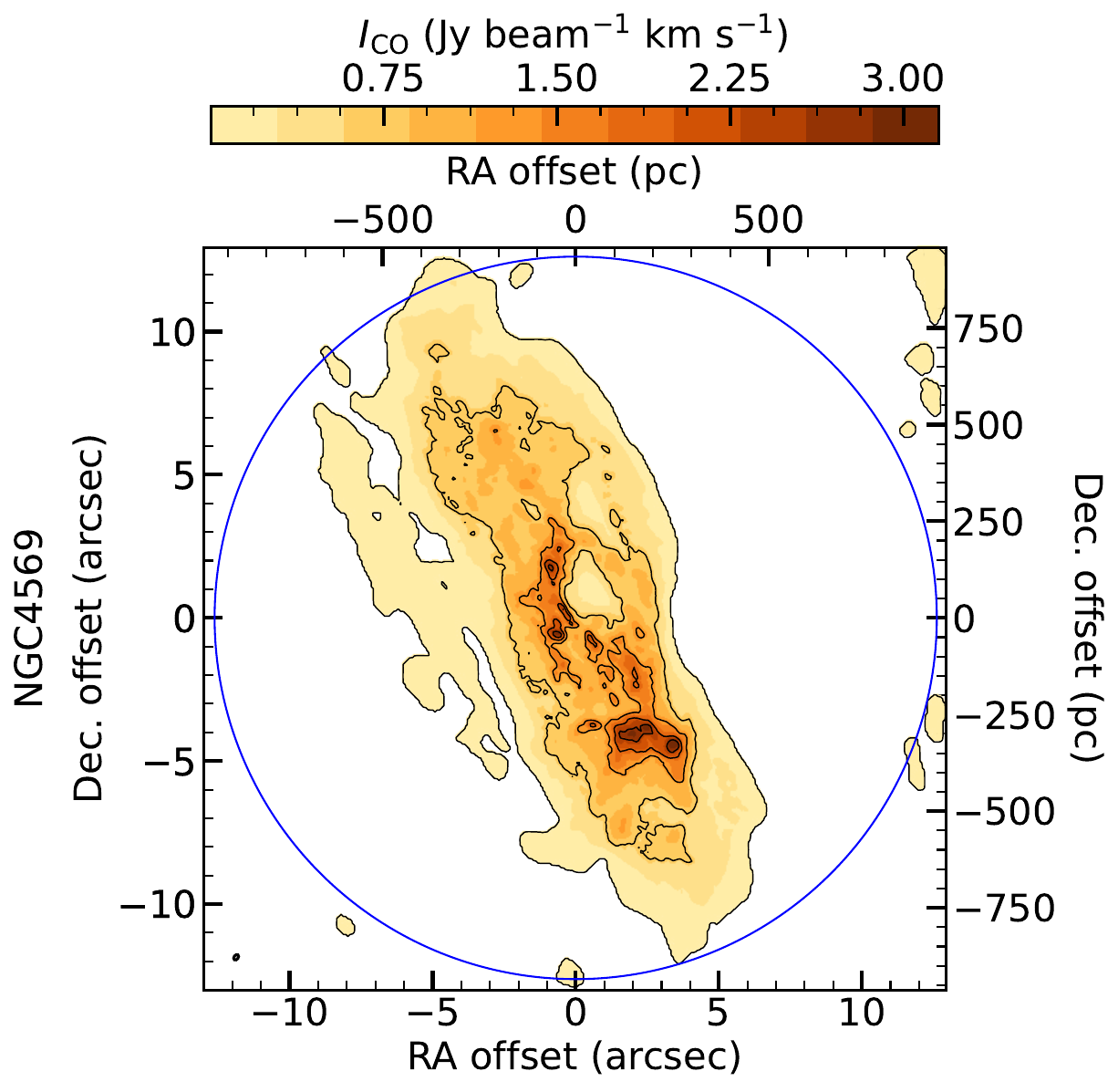}
    \end{subfigure}
    \begin{subfigure}{.32\linewidth}
    \centering
    \includegraphics[height=6.0cm]{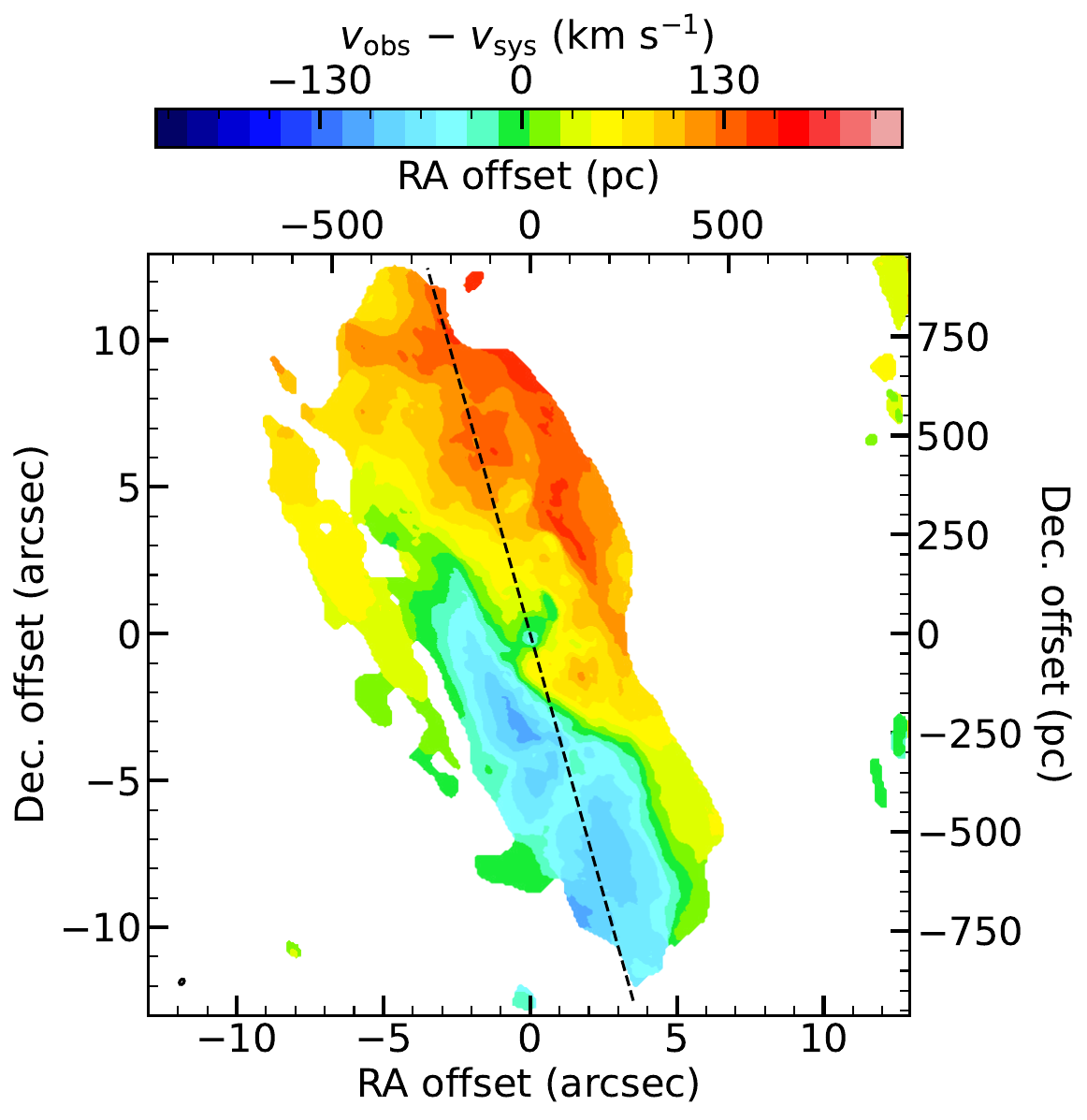}
    \end{subfigure}
    \begin{subfigure}{.32\linewidth}
    \centering
    \includegraphics[height=6.0cm]{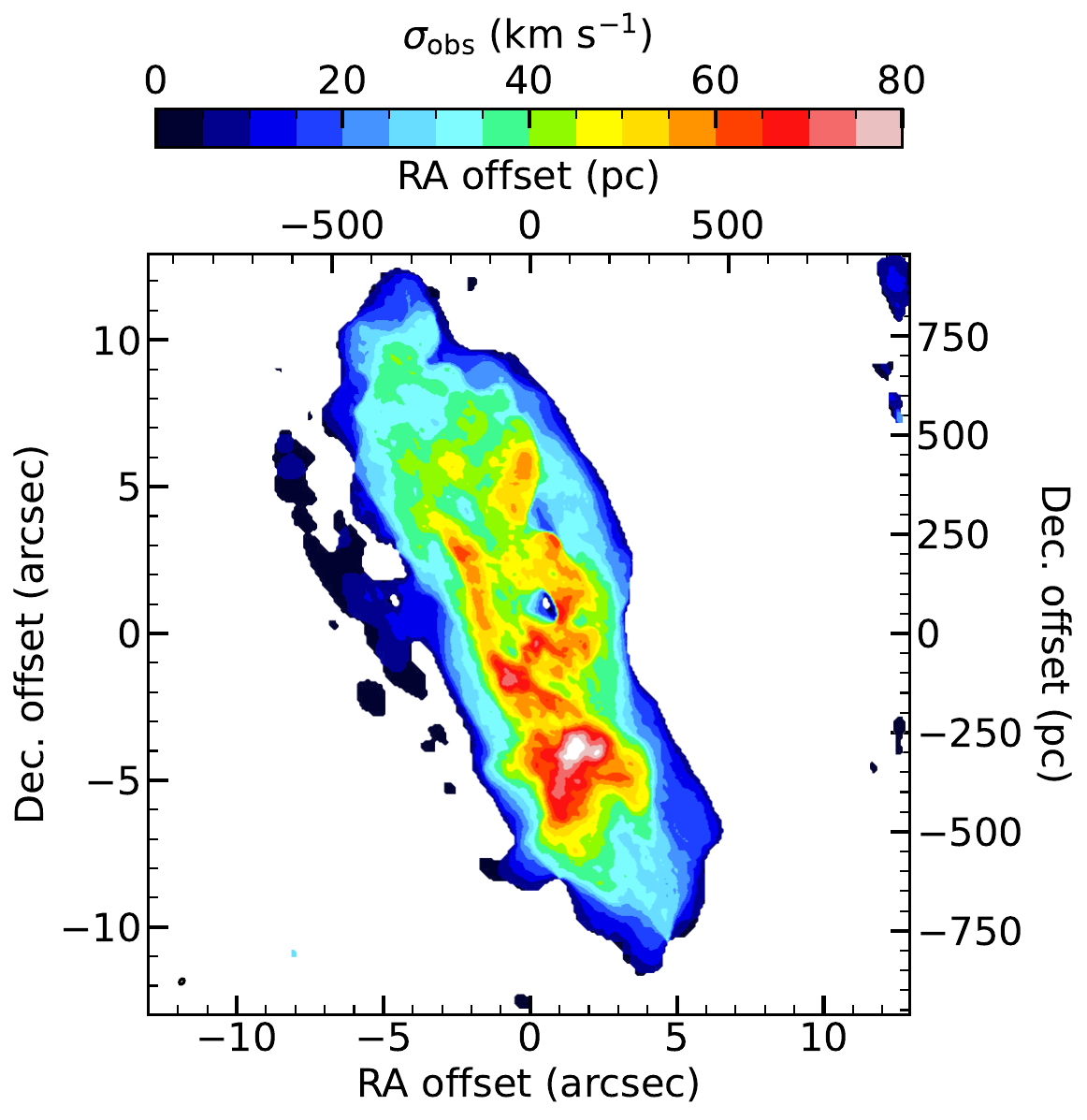}
    \end{subfigure}
    \\
    \begin{subfigure}{.34\linewidth}
    \centering
    \includegraphics[height=5.2cm]{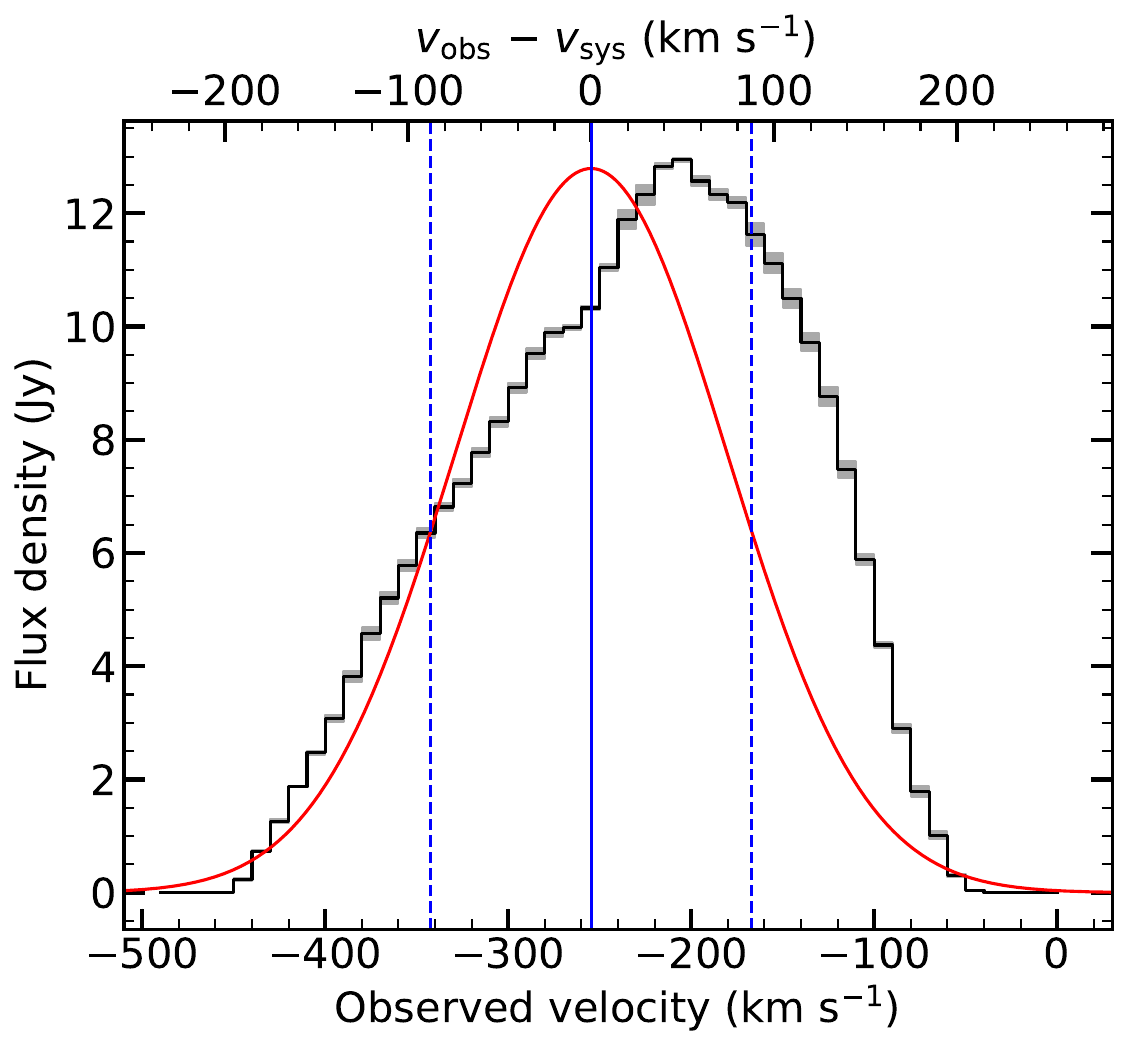}
    \end{subfigure}
    \begin{subfigure}{.32\linewidth}
    \centering
    \includegraphics[height=4.8cm]{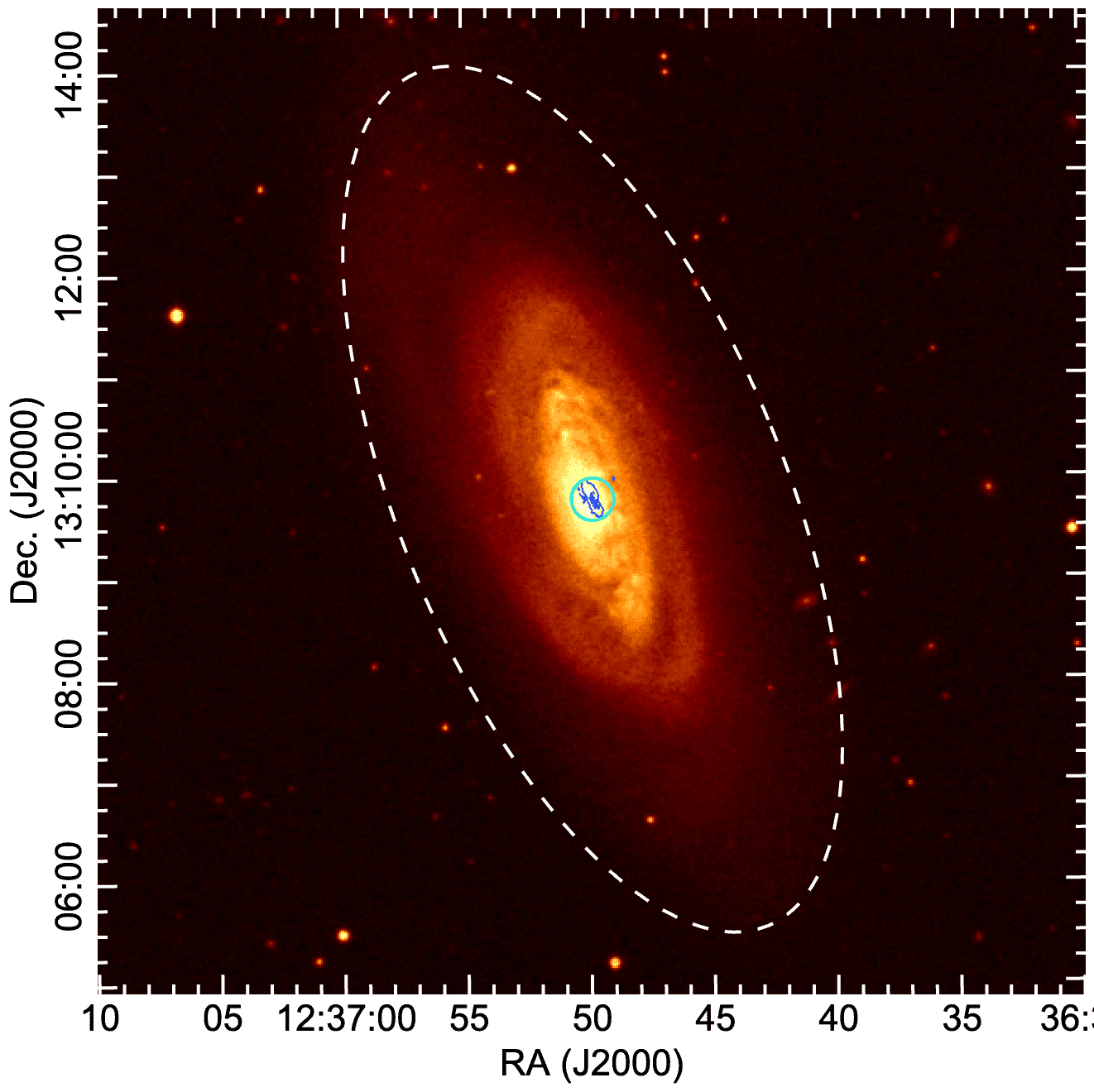}
    \end{subfigure}
    \begin{subfigure}{.33\linewidth}
    \centering
    \includegraphics[height=5.2cm]{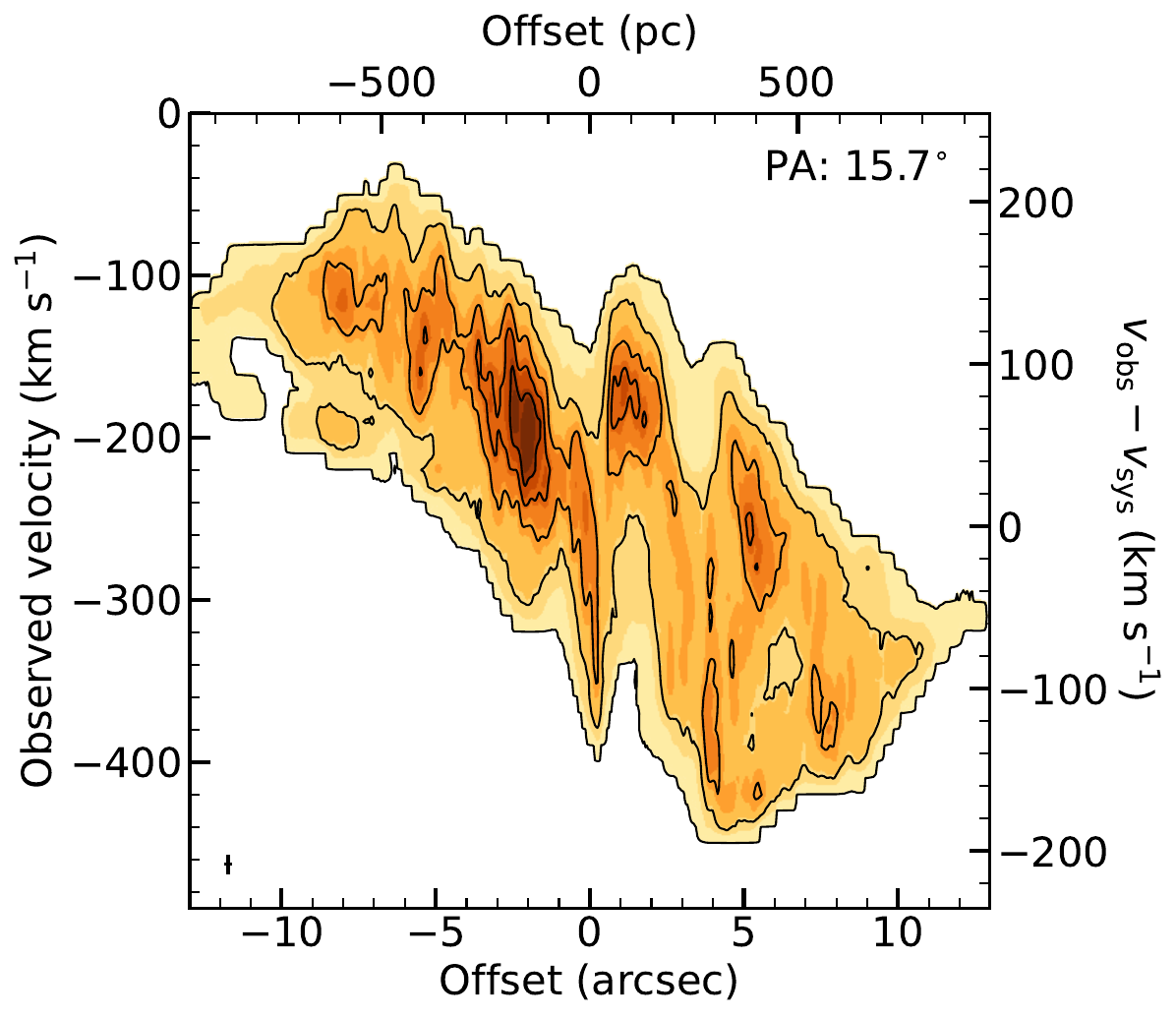}
    \end{subfigure}
    \begin{subfigure}{.34\linewidth}
    \centering
    \includegraphics[height=6.0cm]{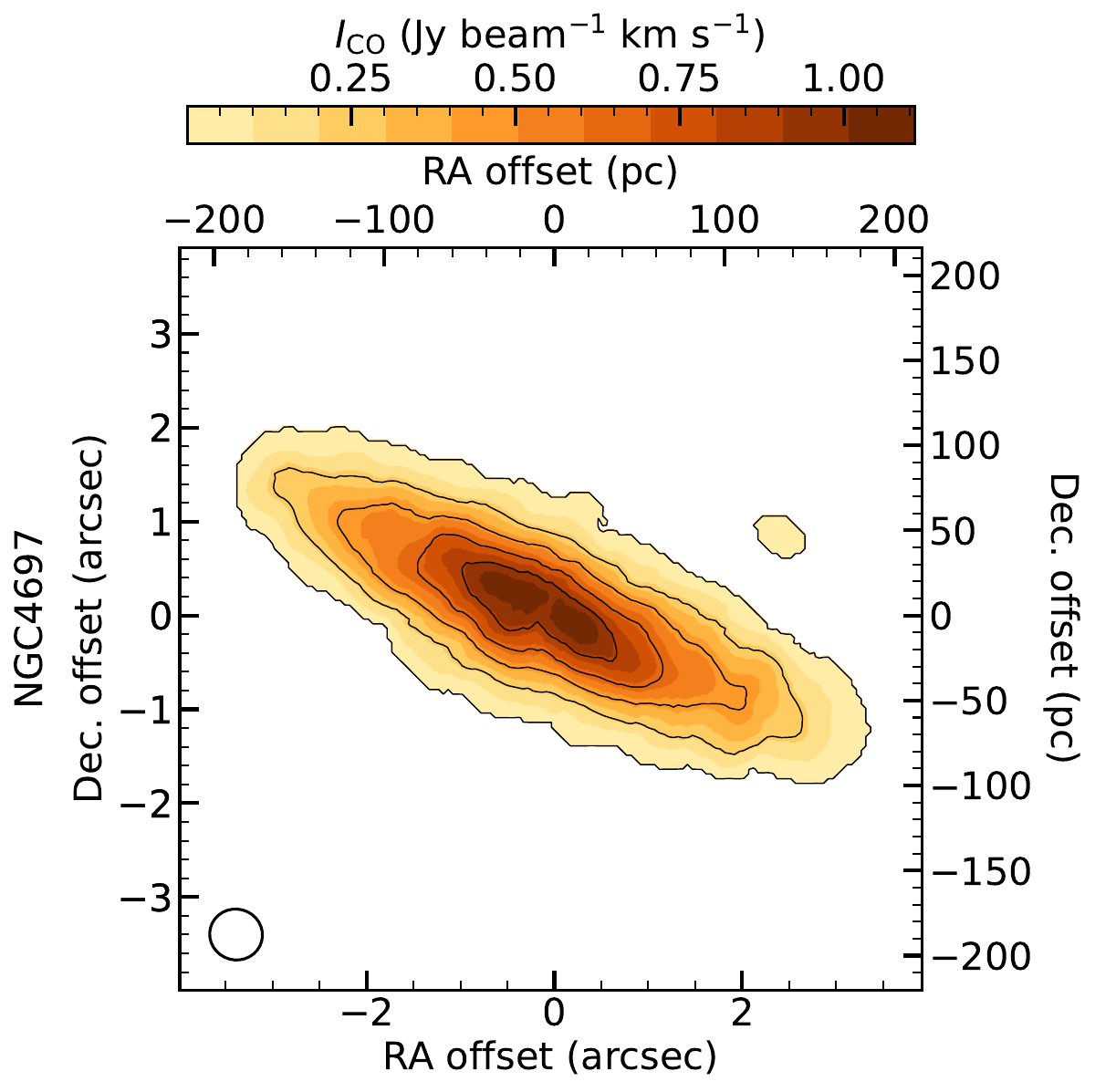}
    \end{subfigure}
    \begin{subfigure}{.32\linewidth}
    \centering
    \includegraphics[height=6.0cm]{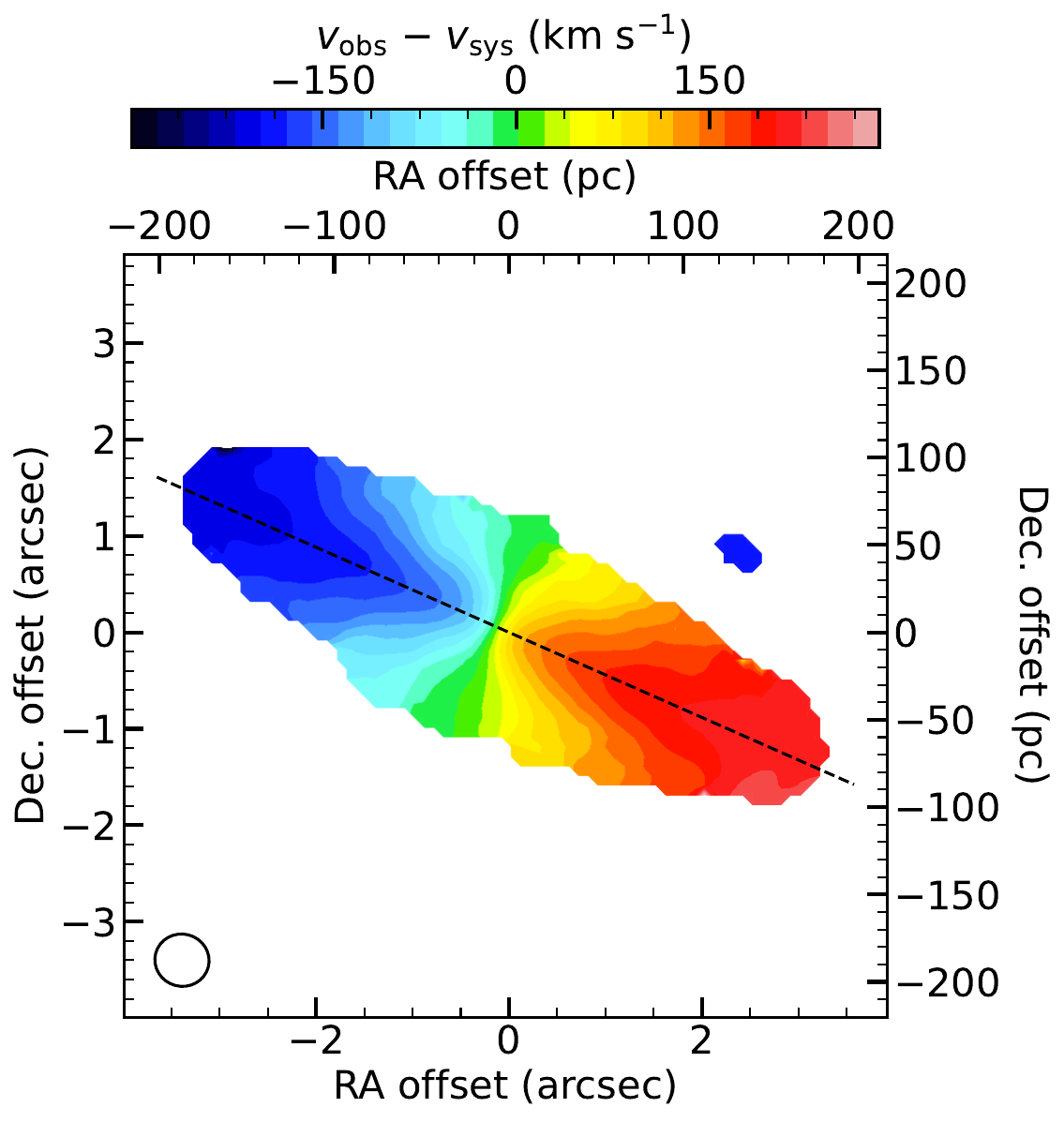}
    \end{subfigure}
    \begin{subfigure}{.32\linewidth}
    \centering
    \includegraphics[height=6.0cm]{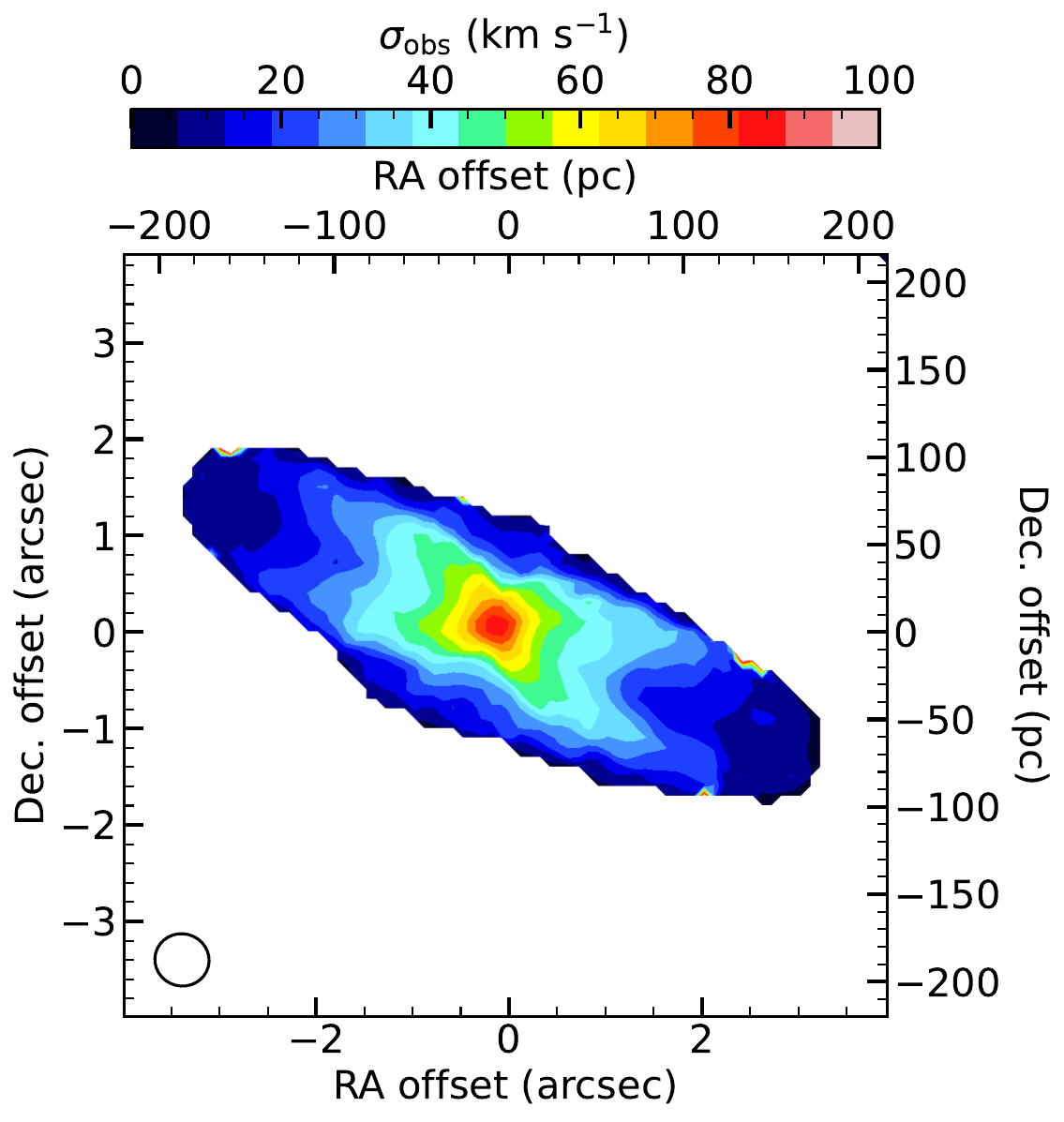}
    \end{subfigure}
    \\
    \begin{subfigure}{.34\linewidth}
    \centering
    \includegraphics[height=5.2cm]{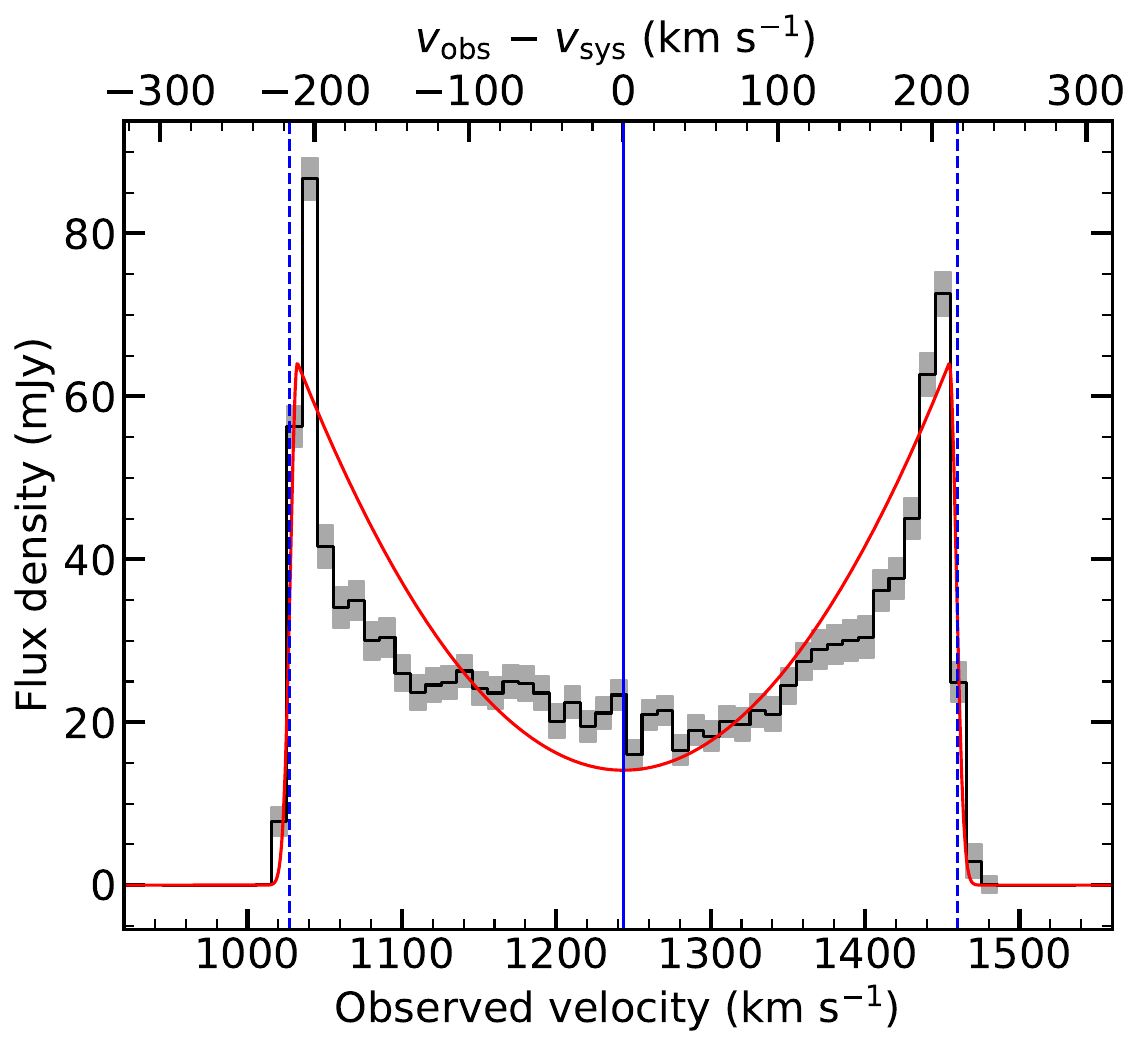}
    \end{subfigure}
    \begin{subfigure}{.32\linewidth}
    \centering
    \includegraphics[height=4.8cm]{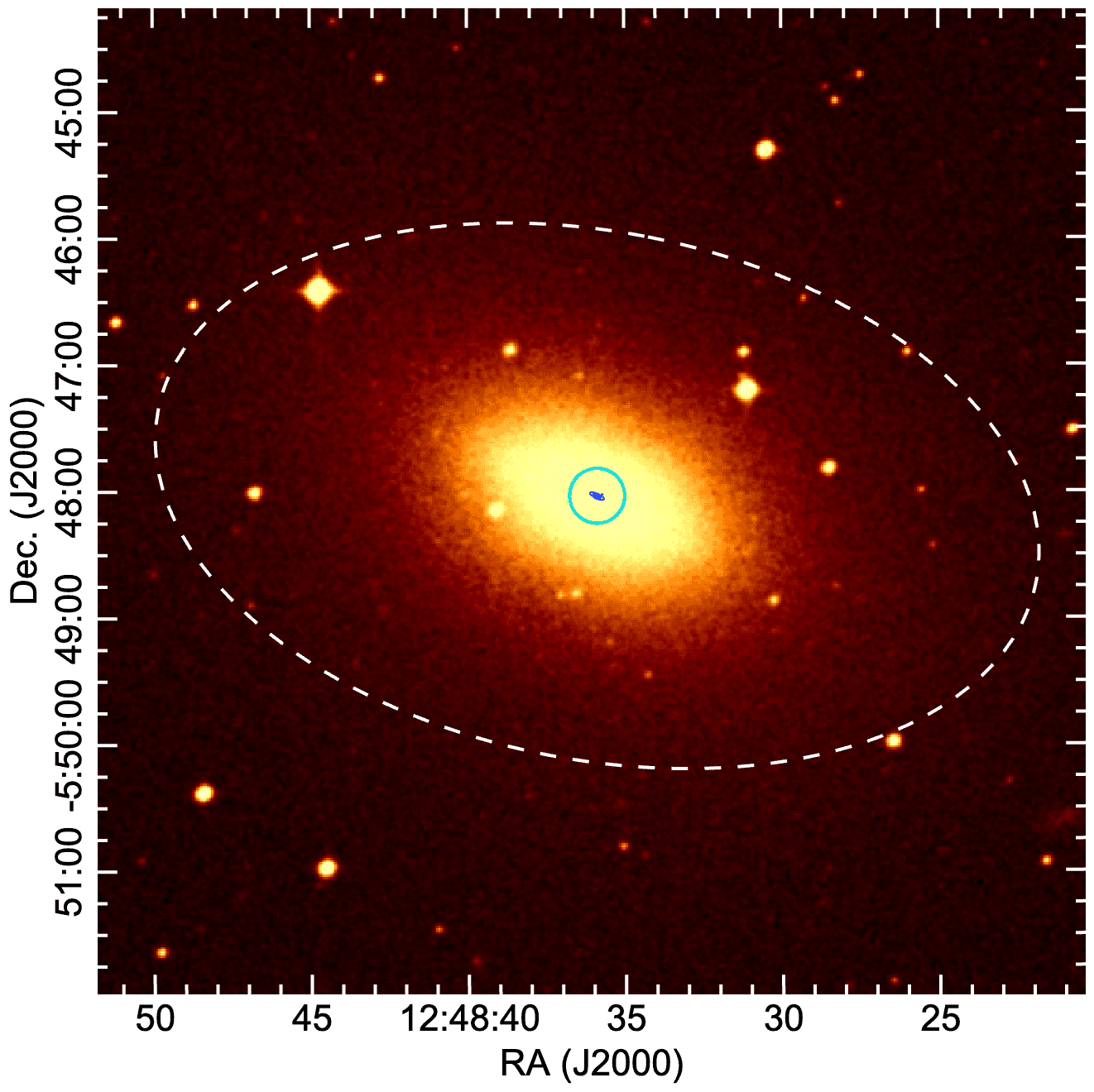}
    \end{subfigure}
    \begin{subfigure}{.33\linewidth}
    \centering
    \includegraphics[height=5.2cm]{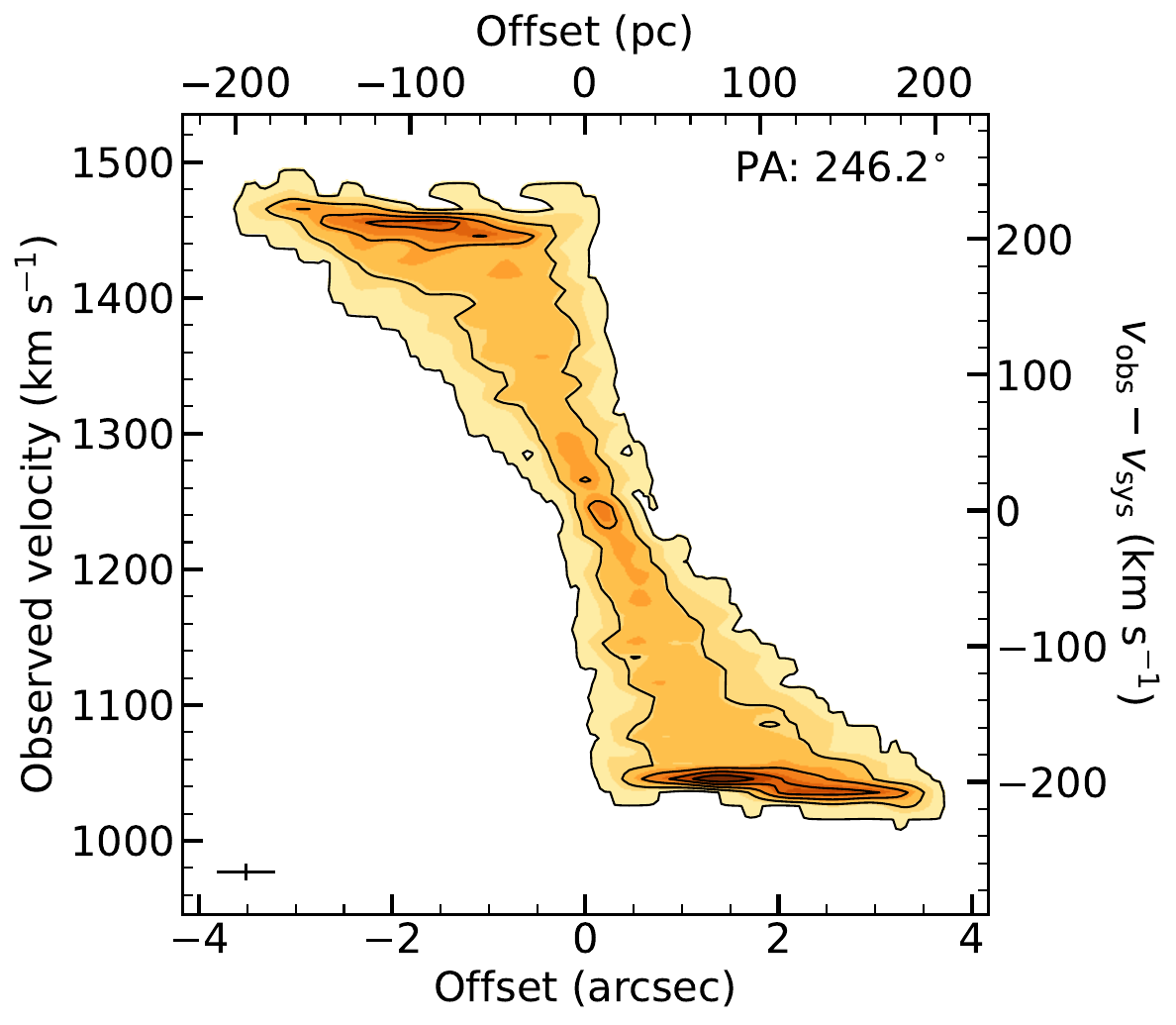}
    \end{subfigure}
    \caption[]{continued.}
\end{figure*}

\begin{figure*}
    \ContinuedFloat
    \begin{subfigure}{.34\linewidth}
    \centering
    \includegraphics[height=6.0cm]{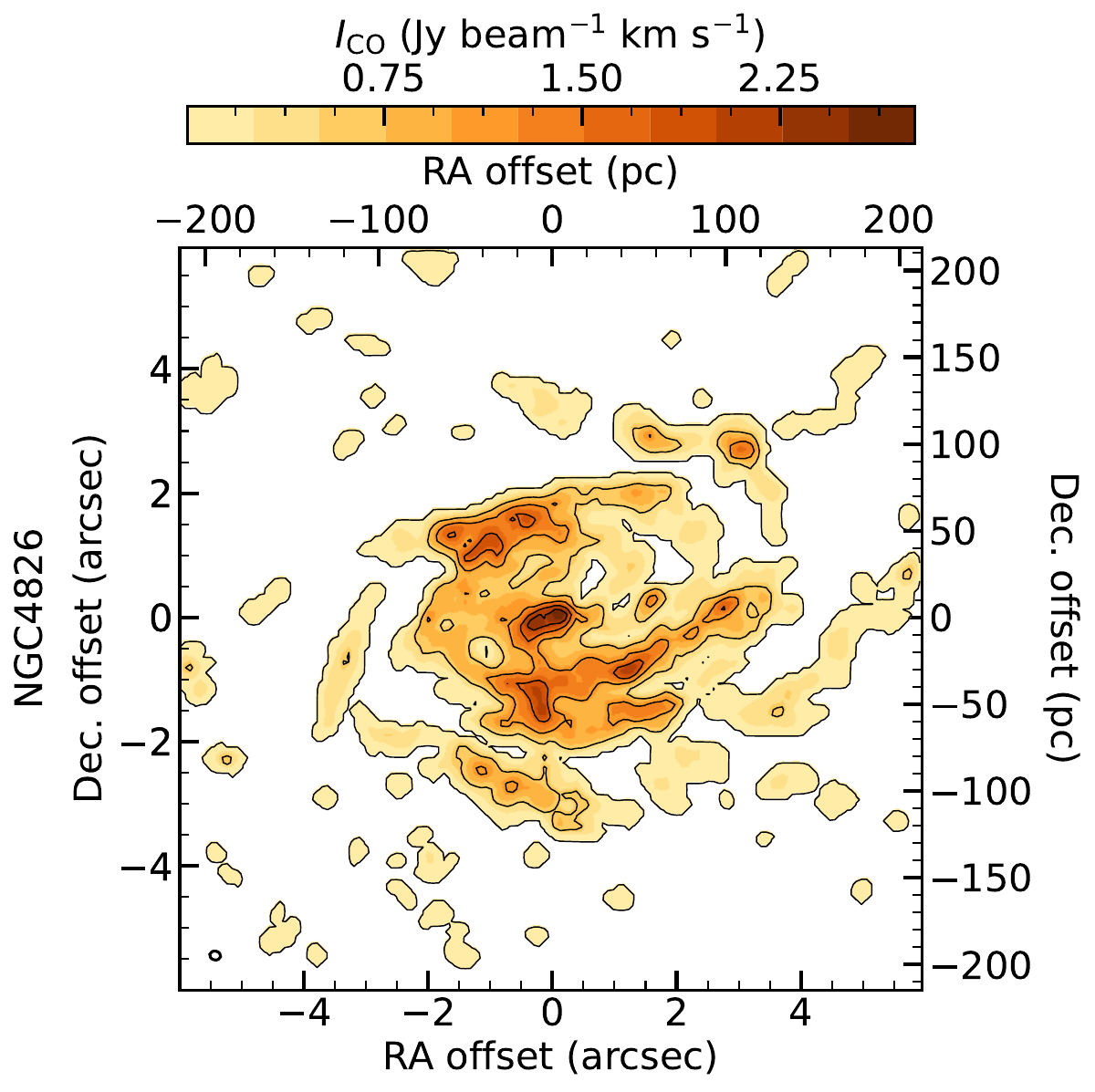}
    \end{subfigure}
    \begin{subfigure}{.32\linewidth}
    \centering
    \includegraphics[height=6.0cm]{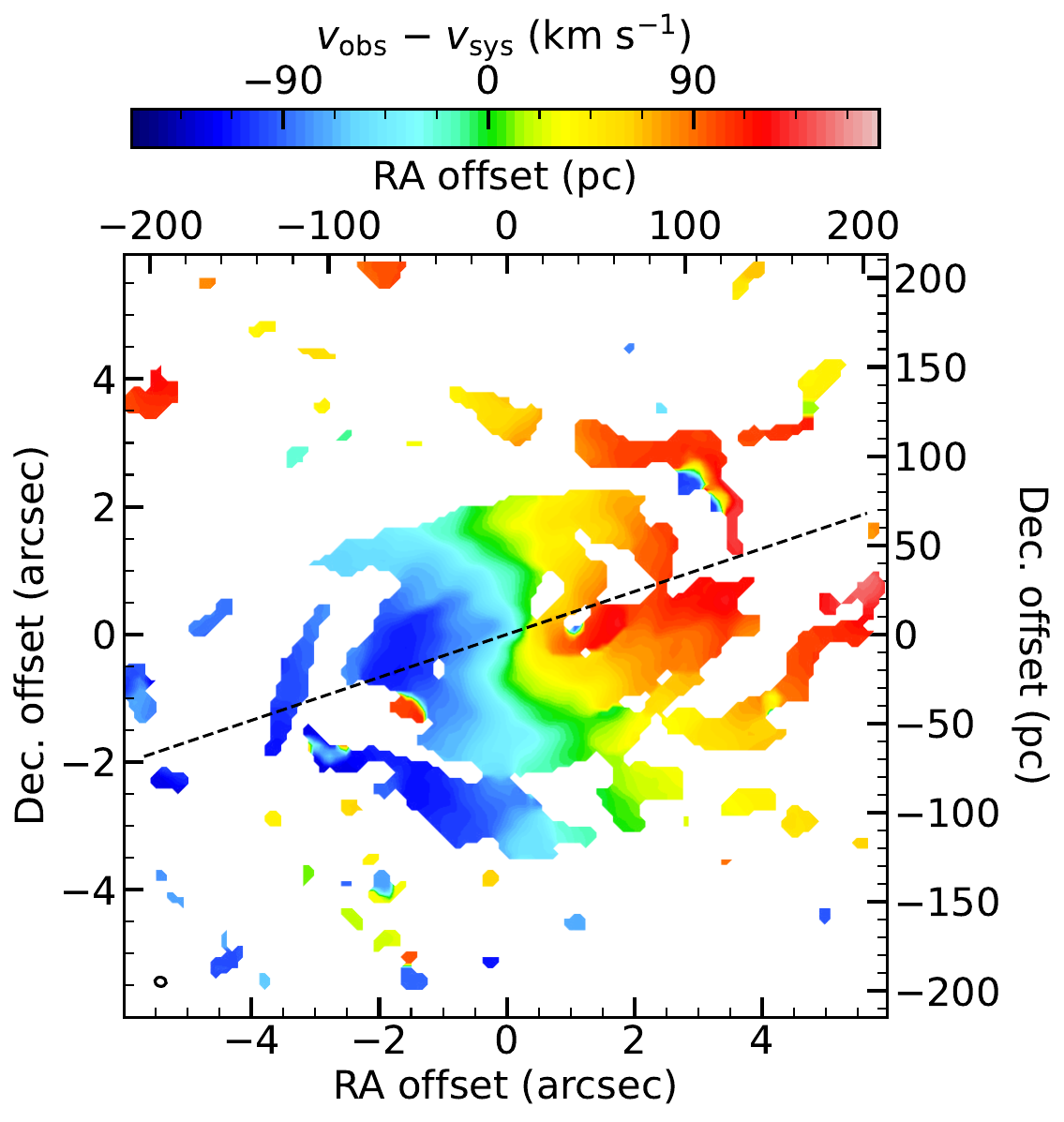}
    \end{subfigure}
    \begin{subfigure}{.32\linewidth}
    \centering
    \includegraphics[height=6.0cm]{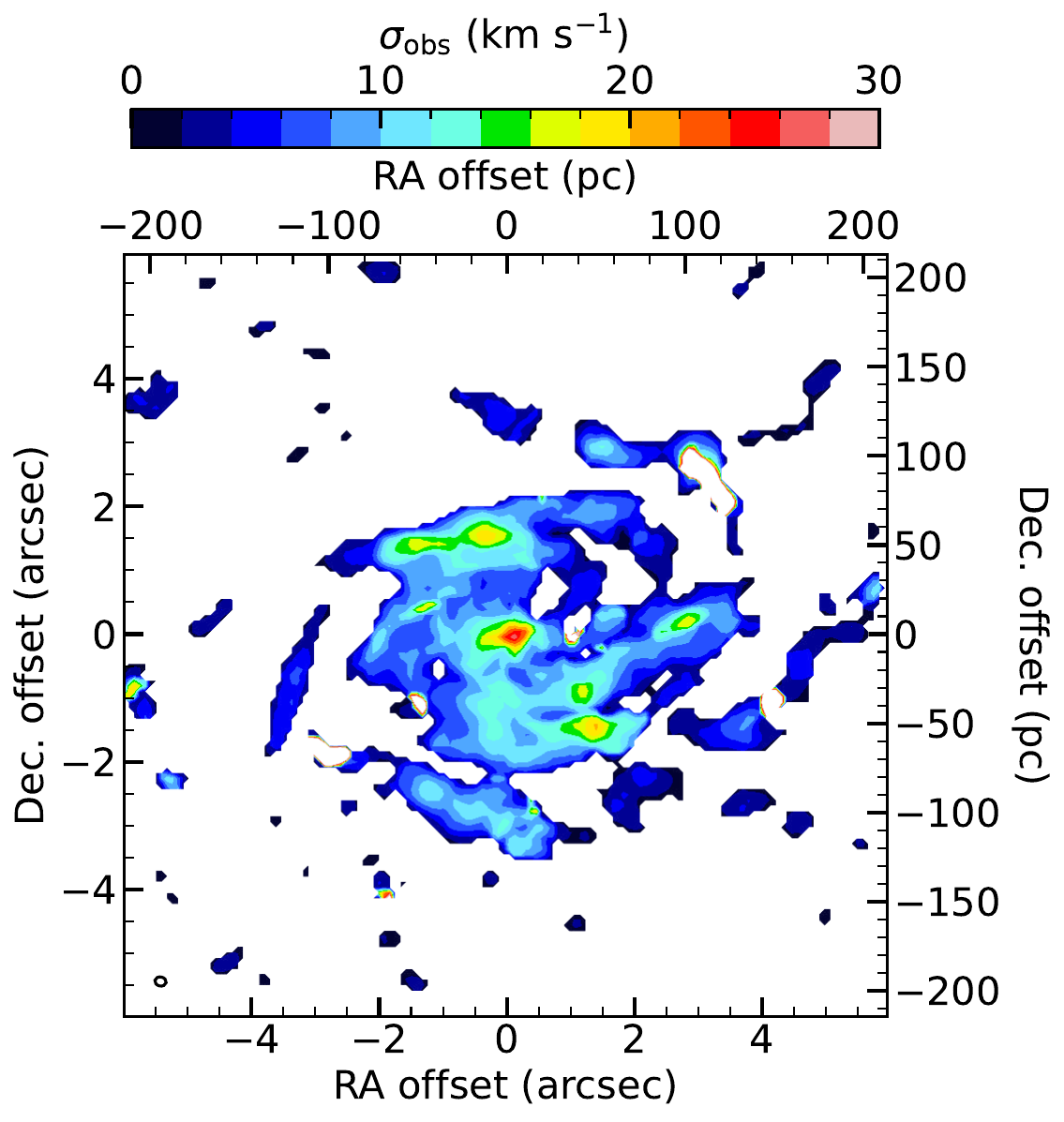}
    \end{subfigure}
    \\
    \begin{subfigure}{.34\linewidth}
    \centering
    \includegraphics[height=5.2cm]{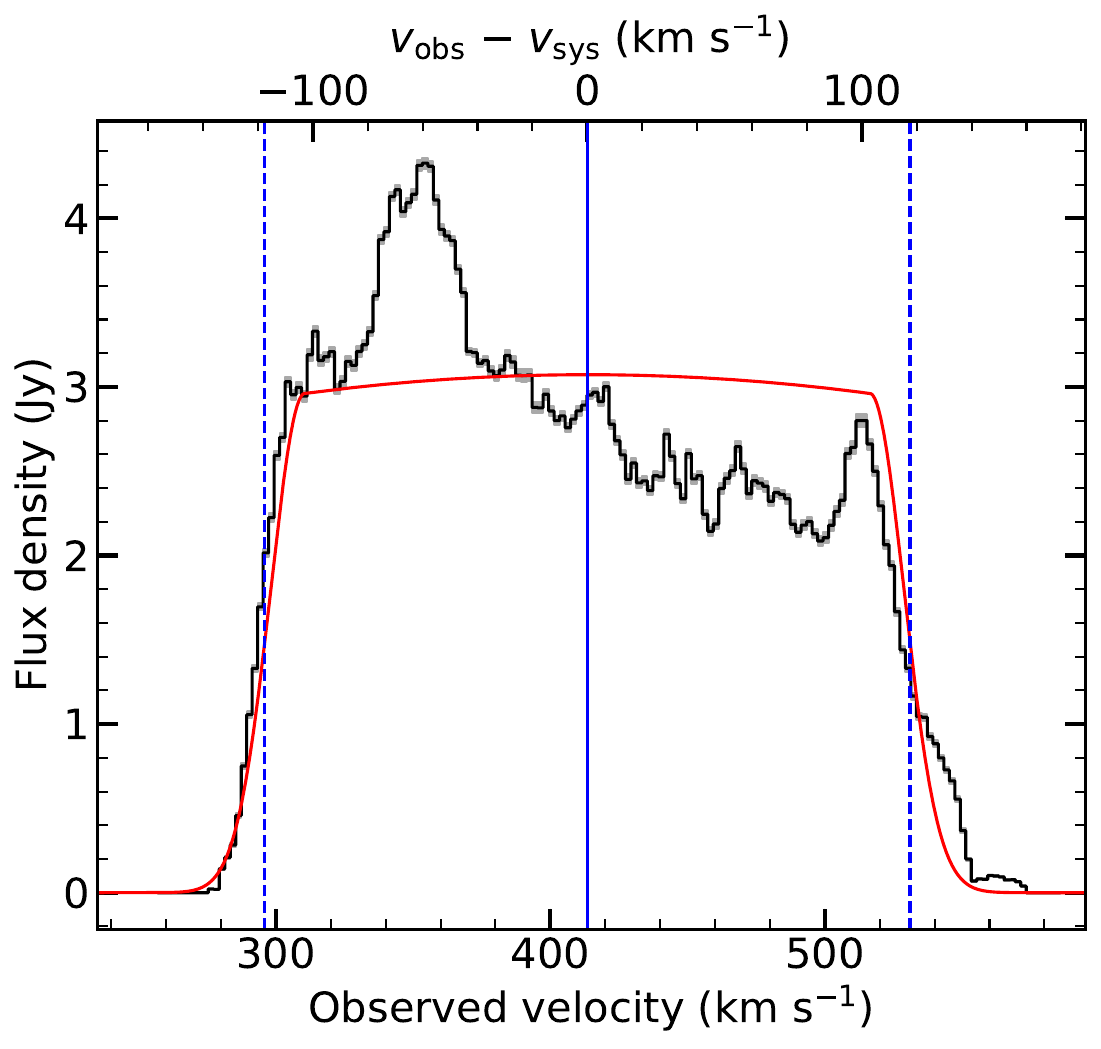}
    \end{subfigure}
    \begin{subfigure}{.32\linewidth}
    \centering
    \includegraphics[height=4.8cm]{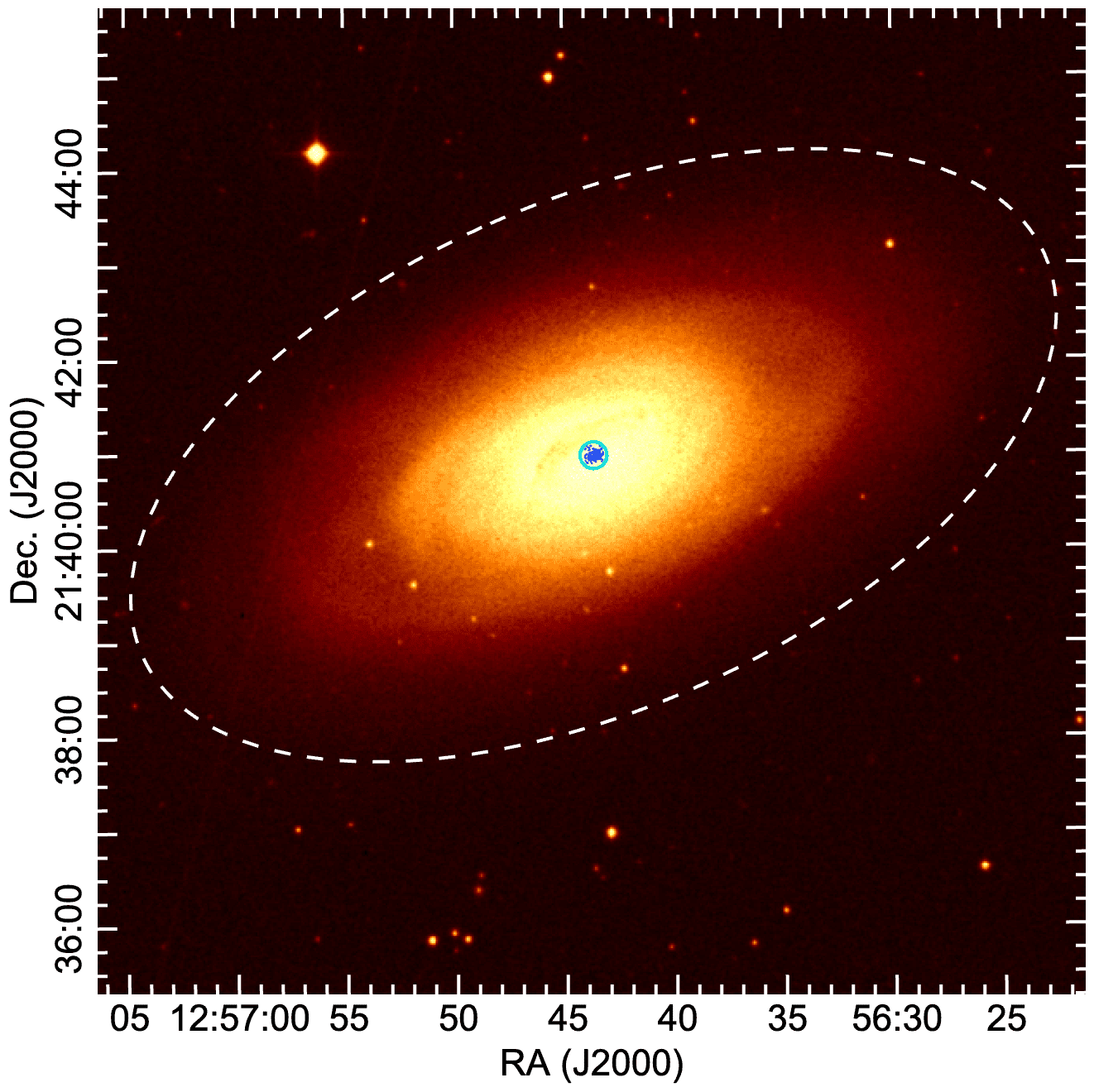}
    \end{subfigure}
    \begin{subfigure}{.33\linewidth}
    \centering
    \includegraphics[height=5.2cm]{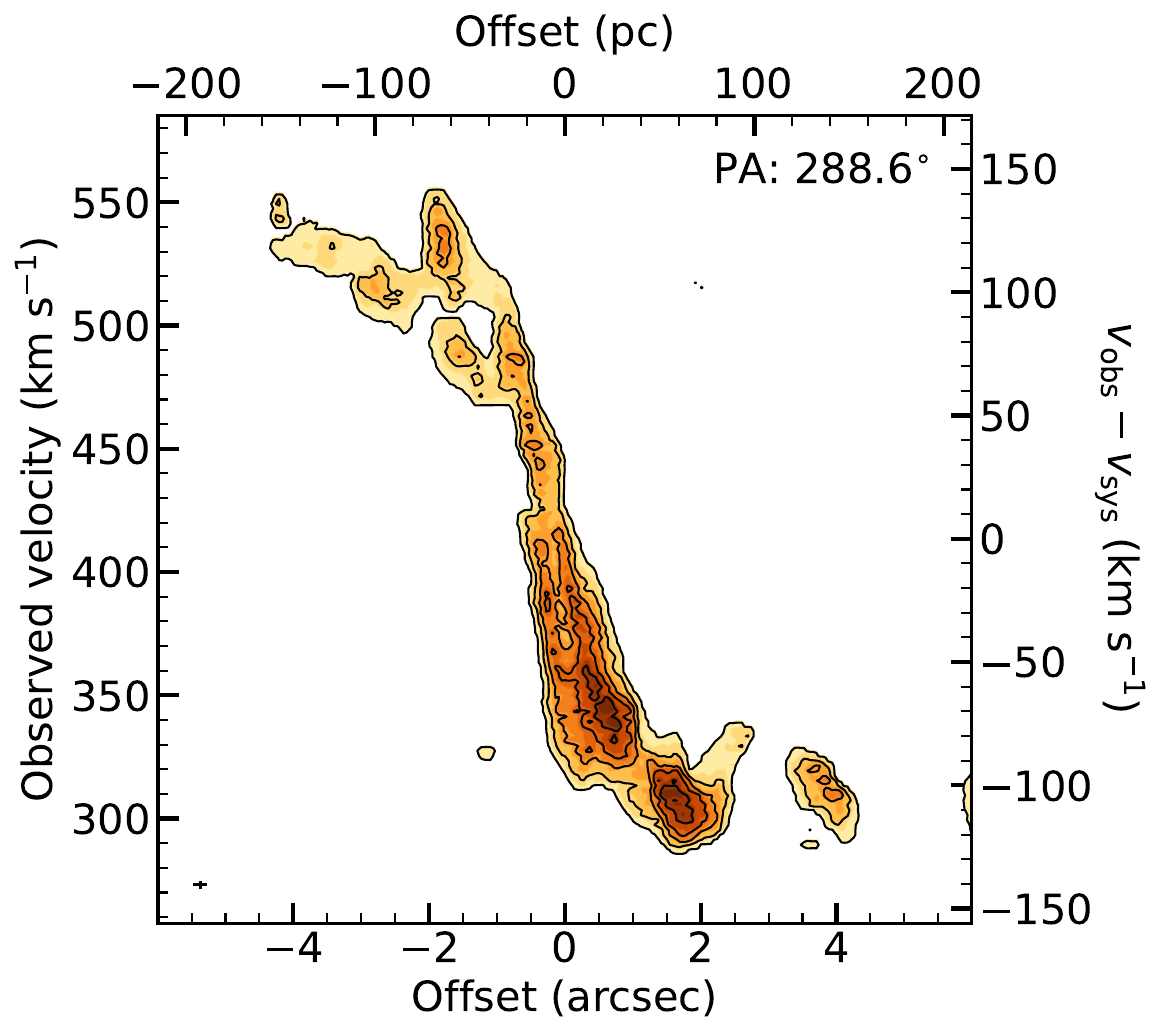}
    \end{subfigure}
    \begin{subfigure}{.34\linewidth}
    \centering
    \includegraphics[height=6.0cm]{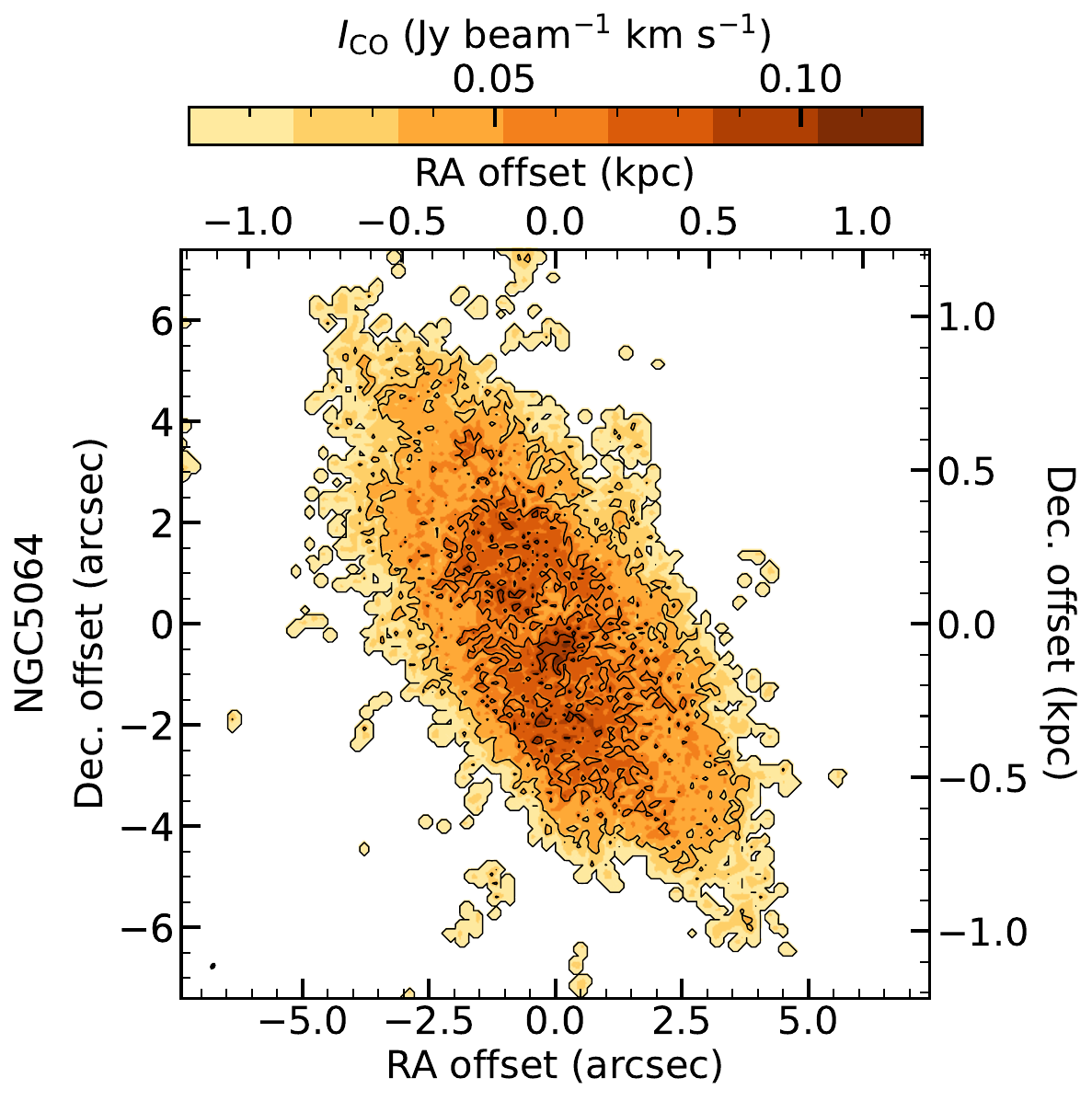}
    \end{subfigure}
    \begin{subfigure}{.32\linewidth}
    \centering
    \includegraphics[height=6.0cm]{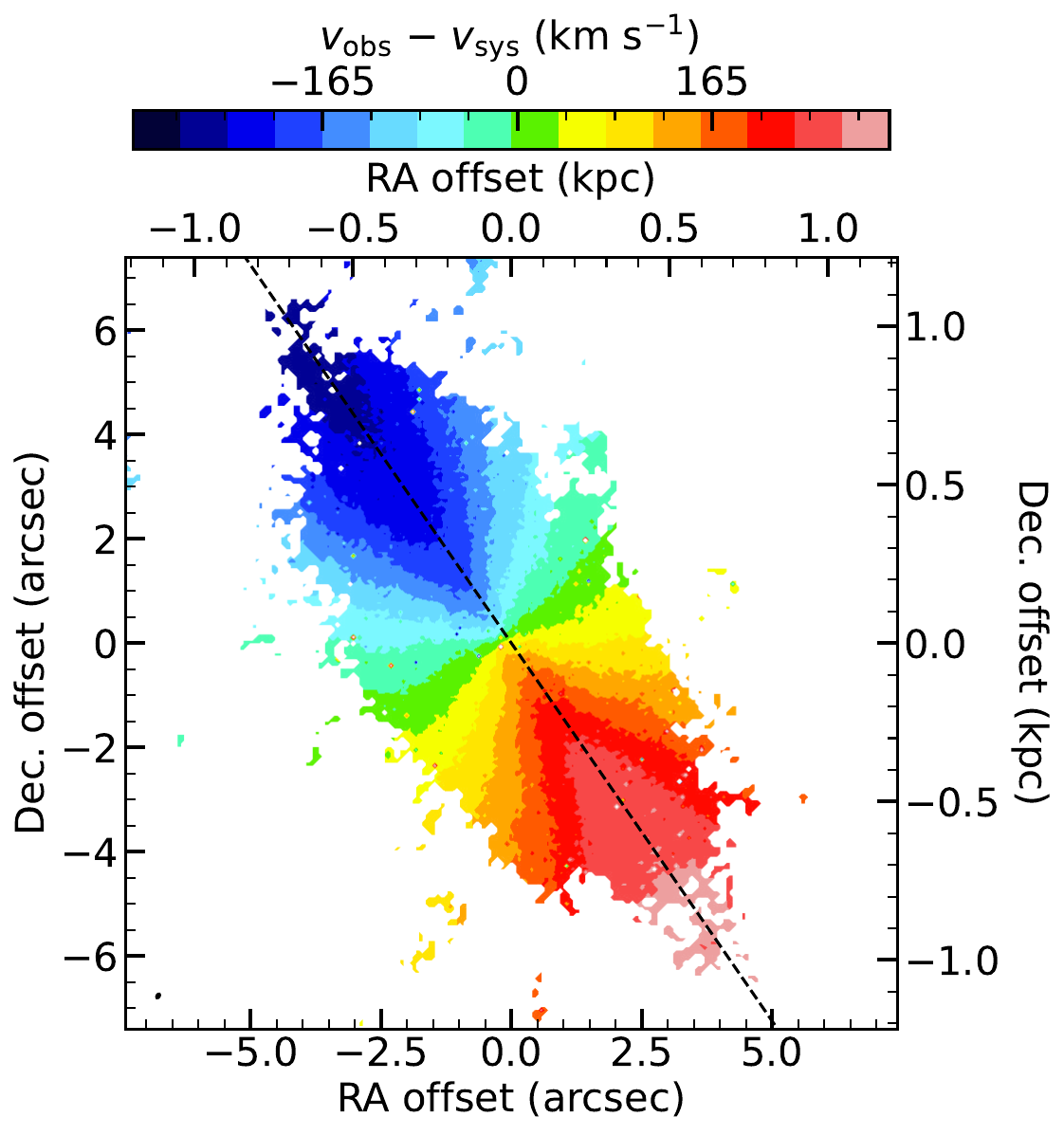}
    \end{subfigure}
    \begin{subfigure}{.32\linewidth}
    \centering
    \includegraphics[height=6.0cm]{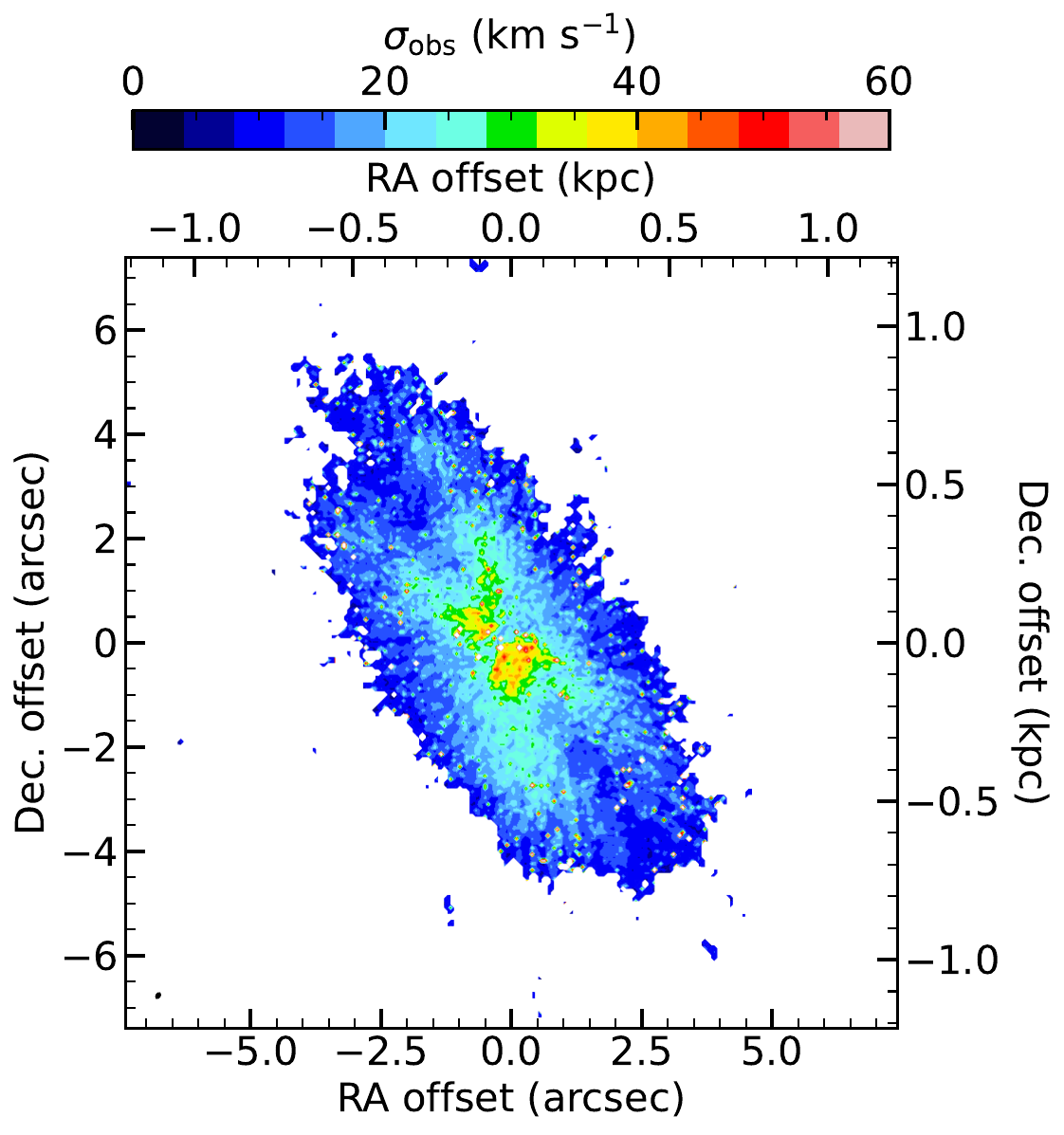}
    \end{subfigure}
    \\
    \begin{subfigure}{.34\linewidth}
    \centering
    \includegraphics[height=5.2cm]{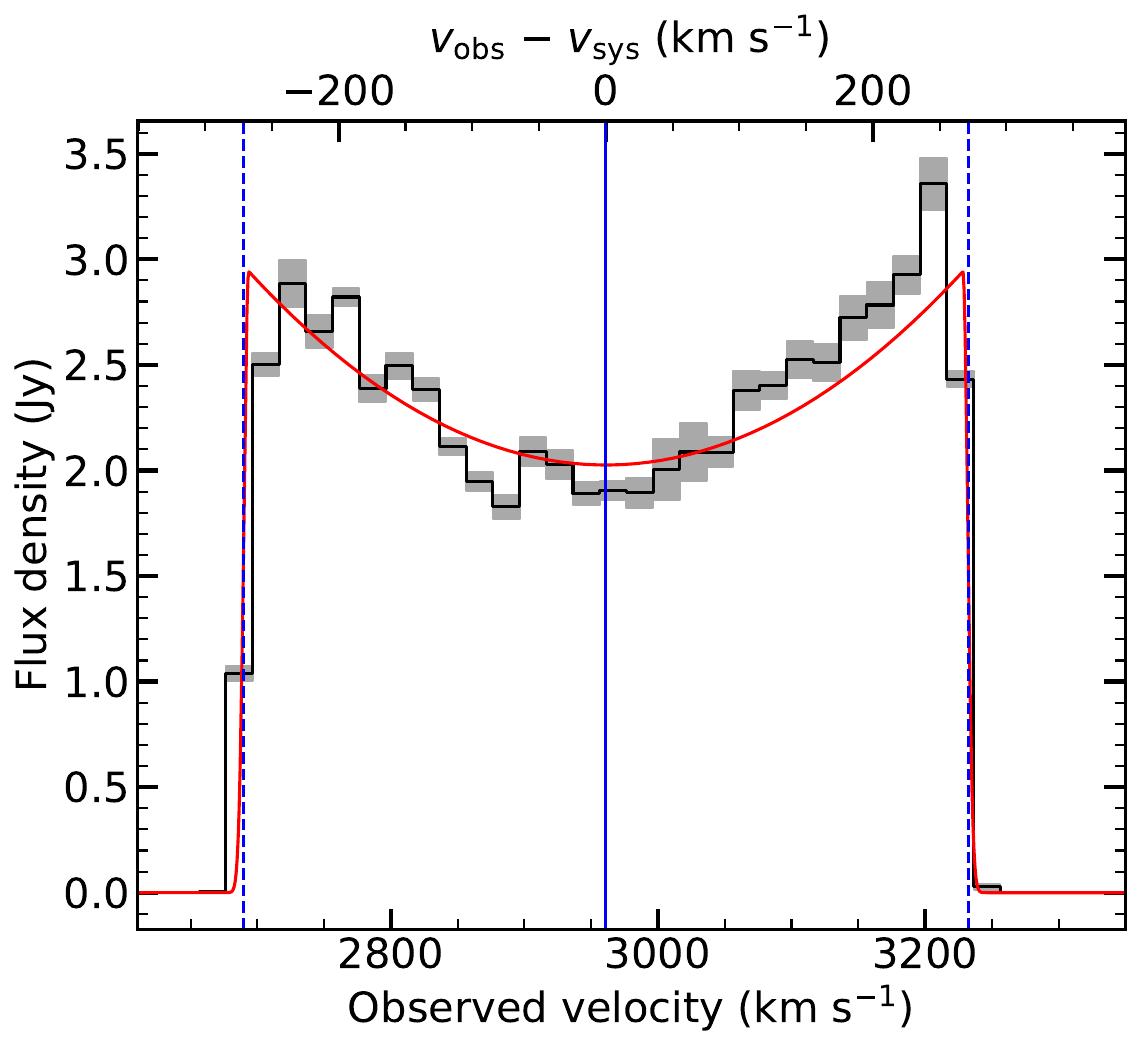}
    \end{subfigure}
    \begin{subfigure}{.32\linewidth}
    \centering
    \includegraphics[height=4.8cm]{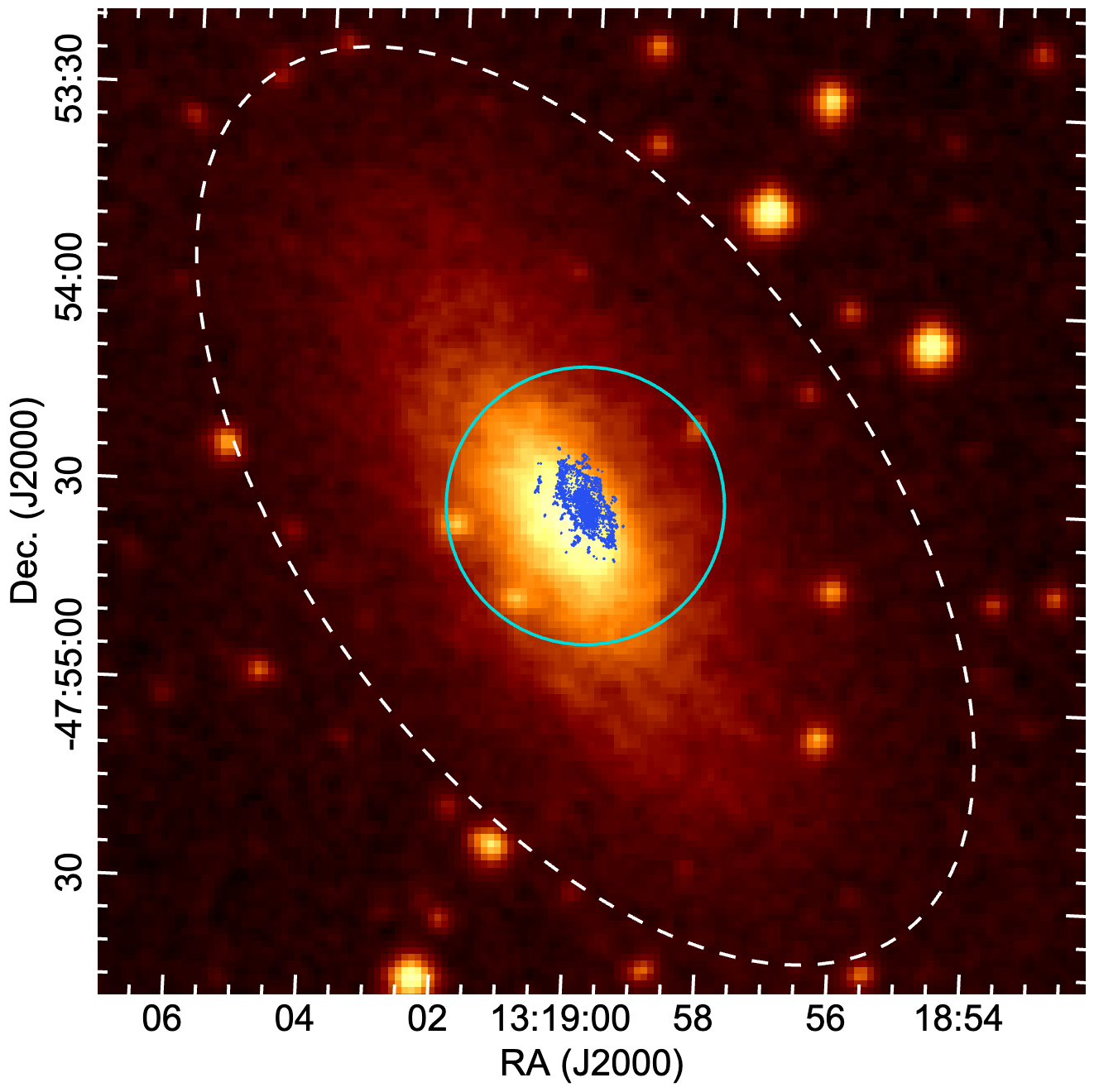}
    \end{subfigure}
    \begin{subfigure}{.33\linewidth}
    \centering
    \includegraphics[height=5.2cm]{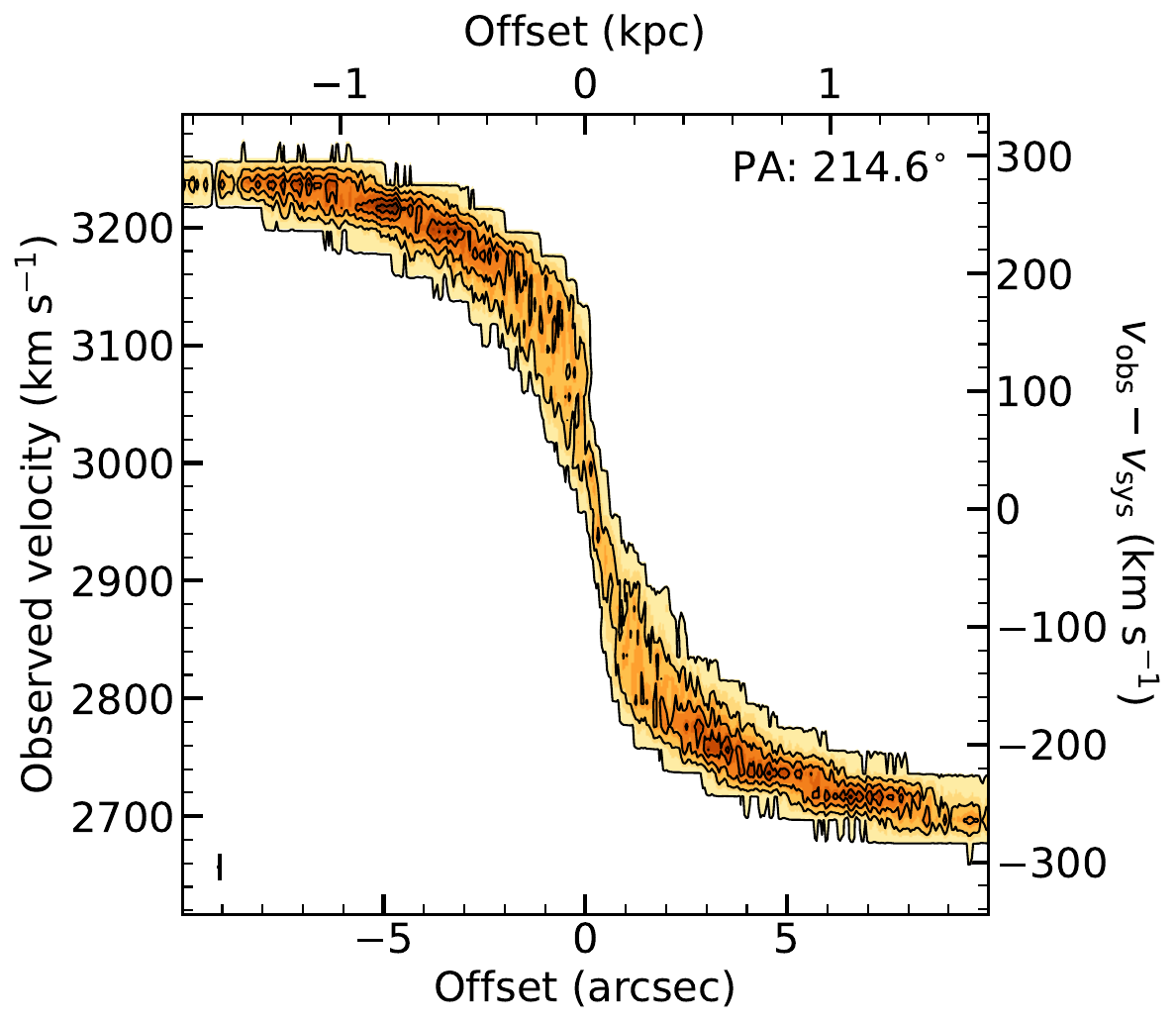}
    \end{subfigure}
    \caption[]{continued.}
\end{figure*}

\begin{figure*}
    \ContinuedFloat
    \begin{subfigure}{.34\linewidth}
    \centering
    \includegraphics[height=6.0cm]{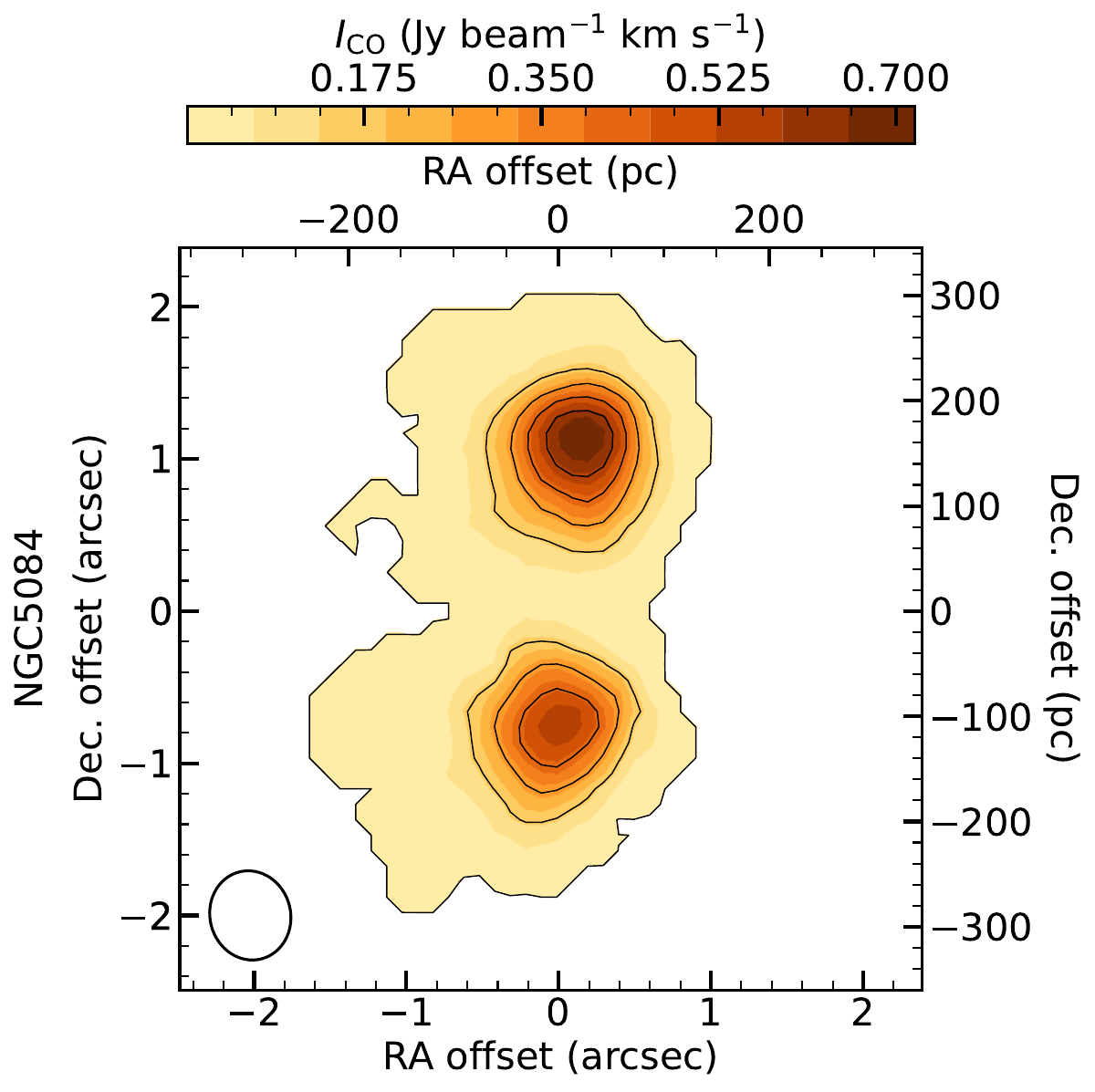}
    \end{subfigure}
    \begin{subfigure}{.32\linewidth}
    \centering
    \includegraphics[height=6.0cm]{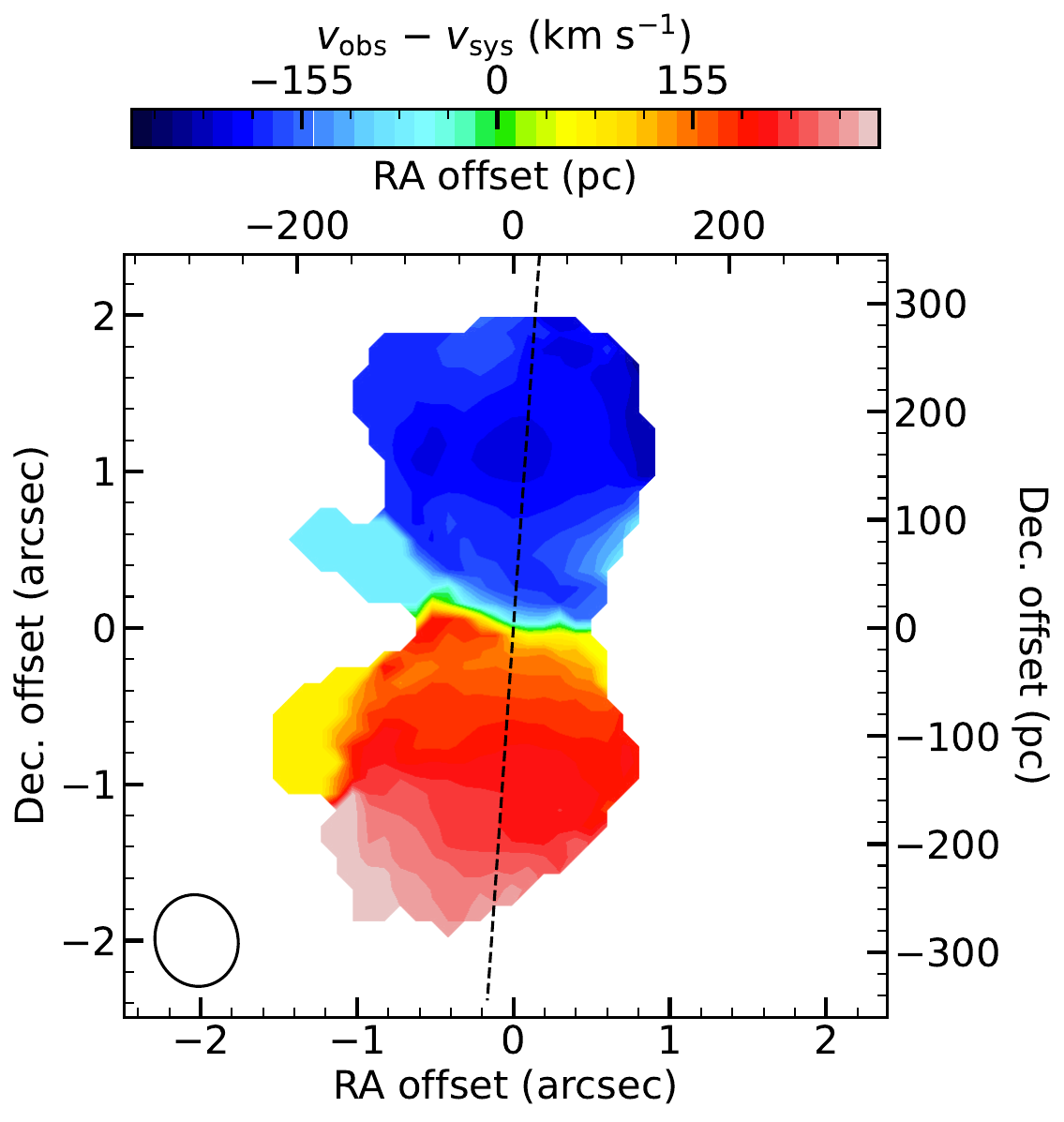}
    \end{subfigure}
    \begin{subfigure}{.32\linewidth}
    \centering
    \includegraphics[height=6.0cm]{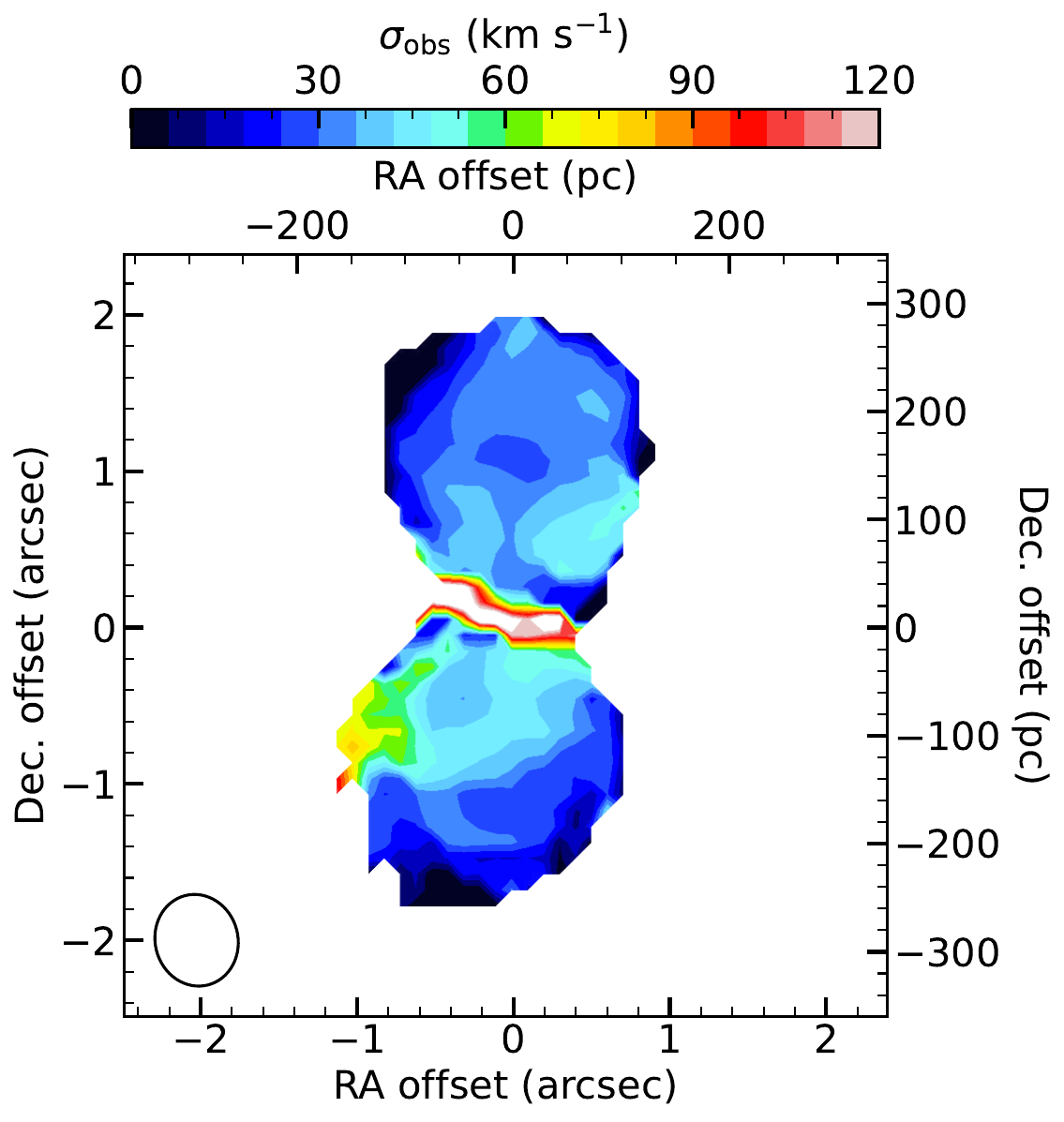}
    \end{subfigure}
    \\
    \begin{subfigure}{.34\linewidth}
    \centering
    \includegraphics[height=5.2cm]{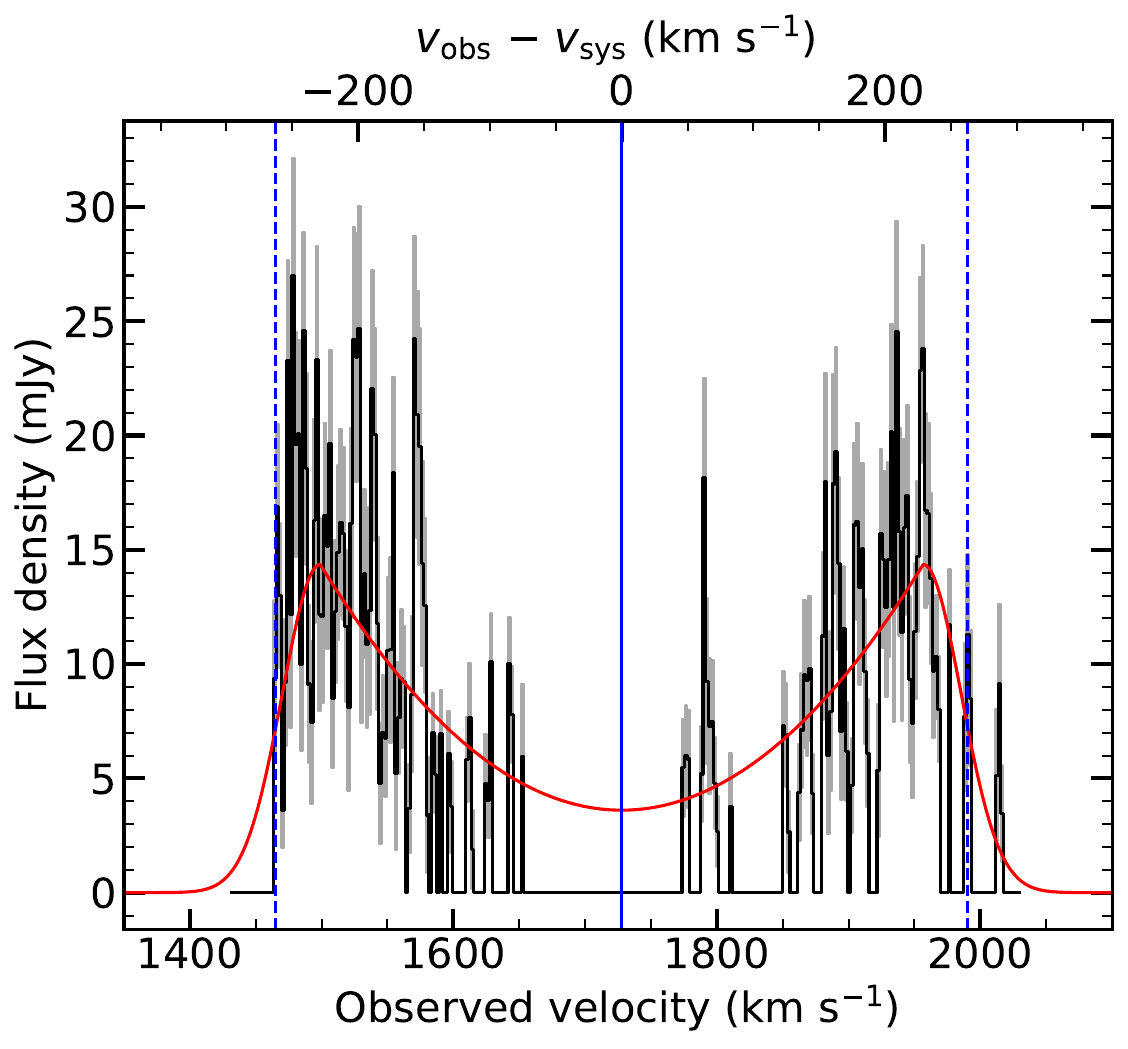}
    \end{subfigure}
    \begin{subfigure}{.32\linewidth}
    \centering
    \includegraphics[height=4.8cm]{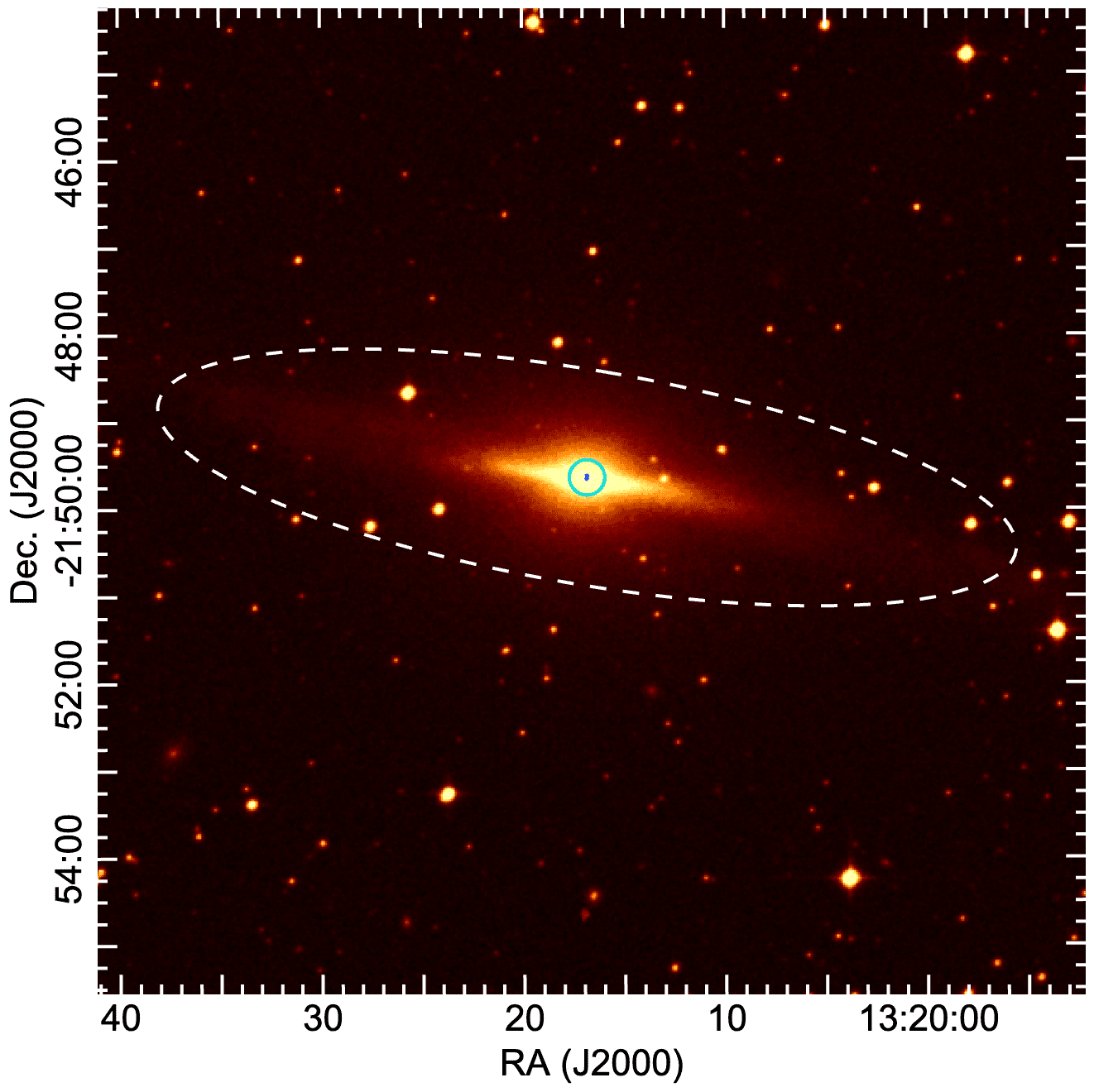}
    \end{subfigure}
    \begin{subfigure}{.33\linewidth}
    \centering
    \includegraphics[height=5.2cm]{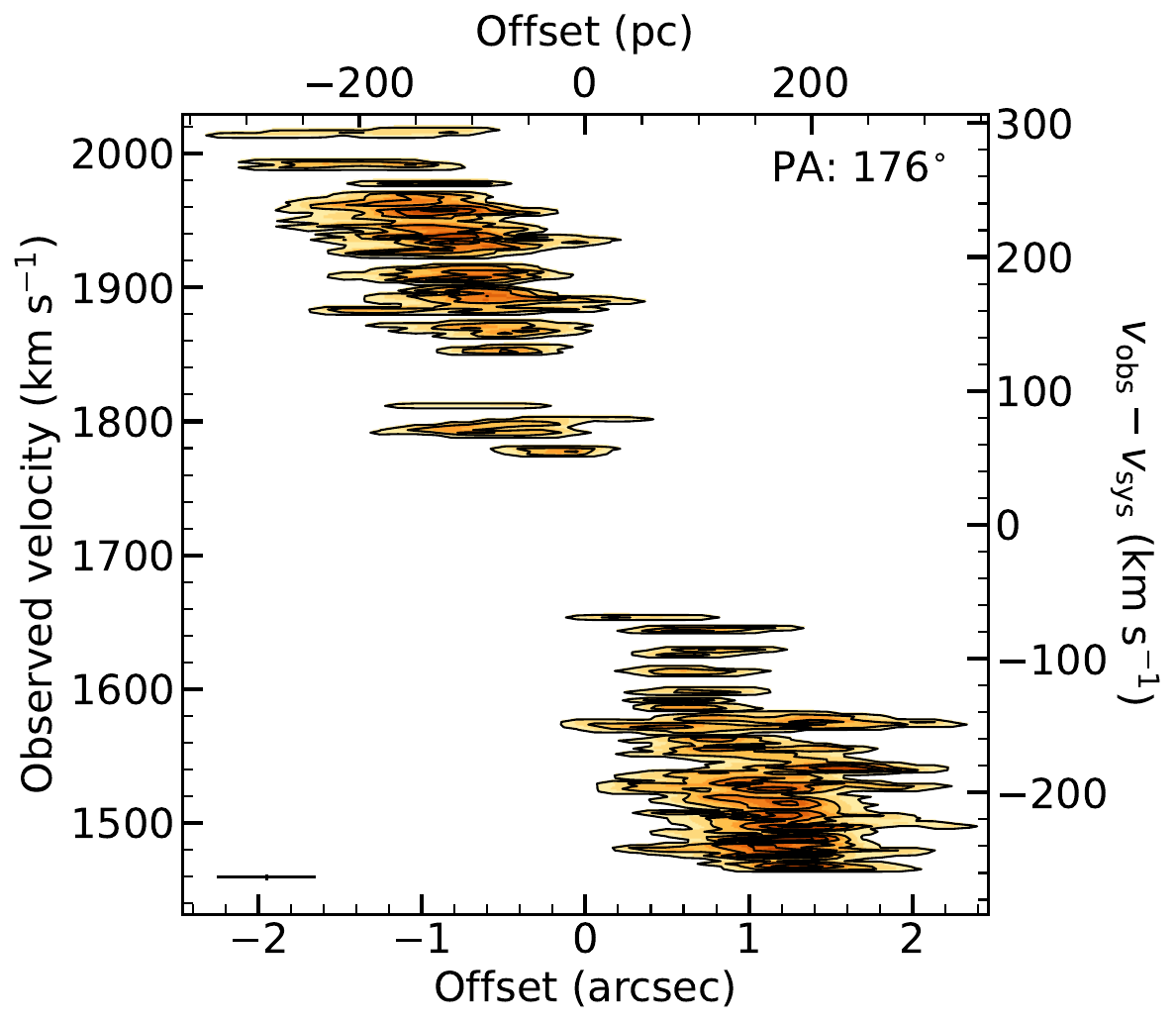}
    \end{subfigure}
    \begin{subfigure}{.34\linewidth}
    \centering
    \includegraphics[height=6.0cm]{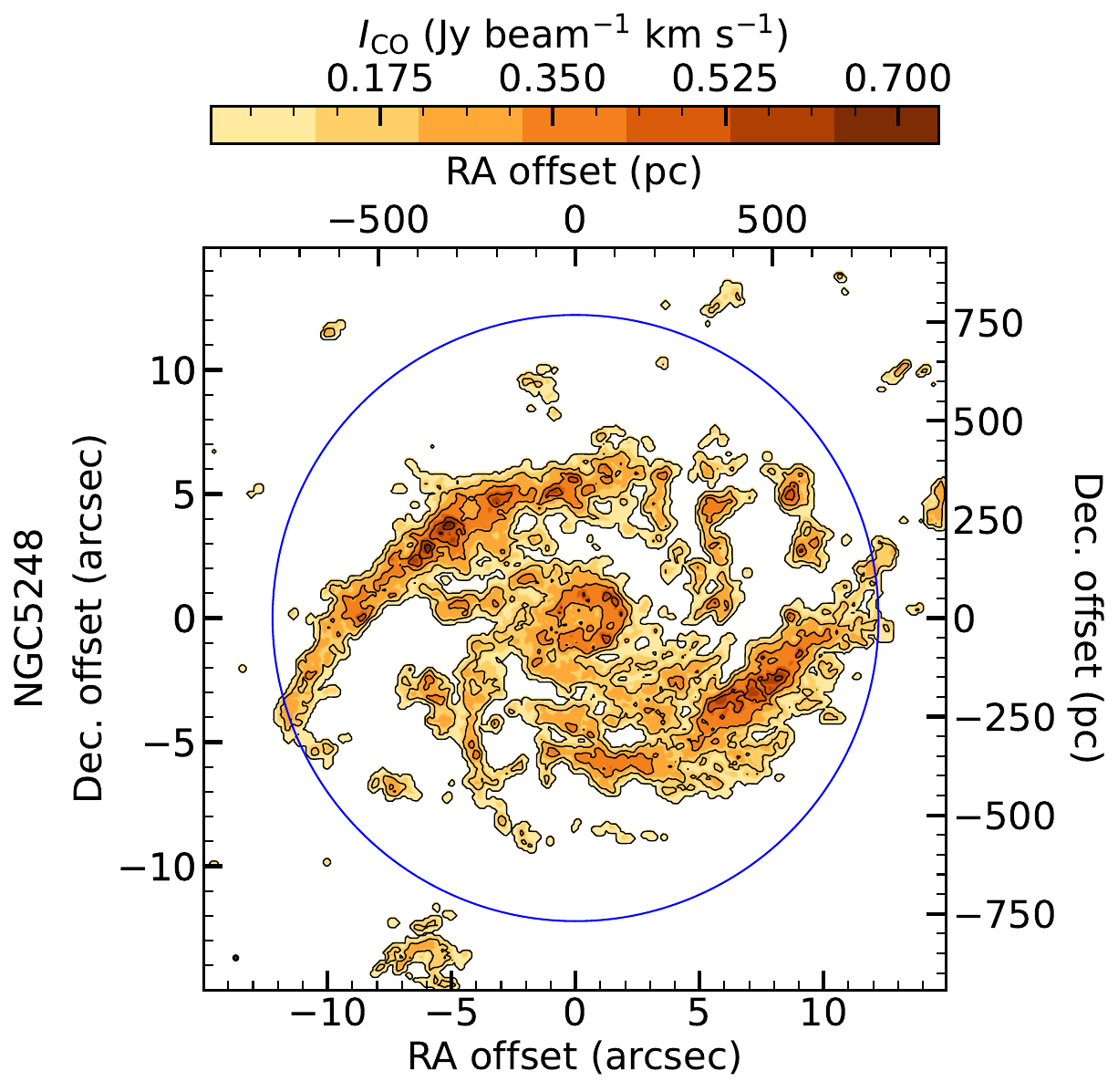}
    \end{subfigure}
    \begin{subfigure}{.32\linewidth}
    \centering
    \includegraphics[height=6.0cm]{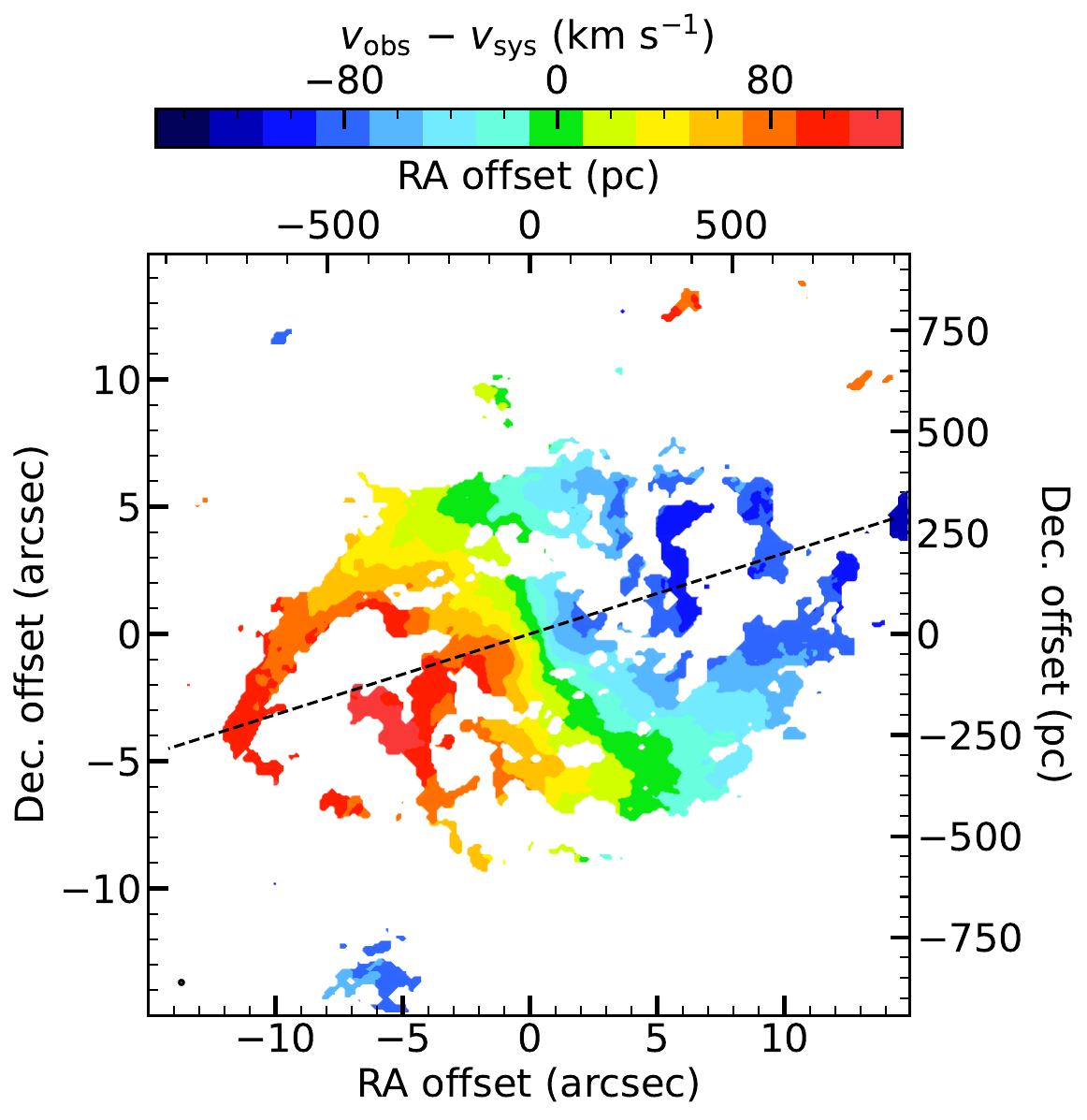}
    \end{subfigure}
    \begin{subfigure}{.32\linewidth}
    \centering
    \includegraphics[height=6.0cm]{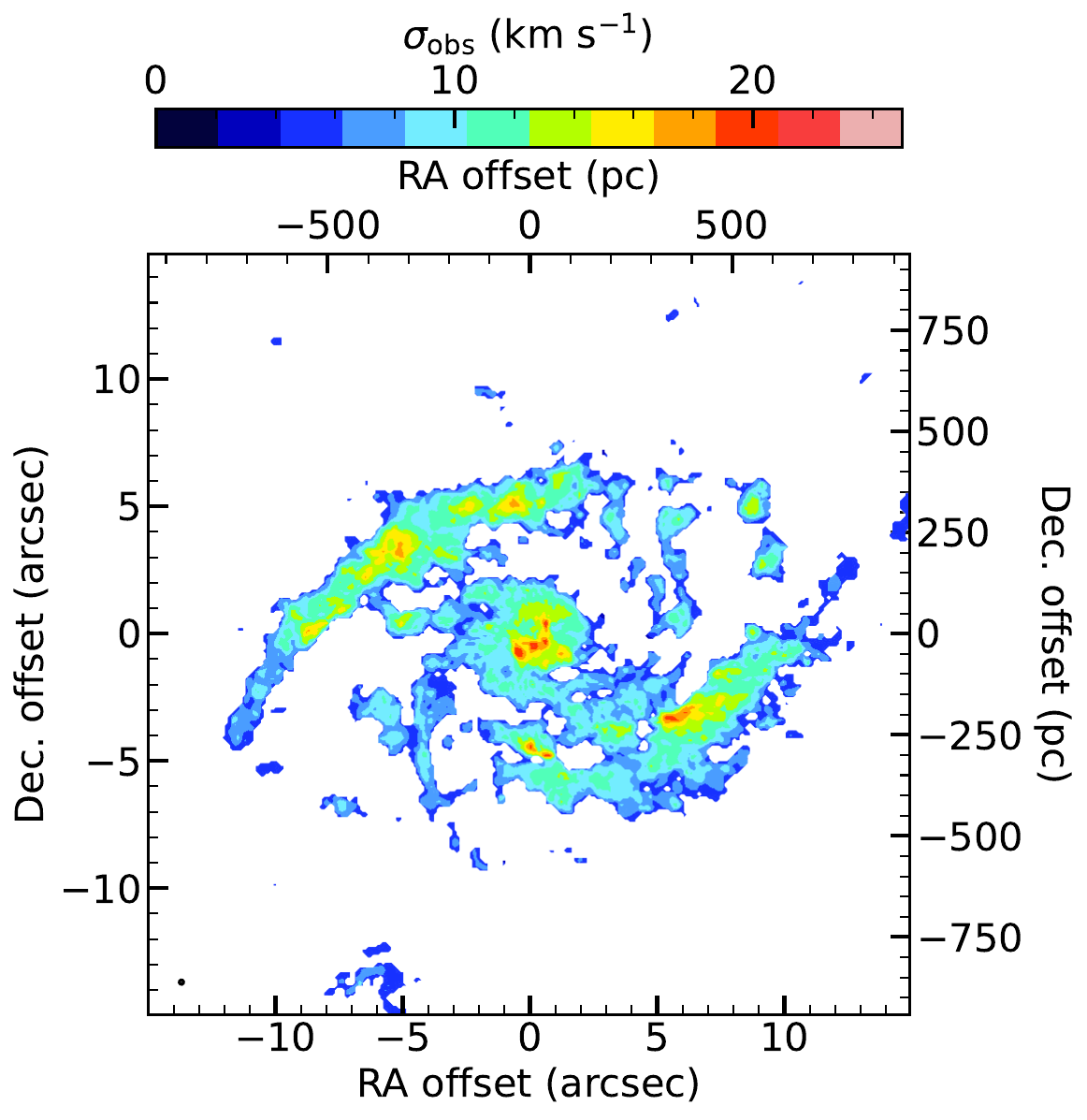}
    \end{subfigure}
    \\
    \begin{subfigure}{.34\linewidth}
    \centering
    \includegraphics[height=5.2cm]{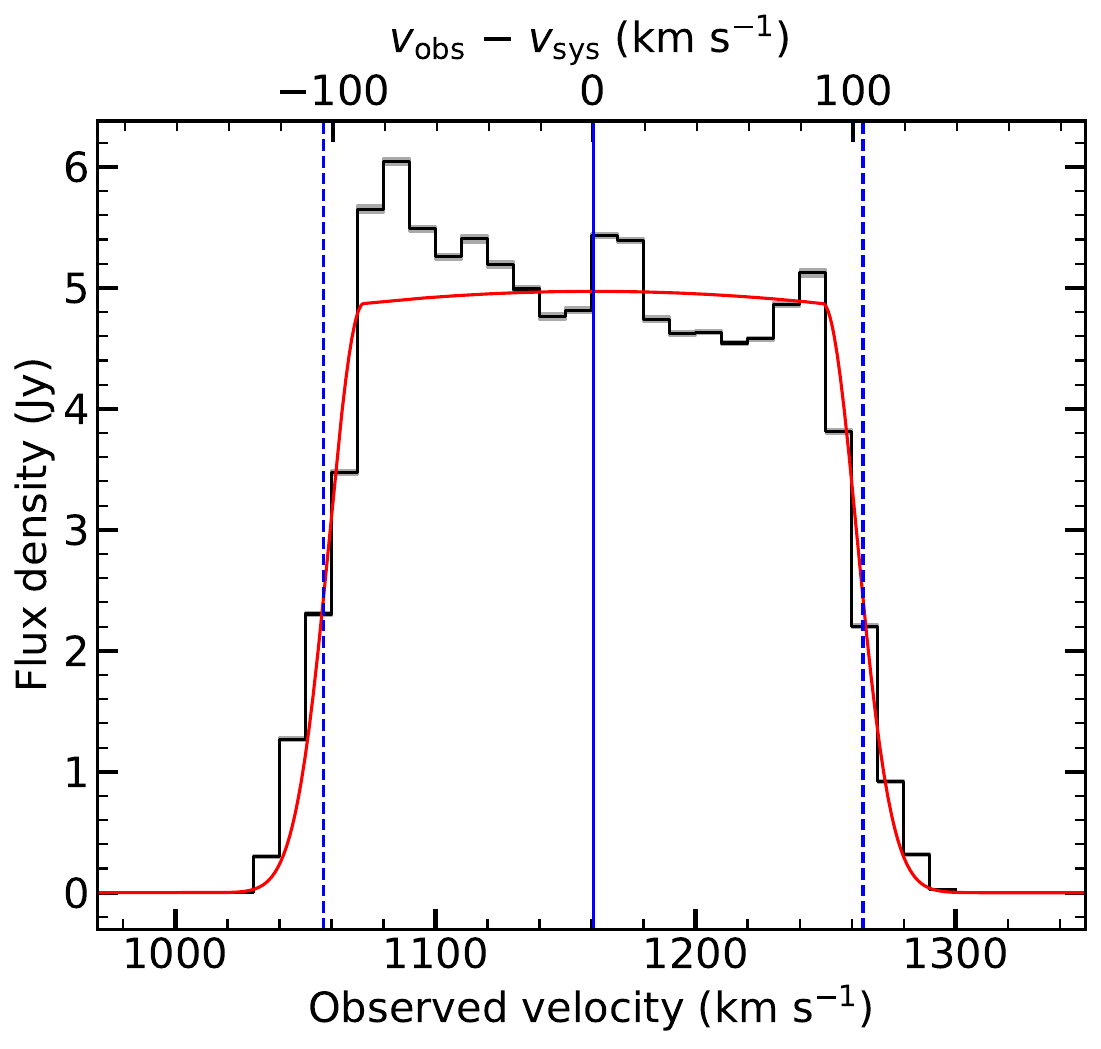}
    \end{subfigure}
    \begin{subfigure}{.32\linewidth}
    \centering
    \includegraphics[height=4.8cm]{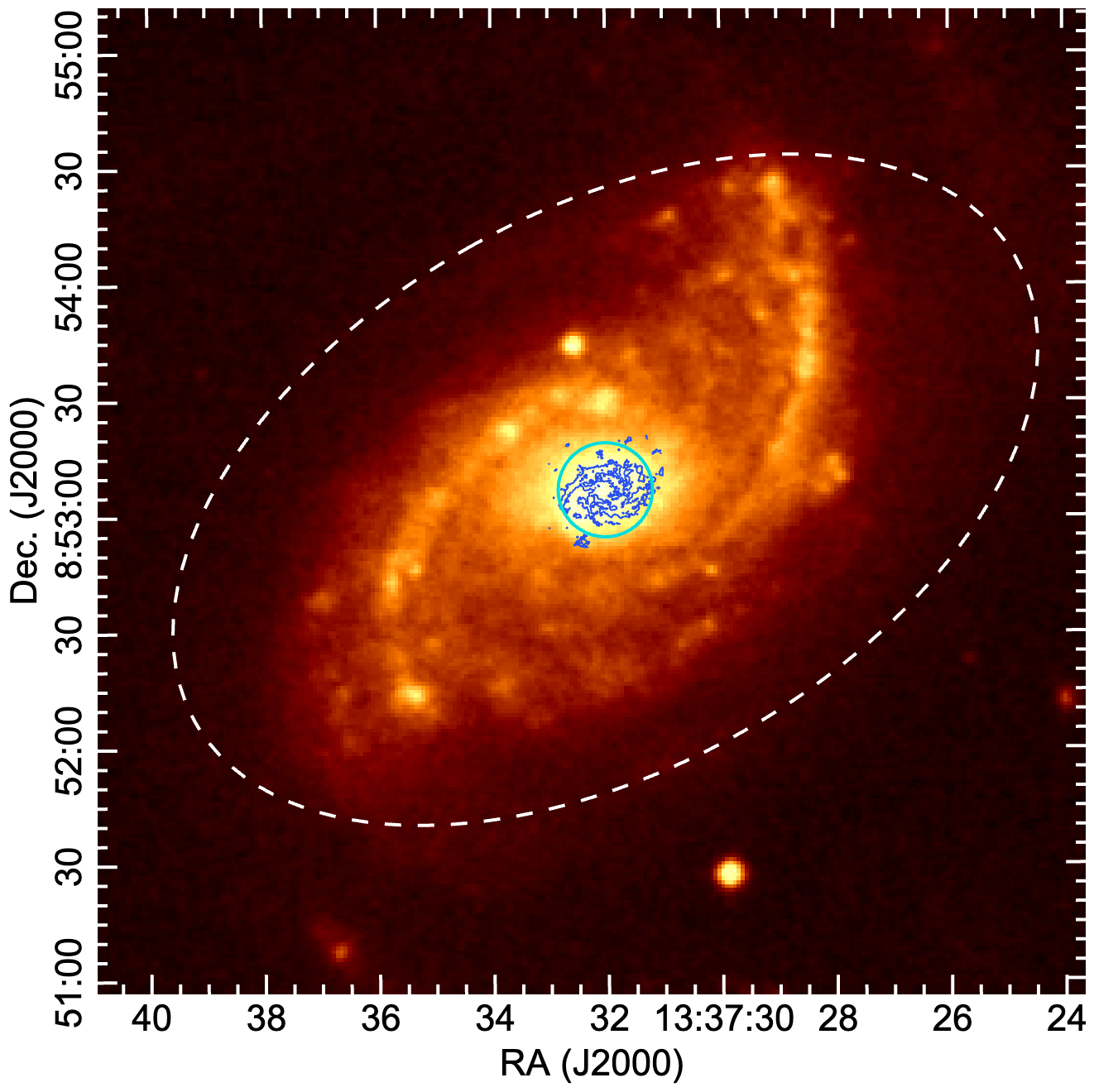}
    \end{subfigure}
    \begin{subfigure}{.33\linewidth}
    \centering
    \includegraphics[height=5.2cm]{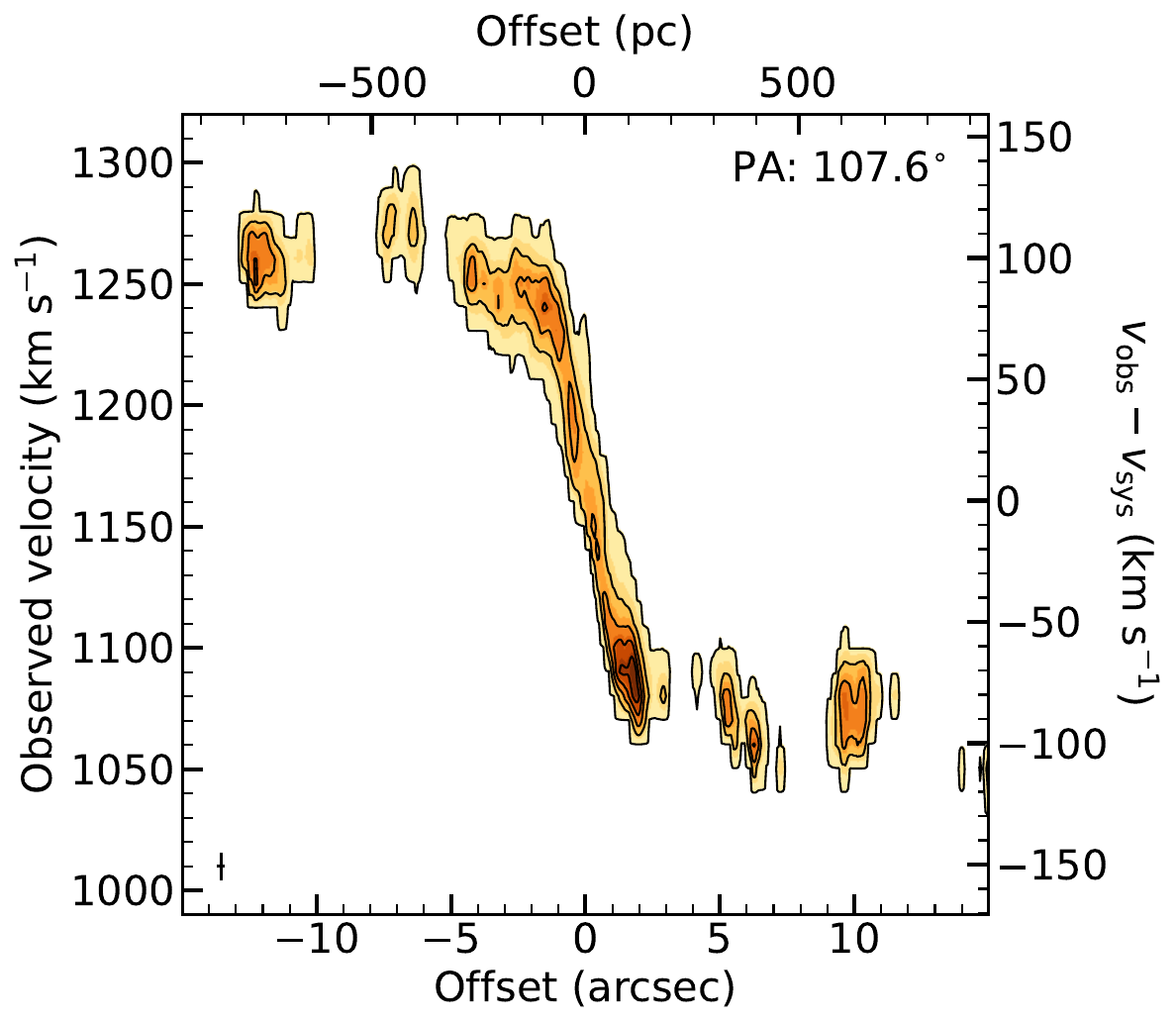}
    \end{subfigure}
    \caption[]{continued.}
\end{figure*}

\begin{figure*}
    \ContinuedFloat
    \begin{subfigure}{.34\linewidth}
    \centering
    \includegraphics[height=6.0cm]{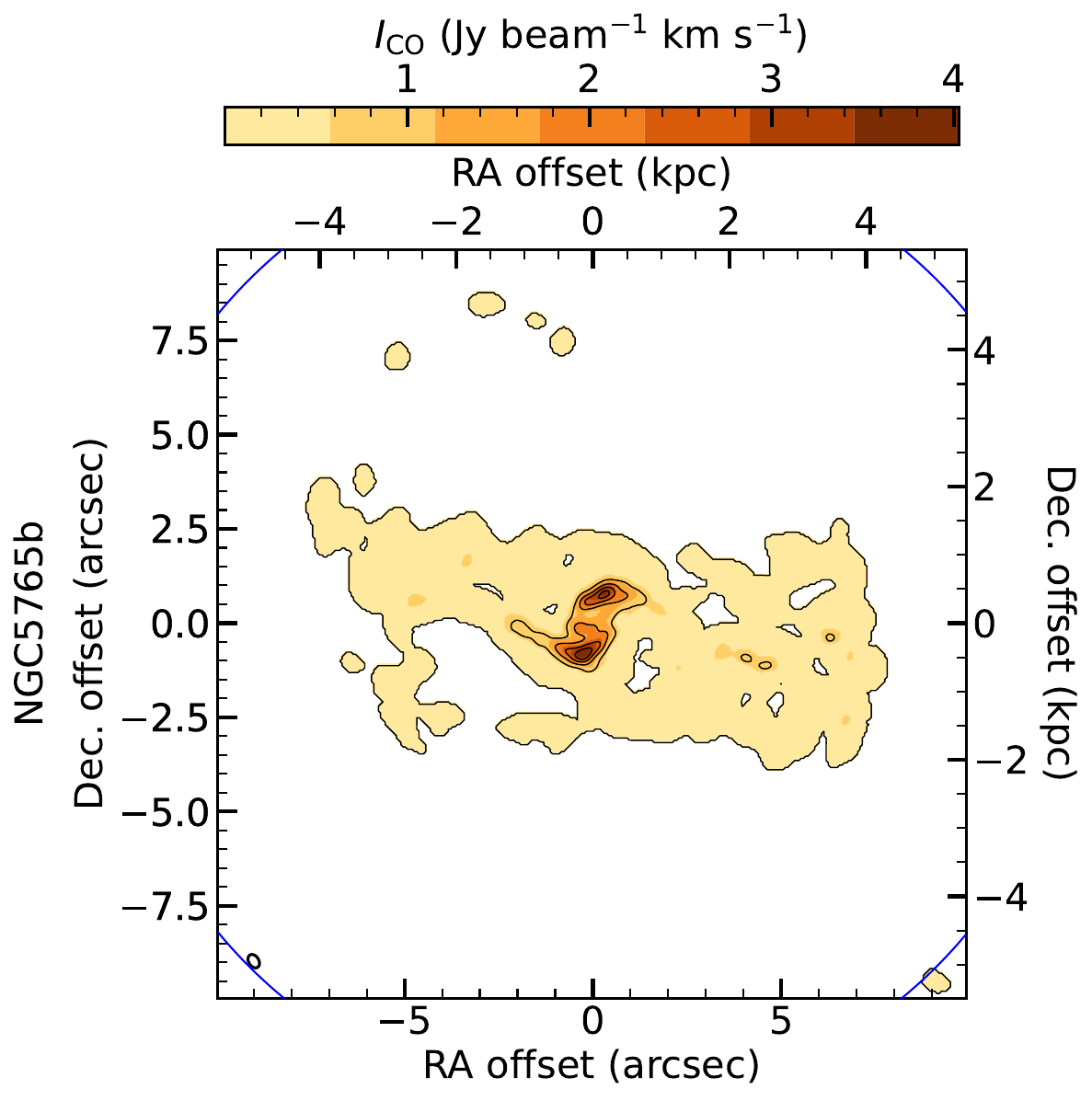}
    \end{subfigure}
    \begin{subfigure}{.32\linewidth}
    \centering
    \includegraphics[height=6.0cm]{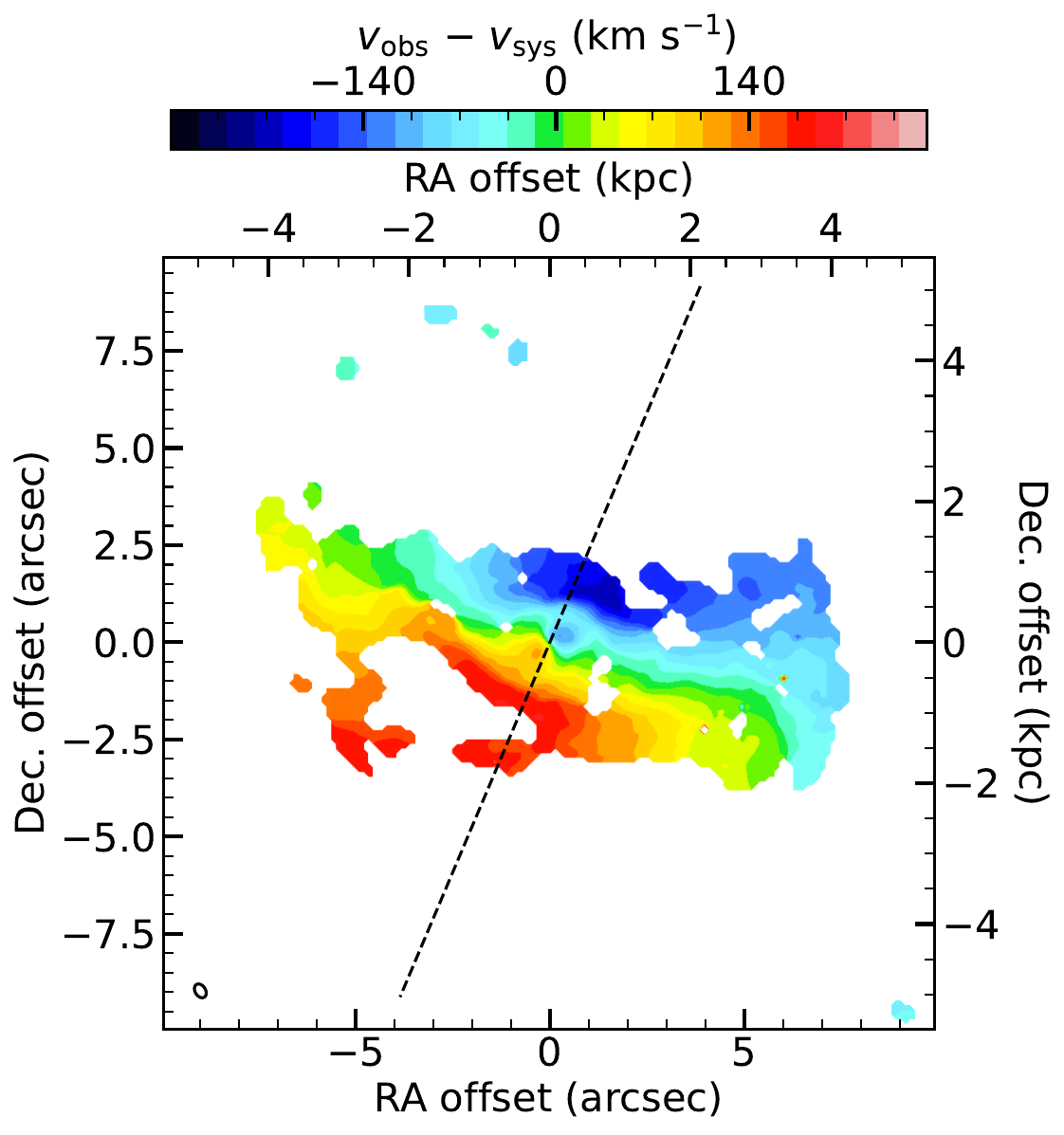}
    \end{subfigure}
    \begin{subfigure}{.32\linewidth}
    \centering
    \includegraphics[height=6.0cm]{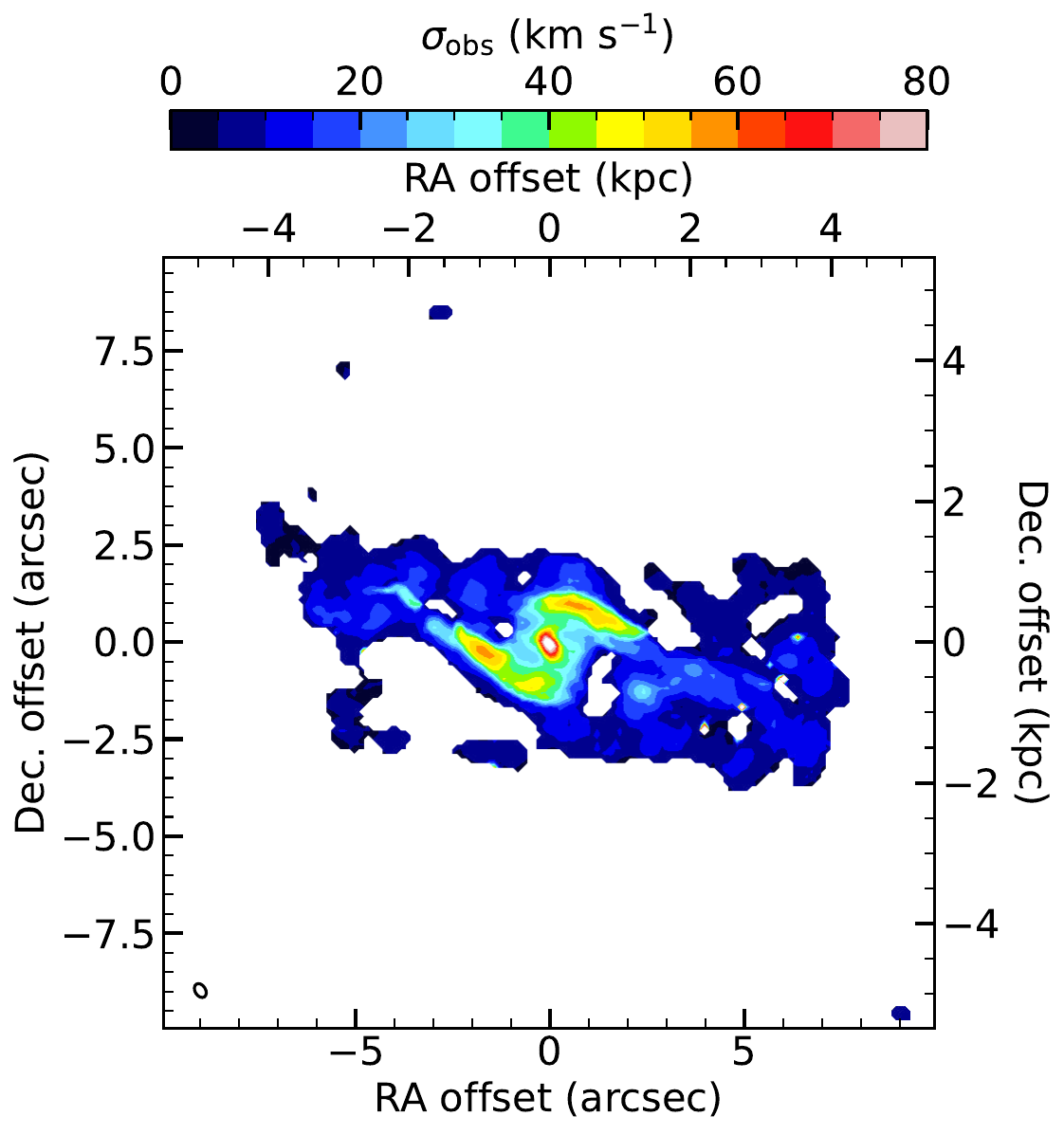}
    \end{subfigure}
    \\
    \begin{subfigure}{.34\linewidth}
    \centering
    \includegraphics[height=5.2cm]{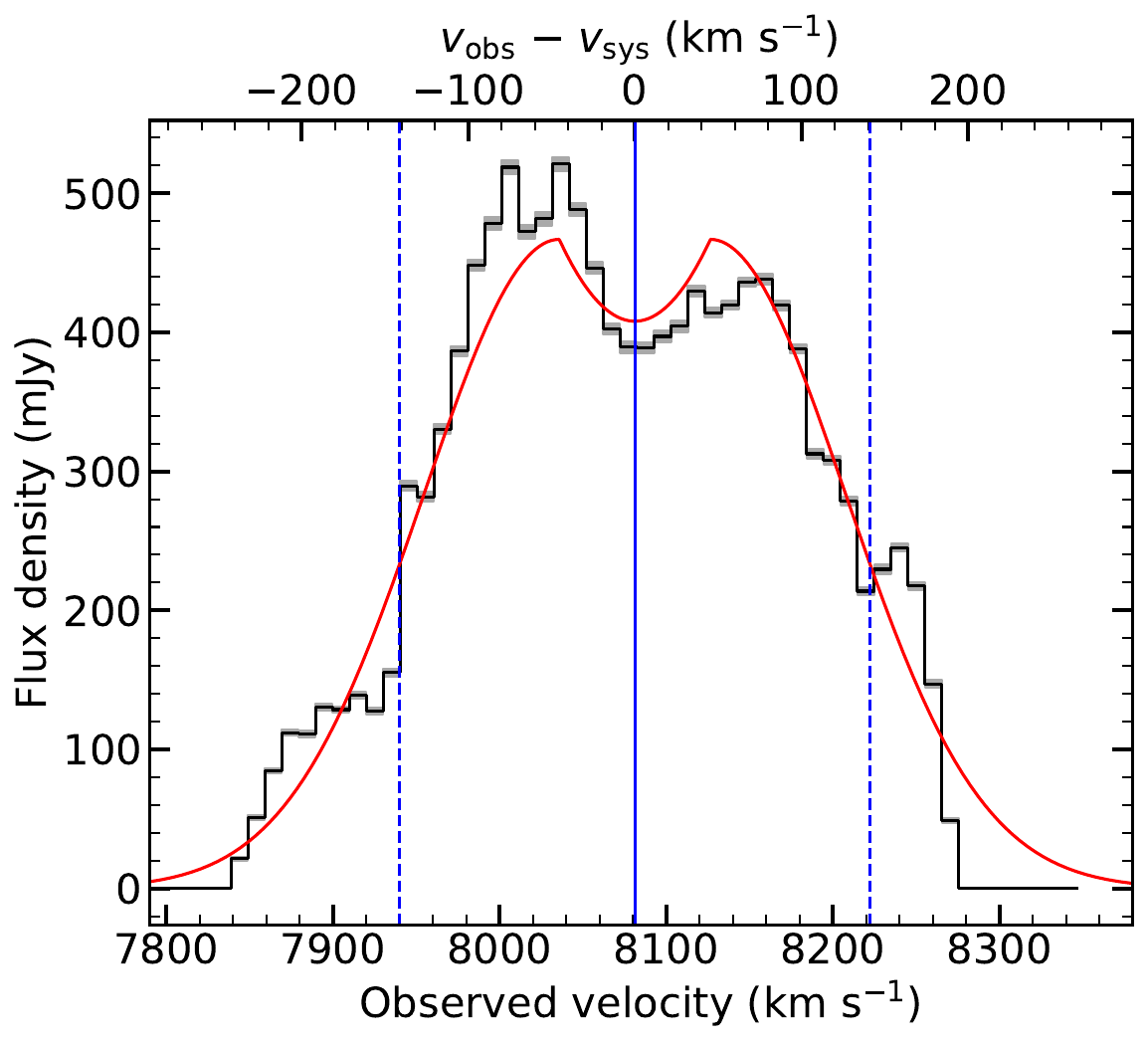}
    \end{subfigure}
    \begin{subfigure}{.32\linewidth}
    \centering
    \includegraphics[height=4.8cm]{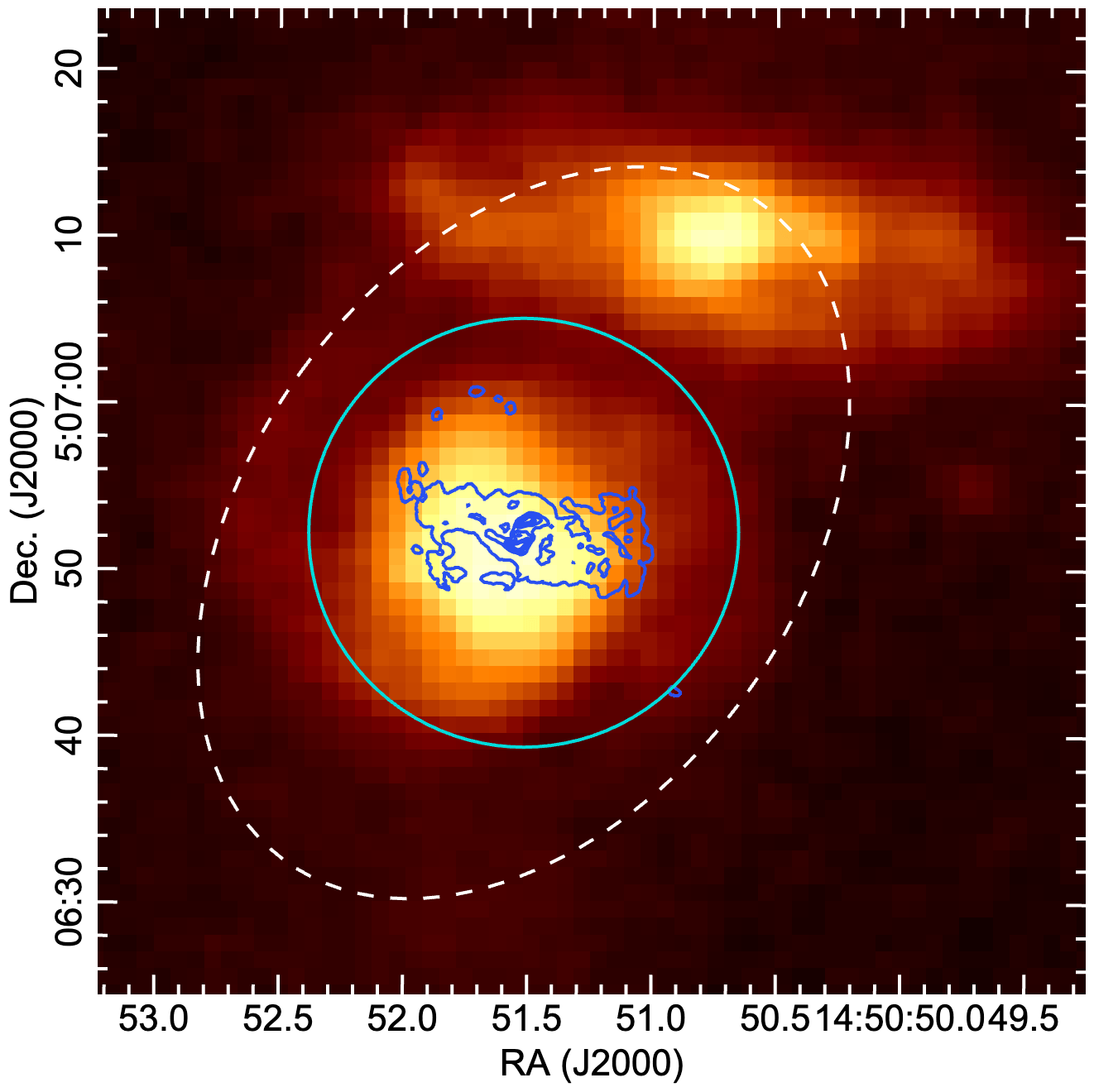}
    \end{subfigure}
    \begin{subfigure}{.33\linewidth}
    \centering
    \includegraphics[height=5.2cm]{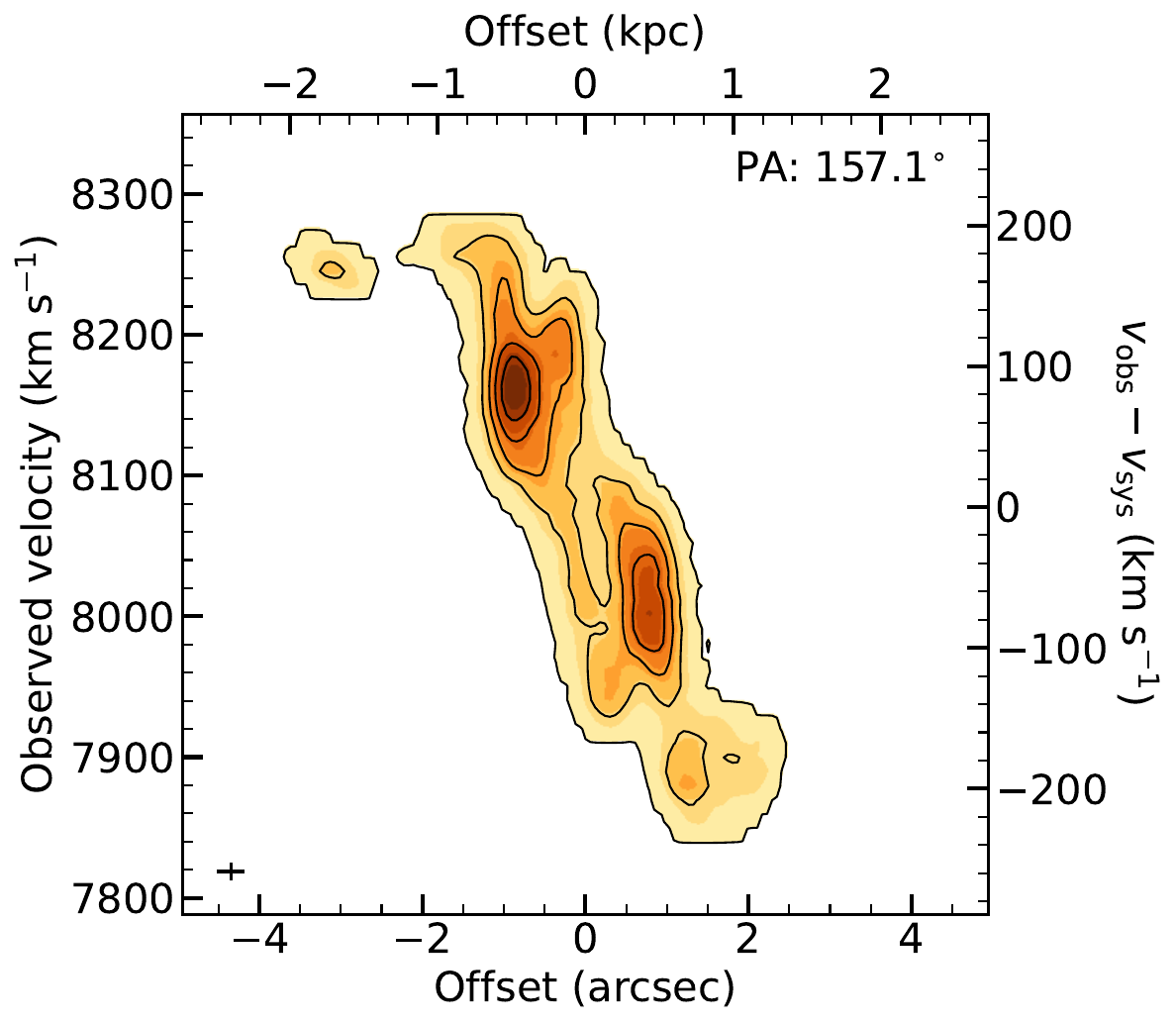}
    \end{subfigure}
    \begin{subfigure}{.34\linewidth}
    \centering
    \includegraphics[height=6.0cm]{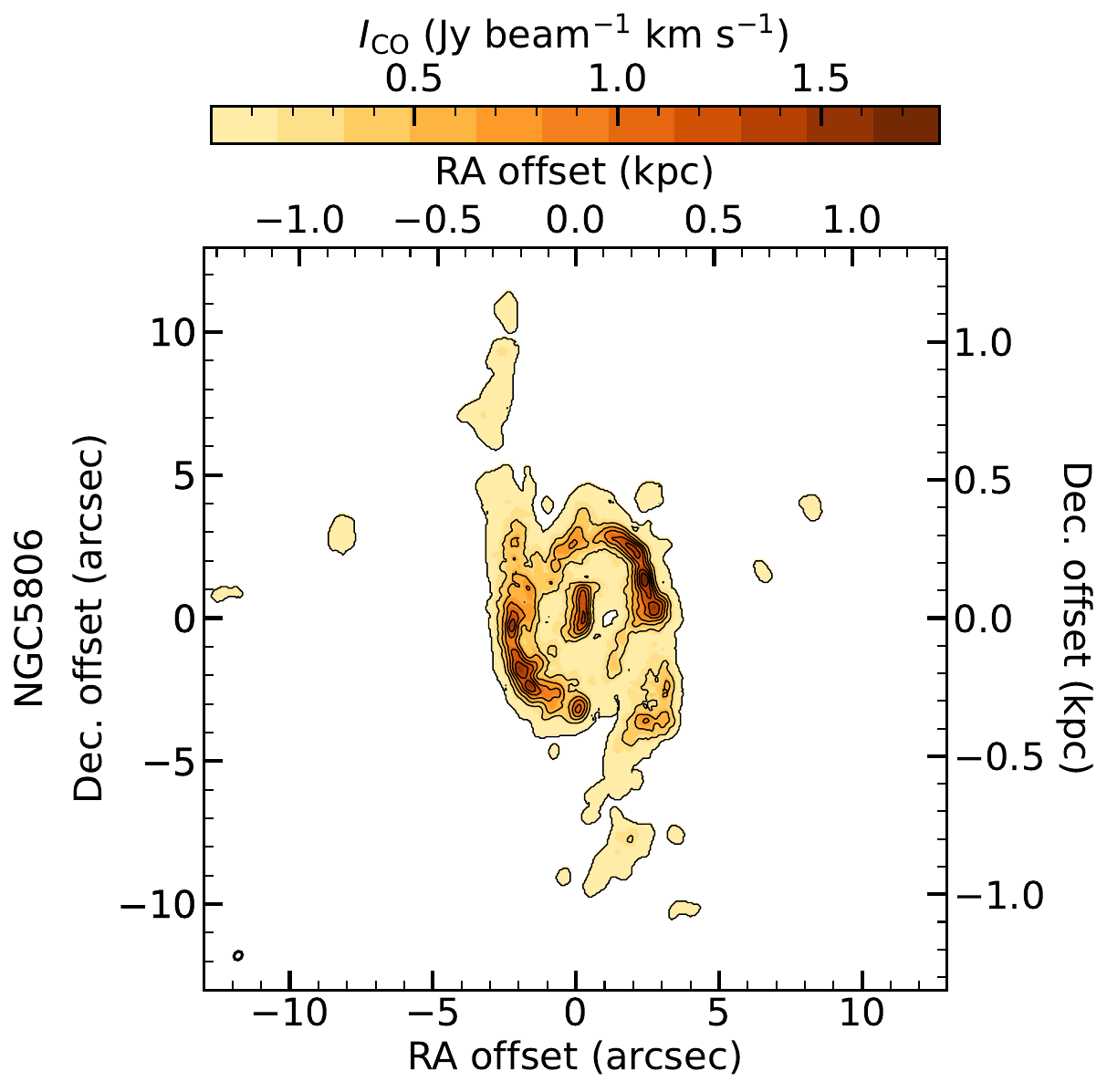}
    \end{subfigure}
    \begin{subfigure}{.32\linewidth}
    \centering
    \includegraphics[height=6.0cm]{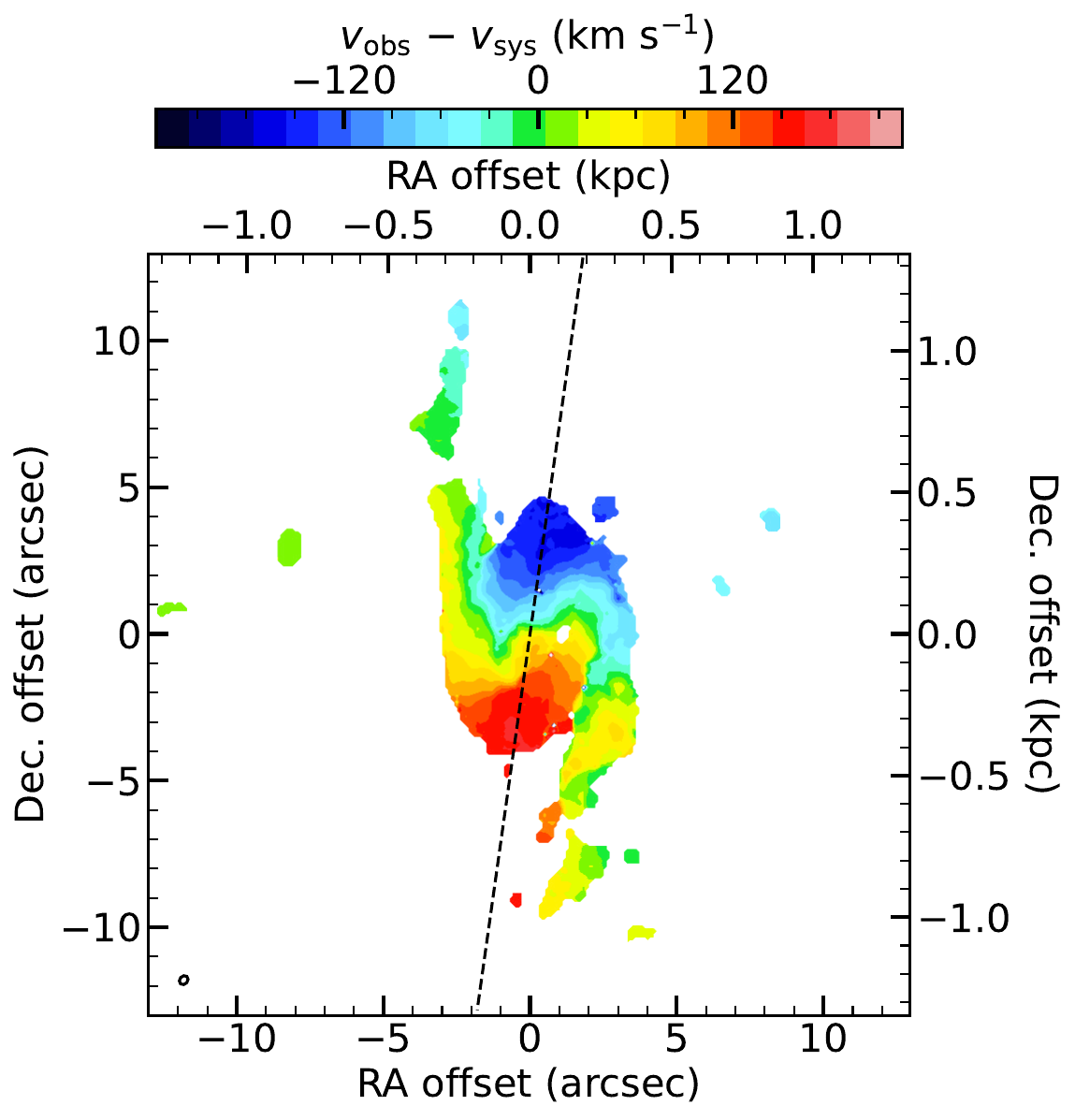}
    \end{subfigure}
    \begin{subfigure}{.32\linewidth}
    \centering
    \includegraphics[height=6.0cm]{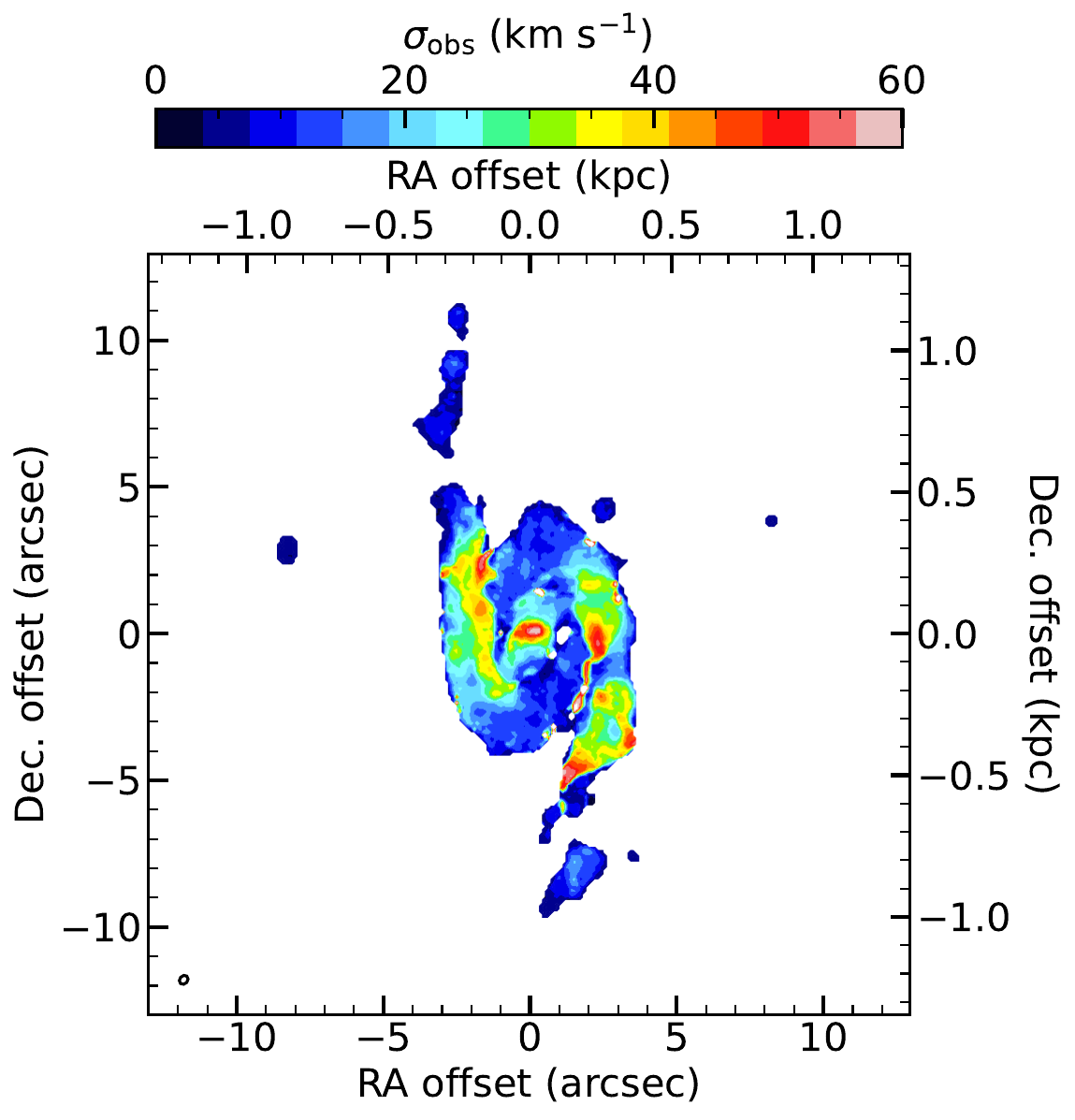}
    \end{subfigure}
    \\
    \begin{subfigure}{.34\linewidth}
    \centering
    \includegraphics[height=5.2cm]{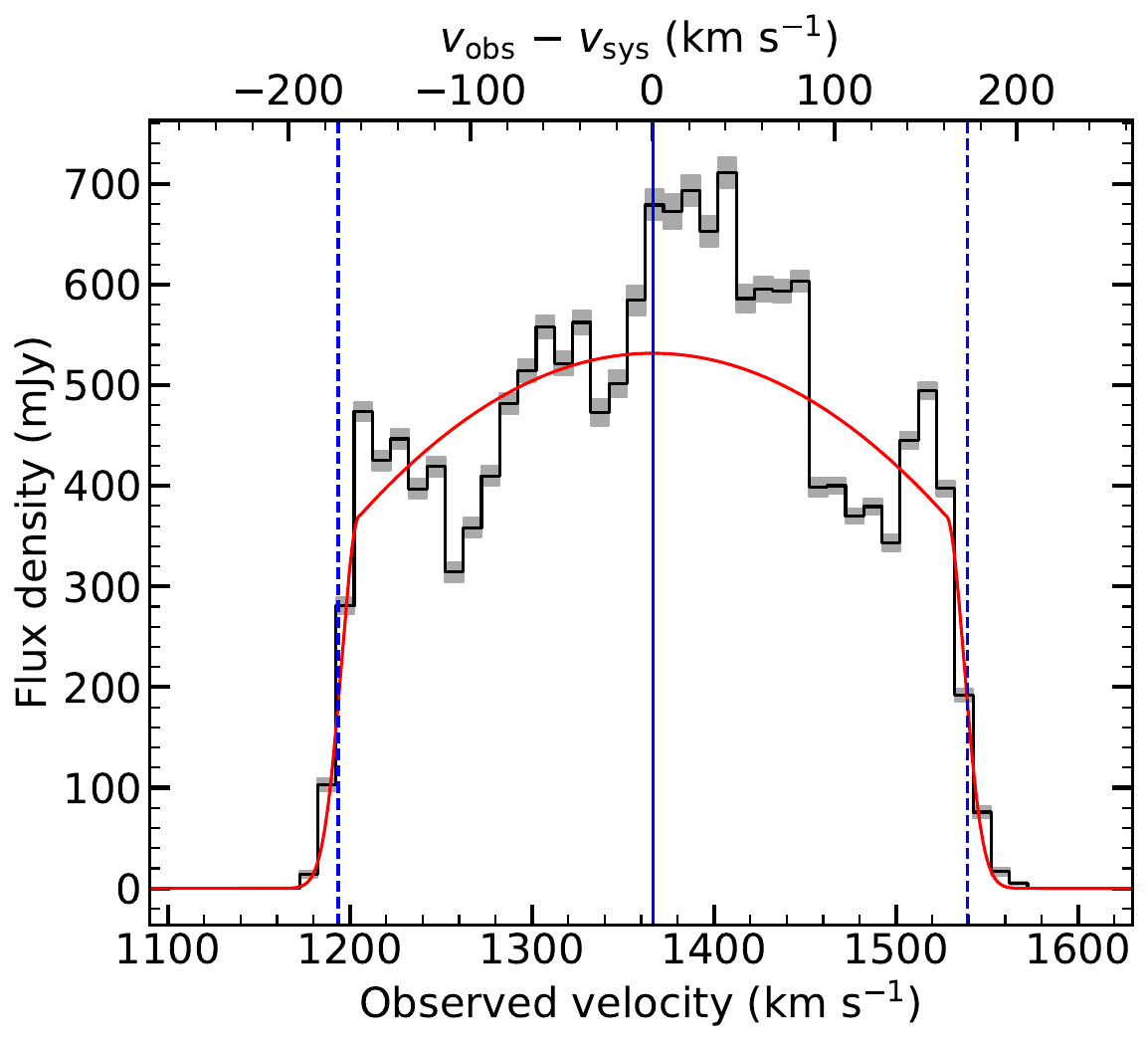}
    \end{subfigure}
    \begin{subfigure}{.32\linewidth}
    \centering
    \includegraphics[height=4.8cm]{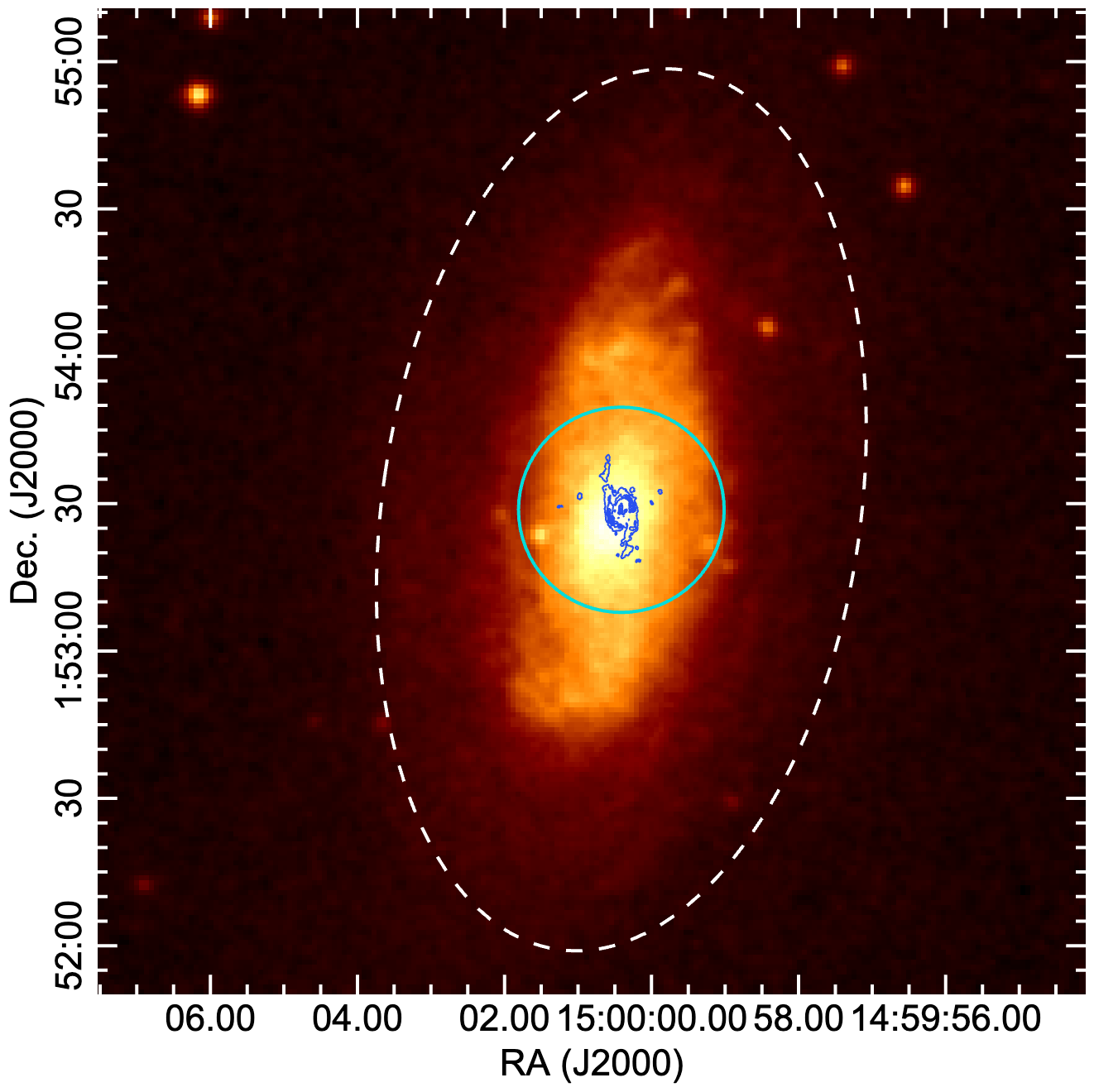}
    \end{subfigure}
    \begin{subfigure}{.33\linewidth}
    \centering
    \includegraphics[height=5.2cm]{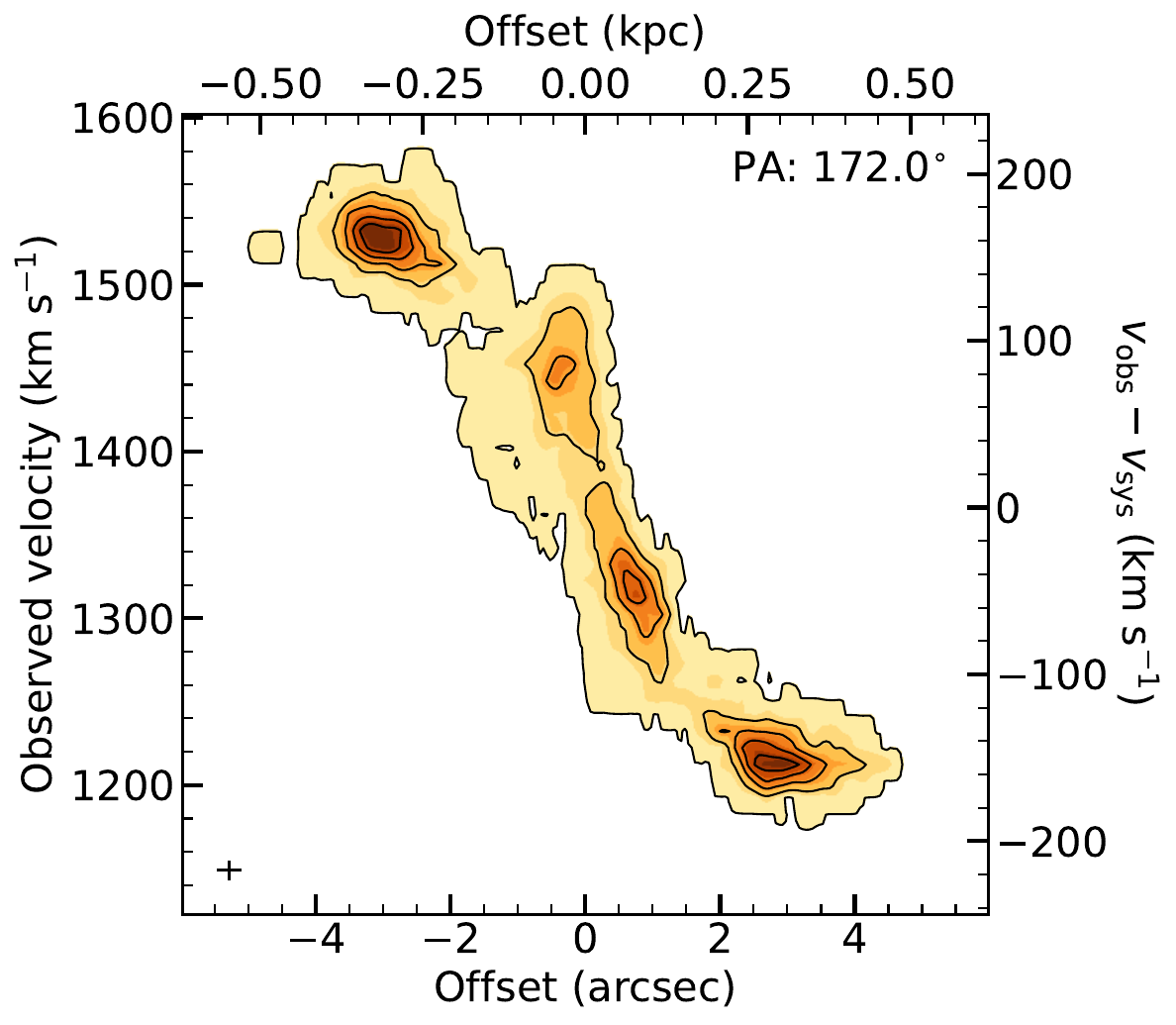}
    \end{subfigure}
    \caption[]{continued.}
\end{figure*}

\begin{figure*}
    \ContinuedFloat
    \begin{subfigure}{.34\linewidth}
    \centering
    \includegraphics[height=6.0cm]{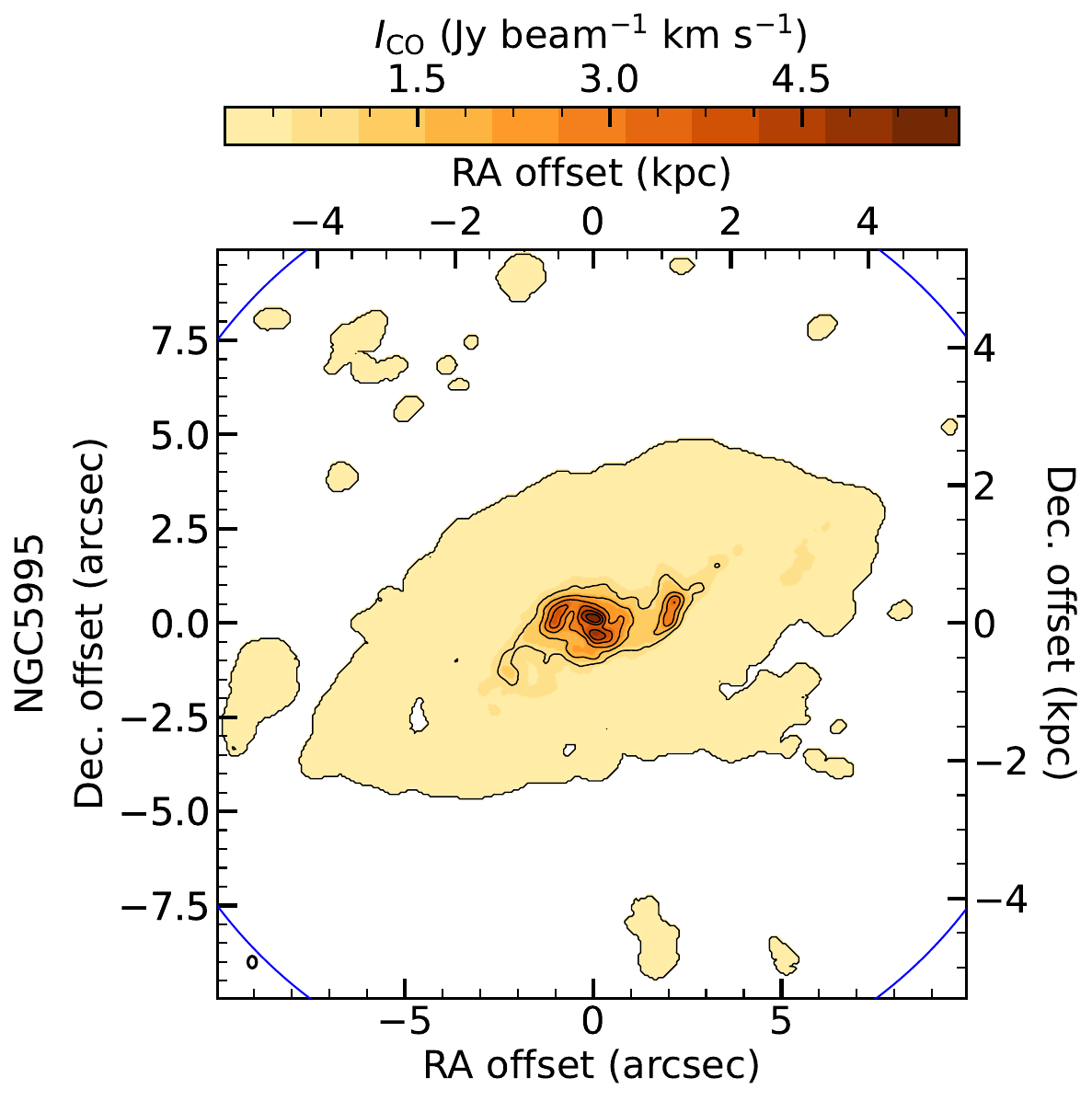}
    \end{subfigure}
    \begin{subfigure}{.32\linewidth}
    \centering
    \includegraphics[height=6.0cm]{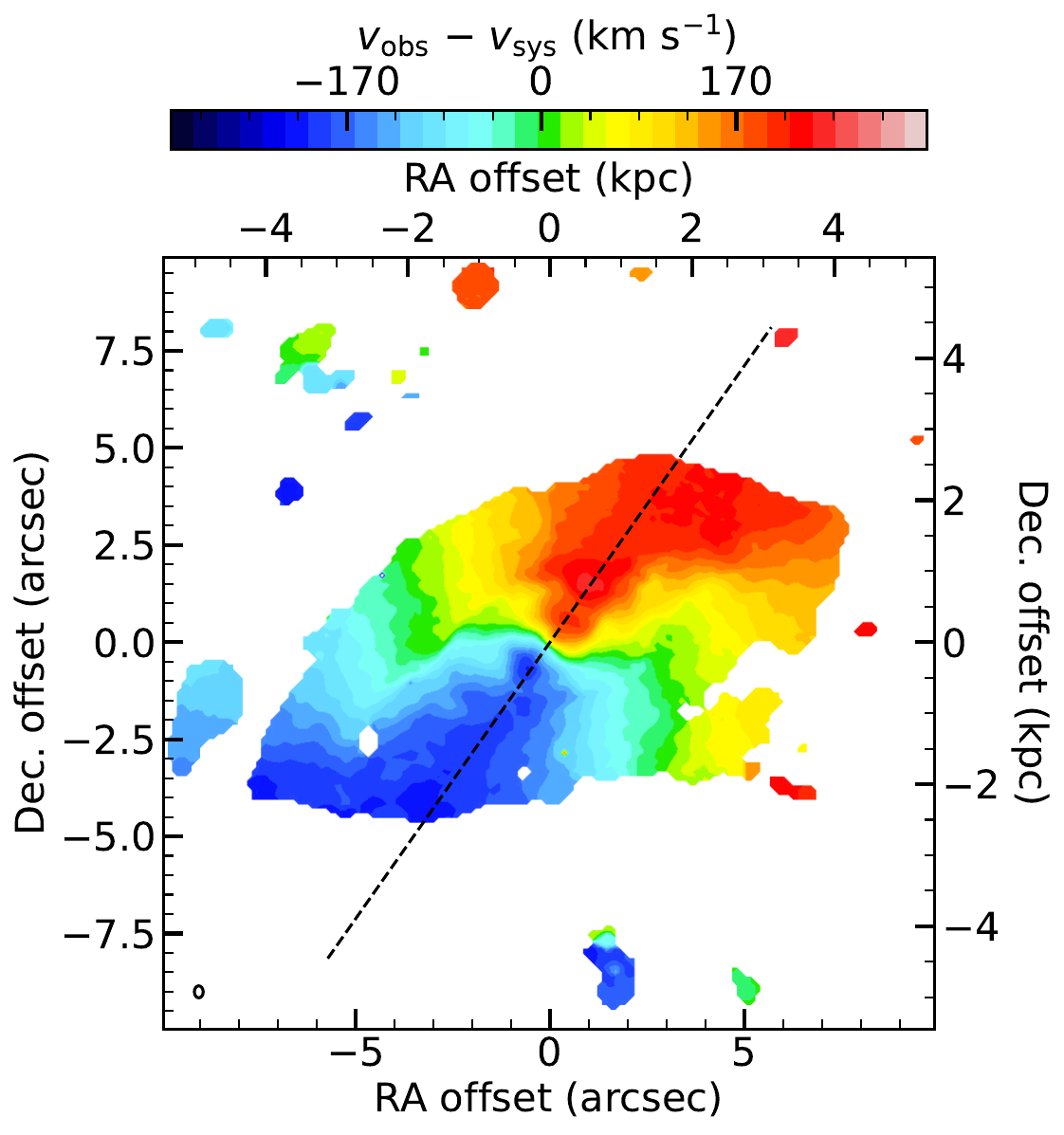}
    \end{subfigure}
    \begin{subfigure}{.32\linewidth}
    \centering
    \includegraphics[height=6.0cm]{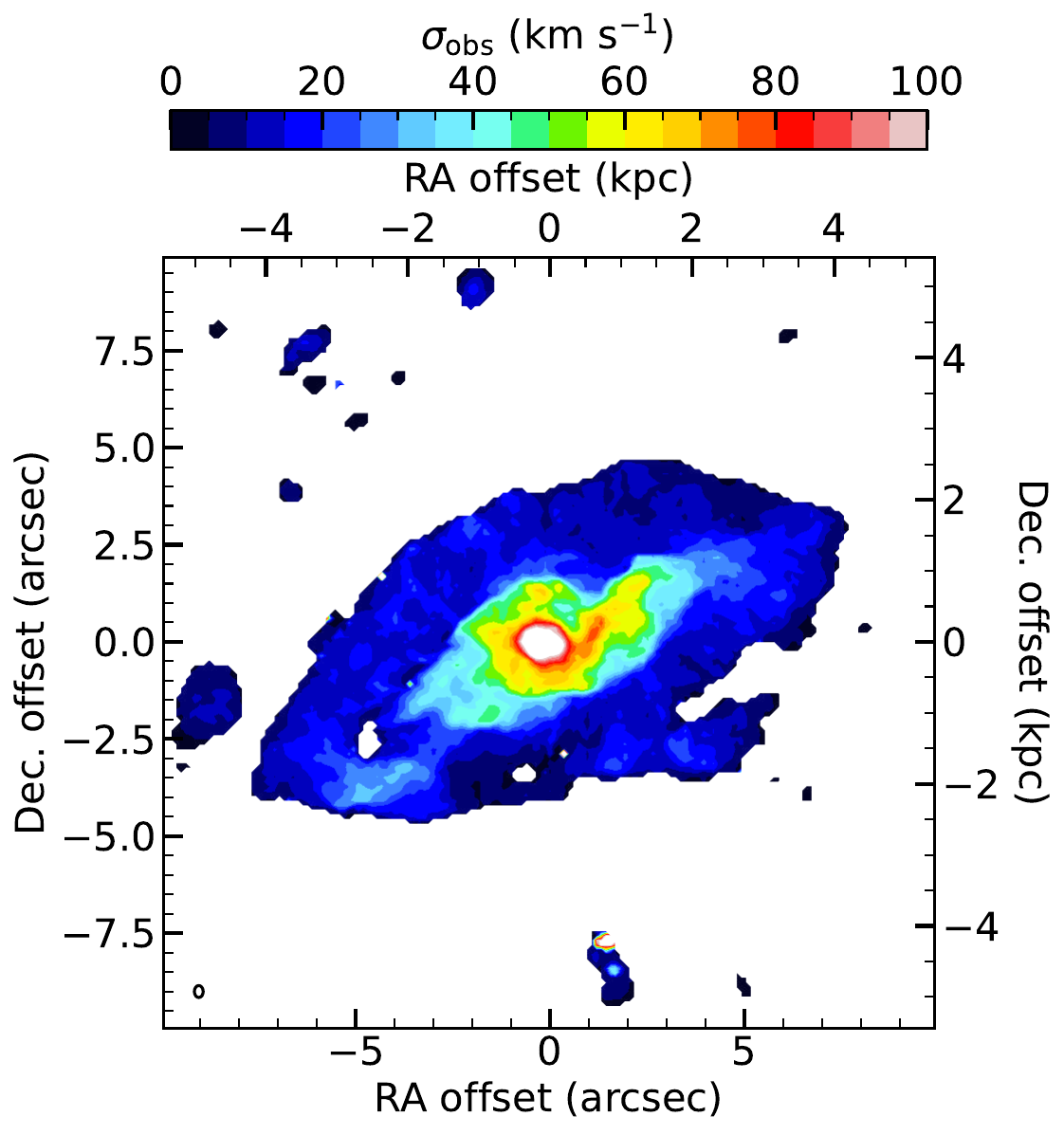}
    \end{subfigure}
    \\
    \begin{subfigure}{.34\linewidth}
    \centering
    \includegraphics[height=5.2cm]{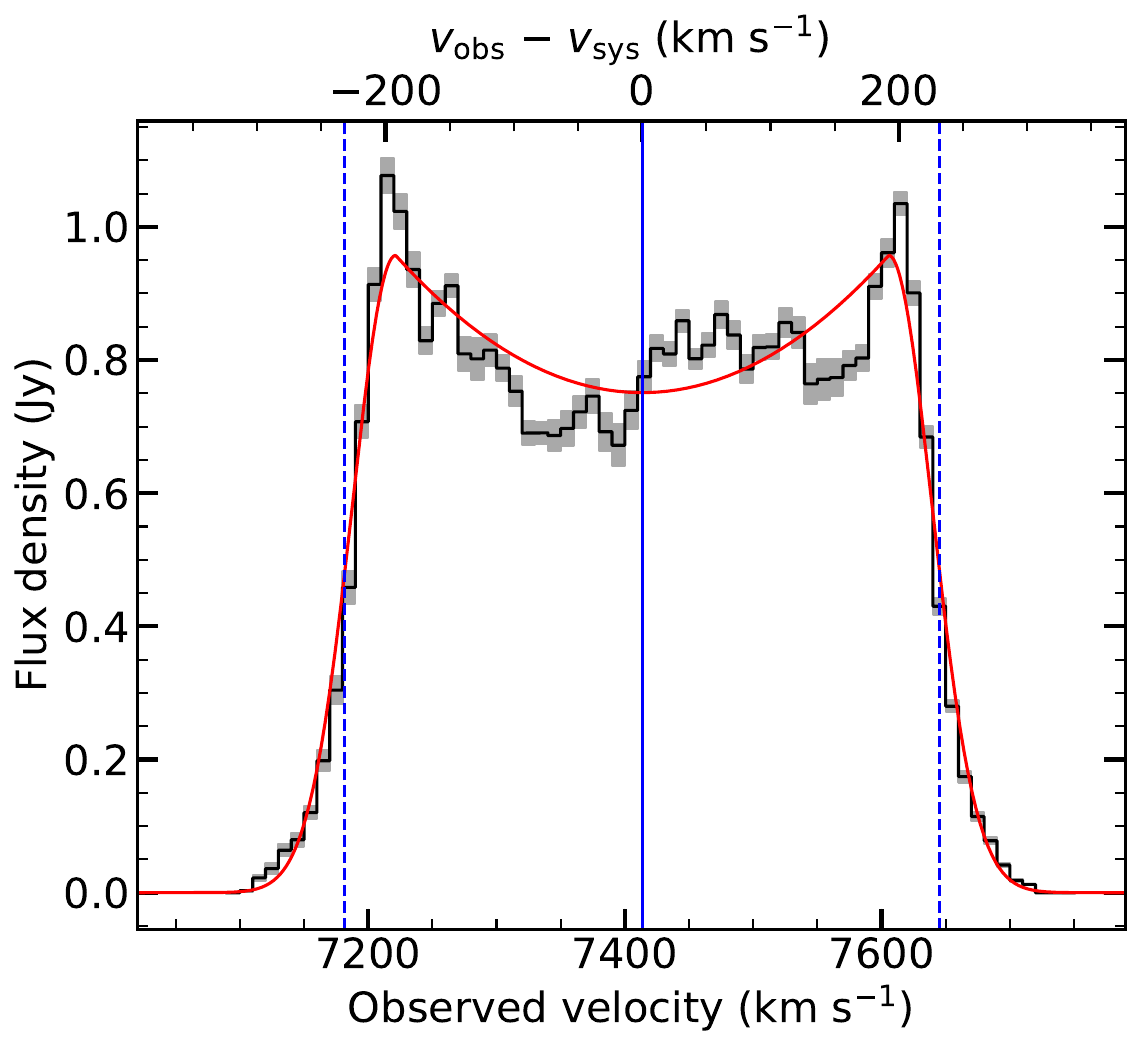}
    \end{subfigure}
    \begin{subfigure}{.32\linewidth}
    \centering
    \includegraphics[height=4.8cm]{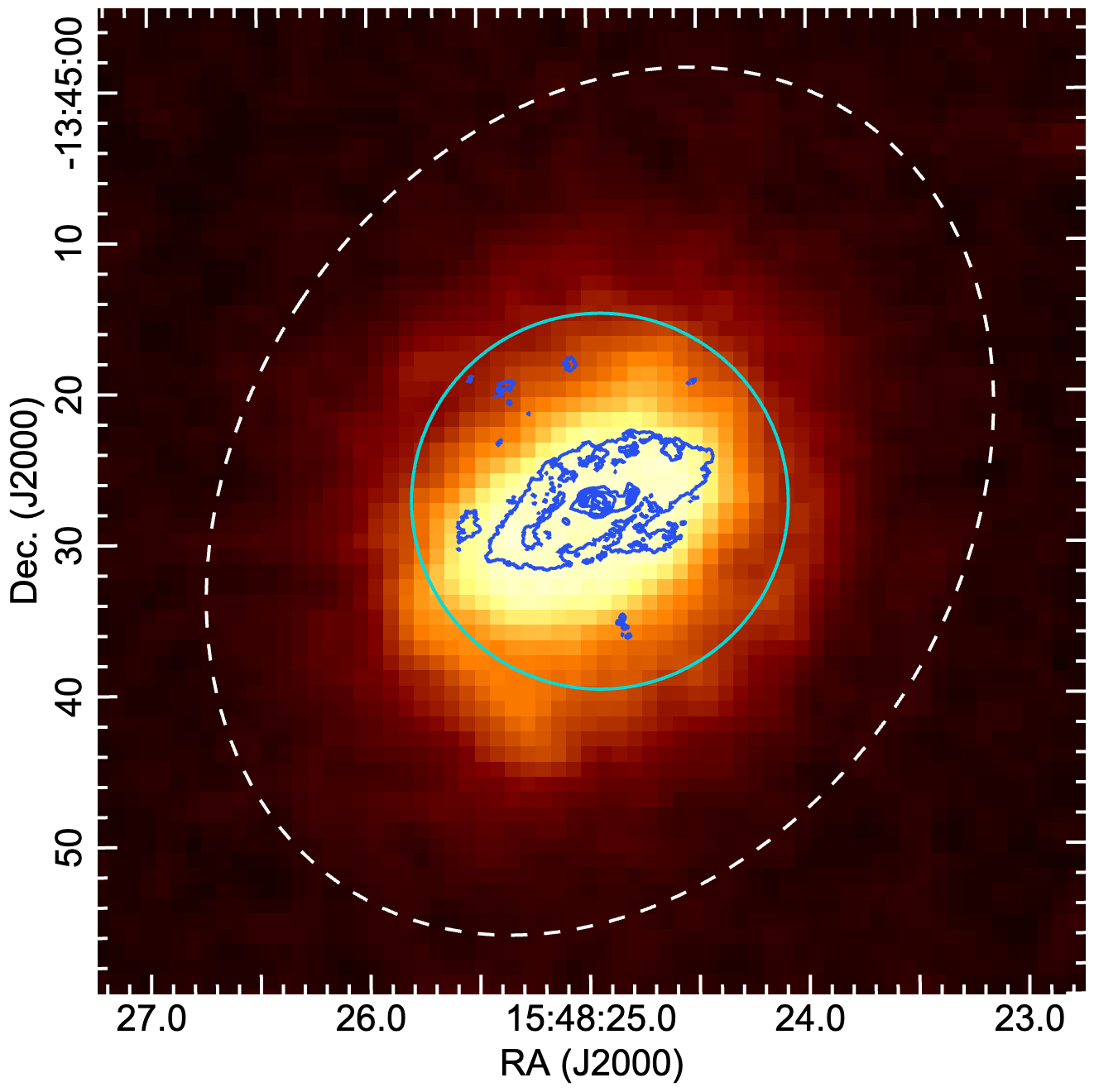}
    \end{subfigure}
    \begin{subfigure}{.33\linewidth}
    \centering
    \includegraphics[height=5.2cm]{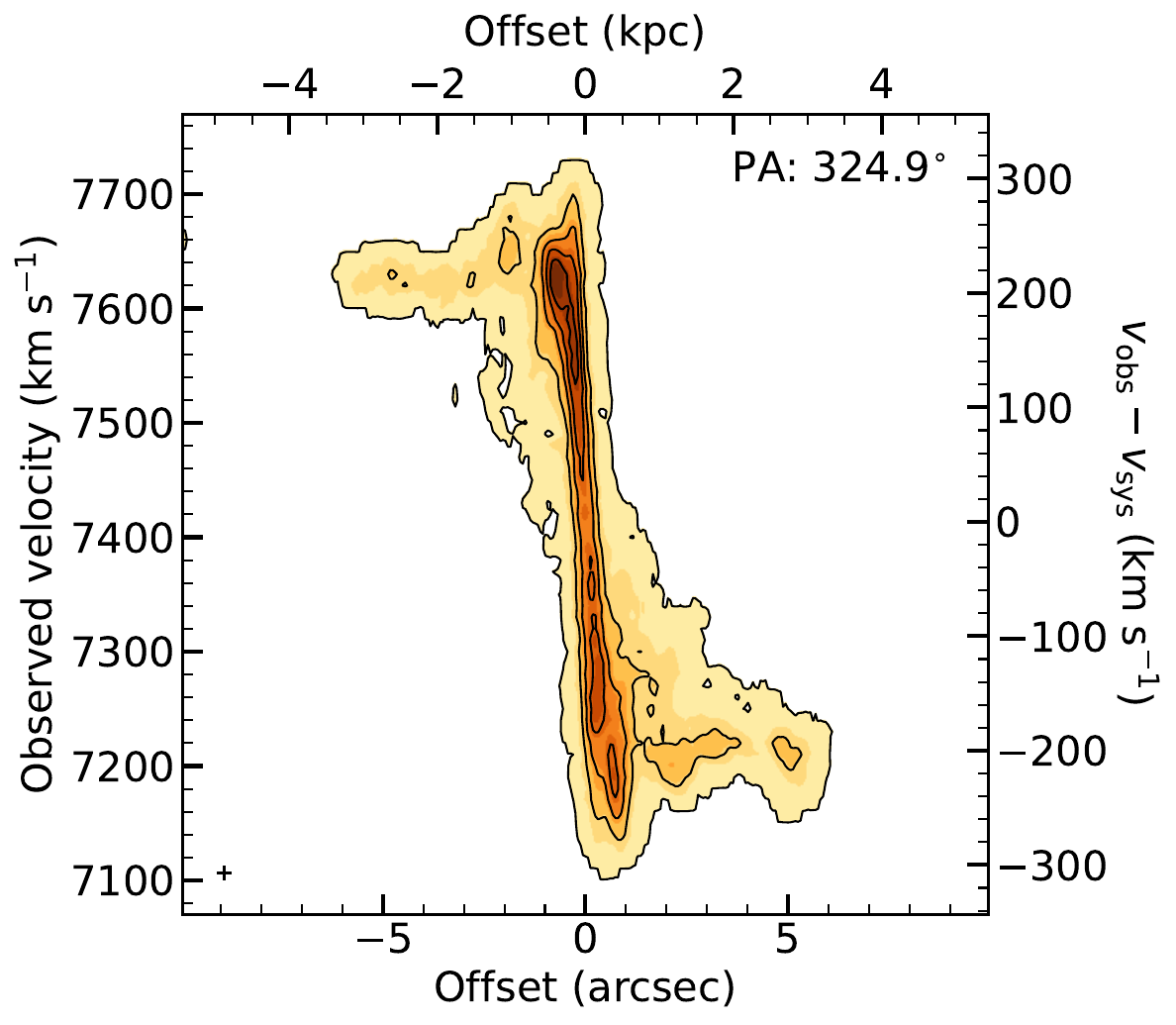}
    \end{subfigure}
    \begin{subfigure}{.34\linewidth}
    \centering
    \includegraphics[height=6.0cm]{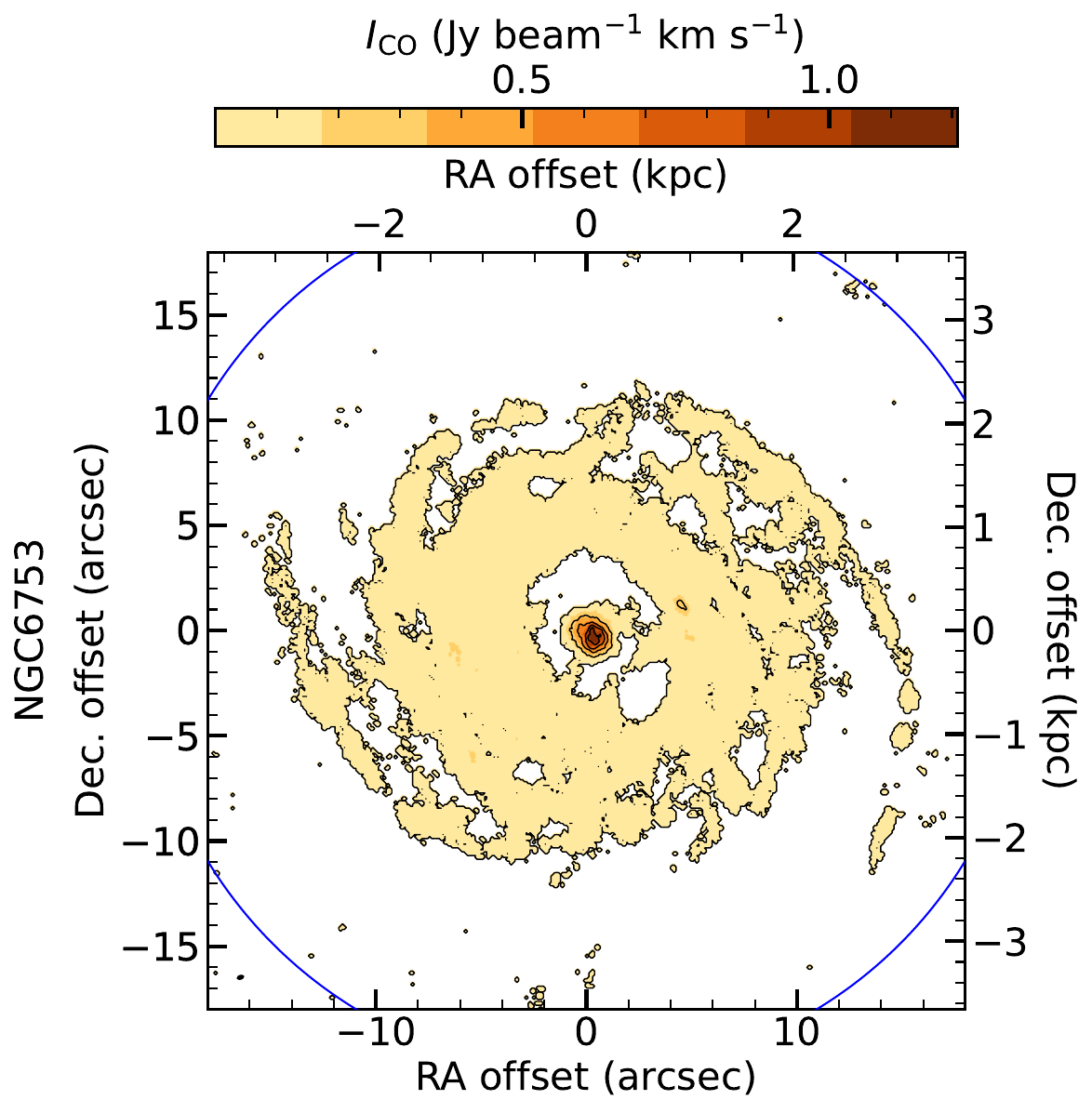}
    \end{subfigure}
    \begin{subfigure}{.32\linewidth}
    \centering
    \includegraphics[height=6.0cm]{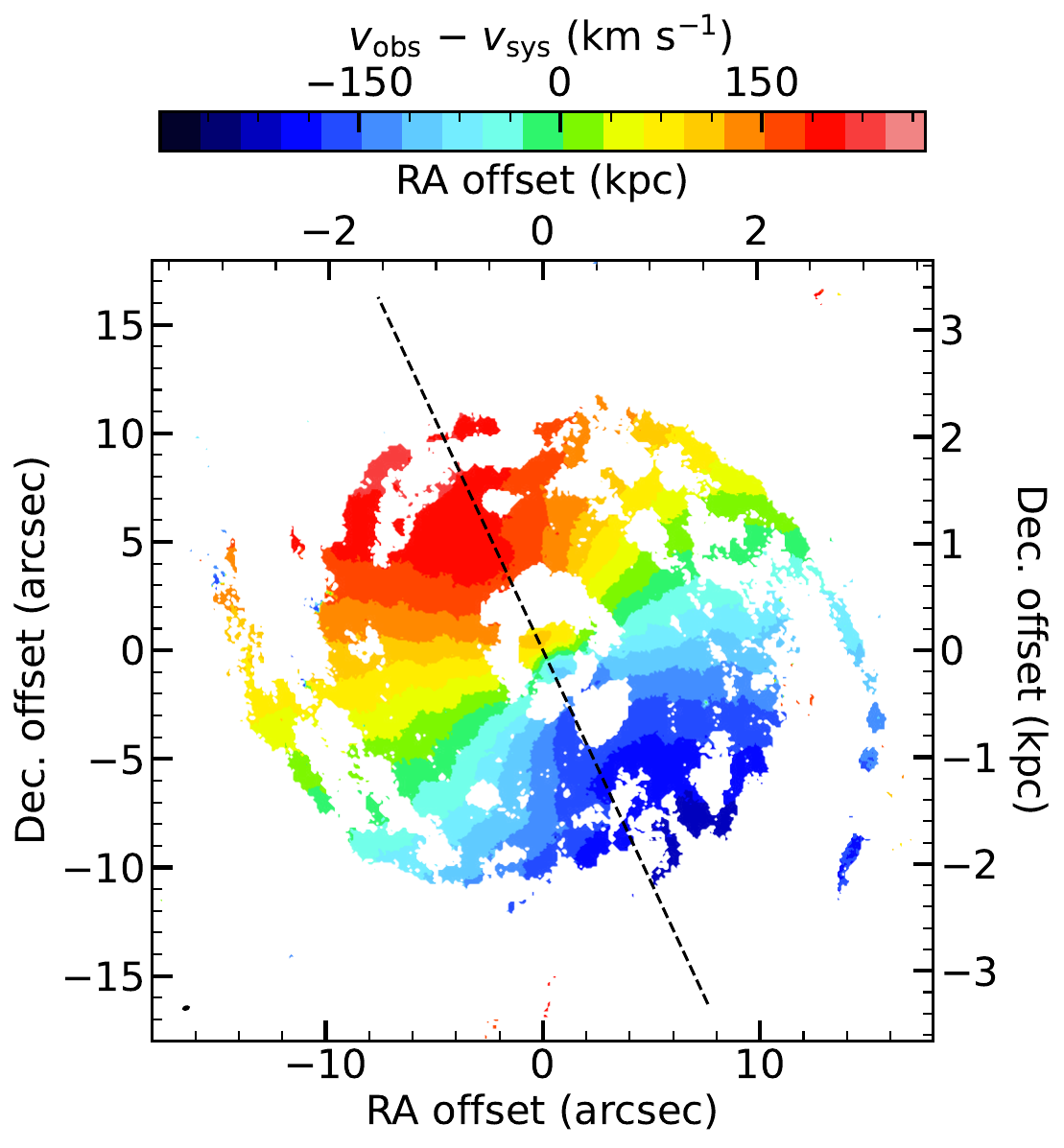}
    \end{subfigure}
    \begin{subfigure}{.32\linewidth}
    \centering
    \includegraphics[height=6.0cm]{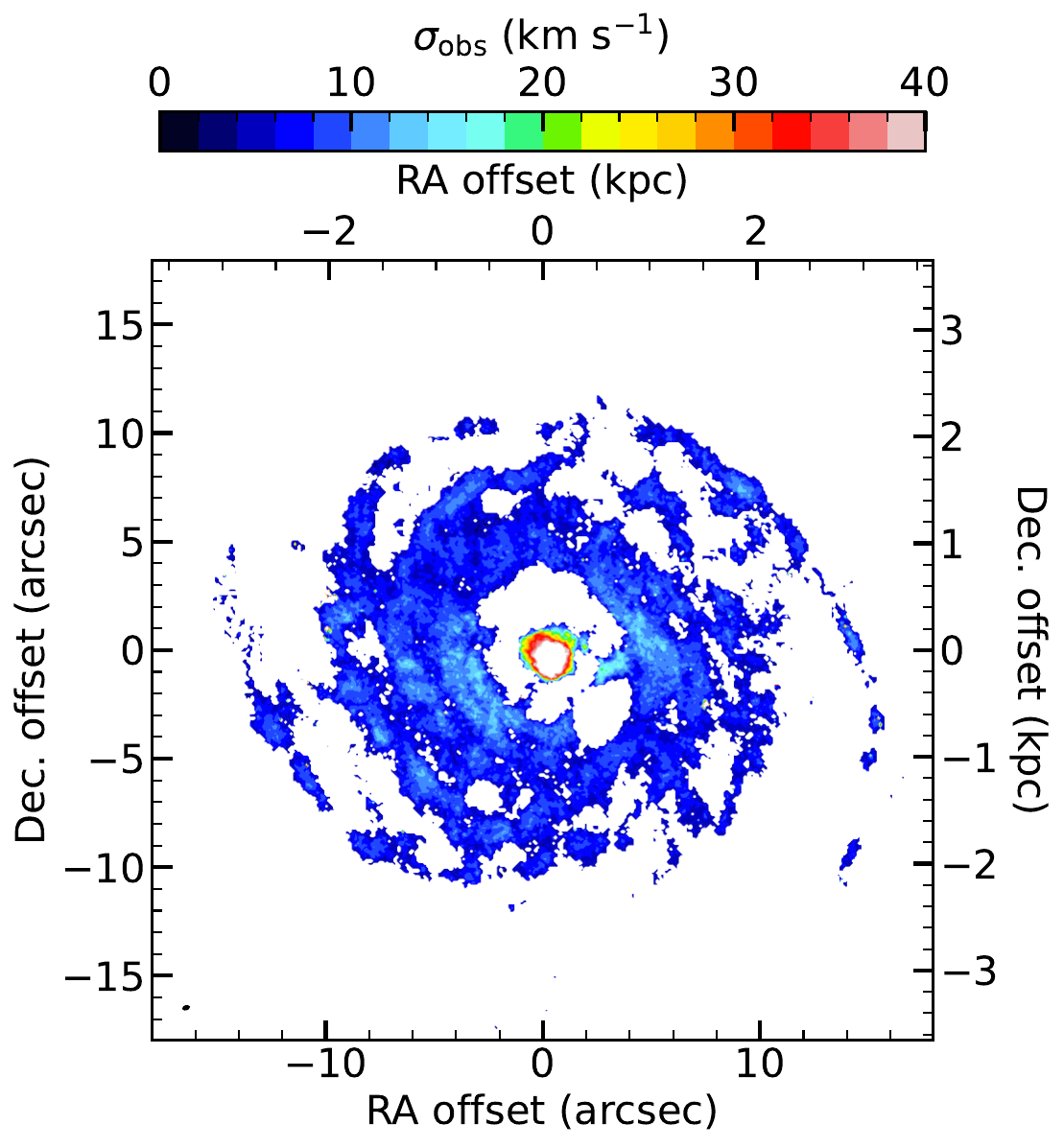}
    \end{subfigure}
    \\
    \begin{subfigure}{.34\linewidth}
    \centering
    \includegraphics[height=5.2cm]{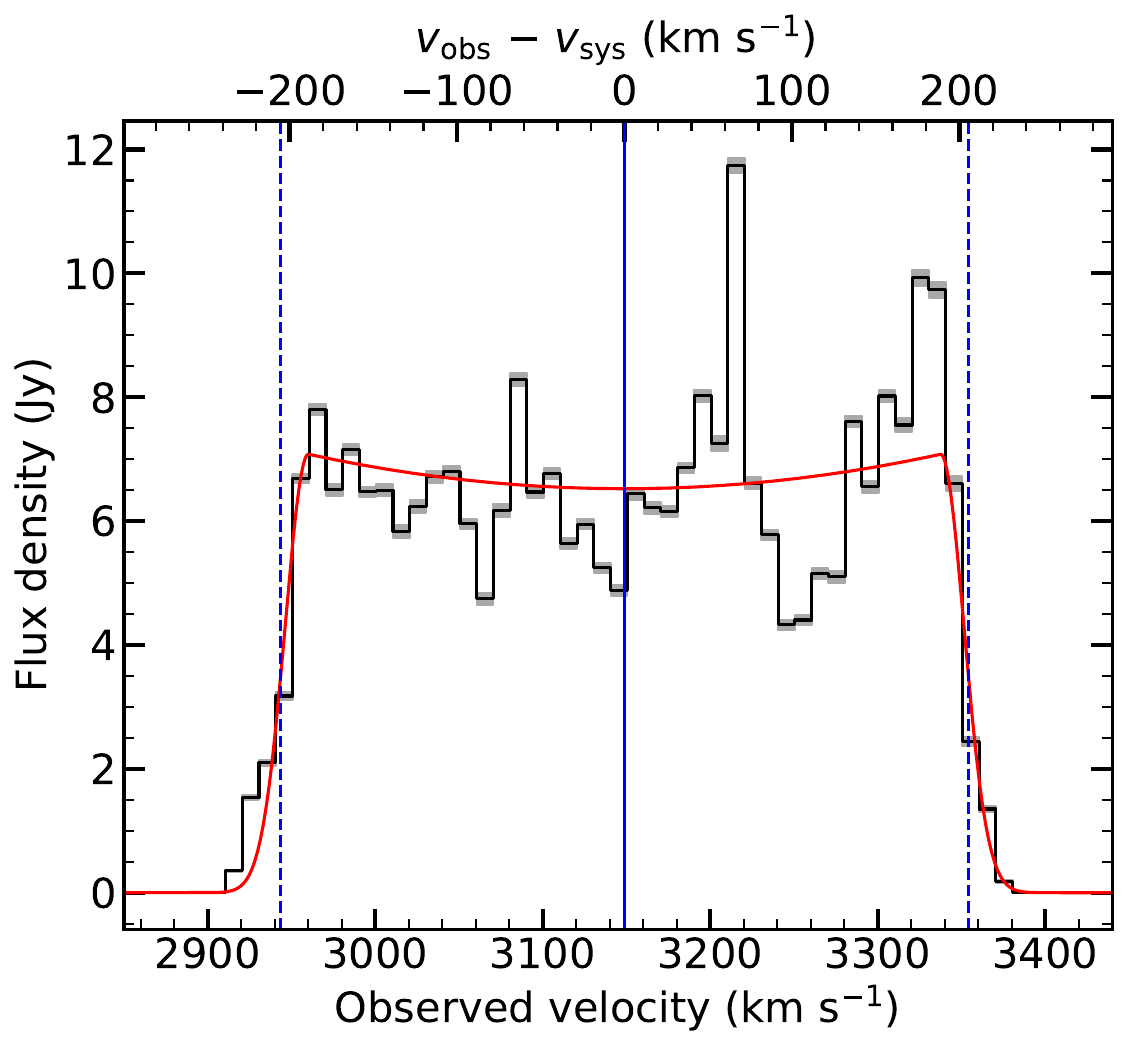}
    \end{subfigure}
    \begin{subfigure}{.32\linewidth}
    \centering
    \includegraphics[height=4.8cm]{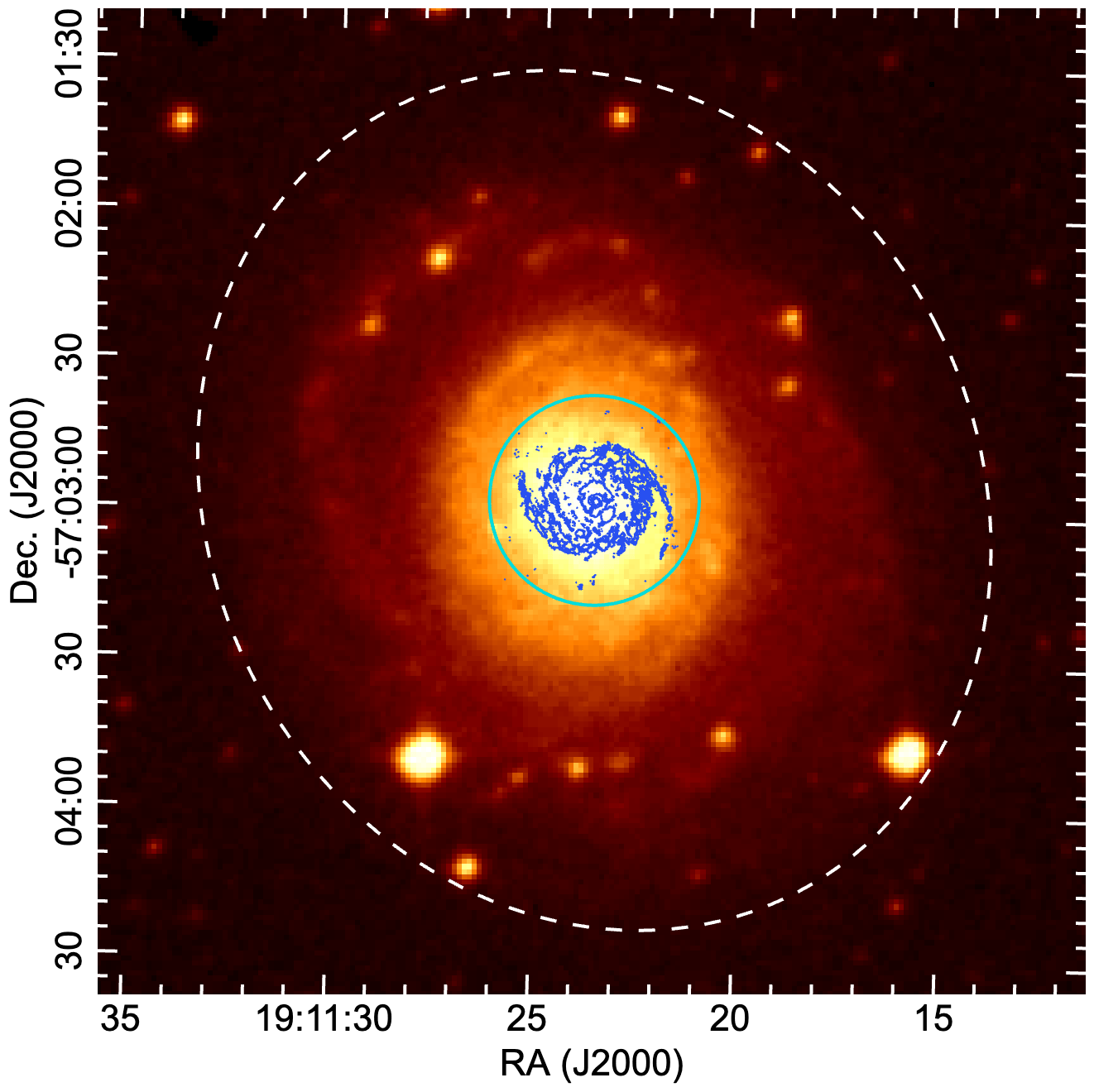}
    \end{subfigure}
    \begin{subfigure}{.33\linewidth}
    \centering
    \includegraphics[height=5.2cm]{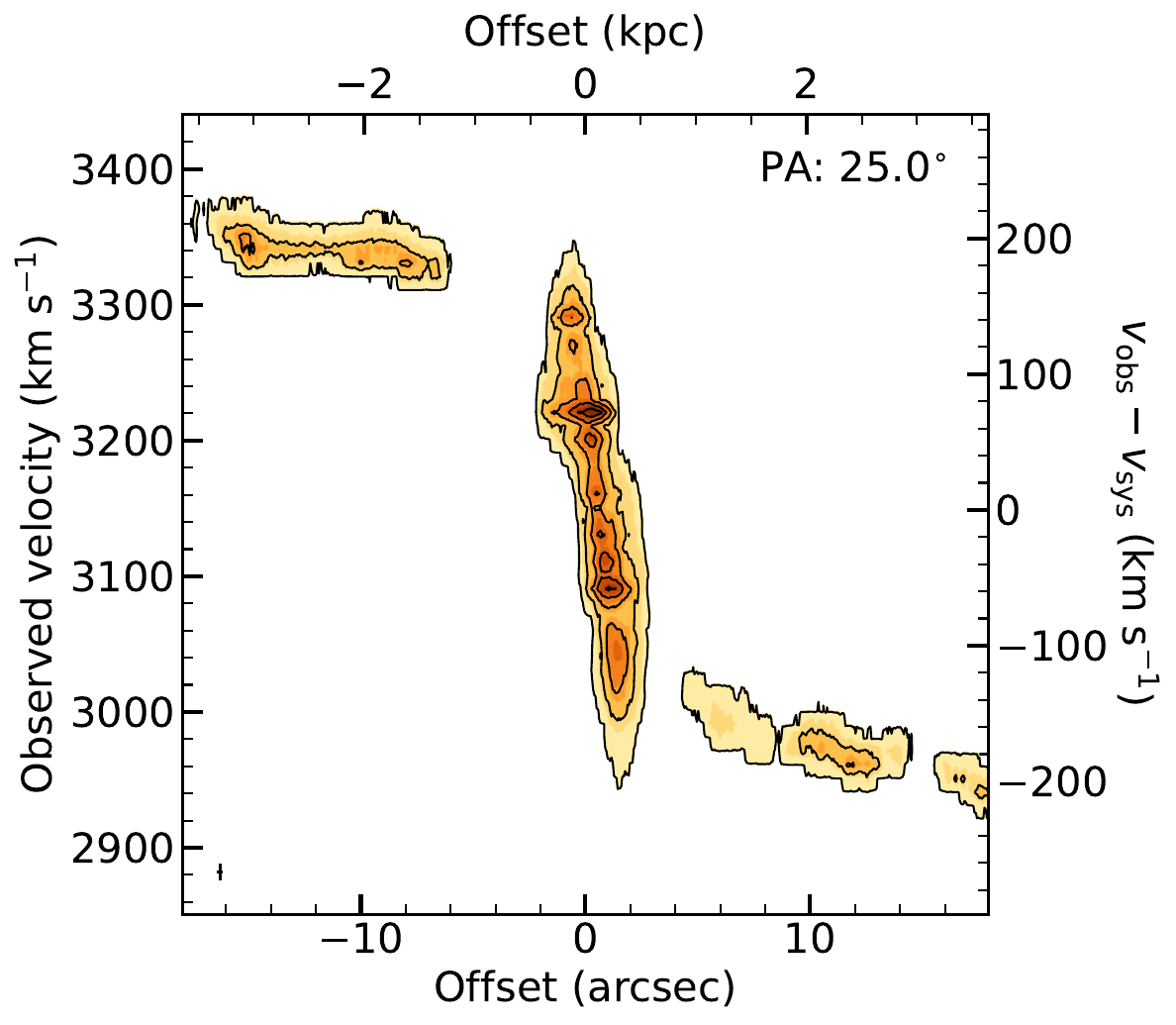}
    \end{subfigure}
    \caption[]{continued.}
\end{figure*}

\begin{figure*}
    \ContinuedFloat
    \begin{subfigure}{.34\linewidth}
    \centering
    \includegraphics[height=6.0cm]{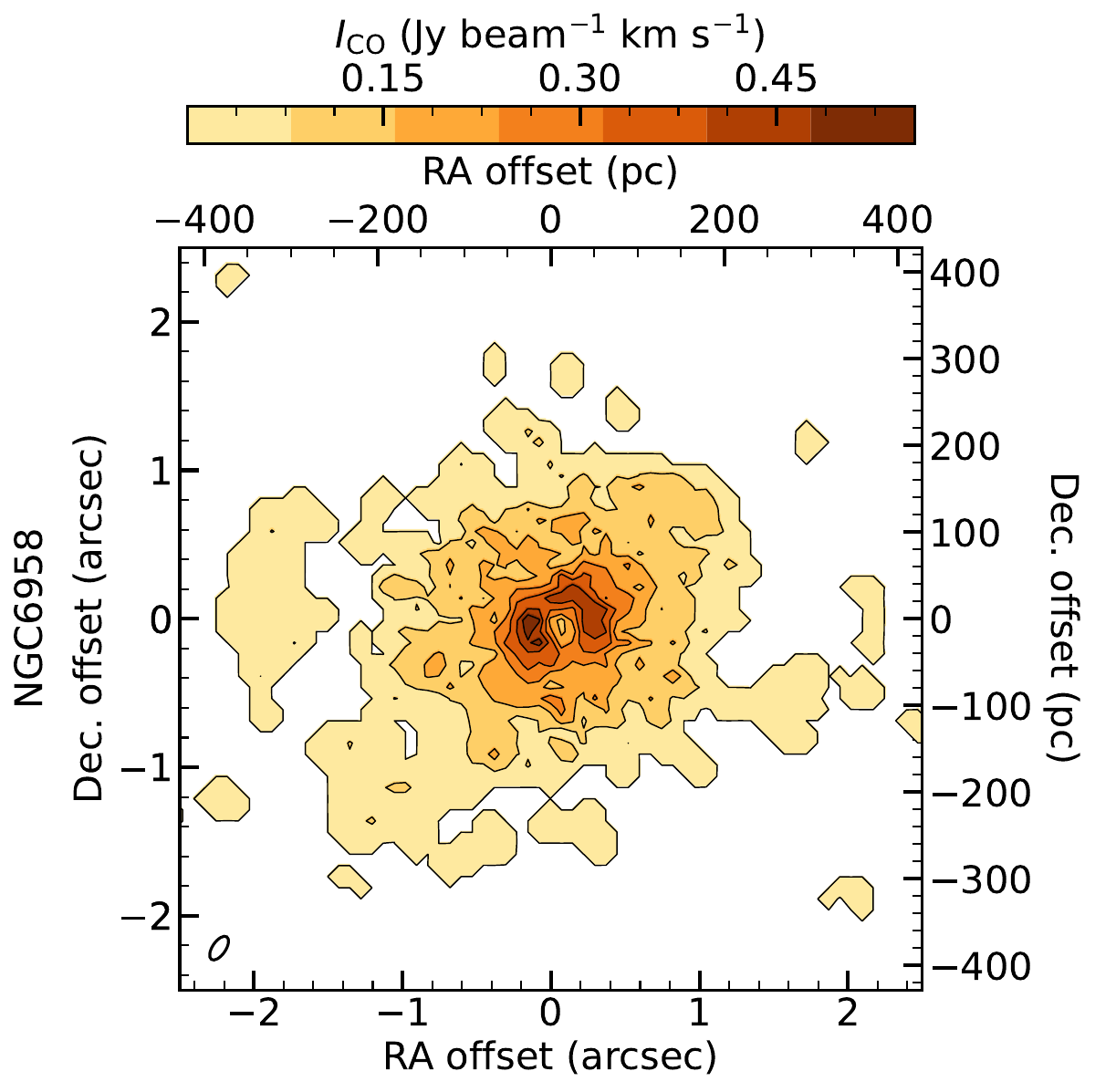}
    \end{subfigure}
    \begin{subfigure}{.32\linewidth}
    \centering
    \includegraphics[height=6.0cm]{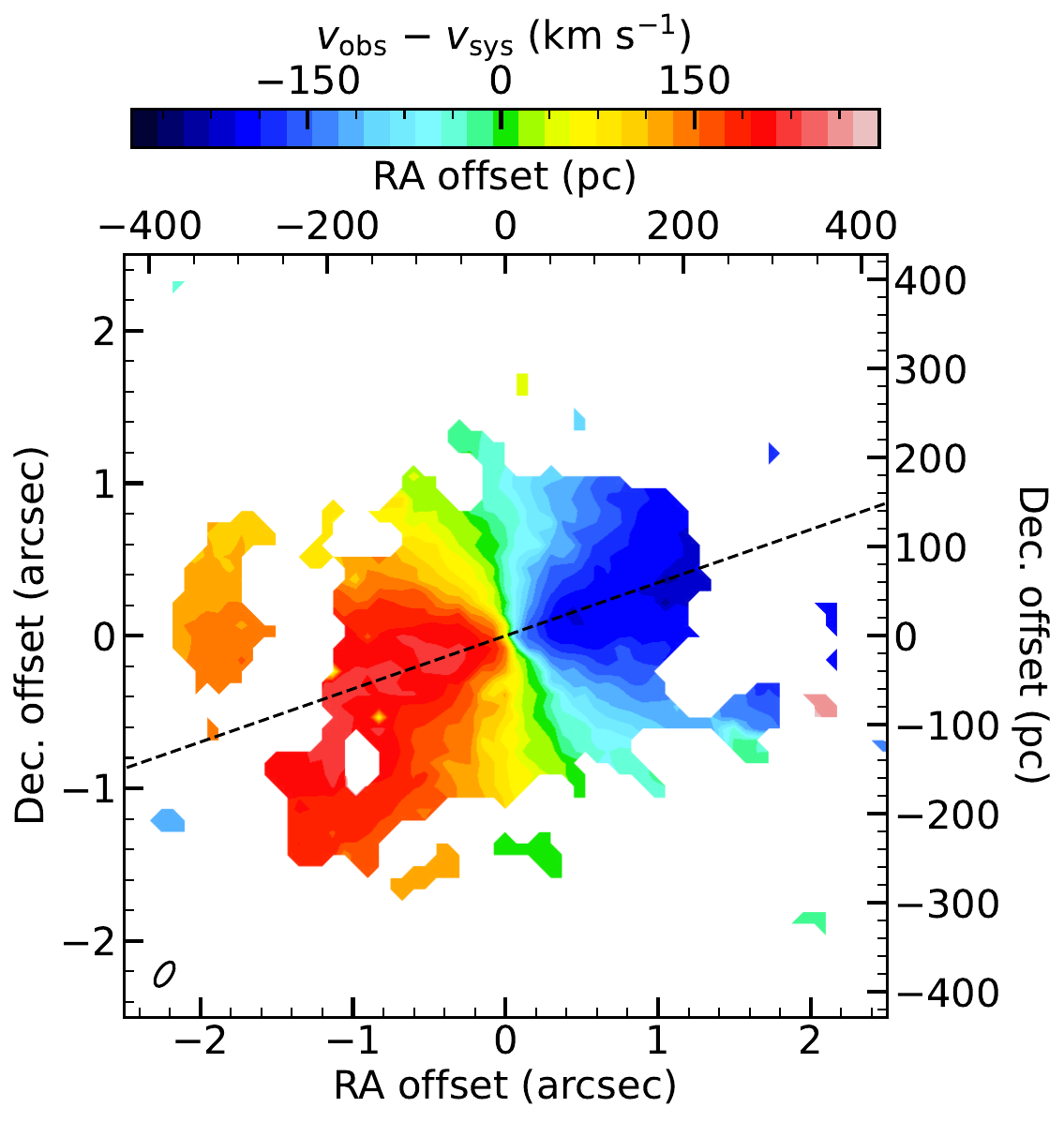}
    \end{subfigure}
    \begin{subfigure}{.32\linewidth}
    \centering
    \includegraphics[height=6.0cm]{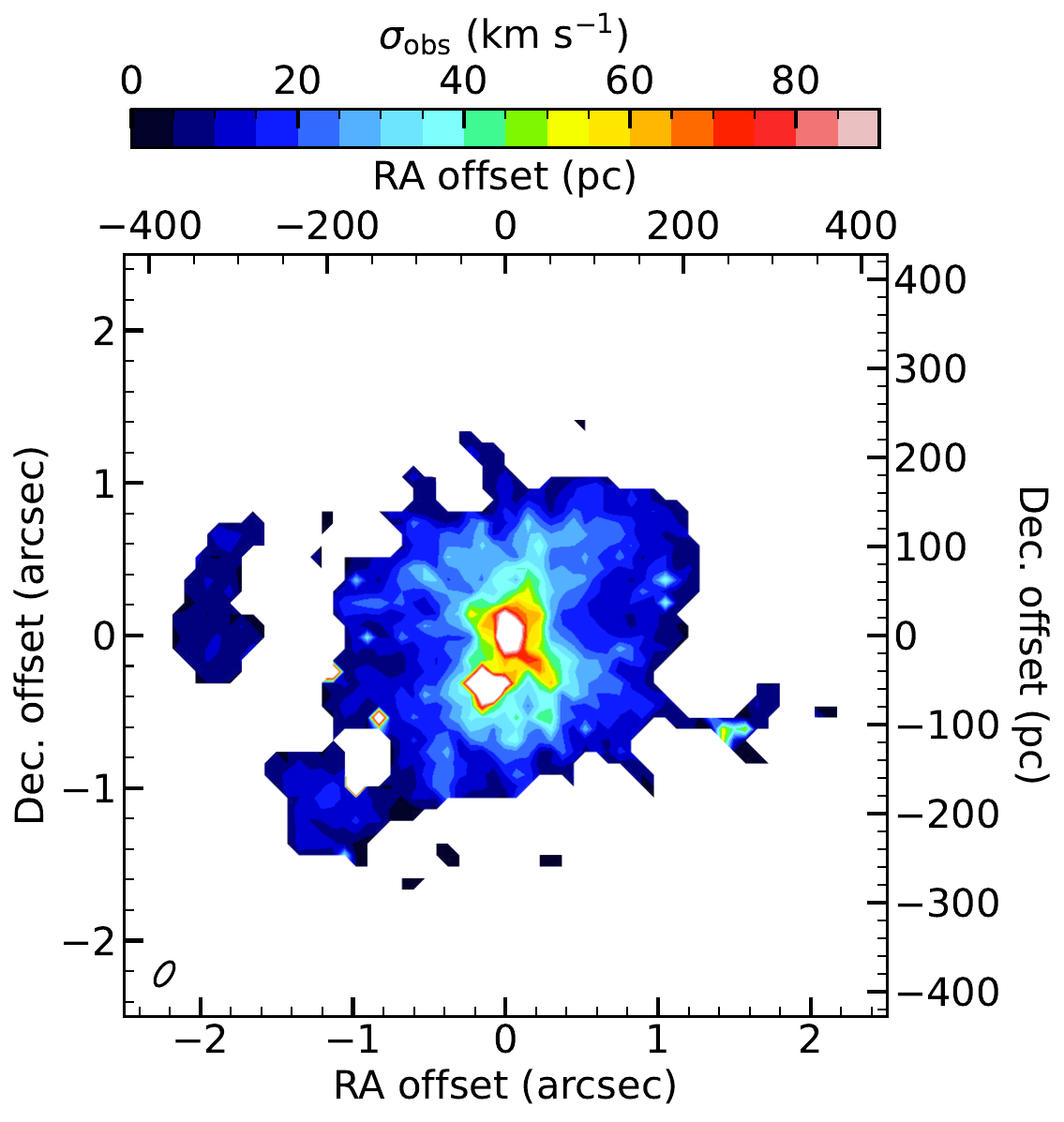}
    \end{subfigure}
    \\
    \begin{subfigure}{.34\linewidth}
    \centering
    \includegraphics[height=5.2cm]{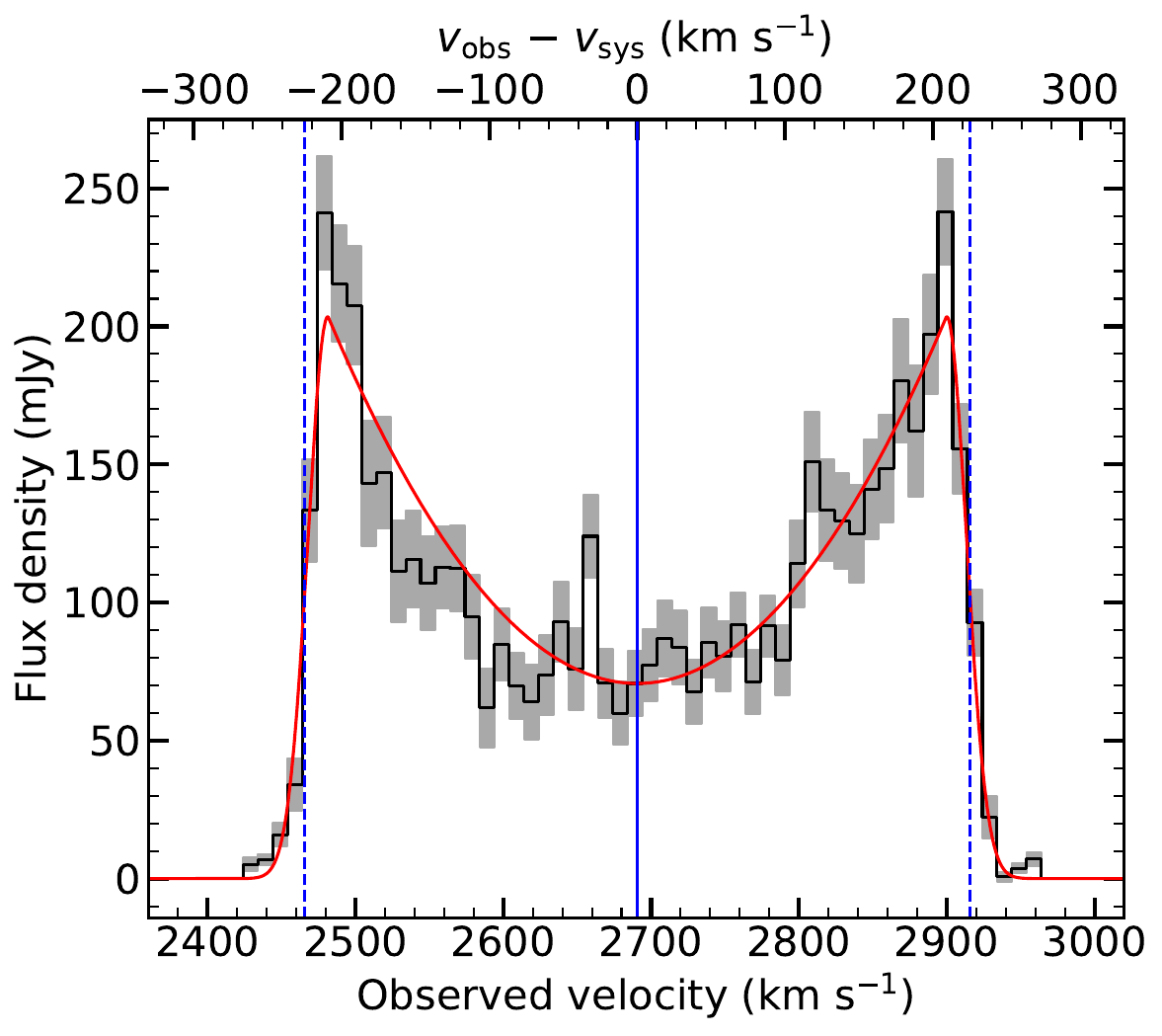}
    \end{subfigure}
    \begin{subfigure}{.32\linewidth}
    \centering
    \includegraphics[height=4.8cm]{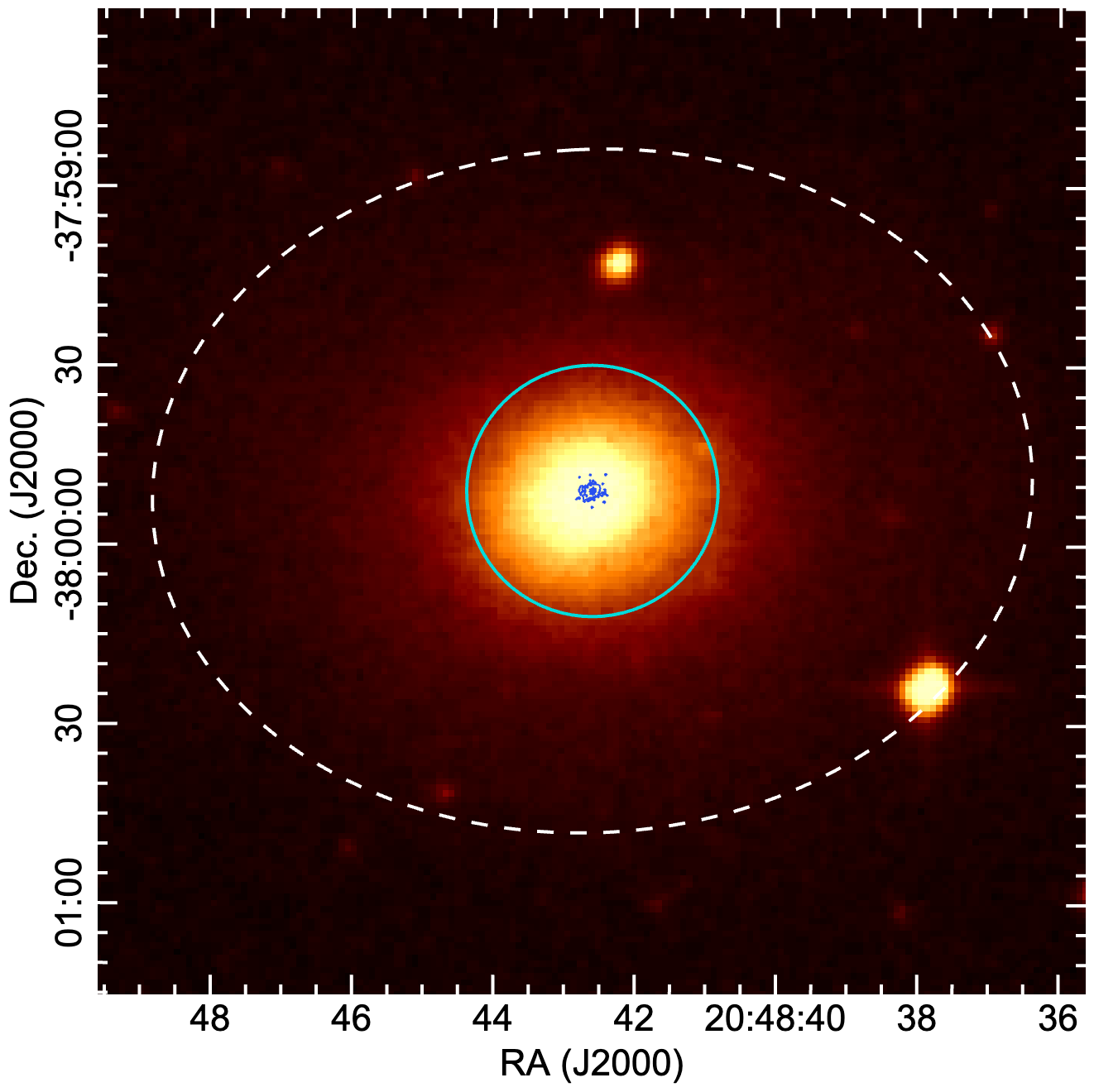}
    \end{subfigure}
    \begin{subfigure}{.33\linewidth}
    \centering
    \includegraphics[height=5.2cm]{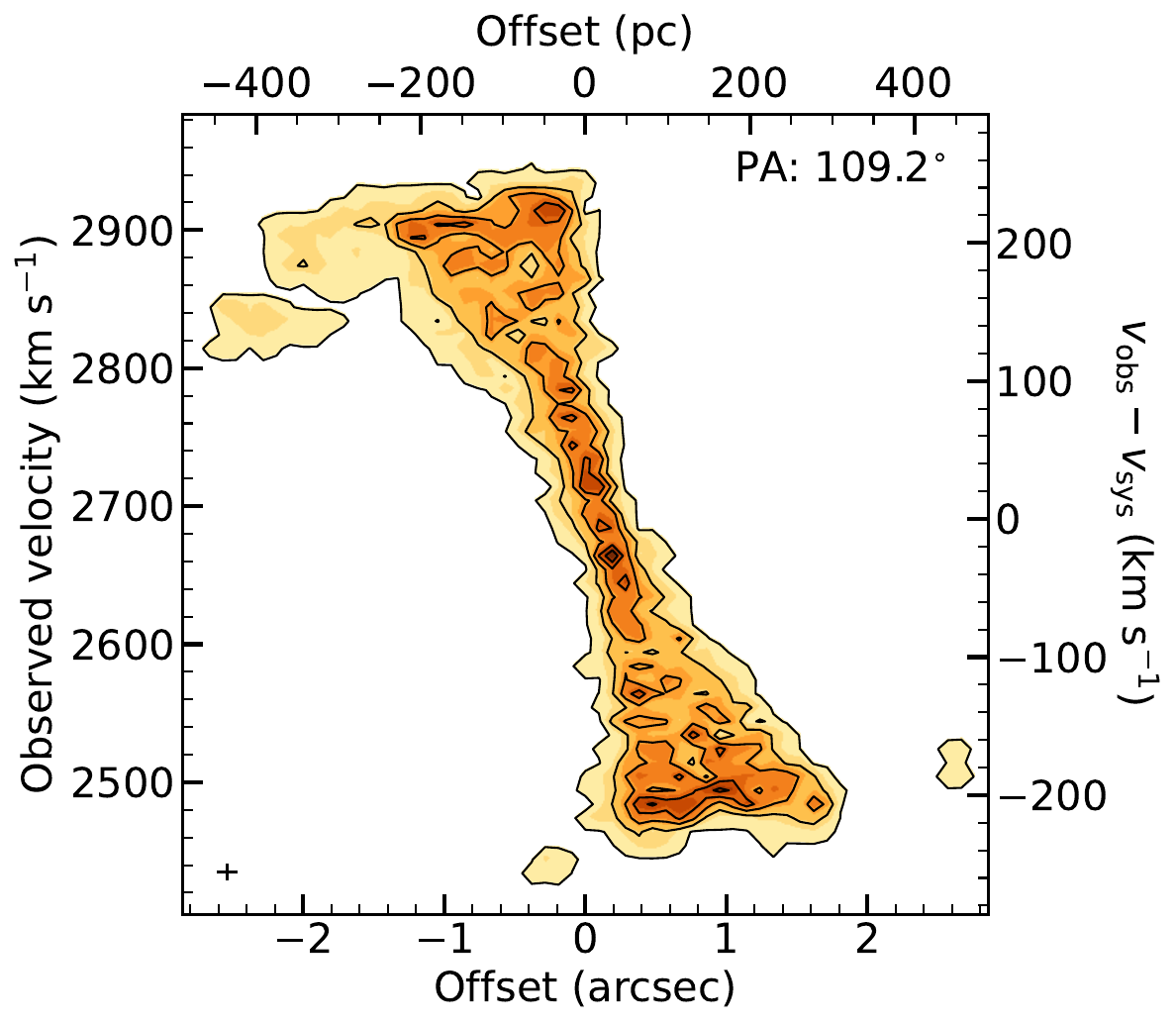}
    \end{subfigure}
    \\
    \begin{subfigure}{.34\linewidth}
    \centering
    \includegraphics[height=6.0cm]{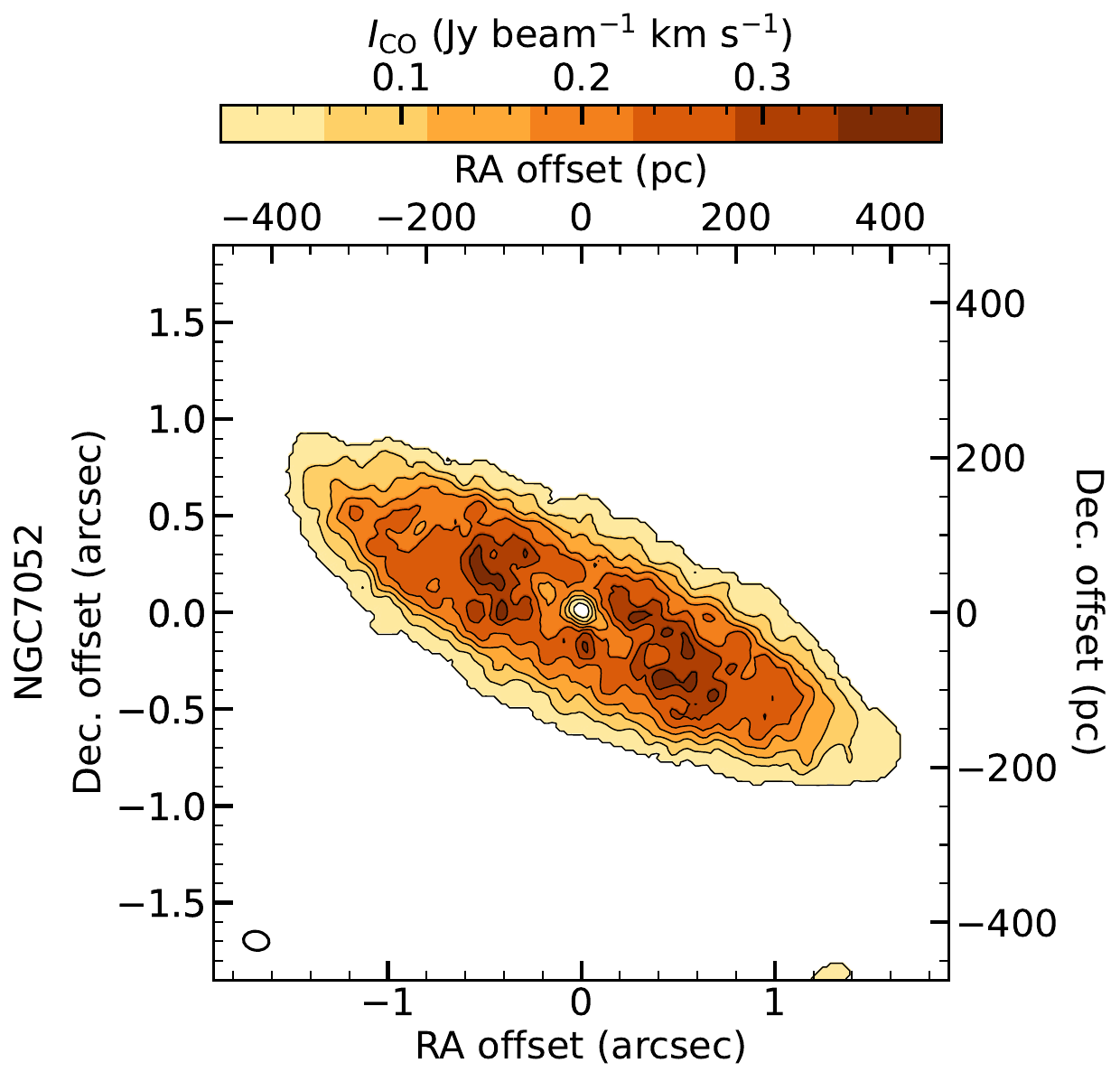}
    \end{subfigure}
    \begin{subfigure}{.32\linewidth}
    \centering
    \includegraphics[height=6.0cm]{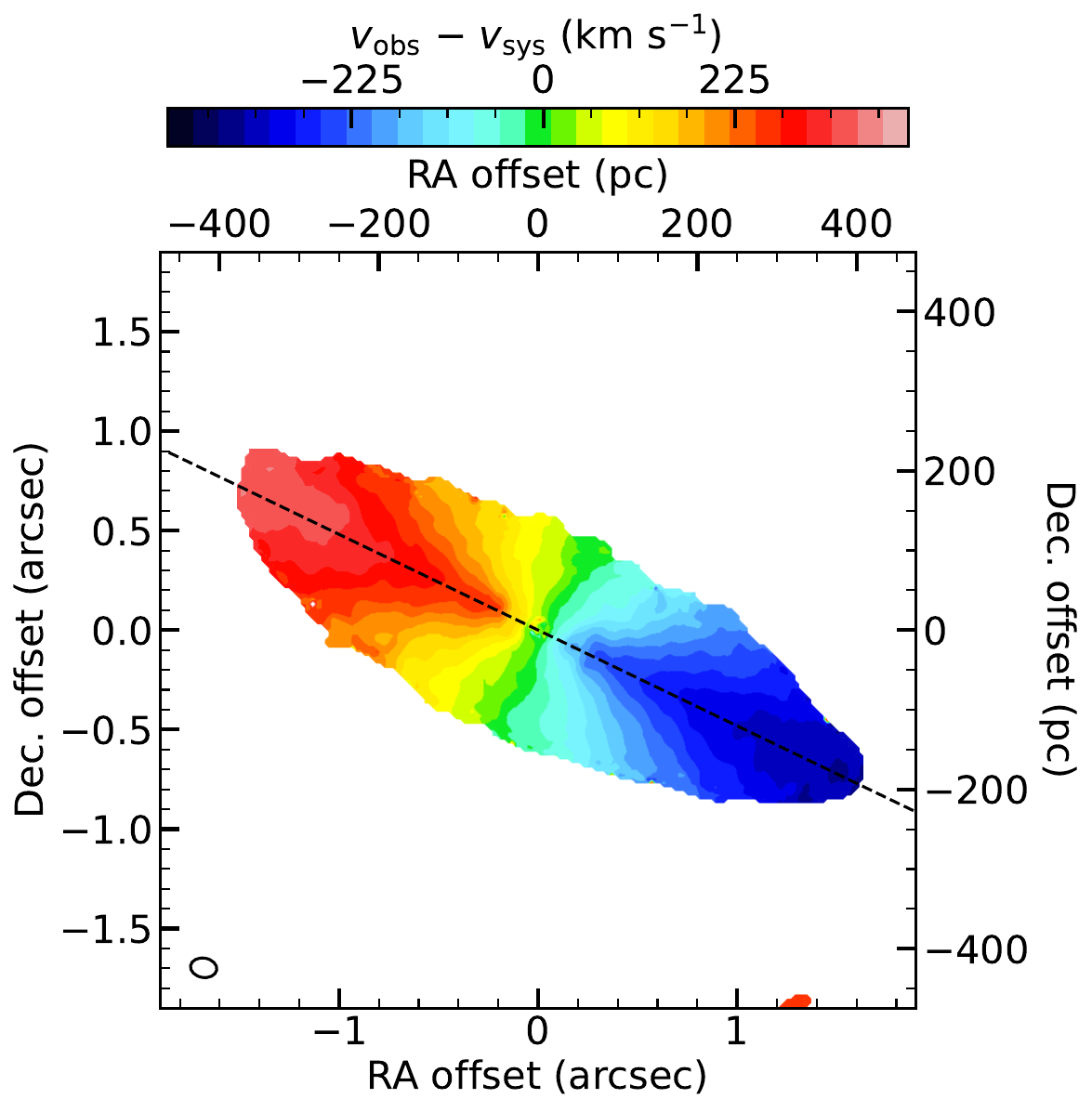}
    \end{subfigure}
    \begin{subfigure}{.32\linewidth}
    \centering
    \includegraphics[height=6.0cm]{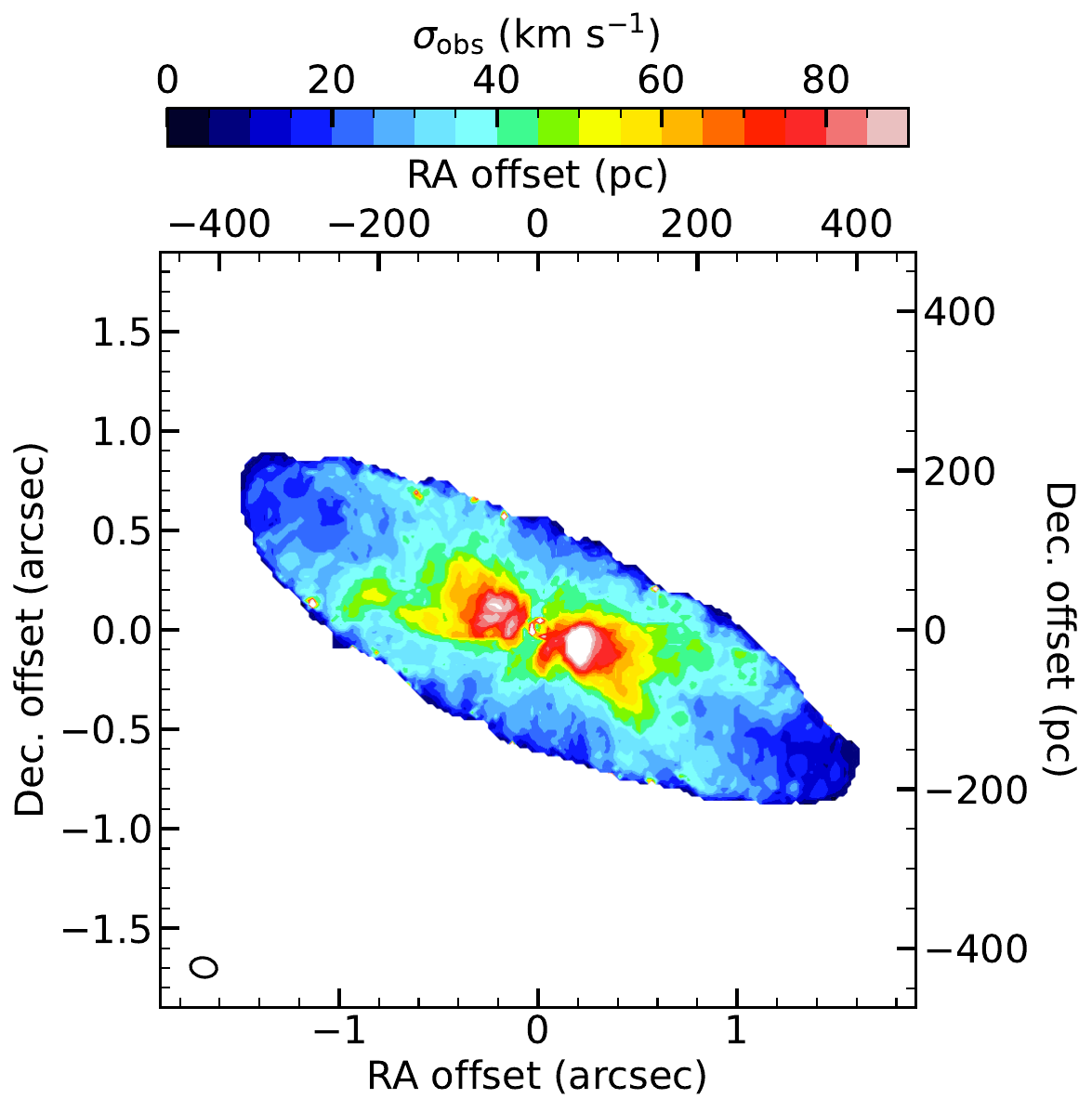}
    \end{subfigure}
    \\
    \begin{subfigure}{.34\linewidth}
    \centering
    \includegraphics[height=5.2cm]{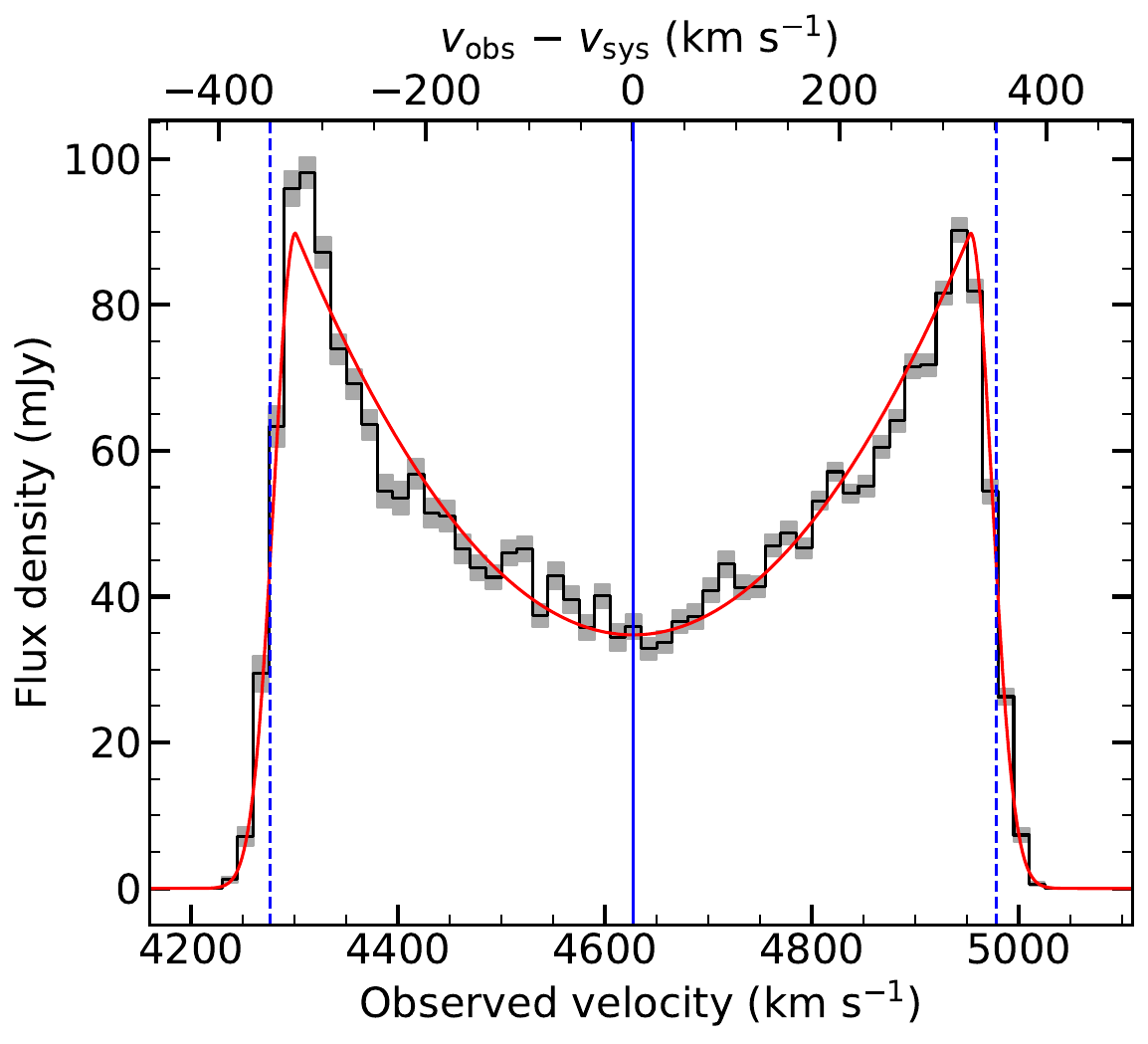}
    \end{subfigure}
    \begin{subfigure}{.32\linewidth}
    \centering
    \includegraphics[height=4.8cm]{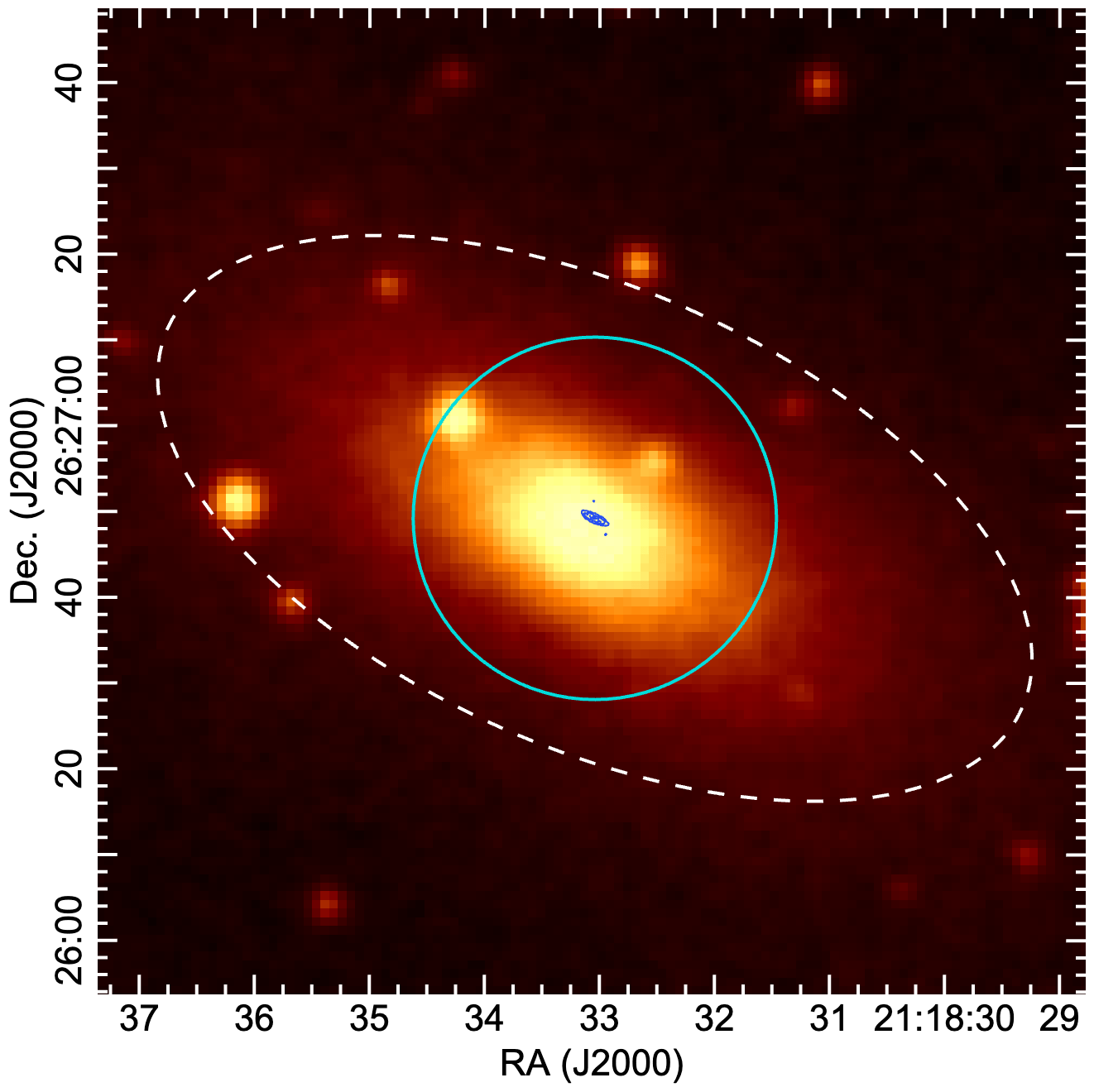}
    \end{subfigure}
    \begin{subfigure}{.33\linewidth}
    \centering
    \includegraphics[height=5.2cm]{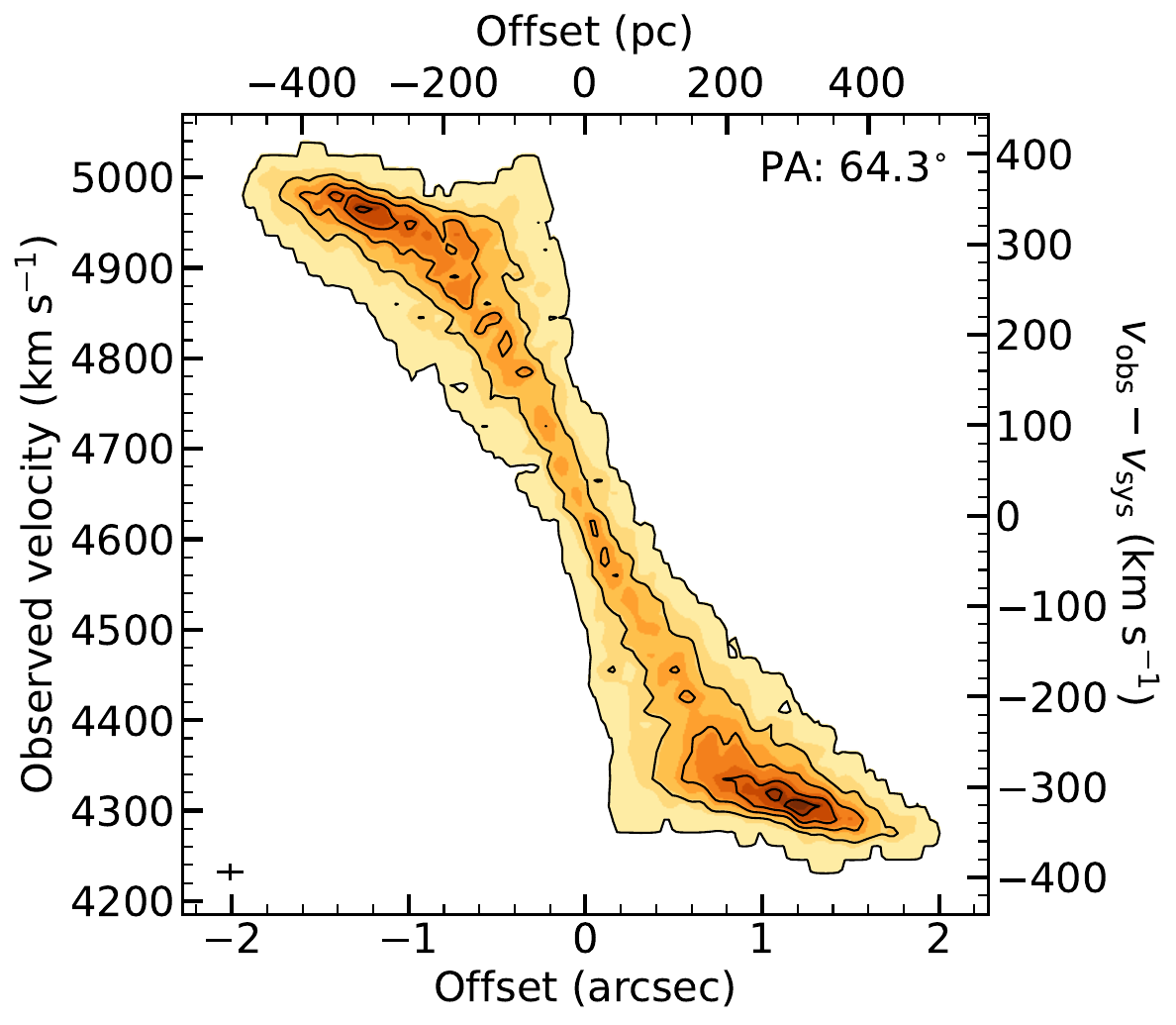}
    \end{subfigure}
    \\
    \caption[]{continued.}
\end{figure*}

\begin{figure*}
    \ContinuedFloat
    \begin{subfigure}{.34\linewidth}
    \centering
    \includegraphics[height=6.0cm]{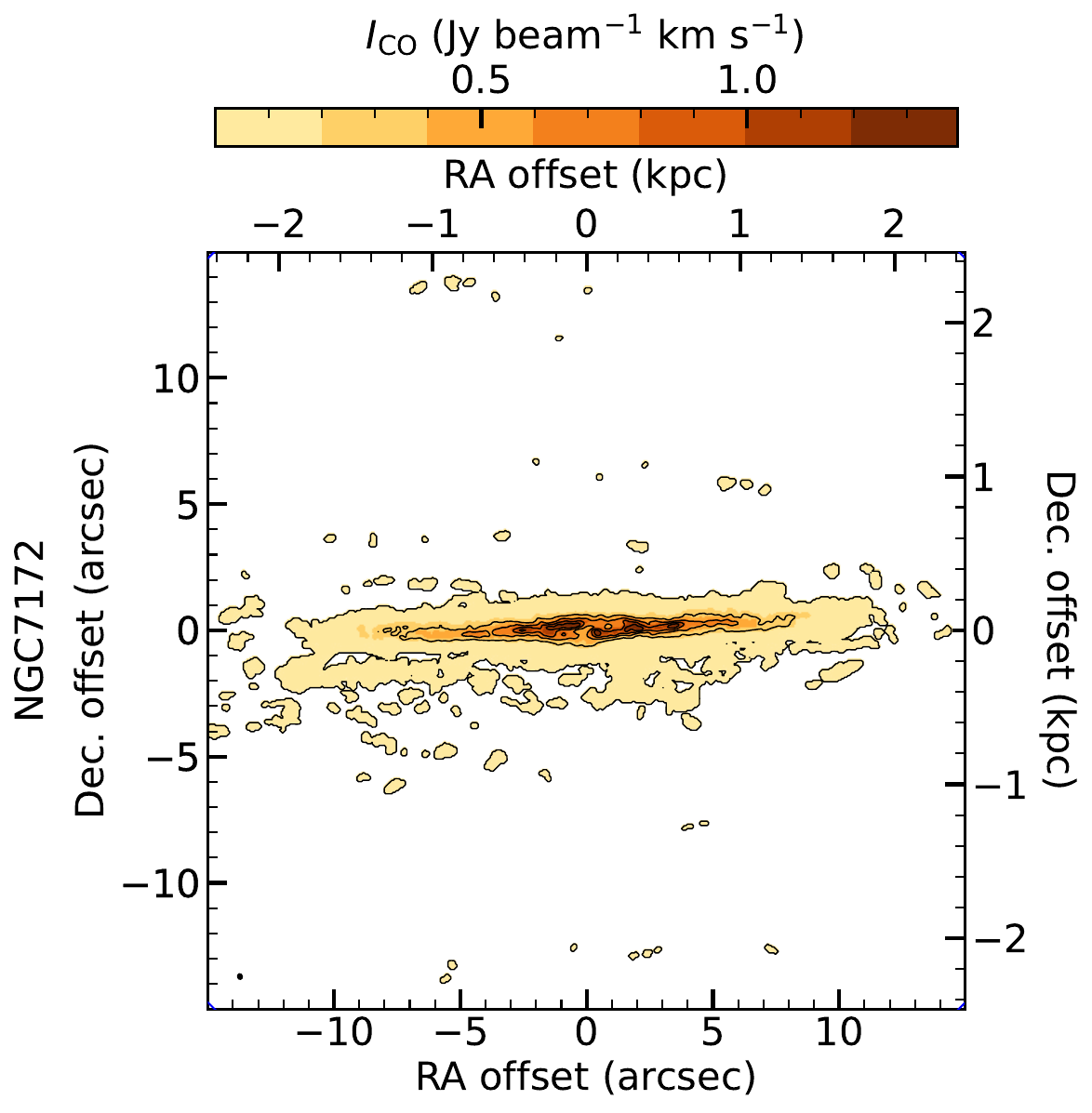}
    \end{subfigure}
    \begin{subfigure}{.32\linewidth}
    \centering
    \includegraphics[height=6.0cm]{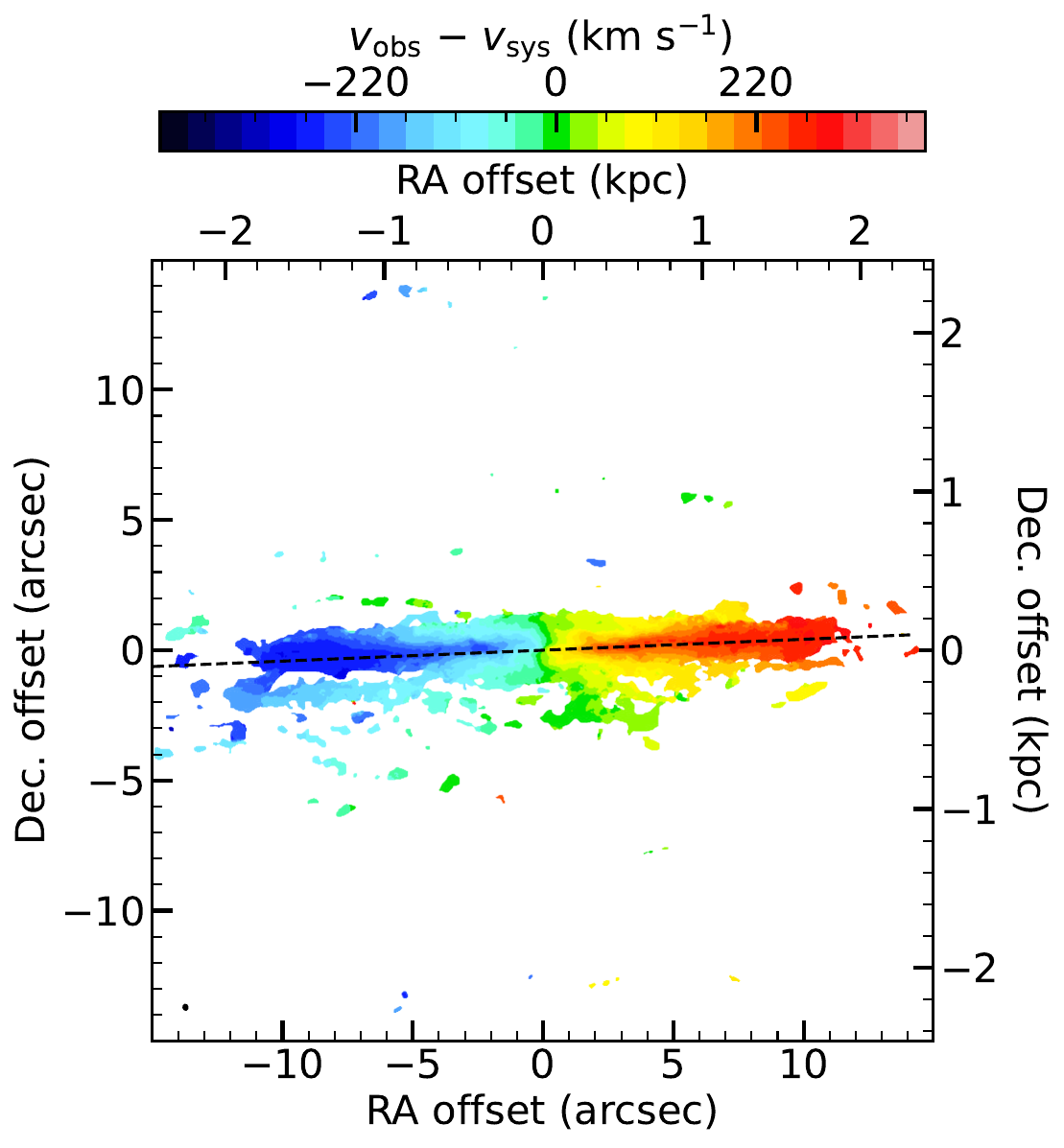}
    \end{subfigure}
    \begin{subfigure}{.32\linewidth}
    \centering
    \includegraphics[height=6.0cm]{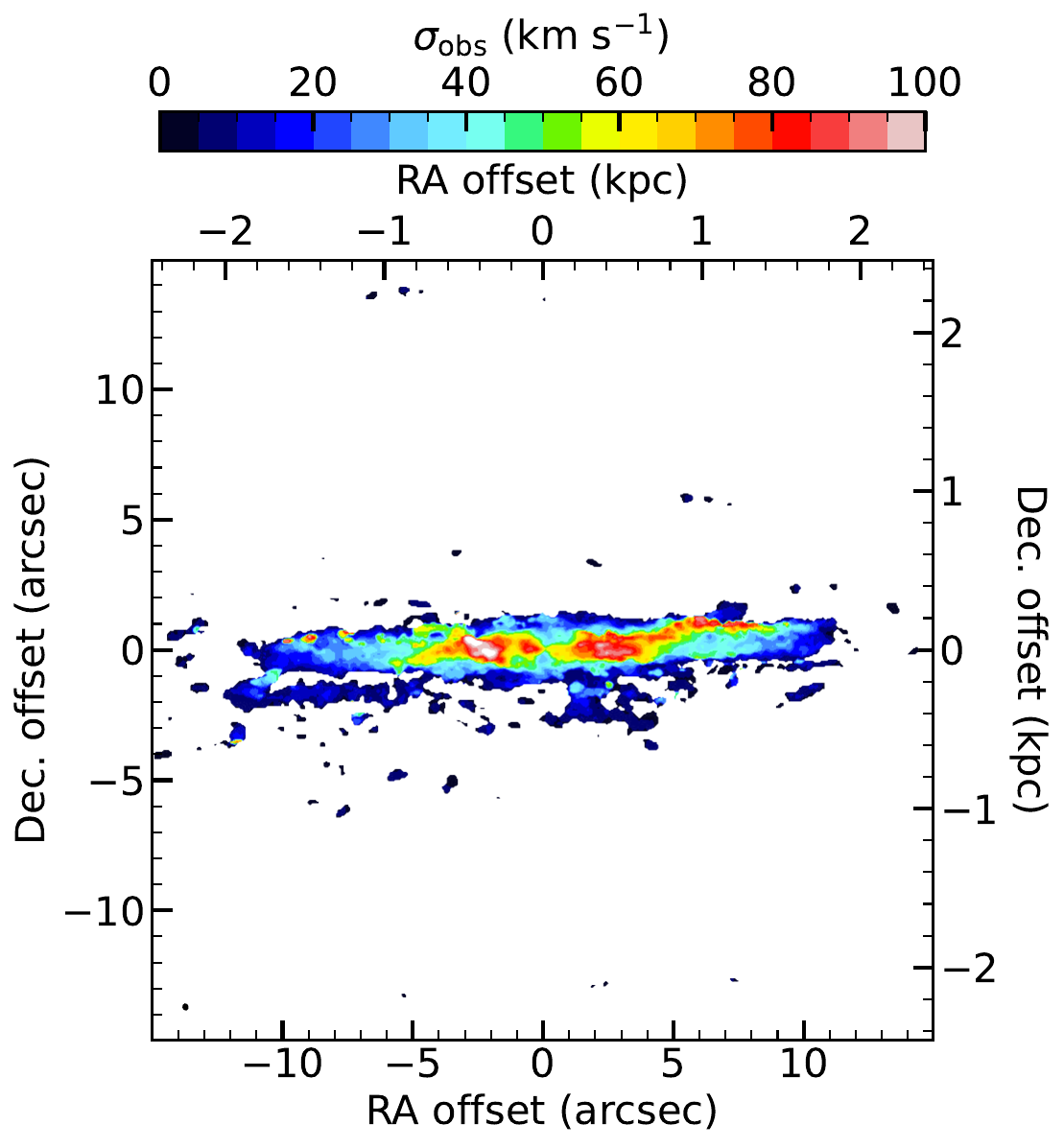}
    \end{subfigure}
    \\
    \begin{subfigure}{.34\linewidth}
    \centering
    \includegraphics[height=5.2cm]{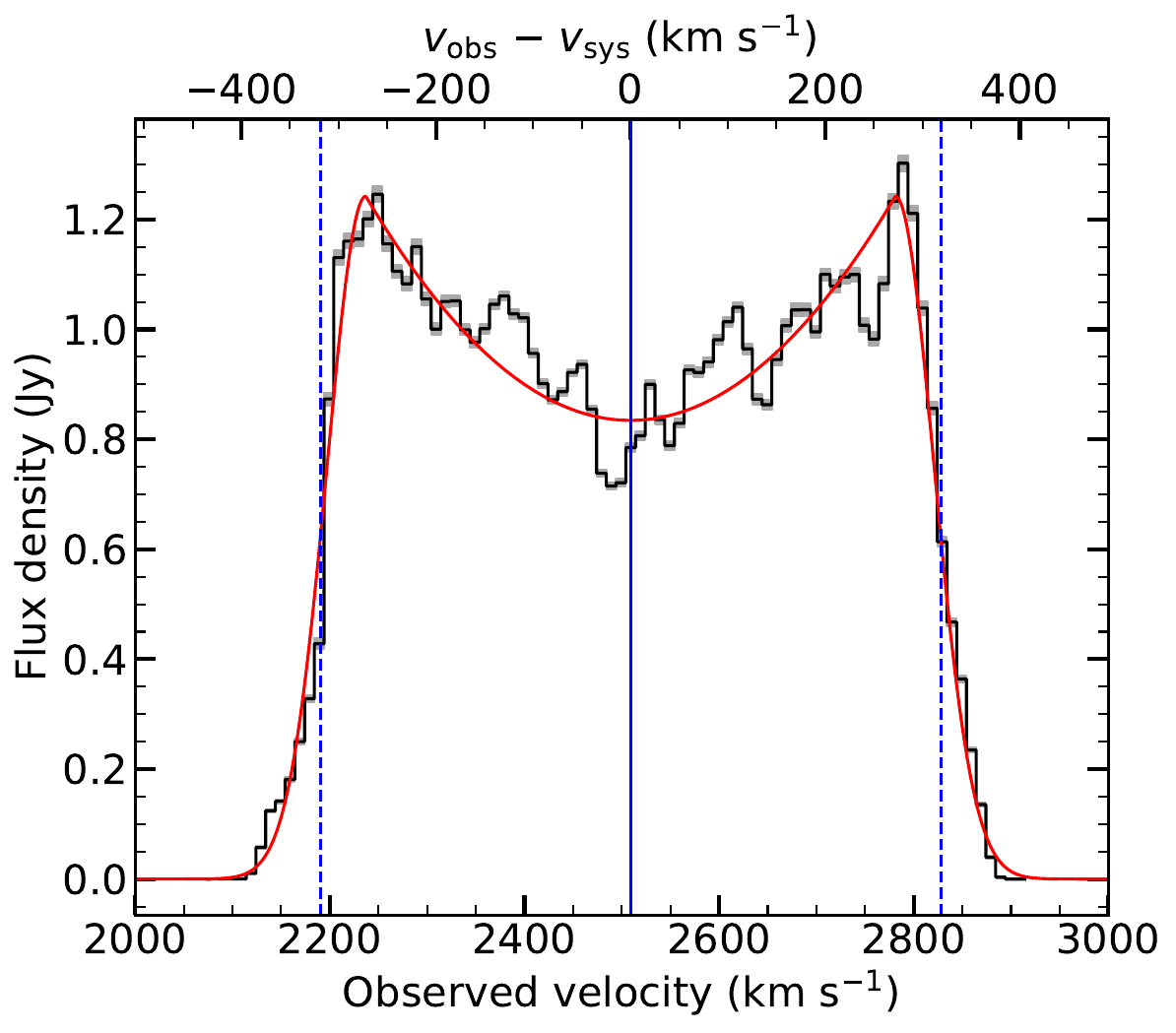}
    \end{subfigure}
    \begin{subfigure}{.32\linewidth}
    \centering
    \includegraphics[height=4.8cm]{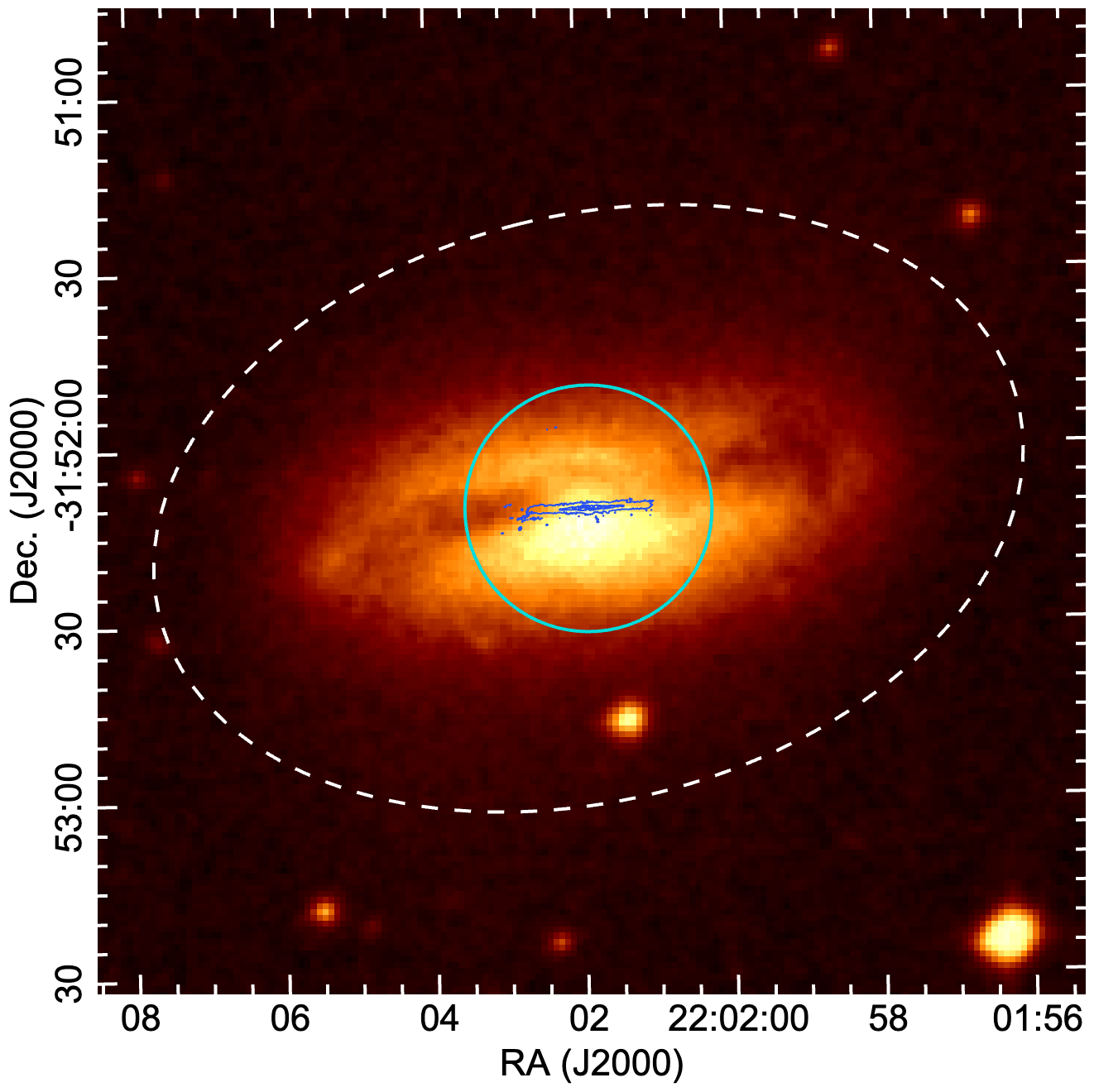}
    \end{subfigure}
    \begin{subfigure}{.33\linewidth}
    \centering
    \includegraphics[height=5.2cm]{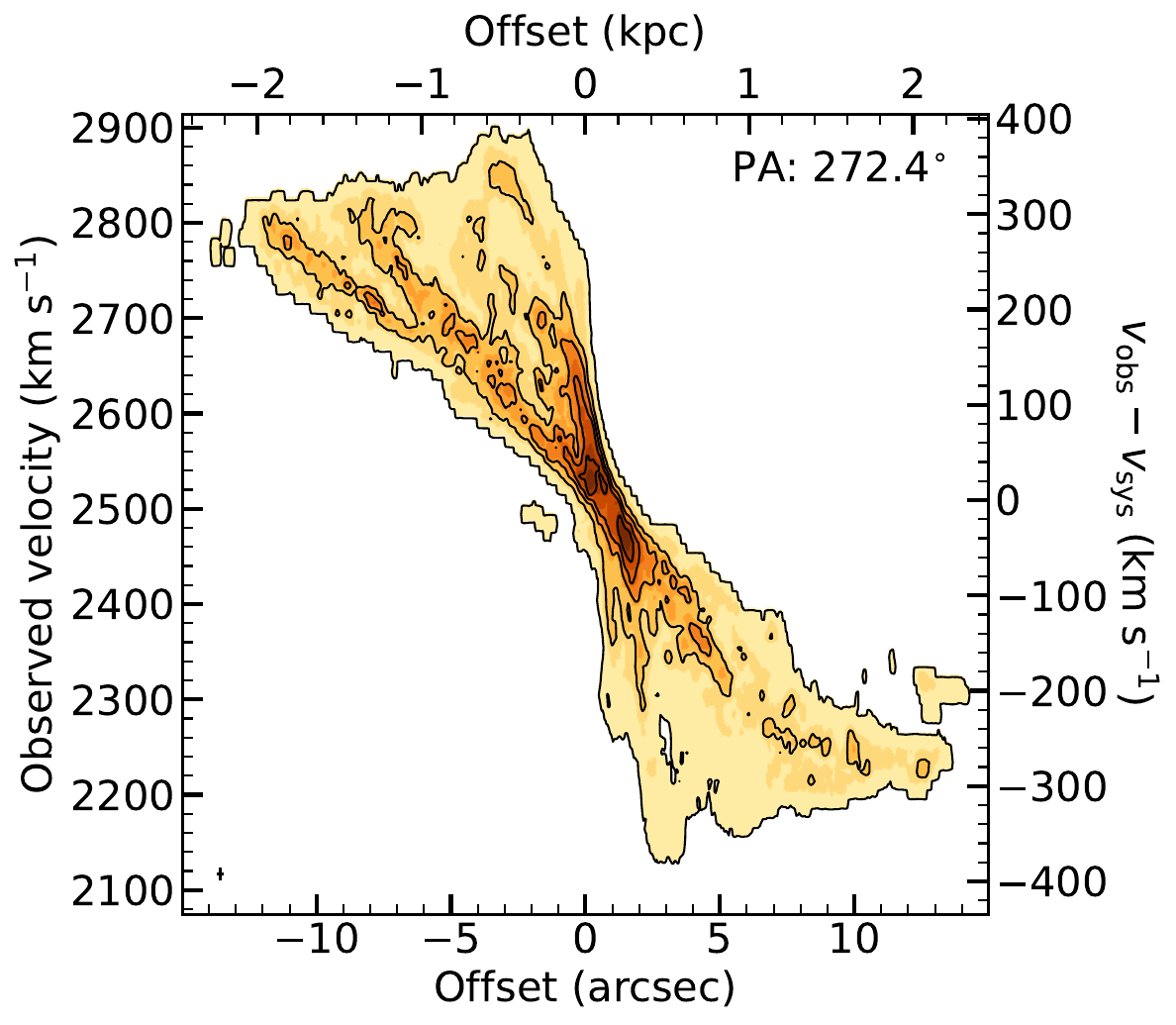}
    \end{subfigure}
    \\
    \begin{subfigure}{.34\linewidth}
    \centering
    \includegraphics[height=6.0cm]{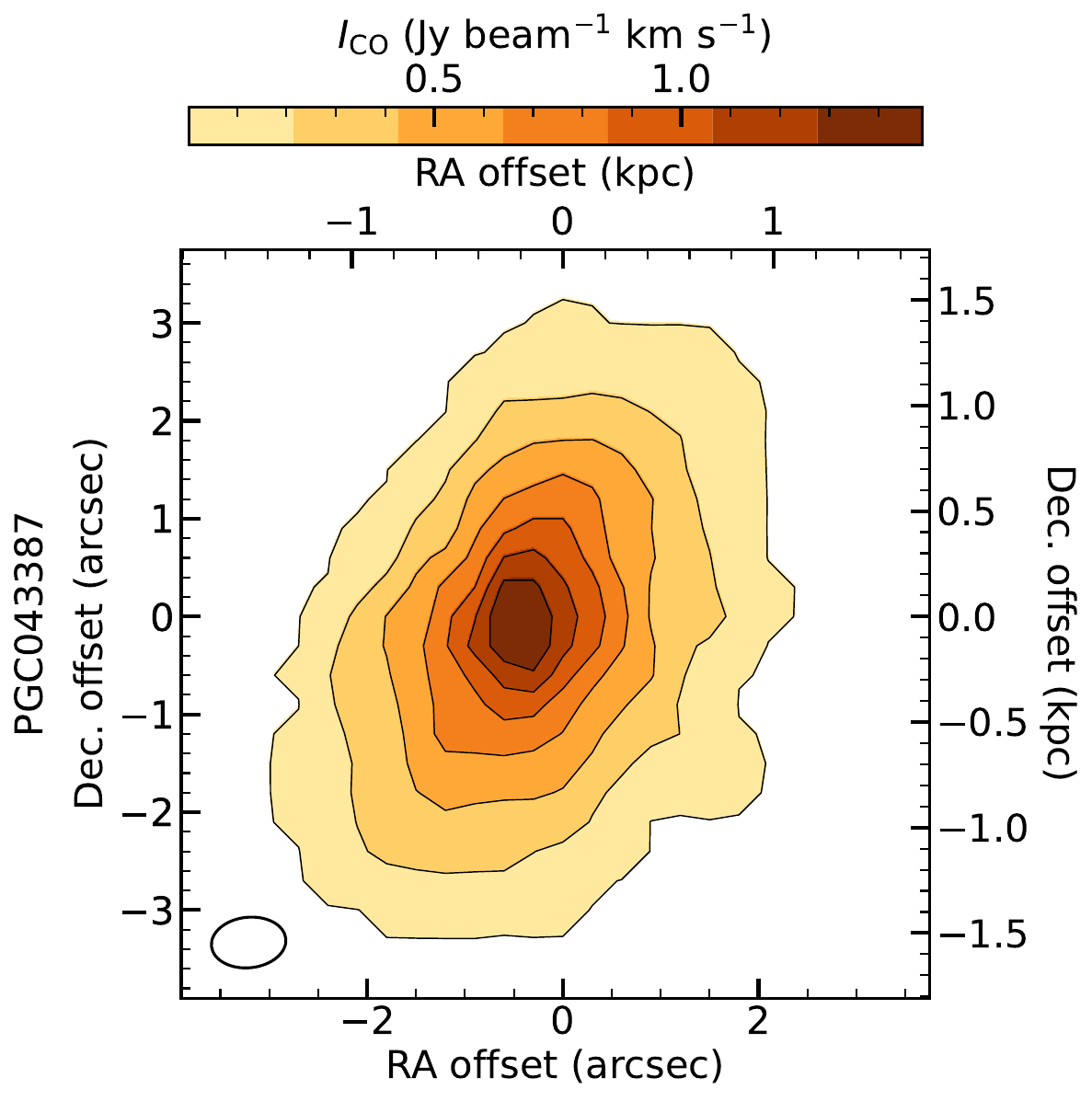}
    \end{subfigure}
    \begin{subfigure}{.32\linewidth}
    \centering
    \includegraphics[height=6.0cm]{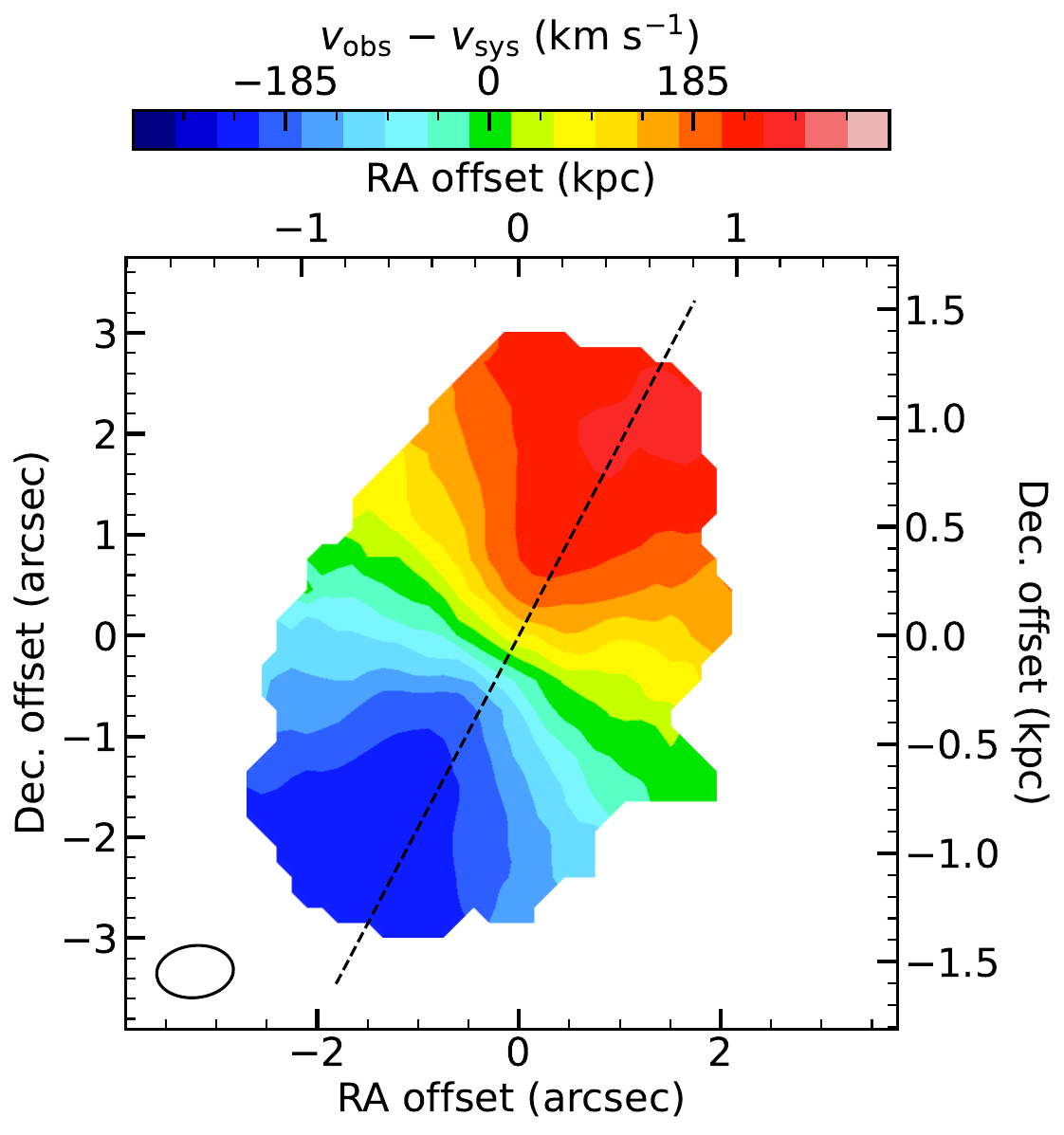}
    \end{subfigure}
    \begin{subfigure}{.32\linewidth}
    \centering
    \includegraphics[height=6.0cm]{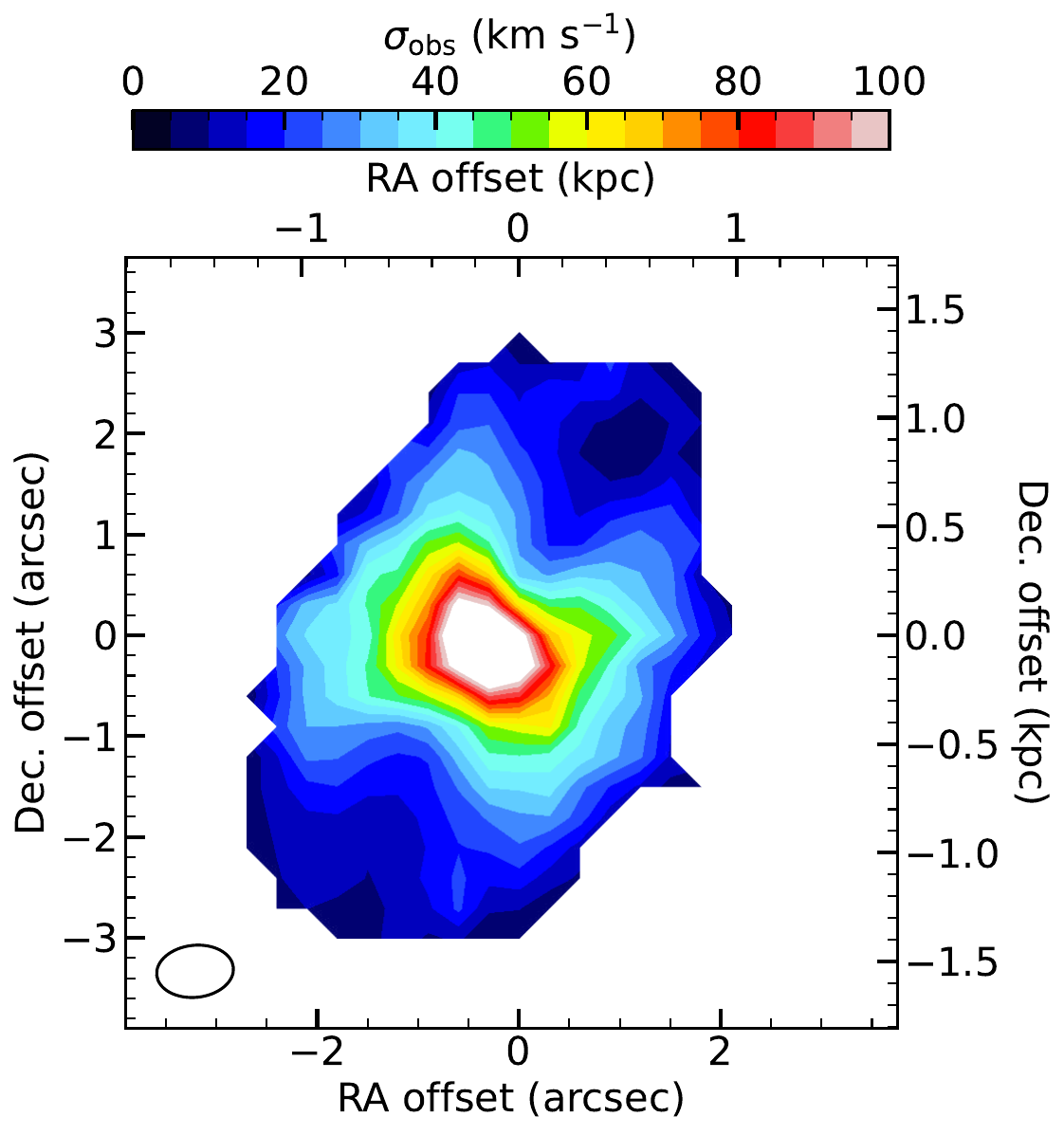}
    \end{subfigure}
    \\
    \begin{subfigure}{.34\linewidth}
    \centering
    \includegraphics[height=5.2cm]{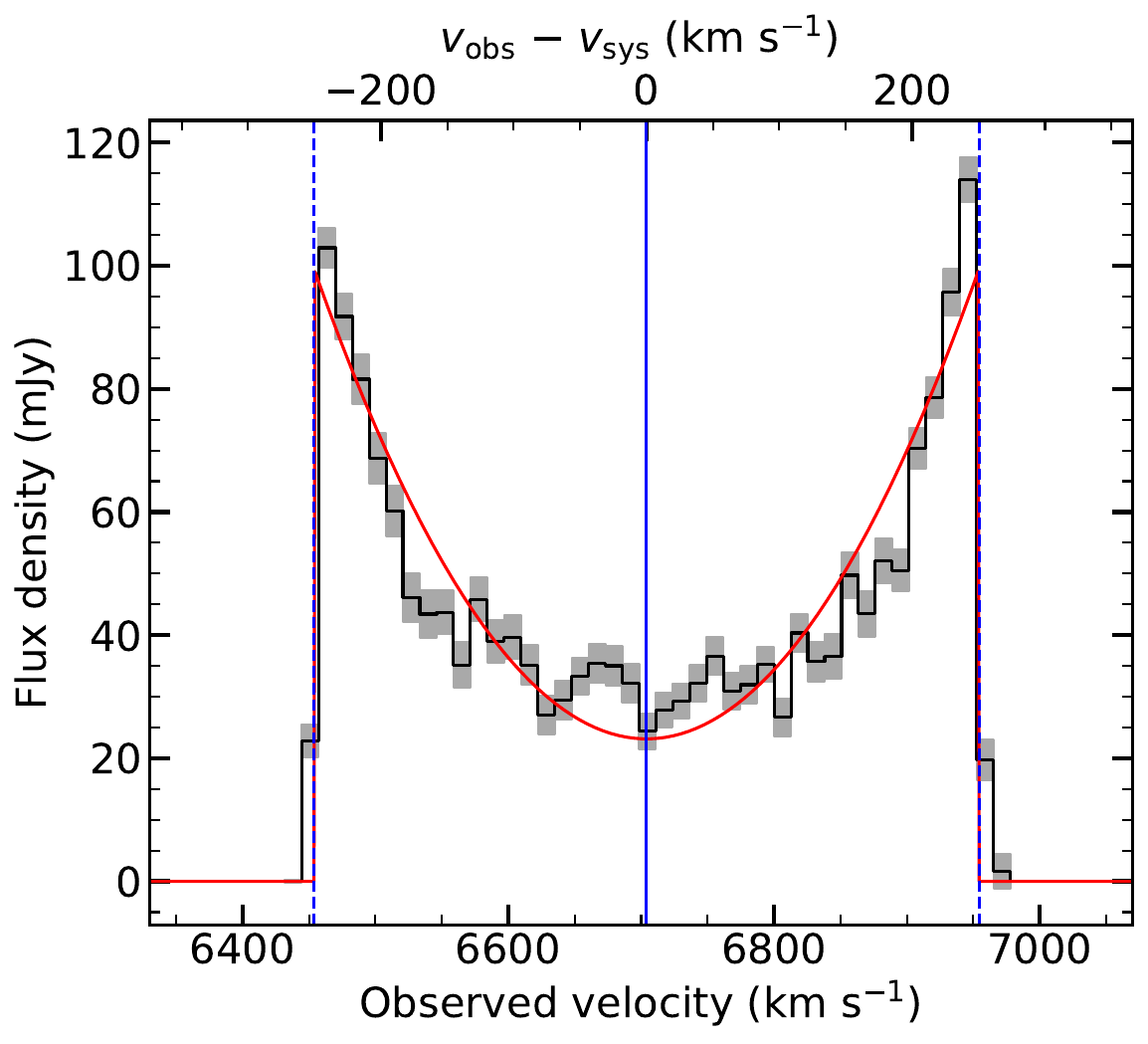}
    \end{subfigure}
    \begin{subfigure}{.32\linewidth}
    \centering
    \includegraphics[height=4.8cm]{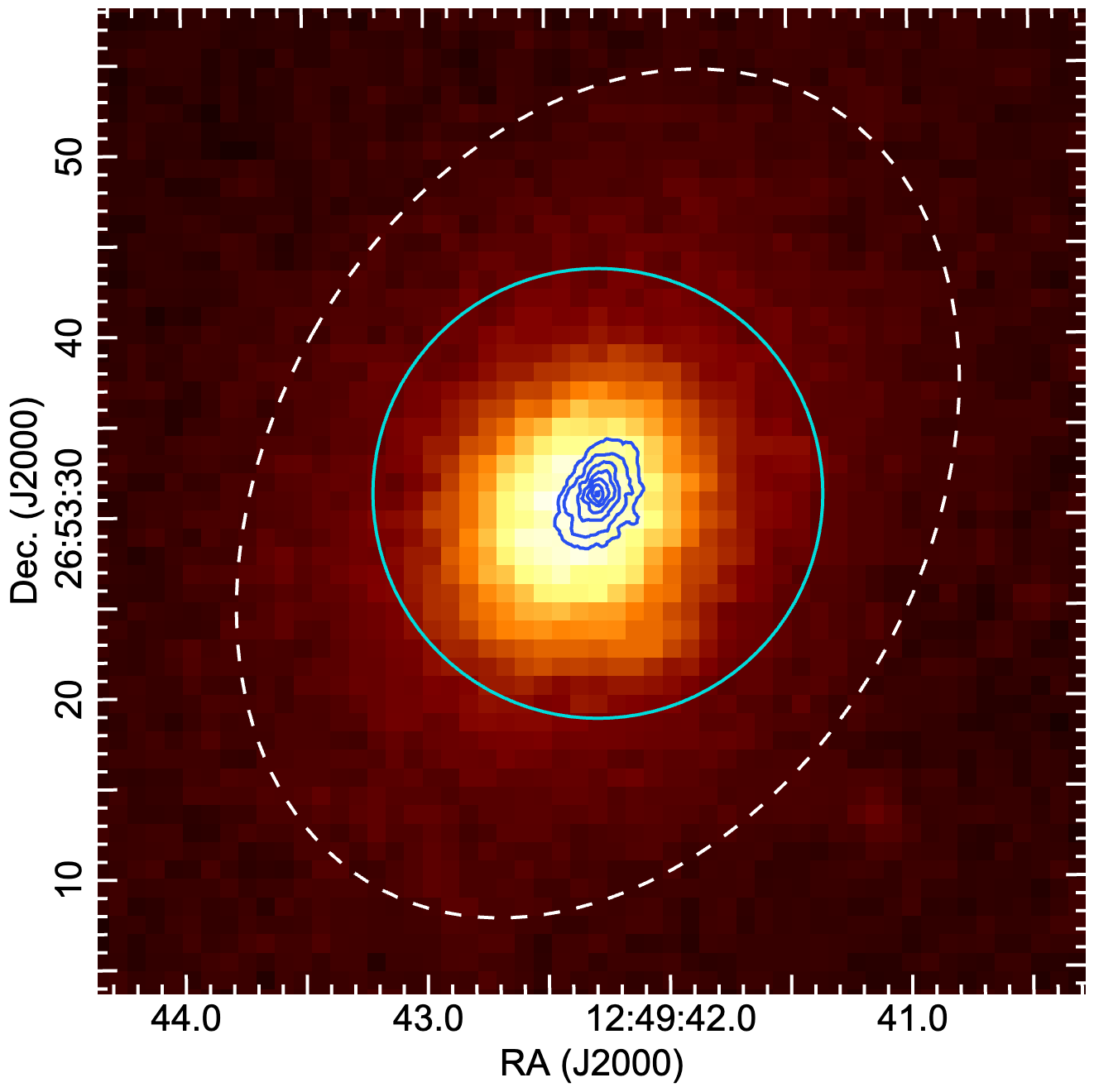}
    \end{subfigure}
    \begin{subfigure}{.33\linewidth}
    \centering
    \includegraphics[height=5.2cm]{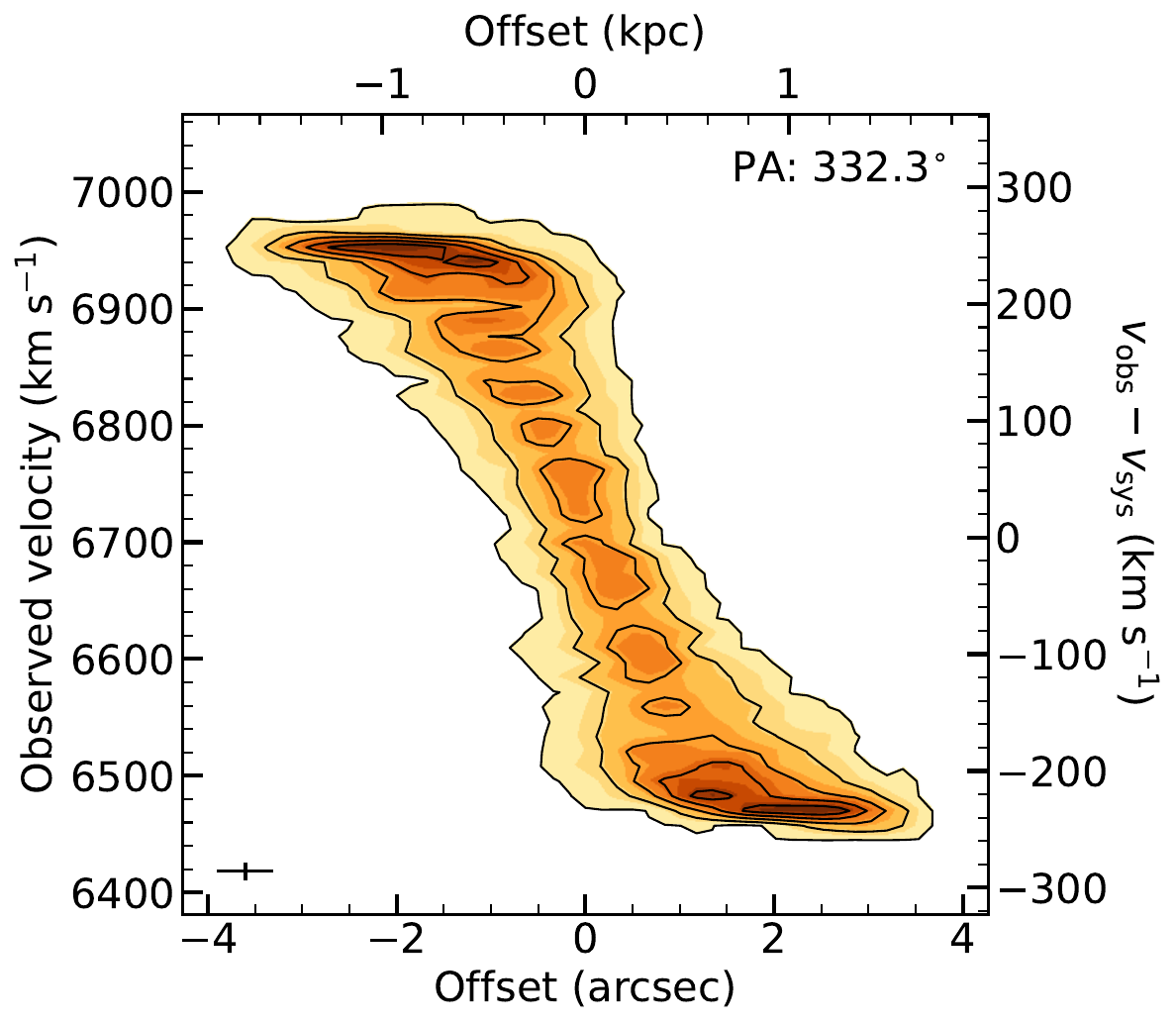}
    \end{subfigure}
    \\
    \caption[]{continued.}
\end{figure*}


\bsp	
\label{lastpage}

\end{document}